%% file: acfarep.tex
\documentclass[12pt]{report}
\usepackage{epsf, acfarep}
\begin{document}
\title{{\Huge Particle Physics Experiments\\
at\\
JLC}\\
\vglue 12pt
\centerline{\epsfbox{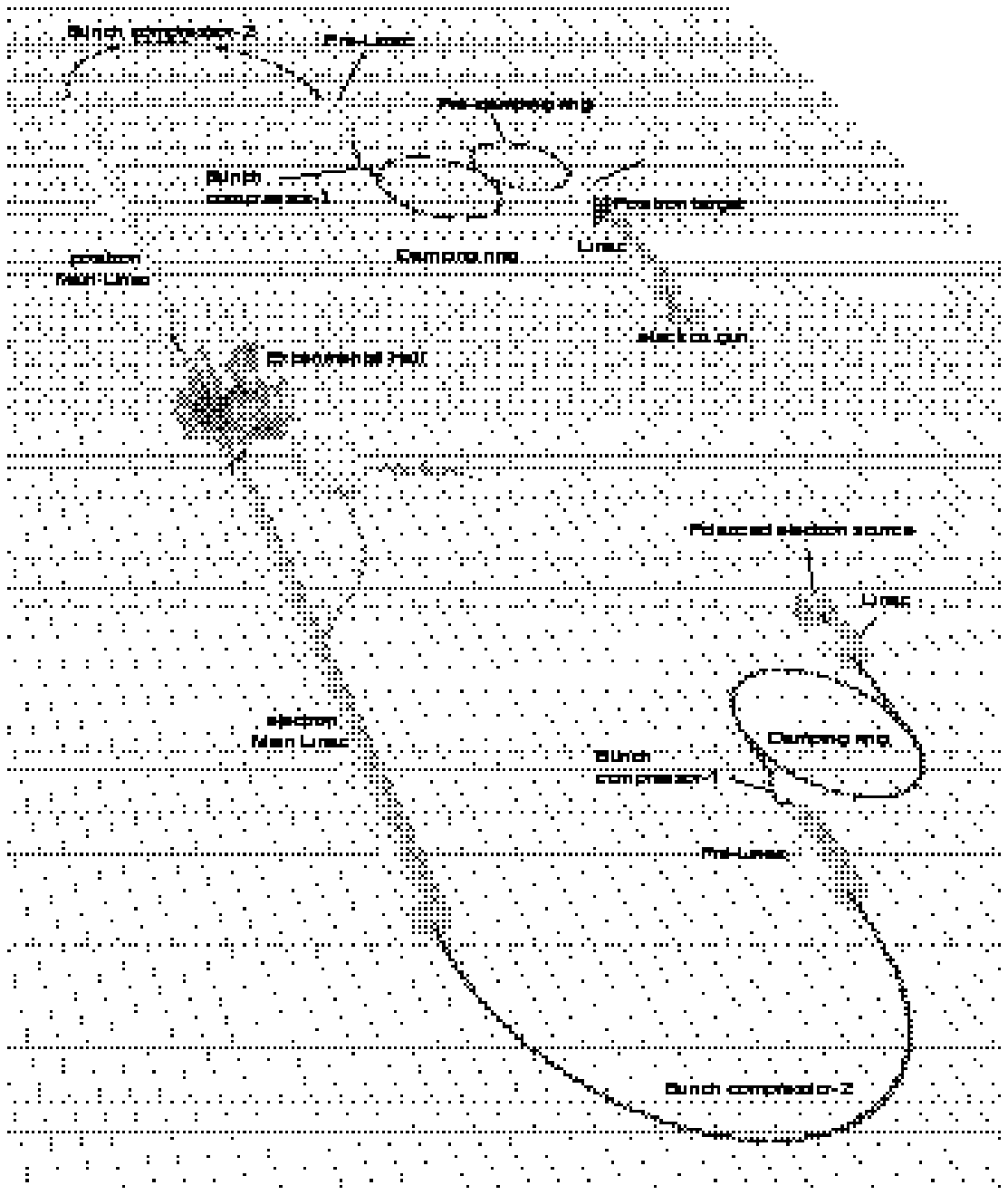}}
}
\author{{\Large ACFA Linear Collider 
Working Group Report}}
\date{}
\maketitle
\input headers/authorlist.tex
\newpage
\input headers/preface.tex

\tableofcontents

\clearpage
\pagestyle{headings}
\pagenumbering{arabic}
\setcounter{page}{1}
\part{Project overview}
\label{part-overview}
\input headers/overview.tex
\part{Physics}
\label{part-physics}
\input physhiggs/main.tex

\input physsusy/main.tex
\input phystop/main.tex

\input physqcd/main.tex

\input physew/main.tex

\part{Detectors}
\label{part-detector}
\input detir/main.tex
\input dettrk/main.tex

\input detcal/main.tex
\input detmuon/main.tex

\input detmag/main.tex

\input detsim/main.tex
\part{Options}
\label{part-options}
\input options/main.tex

\newpage
\pagestyle{empty}
\addcontentsline{toc}{part}{Acknowledgements}
\input headers/acknowledgements.tex
\newpage
\pagestyle{headings}
\appendix
\addcontentsline{toc}{part}{Appendix}
\input headers/acfastatement.tex

\end{document}

%% file: headers/authorlist.tex
\noindent
\begin{center}
\noindent
Koh~Abe$^{67}$,
Koya~Abe$^{53}$,
Toshinori~Abe$^{29}$,
Andrew~G.~Akeroyd$^{15}$,
Kazuaki~Anraku$^{66}$,
Mamoru~Araya$^{56}$,
Abdesslam~Arhrib$^{33}$,
Dennis~C.~Arogancia$^{26}$,
Shoji~Asai$^{66}$,
Eri~Asakawa$^{37}$
% ...
\end{center}
\renewcommand{\thefootnote}{\fnsymbol{footnote}}
\begin{center}
(ACFA Linear Collider Working Group\footnote{Group information is available at http://acfahep.kek.jp/.})
\end{center}
\renewcommand{\thefootnote}{\arabic{footnote}}
\footnotesize
Postal address to contact:\\
% ...
$^{15}$ KIAS, 207-43 Cheongryangri-dong, Dongdaemun-gu, Seoul 130-012, Korea\\
% ...
$^{26}$ Mindanao State University, Iligan city, Philippines\\
% ...
$^{29}$ Nagoya University, Furo-cho, Chikusa-ku, Nagoya 464-8601, Japan\\
% ...
$^{33}$ National Taiwan University, Taipei 10617, Taiwan\\
% ...
$^{37}$ Ochanomizu University, 1 Ohtsuka 2-1, Bunkyo-ku, Tokyo 112, Japan \\
% ...
$^{53}$ Tohoku University, Aoba, Aramaki, Aoba-ku, Sendai 980-8578, Japan\\
% ...
$^{56}$ Tokyo A\&T, Nakacho 2-24-16, Koganeishi, Tokyo 184-8588, Japan\\
% ...
$^{66}$ University of Tokyo, ICEPP, 7-3-1 Hongo, Bunkyo-ku, Tokyo 113-0033, Japan\\
$^{67}$ University of Tokyo, School of science, 7-3-1 Hongo, Bunkyo-ku, Tokyo 113-0033, Japan\\
% ...

\normalsize

%% file: headers/preface.tex
\begin{center}
\Large Preface
\normalsize
\end{center}
\vglue 12pt

The electron-positron linear collider project 
stems from the Japanese High Energy Committee's
recommendation made back in 1986 as the post-TRISTAN program for
energy-frontier physics. 
Following the recommendation a 5-years R\&D program was set up
to address key issues in basic component technologies and to
formulate the project as a whole. 
The R\&D program crystallized into the first
project design in 1992 which elucidated physics prospects and novel detector
concepts matching the opportunities, and outlined the accelerator complex
including its application to X-ray laser production.

The basic physics that motivated the project is intact and  
the principal guideline shown there for the project promotion remains
unchanged or even enforced especially in the necessity for further
internationalization of the project. 
In this respect, the 1997 endorsement of the JLC as one 
of the major future facilities in the Asia-Pacific region
is a remarkable milestone made by the Asian Committee for Future
Accelerators (ACFA) and was an important step towards its realization in
this region.
The inter-regional cooperation also became more important than ever, which
is reflected in recent close cooperation with European and North American
regions to promote R\&D's for the JLC facility which is, in its first stage,
to cover center of mass energies up to about 500 GeV with a luminosity 
in the  $ 10^{34} {\rm cm}^{-2} {\rm sec}^{-1}$ range.

The JLC, being an unprecedentedly high luminosity $e^+e^-$ collider,
is capable of producing
more than 0.5 million light Higgs bosons, if any, and top quarks, for sure,
within 5-6 years, and naturally
serves as a Higgs and top factory.
In addition it has a remarkable physics
potential as a $W$ and $Z$ factory equipped with highly polarized beams and a
two-orders of magnitude higher luminosity than those of existing facilities.
Precision study of these particles at the JLC
is a crucial step to establish the standard model
and will pave the way to go beyond it.
\\

In response to the ACFA statement of the linear collider project, a study
group, Joint Linear Collider Physics and Detector Working Group, has been
set up under ACFA. The charge of the working group is to elucidate physics
scenario and experimental feasibilities. 

The working group has been
subdivided, according to physics topics and detector components and many
institutions in Asia are participating in various subgroups, expanding their
research activities at home institutes. 
The subgroup activities in different
part of Asia are discussed and exchanged over Internet and have been
summarized annually at a series of ACFA workshops.

Taking into account the scale of and the worldwide interests in linear
collider projects, it is highly desirable that actual studies be carried out
in a more global scope in spite of the regional nature of ACFA's initiative.
In this respect, it should also be noted that, although the body of the JLC
project promotion lies in Asia, there have always been active participants
from Europe and North America in the framework of the inter-regional
cooperation. 
Reciprocally Asian colleagues have been contributing to the
workshops of the other regions, and all of these regional activities are
reported and discussed at a series of worldwide linear collider workshops.
\\

This report is a result from the past few years of the working
groups' activities,
and is intended to bring back and further
enforce the importance of the physics that motivates the project,
as well as to detail the technical feasibility of the experimentation
that is involved in it.

The report is organized in the following format: in Chapter 1,
we first detail the mission imposed upon
the working group, and 
overview physics prospects and the detector model.
We then list up 
basic sets of parameters of the JLC machine, 
on which all the studies are based.
The subsequent chapters are grouped into three parts.
First two of them are for  
physics and detectors, which are subdivided,
according to the subgroups formation, and
devoted to summaries of subgroups' activities.
The last chapter describes optional experiments using
$\gamma\gamma/e^-\gamma/e^-e^-$ collisions.
\\

We are convinced that the JLC is a reasonable next step 
for the Asian high energy physics
community well motivated by our present knowledge of particle physics. 
We have been integrating outcomes from the accelerator, physics, and detector
studies to realize the project in Asia through healthy collaboration among
Asian laboratories and universities.
\\

We are willing to host this challenging large-scale inter-regional
facility in the 21st century and we believe 
the project would be a model case for the
promotion of accelerator-based science in Asia.

%% file: headers/overview.tex
\chapter{Overview}
\label{chapter-headers}
\input headers/mission.tex
\input headers/physics.overview.tex

\input headers/detector.overview.tex
\input headers/machine.overview.tex

\input headers/bib.tex

%% file: headers/mission.tex
\section{Statement of the Mission}

Clearly our goal is to build a linear $e^+e^-$ collider 
and to open up a new era of high energy physics through 
the experiments there-at. 
As a necessary step toward this goal, the ACFA requested us to prepare 
a written report on the physics and detectors at the collider\cite{acfastatement,ACFAWorkshops}.
The document should identify important physics targets and, by doing so, 
clarify required machine parameters such as beam energy and its spread, 
beamstrahlung, luminosity, background, etc. and detector parameters 
such as momentum resolution for tracking, energy resolution for calorimetry, 
impact parameter resolution for vertexing, minimum veto angle, and so on. 
\\

The best map of the world for the linear 
collider to explore is called the standard model Lagrangian, 
consisting of three parts: gauge, Higgs, and Yukawa sectors.
While the gauge sector has been investigated in depth
by experiments in the last century,
only a very little is known about 
the remaining two, the Higgs and the Yukawa sectors,
reflecting the fact that there are two particles, 
the Higgs boson and the top quark, 
of which we do not know very much yet. 
The situation was more or less the same back in 1992 with the exception 
that there was no top quark discovered at that time. 

The most important parameters that determine the overall scale of the project 
are naturally the masses of these two particles, since they decide the required 
beam energy for their direct productions. 
Although we had, already at that time, a fairly good estimate of 
the top quark mass, its recent direct measurement at Tevatron 
is far much better and gives us confidence to set our initial target
machine energy to cover 350 to 400 GeV in the center of mass frame.
As for the Higgs mass, we now have an indirect measurement, $M_h < 215$ GeV
at the 95\% confidence level, thanks to the direct top mass measurement 
at Tevatron and various precision electroweak measurements 
at LEP and SLC, in particular. 
This again points us to the energy range: 300 to 400 GeV. 
It is thus very important for us to elucidate all the conceivable physics 
in this initial energy range of the project. 
The next key issue to constrain the machine is the luminosity requirement. 
Past studies showed that a luminosity of 
$5 \times 10^{32} {\rm cm}^{-2}{\rm sec}^{-1}$ is enough for most 
discovery physics, and at least $5 \times 10^{33} {\rm cm}^{-2}{\rm sec}^{-1}$
is needed for precision studies. 
This statement is still valid, but we definitely need more 
if we are to define the JLC as a Higgs/Top/$W$/$Z$ factory. 
How much more should be answered in the report. 

As a working assumption, this report assumes that the JLC
can cover an energy range of $250 < E_{cm} < 500$ GeV
with a luminosity up to $1.5 \times 10^{34} {\rm cm}^{-2}{\rm sec}^{-1}$.
\\

The detector parameters should be decided so as to make maximal use of
the potential of the collider. 
It is thus very important to understand new features expected for the 
future linear collider experiments. 
Firstly, since jets become jettier and calorimetric resolutions improve
with energy, we will be able to identify heavy "partons" such as 
$W$/$Z$ bosons 
and $t$-quarks by reconstructing their hadronic decays 
with jet-invariant-mass method. 
Full reconstruction of final states in this way is possible only 
in the clean environment of an $e^+e^-$ collider and will allow us 
to measure the four-momenta of final-state "partons", 
where "partons" include light quarks, charm, bottom, and top quarks, 
charged leptons, neutrinos as missing four-momenta, photons, 
$W$, $Z$, and gluons. 
For the heavy "partons" which involve decays, 
we may even be able to measure their spin polarizations. 
This is perhaps the most important new feature which makes 
unique the future linear collider experiments: 
the Feynman diagrams behind a reaction become 
almost directly observable. 
The detector must take advantage of this and, therefore, 
has to be equipped with high resolution tracking and calorimetry 
for the jet-invariant-mass method, 
high resolution vertexing for heavy-flavor tagging, 
and hermeticity for indirect neutrino detection as missing momenta. 

Another remarkable advantage of the future linear collider experiments 
is the availability of a highly polarized electron beam. 
For instance, let us consider the reaction: $e^+e^-$ to $W^+W^-$. 
In the symmetry limit where the gauge boson masses are negligible,
we can treat this process in terms of weak eigen states. 
This process then involves only two diagrams: an $s$-channel diagram 
with a triple gauge boson coupling and a $t$-channel diagram with 
a neutrino exchange. 
Notice that only $W^0$ (the neutral member of $SU(2)_L$ gauge bosons) 
can appear in the $s$-channel since $B$, belonging to $U(1)_Y$, 
has no self-coupling. 
Because of this the cross section for this process will be highly 
suppressed at high energies, when the electron beam is polarized 
right-handed. 
We can say that Feynman diagrams are not only observable, but also selectable. 
\\

There are, however, some drawbacks inherent in linear collider experiments. 
Since the collider operates in a single pass mode, 
one has to squeeze the transverse size of its beam bunches 
to a nano-meter level to achieve the luminosity 
we required above. 
Such high density beam bunches produce an electro-magnetic field 
which is strong enough to significantly bend the particles 
in the opposing bunches and thereby generating bremsstrahlung photons. 
This phenomenon is called beamstrahlung. 
Because of this, electrons or positrons in the beam bunches lose 
part of their energies before collision. 
When plotted as a function of the effective center of mass energy,
the differential luminosity distribution thus 
shows a long tail towards the low energy region
in addition to a sharp peak (delta-function part) 
at the nominal center of mass energy, 
corresponding to collisions without beamstrahlung.
The existence of this sharp peak implies that for most physics programs
we can benefit from the well-defined initial state energy, 
a good tradition of $e^+e^-$ colliders. 
In some particular cases, however, the finite width of the 
delta-function part, 
which is determined by the natural beam energy spread 
in the main linacs, and the beamstrahlung tail can be
a potential problem. 
Top pair production at threshold is a typical example. 

There is yet another potentially serious problem inherent in 
linear collider experiments, which is new kinds of background 
induced by beam-beam interactions. 
The beam-beam background includes low energy $e^+e^-$ pairs 
and so called mini-jets.
One of the most important tasks of the working group
is to carefully design background mask system and 
detectors with good time resolution. 
\\

Any detector design should take these new features into account. 
We can summarize the performance goals as: 
\begin{itemize}
\item	Efficient and high purity $b/c/(u,d,s,g)$ tagging for top and higgs studies.
\item	Recoil mass resolution limited by natural beam energy spread 
        but not by trackers for the reaction: $e^+e^-$ to $Zh$ 
        followed by $Z$ to $l^+l^-$. 
        This is necessary to confirm the narrowness of the higgs width.
\item	2-jet invariant mass resolution comparable with the natural 
        widths of $W$ and $Z$ for their separation in hadronic final states.
\item	Hermeticity to indirectly detect invisible particles such as neutrinos, LSP, etc.
\item	A well designed BG masking system and time stamping capability
\end{itemize}

The first of these requirements sets a performance goal for a vertex detector, 
while the second imposes the most stringent constraint on the tracking system. 
It should be emphasized that the third point not only requires a good 
calorimeter, 
but also a good track-cluster matching capability to enable good energy flow 
measurements. 

The possible detector system that we proposed in 1992\cite{JLCI-report}
was designed to satisfy the above requirements, and
contains both the central tracking chamber (CDC) 
and the calorimeter (CAL) in a solenoidal magnetic field of 2 Tesla 
to achieve good resolution and hermeticity.
The design also required that final focus quadrupole magnets 
and a background mask system 
be supported by a support cylinder installed in the detector.
The final focus magnets and the mask system 
should, thus, be considered as part of the detector system. 
Although there is no immediate need to change the design principle of 
the detector system, this design is almost 8 years old now and parameters 
of each detector component should be reexamined carefully, 
taking into account achievements in the past detector R\&D's. 
3 Tesla option is definitely one of the most important study items 
for the working group.

%% file: headers/physics.overview.tex
\section{Physics Overview}

\subsection{The Standard Model}

The goal of elementary particle physics is to identify the ultimate
building blocks of Nature and the interactions among them, and find
their simple description.
Primary means of this endeavor is high energy accelerators.
Advance of accelerator technology
has been enabling us to probe ever-higher energy and thus
ever-shorter distance,
thereby leading us to deeper understanding of Nature.
Over the past decades, we have learned that
Nature consists of a small number of matter particles
and among these matter particles lies
a beautiful symmetry that is
deeply connected with their interactions,
and that the microscopic world of these elementary particles
can be described by quantum field theory.
The matter particles here are two kinds of spin 1/2
fermions, quarks and leptons, and the symmetry
here is called gauge symmetry.

Conversely, we can start from the gauge symmetry and
demand that any matter particle has to belong to
some multiplet that is allowed by the gauge symmetry.
A set of particles that comprise a multiplet
mutually transform each other by gauge transformations
and thus have to be regarded as different states of 
a single particle.
Their distinction thus loses its absolute meanings, 
which naturally leads us to demand invariance of physics
by any gauge transformations made independently at
different points in space-time.
The key point here is that 
this requirement of local gauge invariance
forces us to introduce spin 1 gauge particles (gauge bosons)
and, moreover, it dictates the form of the interaction 
mediated by them.
This is called gauge principle.

The first and most important question in any model
building guided by the gauge principle is the choice
of the gauge symmetry (or corresponding gauge group).
The gauge field theory that is based on the
$SU(3)_C \otimes SU(2)_L \otimes U(1)_Y$ is the
Standard Model\cite{WS}.
The standard model succeeded in
describing all but one of the four known 
interactions---electromagnetic,
weak, and strong---and has been tested to a great precision
in particular in the last decade mainly through
collider experiments.
The test has reached a quantum level and firmly
established the gauge principle.
It is remarkable in this respect that the top quark, which
was missing when the first JLC project design was drawn 
in 1992\cite{JLCI-report},
was discovered\cite{CDF:top} 
in the mass range that had been
predicted\cite{LEP:top}
through the analysis of the quantum corrections.
This filled the last empty slot of the matter multiplet
of the Standard Model.
Although the recent discovery of neutrino oscillation\cite{Ref:superK}
requires extension of its particle contents\cite{NEUTRINO},
the other part of the Standard Model is still intact.

Nevertheless, there is a good reason for the Standard Model
still being called a model.
This is primarily because its core ingredient,
the mechanism that is responsible for
the spontaneous breaking of the gauge symmetry\cite{HIGGS}
hence for the generation of the masses of
otherwise massless matter and force carrying particles,
is left untested.
In the Standard Model, a fundamental scalar 
(Higgs boson) field plays this role.
Because of a new self-interaction (a four-point self-coupling
hereafter called Higgs force),
a Higgs field condenses in the vacuum
and spontaneously breaks the $SU(2)_L \otimes U(1)_Y$
gauge symmetry.
The masses of the matter and force carrying particles
are generated through their interactions
with the Higgs field condensed in the vacuum,
and are consequently proportional to
their coupling strengths to and
the density (the vacuum expectation value)
of the condensed Higgs field.
The masses of the gauge bosons, $W$ and $Z$, could be predicted,
since their interaction with the Higgs field that
is responsible for the mass generation
is the universal gauge interaction.
The discovery of $W$ and $Z$ at the predicted
masses\cite{UA1} is a great triumph of the Standard Model.
On the other hand, the masses of quarks and leptons
are generated through yet another new interaction 
(hereafter called Yukawa force)
that is arbitrarily put in by hand to parametrize the observed
mass spectrum and mixing of the matter particles;
more than half of the 18 parameters 
of the Standard Model are thus used for this parametrization.

We can summarize the current situation as follows.
There is no doubt about the gauge principle on which
the Standard Model is based and thus its breaking
has to be spontaneous and caused by "something"
that condenses in the vacuum.
But the nature of this "something" still remains mysterious.
Without revealing its nature, 
it will be difficult to understand
real implications of the data on $CP$ violation
and flavor mixing to be accumulated 
at various laboratories in this decade.
It should be stressed that, since the coupling
of this "something" with a matter particle
is proportional to the mass of the matter particle,
the heaviest matter fermion found so far, 
the top quark, might hold the key to uncover the
nature of this "something".
In order to understand the Higgs and
the Yukawa forces, therefore,
we need not only to find the Higgs boson but also
to study both the Higgs boson and the top quark in detail.

It is remarkable that just like the analysis of
the quantum corrections enabled us to predict
the mass range of the top quark before its discovery,
the advance of the precision measurements
in the last decade now allows us to indirectly
measure the mass of the Higgs boson in the framework
of the Standard Model.
The data tell us that the mass of the Standard Model
Higgs boson is less than 215 GeV at the 95\% 
confidence level\cite{HIGGSMASSUPPERBOUND}.
Recall that the mass of the Higgs boson is related to 
its four-point self-coupling, 
which becomes stronger at higher energies.
This upper bound is surprisingly consistent
with the picture that the Higgs self-coupling
stays perturbative up to very high energy
near the Planck scale and also with
the recent indication of a possible Higgs signal 
at LEP\cite{HIGGSCANDIDATE}.
Such a light Higgs boson lies well within the reach of the JLC
in its startup phase and can be studied in great detail
as we shall see in Chapter \ref{chapter-physhiggs}.
We will be able to verify its quantum numbers and its couplings
as well as to precisely determine its mass.
If the Higgs boson mass is less than 150 GeV,
we should be able to test
the mechanism of the fermion mass generation. 
The top quark threshold region is
sensitive to the Yukawa potential due to the Higgs boson exchange
and at higher energies we will be able to
measure the top Yukawa coupling directly
as discussed in Chapter \ref{chapter-phystop}.
The study of the Higgs boson branching ratios\cite{Ref:kawagoe.etal} can
also tell us if the Yukawa coupling constants are proportional 
to the fermion masses.
Such tests can be performed only at $e^+ e^-$ colliders 
with a clean environment.  
In this way, the JLC is able to thoroughly establish the Standard Model.

\subsection{Problems with the Standard Model}

Once the $SU(3)_C \times SU(2)_L \times U(1)_Y$
gauge structure and the mass generation mechanism
are established this way,
we may start seriously asking many unresolved questions 
within the Standard Model.
Why do the electric charges of electron and proton exactly balance? 
Why are the strengths of the gauge interactions so different? 
Why is the number of generations three? 
Why do the seemingly independent anomalies from the quark
sector and lepton sector cancel? 
Where do the fermion masses come from? 
Why is the $CP$ invariance broken? 
And many others.  
Among them, the far most important question is: 
Why is the electroweak symmetry broken, and why at
the scale $\langle H \rangle = 246$~GeV?

The Standard Model cannot answer any of these questions.
This is exactly why we believe that
there lies a more fundamental physics at a higher energy scale 
which leads to the unanswered characteristics of the Standard Model. 
Then all the parameters and quantum numbers in the Standard Model 
can be {\it derived}\/ from the more fundamental description of 
Nature, leading to the Standard Model as an
effective low-energy theory.
In particular, the weak scale itself $\langle H
\rangle = 246$~GeV should be a {\it prediction}\/ of the deeper theory. 
The scale of the fundamental physics can be regarded as 
a cutoff to the Standard Model. 
Above this cutoff scale, the Standard Model ceases to be valid and the
new physics takes over.

The mass term of the Higgs field is of the order of the weak scale,
whereas the natural scale for the mass term is, however, 
the cutoff scale of the theory, since the quantum correction to the
mass term is proportional to the cutoff scale squared because 
of the quadratic divergence. 
This problem, so-called the naturalness problem, is one of the main
obstacles we encounter, when we wish to construct realistic models of the
``fundamental physics'' beyond the Standard Model. 
If the cutoff scale of the Standard Model is near the Planck scale, 
one needs to fine-tune the bare mass term of the Higgs potential 
to many orders of magnitude to keep the weak scale
very tiny compared to the Planck scale. 
There are only two known possibilities
to solve this problem. 
One is to assume that the cutoff scale of the Standard Model 
lies just above the weak scale and
the other is to introduce a new symmetry 
to eliminate the quadratic divergence: supersymmetry.
In the former scenario the Higgs boson mass tends to be heavy, if any,
while in the latter it is expected to be light.

\subsection{Supersymmetry}

The existence of a light Higgs boson below the experimental
upper bound mentioned above thus makes the
latter possibility more plausible.
Supersymmetry (SUSY)\cite{SUSY} is a symmetry between bosons and fermions 
and imports chiral symmetry that protects fermion masses
from divergence into scalar fields, thereby eliminating
the quadratic divergence of scalar mass parameters.
Since the principal origin of the naturalness problem in the
Standard Model is the quadratic divergence of the Higgs mass parameter, 
its absence in the supersymmetric models allows us to push the cutoff 
up to a very high scale\cite{Veltman}.
This possibility, that the cutoff scale may be very high, 
provides us an exciting scenario, that all the weak scale parameters
are determined directly from those at the very high scale
where the supersymmetry is naturally understood in the context 
of supergravity.
Stated conversely, we can probe the physics at the
very high scale from the experiments at the weak scale. 

Supersymmetry is, however, obviously broken,
if it exists at all.
Nevertheless, it should not arbitrarily be broken,
as long as it is meant to solve the naturalness problem:
only Soft Supersymmetry Braking (SSB) terms are allowed
and the mass difference between any Standard Model
particle and its superpartner should not 
exceed $O(1)$~TeV.
Most of more detailed analyses along this line put 
upper mass bounds on some SUSY particles that
make their pair productions at the JLC
possible\cite{NATURALNESSBOUNDS}.
Turning our attention to the theoretical aspect 
of the above restriction to the SUSY breaking mechanism,
we can classify phenomenologically viable models
in terms of how the SSB takes place and
how it is transmitted to our observable sector.
Various SSB parameters at the high scale 
of SUSY breaking are determined by the choice of the 
SSB mechanism and the mediation mechanism. 

In the early days of SUSY model building
there existed essentially only one class of models, 
where the SSB is transmitted via gravity to the
low energy world. 
While these models still being valid, 
the past few years changed the situation drastically
and now we have a set of different models that include
the aforementioned gravity-mediated models, 
Anomaly-Mediated SUSY Breaking (AMSB) models\cite{AMSB},
Gauge-Mediated SUSY Breaking (GMSB) models\cite{GMSB},
and models where the SSB mediation is dominated by
gauginos\cite{GAUGINOMSB}.
In any case, the low energy values of the SSB parameters
are derived from those at the high scale by
evolving them down to the weak scale via 
renormalization group equations.
Consequently the SUSY particle masses and, 
in cases where mixing occurs, 
even their couplings are given in terms of
the SSB parameters at the high scale,
depending on the SSB mechanism.
As we shall see in Chapter \ref{chapter-physsusy},
once the first SUSY particle is found,
we can carry out a consistent SUSY study program
to step through the spectrum of SUSY particles
and measure their properties in detail
with the help of the clean environment and
the powerful polarized beam that is available
only at the $e^+e^-$ linear colliders\cite{Ref:tsukamoto.etal,Ref:nojiri.etal}.
Once these low energy values of SUSY particle properties are measured, 
we can then in principle point towards the
physics at the high scale 
and hence at the SUSY breaking mechanism.

The idea that the weak scale parameters are directly determined 
from a very high energy scale has naturally led us 
to the concept of the grand unified theory (GUT)\cite{Georgi-Glashow}.
It is well known that the three gauge coupling constants
of the Standard Model do not unify, when extrapolated to the
high scale using the Renormalization Group Evolution,
while they do to a good approximation,
if the Standard Model is supersymmetrized.
The supersymmetric GUT models thus quantitatively explain
the relative strengths of the three gauge coupling constants
of the Standard Model.
Furthermore, the baroque structure of the fermion quantum
numbers in the Standard Model can be naturally embedded 
into a GUT gauge group, leading to the exact quantization 
of the electric charge and the precise cancellation of the anomalies.
It is also worth mentioning that the heavy top quark
naturally fits in the supersymmetric models,
since its Yukawa coupling can drive the 
Higgs boson mass squared to negative at the weak scale
thereby radiatively breaking the $SU(2)_L \otimes U(1)_Y$
as needed\cite{radiative}.

\subsection{Alternative Scenarios}

Although the experimental data prefer the 
former light Higgs scenario, 
there is logical possibility that Nature had
taken the latter path.
Technicolor scenario, which is based on a new strong interaction,
belongs to this class, solving the
naturalness problem by setting the cutoff just above the TeV scale.  
The elementary Higgs field is replaced by the Nambu-Goldstone bosons 
associated with a dynamical chiral symmetry breaking 
in the techni-fermion sector\cite{technicolor}.
Unlike the early models of supersymmetry, however,
all the early models in this category have been
excluded experimentally and
the remaining models have lost most of the original
beauty that motivated the scenario.
Nevertheless, if no light Higgs boson is found, we will
have to abandon the former scenario and seriously
confront this latter possibility.
In this sense, this is the crucial branch point
to decide the future direction of high energy physics
and the JLC can clearly show us which way to take by 
unambiguously testing the existence of the
light Higgs boson as described in Chapter \ref{chapter-physhiggs}.
If we are to take the latter path, 
we will need to scrutinize 
the $W$ and $Z$ bosons as well as the top quark
in great detail to spot any deviation from the Standard
Model in order to get insight into the underlying 
dynamics that is responsible for the spontaneous
breaking of the electroweak gauge symmetry.
The JLC's ability in such measurements are elucidated in
Chapters \ref{chapter-phystop} and \ref{wz-sect}.

The study of the top quark has, however,
fundamental importance in its own right
as described in Chapter \ref{chapter-phystop}.
First of all, a precise measurement of its mass, 
$\Delta m_t \lsim 100$~MeV, 
is found possible at the $t\bar{t}$ threshold
thanks to the recent progress 
in the nonrelativistic QCD.
The large width of the top quark acts as an
infrared cutoff to the QCD interaction\cite{Ref:fadin.khoze}, 
allowing us to make definite
theoretical predictions using perturbative QCD. 
This remarkable feature provides a new and 
clean test of perturbative QCD
as well as a precise measurement of the strong coupling constant 
$\alpha_s$\cite{Ref:fujii.etal}. 
Both of these measurements together with the $W$ mass
determination discussed in Chapter \ref{wz-sect}
will be indispensable when one tries to
probe the physics beyond the Standard Model from
the analysis of the radiative corrections.
Search for possible $CP$ violation in the top quark system
also deserves special mention since its discovery
immediately signals physics beyond the Standard Model.

Finally, speaking of the physics beyond the Standard Model
that may put the cutoff scale just above the weak scale,
we cannot but mention the recent remarkable proposals that 
have literally added extra dimensions to 
the possible scenarios\cite{EXTRADIM}.
These extra spatial dimensions
may give rise to new states that would appear as $Z'$ bosons
or spin-2 resonances, depending on the models.
In the brane world scenarios, which embed
our world as a four-dimensional membrane in
a space with higher dimensions,
gravity would become strong at TeV energies.
In these models, gravitons could be radiated into the bulk
and would leave missing energy signals for
the process like $e^+e^- \to \gamma G$.
It is also possible that the effect of
the virtual graviton exchange would show up
as a deviation from the Standard Model.
The sensitivity of the 500 GeV linear collider
is expected to reach a few TeV\cite{EXTRADIMLIMITS},
which seems enough to find some positive signal
if the extra dimensions are somehow related to the
naturalness problem.

We have seen above that there are many possibilities
and we do not know which way to take for sure, but
one thing is clear.
Whatever new physics lies beyond the Standard Model,
the key to open it is the understanding of
the electroweak symmetry breaking, and
the clean environment, high luminosity, and the large beam 
polarization at the JLC will give us a definite
answer for it, thereby leading us
to make an entirely new step towards the deeper understanding of
Nature.

%% file: headers/detector.overview.tex
\section{Detector Model}

The first design of the JLC detector was made in 1991
as a part of the JLC-I study report\cite{JLCI-report}.
It was a large-volume general-purpose detector aiming at the 
highest performance with rather conventional technology.
The design criteria adopted that time are still valid
even now since motivating physics has not been changed
up to now.

\vspace{0.5cm}

The design criteria in terms of event-reconstruction performance
are as follows:
\begin{itemize}
\item Two-jet mass resolution be as good as natural widths of weak bosons;
\item Mass resolution of the recoil system to a muon pair be as good as 
beam-energy spread for Higgs-strahlung events
followed by $Z$-decay to muons;
\item Vertex resolution be capable of reconstructing cascade decays.
\end{itemize}
\noindent
How these criteria are translated into performance of each
sub-detector under linear-collider environment
is described in the detector chapters (Part \ref{part-detector} 
of this volume) in detail.

\begin{figure}[bht]
\centerline{
\epsfysize=12cm \epsfbox{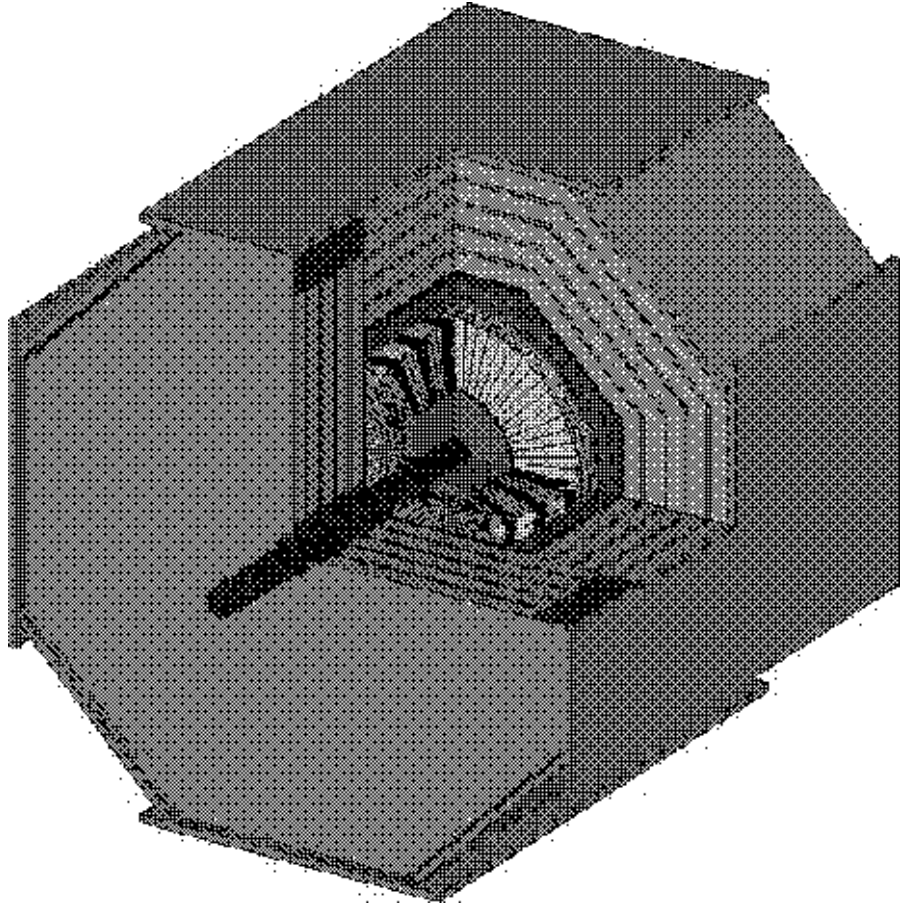}}
\caption{\label{det3T}
{\sl Configuration of the baseline JLC detector.}}
\end{figure}

\vspace{0.5cm}

There has been a great progress in evaluating backgrounds
to the tracking detectors these years.
As a result, the magnetic field of 2~Tesla in the original
design is now thought to be a little 
too weak to suppress backgrounds to the vertex detector (VTX)
and the central drift chamber (CDC) in the case of high-luminosity
operation (parameter set {\bf Y}) of $2.5 \times 10^{34}$cm$^2$sec$^{-1}$.
We therefore started studies for a new detector design with 
3~Tesla magnetic-field, and consequently with smaller dimensions.
This is why, for some sub-detectors described
in the detector chapters, 
both the 2~Tesla- and the 3~Tesla-designs are
discussed in parallel.
Inner detectors, on the other hand, have common designs,
regardless of the field choice,
as far as the aforementioned two options are concerned.

As a matter of fact,
there has been a proposal to go to much higher magnetic field
and much smaller size. 
However, studies on this compact-detector option are yet very limited,
and thus are omitted in this report.

In designing the 3~Tesla-version,
some detector parameters have been revised in accordance with
progress in each sub-detector R\&D and of simulation studies.
There has also been significant advance in detector
technologies as consequences of SSC, Tevatron, and LHC studies. 
Some sub-detector designs have been revised to utilize 
these new technologies.

\vspace{0.5cm}

Fig.~\ref{det3T} shows a cut view of the
present (revised) JLC detector of the 3~Tesla design.
From inside to outside in the radial direction, 
it is composed of VTX, an Intermediate Silicon Tracker (IT), 
the CDC, a Calorimeter (CAL), a Superconducting Solenoid, 
and Muon Counters (MU) interleaved with flux-return iron yokes.
There is a supporting tube between the IT and the CDC,
which supports inner detectors and Interaction Region (IR) devices
such as VTX, the IT, Luminosity and Pair Monitors (LM and PM), 
final Q-magnets, conical masks with active energy taggers (AM) on
their tips, and compensating solenoids. 
There is no design for forward trackers (FT) yet.
The overall dimension of the 3~Tesla design is about 
$14{\rm m} \times 14{\rm m} \times 13{\rm m}$, 
and weighs about 13 kilo-tons.
This volume is slightly smaller than that of the CMS detector at LHC.

Key parameters and expected performances of the sub-detectors are
summarized in Table~\ref{dettable}.
Notation of sub-detectors in the table follows 
the abbreviations given above.

\begin{table}[p]
\caption{\label{dettable}
{\sl Parameters and Performances of the baseline JLC detector.}}
\vspace*{6pt}
\centerline{
\epsfxsize=16cm \epsfbox{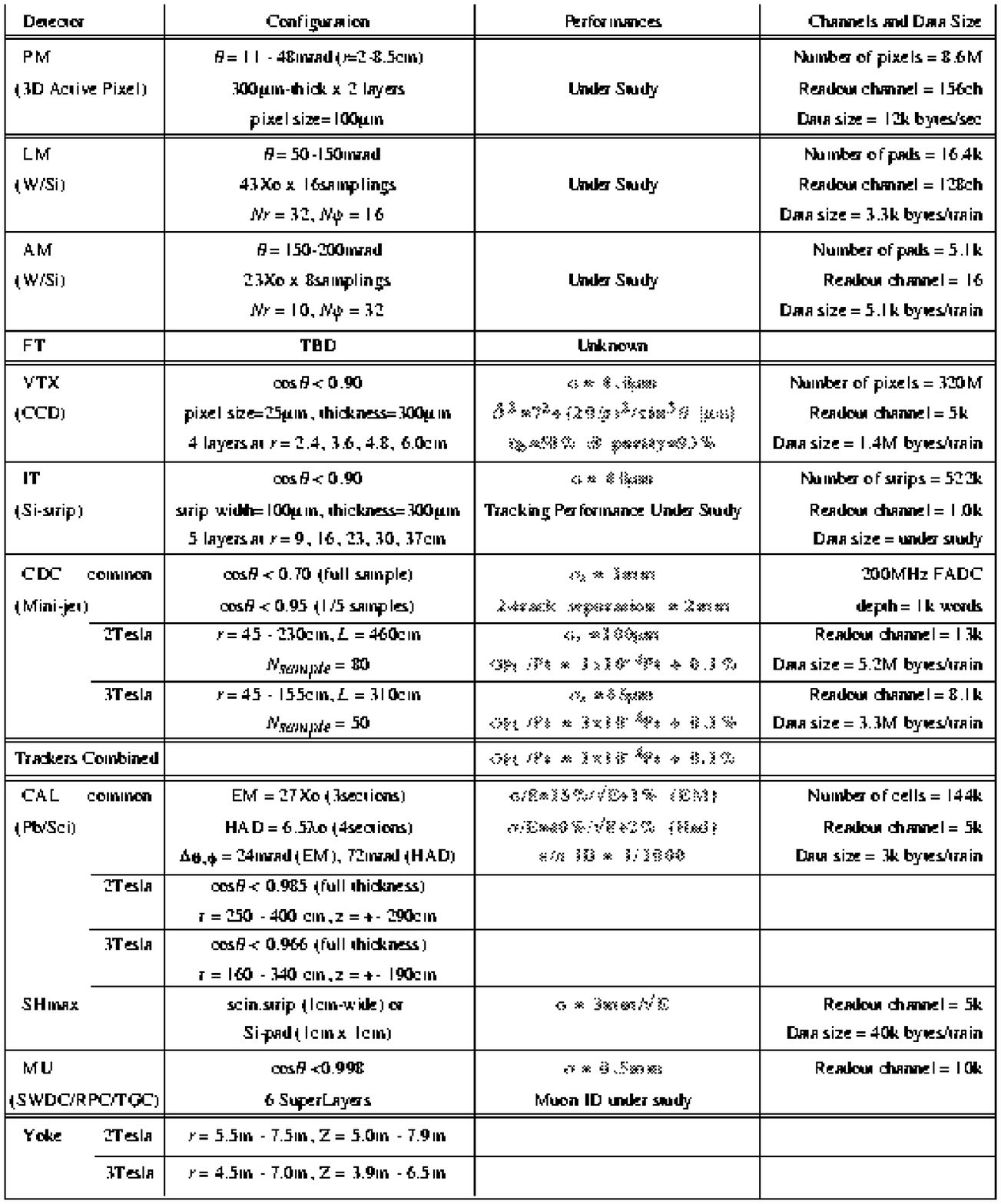}}
\end{table}

\vspace{0.5cm}

\noindent
Key points of the sub-detector R\&D's are very briefly listed below.

\vspace{0.5cm}

\noindent
{\bf IR}

The most important tasks of the IR study are:
\begin{itemize}
\item to design beam line devices which minimize backgrounds to the detectors,
\item to estimate the backgrounds to the detectors reliably,
\item to design a radiation-resistant detector for the pair monitor, and
\item to design a stable support system for the IR devices.
\end{itemize}
Though there remains some technical issue to be solved,
most of these tasks are satisfactorily achieved 
as described in Chapter~\ref{chapter-detir}.

\vspace{0.5cm}

\noindent
{\bf Trackers}

The most crucial R\&D item on the VTX detector is radiation hardness of CCD.
The most up-to-date technologies such as two-phase operation and
notch structure seem promising to enable operation
of a few years to several years.
However, further establishment is needed.
Other R\&D items such as position resolution,
room-temperature operation, and fast readout are shown to
be of no problem.
Detail will be presented in Chapter~\ref{chapter-dettrk}. 

\vspace{0.3cm}

Many R\&D items of the CDC are related to the feasibility of a
4.6m-long chamber: wire creep, wire sag, stereo-cell geometry,
or a thin mechanical structure which can support huge wire tension. 
These problems may disappear in the case of the 3~Tesla design.
However, the 3~Tesla-field raised a new problem: large Lorentz angle.
Some of the problems above are almost solved,
while others need further studies.

Generic studies such as gas gain, position resolution, and
two-track separation, on the other hand, are straightforward
tasks, and seem to be almost established.
These items will be described in Section~\ref{section-cdc}.
 
\vspace{0.3cm}

Studies on the IT are yet rather limited.
Simulation studies on its performance are in progress 
with a 1st trial set of parameters.
Hardware design studies, however, are very primitive.
The description of the IT given in Section~\ref{section-it} is, therefore,  
mostly on the simulation results.

Studies on the forward trackers have not yet started.

\vspace{0.5cm}

\noindent
{\bf PID}

Studies are rather limited up to now about dedicated detectors
for particle identification, especially for $\pi/K$ separation.
Though it was reported that $dE/dx$ measurement with 
pressurized gaseous tracking detectors 
can provide good $\pi/K$ separation
for momentum region below 30~GeV \cite{SitgesYama},
this momentum region may not be high enough to improve
$W$-charge determination via jet-mode at the highest energy.
Furthermore, in our current scope,
the JLC-CDC will not be pressurized.
In such cases, challenging detectors such as a
focusing DIRC \cite{SitgesWilson} would be helpful.
However, since there have been almost no systematic studies about
possible impact of such detectors on physics sensitivity, 
no dedicated chapter for particle identification
is given in this report.

\vspace{0.5cm}

\noindent
{\bf CAL}

Calorimeter design has been changed significantly since the
JLC-I study report.
Finer granularity is aimed at while keeping the best energy
resolution.
The latter has been established by series of test beam measurements,
while optimization of granularity is still underway.
Studies on photo-detectors and engineering issues are also in progress.
The detail of the study including historical review is 
presented in Chapter~\ref{chapter-detcal}.

\vspace{0.5cm}

\noindent
{\bf MU}

Requirements on muon detectors are neither so severe
nor unique under linear collider environment.
We therefore think that muon detectors developed for B-factories
or for LHC experiments will be usable for linear collider experiments.
Applicability of
such muon detectors is examined in Chapter~\ref{chapter-detmuon}.

\vspace{0.5cm}

\noindent
{\bf DAQ}

The number of read-out channels and average data size are listed in 
Table~\ref{dettable}.
Zero-suppression-on-board is assumed for the data-size estimation.

The CCD-VTX detector was once thought to be too slow to finish readout
within the bunch-train crossing-interval of 6~msec, 
and various pre-processing schemes were investigated.
Recent studies have shown that it is feasible to
read-out all the CCD data during the train-crossing interval.
Therefore all the data from all the sub-detectors can directly be 
transfered to a CPU farm for judgement. 
Since there is no complicated architecture in the DAQ design,
dedicated chapter is not given in this report.

\vspace{0.5cm}

\noindent
{\bf Structure}

In Chapter~\ref{chapter-detmag}, design of the solenoid structure and its
engineering studies on mechanical and magnetic properties are described.
Because of the huge magnetic force on the endcaps
we may end up with reducing
the number of super-layers of the muon detector interleaved with
the endcap iron yokes in the case of the 3~Tesla design.
Otherwise there is no foreseen difficulty, and
one possible solution, though not yet final, is presented.

\vspace{0.5cm}

\noindent
{\bf Simulation Tools}

Two kinds of detector simulators,
QuickSim and JIM, have been used for our studies.
QuickSim is a simple but fast simulator
mostly used for physics studies.  JIM is a full detector simulator
based on the Geant3 packages and used for studies of detector performance.
These tools are described in Chapter~\ref{chapter-detsim},
together with tools for event generations.
Recently, we started developing a new full detector simulator, JUPITER,  
using object oriented technology based on Geant4.  Short report
on its status is also given there.

%% file: headers/machine.overview.tex
\section{Accelerator Overview}
\subsection{Accelerator Complex}
A detailed report on the R\&D status as of 1997 was published as
`JLC Design Study'\cite{JLCDS} and recent updates are described in
`ISG Progress Report'\cite{ISG}. Look at the web pages\cite{JLCaccWeb}
for the latest status. 
Here, we shall summarize the
machine aspects very briefly.
\begin{figure}
\centerline{\epsfxsize=\textwidth \epsfbox{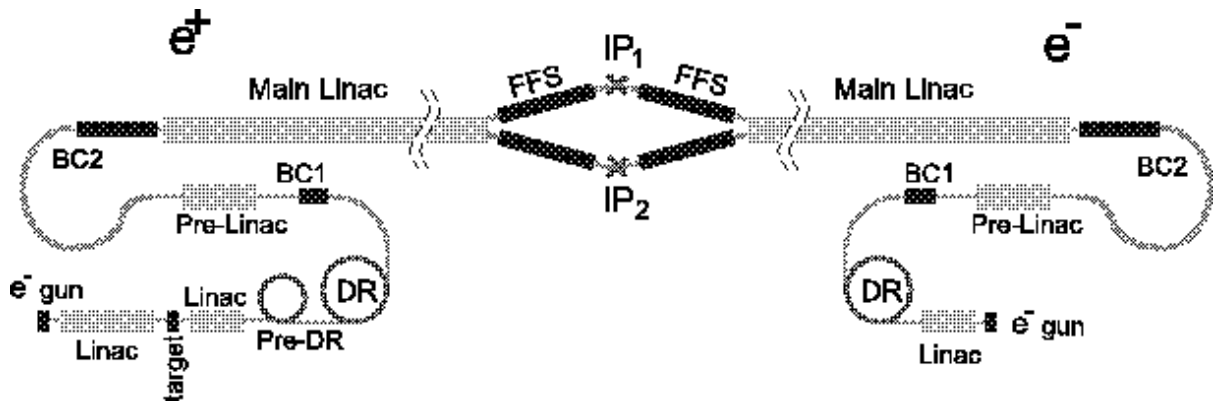}}
\begin{center}\begin{minipage}{\figurewidth}
\caption{\sl Schematic Layout of JLC. Not to scale. 
DR: Damping Ring, BC: Bunch Compressor, FFS: Final Focus System,
IP: Interaction Point.
\label{fig:JLClayout}}
\end{minipage}\end{center} 
\end{figure}

The whole accelerator complex
is schematically depicted in Figure~\ref{fig:JLClayout}.

\subsubsection{Injectors\protect\footnote{Here we give parameters based on X-band main
linac. There are minor differences for C-band up to a factor of 2.}}

The e$^+$e$^-$ beams to be injected to the main linacs have the
following properties:
\begin{itemize} \itemsep 0mm
\item The beam energy 10~GeV.
\item One pulse consists of 95 bunches with the separation 2.8~nsec in-between
(190 bunches $\times$ 1.4~nsec when upgraded).
\item Each bunch contains $\sim10^{10}$ particles. The r.m.s.~bunch length
is $\sim 100~\mu$m.
\item The pulse is repeated at 150~Hz.
\end{itemize}
\begin{flushright}
\centerline{\epsfxsize=12cm \epsfbox{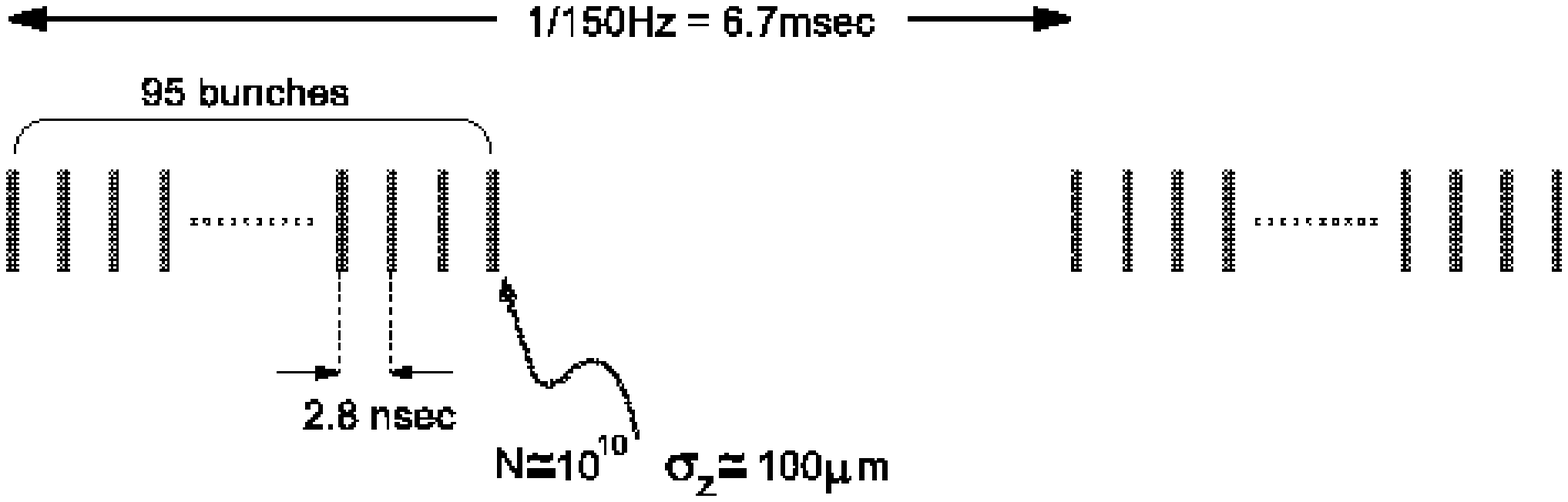}}
\end{flushright}
\noindent The electron beam is created in the following scenario:
\begin{itemize} \itemsep 0mm
\item The beam generated by the electron gun is accelerated to 1.98~GeV
by an S-band linac and cooled to the required emittance in the damping ring (DR).
\item The bunch length is $\sim$5~mm when extracted from DR. It is compressed
by the first bunch compressor BC1 to $\sim500~\mu$m.
\item Accelerated to 10~GeV by another S-band linac (pre-linac).
\item The bunch is compressed to the desired length in the second bunch
compressor BC2. BC2 is a fairly large structure, consisting of a large arc
($\sim$300~m long), RF cavities ($\sim$300~m long), and a chicane.
This makes a 180-degree turn so that the pre-linac and the main linac
can be accommodated in the same tunnel.
\end{itemize}
\noindent For the positron beam some more facilities are needed:
\begin{itemize} \itemsep 0mm
\item A high intensity electron beam is accelerated to $\sim$10~GeV
and is led to a target to generate positrons.
\item This positron beam is collected and accelerated to 1.98~GeV.
\item Since the emittance is much larger than that of the electron beam
from the gun, this positron beam is once cooled in the pre-damping ring
before it is injected to DR.
\end{itemize}

\newcommand{\multcolb}[1]{\multicolumn{2}{c|}{#1}}
\newcommand{\multcolc}[1]{\multicolumn{3}{c|}{#1}}
\newcommand{\multcold}[1]{\multicolumn{4}{c|}{#1}}
\newcommand{\multcole}[1]{\multicolumn{5}{c|}{#1}}
\newcommand{\multcolh}[1]{\multicolumn{8}{c|}{#1}}
\newcommand{\itemheader}[1]{\multicolumn{4}{|l|}{{\bf #1}}}
\newcommand{\avr}[1]{\left\langle{#1}\right\rangle}

\begin{table}
\begin{center}
\caption{\sl Main Linac RF System. \label{tab:RFparam}}
\begin{tabular}{|l|l|r|l|} 
\hline 
\itemheader{Overall parameters} \\
\hline
 Unloaded gradient    & $G_0$ & 72 & MV/m \\
 RF System Efficiency  & $\eta_{AC\rightarrow RF}$ & 38 & \% \\ 
 Repetition rate     & $f_{rep}$ & 100 or 150 & Hz \\
\hline
\itemheader{Modulator} \\
\hline
 Efficiency       & $\eta_{mod}$ & 75  & \% \\
\hline
\itemheader{Klystron} \\
\hline
 Klystron Peak Power &  & 75 & MW \\ 
 Klystron Pulse Length &  & 1.5 & $\mu$s \\ 
 Efficiency      & $\eta_{kly}$ & 60  & \% \\
\hline
\itemheader{Pulse Compressor} \\
\hline
Pulse compression method   &  & \multicolumn{2}{c|}{2 mode 4/4 DLDS} \\ 
Pulse compression power gain & & 3.4 & \\ 
Efficiency      & $\eta_{compr}$ & 85  & \% \\
\hline
\itemheader{Accelerating Structure} \\
\hline
Structure type    &  & \multicolumn{2}{c|}{RDDS $2\pi/3$ mode} \\ 
Structure length   & $L_s$   & 1.80 & m \\ 
Number of cells     &    & 206 & \\
Average iris radius    & $\avr{a/\lambda}$   & 0.18 & \\ 
Attenuation parameter & $\tau$   & 0.47 &  \\ 
Shunt impedance   & $r_s$ & 90 & M$\Omega$/m \\ 
Fill Time    & $T_f$  & 103 & ns \\ 
Q-factor        & $Q$    & 7800 & \\ 
\hline 
\end{tabular}
\end{center}
\end{table} 

\subsubsection{Main Linac}
 The acceleration scheme of the main linac is a conventional one:
The electric power from commercial line is converted to a high-voltage
(a few hundred kilo-volts), short (a few microseconds) pulse
by klystron modulators. This pulse is converted into microwave
by high-power klystrons and is led to normal-conducting accelerating 
structures. 
What is not conventional is that a very high accelerating gradient is
required to make the whole system reasonably short.
Generally speaking, a higher accelerating frequency of microwave is better 
for higher gradient but is more difficult technologically.
 There are at present two possible choices of the
main accelerating frequency, X-band (11.424~GHz) and C-band (5.712~GHz),
both being higher than conventional frequencies for linacs.
The latter is considered to be a backup scheme in case
the X-band R\&D would delay or fail.

The development of high-power klystrons is going well but 
the technological limit of the klystron peak power is far below the
value needed to reach the desired accelerating gradient (over 50~MV/m for
X-band and over 30~MV/m for C-band).
On the other hand it is relatively easy to obtain a long klystron pulse.
Therefore, one should compress the pulse to obtain a higher peak power with shorter
length.
The pulse compression scheme is different between X- and C-band designs.
The X-band design adopts the DLDS
(Delay Line Distributed System) as the effective pulse compression method.
The output microwaves (1.5~$\mu$sec long) from 8 klystrons are combined
and cut into four in time. Each time slice is delivered to different
accelerating structures upstream. The C-band design adopts 
a disk-loaded structure made of 3-cell coupled-cavity. The power efficiency is
lower than the DLDS but the system is much more compact.
 In the following we shall describe the X-band design.
The parameters related to the X-band RF system are summarized in Table~\ref{tab:RFparam}.

\subsection{Overview of JLC Parameters}
\label{headers_movv_param}
The latest parameter sets are available 
in the ISG (International Study Group) Progress Report\cite{ISG}.
Here, we shall briefly summarize the major points and add some more
detailed description on the issues related to the beam properties
at the collision point.

\subsubsection{Standard Parameter Sets}
  During the ISG study we have put emphasis on the cost and power
minimization and the relaxation of the tolerances against various
errors. As a result we decided not to give one single parameter set
but to give a range of parameters in the form of three different sets
{\bf A}, {\bf B}, and {\bf C}. These do not mean three different designs 
but three different operation modes of the same machine. 
Basically, {\bf A} adopts a low-current and small-emittance beam and
{\bf C} a high-current and large-emittance (i.e., accepts larger
emittance growth). We demand that any of
these operation modes (actually continuously from {\bf A} to {\bf C})
can be realized. For example the injector system should be able to deliver the highest
current assigned for {\bf C} and the bunch compressors can produce
the shortest bunch for {\bf A}.

Table~\ref{JLCparameters} shows the parameter sets {\bf A}, {\bf B}, and {\bf C}
at the center-of-mass energy $W_{cm}\simeq 500$~GeV.\footnote{
Since the three sets of parameters refer to the same machine length with
different loading, the center-of-mass energies are not exactly the same.
Several typos in \cite{ISG} have been fixed. There are slight differences 
from \cite{ISG} in the
number of beamstrahlung photons, luminosity, etc., 
because here we used computer simulations
of the beam-beam interaction for beamstrahlung, pinch-enhancement of the 
luminosity, etc.,
instead of using simplified analytic formulas.}

\medskip
The luminosity values in Table~\ref{JLCparameters} do not include the crossing angle
at the collision point. To avoid background events we are thinking of a
(full) crossing angle $\phi_c=7$~mrad, which will cause a luminosity 
reduction of about a factor 0.6. 
This reduction can be avoided by introducing the so-called crab
crossing.

\medskip
The SLC has provided a polarized electron beam ($\sim$80\%). This will also
be possible in JLC. The polarized electron gun for multi-bunch operation
is not ready yet but is expected to be feasible by the time of JLC completion.
The depolarization by the beam-beam interaction will be a few percent.

\subsubsection{Upgrade of Luminosity}
 When the machine is well tuned after several years of operation,
we may hope that the machine can be operated at a high current like {\bf C}
with a small emittance like {\bf A} giving a very high luminosity. This
parameter set is not consistent in that (1) the beamstrahlung would be too strong
and (2) the alignment tolerance too tight. However, we can overcome these
difficulties if we split the charge of each bunch into two bunches.
We cannot change the total train length because of the pulse compression 
system already built.
Thus, we have to halve the distance between bunches keeping the
total train length. Thus, we come to the
parameter set {\bf X} shown in Table~\ref{JLCparameters}. The changes
from {\bf A} to {\bf X} are to
\begin{itemize} \itemsep 0mm
\item  Increase the number of bunches from 95 to 190 and
    halve the bunch distance to 1.4~nsec. (To do so an absolute 
    constraint in the design stage is that the RF frequencies lower
    than 714~MHz must not be used.)
\item  Set the bunch charge to 0.55$\times 10^{10}$ so that the total
    charge in a train is the same as in {\bf C}.
\item  Improve the vertical emittance from the damping ring
    by a factor 2/3 and keep the same emittance blowup ratio 
    in the main linac as in {\bf A}
    (absolute blowup is smaller).
\item  Slightly shorten the bunch length.
\item  Slightly improve the beta functions at the IP, which is
    expected to be feasible at energies much lower than the highest
    design energy (1 to 1.5~TeV).
\end{itemize}
The beamstrahlung and alignment tolerance of accelerating structures
are better than in the standard sets owing to the greatly reduced bunch charge.
Thus, a further upgrade of luminosity will be possible if we can increase
the bunch charge back to that in {\bf A}, resulting in the parameter set {\bf Y}.
(Note that the linac must slightly be lengthened to achieve {\bf Y} due to the
lower loaded gradient.)
However, this parameter set demands very high beamloading everywhere in the system.
Also, a higher production rate of positrons (factor 1.27 over {\bf C})
is required.

\renewcommand{\itemheader}[1]{\multicolumn{8}{|l|}{{\bf #1}}}
\begin{table}
\begin{center}\begin{minipage}{\figurewidth}
\caption{\sl Parameters for the JLC at $E_{CM}\simeq$500~GeV \label{JLCparameters}}
\end{minipage}\end{center}
\medskip
\begin{small}\begin{center}
\begin{tabular}{|l|l|ccc|cc|l|} 
\hline
 & & A & B & C & X & Y & \\ 
\hline 
\itemheader{Beam parameters} \\
\hline 
 Center-of-mass energy & $E_{CM}$ & 535 & 515 & 500 & 497 & 501 & GeV \\ 
 Repetition rate & $f_{rep}$  & \multcole{150} & Hz \\ 
 Number of particles per bunch & $N$ 
    & 0.75 & 0.95 & 1.10 & 0.55 & 0.70 & $10^{10}$ \\ 
 Number of bunches/RF Pulse & $n_b$ & \multcolc{95} & \multcolb{190} & \\ 
 Bunch separation & $t_b$  & \multcolc{2.8} & \multcolb{1.4} & ns \\ 
 R.m.s. bunch length & $\sigma_z$   & 90 & 120 & 145 & 80 & 80 & $\mu$m \\ 
 Normalized emittance at DR exit & $\gamma\varepsilon_x$ 
 & \multcolc{300} & \multcolb{300} & $10^{-8}$m$\cdot$rad \\ 
 & $\gamma\varepsilon_y$ 
 & \multcolc{3.0} & \multcolb{2.0} & $10^{-8}$m$\cdot$rad \\ 
\hline
\itemheader{Main Linac}\\
\hline
 Effective Gradient$^{1)}$ & $G_{eff}$ 
   & 59.7 & 56.7 & 54.5 & 54.2 & 50.2 & MV/m \\
 Power/Beam & $P_B$  & 4.58 & 5.58 & 6.28 & 6.24 & 7.99 & MW \\
 Average rf phase & $\phi_{rf}$  & 10.6 & 11.7 & 13.0 & &  & deg. \\ 
 Linac Tolerances & $y_{c}$ & 16.1 & 15.2 & 14.6 & 18. & 14. & $\mu$m \\ 
 \hline 
 Number of DLDS nonets   &     & \multcold{23} & 25 & \\ 
 Number of structures per linac  &   & \multcold{2484} & 2700 & \\ 
 Number of klystrons per linac & & \multcold{1656} & 1800  & \\ 
 Active linac length &  & \multcold{4.47} & 4.86  & km \\
 Linac length &     & \multcold{5.06} & 5.50  & km \\
 Total AC power & $P_{AC}$  & \multcold{118} & 128 & km \\ 
\hline
\itemheader{IP Parameters}\\
\hline
 Normalized emittance at IP & $\gamma\varepsilon_x$ & 
     400 & 450 &  500 & 400 & 400 & $10^{-8}$m$\cdot$rad \\ 
 & $\gamma\varepsilon_y$ & 
    6.0 & 10 & 14 & 4.0 & 4.0  & $10^{-8}$m$\cdot$rad \\
 Beta function at IP & $\beta_x$
   & 10 & 12 & 13 & 7 & 7  & mm \\ 
 & $\beta_y$ & 0.10 & 0.12 & 0.20 & 0.08 & 0.08 & mm \\
 R.m.s. beam size at IP & $\sigma_x$
   & 277 & 330 & 365 & 239 & 239  & nm \\ 
 &  $\sigma_y$ & 3.39 & 4.88 & 7.57 & 2.57 & 2.55  & nm \\ 
 Disruption parameter & $D_x$ 
   & 0.0940 & 0.117 & 0.136 & 0.0876 & 0.112 & \\
  & $D_y$ & 7.67 & 7.86 & 6.53 & 8.20 & 10.43 & \\
 Beamstrahlung param & $\avr{\Upsilon}$ 
    & 0.14 & 0.11 & 0.09 & 0.127 & 0.163 & \\
 Beamstrahlung energy loss & $\delta_B$ 
    & 4.42 & 4.09 & 3.82 & 3.49 & 5.22 & \% \\
 Number of photons per $e^-/e^+$ & $n_\gamma$ 
    & 1.10 & 1.20 & 1.26 & 0.941 & 1.19 & \\
 Nominal luminosity  & ${\cal L}_{00}$ & 6.82 & 6.41 & 4.98 & 11.15 & 18.20 &
  $10^{33}\mbox{cm}^{-2}\mbox{s}^{-1}$ \\
 Pinch Enhancement$^{2)}$ & $H_D$ 
    & 1.444 & 1.392 & 1.562 & 1.389 & 1.483 & \\
 Luminosity w/ IP dilutions & ${\cal L}$ 
   & 9.84 & 8.92 & 7.77 & 15.48 & 27.0 &
  $10^{33}\mbox{cm}^{-2}\mbox{s}^{-1}$ \\ 
\hline
\end{tabular}
\end{center}
\medskip
1) Effective gradient includes rf overhead (8\%) and 
average rf phase $\avr{\cos\phi_{rf}}$. \\
2) $H_D$ includes geometric reduction (hour-glass) and dynamic enhancement.
The focal points of the two beams are made separated to each other by about $1\sigma_z$
for higher $H_D$ ($\sim$10\%)
\end{small}
\end{table} 

\subsubsection{Scaling to Lower Energies}
\begin{figure}
\centerline{\epsfxsize=12cm \epsfbox{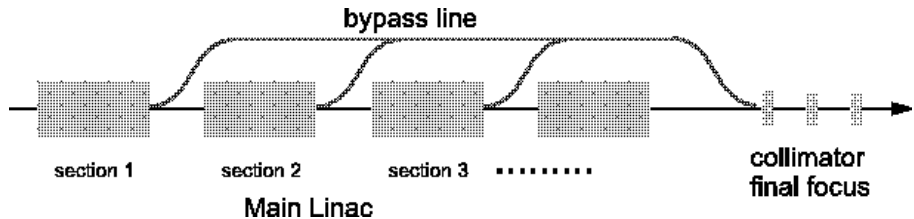}}
\caption{\sl Schematic layout of the bypass \label{fig:bypass}}
\end{figure}

For given normalized emittance $\gamma\epsilon_x$ and $\gamma\epsilon_y$,
the beam size at the IP scales as $1/\sqrt{W_{cm}}$ so that the luminosity
is simply proportional to $W_{cm}$. This is not correct, however, if one
lowers the energy by reducing the accelerating gradient. The beam loading
effect is characterized by the ratio of the beam-induced field to the
external field (by klystron). Therefore, when one reduces the accelerating
gradient, one has to reduce the charge proportionally for avoiding too
large beam loading. 
Thus, the luminosity
would scale as $W_{cm}^3$, which is not acceptable for low energy experiments.
For high luminosities we have to keep the gradient as high as possible. One might think
that one can accelerate the beam to the desired energy and let the accelerated
beam go through the empty accelerating cavities. This does not work 
because the beam-induced
field in the empty cavities will destroy the beam quality.
Thus, to get the luminosity scaling $\propto W_{cm}$, one has either
\begin{itemize} \itemsep 0mm
\item to perform low energy experiments during the construction stage
   when the unnecessary accelerating structures have not yet been installed
\item or to construct a bypass system such that the beam accelerated to
   a desired energy skips the rest of the linac and is directly transported
   to the final focus system. (See Figure~\ref{fig:bypass})
\end{itemize}
Obviously neither of these can be continuous: the construction stage
is discrete and the bypass line as well.
Thus, large changes of energy should be made by the above ways and
fine adjustment by changing the gradient.
Table~\ref{headers-movv-scale} 
shows the scaling of parameters 
(for each of {\bf A,B,C,X,Y}) under the two conditions: (a) constant gradient
and (b) constant active linac length. For example, if one wants $W_{cm}$=250~GeV
but only a bypass for 300~GeV is available, the luminosity will be
$(300/500)\times(250/300)^3$ times the luminosity at 500~GeV.
We do not have a design of the bypass system yet.

\begin{table}
\begin{center}\begin{minipage}{\figurewidth}
\caption{\sl\label{headers-movv-scale}
Energy scaling of parameters for (a) 
constant gradient and (b) constant active linac length.
The numbers are the power of $W_{cm}$.
The luminosity($\dagger$) is not rigorous due to the 
change of pinch enhancement factor.
Beamstrahlung energy loss and the number of beamstrahlung 
photons($\ddagger$) are not rigorous due to the change of 
$\Upsilon$.
}
\end{minipage}\end{center}
\vspace{6pt}
\begin{center}
\begin{tabular}{|lrr|}
 \hline
                  & (a)  & (b) \\
 \hline
 Luminosity ${\cal L}$ $^(\dagger)$  &  1  &  3 \\
 Bunch charge $N$         &  0  &  1 \\
 Bunch length $\sigma_z$     &  0  &  0 \\
 IP beam size $\sigma_x$, $\sigma_y$ &  -0.5 & -0.5 \\
 Disruption parameters $D_x$, $D_y$  & 0 & 1 \\
 Upsilon parameter $\Upsilon $  &  1.5  & 2.5 \\
 Beamstrahlung energy loss (relative) $^(\ddagger)$ & 2 & 4 \\
 Number of beamstrahlung photons $^(\ddagger)$ & 0.5 & 1.5 \\
 \hline
\end{tabular}
\end{center}
\end{table}

\subsubsection{Beam Energy Spread}
The beam energy spread before collision is normally a small fraction
of a percent and is much smaller than the effect of the beamstrahlung.
Nonetheless, the beam energy spread can be important, depending on the 
type of experiments, because the high-energy spectrum edge still remains
under the beamstrahlung but will be blurred by the energy spread of the
input beam.

The beam energy spread within a bunch comes from
\begin{itemize} \itemsep 0mm
\item effect of the short-range longitudinal wake (beam-induced field within
    a bunch)
\item energy spread before the main linac
\end{itemize}
The latter is small ($\leq 2\mbox{\%}\times \mbox{10~GeV}/E$)
except for experiments at very low energies.
We can minimize the former, which causes more energy loss
at the bunch tail than at the head, by placing the bunch
at the optimum phase of the sinusoidal curve of the accelerating field.
The phase angle is also used for other purpose (to control the
transverse instability) but a control of the energy spread can still
be possible by changing the phase in upstream and downstream parts
of the linac separately, since the instability is severer in the
low energy part.
 In general, an under-correction (small phase) causes
a doubly-peaked distribution with a short tail and an overcorrection causes
a sharp single peak with a long tail.
\begin{figure}
\centerline{\epsfxsize=12cm \epsfbox{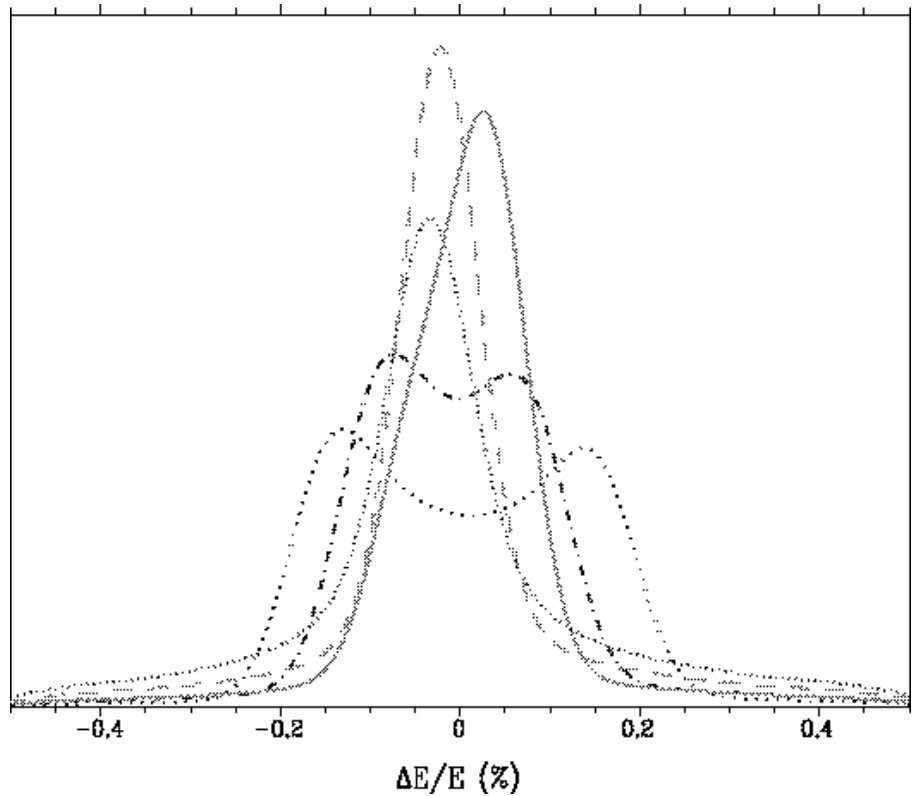}}
\begin{center}\begin{minipage}{\figurewidth}
\caption{An example of beam energy spread at 250~GeV changing the RF phase.
An initial spread of 1.5\% (rms) at 10~GeV has been added. \label{fig:EnergySpread}}
\end{minipage}\end{center}
\end{figure}
Figure~\ref{fig:EnergySpread} shows an example of the energy spectrum 
at various RF phases for the parameter set {\bf A} at the beam energy 250~GeV. 
The vertical scale is normalized for each curve.
An initial energy spread of 1.5\% (rms) at 10~GeV has been added in this figure. 

In addition to the above spread within a bunch,
there will be a fluctuation of average
energy from bunch to bunch and from pulse to pulse. The possible reasons
are:
\begin{itemize} \itemsep 0mm
\item miss-compensation of beam loading
\item jitter of the klystron power and phase
\item change of beam loading due to the fluctuation of the pulse charge
\item jitter of the longitudinal bunch position when the bunch enters the
linac.
\end{itemize}
We shall try to keep them within $\pm$0.1\%. The measurement of the
average beam energy over each pulse 
will be possible at the linac exit with an accuracy better than 0.1\%. 
It will also be possible to measure the energy of each bunch by using 
improved fast position monitors. Then the effects of the energy fluctuation
listed above (but not the spread like the wake effect)
can effectively be eliminated in the software level 
(i.e., the collision energy of each bunch is known although it fluctuates).

An energy feedback is possible
if the fluctuation is slow enough (lower than $\sim$10~Hz)
but the energy spread within a bunch cannot be corrected.

\subsubsection{Luminosity Spectrum}
The electrons(positrons) emit synchrotron radiation during the collision
due to the electromagnetic field (several kilo Tesla) created by the 
opposing beam. This phenomenon
is called `beamstrahlung'. Due to the beamstrahlung the particles loose
a few percent of their energies on the average. This causes a spread 
in the center-of-mass energy in addition to the initial beam energy spread 
from linacs.

\begin{figure}
\centerline{\epsfxsize=12cm \epsfbox{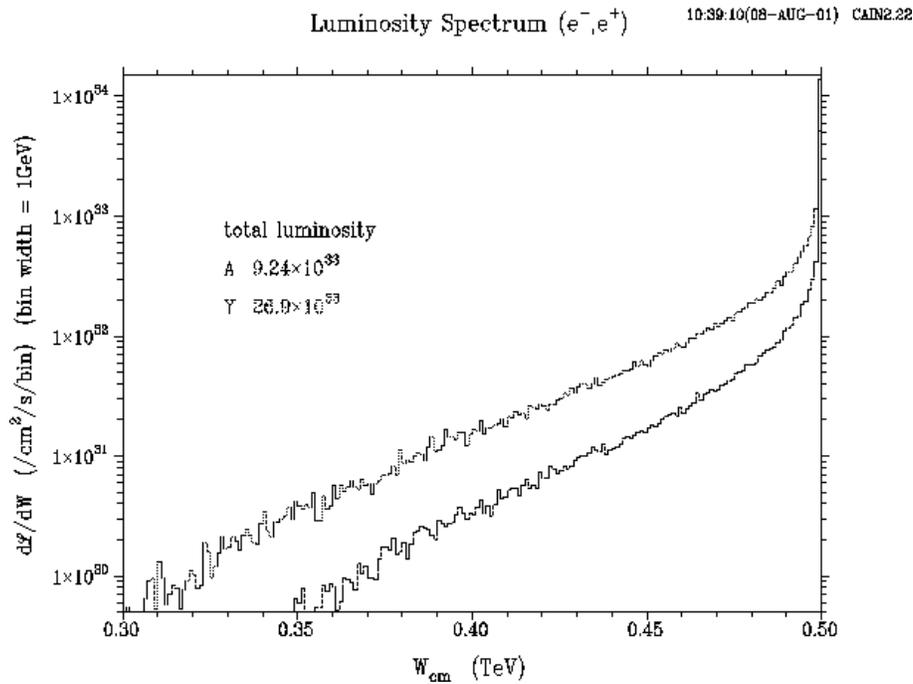}}
\begin{center}\begin{minipage}{\figurewidth}
\caption{\sl Luminosity spectrum at $W_{cm}$=500~GeV for the parameter set 
{\bf A} (thin solid) and {\bf Y}
(solid). The beam energy spread before collision is not included.
\label{fig:lumspec}}
\end{minipage}\end{center}
\end{figure}

\begin{figure}
\centerline{\epsfxsize=12cm \epsfbox{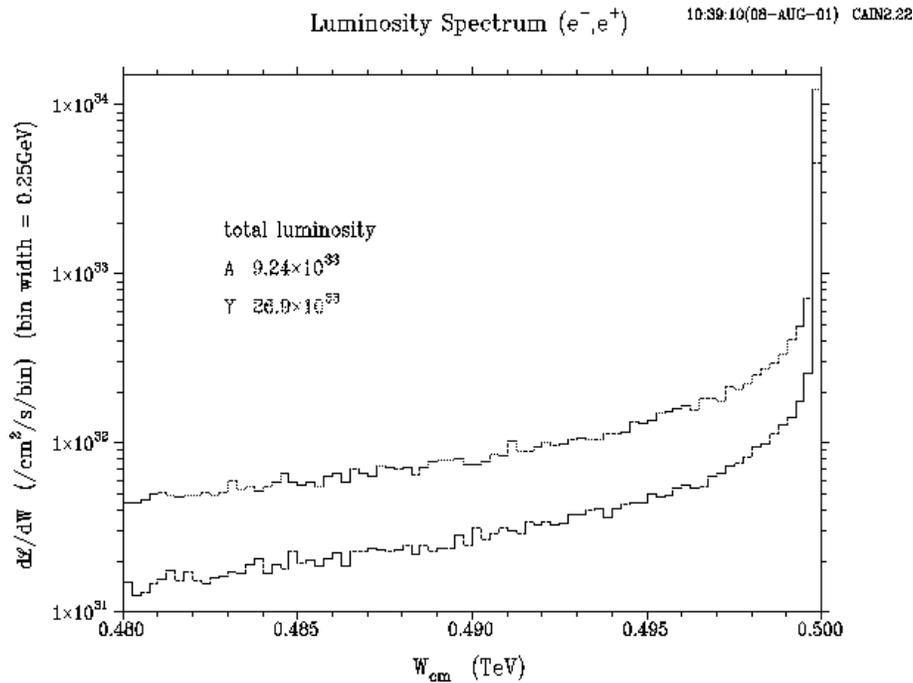}}
\begin{center}\begin{minipage}{\figurewidth}
\caption{\sl High-energy end of the luminosity spectrum at $W_{cm}$=500~GeV for the parameter set 
{\bf A} (thin solid) and {\bf Y}
(solid). The beam energy spread before collision is not included.
\label{fig:lumspec2}}
\end{minipage}\end{center}
\end{figure}

 The expected luminosity spectrum $d{\cal L}/dW_{cm}$
is plotted in Figure~\ref{fig:lumspec}
for {\bf A} and {\bf Y}. The high-energy end is shown in Figure~\ref{fig:lumspec2}.
For these figures $W_{cm}$ is adjusted to 500~GeV 
for the convenience of
comparison so that the luminosities are slightly different from those
in Table~\ref{JLCparameters}. (Linear $W_{cm}$ scaling rather than $W_{cm}^3$ is used here.)
The beam energy spread before collision is not included.
In Figure~\ref{fig:lumspec2} the highest bin ($499.75\leq W_{cm} \leq 500$~GeV)
contains 49\% (46\%) for {\bf A} ({\bf Y}) of the total luminosity.

 Both the luminosity and the average loss by beamstrahlung is
proportional to the bunch population squared. When the parameter set {\bf A}
is achieved and if a much narrower energy spread is desired,
one can reduce the beamstrahlung loss by a factor of four by making the
bunch population half but with pulse structure the same as in {\bf Y}.
The reduction of luminosity from {\bf A} is only a factor of two rather than four. 
This is technically even easier than {\bf A} (same loading, relaxed tolerances).


%% file: physhiggs/main.tex
\chapter{Higgs}
\label{chapter-physhiggs}
\newcommand{\ee}{\mbox{${\mathrm{e}}^+ {\mathrm{e}}^-$}}
\newcommand{\tautau}{\mbox{$\tau^+\tau^-$}}
\newcommand{\mm}{\mbox{$\mu^+\mu^-$}}
\newcommand{\ellell}{\mbox{$\ell^+\ell^-$}}
\newcommand{\qq}         {\mbox{$\mathrm{q}\bar{\mathrm{q}}$}}
\newcommand{\bb}         {\mbox{$\mathrm{b}\bar{\mathrm{b}}$}}
\newcommand{\toptop}         {\mbox{$\mathrm{t}\bar{\mathrm{t}}$}}
\newcommand{\cc}         {\mbox{$\mathrm{c}\bar{\mathrm{c}}$}}
\newcommand{\ff}         {\mbox{$\mathrm{f}\bar{\mathrm{f}}$}}
\newcommand{\nunu}       {\mbox{$\nu\bar{\nu}$}}
\newcommand{\mZ}         {\mbox{$m_{\mathrm{Z}}$}}
\newcommand{\mH}         {\mbox{$m_{\mathrm{H}}$}}
\newcommand{\mh}         {\mbox{$m_{\mathrm{h}}$}}
\newcommand{\mA}         {\mbox{$m_{\mathrm{A}}$}}
\newcommand{\czcz}{\mbox{$\chi^0_1\chi^0_1$}}
\newcommand{\cz}{\mbox{$\chi^{0}$}}
\newcommand{\co}{\mbox{${\tilde{\chi}_1^0}$}}
\newcommand{\ct}{\mbox{${\tilde{\chi}_2^0}$}}
\newcommand{\coct}{\mbox{$\chi^0_1\chi^0_2$}}
\newcommand{\ctcoz}{\ct\ra\Zs}
\newcommand{\ctcog}{$\ct\ra\co\gamma$}
\newcommand{\hczcz}{\ho\ra\co\co}
\newcommand{\hcoct}{\ho\ra\co\ct}
\newcommand {\Ho}        {\mbox{$\mathrm{H}^{0}$}}
\newcommand {\Hpm}       {\mbox{$\mathrm{H}^{\pm}$}}
\newcommand {\Ao}        {\mbox{$\mathrm{A}^{0}$}}
\newcommand {\ho}        {\mbox{$\mathrm{h}^{0}$}}
\newcommand {\Zo}        {\mbox{$\mathrm{Z}^{0}$}}
\newcommand{\Zs}         {\mbox{${\mathrm{Z}}^{*}$}}
\newcommand{\Zgs}        {\mbox{$\mathrm{(Z/\gamma)}^{*}$}}
\newcommand {\Wpm}       {\mbox{$\mathrm{W}^{\pm}$}}
\newcommand{\MH}{M_{\mathrm H}}
\newcommand{\MZ}{M_{\mathrm Z}}
\newcommand{\qqbar}{{\mathrm q}\bar{\mathrm q}}
\newcommand{\nn}{\nu \bar{\nu}}
\newcommand{\gaga}       {\mbox{$\gamma\gamma$}}
\newcommand{\WW}         {\mbox{$\mathrm{W}^+\mathrm{W}^-$}}
\newcommand{\WstarWstar} {\mbox{$\mathrm{W}^{+*}\mathrm{W}^{-*}$}}
\newcommand{\bquark}{\mbox{${\rm b}$}}
\newcommand{\electron}{\mbox{${\rm e}$}}
\newcommand{\epem}{\mbox{$\electron^{+}\electron^{-}$}}
\newcommand{\muon}{\mbox{$\mu$}}
\newcommand{\micron}{\mbox{$\mu{\rm m}$}}
\newcommand{\ra}        {\mbox{$\rightarrow$}}   
\section{Introduction}

We report Higgs physics at JLC
based on the recent progress in the experimental
and theoretical studies.
We focus on the discoveries and measurements 
expected to be achieved at the first phase 
of the JLC project $\sqrt{s}=$250--500 GeV, especially at the early stage
of low energies at $\sqrt{s}\approx 300$ GeV.
The key measurements and the problems to be solved before the experiment
are discussed. 
We first start from the astonishing Higgs discovery sensitivity at JLC,
followed by discussions how to define the quantum number for the found new 
particle in order to confirm it is indeed Higgs particle. Next step is to
measure the Gauge coupling. The measurements of Yukawa couplings with
leptons and quark flavours including top are one of the
highlights of the Higgs study. 
The experimental studies are based on detector performance expected in 
the JLC-I detector. 
Based on the results obtained so far in the simulation studies, 
impacts on the physics and models are discussed from theoretical view points.
\vspace{12pt}
The studies of the Higgs boson~\cite{Higgs} is the key to open the
future generation of the particle physics. 
Our understanding of nature based on the local gauge principle demands
a scheme to make gauge boson and quark/leptons massive. The origin of the
mass is either fundamental scaler particle, Higgs boson, or new interaction
such as technicolor. 

The existence of the Higgs boson
is of course the first question to be answered by experiments.
If it is found to exist,
the next essential questions are how many Higgs bosons, 
and how the relations
are between couplings and masses of quark/lepton/gauge-bosons.
The Standard Model (SM)~\cite{sm} 
assumes a single Higgs field doublet, hence the
single physical state $\Ho$ generates masses of quark/lepton and Z/W with
completely defined couplings except for the Higgs boson mass itself.
There are also varieties of models in extensions of the SM in the Higgs 
sectors, and nobody knows the answer to the questions.

As one knows the SM Higgs sector has various problematic issues inside 
such as the fine-tuning problem.
The Higgs couplings and their evolutions are significantly
sensitive to the fundamental structure of interaction and contents 
of particles up to GUT scale ($\sim 10^{16}$ GeV) via loop effects, 
which becomes one of
the strong motivations to introduce the supersymmetry (SUSY).
Even in the Standard Model, the mass of the Higgs has both upper and
lower values by vacuum stability and self-coupling evolution depending on
the cutoff scale. 
Hence the Higgs study is a strong tool to look into physics up to GUT scale.

Currently running B-factories in Japan and US measure the CKM matrix 
parameters in order to investigate the origin of CP violation. 
Also neutrino physics studied at SuperKamiokande together with
K2K project in Japan look into the mass and mixing in the lepton sector.
The mass matrix in quark and lepton sectors is an obvious key,
hence the origin of the mass of the fundamental particles investigated
at JLC, together with
the results to be obtained at B-factory and SuperKamiokande/K2K, would
give us a comprehensive views for the origin of the flavour mixing, 
CP violation, and a road to understand why 3 generations exist in our Nature.

Possible scenario for the first discovery depends on the Higgs in models.
At LEP in 2000, an indication of possible signal has already obtained 
at a mass of 115 GeV~\cite{higgs_lep}.
If this signature indeed comes from a Higgs boson, 
it would be confirmed at Tevatron~\cite{RUN-II}
and/or LHC~\cite{LHC} in years around 2007 since it is the SM-like Higgs 
boson. 
So far the lower mass bound of the SM Higgs is
around 113 GeV and about 90 GeV for the Higgs in
the minimal supersymmetric extension of the SM (MSSM)~\cite{mssm} 
(h$^{0}$, A$^{0}$)
at the 95 \% CL from LEP-II~\cite{higgs_lep,LEP-II}.
Tevatron run2 in the following years is expected to be sensitive 
to discover the SM Higgs of its mass up to 120 GeV or more~\cite{RUN-II} 
provided the integrated luminosity exceeds a few fb$^{-1}$. 
At forthcoming LHC, there are varieties of channels in the SM and MSSM 
in which we have large discovery potential with significance 5--10$\sigma$
in wide range of the mass of the Higgs bosons~\cite{LHC}, 
while the detectable channels are limited due to
huge backgrounds in the $pp$ collisions and
also the production yield and detection efficiency could be much lower 
if nature is further beyond the MSSM.
Note that the experimental boundary in the extension of the SM in general
two Higgs field doublet models (2HDM) is much weaker~\cite{2hdm} 
than that of the SM or MSSM Higgs. 
For the charged Higgs boson predicted in the 2HDM
has been looked for only up to around 80 GeV~\cite{LEP-II}.

The Higgs discovery itself is significantly easier at LC than Tevatron/LHC,
provided that we establish JLC with luminosity of order of
10$^{34}$ cm$^{-2}$s$^{-1}$ at the beam energy high enough 
to produce Higgs boson kinematically, $\sqrt{s}>\mZ + \mH$.
We need only ``one day'' running to discover the SM-like Higgs, 
which is extremely different situation
from other experiments like LHC. 
Even for the SUSY models beyond the MSSM, 
there always exist the lower limits 
of the production cross-section for at least one CP-even Higgs.
Extremely less and well defined background at JLC 
compared to hadron colliders enables us to discover Higgs boson,
even if the cross-section is 10 times smaller than the lower limits
predicted by SUSY models.

The possible mass regions of the lightest Higgs boson differ model by model.
There are, however, strong indications that the mass is lower than 250 GeV
where the first phase of JLC project covers well without questions.
The indirect measurements of the electroweak parameters
at LEP/SLC/Tevatron with recent measurements of hadronic cross-section
at BEPC~\cite{BEPC} in China give an upper mass bound of about 215 GeV 
at the 95\% CL~\cite{EW} for the SM Higgs.
When we assume the MSSM, the mass must be lower than 140 GeV.
With an additional SU(2) singlet to the MSSM (NMSSM), the upper bound is 
about 140 GeV assuming the finite Higgs coupling 
up to GUT scale~\cite{okada}, 
and even in more general scheme in SUSY
the bound does not exceed 210 GeV~\cite{xmssm}.
Hence the observation of at least one Higgs is guaranteed 
unless our understanding of the nature is completely wrong.
One can say, in other words, it is really a ``big discovery'' when we 
find no Higgs at JLC phase-I. 

At the first phase of JLC, apparently the target Higgs mass is
from 100 GeV (just above the current LEP boundary) to 250 GeV. 
Our purpose of the
experiment at JLC is not only to judge the existence of the light Higgs
predicted in the SM or SUSY, while
we may have an opportunity to reject some of above models of
Higgs sectors only from the observed mass of the Higgs
as are discussed above.

The real Higgs physics starts after the discovery, in order 
to define the Lagrangian, and
to scope the physics up to GUT scale.
The important feature in JLC is its well defined background, and high purity
of the signal.
We expect to select more than 10$^{5}$ Higgs events in 3--5 years running 
with background contamination less than 20\%.
The JLC gives us the unique opportunity to 
measure the cross-section and the branching ratio
in percent order or even per-mill level in its error, which results in
the precise measurements of the Higgs gauge coupling,
Yukawa-coupling, and furthermore derivation of Higgs self-couplings 
in multi-Higgs production. All of these measurements can be made in model
independent fashion. And then, relations among the couplings 
predicted in models can be cross-checked in various measurements.
The scenario of the CP-even Higgs hunting/measurements at JLC is as follows;
the first we start from model independent measurements:
\begin{enumerate}
\setlength{\itemsep}{-3pt}
\item If Tevatron or LHC missed it, JLC discovers Higgs. 
\item Mass measurement.
\item Cross-section measurement.
\item CP, spin determination by the angular distributions, 
      energy scan with/without polarized beam option.
\item Establishment of ZZH coupling.
\item Gauge coupling (F$_{\rm{ZZH}}$, F$_{\rm{WWH}}$) measurements.
\item Yukawa coupling ($\lambda_{\tau}$, $\lambda_{b}$, $\lambda_{c}$, 
$\lambda_{t}$) measurements.
\item Natural width ($\Gamma_{H}$) measurement.
\item Self-coupling ($\lambda$) measurement.
\end{enumerate}
Even after the discovery of a Higgs particle, many unexpected discovery
which means further beyond SM/SUSY might yet wait for us at JLC.
In the SM and SUSY models, there always have numbers of relations
in couplings. We ask; if coupling to Z and W relates as expected from
SU(2)$\times$U(1) symmetry breaking?;
if ratio of Yukawa coupling for tau and b 
is really the ratio of these masses?; if all 3 generations of 
up (down) type quarks have the same coupling normalized with mass?;
and if CP is conserved in Higgs sector?

These precisely measured values are also used to severely check  
the SM, SUSY models and others.
In the SM, the couplings are completely determined by the measured SM
parameters. Once we put the mass of the Higgs, all are calculable.
Thanks to the complete initial state of $\ee$ collision at JLC,
the cross-section for the SM is theoretically known within less 
than 1\% accuracy.
Yukawa coupling for tau has been constrained within per-mill.
Yukawa coupling for b-decay is known within a few \%, 
and the accuracy will increase furthermore by data from the B-factories.
If one measures the couplings within percent level, we can severely check
the SM at the level of the virtual effect from other Higgs or SUSY particles.
We also translate the precisely measured model-independent values to
other model parameters such as tan$\beta$ or $\alpha$ for MSSM as an example.
We check the internal consistency (universality) in the translated 
parameters from various measurements. 
If we have other Higgs mass lighter than TeV, we would have such
effects at percent level or more. In other words we can measure the mass of 
the other Higgs depending on the models. 
At next phase of JLC, LHC or even already in JLC phase-I,
we look into the predicted 
mass, and can determine the final Lagrangian of the Higgs sector. 
The couplings and other physics parameters are finally examined in the models 
of the Grand Unification using renormalization group evolution. 
Precise measurements of Higgs properties at JLC are essential tool to give 
definite answers to physics models and to determine the fundamental 
structure of the interaction in nature up to GUT scale.

\vspace*{0.4cm}

This report discuss the recent Higgs studies at JLC which have been
made in ACFA Higgs working group~\cite{acfahiggs,acfa-taipei} and
as a regional studies for the world-wide LC 
Higgs cooperation~\cite{wwlchiggs}.
The studies in experimental feasibility does depend significantly
on the accelerator and detector performance. Hence in this report
as a first priority we made following direction:
\begin{itemize}
\setlength{\itemsep}{-3pt}
\item Studies based on recent JLC-I~\cite{JLC-I}  
detector design (see Part~\ref{part-detector})
as a model detector, using the fast~\cite{JSF} and full~\cite{JIM}
detector simulation softwares (see Chapter~\ref{chapter-detsim}).
\item Recent accelerator parameters, such as 
beamstrahlung~\cite{beam-beam,UEDA}.
\item Concentrate on the JLC phase-I energy (250--500 GeV), especially
for the energy at early stage ($\sqrt{s} \le 350$ GeV). 
\item Studies on light Higgs ($\mH<210$ GeV) predicted in the SM and 
SUSY models, especially for 120--140 GeV region where the variety of
decay branch can be simultaneously studied.
\item Emphasize on the measurements independent to models, since nobody
knows the true model.
\item Simulation tests in the realistic experimental 
environment~\cite{SY-KANZAKI}.
\end{itemize}

The important but missing ingredients in this report
for the Higgs studies at JLC program
such as multi-Higgs production, Heavy Higgs, 
measurements of coupling with SUSY particles, 
precise measurement of top-Yukawa coupling,
self-coupling and its running measurements, are only simply described.
The direction and feasibility of these studies depends on the physics results
at the initial stage of the JLC phase-I. 
The energy and luminosity should be optimized for these studies;
either quick upgrade of the energy, or concentrate on the lower energy and 
accumulate the integrated luminosity furthermore.
The studies are open for the next ACFA report.

\section{Theoretical Overview}

In the minimal SM, we introduce only one Higgs doublet field. The 
Higgs potential is given by
\begin{equation} 
V=-\mu^2|H|^2 + \lambda|H|^4.
\end{equation}
Through the electroweak symmetry breaking, the gauge bosons, fermions
and the Higgs boson receive their masses. At the tree level, the mass 
formulae are given by $m_W= \frac{1}{2}g v$ for the $W$ boson,
$m_Z= \frac{1}{2}\sqrt{g+g'}v$ for the $Z$ boson,
$m_f= \frac{y_f}{\sqrt2}v$ for fermions and $m_H= \sqrt{\lambda}v$,
where $v (\simeq $ 246 GeV) is the Higgs vacuum expectation value,
$g$ and $g'$ are SU(2) and U(1) gauge coupling constants, and 
$y_f$ is the Yukawa coupling constant for the fermion $f$. These 
expressions imply that a particle mass is determined by 
the strength of its interaction to the Higgs field, therefore the 
measurement of the coupling constants related to the Higgs boson
is an important check of the mass generation mechanism  
in the SM.

In particular, the formula $m_H= \sqrt{\lambda}v$ suggests that
the mass of the Higgs boson reflects the strength of 
the electroweak symmetry breaking dynamics. The heavy Higgs boson 
implies the strongly-interacting dynamics and the light Higgs boson
is consistent to the weakly interacting scenario such as
grand unified theory (GUT) or supersymmetric (SUSY) unified 
models. Although the Higgs boson mass is a free parameter within the 
minimal SM, we can determine the upper and lower bound of its mass,
if we require that the SM is valid up to some cut-off scale, beyond 
which the SM should be replaced by a more fundamental theory.  
If the cut-off scale is taken to be the Planck scale $(\sim 10^{19})$
GeV, the possible mass range is 135 - 180 GeV. For a larger mass 
the Higgs self-coupling constant blows up below the Planck scale,
and for a lower mass the vacuum stability is not guaranteed.  

There are many possibilities to extend the Higgs sector of the minimal 
SM. Because the $\rho$ parameter determined from the electroweak 
measurements is close to unity, the dominant contribution 
to the electroweak breaking should come from weak-doublet fields.
Two Higgs doublet model is one of the simplest extensions. In this case,
there are two CP-even Higgs bosons $(h, H)$,
a CP-odd Higgs boson $(A)$ and a pair of charged Higgs bosons 
$(H^\pm$). If we require the two Higgs doublet model is valid 
up to the cut-off scale, the mass range of the 
lighter CP-even Higgs boson $(h)$ is determined in a similar
way to the SM case. For the case that the Planck scale is
the cut-off scale, the upper-bound is 180 GeV, just like the SM. 
The lower bound can be smaller than the corresponding mass bound 
in the SM. In particular, the lower-bound is about 100 GeV in the case 
that the only one SM-like Higgs boson becomes light compared to
other Higgs states.
 
Since the present electroweak measurements are precise enough to be 
sensitive to virtual loop effects, we can put a useful bound  
on the Higgs boson mass, independent of the theoretical assumption
on the validity of the theory up to some high energy scale. 
Within the minimal SM, the 95 \% upper bound on the Higgs boson mass
is about 210 GeV\cite{EW} using the direct measurements of 
the $W$ boson and top quark mass in  addition to various electroweak data
at LEP and SLD experiments. For more general model, for example 
in the two Higgs doublet model, this kind of strong bound is not 
obtained only from the electroweak data.

Since the end of 1970's, it has been known that supersymmetry (SUSY) 
can be a solution of the naturalness problem in the SM. If we consider
the cut-off scale of the SM is the Planck or the GUT scale, we need 
extremely precise fine-tuning in the renormalization of the
Higgs mass term to keep the weak scale much smaller
than the cut-off scale. There is no such problem in SUSY models 
because the problematic quadratic divergence in the scalar
mass renormalization is absent. Motivated by this observation,
SUSY extensions of the SM and the GUT were proposed, and many
phenomenological studies have been done. For the last ten years,
SUSY models has become the most promising candidate of physics beyond 
the SM, because the gauge coupling constants determined at LEP and SLD 
experiments turned out to be consistent with the prediction of the SUSY
GUT scenario.

The simplest SUSY extension of the SM is called the minimal
supersymmetric standard model (MSSM). The Higgs sector of the
MSSM is the type-II two Higgs doublet model, where Higgs doublet fields
$H_1$ and $H_2$ are introduced for the down-type quark/lepton Yukawa coupling
and the up-type quark Yukawa coupling, respectively. We define two 
angle variables to parameterize the Higgs sector. One is the vacuum 
mixing angle given by $\tan{\beta}\equiv <H_2^0>/<H_1^0>$.
The other is the mixing angel between two CP-even Higgs bosons
$h$ and $H$, $(m_h<m_H)$.  
\begin{eqnarray}
Re H_1^0&=&(v \cos{\beta} - h \sin{\alpha} + H \cos{\alpha})/\sqrt{2},\\ 
Re H_2^0&=&(v \sin{\beta} + h \cos{\alpha} + H \sin{\alpha})/\sqrt{2},
\end{eqnarray}  
where $H_1^0$ and $H_2^0$ are the neutral components of two Higgs 
doublet fields.
The ratio of various tree level coupling constants in the MSSM and
those in the SM are listed in table \ref{tab:mssmcpl}.
\begin{table}
\begin{center}\begin{minipage}{\figurewidth}
\caption[]{\sl
The coupling constant in the MSSM normalized by the
corresponding coupling constant in the SM. In the Yukawa coupling
for the CP-odd Higgs boson $(A)$, the fermion current
is of the pseudo-scalar type.
\label{tab:mssmcpl}}
\end{minipage}\end{center}
\begin{center}
\begin{tabular}{|c|c|c|c|c|c|c|c|c|}
\hline
$hWW, hZZ$&$HWW, HZZ$&$ht\bar{t}$&$hb\bar{b}, h\tau\bar{\tau}$
&$Ht\bar{t}$&$Hb\bar{b}, H\tau\bar{\tau}$&$At\bar{t}$
&$Ab\bar{b}, A\tau\bar{\tau}$\\
\hline
$\sin{(\beta-\alpha)}$&$\cos{(\beta-\alpha)}$
&$\frac{\cos{\alpha}}{\sin{\beta}}$
&$-\frac{\sin{\alpha}}{\cos{\beta}}$
&$\frac{\sin{\alpha}}{\sin{\beta}}$
&$\frac{\cos{\alpha}}{\cos{\beta}}$
&$\cot{\beta}$&$\tan{\beta}$\\
\hline
\end{tabular}
\end{center}
\end{table}

\begin{figure}
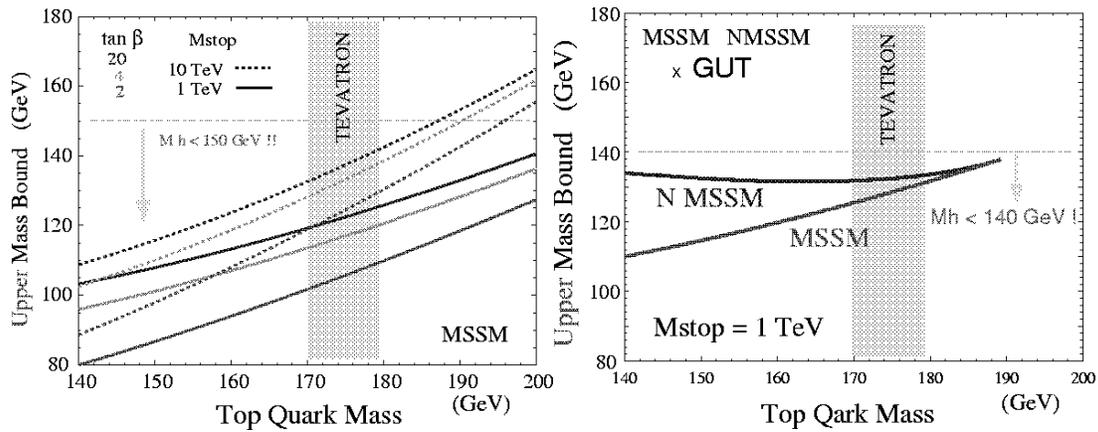

\centerline{
\epsfxsize=7.2cm \epsfbox{physhiggs/okada_limit.epsi}
\epsfxsize=7.2cm \epsfbox{physhiggs/okada_nmssm.epsi}
}
\begin{center}\begin{minipage}{\figurewidth}
\caption[]{\sl
The upper bounds of the mass of the lightest CP-even Higgs boson
in MSSM and NMSSM calculated by Y.~Okada et al~\cite{okada}.
(Left) Upper bounds as a function of the top quark mass in the MSSM, with
various MSSM parameters as shown in the plot. 
(Right) Upper bound for MSSM and NMSSM with additional constraints
of the finite Higgs coupling up to GUT scale. The top quark mass measured
at Tevatron is indicated by the grey bands.
\label{fig:okada1}}
\end{minipage}\end{center}
\end{figure}

In the MSSM, we can derive the upper-bound of the the lightest CP-even 
Higgs boson mass $(m_h)$ without reference to the cut-off scale of the theory.
This is because the self-coupling of the Higgs field is completely 
determined by the gauge coupling constants at the tree level. 
Although $h$ has to be lighter than Z$^{0}$ boson at the
tree level, contribution from the loop effects by the top quark and stop 
squark can extend the possible mass region~\cite{yanagida}.
Taking into account the top and stop one-loop corrections,  
$m_h$ is given by 
\begin{equation}
m_h^2 \leq m_Z^2 \cos^2{2\beta}+\frac{3}{2\pi^2}\frac{m_t^4}{v^2}
ln{\frac{m_{stop}^2}{{m_t}^2}},
\end{equation}
where we assume that two stop mass states have the same mass.
More precise formula is available in the literature. It is concluded 
that $m_h$ is bounded by about 130 GeV, even if we take the stop mass 
to be a few TeV. The upper-bound of $m_h$  is shown in Fig \ref{fig:okada1}. 

\begin{figure}
\centerline{
\epsfxsize=3.5in \epsfbox{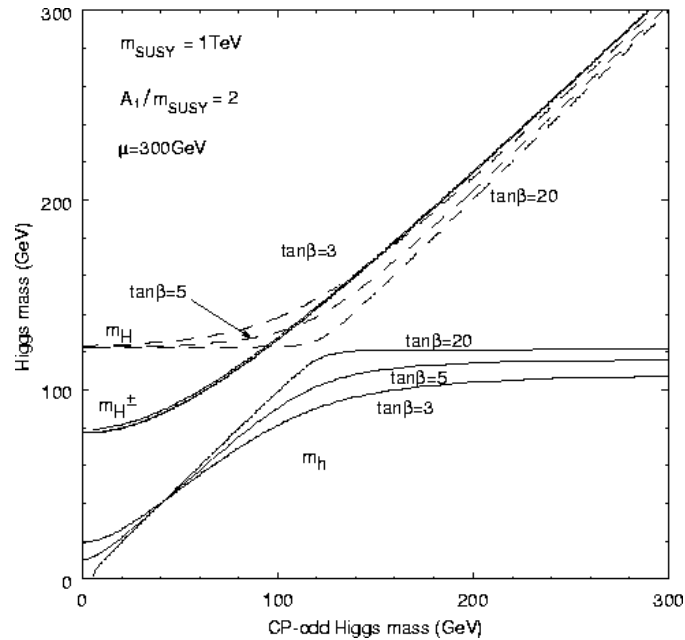}
}
\vspace{10pt}
\begin{center}\begin{minipage}{\figurewidth}
\caption{\sl
Higgs boson masses as a function of the CP-odd Higgs boson mass 
in MSSM.\label{fig:mssmmass}}
\end{minipage}\end{center}
\end{figure}

The radiative correction can also change the mass formulas of the heaver
Higgs bosons. The Higgs potential is parametrized by 
three mass parameters, gauge coupling constants and the parameters of the
top and stop sector through the one-loop correction. The independent 
parameters can be taken as $\tan{\beta}$, and one of Higgs boson masses, 
which is usually taken as the CP-odd Higgs boson mass $(m_A)$ and the top 
and the stop masses.  More precisely, the parameters appearing in the 
formula of the radiative correction 
are not just one stop mass, but two stop masses, 
trilinear coupling constant for stop sector $(A_t)$, the higgsino mass 
parameter $\mu$, and sbottom masses, 
etc. (If we use more precise formula, we need to specify more input
parameters.) Once these parameters are specified, we can calculate the mass
and the mixing of the Higgs sector. The Higgs boson masses are shown as 
a function of the CP-odd Higgs mass in Figure~\ref{fig:mssmmass}.
It is important to distinguish two regions in this figure.
Namely, when $m_A$ is much larger than 150 GeV, $H, A$ and $H^{\pm}$
states become approximately degenerate and the mass of $h$ approaches
to its upper-bound value for each $\tan{\beta}$. This limit is
called the decoupling limit. In this limit, $h$ has properties similar 
to the SM Higgs boson, and the coupling of the heavy Higgs bosons to
two gauge-boson states is suppressed. On the other hand, if $m_A$
is less than 150 GeV, the lightest CP-even Higgs boson has sizable
components of the SM Higgs field and the other doublet field.

Besides the Higgs potential, there is a case where radiative correction
becomes potentially important for the MSSM Higgs boson phenomenology. 
For a large value of $\tan{\beta}$, SUSY correction can generate  
contributions to the bottom-higgs Yukawa coupling which is not present 
at the tree level\cite{babu99}.  The top and bottom Yukawa couplings
with the neutral Higgs fields are given by
\begin{equation}
L_{Yukawa}=y_t\bar{t_L}t_R H_2^0 + y_b\bar{b_L}b_R H_1^0 +
\epsilon_b y_b\bar{b_L}b_R H_2^{0*} + h.c.
\nonumber
\end{equation} 
The $\epsilon_b$ term is induced by loop diagrams with internal 
sbottom-gluino and stop-chargino. The bottom mass is then 
expressed by $m_b=y_b(1+\epsilon_b\tan{\beta})v\cos{\beta}/\sqrt{2}$.
Although $\epsilon_b$ is typically $O(10^{-2})$,
the correction to $m_b$ enters with a combination of 
$\epsilon_b\tan{\beta}$ which can be close to $O(1)$ for a large
value of $\tan{\beta}$. 
Because of this correction, the ratio of 
$B(h \rightarrow \tau^+\tau^-)$ and $B(h \rightarrow b\bar{b})$ 
is modified to
\begin{equation}
R_{\tau \tau/bb}\equiv  \frac{B(h \rightarrow \tau^+\tau^-)}
{B(h \rightarrow b\bar{b})}=\left( \frac{1+\epsilon_b \tan{\beta}}
{1-\epsilon_b/\tan{\alpha}} \right)^2 R_{\tau \tau/bb}(SM),
\end{equation}
where $R_{\tau \tau/bb}(SM)$ is the same ratio evaluated in the SM.
$R_{\tau \tau/bb}$ is the same as $R_{\tau \tau/bb}(SM)$ if
the SUSY loop effect to the $b\bar{b}H_2^{0*}$ vertex is negligible.
Notice that in the decoupling limit with $m_A \rightarrow \infty$,
$\tan{\alpha}$ is approaching to $-1/\tan{\beta}$, so that the ratio 
reduces to the SM prediction.
In actual evaluation, however, the approach to the asymptotic 
form is slow for large $\tan{\beta}$, so that deviation
from the SM prediction can be sizable for $\tan{\beta}\gsim 30$.

In extended versions of SUSY model, the upper-bound of the lightest 
CP-even Higgs boson can be determined only if we require that any of
dimensionless coupling constants of the model does not blow up below some 
cut-off scale. For the SUSY model with an extra gauge singlet Higgs field,
called the next-to-minimal supersymmetric standard model (NMSSM),
the bound is a slightly larger than the upper-bound for the MSSM case. 
Because there is a new tree level contribution
to the Higgs mass formula, the maximum value corresponds to a lower 
value of $\tan{\beta}$, which is quite different from the MSSM case where
the Higgs mass becomes larger for large $\tan{\beta}$. 
The upper-bound for the lightest CP-even Higgs boson in the NMSSM 
is shown in Fig \ref{fig:okada1}.
In Ref.~\cite{xmssm} the upper-bound of the lightest CP-even Higgs 
boson was calculated for SUSY 
models with gauge-singlet or gauge-triplet Higgs field and the maximal
possible value was studied in those extensions of the MSSM. It was 
concluded that the mass bound can be as large as 210 GeV for a specific 
type model with a triplet-Higgs field for a stop mass of 1 TeV. The mass 
bound was also studied for the SUSY model with extra matter fields. In this 
model the upper-bound becomes larger due to loop corrections of extra matter
multiplets. If the extra fields have $\bar{5}+10+5+\bar{10}$ representations
in SU(5) GUT symmetry, the maximum value of the lightest CP-even Higgs 
boson mass becomes 180 GeV for the case that the squark mass is 1 TeV 
\cite{moro92}.   

As we show above, it is very likely that the scalar boson 
associated with the electroweak symmetry breaking exists 
below 200 GeV, as long as we take a scenario that the Higgs sector
remains weakly-interacting up to the GUT or the Planck scale, where
the unification of gauge interactions, or gauge and gravity 
interactions may take place. In particular, there is a strict 
theoretical mass bound for the lightest CP-even Higgs boson in
the MSSM. The precise determination on properties of 
a light Higgs boson is one of the most important tasks of the 
LC experiment. 

\section{Higgs Production, Decay, Background Processes}
\subsection{Higgs Production}

\begin{figure}
\centerline{
\epsfxsize=12.8cm \epsfbox{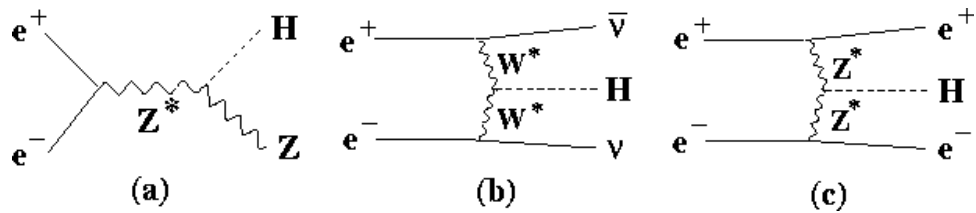}
}
\begin{center}\begin{minipage}{\figurewidth}
\caption[]{\sl
Diagrams of CP-even higgs production expected at JLC phase-I in
(a) Higgs-strahlung (Bijorken) $\ho\Zo$ production, 
(b) WW-fusion and (c) ZZ-fusion process.
\label{fig:gra_zh_ww}}
\end{minipage}\end{center}
\end{figure}

In the $\ee$ collision, 
the CP-even Higgs (h$^0$), either in SM or other models such as SUSY, 
can be produced via Higgsstrahlung (Bijorken) process, 
$\ee\ra\Zo ^{*} \ra\ho\Zo$, 
and WW(ZZ)-fusion process~\cite{wwfusion}, 
$\ee\ra\nn {\mathrm{W}}^* {\mathrm{W}}^* \ra\nn\ho$
($\ee\ra\ee {\mathrm{Z}}^* {\mathrm{Z}}^* \ra\ee\ho$)
as shown in Figure~\ref{fig:gra_zh_ww}.
The ZZH coupling generates Higgsstrahlung and ZZ-fusion processes, 
and WWH coupling generates WW-fusion process.
These processes yields final state of Higgs decay products and
a fermion pair either from the associated Z boson's decay 
or $\nn$ ($\ee$) in WW(ZZ)-fusion.
In general, the coupling between ZZH and WWH relates in SU(2)$\times$U(1)
symmetry breaking which is one of the experimental target to verify at 
the early stage of JLC phase-I.
The ZZ-fusion process is suppressed by one order of magnitude 
compared to WW-fusion, mainly due to ratio between neutral and charged 
currents, 16cos$^{4}\theta_{W}$. However the channel is important in 
electron-electron collider option at the
JLC where we have significantly less backgrounds similar to the signal
while one order of magnitude less luminosity is expected due to the repulsive
force between two electron beams.

The Higgsstrahlung process is dominant at lower energy.
In the Higgsstrahlung process, there are 4 modes 
according to the final state of the associated Z boson, 
namely $\qq$ ($\sim$70\%), $\nn$ ($\sim$20\%), $\tautau$ ($\sim$3\%), 
$\ee$ or $\mm$ ($\sim$6\%).
The signature of the Higgsstrahlung process 
is the invariant mass of $\qq$, lepton
pairs (or the missing mass due to $\nn$) around Z-pole, and invariant mass
of Higgs decay products to a mass of the Higgs. This process is cleanly
identified at JLC thanks to the well-defined initial states, and
hence, strong kinematic constraints are available such as recoil mass 
of the $\qq$ or leptons from Z to be equal to the invariant mass of 
the Higgs decay products.
The Higgsstrahlung process can be isolated from WW-fusion if we select 
H$\qq$, H$\tautau$ or H$\mm$.
In the final state of H$\nn$ and H$\ee$, both Higgsstrahlung and the 
fusion processes contribute with non-negligible interference. 
The WW-fusion process becomes 
important at high energy for light Higgs. 
In the process, in principle,
the kinematic boundary of the Higgs boson mass to be produced can be extended
up to the $\ee$ center-of-mass energy. This production has unique feature,
since it provides us the direct and precise measurement of the coupling 
of the Higgs with W-boson. 

One more important channel is the Higgs production radiated off
fermions with Yukawa coupling, e$^{+}$e$^{-}\ra\ff\ra\ff\ho$.
The channel is especially important at JLC to measure top-quark Yukawa 
coupling.

\begin{figure}
\centerline{
\epsfxsize=7.84cm \epsfbox{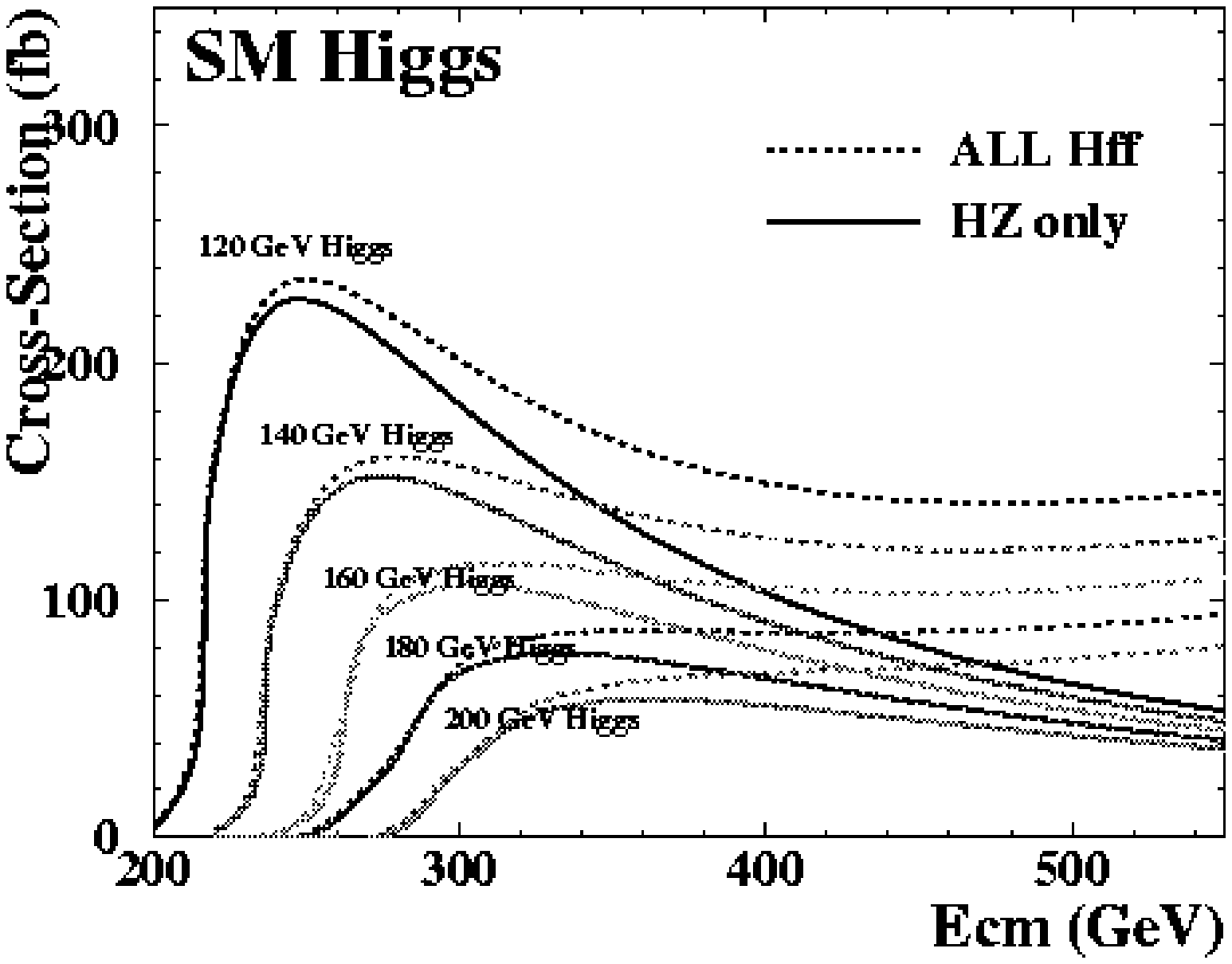}
\epsfxsize=7.84cm \epsfbox{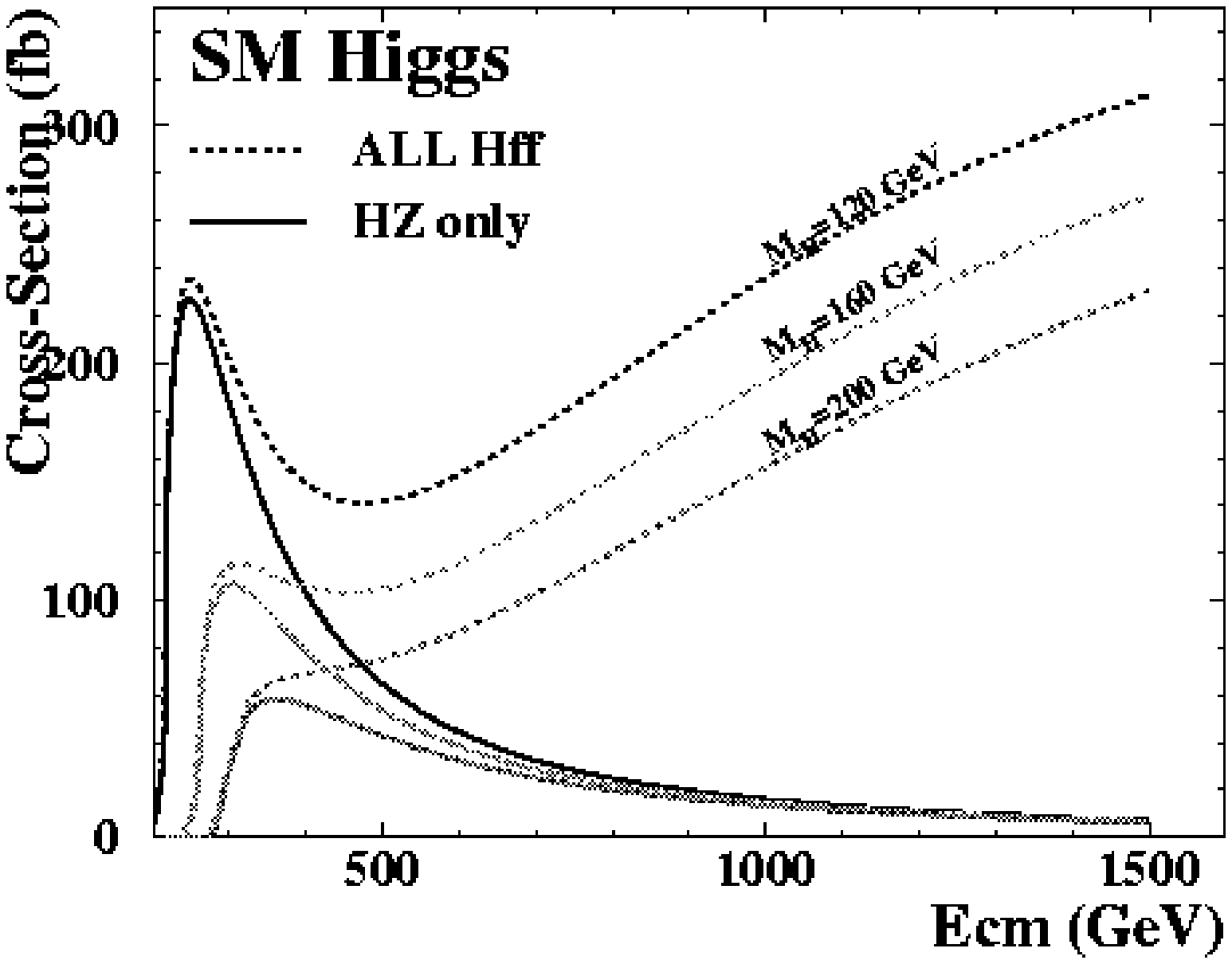}
}
\begin{center}\begin{minipage}{\figurewidth}
\caption[]{\sl 
Production cross-section of the SM Higgs. 
(Left) The production cross-section of $\ho\Zo$ process (solid lines),
and all $\ho\ff$ process including fusion processes (dotted lines) 
for 120--200 GeV SM Higgs 
as a function of the center-of-mass energy in the range for the JLC phase-I. 
(Right) The cross-section for wider range in center-of-mass energies.
\label{fig:xsec}}
\end{minipage}\end{center}
\end{figure}

In the SM Higgs, the production cross-section is completely determined 
for a given Higgs boson mass. No free parameter exists 
except for the Higgs boson mass itself.
Figure~\ref{fig:xsec} shows the production cross-section.
With a naive calculation we have approximately $10^{5}$ Higgs events
in 5 years for 120 GeV Higgs
if we have the luminosity of 10$^{34}$ cm$^{-2}$s$^{-1}$, 
with 10$^{7}$ seconds running per year. Note that the current JLC
design expects to have much larger duty factor (live running-time).

\begin{figure}
\centerline{
\epsfxsize=9.6cm \epsfbox{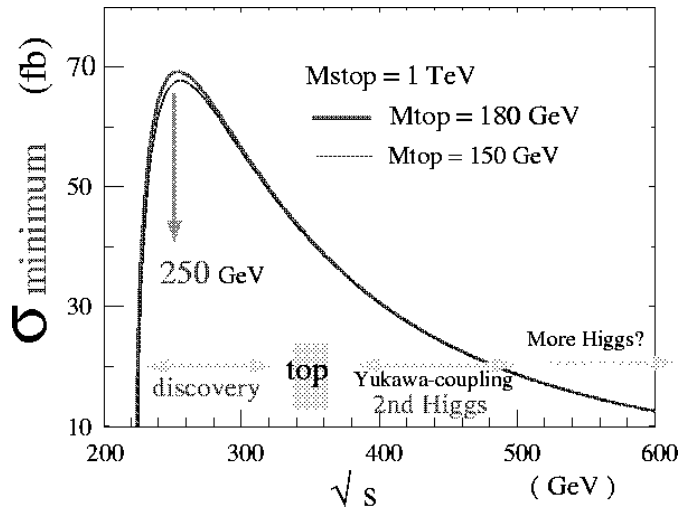}
}
\begin{center}\begin{minipage}{\figurewidth}
\caption[]{\sl
The minimum cross-section of the CP-even Higgs production 
in the NMSSM calculated by Y.~Okada et al~\cite{okada}. The two lines
correspond to the top quark mass of 150 and 180 GeV.
\label{fig:minxsec}}
\end{minipage}\end{center}
\end{figure}

For the MSSM, the production cross-section of the $\ho\Zo$ process
$\sigma_{\mathrm{hZ}}^{\mathrm{MSSM}}$ can be written as 
$\sigma_{\mathrm{hZ}}^{\mathrm{MSSM}} = {\mathrm{sin}}^{2} (\beta - \alpha ) \times \sigma_{\mathrm{HZ}}^{\mathrm{SM}}$,
where $\sigma_{\mathrm{HZ}}^{\mathrm{SM}}$ is the SM Higgs cross-section. 
The $\beta$ is defined by
tan$\beta$=$v_{1}/v_{2}$, 
the ratio of the vacuum expectation values of two Higgs field 
doublets. The $\alpha$ is the mixing angle of the two CP-even Higgs states.
Note that the $\alpha$ and $\beta$ are related in the MSSM in terms of
CP-odd Higgs boson (A$^{0}$) mass and other SUSY parameters.   
In more general SUSY Higgs models such as NMSSM, the cross-section depends
on more parameters. However, there exist absolute lower limits of the
cross-section ($\sigma_{\mathrm{minimum}}$). 
Figure~\ref{fig:minxsec} 
shows the $\sigma_{\mathrm{minimum}}$
as a function of center-of-mass
energy of $\ee$ collision for the NMSSM, taken from Ref.~\cite{okada}, 
which is valid also for the MSSM Higgs.

The other Higgs expected in the multi Higgs models such as MSSM 
might be discovered even in the first
phase of JLC project at $\sqrt{s}$ lower than 500 GeV.
The one of the important measurements would be
$\Ao$ ($\Ho$) production radiated off the $b$ quark or top quark,
for example $\ee\ra\toptop\ra\toptop\Ao$, via Yukawa-coupling. 
Also other channels with multi-Higgs production
could be observed via Higgs self-coupling. The strategy and scenario at the
next step depend on the results of the first phase. 
For MSSM (or general two Higgs doublets scenario), 
there is an additional important channel, which is $\ee\ra\Ao\ho$. 
However, if we stick to the MSSM, the production cross-section of the
$\Ao\ho$ process is expected to be considerably small,
if the $\ho$ mass exceed 110 GeV, where LEP-II sensitivity ends.
The heavier Higgs such as $\Ao$ and $\Ho$ would be the essential
subjects to be investigated at later stage 
of JLC phase-I ($\sqrt{s}>350$ GeV), 
the second phase of the JLC at $\sqrt{s}$
higher than 500 GeV, and/or $\gamma\gamma$ option of the JLC 
(see Chapter~\ref{part-options}).

\subsection{Higgs Decay}

\begin{figure}
\centerline{
\epsfxsize=7.68cm \epsfbox{physhiggs/hbr_1.epsi}
\epsfxsize=7.68cm \epsfbox{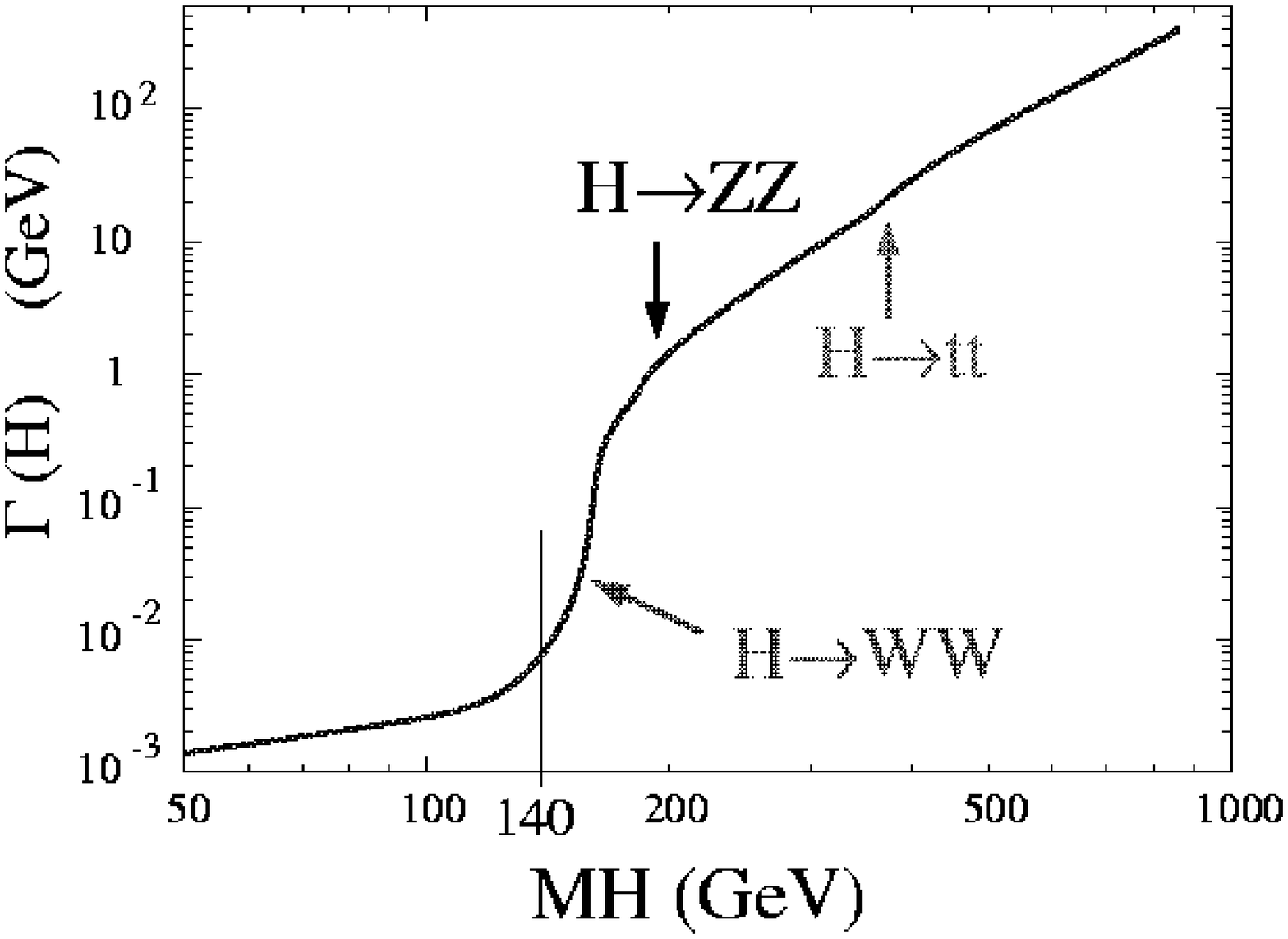}
}
\begin{center}\begin{minipage}{\figurewidth}
\caption[]{\sl
The branching ratio (Left) and the total decay width (Right)
of the SM Higgs as functions of the Higgs mass.
\label{fig:smhbr}}
\end{minipage}\end{center}
\end{figure}

The decay branching ratio of the SM Higgs is shown in figure~\ref{fig:smhbr}.
The Higgs boson couples to 
mass of the particle (coupling determines mass of the 
particle via Higgs mechanism).
In the SM, coupling with fermion, namely Yukawa coupling, is m$_{f}/v$ and 
M$_{g}^{2}/v$ with weak gauge bosons where m$_{f}$, M$_{g}$ are fermion
mass and W or Z masses, respectively, and $v$ is the
vacuum expectation value $v\approx$246 GeV.
If the Higgs is of the SM like, 
the dominant decay branch is $\bb$ for Higgs mass of lighter than
130 GeV. However, once the decay channel into W pair open, the most of the
Higgs is expected to decay to W pair. 
The decay of the MSSM Higgs depends also on $\alpha$ and $\beta$.
For the first target mass region between 100 and 140 GeV 
where we expect the lightest
Higgs in MSSM and NMSSM, again the $\bb$ branch dominates in
wide range of the parameter space.
A good b-flavour tagging is one of the essential techniques 
to discover and precisely measure the Higgs boson.

\subsection{Physics Backgrounds}

\begin{figure}
\centerline{
\epsfxsize=8.8cm \epsfbox{physhiggs/bkg.epsi}
\epsfxsize=7.0cm \epsfbox{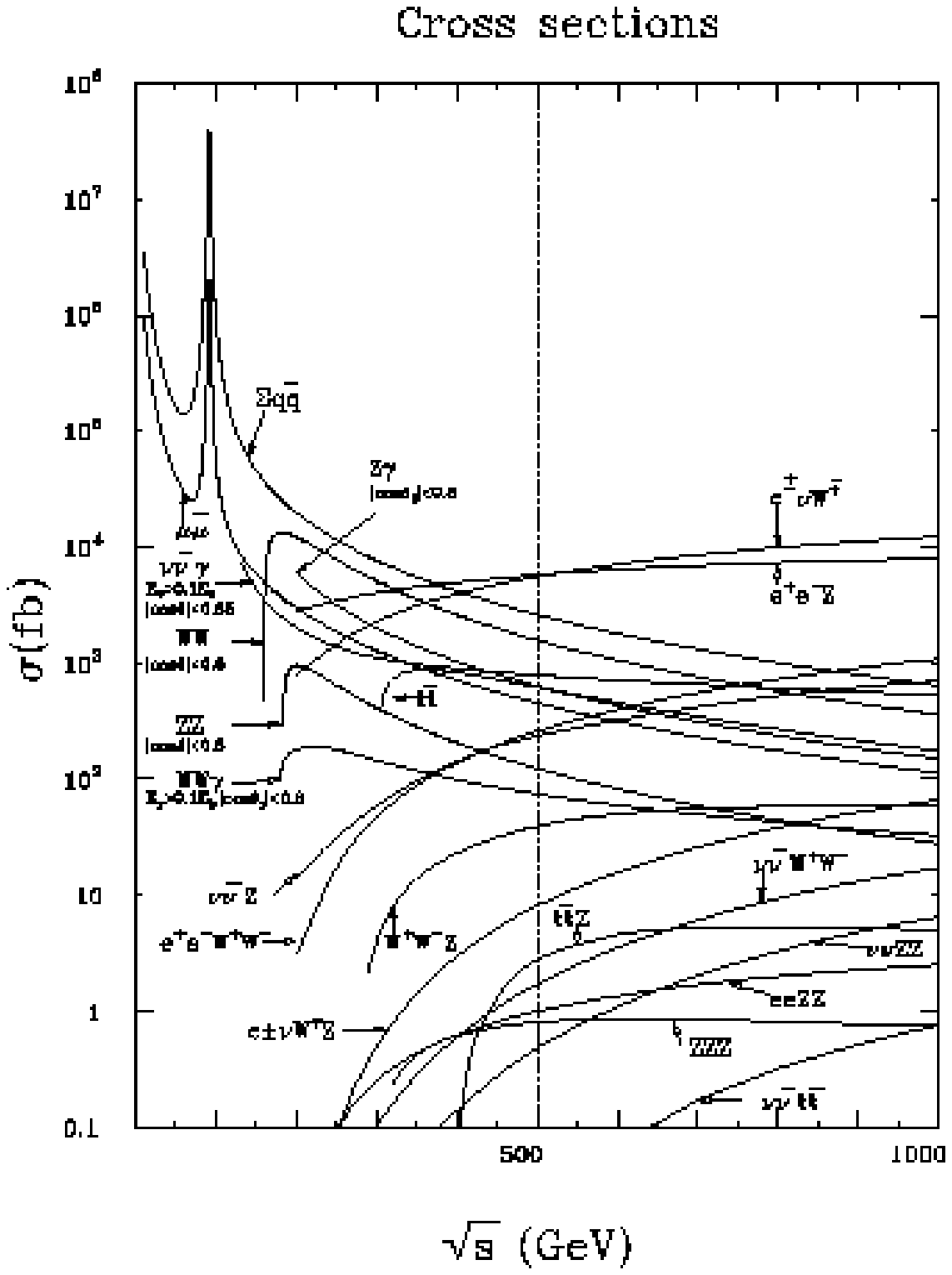}
}
\begin{center}\begin{minipage}{\figurewidth}
\caption[]{
\sl
Cross-section of the physics background as a function of 
center-of-mass energy up to 1 TeV.
\label{fig:backgrd}}
\end{minipage}\end{center}
\end{figure}

Figure~\ref{fig:backgrd}
shows the cross-section of the SM background
processes at energies of LEP-II and JLC region.
For example, at $\sqrt{s}=300$ GeV, the dominant backgrounds are;
$\qq (\gamma)$ (35 pb), $\WW$ (17 pb), $\Zo\Zo$ (1.0 pb)
We$\nu$ (2.0 pb), and $\Zo\ee$ (1.2 pb). Once the center-of-mass energy
exceeds the top-quark pair threshold, the process $\ee\ra\toptop$ 
also contributes as one of the main background. At $\sqrt{s}\approx 350$ GeV,
$\ee\ra\toptop$ cross-section is approximately 0.7 pb 
(see Chapter\ref{chapter-phystop}).
  
The $\qq (\gamma)$ background is relatively
easy to be reduced by analyses in the event topology.
The most severe background is the W pair production. In case mass of 
the Higgs boson is close to $\mZ$, the $\ee\ra\Zo\Zo$ process 
with heavy flavour or tau lepton in the final states could be 
irreducible backgrounds while angular distribution can help to reduce
the contamination. 
The situation is much better than LHC where several order of magnitude
higher background level is expected and the systematical 
control of the background is difficult. 
However our ultimate purpose of the Higgs study is
not only the Higgs discovery, but also to measure the Higgs properties 
precisely, in order to determine the direction of the future physics. 
The requirements of the purity and accuracy are far beyond those at LHC. 
Note that the physics background can be strongly suppressed furthermore 
by the polarized beam option~\cite{polbeam}, 
which is one of the big advantages of the linear collider.

\section{Experimental Considerations}

At JLC phase-I, the numbers of essential programs are waiting for us
in the Higgs studies. In order to evaluate the experimental feasibilities,
we use a model detector, i.e. JLC-I design.
The details of JLC-I detector and its simulation
methods are found elsewhere in this report.
The detector has the expected performance high enough to
satisfy the most of the requirements of the Higgs measurements as follows;
\begin{itemize}
\setlength{\itemsep}{-3pt}
\item Good energy measurements of charged tracks, neutral particles, and jets,
in order to reconstruct Higgs boson mass.
\item Precise positioning of the primary and the secondary vertices 
in order to tag the b-quark and c-quark from Higgs decay.
The primary vertex resolution is good enough at LC 
since the beam size of less than $\mu$m is used in 
the primary vertex fitting.
To establish the Higgs tagging with more than 80 \% for $\bb$ decay 
with reasonable purity against background process 
we need to tag the secondary vertex with its decay length 
less than 1 mm. Furthermore, the c-quark tagging might require more 
accuracy. 
The dominant part of the charged particle from
the b-hadron decay have its momentum less than 3 GeV. The detector
is required to have less materials in vertex detectors.
\item Good hermeticity of the detector. It is essential especially for
the neutrino channel, $\ee\ra\ho\nn$. 
It is also important for detection of
Higgs decay to W-pair for the 
$\ho\ra$WW$^{(*)}\ra${\it{l}}$\nu$q$\bar{\mathrm{q}}$ events.
\item Particle identification especially for leptons. 
The $\tau$-identification performance is one of the keys for the Higgs
studies. A fine segmentation of the calorimetry with enough 
neutral hadron/$\gamma$ discrimination might help.
\end{itemize}

Here we review the detector performance expected at JLC phase-I based on the
simulation studies of the detector.
We show energy-momentum measurement for jets and leptons, 
and 
heavy flavour tagging such as b-tagging, which are crucial in the light Higgs
studies, based on the preliminary simulation studies with the current 
JLC-I detector as a basic model.

\subsection{Jet Energy Momentum Measurements}

The one of the keys of the Higgs studies is the jet energy measurements.
In most of the cases, the Higgs is expected to decay hadronically.
Jet energies are measured by the central tracker system and calorimeters.
The $\gamma$'s mainly from $\pi^{0}$ decay and 
long-lived neutral hadrons such as neutrons and K$_L^0$ are measured
by the calorimetry. 
Charged particles are detected both by trackers and calorimetry.
In order to obtain the good resolution, the double counting in the
detectors are suppressed using the information of the
geometrical matching between tracks and clusters in the calorimetry systems.
The double counting suppression is the one of the essential techniques for the
Higgs study. The method to achieve the best jet-energy resolution
(energy-flow calculation) is under development and open for 
the future studies. 

Figure~\ref{fig:nn-mass} shows the reconstructed mass distribution 
with JLC-I simulation with the current energy-flow calculation
for Higgs signals in $\ee\ra\nn\ho\ra\nn\cc$, $\nn gg$(gluons) and $\nn\bb$.
For the plot, the visible masses without rescaling or kinematic fit
are used. 
The bottom lego plot shows the correlation between 
missing mass, which corresponds to $\Zo$ mass due to missing neutrino from
Z decay, and the visible mass corresponding to Higgs boson mass.
The reason why the $\bb$ decay has wider resolution is mainly due to
additional neutrinos in b-hadron semi-leptonic decays.
\begin{figure}
\centerline{
\epsfxsize=5.2cm \epsfbox{physhiggs/hnn_cc.epsi}
\epsfxsize=5.2cm \epsfbox{physhiggs/hnn_gg.epsi}
\epsfxsize=5.2cm \epsfbox{physhiggs/hnn_bb.epsi}
}
\vspace{18pt}
\centerline{
\epsfxsize=12.0cm \epsfbox{physhiggs/hnn_2dim.epsi}
}
\begin{center}\begin{minipage}{\figurewidth}
\caption[]{
\sl
The visible mass distribution of the process
$\ee\ra\nn\ho\ra\nn\cc$ (a), $\nn\ho\ra\nn gg$ (b)
and $\ee\ra\nn\ho\ra\nn\bb$ (c)
reconstructed in detector simulation for 120 GeV Higgs.
The bottom plot shows the correlation between visible mass and the
missing mass.
\label{fig:nn-mass}}
\end{minipage}\end{center}
\end{figure}

\subsection{Lepton Momentum Measurements}

\begin{figure}
\centerline{
\epsfxsize=9.6cm \epsfbox{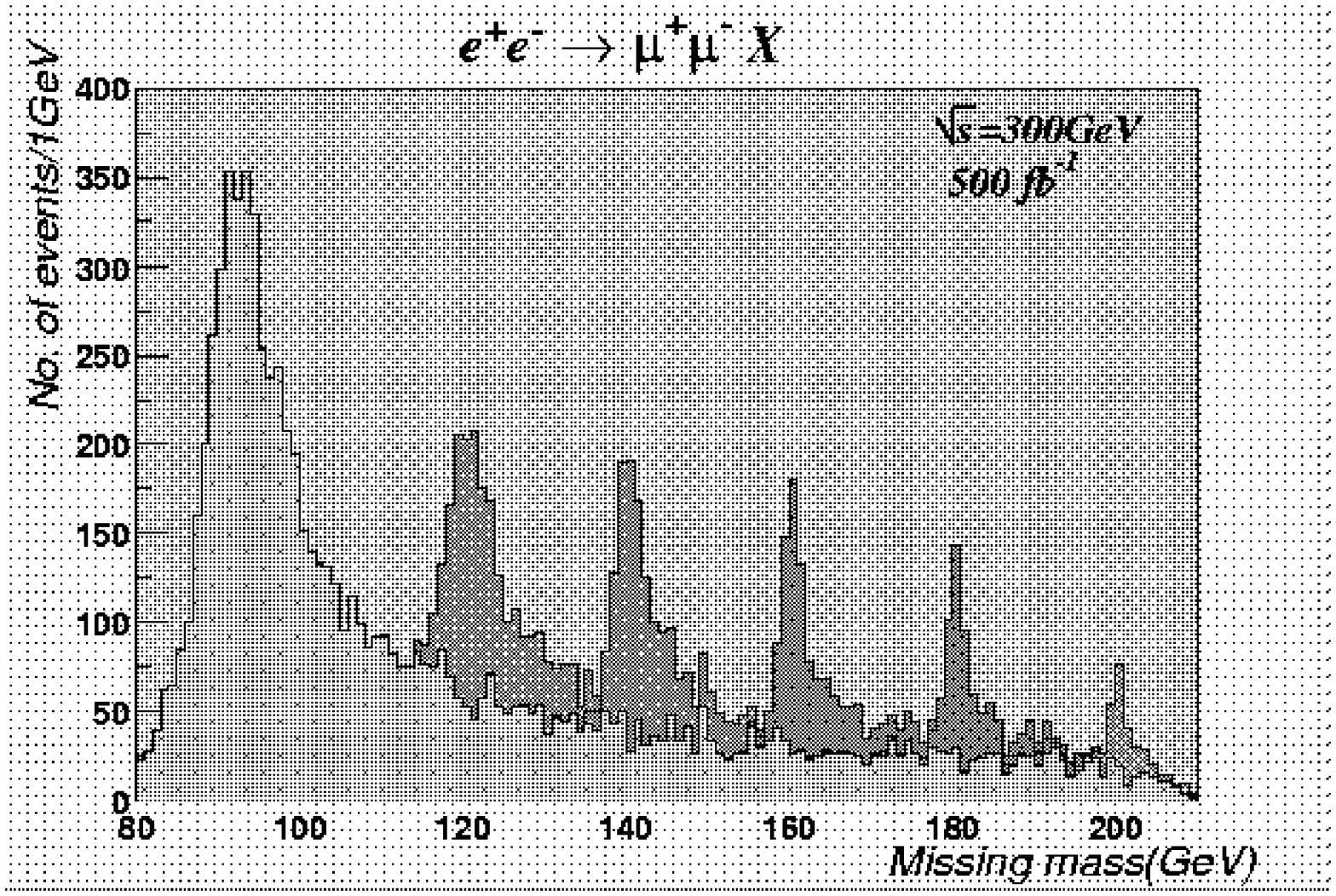}
}
\begin{center}\begin{minipage}{\figurewidth}
\caption[]{\sl
The distribution of the recoil mass of $\mm$ pair in $\ee\ra\mm$X 
normalized to 500 fb$^{-1}$ at $\sqrt{s}=300$ GeV.
The background of Z-pair production process, $\ee\ra\Zo\Zo\ra\mm$X, and
the SM Higgs boson signals of 120, 140, 160, 180 and 200 GeV are shown.
The full simulation program JIM (see Chapter~\ref{chapter-detsim}) is
uses for the simulation.
\label{fig:mmrecoil}}
\end{minipage}\end{center}
\end{figure}

In the Higgs-strahlung process, the leptonic decay ($\ee$ or $\mm$) 
in the associated Z boson plays a special role. 
The channels are the most clean ones in all final states
of $\ho\Zo$. Event-topology for the signal is high energy $\ee$ or $\mm$
with the invariant mass of lepton pair similar to Z.
Although the branching ratio in the Z decay to $\ee$ or $\mm$
is limited ($\sim 6$\%), the background similar to the signal is only those
of the $\ee\ra\Zo\Zo\ra\ee$X or $\mm$X. Recoil mass of the two leptons
corresponds to the Higgs mass without dilution by the Z natural width.
The recoil mass distributions of $\ee\ra\mm$X simulated 
with JLC-I detector for background and signal is found in 
Figure~\ref{fig:mmrecoil}. 

\subsection{Heavy Flavour Tagging}
\begin{figure}
\centerline{
\epsfxsize=9.6cm \epsfbox{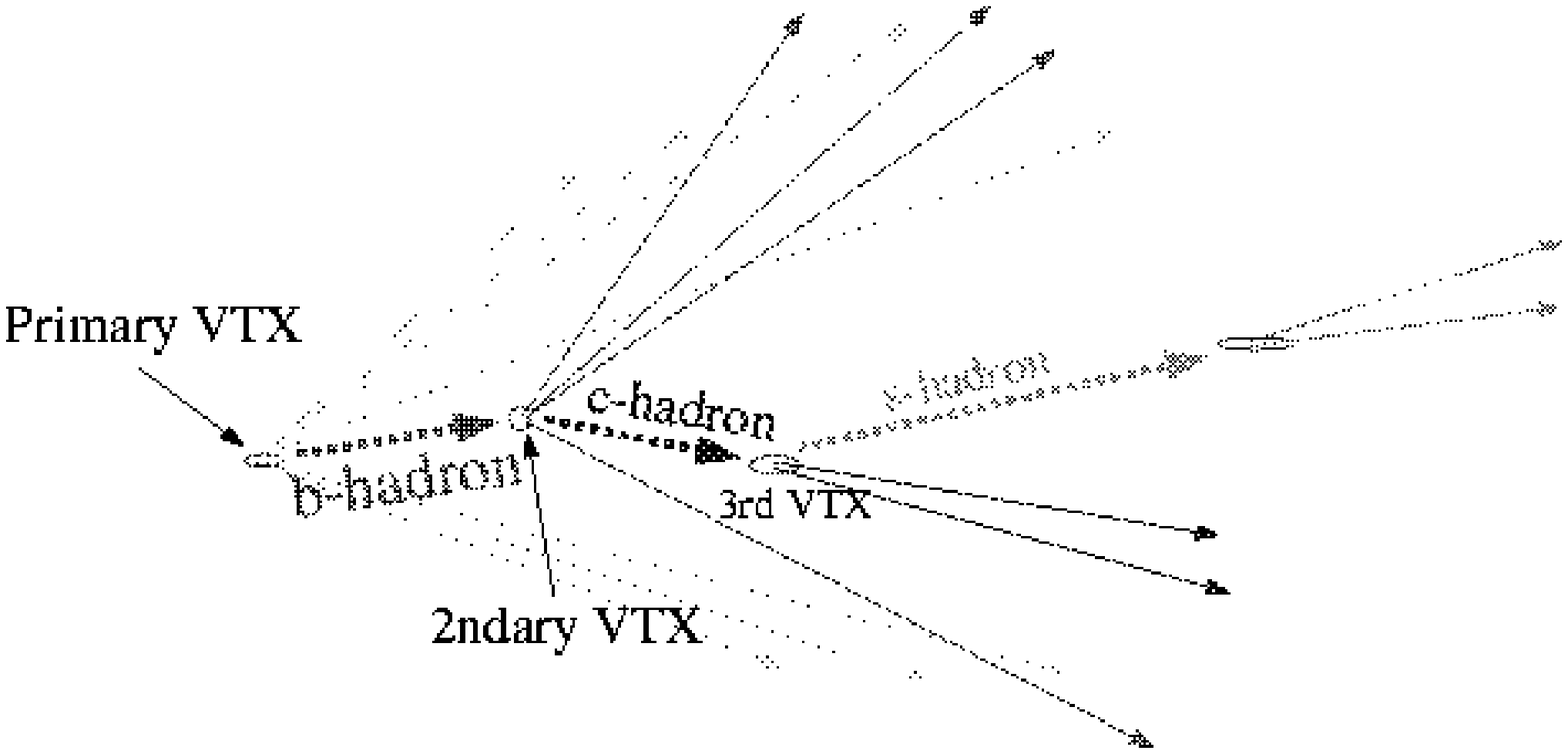}
}
\begin{center}\begin{minipage}{\figurewidth}
\caption[]{\sl
Schematic view of the cascade decay of the b-hadron (Left) and the
scheme of the definition of the impact parameters (Right) 
for the heavy-flavour tagging.
\label{fig:btagscheme}}
\end{minipage}\end{center}
\end{figure}

  As we have discussed, if the Higgs boson has mass of less than 140 GeV,
  the dominant decay mode is expected to be $\bb$. Hence the b-flavour 
  tagging is the essential technique in the light Higgs study.
  Figure~\ref{fig:btagscheme} shows a schematic view of the cascade decay of
  the b-hadron. The b-hadron tends to decay sequentially to c-hadron and to
  hadrons with strangeness. The lifetime of the b-hadron is long enough
  ($\beta\gamma c\tau \sim$3--5 mm) to be observed 
  as a displaced vertex in the detector. 

\begin{figure}
\centerline{
\epsfxsize=7.2cm \epsfbox{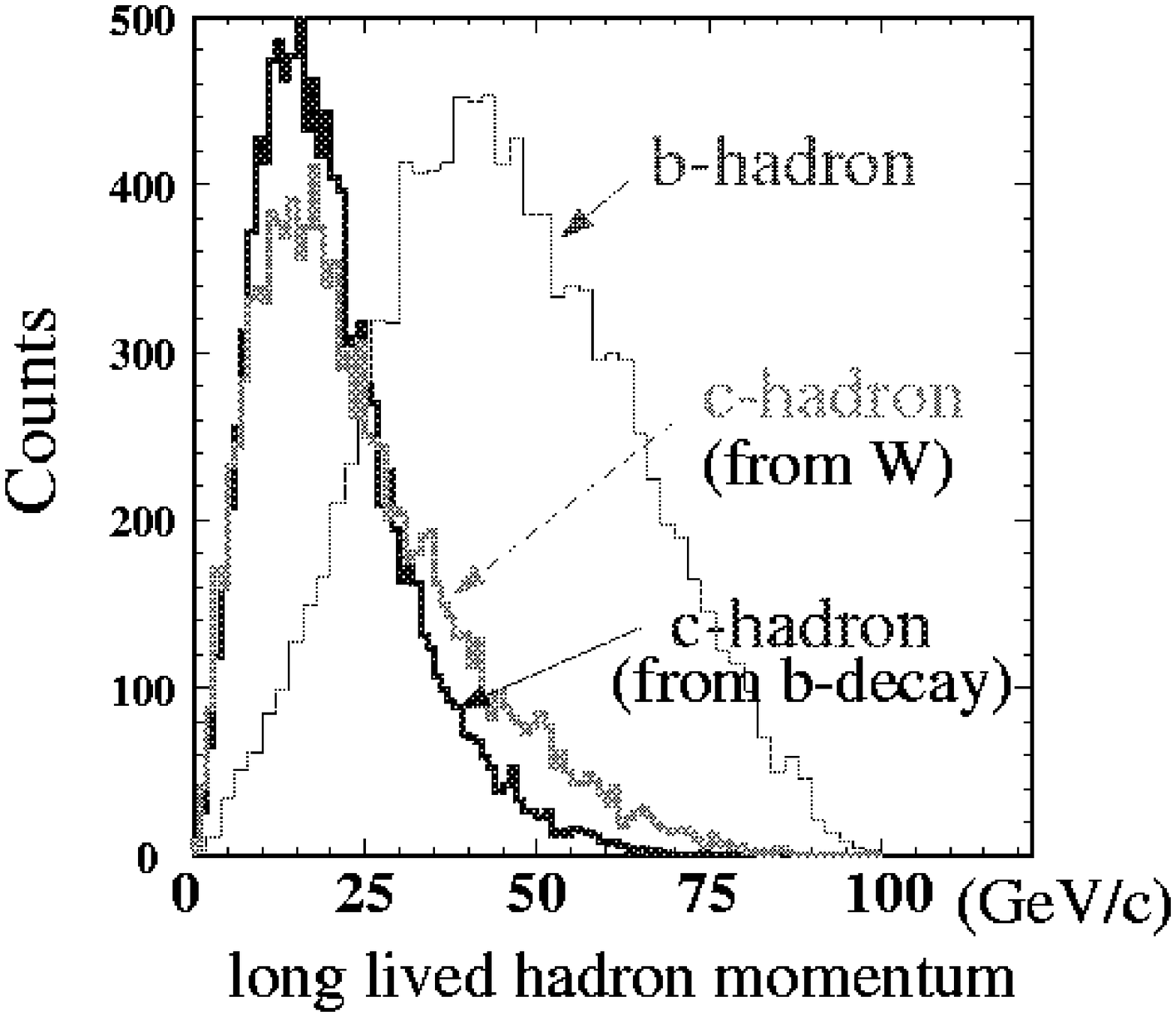}
\epsfxsize=6.4cm \epsfbox{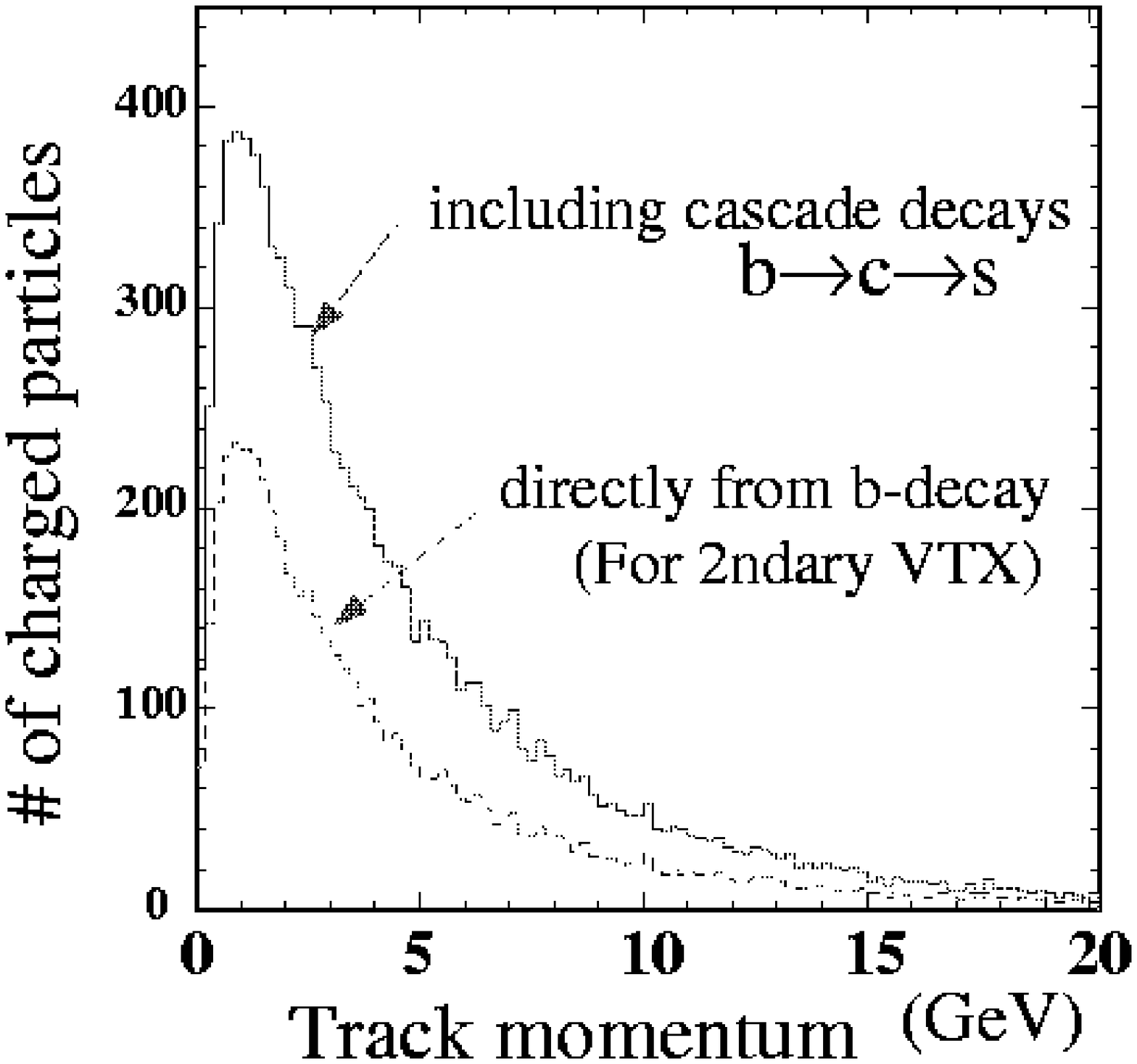}
}
\centerline{
\epsfxsize=7.2cm \epsfbox{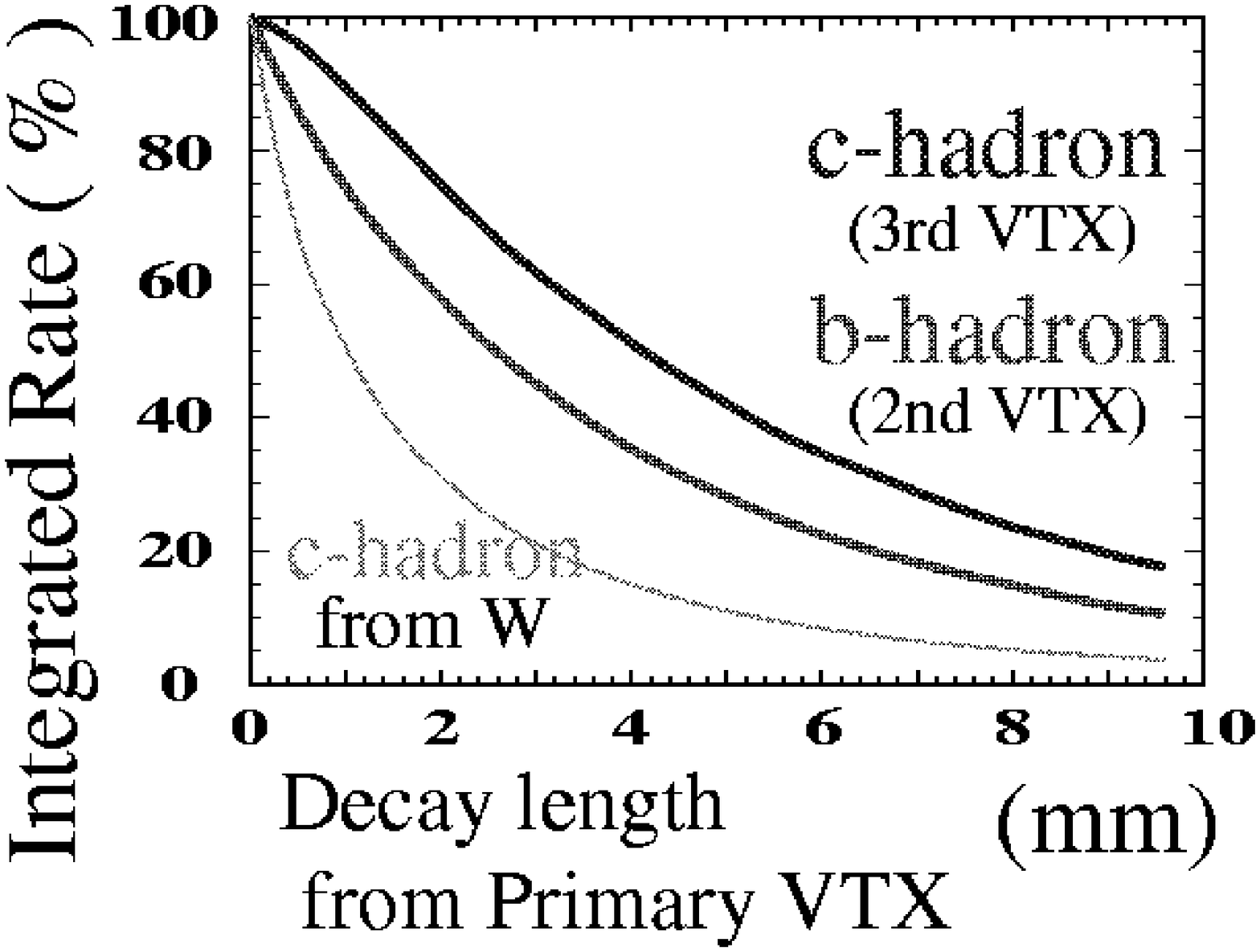}
\epsfxsize=5.6cm \epsfbox{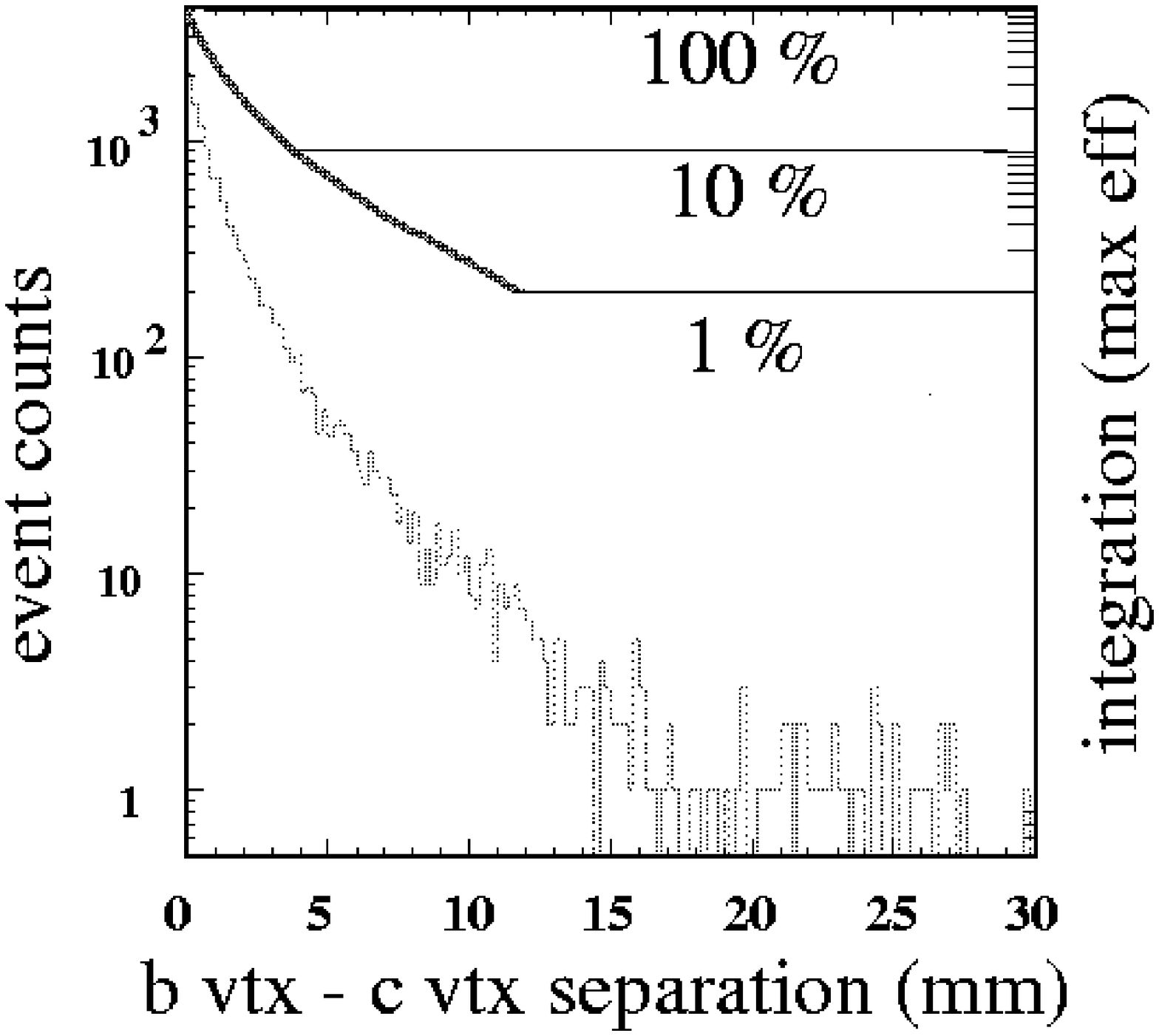}
}
\begin{center}\begin{minipage}{\figurewidth}
\caption[]{\sl
Properties of the b and c-hadrons 
and decay products from Higgs boson
and background W pair production: (a) The momentum distribution of the
heavy-flavour hadrons. (b) Distributions of the track momentum originating
the b-hadron decay. (c) Distributions of decay length (the distance from
the event production point to the decay point. (d) Rate (tagging efficiency) 
in integration of events as a function of the cut position of the decay 
length from the event production point.
\label{fig:bprop1}}
\end{minipage}\end{center}
\end{figure}

At JLC, the position resolution of the initial production of an event 
(primary vertex) is expected to be
much better than that of LEP, since the beam size of less than $\mu$m
at JLC can be used as constraints of the primary vertex fitting.
Several properties of the $\bb$ decay of the Higgs signal 
can be seen in Figure~\ref{fig:bprop1}. 
The dominant part of the charged particle from
the b-hadron decay have its momentum less than 3 GeV. Hence the
low momentum track is important for the efficient Higgs tagging. 
One can see, to establish the Higgs tagging with
more than 80 \% for $\bb$ decay, we need to tag 
the secondary vertex for that with decay length less than 1 mm. 
In order to purify the b against c-quarks, the vertex mass and
tagging of the 3rd vertex are powerful. The distance between b and c
decay in a decay sequence is also less than 1 mm as shown in the figure.
Kinematics of the jet and vertex is also the discriminatories
of the b-hadron against lighter flavours. The b-hadron tends to decay with
large multiplicity and large jet invariant mass due to heavy mass of the
b-quark. 

The distributions of the obtained normalized impact parameter, 
impact parameter divided by the expected resolution, 
are shown in the bottom-left plot. 
In the plot, the distributions for the tracks from the b-hadrons decay 
in the Higgs event and those from the charm hadron in the W-decay
are seen, together with the distribution for the tracks produced at
the primary vertex. 
Note that the track fitting has been made with the correct association of the
hits, hence no effects of the mis-assignment of the hits are included.

There are many
methods for efficient b-tagging 
developed and used in the experiments so far.
SLD experiment at SLC extensively uses the combined information of the
vertex decay length and its mass (SLD mass tagging). At LEP, sophisticated
methods combining several discriminant variables are used. For example 
OPAL experiment uses tagging based on long lifetime
combined in a likelihood method with high $p_{t}$ lepton tagging 
and jet kinematics.  
The feature of the tagging methods can be seen in Ref.~\cite{acfahiggs}.

\begin{figure}
\centerline{
\epsfxsize=6.72cm \epsfbox{physhiggs/IP-norm.epsi}
\epsfxsize=6.72cm \epsfbox{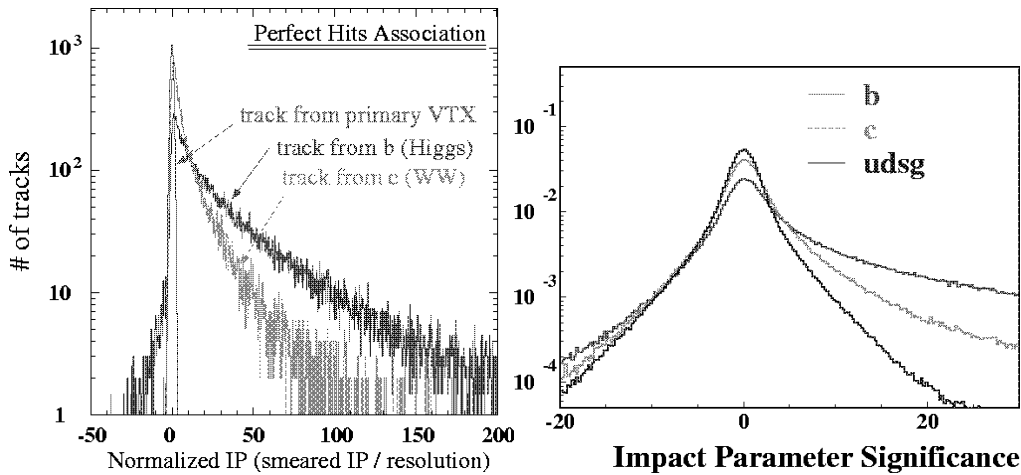}
}
\begin{center}\begin{minipage}{\figurewidth}
\caption[]{\sl
Performances obtained in the smearing simulation.
\label{fig:preres}}
\end{minipage}\end{center}
\end{figure}

  Figure~\ref{fig:preres} shows the performances expected with the JLC-I
  detector with simulations. The left plot shows the expected 
  impact parameter distribution for the signal process and background process
  of W-pair production.
  The right plot
  shows the results obtained for the impact parameter resolution
  for each flavour of the decay from the Higgs.

\section{Experimental Overview; Higgs Detection and Measurement}

The essential measurements of the Higgs is the mass and the coupling
to other particles and to Higgs particle itself. These can be achieved by
the precise measurements of the Higgs production and the decay.
The key measurements at the first phase of the JLC would be as follows;
\begin{itemize}
\item Mass of the Higgs.
\item Production and decay angular distribution and their correlations.
\item Cross-section in $\Zo\ho$ and WW-fusion as a function of
$\sqrt{s}$ with/without polarized beam.
\item Branching ratio to $\bb$ (Br($\ho\ra\bb$)).
\item Br($\ho\ra\WW$).
\item Br($\ho\ra\tautau$)/Br($\ho\ra\bb$).
\item Br($\ho\ra\cc$)/Br($\ho\ra\bb$), Br($\ho\ra gg$)/Br($\ho\ra\bb$), 
Br($\ho\ra\cc +gg)$)/Br($\ho\ra\bb$).
\end{itemize}

The important thing is that we can establish/measure the properties
of the Higgs particle model-independently in following things:
\begin{itemize}
\item Mass, which is final inputs in the SM.
\item Establishment of quantum numbers (CP, spin) and s-channel production
of $\ho\Zo$ final states through angular distributions and/or 
energy-scan
\item Establish the Z mediation in $\ho\Zo$ production
with polarized beam option
\item SU(2)$\times$U(1) universality of HZZ and HWW comparing
$\ho\Zo$ production and $\ho\nn$ production which contains WW-fusion
process.
\item Absolute strength of gauge coupling to Z and to W via cross-section measurements.
\item Total decay width of Higgs which is significantly sensitive to SUSY via
Higgs decay branching ratio into W or Z together with the measured absolute
strength of the coupling between Higgs and weak bosons.
\item Absolute strength of Yukawa couplings to quarks and leptons
via branching ratio to fermions together with the measured total decay width.
The feasible flavour is b, c quarks, and tau. 
\item Self-coupling via multi Higgs production either in $\ee\ra\Zo\ho\ho$
and/or $\ee\ra\nn\ho\ho$.
\end{itemize}

In addition to these, top Yukawa coupling is
measured directly from a measurement of the process 
$\ee\ra\toptop\ra\toptop\ho$.
In case we neglect the new particle (such as SUSY) loops, 
the decay branching to gluons are
also used to determine the top Yukawa coupling assuming the dominant
contribution is a top quark loop. 
In other words the comparison between top Yukawa coupling and gluonic decay
measures new particles with QCD color charge.

\subsection{Discovery Potential and Sensitive Signal Cross-Section}

In order to demonstrate the clean experimental environment compared
to Hadron colliders, we first describe the Higgs discovery potential.

\subsubsection{$\ho\Zo$ Production}

In the e$^{+}$e$^{-}$ collider, the Higgs signal is cleanly identified.
The Higgsstrahlung production process is the most clean channel to be sought.
In the $\Zo\ho$ production, the event topologies are categorized
by the final state of the associated $\Zo$ boson, namely 
$\qq$, $\tautau$, $\nn$, $\ee$ and $\mm$. 

\begin{figure}
\centerline{
\epsfxsize=7.08cm \epsfbox{physhiggs/eeqq1.epsi}
\epsfxsize=6.88cm \epsfbox{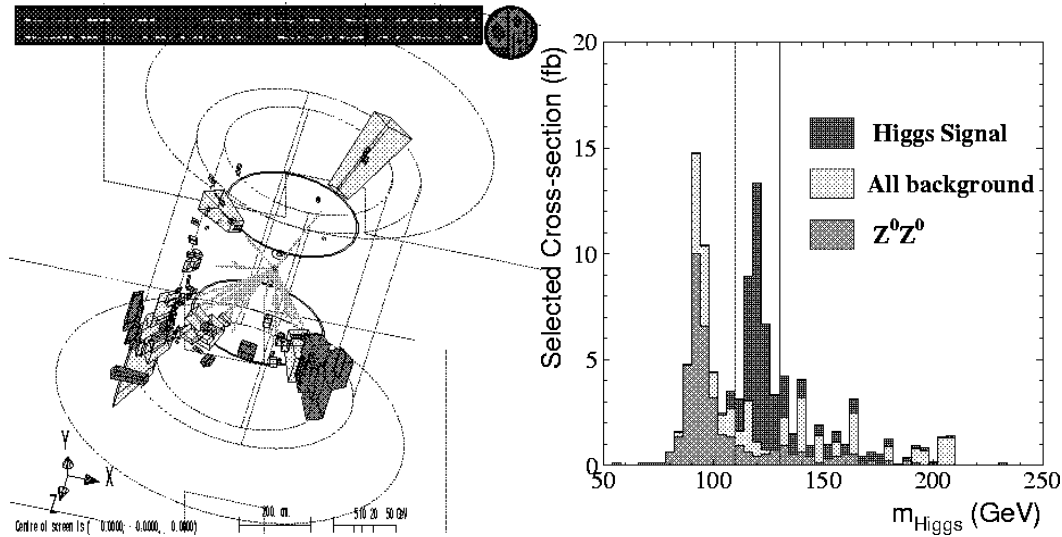}
}
\begin{center}\begin{minipage}{\figurewidth}
\caption[]{\sl
Demonstration of Higgs hunt with OPAL detector at $\sqrt{s}=250$ GeV.
The SM Higgs of 120 GeV is assumed.
\label{fig:opaldemo}}
\end{minipage}\end{center}
\end{figure}

One of the realistic estimates of the detector effects for Higgs hunt 
would be made by a simulation study of existing detector known to precisely
follow the detector responses, 
while the expected performance of the detector at JLC is 
much better than that of the old detectors. 
Full simulation code of the OPAL detector at LEP was used 
to simulate the SM physics at center-of-mass energy of
250 GeV together with 120 GeV Higgs.
The tagging of the Higgs event uses the energy-flow and
b-tagging algorithms actually used in the OPAL physics at LEP-II including 
Higgs searches. 
The figure~\ref{fig:opaldemo} shows the
reconstructed mass distribution for the events after selection.
The effective cross-section after the selection is 28.2 fb for the signal
in the mass window indicated in Figure~\ref{fig:opaldemo}.
The background has been reduced to 7.2 fb after the selection.
At a glance, even with the integrated luminosity of 1 fb$^{-1}$, 
which corresponds to 
only about one day (or shorter) running of JLC, 
we can discover Higgs easily, 
even with the performance of the detector at LEP.

At JLC, in order to definitely answer to the existence of the
light Higgs predicted by the SM and SUSY, we must have the sensitivity
down to $\sigma\sim10$ fb level which is one order of magnitude less
than that of the SM Higgs.
Once we know the efficiency and recoil mass distribution 
after a selection for backgrounds and a signal at certain mass, 
we can easily estimate the discovery sensitivity.

\begin{figure}
\centerline{
\epsfxsize=9.0cm \epsfbox{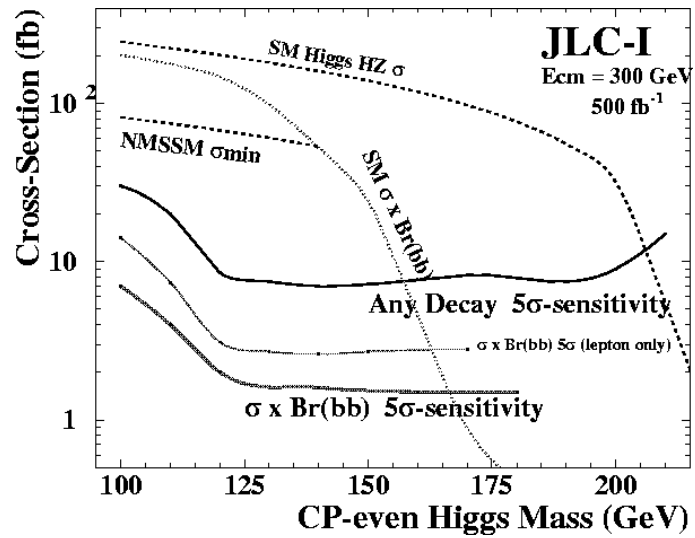}
}
\begin{center}\begin{minipage}{\figurewidth}
\caption[]{\sl
JLC-I sensitivity for the CP-even Higgs: 
Production cross-section sensitive at JLC-I at 300 GeV with 500 fb$^{-1}$
at more than 5$\sigma$ are shown as a function of the Higgs mass,
with and without b-flavour tagging for the Higgs decay products.
The line indicated by ``Any decay 5$\sigma$ sensitivity'' shows the
JLC-I sensitivity only with leptonic channel $\ee\ra\ellell\ho$ independent
to decay of the Higgs.
The cross-section for the SM Higgs, minimum cross-section for NMSSM
are also shown to be compared. 
To compare the sensitivity for exclusive channel
$\ho\ra\bb$ for $\sigma\times$Br($\ho\ra\bb$), 
the SM Higgs $\sigma\times$Br($\ho\ra\bb$), and
expected sensitivity at JLC are also shown.
\label{fig:disc0}}
\end{minipage}\end{center}
\end{figure}

As an example the sensitivity is studied at 300 GeV, which is the early
stage of the JLC phase-I using the JLC quick simulator.
Figure~\ref{fig:disc0} shows the discovery sensitivity for 500 fb$^{-1}$
in channels.

We first check the simplest mode, $\ee\ra\ho\Zo$ and $\Zo\ra\ee$ or $\mm$.
These channels are only 6\% of the produced $\ho\Zo$ due to the branching ratio
of the Z.
The $\Zo\Zo$ process is almost all the background. 
Only the recoil mass is used in the analysis in order to have the sensitivity
independent to the decay of the Higgs.
As shown in the figure, we are sensitive up to more than 200 GeV
for the CP-even Higgs with cross-section similar to the SM model
even at $\sqrt{s}=300$ GeV only from the lepton spectra
independent to decay of the Higgs. 

When we utilize the b-tagging, we can suppress the huge background from
$\WW$ production, which enables us to use hadronic decay of Z,
$\ho\Zo\ra\ho\qq$, which carries 70\% (Br($\Zo\ra\qq$))
of the $\ho\Zo$ signal process.
While the signal of $\ho\Zo\ra\bb\Zo$ 
also reduces the efficiency,  
the sensitivity to $\sigma\times$Br($\bb$) reaches down to less than 2 fb
as shown in the figure. 

\begin{figure}
\centerline{
\epsfxsize=9.0cm \epsfbox{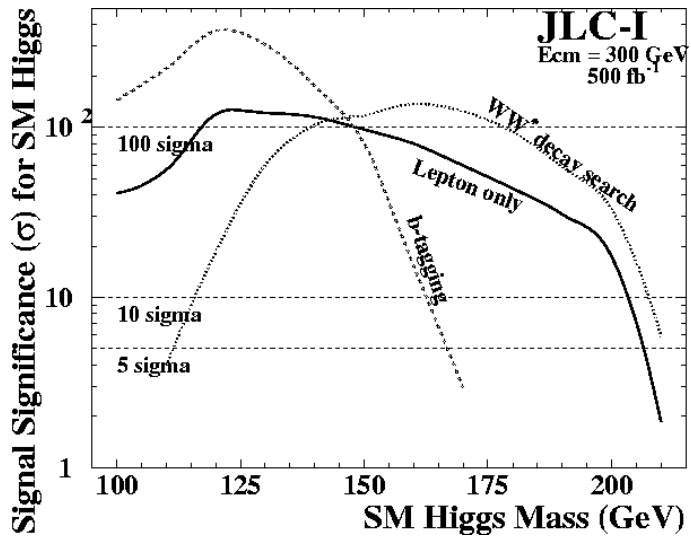}
}
\begin{center}\begin{minipage}{\figurewidth}
\caption[]{\sl
Higgs signal significance for the SM Higgs in $\Ho\Zo$ channels: 
Three methods are shown; 1) only lepton channel independent to Higgs decay,
2) using b-tagging enrich $\ho\ra\bb$ signal expected for light Higgs, and
3) analysis optimized for $\ho\ra\WW^{*}$ decay.
\label{fig:smsig}}
\end{minipage}\end{center}
\end{figure}

\begin{figure}
\centerline{
\epsfxsize=9.0cm \epsfbox{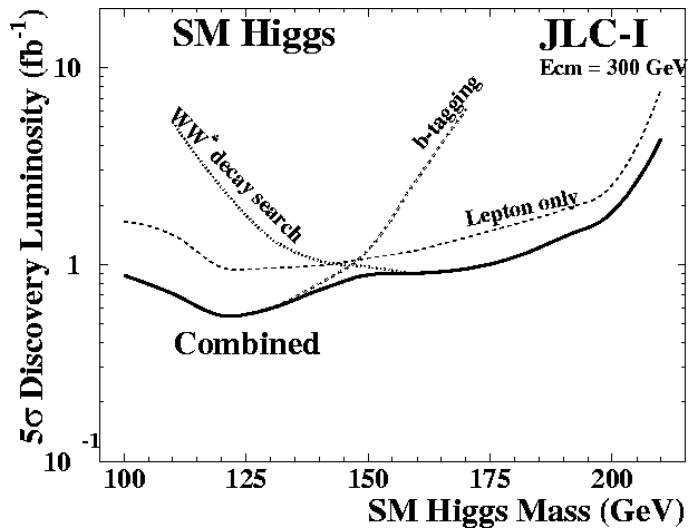}
}
\begin{center}\begin{minipage}{\figurewidth}
\caption[]{\sl
Integrated luminosity necessary to 
discover (5$\sigma$) the SM Higgs 
at JLC phase-I $\sqrt{s}=300$ GeV. Luminosity of 1 fb$^{-1}$ correspond to
about ``one day'' running at JLC.
\label{fig:smdisclumi}}
\end{minipage}\end{center}
\end{figure}

The luminosity necessary to discover the SM Higgs is shown
in Figure~\ref{fig:smdisclumi} as a function of the SM Higgs mass.
Note that the data correspond to 1 fb$^{-1}$ 
is expected to be accumulated in less than one day
at the JLC.

\subsubsection{$\Ao\ho$ Production}
Let's consider the two Higgs field doublets (2HDM) type-II including MSSM.
We assume here CP conservation in the Higgs sector.
We have 5 physical states, two CP-even $\ho$ and $\Ho$, CP-odd $\Ao$
and positively and negatively charged Higgs. 
$\Ao$ mass is unknown and possible mass is from 90 to 2 TeV 
or more in the MSSM. 
CP-odd nature of $\Ao$ inhibits the decay to pair of gauge bosons.
In general $\Ao$ decay to heavy quarks in wide range of 2HDM parameters. 
Kinematical accessible mass is $\mA < \sqrt{s}-\mh$, hence we have
a potential to produce $\Ao\ho$ in the case such as 
$\mh=120$ GeV, $\mA=300$ GeV at $\sqrt{s}=500$ GeV.
At that time, the CP-even light Higgs is measured already by $\ho\Zo$ 
production at JLC. For the branching ratio of $\ho$ to $\bb$, we assume 70\%.
In this benchmark case, $\Ao$ cannot decay to top quark, and we assume here
the decay to $\bb$ has branching ratio more than 80\%.
The production cross-section depends on tan$\beta$ and $\alpha$.
The analysis can be made in two steps; the first clean four jet topology
with b quark signature in all jets to be selected, then
next choose a jet-pair which corresponds to CP-even Higgs mass.
The other jets are combined to measure $\Ao$ mass.

The dominant backgrounds are; 1) $\Zo\Zo$ process with b-quark decay of both Z,
2) $\bb$ production together with multi hard gluon radiation followed by
b production, 3) top-quark pair production followed by 
$\toptop\ra$bWbW$\ra$bqqbqq, and 
4) production of $\ho$ signal itself in the Higgsstrahlung process.
The top quark pair and $\bb$ backgrounds are 
suppressed by its event topology and four b quark requirement.
The $\Zo\Zo\ra\bb\bb$ background is further suppressed 
by a veto for the events which corresponds to ZZ in a jet combination. 
Finally the jet pair mass is checked if the accumulation exist 
compared to background.
The selection efficiency for the process is 20\% with selected 
background cross-section of about 0.05 fb 
from non-Higgs background. The most sever background might be Higgsstrahlung
signal itself. It corresponds to about 0.1 fb
if the $\ho$ is of SM-like (sin$^{2}(\alpha-\beta)\sim 1$).
At 1000 fb$^{-1}$, we have 50 event non-Higgs background and 
100 events from Higgsstrahlung in the mass window. 
It means 5$\sigma$ corresponds to signal cross-section of about 0.4 fb.

If we have a sensitivity down to 0.4 fb, we can estimate
the sensitivity to cos$^{2}$($\alpha-\beta$) in $\mh$-$\mA$ plane.
In all cases, $\ho$ is assumed to be SM-like in order to check the
sensitivity to cos$^{2}$($\alpha-\beta$), hence SM Br is used for $\ho$.

\subsubsection{Yukawa Production of $\ho$, $\Ao$}

The cross-section for the SM Higgs is shown in figure??? for the processes
$\ee\ra\toptop\ra\toptop\Ho$ and $\ee\ra\bb\ra\bb\Ho$ 
for $\sqrt{s}=$500, and 1000 GeV as a function of the SM Higgs mass.
The measured error $\Delta\sigma / \sigma$ 
linearly depend on the inverse of the square root of the cross-section 
times the integrated luminosity,
in case the background is negligible.
The sensitive cross-section is about 0.1 fb for 1000 fb$^{-1}$.

\subsection{Mass Measurements}

The Higgs boson mass is the fundamental value which might give us information
up to GUT scale as we discussed above. 
A precise measurement of the Higgs mass with an accuracy of about 
100 MeV or less is necessary 
in order to obtain fruitful results from the other
measurements such as branching ratio especially for Higgs 
around 120--150 GeV. 
The Higgs branching ratio
is a strong function of its mass mainly due to kinematic phase
space factor of Higgs decay into W-pair.

\begin{figure}
\centerline{
\epsfxsize=5.0cm \epsfbox{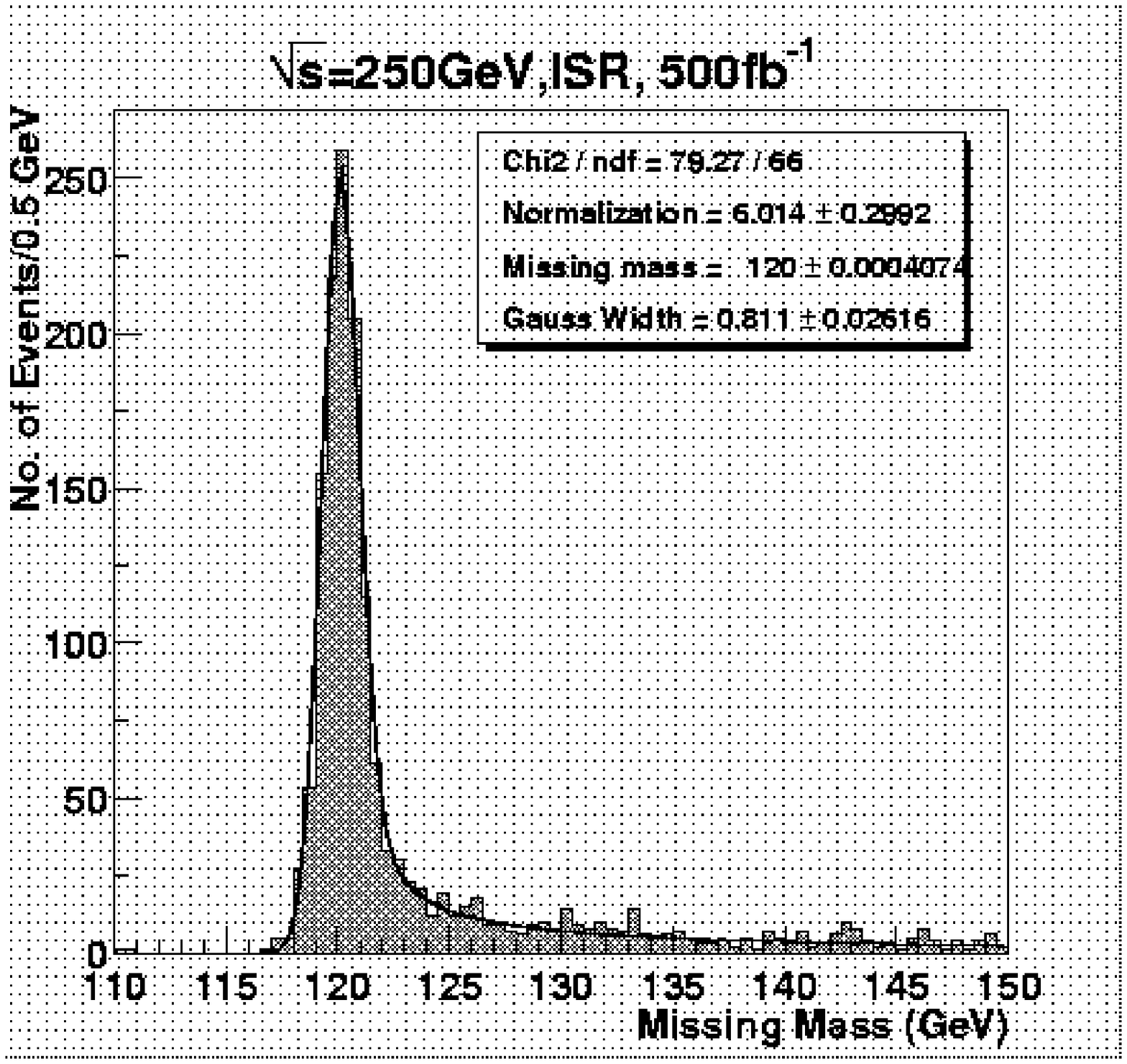}
\epsfxsize=5.0cm \epsfbox{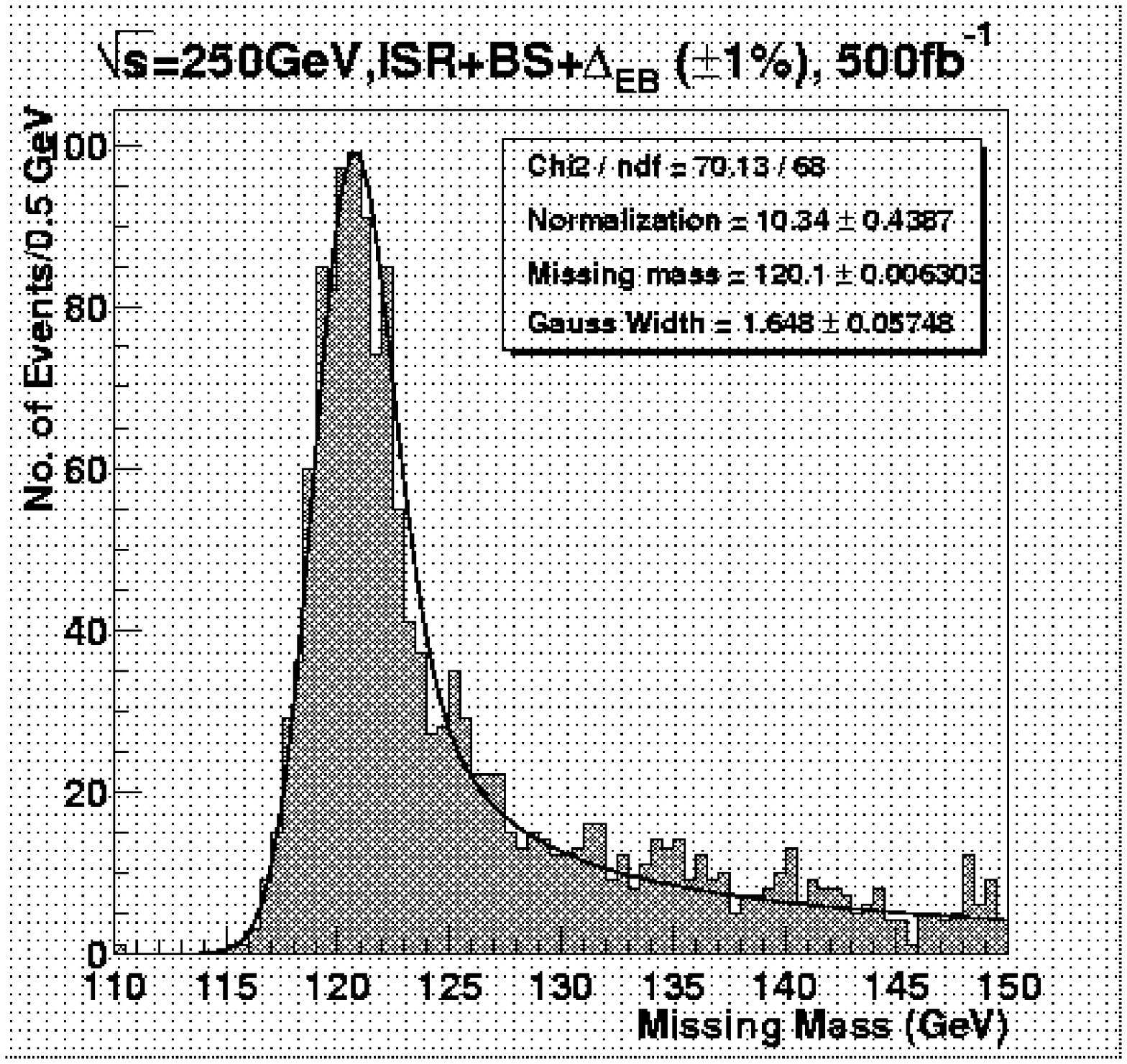}
}
\centerline{
\epsfxsize=5.0cm \epsfbox{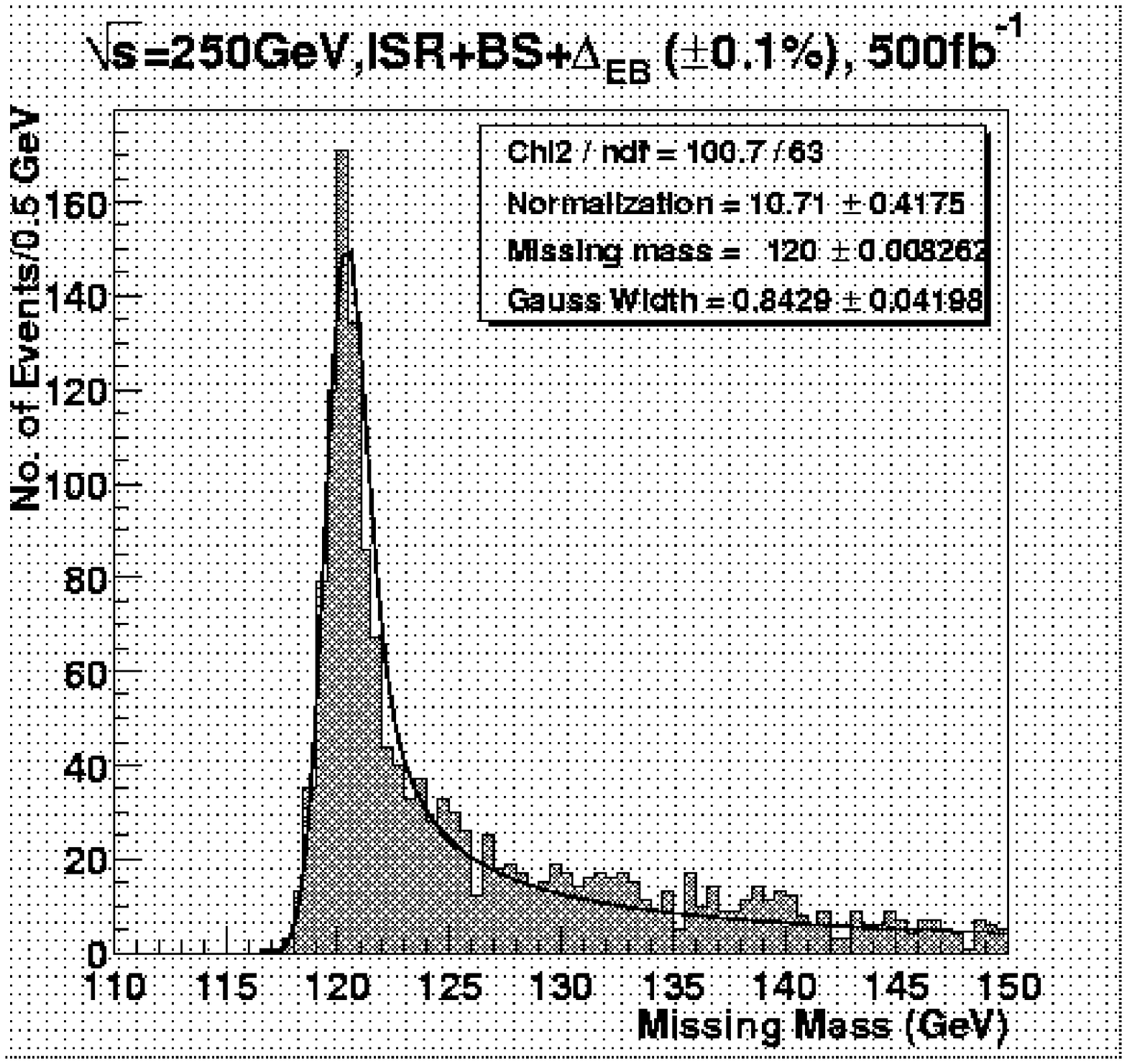}
\epsfxsize=5.0cm \epsfbox{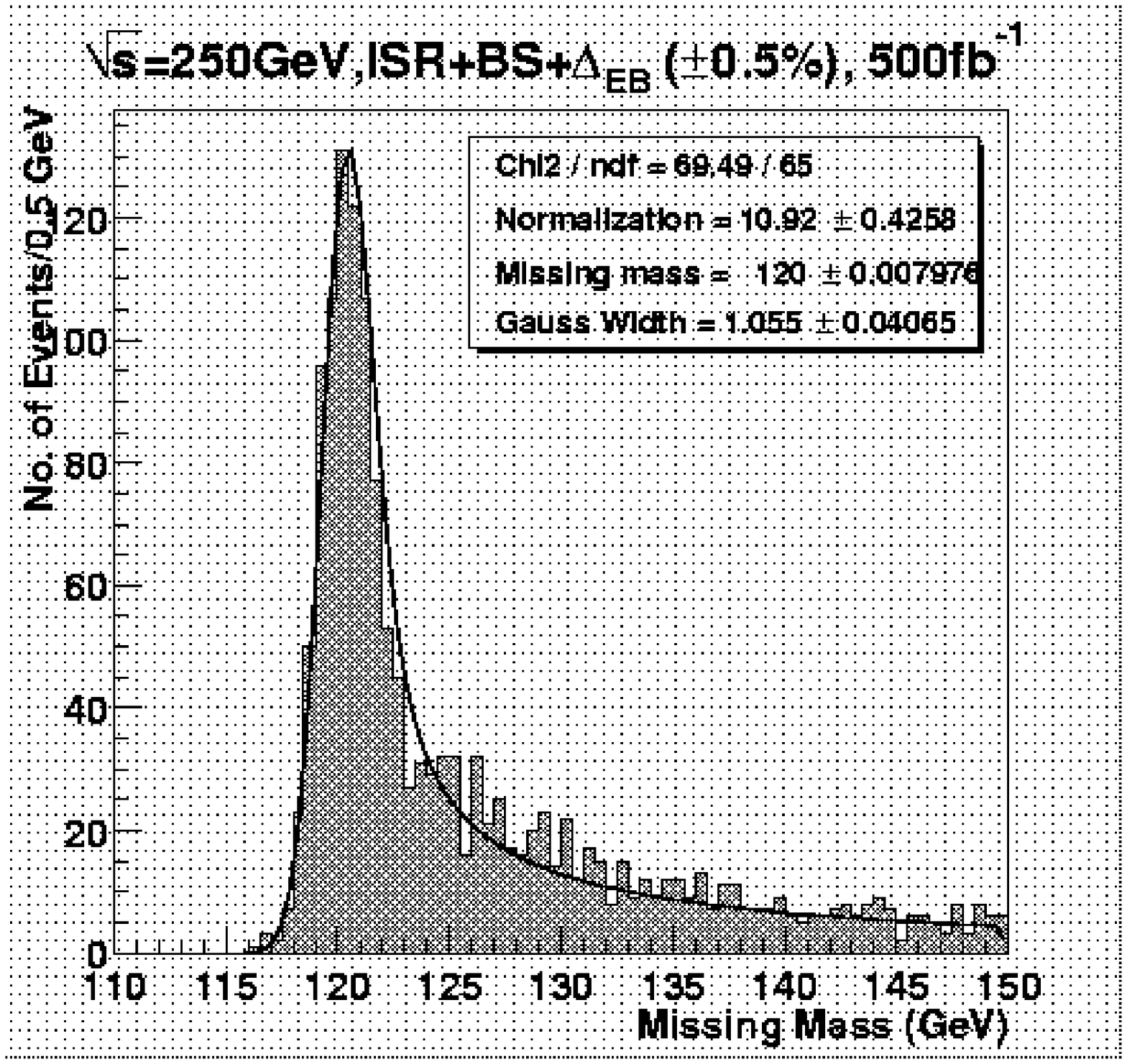}
}
\begin{center}\begin{minipage}{\figurewidth}
\caption[]{\sl
The distribution of the recoil mass of $\mm$ pair in $\ee\ra\mm\ho$
for the 120 GeV SM Higgs normalized to 500 fb$^{-1}$ at $\sqrt{s}=250$ GeV.
Different beam parameters are used.
\label{fig:mmisr}}
\end{minipage}\end{center}
\end{figure}

The Higgs mass can be obtained either with direct reconstruction of the
jets from Higgs, or with recoil mass in case the associated Z$^{0}$ decays
to e$^{+}$e$^{-}$ or $\mu^{+}\mu^{-}$, or with all information combined with
fitting with energy-momentum (and $\Zo$ mass) constraints.
It can be compared to the W boson mass measurement ($\Delta m \sim 40$ MeV)
at LEP using the similar technique of the event-by-event direct 
reconstruction with about 50000 WW events in LEP which is
approximately the same order as the expected Higgs signal.
What is different with WW measurements at LEP and Higgs mass measurement at
JLC are;
\begin{itemize}
\item wider beam energy spread ($\sim0.5$\%)
\item better detector resolution
\item smaller (negligible) natural width of Higgs
\item precise knowledge of Z mass can be used
\end{itemize}
Precise calibration of the detector and beam energy
measurement are essential as of W mass measurement. 
Calibration runs at $\Zo$ pole 
might be helpful~\cite{omori}. 
It should be noted that the actual e$^{+}$e$^{-}$ collision energy
has a relatively large spread in linear colliders compared to circular
colliders like LEP. Also there we expect
significant beamstrahlung effect~\cite{beam-beam}
in exchange for the high luminosity.
These make a distortion of the reconstructed Higgs mass as well as 
the background processes when we use the beam energy constraints. 
In case we use the recoil mass or constraint fitting, also the
ISR at the physics collision affects the mass resolution if 
the center-of-mass energy is significantly higher than $\mH + \mZ$.
These effects are seen in Figure~\ref{fig:mmisr} for $\ee\ra\Zo\ho\ra\mm\ho$
for the SM Higgs with its mass of 120 GeV at $\sqrt{s}=250$ GeV.

The simplest way to determine the Higgs mass is to use lepton final
state of the associated Z (lepton channel). 
In this case, we measure the mass of the
Higgs as the recoil mass of the leptons. For the light Higgs of
120--140 GeV, our first target, 
we have production cross-section more than 100 fb$^{-1}$
at JLC phase-I for the SM-type Higgs.
The selected event rate for the lepton channel is about 2000 events 
per 500 fb$^{-1}$, and the mass resolution per event is $\sim$2 GeV 
at 300 GeV.
With design beam parameter at JLC, and including ISR, we expect
80 MeV for the precision of the mass from the lepton channels. 
Note that the precise
calibration for the electron energy scale in detector response
is important for electron channel. 
In order to avoid the possible bias in the energy scale, 
we may use the kinematic fitting.

If we utilize other channels, the statistics of the Higgs signal
jumps up. In the missing energy channel, the visible mass correspond to
Higgs boson. JLC detector has jet-jet invariant mass resolution of around
3 GeV for 120 GeV Higgs at 300 GeV.  Hence if we know the detector 
energy scale precisely enough we may measure the Higgs in this channel 
with the similar level as leptonic channel or even less. 
The precision is determined by the error in the energy scale in detector. 
The constraints fit of the missing mass to Z does not gain. 
If we calibrate the detector response in jet energy scale either 
using Z$^{0}$ calibration and/or energy scaling by visible energy peak
to $\Zo$ mass using $\Zo\Zo\ra\qq\nn$ especially for $\Zo\Zo\ra\bb\nn$,
we may achieve the necessary precision. 

Also we can use the four quark final states $\ee\ra\Zo\ho\ra\qq\bb$.
In this channel, the Higgs mass is in principle deducible from the
jet-jet invariant mass for jets assigned to Higgs, and/or recoil mass
of the jet pair associated to Z. However the jet clustering of decay
products from Higgs and Z mixes each other. A possible way is to use
the constraints fit or scaling with energy-momentum constraints.
In this case the jet direction, which is not affected much by the mixing,
can be used strongly. The mass of the Higgs is assigned event by event
as $M_{12}+M_{34}-\mZ$ where $M_{ij}$ is the jet-pair invariant mass after
the fitting. It is almost equivalent to 5-constraints fitting
(Z$^{0}$ constraints for a jet pair in addition to 
the energy-momentum conservation).
The mass resolution is governed 
mostly by the $\Zo$ natural width ($\sim 2$ GeV) and the beam energy spread.
In this case, again the knowledge of the absolute central beam energy 
and spread is essential. 
Also the effect of the ISR has to be controlled theoretically.

Through these measurements, we expect to be able to measure the Higgs mass
within accuracy of 40 MeV assuming the central beam energy is known 
to less than the error, 
which is roughly the same size as LEP WW mass as we expected. 

\begin{figure}[htbp]
\centerline{
\epsfxsize=5.0cm \epsfbox{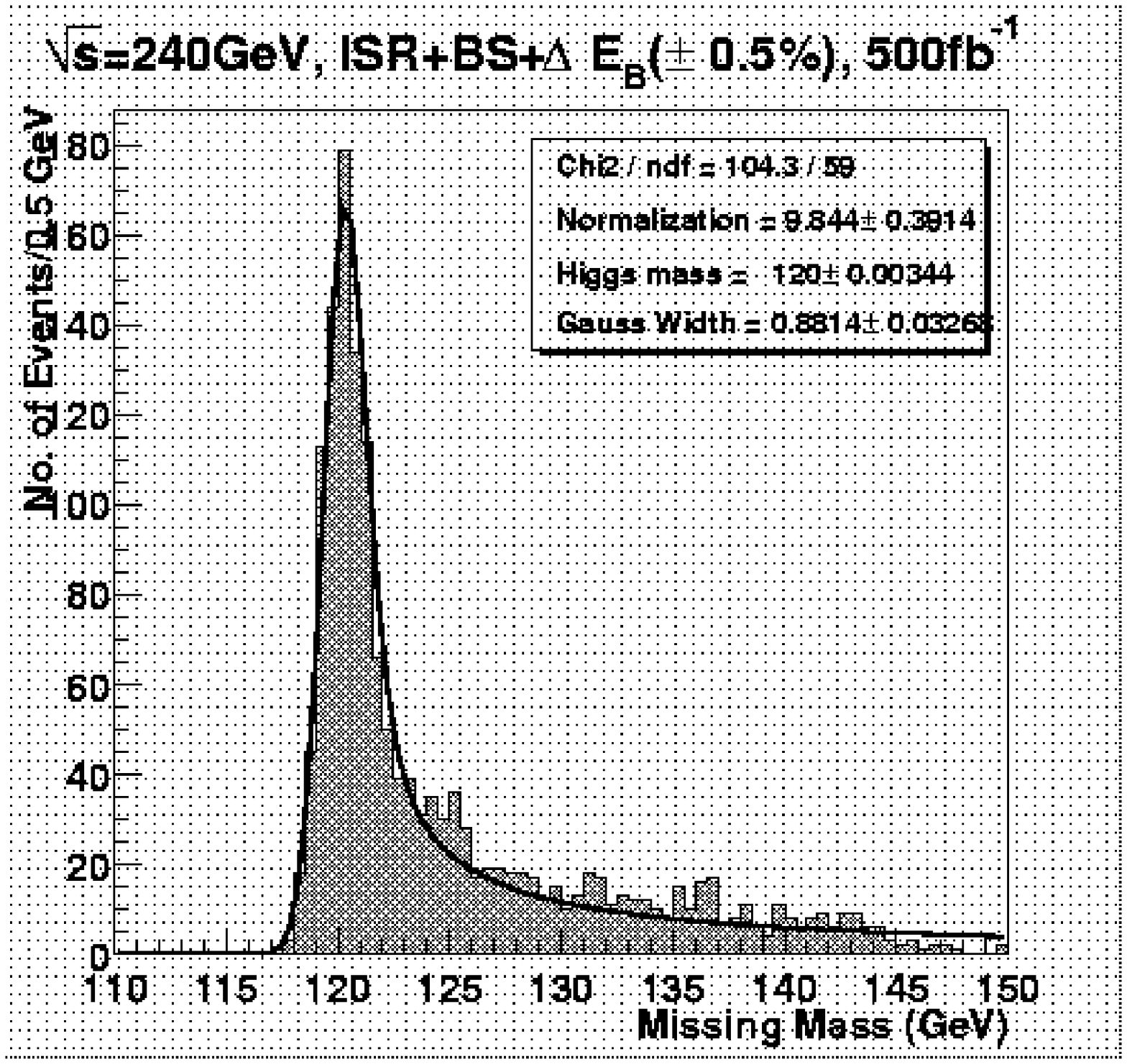}
\epsfxsize=5.0cm \epsfbox{physhiggs/MMNEW.25sb5m-500fb.eps}
\epsfxsize=5.0cm \epsfbox{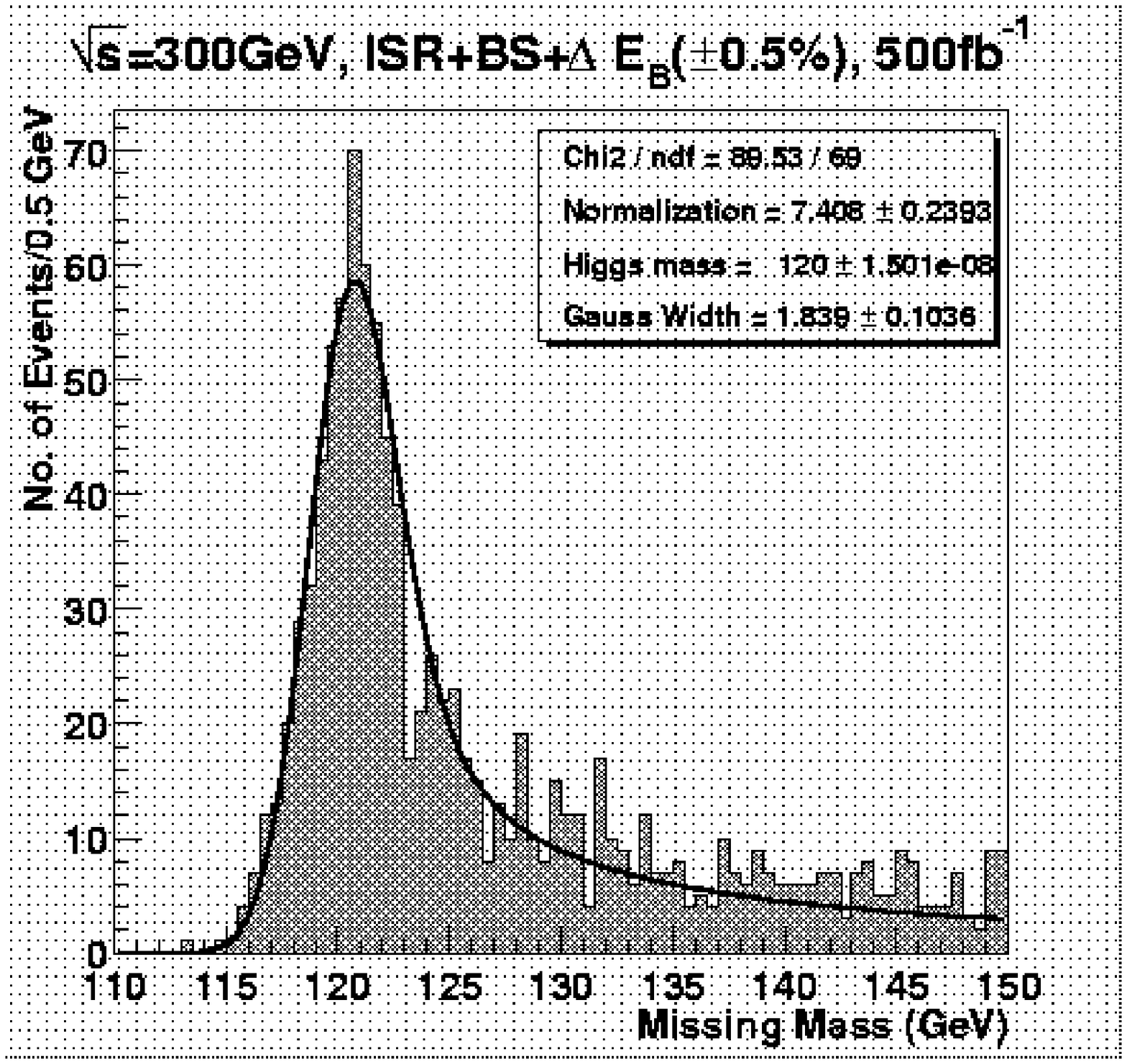}
}
\begin{center}\begin{minipage}{\figurewidth}
\caption[]{\sl
The distribution of the recoil mass of $\mm$ pair in $\ee\ra\mm\ho$
for the 120 GeV SM Higgs normalised to 500 fb$^{-1}$ at $\sqrt{s}=240$,
250 and 300 GeV. 
\label{fig:mmrmass_ecm}}
\end{minipage}\end{center}
\end{figure}

Note that the effects of the ISR and beamstrahlung on the mass measurement
significantly depend on the center-of-mass energy. Cross-section and
the momentum resolution also depend on the center-of-mass energy.
The dependence of the center-of-mass energies for the Higgs mass
reconstruction is shown in Figure~\ref{fig:mmrmass_ecm} for the muon channel. 
As shown in the figure, for 120 GeV Higgs, 
we have a recoil mass resolution of
about 0.8, 1.0 and 1.8 GeV for $\sqrt{s}=240$, 250 and 300 GeV, respectively.
Once we discover Higgs, beam parameters should be re-optimized.

\subsection{Quantum Numbers and Verification of ZZH Coupling}

It is essential to verify the particle found is produced via
ZZH coupling. If we establish the coupling, and verify the spin 0 scalar
particle, then it is a direct evidence that the particle should have a 
vacuum expectation value from the requirement of gauge coupling.

The final state of ZH is considered.
The verification is easily made via three steps;
\begin{enumerate}
\item s-channel production is verified from the production angle measurement.
\item The particle spin and parity is established also from the production
angle together with the decay angle, and also from the energy scan.
\item Then finally the question if the intermediated particle is Z or not
is verified for example with polarized beam option. 
\end{enumerate}

\subsection{Higgs strahlung and WW(ZZ)-fusion Cross-section Measurements}

The cross-section measurements of the $\Zo\ho$ 
(and WW-fusion) process mean the direct determination of the coupling 
between Higgs and weak bosons. 
The measurement is significantly important. If we find the cross section
even slightly different from what we expect from the SM, it means
we have more than one Higgs in the Universe.
In MSSM or type-II 2HDM, the cross-section normalized to the SM value 
is sin$^{2}$($\alpha-\beta$). If heavier Higgs is found in JLC or LHC,
we are already able to check the MSSM using the relation between
light Higgs mass, heavier Higgs mass and $\alpha-\beta$ measured above. 
The ZZh coupling is directly measured with 
Higgs-strahlung process
using $\ee\ra\qq\ho$, $\mm\ho$, $\tautau\ho$ or $\ee\ho$ 
(with small contribution from ZZ-fusion).
The neutrino channel $\ee\ra\ho\nn$ is important to measure
WWh coupling via WW-fusion process~\cite{Ishii}. 
The comparison between ZZh and WWh
couplings verify the SU(2)$\times$U(1) structure, which is
essential to establish the Higgs mechanism.
Once the SU(2)$\times$U(1) coupling is established, all channels are
combined to precisely determine both ZZh and WWh couplings. 

\subsubsection{HZ Cross-section measurement}

As we've described above, in the $\Zo\ho$ production, 
the event topologies are categorized
by the final state of the associated $\Zo$ boson, namely 
$\qq$, $\tautau$, $\nn$, $\ee$ and $\mm$. 
In order to measure the cross-section independently of the Higgs decay
branching, the simplest way is to use 
$\ee\ra\Zo\ho\ra\mm\ho$ channel in the recoil mass 
distribution of the muon pairs.
We assume HZ cross section of 200 fb which is similar to SM optimizing the 
beam energy, and 500 fb$^{-1}$ integrated luminosity. 
We have about 1500 selected 
signal events in the mass window with background of 400 events for 120 GeV.
It corresponds to about 3\% error on $\Delta\sigma/\sigma$.
We add the statistics with electron channel which includes small contribution
from the ZZ-fusion to be corrected. 
These two lepton channels already reach to 2\% level. 
Systematic errors are the uncertainties in the selection efficiency
of the signal due to acceptance such as isolation requirement 
of lepton and jets, that in the background and signal shapes especially
due to uncertainty of the tail in the beam energy spectrum. 
These errors can be controlled in the data itself using similar event
topology like ZZ process for which the cross-section is known already at
the 1--2 \% level at present, and theoretical works are in progress.

These leptonic channels have small branching compared
to all $\Zo\ho$ production due to only 3 \% of the branching 
from $\Zo$ to each lepton flavour.
Hence the inclusion of the other channels 
may be helpful. 
However the current analysis for other channels needs requirements
to the Higgs decay products in order to purify the signal.
Once we measure the branching ratio in dominant modes such as 
to $\bb$ (or $\WW$ for higher mass), 
the other channels might also be used to determine the 
cross-section more precisely. 
In this case, however, the precision of the cross-section 
measurement is determined by that of the Br-measurements in this case,
which is expected to be the same level. 
See below.

\subsubsection{WW-fusion Cross-section measurement}

The WW-fusion process is a key to establish the universality between
WWH and ZZH coupling in SU(2)$\times$U(1) symmetry breaking.
The WW-fusion happens for final state of $\ho\nn$ for the electron neutrino.
The interference between fusion and Higgsstrahlung is not negligible, and
the cross-section for the WW-fusion process increase quite rapidly 
as a function of the beam energy.
In this final state, $\ho\nn$, all the visible particles are the decay
products of the Higgs.
The first signature of this final state is the large missing energy due
to the escaping neutrino.
In order to free from the Z-mediated or WW-fusion,
we do not require the missing mass to be Z mass.
The main background at $\sqrt{s} < 350$ GeV is the WW background
with leptons from a W untagged by the detector. One of the
case is the lepton going inside beam pipe direction, and
in case the lepton is tau with its decay products close to other jet.
Detector hermeticity, lepton tagging even inside the jets, and tau tagging
are essential.
The visible mass of the event is the most essential.

There are many ways to investigate the $\ho\nn$ final state for the WW fusion.
\begin{itemize}
\item simple counting experiment with tight selection
\item shape fits in the missing mass and/or production angle
\item energy-scan
\item polarized beam
\end{itemize}

So far we have studied
the $\ho\nn$ signal detection in the JLC-I simulation in a simple 
counting experiment, 
using a loose b-quark tagging based on the impact parameter 
to further suppress the background, especially WW process. 
First attempt was made neglecting the interference between WW-fusion and
Higgsstrahlung.
The center-of-mass energy and Higgs mass are set to 
$\sqrt{s}$=300 GeV and $\mh$=120 GeV.
The analysis uses a standard set of cuts
like acoplanarity, multiplicity.
Finally a rather wide range of the mass from 100--130 GeV are sought
to count the number of events.

The expected number of signal and background events are 
summarized in the table~\ref{res-wwh}.
A total signal selection efficiency is estimated to be $\sim$ 50 \%.
A statistical error of the measurement of the Higgs production cross-section 
times branching ratio to $\bb$ in $\ho\nn$ final states
including HZ and WW fusion process is found to be
$\delta\sigma / \sigma$ = 1.2\% 
at an integrated luminosity 
of 500 pb$^{-1}$. 

\begin{table}
\begin{center}\begin{minipage}{\figurewidth}
\caption{\sl
The expected number of signal and background events
at the center of mass energy of 300 GeV, 
Higgs mass of 120 GeV and an integrated
luminosity of 500 fb$^{-1}$ which corresponds to several years JLC running.
\label{res-wwh}}
\end{minipage}\end{center}
\begin{center}
\begin{tabular}{|l|c|} 
\hline
Process & Selected Number of Events\\ \hline
HZ($\nu\nu$bb) & 7015 \\
WW fusion($\nu\nu$bb) & 3580 \\ \hline\hline
Other Higgs signal & 231.5 \\ \hline
$\ee\ra\qq(\gamma)$  & 348.5 \\
WW background & 998 \\
ZZ background & 4990 \\
We$\nu$ background & 245 \\
Zee background & 150.5 \\ \hline
\end{tabular}
\end{center}
\end{table}

An attempt was also 
made to suppress the Higgs strahlung with a requirement on the
missing mass to be larger than 140 GeV.
The 2480 events of WW fusion process are expected to pass the additional cut, 
while the other Higgs process events and the total backgrounds 
are suppressed to be 1030 and 2475, respectively.
An error of the measurement of the Higgs production cross-section
for WW fusion process times b-quark decay branching 
is obtained to be about 3.1\% in this case.

The better selection using likelihood or neural network may help.
In order to deduce the
absolute cross-section independent to branching ratio, we again need the
accurate measurement of the branching ratio to $\bb$ in this case
as well as the accurate estimate of the signal and background selection
efficiency. Fitting in shape of the missing mass and production angle
may be much powerful than the method described above in order to
check the contribution of the WW-fusion in $\ho\nn$ process, since the
absolute strength of the HZ process can be normalized to what we obtained
by the HZ cross-section measurement in other channels mentioned above.
We also have many other ways to investigate the contribution from the WW-fusion
process as are mentioned above. 

\subsection{Branching Ratio Measurements}

\subsubsection{Outlook}

The measurements of the branching ratio (Br-measurement) 
of the Higgs is one of the keys to distinguish
the SM and other models such as SUSY.
Once we determine the Higgs mass, the branching ratio
of the SM Higgs is completely determined by the known SM parameters.
On the other hand, for SUSY Higgs, the branching can be affected by
other SUSY parameters such as $\Ao$ mass or tan$\beta$, 
which result in the different
branching ratio from the SM case. 

The measurements of the branching ratio (Br) 
of the Higgs is one of the keys to distinguish
the SM from other models such as SUSY.
Once we determine the Higgs mass, the branching ratios
of the SM Higgs are calculable with the known SM parameters.
On the other hand, for SUSY Higgs or other Higgs models, 
the branching can be affected by
other model parameters such as $\Ao$ mass or tan$\beta$, 
which result in the different
branching ratio from the SM. 

When we assume heavy ($\sim 2$ TeV) mass for $\Ao$ and other SUSY particles,
the branching ratios in MSSM are very similar to those of the SM by the
decoupling theorem.
There are a lot of Br-measurements. Of these the most interesting are
the Br-measurement of Higgs decaying into $\bb$, $\tautau$ and $\WW$. 

When we combine the Br($\ho\ra\WW$) with the cross-section measurement,
the total width of the Higgs can be deduced. 
The absolute coupling of Higgs and
W boson can be determined by the cross-section measurement 
either of $\Zo\ho$ production
with SU(2)$\times$U(1) coupling assumed, or of the WW-fusion process. 
The partial width of the Higgs to $WW$, $\Gamma_{\ho\ra\WW}$, is calculable
from the coupling strength obtained by the cross-section measurement.
It can be written as 
\begin{equation}
{\Gamma_{\ho\ra\WW} = \Gamma_{\ho\ra\WW}^{\mathrm{SM}} \times 
\frac{\sigma_{\ho\Zo}}{\sigma_{\ho\Zo}^{\mathrm{SM}}}}~~~~, 
\end{equation}
where $\Gamma_{\ho\ra\WW}^{\mathrm{SM}}$ is the partial width of 
the $\ho\ra\WW$ in the SM, $\sigma_{\ho\Zo}$ is the production 
cross-section obtained in the experiment, 
and the $\sigma_{\ho\Zo}^{\mathrm{SM}}$ is that for the SM.
Finally the total width of the Higgs $\Gamma_{\ho}$ is calculated as
\begin{equation}
{\Gamma_{\ho} = \frac{\Gamma_{\ho\ra\WW}}{{\mathrm{Br}}(\ho\ra\WW)} =
\Gamma_{\ho}^{\mathrm{SM}} \times 
\frac{{\mathrm{Br}}^{\mathrm{SM}}(\ho\ra\WW)}{{\mathrm{Br}}(\ho\ra\WW)}
\times\frac{\sigma_{\ho\Zo}}{\sigma_{\ho\Zo}^{\mathrm{SM}}}}~~~,
\end{equation}
where $\Gamma_{\ho}^{\mathrm{SM}}$ is the total width for the SM Higgs.

In a similar manner, 
the absolute strength of the Yukawa-couplings $\lambda_{f}$ of 
the Higgs boson to fermions can be deduced in a model independent fashion,
using the relative branching ratio 
Br($\ho\ra\ff$)/Br($\ho\ra\WW$) together with
the measured cross-section.

\begin{equation}
\lambda_{f}^{2} = \lambda^{2}_{f,SM}\times
\frac{\mathrm{BR}(\ho\ra\ff)\cdot\sigma_{\rm HZ}\cdot\mathrm{BR}^{SM}(\ho\ra\WW)}{\mathrm{BR}^{SM}(\ho\ra\ff)\cdot\sigma^{SM}_{\rm HZ}\cdot\mathrm{BR}(\ho\ra\WW)}~~~~, 
\end{equation}

The ratio Br($\ho\ra\tautau$)/Br($\ho\ra\bb$) is also the one of the
essential measurements. In SM, various SUSY, and also in more general
type-II two Higgs doublets model, an universal coupling to the charged
lepton and down-type quarks are assumed at tree level, \\
Br($\ho\ra\tautau$)/Br($\ho\ra\bb$)=$m^{2}_{\tau}/m^{2}_{b}$. 
While the different absolute values 
of the Yukawa-coupling strength are expected in models, 
the ratio Br($\ho\ra\tautau$)/Br($\ho\ra\bb$) is the
same in the SM and type-II models. 
This is the key measurement to make a definite answer in the origin
of the mass in the lepton sector.
On the other hand, if we assume 
the Type-II coupling like SM or MSSM,
the measurement serves as the precise measurement of the running
b-quark mass at the Higgs mass.
Note that the loop effects by SUSY particles especially scalar top/bottom 
quarks and gluinos could make a significant
correction in the decay branch into quarks. Hence the measurement is one of
the key to determine SUSY parameters as described later
if we live in the SUSY world.

The measurements of 
Br($\ho\ra\cc$)/Br($\ho\ra\bb$) is also the important to measure
the Yukawa-coupling for up-type quarks.
Br($\ho\ra gg$)/Br($\ho\ra\bb$) or Br($\ho\ra\cc + gg)$)/Br($\ho\ra\bb$)
are also to be used to determine the
Yukawa-coupling for up-type quarks since 
the gluonic decay occurs mainly via the top quark loop in most of the models.
  
\begin{figure}
\centerline{
\epsfxsize=8.6cm \epsfbox{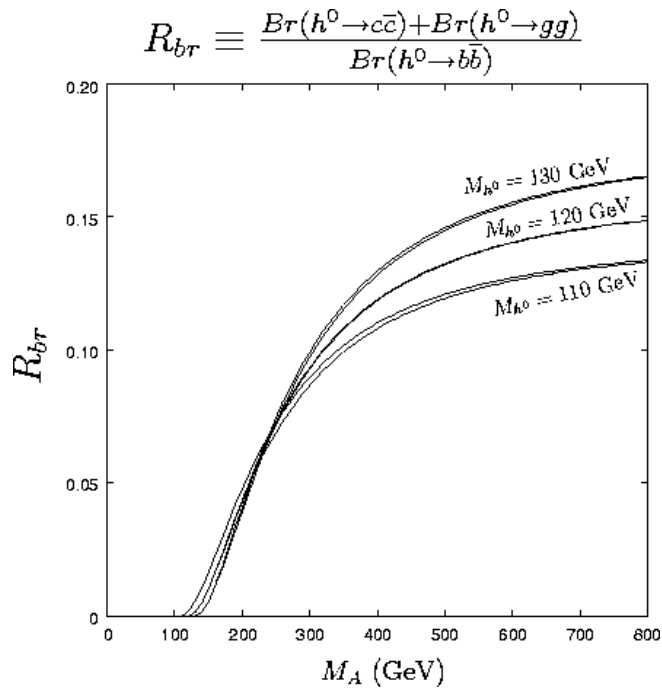}
}
\begin{center}\begin{minipage}{\figurewidth}
\caption[]{\sl
The ratio Br($\ho\ra\cc +gg$)/Br($\ho\ra\bb$) predicted by 
J.~Kamoshita et al~\cite{kamoshita} in MSSM, as a function of the
mass of the CP-odd Higgs ($\Ao$). From the top, the set of parameters
$(m_h, M_{SUSY}, A_t, \mu)$ are (130, 5000, 10000, 500),
(130, 10000, 20000, 500), (120, 1000, 2000, 300), (120, 5000, 5000, 300),
(110, 1000, 0, 300), and (110, 500, 750, 300) in GeV, and $\tan{\beta}$ is 
 solved to fix $m_h$.
\label{fig:kamoshita}}
\end{minipage}\end{center}
\end{figure}

In SUSY,
the gluonic decay can happen also via scalar top or scaler bottom loops, 
hence comparison between gluonic decay branching and Yukawa coupling to be
directly measured at LHC and/or by the $\ee\ra\toptop\ra\toptop\ho$ at JLC 
could give us information in the scaler quark sector.
The ratios relative to those of the SM are proportional to 
tan$^{-2}\alpha\times$tan$^{-2}\beta$ which can be approximated as
$[(\mA^{2}-\mh^{2})/(\mA^{2}-\mZ^{2})]^{2}$ 
for large tan$\beta$, which leads the indirect estimation of
CP-odd Higgs $\Ao$ mass~\cite{kamoshita,kawagoe}.
However one should note that the theoretical ambiguities is large
due to uncertainty of $\alpha_s$ and running c-quark mass.
Theoretical and experimental improvements are awaited.

The measurements of 
Br($\ho\ra\cc$)/Br($\ho\ra\bb$), Br($\ho\ra gg$)/Br($\ho\ra\bb$) or
Br($\ho\ra\cc + gg$)/Br($\ho\ra\bb$) have interesting feature.
The gluonic decay occurs mainly due to the top quark loop, and it has similar
SUSY-parameter dependence to the Higgs decay to c-quarks.
J.~Kamoshita et al~\cite{kamoshita} calculated the 
effects of the SUSY-parameters in MSSM as shown in Fig~\ref{fig:kamoshita}.

The expected value of Br($\ho\ra\cc + gg$)/Br($\ho\ra\bb$) for 110 GeV Higgs
in the SM is about 15 \%.
The ratios are sensitive to $\Ao$ mass in MSSM, with little
dependence on other SUSY parameters such as scalar top mass.
The figure indicates that we can discriminate SM and MSSM in case
$\Ao$ mass less than 500 GeV with the Br-measurement accuracy of 10 \%.
Preliminary expectation of the experimental sensitivity of the 
Br-measurement
are found in Ref~\cite{suneokun} and ~\cite{kawagoe}. 
However one should note that the line in the figure have, in relative, 
about 10 \%
ambiguity due to the uncertainty of the c-quark mass and strong
coupling constant which affect the gluon radiation rate, while
we expect more precise measurement of the $\alpha_{s}$ at JLC. 
Theoretical improvements are awaited.

When we assume a model such as MSSM, the results of the various 
decay modes can be combined to estimate the masses of the 
other Higgs and other model parameters in the model assumed, 
like $\mA$, $\mH$ and tan$\beta$ of the MSSM.
These estimated values would be examined in comparison with
results from LHC and/or those from the next stage of the JLC,
from which we could obtain the answer for the physics up to GUT scale.

\subsubsection{Current Analysis Procedure}

Here we discuss the results of the simulation
studies, which are very preliminary in the sense that the analysis 
is not optimized yet,
that the b-tagging, tau, and other flavour tagging 
is simple and very primitive ones without using vertex reconstruction
which is known to be much powerful than those of simple impact parameter
used in the present analyses.
Also the helpful signal contribution from the WW-fusion process
at higher energy is neglected. Hence the results is to be considered as
a conservative estimates in the analyses procedure, while
the degradation of the performance due to accelerator noise and
two-photon~\cite{twophoton} events overlap are neglected. 

The selection of the events are made using a simple likelihood method
combining the kinematical information and flavour-tagging.
The most major background is ZZ process in $\bb$ decay.
For the rare decay process such as $\ho\ra\cc$ or gluons, the dominant
background is the Higgs signal itself with different decay modes. 

As an initial studies, the benchmark Higgs mass of 120 and 140 GeV are
considered. The HZ production processes as well as the background
processes are simulated using JSF quick simulator.
The analysis start from the event topology classification, followed by
the Higgs tagging. Since we can use the mass determined by JLC itself,
the Higgs mass can be used as an additional constraints in the selection.

The W decay of the Higgs was studied in the sequence
$\ee\ra\Zo\ho\ra\qq WW^{*}\ra\qq l\nu\qq$, 
$\ee\ra\Zo\ho\ra\ellell WW^{*}\ra\ellell l\nu\qq$.
The most important key in the analysis is the good energy-momentum
measurements of the jets and an event.
In often cases, one of the W from Higgs decay has mass close to the
W-pole ($\sim$on-shell). 
In case the W decays hadronically, the jet-jet invariant mass
is to be requested to be close to W mass. For case the W decays to
lepton and neutrino, we can also demand similarly since the momentum
carried by the neutrino can be measured as the missing one in the events
in the final state above. 
In these channels, the Higgs mass is calculated by the invariant mass of
the four momenta of the two W's. We may improve the resolution by a kinematic
fitting requiring the total energy-momentum, negligible neutrino mass, 
W mass for one of the W's, etc.. This is also the next subject to study. 

\begin{figure}
\centerline{
\epsfxsize=5.92cm \epsfbox{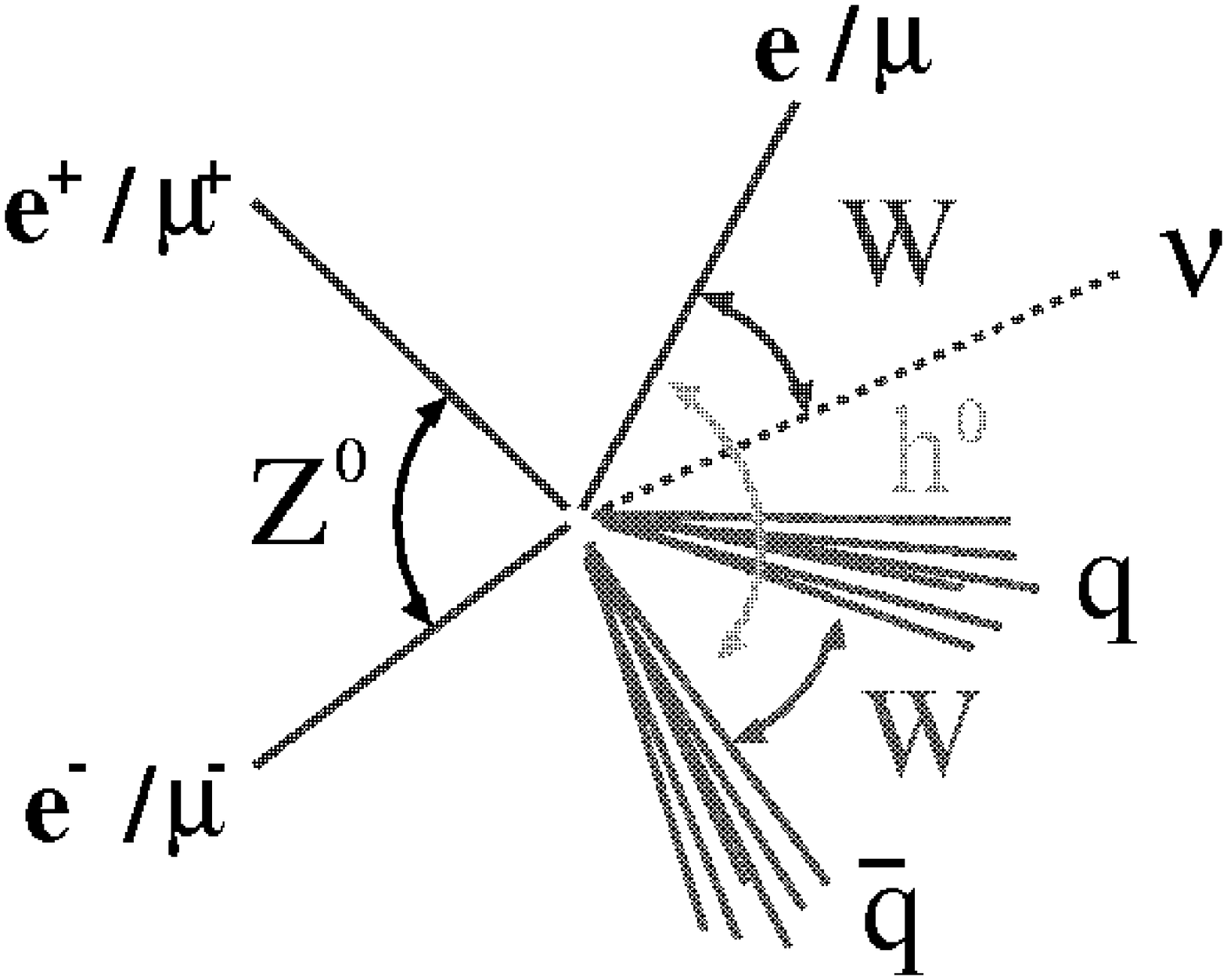}
\epsfxsize=6.72cm \epsfbox{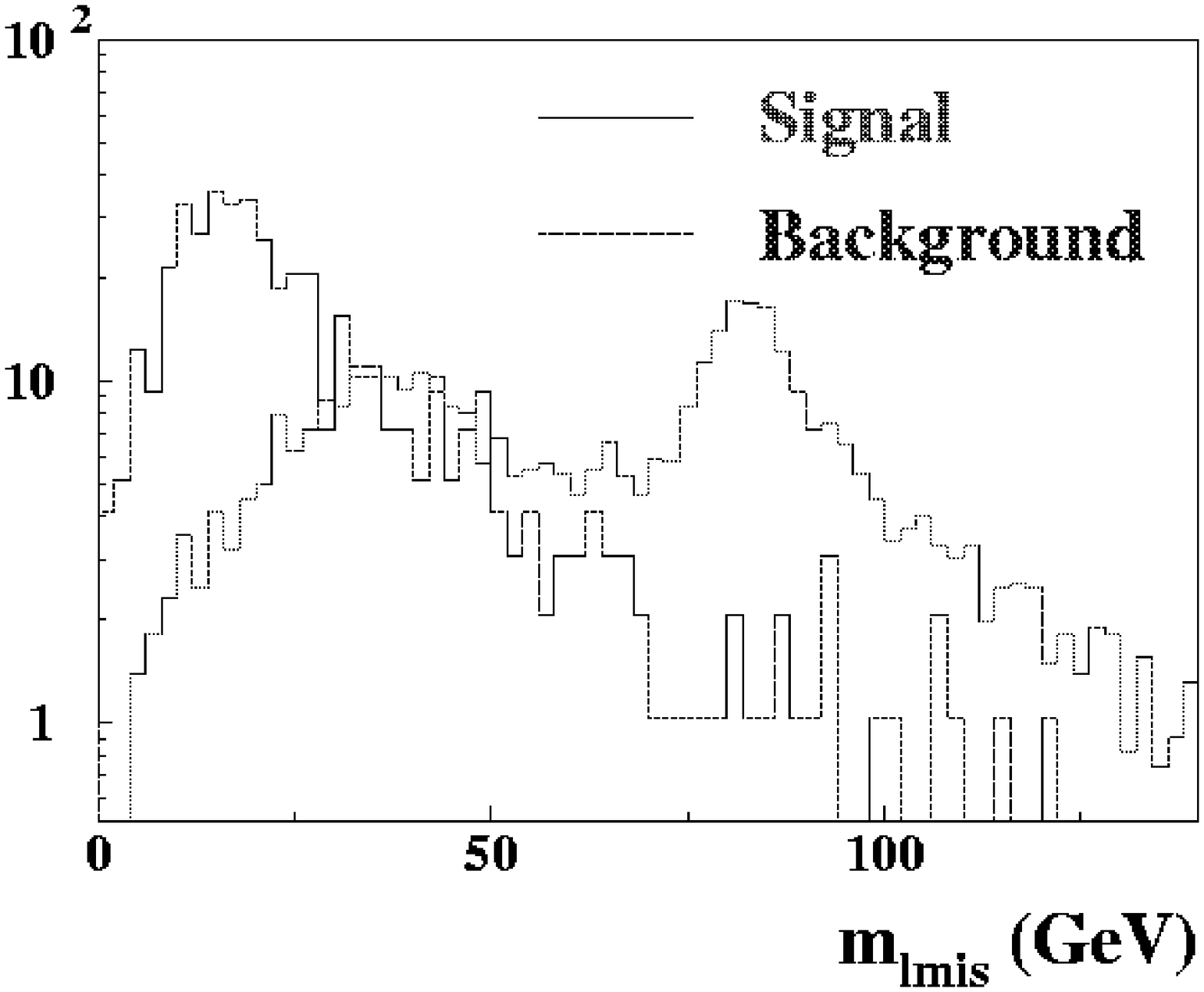}
}
\begin{center}\begin{minipage}{\figurewidth}
\caption[]{\sl
$\ho\ra\WW$: (Left) $\ee\ra\ho\Zo\ra\WW\ellell$ topology
in leptonic channel ($\Zo\ra\ee$ or $\mm$).
(Right) Invariant mass of 3rd lepton and $p_{miss}$ normalized 
to 10 fb$^{-1}$ (about 10 days running at JLC). 
For signal, plots for 100\% decay of the $\ho\ra\WW$ for
120 GeV Higgs with production cross-section of the SM Higgs.
\label{fig:schemeww}}
\end{minipage}\end{center}
\end{figure}

Note that the recent studies by A. L. C. Sanchez, J. B. Magallanes et al 
in channel $\ee\ra\Zo\ho\ra\qq WW^{*}\ra\qq\qq\qq$, is also 
on going~\cite{phillipin-acfa}.
The important other channel
$\ee\ra\Zo\ho\ra\nn WW^{*}\ra\nn\qq\qq$
should also be studied. In this case the 4 jet topology with huge missing mass
should be the initial request, and one of the jet-pair invariant mass
close to W may be added. An anti-b tagging may also be helpful to suppress the
background from $\bb$ decay of the Higgs with multi hard gluon radiation. 
Also the process, $\ee\ra\Zo\ho\ra\ellell WW^{*}\ra\ellell\qq\qq$, should be
studied.

For the $\qq$ or gluonic decay of the Higgs,
the selection were made imposing the $\mZ$ and known Higgs mass.
Finally the flavour of the jets assigned to the Higgs are examined and
measure the relative branching ratio for the Higgs decay into quarks or gluons.
The signature of the gluonic decay is the rather spherical shape of the
jets and high multiplicity, while the c-quark decay tends to have sharp jets
with much less multiplicity.   

In case the Z decays to muons, electrons or quarks, the tau decay
is feasible to identify. 
The tau identification is the key of the measurement.
Recoil mass of the leptons or jets originating from the Z also
represents the Higgs mass. 
The direction of the decay products
of the taus well represents the direction of the taus. Hence 
either constraint fits or scaling can be used to have better mass resolution
of the invariant mass of the tau-pairs which is to be requested close to Higgs
mass. There are many rooms to improve the analyses.

\begin{table}
\begin{center}\begin{minipage}{\figurewidth}
\caption[]{\sl
The simulation results for selection optimized for
$\ho\ra\bb$ at JLC $\sqrt{s}=300$ GeV: 
Number of events selected for 120 GeV. The SM Higgs cross-section
and the branching ratio is used as an input. Event counts are
normalized to ${\cal{L}}$=500 fb$^{-1}$. Fusion process, which is
helpful to increase the signal statistics, is not included
in the analysis.
\label{tab:bbanal}}
\end{minipage}\end{center}
\begin{center}
\begin{tabular}{|l|c|c|c||c|}
\hline
 Selection    &  $\qq\qq$  &  $\nn\qq$ &  $\ellell\qq$ & Total \\
\hline
\hline
$\ho\ra\bb$         &  6880   & 1580 & 1129 & 9590 \\\hline\hline
$\ho\ra\cc$         &    95   &   28 &   15 &  139\\
$\ho\ra$gg          &    63   &   29 &   12 &  104\\
$\ho\ra\WW$         &    67   &   31 &   18 &  116\\
$\ho\ra\tautau$     &     0   &    0 &    0 &    0\\\hline
non-Higgs SM Bkg       &   869   &  374 &  134 & 1734\\\hline
\end{tabular}
\end{center}
\end{table}

\begin{table}
\begin{center}\begin{minipage}{\figurewidth}
\caption[]{\sl
The simulation results for selection optimized for
$\ho\ra\WW$ at JLC $\sqrt{s}=300$ GeV: 
Number of events selected for 120 GeV. The SM Higgs cross-section
and the branching ratio is used as an input. Event counts are
normalized to ${\cal{L}}$=500 fb$^{-1}$. 
\label{tab:wwanal}}
\end{minipage}\end{center}
\begin{center}
\begin{tabular}{|c|c|c||c|}
\hline
$\ho\Zo\ra\WW\Zo\ra$ &  $\ell\nu\qq\qq$ & $\ell\nu\qq\ellell$ & Total \\
\hline
\hline
Efficiency for $\ho\Zo\ra\WW\qq$ & 5.8\% & --- &  \\\hline\hline
Efficiency for $\ho\Zo\ra\WW\ellell$ & --- & 17.3\% &  \\\hline\hline
Selected Signal &  497   &   142 &   689 \\\hline\hline
Bkg from other Higgs decay & 44 & 40 & 84 \\
Bkg from non-Higgs SM      & 397 & 77 & 474 \\\hline
Total Bkg &  440 & 117 & 557 \\\hline
\end{tabular}
\end{center}
\end{table}

\subsubsection{Results of the preliminary simulation studies}

\begin{figure}
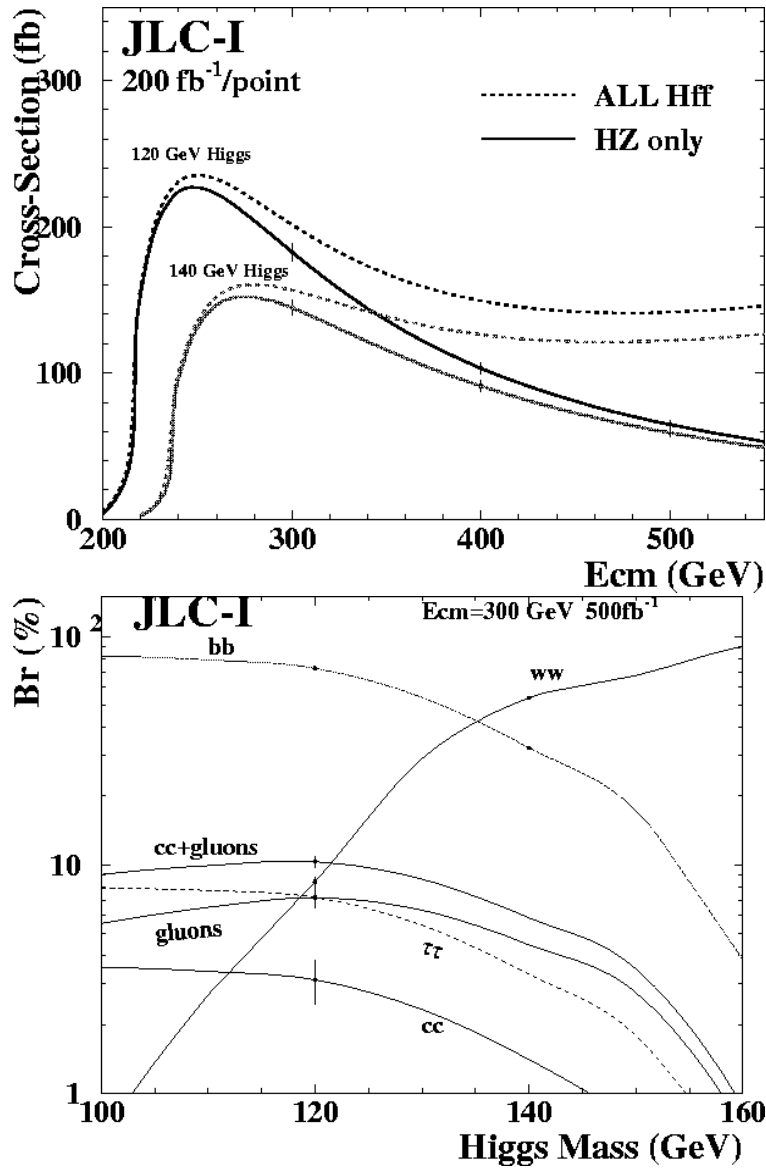

\centerline{
\epsfxsize=10.0cm \epsfbox{physhiggs/jlc_xsec.epsi}
}
\centerline{
\epsfxsize=10.0cm \epsfbox{physhiggs/jlc_br.epsi}
}
\begin{center}\begin{minipage}{\figurewidth}
\caption[]{\sl
(Top) Production cross-section 
(for M$_{\mathrm{H}}$=120 and 140 GeV) as a function of $\sqrt{s}$.
The accuracy in the cross-section measurement in $\ho\Zo$ process
are shown at 300, 400 and 500 GeV as error bars, assuming 200 fb$^{-1}$ 
at each $\sqrt{s}$ point.
(Bottom) Branching ratio of the SM Higgs as a function of the
Higgs mass. The measurements errors obtained in the simulation studies 
done so far are also shown in error bars.
Note that the theoretical errors are not shown in this plot.
\label{fig:jlcxsec}}
\end{minipage}\end{center}
\end{figure}

\begin{figure}
\centerline{
\epsfxsize=10.0cm \epsfbox{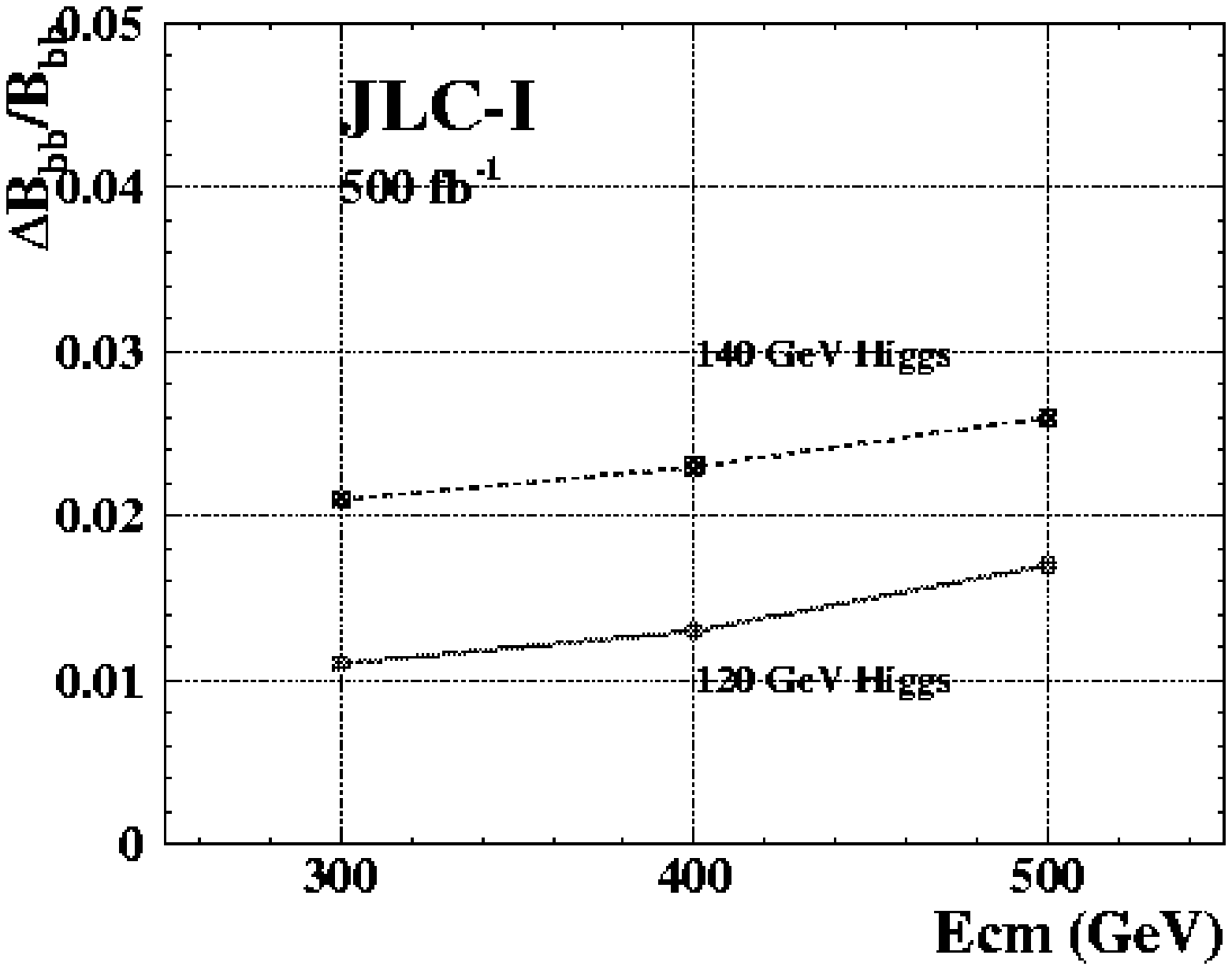}
}
\centerline{
\epsfxsize=10.0cm \epsfbox{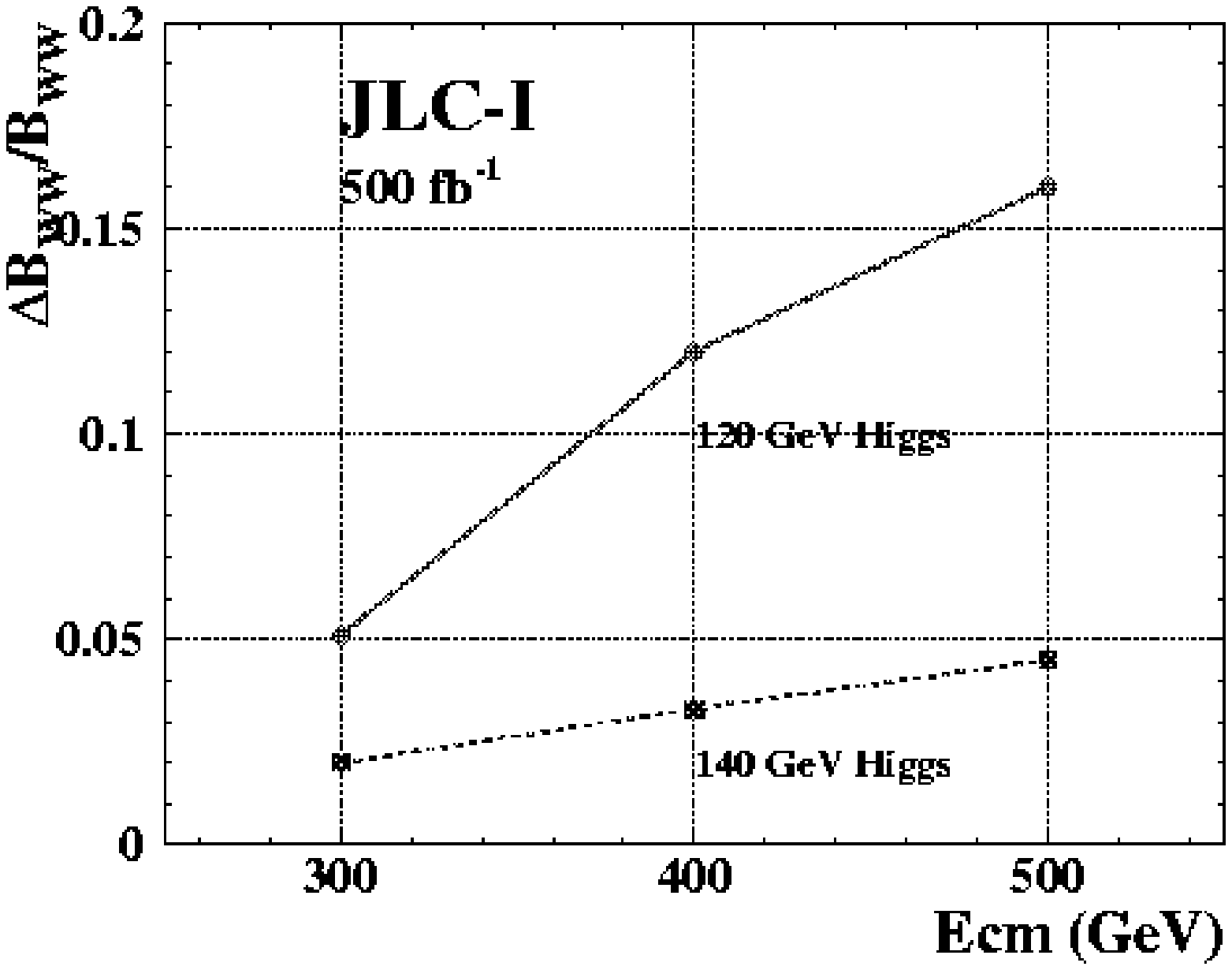}
}
\begin{center}\begin{minipage}{\figurewidth}
\caption[]{\sl
Expected measurement error $\Delta$Br/Br in Br($\ho\ra\bb$) and
Br($\ho\ra\WW$), for 120 and 140 GeV SM Higgs as a function of $\sqrt{s}$. 
Luminosity of 500 fb$^{-1}$ ($\sim$3 years JLC-I running) 
is used:
Note that the huge contribution from the WW-fusion especially at the high
energy is neglected in this analysis. 
Including the process, the 
sensitivity at 500 GeV is expected to be similar to that of 300 GeV. 
\label{fig:brw}}
\end{minipage}\end{center}
\end{figure}

\begin{figure}
\centerline{
\epsfxsize=10.0cm \epsfbox{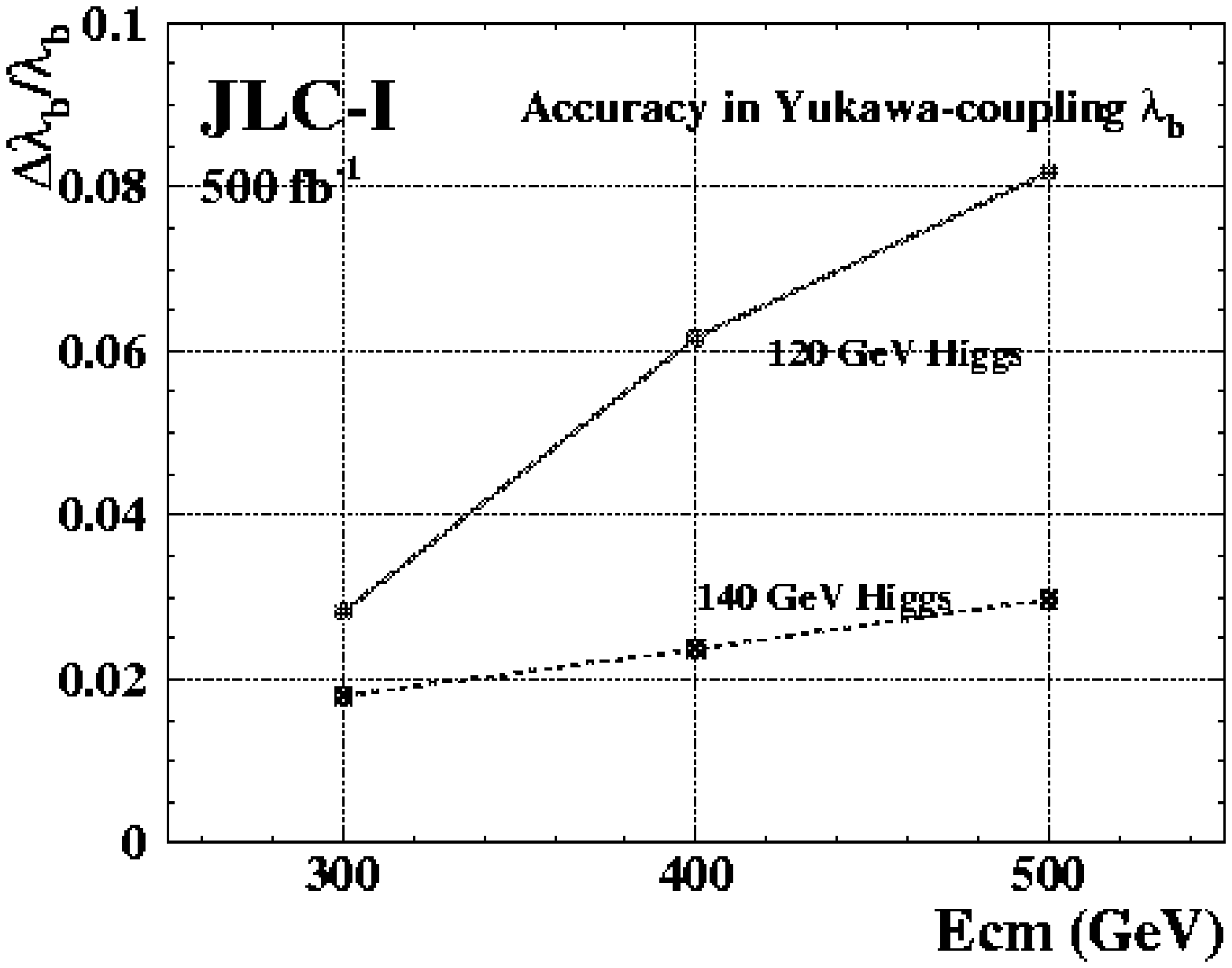}
}
\centerline{
\epsfxsize=10.0cm \epsfbox{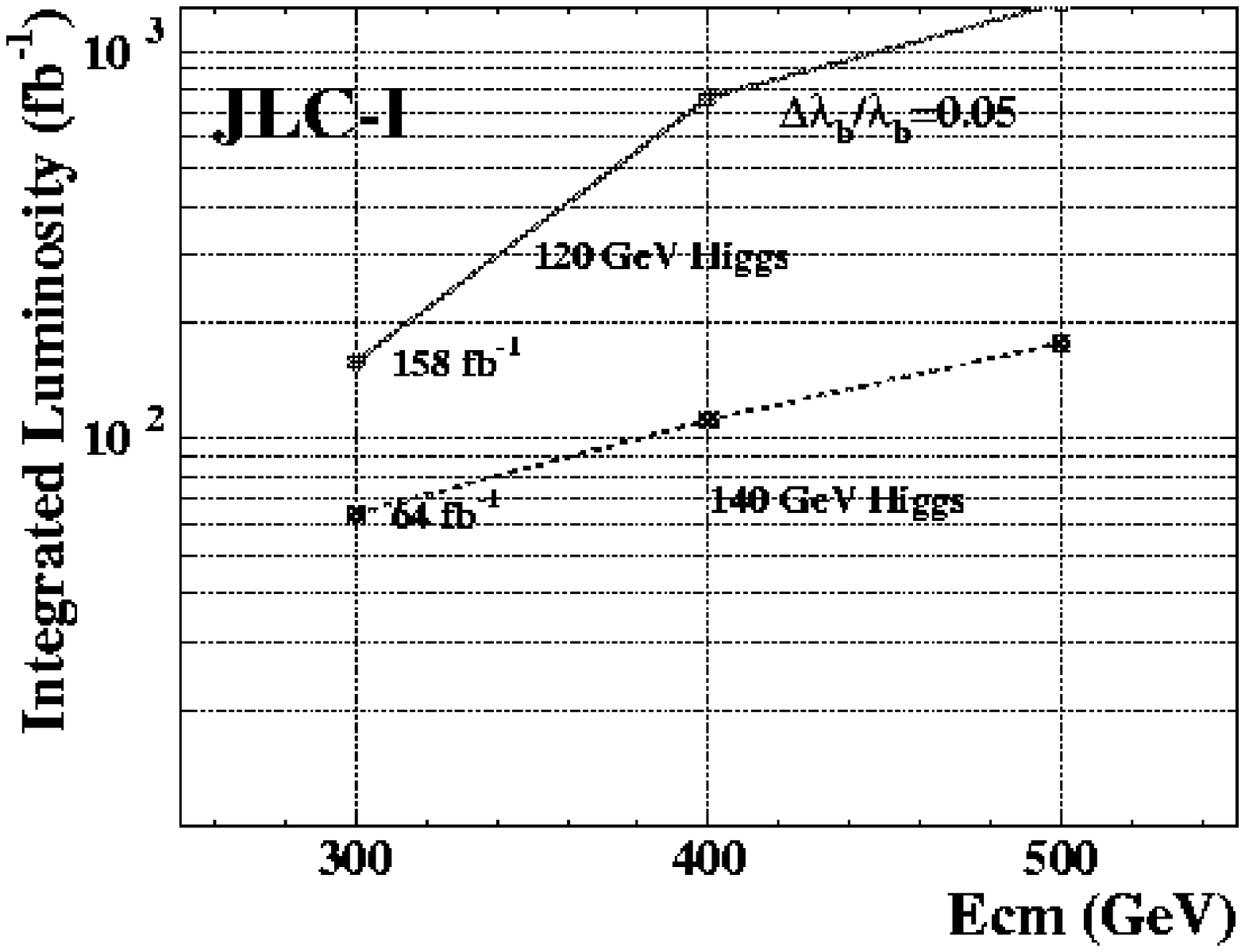}
}
\begin{center}\begin{minipage}{\figurewidth}
\caption[]{\sl
(Top) expected measurement error in Yukawa-coupling to b-quark, 
$\Delta\lambda_b/\lambda_b$,
for 120 and 140 GeV SM Higgs as a function of $\sqrt{s}$. Luminosity of 500
fb$^{-1}$ ($\sim$3 years JLC running) is used:
(Bottom) necessary luminosity for measurement of
$\lambda_b$ to achieve 5\% error which is similar to current knowledge of the
b-quark mass.
The luminosity of less than 160 fb$^{-1}$ is necessary to achieve the
5\% accuracy for 120--140 GeV Higgs.
Note that the huge contribution from the WW-fusion especially at the high
energy is neglected in this analysis. 
Including the process, the 
sensitivity at 500 GeV is expected to be similar to that of 300 GeV. 
\label{fig:brb}}
\end{minipage}\end{center}
\end{figure}

Many studies have been done for varieties of decay channels 
with simulation with JLC-I model detector, at the typical benchmark
points in center-of-mass energies and Higgs masses. Some of the results
are summarized in Tab~\ref{tab:summary1} and \ref{tab:summary140}. 
Figure~\ref{fig:jlcxsec} shows the expected accuracies of 
the cross-section and branching ratio measurements.
The dependence of the center-of-mass energy on the 
sensitivity of the branching ratio and Yukawa-coupling 
are shown in figures~\ref{fig:brw}
and \ref{fig:brb}, respectively. Note that we neglect 
the huge signal contribution
especially at higher energy from the WW-fusion process. 
Hence the signal events used in these analyses are considerably 
less at higher energies (400 and 500 GeV) than 300 GeV, which leads
the worse sensitivity at higher energies. 
When we include the WW-fusion process, the sensitivity is expected
to be similar at 300--500 GeV.
Table~\ref{tab:summarysummary} summarizes the accuracies of the
measurements obtained in these studies so far for $\sqrt{s}=$300, 400 and
500 GeV, for a Higgs boson mass of 120 GeV, with $\cal{L}=$500 fb$^{-1}$.

\begin{table}
\begin{center}\begin{minipage}{\figurewidth}
\caption[]{\sl
Accuracy in the branching ratio obtained in the 
analysis for ${\cal{L}}$=500 fb$^{-1}$, and for 120 GeV Higgs. Note that
the WW-fusion process, which increases the signal statistics significantly
at higher energy, is not used in the analysis so far.
\label{tab:summary1}}
\end{minipage}\end{center}
\begin{center}
\begin{tabular}{|c|c|c|c|}
\hline
   $\sqrt{s}$    &  300 GeV &  400 GeV &  500 GeV \\
\hline
\hline
$\Delta$Br/Br & & & \\
$\ho\ra\bb$         &  1.1\%   & 1.3\% & 1.7\% \\
$\ho\ra\WW$         &  5.1\%   & 12\%  & 16\% \\
$\ho\ra\tautau$     &  4.4\%   & ---   & --- \\
$\ho\ra\cc$+gg      &  6.3\%   & ---   & --- \\
$\ho\ra\cc$         &  22\%    & 23\%  & 27\% \\
$\ho\ra$gg          &  10\%    & 11\%  & 13\% \\
$\ho\ra\gamma\gamma$& ---      & --- & --- \\
$\ho\ra\Zo\gamma$   & ---      & --- & --- \\
\hline
\end{tabular}
\end{center}
\end{table}

\begin{table}
\begin{center}\begin{minipage}{\figurewidth}
\caption[]{\sl
Accuracy in the branching ratio obtained in the 
analysis for ${\cal{L}}$=500 fb$^{-1}$, and for 140 GeV Higgs. Note that
the WW-fusion process, which increases the signal statistics significantly
at higher energy, is not used in the analysis so far.
\label{tab:summary140}}
\end{minipage}\end{center}
\begin{center}
\begin{tabular}{|c|c|c|c|}
\hline
   $\sqrt{s}$    &  300 GeV &  400 GeV &  500 GeV \\
\hline
\hline
$\Delta$Br/Br & & & \\
$\ho\ra\bb$         &  2.1\%   & 2.3\% & 2.6\% \\
$\ho\ra\WW$         &  2.0\%   & 3.3\% & 4.5\% \\
\hline
\end{tabular}
\end{center}
\end{table}

\begin{table}
\begin{center}\begin{minipage}{\figurewidth}
\caption[]{\sl
Accuracy at $\sqrt{s}=$300, 400 and 500 GeV 
with ${\cal{L}}$=500 fb$^{-1}$ for 120 GeV CP-even Higgs at JLC. 
The Higgs boson of SM-like is used as an
input.
}
\label{tab:summarysummary}
\end{minipage}\end{center}
\begin{center}
\begin{tabular}{|c|c|c|c|}
\hline
   $\sqrt{s}$    &  300 GeV &  400 GeV &  500 GeV \\
\hline
\hline
$\Delta\mh$ (lepton-only) & 80 MeV & --- & --- \\
$\Delta\mh$               & 40 MeV & --- & --- \\
\hline
$\Delta\sigma/\sigma$ (lepton-only) & 2.1\% & 2.5\% & 2.9\% \\
$\Delta\sigma/\sigma$               & 1.3\% & ---   & ---   \\
$\Delta(\sigma_{h\nn}\cdot$Br($\bb$)& 2.0\% & ---   & ---   \\
ZZH-coupling $\Delta$ZZH/ZZH  & 1.1\% & 1.3\% & 1.5\% \\
WWH-coupling $\Delta$WWH/WWH  & 1.6\% & --- & --- \\
\hline
$\Delta\Gamma_{\ho}/\Gamma_{\ho}$  &  5.5\%   & 12\%  & 16\% \\
\hline
Yukawa coupling $\Delta\lambda/\lambda$ & & & \\
$\lambda_{b}$       &  2.8\%   & 6.1\% & 8.1\% \\
$\lambda_{\tau}$    &  3.5\%   & ---   & --- \\
$\lambda_{c}$       & 11.3\%   & 13\%  & 15\% \\\hline
$\lambda_{b}/\lambda_{\tau}$  &  2.3\% & --- & --- \\
$\lambda_{b}/\lambda_{c}$  &  11\%   & 12\%   & 14\% \\\hline
$\lambda_{up-type}$ &  4.1\%   & ---   & --- \\
$\lambda_{down-type}/\lambda_{up-type}$ &  3.2\%  & ---   & --- \\
\hline
$\Delta(\sigma\cdot$Br)/$(\sigma\cdot$Br)  & & & \\
$\ho\ra\bb$         &  1.1\%   & 1.3\% & 1.7\% \\
$\ho\ra\WW$         &  5.1\%   & 12\%  & 16\% \\
$\ho\ra\tautau$     &  4.4\%   & ---   & --- \\
$\ho\ra\cc$+gg      &  6.3\%   & ---   & --- \\
$\ho\ra\cc$         &  22\%    & 23\%  & 27\% \\
$\ho\ra$gg          &  10\%    & 11\%  & 13\% \\
$\ho\ra\gamma\gamma$& ---      & --- & --- \\
$\ho\ra\Zo\gamma$   & ---      & --- & --- \\
\hline
\end{tabular}
\end{center}
\end{table}

\section{Physics Outputs and Impacts in Models}
In this section, we discuss impacts of the Higgs boson studies at
JLC. Using the production cross section and the branching ratios
of the Higgs boson, we show how and to what extent the SM
is distinguished from other models like two Higgs doublet model
and MSSM.  Within the context of the MSSM, we can determine 
the parameters of the model, especially the heavy Higgs boson mass
from the properties of the lightest Higgs boson. We also discuss
the implications of the Higgs study when the heavy Higgs bosons 
or SUSY particles are discovered at LHC and JLC experiments.

\subsection{Model-independent Analysis for the Higgs Boson Couplings}
In order to distinguish various models, we first introduce 
a model-independent parameterization for various couplings to
Higgs boson. The following four parameters represent the multiplicative
factors in the Higgs-boson coupling constants with down-type quarks,
up-type quarks, charged-leptons and gauge bosons. 
\begin{eqnarray}
L&=& x \frac{m_b}{v}h\bar{b}b +y( \frac{m_t}{v}h\bar{t}t
+ \frac{m_c}{v}h\bar{c}c)
\nonumber\\
&&+z \frac{m_{\tau}}{v}h\bar{\tau}\tau 
+u(g m_W hW_{\mu}W^{\mu}+\frac{\sqrt{g^2+g'^2}}{2}m_Z hZ_{\mu}Z^{\mu}),
\end{eqnarray}
where the coupling with light quarks and leptons are suppressed.
This is, of course, not general way to parameterize the coupling 
constants. We can introduce different parameters
for charm and top quarks, and $W$ and $Z$ bosons, and 
flavour-changing couplings are also possible. The above
parameterization is still useful because models
such as MSSM, NMSSM and multi-Higgs doublet model without tree-level 
flavour changing neutral current fall into this category.

\begin{figure}
\centerline{
\epsfxsize=9cm \epsfbox{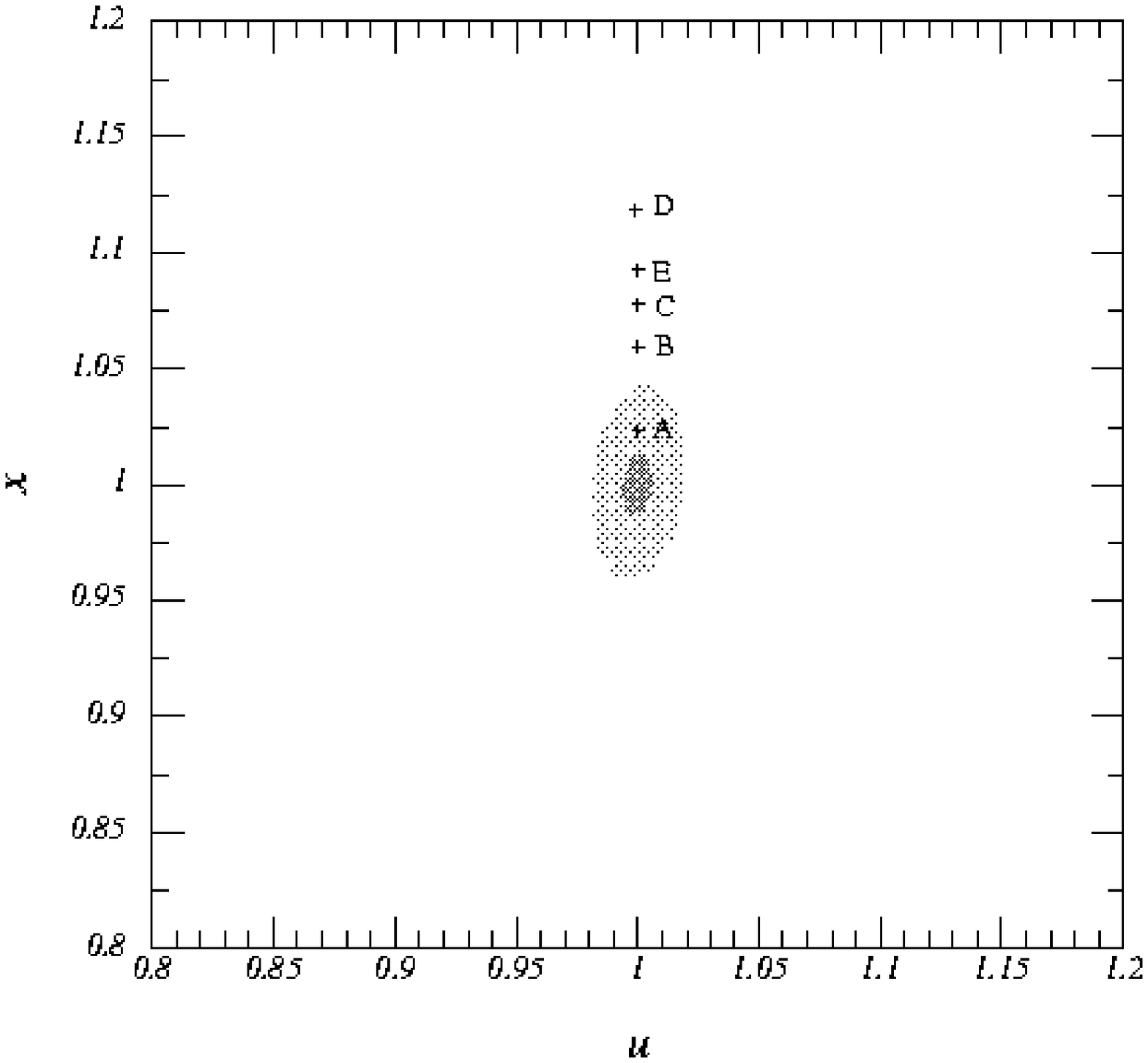}
\epsfxsize=9cm \epsfbox{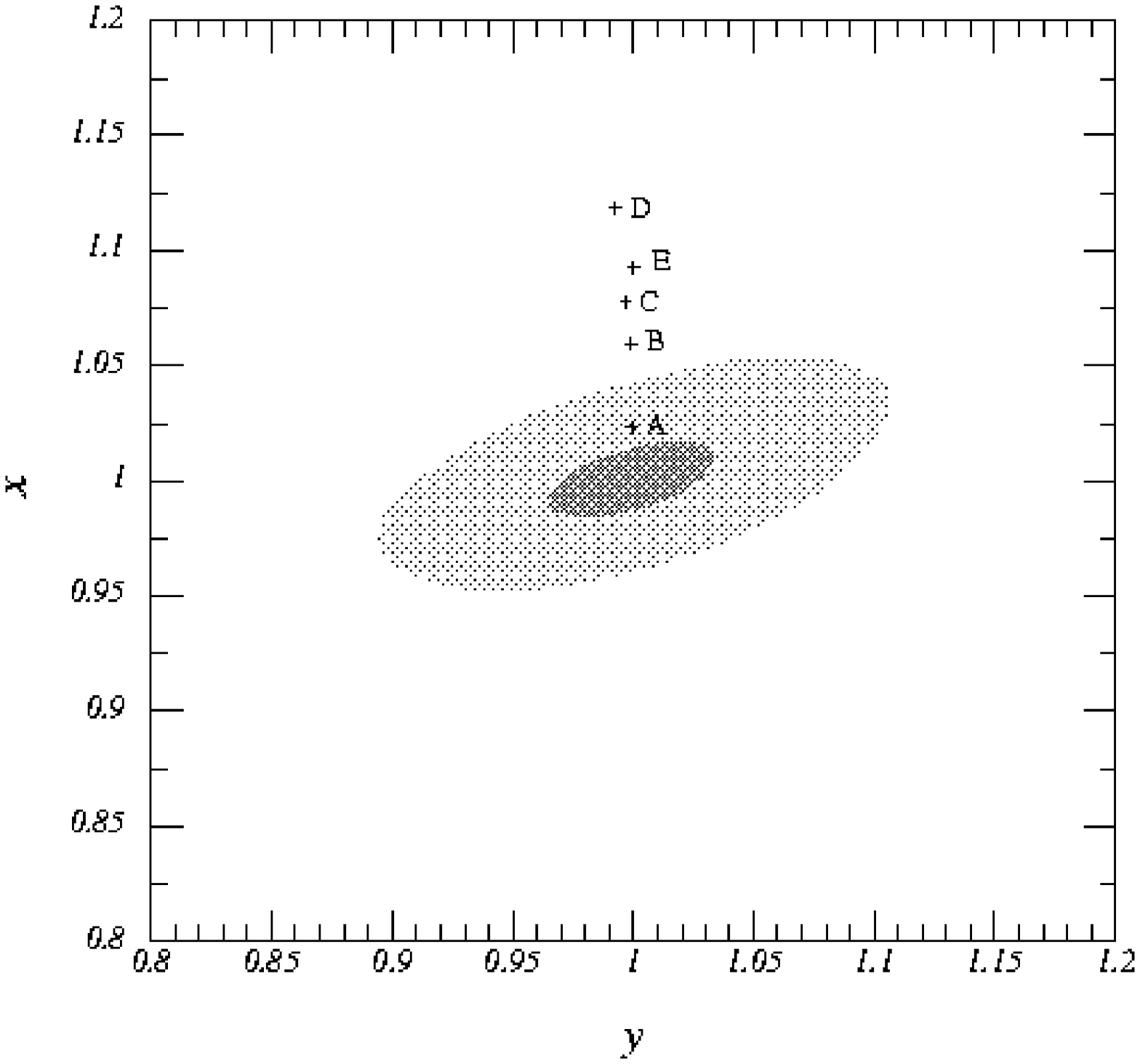}
}
\centerline{
\epsfxsize=9cm \epsfbox{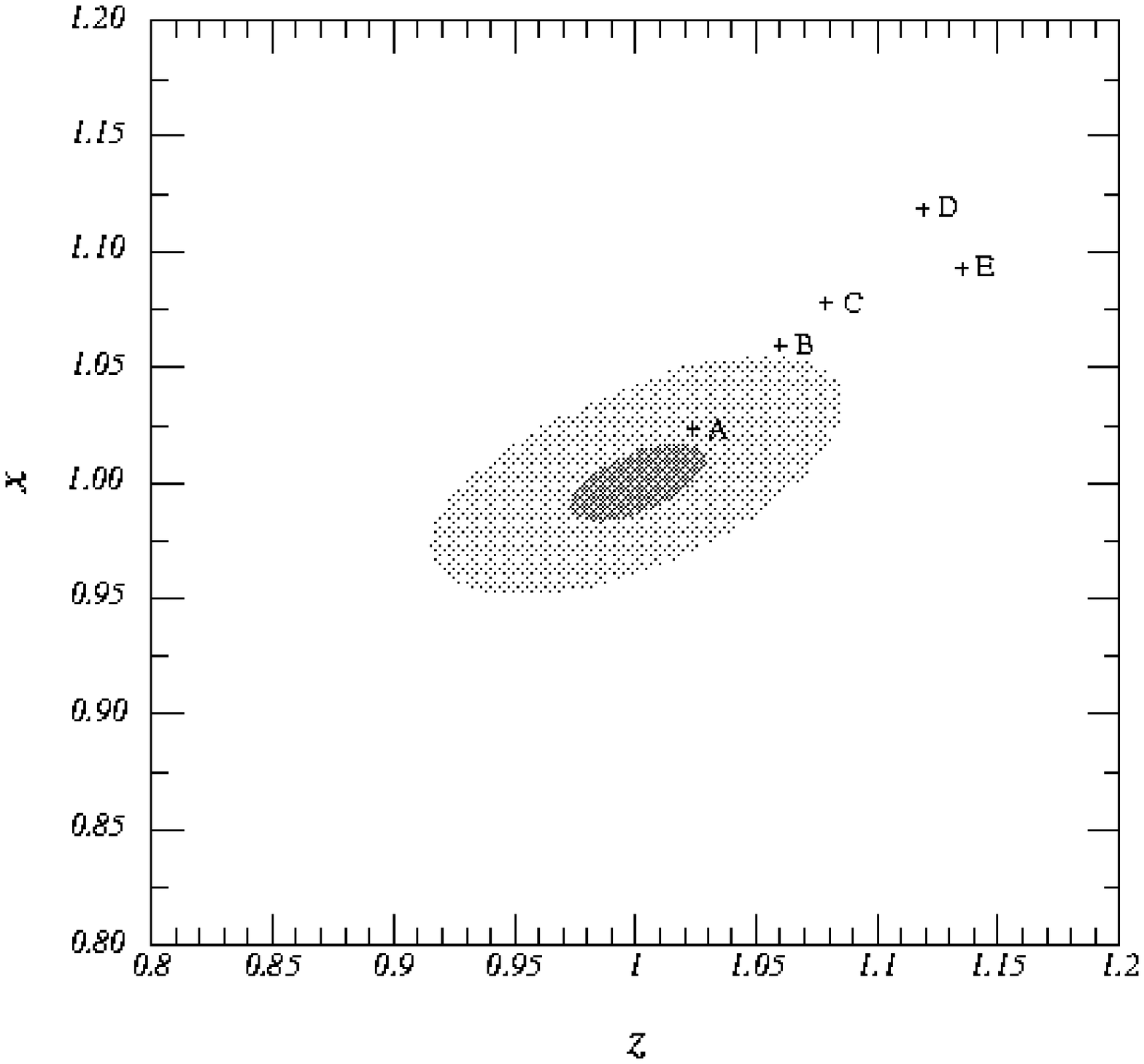}
}
\vspace{10pt}
\begin{center}\begin{minipage}{\figurewidth}
\caption{\sl
Experimental statistical accuracy of the parameter determination 
in $u - x$, $y - x$, and $z - x$ space at the JLC experiment with
an integrated luminosity of 500 fb$^{-1}$ and  $ \sqrt{s}= 300$ GeV.
The inner (outer) contour
is the 1 $\sigma$ (95\% CL) curve. The input point is the SM 
point $(x=y=z=u=1)$. The Higgs boson mass is 120 GeV.  
The points  A - E correspond to the projections on each plane of 
the $x, y, z,$ and $u$ values evaluated in the following parameter sets 
of the MSSM. ($m_A$ (GeV), $M_S$ (GeV), $X_t$, $\tan{\beta}$) are 
(1000, 500, 0, 10), (600, 550, -2, 10), (550, 1430, 2, 5),
(450, 25430, 0, 4) and (500, 4810, 2, 30) for A - E, respectively.
We also take the gluino mass $(M_3)$ and the higgsino mass parameter
$(\mu)$ as 300 GeV. 
\label{fig:xyzu}}
\end{minipage}\end{center}
\end{figure}

In Figure~\ref{fig:xyzu}, we show the expected accuracy of the
parameter determination in four dimensional space at the JLC
experiment with an integrated luminosity of 500 fb$^{-1}$. 
We use the production cross section and branching ratios
listed in Table~\ref{tab:summary1} for the case of $ \sqrt{s}= 300$ GeV and 
$m_h =$ 120 GeV. The input point is taken to be the SM point. We can see 
that the $u$ and $x$ parameters are determined to a few \% level, and $y$ 
and $z$ are constrained to less than 10 \%. 
$x$, $y$, $z$, and $u$ in various models are given by model parameters
as listed in Table~\ref{tab:xyzu}. In the MSSM model,
in addition to the tree level coupling constants, we take into account 
the correction to the $h b \bar{b}$ coupling constant denoted by
$\epsilon_b$ defined in section 2.1, whose effect becomes significant
only for large values of $\tan{\beta}$. We also assume that the 
coupling of the Higgs boson with two gluons is dominated by the internal
top-quark loop diagram, although the SUSY loop corrections become sizable
in some parameter space. In the figure we also show points corresponding 
to several input parameters in the MSSM.  From the correlation of 
the four parameters determined at the LC experiment,
it is possible to distinguish various models. 
For example,  Type-I Higgs doublet model and MSSM have different relation 
in the $x-y$ space. For a large $\tan{\beta}$ value, the allowed range
of the $x-z$ plane can deviate from the $x=z$ line for the MSSM
because of the SUSY correction to the  $h b \bar{b}$ vertex, as shown
for the point E. 

\begin{table}
\begin{center}\begin{minipage}{\figurewidth}
\caption[]{\sl
The parameters $x$, $y$, $z$, and $u$ in various
models. Two choices of the parameters are shown for the Type I 
two Higgs doublet model. 
\label{tab:xyzu}}
\end{minipage}\end{center}
\begin{center}
\begin{tabular}{|c|c|c|c|c|}
\hline
model & $x$ & $y$ & $z$ & $u$\\
\hline
SM & 1 & 1 & 1& 1\\
\rule[-12pt]{0pt}{30pt}
\begin{minipage}[c]{3.5cm}
\begin{center}
Type-I Two Higgs\\
doublet Model
\end{center}
\end{minipage} &
$-\frac{\sin{\alpha}}{\cos{\beta}}$($\frac{\cos{\alpha}}{\sin{\beta}}$)&
$-\frac{\sin{\alpha}}{\cos{\beta}}$($\frac{\cos{\alpha}}{\sin{\beta}}$)&
$-\frac{\sin{\alpha}}{\cos{\beta}}$($\frac{\cos{\alpha}}{\sin{\beta}}$)&
$\sin{(\beta-\alpha)}$\\
\rule[-12pt]{0pt}{30pt}
\begin{minipage}[c]{3.5cm}
\begin{center}
Type-II Two Higgs\\
doublet Model
\end{center}
\end{minipage} &
$-\frac{\sin{\alpha}}{\cos{\beta}}$&
$\frac{\cos{\alpha}}{\sin{\beta}}$&
$-\frac{\sin{\alpha}}{\cos{\beta}}$&
$\sin{(\beta-\alpha)}$\\
MSSM&
$-\frac{\sin{\alpha}}{\cos{\beta}}
\frac{1-\epsilon_b/\tan{\alpha}}{1+\epsilon_b \tan{\beta}}$&
$\frac{\cos{\alpha}}{\sin{\beta}}$&
$-\frac{\sin{\alpha}}{\cos{\beta}}$&
$\sin{(\beta-\alpha)}$\\
\hline
\end{tabular}
\end{center}
\end{table}

\subsection{Determination of Heavy Higgs Boson Mass in MSSM}
As we discuss in section 2.5.5 we can put constrains on the 
CP-even  Higgs boson mass ($m_A$) from the branching ratios
of the lightest CP even Higgs boson in the MSSM. 
Within the approximation that the stop mixing is neglected
in the one-loop Higgs potential and the $h b \bar{b}$ vertex 
correction is small, we can derive the following formulas
for $y^2/x^2$, $y^2/z^2$, $u^2/x^2$, and $u^2/z^2$, which 
corresponds to $(B(h \rightarrow c \bar{c})+B(h \rightarrow gg))
/B(h \rightarrow b \bar{b})$,
$(B(h \rightarrow c \bar{c})+B(h \rightarrow gg))
/B(h \rightarrow \tau^+ \tau^-)$,
$B(h \rightarrow W^{(*)} W^{(*)})
/B(h \rightarrow b \bar{b})$, and 
$B(h \rightarrow W^{(*)} W^{(*)})
/B(h \rightarrow \tau^+ \tau^-)$, respectively.
\begin{eqnarray}
y^2/x^2 &= &y^2/z^2 = \left( \frac{m_A^2-m_h^2}{m_A^2+m_Z^2}\right) ^2
\left( 1-\frac{m_h^2-m_Z^2}{m_A^2-m_h^2} \frac{1}{\tan^2{\beta}}\right)^2,\\
u^2/x^2 &= &u^2/z^2 =  \left( \frac{m_A^2-m_h^2}{m_A^2+m_Z^2}\right) ^2
\left( 1+\frac{2 m_Z^2}{m_A^2-m_h^2}\frac{1}{1+\tan^2{\beta}}\right)^2.
\end{eqnarray} 
For $m_A \gsim 200$ GeV, the first equation shows that $ y^2/x^2$ and 
$y^2/z^2$ are approximately given by 
$(m_A^2-m_h^2)^2/(m_A^2+m_Z^2)^2 $
because of the small factor of $(m_h^2-m_Z^2)/(m_A^2-m_h^2)$. 
From the second equation, we see that $ u^2/x^2$ and $u^2/z^2$ are 
also given by the same quantity if $\tan{\beta}$ is not small.
This should be a good approximation because the Higgs boson search 
at LEP already excluded the region of $\tan{\beta}\lsim 2$.
If we take into account the stop mixing effects in the one-loop 
Higgs potential and the $h b \bar{b}$ vertex correction denoted by 
$\epsilon_b$, the above approximate formulas receive corrections.
\begin{figure}
\centerline{
\epsfxsize=1.0\textwidth \epsfbox{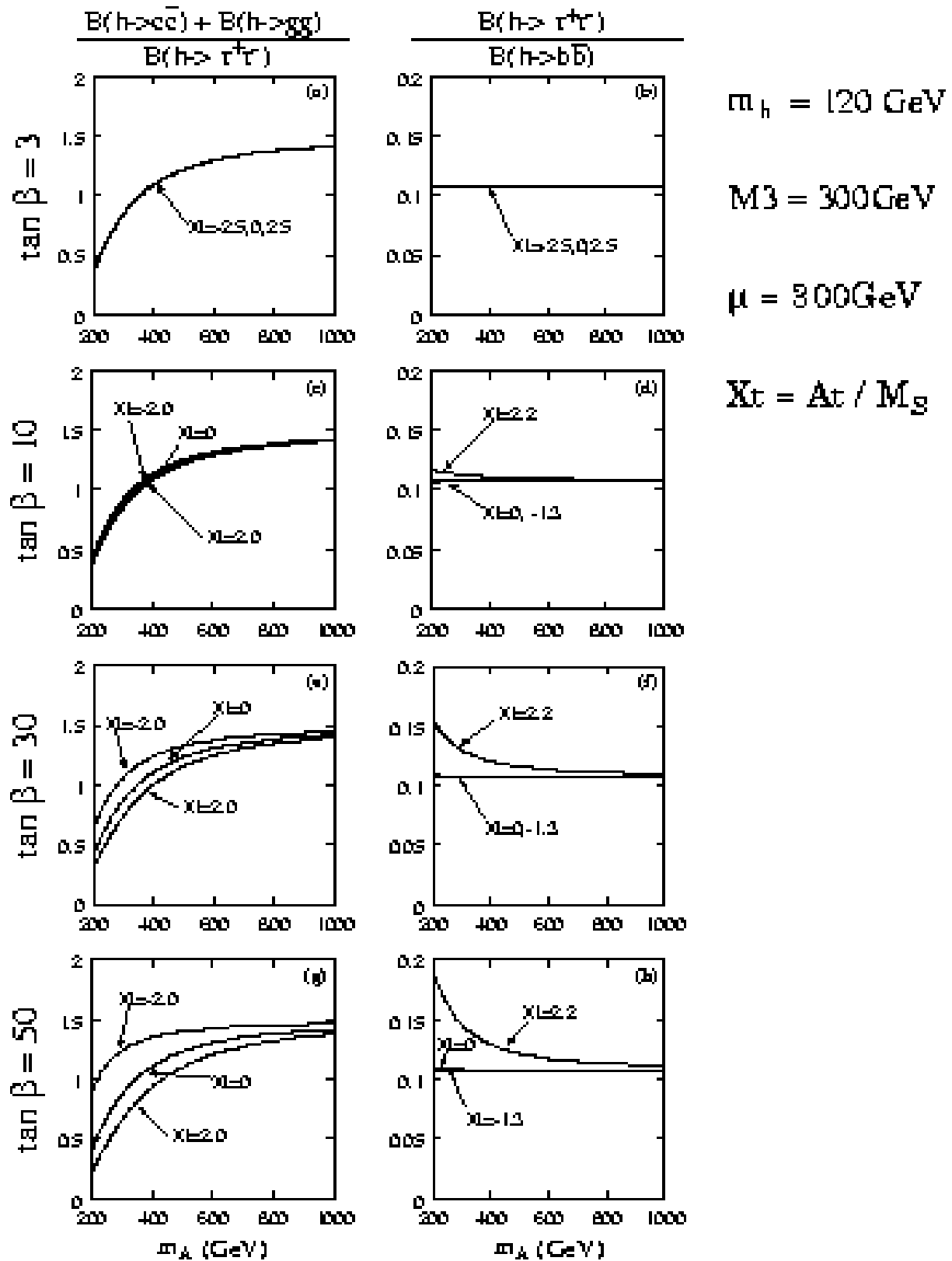}
}
\vspace{10pt}
\begin{center}\begin{minipage}{\figurewidth}
\caption{\sl
$(B(h \rightarrow c \bar{c})+B(h \rightarrow gg))
/B(h \rightarrow \tau^+ \tau^-)$ and 
$B(h \rightarrow \tau^+ \tau^-)/B(h \rightarrow b \bar{b})$
of the 120GeV
lightest CP-even Higgs boson for $\tan\beta=3,10,30,$ and $50$.
\label{fig:2x4_mu300}}
\end{minipage}\end{center}
\end{figure}
\begin{figure}
\centerline{
\epsfxsize=1.0\textwidth \epsfbox{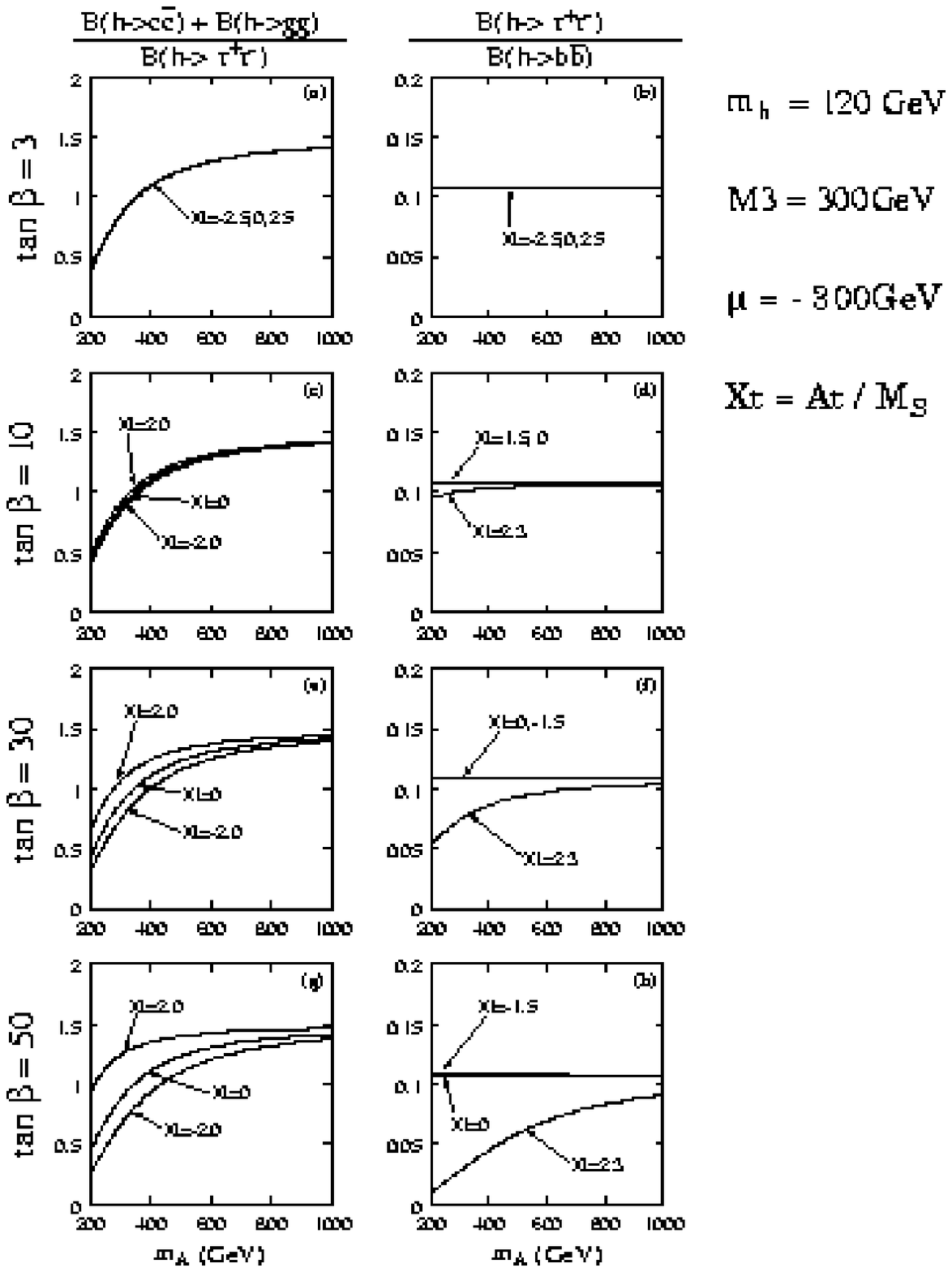}
}
\vspace{10pt}
\begin{center}\begin{minipage}{\figurewidth}
\caption{\sl
The same as Figure~\ref{fig:2x4_mu-300}, except $\mu=-300$ GeV.
\label{fig:2x4_mu-300}}
\end{minipage}\end{center}
\end{figure}

In order to show numerical impacts of  these corrections we present 
$(B(h \rightarrow c \bar{c})+B(h \rightarrow gg))
/B(h \rightarrow \tau^+ \tau^-)$ and
$B(h \rightarrow \tau^+ \tau^-)/B(h \rightarrow b \bar{b})$
for different choices of $\tan{\beta}$ in Figure~\ref{fig:2x4_mu300}
(Figure~\ref{fig:2x4_mu-300}).
Here we introduce a SUSY breaking scale $M_S$ as
$m^2_{\tilde{q_3}_L}=m^2_{\tilde{t}_R}=m^2_{\tilde{b}_R}= M_S^2$
and the squark mixing parameter $A_t=X_t M_S$ and $A_b=0$.
We set the gluino mass $(M_3)$ as 300 GeV and the higgsino mass parameter
$(\mu)$ as 300 GeV (-300 GeV) and fix the Higgs boson mass as 120 GeV
and solve $M_S$ in term of $\tan{\beta}$, $X_t$ and $m_A$. The
possible range of the ratios are shown as a function of $m_A$ by changing the 
parameter $X_t$.
Notice that  $B(h \rightarrow \tau^+ \tau^-)/B(h \rightarrow b \bar{b})$
is the same as the SM prediction and therefore independent of $m_A$ if
$\epsilon_b$ vanishes.
In Figure~\ref{fig:2x4_mu300} and Figure~\ref{fig:2x4_mu-300},
the plots (b) and (d) show that the effect of $\epsilon_b$ on this
ratio is tiny for $\tan \beta \lsim 10$. On the other hand, for large
$\tan \beta$, e.g. the plot (h), the deviation of this ratio from the SM
value can be significant.  Furthermore, by comparing 
Figure~\ref{fig:2x4_mu300}(h) and Fig~\ref{fig:2x4_mu-300}(h), we can see
that the effect of $\epsilon_b$ enhances or suppresses the ratio
depending on the sign of $\mu$.
In the plots (e) and (g) of Figure~\ref{fig:2x4_mu300} and
Figure~\ref{fig:2x4_mu-300}, we can see that the deviation of
$y^2/z^2$ from the above approximate formula is large for
$\tan{\beta} \gsim 30$. Since $y$ and $z$ are independent of
$\epsilon_b$, this deviation is due to the stop
mixing effects on the one-loop Higgs potential.
On the other hand, the correction is small for $\tan{\beta} \lsim 10$.
Because it is likely that the only one light Higgs boson is
found in the LHC experiment for $3 \lsim tan{\beta} \lsim 10$, 
the indirect determination on $m_A$ could become very important
in such parameter space. Note, however, that the deviation from the
approximate formula depends on the magnitude of $\mu$. For larger
$|\mu|$ the deviation can be significant even for $\tan \beta \lsim 10$.

\begin{figure}
\centerline{
\epsfxsize=0.9\textwidth \epsfbox{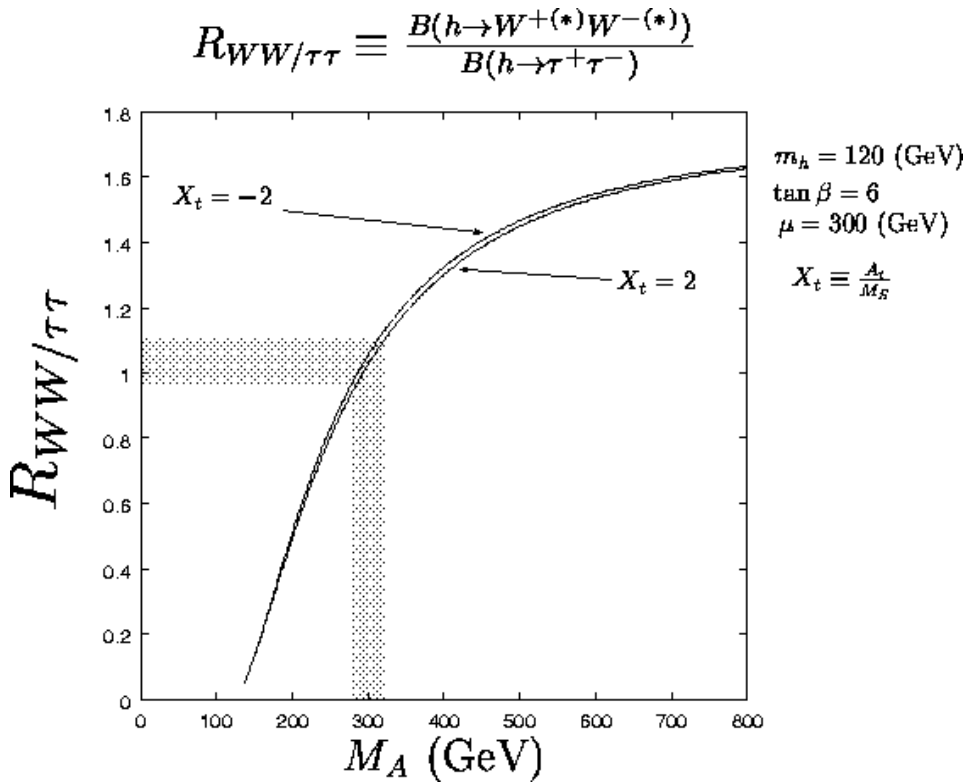}
}
\begin{center}\begin{minipage}{\figurewidth}
\caption{\sl
The ratio $B(h \rightarrow W^{+(*)}W^{-(*)})/B(h \rightarrow
\tau^+ \tau^-)$ of the 120 GeV lightest Higgs boson for
$\tan\beta=6$, $\mu=300$ GeV, and $X_t=\pm2$ as a function of $m_A$.
The shaded area shows the $m_A$ determination from the ratio,
obtained by the assumption that $B(h \rightarrow W^{+(*)}W^{-(*)})/B(h
\rightarrow \tau^+ \tau^-)$ is determined in 6.7\% accuracy (see
Table\ref{sm_error}).
\label{fig:Rwwtautau_ma}}
\end{minipage}\end{center}
\end{figure}

\begin{figure}
\centerline{
\epsfxsize=0.7\textwidth \epsfbox{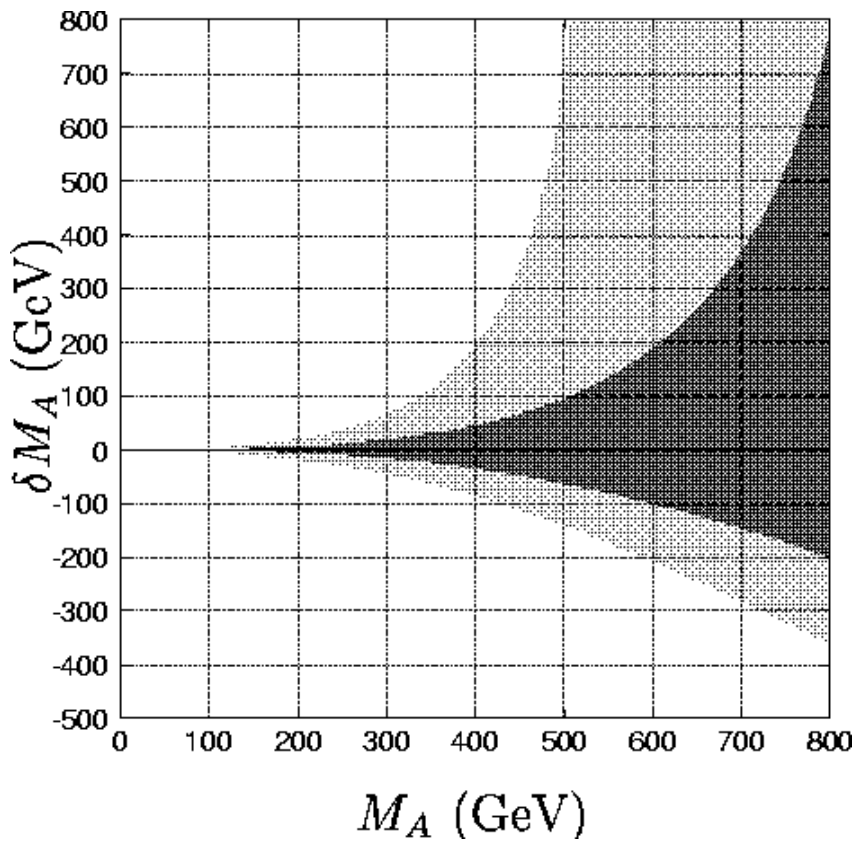}
}
\begin{center}\begin{minipage}{\figurewidth}
\caption{\sl
Accuracy of the $m_A$ determination 
as a function of $m_A$ from branching ratio measurements.
The dark area corresponds to the error of $m_A$ from
$B(h \rightarrow c \bar{c})+B(h \rightarrow gg)$,
$B(h \rightarrow \tau^+ \tau^-)$, and  $B(h \rightarrow W^{(*)} W^{(*)})$
measurements at the JLC experiment. The light area is obtained by the 
assumption that $\Gamma ( h \rightarrow W^{(*)} W^{(*)}) /
\Gamma (h \rightarrow \tau^+ \tau^-)$ is determined in 15\% accuracy,
which corresponds to an estimated statistical error at the LHC experiment.
\label{fig:deltama}}
\end{minipage}\end{center}
\end{figure}

\begin{table}
\begin{center}\begin{minipage}{\figurewidth}
\caption{\sl
The errors of the double ratios for the 120 GeV SM Higgs
boson due to the theoretical uncertainties of input parameters 
and experimental statistical errors (\%) for JLC with an integrated
luminosity of 500 fb$^{-1}$ and $\sqrt{s}$ = 300 GeV.
For theoretical errors, we take into account the uncertainty from the strong 
coupling constant and the bottom and charm quark masses as 
${\alpha}_s(m_Z)=0.1181 \pm 0.002,
{m}^{\overline{MS}}_b(m_b)=4.20 \pm 0.13 ~\mbox{GeV} ~(\pm3\%),
{m}^{\overline{MS}}_c(m_c)=1.25 \pm 0.06 ~\mbox{GeV} ~(\pm5\%).$
\label{sm_error}}
\end{minipage}\end{center}
\begin{center}
\begin{tabular}{l|ccc}
\hline
 & Theoretical error & Experimental error
 & Total error \\
\hline
$\frac{Br(h \rightarrow c\bar{c})+Br(h \rightarrow gg)}
{Br(h \rightarrow b\bar{b})}$ &
  8.6 & 6.4 & 10.7 \\
$\frac{Br(h \rightarrow W^{(*)}W^{(*)})}
{Br(h \rightarrow b\bar{b})}$ &
  7.6 & 5.2 & 9.2 \\
$\frac{Br(h \rightarrow c\bar{c})+Br(h \rightarrow gg)}
{Br(h \rightarrow \tau^+ \tau^-)}$ &
  4.1 & 7.7 & 8.7 \\
$\frac{Br(h \rightarrow W^{(*)}W^{(*)})}
{Br(h \rightarrow \tau^+ \tau^-)}$ &
  0 &  6.7 & 6.7 \\
\hline
\end{tabular}
\end{center}
\end{table}

The ratio of the branching ratios is useful to constrain $m_A$ as
shown in Figure~\ref{fig:Rwwtautau_ma}.
In Figure~\ref{fig:deltama},
we show the precision of the indirect determination 
on $m_A$ from the above branching ratios at the JLC experiment
with integrated luminosity of 500 fb$^{-1}$ at $ \sqrt{s}= 300$ GeV. 
The Higgs boson mass is taken to be 120 GeV. The theoretical uncertainty 
of the branching ratio calculation in the SM and the estimated experimental 
statistical errors are summarized in Table \ref{sm_error}.
The combined error to determine $m_A$ from 
$(B(h \rightarrow c\bar{c})+B(h \rightarrow gg))
/B(h \rightarrow b\bar{b})$ and $B(h \rightarrow W^{(*)}W^{(*)})
/B(h \rightarrow b\bar{b})$ is 7.0 \%, and that from 
$(Br(h \rightarrow c\bar{c})+Br(h \rightarrow gg))
/Br(h \rightarrow \tau^+ \tau^-)$ and 
$Br(h \rightarrow W^{(*)}W^{(*)})/Br(h \rightarrow \tau^+ \tau^-)$ 
is 5.3\%. In order to draw the upper and lower limits of $m_A$,
we use the 5.3\% error and assume that these ratios normalized by
the SM values are given by $(m_A^2-m_h^2)^2/(m_A^2+m_Z^2)^2 $. 
Notice that for the ratios normalized by $Br(h \rightarrow \tau^+
\tau^-)$, there is no correction from the $\epsilon_b$ term.
We can see that the useful upper bound of the heavy 
Higgs boson mass is obtained, if the true value of $m_A$ is less than 600 GeV. 
Even if the branching ratio is consistent to the SM value, the lower 
bound will be determined.
This is compared with the precision of the ratio of partial decay widths 
$\Gamma ( h \rightarrow W^{(*)} W^{(*)}) /
\Gamma (h \rightarrow \tau^+ \tau^-)$ at LHC. Using the weak boson fusion 
process for Higgs boson production, the statistical error is estimated 
to be 15\% for the integrated luminosity of 200 fb$^{-1}$\cite{zepp00}.

\begin{figure}
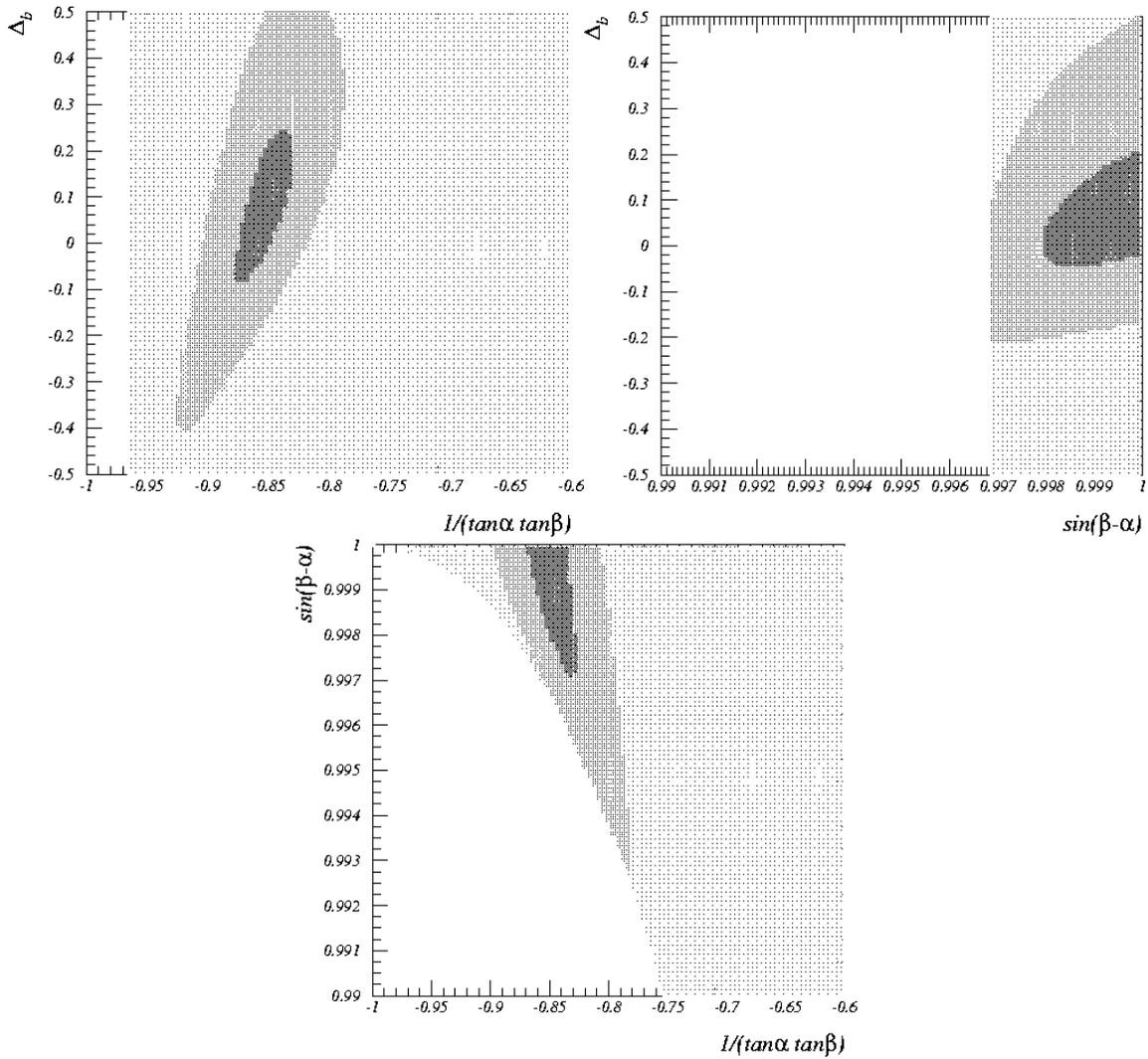

\centerline{
\epsfxsize=7.7cm \epsfbox{physhiggs/chi2_SUSY_QD_tb10.epsi}
\epsfxsize=7.5cm \epsfbox{physhiggs/chi2_SUSY_UD_tb10.epsi}
}
\centerline{
\epsfxsize=7.5cm \epsfbox{physhiggs/chi2_SUSY_UQ_tb10.epsi}
}
\vspace{10pt}
\begin{center}\begin{minipage}{\figurewidth}
\caption{\sl
Experimental statistical accuracy of the 
$1/(\tan{\alpha}\tan{\beta})$,
$\sin{(\beta-\alpha)}$, and $\Delta_b$ determination
for $\tan{\beta}=10$ in the MSSM. The darkest (second darkest) area 
corresponds to the 1$\sigma$ (95\% CL) region. No solution is obtained
in the white area. The input point
is $1/(\tan{\alpha}\tan{\beta})= -0.852$,
$\sin{(\beta-\alpha)}=0.99985$, and $\Delta_b
(\equiv \epsilon_b \tan{\beta})=0.08$, 
corresponding to SUSY parameters, $m_A$=400 GeV, $M_S$=654 GeV, $X_t$=2,
$\mu$ =300 GeV, and $M_3$ = 300 GeV.
\label{fig:chi210}}
\end{minipage}\end{center}
\end{figure}

\begin{figure}
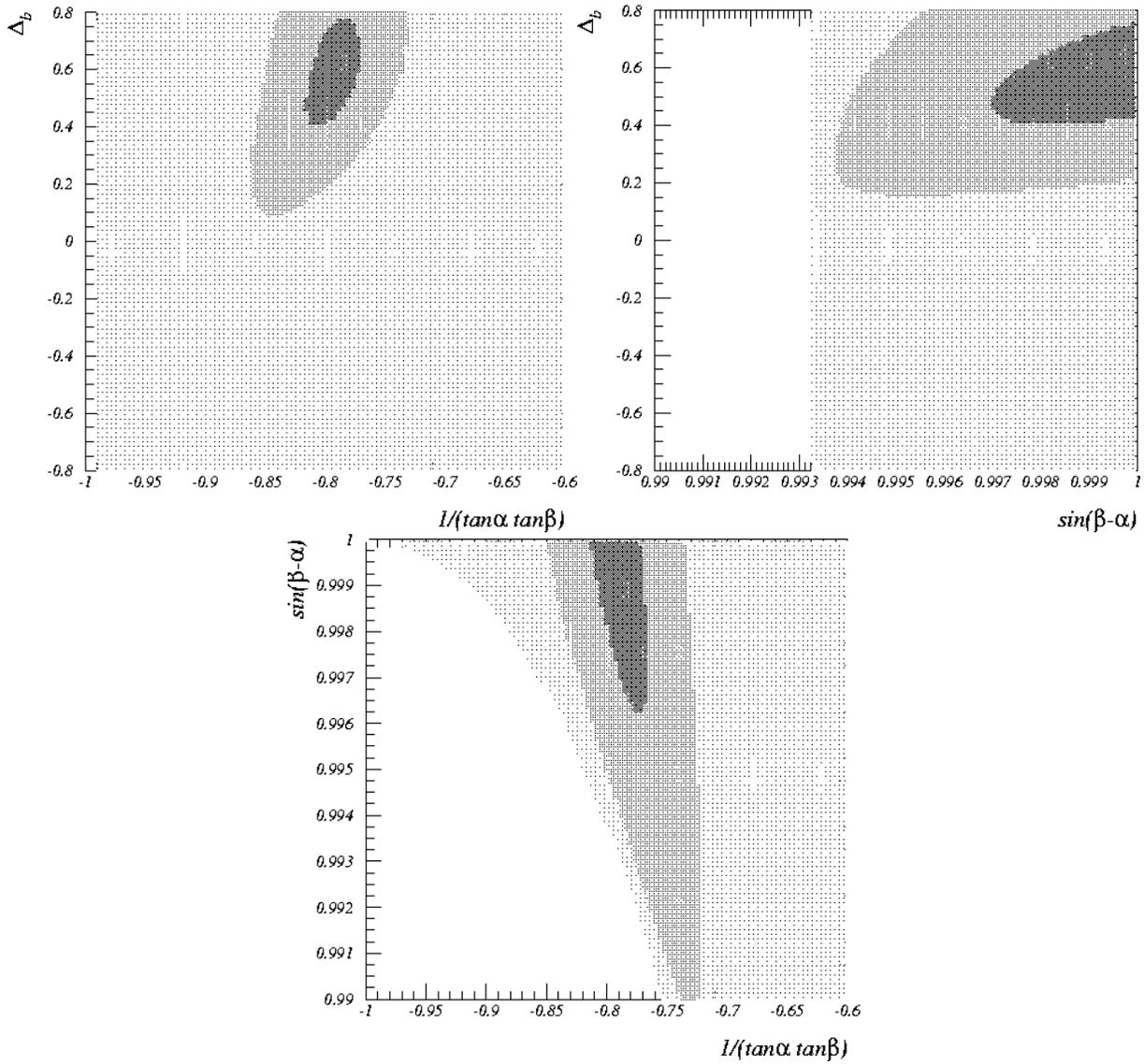

\centerline{
\epsfxsize=8.1cm \epsfbox{physhiggs/chi2_SUSY_QD_tb50.epsi}
\epsfysize=7.5cm \epsfbox{physhiggs/chi2_SUSY_UD_tb50.epsi}
}
\centerline{
\epsfysize=7.5cm \epsfbox{physhiggs/chi2_SUSY_UQ_tb50.epsi}
}
\vspace{10pt}
\begin{center}\begin{minipage}{\figurewidth}
\caption{\sl
The same figure as Figure~\ref{fig:chi210}
for the case of  $\tan{\beta}=50$.
The input point
is $1/(\tan{\alpha}\tan{\beta})= -0.793$,
$\sin{(\beta-\alpha)}=0.999986$, and $\Delta_b
(\equiv \epsilon_b \tan{\beta})=0.58$, 
corresponding to SUSY parameters, $m_A$=400 GeV, $M_S$=462 GeV, $X_t$=2,
$\mu$ =300 GeV and $M_3$ = 300 GeV.
\label{fig:chi250}}
\end{minipage}\end{center}
\end{figure}

When $tan{\beta}$ is large, we may be able to determine the
parameter $\Delta_b \equiv \epsilon_b \tan{\beta}$ from production cross 
section and branching ratios. As an example we show the experimental 
statistical accuracy of the  MSSM parameter determination including
$\Delta_b$.
We present the $\chi^2$ contour in $1/(\tan{\alpha}\tan{\beta})$,
$\sin{(\beta-\alpha)}$, and $\Delta_b$ space for $\tan{\beta}= 10$
(Figure~\ref{fig:chi210}) 
and $\tan{\beta}= 50$ (Figure~\ref{fig:chi250}). 
We can see that non-zero $\Delta_b$ can be clearly distinguished
in the case of  $\tan{\beta}=50$. Establishing the non-vanishing $\Delta_b$ is 
very interesting because it is induced by virtual correction due to SUSY 
particles.

\subsection{Possible Scenarios of SUSY Parameter Determination}
If the heavy Higgs bosons or SUSY particles are found in 
the LHC experiment and the JLC experiment, we can perform
many consistency checks, coupling constant determinations, and 
model discrimination with help of the Higgs boson studies. 
Possible scenarios are as follows:
\begin{itemize}    
\item If the heavy Higgs bosons $(A$ and $H)$ are found,
we can compare their masses with the value determined
from branching measurement of the lightest Higgs boson.
This is an important test of the MSSM because two values
are not necessarily the same for two Higgs doublet model in general.
\item If the stop sectors are determined from the direct SUSY search
in addition to the information on $m_A$ from either direct discovery
of the heavy Higgs bosons or indirect determination from the branching 
measurements, we can determine $\tan{\beta}$ from the mass formula
of the lightest CP-even Higgs boson in the MSSM. In order to do so,
it is important to determine two masses and the left-right mixing 
of the stop sector, because the mixing parameter gives a sizable
correction to the lightest Higgs boson mass.
\item If $\tan{\beta}$ value is obtained from other sectors of the 
SUSY model, we can compare it with $\tan{\beta}$ inferred from the 
studies on the lightest Higgs boson. For possibilities of other 
$\tan{\beta}$ measurements, we can consider the stau decay\cite{nojiri}, 
the heavy Higgs production and decay\cite{feng}, 
and the chargino and neutralino production\cite{tsukamoto}. 
The consistency of $\tan{\beta}$ determination is in fact the direct 
confirmation of the MSSM Higgs mass formula, and tells us that the 
Lagrangian of the Higgs sector is likely to be that of the MSSM.
If there is a deviation we have to extend the Higgs sector. For
example, we may be able to give a constraint on the additional 
coupling constant in the mass formulas of the NMSSM Higgs sector,
which is given by  
\begin{equation}
m_h^2 \leq \frac{1}{2}\lambda^2v^2\sin^2{2\beta}
+m_Z^2 \cos^2{2\beta}+\frac{3}{2\pi^2}\frac{m_t^4}{v^2}
ln{\frac{m_{stop}^2}{{m_t}^2}},
\end{equation}
where $\lambda$ is a new coupling constant associated with a
singlet Higgs field. It is conceivable that the upper bound
is saturated for the lightest CP-even Higgs boson and therefore
its properties are not much different from the SM Higgs boson. 
From precise determination of the Higgs boson properties, the 
stop sector, and $\tan{\beta}$ from other processes, we may be 
able to distinguish NMSSM from MSSM and  determine the parameter 
$\lambda$.
\end{itemize}  
In this way, studies of the Higgs boson properties are 
useful to determine the Lagrangian of the SUSY model.
Through the Higgs mass formula, we can obtain information
on various SUSY parameters. The precision studies 
of the Higgs boson, therefore, play an important role in distinguishing 
and establishing different models, even after the discovery of  
SUSY particles.

\section{Further Experimental Considerations}

In studies of the Higgs particles at JLC,
accurate tracking and event kinematic reconstructions are essential. 
One needs to investigate the performances of such key techniques under
very realistic experimental environments.
The extremely high luminosity of order of
$10^{34}$ cm$^{-2}$s$^{-1}$, which is about 2 order higher than that at
LEP, and the beam structure necessary for the
acceleration in the linear collider JLC,
make the experimental situation more severe 
than that at LEP and SLC in the following points;

\begin{itemize}
\item Large number of noise hits on the vertex detectors due to
huge number of \ee\ pairs created by beam-beam interactions~\cite{beam-beam}.
It makes proper associations of the
hits for the tracking difficult especially for the low momentum tracks,
which are rather important for \bquark\ tagging,
due to multiple Coulomb scatterings and energy losses in
detector materials;
\item Overlap of two or more physics 
events, especially two-photon backgrounds~\cite{twophoton}, in a same bunch 
collision with a signal or other hard scattering event
due to very high integrated luminosity per bunch.
It might reach to 1 $\mu$b$^{-1}$ level at JLC
which is 2--3 order higher than that at LEP/SLC.
The average rate of the overlap of the two photon process on signal events
might reach to about 10 \% or more
\item Degradation of the collision energy due to beamstrahlung.
\end{itemize}

The studies have been made with the current design of the detector at 
JLC~\cite{JLC-I} as a model detector.

Studies of the Higgs measurements with such expected experimental
condition have been attempted~\cite{SY-KANZAKI}.
The most inner layer of the vertex detector might have the noise hit
density of about 1 hits/mm$^{2}$, which makes proper associations of the
hits difficult especially for the low momentum tracks. 
For the event overlapping,
there are two major effects in the measurements. One is the effect on the
kinematics such as reconstructed mass of the Higgs and the selection bias.
The other effect is on the flavour tagging which
relies on the tracks displaced from the event primary vertex.  

\subsection{Noise Hits on Vertex Detector due to Beam-beam Interaction}

The beam at JLC has 50--100 of bunches in a beam train with bunch spacing
of 1.4 to 5.6 nsec and the repetition rate of 50--150 Hz. 
Since the present design uses CCD readout for the vertex detector,
the hits from multi-bunches in a train overlap each other. 
In single train collision, the expected rate of the noise hits due to
the beam-beam interaction is about 30--100 hits/cm$^{2}$ on the inner 
most layer of the vertex detectors.
Studies have been made with full simulation program JIM~\cite{JIM}
based on the GEANT3 package.
Tracking devices in the JLC detector
are: central tracker (CT), intermediate tracker (IT) and vertex detector (VX).
Detector parameters are summarized in
Tab~\ref{tab:detpar}.

\begin{table}
\begin{center}\begin{minipage}{\figurewidth}
\caption{\sl
Detector parameters which are relevant to this analysis.
Magnetic field of 2 T is used.
\label{tab:detpar}}
\end{minipage}\end{center}
\begin{center}
\begin{tabular}{|l|l|l|l|l|l|l|} \hline
Detector & No. of & \multicolumn{2}{c|}{Radius of layers [cm]} 
& Half-$z$  & \multicolumn{2}{c|}{Resolution [\micron]} \\
\cline{3-4}\cline{6-7} 
Name & Layers & Inner & Outer & Length [cm] & $r$-$\phi$ & $z$ \\
\hline 
CT & 100 & 45 & 230 & 125 & 100 & 2000 \\ \hline
IT & 5 & 10 & 38 & 21-79 & 20 & 20 \\ \hline
VX & 4 & 2.4 & 6.0 & 5-12.5  & 4 & 4 \\ \hline
\end{tabular}
\end{center}
\end{table}
Single \muon\ track generated with flat distribution in polar angle
in cos$\theta\pm$0.8 with momentum ranging from 0.5
to 50 GeV, are used as a ``signal'' track 
to study the proper track finding efficiency.
Two kinds of backgrounds are considered in this analysis.
One is the low energy \epem\ pairs induced by the beam-beam interactions.
This background is simulated by the CAIN code~\cite{ABEL}, and generated \epem\
pairs are also processed through the same full detector simulation program.
The simulated detector response from the \epem\ pairs,
corresponding to numbers of bunch collisions 
(100 bunches as a default), are overlapped to each signal event. 
Another background source added are low
momentum charged tracks from two-photon process (corresponding to
10 events as a default).

First step of the analysis starts from the reduction of the
noise hits with cluster shape analyses.
Almost all the background hits in the VX detector are generated
by very low momenta \epem's which are curling as a helix with small radius
within the VX detector region.
Thus the shape of the cluster of the noise hits on the VX tends to 
be longer than that of hits generated by ``signal'' charged tracks.
By applying cuts on shapes of VX hit cluster, number of background hits 
has been reduced to 1/3 
keeping the signal hit efficiency to be more than 98 \%.

We start the tracking from the fitting of the hits in the CT. 
Then the hits in IT are associated to these tracks.
Since most of the problematic \epem pairs have very low momenta, 
the hit density in IT region is already low enough. 
Hence the association of the hits on the IT is rather easy.
Then analysis proceeds to the association of the tracks 
obtained from CT and IT (CT+IT track) 
to the clusters found in the VX.
We have compared the following two methods for the association;
\begin{enumerate}
\item a conventional ``Road Method'':
The positions predicted by the CT+IT track are compared with those
of VX clusters in each VX layer, and nearest clusters are retained.
\item ``Local Tracking'': first, candidates of track segments are locally made
only with the VX clusters. 
In order to reduce the combinatorial associations, 
accumulated distribution of VX hits in all layers 
in the $\phi$-$z$ plane is used, and
peaked clusters in the plane are defined as a local track segments.
Then the track parameters of the local segments are compared with
those of CT+IT tracks, 
and associated to the most appropriate tracks from the CT+IT.
Finally re-fits are made globally for all hits found in CT, IT and VX.
\end{enumerate}

\begin{figure}
\centerline{
\epsfxsize=9.0cm \epsfbox{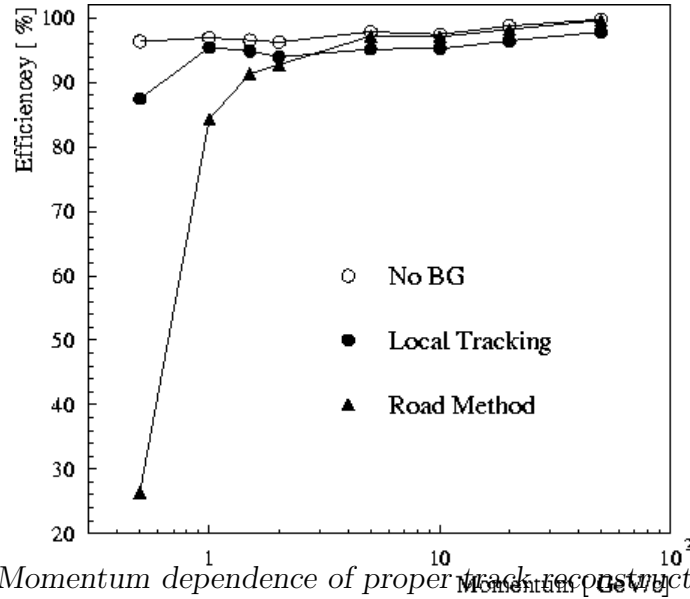}
}
\vspace*{-0.8cm}
\begin{center}\begin{minipage}{\figurewidth}
\caption{\sl
Momentum dependence of proper track reconstruction efficiency.
\label{fig:vx-eff-pdep}}
\end{minipage}\end{center}
\end{figure}

\begin{figure}
\centerline{
\epsfxsize=9.5cm \epsfbox{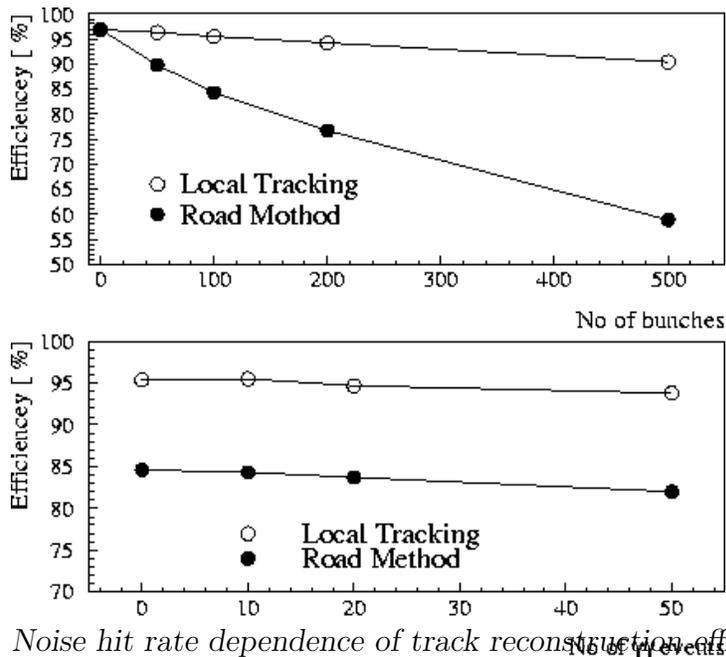}
}
\vspace*{-0.8cm}
\begin{center}\begin{minipage}{\figurewidth}
\caption{\sl
Noise hit rate dependence of track reconstruction efficiency for
1 GeV single muon.
\label{fig:vx-eff-bgdep}}
\end{minipage}\end{center}
\end{figure}

To check the tracking performance, we define the proper tracking
as the track having correct hit association for all layers of VX. 
Dependence of the proper tracking efficiency on the 
momentum of the signal \muon\ is shown in Fig~\ref{fig:vx-eff-pdep}.
As a default, the \epem\ pairs from 100 bunch-interactions and charged
tracks from 10 two-photon events are overlapped.
Especially for charged tracks with momentum below 1 GeV, difference
of efficiencies between two methods mentioned above are fairly large. 
Fig\ref{fig:vx-eff-bgdep} shows the proper track
reconstruction efficiencies as functions of the 
amount of overlapped background of; 
(upper) \epem\ pairs indicated by the number of bunches in a train, 
and (lower) two-photon processes. 
The signal \muon\ momentum is fixed to 1 GeV.
For both cases, amount of the other background is
fixed to the default values.

\noindent

Efficiencies obtained by two track finding methods do not depend on
amount of overlapped two-photon background significantly.
The degradation of the efficiency obtained by the ``Local Tracking'' method 
is found to be small even at the extremely high background rate,
while
the efficiency by the simple ``Road method'' decreases 
rapidly as amount of background from \epem\ pairs grows.
In order to perform a good tracking, a self-fit capability in vertex detector
is found to be very helpful. 
Parameters in this method should be optimized further. 
Also an interpolation of tracks taking into account energy losses and
multiple-scatterings should be used, 
especially for lower momentum charged tracks found in the CT detector.
These improvements are addressed in the future studies.

\subsection{Overlap with Two-photon Processes}

The effective cross-section $\sigma_{eff}$ of the overlapping 
events~\footnote{
It has to be modified with Poisson probability for
only one two-photon overlap, two two-photon overlap, three or more etc..
for large overlapping rate. 
}
can be written as
$\sigma_{eff} = \sigma_{1} \times \sigma_{2} \times {\cal{L}}_{bunch}$,
where $\sigma_{1,2}$ are the production cross-section of each
physics process, and ${\cal{L}}_{bunch}$ is 
the integrated luminosity in a bunch (luminosity per bunch).
At the future linear collider,
the luminosity per bunch can reach up to 1 $\mu$b$^{-1}$ level,
which is 2--3 order of magnitude higher than that at LEP/SLC.
The most severe overlap comes from two-photon backgrounds.
Roughly speaking, its cross-section is around 100 nb, 
while the cross-section depends on the
cut in the kinematics, and the calculated cross-section has large 
uncertainty by nearly factor two~\footnote{
Note that the two-photon processes include ``normal'' two-photon process
in which the two virtual photon interacts, and interaction between
a virtual photon and real photon induced by the beamstrahlung effect,
which may roughly doubles the cross-section.}.
Hence the average overlap rate could reach to 10 \% or more for
the luminosity per bunch of 1 $\mu$b$^{-1}$, even if we assume that
we have perfect identification of the bunch of the event. 
The assumption in the bunch identification may have to be relaxed
when we have short bunch spacing of 1.4 nsec at JLC.

\noindent
\begin{figure}
\hspace*{-0.1cm}
\centerline{\epsfxsize=12.8cm \epsfbox{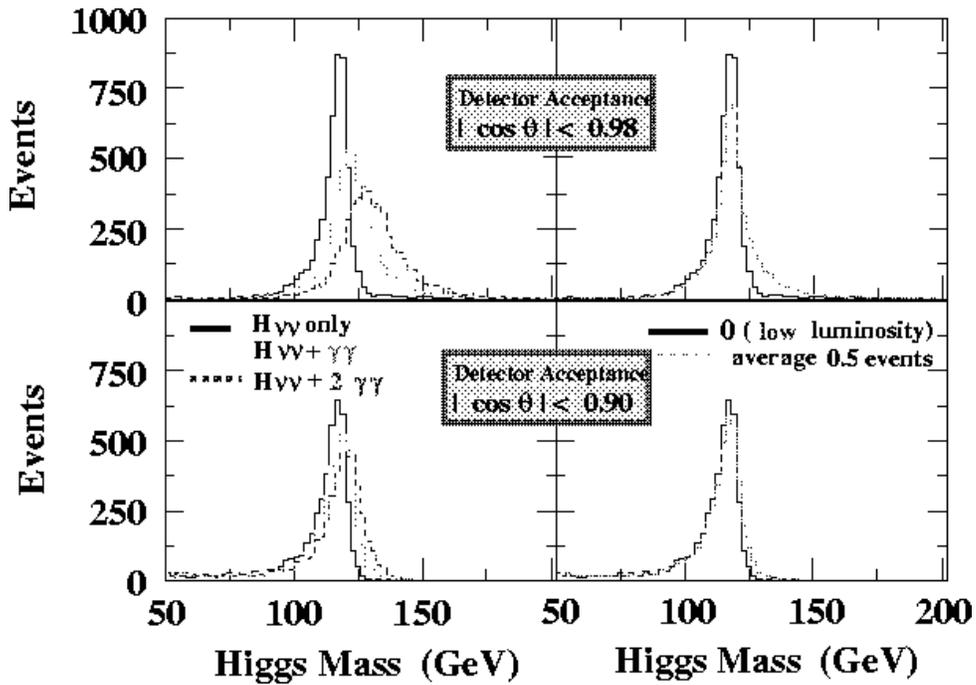}}
\begin{center}\begin{minipage}{\figurewidth}
\caption[]{\sl
Distributions of the reconstructed mass of the Higgs in
$\ee\ra\nn\ho$ process. Upper plots are for the mass with hits 
observed in the detector acceptance $| {\mathrm{cos \theta}} | \leq 0.98 $,
and bottom plots are for $\leq 0.90 $. 
Left plots are those for none, one, two two-photon events overlapped.
The dotted lines in the right side plots are
the reconstructed mass in case 0.5 events are expected in average for
the overlap.
\label{fig:nnmass}}
\end{minipage}\end{center}
\end{figure}

In case two or more events happen in
single bunch collision, the separation of the events is extremely difficult.
The separation is almost
impossible for the neutral particles like $\gamma$ or K$_{L}$.
Some of the charged tracks might be rejected looking into the
production point in beam direction, however such rejection
might make significant bias in the Higgs
precise measurements since
the displaced vertex may come from the heavy quark decay of the Higgs.

Several simulation studies have been attempted. The simulation overlaps
the hard scattering events of signal or background
processes, with one or more two-photon events.
We have used the JSF Quick-simulator~\cite{JSF} for this purpose.
The data base are separately generated for the hard processes 
and the two-photon backgrounds. Afterward they are overlapped each other
with event production points displaced according to the assumed beam spread
at the interaction point. One of the significant effect is found in
the reconstructed mass of the Higgs in $\ee\ra\nn\ho$ process.

Figure~\ref{fig:nnmass} shows the reconstructed 
mass with/without two-photon overlap. 
No event selection is made for the plots. 
The distributions are shown for the different detector 
acceptance to be used in the mass reconstruction. 
Since the two-photon background tends to make energy-flow in the
forward region, 
the bias in the mass reconstruction could be reduced if we use the detector
information only from the central part, while the signal tagging efficiency
in a mass window is reduced as shown in the figure.

\begin{figure}
\hspace*{-0.5cm}
\centerline{
\epsfxsize=10.4cm \epsfbox{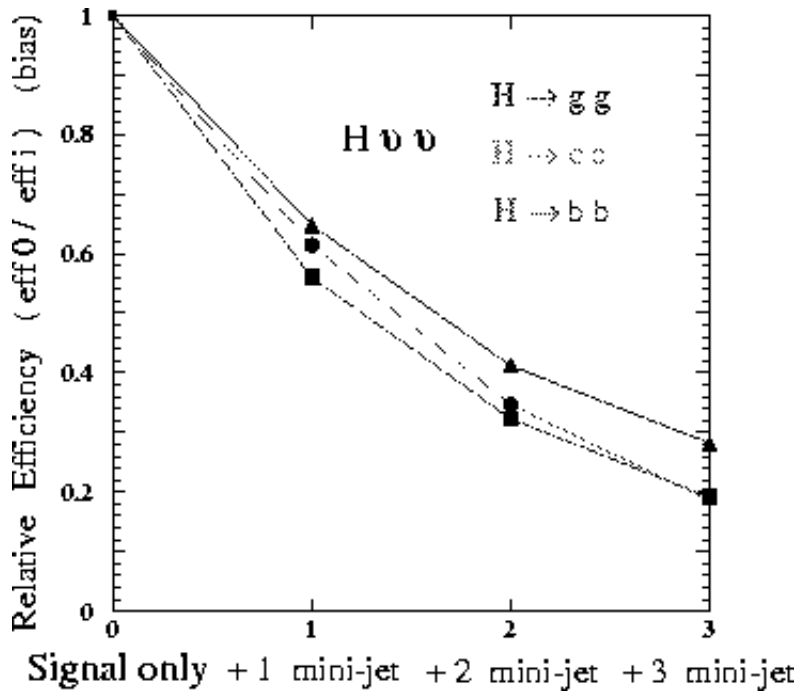}
}
\begin{center}\begin{minipage}{\figurewidth}
\caption[]{\sl
Degradation of the detection efficiency for the
$\ee\ra\ho\nn$ process as a function of the number of two-photon
processes overlapped to the signal event. Three lines correspond to
different decay of the Higgs. 
\label{fig:hggeff}}
\end{minipage}\end{center}
\end{figure}

The effect on the signal tagging efficiencies is shown 
in Fig\ref{fig:hggeff}. The vertical axis is the efficiency normalized to
that without two-photon overlap. Significant degradation of the tagging
efficiency is found. The main reason of the loss of the efficiency 
is the distorted mass distribution with overlapping two-photon events.
Also the other kinematic values such as missing momentum direction
are affected significantly. Furthermore the background contamination
especially We$\nu$ events
increases rapidly as two-photon overlap rate grows.

The other effect is that in the flavour tagging which
relies on the tracks displaced from the event primary vertex.  
The overlapping event has, in general, 
its event-production-point different from
the original event, since the beam has a finite width. 
If we use the information only in r-$\phi$ projection, the effect can be
minimized thanks to the tiny width of the beam in the projection.
However the beam spread is larger in the beam-direction.
Simulation studies have been made also for the flavour tagging.
Possible bias has been checked with
different assumptions for the beam spread in beam direction.
As we have expected,
the effects are found to be negligible in r-$\phi$ projection.
The results obtained in the 3-dimensional tagging 
show a small effects for 80 $\mu$m spread 
in z-direction at the interaction point which is the current design value
at JLC. When we increase the beam spread, the bias increases almost linearly.

Although the studies have just started and the analyses are not optimized for
the two-photon overlapping, the results indicate the several important
issues for the accelerators, detectors and monitoring systems.
Note that our aimed precisions for the Higgs measurements are 
percent order or less.
We need good ``time-stumps'' for each hits in the detectors to identify
the corresponding bunch. Both trackers and calorimetry are required to have
good time resolution.
Since the overlapping rate depends on the luminosity per bunch, 
the operation of the accelerators are required to be
stable bunch by bunch, and the monitoring of the luminosity per bunch
might be necessary. One may use the data at low luminosity to investigate
the effects in real data, overlapping the hits from hard scattering events 
with minimum-bias events including two-photons. 
Many studies are addressed in the future. 

Here we summarize the conclusion obtained so far.
\begin{itemize}
\item The effect of the two photon background can be minimized if we use
the central part of the detector in order to avoid the bias due to
two photon yields. 
\item The systematics can be minimized by the studies in data itself
using the overlapping events in hit level for events obtained
at lower luminosity data.
\item The effects in flavour tagging needs care. For small beam size
in beam direction in JLC the current design is good in order to
avoid the bias. If we make larger $\sigma_{z}$, we may need 
special tricks to reject tracks from two-photon. 
\end{itemize}

\subsection{Requests to Detector}

Through above discussion, here we list requests to 
detector design optimizations. 
\begin{enumerate}
\item Vertex detector should have at least 4 layers, each with r-$\phi$ and
z coordinates, in order to assure the local tracking capability.
\item To deal with two-photon overlap, the central tracker (CT) should have the
bunch identification (time stamp) performance of 1 nsec.
\item It would be very helpful to have a fast detector 
for the intermediate tracker (IT)
to help the bunch identification.
\item Calorimeter should have a time resolution of less than 1 nsec 
for a given hit. This request is very important especially for the endcap 
region. Granularity of the calorimeter also should be optimized. 
\end{enumerate}

Above requests are serious especially for high luminosity option of the
JLC accelerator.

\section{Future Studies to be Made}

Here we summarize the the high priority studies which should be made
in ACFA for JLC projects.

\paragraph{Physics channels\\}
\noindent
In this report we focus on the physics at JLC-I especially the SM Higgs
and the lightest CP-even Higgs. Important channels which are not described
in detail in this report are as follows.
\begin{itemize}
\setlength{\itemsep}{-3pt}
\item The Yukawa processes
such as $\ee\ra\toptop\ra\toptop\ho$,
and also heavier Higgs such as $\toptop\Ao$ and $\toptop\Ho$,
have to be studied in detail.      
\item Multi-Higgs production such as $\ho\ho\Zo$ and direct measurement of
the self-coupling strength. 
\item The heavy Higgs pair production, $\Ao\Ho$. 
\item Charged Higgs in $\ee\ra$H$^{+}$H$^{-}$ and the single production
such as $\ee\ra\tau\nu\Hpm$~\cite{singlech}.
\end{itemize}
These studies should be extended also at higher energies 
in a TeV and multi-TeV
regions which is the one of the options at later stages of the JLC project.

\paragraph{Analyses Techniques\\}
\noindent
The sensitivity to Higgs bosons are determined not only by the
accelerators and detectors but also the essential analyses techniques.
The essential techniques are;
\begin{itemize}
\setlength{\itemsep}{-3pt}
\item Energy-flow calculation.
\item Flavour tagging.
\item tau tagging.
\end{itemize}
The current studies uses rather simple and old-fashion procedures.
For example, in this study we use the simple likelihood calculation for the
flavour-tagging based on the impact parameters of the tracks.
At recent experiments like LEP/SLC, various more sophisticated methods with
much higher performance have been developed using more information
such as secondary vertex reconstruction. Multi-variables have been combined
using neural-network techniques or likelihood combinations. These experience
should be applied for the next steps. 
Above these crucial issues are under development and 
open for the future studies.

\paragraph{More studies in the real experimental situation\\}
\noindent
As we have described above, the experimental situation at JLC is much cleaner
than the hadron colliders. 
We have described the effects of the beam-beam interaction and
mini-jet (two-photon) event overlaps, which are major issues at high luminosity
operation at JLC. As is discussed, these are expected not to be
serious at JLC. In order to obtain better performance and feed-back to
final designs of the accelerator and detector, further studies are addressed.

\section{Summary}

The future linear $\ee$ collider, JLC, gives us fantastic occasions
to investigate the physics up to GUT scale through studies of Higgs
production and decay properties. 

In this report we studied varieties of Higgs physics based on the
recent progress in theoretical and experimental sides.

The experimental feasibilities and sensitivities are investigated 
especially for physics at early stage of the JLC phase-I.
We discussed the sensitive signal cross-section,
model-independent measurements, and expected accuracy in physics parameters
such as Yukawa-coupling.  
Full simulation studies for the tracking as well as the effect of the
event overlapping expected at JLC has been started already. 

The existence of the light Higgs predicted in SUSY models is first 
to be clarified.
At JLC, we may need only a few days to discover (at least one of) 
the Higgs bosons if the current design luminosity value 
of JLC is realized. 
There is always a potential for us 
to find unexpected ``big things'' through Higgs hunting and measurements even
at very early stage of the JLC experiment.

We have enough sensitivity even to the worst case with the lowest 
cross-section in SUSY models.
The physics background processes which have event topologies similar
to those of Higgs signals have cross-section just one or two orders of
magnitude higher than the signal. 
The backgrounds are well under control 
thanks to the well defined initial states,
which is necessary for the precise measurement of the Higgs and 
in order to be sensitive to new particle production with tiny cross-section.
We are sensitive to a cross-section down to 1 fb level.
One can say, in other words, it is really a ``big discovery'' when we 
find no Higgs at JLC phase-I. 

Varieties of discoveries are yet waiting for us. 
For instance, if Br$(\ho\ra\bb)$/Br$(\ho\ra\tau\tau)$ is found to be
different from $m_{b}^2/m_{\tau}^2$, all models belonging to 
SM or type-II 2HDM, such as SUSY models, are excluded. 
If Yukawa couplings of
top and charm normalized to its mass are found to be different, we need 
completely new theory for the fermion generation.
If the measured cross-section is smaller than
the minimum cross-section~\cite{okada,xmssm} expected in SUSY models, 
it means new physics
further beyond SUSY between EW and GUT scale.
We may discover CP-mixing in Higgs sector. 
We also may discover other Higgs with tiny cross-section at JLC
even at phase-I.

Our purpose of the
experiments at JLC is not only to judge the existence of the light Higgs,
but also to measure the Higgs properties precise enough if it exists.
We would measure the cross-section, the branching ratio, natural width
in percent order or less in its error at JLC based on 
more than 10$^5$ Higgs events in a few years running 
which results in
the precise measurements of the Higgs gauge coupling and
Yukawa-coupling, and furthermore derivation of Higgs self-couplings 
in multi-Higgs production. These can be made model independent fashion.

From these measurements, we further measure various model parameters,
and test in the internal consistency.
The precise values of the Higgs properties, we are also sensitive
to a loop effect of new particles such as scaler top quark.
The mass of the other Higgs and properties are indirectly measured
if we missed it at JLC phase-I, and can be tested at the next step of
the JLC. The information from LHC are also helpful. 
All of those,
which are expected to be done only with the next e$^{+}$e$^{-}$
Linear Collider, JLC,
are essential to give definite answers to physics models 
and to determine the fundamental structure of 
the interaction in nature up to GUT scale.
JLC is a big step to definitely answer the question whether we live 
in SUSY world or SM-like world, or completely new unexpected world. 

\begin{table}
\begin{center}\begin{minipage}{\figurewidth}
\caption[]{\sl
Accuracy at $\sqrt{s}=$300, 400 and 500 GeV 
with ${\cal{L}}$=500 fb$^{-1}$ for 120 GeV CP-even Higgs at JLC. 
The Higgs boson of SM-like is used as an
input.
\label{tab:summarysummary2}}
\end{minipage}\end{center}
\begin{center}
\begin{tabular}{|c|c|c|c|}
\hline
$\sqrt{s}$    &  300 GeV &  400 GeV &  500 GeV \\
\hline
\hline
$\Delta\mh$ (lepton-only) & 80 MeV & --- & --- \\
$\Delta\mh$               & 40 MeV & --- & --- \\
\hline
$\Delta\sigma/\sigma$ (lepton-only) & 2.1\% & 2.5\% & 2.9\% \\
$\Delta\sigma/\sigma$               & 1.3\% & ---   & ---   \\
$\Delta(\sigma_{h\nn}\cdot$Br($\bb$)& 2.0\% & ---   & ---   \\
ZZH-coupling $\Delta$ZZH/ZZH  & 1.1\% & 1.3\% & 1.5\% \\
WWH-coupling $\Delta$WWH/WWH  & 1.6\% & --- & --- \\
\hline
$\Delta\Gamma_{\ho}/\Gamma_{\ho}$  &  5.5\%   & 12\%  & 16\% \\
\hline
Yukawa coupling $\Delta\lambda/\lambda$ & & & \\
$\lambda_{b}$       &  2.8\%   & 6.1\% & 8.1\% \\
$\lambda_{\tau}$    &  3.5\%   & ---   & --- \\
$\lambda_{c}$       & 11.3\%   & 13\%  & 15\% \\\hline
$\lambda_{b}/\lambda_{\tau}$  &  2.3\% & --- & --- \\
$\lambda_{b}/\lambda_{c}$  &  11\%   & 12\%   & 14\% \\\hline
$\lambda_{up-type}$ &  4.1\%   & ---   & --- \\
$\lambda_{down-type}/\lambda_{up-type}$ &  3.2\%  & ---   & --- \\
\hline
$\Delta(\sigma\cdot$Br)/$(\sigma\cdot$Br)  & & & \\
$\ho\ra\bb$         &  1.1\%   & 1.3\% & 1.7\% \\
$\ho\ra\WW$         &  5.1\%   & 12\%  & 16\% \\
$\ho\ra\tautau$     &  4.4\%   & ---   & --- \\
$\ho\ra\cc$+gg      &  6.3\%   & ---   & --- \\
$\ho\ra\cc$         &  22\%    & 23\%  & 27\% \\
$\ho\ra$gg          &  10\%    & 11\%  & 13\% \\
$\ho\ra\gamma\gamma$& ---      & --- & --- \\
$\ho\ra\Zo\gamma$   & ---      & --- & --- \\
\hline
\end{tabular}
\end{center}
\end{table}

%% file: physsusy/main.tex
\chapter{Supersymmetry}
\label{chapter-physsusy}
\input physsusy/overview.tex

\input physsusy/scenarios.tex

\input physsusy/mssm.tex
\input physsusy/otherscenarios.tex

\input physsusy/detector.tex
\input physsusy/highscale.tex

\input physsusy/ref.tex

%% file: physsusy/overview.tex
\section{Introduction}

The existence of a light neutral Higgs boson strongly indicates
supersymmetry but it is by no means a solid proof. 
It is definitely necessary to discover {\it at least} one 
supersymmetric particle to prove it.
We have emphasized that there are a lot of chances to discover at least 
one SUSY particle at the JLC\cite{c2s4_REFSUSYTH}.
Although which is the first to be discovered is model-dependent, the SUSY
search methods to be invoked at the JLC is largely model-independent and,
once we find one SUSY particle, it will guide us to discover the next.
More important than discoveries are the precision measurements
of the masses and couplings of these SUSY particles, which
can be carried out model-independently, thereby allowing
us to test model assumptions.
It should also be noted that 
the polarized electron beam will play an essential role,
in the course of the SUSY searches and studies.

Basically, 
these statements made back in 1992\cite{chapter-physsusy:Ref:jlc1} 
are still valid except for the fact that the only SUSY breaking 
scheme available at that time was the one exploiting
gravity to transmit the spontaneous breaking of supersymmetry
in some hidden sector to our world.
In the past few years, there has been tremendous progress
and now we have a set of different models for such
SUSY breaking transmission.
The subgroup's first mission is thus to list up
currently available SUSY breaking schemes
and clarify their experimental implications.
For theorists, this means to try to exhaust
possible SUSY breaking mechanisms and 
examine their characteristic features
in terms of collider phenomenology and,
for experimentalists, this means
to identify possible new signatures of SUSY
and investigate their detectability at the JLC\cite{chapter-physsusy:Ref:rohini1}.
The other lines of studies carried out in the ACFA studies
include reexamination of the simplifying assumptions we made
in our past studies:
studies of $R$-parity breaking phenomena 
and special corners of the parameter space leading
to some experimentally difficult situation such
as the lighter chargino decaying into a stau plus a neutrino\cite{chapter-physsusy:Ref:rohini2}.

As already stated, even more interesting than the mere
discovery of SUSY particles are precision measurements
of their masses and couplings. 
The past JLC studies assumed an integrated luminosity of
a couple of tens of ${\rm fb}^{-1}$.
Recent advance in accelerator physics has, however,
opened up possibility of some really high luminosity
which was unconceivable at the time of the green book
being written\cite{chapter-physsusy:Ref:jlc1}.
The subgroup's next task is thus to clarify
what we can learn from first a couple of SUSY particles,
given a real high integrated luminosity such as
$1~{\rm ab}^{-1}$, 
quantify required precisions for model discrimination
for various observables, and translate them into
required detector performance.

In this chapter, we first quickly review the phenomenological
consequences of various SUSY breaking scenarios
and experimental expectations from the next generation
colliders.
We then perform simulated experiments in the framework
of the minimal supersymmetric standard model,
in order to demonstrate basic experimental 
techniques to study SUSY particles at the JLC:
measurements of masses, mixings, and quantum numbers
of SUSY particles.
This will be followed by discussions on
possible complications and ways out,
expected for some particular corner of the parameter space
or from the other scenarios including
gauge or anomaly mediated SUSY breaking
as well as $R$-parity violation.
We then examine the physics demands on the detector 
performance.
The precision measurements of SUSY particles at the JLC
will serve as a telescope to look into
physics at really high scale.
We conclude this chapter by showing how and to what extent
we can carry out this very ambitious program.

%% file: physsusy/scenarios.tex
\section{SUSY Breaking Scenarios}
\label{Chap:susy:Sec:scenarios}

It is clear that the SUSY is a broken symmetry,
if it exists at all.
It should not arbitrarily be broken, however,
as long as it is meant to solve the naturalness problem:
only Soft Supersymmetry Braking (SSB) terms are allowed
to tame the quadratic divergence of the Higgs mass correction\footnote{
Of course, `warped large' extra dimensions~\cite{RS1} 
might obviate the hierarchy problem completely.
}.
Phenomenologically viable models can thus be
classified in terms of how the SSB takes place and
how it is transmitted to our observable sector.
In almost all of the models, SUSY is broken dynamically 
at a high scale and then this breaking is mediated to our 
low energy world. 
Various SSB parameters at the high scale 
of SUSY breaking are determined by the choice of the SSB mechanism and 
the mediation mechanism. 
Various theoretical and experimental considerations restrict the scale 
to a rather big range $10^4$ GeV $<$ $M_{SSB}$ $<$ $M_{Pl}$. 
The low energy values of the SUSY breaking
parameters are then decided by evolving them back
to the weak scale via renormalization group equations. 
Consequently the sparticle masses and, in cases where mixing occurs, 
even their couplings depend on the SSB mechanism. 
Once these low energy values of sparticle properties
are measured, therefore, we can in principle point towards the
physics at high scale and hence at the SUSY breaking mechanism.

As already mentioned,
in the early days of SUSY model building there existed essentially only
one class of models where the SSB is transmitted via gravity to the
low energy world. 
The past few years changed the situation drastically
and now we have a set of different models that include
the following:
\begin{description}
\item[1)] 
Gravity mediated models that include (minimal) SUGRA
(mSUGRA), (constrained) MSSM (cMSSM), etc., where
supergravity couplings of the fields in the hidden sector
with the SM fields are responsible for the SSB terms.
The difference between mSUGRA and cMSSM lies in the fact that
the former fixes the Higgsino mixing mass parameter $\mu$
by demanding the radiative breaking of the EW symmetry,
while the latter leaves it as a free parameter.
Both assume universality of the gaugino and sfermion 
masses at the high scale.  
These models always have extra scalar mass parameter 
$m_0^2$ which needs fine tuning so that the sparticle exchange does not 
generate FCNC effects, at an unacceptable level.  
\item[2)]
Anomaly Mediated Supersymmetry Breaking (AMSB) models, for which 
supergravity couplings that induce mediation are absent and the 
SSB is caused by loop effects. 
The conformal anomaly, which is always present, generates the 
SSB terms and the sparticles acquire masses due to the
breaking of scale invariance.
Note that this contribution exists
even in the case of mSUGRA/MSSM, but is much smaller
in comparison with the tree level terms that exist in those models. 
This mechanism becomes a viable one for solely generating the SSB 
terms, when the quantum contributions to the gaugino masses
due to the `superconformal anomaly' can be large~\cite{RS2,GR}, 
hence the name Anomaly mediation for them.  
The slepton masses in the simplest model of this kind are 
tachyonic and require some other SUSY breaking mechanism
to obtain phenomenologically acceptable mass spectrum.
One way to fix this problem is to introduce
a scalar mass parameter $m_0^2$.
\item[3)] 
Gauge Mediated Supersymmetry Breaking (GMSB) models~\cite{gmsb_rev},
where the SSB is transmitted to the low energy world 
via a messenger sector through messenger fields which have 
gauge interactions.
These models have no problems with the FCNC 
and do not involve any scalar mass parameter.
\item[4)]  
Models where the SSB mediation
is dominated by gauginos~\cite{gaumsb}.
These models are based on the {\it brane world scenarios},
where the brane (our world) on which
the matter particles and their superpartners
live is separated in the bulk from
the one that is responsible for the SUSY breaking.
Consequently, the wave functions of the matter particles 
and their superpartners on the SUSY breaking brane are suppressed, 
whereas those of the gauginos are substantial, 
due to the fact that the gauge superfields live in the bulk. 
Hence the matter sector feels the effects of SUSY breaking
dominantly via gauge superfields. 
As a result, in these scenarios, one expects 
$m_0 \ll m_{1/2}$, reminiscent of the `no scale' models.
\end{description}
All of these models clearly differ in their specific predictions for various 
sparticle spectra, features of some of which are summarized in 
Table~\ref{T:rgplen:1}\cite{peskin_talk}, where the usual
messenger scale parameter $\Lambda$ had been traded for $M_2$ for ease of
comparison. 
\begin{table}[htb]
\begin{center}\begin{minipage}{\figurewidth}
\caption{ \sl
Predictions of different types of SUSY breaking
models for gravitino, gaugino, and scalar masses.
$\alpha_{i} = {g_{i}^{2}}/{4 \pi}$ 
(i=1,2,3 corresponds to $U(1)$, $SU(2)$ and $SU(3)$, respectively), 
$b_{i}$ are the coefficients of the
${-g_{i}^{2}}/{(4 \pi)^{2}}$ in the expansion of the $\beta$ functions 
$\beta_{i}$ for the coupling $g_i$ and $a_i$ are the coefficients 
of the corresponding expansion of the anomalous dimension. 
The coefficients 
$G_i$ are the squared gauge charges multiplied by various factors which 
depend on the loop contributions to the scalar masses in the different models.
\label{T:rgplen:1}}
\end{minipage}\end{center}
\begin{center}
\small
\begin{tabular}{cccc}
 Model & $m_{\tilde{G}}$ & $(mass)^2$ for gauginos & $(mass)^2$ for scalars\\
\hline
&&&\\
mSUGRA & ${{M_{SSB}^{2}} / {\sqrt{3} M_{pl}}}$ $\sim $ TeV &
$({\alpha_{i}}/{\alpha_{2}})^{2}$ $M_{2}^{2}$ & $m_{0}^{2} + \sum_{i}
G_{i} M_{i}^{2}$ \\
cMSSM & $M_{SSB} \sim 10^{10} - 10^{11}$ GeV & $\mbox{ }$ & $\mbox{  }$ 
\\
&&&\\
\hline
&&&\\
GMSB & $({\sqrt{F}}/{100TeV})^{2}$ eV &
$({\alpha_{i}}/{\alpha_{2}})^{2} M_2^2$ & $\sum_{i} G_{i}^{'} M_{2}^{2}$ \\
$\mbox { }$  & 10 $<  \sqrt{F} < 10^4 $ TeV &  & \\ 
&&&\\
\hline
&&&\\
AMSB & $\sim $ 100 TeV & 
$({\alpha_{i}}/{\alpha_{2}})^{2} ({b_{i}}/{b_{2}})^{2} M_2^2$ & 
$\sum_{i} 2 a_{i} b{i} ({\alpha_{i}}/{\alpha_{2}})^{2} M_2^2$ \\ 
&&&\\
\hline
\end{tabular}
\end{center}
\end{table}

As one can see, the expected gravitino mass 
varies widely in different models. 
The SUSY breaking scale
$\sqrt{F}$ in GMSB model is restricted to the range shown 
in the table by cosmological considerations. 
Since $SU(2)$ and $U(1)$ gauge groups are not
asymptotically free, {\it i.e.}, $b_i$ are negative, 
the slepton masses are tachyonic in the AMSB model, 
without a scalar mass parameter, as can be seen from 
the third column of the table. 
The minimal cure to this is, as mentioned before, to add an additional 
parameter $m_{0}^{2}$, not shown in the table, which however spoils
the RG invariance.  
In the gravity mediated models like mSUGRA, cMSSM, and 
most of GMSB models, 
gaugino masses unify at high scale, whereas 
in the AMSB models the gaugino masses are given by RG invariant equations 
and hence are determined completely by the values of the couplings at 
low energies and become ultraviolet insensitive. 
Due to this very different scale dependence,
the ratio of gaugino mass parameters at the weak scale in the two sets of
models are quite different: 
gravity mediated models and GMSB models have $M_1 : M_2 : M_3$ = 1 : 2 : 7,
whereas the AMSB model has $M_1 : M_2 : M_3$ = 2.8 : 1 : 8.3.
The latter therefore, has the striking prediction that the lightest chargino
$\tilde{\chi}^\pm_1$ and the lightest supersymmetric particle (LSP)
$\tilde{\chi}^0_1$, are almost  pure
SU(2) gauginos and are almost mass-degenerate.
The expected sparticle spectra in
any given model can vary a lot. 
But still one can make certain general
statements, {\it e.g.} the ratio of squark masses to slepton masses 
is usually larger in the GMSB models as compared to mSUGRA.
In mSUGRA one expects the sleptons to be lighter than the first two
generation squarks, the LSP is expected mostly to be a bino 
and the right-handed sleptons are lighter than the left-handed sleptons. 
On the other hand, in the AMSB models, the left- and right-handed sleptons 
are almost degenerate. 
The above mentioned degeneracy between $\tilde{\chi}_{1}^{\pm}$
and $\tilde{\chi}_{1}^{0}$ is lifted by the loop effects~\cite{extra1}. 
For $\Delta M$ = $m_{\tilde{\chi}_{1}^{\pm}}$ - $m_{\tilde{\chi}_{1}^{0}}$
$<$ 1 GeV, the phenomenology of the sparticle searches 
in AMSB models will be strikingly different from that in mSUGRA, MSSM, etc.
In the GMSB models, 
the LSP is gravitino and is indeed `light' for the range of the values of
$\sqrt{F}$ shown in Table~\ref{T:rgplen:1}. 
The candidate for the next lightest sparticle, 
the NLSP, can be $\tilde{\chi}^0_1$, $\tilde{\tau}_1$, or
$\tilde{e}_R$ depending on model parameters. 
The NLSP life time
and hence the decay length of the NLSP in lab is given by 
$ L = c\tau \beta \gamma \propto \frac{1}{(M_{NLSP})^5}$ $(\sqrt{F})^{4}$. 
Since the theoretically allowed 
values of $\sqrt{F}$ span a very wide range as shown in Table~\ref{T:rgplen:1}, 
so do those for the expected life time and this range is given by 
$10^{-4}$ $<$ c$\tau \gamma \beta$ $<$ $10^{5}$ cm.
Since the crucial differences in different models exist in the slepton
and the chargino/neutralino sector,  
it is clear that the leptonic 
colliders which can study these sparticles with the EW interactions, 
with great precision, 
can really play a crucial role in model discrimination.

The above discussion, which illustrates the wide `range' of predictions of
the SUSY models, also makes it clear that a general discussion of the  
sparticle phenomenology at any collider is far too complicated. 
This makes it even more imperative 
that we try to extract as much model independent information
as possible from the experimental measurements. 
This is one aspect where the JLC can really play an extremely important role.

\subsection{SUSY Questions to be Answered by Next Generation Colliders}

\begin{figure}[htb]
\centerline{
\epsfxsize=10cm 
\epsfbox{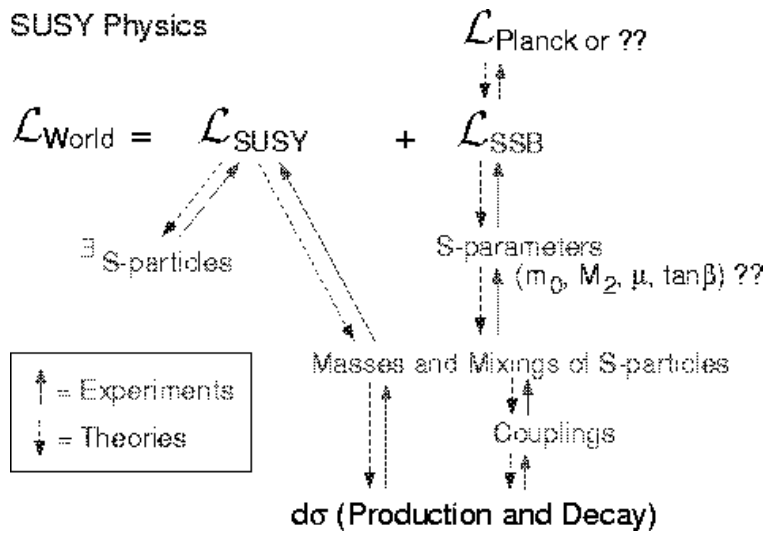}}
\begin{center}\begin{minipage}{\figurewidth}
\caption[Fig_Lworld]{\label{Fig_Lworld} \sl 
A schematic diagram showing the relation between the SUSY studies at
the JLC and physics at high scale.
The dashed arrows indicate the logical flow of theoretical implications,
while the solid arrows show that of experimental inputs.
}
\end{minipage}\end{center}
\end{figure}

The existence of a sparticle is a qualitative evidence for
the existence of a supersymmetric part (${\cal L}_{SUSY}$)
in the Lagrangian of the world (${\cal L}_{world}$).
The Lagrangian of the world must
contain, however, a soft SUSY breaking
part (${\cal L}_{SSB}$), which
determines masses and mixings of sparticles.
As stressed above, this SSB part is believed to be determined by 
physics at high scale: such as GUT or
Planck scale physics, 
and its studies will enable us to make
a first realistic step towards the ultra-high energy physics.
As illustrated in Fig.~\ref{Fig_Lworld},
experimentalists' tasks can, therefore, be summarized as follows:
(i) search for a sparticle to qualitatively 
prove the existence of ${\cal L}_{SUSY}$,
and 
(ii) determine masses and mixings of sparticles through the
measurements of various differential cross sections,
thereby testing SUSY quantitatively and uncovering the
structure of ${\cal L}_{SSB}$.
More specifically, we need to
\begin{itemize}
\item[1)] 
{\bf Find the sparticles} and establish their quantum numbers
by checking their interactions.
\item[2)] 
Establish coupling equalities implied by the SUSY, thereby
proving SUSY quantitatively.
\item[3)] 
Determine the {\bf scalar masses, gaugino masses}, and gaugino-higgsino
mixing.
\item[4)] 
Measure the properties of the third generation sfermions including the
L-R mixing.
\end{itemize}
The measurements mentioned in (3) above can give information about $\mu$,
$\tan\beta$, and some of the soft SUSY breaking parameters, 
whereas (4) above can
further add to the determination of $\mu$, tan$\beta$, trilinear A
parameters, and the scalar mass parameters. 

The LHC will be able to achieve
the goals given in `bold face' in the list above; for the remaining tasks we
need the clean environment of the $e^+e^-$ collider.

\subsection{What LHC Can Do}

Let us start with a summary of major 
hopes~\cite{cms_rep,atlas_tdr,lhc_susy,extra2} from the LHC for SUSY enthusiasts.
Various versions of `naturalness' arguments~\cite{barb,anderson,moroi} 
indicate that if theories are `natural', at least some of the sparticles, 
notably the gauginos/higgsinos, must be accessible at the LHC. Thus if SUSY 
is realized in  nature, the LHC should be able to provide some proof for it. 

Being a hadronic collider, the LHC is best suited
for the search of strongly interacting particle sector. 
The heavier `strongly
interacting' sparticles will be produced first and the lighter sparticles
with EW interactions only in the decay. 
The very high rates~\cite{tata1}
({\it e.g.}, even for a gluino mass  of 2 TeV, the expected cross-section 
is $\sim 10$ fb, giving about 1000 events for the high luminosity option)
make discovery easy. 

Methods have been developed to make accurate measurements
of different sparticle masses;
a nontrivial task as the worst background for 
SUSY searches is SUSY itself~\cite{paige1}. 
Depending on the point in mSUGRA parameter space chosen for analysis, a 
determination of $m_{\tilde{q}_L}, m_{\tilde{g}}$ up to an accuracy of 
$5-7 \% $ is possible, 
whereas the  masses $m_{\tilde{\chi}^0_1}, m_{\tilde{\chi}^0_2}$ 
can be determined 
with $ < 10\%$ accuracy~\cite{atlas_tdr,lhc_susy,tata1,paige1}. 
For some of the 
points chosen for studies high accuracies $\sim 1-2 \%$ are also possible
for neutralino mass determination.
The heavier gauginos are, however, inaccessible in general since
the rates for direct EW production are very low. 
The reach for sleptons at the LHC is also limited 
as compared to that for the strongly interacting 
particles and is $m_{\tilde{l}}$ $\leq$ 360 GeV unless it is produced
in cascades of squarks; a model dependent fact.
Their mass measurements are thus difficult, if not impossible. 
Ingenious methods have been developed to get an idea of the effective 
SUSY breaking scale~\cite{paige1}. 
However, accurate information about
the SUSY breaking scale {\bf and} mechanism generally does not seem easily
extractable. 
Further, a direct determination of quantum numbers and couplings
of the sparticles is not possible.
It has been shown that many
SUSY model parameters such as $\mu$, tan$\beta$, $M_2$, $M_3$ can be
determined with an accuracy of a percent level~\cite{paige1,atlas_tdr1},  
within a model. 
However, model independent analyses do not yet promise 
similar accuracy~\cite{paige2}. 

There is another point to make.
The lightest SUSY particle is a good candidate of
the cold dark matter in the universe.
It is thus desirable that collider measurements 
provide us with a clue about the nature of the dark matter.
Such an analysis
essentially needs determinations of the chargino/higgsino content of
$\tilde{\chi}_{i}^{0}$ and the slepton masses. 
At the LHC, however, this is, so far, only shown to be possible 
for $m_{\tilde{l}_{R}}$ $<$ $m_{\tilde{\chi}_{2}^{0}}$\cite{man-mih}.
On the other hand, a lepton collider can provide
the necessary information on these
sparticles, once produced, thereby making 
very crucial contributions. 
As a matter of fact, this information, 
if available, can play a very useful role in LHC analyses, too. 
Thus information obtained from the JLC can feed back 
into the LHC analyses.

\subsection{What We Expect from JLC}

The above discussion identifies the areas in which we
expect significant contributions from the linear collider:
\begin{description}
\item[1)] 
The JLC should provide {\bf precision} measurement of sparticle 
masses and mixing.
Since the machine energy limits
the sparticle spectrum that we can cover,
it is desirable that the JLC shall eventually reach
the TeV region.
\item[2)] 
The linear collider should be able to pin down quantum numbers such as 
spin, hypercharge and establish the equality of couplings predicted by SUSY.
\item[3)] 
Information from the LHC, along with measurements in (2) can then be 
used to get information about the SUSY breaking at high scale.
\end{description}
As seen before, the LHC can achieve the first goal only partially and the second
one only indirectly.
The information on sparticle masses obtained from the LHC
can serve as an important input to guide energy upgrade
strategy of the JLC.
The tunable energy of the $e^+e^-$ linear collider allows 
for sequential production of
various sparticles and hence a better control of the possible SUSY
background to SUSY searches.

\begin{figure}[ht]
\centerline{
\epsfxsize=2.8in\epsfysize=2.5in\epsfbox{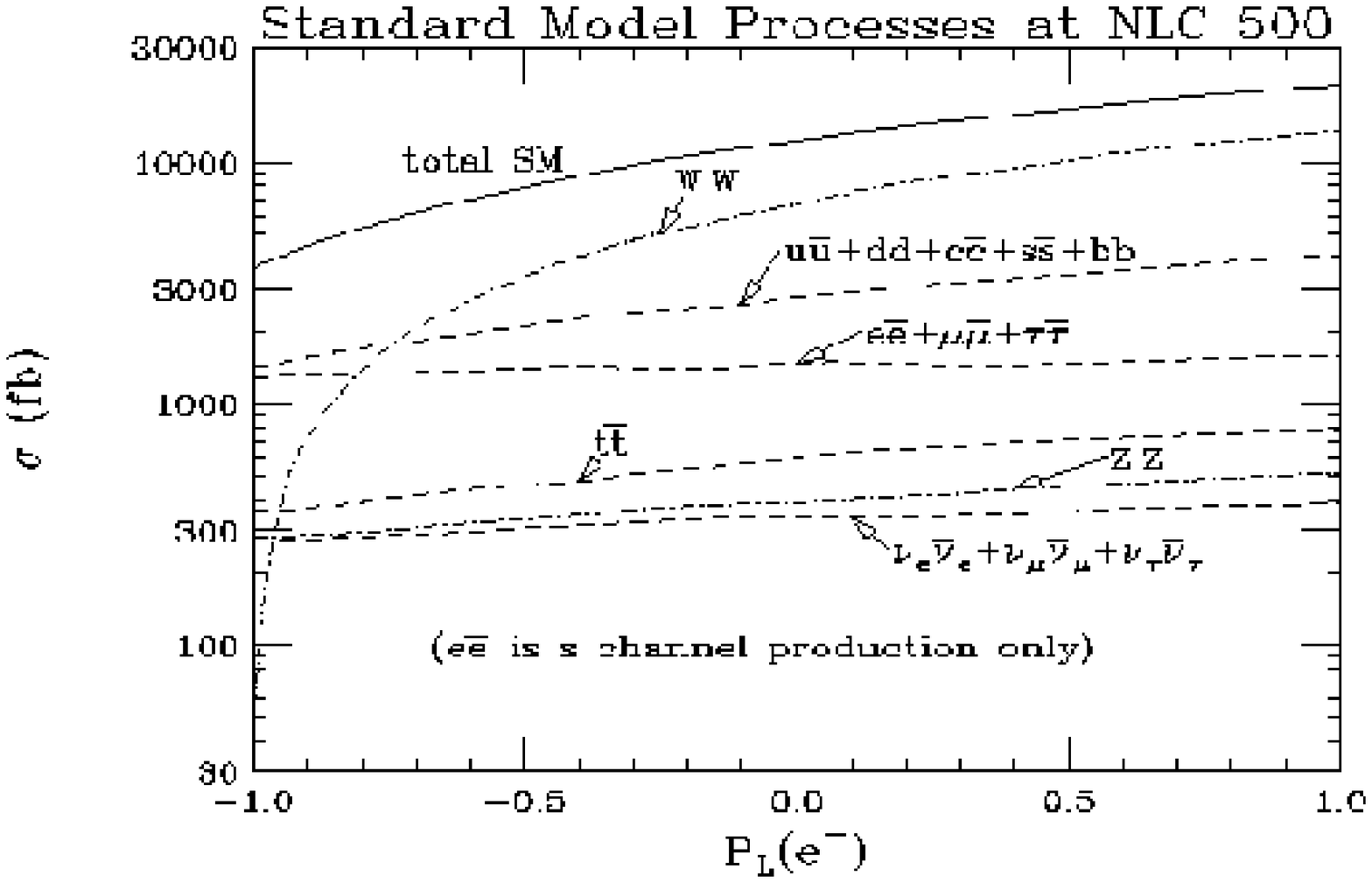}
\epsfxsize=2.8in\epsfysize=2.5in\epsfbox{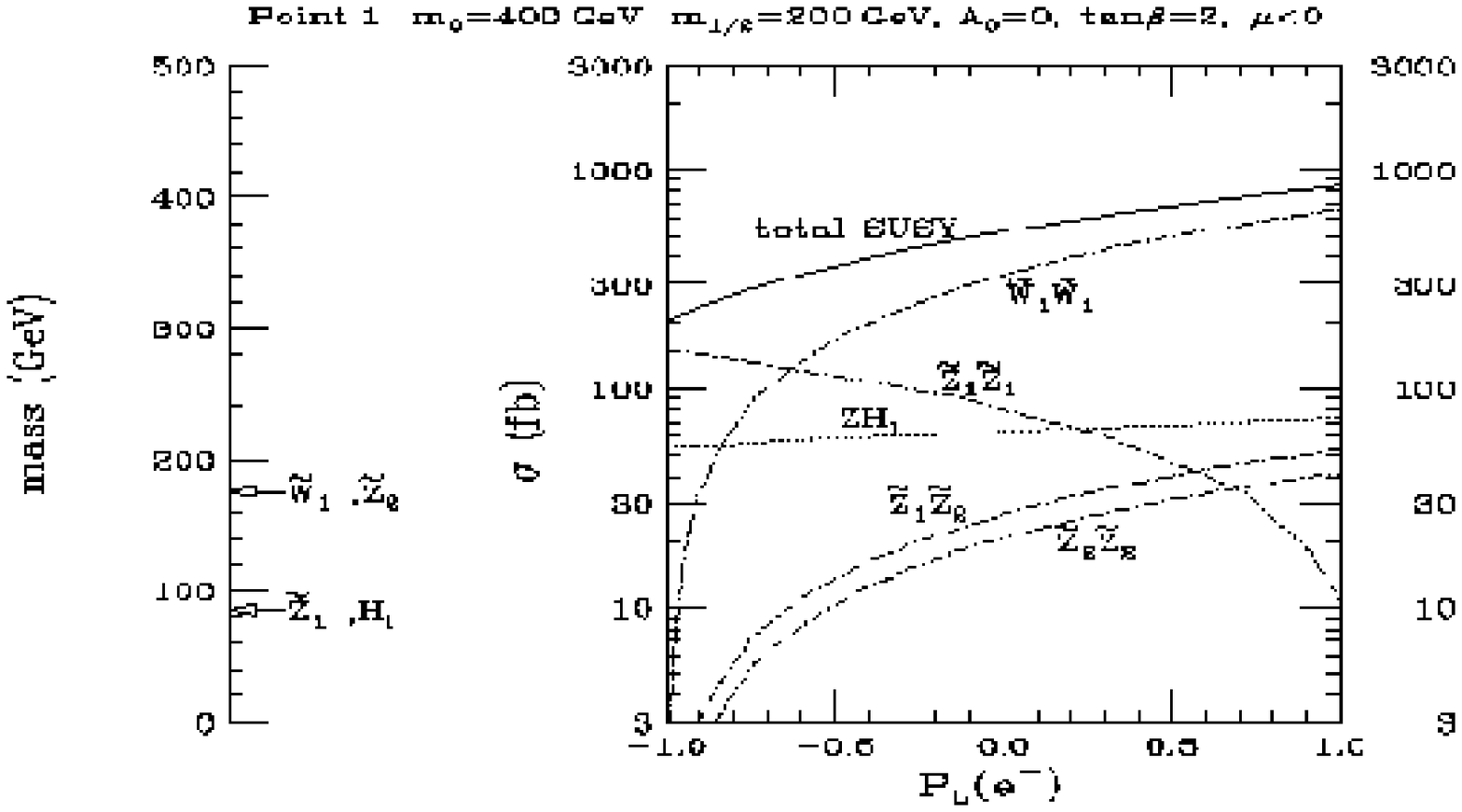}
}
\begin{center}\begin{minipage}{\figurewidth}
\caption{\sl 
Cross sections of different SM processes as well as the 
chargino/neutralino production. Model parameter values
are shown in the figure.
\label{F:rgplen:1}}
\end{minipage}\end{center}
\end{figure}

Since SUSY involves chiral fermions and their super-partners, 
The polarization of the initial $e^+$/$e^-$ beams can be used very
effectively to project out information about 
sparticle spectra and couplings.
Appropriate choice of the polarization can also reduce effectively the 
background due to $W^+W^-$  production which has a very high rate.
Fig.~\ref{F:rgplen:1} taken from Ref.~\cite{tata2} shows the
cross-sections for different SM processes and the corresponding ones for the
SUSY model dependent chargino/neutralino pair production, at a chosen  point 
in the mSUGRA parameter space. 
From the figure it is clear that with a judicious
choice of polarization of $e^-$/$e^+$ beam, the SM background can be handled
and precision measurements of chargino/neutralino sector are possible. 
The $e^+e^-$ collider produces democratically all the sparticles that have EW 
couplings. 
Hence it is better suited than the LHC to study the 
gauginos/higgsinos and sleptons and will complement the information in
these sectors from the LHC very effectively. 
The correlation between properties of gluino that 
will be obtained from the LHC and those of
chargino/neutralino sector from LHC/JLC can disentangle the various gaugino
mass parameters $M_i$ at weak scale. 
For reasons outlined above
knowledge about the relative values of $M_i (i=1,3)$ at the weak scale, 
from independent sources, contains crucial clues to the physics at high
scale. 
This also shows how truly the LHC and the JLC are complementary to each
other and thus how necessary both are to solve the puzzle of EWSB.

%% file: physsusy/mssm.tex
\section{Simulated Experiments}
\subsection{Model Assumptions}

To be specific,
we will work within the framework of supergravity (SUGRA) models 
with the ``GUT-condition''s,
although the search and study methods are largely 
model-independent. 
These models involve, in general, the following parameters:
($m_0$, $M_2$, $\mu$, $\tan \beta$),
which determine the mass spectra and the interactions of 
supersymmetric particles\cite{c2s4_REFHIKASA}.
$\mu$ is the higgsino mixing mass parameter and $\tan\beta$ is the
ratio of the two vacuum expectation values of two Higgs doublets.
$m_0$ is the scalar mass parameter, 
which is common to all the scalar particles 
in the case of SUGRA models.
$M_2$ is the $SU(2)$ gaugino mass parameter, 
which is related to the $SU(3)$ and
$U(1)$ gaugino mass parameters:
$$
{M_1 \over { 5 \over 3 } \alpha /\cos^2\theta_W }
={M_2 \over \alpha/\sin^2\theta_W } 
={M_3 \over \alpha_s }
$$
under the GUT conditions. 
As already mentioned in the previous section,
this means, numerically,
\begin{equation}\begin{array}{lll}
M_1:M_2:M_3 \simeq 1:2:7,
\end{array}\end{equation}
which implies the following inequality between the
lighter chargino and the gluino masses:
\begin{equation}\begin{array}{lll}
\label{Eq_mi}
m_{\tilde{\chi}^\pm_1} \lsim \frac{1}{3} m_{\tilde{g}}.
\end{array}\end{equation}
As demonstrated later, these SUGRA-GUT assumptions can be
tested to a high precision.

On the other hand, sfermion masses can be written in the form:
\begin{equation}\begin{array}{lll}
\label{Eq_mf}
m_{\tilde{f}}^2 = m_0^2 + G_{\tilde{f}} M^2 
+ D_{\tilde{f}} m_Z^2,
\end{array}\end{equation}
where the first term on the right-hand side is the
contribution from the common scalar mass ($m_0$).
The coefficient of the gauge term ($G_{\tilde{f}}$) is controlled
by the size of the gauge group to which the sfermion belongs and the
coefficient of the $D$-term ($D_{\tilde{f}}$) depends on $\tan \beta$ and
is of $O(1)$ or less. 
When $M_2^2 \gg m_Z^2$, the sfermion
mass spectrum is thus largely determined by the gauge term
in the gravity-mediated SUSY breaking models.
Taking into account
the above two mass relations (Eqs.(\ref{Eq_mi}) and (\ref{Eq_mf})),
we can conclude that colored sparticles are heavier than
colorless ones and right-handed sfermions are
lighter than left-handed ones, in the gravity-mediated
SUSY breaking models.

In order to avoid unnecessary complications, 
we will make the following simplifying
assumptions: a) the $R$-parity is exactly conserved
which implies that SUSY particles can only be pair-produced 
and the lightest SUSY
particle (LSP) is absolutely stable and   
b) the LSP is the lightest neutralino to be consistent with cosmology.
Complications expected when these simplifying
assumptions are lifted and possible ways out
will be discussed later.
Our first SUSY particle (FSP) candidates are thus the lighter chargino 
($\tilde{\chi}^{\pm}_1$) 
or the right-handed sleptons ($\tilde{l}^{\pm}_R$),
except for the light third generation case.
The low lying SUSY particles accessible for the JLC usually have 
a reasonable branching fraction for direct decays into 
the LSP\cite{c2s4_REFSUSYEXP}. 
Therefore, the signal for SUSY particle
productions is a missing transverse momentum or a large acoplanarity.

Taking these into account,
we will focus our attention on chargino and slepton pair productions.
In any case, it is usually straightforward to discover these SUSY  
particles at the JLC, once their thresholds are crossed. 
Moreover, we will be able to study their properties in detail. 
It should be emphasized that detailed studies of the first 
SUSY  particle alone can teach us a lot about the model parameters 
and will guide us to the discovery of the next.       

In what follows we will demonstrate how
a typical SUSY study program goes\cite{chapter-physsusy:Ref:tsukamoto.etal},
taking a sample case:
$$
(m_0,M_2,\mu,\tan\beta)=(70~{\rm GeV},250~{\rm GeV},400~{\rm GeV},2)
$$
unless otherwise stated\footnote{
The value of $\tan\beta$ has been excluded by the current
higgs mass bound, but increasing it will not alter the
following discussions in any significant manner.
}.
This parameter set gives the following sparticle mass spectrum:
$$
( m_{\tilde{\chi}^0_1},
m_{\tilde{\chi}^0_2},
m_{\tilde{\chi}^\pm_1},
m_{\tilde{l}_R},
m_{\tilde{l}_L},
m_{\tilde{\nu}_L} ) 
= (118, 222, 220, 142, 236, 227)~{\rm GeV}.
$$
Thus the first SUSY particle (FSP) in this case 
is the right-handed sleptons
pair-produced in $e^+e^-$ collisions.
The third generation sleptons will be treated separately, since
for them we expect significant left-right mixing.

\subsection{Study of $\protect \mathbold{\tilde{\it l}^{\pm}_{\it R} 
({\it l}\ne\tau)}$}
\subsubsection{Slepton Signature}

\begin{figure}[htb]
\centerline{
\epsfxsize=12cm\epsfbox{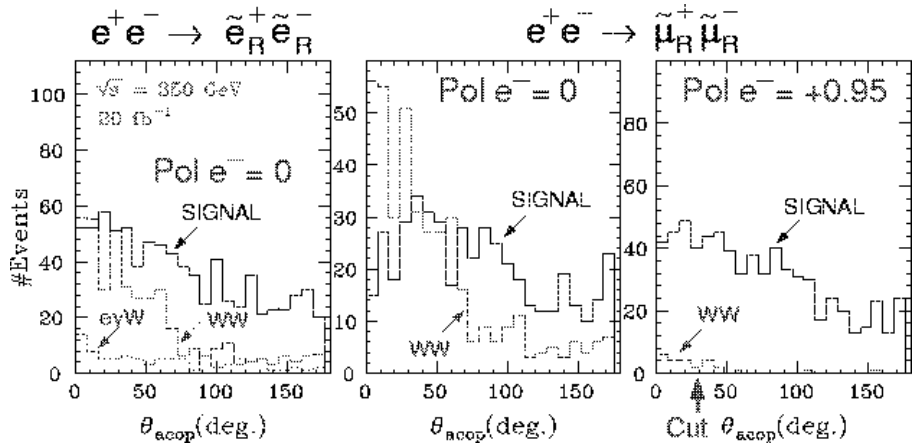}}
\begin{center}\begin{minipage}{\figurewidth}
\caption[ACOPSLEPTON]{\label{ACOPSLEPTON} \sl
Examples of acoplanarity distributions for
(a) selectron pair and (b) smuon pair productions. 
The Monte Carlo data correspond to an integrated luminosity of
$20~{\rm fb}^{-1}$ at $\sqrt{s} = 350~{\rm GeV}$ 
with an unpolarized electron beam.
The solid histograms are for the signal events, while the dashed
histograms are the background from $W^+W^-$ productions.
The effect of the right-handed electron beam is demonstrated in (c).
}
\end{minipage}\end{center}
\end{figure}

Since a right-handed slepton decays directly into a lepton plus
an LSP, the signal to look for is an acoplanar lepton pair:
$e^+e^- \to \tilde{l}^+_R\tilde{l}^-_R 
\to l^+\tilde{\chi}^0_1 l^-\tilde{\chi}^0_1$.
On the other hand, the background to this reaction is standard model (SM)
processes with neutrinos which mimic the LSP.
The key point here is the power of a highly
polarized electron beam available only at linear colliders.
For instance, one can eliminate the largest
SM background ($e^+e^- \to W^+W^-$ with $W \to l\nu$) very
effectively, using a right-handed electron beam:
the transverse $W$-pair production vanishes in the symmetry limit
since the $s$-channel diagram involves a $W_3$,
the third component of the $SU(2)_L$ gauge multiplet,
and the $t$-channel
diagram exchanges a $\nu_e$, both of which only couple to
left-handed electrons.
On the other hand, the signal cross section will be enhanced
because of the weak hypercharge difference between $e^-_L$ and $e^-_R$:
$\sigma_R = 4 \sigma_L$ in the symmetry limit.
Figs.~\ref{ACOPSLEPTON}-a) 
and -b) are examples of acoplanarity distributions for selectron and 
smuon pair productions, respectively, after the accumulation of 
$20~{\rm fb}^{-1}$ at $\sqrt{s} = 350~{\rm GeV}$.
We can see that the signals can be effectively enhanced over the background 
mainly consisting of $W$ pairs, by applying a cut on the acoplanarity angle. 
A detection efficiency of $\epsilon \gsim 50~\%$  is easily achieved with
a signal to background ratio of $S/B \gsim 1.6$ even for smuons.
As stressed above, the signal to background ratio can be 
further improved by using a right-handed electron beam 
as shown in fig.~\ref{ACOPSLEPTON}-c),
resulting in a very clean event sample
suitable for precision studies.

\subsubsection{Mass Determination}

Using this clean sample, we can determine the masses of the LSP and the
right-handed slepton through the measurement of the final-state lepton
energy distribution.
The two-body decay of any spinless particle gives a
flat energy distribution with its end points kinematically fixed
by the masses of the parent and the daughter particles:
in the case of the slepton decay, the slepton
and the LSP masses:
\begin{eqnarray}
E_{min} & = & \frac{m_{\tilde{\mu}_R}}{2} 
\left( 1 - \frac{m_{\tilde{\chi}^0_1}^2}{m_{\tilde{\mu}_R}^2} \right)
\gamma (1 - \beta) \cr
E_{max} & = & \frac{m_{\tilde{\mu}_R}}{2} 
\left( 1 - \frac{m_{\tilde{\chi}^0_1}^2}{m_{\tilde{\mu}_R}^2} \right)
\gamma (1 + \beta),
\end{eqnarray}
where $\gamma = E_{beam}/m_{\tilde{\mu}_R}$ and 
$\beta = \sqrt{1-(m_{\tilde{\mu}_R}/E_{beam})^2}$.

\begin{figure}[htb]
\centerline{
\epsfxsize=13cm\epsfbox{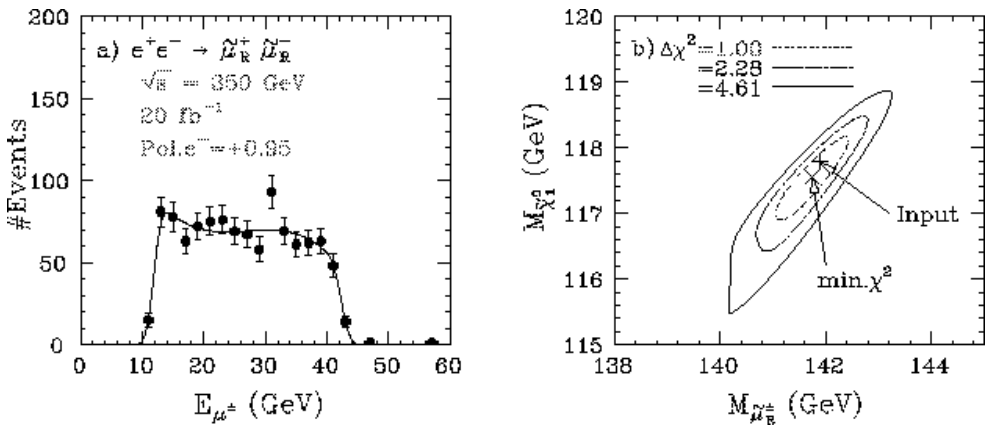}}
\begin{center}\begin{minipage}{\figurewidth} 
\caption[ELSMU]{\label{ELSMU} \sl 
(a) The energy distribution of muons from smuon decays 
for the same Monte Carlo parameters as with Fig.\ref{ACOPSLEPTON}-b).
The solid line corresponds to the best fit curve,
letting $m_{\tilde{\mu}_R}$ and $m_{\tilde{\chi}^0_1}$ move freely.
(b) The contours in the $m_{\tilde{\mu}}$-$M_{\tilde{\chi}^0_1}$
plane obtained from the fit to the energy distribution.
}
\end{minipage}\end{center}
\end{figure}

Fig.~\ref{ELSMU}-a) is the energy distribution of muons 
from smuon decays for an integrated luminosity
of $20~{\rm fb}^{-1}$ with a right-handed electron beam. 
Though the distribution is a little bit different from the expected
rectangular shape due, primarily, to selection cuts, 
the lower and the higher edges are sharp enough. 
These end points of the energy spectrum determine $m_{\tilde{\mu}}$ and
$M_{\tilde{\chi^0_1}}$. 
Fig.~\ref{ELSMU}-b)
shows the contours obtained from the fit to the Monte
Carlo data.
We can determine the smuon and the LSP masses 
to a 1\% level.

\subsubsection{Test of Generation Independence of Sfermion Masses}

With the right-handed selectron and smuon masses determined this way,
we can make a very important test of the SUSY breaking sector,
which is a test of generation independence of sfermion masses.
\begin{figure}[htb]
\centerline{
\epsfxsize=7cm 
\epsfbox{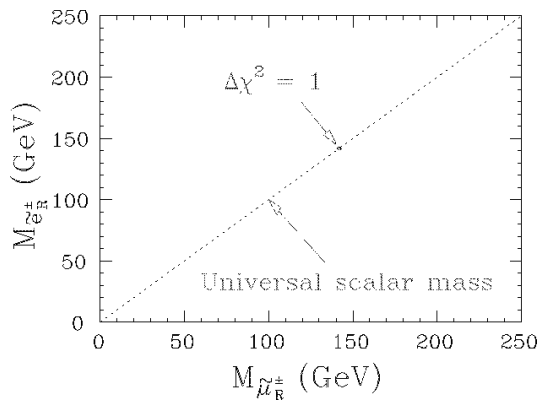}}
\begin{center}\begin{minipage}{\figurewidth}
\caption[Fig_slr_msesm]{\label{Fig_slr_msesm} \sl
The expected $\Delta \chi^2=1$ contour 
in the $m_{\tilde{\mu}_R}$-$m_{\tilde{e}_R}$ plane.
}
\end{minipage}\end{center}
\end{figure}
The gravity-mediated SUSY breaking implies the universality of
scalar masses at the GUT scale, which leads to the mass degeneracy
of the first and the second generations.
Fig.~\ref{Fig_slr_msesm} is an example of the test of the
generation independence, demonstrating potential precision
available at the JLC.

\subsubsection{Measurements of Differential Cross Sections}

Now that we know the masses of the LSP and the sleptons, we can
solve kinematics up to 2-fold ambiguities.
\begin{figure}[htb]
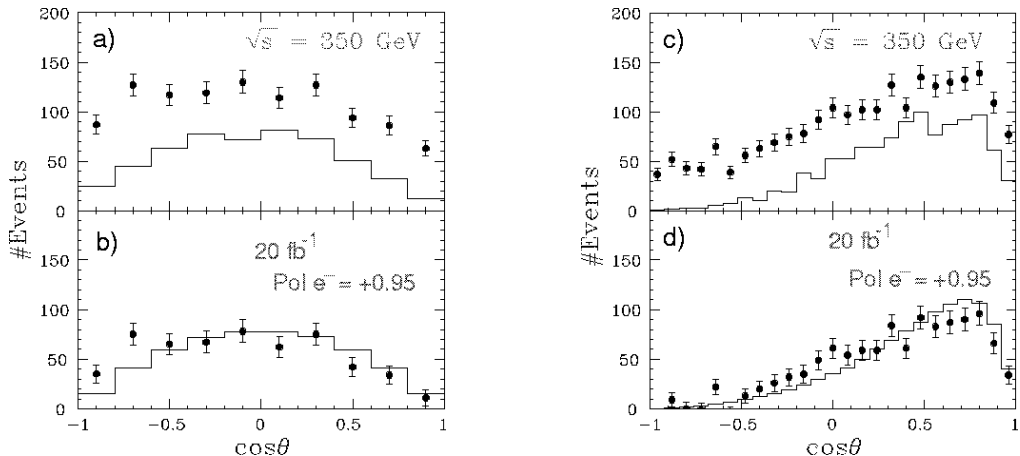

\centerline{
\epsfxsize=5.8cm 
\epsfbox{physsusy/figs/02_slr_xsm.epsf}
\hskip 1.8cm
\epsfxsize=5.8cm 
\epsfbox{physsusy/figs/02_slr_xse.epsf}
}
\begin{center}\begin{minipage}{\figurewidth}
\caption[Fig_slr_xsm]{\label{Fig_slr_xsm} \sl
(a) Production angle distribution of $\tilde{\mu}^-_R$ with respect to
the electron beam axis. The points with error bars are the
distribution of the two solutions reconstructed from the selected
sample corresponding to Fig.~\ref{ELSMU}-a).
The histogram is the generated $\cos\theta$ distribution for the
selected sample.
(b) Production angle distribution after the background subtraction compared 
with the scaled generated distribution before selection cuts.
No acceptance correction is applied to the reconstructed distribution.
(c) and (d) are similar plots to (a) and (b) for
$e^+e^- \to \tilde{e}^+_R \tilde{e}^-_R$.
}
\end{minipage}\end{center}
\end{figure}
Fig.~\ref{Fig_slr_xsm}-a) plots the two solutions for the
cosine of the smuon production angle.
Comparison of this with the corresponding generated 
one (histogram) suggests that the wrong solution makes
a flat background.
After the background subtraction, we get the plot
in Fig.~\ref{Fig_slr_xsm}-b), which shows a $\sin^2\theta$
distribution characteristic of $s$-channel pair productions of
a spinless particle.
In this way we can confirm that our smuon is really a scalar
particle.
 
On the other hand, the right-handed selectron pair production has
$t$-channel neutralino-exchange diagrams in addition.
Notice that only their bino ($\tilde{B}$) components
contribute here, since the wino component does not
couple to the right-handed sleptons and the higgsino
couplings are proportional to the electron mass.
The $t$-channel diagram, if active,
will produce a forward peak if significant.
Figs.~\ref{Fig_slr_xsm}-c) and -d) are similar plots to
Figs.~\ref{Fig_slr_xsm}-a) and -b) for the
right-handed selectron pair production.
The forward peak indicates that the LSP in our sample case 
is bino-dominant.
The JLC's polarized electron beam will, again, play a crucial role here 
since the $t$-channel diagrams exist only for the selectron 
with the same chirality as the beam electron. 

\subsubsection{Test of Supersymmetry}
\label{Chap:susy:susytest}

Up to now, we have been implicitly assuming that
the discovered sparticles obey supersymmetry.
In this subsection, we turn our attention to
how we can quantitatively test supersymmetry.

Supersymmetry not only demands existence of a
super-partner for each of the standard model particles
but also requires the supersymmetric relations
among couplings that are related by supersymmetry.
Because of this, the fermion-sfermion-gaugino coupling,
for instance, has the following relation with
the fermion-fermion-gauge coupling
in the lowest order of perturbation theory:
\begin{eqnarray}
g_{\tilde{B}\tilde{e}_R e_R}&= \sqrt{2}g\tan\theta_W=\sqrt{2}g',\\
\end{eqnarray}
If we can test such a SUSY relation,
we can quantitatively prove that
there is indeed supersymmetry in nature.
Here, we discuss to what extent we can
test the SUSY relation for 
the $e$-$\tilde{e}_R$-$\tilde{B}$ coupling,
using the selectron pair production.

Let us parametrize the deviation of this coupling
from the gauge coupling as
$$
g_{\tilde{B}\tilde{e}_R e_R} = \sqrt{2}g' Y_{\tilde{B}},
$$
and consider the constraint on $Y_{\tilde{B}}$
from the $\tilde{e}_R\tilde{e}_R$ production.
In the limit that the beam energy is high enough
and the neutralino mixing is negligible: 
$\sqrt{s} \gg m_Z$ and $M_1 \ll \mu$,
The $\tilde{e}_R$ pair production amplitude
can be written in the form: 
$$
{\cal M}\propto\sin\theta\left[
1 - \frac{4Y^2_{\tilde{B}}}{1-2\cos\theta\beta_f +\beta_f^2 4M_1^2/s}
\right],
$$
where $\theta$ is the selectron production angle measured
from the electron beam axis.
Notice that the $s$($t$)-channel particle that contributes to the
right-handed selectron pair production
is only $B$ ($\tilde{B}$),
and consequently the cross section only depends on
the hyper-charge.

\begin{figure}[htb]
\centerline{
\epsfxsize=7cm
\epsfbox{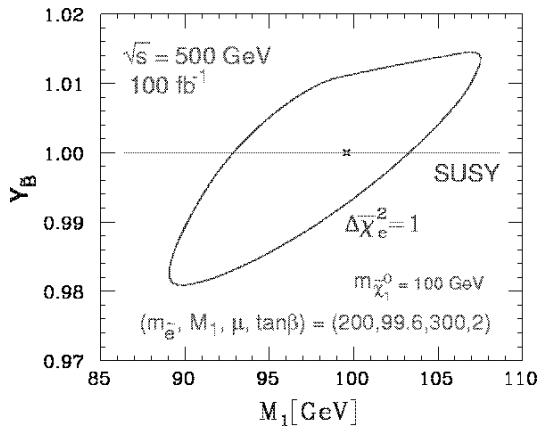}
}
\begin{center}\begin{minipage}{\figurewidth}
\caption[physics:susy:stau:susytest]{\sl \label{physics:susy:stau:susytest}
The $\Delta\chi^2=1$ contour in
the $M_1$-$Y_{\tilde{B}}(\equiv g_{\tilde{B}\tilde{e}_Re}/g'\sqrt{2})$
plane for pairs of $\tilde{e}_R$ with a mass of 200 GeV generated at
$\sqrt{s}=500$ GeV with $\int L dt=100 fb^{-1}$.
Input parameters are $\mu=300$ GeV, $M_1=99.57$, and $\tan\beta=2$. 
}
\end{minipage}\end{center}
\end{figure}

The formula tells us that we can constrain
$Y_{\tilde{B}}$ by measuring the differential cross section:
$d\sigma(e^+e^-\rightarrow \tilde{e}^+_R \tilde{e}^-_R)/d\cos\theta$
(we have already demonstrated that the selectron production angle
can be reconstructed from the daughter $e^+$ and $e^-$
momenta.
Fig.\ref{physics:susy:stau:susytest} is the constraint
shown in the $M_1$-$Y_{\tilde{B}}$ plane from
Monte Carlo data sample corresponding to 100 fb$^{-1}$.
We used the information on the selectron and the neutralino
masses, and the selectron differential cross section,
and assumed the GUT relation between $M_1$ and $M_2$.
Notice that once we find
the charginos or the other neutralinos,
we can use them to
constrain the ino mass matrices,
and further improve the constraint on $Y_{\tilde{B}}$
and other couplings.

On the other hand, $Y_{\tilde{B}}=1$ holds only at the
lowest order of perturbation and is expected to
deviate from $1$ through radiative corrections.
In the case of $m_{\tilde{q}}\gg m_{\tilde{l}}$,
the sizes of the radiative corrections are
\begin{eqnarray}
\delta Y_{\tilde{B}\tilde{e}_Re}&\sim 
0.007 \log_{10} (m_{\tilde{q}}/m_{\tilde{l}})
\\
\delta Y_{\tilde{W}\tilde{\nu}e}&\sim 0.02 \log_{10}
(m_{\tilde{q}}/m_{\tilde{\nu}})
\end{eqnarray}
and their effects on the cross sections are expected to be
2.8\% and 8.2\%, respectively, for
$m_{\tilde{q}}/m_{\tilde{\nu}}\sim 10$.
If we have enough statistics for
$\tilde{e}_{L,R}$, $\tilde{\nu}_L$
we can thus put some limit on $m_{\tilde{q}}$
through the measurements of the sizes of the radiative corrections.

\subsection{Study of $\protect \mathbold{\tilde{\chi}^{\pm}_1}$}

The charginos $\chi_1^\pm$, $\chi_2^\pm$ are 
the mixtures of charged wino $\tilde{W}^\pm$ 
and charged higgsino $\tilde{H}^\pm$ 
obtained from the diagonalization of the mass matrix
\begin{equation}                                                               
\label{Eq:M_chic}
{\cal L}_{\rm mass} = (\tilde{W}^+ ~~ \tilde{H}^+) 
\pmatrix{ M_2 & \sqrt{2}m_W\cos\beta \cr
\sqrt{2}m_W\sin\beta & \mu  \cr }
\pmatrix{ \tilde{W}^- \cr \tilde{H}^- \cr }.
\end{equation}
Notice that this mass matrix is given in terms of 
the three SUSY parameters: ($M_2$, $\mu$, $\tan\beta$) 
of which $M_2$ and $\mu$ mainly control the mixing
between the weak eigenstates.
When the splitting between $M_2$ and $\mu$ is large, 
the mixing is small and the pure states, the charged wino and higgsino, 
essentially become mass eigenstates.

From the slepton studies, we can determine the mass of
the LSP($\tilde{\chi}^0_1$) to better than 1~\%.
From this and the mass formula above,
we can set an upper limit on the lighter chargino mass,
assuming the GUT relation among the gaugino mass parameters.
This upper limit corresponds to the gaugino-dominant case
($m_{\tilde{\chi}^\pm_1} \simeq 2 \times m_{\tilde{\chi}^0_1}$),
while the LSP and the lighter chargino will be almost
mass-degenerate in the higgsino-dominant case.
The upper limit is plotted in Fig.~\ref{Fig_chic1_01}
as a function of the LSP mass. 

\subsubsection{Chargino Signature}

We thus set our center of mass energy just above $4 \times m_{LSP}$ and
look for $\tilde{\chi}^\pm_1$ pair productions.
In our sample case, the chargino decays into a real $W$ and an LSP
($\tilde{\chi}^\pm_1 \to W^\pm \tilde{\chi}^0_1$).
\begin{figure}[htb]
\hspace{0.5cm}
\begin{minipage}[t]{7cm}
\centerline{
\epsfysize=5.5cm 
\epsfbox{physsusy/figs/03_chic1_01.epsf}}
\caption[Fig_chic1_01]{\label{Fig_chic1_01} \sl
Upper limit on the lighter chargino mass as a function of
the LSP mass, assuming the GUT relation.
}
\end{minipage}
\hfill
\begin{minipage}[t]{7cm}
\centerline{
\epsfysize=5.5cm 
\epsfbox{physsusy/figs/03_chic1_acop.epsf}}
\caption[Fig_chic1_acop]{\label{Fig_chic1_acop} \sl
Acoplanarity angle distribution for 
$e^+e^- \to \tilde{\chi}^+_1 \tilde{\chi}^-_1$
together with major standard-model backgrounds.
}
\end{minipage}
\hspace{0.5cm}
\end{figure}
The signal to look for will thus be an acoplanar $W$ pair 
in 4-jet final states.
Fig.~\ref{Fig_chic1_acop} shows the acoplanarity angle 
distribution for the signal events (solid histogram) together
with major standard-model backgrounds:
the $W^+W^-$ (dash),
the $e^+e^-W^+W^-$ (dot), and the sum of
the $e^\pm \mathop{\nu_e}\limits^{\mbox{\tiny $(-)$}} W^\mp Z^0$,
$W^+ W^- Z^0$, and $\nu_e \bar{\nu}_e W^+ W^-$ 
(dot-dash) productions,
after demanding two $W$ candidates in the final states.
A cut at $\theta_A = 30^\circ$ will give us a fairly clean sample
with a selection efficiency in excess of $10~\%$ including
the branching fraction to the 4-jet final states.

\subsubsection{Mass Determination}

As we did to the sleptons, we can first use this clean sample to determine
the chargino mass.
\begin{figure}[htb]
\centerline{
\epsfxsize=14cm 
\epsfbox{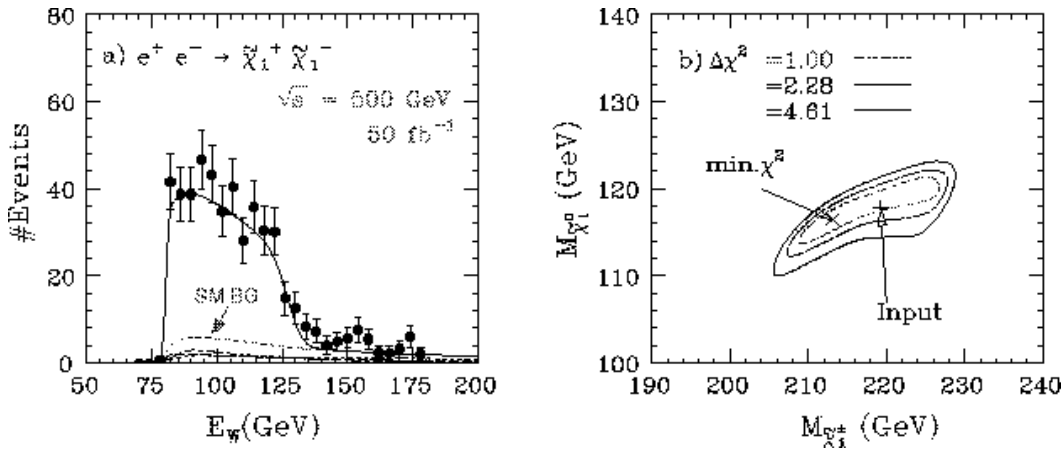}}
\begin{center}\begin{minipage}{\figurewidth}
\caption[Fig_chic1_cmass]{\label{Fig_chic1_cmass} \sl
(a) Energy distribution of final-state $W$'s from 
chargino decays: $\tilde{\chi}^\pm_1 \to W^\pm \tilde{\chi}^0_1$
(points with error bars) for the sample shown in
Fig.~\ref{Fig_chic1_acop} after the acoplanarity angle cut
at $\theta_A = 30^\circ$.
The solid curve is the best-fit curve to determine
$m_{\tilde{\chi}^\pm_1}$ and $m_{\tilde{\chi}^0_1}$.
The other curves correspond to the histograms in Fig.~\ref{Fig_chic1_acop}.
(b) Resultant contours from the 2-parameter fit.
}
\end{minipage}\end{center}
\end{figure}
Fig.~\ref{Fig_chic1_cmass}-a) plots the expected
energy distribution of the final-state $W$'s from the chargino decays.
The solid curve is the best-fit curve from a 2-parameter fit
letting the $m_{\tilde{\chi}^\pm_1}$ and $m_{\tilde{\chi}^0_1}$
move freely. 
Notice that the fit includes the
standard-model backgrounds
shown as dashed, dot-dashed, and dotted curves,
corresponding to those in Fig.~\ref{Fig_chic1_acop}.
Fig.~\ref{Fig_chic1_cmass}-b) is the resultant contour plot in
the $m_{\tilde{\chi}^\pm_1}$-$m_{\tilde{\chi}^0_1}$ plane,
which tells us that we can determine
the chargino mass to $\Delta m_{\tilde{\chi}^\pm_1} \simeq 8~{\rm GeV}$.
If we use the LSP mass constraint from the slepton study,
we can improve this to $\Delta m_{\tilde{\chi}^\pm_1} \simeq 5~{\rm GeV}$.

\subsubsection{Measurement of Differential Cross Section}

\begin{figure}[h]
\centerline{
\epsfxsize=7cm\epsfbox{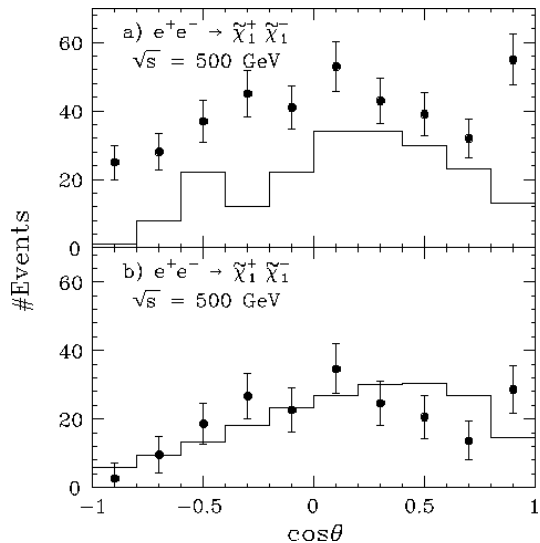}}
\begin{center}\begin{minipage}{\figurewidth}
\caption[Fig_dsig_chic]{\label{Fig_dsig_chic} \sl
(a) An example of the reconstructed production angle distribution
(data points) compared with the generated angle distribution
of the selected sample (histogram). 
The Monte Carlo events were generated at
$\sqrt{s} = 500~{\rm GeV}$ and correspond to an integrated luminosity
of $50~{\rm fb}^{-1}$.
We have assumed a $100 \%$ charge ID efficiency for $W$'s here.
The two solutions were plotted in the same figure.
(b) Same as (a) but after the subtraction of the background due
to wrong solutions.
The histogram is the scaled generated angle distribution
for the initial sample before selection cuts, while the
data points are the reconstructed without acceptance correction.
}
\end{minipage}\end{center}
\end{figure}

The clean sample can also be used to study
the angular distribution of $\tilde{\chi}_1^\pm$'s
as has been done to
the right-handed slepton-pair production.
One can determine the four-momenta of the charginos 
up to a two-fold ambiguity, once their masses are known. 
Fig.\ref{Fig_dsig_chic}-a) plots the angular distribution with the two-fold
ambiguity unresolved.
The ``wrong'' solutions again give an almost flat distribution, 
which can be subtracted to reproduce the real angular distribution
as shown in Fig.\ref{Fig_dsig_chic}-b)\footnote{
We assumed here that we can determine the charge of 
at least one $W$ candidate in a reconstructed event by using, 
for instance, the charge of a lepton from
charm decay or the reconstruction of a charmed meson or both.
This is one of the most important cases in which
a good particle ID system is essential. 
}.
The event excess in the forward region suggests that
there is some diagram with $t$-channel particle exchange. 
The only particle exchanged in the $t$-channel here 
is the electron sneutrino $\tilde{\nu}_{e_L}$. 

\subsubsection{Test of SUGRA-GUT Relation}

More interesting is the measurement of the production cross section
for the polarized electron beam.
Notice that, for the right-handed electron beam, 
only higgsino components contribute to the chargino pair production
in the symmetry limit, since the gauge boson exchanged in
the $s$-channel is $B$ ($U(1)_Y$ gauge boson),
while the $t$-channel sneutrino exchange diagram is absent.
By measuring the production cross section for the right-handed
electron beam, we can thus determine the composition of
$\tilde{\chi}^\pm_1$.

\begin{figure}[hbt]
\begin{minipage}[t]{7.5cm}
\centerline{
\epsfysize=5.3cm 
\epsfbox{physsusy/figs/03_chic1_gut.epsf}
}
\caption[Fig_chic1_gut]{\label{Fig_chic1_gut} \sl
Constant $\chi^2$
contours in the $M_2$-$M_1$ plane obtained from the global fit
explained in the text.
The dotted line shows the GUT relation.
}
\end{minipage}
\hfill
\begin{minipage}[t]{7.5cm}
\begin{center}
\epsfysize=5.3cm\epsfbox{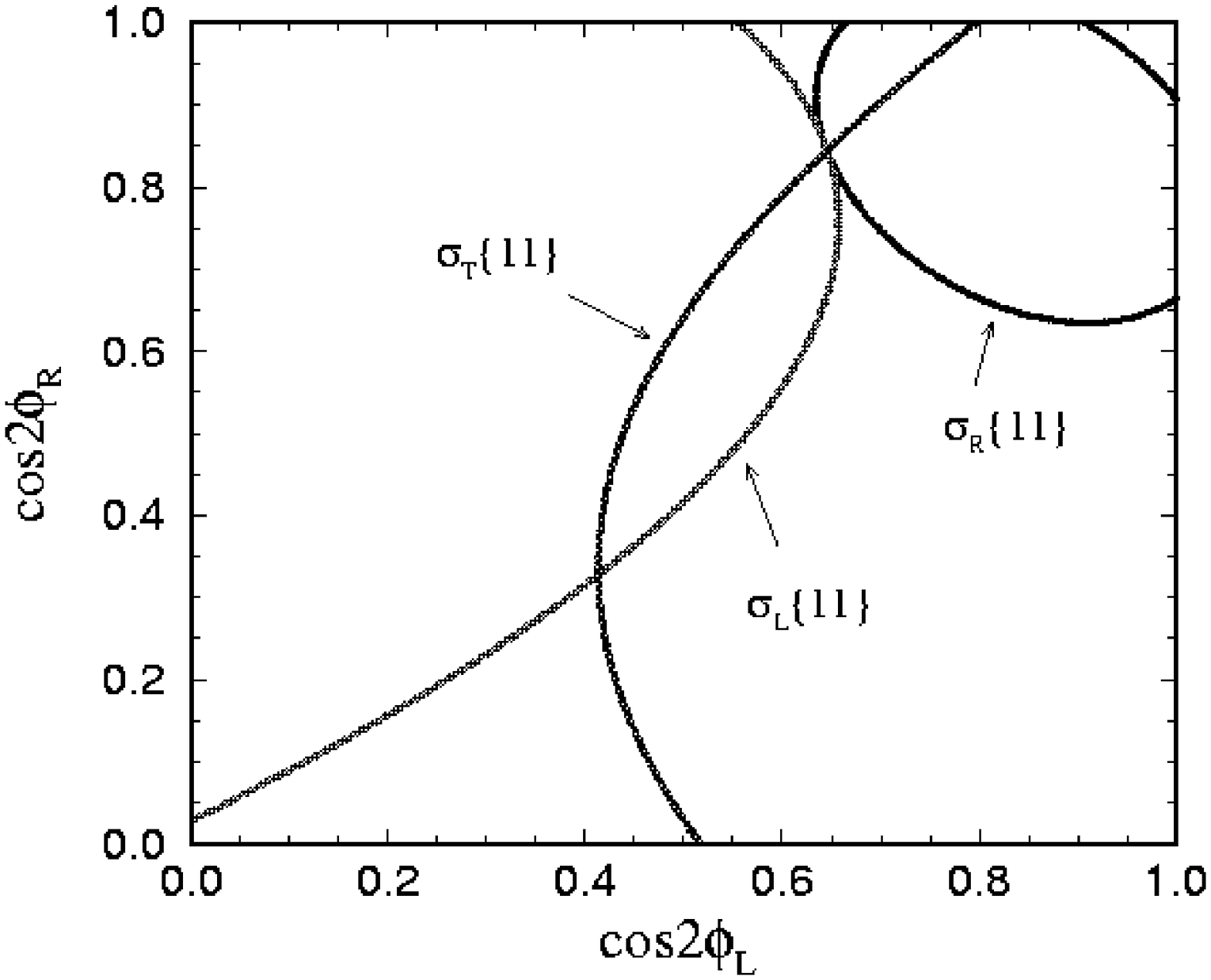}
\end{center}
\vspace{-26pt}
\caption[Fig:choi_cosL_cosR_1]{\label{Fig:choi_cosL_cosR_1} \sl 
Contours of the cross sections $\sigma_L\{11\}$,~$\sigma_R\{11\}$ 
and $\sigma_T\{11\}$ in the $\cos2\phi_L$-$\cos2\phi_R$ plane 
for $\tan\beta=3, m_0=100~{\rm GeV}, M_{1/2}=200~{\rm GeV}]$ 
at $\sqrt{s} = 400$~GeV\cite{Ref:Choi}.
} 
\end{minipage}
\end{figure}

We now know the LSP ($\tilde{\chi}^0_1$)
and the chargino ($\tilde{\chi}^\pm_1$) masses,
which constrain the chargino and neutralino mass matrices,
the production cross section for $e^+e^-_R \to \tilde{e}^+_R \tilde{e}^-_R$,
which depends on the bino component of the LSP, and
the production cross section 
for $e^+e^-_R \to \tilde{\chi}^+_1 \tilde{\chi}^-_1$,
which provides information on the chargino composition.
Combining these measurements together, we can
carry out a global fit to determine
SUSY breaking parameters: $(M_1, M_2, \mu, \tan\beta)$.
Notice that we did not assume the GUT relation 
between $M_1$ and $M_2$ here, thereby
testing it to get insight into the SUSY breaking mechanism.
Fig.~\ref{Fig_chic1_gut} is the result from such a global
fit to Monte Carlo data generated with the GUT relation
indicated as a dashed line.
In this way, we can test the GUT relation.

\subsubsection{Full Reconstruction of the Chargino Mass Matrix}
\label{Chap:susy:Sec:chic:massmatrix}

The analysis presented above can be worked out 
further\cite{Ref:Choi}
to fully reconstruct the chargino mass matrix given
by Eq.~(\ref{Eq:M_chic}), which we repeat here for
convenience:
$$
{\cal M}_C = \pmatrix{ M_2 & \sqrt{2}m_W\cos\beta \cr
\sqrt{2}m_W\sin\beta & \mu  }
$$
Notice that the mass matrix, being asymmetric, requires
two unitary matrices for diagonalization.
We thus have two (real) mixing angles, $\phi_L$ and $\phi_R$,
which are given in terms of $M_2$, $\mu$, and $\tan\beta$.
These SUSY parameters are, in general,
complex, implying a possible non-trivial $CP$ phase
which cannot be removed by field redefinitions.
Without loss of generality, we can attribute this phase
to the $\mu$ parameter.

As mentioned above, only the $s$-channel
$B$ exchange diagram contribute to
$e^+e^-_R \to \tilde{\chi}^+_1 \tilde{\chi}^-_1$,
thereby singling out the higgsino component of
the chargino.
Because of this the corresponding cross section
($\sigma_R\{11\}$) is symmetric with respect to
$\cos 2\phi_L$ and $\cos 2\phi_R$ and is
independent of the mass
of the sneutrino that could be exchanged in the
$t$-channel if the electron beam were left-handed\footnote{
We assume here that the chargino mass is known either
from the end-point method as explained above or
by threshold scan, which can pin down the
gaugino mass to per mil level 
for $100~{\rm fb}^{-1}$\cite{Ref:Martyn}.}
On the other hand, the left-handed ($\sigma_L\{11\}$)
and the transverse ($\sigma_T\{11\}$) cross sections
vary with the sneutrino mass and the $e\tilde{\nu}\tilde{W}$
Yukawa coupling ($g_{e\tilde{\nu}\tilde{W}}$).
When mapped into the $\cos 2\phi_L$-$\cos 2\phi_R$ plane,
these cross sections comprise elliptic or parabolic contours,
which may cross each other up to four times.
Remember that $\sigma_R\{11\}$ is invariant
against any change of $m_{\tilde{\nu}_e}$ 
or $g_{e\tilde{\nu}\tilde{W}}$,
while $\sigma_L\{11\}$ and $\sigma_T\{11\}$ move with them.
Since the three cross section measurements meet at a
single point in the $\cos 2\phi_L$-$\cos 2\phi_R$ plane
only for the correct solution,
we can thus decide the mixing angles by changing
$m_{\tilde{\nu}_e}$ provided that the $e\tilde{\nu}\tilde{W}$
Yukawa coupling is identified with the gauge coupling
as dictated by the supersymmetry.
Fig.~\ref{Fig:choi_cosL_cosR_1} demonstrate this.

Instead of using beam polarizations, we can
also use the final-state polarization of charginos,
which can be extracted by measuring the distributions
of their decay daughters.
It has been shown that we can reconstruct
the polarization vectors of the charginos
and their spin-spin correlation tensor
dynamics-independently, and
use them to decide the mixing angles\cite{Ref:Choi}.
Note also that these additional measurements,
including the differential cross section measurements, 
can be used,
in combination with the beam polarizations,
to check the equality of the $e\tilde{\nu}\tilde{W}$
Yukawa coupling to the gauge coupling
(a quantitative test of supersymmetry)\cite{Ref:susy:feng}.

\begin{figure}[htb]
\centerline{
\epsfxsize=13cm\epsfbox{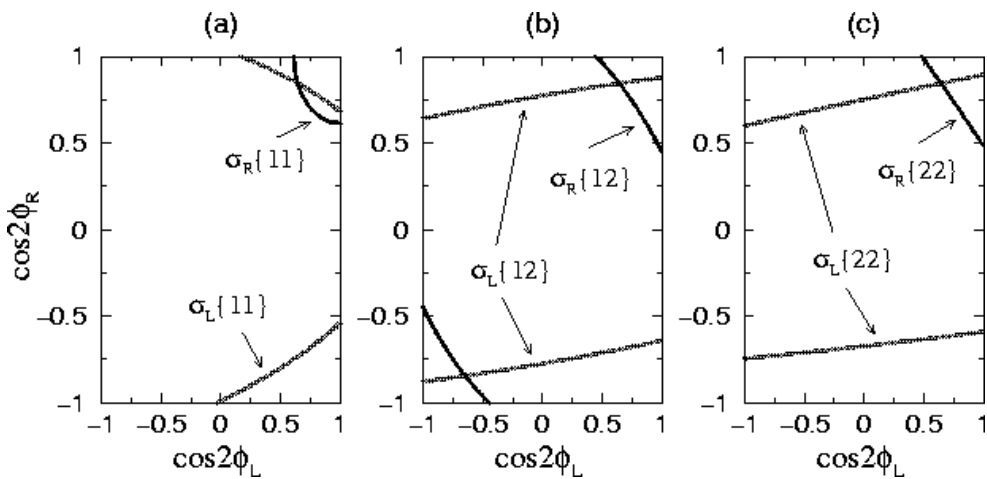}}
\begin{center}\begin{minipage}{\figurewidth}
\caption[Fig:choi_cosL_cosR_2]{\label{Fig:choi_cosL_cosR_2}\sl
Contours of the cross sections (a) $\sigma_{L/R}\{11\}$,
(b) $\sigma_{L/R}\{12\}$, and (c) $\sigma_{L/R}\{22\}$
in the $\cos2\phi_L$-$\cos2\phi_R$ plane 
for $\tan\beta=3, m_0=100~{\rm GeV}, M_{1/2}=200~{\rm GeV}]$ 
at $\sqrt{s} = 800$~GeV\cite{Ref:Choi}.
} 
\end{minipage}\end{center}
\end{figure}

When the machine energy reaches the pair production
threshold for the heavier chargino,
we will be able to study all of the three combinations:
\begin{eqnarray}
\nonumber
e^+e^- & \to & \tilde{\chi}^+_1 \tilde{\chi}^-_1 \cr
~  & \to & \tilde{\chi}^+_1 \tilde{\chi}^-_2 \cr
~  & \to & \tilde{\chi}^+_2 \tilde{\chi}^-_2.
\end{eqnarray}
This will allow us to measure the masses of the two charginos,
the production cross sections for the left- and right-handed
electron beams for the three processes, and
various decay distributions.
These measurements can then be use 
to unambiguously determine all of the
chargino mass matrix parameters, including $CP$ phase.
Figure~\ref{Fig:choi_cosL_cosR_2}
shows the cross section contours in the $\cos2\phi_L$-$\cos2\phi_R$ plane
for the three processes with different beam polarizations.
We can see that the three figures uniquely select a single
point $(\cos\theta_L,\cos\theta_R) = (0.645,0.844)$
in the mixing angle plane.
In this way, we can thus decide the mixing angles with
a statistical error of a percent level or better,
given an integrated luminosity of $1~{\rm ab}^{-1}$.
Notice also that the mass matrix is clearly over-constrained
by the measurements, we can relax the supersymmetry
condition on the $e\tilde{\nu}\tilde{W}$ Yukawa coupling,
and test it instead of assuming it.
The ratio of the $e\tilde{\nu}\tilde{W}$ Yukawa coupling
to the gauge coupling can thus be tested
at per mil level.
Table \ref{tab:tab1} compares the input and output values
of the masses, mixing angles, and the Yukawa coupling
for $1~{\rm ab}^{-1}$\cite{Ref:Choi}.
\begin{table}[htb]
\begin{center}\begin{minipage}{\figurewidth}
\caption{ \sl
Comparison of the input and output values of the mixing angles in 
the chargino sector extracted from different  chargino measurements for
$\int {\cal L} dt = 1 {\rm ab}^{-1}$\cite{Ref:Choi}.
\label{tab:tab1}
}
\end{minipage}\end{center}
\begin{center}
\begin{tabular}{|c|c|}
\hline
 Input  &  Extracted \\
\hline
$m_{\tilde{\chi}_1^{\pm}}$ = 128 GeV,&
$m_{\tilde{\chi}_1^{\pm}}$ = 128 $\pm 0.04 $GeV, \\
$m_{\tilde{\chi}_2^{\pm}}$ = 346 GeV.& $m_{\tilde{\chi}_2^{\pm}} = 346$
$\pm 0.25$ GeV. \\ 
\hline
$\cos 2 \phi_{L} = 0.645$,& $\cos 2 \phi_{L} = 0.645 \pm 0.02,$\\
$\cos 2 \phi_{R} = 0.844.$ & $\cos 2 \phi_{R} = 0.844 \pm 0.005 $. \\ 
\hline
$g_{e \tilde {\nu} \tilde {W}} / g_{e \nu W}  =  1$ &
$g_{e \tilde {\nu} \tilde {W}} / g_{e \nu W}  =  1 \pm 0.01 $ \\ 
\hline
\end{tabular}
\end{center}
\end{table}
On the other hand Table \ref{tab:tab2} summarizes
the input and extracted values of the fundamental parameters
of the chargino mass matrix.
\begin{table}[t]
\begin{center}\begin{minipage}{\figurewidth}
\caption{ \sl
The input and output values of $M_2,\mu $ and 
$\tan\beta$ for two input points for an integrated luminosity 
$1 ab^{-1}$\cite{Ref:Choi}.
\label{tab:tab2}
}
\end{minipage}\end{center}
\begin{center}
\begin{tabular}{|c|c|c|c|c|}
\hline
parameter & Input & Extracted & Input & Extracted \\ \hline
$M_{2}$ & 152 & $152 \pm 1.75$ & 150 & $150 \pm 1.2 $ \\ \hline
$\mu$ & 316 & $316 \pm 0.87 $ & 263 & $ 263 \pm 0.7 $ \\ \hline
$\tan \beta$ & 3 &  3 $\pm$ 0.69 & 30 & $ > $ 20.2 \\ \hline
\end{tabular}
\end{center}
\end{table}
Note that the expected precisions are remarkable except
for the large $\tan\beta$ case.
This is easy to understand since 
all the chargino variables are proportional
to $\cos 2\beta$ whose dependence on $\beta$ becomes flat
as $2\beta$ goes to $\pi$ ({\it i.e.} $\tan\beta$ increases). 

\subsubsection{Extraction of Sneutrino Mass}

\begin{figure}[hbt]
\centerline{
\epsfysize=5.3cm 
\epsfbox{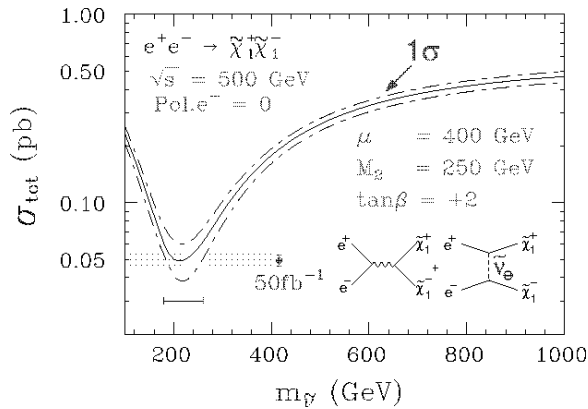}
}
\begin{center}\begin{minipage}{\figurewidth}
\caption[Fig_chic1_snu]{\label{Fig_chic1_snu} \sl
Lighter chargino pair production cross section as a function of
$m_{\tilde{\nu}_e}$.
The dot-dashed curves indicate the 1-$\sigma$ bound from
the global fit (see the text).
}
\end{minipage}\end{center}
\end{figure}

The chargino pair production for a left-handed electron beam
involves a $t$-channel exchange of $\tilde{\nu}_e$, which destructively
interferes with the $s$-channel gauge boson exchange diagrams.
Since we already know the lighter chargino mass and
composition, the only unknown parameter in the cross section is
the mass of $\tilde{\nu}_e$.
The solid curve in Fig.~\ref{Fig_chic1_snu} shows
the cross section 
for the unpolarized electron beam as
a function of the mass of $\tilde{\nu}_e$,
while the dot-dashed lines indicate the 1-$\sigma$
bound from the global fit explained above.
We can see that the total cross section measurement
constrains the sneutrino mass fairly well.
Using this and the following model-independent inequality
\begin{equation}\begin{array}{lll}
 m_{\tilde{e}^\pm_L}^2 \le m_{\tilde{\nu}_e}^2 + 0.77 m_Z^2,
\end{array}\end{equation}
we can set an upper limit on the left-handed selectron mass.
We can thus go and look for the left-handed selectron
in $e^+e^-_R \to \tilde{e}^+_L \tilde{e}^-_R$\footnote{
Another strategy is to go directory to 
the energy just above the pair production
thresholds for $\tilde{\nu}_e$ or $\tilde{e}_L$.
Their cross sections are large, in particular when
the gaugino mass is small and thus the $t$-channel
diagram dominates.
We can then carry out similar analyses as we have done
to the right-handed sleptons 
(see, for instance, Ref.\cite{Ref:tata}).
}.

\subsection{$\protect \mathbold{\tilde{\it e}^\pm_{\it L}}$ Study}

\subsubsection{$\protect \mathbold{\tilde{\it e}^\pm_{\it L}}$ Signature}

\begin{figure}[htb]
\centerline{
\epsfysize=6.cm 
\epsfbox{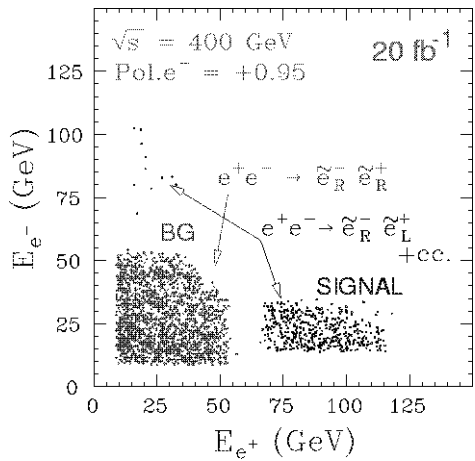}
}
\begin{center}\begin{minipage}{\figurewidth}
\caption[Fig_sel_Ee]{\label{Fig_sel_Ee} \sl
Distribution of
final-state electron energy against that of the positron for
$e^+e^-_R \to \tilde{e}^+_L\tilde{e}^-_R$ (signal) and
$\tilde{e}^+_R\tilde{e}^-_R$ (background)
at $\sqrt{s}=400~{\rm GeV}$ with an integrated luminosity of
$20~{\rm fb}^{-1}$.
}
\end{minipage}\end{center}
\end{figure}

The associate production of $\tilde{e}_L$ with $\tilde{e}_R$
proceeds via $t$-channel exchange of neutralinos, to which
only the $\tilde{B}$ components contribute.
The decay of the left-handed slepton may be a little
complicated, since they tend to decay in cascade as
the scalar mass becomes heavier.
Nevertheless, the branching fraction for the direct
two-body decay into the LSP: $\tilde{l}_L^\pm \to l^\pm \tilde{\chi}_1^0$
usually exceeds $10~\%$.
For our parameter set, 
the $\tilde{l}_L^\pm \to l^\pm \tilde{\chi}_1^0$ decay
is actually dominant.
The signal for this process is therefore
an acoplanar $e^+e^-$ pair.
When we use the right-handed electron beam to suppress
standard-model backgrounds, the major background is our
previous signal: $e^+e^-_R \to \tilde{e}^+_R \tilde{e}^-_R$.
Notice that the use of the right-handed electron beam selects
the $e^+$ in the final state as a carrier of the
$\tilde{e}_L$ information.
Fig.~\ref{Fig_sel_Ee} plots the final-state electron energy against
that of the positron, clearly demonstrating this fact.

\subsubsection{Mass Determination}

\begin{figure}[htb]
\centerline{
\epsfxsize=14cm 
\epsfbox{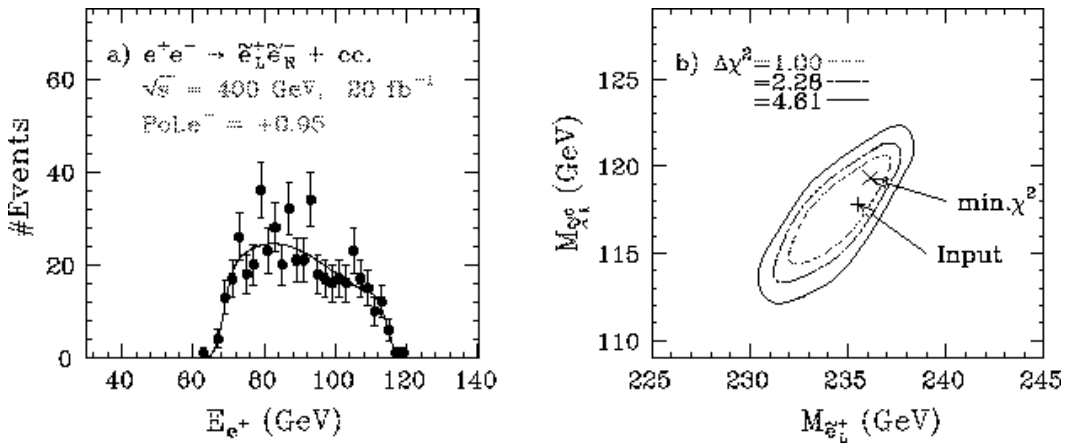}}
\begin{center}\begin{minipage}{\figurewidth}
\caption[Fig_sel_mass]{\label{Fig_sel_mass} \sl
(a) Energy distribution of positrons from $\tilde{e}^+_L$ decays 
for the same Monte Carlo parameters as with Fig.\ref{Fig_sel_Ee}. 
The solid line corresponds to the best fit curve,
letting $m_{\tilde{e}_L}$ and $m_{\tilde{\chi}^0_1}$ move freely.
(b) Contours in the $m_{\tilde{e}_L}$-$m_{\tilde{\chi}^0_1}$
plane obtained from the fit to the energy distribution.
}
\end{minipage}\end{center}
\end{figure}

Projecting Fig.~\ref{Fig_sel_Ee} to the $E_{e^+}$ axis, we obtain
the distribution shown in Fig.~\ref{Fig_sel_mass}-a), from
which we can extract the sparticle masses as before.
The contours from a 2-parameter fit are shown in Fig.~\ref{Fig_sel_mass}-b),
which tells us that we can determine the left-handed selectron mass
to an accuracy better than $1~\%$ with $20~{\rm fb}^{-1}$.

\subsubsection{Test of Left-Right Universality of Scalar Masses}

\begin{figure}[htb]
\centerline{
\epsfxsize=7cm 
\epsfbox{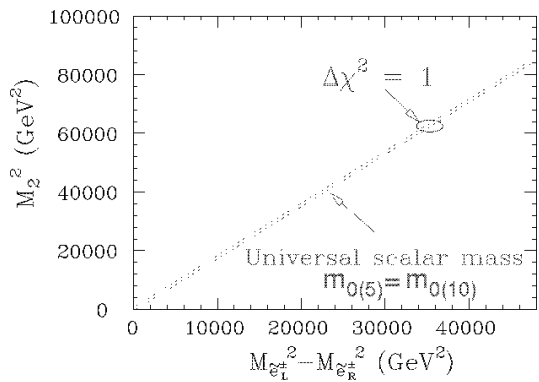}
}
\begin{center}\begin{minipage}{\figurewidth}
\caption[Fig_sel_usm]{\label{Fig_sel_usm} \sl
$\Delta \chi^2=1$ contour from the selectron mass measurements,
compared with the prediction from the universal scalar mass
hypothesis, indicated as dotted lines for $\tan\beta=0$ and $30$.
}
\end{minipage}\end{center}
\end{figure}

Knowing both the right-handed and left-handed selectron masses
enables us to make another test of the universal scalar mass
hypothesis.
The squared mass difference of the right-handed 
(belonging to $\mathbold{5^*}$)
and left-handed (belonging to $\mathbold{10}$)
selectrons is related through the following scalar mass
formula:
\begin{equation}\begin{array}{lll}
m_{\tilde{e}_L}^2 - m_{\tilde{e}_R}^2
= m_{0({\bf 5})}^2 -  m_{0({\bf 10})}^2
+ 0.5 M_2^2 - 0.04 \cdot m_Z^2 \cos 2\beta.
\end{array}\end{equation}
The universal scalar mass hypothesis implies the
representation independence ($m_{0({\bf 5})} = m_{0({\bf 10})})$,
which can be tested as shown in Fig.~\ref{Fig_sel_usm}:
compare the $\Delta \chi^2=1$ contour with the prediction
of the universal scalar mass hypothesis (dashed lines
for $\tan\beta = 0$ and $30$).
Notice that the last term ($\tan\beta$-dependent term)
of the above equation is small and only makes slight difference,
allowing us a clean test of the representation independence.

Once the GUT relation is confirmed and the universal scalar mass
hypothesis is verified in the slepton sector, we can set,
with confidence, the next target energy for
$\tilde{q}$ pair production.

\subsection{ Study of 
$\protect \mathbold{\tilde{\it q}_{\it G} ({\it G} \ne 3)}$}

\subsubsection{Squark Signature}

A squark may decay into a quark plus an LSP or into
a quark plus a chargino or heavier neutralino, depending on
the SUSY breaking parameters.
In the former case, the signal for squark pair production is
an acoplanar 2-jet final state, while in the latter case
it is a final state consisting of two jets plus one or two
$W/Z$ bosons from the cascade decays of 
the charginos or neutralinos, respectively.
Potential backgrounds to this process thus include
$e^+e^- \to W^+W^-$, 
$e^\pm\mathop{\nu_e}\limits^{\mbox{\scriptsize$(-)$}} W^\mp$,
$\nu\bar{\nu}Z$,
$\tilde{\chi}^+_1\tilde{\chi}^-_1$, and
$\tilde{\chi}^0_i\tilde{\chi}^0_j$.
Notice that the 2-jet systems in the final states of
these background processes are all from $W^{(*)}/Z^{(*)}$'s,
thereby having invariant masses smaller than $100~{\rm GeV}$.
We can therefore very effectively eliminate them
by requiring $m_{q\bar{q}} > 100~{\rm GeV}$,
together with usual acoplanar 2-jet selection criteria like
$\sla{p}_T > 35~{\rm GeV}$, and $\theta_A > 30^\circ$.
We can thus use both $e^-_L$ and $e^-_R$ beams without
suffering from these would-be serious backgrounds.
It should also be emphasized that the chirality selection of
the final-state squarks works better than the slepton case:
$\sigma_R(\tilde{q}_R \bar{\tilde{q} }_R) : 
\sigma_R(\tilde{q}_L\bar{\tilde{q}}_L)
\simeq
\sigma_L(\tilde{q}_L\bar{\tilde{q}}_L) :
\sigma_L(\tilde{q}_R\bar{\tilde{q}}_R)
\simeq 9 : 1$,
if $m_{\tilde{q}_L} \simeq m_{\tilde{q}_R}$.
By controlling the electron beam polarization, we can
thus select the chirality of the final-state squarks and,
consequently, study their properties separately.

\subsubsection{Mass Determination}

\begin{figure}[htb]
\centerline{
\epsfxsize=7cm\epsfbox{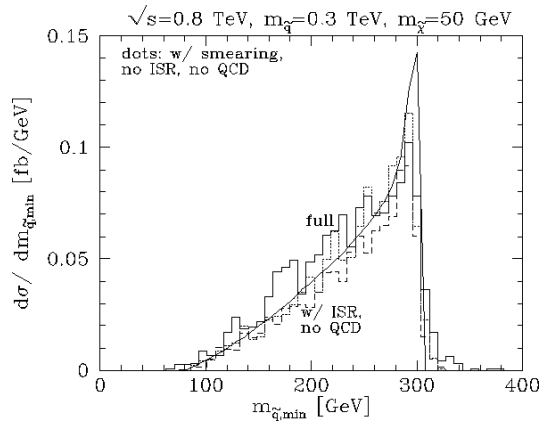}
}
\begin{center}\begin{minipage}{\figurewidth}
\caption{\label{Fig:sqmass_radcorr}\sl
Accuracy of determination of the squark mass using the 
$m_{\tilde{q}}^{min}$ estimator.
The input squark mass is 300 GeV and Monte Carlo data
were generated at $\sqrt{s} = 0.8$~TeV and 
correspond to $50~{\rm fb}^{-1}$
}
\end{minipage}\end{center}
\end{figure}

Let us first consider the mass determination in the case of the
direct decay.
In this case, we can use the end-point method as with
$\tilde{l}^\pm$ and $\tilde{\chi}^\pm_1$.
There is, however, a better quantity called
minimum squark mass ($m_{\tilde{q}}^{min}$)\cite{Ref_Feng2}.
defined by
\begin{equation}\begin{array}{lll}
\left(m_{\tilde{q}}^{min}\right)^2 = E_{beam}^2 - |{\bf p}_1|^2_{max}
= E_{beam}^2 - |{\bf p}_2|^2 - |{\bf p}_3|^2
+ 2 |{\bf p}_2| |{\bf p}_3| \cos(\delta+\gamma),
\end{array}\end{equation}
where the momenta and angles are defined in Fig.~\ref{Fig_sq_msqmdef}.
\begin{figure}[htb]
\hspace{0.5cm}
\begin{minipage}[t]{7cm}
\centerline{
\epsfysize=5.5cm 
\epsfbox{physsusy/figs/05_sq_msqmdef.epsf}
}
\caption[Fig_sq_msqmdef]{\label{Fig_sq_msqmdef} \sl
Definitions of the momenta and angles used to
define the minimum squark mass.
The LSP mass is assumed to be known precisely.
}
\end{minipage}
\hfill
\begin{minipage}[t]{7cm}
\centerline{
\epsfysize=5.5cm 
\epsfbox{physsusy/figs/05_sq_msqm.epsf}
}
\caption[Fig_sq_msqm]{\label{Fig_sq_msqm} \sl
Expected $m_{\tilde{q}}^{min}$ distributions for
left- and right-handed squarks cited from
Ref.\cite{Ref_Feng2}.
}
\end{minipage}
\hspace{0.5cm}
\end{figure}
Notice that the two angles $\delta$ and $\gamma$ and the
magnitude of the LSP momentum are calculable, provided that
the LSP mass is known. 
Notice also that $m_{\tilde{q}}^{min}$ makes use of the
information contained in the relative configuration of
the final-state $q$ and $\bar{q}$ unlike in the end-point
method.
Fig.~\ref{Fig_sq_msqm} shows an example of expected
$m_{\tilde{q}}^{min}$ distributions for parameters
shown in the figure.
We can see that the Jacobian peaks are good measures of the
squark masses and allow their precision measurements:
$\Delta m_{\tilde{q}} \simeq 1~{\rm GeV}$ at the $95~\%$
confidence level.
The effects of the radiative corrections to 
production and decay as well as the effects of the ISR
have been studied separately\cite{DEKG}
and found to be harmless:
the peak of the $m_{\tilde{q}}^{min}$ distribution
is essentially unaffected as shown in Fig.~\ref{Fig:sqmass_radcorr}
and still allows us to measure the squark mass
to a percent level with $50~{\rm fb}^{-1}$.

If cascade decays dominate the direct decay,
we select final-states with two jets plus leptons and
pretend to be blind to the leptons.
Notice that these leptons are end-products of the chargino 
($\tilde{\chi}^\pm_1$) or
heavier neutralino ($\tilde{\chi}^0_2$),
whose masses are almost degenerate and presumably known from
the earlier measurements.
We can thus change the role of
the LSP in the direct decay case by that of the chargino or 
second neutralino, by ignoring the final-state leptons.
In this way, we can again calculate the minimum squark mass for each event
and carry out the squark mass measurement at a $1~\%$ level,
provided that $m_{\tilde{\chi}^\pm_1}$ and $m_{\tilde{\chi}^0_1}$
are known.

\subsection{Study of Third Generation Sleptons}

For the first and second generation, we could
ignore the left-right mixing.
This is due to the fact that
the off-diagonal element of the mass matrix:
\begin{equation}\label{e6}
{\cal M}^2_{\tilde{f}} 
= \left( \begin{array}{cc} 
m_{\tilde{f}_L}^2 & -m_{f}(A_{f} + \mu \tan\beta) \cr 
-m_{f}(A_f + \mu \tan\beta) &  m_{\tilde{f}_R}^2
\end{array}\right),
\end{equation}
which is responsible for the left-right
mixing is proportional to the corresponding
fermion mass.
For the third generation, since the tau mass ($m_{\tau}$)
is much heavier than the other leptons,
this off-diagonal element can be substantial,
if the soft SUSY-breaking trilinear scalar coupling
($A_{\tau}$) or $\mu\tan\beta$ or both are large.
The mass eigen states therefore become mixtures of 
$\tilde{\tau}_L$ and $\tilde{\tau}_R$:
\begin{eqnarray}
\pmatrix{ \tilde{\tau}_1 \cr
\tilde{\tau}_2 }
= \pmatrix{ \cos\theta_\tau & \sin\theta_\tau \cr
-\sin\theta_\tau & \cos\theta_\tau }	
\pmatrix{ \tilde{\tau}_L \cr
\tilde{\tau}_R },
\end{eqnarray}
where 
$m_{\tilde{\tau}_1} < m_{\tilde{\tau}_2}$
by definition.

Note also that the $\tau$ lepton has a Yukawa coupling: 
\begin{eqnarray}
\nonumber 
Y_{\tau}=-gm_{\tau}/(\sqrt{2}m_W\cos\beta),
\end{eqnarray}
which becomes non-negligible for a large $\tan\beta$.
The correction to the diagonal elements is proportional to
$Y_{\tau}^2\log(M_{GUT}/M_W)$ and
makes $m_{\tilde{\tau}_{L(R)}}$
smaller than those of the first and second generations.
If $\tan\beta$ is large, $\mu\tan\beta$
that controls the off-diagonal element becomes substantial
and consequently lowers the lighter stau mass further.
In SUGRA, we thus expect that the lighter
stau is lighter than the other sleptons.

Conversely, once the masses of the two
staus, their mixing angle, and the tau
Yukawa coupling are determined,
we can calculate $m_{\tilde{\tau}_{L(R)}}$
thereby testing the prediction of SUGRA.
It has been pointed out that
the determination of $m_{\tilde{\tau}_{L(R)}}$
is crucial for testing grand unified theories:
in GUTs a heavy colored Higgs boson with a mass
of $O(M_{GUT})$ couples to the stau leptons
via strong Yukawa interactions and further reduces
$m_{\tilde{\tau}_{L,R}}$.

In this way, the stau can be the first slepton
to be discovered at the JLC and, once it is found,
its detailed studies may provide us
an evidence of grand unification at high scale.

\subsubsection{Stau Signature}

The lighter stau $\tilde{\tau}_1$ decays into
a chargino ($\chi^+_i$) and a tau neutrino ($\nu_{\tau}$)
or into a neutralino ($\chi^0_i$) and a tau ($\tau$).
The latter decay mode leaves an acoplanar tau pair
in the final state and provides a signature
of the stau production.
In particular when $\tilde{\tau}_1$ is the
lightest charged SUSY particle,
it decays into the lightest neutralino and a tau.
The taus in the final state further decay
into $e(\mu)\bar{\nu}_{e(\mu)}\nu_{\tau}$,
$\pi \nu_{\tau}$,
$\rho \nu_{\tau}\rightarrow \pi^+ \pi^0 \nu_{\tau}$, etc.
Consequently, the stau pair production
results in an acoplanar event consisting of
a pair of leptons or low multiplicity hadron jets.
The most serious background in the search of
the stau pair production is
the two-photon process:
$e^+e^-\rightarrow e^+e^-\tau^+\tau^-$,
which mimics the stau pair production
when the final-state $e^+$ and $e^-$
escape from detection into the beam-pipe.
Note that for this background, the missing
transverse momentum ($P_T$) cut is not that effective,
because of the neutrinos from the tau decays
always carrying away significant fraction of $P_T$.
Simultaneous cuts on the polar angles of the jets,
$\cos\theta_{jet}$, and
the acoplanarity, $\theta_{acop}$,
are effective but still not enough to reduce
the huge background
for signal events with a missing $P_T$
that is similar to the $P_T$'s carried
away by the escaping $e^+$ and $e^-$.
It is essential to be able to
veto low angle electrons and positrons
to reduce the two-photon background.
We will return to this point later.

\subsubsection{Polarization Measurements}

\begin{figure}[htbp]
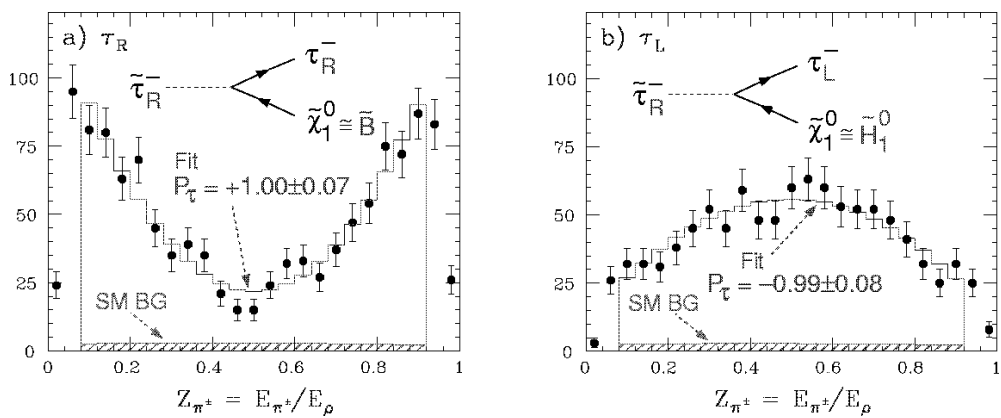

\centerline{
\epsfxsize=6cm
\epsfbox{physsusy/figs/03_zc_a.epsf}
\hspace {1cm}
\epsfxsize=6cm
\epsfbox{physsusy/figs/03_zc_b.epsf}
}
\begin{center}\begin{minipage}{\figurewidth}
\caption[physics:susy:stau:zc]{\sl \label{physics:susy:stau:zc}
$z_c=E_{\pi^{\pm}}/E_{jet}$ distribution of
the selected candidates of the cascade decay:
$\tilde{\tau}_1\rightarrow \tau\tilde{\chi}^0_1$ 
followed by 
$\tau\rightarrow \rho\nu$
for $10^4$ Monte Carlo stau pair events.
The histograms are the results of the fits to determine
the tau polarization.
}
\end{minipage}\end{center}
\end{figure}

Since the stau decays in cascade,
its analysis is complicated.
Its detailed study, however, provides
us with another handle to uncover the nature
of the LSP.
Namely, we can study the polarization of the tau leptons
by measuring the energy distribution of 
the tau's decay products\cite{Ref:susy:stau}.
Fig. \ref{physics:susy:stau:zc}
plots the distribution of $E_{\pi^{\pm}}/E_{\rho}$
for the cascade decay: 
$\tilde{\tau}_1\rightarrow \tau\tilde{\chi}^0_1$
followed by 
$\tau\rightarrow \rho\nu_{\tau}$, $\rho\rightarrow \pi^{\pm} \pi^0$,
when the intermediate tau has
the right-handed or left-handed polarization.
We can see that the right-handed tau tends to give most of its energy to
one of the $\pi^+$, $\pi^0$, while
the left-handed tau tends to equally distribute
its energy to both of them.
Given $10^4$ stau pairs, 
the Monte Carlo simulations tells us that
the tau polarization can be measured
to an accuracy of $8\%$.

The tau polarization provides a very important
piece of information for the determination of
the stau's Yukawa coupling.
Let us now consider the 
$\tilde{\tau}_{R(L)}$- $\chi^0_1$-$\tau$
coupling which is relevant to the stau decay.
The neutralinos
($\tilde{\chi}^0_1$, 
$\tilde{\chi}^0_2$, $\tilde{\chi}^0_3$, $\tilde{\chi}^0_4$)
are the mixtures of
the bino ($\tilde{B}^0$), the wino ($\tilde{W}^0_3$),
and the two higgsinos
($\tilde{H}^0_1$, $\tilde{H}^0_2$),
whose mass matrix is given by
\begin{eqnarray}
\nonumber
{\cal M}_N & = &
\left(\begin{array}{cccc}
M_1 &0&-m_Z\sin\theta_W\cos\beta&m_Z\sin\theta_W\sin\beta\\
0 &M_2& m_Z\cos\theta_W\cos\beta&-m_Z\cos\theta_W\sin\beta\\
-m_Z\sin\theta_W\cos\beta&m_Z\cos\theta_W\cos\beta&0&-\mu \\
m_Z\sin\theta_W\sin\beta&-m_Z\cos\theta_W\sin\beta&-\mu & 0
\end{array}\right)
\end{eqnarray}
and its mass eigen states can be expressed in the form:
$$
\tilde{\chi}^0_i= N_{i1}\tilde{B}^0+N_{i2}\tilde{W}^3+
N_{i3}\tilde{H}^0_1+N_{i4}\tilde{H}^0_2.
$$

\begin{figure}[htbp]
\centerline{
\epsfxsize=10cm
\epsfbox{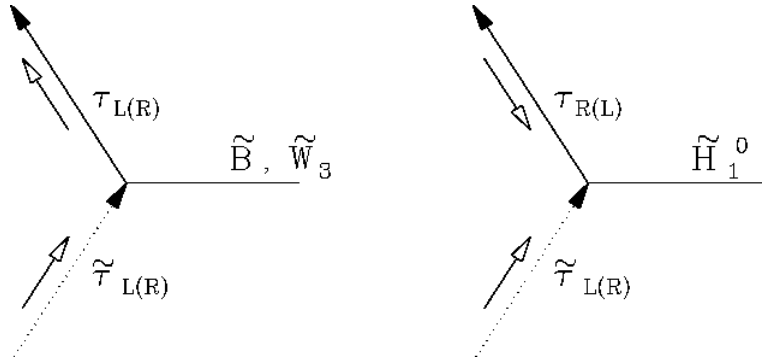}
}
\begin{center}\begin{minipage}{\figurewidth}
\caption[physics:susy:stau:feyn]{\sl \label{physics:susy:stau:feyn}
Feynmann diagarams for
$\tilde{\tau}_R$ decaying into a tau and a bino
($\tilde{B}^0$) or a higgsino ($\tilde{H}^0_1$).
The white arrows indicate chirality flow.
}
\end{minipage}\end{center}
\end{figure}

Fig.~\ref{physics:susy:stau:feyn}
illustrates the right-handed stau's interaction with
the bino and higgsino components.
Notice that the interaction differs not only in its
strength but also in the resultant polarization
of the final-state tau.

In general, interaction of matter fermions with gauge bosons
preserves chirality, while
that with the Higgs bosons flips it.
By supersymmetry, therefore,
$\tilde{\tau}_{R(L)}$
interacts with a gaugino (higgsino)
and turns into a $\tau_{R(L)}$($\tau_{L(R)}$).
The polarization of the daughter tau
is determined by the neutralino mixing angles,
$N_{ij}$, and the stau mixing angle, $\theta_{\tilde{\tau}}$,
and, in the limit that
the neutrarino is a pure bino and
the stau is a weak eigen state, {\it i.e.}
either $\tilde{\tau}_R$ or $\tilde{\tau}_L$,
is expressed as
\begin{eqnarray}\label{e8}
\nonumber
P_{\tau}\left(\tilde{\tau}_1\rightarrow\tilde{B}\tau\right)&=&
\frac{4\sin^2\theta_{\tilde{\tau}} - \cos^2\theta_{\tilde{\tau}}}
{4\sin^2\theta_{\tilde{\tau}} + \cos^2\theta_{\tilde{\tau}}} \label{e8a}\\
\nonumber
P_{\tau}\left(\tilde{\tau}_R\rightarrow\tilde{\chi}^0_1\tau\right)&=&
\frac{\left(\sqrt{2}N_{11}\tan\theta_W\right)^2 -
\left(Y_{\tau}N_{13}\right)^2}
{\left(\sqrt{2}N_{11}\tan\theta_W\right)^2
+\left(Y_{\tau}N_{13}\right)^2} \label{e8b}\\
\nonumber
P_{\tau}\left(\tilde{\tau}_L\rightarrow\tilde{\chi}^0_1\tau\right)&=&
\frac{\left(\sqrt{2}Y_{\tau}N_{13}\right)^2-g^2
\left(N_{12}+N_{11}\tan\theta_W\right)^2}
{\left(\sqrt{2}Y_{\tau}N_{13}\right)^2+
g^2\left(N_{12}+N_{11}\tan\theta_W\right)^2}.
\end{eqnarray}
Notice that, unless the neutralino
is a pure bino,
the polarization depends on the Yukawa coupling.
We shall discuss later how
this can be exploited to 
determine the Yukawa coupling,
when the neutralino and stau mixing angles can
be constrained.

\subsubsection{Mass Determination}

\begin{figure}[htbp]
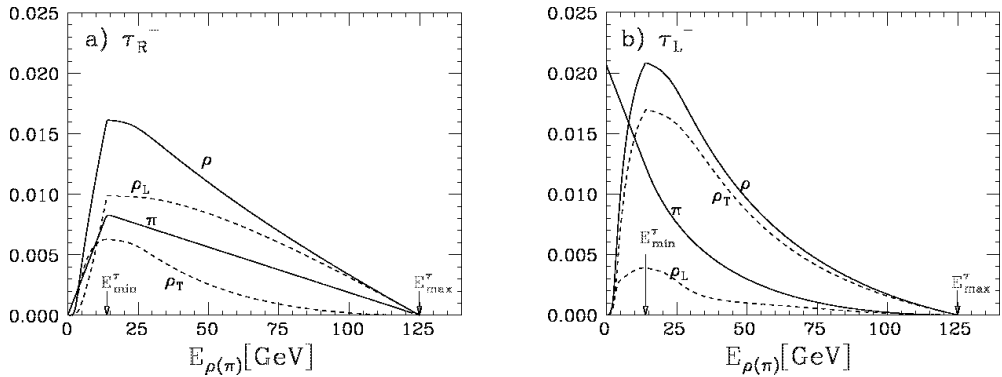

\centerline{
\epsfxsize=6cm
\epsfbox{physsusy/figs/05_ej_a.epsf}
\hspace {1cm}
\epsfxsize=6cm
\epsfbox{physsusy/figs/05_ej_b.epsf}
}
\begin{center}\begin{minipage}{\figurewidth}
\caption[physics:susy:stau:ej]{\sl \label{physics:susy:stau:ej}
Energy distributions of $\rho(\pi)$'s
from 
$\tilde{\tau}\rightarrow\tau_{L(R)}$ 
$\rightarrow\rho(\pi)$ decays
for $m_{\tilde{\tau}}=150$ GeV, 
$m_{\tilde{\chi}^0_1}=100$ GeV, and $\sqrt{s}=500$ GeV.
$E^{\tau}_{max}$ and $E^{\tau}_{min}$
are the upper and lower limits of the tau energy.
}
\end{minipage}\end{center}
\end{figure}

The masses of the first and second generation sleptons
and the lightest neutralino could be determined
by measuring the end points of
the square-shaped energy distribution of the
daughter leptons from the slepton decays.
For the stau, however, since the daughter tau
from its decay further decays into
a neutrino and hadrons,
the mass determination must rely
not merely on the end points but 
the shape of the distribution of
the energy of the decay products from the tau decay.
Fig. \ref{physics:susy:stau:ej} shows 
the energy distributions
of the final-state hadrons
for the $\tau \rightarrow \nu_{\tau} + \pi^-/\rho^-$ decays.
The higher end point of the distribution
is equal to that of the tau energy distribution,
while the peak of the energy distribution
of the $\rho$'s corresponds 
to the lower limit of the tau energy distribution.
If we can determine these two values,
we can thus obtain the stau and the neutralino
masses.
The energy distribution depends on the tau polarization,
but it can be measured as explained above.

\begin{figure}
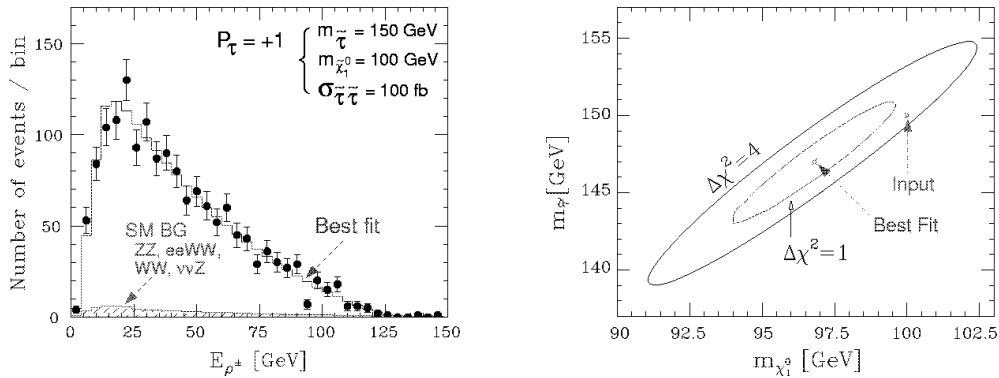

\centerline{
\epsfxsize=6cm
\epsfbox{physsusy/figs/06_massfit_a.epsf}
\hspace {1cm}
\epsfxsize=6cm
\epsfbox{physsusy/figs/06_massfit_b.epsf}
}
\begin{center}\begin{minipage}{\figurewidth}
\caption[physics:susy:stau:massfit]{\sl \label{physics:susy:stau:massfit}
a) energy distribution of the final-state hadrons selected as $\rho$'s
from stau decays shown together with the best fit,
when $10^4$ $\tilde{\tau}_1$ stau pairs
are produced and decay in the $\tilde{\chi}^0_1\tau_R$ mode.
b) $\Delta\chi^2=1,4$ contour in the
$m_{\tilde{\chi}^0_1}$-$m_{\tilde{\tau}_1}$ plane. 
}
\end{minipage}\end{center}
\end{figure}

\noindent
Fig.\ref{physics:susy:stau:massfit}-a)
shows the tau energy distribution
for Monte Carlo data with background events,
when $10^4$ stau pairs are produced.
The expected resolution for 
the stau mass is 2.6\%, while
that for the $\tilde{\chi}^0_1$ mass is
2.8\%.
These accuracies are worse than what we expect
for the first and second generation sleptons,
which is due to the fact that the tau energy
end points are more difficult to measure.
Note that, if the other sleptons are
already found, we can use the $\chi^0_1$ mass
from the end point measurements for them to
improve the measurement.

\subsubsection{Determination of Mixing Angle}
\label{Chap:susy:Sec:stau:mixing}

\begin{figure}
\centerline{
\epsfxsize=7.5cm
\epsfbox{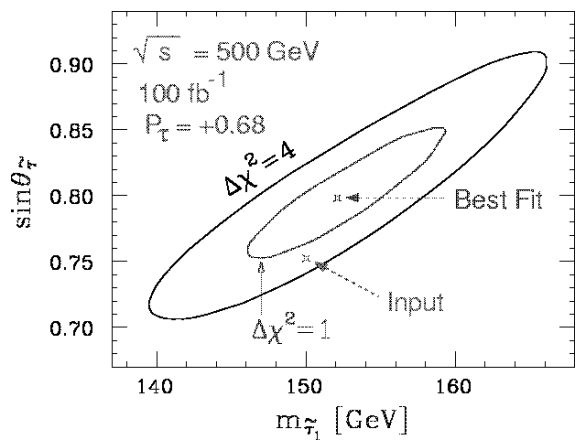}
}
\begin{center}\begin{minipage}{\figurewidth}
\caption[physics:susy:stau:theta_mix]{\sl \label{physics:susy:stau:theta_mix}
The $\Delta\chi^2=1,4$ contour in the
$m_{\tilde{\tau}_1}$-$\sin\theta_{\tilde{\tau}}$
plane obtained from a fit of 5000 Monte Carlo stau pair events
generated at $\sqrt{s}=500$ GeV
for $m_{\tilde{\tau}}=150$ GeV,
$m_{\tilde{\chi}^0_1}=100$ GeV,
$P_{\tau}=0.6788$, and 
$\sin\theta_{\tilde{\tau}}=0.7526$.
The Monte Carlo statistics is
$\int {\cal L} dt=100 {\rm fb}^{-1}$ equivalent.
The tau polarization corresponds to
the case in which the stau decays into a tau and a bino.
}
\end{minipage}\end{center}
\end{figure}

Once the cross section and the stau mass are measured,
we can determine the stau mixing angle.
Fig.\ref{physics:susy:stau:theta_mix} shows the
expected contour of $\Delta\chi^2=1,4$
in the $m_{\tilde{\tau}_1}$-$\sin\theta_{\tilde{\tau}}$
plane for $m_{\tilde{\tau}}=150$ GeV and
$m_{\tilde{\chi}^0_1}=100$ GeV,
when the production cross section is 50~fb
(corresponding to $\sin\theta_{\tilde{\tau}}=0.7526$)
and an integrated luminosity of 100 ${\rm fb}^{-1}$
is accumulated at $\sqrt{s}=500$ GeV.
In this case, $\sin\theta_{\tilde{\tau}}$ can be
determined to 6.5\%.

\subsubsection{Determination of Yukawa Coupling}

As mentioned above, the tau polarization
carries information on the  stau's Yukawa coupling.
In order to extract the information on the stau's Yukawa coupling,
we need to determine the mixing angles of the stau and
the neutralino.
The stau's mixing angle can be determined
by measuring the production cross section.
Combining the information from the productions
of the other sleptons, we can further
decide the neutralino mixing, and
consequently obtain the information 
on the stau's Yukawa coupling. 

\begin{figure}
\centerline{
\hspace{0.2cm}
\epsfxsize=6.7cm
\epsfbox{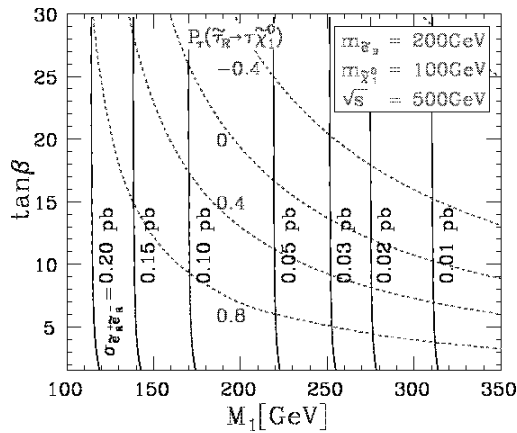}
}
\begin{center}\begin{minipage}{\figurewidth}
\caption[physics:susy:stau:pol_mix]{\sl \label{physics:susy:stau:pol_mix}
The neutralino-mixing dependence of
the tau polarization in
the $\tilde{\tau}_1\rightarrow\tau\tilde{\chi}^0_1$
decay.
The neutralino mixing angle is determined by
$(M_1,M_2,\mu,\tan\beta)$ in general.
If we assume the GUT relation,
$M_1=5/3\cdot\tan^2\theta_W M_2$, 
and $m_{\tilde{\chi}^0_1}=100$ GeV,
we can eliminate two of the four parameters.
The tau polarization is thus shown in the
$M_1-\tan\beta$ plane.
}
\end{minipage}\end{center}
\end{figure}

Fig. \ref{physics:susy:stau:pol_mix} shows the tau
polarization in the $M_1$-$\tan\beta$ plane,
when the $\tilde{\tau}_R$ decays into
a tau and a 100 GeV neutralino.
We assumed the GUT relation for $M_1$ and $M_2$.
Notice that the neutralino mass matrix is 
completely determined by
$M_1$, $M_2$, $\mu$, and $\tan\beta$.
Once the neutralino mass is given, therefore,
the neutralino properties are completely
determined by the two parameters. 

\begin{figure}
\centerline{
\epsfxsize=7cm
\epsfbox{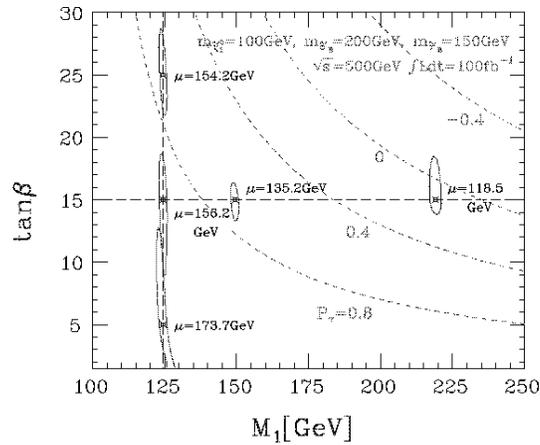}
}
\begin{center}\begin{minipage}{\figurewidth}
\caption[physics:susy:stau:m1_tanb]{\sl \label{physics:susy:stau:m1_tanb}
The $1\sigma$ error contours in the
$M_1$-$\tan\beta$ plane expected from the measurements of
the selectron and stau productions.
The contours are plotted for
$m_{\tilde{\tau}_1}=150$ GeV, 
$\sin\theta_{\tilde{\tau}}=1$ ($\tilde{\tau}_1=\tilde{\tau}_R$),
$m_{\tilde{\chi}^0_1}=100$ GeV($\mu>0$), 
and $m_{\tilde{e}_R}=200$ GeV.
}
\end{minipage}\end{center}
\end{figure}

Fig. \ref{physics:susy:stau:m1_tanb}, on the other hand,
demonstrates how the tau polarization and
the selectron and stau cross section measurements
constrain $\tan\beta$ in the $M_1$-$\tan\beta$ plane,
where we assumed that the measurement errors are
dominated by statistical ones and that
the GUT relation between $M_1$ and $M_2$ holds
\footnote{As mentioned above, it is possible to
determine $M_2$ and $M_1$ separately by studying
chargino pair productions. 
Here, however, we assumed the GUT relation for simplicity.}.
The points in the figure indicate the input
model parameter sets and the curves surrounding them
are the expected $\Delta\chi^2 =1$ contours.
The neutralino, stau, and selectron masses are
fixed at 100 GeV, 150 GeV, and 200 GeV,
respectively.
Notice that $\tan\beta$ determination becomes less accurate
as $M_1$ decreases.
This is because $\tilde{\chi}^0_1$ becomes
more and more $\tilde{B}$-like, as
$M_1$ approaches $m_{\tilde{\chi}^0_1}$,
and consequently loses the $\tan\beta$ dependence
of the tau polarization.
On the other hand, in the 
large $M_1$ and $\tan\beta$ region,
we can determine $\tan\beta$ reasonably well.

\begin{figure}
\centerline{
\epsfxsize=6.7cm
\epsfbox{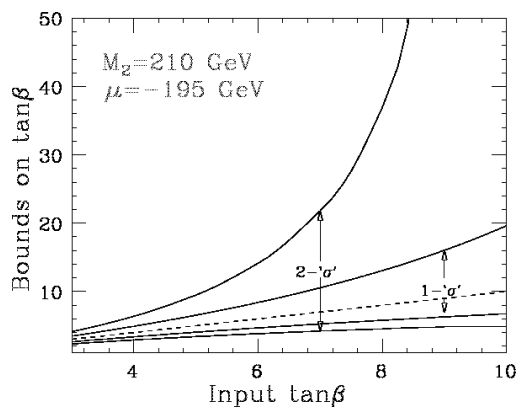}
}
\begin{center}\begin{minipage}{\figurewidth}
\caption[physics:susy:stau:tanb_xc]{\sl \label{physics:susy:stau:tanb_xc}
The constraint on $\tan\beta$ from the measurement
of the forward-backward asymmetry of the
chargino production for
$M_2=210$ GeV, $\mu=-195$ GeV, and $m_{\tilde{\nu}_e}=500$ GeV.
The error on the two chargino masses is assumed
to be common and 2\%.
}
\end{minipage}\end{center}
\end{figure}

There are other possibilities to determine $\tan\beta$.
The $\tan\beta$ measurement using the
stau production is, however, particularly useful when
$\tan\beta > 10$.
On the other hand, the determinations using
the $\tilde{e}_L$-$\tilde{\nu}_L$ mass difference
or the forward-backward asymmetry of the chargino
pair production are very effective
for $\tan\beta\lsim 5$.
The Higgs production is also effective
for $\tan\beta \lsim 10$
(See Fig.\ref{physics:susy:stau:tanb_xc}).

%% file: physsusy/otherscenarios.tex
\section{Other Scenarios}

Up to now we have been concentrating 
our attention on typical SUGRA scenarios.
We now turn to other scenarios and examine
expected complications and possible ways out.
We will first study a particular case
where the lighter chargino decays into $\tilde{\tau}_1 + \nu_\tau$
and therefore the end-point method we used for
the $\tilde{\chi}_1^\pm \to W^\pm \tilde{\chi}_1^0$ decay
is not applicable.
Effects of possible $CP$ phases will be described
after this.
We then move on to examine other SUSY breaking schemes:
Gauge Mediated Supersymmetry Breaking (GMSB) and
Anomaly Mediated Supersymmetry Breaking (AMSB),
which will be followed by discussions on
$R$-parity violation.

\subsection{Chargino Decaying into Stau plus Neutrino}
\label{Chap:susy:Sec:otherscenarios:kato}

\begin{figure}[h]
\centerline{
\epsfxsize=6.5cm\epsfbox{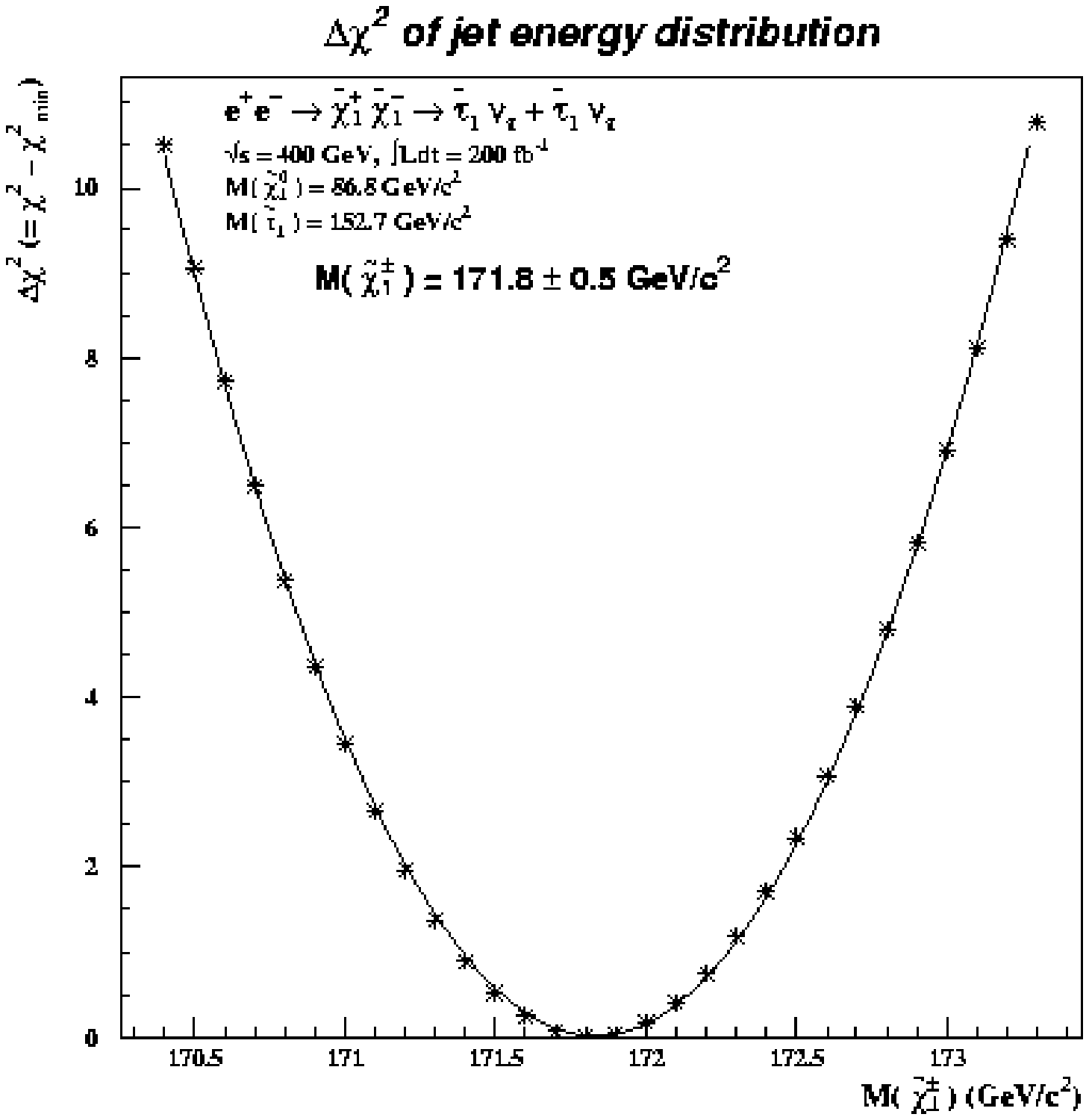}
\hspace{1cm}
\epsfxsize=6.5cm\epsfbox{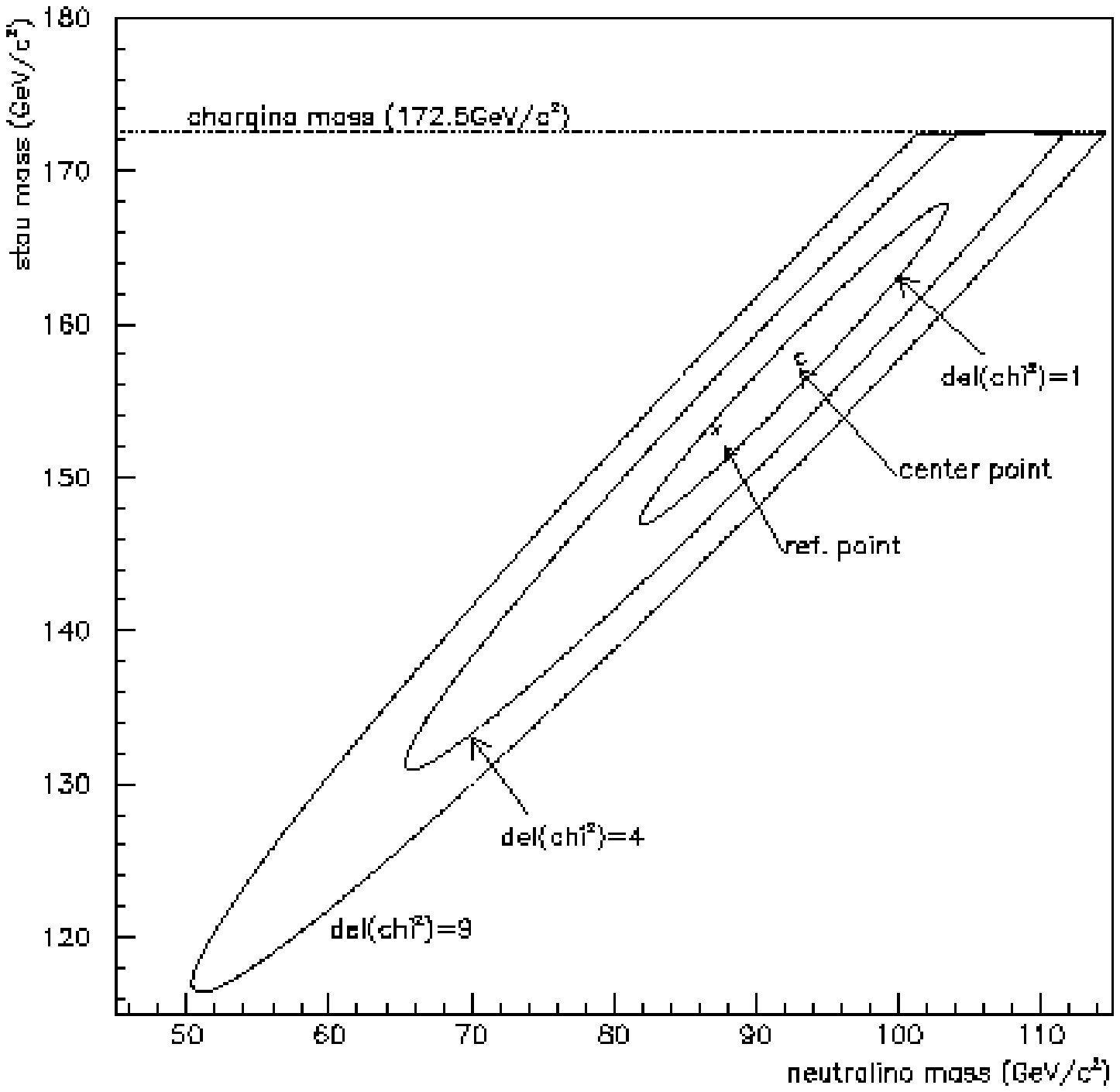}}
\begin{center}\begin{minipage}{\figurewidth}
\caption{\label{fig:delchi} \sl
(a) $\Delta\chi^2$ for the energy distribution of the $\tau$ jets from
$\tilde{\chi}_1^\pm \to 
\tilde{\tau}_1^\pm \mathop{\nu_\tau}\limits^{\mbox{\scriptsize$(-)$}} 
\to \tau^\pm \tilde{\chi}_1^0 
\mathop{\nu_\tau}\limits^{\mbox{\scriptsize$(-)$}}$, 
(b) contours of the $\Delta\chi^2$,
when the stau and the neutralino masses are changed.
Input masses are $m_{\tilde{\chi}_1^+}$ = 172.5 GeV,
$m_{\tilde{\tau}_1}$ = 152.7 GeV,
and $m_{\tilde{\chi}_1^0}$ = 86.8 GeV.
An integrated luminosity of $200~{\rm fb}^{-1}$ is assumed.
}
\end{minipage}
\end{center}
\end{figure}

As we have seen above, if $\tan\beta$ is large,
it is quite possible that the
stau is the first SUSY particle to be discovered at JLC.
If the mass hierarchy (expected
at large $\tan \beta$ in (M)SUGRA scenario as well) is such that
$m_{\tilde{l}} > m_{\tilde{\chi}_1^\pm} > 
m_{\tilde{\tau}_1} > m_{\tilde{\chi}_1^0}$, 
then the lighter chargino almost 100~\% decays into 
$\tilde{\tau}_1 + \nu_\tau$,
through the left-handed component
due to the stau's left-right mixing.
The signature of the chargino pair production
becomes a pair of acoplanar tau leptons.
The background processes thus include the former signal process:
$e^+e^- \to \tilde{\tau}_1^+ \tilde{\tau}_1^-$,
and the standard model ones: $ee\tau\tau$, $WW$, and $ZZ$.
Nevertheless, it has been shown that we can determine 
the chargino mass by measuring the $\tau$-jet energy.
In this scenario, the chargino mass determination
based on the end-point measurement of the
daughter $W$'s from 
the $\tilde{\chi}_1^\pm \to W^\pm \tilde{\chi}_1^0$ decay
will be useless.
For the mass hierarchy given above, however,
we can assume that $m_{{\tilde{\tau}}_1}$ and
$m_{\tilde{\chi}_1^0}$ to be known from 
the studies of $\tilde{\tau}_1^+ \tilde{\tau}_1^-$ production. 
Using the $E_{jet}$ distribution ($\tau$ is 
detected as a low multiplicity jet), therefore,
we can still determine the chargino mass:
for an integrated luminosity of $200~{\rm fb}^{-1}$, 
with inputs $m_{\tilde{\chi}_1^+}$ = 172.5 GeV,
$m_{\tilde{\tau}_1}$ = 152.7 GeV,
and $m_{\tilde{\chi}_1^0}$ = 86.8 GeV, 
the fit to the $E_{jet}$ distribution yielded
$m_{\tilde{\chi}_1^+} = 171.8 \pm 0.5$ GeV
(see Fig.~\ref{fig:delchi}-(a) for the $\Delta \chi^2$ curve).
This value is obtained, assuming that the errors on
the stau and the LSP masses are negligible. 
In order to estimate their effects on the chargino
mass determination, their $\chi^2$ contribution
to the $E_{jet}$ fit has been estimated,
which is shown in Fig.~\ref{fig:delchi}-(b).
which is shown in Fig.~\ref{fig:delchi}-(b).
If the error on these masses are less than 3~GeV,
as expected from Fig.~\ref{physics:susy:stau:massfit},
the effect will not be a problem.

\subsection{$\protect \mathbold{CP}$ Violating Phases}

As already exemplified in the previous section 
(\ref{Chap:susy:Sec:chic:massmatrix}),
the new mass parameters introduced in association with
supersymmetry contain $CP$ phases that 
cannot be eliminated by field redefinitions.
We have not yet examined their effects in any detail
and that is what we will turn our attention to here.
Firstly, it should be noted that the $CP$ phases can modify
not only the couplings but also
the sparticle mass spectrum significantly\cite{Ref:BK}.
Secondly, although there are potentially many
$CP$ violating phases in the full MSSM Lagrangian,
it is known that only two $CP$-odd rephase-invariant phases:
$$
\mu = \left| \mu \right| \exp^{i\phi_\mu} 
~~~~~\mbox{and}~~~~~~ 
M_1 = \left| M_1 \right| \exp^{i\phi_1},
$$
stemming from the chargino and neutralino mass matrices,
take part in the chargino and neutralino 
production processes.

\subsubsection{$\protect \mathbold{e^+e^- \to \tilde{\chi}_1^0 \tilde{\chi}_2^0
\to
\tilde{\chi}_1^0  \tilde{\chi}_1^0 l^+ l^-}$}

\begin{figure}[h]
\centerline{
\epsfxsize=13cm\epsfbox{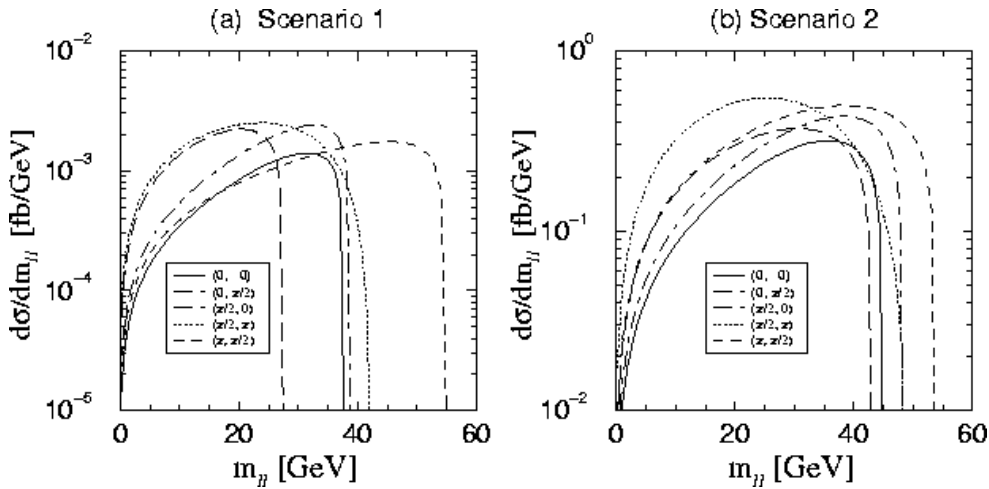}
}
\begin{center}\begin{minipage}{\figurewidth}
\caption{\sl  \label{Fig:cpv:mll}
Dilepton invariant mass distributions for the process:
$e^+e^- \to \tilde{\chi}_1^0 \tilde{\chi}_2^0
\to
\tilde{\chi}_1^0  \tilde{\chi}_1^0 l^+ l^-$
in (a) scenario ${\cal S}1$ and (b) ${\cal S}2$
for five combinations of the values of two $CP$ phases:
$\phi_\mu$ and $\phi_1$. 
\cite{Ref:Choi2}.
}
\end{minipage}
\end{center}
\end{figure}

The good place, where we can study the effects of the
$CP$-violating phases, is the following associated
neutralino production process:
$e^+e^- \to \tilde{\chi}_1^0 \tilde{\chi}_2^0$,
followed by
$\tilde{\chi}_2^0 \to \tilde{\chi}_1^0 l^+ l^-$\cite{Ref:Gooty,Ref:Choi2},
since it is likely that this is the
first process to access the gaugino sector,
and that the final-state leptons allow us to
construct clean and useful observables for 
the study of the $CP$ phases.
First such an observable is the dilepton invariant mass ($m_{ll}$).
It is note worthy that the distribution shape of
this observable is independent of
the production mechanism of the decaying neutralino,
which is because the invariant mass does not involve any
angular variables describing the decays so that the
polarization of the decaying neutralino is irrelevant.
In Ref.\cite{Ref:Choi2}, two scenarios 
(${\cal S}1$ and ${\cal S}2$) are considered that satisfy
the electron EDM constraints:
$$
\begin{array}{lllll}
{\cal S}1: & ~ & M_2 = 100~{\rm GeV}, &
		 \left| \mu \right| = 200~{\rm GeV}, &
		 m_{\tilde{e}} = 10~{\rm TeV} \cr
{\cal S}2: & ~ & M_2 = 100~{\rm GeV}, &
		 \left| \mu \right| = 700~{\rm GeV}, &
		 m_{\tilde{e}} = 200~{\rm GeV}.
\end{array}
$$
Figs.\ref{Fig:cpv:mll}-a) and -b) show the invariant mass
distributions for the two cases.
We can see that the $m_{ll}$ distribution
is sensitive to the $CP$ phases.

The final-state lepton angular distribution is
also known to be useful.
Unlike the dilepton invariant mass
this variable is crucially dependent on
the production-decay spin correlations,
and its distribution shape
sensitively reflects the slepton mass
and the neutralino mixing\cite{Ref:Gooty}.

\begin{figure}[h]
\centerline{
\epsfxsize=7cm\epsfbox{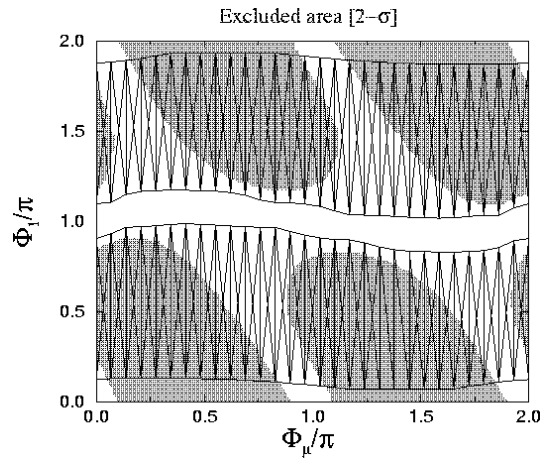}
}
\begin{center}\begin{minipage}{\figurewidth}
\caption[physics:susy:cpv:ot]{\sl \label{Fig:cpv:ot}
Expected $95\%$-confidence-level
excluded regions of the $\phi_\mu$-$\phi_1$ plane
by the EDM constraints (shaded) and by the ${\cal O}_T$
measurement at $\sqrt{s} = 500~{\rm GeV}$ with $200~{\rm fb}^{-1}$.
}
\end{minipage}\end{center}
\end{figure}

The two observables discussed above are
$CP$-even and only indirectly depends on the $CP$ phases.
It is possible, however, to construct a $T$-odd,
thus $CP$-odd, observable:
$$
{\cal O}_T = {\bf p}_e \cdot ( {\bf p}_{l^-} \times {\bf p}_{l^+} ),
$$
whose expectation value can be sizable in scenario ${\cal S}2$,
thereby directly signaling the $CP$ phases at work\cite{Ref:Choi2}.
Fig.~\ref{Fig:cpv:ot} is an expected exclusion plot
for the $CP$ phases from the measurement of ${\cal O}_T$
at $\sqrt{s} = 500~{\rm GeV}$ with an integrated luminosity
of $200~{\rm fb}^{-1}$.
One can see that the ${\cal O}_T$ measurement is
complementary to the EDM measurement shown as shaded regions.

\subsubsection{$\protect \mathbold{e^+e^- \to \tilde{t}_1 \tilde{t}_1^*h}$}

In the light of possible $CP$ phases that may appear
in the scalar sector,
$e^+e^- \to \tilde{t}_1 \tilde{t}_1^*h$ 
is interesting, since the stop can be light enough
to make this process useful 
already at $\sqrt{s} = 500~{\rm GeV}$.
The $CP$ violating phases in $A_t$ and $\mu$ 
affects not only the masses and the mixings of the stops and
the Higgs bosons but also the $\tilde{t}\tilde{t}^*h$ coupling,
which has been studied in Ref.\cite{Ref:Bae}.
In order to satisfy the neutron electric dipole constraints 
as well as cosmological ones\cite{Ref:pilaftsis}, the
following conditions are imposed in the study:
i) Arg($\mu) < 10^{-2}$,  ii) $m_{\tilde{g}} > 400$ GeV,
iii) $ |A_{e}|, |A_{u,c}|, |A_{d,e}| < 10^{-3} |\mu|$,
and iv) maximal mixing in the stop sector $|A_{t}| = |\mu \cot\beta |$.
\begin{figure}[htb]
\centerline{
\epsfxsize=7cm\epsfbox{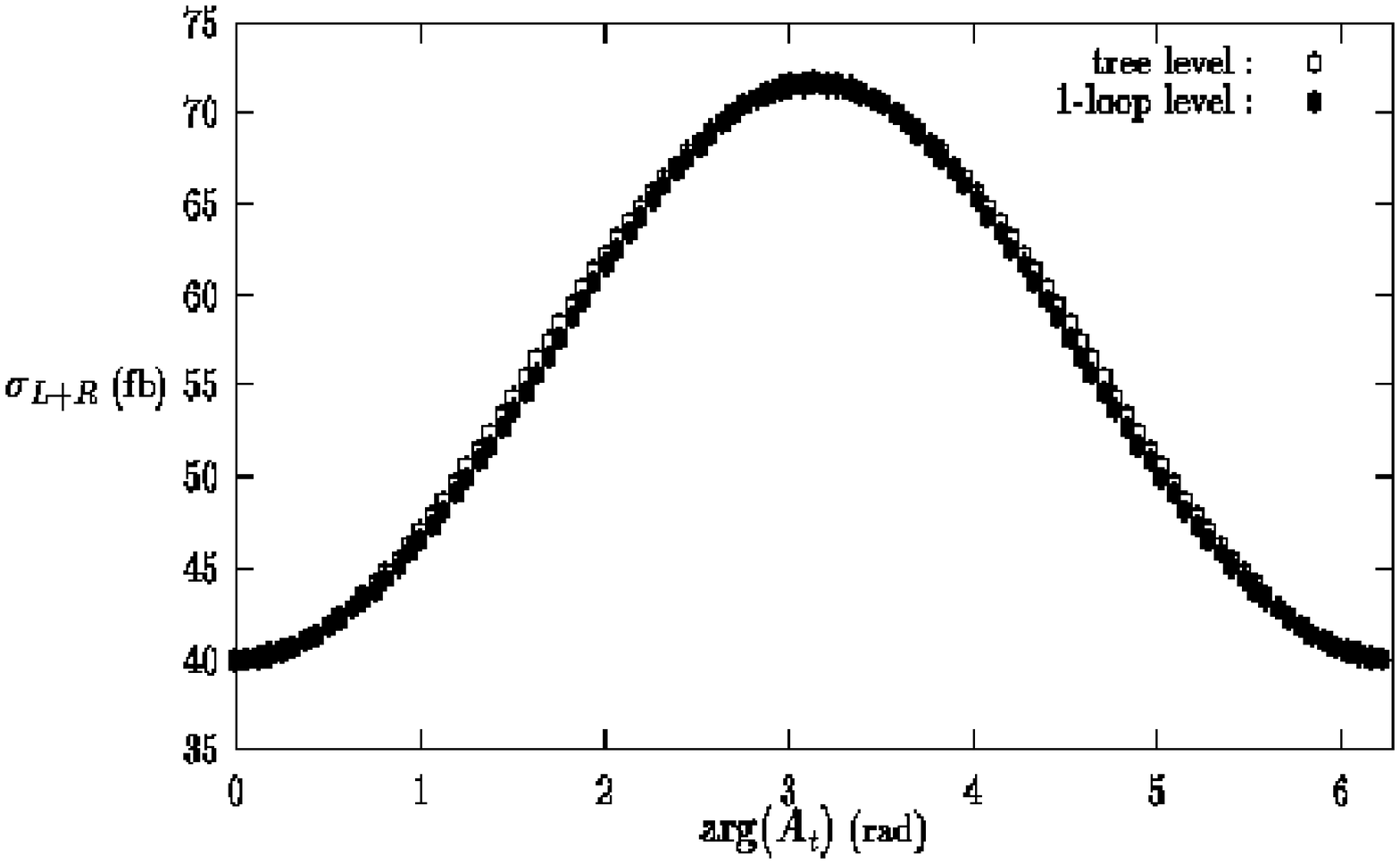}
\hspace{0.7cm}
\epsfxsize=7cm\epsfbox{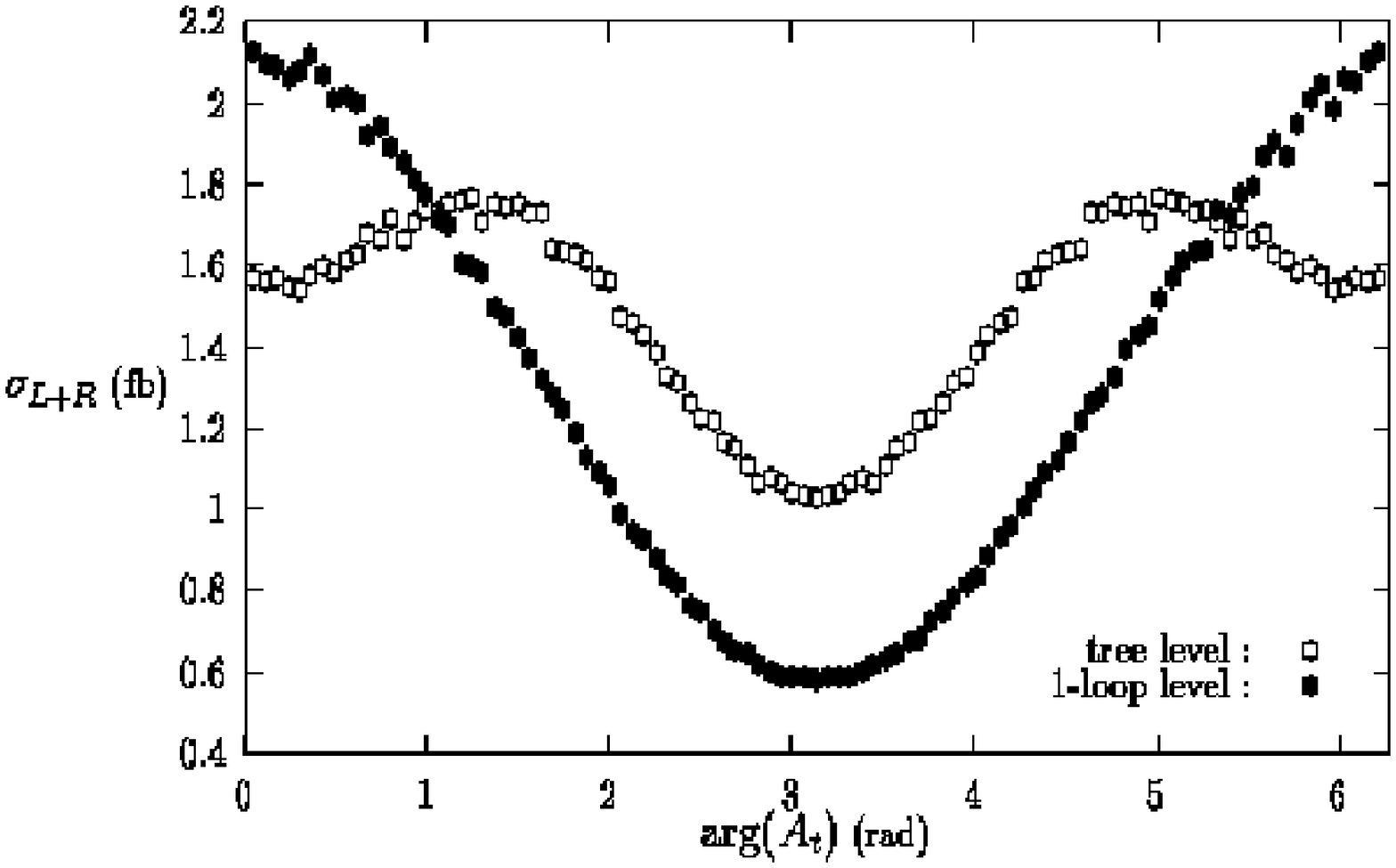}
}
\begin{center}\begin{minipage}{\figurewidth}
\caption{\sl \label{Fig:cpv:stop} 
Effect of loop corrections on 
$\sigma(\tilde{t}_1 \tilde{t}_1^*)$ (left) and 
$\sigma(\tilde{t}_1 \tilde{t}_1^*h)$ (right) 
as a function of Arg$(A_t)$ for 
$\sqrt{s} = 500$ GeV, $\mu = 500$ GeV, $|A_t| = 250$ GeV, 
$M_A \simeq 194$ GeV, $\tan \beta = 2$, 
and $M_{\rm SUSY} = 500 $ GeV\cite{Ref:Bae}.
}
\end{minipage}
\end{center}
\end{figure}
Fig.~\ref{Fig:cpv:stop} shows 
$\sigma_{L+R}(e^+e^- \to \tilde{t}_1 \tilde{t}_1^*)$ 
and 
$\sigma_{L+R}(e^+e^- \to \tilde{t}_1 \tilde{t}_1^* h)$
in the left and the right panels, respectively. 
We see that though the loop effects are minimal 
in $\tilde{t}_1 \tilde{t}_1^*$ production, 
the dependence on Arg($A_{t}$) is quite strong.
This strong dependency comes from the strong dependence of
the stop mass $m_{\tilde{t}_1}$ on Arg($A_{t}$).
On the other hand 
for $e^+e^- \to \tilde{t}_1 \tilde{t}_1^* h$ 
the cross sections are rather small (a few fb level) but loop effects
are substantial and can be as much as 100 \%. 
This is because the light Higgs boson mass can be strongly
affected by loop corrections with the complex trilinear coupling $A_t$.
though the neutral Higgs boson mixing can hardly be changed 
significantly.
It is in principle possible to extract $|A_{t}|$, Arg$(A_{t})$ 
by combining the measurements of these cross-sections 
with the knowledge of higgs masses\cite{Ref:Bae}.

\subsection{Gauge Mediated Supersymmetry Breaking}

As a result of the super-Higgs mechanism, the gravitino
acquires a SUSY-breaking mass:
\begin{eqnarray}
\nonumber
m_{\tilde{G}}  =  \frac{F}{\sqrt{3} M_{Pl}} 
 \simeq  \left( \frac{\sqrt{F}}{100~{\rm TeV}} \right) {\rm eV},
\end{eqnarray}
where $M_{Pl}$ is the reduced Planck mass given by
$M_{Pl} = (8\pi G_N)^{-1/2}$.
As already mentioned in Section~\ref{Chap:susy:Sec:scenarios},
the GMSB thus predicts a very light gravitino mass
for the theoretically allowed range of the SUSY breaking scale:
$10 < \sqrt{F} < 10^4~{\rm TeV}$
shown in Table~\ref{T:rgplen:1}.
In the GMSB models, therefore, the LSP is always the gravitino.
Since the longitudinal component of the now massive gravitino,
being the goldstino originally, has an effective coupling with
the NLSP that is proportional to 
$$
\left(\frac{E_{\tilde{G}}}{M_{Pl}} \right).
\left(\frac{E_{\tilde{G}}}{m_{\tilde{G}}} \right)
\simeq 
\left(\frac{E_{\tilde{G}}^2}{F} \right)
\simeq
\left(\frac{M_{NLSP}}{\sqrt{F}} \right)^2,
$$
which can be much larger than one would expect
for gravitational interaction\footnote{
This is in contrast to the gravity-mediated 
soft SUSY breaking scenarios,
where $\sqrt{F} \simeq 10^{10}$-$10^{11}~{\rm GeV}$,
which makes the gravitino interaction with other
MSSM fields phenomenologically irrelevant.
} 
This makes the NLSP decay into the gravitino
relevant to collider phenomenology.
The decay length is given by
$$
L = c\tau \beta \gamma \propto \frac{1}{(M_{NLSP})^5} (\sqrt{F})^{4}
$$
and depends on the model parameters.
Since $\sqrt{F}$ spans a very wide range (see Table~\ref{T:rgplen:1}), 
so does the expected decay length:
$$
10^{-4} < L < 10^{5}~{\rm cm}.
$$
Since the GMSB models preserve the GUT relation among the gauginos
and usually predict light sleptons lighter as compared with
the SUGRA models,
the NLSP will be $\tilde{\chi}_1^0$, or $\tilde{\tau}_1$,
or $\tilde{l}_R (l \ne \tau)$, depending on the model parameters.
There are thus logically nine cases to be considered:
\begin{equation}
\pmatrix{ NLSP = \tilde{\chi}^0_1 \cr 
NLSP = \tilde{\tau}_1 \cr 
NLSP = \tilde{l}_R
} \otimes
\pmatrix{ L \ll R \cr 
L \lsim R \cr 
L \gg R 
} 
\end{equation}
where $R$ is the detector radius. 
If the decay length is much bigger than the detector radius
($L \gg R$), the signatures of the sparticle productions
will stay the same as with the gravity-mediated models.

If the decay length is much
shorter than the detector radius ($L \ll R$),
the final-state NLSP's decay 
and leave a missing $P_T$ signal
with two isolated energetic photons
in the case of the $\tilde{\chi}^0_1$ NLSP,
or two isolated energetic taus or electrons or muons
in the cases of the $\tilde{\tau}_1$ NLSP or the 
$\tilde{l}_R$ NLSP.
In the case of the $\tilde{\chi}^0_1$ NLSP,
the signatures of the sparticle productions
are then missing $P_T$ and extra isolated energetic
photons in addition to those
final-state particles expected 
for the gravity-mediated models, if any.
It should also be noted that we can now detect
the $\tilde{\chi}^0_1$ pair production as an
acoplanar photon pair.
On the other hand, the slepton NLSP cases will
have the same signatures as
the massless $\tilde{\chi}^0_1$ limit 
of the gravity-mediated cases.
In such a case, the study of the chargino will be similar
to the one discussed above 
in Subsection \ref{Chap:susy:Sec:otherscenarios:kato}.

If the decay length is comparable with the detector size
($L \lsim R$), one or both of the NLSP's can decay
inside the detector volume after flying over finite
lengths.
In the case of the $\tilde{\chi}^0_1$ NLSP,
this will result in
one or two extra energetic photons
which do not point to the interaction point (IP).
The $\tilde{\chi}^0_1$ pair production can then be detected 
as an acoplanar photon pair with displaced vertices.
In the case of the slepton NLSP,
we might even be able to see their tracks,
and possibly kinks also, in the tracking detector. 
They will appear as heavily ionizing stable particles,
provided that $dE/dx$ measurement is available\cite{Ref:HH}.
The mass determination for the NLSP can be carried out
by measuring the high end of the photon or lepton
energy distribution by virtue of the simple kinematics
of the two-body decay into two practically massless particles.
It has been shown that $\tilde{\chi}^0_1$ mass
can be determined to $0.2~\%$ for $m_{\tilde{\chi}^0_1} = 200~{\rm GeV}$
with an integrated luminosity of 
$200~{\rm fb}^{-1}$\cite{Ref:AB}.
Note also that when the displaced vertices are detectable,
the so determined mass, together with the decay length
measured, will constrain the SUSY breaking scale $F$.

Once the NLSP is identified and its decay pattern established,
we can carry out analyses 
similar to those described above for the MSSM case,
by ignoring decay products from the NLSP's in the
final states.

\subsection{Anomaly Mediated Supersymmetry Breaking}

Anomaly mediation is a special case of gravity mediation
where there is no direct tree level coupling that
transmits the SUSY breaking in the hidden sector
to the observable one.
In this case the masses of the gauginos are
generated at one-loop, while
those of the scalars are generated at two-loop level,
because of the superconformal anomaly
that breaks scale invariance\cite{RS2,GR}.
Since the anomaly is topological in origin, it
naturally conserves flavor, thereby inducing
no new FCNC amplitudes.
AMSB models thus preserve virtues of the gravity-mediated 
models with the FCNC problem resolved.
The scale invariant one-loop gaugino mass expressions are
\begin{eqnarray}
\nonumber
M_i & = & b_i \left( \frac{\alpha_i}{4\pi} \right) m_{3/2}
 ~~=~~  \left( \frac{b_i}{b_2} \right) 
          \left( \frac{\alpha_i}{\alpha_2} \right),
\end{eqnarray}
where $b_i$'s are related to the one-loop beta functions through
$\beta_i = - b_i g_i^3/(4\pi)^2$ and numerically
$b_1 = 33/5$, $b_2 = 1$, $b_3 = -3$.
In the minimal AMSB model, however, the sleptons
become tachyonic, which necessitated, for instance,
introduction of a universal scalar mass parameter
$m_0^2$:
\begin{eqnarray}
\nonumber
m_{\tilde{f}}^2 & = & 
		C_{\tilde{f}} \frac{m_{3/2}^2}{(16\pi^2)^2} + m_0^2
    ~~=~~ \sum 2a_{\tilde{f},i} b_i 
         \left( \frac{\alpha_i}{\alpha_2} \right)^2 M_2^2 
         + m_0^2,
\end{eqnarray}
where $a_{\tilde{f},i}$'s are related to the RG functions
through $\gamma_{\tilde{f}} = - a_{\tilde{f},i} g_i^2/(4\pi)^2$.

\begin{figure}[h]
\centerline{
\epsfxsize=6cm\epsfbox{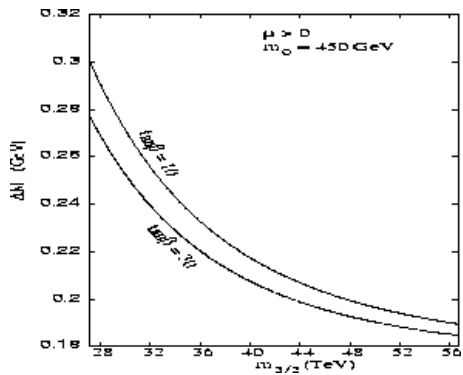}
}
\begin{center}\begin{minipage}{\figurewidth}
\caption[physics:susy:amsb:delm]{\sl \label{Fig:delm}
Mass difference 
$\Delta M \equiv m_{\tilde{\chi}_1^\pm} - m_{\tilde{\chi}_1^0}$
as a function of the gravitino mass for $\tan\beta = 10$ (upper curve)
and $\tan\beta = 30$ (lower curve), $\mu > 0$, and $m_0 = 450~{\rm GeV}$.
}
\end{minipage}\end{center}
\end{figure}

\begin{figure}[h]
\centerline{
\epsfxsize=7cm\epsfbox{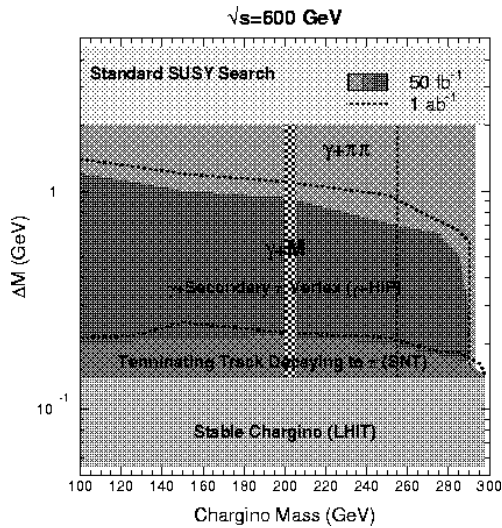}
}
\begin{center}\begin{minipage}{\figurewidth}
\caption[physics:susy:amsb:regions]{\sl \label{Fig:regions}
Viable search modes in different regions
in the $m_{\tilde{\chi}_1^\pm}$-$\Delta M$
($\equiv m_{\tilde{\chi}_1^\pm} - m_{\tilde{\chi}_1^0}$) plane.
Discovery reach is given
for $50~{\rm fb}^{-1}$ and also for $1~{\rm ab}^{-1}$
accumulated at $\sqrt{s} = 600~{\rm GeV}$.
The  vertical band and vertical line (the band for $50~{\rm fb}^{-1}$ 
and the line for $1~{\rm ab}^{-1}$)
is the reach of the $\gamma+\sla{M}$ detection mode, which
is relevant only if the $\pi$'s are too soft to detect\cite{Ref:Gunion}.
}
\end{minipage}\end{center}
\end{figure}

As already stressed in Section \ref{Chap:susy:Sec:scenarios},
the first and the most important message from 
the gaugino mass formula
above is the hierarchy:
$$
  M_1 : M_2 : M_3 = 2.8 : 1 : 8.3
$$
as opposed to 
$M_1 : M_2 : M_3 = 1 : 2 : 7$
that is expected for the gravity- or gauge-mediated models.
This implies that the lightest neutralino ($\tilde{\chi}_1^0$) 
and the lighter chargino ($\tilde{\chi}_1^\pm$) 
are almost pure winos and consequently
mass-degenerate\footnote{
The lighter chargino and the lightest neutralino can
be almost mass-degenerate also in the MSSM models.
In such a case, however, they are almost
pure higgsinos instead of winos.
}.
This degeneracy is lifted by 
the tree-level gaugino-higgsino mixing
and loop corrections, resulting in a small
but finite mass splitting as depicted in
Fig.~\ref{Fig:delm}
for $\tan\beta = 10, 30$, $\mu > 0$, and $m_0 = 450~{\rm GeV}$.
The lighter chargino thus decays mostly (96-98~\%)
into the lightest neutralino and a soft $\pi^\pm$,
possibly with a visibly displaced vertex.
Ref.~\cite{Ref:Gunion} discusses search strategies for
the chargino pair production:
$e^+e^- \to \tilde{\chi}_1^+  \tilde{\chi}_1^- (\gamma)$,
where the additional photon will be very useful to
suppress the huge two-photon $\pi\pi$ background
expected for the $\Delta M$ range shown above, or to
tag the chargino pair production when the final-state pions are
too soft to be detected.
When the decay length is comparable or larger than the detector
size, the charginos will appear as heavily ionizing particles
as in the case of the slepton LSP in the GMSB models.
If the daughter pions are too soft to be detected, one may
observe abruptly terminating tracks in the central tracker.
Fig.~\ref{Fig:regions} summarizes the search methods and
corresponding discovery limits shown in 
the $m_{\tilde{\chi}_1^\pm}$-$\Delta M$ plane.
The figure tells us that,
with all the methods combined,
one can cover the parameter space almost to
the kinematical limit.

The scalar mass formula also contains a phenomenologically
important message that is the near degeneracy of the
left- and right-handed sleptons, which
means that left- and right-handed sleptons can be
produced simultaneously, with relative rates
controlled by the beam polarization.
Slepton productions are studied\cite{Ref:Ghosh1,Ref:Ghosh2}
in the context of the AMSB models
assuming $\sqrt{s} = 1~{\rm TeV}$\footnote{
The sleptons are expected to be heavy on the basis of
the required absence\cite{Ref:Datta} of
charge and color violating minima in the
one-loop effective potential.
}.

\begin{figure}[h]
\centerline{
\epsfxsize=13cm\epsfbox{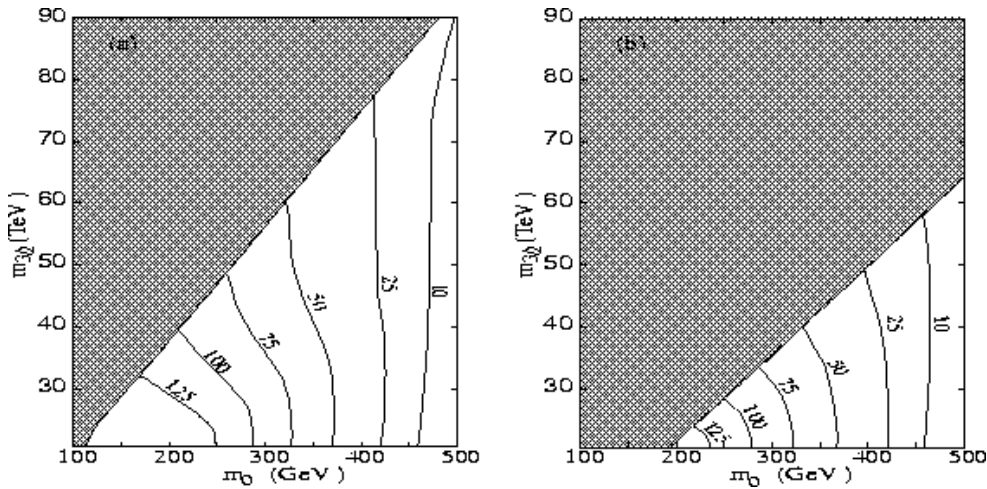}
}
\begin{center}\begin{minipage}{\figurewidth}
\caption{\sl  \label{Fig:amsb:m0m32}
Effective cross section contours in the $m_{3/2}$-$m_0$ plane
expected for the 
$e^\pm \tilde{\chi}_1^0 \mathop{\nu_e}\limits^{\mbox{\scriptsize$(-)$}} 
\tilde{\chi}_1^0 \pi^\mp$ signal from
$e^+e^- \to \tilde{e}_L^+ \tilde{e}_L^-$ at $\sqrt{s} = 1~{\rm TeV}$:
(a) $\tan\beta = 3$, and (b) $\tan\beta = 30$.
The shaded regions are ruled out by various constraints\cite{Ref:Ghosh2}.
}
\end{minipage}
\end{center}
\end{figure}

For instance, let us consider the left-handed selectron
pair production:$e^+e^- \to \tilde{e}_L^+ \tilde{e}_L^-$
followed by the mixed decays:
$\tilde{e}_L^\pm \to e^\pm \tilde{\chi}_1^0$
and
$\tilde{e}_L^\mp \to \mathop{\nu_e}\limits^{\mbox{\scriptsize$(-)$}} 
\tilde{\chi}_1^\mp$
with 
$\tilde{\chi}_1^\mp$ decaying slowly into
$\tilde{\chi}_1^\mp \to \tilde{\chi}_1^0 + \pi^\mp$.
This results in a final state:
$e^\pm \tilde{\chi}_1^0 
\mathop{\nu_e}\limits^{\mbox{\scriptsize$(-)$}} \tilde{\chi}_1^0 \pi^\mp$.
The signal is a fast $e^\pm + \sla{E}_T$ and a soft $\pi^\mp$. 
The soft $\pi$ can give rise to a visibly displaced vertex,
if the decay length of the chargino is less than 3cm. 
If it is longer than 3cm, 
then one sees a heavily ionizing track of the chargino
as discussed above.
Fig.~\ref{Fig:amsb:m0m32} shows the effective cross sections
after cuts to select the signal events
as contours in the $m_{3/2}$-$m_0$ plane.
Note that the effective cross sections include
branching fraction as well as the selection efficiency
due to the selection cuts.
We can see that large regions, allowed by all the
current constraints, have large effective cross sections 
$\sim 10-100$ fb at $\sqrt{s}$ = 1000 GeV.

\subsection{$\protect \mathbold{\it R}$-Parity Violation}

Although we have been assuming exact conservation of $R$-parity ($R_p$)
in the discussions so far,
this symmetry is not automatic in the MSSM.
Without $R$-parity conservation, the superpotential
may contain $R$-parity violating terms:
\begin{eqnarray}
\nonumber
W_{\sla{R}_p} & = &   \lambda_{ijk} L_i L_j e^c_k
+ \lambda'_{ijk} L_i Q_j d^c_k
+ \lambda''_{iji} u^c_i d^c_j d^c_k 
+ \mu_i H_u L_i.
\end{eqnarray}
The $\lambda$- and $\lambda'$-terms violate lepton number ($L$), 
while the $\lambda''$-terms break baryon number ($B$).
In order to avoid unacceptably rapid proton decay,
the $\sla{B}$ and the $\sla{L}$ terms should not be made finite
simultaneously.
The $\mu_i$ terms are also lepton-number violating, but
can be rotated away at tree level.

Here, we assume that these $R$-parity violating couplings are 
strong enough to make the LSP ($\tilde{\chi}_1^0$) decay
promptly in the detector volume, but other than that
they are too weak to participate in the observable phenomena\footnote{
When the $R$-parity violating terms are large,
we may expect new SUSY signals that are absent from the $R$-parity
conserving scenarios: for instance, if the $\lambda$-coupling is sizable
we may see a spectacular $s$-channel resonance production of
a sneutrino.
}.
Let us then consider $e^+e^- \to \tilde{\chi}_{i}^{+} \tilde{\chi}_{j}^{-}$
and $e^+e^- \to \tilde{\chi}_{i}^{0} \tilde{\chi}_{j}^{0}$\cite{Ref:Ghosh3}. 
Once the LEP constraints on
$m_{\tilde{\chi}_{1}^{\pm}}$ are imposed, 
it is found that over a large part of 
the region of parameter space which allows 
$\tilde{\chi}_{1}^{+}$ within the 
reach of a 500 GeV linear collider, 
$\tilde{\chi}_{3}^{0}, \tilde{\chi}_{4}^{0}$, 
and $\tilde{\chi}_{2}^{+}$ are almost always beyond its reach, 
at least in the framework of the (c)MSSM.
Hence it is sufficient to consider 
i) $e^+e^- \to \tilde{\chi}_{1}^{0} \tilde{\chi}_{2}^{0}$,
ii)$e^+e^- \to \tilde{\chi}_{2}^{0} \tilde{\chi}_{2}^{0}$,
and iii) $e^+e^- \to \tilde{\chi}_{1}^{+} \tilde{\chi}_{1}^{-}$. 
Further, using the
approximate degeneracy of $\tilde{\chi}_{1}^{\pm}$ 
and $\tilde{\chi}_{2}^{0}$, 
the number of decay chains to be considered are reduced to
manageable numbers.
The final-state $\tilde{\chi}_1^0$'s decay via
$R$-parity violating terms:
\begin{eqnarray}
\nonumber
L L e^c (\lambda) & : & 
	\tilde{\chi}_1^0 \to l_1^+ l_2^- + \sla{E}_T \cr
\nonumber
L Q d^c (\lambda') & : &
	\tilde{\chi}_1^0 \to l^\pm + \sla{E}_T 
		~~\mbox{or}~~ l^\pm + 2~\mbox{Jets} \cr
\nonumber
u^c d^c d^c (\lambda'') & : &
	\tilde{\chi}_1^0 \to 3~\mbox{Jets},
\end{eqnarray}
where $\sla{E}_T$'s are due to neutrinos. 
For the $\sla{L}$ $\lambda$-couplings the final state will
consist of $m$ leptons and $\sla{E}_T$; 
for the $\sla{L}$ and $\sla{B}$
$\lambda'$-couplings it will consist of $m$ leptons, 
$n$ jets and $\sla{E}_T$,
whereas $\sla{B}$ $\lambda''$-couplings 
give rise to a final state with only jets.
Potential standard model backgrounds include:
$e^+e^- \to W^+W^-$,
$e^\pm\mathop{\nu_e}\limits^{\mbox{\scriptsize$(-)$}} W^\mp Z$,
$ZZ$, $t\bar{t}(g)$,
$W^+W^-Z$, $e^+e^-Z$, etc.
\begin{table}[htb]
\begin{center}\begin{minipage}{\figurewidth}
\caption{\sl Contributions (in fb) of different (light)
chargino and neutralino production modes to multi-lepton signals
in the case of $\lambda$-couplings for
point {\bf A} of the five points studied\cite{Ref:Ghosh3}:
$M_2 = 100~{\rm GeV}$,
$\mu = -200~{\rm GeV}$,
$\tan\beta = 2$, and
$M_{\tilde{e}_L} = M_{\tilde{e}_R} = 150~{\rm GeV}$.
The last column shows the SM background.
\label{tab:tab3}}
\end{minipage}\end{center}
\begin{center}
\begin{tabular}{|c|l|cccc|c|c|}
\hline  
&&&&&&&\\
&Signal       
&$  \tilde{\chi}^0_1 \tilde{\chi}^0_1 $ & $ \tilde{\chi}^0_1 \tilde{\chi}^0_2$  & $ \tilde{\chi}^0_2 \tilde{\chi}^0_2$  & $\tilde{\chi}^+_1 \tilde{\chi}^-_1$ &  
{\bf Signal} & {\bf Bkgd.}  \\
&&&&&& (fb) & (fb)\\
\hline 
{\bf A} 
&$1\ell + \sla{E}_T$ & $1.1$ & $0.4$ & $0.2$ & $1.5$   &$3.2$ & $8272.5$
\\ \cline{2-8}
& $2\ell + \sla{E}_T$ & $ 14.9$ & $5.2$ & $1.8$ & $15.3$   &$ 37.2$ &$ 2347.4$ 
\\ \cline{2-8}
& $3\ell + \sla{E}_T$ & 91.7 & 25.3 & 7.2 & 71.6   & 195.8 & 1.5
\\ \cline{2-8}
& $4\ell + \sla{E}_T$ &  212.8 & 49.6 &13.6 & 152.8   & 428.8 & 0.4 
\\ \cline{2-8}
&$5\ell + \sla{E}_T$ &  0.0 & 37.8 & 19.3 & 113.5   & 170.6 &  -
\\ \cline{2-8}
& $6\ell + \sla{E}_T$ &  0.0 & 39.6 & 21.6 & 26.9   & 88.0 &  -
\\ \cline{2-8}
& $7\ell + \sla{E}_T$ & 0.0 & 0.0 & 11.9 & 0.0   & 11.9 &  -
\\ \cline{2-8}
& $8\ell + \sla{E}_T$ &0.0 &0.0 &8.0 &0.0   &8.0 &  -
\\ \hline
\end{tabular}
\end{center}
\end{table}
Table~\ref{tab:tab3} shows, for a particular point (point {\bf A}),
results of the Monte 
Carlo analysis for the case of $\lambda$-couplings.
We can see that the requirements of three or more isolated
energetic leptons in the final states retains a large fraction
of signal events, while suppressing the background
to a negligible level.
In the case of the $\lambda'$-couplings, we can still
use the same strategy but with significant loss of
signal events due to the smaller expected number of 
leptons in the final states. 
Although detailed simulations are yet to be done,
the cleanness of the final sample seems to allow
mass determinations:
we can carry out similar analyses as with the $R$-parity
conserving cases by ignoring the decay products of the LSP
or, if there is no missing neutrino in the LSP decay,
we can directly reconstruct the masses involved in the
decay chains.

\begin{figure}[htb]
\centerline{
\epsfxsize=12cm\epsfbox{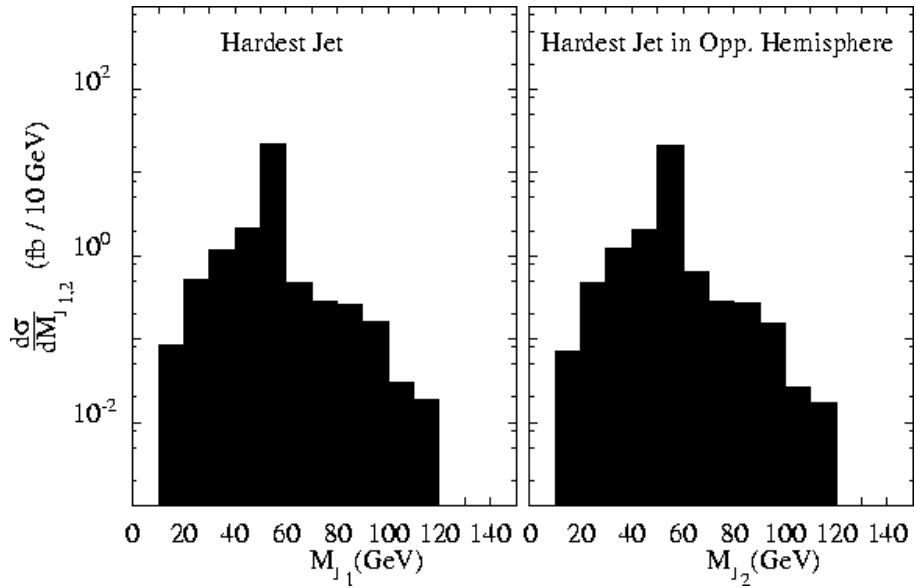}
}
\begin{center}\begin{minipage}{\figurewidth}
\caption{\sl \label{Fig:rpv:mj}
Distribution of invariant mass
reconstructed from (a) the hardest jet and all jets in the same
hemisphere, and (b) the hardest jet in the opposite hemisphere and all
remaining hadronic jets, when all contributions are summed over (signal
only), for a chosen point in the parameter space\cite{Ref:Ghosh3}.
}
\end{minipage}
\end{center}
\end{figure}

In the case of $\lambda''$-couplings,
we can no longer rely on leptons. 
The final states involve
a large number of partons with or without leptons. 
Some of these partons may merge together in jet definitions,
removing the connection between the jet multiplicity 
and the number of initial partons in the final state. 
Nevertheless, it is pointed out that kinematic mass reconstructions
can be used to study the multi-jet events and identify 
those coming from $\sla{B}$ couplings. 
Fig.~\ref{Fig:rpv:mj} shows the distributions of the invariant mass
constructed from the hardest jet and all the other jets in the same
hemisphere, and the same for the hardest jet in the opposite
hemisphere.
The distribution shows clear peaking at $m_{\tilde{\chi}_{1}^{0}}$ 
as well as a sharp cutoff 
at $m_{\tilde{\chi}_{1}^{\pm}} \approx m_{\tilde{\chi}_{2}^{0}}$. 
Thus kinematic distributions can be used quite effectively 
even for the case of the multi-jet events.

%% file: physsusy/detector.tex
\section{Requirements to Machine and Detector}

Most of the studies presented above assumed
an integrated luminosity ranging from $20$ 
to $100~{\rm fb}^{-1}$,
where measurement accuracy is largely limited by
statistical errors.
Recent progress of machine technology, however, allows us
to think about much higher luminosity than
one could hope at the time of the green book
writing\cite{chapter-physsusy:Ref:jlc1}.
In this context, we will examine here how the
detector resolution or the machine energy uncertainty
might affect the SUSY study scenarios we discussed so far. 

Major roles of the detector components are listed below:
\begin{itemize}
\item[VTX] 
Heavy flavor ID. Particularly important, if one wants to
fully reconstruct chargino pair production in the 4-jet
mode in combination with particle ID information.
It is also important for detecting visibly displaced decay
vertices of the LSP as expected in GMSB models.
\item[CDC]
Good end-point energy resolution for leptons.
The central tracker will also be essential to achieving
good energy flow measurement for end-point energy resolution for jets.
Energy loss measurement for GMSB staus.
Some care must be taken for triggering such heavy and therefore
slowly moving stable particles.
\item[CAL]
Good energy flow measurement for end-point energy resolution for jets.
Vertex pointing resolution for the GMSB neutralino.
Low angle coverage to push $\theta_{min}$ as much as possible
down to the beam line.
\item[P-ID]
$K/\pi$ separation might be important to decide the charge
of a $c(\bar{c})$-jet from, for instance, $W$ bosons
produced in the chargino decays:
full reconstruction of the chargino pair production then
becomes possible.
\end{itemize}

Taking into account that the SUSY studies we have discussed above 
are largely based on the precision measurements of the masses
and polarization-dependent cross sections (mixings) 
of the sparticles, we focus, in what follows, our attention 
on effects of the detector and machine performance
on these measurements.

\subsection{End-Point Measurements}

Let us begin with the end-point measurement,
since it is not only one of the most important experimental
techniques but also a typical place where detector
and machine energy resolutions might play some
important role, once the integrated luminosity 
well exceeds $100~{\rm fb}^{-1}$.

\begin{figure}[h]
\centerline{
\epsfxsize=7cm\epsfbox{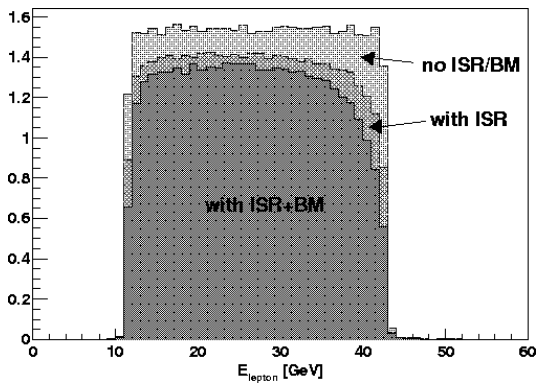}
}
\begin{center}\begin{minipage}{\figurewidth}
\caption[physics:susy:detect:epbmeff]{\sl \label{Fig:epbmeff}
Muon energy distributions for 
$e^+e^- \to \tilde{\mu}_R^+ \tilde{\mu}_R^-$ followed by
$\tilde{\mu}_R^\pm \to \mu^\pm \tilde{\chi}_1^0$
(dotted) with neither ISR nor beam effects (BM),
(dot-dashed) with ISR only,
and (solid) with both ISR and beam effects.
The Monte Carlo data correspond to Fig.~\ref{ELSMU}
and were generated at $\sqrt{s} = 350~{\rm GeV}$
with $m_0 = 70~{\rm GeV}$, $M_2 = 250~{\rm GeV}$, $\mu = 400~{\rm GeV}$,
and $\tan\beta=2$.
}
\end{minipage}\end{center}
\end{figure}

First of all, it should be noted that initial-state 
radiation (ISR) and beamstrahlung do not change
the end-point locations, 
since the reduction of the effective beam energy,
if sizable, makes the higher end-point ($E_+$) lower
and the lower end-point ($E_-$) higher.
Fig.~\ref{Fig:epbmeff} clearly demonstrates this fact.
Consider a two-body decay of a particle with mass $M$
to a lepton (muon in our case) and
a massive daughter with mass $m$.
Ignoring the lepton mass, $M$ and $m$
can be expressed in terms of
the end points, $E_+$ and $E_-$ as
\begin{eqnarray}
\nonumber
\left\{
\begin{array}{lll}
M & = & 2E \frac{\sqrt{E_+ E_-}}{E_+ + E_-} \cr
&& \cr
\frac{m}{M} & = & \sqrt{ \frac{E - (E_+ + E_-)}{E} } ,
\end{array}
\right.
\end{eqnarray}
where $E$ is the beam energy.
This leads, for instance, to the following formula for the relative
error on the parent mass:
\begin{eqnarray}
\nonumber
\sigma_M & = & \left( \frac{M}{2} \right)
\left( \frac{E_+ - E_-}{E_+ + E_-} \right)
\sqrt{
\left( \frac{\sigma_{E_+}}{E_+} \right)^2
+ \left( \frac{\sigma_{E_-}}{E_-} \right)^2
}.
\end{eqnarray}
Since the lepton energy will be measured by the central tracker
that has energy (momentum) resolution that is proportional to the
energy:
$\sigma_E/E = a E$,
The above equation becomes:
\begin{eqnarray}
\nonumber
\sigma_M & = & \left( \frac{M}{2} \right)
\left( \frac{E_+ - E_-}{E_+ + E_-} \right)
a \sqrt{E_+^2 + E_-^2}.
\end{eqnarray}
Using this formula, we can translate 
the current performance goal of the central tracker,
$a \simeq 5 \times 10^{-5}~{\rm GeV}^{-1}$,
to the $\sigma_M$ of about $0.1~{\rm GeV}$
for the example shown in Fig.~\ref{ELSMU} or in Fig.~\ref{Fig:epbmeff}. 
This value should be compared with 
$\sigma_M \simeq 0.7~{\rm GeV}$ expected for $20~{\rm fb}^{-1}$
(see Fig.~\ref{ELSMU}).
Assuming that this statistical error scales as $1/\sqrt{N}$,
we expect $\sigma_M \simeq 0.1~{\rm GeV}$ 
for an integrated luminosity of $1~{\rm ab}^{-1}$,
which is comparable with the detector resolution.
This means that the tracker resolution is good enough
for the measurement, provided that we understand
the detector performance well\footnote{
Note also that if we reach this level of precision,
we may start worrying about the effect of the finite
width of the parent particle.
As a matter of fact, the width of the right-handed
smuon in our example is about $0.7~{\rm GeV}$,
which is indeed comparable with 
the ultimately expected precision.
}.

For the chargino mass determination, we
need to retain the $W$ mass in the end-point formula:
\begin{eqnarray}
\nonumber
M&=&\sqrt{2} E \frac{ \sqrt{ E_+ E_- + m_W^2
+\sqrt{(E_+^2 - m_W^2)(E_-^2 - m_W^2)} 
}
}
{E_+ + E_-},
\end{eqnarray}
from which we obtain
\begin{eqnarray}
\nonumber
\sigma_M^2 & = & \left(\frac{\partial M}{\partial E_+}\right)^2 \sigma_{E_+}^2
+\left(\frac{\partial M}{\partial E_-}\right)^2 \sigma_{E_-}^2.
\end{eqnarray}
with
\begin{eqnarray}
\nonumber
\left(\frac{\partial M}{\partial E_\pm}\right)^2
&=&\left( \frac{M}{E_+ + E_-} \right)
\left[ \left(\frac{E}{M}\right)^2
\frac{E_\mp + E_pm \sqrt{\frac{E_\mp^2 - m_W^2}{E_\pm^2 - m_W^2}}}
{E_+ + E_-} - 1
\right] .
\end{eqnarray}
Notice that,
since we need to rely on the energy flow measurement for jets,
we need to use
\begin{eqnarray}
\nonumber
\sigma_{E_\pm}/E_\pm & = & b / \sqrt{E_\pm}.
\end{eqnarray}
The detector resolution is already important
for the integrated luminosity of $50~{\rm fb}^{-1}$
assumed in Fig.~\ref{Fig_chic1_cmass},
even if we assume currently conceivable best energy
flow resolution: $b \simeq 0.1~{\rm GeV}^{-1/2}$
corresponding to $\sigma_M \simeq 8~{\rm GeV}$.
This means that we need to understand the resolution
function to the accuracy that stands up to the expected
statistical error.

\begin{figure}[h]
\centerline{
\epsfxsize=7.6cm\epsfbox{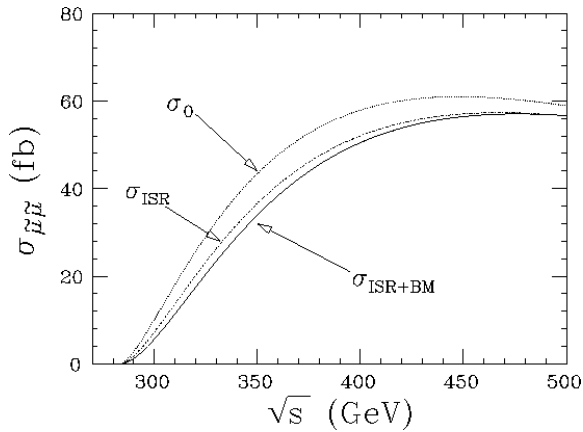}}
\begin{center}\begin{minipage}{\figurewidth}
\caption[physics:susy:detect:sgbmeff]{\sl \label{Fig:sgbmeff}
Production cross sections for
$e^+e^- \to \tilde{\mu}_R^+ \tilde{\mu}_R^-$
(dotted) with neither ISR nor beam effects (BM),
(dot-dashed) with ISR only,
and (solid) with both ISR and beam effects.
The SUSY parameters are the same as with Fig~\ref{Fig:epbmeff}:
with $m_0 = 70~{\rm GeV}$, $M_2 = 250~{\rm GeV}$, $\mu = 400~{\rm GeV}$,
and $\tan\beta=2$.
}
\end{minipage}\end{center}
\end{figure}

On the other hand, the mass determination by means of
threshold scan will be rather insensitive to the detector
resolution and potentially more accurate in particular
for spin 1/2 particles with rapidly rising cross 
sections\cite{MB1}.
This technique might, however, be affected by the natural beam
energy spread and beamstrahlung.
Fig.~\ref{Fig:sgbmeff} shows the effects of ISR 
and the beam effects on the threshold shape of the
$e^+e^- \to \tilde{\mu}_R^+ \tilde{\mu}_R^-$ process.
If we are to carry out threshold scan 
we need to understand the energy distribution
of the collider very well.
It should also be noted that the branching fraction
to the final states to be used for a threshold scan
is not {\it a priori} known, which forces us to make
the overall normalization a free parameter in the fit.
Moreover, the cross sections in particular for the second and
the third-generation sneutrinos are not necessarily large enough
to carry out their threshold scan with a practicable 
luminosity\cite{Ref:Mizukoshi}.

\subsection{Beam Polarization}

\begin{figure}[h]
\centerline{
\epsfxsize=7cm
\epsfbox{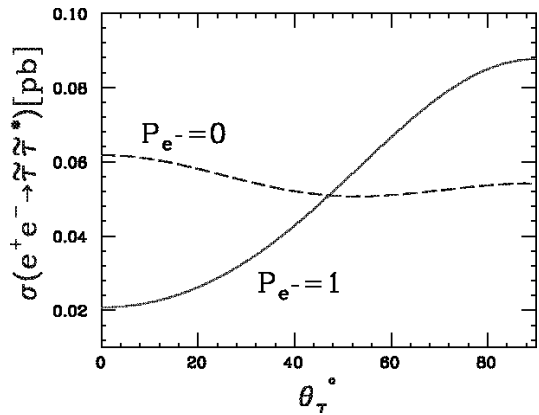}
}
\begin{center}\begin{minipage}{\figurewidth}
\caption[physics:susy:stau:sg]{\sl \label{Fig:physics:susy:stau:sg}
The mixing angle ($\theta_{\tau}$) dependence
of the production cross section of
$\tilde{\tau}_1\tilde{\tau}_1$
for $m_{\tilde{\tau}}=150$ GeV and $\sqrt{s}=500$ GeV.
The cross section is heavily dependent on
$\theta_\tau$ for the right-handed electron beam
($P_{e}=1$).
}
\end{minipage}\end{center}
\end{figure}

It has been stressed many times that the beam polarization
plays a crucial role in sorting out the sparticle mixings.
For instance, let us recall the mixing angle determination
for the stau.
The production cross section 
for the $e^+e^- \to \tilde{\tau}_1^+ \tilde{\tau}_1^-$
process depends on the electron beam polarization,
as well as on the stau mixing angle $\theta_\tau$
as shown in Fig.~\ref{Fig:physics:susy:stau:sg}.
This suggests that, even if we perfectly understand the effective
center of mass energy distribution (beam effects) of the machine, 
there can still be some uncertainty due to
the error in the beam polarization measurements.
The expected error on the mixing angle was estimated
to be $\Delta \sin\theta_\tau / \sin\theta_\tau \simeq 0.065$
(see Section \ref{Chap:susy:Sec:stau:mixing}) 
for an integrated luminosity of $100~{\rm fb}^{-1}$. 
This implies that for $1~{\rm ab}^{-1}$,
the statistical error reaches $2~\%$ level.
The error on the beam polarization must thus be
controlled down to $1~\%$ level or less.
It should also be pointed out that 
the higher the beam polarization,
the easier the mixing angle
determinations through polarization-dependent 
cross section measurements.
In this respect, the positron beam polarization is useful. 
The power of the positron beam polarization is, however,
further enhanced for the SUSY processes with
$t$-channel neutralino exchanges, since the neutralinos,
being of Majorana type, do not respect
the automatic helicity selection rule between the
electron and the positron beams.
As we have seen in the $CP$ violation studies,
transverse beam polarization is also a useful
option to consider.

\subsection{Forward/Backward Detector to Veto Low Angle 
$\protect \mathbold{e^+e^-}$}

For detecting missing $P_T$ signals
of sparticle productions,
it is essential to be able to veto low angle
$e^+/e^-$'s to suppress two-photon backgrounds.
Fig. \ref{physics:susy:stau:pt}
shows the $P_T$ distributions
of the signal ($e^+e^- \to \tilde{\tau}_1^+ \tilde{\tau}_1^-$) 
and the background ($e^+e^-\rightarrow e^+e^-\tau^+\tau^-$), 
for $\theta_{acop}>30^{\circ}$ and $\vert\cos\theta_{jet}\vert<0.8$.
When the minimal veto angle
($\theta^{veto}_e$) for forward and backward
electrons and positrons is 150 mrad,
the background rate is intolerable high
up to $P_T<35$ GeV.
\begin{figure}[h]
\begin{minipage}[t]{7.5cm}
\centerline{
\epsfxsize=7cm
\epsfbox{physsusy/figs/02_pt.epsf}
}
\caption[physics:susy:stau:pt]{\sl \label{physics:susy:stau:pt}
$P_T$ distributions for the signal and
the background $e^+e^-\tau^+\tau^-$ events
at $\sqrt{s}=500$ GeV after the cuts:
$\vert\cos\theta_{jet}\vert<0.8$ and $\theta_{acop}>30^\circ$,
for different minimum veto angles.
$10^{4}$ signal events were generated with
$m_{\tilde{\tau}}=150$ GeV and
$m_{\tilde{\chi}^0_1}=100$ GeV,
corresponding
to $\int L dt= 100$ $fb^{-1}$.
}
\end{minipage}
\hfill
\begin{minipage}[t]{7.5cm}
\centerline{
\epsfxsize=6.8cm
\epsfbox{physsusy/figs/07_ej_pt.epsf}
}
\caption[physics:susy:stau:ej_pt]{\sl \label{physics:susy:stau:ej_pt}
$P_T$ cut dependence of the energy distribution
of the final-state hadrons from the stau decays.
The $P_T$ cut at 35 GeV significantly distorts the
energy distribution near its peak. 
}
\end{minipage}
\end{figure}

If $\theta^{veto}_e = $ 150 mrad
we thus need to require
$P_T>35$ GeV, which distorts the energy distribution
significantly and reduces the reliability of
the mass determination 
in this method (see Fig. \ref{physics:susy:stau:ej_pt}).
On the other hand, for $\theta^{veto}_e$ = 50 mrad,
the cut on $P_T$ can be lowered to about 15 GeV,
which makes the resultant energy distribution
much less affected.
It is thus essential to push the minimum veto angle
down to at least $50$~mrad.

%% file: physsusy/highscale.tex
\section{Tests of SUSY Breaking Mechanisms}

We have seen above that the precision measurements of the masses 
and the mixings in the sfermion and the chargino/neutralino sector 
will certainly allow us to quantitatively test various
supersymmetry relations, thereby firmly establishing
the existence of supersymmetry as a dynamical symmetry 
of particle interactions.
More importantly, the precision we expect for these measurements 
at the JLC is so high that we will be able to
infer the SUSY breaking scale and 
the values of the SUSY breaking parameters at this high scale.
Since these SUSY breaking parameters are presumably determined
by the physics at the high scale,
this will open up a first realistic way to systematically 
study the physics at this high scale experimentally.

Some examples of such studies are already mentioned above,
including tests of universality among scalar masses and
gaugino mass unification expected in the SUGRA models.
These tests can be carried out step by step in the course of the 
SUSY discovery and study scenarios.
We can hope that, eventually, we will be able to
determine all the masses and the mixings
of the sparticle spectrum,
thereby fixing all of the Lagrangian
parameters such as
$M_1,M_2,\mu,\tan \beta$ at the {\it electroweak (EW) scale}.
In this section, we review what and to what extent
we can learn then from these precision
measurements of the masses and the mixings.

\begin{figure}[htb]
\centerline{
\epsfysize=9cm\epsfbox{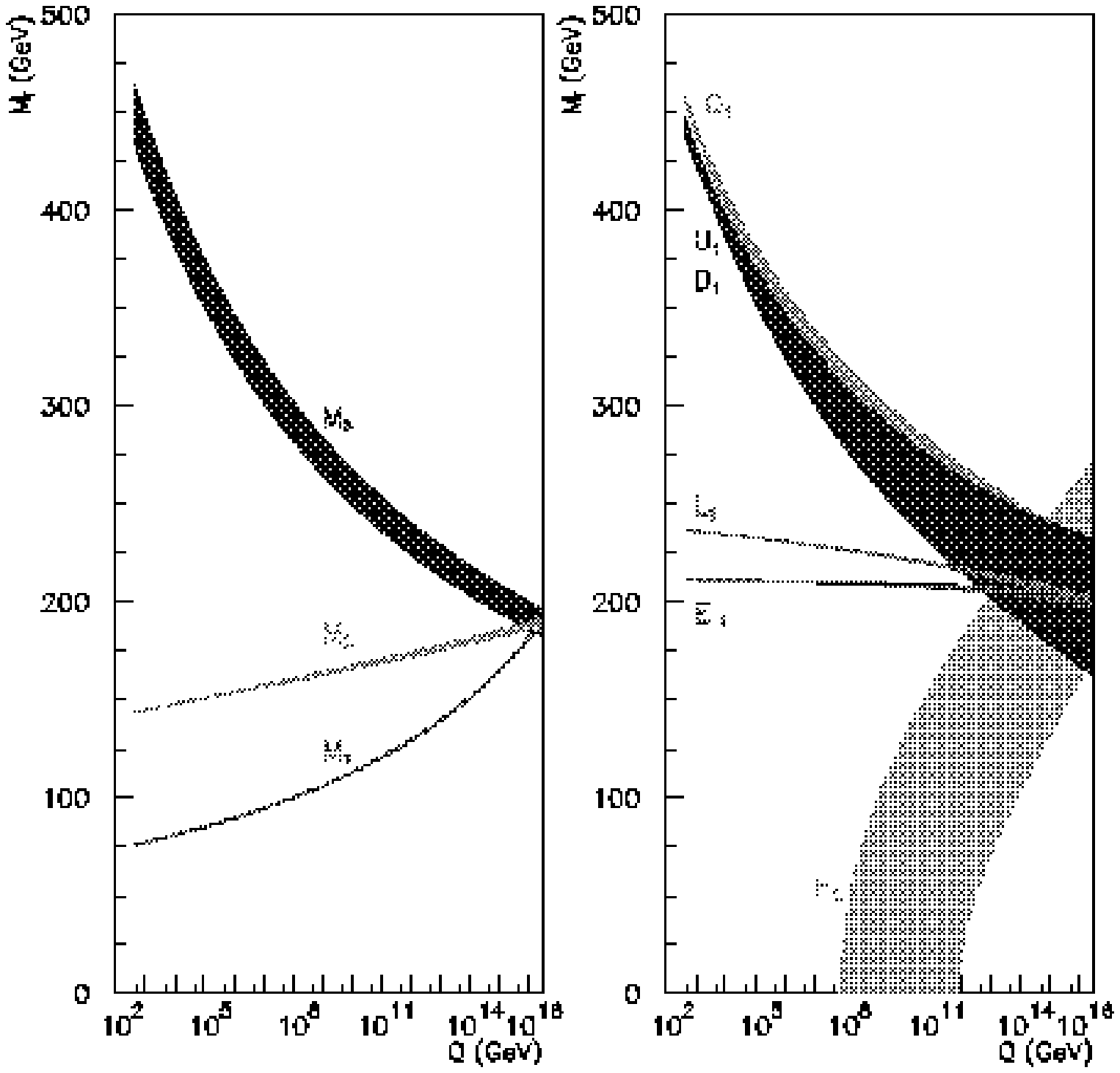}
\epsfysize=9cm\epsfbox{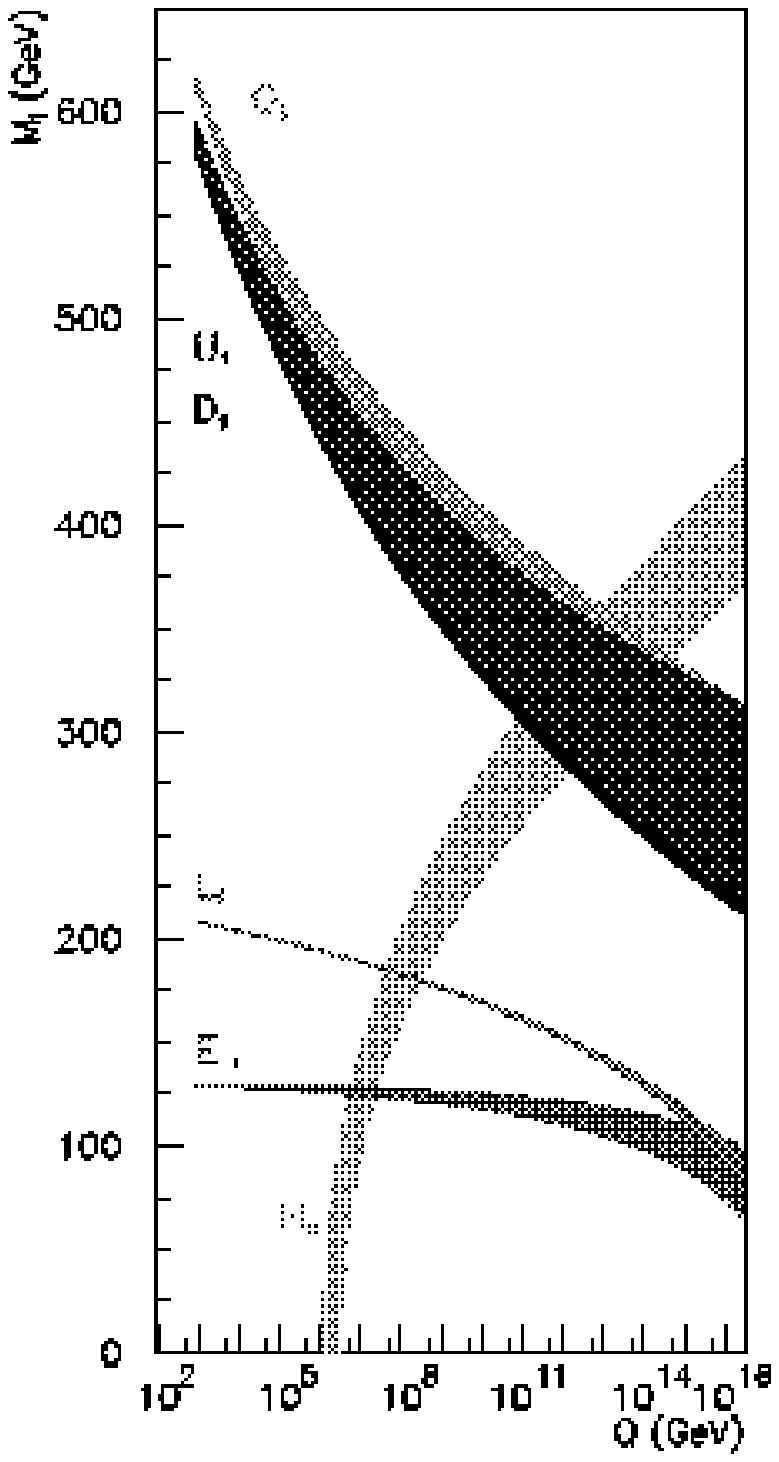}
}
\begin{center}\begin{minipage}[t]{\figurewidth}
\caption{\sl \label{F:rgplen:14}
Bottom up approach of the determination of the sparticle mass
parameters for mSUGRA and GMSB\protect\cite{PZB}. 
Values of the model parameters as given there.
}
\end{minipage}\end{center}
\end{figure}

One can use these accurately determined Lagrangian parameters 
at the EW scale to fit their values at the high scale.
It has been shown~\cite{MB1} in this approach, 
that the projected accurate measurements of the 
various sparticle masses through an energy scan
for each sparticle's threshold, using
ten energy points with $10$\ ${\rm fb}^{-1}$ each,
allows one to determine
$M_1, M_2,m_0, \mu$, and $\tan\beta$ at the high scale
to an accuracy better than $1 \%$\footnote{
It has recently been pointed out\cite{Ref:Mizukoshi} 
that the quoted accuracy in Ref.\cite{MB1} might
be too optimistic in particular for the third generation
sneutrino, though
to what extent this will affect the 
results including those in Tables~\ref{T:rgplen:3},
\ref{T:rgplen:4}, and Fig.~\ref{F:rgplen:14}
is still an open question.
}. 
The accuracy is much worse, however, for
higher values of $\tan \beta$.
The expected accuracy for the trilinear term is rather poor 
as shown in Tables \ref{T:rgplen:3} 
and \ref{T:rgplen:4} taken from Ref.~\cite{MB1},
which is due to 
the fact that most of 
the physical observables are rather insensitive 
to the parameters $A_k$.
\begin{table}
\noindent\begin{minipage}{0.5in}
$\mbox{ }$
\end{minipage}
\noindent\begin{minipage}{2in}
\caption{\sl Reconstruction of SUGRA parameters assuming universal masses.
\label{T:rgplen:3}}
\centering
\begin{tabular}{ccc}
$\mbox{ }$ & True value & Error \\
\hline
$m_0$ & 100 & 0.09 \\
$m_{1/2}$ & 200 & 0.10 \\
$A_0$ & 0 & 6.3 \\
tan$\beta$ & 3 & 0.02 \\
$\mbox{ }$ & $\mbox{ }$ & $\mbox{ }$ \\
\end{tabular}
\end{minipage}
\hfill
\noindent\begin{minipage}{2in}
\caption{\sl Reconstruction of SUGRA parameters with nonuniversal gaugino masses.}
\label{T:rgplen:4}
\centering
\begin{tabular}{ccc}
$\mbox{ }$ & True value & Error \\
\hline
$m_0$ & 100 & 0.09 \\
$M_1$ & 200 & 0.20 \\
$M_2$ & 200 & 0.20 \\
$A_0$ & 0 & 10.3 \\
tan$\beta$ & 3 & 0.04 \\
\end{tabular}
\end{minipage}
\noindent\begin{minipage}{0.5in}
$\mbox{ }$
\end{minipage}
\end{table}
In this approach model selection will be made
on the basis of goodness of the fit.

There is, however, a totally different approach to
test the SUSY breaking mechanisms,
the bottom up approach\cite{PZB},
where one starts with these Lagrangian parameters extracted 
at the EW scale and extrapolates them to the high scale
using the renormalization group evolution (RGE).
As explained in Section \ref{Chap:susy:Sec:scenarios},
different SUSY breaking mechanisms predict
different relations among these parameters 
at the high scale.
In the bottom up approach,
one can test these relations 
{\it directly} by reconstructing them
from their low energy values using the RGE.
In Ref.\cite{PZB},
{\it experimental} values of the various sparticle masses 
are generated in a given scenario (mSUGRA, GMSB etc.), 
by starting from the universal parameters at the high scale 
appropriate for the model under consideration
and evolving them down to the EW scale. 
These masses are then endowed with experimental errors 
expected to be reached eventually in the combined analyses 
of data from LHC and a TeV scale linear collider 
with an integrated luminosity of $1 {\rm ab}^{-1}$. 
These {\it measured} values are then evolved 
back to the high scale. 
The left two plots in Fig.~\ref{F:rgplen:14}
show results of such an exercise for the gaugino 
and sfermion masses, respectively, for the mSUGRA case,
while the one on the right for sfermion masses in GMSB. 
The width of the bands indicates $95 \%$ C.L. limits.

We can see that with the projected 
accuracies of measurements, 
the unification of the gaugino masses will be indeed demonstrated 
very clearly in the mSUGRA case. 
The extrapolation errors for the evolution of the slepton masses  
are also small, since only the EW gauge couplings contribute to it,
and allows one to test the unification of the slepton masses 
at the high scale.
This is in contrast to the squark and higgs masses,
for which the extrapolation errors are rather large.
In the case of the Higgs mass parameters this insensitivity to 
the common scalar mass $m_0$ is due to an accidental cancellation 
between different contributions in the loop corrections 
to these masses which in turn control the RG evolution.
In the case of the squarks the extrapolation errors
are caused by the stronger dependence of the radiative corrections 
on the common gaugino mass, 
due to the strong interactions of the squarks. 
As a result of these the small error at the EW scale
expands rapidly, when extrapolated to the unification scale.
It should also be noted that
the trilinear $A$ coupling for the top shows a pseudo 
fixed point behavior, 
which also makes the EW scale value 
insensitive to $m_0$. 
If the universal gaugino mass $m_{1/2}$ is larger than the $m_0$ 
then this pseudo fixed point behavior increases the errors 
in the determination of the third generation squark mass 
at the EW scale.  
This picture shows us clearly the extent to which the unification 
at high scale can  be tested.
If we compare this with the results of Tables~\ref{T:rgplen:3} 
and \ref{T:rgplen:4}, we see that with the bottom up approach 
we have a much clearer representation of the situation.
It should be stressed that 
the $95 \%$ C.L. bands on the squark and the higgs mass parameters 
become much wider if one only assumes the accuracies expected 
to be reached at the LHC\cite{BWS}. 
The linear collider's precision measurements are indispensable to
get a clearer picture of the SUSY breaking mechanism at
the high scale. 

The right-most plot in Fig.~\ref{F:rgplen:14} 
shows the results of a similar 
exercise but for the GMSB model, 
where with the assumed values of the model parameters, 
one would need to have a $1.5$ TeV linear collider 
to access the full sparticle spectrum. 
In this case the doublet slepton mass and the Higgs mass 
parameter are expected to unify at the messenger scale, 
which the {\it data} show quite clearly.
It is remarkable that the extrapolation
of the reconstructed slepton and squark masses 
to the high scale is stable enough to reveal the
entirely different unification patterns expected 
in this case as opposed to the mSUGRA case.
The bottom up approach of testing the structure 
of the SUSY breaking sector definitely requires
the high accuracy that can be achieved only with
the linear collider\cite{BWS}.

All of these discussions assume that most of the sparticle spectrum
are accessible jointly by the LHC and a TeV scale
linear collider. 
If we are unlucky and the squarks are super-heavy\footnote{
In the focus point SUSY scenarios the entire scalar sector might be 
beyond a few TeV\cite{fengbagger,moroi}.
}, 
then perhaps the only clue to their existence 
can be obtained through the analogue of precision
measurements of the oblique correction to 
the SM parameters at the $Z$ pole. 
These super-oblique corrections~\cite{MIHO3} modify the tree-level 
supersymmetry relations between various couplings as already mentioned
in Section \ref{Chap:susy:susytest}.
These modifications arise if there is a large mass splitting between 
the sleptons and the squarks. 
The expected radiative corrections imply
\begin{eqnarray}
{{\delta g_Y} \over g_Y}  & \simeq &  {{11 g_Y^2} \over {48 \pi^2}} 
ln \left({m_{\tilde{q}}} \over {m_{\tilde l}}\right),
\end{eqnarray}
which amounts to about $0.7 \%$,
if the mass splitting is a factor of 10.
We have shown in the previous sections that it might well 
be possible to make a measurement with such accuracy at the JLC.

%% file: phystop/main.tex
\newcommand \bra[1]{\left< {#1} \,\right\vert}
\newcommand \ket[1]{\left\vert\, {#1} \, \right>}
\newcommand \braket[2]{\hbox{$\left< {#1} \,\vrule\, {#2} \right>$}}
\newcommand{\bea}{\begin{eqnarray}}
\newcommand{\eea}{\end{eqnarray}}
\newcommand{\simgt}{\hbox{ \raise3pt\hbox to 0pt{$>$}\raise-3pt\hbox{$\sim$} }}
\newcommand{\simlt}{\hbox{ \raise3pt\hbox to 0pt{$<$}\raise-3pt\hbox{$\sim$} }}
\newcommand \vc[1]{{\bf {#1}}}
\newcommand{\clfn}{\setcounter{footnote}{0}}

\chapter{Top Quark Physics}
\label{chapter-phystop}

\section{Introduction}

Among all the fermions included in the Standard Model (SM),
the top quark plays a very unique role.
The mass of the top quark is by far the largest and approximates the 
electroweak (EW) symmetry breaking scale.
The recent reported value from the CDF and D0 collaborations reads
\bea
m_t = 
174.3 \pm 3.2 (stat) \pm 4.0 (syst) ~\mbox{GeV}  
~~~~~
\mbox{(CDF+D0 combined\cite{moriond2000})}
.
\eea
In fact the top quark is the heaviest of all the 
elementary particles discovered up to now.
It means that in the SM Lagrangian the top quark mass
term breaks the $SU(2)_L \times U(1)_Y$ symmetry maximally.
This fact suggests that the top quark couples strongly to the
physics that breaks the EW symmetry.
It is therefore important to investigate properties of the
top quark in detail, for the purpose of probing the 
symmetry breaking physics as well as
to gain deeper understanding of the origin of the flavor structure.
The standard procedures for investigating top quark properties
are: measurements of fundamental quantities such as its mass
and decay width; 
detailed examinations of various interactions of top quark 
to see if there are signs of new physics.

Studies of various properties of the top quark 
have been started at Tevatron.
For example, the polarization of $W$ bosons from the decay
of top quarks has been measured.
More detailed properties will be investigated further
in future experiments at Tevatron Run II,
at LHC and at JLC.
In particular, experiments at 
JLC will be able to uncover detailed
properties through 
precision measurements
of various top quark interactions.
For example,
another fundamental parameter of the top quark, its decay width
$\Gamma_t$, can be measured accurately.
Within the SM the top quark decays almost 100\% to
the $b$ quark and $W$.
The width can be predicted as a function of $m_t$,
$\alpha_s$, $M_W$, etc.
Already a fairly precise theoretical prediction
at the level of a few percent accuracy is available \cite{jk1}.
Here, we only note that $\Gamma_t \simeq 1.5$~GeV for the
above top quark mass range.
This value deviates with simple extensions of the SM:
e.g.\ $\Gamma_t \propto |V_{tb}|^2$ 
can be smaller if there is a fourth generation, whereas it
will be greater if there are additional decay modes such as $t \to b H^+$
or $t \to \tilde{t} \tilde{\chi}$.
Thus, it is important to measure $\Gamma_t$ precisely.
An important property of the top quark, besides its heaviness,  
is that it decays so quickly that
no top-hadrons will be formed.
Therefore all the spin information of the top quark
will be transferred to its decay daughters in its decay 
processes \cite{kuehn}, and
the energy-angular distributions of the decay products are
calculable as purely partonic processes.
In fact we may take full advantage of the spin information 
in studying the
top quark properties through its decay processes \cite{peskin}.

In this chapter we review the studies of 
top quark physics at JLC.
We will first focus on the prospects in the $t\bar{t}$ threshold
region in three steps:
In Subsection \ref{mandg} we clarify the physics motivation and
goals;
In Subsection \ref{topmass} we discuss the precision measurements
of the top quark mass;
In Subsection \ref{dyn} we discuss the relevant observables;
Subsection \ref{topcoupling} is devoted to the sensitivity
studies on various top quark couplings.
In the rest, we will discuss the physics in the open top
region.
In Section \ref{dilepton} we discuss a measurement of $m_t$
from the dilepton events.
In Section \ref{yukawa} measurement of the
top-Yukawa coupling is studied.
In Section \ref{ttH} anomalous couplings of the top quark with the
Higgs boson are studied.
In Section \ref{anom} anomaly in the Standard-Model
top quark interactions is studied through the decay processes of the
top quark.
In Section \ref{CPv-opentop} we study how to probe $CP$ violation in
the open top region, using both $e^+e^- \to t\bar{t}$ and
$\gamma \gamma \to t\bar{t}$ processes.
On the other hand, Section \ref{R-v} examines the possibility to
observe $R$-parity violating SUSY interactions using the top quark.
We summarize our studies in Section \ref{phystop-summary}.

\section{Top Quark Threshold Region}

An experiment in the top quark threshold region ($\sqrt{s} \approx 2 m_t$)
is considered as a serious candidate for the first stage operation
of JLC.
In this section we review the subjects that can be covered in experiments
in the $t\bar{t}$ threshold.
We will be able to extract rich physics and test properties of the
top quark in detail.

\subsection{Physics Motivations and Goals}
\label{mandg}

The major three goals that should be achieved in the experiment
in the $t\bar{t}$ threshold region
will be as follows:
\begin{itemize}
\item[1)]
A precision measurement of the top quark mass.
\item[2)]
To elucidate dynamics involved in formations and decays 
of (remnants of) the toponium resonances.
\item[3)]
Tests of the various top quark interactions:
search for anomalies in the
$t\bar{t}\gamma$, $t\bar{t}Z$, $tbW$, $ttg$ and $ttH$ couplings
and possibly for other non-SM interactions.
\end{itemize}
Here, we briefly explain the physics motivation 
for each of these goals.

\subsubsection*{{ 1) Top Mass Determination}}
\vspace{3mm}

The top quark mass $m_t$ is one of the fundamental parameters of the 
SM.
It is conceivable that, in the future,
to know its precise value will be crucial
to achieve a high predictive power in precision physics.
In fact some important observables depend on $m_t$ sensitively through
radiative corrections,
when $m_t$ acts as a non-decoupling parameter.
Then it is required that $m_t$ should be known more precisely than
other parameters of the theory.
For example, suppose that the Higgs boson is discovered in the future, and
suppose that 
we want to test predictions of its mass by the Minimal Supersymmetric Model
within 50~MeV accuracy.
Then we need to know the top quark mass with a similar accuracy due to
the large dependence of the radiative correction on the top quark mass:
$\delta m_H^2 \sim G_F m_t^4 \ln ( m_{\tilde{t}_1} m_{\tilde{t}_2}/m_t^2 )$.

\subsubsection*{2) Testing Dynamics of $t\bar{t}$ Resonances}
\vspace{3mm}

On the one hand,
we will be able to observe intriguing phenomena of the
(remnants of) $t\bar{t}$ resonances
which are quite different
from the phenomenology of charmonium and bottomonium.
This is because: (1) The top quarks decay via the EW interaction very
quickly, and the decay daughters carry the information on the dynamics
of the top quarks inside the resonances.
(2) The interplays of the QCD and EW interactions induce unique 
phenomena.
On the other hand, due to the large
mass and the large decay width of the top quark,
high precision theoretical predictions
based on the first principles are available.
Therefore we will be able to make detailed tests of
the dynamics of the boundstate formation and decays.
This aspect is important since predictability of the theory on the
quarkonium physics has improved dramatically recently.
For example, so far the quarkonium wave functions calculated
in perturbative QCD have never been tested experimentally,
and this will be possible for the $t\bar{t}$ resonances.

\subsubsection*{3) Examinations of Various Top Quark Interactions}
\vspace{3mm}

Studies of various top quark interactions 
will be pursued both in the open top region,
where we can explore more deeply into 
the structures of many of the interactions,
and in the $t\bar{t}$ threshold region.
There are specific advantages in each case.
The advantages in the $t\bar{t}$ threshold region are as follows.
First, to study the decay properties of the top quark, the threshold region
will be optimal.
This is because:
the top quark can be polarized maximally 
by polarizing the electron beam; 
we are guaranteed to be (almost) in the rest frame of the top quark
without need to reconstruct its momentum;
we do not gain resolving power by raising the c.m.\ energy.
In particular, the top quark decay width can be measured
most precisely in the threshold region.
Secondly, we can test not only the EW interactions of the top quark
but also the QCD interactions.
Since the threshold cross section includes an infinite number of
$ttg$ couplings at leading order, it has a high sensitivity also to
the anomalous $ttg$ couplings.

\subsection{Top Quark Mass}
\label{topmass}

\subsubsection*{\underline{How to Determine the Mass}}

\begin{figure}
\centerline{\epsfxsize=7.cm \epsffile{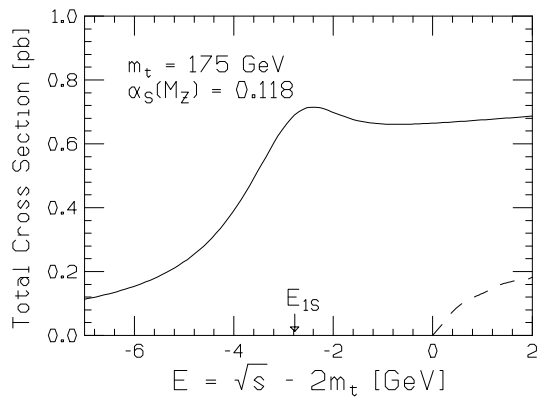}}
\begin{center}
\begin{minipage}[t]{\figurewidth}
  \caption[]{\sl
        \label{totcs}The total cross section vs.\ energy, 
$E = \sqrt{s}-2m_t$.
The solid curve is calculated from the Green function.
The dashed curve shows the tree-level total cross section for
a stable top quark.
}
\end{minipage}
\end{center}
\end{figure}

The top quark mass will be determined to high accuracy from
the shape of the $t\bar{t}$ total production cross section 
in the threshold region \cite{exp}.
If the top quark were stable, the cross section would 
show distinct resonance peaks in this region.
Due to the large decay width of the top quark, $\Gamma_t \simeq 1.5$~GeV,
however, these peaks will be smeared out.
The resonances merge with one another, leading to a broad enhancement
of the cross section over the threshold region; see Fig.~\ref{totcs}.
Still, the location of the sharp rise of the cross section is determined
mainly from the mass of the lowest-lying ($1S$) $t\bar{t}$ resonance.
Since the mass of the resonance can be
calculated as a function of the top quark mass and $\alpha_s$
from perturbative QCD,
we will be able to extract
the top quark mass from the measurement of the $1S$ peak position
of the cross section.

In the beginning of the studies of the threshold cross section, 
people planned to extract the pole-mass of the top quark
from the cross section.
This is because the position of the ``threshold'' is determined as
twice the top quark pole-mass, and it was considered most natural to relate
this mass parameter directly to the threshold cross section or the resonance
spectrum.
On the other hand, existence of the pole-mass of a quark contradicts
the confinement picture of QCD and this mass
is most likely to be ill-defined beyond perturbation theory;
even within perturbative QCD we face a problem in any attempt to define
this quantity to ${\cal O}(\Lambda_{QCD}) = {\cal O}$(300~MeV)
or better.
Recently there has been significant progress in our 
theoretical understanding
of heavy quarkonia such as $\Upsilon$'s and (remnant of) toponium
resonances.
Developments in technologies of higher-order calculations and the
subsequent discovery of renormalon cancellation enabled extractions
of the top quark mass  with high accuracy in future experiments.
It became clear that we have to relate the threshold cross section
to a short-distance mass, such as the $\overline{\rm MS}$ mass,
in order to extract a precise and physically meaningful mass parameter.

The upshot of the theoretical studies up to now is as follows.
The series expansion of the mass of (remnant of) the $1S$ $t\bar{t}$
resonance
in $\alpha_s$ becomes much more convergent when we express the
series in terms of the $\overline{\rm MS}$ mass of the top quark
instead of the pole-mass.
At the present situation, theoretical uncertainty of the 
$\overline{\rm MS}$ mass 
$\overline{m}_t \equiv m_t^{\overline{\rm MS}}(m_t^{\overline{\rm MS}})$
is about 150~MeV when we determine it from the threshold scan.
Also, concerning future prospects,
there seems to be a good chance to reduce the theoretical uncertainty to
about 50~MeV by the time of the JLC operation.

\subsubsection*{\underline{Simulation Studies on Expected Precision}}

There have been a number of simulation studies 
which incorporate realistic
experimental conditions expected at JLC.
Here, we quote the result of the study \cite{exp} on the determination
of the peak position of the $t\bar{t}$ production cross section, 
from which the $\overline{\rm MS}$
mass of the top quark
$\overline{m}_t$
can be extracted.
The effects of the beam energy spread and the beamstrahlung
as well as those of the
initial state radiation are important when studying the threshold
cross section.
In particular the beam energy spread smears out the 
$1S$ peak of the cross section.
In order to determine the peak position of the original
cross section,
we convolute the original cross section with the radiator function and
with
the beam energy spectrum, and then we fit it to the measured cross section.
Selection of top quark events 
can be performed with high efficiency and 
good background suppression.
(More details will be explained in 
Subsection~\ref{topcoupling}.)
Taking these into account, a simulation study of
the energy scan of the cross section was
performed.
Using 11 energy points with 1~fb$^{-1}$ each, we can determine
the peak position with a statistical error $\approx 200$~MeV,
which translates to a statistical error on the top quark 
$\overline{\rm MS}$ mass of
$\Delta \overline{m}_t \approx 100$~MeV.
\medbreak

\subsection{Observables}
\label{dyn}

In this subsection we discuss 
the observables which contain interesting information
close to the top quark threshold.
Here, we restrict ourselves to qualitative arguments.
More quantitative analyses will be given in the next subsection.

\subsubsection*{\underline{A) Production Process of Top Quarks}}

\subsubsection*{\underline{Total Cross Section}}

The first observable we measure in the $t\bar{t}$ threshold region
will be the total cross section.
The total cross section can be written as
\cite{fk,sp}
\bea
\sigma_{\rm tot} (e^+e^- \to t\bar{t})
\propto
- {\rm Im}
\sum_n
\frac{|\psi_n(\vc{0})|^2}{E-E_n+i\Gamma_t} .
\label{totcs-f}
\eea
One sees that the energy dependence of the total
cross section is determined by the resonance spectra.
Due to the large width $\Gamma_t$
of the top quark, however, distinct resonance peaks are smeared out.
The resonances merge with one another, leading to a
broad enhancement of the cross section over the threshold region
as seen in Fig.~\ref{totcs}.
(We will show explicitly the resonance spectra below.)
In the same figure, the tree level cross section is also shown as a dashed
curve.
Despite the disappearance of each resonance peak, one sees that 
the cross section is indeed largely enhanced by the QCD interaction,
and that inclusion of
the QCD binding effect is mandatory for a proper account of the
cross section in the threshold region.

\subsubsection*{\underline{Top Momentum Distribution}}

\begin{figure}
\begin{center}
  \begin{minipage}{7cm}\centering
    \hspace*{-0.5cm}
    \epsfysize=5.5cm
    \epsfxsize=7cm\epsffile{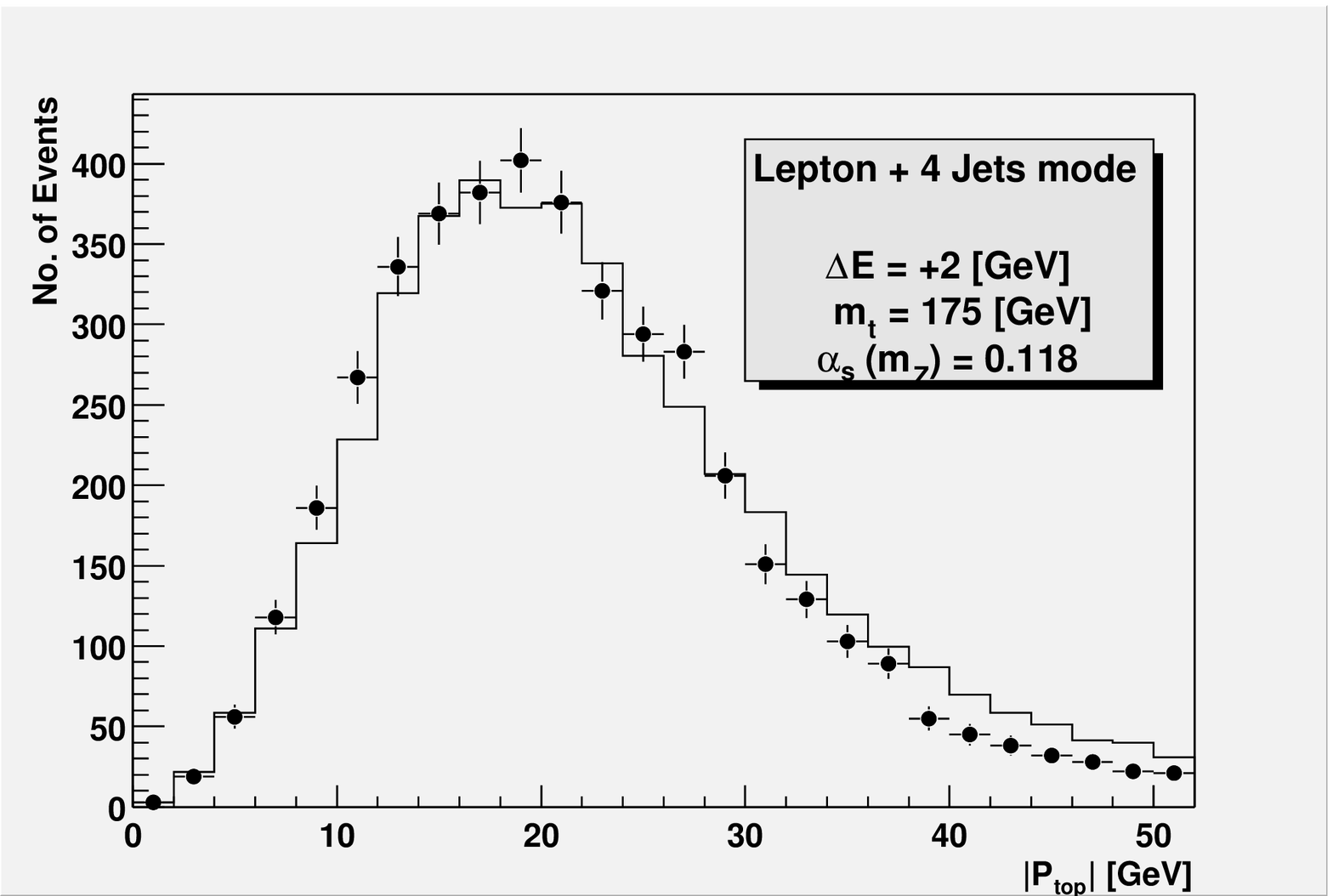}
  \end{minipage}
  \begin{minipage}{6.5cm}\centering
    \vspace{7mm}
    \hspace*{+0.5cm}
    \epsfxsize=6.5cm\epsffile{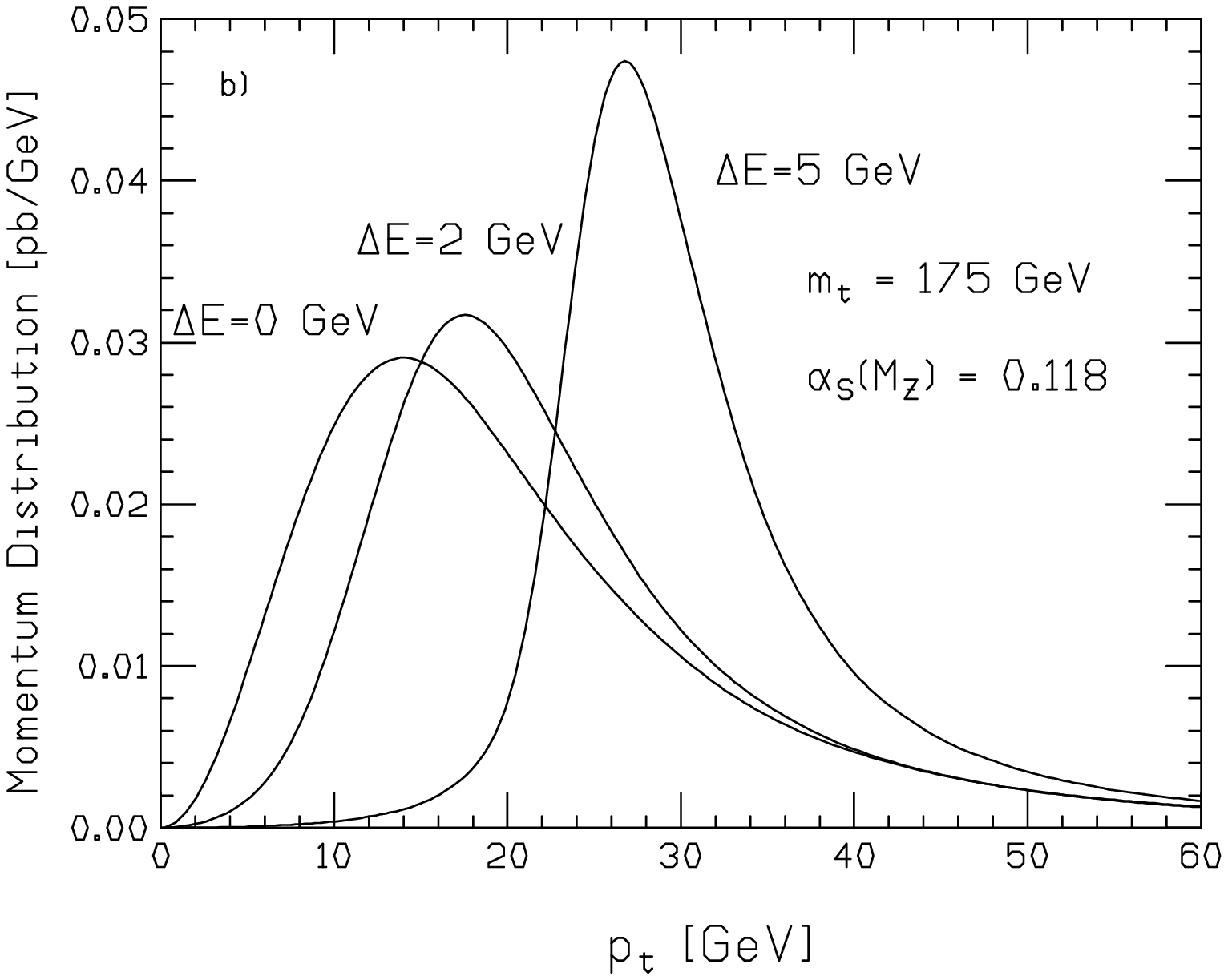}
  \end{minipage}
\\
\end{center}\vspace{-2mm}
\begin{center}\begin{minipage}{\figurewidth}
  \caption[]{\sl
        \label{momdist}
(a) Reconstructed momentum distribution (solid circles) for the 
lepton-plus-4-jet mode, compared with the generated distribution (histogram).
The Monte Carlo events were generated with 
$\alpha_s(M_Z) = 0.12$ and $m_t = 150$~GeV \cite{exp}.
(b) Top-quark momentum distributions $d\sigma/d|\vc{p}_t|$
for various c.m.\ energies measured from
the lowest lying resonance, $\Delta E = \sqrt{s} - M_{1S}$,
taking $\alpha_s(M_Z) = 0.118$ and $m_t = 175$~GeV.
}
\end{minipage}\end{center}
\end{figure}

Next we consider the top-quark momentum ($|\vc{p}_t|$) distribution
near $t\bar{t}$ threshold \cite{sfhmn,jkt}.
It has been shown that experimentally it will be possible to reconstruct
the top-quark momentum $\vc{p}_t$ from its decay products with
reasonable resolution and detection efficiency.
Fig.~\ref{momdist}(a) 
shows a comparison of reconstructed top momenta (solid circles)
with that of generated ones (histogram), 
where the events are generated by a Monte Carlo generator
and are reconstructed after going through detector simulators
and selection cuts; see Subsection \ref{topcoupling} for
details.
The figure demonstrates that the agreement is fairly good.

Theoretically, the top-quark momentum distribution is given by
\bea
\frac{d\sigma}{d|\vc{p}_t|} &\propto&
\biggl| \sum_n
\frac{\phi_n (\vc{p}_t)\psi_n^* (\vc{0})}{E-E_n+i\Gamma_t}
\biggl|^2  + \mbox{(sub-leading)} .
\label{msgf}
\eea
The $|\vc{p}_t|$-distribution is thus governed by the momentum-space
wave functions of the resonances. 
By measuring the momentum distribution, essentially we measure 
(a superposition of) the wave functions of the toponium resonances.
Shown in Fig.~\ref{momdist}(b) are the top momentum distributions for various
energies.
One may also vary the magnitude of $\alpha_s$ and confirm that the
distribution is indeed sensitive to the resonance wave functions
\cite{sfhmn,jkt}.
Hence, the momentum distribution provides information independent of that
from the total cross section.

\subsubsection*{\underline{Forward-Backward Asymmetry}}

Another observable that can be measured experimentally
is the forward-backward asymmetry of the top quark \cite{ms}.
Generally in a fermion pair production process, a forward-backward asymmetric
distribution originates from an interference of
the vector and axial-vector $f\bar{f}$ production vertices at tree level
of electroweak interaction.
One can show from the spin-parity
argument that in the threshold region the $t\bar{t}$ vector vertex 
creates S-wave resonance states, while the 
$t\bar{t}Z$ axial-vector vertex creates P-wave states.
Therefore, by observing the forward-backward asymmetry of the top quark,
we observe an interference of the S-wave and P-wave states.

\begin{figure}
\centerline{    \epsfxsize=7cm\epsffile{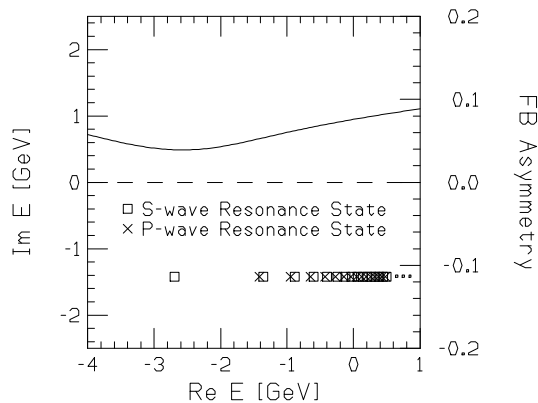}}
\begin{center}\begin{minipage}{\figurewidth}{
  \caption[]{\sl
        \label{fbasym}The positions of poles of the S-wave and P-wave
states on the complex energy plane, together with the forward-backward
asymmetry as a function of the energy, taking
$\alpha_s(M_Z) = 0.118$ and $m_t = 175$~GeV.
The right-axis is for the forward-backward asymmetry.
}
}
\end{minipage}\end{center}
\end{figure}

In general, S-wave resonance states and P-wave resonance states have
different energy spectra.
So if the c.m.\ energy is fixed at some resonance in 
either one of the spectra, there would be no
contribution from the other.
However, the widths of resonances are large for the toponium
in comparison to their level splittings,
which permit sizable interferences of the $S$-wave and $P$-wave 
states.
Fig.~\ref{fbasym} 
shows the pole position $E_n \! - \! i\Gamma_t$ of these states
on the complex energy plane.
One sees that the widths of the resonances are comparable to the
mass difference between the lowest lying S-wave and P-wave states, and
exceeds by far the level spacings between higher S-wave and P-wave states.
This gives rise to a forward-backward 
asymmetry even below threshold, and provides
information on the resonance level structure which is concealed in the total
cross section.
Shown on the same figure is the forward-backward asymmetry as a function
of the energy.
It is seen that the asymmetry takes its minimum value
at around the lowest lying S-wave state, where the interference is
smallest, and
increases up to $\sim 10$\% with energy as the resonance spectra appear
closer to each other.
One may also confirm that essentially the forward-backward asymmetry 
measures the degree of overlap of the S-wave and P-wave states
by varying the coupling constant $\alpha_s$ or the top quark
decay width $\Gamma_t$.

\subsubsection*{\underline{B) 
Decay of Top Quarks and Final-State Interactions}}

We turn to decay processes of the top quark near threshold.
The top quarks produced via $e^+e^- \to t\bar{t}$ in the
threshold region will be highly polarized \cite{kuehn}.
Therefore, the threshold region can be an ideal place
for studying properties of the top quark through
decay processes, using the highly polarized
top quark sample.

\subsubsection*{\underline{Decay of Free Polarized Top Quarks}}

Detailed studies of the decay of {\it free} polarized top quarks have
already been available including the full ${\cal O}(\alpha_s)$ corrections
\cite{r44,r27,r28}.
The best example is the energy-angular distribution of charged
leptons $l^+$ in the semi-leptonic decay of the top quark.
At leading order, the $l^+$ distribution
has a form where the energy and angular dependences are factorized
\cite{langdist,r43}:
\bea
\frac{d\Gamma_{t \to bl^+\nu}
(\vc{S})}
{dE_l d\Omega_l} 
= h(E_l) \, ( 1 + |\vc{S}| \cos \theta_l ) ~~ + ~~
[ \, {\cal O}(\alpha_s)~\mbox{correction} \, ] .
\label{leneangular}
\eea
Here $E_l$, $\Omega_l$, and $\theta_l$
denote, respectively, the $l^+$ energy, the solid angle of $l^+$, 
and the angle between the $l^+$ direction and the 
top polarization vector $\vc{S}$, all of which 
are defined in the top-quark rest frame.
Hence, we may measure the top-quark polarization 
with maximal sensitivity using the $l^+$ angular
distribution.

\subsubsection*{\underline{Effects of Final-State Interactions}}

Close to threshold, the above precise
analyses of the free top-quark decays do not apply directly
because of the existence of corrections unique to this region.
Namely, these are 
the final-state interactions due to gluon exchange between $t$
and $\bar{b}$ ($\bar{t}$ and $b$) or between $b$ and $\bar{b}$.
The size of the corrections is at the 10\% level 
in the threshold region, hence it is 
necessary to incorporate their
effects in precision studies of top-quark production and decay near
threshold.

Here, we present the formula for the charged 
lepton energy-angular distribution
in the decay of top quarks that are produced via $e^+e^- \to t\bar{t}$
near threshold.
First, without including the final-state interactions,
the differential distribution of $t$ and $l^+$ has a form where the
production and decay processes of the top quark
are factorized:
\bea
\frac{d\sigma(e^+e^- \! \to t\bar{t} \to bl^+\nu\bar{b}W^-)}
{d^3\vc{p}_t dE_l d\Omega_l}
=
\frac{d\sigma(e^+e^- \! \to t\bar{t})}
{d^3\vc{p}_t} \times
\frac{1}{\Gamma_t} \,
\frac{d\Gamma_{t \to bl^+\nu}
(\vc{S}) }
{dE_l d\Omega_l} .
\label{born}
\eea
Namely, the cross section is given as a product of the production
cross section of unpolarized top quarks and the differential decay
distribution of $l^+$ from polarized top quarks.
The above formula holds even including all
${\cal O}(\alpha_s)$ corrections other than the final-state
interactions.

Including the final-state interactions,
the factorization of production and decay processes is destroyed.
The formula including the full ${\cal O}(\alpha_s)$ corrections
is given by \cite{ps}
\bea
\frac{d\sigma(e^+e^- \! \to t\bar{t} \to bl^+\nu\bar{b}W^-)}
{d^3\vc{p}_t dE_l d\Omega_l}
=
\frac{d\sigma(e^+e^- \! \to t\bar{t})}
{d^3\vc{p}_t} \times
( 1 + \delta_0 + \delta_1 \cos \theta_{te} )
\nonumber \\
\times
\frac{1}{\Gamma_t} \,
\frac{d\Gamma_{t \to bl^+\nu}
(\vc{S}+\delta \vc{S}) }
{dE_l d\Omega_l} 
\times \left[ 1 +
\xi (|\vc{p}_t|,E,E_l,\cos\theta_{lt}) \right] .
\nonumber \\
\label{full}
\eea
Here, the first line on the right-hand-side shows that 
there are corrections to the top-quark
production cross section,
while the second line shows that the correction to the 
decay distribution of $l^+$ is
accounted for by a modification of the parent top-quark polarization vector,
and finally there is a non-factorizable correction $\xi$ which cannot be
assigned either to the production or the decay process alone.

The top momentum distribution
is modified by $\delta_0$ to take a lower average momentum.
The forward-backward asymmetric distribution 
and the top polarization
vector get corrections as
\bea
&&
\delta_1 \cos \theta_{te} =
\kappa S_\| \cos \theta_{te} \times \frac{1}{2}\psi_{\rm _R} ,
\label{del1}
\\
&&
\delta \vc{S} =
\left[ 1 - (S_\|)^2 \right] \times
\kappa \cos \theta_{te} \times {\frac{1}{2}} \psi_{\rm _R}
\cdot \hat{\vc{n}}_\| 
\label{delS}
\eea
with
\bea
\psi_{\rm _R}(|\vc{p}_t|,E) &=& - \, C_F \! \cdot \! 4 \pi \alpha_s \,
\nonumber \\ && \times \,
\mbox{Pr.}\int \frac{d^3\vc{q}}{(2\pi)^3} \,
\, \frac{1 \,}{|\vc{q}\! -\! \vc{p}_t|^3} \,
\frac{\vc{p}_t\! \cdot \! (\vc{q}\! -\! \vc{p}_t)}{|\vc{p}_t|\, 
|\vc{q}\! -\! \vc{p}_t|}
\,
2 \,
\mbox{Re}
\biggl[ \frac{\tilde{G}(\vc{q};E)}{\tilde{G}(\vc{p}_t;E)} \biggl] .
\nonumber \\
\eea

\begin{figure}
\centerline{ \epsfxsize=7cm\epsffile{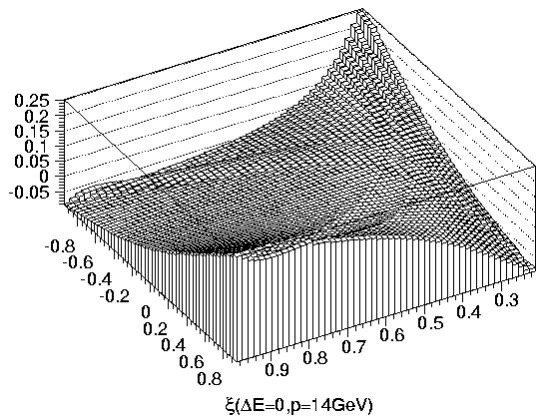}}
\begin{center}\begin{minipage}{\figurewidth}{
  \caption[]{\sl
        \label{xi}
A three-dimensional plot of $\xi$ as a function of $2E_l/m_t$ ($x$-axis) and
$\cos\theta_{te}$ ($y$-axis).
}}
\end{minipage}\end{center}
\end{figure}

We show the $\cos \theta_{lt}$ and $E_l$ dependences of the
non-factorizable correction $\xi$ as a 3-dimensional plot in
Fig.~\ref{xi}.
One can see that $\xi$ takes comparatively large
positive values for
either ``small $E_l$ and $\cos \theta_{lt} \simeq -1$'' or
``large $E_l$ and $\cos \theta_{lt} \simeq +1$''.
Oppositely,
in the other two corners of the $E_l$--$\cos \theta_{lt}$ plane
$\xi$ becomes negative.
The typical magnitude of $\xi$ is 10--20\%.

\subsection{Measurements of Top Quark Couplings}
\label{topcoupling}
\clfn

As we have seen so far,
studying various top quark properties in the
$t\bar{t}$ threshold region is promising and interesting.
Regarding the energy upgrading scenario of JLC,
it is likely that the machine will
operate first in the $t\bar{t}$ threshold region for a significant
amount of time,
while measuring the top quark mass precisely, etc., before
the beam energies are increased to the open-top region.
Therefore it is desirable that measurements of the various top quark
couplings can be performed concurrently with 
other unique measurements near threshold,
with sensitivities comparable to those in the open-top region.

Thus, systematic
simulation studies on sensitivities to all the top quark couplings
or the form factors in the $t\bar{t}$ threshold region
are requested, which are not yet complete.
Up to now, sensitivity studies in the threshold region were performed 
on the top quark width \cite{exp}, on
the top-Higgs Yukawa coupling \cite{exp,hjk}, 
as well as on the strong coupling constant $\alpha_s$.
Also, there exists a theoretical study on the $CP$ violating couplings
of the top quark.
We review the present status of these studies below.

\subsubsection*{\underline{A) 
Measurements of $\Gamma_t$, $g_{tH}$, $\alpha_s$}}

Among various observables in the top quark threshold region,
the height (normalization) of the $1S$ peak of the threshold cross section 
is most sensitive to the top quark decay width $\Gamma_t$, the
top-Higgs Yukawa coupling $g_{tH}$, and the strong coupling constant
$\alpha_s$.
This is because:
(1) according to Eq.~(\ref{totcs-f}), the peak height is inversely proportional
to $\Gamma_t$ in the narrow width limit;
(2) the effect of Higgs exchange enhances the cross section;
(3) the wave function at origin $|\psi_n(\vc{0})|^2$ is proportional to
$\alpha_s^3$; when we take into account 
the smearing effect by the large decay width the
sensitivity reduces to $\sim \alpha_s^2$.
Nevertheless, extracting these quantities requires a precise 
theoretical prediction for the 
normalization of the threshold cross section.
In view of the present large theoretical uncertainty, 
accurate measurements of $g_{tH}$ and $\alpha_s$
seem to be difficult, whereas we will still be able to measure $\Gamma_t$
relatively accurately.
Also, the top momentum distribution and forward-backward asymmetry
of the top quark depend on $\Gamma_t$ and $\alpha_s$.
Theoretical predictions for these observables are expected to be 
more stable, but 
some more calculations are needed for accurate predictions.

\subsubsection*{\underline{Total $t\bar{t}$ Cross Section}}

Let us first discuss the results of the simulation studies, which 
set a benchmark of measurements in future experiments.

It is well known that the initial state radiation has to be properly
taken into account in the vicinity of a resonance state.
Also, in order to study experimental feasibilities in a realistic
environment, we need to include the effects due to the beam energy spread
and beamstrahlung (a
bremsstrahlung from beam particles due to the strong electromagnetic
fields produced by opposing beam bunches).
These effects can be incorporated by convoluting the original cross
section with the radiator function and the beam spectrum.
The beam spectrum is machine dependent, and in our study below we use the
parameters given in \cite{exp}.

\begin{figure}[tbh]
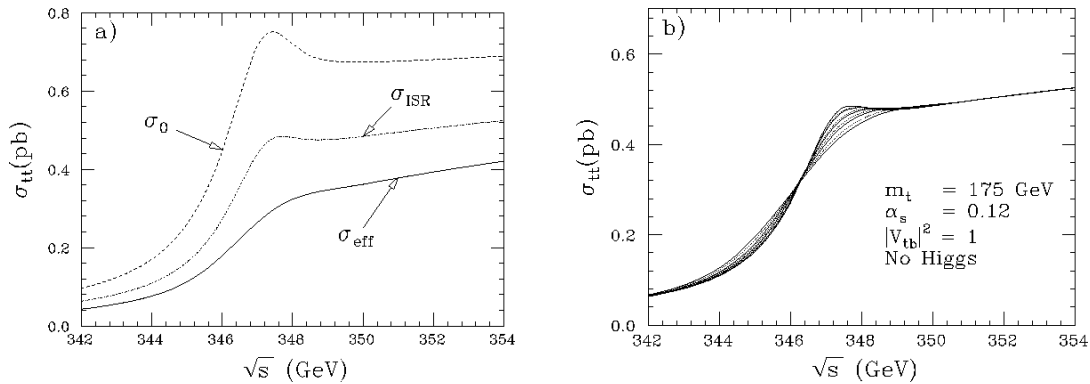

\begin{center}
\vspace*{0.5cm}
\begin{minipage}[t]{7.5cm}
    \epsfxsize=6.8cm\epsffile{phystop/figs/sgttbmeffa.epsf}
\end{minipage}
\hspace{0.5cm}
\begin{minipage}[t]{7.5cm}
    \epsfxsize=6.8cm\epsffile{phystop/figs/sgttbmeffb.epsf}
\end{minipage}
\end{center}

\vspace*{-1cm}
\begin{center}\begin{minipage}{\figurewidth}
\caption{\sl
(a) Effects of the initial state radiation (ISR)
and the beam energy spread and 
beamstrahlung on the threshold shape for $m_t=175$~GeV and 
$\alpha_s(M_Z)=0.12$: with no ISR and no beam effects (dash), with
ISR but without beam effects (dotdash), and with both ISR and beam
effects (solid).
(b) Threshold cross section including the effects of ISR and natural
beam energy spread but leaving out the effect of beamstrahlung.
Two kinds of spectra, flat-top (dot) and double-peaked (solid), which
is more realistic, are examined for various beam energy spread:
0.1\%, 0.4\%, 0.7\%, 1.0\%, 1.4\%.
\label{beameff}
}
\end{minipage}
\end{center}
\end{figure}

The effects on the cross section can be summarized as follows
(see Figs.~\ref{beameff}).
(i) The initial state radiation and the 
beamstrahlung reduce usable luminosity in the threshold
region. 
The usable part is essentially restricted to the 
$\delta$-function part of the beam spectrum.
Therefore, we need to know the height of the $\delta$-function part
accurately for precision measurements.
(ii) The beam energy spread is the major source of the peak smearing.
When the energy spread is less than 0.4\%, the cross section shape
is practically independent of the structure inside the $\delta$-function part.
If we cannot achieve such a narrow band beam, it is important to measure
the spectrum inside the $\delta$-function part with a high-energy
resolution.

The signature of a $t\bar{t}$ pair production is two $b$ quarks and
two $W$ bosons in the final state.
The two $W$ bosons decay into either $q\bar{q}'$ or $\ell \nu$.
Therefore the final state configurations are 
(i) two $b$ jets and four jets from $W$'s (45\%),
(ii) two $b$ jets, two jets, and one charged lepton (44\%), and
(iii) two $b$ jets and two charged leptons (11\%).
All three cases can be used for the measurement of the total cross section.
Case (i) and (ii) are useful for the measurements of the top momentum
distribution and top quark forward-backward asymmetry.\footnote{
A recent study shows that the charge of $b$ can be identified efficiently
from a measurement of the vertex charge of $b$ jets.
In this way $t$ and $\bar{t}$ can be distinguished even in case
(i) \cite{iwasaki}.
}
The basic cuts used in the event selection can be classified into
four groups:
(a) event shape cuts such as those on the number of charged particles,
the number of jets, and thrust,
(b) mass cuts to select $W$'s and $t$'s by jet-invariant-mass method,
(c) requirements of leptons in cases (ii) and (iii), and
(d) $b$ taggings.

In order to measure the total $t\bar{t}$ cross section reliably, we need to
select $t\bar{t}$ events with a good signal-to-background
ratio.
The largest background comes from $W^+W^-$ pair productions, whose
cross section is larger by a factor $\sim 10^{3}$.
We take case (i) as an example and show the result of a Monte
Carlo simulation study.
We require 20 or more charged tracks and impose cuts on the visible energy
(the energy sum of all the detected particles including neutral ones) and the
total transverse momentum:
$E_{\rm vis} \geq 250$~GeV and $p_T \leq 50$~GeV.
The remaining events must contain six or more hadronic jets after jet
clustering with $y_{\rm cut} = 5 \times 10^{-3}$, where $y_{\rm cut}$ is
the cut on jet-invariant-masses normalized by the visible energy.
If there are more than six jets, we increase the $y_{\rm cut}$ value
so as to make the event yield exactly six jets.
Two pairs of jets out of these six jets must have 2-jet invariant masses
consistent with $M_W$.
These two $W$ candidates are then required to make invariant masses 
consistent with $m_t$, when combined with one of the remaining two jets.
Finally we impose a cut on the event thrust: ${\rm thrust} \leq 0.75$.
After the final cut, the detection efficiency is 29\%, while the 
signal-to-background ratio is greater than 10.
This detection efficiency translates to about 63\% when the branching
ratio for $W \to q \bar{q}'$ of 67\% is taken into account.
We can select events with
similar efficiencies and signal-to-background ratios for
cases (ii) and (iii) as well.

Using the $t\bar{t}$ samples so obtained, we can determine the
parameters $\Gamma_t$, $g_{tH}$, $\alpha_s$.
11 points with 1~fb$^{-1}$ each were generated with the inputs
$m_{t,{\rm pole}} = 170$~GeV,
$\alpha_s(M_Z)=0.12$, $m_H = \infty$ (no Higgs effects) and the
SM value for $\Gamma_t$.
By fitting these data points to
the theoretical
prediction for the total $t\bar{t}$ production cross section, convoluted
with the beam spectrum,
we obtained the following estimates of the statistical errors on the
above parameters:
$\Delta \Gamma_t / \Gamma_t^{\rm SM} = 0.18$,
$\Delta g_{tH}/g_{tH}^{\rm SM} = 0.3$ in the case $M_H = 115$~GeV,
and $\Delta \alpha_s(M_Z) = 0.002$.
Each error is given for a measurement of each parameter, assuming
that the other parameters are known from other sources.

\begin{figure}[t]
\begin{center}
\begin{minipage}{9.0cm}\centering
\epsfxsize=10cm\epsffile{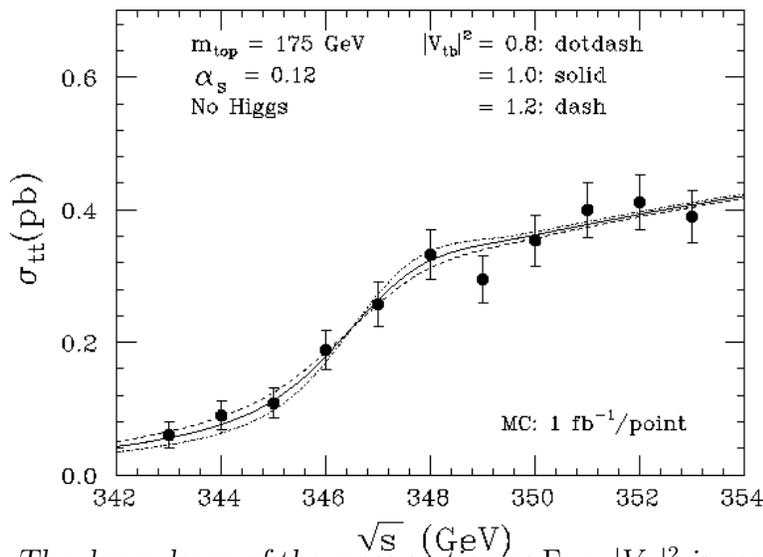}
\vspace*{-1cm}
\end{minipage}\\
\end{center}
\begin{center}\begin{minipage}{\figurewidth}{
\caption{\sl
The dependence of the cross section as
$\Gamma_t \propto |V_{tb}|^2$ is varied.
An example of energy scan is superimposed, where each point 
corresponds to 1~fb$^{-1}$.
      \label{widtheff}
}}
\end{minipage}
\end{center}
\end{figure}

While the above errors serve as measures for 
potential sensitivities of the threshold measurements, as long as the
theoretical uncertainty on the normalization of the cross section remains
at the present level, theoretical errors on these parameters can be 
much larger.\footnote{
An attempt at solving this problem has been given recently in \cite{rg}.
}
For example, the induced theoretical error on the strong coupling constant
is $\Delta \alpha_s(M_Z) \simeq 0.01$.
As for $\Gamma_t$, we may consider a strategy to determine it to
a good accuracy, nonetheless.
Since the effect of the variation of $\Gamma_t$ appears only in the
vicinity of the $1S$ peak\footnote{
In the limit $\sqrt{s} \gg 2 m_t$, the normalization of the cross section
becomes independent of $\Gamma_t$.
} (see Fig.~\ref{widtheff}), we may use, for instance, the ratio of the
$t\bar{t}$ total cross section at 
$\sqrt{s} \simeq M_{1S}$ and at $\sqrt{s} \simeq M_{1S}+3$~GeV.
This ratio is much less affected by the normalization uncertainty,
and the statistical error on $\Delta \Gamma_t/\Gamma_t$ will be determined
by the luminosity we invest at $\sqrt{s} \simeq M_{1S}$ and 
$\sqrt{s} \simeq M_{1S}+3$~GeV.
We expect that an accuracy better than 15\%
(both theoretical and statistical) would be achievable
with a moderate integrated luminosity.
On the other hand, extraction of the Yukawa coupling 
requires, at the present status, a more accurate theoretical
prediction.
This is because its effect is to enhance the cross section almost
energy independently in the threshold region, and disentanglement from
the normalization uncertainty is difficult.
To match the above statistical accuracy of the  Yukawa coupling,
the theoretical accuracy should become smaller than 4\%.
For the same reason,
a precise measurement of $\alpha_s$ will also be difficult.
A better strategy would be to determine the value of $\alpha_s(M_Z)$
in other experiments and to
use it as an input of the theoretical predictions in the top 
quark threshold region.

\subsubsection*{\underline{Top Momentum Distribution}} 

By virtue of the large top quark width, we can measure the top quark
momentum by reconstructing the 3-jet decay of a top quark, even below
the $t\bar{t}$ threshold.
This is in contrast to the charmonium and bottomonium cases,
where the annihilation modes dominate.
We assume that the program of JLC is such that
the machine first scans the threshold shape with a relatively low
integrated luminosity and determines the top quark mass accurately, then
the c.m.\ energy is fixed at some point in the threshold
region and we invest more luminosity to perform the measurements of the
top momentum distribution, etc.
The choice of the c.m.\ energy should be made from the viewpoint of 
insensitivity to theoretical ambiguities and beam effects and of
sensitivities to physical parameters to be determined.
Since the momentum distribution depends heavily on the c.m.\ energy,
the initial state radiation and beam effects 
must be properly taken into account.
From the study in \cite{exp}, we find that the position of the
peak momentum $|\vc{p}|_{\rm peak}$ is fairly stable against these
effects, although the distribution itself is quite affected.
An optimal c.m.\ energy, which maximizes the sensitivity to
$\alpha_s(M_Z)$ and $\Gamma_t$, while keeping small the beam effects
and theoretical uncertainties, is chosen as 
2~GeV above the 1S peak energy,
$\Delta E \equiv \sqrt{s} - M_{1S} = 2$~GeV.
For simplicity, we assume that $M_{1S}$ is known in what follows.

In order to measure the top momentum, we need to select top quarks
which are well-reconstructed as 3-jet systems from $t \to b W$ decays.
This requires final states with at least one $t$ or $\bar{t}$ quark
decaying into three jets:
the configurations (i) and (ii) discussed above.
The 6-jet mode [case (i)] is advantageous from the statistical point of
view, since both $t$ and $\bar{t}$ quarks in a single event can be used
for the momentum measurement.
This mode, however, tends to suffer from combinatorial background:
even if we have a perfect $b$-tagging capability, some
ambiguity would still remain in assigning other four jets to
two $W$'s.
The lepton-plus-4-jet mode [case (ii)] is cleaner in this respect.

The selection of top quark events goes similarly to the one discussed in 
the total cross section analysis.
In the present case, however,
we have to make sure that the momentum of a reconstructed top quark
is well measured.
Rejection of events with energetic missing neutrinos is thus
necessary.
The main differences are the cut values which must be much tighter
to require four-momentum balance, well reconstructed $W$'s and
$t$'s.
The $b$ tagging and the cut on the $b$-$W$ angle are necessary in addition.
(See \cite{exp} for details.)
In case (ii)
Monte Carlo simulations resulted in a detection efficiency
of 4\% including the branching fraction to the
lepton-plus-4-jet mode.
The systematic shift between the peak momentum of the generated events
and that of the reconstructed events is less than 1~GeV. 
The momentum resolution is 3--4~GeV in the relevant range, 
which is much smaller than the width of the momentum distribution.
In case (i) the detection efficiency is
2\%, including the branching fraction to the 6-jet mode.
Since both $t$ and $\bar{t}$ can be used for the momentum measurement,
effective selection efficiency is twice this value.
The agreement of the reconstructed momentum distribution with
the generated one is not as good as in the lepton-plus-4-jet case,
because of the combinatorial background.
A further study to gain in the detection efficiency using 
a kinematical fitting procedure is now underway.

We can estimate the statistical error on $|\vc{p}|_{\rm peak}$
from the width of the expected momentum distribution and the expected
number of reconstructed top quarks.
Using the events reconstructed from the lepton-plus-4-jet mode
at $\Delta E = 2$~GeV with $m_{t,{\rm pole}}=170$~GeV,
the $1\sigma$ bound corresponding to an integrated luminosity of
100~fb$^{-1}$ is
$\Delta | \vc{p} |_{\rm peak} \simeq 200$~MeV.
This value translates to sensitivities to 
$\Gamma_t$ and $\alpha_s(M_Z)$ of
$\Delta \Gamma_t / \Gamma_t = 0.03$ and
$\Delta \alpha_s(M_Z) = 0.002$, when
each parameter is varied while the other is fixed
and assuming $M_{1S}$ is known.

\subsubsection*{\underline{Forward-Backward Asymmetry}}

As we have seen in Subsection~\ref{dyn}, interference of the
$S$- and $P$-wave resonances produces a forward-backward asymmetry
of the top quark.
The spectrum of the would-be toponium resonances is such that the
lowest-lying $S$-wave resonance ($1S$ state) stands alone, while the
higher $S$-wave resonances are accompanied by nearly degenerate
$P$-wave resonances.
The level spacings are determined by $\alpha_s$ while the resonance
widths are determined mainly by the top quark width.
The relative size of the two determines the size of the interference.
Therefore, we expect that the forward-backward asymmetry provides
additional information on the top quark width.

The c.m.\ energy is chosen in the same strategy as in the momentum
distribution measurement.
The asymmetry depends fairly strongly on the c.m.\ energy.
As a consequence, the beam effects are sizable in the entire threshold 
region, and a 
good understanding or control of these effects will be mandatory in the 
asymmetry measurement.
In consideration of the sensitivity to $\Gamma_t$, we chose the
c.m.\ energy as
$\Delta E = 1$~GeV.
Since the asymmetry is quite small, the systematic error  due to
the detector asymmetry must be carefully studied in actual experiments.
We expect collider operations at the $Z$ pole (Giga $Z$) will be 
very useful, since the experimental limit on the measurable asymmetry
will be determined by the number of events available for the
detector calibration.
We assume that a sample of 40,000 reconstructed events is given
and estimate the expected statistical errors on $\alpha_s$
and $\Gamma_t$.
The sample corresponds to 200~fb$^{-1}$ even if the detection efficiency
is as high as 40\%.
For $m_{t,{\rm pole}}=170$~GeV, the sensitivity to the strong coupling
constant is $\Delta \alpha_s(M_Z)=0.004$ if $M_{1S}$ and $\Gamma_t$ are
known.
On the other hand, the sensitivity to the top quark width is
$\Delta \Gamma_t / \Gamma_t = 0.06$ if $M_{1S}$ and $\alpha_s(M_Z)$ are
known.

\subsubsection*{\underline{B) $CP$ Violating Couplings}}

Here, we review  the recent theoretical study on
probing the anomalous $CP$ violating interactions of the top quark
in the threshold region \cite{jns}.
Besides its own
interest, it also serves as a case study for the analyses of
top quark properties using its production-decay chain.

Among various interactions of the top quark, 
testing the {\it CP}-violating interactions
is particularly interesting due to 
following reasons:
\begin{itemize}
\item
Within the SM,
{\it CP}-violation in the top quark sector is extremely small.
If any {\it CP}-violating effect is detected in the top quark sector 
in a near-future experiment, it immediately signals new physics.
\item
There can be many sources of {\it CP}-violation
in models that extend the SM.
Besides, the observed baryon asymmetry in the Universe suggests 
existence of {\it CP} violating mechanisms beyond the SM.
\item
In relatively wide class of models
beyond the SM, {\it CP} violation 
emerges especially sizably in the top quark sector.
\end{itemize}
In this analysis we consider 
{\it CP}-violating interactions of top quark with
$\gamma$, $Z$, and $g$.
In particular, we consider
the lowest dimension {\it CP}-odd effective operators:
\bea
{\cal L}_{\mbox{\scriptsize {\it CP}-odd}} &=&
- \frac{e d_{t\gamma}}{2m_t} 
( \bar{t} i \sigma^{\mu\nu} \gamma_5 t )
\partial_\mu A_\nu
- \frac{g_Z d_{tZ}}{2m_t} 
( \bar{t} i \sigma^{\mu\nu} \gamma_5 t )
\partial_\mu Z_\nu
\nonumber \\ &&
- \frac{g_s d_{tg}}{2m_t} 
( \bar{t} i \sigma^{\mu\nu} \gamma_5 T^a t )
\partial_\mu G_\nu^a ,
~~~~~~
 \sigma^{\mu\nu} \equiv {\textstyle \frac{i}{2}}
[ \gamma^\mu , \gamma^\nu ] ,
\label{effop}
\eea
where 
$e = g_W \sin \theta_W$ and
$g_Z = {g_W}/{\cos \theta_W}$.
These represent the interactions of $\gamma$, $Z$, $g$ with
the EDM, $Z$-EDM, 
chromo-EDM of top quark, 
respectively.\footnote{
The magnitudes of these EDMs are given by ${e d_{t\gamma}}/{m_t}$,~
${g_Z d_{tZ}}/{m_t}$,~ ${g_s d_{tg}}/{m_t}$,
respectively.
$d_{t\gamma}=1$ corresponds to $e/m_t \sim 10^{-16}~e \, {\rm cm}$,
etc.
}
As stated, the SM contributions to these couplings are extremely small,
$d_{t\gamma}^{\rm (SM)}, d_{tZ}^{\rm (SM)}, d_{tg}^{\rm (SM)} \sim 10^{-14}$.
Our concern is in the anomalous couplings which are induced from
some new physics.
We assume that generally the couplings
$d_{t\gamma}$, $d_{tZ}$, $d_{tg}$
are complex.

Which {\it CP}-odd observables are sensitive
to the {\it CP}-violating couplings $d_{t\gamma}$, $d_{tZ}$, $d_{tg}$?
For the process $e^+e^- \to t\bar{t}$, 
we may conceive of the following 
expectation values of kinematical variables for
{\it CP}-odd observables:
\bea
&
\left< \,
( \vc{p}_{e} - \bar{\vc{p}}_{e} ) \cdot
( \vc{s}_t - \bar{\vc{s}}_t ) 
\, \right> ,
&
\nonumber \\
&
\left< \,
( \vc{p}_t - \bar{\vc{p}}_t ) \cdot
( \vc{s}_t - \bar{\vc{s}}_t )
\, \right> ,
&
 \\
&
\left< \,
[ ( \vc{p}_{e} - \bar{\vc{p}}_e ) \times
( \vc{p}_t - \bar{\vc{p}}_t ) ] \cdot
( \vc{s}_t - \bar{\vc{s}}_t )
\, \right> ,
&
\nonumber
\eea
where the spins and momenta are defined in the
c.m.\ frame.
The above quantities are the three components of the
difference of the $t$ and $\bar{t}$ spins.
All the other {\it CP}-odd observables for $e^+e^- \to t\bar{t}$
are bilinear in $\vc{s}_t$ and $\bar{\vc{s}}_t$.
Since analyses of spin correlations are complicated, we focus on
the difference of the
polarization vectors.

Practically we can measure the $t$ and $\bar{t}$ polarization
vectors efficiently using $\ell^\pm$ angular distributions.
It is known that the
angular distribution of the charged lepton $\ell^+$ from the
decay of top quark is
maximally sensitive to the top quark polarization
vector.
In the rest frame of top quark,
the $\ell^+$ angular distribution is given by \cite{langdist}
\bea
\frac{1}{\Gamma_t} \,
\frac{d \Gamma ( t \to b \ell^+ \nu )}{d \cos \theta_{\ell^+}}
= \frac{1 + S \cos \theta_{\ell^+}}{2} 
\label{langdistr}
\eea
at tree level,
where $S$ is the top quark polarization and
$\theta_{\ell^+}$ is the angle of $\ell^+$ measured from the direction
of the top quark polarization vector, cf.\ Eq.~(\ref{leneangular}).
Indeed the $\ell^+$ distribution is ideal for extracting
{\it CP}-violation in the $t\bar{t}$ {\it production process};
the above angular distribution is unchanged even if anomalous
interactions are included in the $tbW$ decay vertex,
up to the terms linear in the decay anomalous couplings and within the
approximation $m_b = 0$ \cite{GH}.
Therefore, if we consider the average of the lepton direction,
for instance, we may extract the top quark
polarization vector efficiently:
\bea
\left< \vc{n} \cdot \vc{n}_\ell \right>_{\rm Lab}
\simeq \frac{1}{3} \, \vc{n}\cdot \vc{S} .
\eea
The average is to be taken at the top quark rest frame, but
in the threshold region, we may take the average in the
laboratory frame barely without loss of sensitivities to the
anomalous couplings.

From rough estimates, we estimated the statistical errors to
the anomalous 
{\it CP}-violating couplings of the top quark with $\gamma$,
$Z$ and $g$ in the $t\bar{t}$ threshold region:
$\delta d_{t\gamma}$, $\delta d_{tZ}$, $\delta d_{tg} \sim 10\%$
corresponding to an integrated luminosity of 50~fb$^{-1}$.
In Table~\ref{top-tab1} we summarize the results of 
the sensitivity studies performed
so far, including the results of our present study.
\begin{table}[t]
\begin{center}\begin{minipage}{\figurewidth}
\caption{\sl 
\label{top-tab1}
The results of studies of sensitivities to the anomalous couplings
expected in future
experiments.
For $e^+e^-$ linear colliders (LC),
``open top'' denotes the results of studies performed at $\sqrt{s}=500$~GeV.
}
\end{minipage}\end{center}
\vspace{4pt}
\begin{center}
\small
\begin{tabular}{l|l||c|c|c}
\hline
\multicolumn{2}{l||}{}
& $\delta d_{tg}$ & $\delta d_{t\gamma}$ & $\delta d_{tZ}$
\\ \hline 
\multicolumn{2}{l||}{LHC ~~~ (10~fb$^{-1}$)}
& $10^{-2}$ -- a few$\times 10^{-3}$ & - & -
\\
\hline
$e^+e^-$ LC & open top &
${\cal O}(1)$ & $10^{-1}$ -- a few$\times 10^{-2}$ &
$10^{-1}$ -- a few$\times 10^{-2}$
\\
\cline{2-5}
(50 fb$^{-1}$) & $t\bar{t}$ threshold &
$10^{-1}$ & $10^{-1}$ & $10^{-1}$ 
\\
\hline
\end{tabular}
\end{center}
\end{table}
We may compare the sensitivities of 
experiments in the $t\bar{t}$
threshold region at $e^+e^-$
colliders (JLC) with others.
The sensitivities to $d_{t\gamma}$ and $d_{tZ}$ are comparable
to those attainable in the open-top region at $e^+e^-$
colliders.
The sensitivity to $d_{tg}$ is worse than that expected at 
a hadron collider but
exceeds the sensitivity in the open-top region
at $e^+e^-$ colliders.
(It should be noted, however, that although a hadron collider has
a high sensitivity to this quantity, accuracy of its measurement
is not very good, and {\it vice versa} for $e^+e^-$ colliders.)

Qualitatively, the characteristics of the $t\bar{t}$ region are
summarized as follows.
\begin{enumerate}
\renewcommand{\labelenumi}{(\arabic{enumi})}
\setcounter{enumi}{0}
\item
We can measure the three couplings $d_{t\gamma}$, $d_{tZ}$, $d_{tg}$
simultaneously and we can disentangle each contribution.
\item
Typical sizes of components of the 
difference of the $t$ and $\bar{t}$
polarizations are given by
$$
|\delta {S}_\perp|, \, |\delta {S}_{\rm N}| 
\sim ( \mbox{5--20}\% ) \times ( d_{t\gamma}, d_{tZ}, d_{tg} ) .
$$
They can be extracted efficiently from the directions of
charged leptons from decays of $t$ and $\bar{t}$:
$$
\langle \langle \vc{n} \cdot 
( \vc{n}_{\ell} + \bar{\vc{n}}_{\ell} )  \rangle \rangle _{\vc{p}_t}
\simeq
\frac{2}{3} \, 
\vc{n} \cdot \delta \vc{S} .
$$
\item
We can measure the complex phases of the couplings
$d_{t\gamma}$, $d_{tZ}$, $d_{tg}$.
Since the strong phases can be modulated at our disposal,
a single observable ($\delta {S}_\perp$ or $\delta {S}_{\rm N}$)
probes the phases of the couplings.
\end{enumerate}

We note that if one of the couplings is detected in the future, we 
would certainly want to measure the others 
in order to gain deeper understanding
of the {\it CP}-violating mechanism.
This is because one may readily think of 
various underlying processes which give different contributions
to the individual couplings.

Regarding (1) and (2) above, QCD interaction is used as
a controllable tool for
the detection of the anomalous couplings.
This would be the first trial to use QCD interaction for such a
purpose without requiring any phenomenological inputs.

\section{Top Quark Mass Measurement
   from $ t \bar{t} $  Dilepton events}
\label{dilepton}

 The top quark mass can also be determined from direct reconstruction of
    $$t \bar{t} \rightarrow W^+ b W^- \bar{b}
               \rightarrow \ell^+\ell^-\nu\bar{\nu}b\bar{b} $$
 Dilepton events at JLC, in addition
 to the mass determination from the $t \bar{t} $ threshold scan.
 CDF and D0 have determined the top quark mass to be
    $ 174.3 \pm 5.1$ GeV/c$^2 $
 from direct reconstruction of $ t \bar{t} $ events.
 The top quark mass is expected to be determined to a precision
 of 1 to 2 GeV/c$^2$ at the Tevatron and the LHC.
 For direct reconstruction, the Dilepton channel has the
 smallest systematic uncertainty, in comparison with the Lepton + Jets
 and Multi-jet channels.
 In contrast to the hadron colliders,  the beam energy at the
 Linear Collider provides constraints in the kinematic reconstruction.
 A precision of 200 MeV/c$^2$ on the top quark mass can be achieved
 at JLC \cite{DiLepton}.

\section{Measurement of Top-Yukawa Coupling}
\label{yukawa}

\subsection{Theoretical Background}

At higher energies, where $e^+e^- \to t\bar{t}H$ is open,
we can carry out a more direct measurement
of the top Yukawa coupling.
There are two kinds of diagrams to
the $e^+e^- \to t\bar{t}H$ process at the tree level.
The first kind includes diagrams where a Higgs boson
is radiated off the final-state top or anti-top legs
($H$-off-$t$) and is thus proportional to the top Yukawa coupling.
The second kind consists of the one with a Higgs boson
from the $s$-channel $Z$ ($H$-off-$Z$) and
is independent of the top Yukawa coupling.

\begin{figure}[htb]
\centerline{
\epsfxsize=7cm 
\epsfbox{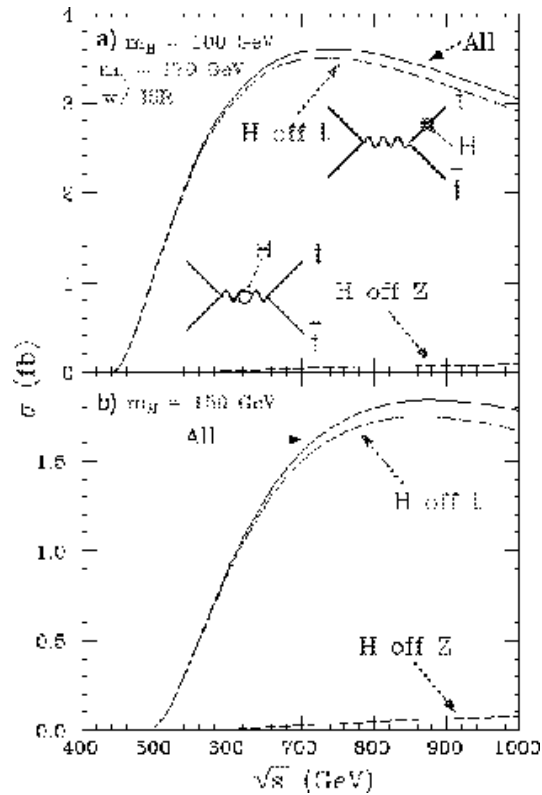}
}
\begin{center}\begin{minipage}{\figurewidth}{
\caption[Fig:sgtth]{\sl\label{Fig:sgtth}
The production cross sections for
$e^+e^- \to t\bar{t}H$ with $m_t = 170~{\rm GeV}$
as functions of
$\sqrt{s}$:
(a) $m_H = 100~{\rm GeV}$ and
(b) $m_H = 150~{\rm GeV}$,
both with initial state radiation, leaving out
the beam effects.
The dot-dashed lines are the contributions from
the $H$-off-$t$ diagrams (signal diagrams),
while the dashed lines are from
the $H$-off-$Z$ diagram (background).
}}
\end{minipage}\end{center}
\end{figure}

Figures \ref{Fig:sgtth}-a) and -b) plot
the contributions of these two kinds of diagrams
in terms of the production cross section
for (a) $m_H = 100~{\rm GeV}$
and (b) $m_H = 150~{\rm GeV}$.
The cross sections are calculated for
$m_t = 170~{\rm GeV}$ with the initial state radiation
but without beam effects which are heavily machine-dependent.
The key point here is that
the $H$-off-$t$ diagrams dominate
the $H$-off-$Z$ diagram, so that
the total cross section is essentially proportional to
$\beta_H^2$.
A simple counting experiment thus serves as a
direct top Yukawa coupling measurement.

\subsection{Event Selection}

The question is how well we can separate signal from background,
since the production cross section is not very large.
The signal for $t\bar{t}H$ production is
2 $W$'s and 4 $b$'s, where the $W$'s decay into either
$q\bar{q}'$ or $l\nu$.
Depending on how the two $W$'s decay, we have three modes:
(1) 8 jets (38\%), (2) a lepton plus 6 jets (37 \%),
and two leptons + 4 jets (the rest).
Here, we consider the modes (1) and (2).
The background, on the other hand, can be
classified into two groups: (i)
irreducible ones containing the $e^+e^- \to t\bar{t}Z$
followed by $Z \to b\bar{b}$ (separable if $m_H$ is far from
$m_Z$) and the $H$-off-$Z$ diagram which is tiny,
(ii) reducible ones of which the biggest is
$e^+e^- \to t\bar{t}$ with gluon emissions.

Basic cuts to select signal events from the background
consist of event shape cuts requiring
8 or 6 jets in the final state
and small thrust (e.g. thrust $< 0.8$),
mass cuts demanding
$m_{2J} \simeq m_W/m_H$ and $m_{3J} \simeq m_t$,
and, in the case of the 8-jet mode,
four-momentum balance: $P_T < 50~{\rm GeV}$ and
$\Delta E_{vis} < 200~{\rm GeV}$.
In addition, the $b$-tagging
requiring more than two $b$-jet candidates in the final state is crucial.
Here, we conservatively assume the
base-line JLC-I values for the tagging efficiencys:
$\epsilon_b = 0.78$ and $\epsilon_c = 0.38$.
The resultant signal-to-noise ratio
heavily depends on the $b$-tagging performance as well as on the Higgs mass
which controls the background from 
the $e^+e^- \to t\bar{t}Z$ process.

\begin{figure}[htb]
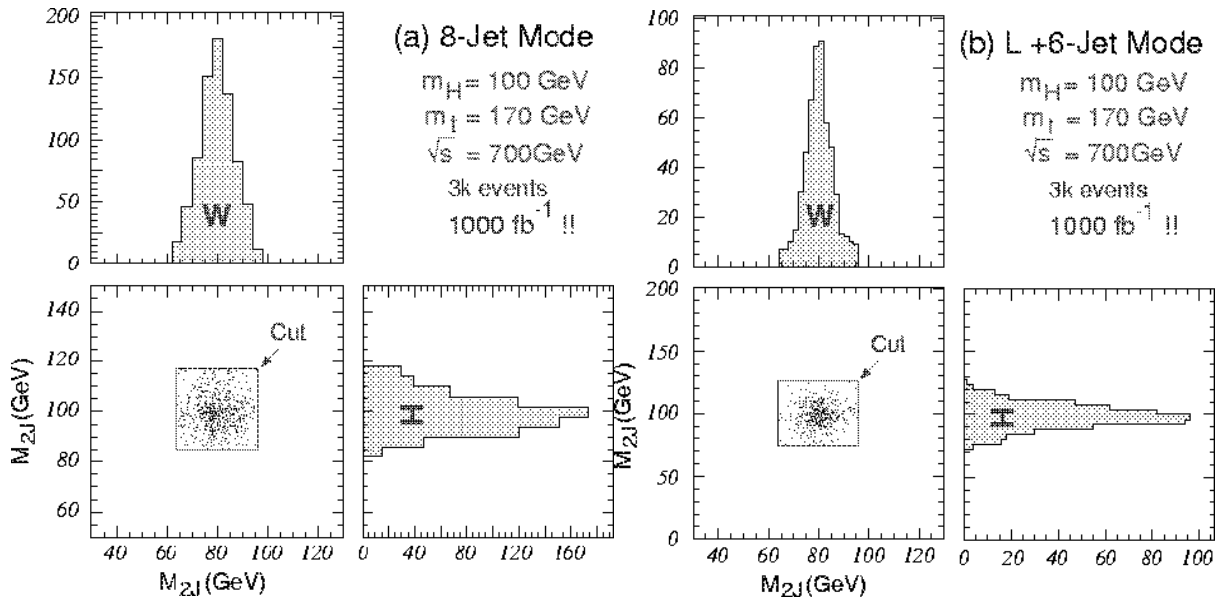

\centerline{
\epsfxsize=8cm 
\epsfbox{phystop/mhmw8j.epsf}
\epsfxsize=8cm
\epsfbox{phystop/mhmwl6j.epsf}
}
\begin{center}\begin{minipage}{\figurewidth}{
\caption[Fig:mhmw]{\sl \label{Fig:mhmw}
A scatter plot of 2-jet invariant masses for
$W$ and $H$ candidates in the signal events
shown together with its projection to
the two axes:
(a) 8-jet mode and (b) lepton-plus-6-jet mode.
}}
\end{minipage}\end{center}
\end{figure}

In order to study the event selection feasibility,
we have generated the $t\bar{t}H$ events
and the $t\bar{t}$ and $t\bar{t}Z$ background events
for $m_t = 170~{\rm GeV}$ and $m_H = 100~{\rm GeV}$
at $\sqrt{s} = 700~{\rm GeV}$,
with the beam width and beamstrahlung
as given in Ref.\cite{Ref:jlc1}.
These events were fed into the
JLC-I detector simulator\cite{Ref:jlc1}
and processed through the event selection
described above.
Scatter plots of 2-jet invariant masses for
$W$ and $H$ candidates
in the signal events after the above selection
is shown in Figs.~\ref{Fig:mhmw}-a) and -b)
for the 8-jet and the lepton-plus-6-jet modes,
respectively.

\begin{figure}[p]
\centerline{
\epsfxsize=8cm 
\epsfbox{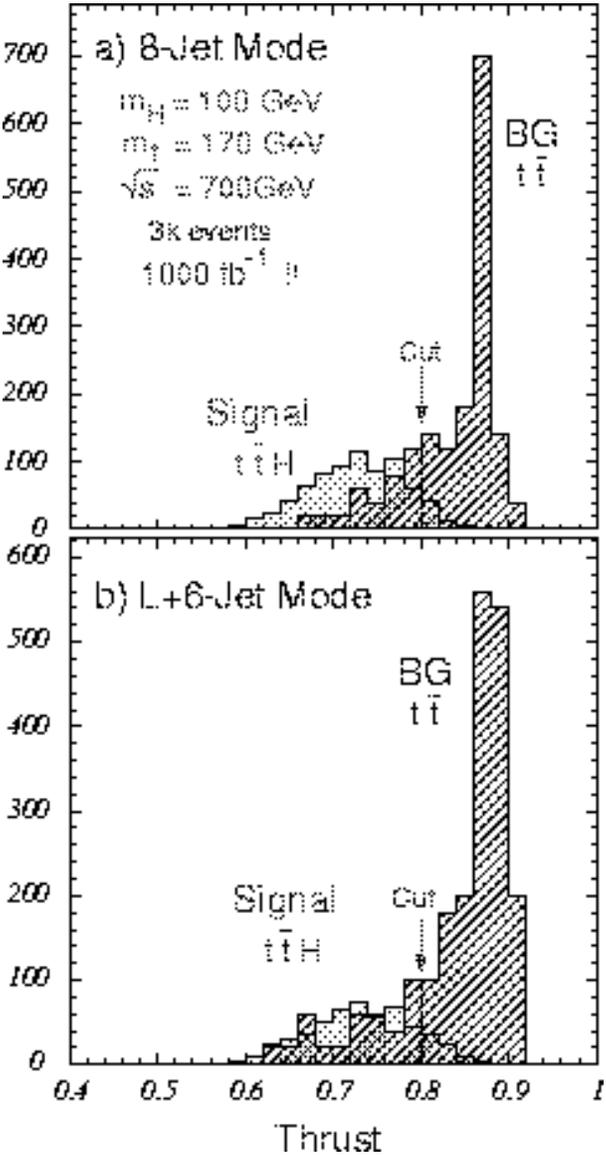}
}
\begin{center}\begin{minipage}{\figurewidth}{
\caption[Fig:thrust]{\sl \label{Fig:thrust}
Thrust distributions for 
the signal $t\bar{t}H$ (solid)
and the background $t\bar{t}$ (hatched) events:
(a) 8-jet mode and (b) lepton-plus-6-jet mode.
}
}\end{minipage}\end{center}
\end{figure}

Next we discuss the S/N ratio.
Figs.~\ref{Fig:thrust}-a) and -b)
are the thrust distributions for
the signal ($t\bar{t}H$) and
the background ($t\bar{t}$) events
for the 8-jet and lepton-plus-6-jet modes,
respectively.
We can see that the signal events are
more spherical than the background
and consequently have smaller thrusts.
The cut at 0.8 thus removes the background
very effectively.
Nevertheless, the $t\bar{t}$
background is still large even after the
rather tight selection described above.
The situation can be improved significantly
by the use of a better vertex detector:
in principle, the requirement of more than
two $b$-jets in the final state reduces
the $t\bar{t}$ background to a negligible
level, as long as the $b$-purity
is near 100\%.

We will, however, assume this S/N
ratio and examine the sensitivity
to the top Yukawa coupling 
in this conservative case.

\subsection{Determination of Top Yukawa Coupling}

With the initial state radiation and beam effects,
the signal cross section
at $\sqrt{s} = 700~{\rm GeV}$ for
$m_t = 170~{\rm GeV}$ and $m_H = 100~{\rm GeV}$
is $\sigma_{t\bar{t}H} \simeq 3.0~{\rm fb}$.
The overall selection efficiencies for the 
8-jet and the lepton-plus-6-jet modes
are 0.23 and 0.16, respectively,
including the branching fractions.
The total selection efficiency thus amounts
to
\begin{eqnarray}
\nonumber
        \epsilon_{t\bar{t}H}=0.23+0.16=0.39.
\end{eqnarray}
On the other hand, the effective background cross sections
are
$\sigma_{t\bar{t}Z} \simeq 4.2~{\rm fb}$
and
$\sigma_{t\bar{t}} \simeq 402~{\rm fb}$,
including the initial state radiation and beam effects.
The corresponding overall detection efficiencies
for the background events are
\begin{eqnarray}
\nonumber
        \epsilon_{t\bar{t}Z} = 0.088({\rm 8J}) + 0.048({\rm L+6J}) = 0.14 \cr
        \epsilon_{t\bar{t}} = 0.0009({\rm 8J}) + 0.0010({\rm L+6J}) = 0.0019.
\end{eqnarray}

The expected number of the signal events for
$100~{\rm fb}^{-1}$ is
$S = 114$ while that for the background
is $B = 133$, resulting in a signal-to-noise ratio
of 0.86.
Making use of the fact that
the number of the signal events is essentially proportional to
the square of the normalized top Yukawa coupling ($\beta_H^2$),
we can translate these numbers to the error on $\beta_H$:
$\Delta \beta_H/\beta_H \simeq 0.14$ for
$100~{\rm fb}^{-1}$ at $\sqrt{s} = 700~{\rm GeV}$.

\section{Probe for anomalous $t\bar{t}H$ couplings at JLC} 
\label{ttH}
In this section, we review the recent theoretical study on probing
the anomalous $t \bar{t} H$ couplings in the process of the Higgs
boson and top-quark associated production \cite{tth}
\begin{equation}
e^+ e^- \to t\bar t H.
\label{tth}
\end{equation}
By scrutinizing this process in detail, one would hope to reveal
the nature of the Higgs and top-quark interactions and hopefully
gain some insight for physics beyond the SM.

\subsection{Effective Interactions} 
As a model-independent approach, anomalous couplings of the 
top quark can be parameterized by a low-energy
effective Lagrangian.
Such an approach has been taken in Ref.~\cite{xinmin}
in a non-linear realization of the gauge symmetry, 
and in Refs.~\cite{linear,Whisnant} in a linear realization with 
an explicit scalar (Higgs boson) field.

In the latter case, the anomalous couplings are parameterized by
a set of 
 higher dimensional operators which contain the SM fields and are
 invariant under the SM gauge group, $SU_c(3)\times SU_L(2)\times U_Y(1)$.
 Below the new physics scale $\Lambda$, the effective Lagrangian 
 can be written as
 \begin{equation}
 \label{eff}
 {\cal L}_{eff}={\cal L}_0+\frac{1}{\Lambda^2}\sum_i C_i O_i
                          +{\cal O}(\frac{1}{\Lambda^4})
                          \end{equation}
  where $\Lambda$ is a cutoff scale above which new physics sets
in, and ${\cal L}_0$ is the SM Lagrangian.
  $O_i$ are dimension-six operators and $C_i$
   represent the coupling strengths of $O_i$ \cite{linear}.  

Following Refs.~\cite{Whisnant,Renard}, we find that there are seven 
dimension-six CP-even operators which give new contributions
to the couplings of $H$ to the top quark, 
\begin{eqnarray}
\label{O1}
O_{t1}&=&(\Phi^{\dagger}\Phi-\frac{v^2}{2})\left [\bar q_L
t_R\widetilde\Phi
+\widetilde\Phi^{\dagger} \bar t_R q_L\right ],\\
O_{t2}&=&i\left [\Phi^{\dagger}D_{\mu}\Phi
-(D_{\mu}\Phi)^{\dagger}\Phi\right ]\bar t_R \gamma^{\mu}t_R,\\
O_{Dt}&=&(\bar q_L D_{\mu} t_R) D^{\mu}\widetilde\Phi
+(D^{\mu}\widetilde\Phi)^{\dagger}(\overline{D_{\mu}t_R}q_L),\\
O_{tW\Phi}&=&\left [(\bar q_L \sigma^{\mu\nu}\tau^I t_R) \widetilde\Phi
+\widetilde\Phi^{\dagger}(\bar t_R \sigma^{\mu\nu}\tau^I
q_L)\right ] W^I_{\mu\nu},\\
O_{tB\Phi}&=&\left [(\bar q_L \sigma^{\mu\nu} t_R) \widetilde\Phi
+\widetilde\Phi^{\dagger}(\bar t_R \sigma^{\mu\nu} q_L)\right ]
B_{\mu\nu},\\
O_{\Phi q}^{(1)}&=&i\left [\Phi^{\dagger}D_{\mu}\Phi
-(D_{\mu}\Phi)^{\dagger}\Phi\right ]\bar q_L \gamma^{\mu}q_L,\\
O_{\Phi q}^{(3)}&=&i\left [\Phi^{\dagger}\tau^I D_{\mu}\Phi
-(D_{\mu}\Phi)^{\dagger}\tau^I\Phi\right ]\bar q_L
\gamma^{\mu}\tau^I q_L ,
\label{O17}
\end{eqnarray}
where $\Phi$ is the Higgs doublet with 
$\widetilde\Phi = i\sigma_2 \Phi^*$,
and $\bar q_L=(\bar t_L, \bar b_L)$.
Similarly, there are seven dimension-six CP-odd 
operators \cite{Yang} which contribute to the couplings of 
$H$ to a top quark,
\begin{eqnarray}
\label{O2}
\overline {O}_{t1}&=&i(\Phi^{\dagger}\Phi-\frac{v^2}{2})
\left [\bar q_L t_R\widetilde \Phi
-\widetilde \Phi^{\dagger} \bar t_R q_L  \right ],\\
\overline {O}_{t2}&=&  \left [\Phi^{\dagger}D_{\mu}\Phi
+(D_{\mu}\Phi)^{\dagger}\Phi  \right ]\bar t_R \gamma^{\mu}t_R,\\  
\overline {O}_{Dt}&=&i  \left [(\bar q_L D_{\mu} t_R) D^{\mu}\widetilde
\Phi
-(D^{\mu}\widetilde\Phi)^{\dagger}(\overline {D_{\mu}t_R}q_L)  \right ],\\
\overline {O}_{tW\Phi}&=&i  \left [(\bar q_L \sigma^{\mu\nu}\tau^I t_R)
\widetilde\Phi
-\widetilde\Phi^{\dagger}(\bar t_R \sigma^{\mu\nu}\tau^I q_L)  \right ]
W^I_{\mu\nu},\\
\overline {O}_{tB\Phi}&=&i\left [(\bar q_L \sigma^{\mu\nu} t_R)
\widetilde\Phi
-\widetilde\Phi^{\dagger}(\bar t_R \sigma^{\mu\nu} q_L)\right ]
B_{\mu\nu},\\
\overline {O}_{\Phi q}^{(1)}&=&\left [\Phi^{\dagger}D_{\mu}\Phi
+(D_{\mu}\Phi)^{\dagger}\Phi\right ]
\bar q_L \gamma^{\mu}q_L,\\
\overline {O}_{\Phi q}^{(3)}&=&\left [\Phi^{\dagger}\tau^I D_{\mu}\Phi
+(D_{\mu}\Phi)^{\dagger}
\tau^I\Phi\right ]\bar q_L \gamma^{\mu}\tau^I q_L .
\label{Olast}
\end{eqnarray}

Operators (\ref{O1})$-$(\ref{Olast}) contribute to both the
three-point coupling $~t\bar tH~$ as well as four-point 
couplings $t\bar t H Z$ and $t\bar t H\gamma$ beyond
the SM. 
Operators $O_{t1}$ and $\overline{O}_{t1}$
give the direct corrections to the top-quark Yukawa coupling 
and is energy-independent.
Only $O_{Dt}$ and $\overline{O}_{Dt}$ contribute
to both three-point and four-point couplings 
and the three-point couplings
are quadratically dependent on energy, 
due to the nature of double-derivative couplings.

\subsection{$t\bar t H$ Production with Non-standard
Couplings }
\begin{figure}
\centerline{
\epsfxsize=13cm\epsffile{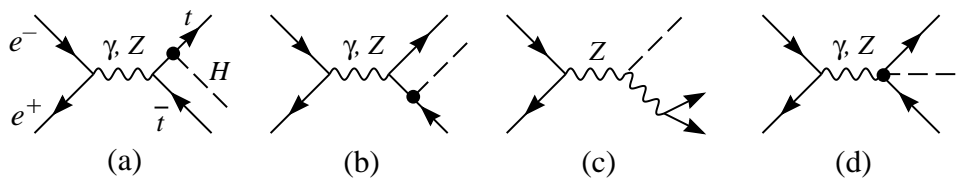}}
\begin{center}\begin{minipage}{\figurewidth}{
\caption{\sl Feynman diagrams for $e^+e^- \to t\bar t H$ production.
(a)-(c) are those in the SM. The dots denote the contribution
from new interactions.
\label{feyn}
}}
\end{minipage}
\end{center}
\end{figure}

The relevant Feynman diagrams for $e^+e^- \to t\bar t H$ 
production are depicted in Fig.~\ref{feyn}, where
(a)$-$(c) are those in the SM and the dots denote the contribution
from new interactions.
For the purpose of illustration, we will only present
results for the operators $O_{t1}$ (energy-independent)
and $O_{Dt}$ (most sensitive to energy scale) and
hope that they are representative to the others with
similar energy-dependence based on the power-counting behavior.
For simplicity, we assume one operator to be non-zero 
at a time in our study.

\begin{figure}
\centerline{
\epsfxsize=7cm\epsffile{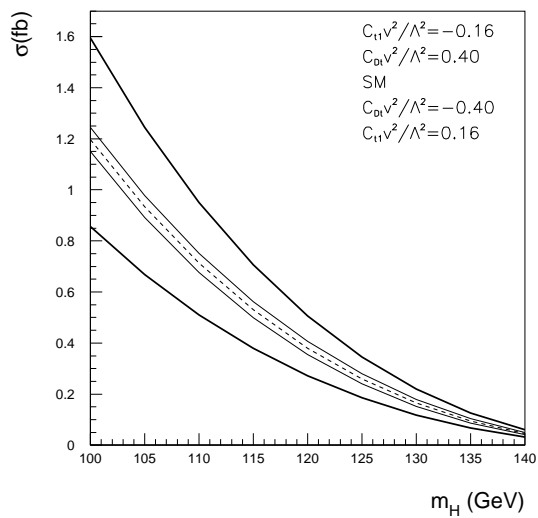}}
\begin{center}\begin{minipage}{\figurewidth}{
\caption{\sl
Total cross section for $e^+e^- \to t\bar t H$ production
versus $m_H$ for $\sqrt s=500$ GeV.
The dashed curves are for the SM expectation.
\label{mhiggs}
}}
\end{minipage}
\end{center}
\end{figure}

\begin{figure}
\centerline{
\epsfxsize=14cm\epsffile{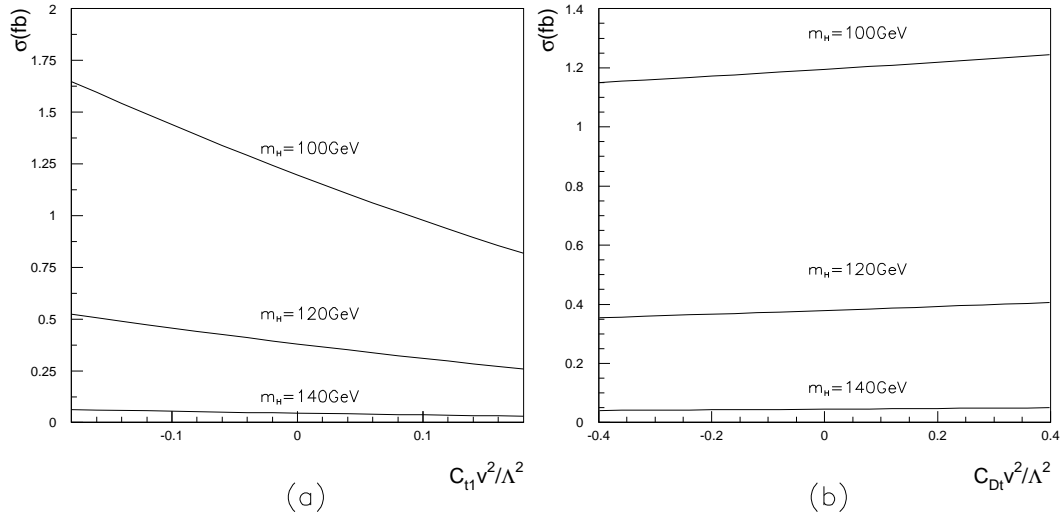}}
\begin{center}\begin{minipage}{\figurewidth}{
\caption{\sl
Total cross section for $e^+e^- \to t\bar t H$ production
versus the couplings 
(a) for $O_{t1}$ and (b) for $O_{Dt}$
at $\sqrt s=500$ GeV with $m_H=100,~120,~140$ GeV.
\label{coup}
}}
\end{minipage}
\end{center}
\end{figure}

In Fig.~\ref{mhiggs} we show the Higgs boson mass 
dependence of the cross section for $\sqrt s=500$ GeV.
A few representative values of the couplings
$C_{t1}$ and $C_{Dt}$ are illustrated. The thick solid curves are
for the operator $O_{t1}$, while the thin solid ones are for $O_{Dt}$.
Fig.~\ref{coup} shows the behavior of the cross sections
as functions of the anomalous couplings.

To establish the sensitivity limits on the non-standard couplings
that may be probed at JLC experiments, 
one needs to consider the identification of the final state
from $t\bar t H$, including the
branching ratios and the detection efficiencies.
For a light Higgs boson of current interest, its leading
decay mode is $H\to b\bar b$. The branching ratio for this
mode is about $80\% \sim 50\%$ for the mass range of
$100\sim 130$ GeV.
To assure a clear signal identification, we require to identify
four $b$-jets in the final state. We assume a 65\% efficiency 
for single $b$-tagging \cite{jackson}.
As for the decays of $W^\pm$ from $t\bar t$, to effectively
increase the signal rate, we include
both the leptonic decay ($e^\pm,\mu^\pm$) \cite{moretti}
and the pure hadronic decay \cite{laura}. These amount to about 85\% 
of the $t\bar t$ events. With the above event selection and 
imposing certain selective acceptance cuts, one expects to 
significantly suppress the QCD and EW background processes
$e^+e^- \to g t\bar t,\ Z t\bar t$ \cite{moretti,laura}.
We estimate an efficiency factor $\epsilon$
for detecting $e^+e^- \to t\bar t H$ to be
$$
\epsilon = 10 - 30\% ,
$$
and a factor $\epsilon '$ for reducing QCD and EW background 
to be
$$
\epsilon '=10\%, 
$$
in our further evaluation.
The background cross sections for QCD,
electroweak and $e^+e^- \to t\bar t H$ 
in the SM at $\sqrt s=500$ GeV without 
branching ratios and cuts included are 
0.84, 0.19, 0.38 $fb$ respectively for $m_H=120$ GeV,
which are consistent with those in \cite{laura}.  
 
To estimate the luminosity ($L$) needed for probing the effects of 
the non-standard couplings, we define the significance of a signal 
rate ($S$) relative to a background rate ($B$) in terms of the
Gaussian statistics, 
\begin{eqnarray}
\label{sb}
\sigma_S = {S\over \sqrt{B}},
\end{eqnarray}
for which a signal at 95\% (99\%) confidence level (C.L.) 
corresponds to $\sigma_S=2\ (3)$. 

\subsection{ CP-even Operators}

\begin{figure}
\centerline{
\epsfxsize=7cm\epsffile{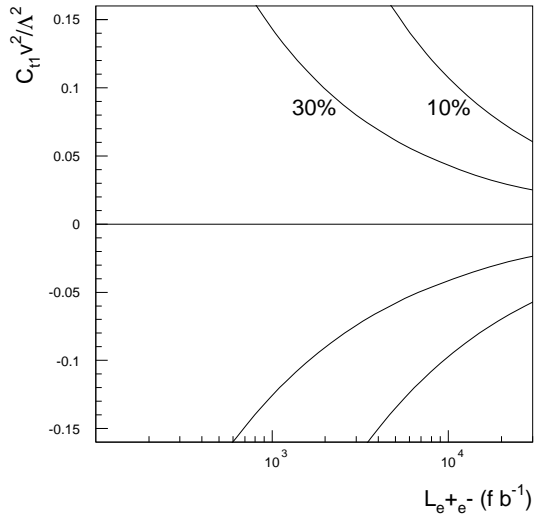}}
\begin{center}\begin{minipage}{\figurewidth}{
\caption{\sl
Sensitivity to the anomalous couplings $O_{t1}$
versus the integrated luminosity for a 95\% confidence level limits
at $\sqrt s=500$ GeV. 
\label{lumi}
}}
\end{minipage}
\end{center}
\end{figure}

In the presence of the CP-even operators, the $t\bar tH$
cross section ($\sigma$) would be thus modified from 
the SM expectation.
The event rates in Eq.~(\ref{sb}) are calculated as
\begin{equation}
S=L(|\sigma-\sigma_{SM}|)\epsilon\ \  {\rm and}\ \  
B=L\left [\sigma_{SM}
\epsilon+(\sigma_{QCD}+\sigma_{EW})\epsilon'\right ].
\end{equation}
We then obtain the luminosity
required for observing the effects of $O_{t1}$ at 95\% C.L. at 500
GeV, which is shown in Fig.~\ref{lumi},
 where the two curves are for 10\% and 30\% 
of signal detection efficiency, respectively.
We see that at a 500 GeV collider, one would need a
high integrated luminosity to reach the sensitivity to
the non-standard couplings.

\subsection{ CP-odd Operators}

If there exist effective CP-odd operators besides the SM 
interaction, then CP will be violated in the Higgs and
top-quark sectors. Similarly to the discussion in the previous
section, one can try to observe the effects of the operators
beyond the SM. 

To unambiguously establish the observation of CP violation, one
needs to examine CP-violating observables. 
The CP-violating effect can be parameterized by a cross section
asymmetry as
\begin{eqnarray}
A_{CP}\equiv \frac{\sigma((p_1\times p_3)\cdot p_4<0)-
\sigma((p_1\times p_3)\cdot p_4>0)}
{\sigma((p_1\times p_3)\cdot p_4<0)+
\sigma((p_1\times p_3)\cdot p_4>0)}
\end{eqnarray}
where $p_1$, $p_3$ and $p_4$ are the momenta of the incoming
electron, top quark and anti-top quark, respectively.
Unfortunately, it requires a higher luminosity to detect the CP-violating
effects.

\section{Probe for Anomalous $t\gamma$, $tZ$, $tW$ Couplings}
\label{anom}
\newcommand{\nc}{\newcommand}
\nc{\postscript}[2] 
{\setlength{\epsfxsize}{#2\hsize}\centerline{\epsfbox{#1}}}
\nc{\bg}{B. Grzadkowski}
\nc{\non}{\nonumber}
\def\dps{\displaystyle}
\def\mib#1{\mbox{\boldmath $#1$}}
\def\sla#1{\mbox{$#1\!\!\scriptstyle{/}$}}
\def\slaq{\mbox{$q\!\!\!\scriptstyle{/}$}\,}
\def\bra#1{\langle #1 |} \def\ket#1{|#1\rangle}
\def\vev#1{\langle #1\rangle}
\def\sst#1{\scriptscriptstyle{#1}}
\def\ssf{\scriptscriptstyle{f}}
\def\ssl{\scriptscriptstyle{\ell}}
\def\ssb{\scriptscriptstyle{b}}
\def\xf{x_{\!\scriptscriptstyle{f}}}
\def\thf{\theta_{\!\scriptscriptstyle{f}}}
\nc{\barx}{\bar{x}}\nc{\pbarn}{\;\hbox {pb}}\nc{\fbarn}{\;\hbox {fb}}
\nc{\hc}{\hbox {h.c.}} \nc{\re}{\hbox {Re}} \def\im{{\rm Im}}
\nc{\mev}{\hbox {MeV}} \nc{\gev}{\;\hbox {GeV}}
\def\gesim{\lower0.5ex\hbox{$\:\buildrel >\over\sim\:$}} 
\def\lesim{\lower0.5ex\hbox{$\:\buildrel <\over\sim\:$}} 

The discovery of the top-quark completed the fermion list
required in the standard EW theory (SM). However it is still an open
question whether this quark interacts with the others the standard
way or there exists any new-physics contribution to its couplings. It
decays immediately after being produced because of the huge mass.
Therefore this process is not influenced by any hadronization effects
and consequently the decay products are expected to tell us a lot
about parent top property. 

Next linear colliders (NLC) of $e^+e^-$ will give us fruitful data
on the top through $e^+e^-\to t\bar{t}$. In particular the final
lepton(s) produced in its semileptonic decay(s) turns out to carry
useful information of the top-quark couplings \cite{SP}. Indeed many
authors have worked on this subject (see the reference list of
Ref.\cite{GH}), and we also have tackled them over the past several
years.

Here we would like to show some of the results of our latest
model-independent analyses of the lepton distributions for arbitrary
longitudinal beam polarizations \cite{GH}, where we have assumed the
most general anomalous couplings both in the production and decay
vertices in contrast to most of the existing works.
What we actually studied are the lepton angular-energy distribution
and the angular distribution, both of which would enable us to
perform interesting tests of the top-quark couplings.

\subsection{Framework}

We can represent the most general covariant $t\bar{t}$ couplings to the
photon and $Z$ boson as
\footnote{
Throughout this report, we use 
simplified expressions. Here, for example, $A_v$ means $A_v +
        \delta\!A_v$ in our original papers.
}
\begin{equation}
{\mit\Gamma}_{vt\bar{t}}^{\mu}=
\frac{g}{2}\,\bar{u}(p_t)\,
\Bigl[\,\gamma^\mu ( A_v-B_v \gamma_5 )
+\frac{(p_t-p_{\bar{t}})^\mu}{2m_t}(C_v-D_v\gamma_5)
\,\Bigr]\,v(p_{\bar{t}})  \label{ff}
\end{equation}
in the $m_e=0$ limit, where $g$ denotes the $SU(2)$ gauge coupling
constant and $v=\gamma/Z$. Among the above form factors, $A_{\gamma,Z}$,
$B_{\gamma,Z}$ and $C_{\gamma,Z}$ are parameterizing
$C\!P$-conserving interactions, while $D_{\gamma,Z}$ is
$C\!P$-violating one. 

On the other hand, we adopted the following parameterization of
the $tbW$ vertex suitable for the $t\to W^+ b$ and $\bar{t}\to W^- \bar{b}$ decays:
\begin{eqnarray}
&&\!\!{\mit\Gamma}^{\mu}_{Wtb}=-{g\over\sqrt{2}}\:
\bar{u}(p_b)\biggl[\,\gamma^{\mu}(f_1^L P_L +f_1^R P_R)
-{{i\sigma^{\mu\nu}k_{\nu}}\over M_W}
(f_2^L P_L +f_2^R P_R)\,\biggr]u(p_t),\ \ \ \ \ \ \label{ffdef}\\
&&\!\!\bar{\mit\Gamma}^{\mu}_{Wtb}=-{g\over\sqrt{2}}\:
\bar{v}(p_{\bar{t}})
\biggl[\,\gamma^{\mu}(\bar{f}_1^L P_L +\bar{f}_1^R P_R)
-{{i\sigma^{\mu\nu}k_{\nu}}\over M_W}
(\bar{f}_2^L P_L +\bar{f}_2^R P_R)\,\biggr]v(p_{\bar{b}}),
\label{ffbdef}
\end{eqnarray}
where $k$ is the $W$-boson momentum and $P_{L/R}=(1\mp\gamma_5)/2$.
This is also the most general form as long as we treat $W$ as an
on-mass-shell particle, which is indeed a good approximation.
It is worth mentioning that these form factors satisfy
$f_1^{L,R}=\pm\bar{f}_1^{L,R}$ and $f_2^{L,R}=\pm\bar{f}_2^{R,L}$,
where upper (lower) signs are those for $C\!P$-conserving
(-violating) contributions.

\subsection{Angular-energy distributions}

After some calculations, we arrived at the following angular-energy
distribution of the final lepton:
\begin{eqnarray}
&&\frac{d^2\sigma}{dx d\cos\theta}
=\frac{3\pi\beta\alpha^2}{2s} B \:
\Bigl[\:S^{(0)}(x, \theta)                        \non\\
&&\ \ \ \ \ \
+\!\!\sum_{v=\gamma,Z}\bigl[\:
 {\rm Re}(\delta\!A_v){\cal F}_{Av}(x, \theta)
+{\rm Re}(\delta\!B_v){\cal F}_{Bv}(x, \theta)
+{\rm Re}(\delta  C_v){\cal F}_{Cv}(x, \theta)    \non\\
&&\ \ \ \ \ \ \ \ \ \ \
+{\rm Re}(\delta\!D_v){\cal F}_{Dv}(x, \theta) \:\bigr]
+{\rm Re}(f^R_2){\cal F}_{2R}(x, \theta)\:\Bigr], \label{A-E}
\end{eqnarray}
where $\beta\:(\equiv\sqrt{1-4m_t^2/s})$ is the top-quark velocity,
$B$ denotes the appropriate branching fraction (=0.22 for $e/\mu$),
$x$ means the normalized energy of $\ell$ defined in terms of its
energy $E$ as $x \equiv 2E\sqrt{(1-\beta)/(1+\beta)}/m_t$,
$\theta$ is the angle between the $e^-$ beam direction and the $\ell$
momentum, all in the $e^+e^-$ CM frame, $S^{(0)}$ is
the SM contribution, $\delta\!A_v \sim \delta\!D_v$ express non-SM
part of $A_v \sim D_v$ (i.e., $\delta\!B_{\gamma}$, $\delta C_v$ and
$\delta\!D_v$ are equivalent respectively to $B_{\gamma}$, $C_v$ and
$D_v$), and ${\cal F}$ are all analytically-expressed functions of
$x$ and $\theta$, which are independent of each other. A similar formula
also holds for the final $b$-quark.

Equation (\ref{A-E}) can be re-expressed as
\begin{equation}
\frac{d^2\sigma}{dx d\cos\theta}
=\frac{3\pi\beta\alpha^2}{2s}B\:
\Bigl[\:{\mit\Theta}_0(x)
+\cos\theta\,{\mit\Theta}_1(x)
+\cos^2\theta\,{\mit\Theta}_2(x) \:\Bigr].
\label{dis1}
\end{equation}
This form directly leads to the angular distribution for $\ell$
through the integration over $x$:
\begin{equation}
\frac{d\sigma}{d\cos\theta}
= \int_{x_-}^{x_+}dx \frac{d^2\sigma}{dx d\cos\theta}=
\frac{3\pi\beta\alpha^2}{2s}B
\left({\mit\Omega}_0+{\mit\Omega}_1
\cos\theta+{\mit\Omega}_2\cos^2\theta\right),   
\label{dis2}
\end{equation}
where ${\mit\Omega}_{0,1,2}
\equiv\int_{x_-}^{x_+}dx\,{\mit\Theta}_{0,1,2}(x)$ and
$x_\pm$ define the kinematical range of $x$.

Surprisingly enough, the non-SM decay part, i.e., $f^R_2$ term completely
disappears through this $x$ integration, and the angular distribution
depends only on the whole production vertex plus the SM decay vertex
\cite{GH,Rin00}.
This never happens in the final $b$-quark distribution.

\subsection{Analyzing the results}

First, we could determine $\delta\!A_v \sim \delta\!D_v$ and $f^R_2$
simultaneously using the angular-energy distribution (\ref{A-E})
via the optimal-observable procedure \cite{opt}, since these anomalous
parameters are all coefficients of independent functions. In the
second paper of Ref.\cite{GH}, we explored the best $e^{\pm}$
polarizations which minimize the expected statistical uncertainty
($1\sigma$) for each parameter.

On the other hand, we can perform another interesting test via
the angular distribution. That is, asymmetries like
\begin{equation}
{\cal A}_{\sst{C\!P}}(\theta)= \Big[\:
{\displaystyle \frac{d\sigma^+(\theta)}{d\cos\theta}-
\frac{d\sigma^-(\pi-\theta)}{d\cos\theta}}
\:\Bigr]\Big/\Bigl[\:
{\displaystyle \frac{d\sigma^+(\theta)}{d\cos\theta}+
\frac{d\sigma^-(\pi-\theta)}{d\cos\theta}}
\:\Bigr]
\end{equation}
or
\begin{equation}
{\cal A}_{\sst{C\!P}}= \frac{
{\displaystyle \int_{-c_m}^{0}\!d\cos\theta
 \frac{d\sigma^{+}(\theta)}{d\cos\theta}
 -\int_{0}^{+c_m}\!d\cos\theta \frac{d\sigma^{-}(\theta)}{d\cos\theta}}}
{{\displaystyle \int_{-c_m}^{0}\!d\cos\theta
  \frac{d\sigma^{+}(\theta)}{d\cos\theta}
 +\int_{0}^{+c_m}\!d\cos\theta \frac{d\sigma^{-}(\theta)}{d\cos\theta}}}, 
\end{equation}

\noindent
where $d\sigma^{\pm}$ are for $\ell^{\pm}$ respectively and $c_m$
expresses an experimental angle cut, are a pure measure of the
$C\!P$-violating anomalous $t\bar{t}\gamma/Z$ parameters.

In Ref.\cite{GH_plb97}, we introduced the following asymmetry
\begin{equation}
A_{\ell\ell}\equiv
\frac
{\dps\int\!\!\int_{x<\bar{x}}dxd\bar{x}\frac{d^2\sigma}{\dps dxd\bar{x}}
 -\int\!\!\int_{x>\bar{x}}dxd\bar{x}\frac{d^2\sigma}{\dps dxd\bar{x}}}
{\dps\int\!\!\int_{x<\bar{x}}dxd\bar{x}\frac{d^2\sigma}{\dps dxd\bar{x}}
 +\int\!\!\int_{x>\bar{x}}dxd\bar{x}\frac{d^2\sigma}{\dps dxd\bar{x}}}
\end{equation}
using the $\ell^{\pm}$ energy correlation $d^2\sigma/dxd\bar{x}$, where
$x$ and $\bar{x}$ are the normalized energies of $\ell^+$ and $\ell^-$
respectively. Generally this is also an asymmetry very sensitive to
$C\!P$ violation. However, when we have no luck and two contributions
from the production and decay vertices cancel each other, we get little
information. This comparison lightens the
outstanding feature of ${\cal A}_{\sst{C\!P}}(\theta)$ and
${\cal A}_{\sst{C\!P}}$ more clearly.

To summarize, we showed here some results of our latest work on the angular and
energy distributions of the lepton ($e$ or $\mu$) produced in $e^+e^-
\to t\bar{t} \to \ell^{\pm}X$. There the most general covariant forms
were assumed both for the $t\bar{t}\gamma/Z$ and $tbW$ couplings, which
makes our analysis fully model-independent.

The angular-energy distribution $d^2\sigma/dx d\cos\theta$ enables
us to determine in principle all the anomalous parameters in the general
$t\bar{t}\gamma/Z$ and $tbW$ couplings simultaneously. Although
extremely high luminosity is required to achieve good precision, it
never means our analysis is impractical. We could get better precision
when we have any other independent information on those anomalous
parameters.

On the other hand, the angular distribution $d\sigma/d\cos\theta$ is
completely free from the non-SM decay vertex. Therefore, once we 
catch any non-trivial signal of non-standard phenomena, it will be
an indication of new-physics effects in $t\bar{t}\gamma/Z$ couplings.
This is quite in contrast to asymmetries using the single or double
energy distributions of $e^+e^-\to t\bar{t} \to \ell^{\pm}X\:/\:
\ell^+ \ell^-X'$, where cancellation between the production and
decay contributions could occur.

\def \eebar{$e^+e^-$}
\def \beq{\begin{equation}}
\def \eeq{\end{equation}}
\def \f{\frac}
\def \cdg {c_d^{\gamma}}
\def \cdz {c_d^Z}
\def \cdgz {c_d^{\gamma,Z}}
\def \el{E_l}

\section{CP violation in the open $t \overline{t}$~region}
\label{CPv-opentop}

Linear colliders can provide a clean environment for the study of CP 
violation in top-quark couplings in the process $e^+e^- \rightarrow$ 
$t\overline{t}$ and also in
$\gamma\gamma\rightarrow $ $t\overline{t}$. 
CP violation in the production process can lead to a definite 
pattern of deviation of $t$ and $\overline{t}$ polarizations from the 
predictions of SM. A specific example is the asymmetry between the rate of 
production of $t_L\overline{t}_L$ and $t_R\overline{t}_R$, where $L,R$ denote
helicities. Since top polarization is measured only by studying distributions
of the decay products, it is advantageous to make predictions for these 
distributions, as far as possible, without reference to details of top 
reconstruction. 

The usual procedure is to study either CP-violating asymmetries or correlations
of CP-violating observables to get a handle on the CP-violating parameters
of the underlying theory. For a given situation, correlations of optimal
CP-violating variables correspond to the maximum statistical sensitivity. It
is however convenient sometimes to consider variable which are simpler and can
be handled more easily theoretically and experimentally.

CP-violation can be studied in $e^+e^-$ collisions as well in the $\gamma\gamma$
collider option. Both these approaches are outlined below.
In either case, it is seen that polarized beams help to increase the 
sensitivity.

\subsection{CP violation studies in $e^+e^- \rightarrow t\overline{t}$}
CP violation in $e^+e^-\rightarrow$ $t\overline{t}$ 
can mainly arise through the couplings of the top
quark to a virtual photon and a virtual $Z$, which are responsible for
$t \overline{t}$ 
production, and the $tbW$ coupling responsible for the dominant decay 
of the top quark in to a $b$ quark and a $W$. The CP-violating couplings of 
a $t \overline{t}$ current to $\gamma$ and $Z$ can be written as $ie\Gamma_\mu^j$,
where
\beq
\Gamma_\mu^j\;=
\;\f{c_d^j}{2\,m_t}\,i\gamma_5\,
(p_t\,-\,p_{\overline{t}})_{\mu},\;\;j\;=\;\gamma,Z,
\eeq
where
$e\cdg/m_t$ and $e\cdz/m_t$ are the electric and ``weak' dipole couplings.
The $tbW$ vertex is parametrized as in eqs. (1.34) and (1.35).

While these CP-violating couplings may be studied using CP-violating 
correlations among momenta and spins which include the $t$ and $\overline{t}$
momenta and spins \cite{lin}, it may be much more useful to study asymmetries and 
correlations constructed out of the initial $e^+/e^-$ momenta and 
the momenta of the decay products, which are more directly observable. 
In addition, the observables using top spin
depend on the basis chosen \cite{parke,lin}, and would require
reconstruction of the basis which has the maximum sensitivity. In studying 
decay distributions, this problem is avoided.

Correlations of optimal CP-violating observables have been studied by Zhou 
\cite{zhou}. Using purely hadronic or hadronic-leptonic variables, limits
on the dipole moment of the order of $10^{-18}$ $e$ cm are shown to be
possible with $\sqrt{s}=500$ GeV and integrated luminosity of 50 fb$^{-1}$.

Examples of CP-violating asymmetries using single-lepton angular distributions
and the lepton energy correlations have been discussed in Sec. 1.6.3. 
In addition, we have
studied, in \cite{pra}, additional CP-violating 
asymmetries which are functions of lepton energy. Using suitable ranges for
the lepton energy, it is possible to enhance the relative contributions of
CP violation in production and CP violation in decay.

One-loop QCD corrections can contribute as much as 30\% to $t \overline{t}$ production
cross section at $\sqrt{s}=500$ GeV \cite{kod}. It is therefore important to include
these in estimates of sensitivities of CP-violating observables. The effect of 
QCD corrections in the soft-gluon approximation in decay lepton distributions in
$e^+e^-\rightarrow$ $t\overline{t}$ were discussed in \cite{sga}. These were incorporated in CP-violating
leptonic angular asymmetries and corresponding limits possible at JLC with 
longitudinal beam polarization were presented in \cite{kek}. These are in the 
laboratory frame, do not need accurate detailed top energy-momentum reconstruction,
and are insensitive to CP violation (or other CP-conserving anomalous effects)
in the $tbW$ vertex. 

Four different asymmetries have been studied in \cite{kek}. In addition to two
asymmetries where the azimuthal angles are integrated over, and which are exactly
the ones defined in \cite{pou}, there are two others which depend on azimuthal
distributions of the lepton. A cut-off $\theta_0$ in the forward and 
backward directions is assumed in the polar angle of the lepton.
 The up-down asymmetry is defined by
\beq
A_{ud}(\theta_0)=\f{1}{2\,\sigma (\theta_0)}\int_{\theta_0}^{\pi-\theta_0}
\left[
\f{d\,\sigma^+_{\rm up} } {d\,\theta_l}
-\f{d\,\sigma^+_{\rm down} } {d\,\theta_l}
+
\f{d\,\sigma^-_{\rm up}} {d\,\theta_l}
-\f{d\,\sigma^-_{\rm down} } {d\,\theta_l}
\right] {d\,\theta_l} ,
\eeq
Here up/down refers to
$(p_{l^{\pm}})_y\;\raisebox{-1.0ex}{$\stackrel{\textstyle>}{<}$}\;0,\;
\:(p_{l^{\pm}})_y$ being the $y$
component of $\vec{p}_{l^{\pm}}$ with respect to a coordinate system
chosen in the $e^+\,e^-$ center-of-mass (cm) frame so that the
$z$-axis is along $\vec{p}_e$, and the $y$-axis is along
$\vec{p}_e\,\times\,\vec{p}_t$.  The $t\bar{t}$ production
plane is thus the $xz$ plane.  Thus, ``up" refers to the range $0<\phi_l<\pi$,
and ``down" refers to the range $\pi<\phi_l<2\pi$. 

The left-right asymmetry is defined by
\beq
A_{lr}(\theta_0)=\f{1}{2\,\sigma (\theta_0)}\int_{\theta_0}^{\pi-\theta_0}
\left[
\f{d\,\sigma^+_{\rm left} } {d\,\theta_l}
-\f{d\,\sigma^+_{\rm right} } {d\,\theta_l}
+
\f{d\,\sigma^-_{\rm left}} {d\,\theta_l}
-\f{d\,\sigma^-_{\rm right} } {d\,\theta_l}
\right] {d\,\theta_l} ,
\eeq
Here left/right refers to
$(p_{l^{\pm}})_x\;\raisebox{-1.0ex}{$\stackrel{\textstyle>}{<}$}\;0,\;
\:(p_{l^{\pm}})_x$ being the $x$
component of $\vec{p}_{l^{\pm}}$ with respect to the coordinate system
system defined above.
Thus, ``left" refers to the range $-\pi /2<\phi_l<\pi /2$,
and ``right" refers to the range $\pi /2<\phi_l<3\pi /2$. 

The simultaneous independent 90\% CL 
limits on the couplings $\cdgz$ which can be obtained 
at a linear collider with $\sqrt{s}=500$ GeV with integrated luminosity 200
fb$^{-1}$, and for $\sqrt{s}=1000$ GeV with integrated luminosity 1000 fb$^{-1}$
and using only $e^-$ longitudinal beam polarization $\pm 0.9$ 
are given in Table \ref{azi}.

\begin{table}[ptb]
\begin{center}\begin{minipage}{\figurewidth}
\caption{\sl\label{azi} Simultaneous limits on dipole couplings combining data from
polarizations $P_e=0.9$ and $P_e=-0.9$, using separately $A_{ud}$ and
$A_{lr}$. Values of
$\sqrt{s}$ and integrated luminosities as in the previous tables.}
\end{minipage}
\end{center}
\vspace{4pt}
\begin{center}
\begin{tabular}{|c|c|c|c|c|c|c|}
\hline
&
\multicolumn{3}{|c|}{$A_{ud}$}&\multicolumn{3}{|c|}{$A_{lr}$}\\
\cline{2-7}
$\sqrt{s}$ (GeV)&  $\theta_0$ & Re$c_d^{\gamma}$ &
 Re$c_d^Z$ &$\theta_0$ & Im$c_d^{\gamma}$ & Im$c_d^Z$ \\
\hline
500 & 25$^{\circ}$ & 0.031& 0.045& $35^{\circ}$ & 0.031& 0.056\\
1000& 30$^{\circ}$ & 0.0085& 0.013 &$60^{\circ}$ & 0.028 & 0.052 \\
\hline
\end{tabular}
\end{center}
\end{table}    

As can be seen from the table, the limits on the dipole couplings 
are of the order of a few times 10$^{-17}$ $e$ cm for $\sqrt{s}=500$ GeV.

\subsection{CP violation studies in $\gamma\gamma \rightarrow t\overline{t}$}

Asakawa et al. \cite{asa} study the possibility of determining completely the 
effective couplings of a neutral Higgs scalar to two photons and to 
$t\overline{t}$ when CP is violated. They study the effects of a neutral Higgs 
boson without definite CP parity in the process $\gamma\gamma \rightarrow 
t\overline{t}$ around the pole of the Higgs boson mass. Near the resonance,
interference between Higgs exchange and the continuum SM amplitude can be
sizeable. This can permit a measurement in a model-independent way of 6 coupling
constant combinations by studying cross sections with initial beam polarizations 
and/or final $t$, $\overline{t}$ polarizations. Using general (circular
as well as linear) polarizations for
the two photons, and different longitudinal polarizations for the $t$ 
and $\overline{t}$, in all 22 combinations 
could be measured, which could be used to determine the 6 parameters of 
the theory. Of these, half are CP-odd, and the remaining are CP-even.
They also consider a specific example of MSSM, where CP-odd measurements are 
sensitive to the case of low $\tan\beta$.

In an earlier work, Ma et al. \cite{ma} discussed the 
CP-violating couplings of 
neutral Higgs in the context of a two-Higgs doublet model. They studied the 
CP-violating asymmetries 
\begin{equation}
\xi_{\rm CP} = \frac{\sigma_{t_L\bar{t}_L} - \sigma_{t_R\bar{t}_R}}
{\sigma_{t\bar{t}}},
\end{equation}
for the case of unpolarized photon beams, and
\begin{equation}
\xi_{\rm CP,1 } = \frac{\sigma^{++}_{t_L\bar{t}_L} 
- \sigma^{--}_{t_R\bar{t}_R}}
{\sigma^{++}_{t_L\bar{t}_L} + \sigma^{--}_{t_R\bar{t}_R}}
\end{equation}  
and
\begin{equation}
\xi_{\rm CP,2 } = \frac{\sigma^{++}_{t_R\bar{t}_R} 
- \sigma^{--}_{t_L\bar{t}_L}}
{\sigma^{++}_{t_R\bar{t}_R} + \sigma^{--}_{t_L\bar{t}_L}}
\end{equation}  
for the case of circularly polarized photon beams, where the superscripts on
$\sigma$ denote the signs of the photon helicities, and the subscripts $L$ and
and $R$ denote left- and right-handed polarizations for the quarks. They found
that asymmetries of the order of 10$^{-4}$ -- 10$^{-3}$ can get enhanced to 
the level of a few percent in the presence of beam polarization, for reasonable
values of the model parameters.
Similar CP-violating observables have been identified in the MSSM by M.-L. Zhou et
al.\cite{mlzhou}.

CP-violating dipole couplings of the top quark to photons can be studied at 
$\gamma\gamma$ colliders. The advantage over the study using $e^+e^-$ collisions
is that the electric dipole moment is obtained independent of the weak dipole
coupling to $Z$. Choi and Hagiwara \cite{choiedm} and Baek et al. 
\cite{baek} have proposed the study of the number asymmetry of top quarks with 
linearly polarized photon beams, and found that a limit of about 10$^{-17}$
$e$ cm can be put on the electric dipole moment (edm) of the top quark with an 
integrated $e^+e^-$ luminosity of 20 fb$^{-1}$ for $\sqrt{s}=500$ GeV. 
Poulose and Rindani \cite{pp} studied asymmetries of charged leptons from 
top decay in $\gamma\gamma\rightarrow$ $t\overline{t}$ with longitudinally polarized photons. These 
asymmetries do not need full reconstruction of the top or anti-top. Limits 
at the 90\% confidence level 
of the order of $2\times 10^{-17}$ $e$ cm on (the imaginary part of) 
the top edm can be obtained with
an $e^+e^-$  luminosity of 20 fb$^{-1}$ and cm energy $\sqrt{s}=500$ GeV and 
suitable choice of electron beam and laser photon polarizations. The
limit can be improved by a factor of 8 by going to $\sqrt{s}=1000$ GeV.
It should be emphasized that the method relies on direct observation of
lepton asymmetries rather than top polarization asymmetries, and hence 
does not depend heavily on the accuracy of top reconstruction.

Poulose and Rindani \cite{pp2} have  also considered the asymmetries discussed
in \cite{pp} to study the 
simultaneous presence of the top edm and an effective CP-violating 
$Z\gamma\gamma$ coupling. By using two different decay-lepton asymmetries,
the top edm coupling and the $Z\gamma\gamma$ coupling can be studied in 
a model-independent manner.

\section{Probe for R-violating top quark decays}
\label{R-v}

In this section we review the possibility of observing exotic 
top quark decays via $R$-Parity violating SUSY interactions 
in $e^{+}e^{-}$ collisions at $\sqrt{s}= 500~GeV$ \cite{top_rv1}. 
We present cross-sections for $t\bar t$ production followed by the 
subsequent decay of either the $t$ or $\overline{t}$
via the $R$-Parity violating interaction while the other undergoes the 
SM decay. We discuss kinematic cuts that allow 
the exotic SUSY decays to be detected over standard model
backgrounds. Discovery limits for $R$-Parity violating couplings 
in the top sector are presented assuming an integrated luminosity 
of $100\,fb^{-1}$. 

\subsection{R-violating top quark decays}
In the minimal supersymmetric model (MSSM), R-parity violation
can induce exotic top decays. As an example, we examine the feasibility 
of detecting $B$ violating $R$-parity interactions 
({\em i.e.} $\lambda^{''}$ couplings only)
in top production and decay in $e^{+}e^{-}$ collisions at $\sqrt{s}= 500~GeV$.
Unlike the case of a hadron collider, where $B$ violating couplings lead to
new $\overline{t}t$ production mechanisms, at an $e^+e^-$ collider the effect of 
$B$ violating couplings has no effect on top pair production.  We thus focus 
on exotic top decay modes induced by $B$ violating couplings.
Furthermore, we assume that the decay of either the $t$ or the 
$\overline{t}$ proceeds via the $R$-parity Violating interactions; 
one quark thus decays via Standard Model channels. With the restriction to 
$B$-Violating couplings only, the possible exotic decay modes are 
\begin{eqnarray}\label{BV}
t\to \tilde{\bar d_i}  \bar d_j,~\tilde{\bar{d}}_j \bar{d}_i
          \to \bar d_i  \bar d_j \tilde \chi^0_1 
\end{eqnarray} 
Among the decay modes which are relatively easy to detect are 
those induced by $\lambda^{''}_{3j3}$. 
To keep the analysis simple we assume that 
either one, but not both, of the tri-linear coupling just mentioned takes a 
non-vanishing value. 
In our analysis we focus on the case of where $\lambda''_{313}$ is 
non-vanishing. As shown in eq.(\ref{BV}), the decay 
$ t\to \bar b  \bar d \tilde \chi^0_1$ can proceed through exchange of
a sbottom ($\tilde b$)  or a down squark ($\tilde d$).  
Since among the down-type squarks the sbottom is most likely to be 
significantly lighter than others, we assume the channel of exchanging 
a sbottom gives the dominant contribution.
Since only a light sbottom is meaningful to our analysis, the dominant 
decay mode of the sbottom is $\tilde b\to b \tilde \chi^0_1$. 
The charged current 
decay mode $\tilde b\to t \tilde \chi^+_1$ is kinematically forbidden for
a light sbottom in our analysis. We do not consider the strong decay mode
$\tilde b\to b \tilde g$ since the gluino $\tilde g$ is likely
to be heavy. 
Note that the LSP ($\tilde \chi^{0}_{1}$) is no longer stable when R-parity is 
violated. In case just one R-violating top quark coupling does not vanish,
the lifetime of the LSP will be very long, depending on the coupling and the
masses of squarks involved in the LSP decay chain. So we assume the LSP decays 
outside the detector.

\subsection{Signal and backgrounds}
The final state in the exotic decay of the $t$ or $\overline{t}$
will consist of two jets accompanied by missing energy. If in addition 
we consider purely hadronic standard model $t(\overline{t})$ decays, 
we will have a very distinctive signal consisting of five jets and missing 
energy. (The inclusion of semi-leptonic
standard model decays not only does not increase the signal by much but 
also complicates the reconstruction of the top pair due to multiple sources
of missing energy.)

The main standard model backgrounds are \newline
\noindent $\bullet$ $W^{+}W^{-}Z$ production 
with the subsequent decay of 
$W^{+}W^{-}$ to five partons and $Z\rightarrow\overline{\nu}\nu$ \newline
\noindent $\bullet$ $Z+5\rm{jets}$ production.

In order to isolate the signal we impose the following phase space cuts:

\noindent $\bullet$ Each jet must have an energy of at least 20 GeV \newline
$\bullet$ There must be a missing $\mbox{p}_{\mbox{t}}$ of at least 20 GeV 
\newline 
$\bullet$ The invariant mass of at least one combination of three jets 
must lie within 10 GeV of 
$m_{t}$, and the invariant mass of two of the these three jets 
must lie within 5 GeV of $m_W$. \newline
$\bullet$ The invariant mass of the remaining two jets and the invisible  
particles must lie within 10 GeV of $m_{t}$. \newline
$\bullet$ The absolute value of the cosine of the angle made by any jet 
with the beam axis must be larger than 0.9. \newline
$\bullet$ For all jets we require the $y_{ij}$ be larger than 0.001 for all
values of $i$ and $j$.
$y_{ij}$ is defined by \begin{displaymath}
\frac{2 \mbox{min}(E_{i}^{2},E_{j}^{2})(1-\cos{\theta_{ij}})}{s}
\end{displaymath} where $i$ and $j$ denote jet indices and run from 1 to 5.

With the cuts listed above, $WWZ$ production gives a background of 
less than one event with a luminosity of 100$fb^{-1}$ and is thus 
small. Estimating the background from (Z +5 jet) production is more tricky due
to the huge number of different graphs involved. Furthermore, large NLO QCD 
corrections may be expected in multi-parton final states \cite{moretti2}. 
Rather than attempt an exact calculation, we will use the numerical results 
for 6 jet production \cite{moretti2} to put an upper bound on this background. 
The cross-section for 6 jet production at $\sqrt{s}$ = 500 GeV with the $y$ cut 
alone is 22 fb, adding the other cuts listed above reduces the phase space by a 
factor of 200. We may 
thus use as an upper limit on the cross-section for (Z + 5 jet) production 
a value of about .15 fb, including a K-factor of 1.5 
to be conservative.
Taking into account the Z branching fraction to neutrino, leads us to a
cross-section of .03 fb, corresponding to an irreducible background of 
3 events with an integrated luminosity of 100 $\mbox{fb}^{-1}$. 
Combining the two backgrounds gives a total of four events.

\subsection{Numerical results} 
For a representative set of 
SUSY parameters: $M=150 {\rm ~GeV}, \mu=300 {\rm ~GeV}, \tan\beta=10$, 
the values of $\lambda^{''}_{313}$ and $m_{\tilde{b}}$ corresponding to 
the discovery level ($5\sigma$)  are displayed in Fig.~1.
For comparison, the results of Tevatron Run 2 (2 fb$^{-1}$), Run 3 
(30 fb$^{-1}$) and LHC (100 fb$^{-1}$) are also presented, which
are taken from \cite{top_rv2}, but renewed by using the new SUSY parameter 
values. 
The current upper bounds on $\lambda''_{313}$, obtained from $Z$ decays 
at LEP I \cite{Zdecay}, are about 0.5 at $1\sigma$ level and 1.0 at 
$2\sigma$ level for squark mass of 100 GeV. For heavier squarks, the bounds 
get weaker because of decoupling property of the MSSM. So one sees from
Fig.~1 that for $0.1 < \lambda^{''}_{313} < 1 $, 
the signal is observable for a sbottom lighter than about 
160 GeV. 
 
\begin{figure}[htb]
\centerline{\epsfxsize=7.cm \epsffile{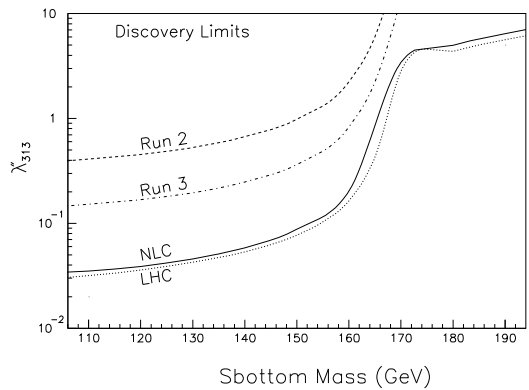}}
\begin{center}\begin{minipage}[t]{\figurewidth}
\caption[]{\sl\label{fig.1}
 The discovery ($5\sigma$) limits of $\lambda''_{313}$ versus 
       sbottom mass. The region above each curve is the 
       corresponding region of discovery. 
}
\end{minipage}\end{center}
\end{figure}

Note that the signal contains like 
sign $b$ quarks, in contrast to the background. In case of a positive
signal, $b$ tagging will present additional evidence for non-standard physics. 
The results can also be applied to the case of the presence of 
$\lambda''_{312}$ with sbottom replaced by strange-squark.

To summarize, we have calculated the cross-section for $R$-parity violating
$t$ decays in $e^{+}e^{-}$ collisions at $\sqrt{s}=500$GeV. The 
standard model backgrounds can be minimized with suitable cuts leading
to discovery bounds about as stringent as at the LHC\cite{top_rv2}.

\section{Summary}
\label{phystop-summary}

In this chapter we have collected the research results of the ACFA
Top Quark Working Group during the period 1998--2001.

Physics studies in the $t\bar{t}$ threshold region will be one of
the major subjects of future experiments at JLC.
Accordingly there have been extensive studies in the working group.
Especially theoretical understandings of the physics in this region
have developed remarkably.
Also, we have initiated detailed Monte Carlo studies focusing in
this energy region and they are still on-going.
The most important outcome up to now is the establishment of a
future prospect that we will be able to measure the top quark 
$\overline{\rm MS}$ mass precisely ($\delta m_t \approx 50$--150~MeV).
Furthermore, we have clarified that a unique study of various properties
of the top quark will be possible.

Studies of the top quark properties in the energy region far above the
threshold extend to wide varieties.
A possibility of the top quark mass determination from the dilepton
channel, under a fairly clean experimental environment, is demonstrated.
The Monte Carlo study on the measurement of the top quark Yukawa coupling
has put a benchmark of achievable accuracy of this coupling.
The theoretical study on probing the anomalous $t\bar{t}H$ coupling
has elucidated an interesting possibility of probing the electroweak
symmetry breaking physics through the top quark interactions with the
Higgs boson.
Even through detailed study of the Standard-Model vertices of the top 
quark, which can be extracted from the decay processes of the top quark,
JLC will enable us to provide hints to new physics.
The study of $CP$-violating interactions of the top quark is interesting
in its own right.
Sensitivity studies to these interactions have been performed and reviewed.
Also, the top quark can be used as a probe for $R$-parity violating SUSY 
interactions.

Thus, we have shown that from the detailed 
studies of the top quark physics at JLC, we will be able to
study deeply into the the Standard-Model physics and beyond.

%% file: physqcd/main.tex
\chapter{QCD and Two Photon Physics}

\section{Introduction}
~Gauge theory is well known from QED and non-Abelian gauge
theories contributed much to the unification of electromagnetic and weak
interactions\cite{physqcd-int}. After the renormalization of unified theory was carried out
by t'Hooft and others, people gave serious considerations to the
non-Abelian gauge theory in the strong interactions of hadrons and finally
found the asymptotic freedom of this theory\cite{physqcd-the}. It has been proved
that only non-Abelian gauge theories are asymptotically free and several
coupling types were studied within the newly introduced color gauge
group. 

~Before the advent of the non-Abelian gauge theory, Gell-Mann
proposed a mechanism for quark imprisonment and the concept of color 
was introduced to explain some difficulties of the quark model. He
named this model the quantum chromodynamics(QCD). Here gluons
mediate the strong force and these gluons can be made to be the
non-Abelian gauge fields. The infrared behavior of the non-Abelian gauge theory attracted
much interests because of the possible relations to quark confinement.
However, in this infrared region, perturbation theory fails and the
problem of infrared slavery is not completely resolved. The non-perturbative
features are closely related to bound state problems formed by strong
forces such as mesons and baryons which are present in the final states
of high energy scattering experiments. In this way, the gluonic behaviors
come to the focus of many theoretical attempts resulting in various
phenomenological models.
 
 In QED, bound state problems can be solved exactly and predictions
can be made explicitly by calculating spin-dependent forces since the
potential form is well-known. On the other hand, the potential form for
bound states is not known in  QCD, and so there exist various potential
models\cite{physqcd-mod}.  Moreover, the conventional perturbation theory in powers of inverse
quark mass and strong coupling constant $\alpha_s$ cannot be applied securely for
low mass hadrons. Therefore the non-relativistic systems of $J/\psi$  and $\Upsilon's$
were important in confirmation of the QCD calculations concerning spin-dependent
forces\cite{physqcd-for}.  Although the explicit form of confining potential cannot be obtained
from a first principle formalism such as Wilson loop, the spin-dependent
corrections to the static energy have been obtained by considering relativistic
propagator expansions. The obtained spin-dependences can be used to classify
and predict hadronic mass spectra, where the variation of $\alpha_s$ can be
confirmed. Of course, the variations of $\alpha_s$ in high energy region have 
been checked by many experiments which are the supporters of the
asymptotic freedom.

 Deep inelastic scatterings in processes such as electro-production,
lepton-hadron, $e^{+} e^{-} \rightarrow $ hadrons and $\nu$ or $\bar{\nu}$ scatterings have shown
scaling phenomena. In order to explain these phenomena, Feynman has
made an induction about the hadronic structure that at extreme energies
hadrons would behave like a composite object of free point-like
particles called partons\cite{physqcd-par}. Theoretically Wilson considered the descriptions
of scale breaking by exploiting the idea of renormalization group and
finally suggested the concept of momentum dependent coupling constant\cite{physqcd-con}.
After the two phenomenological and field theoretical approaches to explain
the scaling behavior were attempted, Lagrangian field theories were examined
to combine the two aspects. The combination of these two aspects can
be done by finding out a Lagrangian which becomes asymptotically free.
A theory is called asymptotically free if it has a fixed point as the
momentum goes to infinity and if the coupling constant vanishes at this
point. Various attempts were made and it became clear that the non-Abelian
gauge theory had to be considered because it contributed much to the
unification of the weak and the electromagnetic interactions. In the long run,
it has been proved that no renormalizable field theory without non-Abelian
gauge fields can be asymptotically free if the Bjorken scaling is accepted
as the evidence for asymptotic freedom. Thus there results the  quark-gluon model for hadrons.

 In a conventional quark parton model, infinite number of quarks are 
embeded in the nucleons and these quarks are separated into valence
and sea parts. This is the dual valence-plus-sea model and infinite
number of uncorrelated quark-antiquark pairs are introduced. However, the
gluons which are known to carry about half of the nucleon momentum and
play essential roles in asymptotically free theories had been simply ignored.
If the non-Abelian guage theories are to be considered more seriously,
the gluons must have some contributions to experimental observations and so
it has been proposed that the vaguely assumed sea of quark-antiquark pairs
have to be replaced by the effects of the gluon partons. This modified
quark parton model assumes that the hadrons are composed of valence quarks
and gluons only and the sea shows up via the pair production of the
gluons. In order to find the gluon  distribution compared with the quark
distributions, it is necessary to consider processes in which contributions
both from gluon and quark components appear explicitly. For example, one photon
inclusive processes in proton-proton collision have the desired feature\cite{physqcd-fea}, and
in this way the shape and the momentum dependences of gluon distributions
have been investigated. The resulting successes of these approaches and the
remaining problems will be discussed in the following sections.
	
\section{Perturbative QCD}
	\subsection{Quark-Gluon Model}
~The quark-gluon model of hadrons can be applied to fix the
initial states of hadronic scatterings and to predict the final states
formed via fragmentations and hadronizations. For initial states,
main researches have been concentrated on the determinations of quark
and gluon distributions in a proton since the proton is the most
suitable one to accelerate. Since the first proposal of partons in
connection with scaling phenomena, many detailed arguments have been
made by Feynman and scattering experiments have been suggested to
test models. The simplest model to give the partons suitable quantum
numbers is the quark model, however, the proton cannot be explained
simply by 3 valence quarks. To describe the proton structure, sea
quarks were introduced rsulting in six distribution functions $u(x), d(x)$,
$s(x), \bar{u}(x), \bar{d}(x), \bar{s}(x)$ in case of 3-quark model\cite{physqcd-del}. These functions are
representing the numbers of quarks and antiquarks in a proton with
momentum between $x$ and $x+dx$. From the condition that proton has
unit charge, $\frac{1}{2}$ isospin and zero strangeness, we can show that the
net number of each kind of quark is just the number of that kind
in conventional 3-quark model. The detailed forms of the dsitribution
functions were derived from experimental comparisons and modified
repeatedly by new data. To test the models of this type, Drell
suggested two kinds of experiments ; lepton pair production in proton-proton 
collisions and the $e^{+} e^{-}$ annihilation into hadrons. Lepton pair
production will be discussed in the next section, and there appears one
more serious contribution. In the quark-parton model analysis, it had
been found that about half of the proton momentum must be carried
by neutral particles, for example, the gluons. These gluon contributions
were thought to be of the same form as the quark contributions,
but concrete calculations were not possible before the advent of
asymptotically free theories of strong interactions. For the non-Abelian
gauge theory, the gluons mediate the strong force, and for sufficiently
high energies the effective gluon coupling constant becomes small enough
so that perturbative calculations can be carried out.

 For quark partons, the $Q^{2}$ dependences of the distribution functions
have been investigated in asymptotically free theories. The results
can be summarized with increasing $Q^{2}$ as\cite{physqcd-as}
\begin{verse}
(1) The momenta carried by valence quarks decrease, and those carried by sea quarks increase.

(2) The average value $\langle x \rangle$ decreases for both  valence and sea quark distributions.

(3) Both valence and sea quark distributions decrease at large $x$. 

(4) There is a strong(weak) increase of the sea(valence) distribution at small $x$.
\end{verse}
~The momentum dependences of gluon distributions have been investigated
by comparing with the quark distributions, and the sea quark contributions
had been shown to be replaced by gluon contributions.
\subsection{Drell-Yan Process}
~Lepton pair can be produced only via a virtual photon when
two protons collide, and it has been suggested that one quark in
one proton and one antiquark in the other annihilate into a
photon which then creates a lepton pair. This mechanism has been
named as Drell-Yan process\cite{physqcd-pro}, however, sea contributions are
inevitable to introduce antiquarks. In quark-gluon model, a nucleon
is taken to be composed of valence quarks and gluons only, and the
sea of quark-antiquark pairs in conventional quark parton model can
be produced by gluonic interactions. In this model, a gluon parton
in one proton and a quark parton in the other produce a virtual
photon via Compton-like scattering, and the produced photon subsequently 
decays into a lepton pair. With appropriate introduction of
gluon and quark distributions in a proton, the calculated results have
been shown to agree very well with experimental data\cite{physqcd-dat}.

~The differences between conventional quark parton model and the
new quark-gluon model exist in the coupling types between constituent
partons. In a conventional model, quark partons are taken to be
free from each other resulting in scaling behavior. As a matter of
fact, scaling behaviors in experimental data led to the idea of
free partonic picture. On the other hand, gluons are interacting with
other gluons and quarks in quark-gluon model, and these interactions
become larger for lower energy scales. These effects have been checked
as logarithmic violation of scaling in the calculation of cross sections for
Drell-Yan process. The scaling violation becomes larger in lower
energy region. These variations are due to the couplings between
partons, and the coupling effects have been analyzed systematically
resulting in the introduction of splitting functions\cite{physqcd-fun}. These splitting
functions are used to define fragmentation functions which play
important roles in predicting hadronic final states in high energy scatterings.

\subsection{Direct Photon Production}
~Since gluons constitute hadrons in quark-gluon model, it is
important to check the gluon distributions in hadrons. The
checking process had been started from a process in which the
gluon distributions can be compared with those of quarks. Direct
photon production processes in proton-proton collisions\cite{physqcd-col} are suitable
ones not only to check the gluon distribution functions but also
to get information about the strong coupling constant $\alpha_s$. By taking
the quark distributions obtained from deep inelastic scatterings,
we can calculate the cross sections for three different final
states $\gamma + \gamma + X$, $ \gamma +$ gluon jet  $+ X$, and $\gamma +$ quark jet $+ X$.
The first two processes have nearly the same form of cross sections
except the fact that instead of one photon in the first one,
a gluon couples with different coupling constant in the second process.
Therefore we can get information about  $\alpha_s$ by comparing
these two processes. The predictions for $\alpha_s$ can be made
conveniently by choosing the special case of both the photon
and the gluon jet coming out at $90^{\circ}$ in the center of mass
system.

~The information about gluon distribution functions can be 
obtained by comparing the gluon jet case with that of quark jet.
The quark jet case corresponds to inverse Compton scattering
with initial gluon so that the ratio of two cross sections including
quark jet and gluon jet depends explicitly on the gluon distribution.
Of course, the functional form of gluon distribution changed from
time to time by the effects of experimental data. As the scattering
energy goes up, the functional form at smaller $x$ values becomes
on the focus and the more precise values of $\alpha_s$ needed. The relation
between $\alpha_s$ and $g(x)$, the gluon distribution function, is well-known,
and we need to analyze evolution equations established by exploiting
the splitting functions introduced to account for scaling violations\cite{physqcd-vio}. The
perturbative features appearing in high energy and large $x$ region gradually
change into non-perturbative ones as the related momentum decreases and
the value of $x$ becomes smaller. The transition region has been
studied extensively resulting in extraction of $g(x)$  for very small $x$
values\cite{physqcd-val}. However, the intrinsic form of $g(x)$ for small $x$ values is unknown
in these perturbative approaches.
\subsection{Hadron Spectra}
~Perturbative QCD calculations are very useful in estimating
the spin splittings of hadronic masses. In order to calculate
hadronic masses, we need to adopt a formalism which can be applied
to the treatment of bound states between quark and antiquark.
The description of bound states between a quark and an antiquark
cannot be done as in QED because the exact form of confining
potential has not been obtained from first principles. In QCD,
the static potential results from an infinite set of graphs through
the interactions of full Yang-Mills couplings, and therefore we have
to introduce an approximation method to treat these infinite graphs.
One method is to calculate directly by computers on suitably 
chosen lattices\cite{physqcd-lat}. This method turns out to be quite successful,
however, there exist restrictions on the number of lattice sites and
critically this theory becomes confining even in QED case.

 Another approximation method to account for the confinement is
to introduce a boundary which forms a bag containing quark,
antiquark, and gluons. In this bag model, the quarks move freely
and relativistically and are described by eigenmodes determined by
the form and the size of the bag. Hadron spectra can be
calculated and fitted fairly well to the observed values, but if
we want to calculate diagrams containing internal propagators such
0as annihilation diagrams, we have to write down the propagators
as sums over bag-model eigenmodes. These propagators result
not only in algebraic complexity but also in some ambiguity
concerning the bound state description connected with the boundary
conditions imposed on the eigenmodes.

 The other method to account for the confining potential is to
assume an appropriate potential form. This method is known to be
convenient for quantitative calculations of mass spectra and decay
processes. The non-relativistic potential model calculations were first
carried out for charmonium system $(c\bar{c})$ just after the $J/\psi$ state
had been observed. Later, the same calculations were applied to the
heavier bottomonium system $(b\bar{b})$ and it was realized that the spin
dependences  between the quark and the antiquark should be accounted for
in a systematic description of quarkonium systems\cite{physqcd-sys}. Although there had
been several attempts to derive spin-dependent forces, the derivations of
Eichten and Feinberg have provided a clear basis for other calculations\cite{physqcd-for}.
They used the Wilson loop formalism and obtained the spin-dependent
potentials up to order $\frac{1}{m^2}$ and $\alpha_s$. These first-order results can
be used to explain nearly all the meson masses except for several
states including the lowest-lying isoscalar-pseudoscalar mesons. The
successful account for spin-dependences can be made possible by considering
gluonic exchanges between quarks with perturbative expansion in $\alpha_s$ .

 We can estimate the values of $\alpha_s$ for different combinations of
quark flavors by comparing observed mass spectra with the theoretical
predictions\cite{physqcd-pre}. The estimated values seem to fulfill the condition of
running in low energy region. However, the related energy scales
could not be easily fixed in bound state problems in contrast to
the high energy scattering cases. One possibility is to use the calculated
effective quark masses which contain averaged momentum values in
bound states for given quark combinations.

 \subsection{Fragmentations and Jets}
~In high energy scattering experiments, the multiparticle distributions
associated with jets have important information about QCD behavior
such as the running of strong coupling constant $\alpha_s$, higher order
contributions, fragmentation schemes, discrimination of gluon jet from
quark jets, and so on. The information can be converted into physical
insights by comparing with theoretical predictions. However, in theoretical
viewpoints, the predictions for QCD processes at large momentum transfer
are given only by factorization scheme like\cite{physqcd-lik}
\begin{equation}
\label{qcd-eq-2-5-1}
d\sigma = d\hat{\sigma}(Q, \mu) \otimes F(\mu, \Lambda_{QCD} ) + O(\frac{\Lambda_{QCD}}{Q}) ,
\end{equation}
where $Q$ is the momentum transfer and $\mu$ is the relevant scale.
Here $d\hat{\sigma}$ can be calculated perturbatively as functions of $\alpha_s$ treating
the scattering process in terms of quark-gluon interactions. The second
factor $F(\mu, \Lambda_{QCD} )$ contains all long distance effects which are essential
in experiments because every measurement is done macroscopically. Of
course $F$ also depends on $\alpha_s$, but in  this case $\alpha_s$ becomes large
enough resulting in non-perturbative situation.

 The perturbative processes have been calculated by considering more
loops and legs with the improvements on scale variations included. For
QCD $\beta$-function and quark mass anomalous dimension, 4-loop calculations
have been carried out by automated computer programs treating as much
as $10^4$  Feynman diagrams\cite{physqcd-dia}. These calculations are closely related to
higher order predictions of jets, but jet calculations become more
difficult because the final state kinematics are complicated for
many jet systems. For example, NLO corrections to $e^+ e^- \rightarrow 4$ jets
and  $e^+ e^- \rightarrow 3$ jets with quark mass effects have been estimated
recently. The NNLO jet calculations are the subjects of forefront
researches dealing with several steps related to the whole picture
of jet phenomena. They provide detailed insights into jet structure
and can be used to reduce the error bar in $\alpha_s$, which are both
important in understanding the factorization scheme.

 For event shape observables such as thrust, jet masses, C parameter,
and so on, we can define average event shape variable $\langle R \rangle$ which
satisfies
\begin{eqnarray}
\langle R(Q) \rangle& = &\langle R_{pert} (\alpha_s (Q)) \rangle_{NLO} \nonumber \\
                                 & + & a_R {\bar{\alpha}}_0 (\mu_{I}) (\frac{\mu_{I}}{Q}) - a_R [{\bar{\alpha}}_0 (\mu_{I} )]_{NLO} (\frac{\mu_{I}}{Q}) + \ldots ,
\end{eqnarray}

where
\begin{equation}
\label{qcd-eq-2-5-3}
\bar{\alpha}_0 (\mu_I) = \int_0^{\mu_I} \frac{dk}{\mu_I} \alpha_s (k) .
\end{equation}
In this way, DELPHI group, for example, obtained\cite{physqcd-obt}
\begin{equation}
\label{qcd-eq-2-5-4}
\alpha_s (M_z) = 0.1173 \pm 0.0023
\end{equation}
by considering 18 shape variables. However, the value of $\alpha_s$ is
somewhat unstable if we fit the value to each $R$, and we need
more study on the non-perturbative features of hadronization processes.

\section{Transition to Non-perturbative Region}
  \subsection{Loop Corrections and Multi-jets}
 ~ Perturbative expansions depend critically on the value of expansion parameter
which is $\alpha_s$ in QCD. Since this parameter is well-known to be running,
we have to fix the momentum scale for each physical process. However, this
procedure cannot be carried out easily because there occur loop corrections
which split the related momentum. Moreover, even in the case of fixed $\alpha_s$,
it is impossible to confirm the convergence of expanded series since the
number of diagrams increases rapidly as the number of corrected loops increases.
These problems are closely related to the transition to non-perturbative
region where $\alpha_s$ becomes large as the related momentum reduces to small
values. This region occupies large part of hadronization processes which
result in observable jets in colliders.

 In order to account for multi-jet configurations, we need to analyze
loop corrections to each basic channels. Let's consider many jet configurations
in turn. The simplest case is that of  2 jets which are in linear
configuration and have no problem of jet definitions or overlapping.
The next 3 jet case becomes planar and one jet is generated by
radiated gluon. The distinction between quark jet and gluon jet is an
important issue to be resolved in analyzing 3 jet events. There occur
some overlapping effects between two nearest jets, that is, usually between
one quark jet and the gluon jet. In general, gluon jet has broader
shape in particle distributions and this broadness sometimes results in
overlapping effects which are related to the so-called string effects\cite{physqcd-eff}.

 The 4 jet configurations become 3-dimensional and the analysis of
final phase space is not trivial. First of all, there exist various ways
of generating 4 jets. In JLC case, 4 jets can be generated from
different initial states such as $q\bar{q}Q\bar{Q}, q\bar{q}gg, W^+ W^-,$ and $ZH$. In order to
predict the directions and the total energies of jets, we need to analyze
the initial creation processes related to each channel. For $q\bar{q}Q\bar{Q}$ channel,
tree diagram correspond to a quark pair creation and one gluon radiation which
results in another quark pair creation, and the resulting 4 quarks are
assumed to generate one jets respectively. To identify Higgs particle, for
example, from 4 jet events via $ZH$, we need to analyze other 4 jet events
in a precise manner. The one-loop corrections to the $e^+ e^-  \rightarrow q\bar{q}Q\bar{Q}$ process
had been calculated in 1997\cite{physqcd-and} and the corrections to $e^+ e^- \rightarrow q\bar{q}gg$ had been
estimated later\cite{physqcd-ter}. The number of diagrams increases as the number of loops
inserted increases. Approximately the number of one-loop diagrams is of order
10 and two-loop diagrams are of order $10^2$. More loop diagrams can be
drawn by computer programs\cite{physqcd-gra} and some 4-loop diagrams of order $10^4$ have
been calculated. Of course, these calculations are useful only in perturbative
region and the series are not convergent. In low energy region, which is
reached in the final stage of fragmentation and hadronization, the situation
becomes non-perturbative mainly due to the non-linear gluonic interactions.
Since we have not solved the problem of QCD with these non-perturbative
features, we have to introduce phenomenological models to describe the
low energy final states. These models are closely related to the descriptions
of jets and we need to study more on many jet systems such as
10 or 20 jets in case of high energy linear collider. In order to describe
such complex situations, we have to devise new method and as one possible
choice we will introduce momentum space flux-tube model\cite{physqcd-mol}.

  \subsection{$\alpha_s$ for  Light  Quarks}
~ The transition from perturbative region to non-perturbative one is
usually parametrized by
\begin{equation}
\label{qcd-eq-3-2-1}
\bar{\alpha}_0 (\mu_I) = \int_0^{\mu_I} \frac{dk}{\mu_I} \alpha_s (k) .
\end{equation}
The value of ${\bar{\alpha}}_0$  turns out to be around $0.5$ at $\mu_I = 2{\rm{GeV}}$
when we compare event shapes in fragmentation with theoretical predictions.
However, the change of $\alpha_s$ into the non-perturbative region
can be explicitly shown by analyzing meson spectra. Since the
spin-dependent forces between quark and antiquark in a meson can
be expanded as functions of $\alpha_s$, we can fit the observed mass
spectra by varying the value of $\alpha_s$. 

 The spin splittings of hadronic spectra result from gluon exchanges which can be calculated by considering
 relativistic propagator corrections in a Wilson loop. 
For each gluon exchange, one strong coupling constant
 $ \alpha_s$ appears, and we can estimate the magnitude of
$\alpha_s$ for each quarkonium system by comparing theoretical
 spin splittings with those observed in experiments.
Although the spin dependences have been calculated to
higher orders in $\alpha_s$,  they cannot be safely applied to the
analyses of quarkonium systems because of the appearance
 of complex problems at second order in $\alpha_s$, such as
mixings, annihilation contributions, and glueball effects.
These effects are closely related to the solution of QCD
which remains as an important problem to be resolved
in the future. In order to extract physically comparable
values of $\alpha_s$, it is better at this point to use the first-order
form of spin-dependent forces comparing different quark
systems composed of various flavor components.

 The first-order spin-dependent potential is given by
\begin{eqnarray}
V_{SD} & = & \frac{1}{m_1 m_2} ( {\bf{s}}_1 + {\bf{s}}_2 ) \cdot {\bf{L}} \frac{4}{3}\alpha_s\frac{1}{ (r + r_q)^3}  \nonumber \\
            &  + & \frac{1}{2}\Big( \frac{{\bf{s}}_1}{m^{2}_{1}} + \frac{{\bf{s}}_2}{m^{2}_{2}} \Big) \cdot {\bf{L}}\Big( -\frac{1}{r}\frac{d\epsilon(r)}{dr} + \frac{8}{3}\alpha_s\frac{1}{(r + r_q)^3 } \Big)\nonumber \\
            & + &\frac{1}{3m_1 m_2}(3{\bf{s}}_1 \cdot \hat{{\bf{r}}}{\bf{s}}_2 \cdot \hat{{\bf{r}}} - {\bf{s}}_1 \cdot {\bf{s}}_2 )\frac{4}{3}\alpha_s\frac{3}{( r + r_q )^3}   \nonumber \\
            & + &\frac{2}{3m_1 m_2}{\bf{s}}_1 \cdot {\bf{s}}_2\frac{4}{3}\alpha_s 4\pi\frac{1}{r^{3}_{0}}exp\Big( -\frac{\pi r^2}{r^{2}_{0}} \Big) ,
\end{eqnarray} 

where ${\bf{s}}_i$ are the quark spins and ${\bf{L}}~ \equiv~ {\bf{r}} \times {\bf{p}}_1 = -{\bf{r}} \times {\bf{p}}_2$.
The parameters $r_q$ and $r_0$ are introduced to smooth the
potential so that we can evaluate the eigenvalues of the
equation
\begin{equation}
\label{qcd-eq-3-2-3}
H\psi = E\psi
\end{equation}
where the Hamiltonian is
\begin{equation}
\label{qcd-eq-3-2-4}
H = \sqrt{m^{2}_{1} + {\bf{p}}^{2}_{1}} + \sqrt{ m^{2}_{2} + {\bf{p}}^{2}_{2}} + \epsilon(r) + V_{SD} .
\end{equation}
Since the form of the spin-independent potential $\epsilon(r)$,
which is related to the unknown solution of QCD, cannot
be deduced from first principles, we just choose a typical
potential
\begin{equation}
\label{qcd-eq-3-2-5}
\epsilon(r) = \frac{r}{a^2} - \frac{4}{3} \alpha_s \frac{1}{r} + b ,
\end{equation}
with two other parameters $a$ and $b$. The linear part
accounts for the confining features in the long-distance
region, and the Coulomb part corresponds to the short
range potential, which has been changed many times and
has been fixed to the above form by many authors.

 Now, we need to approximate the square-root operators
 in Eq.(5.8) in order to avoid the non-locality problem.
In non-relativistic cases, we have
\begin{equation}
\label{qcd-eq-3-2-6}
\sqrt{m^{2}_{i} + {\bf{p}}^{2}_{i}}~ \cong~ m_i + \frac{{\bf{p}}^{2}_{i}}{2m_i} .
\end{equation}
However, in relativistic  cases, such as light quark systems
 or highly excited states of heavy quark systems, it
is better to define the expansion parameter $M$ by
\begin{equation}
\label{qcd-eq-3-2-7}
M = \sqrt{\langle{\bf{p}}^{2}_{i} \rangle + m^{2}_{i}} ,
\end{equation}
including the momentum expectation values. Then, we
can use the expansion
\begin{equation}
\label{qcd-eq-3-2-8}
\sqrt{m^{2}_{i} + {\bf{p}}^{2}_{i}} ~\cong~ \frac{M}{2} + \frac{m^{2}_{i} }{2M} + \frac{{\bf{p}}^{2}_{i}}{2M} .
\end{equation}
Now, the two expanded forms,  Eq. (5.10) and (5.12),
are equivalent if we consider the fact that the spin-independent
 potential $\epsilon(r)$ contains a constant term as 
in Eq.(5.9). We can use the same second-order differential
equation to solve Eq.(5.7) for any system from the lightest
 one to the heaviest one, and the differences between
various systems are represented by the magnitudes of the
momentum expectations in Eq.(5.11).

 The mass parameters $m_1$ and $m_2$ appearing in the
Hamiltonian H cannot be easily fixed because the behaviors
 of quarks in a bound state are not fully understood yet.
 Generally used values of quark masses are deduced
 from some features related to low-lying hadronic
states. For example, current quark masses are introduced
to account for weak interaction features, and constituent
quark masses are used to explain mass spectroscopy and
high energy scattering results. In some cases, the larger
values of dynamical quark masses are used to account
for experimental data. However, in order to estimate
the magnitudes of momenta in quarkonium states, which
are important to fix down the momentum dependences
of $\alpha_s$, it is better to consider the effective quark masses
defined in Eq.(5.11) with the momentum expectation values
 included.  We will take this point of view in this
report so that all the mass parameters in the Hamiltonian
 H are taken to be effective quark masses. When
we deduce the magnitudes of the momenta from the determined
 mass parameters, we need the original quark
masses as in Eq.(5.11) and we will take the original values
as being equal to the constituent quark masses.

 The determination of the free parameters in the Hamiltonian
 H can be carried out by comparing the calculated
 and the observed masses for given combinations
of quarks. Since there exist seven parameters in H, we
have to reduce the number of free parameters by several
methods. Firstly, the two parameters $r_q$ and $r_0$ can be
fixed to some values which do not affect the final results.
We have tested various values for the cases of isosinglet
and isotriplet states, and we will use the same values for
these parameters. The results are
\begin{equation}
\label{qcd-eq-3-2-9}
r_q = 10^{-7}~ {\rm{GeV}}^{-1} , ~~ r_{0} = 1.0~ {\rm{GeV}}^{-1} .
\end{equation}
Secondly, the constant parameter $b$ can be determined
by the mass of 1S triplet state for each combination of
quarks. Then, the remaining four parameters are the
effective quark masses $M_1$ and $M_2$, the potential parameter
 $a$, and the strong coupling constant $\alpha_s$. Best values
for these parameters can be chosen such that the root-mean-square
 mass difference\cite{physqcd-dif}
\begin{equation}
\label{qcd-eq-3-2-10}
\Delta m = \sqrt{\frac{1}{N} \sum_{i} (E^{cal}_{i} - E^{obs}_{i})^2}                                            
\end{equation}
becomes a minimum, where $E^{cal}_{i}$ and $E^{obs}_{i}$ represent  the
calculated and the observed masses and  $N$ is the number
of observed states. However, if all four of these parameters
 are varied freely, the parameter space becomes too
large to be covered easily; therefore, we need to reduce
the number of parameters. One possibility is to consider
the relationship between $M_1$ and $M_2$. Since we are considering
 quarkonium states, the momentum expectation
values can be assumed to satisfy
\begin{equation}
\label{qcd-eq-3-2-11}
\langle {\bf{p}}^{2}_{1} \rangle = \langle{\bf{p}}^{2}_{2}\rangle
\end{equation}
for each given state. Then, we can vary only this magnitude
 of the momentum expectation in $M_1$ and $M_2$ instead
of treating $M_1$ and $M_2$ as independent parameters. For
example, in the case of K meson system, $M_1$ and $M_2$
become
\begin{equation}
\label{qcd-eq-3-2-12}
M_1 = \sqrt{\langle {\bf{p}}^2 \rangle + m^{2}_{s} },~~ M_{2} = \sqrt{\langle {\bf{p}}^2 \rangle + m^{2}_{u} } ,
\end{equation}
where $m_s$ and $m_u$ are the constituent quark masses. We
will assume that the constituent quark masses are
\begin{equation}
\label{qcd-eq-3-2-13}
m_u = m_d = 0.33, ~~ m_s = 0.45,~~  m_c = 1.5,~~  m_b = 4.5
\end{equation}
in units of GeV. In fact, the magnitude of the momentum
expectation varies from state to state, and the assignment
of an average value to a given quarkonium system may
contradict the behavior of variations. However, in this
way, we can reduce the number of parameter by one, and
the remaining three parameters $\langle {\bf{p}}^2 \rangle , a$, and $\alpha_s$ can
be varied to figure out the minimum value of $\Delta m$ .

 Another method for fixing the values of $M_1$ and $M_2$
is to use the effective masses determined from the analyses
 of  isosinglets and isotriplets. In this case, only two
parameters $a$ and $\alpha_s$ are varied, and the minimum $\Delta m$
can be found easily. The used values of effective masses
in units of GeV are
\begin{equation}
\label{qcd-eq-3-2-14}
M_{u,d} = 1.5400, ~~~ M_s = 1.7710,~~~ M_{c} = 1.8777,~~~ M_{b} = 5.4300.
\end{equation}

\begin{figure}[h] 
\centerline{\epsfbox{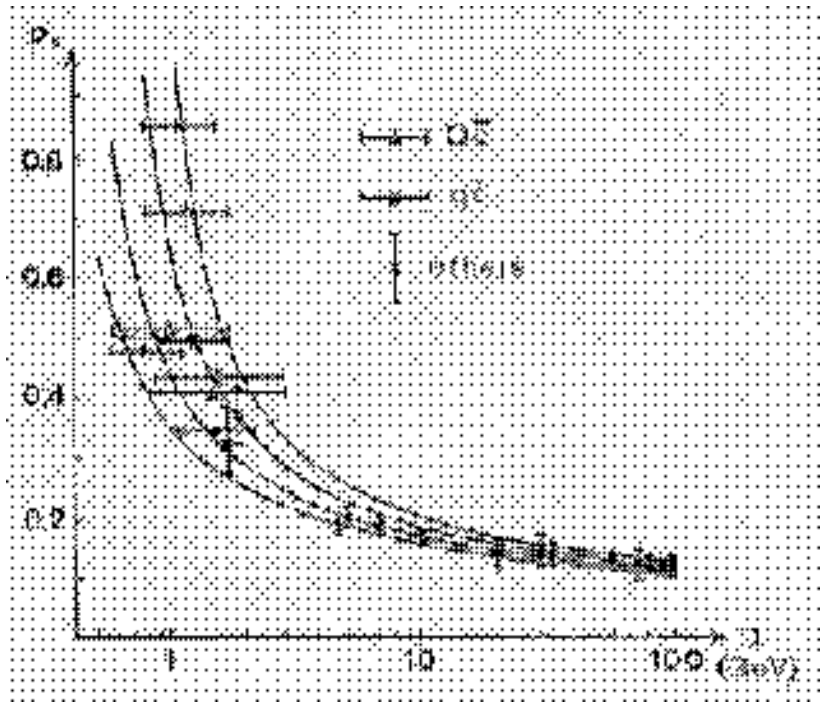}}
\vspace{6pt}
\caption{\sl \label{physqcd-fig5-1} 
$\alpha_s$ values deduced from meson masss spectra.}
\end{figure}

 In this way, we can get $4~ \alpha_s$ values
corresponding to $q\bar{q}$ combinations and $5~ \alpha_s$
values corresponding to $Q\bar{q}$ with
different quark masses. The results are
shown in Fig.~\ref{physqcd-fig5-1}\cite{physqcd-pre}.
\newpage 
 With these results, we can say
safely that $\Delta\alpha_s \stackrel{<}{=} 0.001$ at $ Q=500{\rm{GeV}}$ so that all kinds of QCD calculations
can be carried out within 1\% level at this energy. On the other hand, $\alpha_s$
becomes greater than 0.5 in low energy region corresponding to light quark
hadrons, where perturbative calculations break down. The smooth transition
from perturbative region into non-perturbative one cannot be clearly cut off
at some point, and there remain some uncertainties in defining the related
factorization scales.
\section{Non-perturbative Topics }
 \subsection{Jet Overlapping}
~In order to describe the non-perturbative features of hadronization
processes, we usually assume two generalities ; one is the local
parton-hadron duality and the other is the universality of $\alpha_s$ in low
energy region. With these assumptions we can construct models for
hadronizations resulting in various jet generating algorithms such as
HERWIG from cluster model, JETSET or UCLA from string
model, PYTHIA, PANDORA, ISAJET, SUSYGEN etc. with appropriate
corrections.  For these algorithms, one important point is that there
exist differences between quark jet and gluon jet. These differences
could be critical in analyzing many jet systems resulting from
high energy, $e^{+} e^{-}$ linear collider.
~Another point to be considered in analyzing many jet systems
is jet overlapping. For 2 jet system, there exists no problem
since the configuration of 2 jets is linear.  For 3 jets, the
differences between quark jet and gluon jet turn out to be string
effects with slight overlapping effects. However, for 4 jet system,
the overlapping probability becomes significant. For example, let's
estimate the probability to overlap by assuming jet shapes of cone
structure with subtending solid angle $\frac{\pi}{16}$, which corresponds
approximately to a cone with side angle $\frac{\pi}{6}$. For a fixed cone,
the total solid angle for another cone to overlap becomes $\frac{9}{16}\pi$.
Then the probability for 3 fixed cones to overlap with the
remaining one is $\frac{27}{64}$, which is larger than $\frac{1}{3}$. This
probability becomes $\frac{9}{16}$ for 5 jet system and increases to
$\frac{45}{64}$ for 6 jets. Since $t\bar{t}$ processes in linear collider correspond
to 6 jet system, it becomes problematic to analyze $t\bar{t}$ by just
counting particle trajectories. The situation becomes worse if
we want to get some information about particle polarizations.
Moreover, the $t\bar{t}H$ processes result in 8 jets for which the
overlap probability becomes $\frac{63}{64}$. In our simple estimation, jets
cannot be separately observed in 8 jet system so that it is
meaningless to follow particle trajectories to define jets.

~Since we have to be prepared to account for many jets
such as 10 or 20 jets in high energy linear collider, we need
to construct new theoretical models which can be used to analyze
jet overlapping. One possibility is the momentum space flux-tube
model which will be explained in the following sections. We can
classify flux-tubes and construct topological spaces and then it is
possible to define probability amplitude to have quark pairs which
can be used to predict particle multiplicities in jets. In this
way, we can account for string effects in 3-jets by considering
appropriate differences between quark and gluon jets.

 \subsection{Flux-Tube Model}
~The flux-tube model was originally considered in order to
account for the mass spectra of hadrons and their strong decay processes
which are related to quark pair creations. At first, the creations
of quark pairs were assumed arbitrarily with appropriate operators
resulting in models such as the $^3P_0$ or the $^3S_1$ model. However, it
was thought later that the quark pair creations were controlled by
gluonic degrees of freedom. It is well known that the gluonic degrees
of freedom in bound state problems are not so simple as to be described
by perturbative calculations only. The non-perturbative feature of
gluonic interactions is one of the motives for consideration of the
simple flux-tube formalism.

 The description of gluonic flux-tube was firstly attempted by
a string picture. In quark pair creation model, the created quark
pair breaks a flux-tube with equal probability amplitude anywhere
along the string and in any state of string oscillation. The amplitude
to decay into a particular final state is assumed to depend on the
overlap of original wave functions of quarks and string with the
final two state wave functions separated by the pair creation. For
ground state strings connecting quark and antiquark in mesons, the
amplitude $\gamma(\xi\vec{r}, \vec{\omega})$ to break at point $\frac{\vec{r}} {2} + \vec{\omega}$
was first assumed to be\cite{physqcd-be}
\begin{equation}
\label{qcd-eq-4-2-1}
\gamma(\xi\vec{r}, \vec{\omega}) = \omega_0 e^{-\frac{b}{2} \omega^{2}_{min}}
\end{equation}     
where $\gamma_0$ and $b$ are parameters and
\begin{equation}
\label{qcd-eq-4-2-2}
\frac{\vec{r}} {2} + \vec{\omega} = \vec{r}_{\bar{q}}  +  \xi\vec{r} + \vec{\omega}_{min}
\end{equation}

with $\vec{r} = \vec{r}_q - \vec{r}_{\bar{q}}$ being the difference between the quark and the
antiquark position vectors.  $\vec{\omega}_{min}$ is the vector with the shortest
distance between any point  on the vector $\vec{r}$ and the pair creation
point, and $\xi$ measures the ratio of the distances from the original
antiquark and quark positions to the projection point of the created
quark pair along the vector $\vec{r}$ . The shape of equi-$\gamma$  surface
looks like a cigar which is appropriate to describe a flux-tube
of constant width with end caps.

 For general shapes, we can introduce a $\xi$-dependent factor
 $f(\xi)$ into the exponent of $\gamma$ such that
 \begin{equation}
\label{qcd-eq-4-2-3}
\gamma_{f} = \gamma_0 e^{-\frac{1}{2} f(\xi)b\omega^{2}_{min}} .
\end{equation}

Although it has been found that physical results are nearly independent
of $f(\xi)$ for $\frac{1}{3} < f < 3$,  the arbitrariness of $f(\xi)$ raises the
problem of theoretical basis for the derivation of flux-tube overlap
function $\gamma$. In fact, the form in Eq.(4-2-3)  was derived by
using harmonic oscillator wave functions for discrete string components.
The form was even changed into the spherical one\cite{physqcd-one}
 \begin{equation}
\label{qcd-eq-4-2-4}
\gamma(\vec{r}, \vec{\omega}) = \gamma_0 e^{-\frac{b}{2}\omega^2}
\end{equation}
which is convenient for calculating physical amplitudes expanded in
the harmonic oscillator wave function basis. The changes in the
form of $\gamma$ indicate the fact that no firm theoretical grounds
exist for treating gluonic flux-tubes.

 \subsection{Construction of Topological Spaces}
~In order to deduce flux-tube overlap functions systematically from
a well-defined set of assumptions, we need to devise a framework
for the description of gluonic flux-tubes. Since the QCD bound-state
problem with strong couplings has not been solved as yet, we have
to assume the existence of gluonic flux-tubes and their interactions.
The most fundamental interactions between flux-tubes are their joining
and breaking processes, and these two processes with the existence
assumption make it possible to devise a topological approach to the description
of flux-tubes\cite{physqcd-tub}. To construct topological spaces, we need to classify the
flux-tubes, and it can be done by counting the number of boundary points
which are occupied by a quark or an antiquark behaving as a source
or a sink. The classified sets of flux-tubes can be represented as
$F_{a,\bar{b}}$ which means the flux-tube set with a quarks and b anti-quarks
sitting at boundary points. For convenience, we omit the number if it
becomes zero except for $F_0$ which represents the set with no boundary point,
that is, the flux-tube set corresponding to glueballs.

 We can construct topological spaces on the classified flux-tube sets
with the following assumptions :
\begin{verse}
(1) Open sets are stable flux-tubes.\\
(2) The union of stable flux-tubes becomes a stable flux-tube.\\
(3) The intersection between a connected stable flux-tube and disconnected stable flux-tubes is the reverse operation of the union.\\
\end{verse}
There are two possible senarios distinguished by the condition whether
the set $F_0$ is included or not. If $F_0$  is included, the loop structure
can be added to any flux-tube by the union operation. Then, the loop
can be broken by the intersection operation with one quark-antiquark pair
created as new boundaries. When we exclude four-junction structures
in flux-tubes, inclusion of $F_0$ corresponds to the creation of two
three-junctions. In this case, we can conclude that all kinds of
flux-tube sets are interrelated through the union and intersection
operations if the condition of baryon number conservation is satisfied.
Therefore, topological spaces have to be formed with all these sets under
the restriction of baryon number conservation.

 When $F_0$ is not included in the construction of topological spaces, we
can exclude three-junction creations by assuming the existence of only
non-excited flux-tubes. For this case, we can construct various topological
spaces such as meson, baryon-meson, antibaryon-meson, baryon-meson-baryon
systems, and so on. Let's consider first the meson system $F_{1,\bar{1}}$. A meson
can decay into several mesons by repeated pair creations and if we
represent n meson states by  $F^{n}_{1,\bar{1}}$, the simplest non-trivial topological
space is given by
\begin{equation}
\label{qcd-eq-4-3-1}
T_0~=~\{ \phi, ~F_{1,\bar{1}},~ F^{2}_{1,\bar{1}},~\cdots~,~ F^{n}_{1,\bar{1}},~ \cdots  ~\}
\end{equation}
Since $F_{1,\bar{1}}$ represents the set of all possible flux-tubes with two
boundary points, there is no confusion in writing
\begin{equation}
\label{qcd-eq-4-3-2}
F^{2}_{1,\bar{1}}~=~F_{1,\bar{1}}
\end{equation}
Then $T_0$ becomes
\begin{equation}
\label{qcd-eq-4-3-3}
T_0 = \{ \phi , F_{1, {\bar{1}}} \}
\end{equation}
If we accept that $F_{1,\bar{1}}$ can be multiplied at any time without
violation of baryon number conservation, the topological space for
baryon-meson system can be written as
\begin{equation}
\label{qcd-eq-4-3-4}
T_{1,0}~ \equiv ~T_1 ~= ~\{\phi,~ F_3,~ F_3F_{1,\bar{1}}\}~ =~ \{\phi,~ F_3\}
\end{equation}
and in general, we get
\begin{equation}
\label{qcd-eq-4-3-5}
T_{i,\bar{j}} = \{ \phi, ~F^{i}_{3}F^{j}_{\bar{3}}, ~F^{i-1}_{3}F^{j-1}_{\bar{3}}F_{2,\bar{2}}, ~\cdots ~\}
\end{equation}
\subsection{Connection Amplitude}
~For the defined topological spaces, we can consider the connection
amplitude A for a quark to be connected to another quark or anti-quark
through the given flux-tube open set.  In general, we can assume the
existence of a measure $M$ of the connection amplitude A satisfying the
conditions that
\begin{verse}
(1) $M(A)$ decreases as A increases,\\
(2) $M(A_1) + M(A_2) = M(A_1 A_2)$ when $A_1$ and $A_2$ are independent.\\
\end{verse}
The multiplication of two independent amplitudes $A_1$ and $A_2$ is induced
by the union operation. The above two conditions for the measure of
the connection amplitude lead to the solution
\begin{equation}
\label{qcd-eq-4-4-1}
M(A) = -k \ln\frac{A}{A_0}
\end{equation}
where $A_0$ is a normalization constant and $k$ is an appropriate parameter.
In order to get coordinate dependences, we consider the measure $M$ as
a metric through a flux-tube between two points occupied usually by
quark or antiquark. If we consider the simplest flux-tube type
$F_{1,\bar{1}}$ the metric can be thought of as functions of $|\vec{x} - \vec{y}|$ with
$\vec{x}$ and $\vec{y}$ being the positions of the two boundary points. In general,
the distance function $|\vec{x} - \vec{y}|^\nu$ can be made metric if it satisfies
the triangle inequality
\begin{equation}
\label{qcd-eq-4-4-2}
|\vec{x} - \vec{z}|^\nu + |\vec{z} -\vec{y}|^\nu \stackrel{>}{=} |\vec{x} -\vec{y}|^\nu
\end{equation}
The set of points $\vec{z}$ not satisfying this triangle inequality can
be taken as forming the inner part of the flux-tube where it
is impossible to define a metric from boundary points with given
$\nu$. With this metric condition, we can figure out the shape of
flux-tube, and we can take $|\vec{x} - \vec{y}|^\nu$ as an appropriate measure
to deduce a concrete form for the connection amplitude A. The
lower limit of $\nu$ can be fixed to be 1 because there exists no
point $\vec{z}$ satisfying the triangle inequality with $\nu < 1$ . In order to
account for various possibilities, we need to sum over contributions
from different $\nu$'s.  For a small increment $d\nu$, the product of the two
probability amplitudes for $|\vec{x} - \vec{y}|^\nu$ and $|\vec{x} - \vec{y}|^{\nu + d\nu}$ to satisfy the
metric conditions can be accepted as the probability amplitude for
the increased region to be added to the inner connected region which
is out of the metric condition. Then the full connection amplitude
becomes
\begin{equation}
\label{qcd-eq-4-4-3}
A = A_0 exp\{-\frac{1}{k} \int_1^\alpha F(\nu)r^\nu d\nu \}
\end{equation}
where all possibilities from the line shape with $\nu = 1$ to the arbitrary
shape with $\nu = \alpha$ have been included. The weight factor $F(\nu)$ has
been introduced in order to account for possible different contributions
from different $\nu$'s.  When we consider the case of $\alpha  =  2$,  which
corresponds to a spherical shape flux-tube, and the case of equal
weight $F(\nu) =1$,  we get
\begin{equation}
\label{qcd-eq-4-4-4}
A = A_0 exp\{ -\frac{1}{k} \frac{r^2 -r}{\ln r}\}
\end{equation}
 Slight changes can be made if we replace A into $r^\beta A$,
for which the conditions on $M$ still hold with $\beta  > 0$. Then general
$A$ becomes
\begin{equation}
\label{qcd-eq-4-4-5}
 A = \frac{A_0}{r^\beta} exp\{-\frac{1}{k} \frac{r^2 - r}{\ln r}\}
\end{equation}
\subsection{Gluonic Structures in Hadrons}
~In order to describe the non-perturbative gluonic structures in
hadrons by using flux-tube model, it is necessary to deduce the
relationships between the connection amplitude and the long range
gluonic distributions. The gluonic contents of hadrons can be probed
by estimating the probability amplitudes for quark pair creations
because the quark pairs are created by the gluons inside hadrons.
If we assume that the probabilities for quark pair creations are
proportional to the gluon densities, the gluonic structures of hadrons
can be deduced from the flux-tube overlap functions defined by the
connection amplitudes. The probability amplitude  to have a quark
pair at a given position is taken to be equal to the overlap of
connection amplitudes calculated for the initial and the final boundary
points. The overlap function $\gamma$ can be written as
\begin{equation}
\label{qcd-eq-4-5-1}
 \gamma = A_i A_f
\end{equation}
where $A_i$ and $A_f$ represent the connection amplitudes before and
after quark pair creation.

 Although the gluonic densities are closely related to physically
observable quantities, it is impossible to observe directly the gluonic
densities because gluons are confined. During the formation process
of final hadrons, quark pairs are created by confined gluons. The
confined structures of gluons can be probed only through the created
quark pairs, and the probability amplitude to have a quark pair at
a specified position is proportional to the overlap function $\gamma$ that
can be used to describe gluonic behaviors. We can calculate the
overlap function $\gamma$ for various configurations of boundary points.

 As the first example, let's consider the case of a proton accelerated
in the positive direction of z-axis.  We can assume that electric fields
are applied along positive z direction during the acceleration. Then the
three boundary quarks will be oriented in such a way that the two
positively charged up quarks go ahead pulling the other negatively charged
down quark. If the flux-tubes connecting the three boundary quarks
do not break, the distance between the up quarks and the down quark
will be contracted along the direction of z-axis. The relative positions
of three boundary quarks will change according to the Lorentz boosts,
and the corresponding gluonic structures will change.  If we take
different connection amplitudes with different values of $ \beta$,  the gluonic
structures turn out to be deformed from one to another\cite{physqcd-ano}. These
structures can be Fourier transformed resulting in gluon distribution
functions in momentum space.

 In the case of a neutron, the three boundary points are occupied
by one up quark and two down quarks.  Since the total charge of these
three quarks is zero, a neutron cannot be accelerated as for a proton.
However, the three quarks can be aligned in an electric field, and if
we fix the three boundary points we can calculate the flux-tube overlap
functions.  We can easily check that the shapes of gluon distribution
functions for a proton and for a neutron are different even if we change
the velocity of the neutron.

 For quarkonium mesons, the situation changes into one with only
two boundary points.  For example, if we consider the case of $\pi^+$
accelerated along z direction, the gluon densities along z direction turn
out to be symmetric about the axis through the center point between
the u quark and the $\bar{d}$ quark.  All the other quarkonium mesons
have the similar form.

\subsection{Momentum Space Formulation}
~The necessity to transform coordinate space distributions into
those in momentum space originates from the nature of scattering
processes. In particle scattering processes, the initial and  the final
states are defined by the momenta of each particle and the final
data can be obtained by analyzing particle trajectories.  Because the
quarks cannot exist alone, hadronization processes are essential to
form particles that can generate such trajectories. These hadronization
processes cannot be described by perturbative QCD, and we need to
construct models to account for non-perturbative features of QCD.
In our flux-tube model, we can account for these non-perturbative
features which are related to quark pair creations leading to final
jets observed in high energies. Since the created quark pairs have
to be specified in momentum space for scattering states, we need to
formulate our model in momentum space.

 Because the connection amplitudes have been defined between two
boundary points, it is possible to consider the same amplitudes in
momentum space. The connection amplitude for two boundary points
with momenta $\vec{p_1}$ and $\vec{p_2}$ can be written down as
\begin{equation}
\label{qcd-eq-4-6-1}
 A = A_0 exp\{-\frac{1}{\tau} \int_1^\alpha G(\nu)|\vec{p_1} -\vec{p_2}|^\nu d\nu\}
\end{equation}
where $G(\nu)$ and $\alpha$ are an appropriate weight factor and the boundary
value of $\nu$ respectively. In case of $G(\nu) = 1$ and $\alpha = 2$, we get
\begin{equation}
\label{qcd-eq-4-6-2}
 A(p) = A_0 exp\{-\frac{1}{\tau} \frac{p^2 - p}{\ln p}\} 
\end{equation}
with $p = |\vec{p_1} - \vec{p_2}|$.  We can apply this amplitude to describe fragmentation
processes resulting in jets.

 As an example, we choose the 3-jet events formed from
fragmentation processes of ${Z^0} '$s  produced by $e^+ e^-$ collisions. For $Z^0$
fragmentations, no initial state uncertainties exist, and the final
quark jets and antiquark jets can be identified by lepton-tagging
methods. For non-trivial 3-jet events, the other gluon jet can be
identified clearly and we can compare different properties between quark
jet and gluon jet. The gluonic effects appear as the probability amplitudes
for creation of quark  pairs which generate many hadrons resulting in jets.
For the above $A(p)$, the probability amplitude for a created quark pair
to be connected in momentum to the quark and the gluon boundaries
is proportional to
\begin{equation}
\label{qcd-eq-4-6-3}
C_1 = A^3_0 exp\{-\frac{1}{\tau} (\frac{a^2 - a}{\ln a} + \frac{b^2 - b}{\ln b} + \frac{c^2 - c}{\ln c})\}
\end{equation}
where $a$ and $b$  are the distances in momentum space from the created
quark pair to gluon and quark boundaries, and $c$ is the distance between
gluon and antiquark boundaries. The other amplitude $C_2$  for the created
quark pair to be connected to the antiquark and the gluon boundaries
has the same form with the interchanged roles of quark and antiquark
boundaries. The probability to get particles in a given direction is
\begin{equation}
\label{qcd-eq-4-6-4}
 P = \int_0^h |C_1 + C_2|^2 dR 
\end{equation}
where $h$  is determined by phase space configurations. The shape of
momentum phase space can be determined by using parton model assumptions.
Longitudinal components are taken to be proportional to the total parton
energy which becomes generally the total jet energy. On the other hand,
transverse components are assumed to be small resulting from uncertainties
in momentum specifications. With appropriate choice of phase space, we
can predict the particle multiplicities for jet configurations\cite{physqcd-ons}. For 6 data
samples from the OPAL group, we have fixed parameters to the first
data and then predicted the other 5 samples quite well.
\section{Summary}
~Quantum chromodynamics is taken to be the right theory of strong
interaction, which is responsible for the formation of hadrons. Since the
final states of high energy linear collider will be mostly composed of
hadrons, it is important to check the hadronization processes with the
viewpoints based on QCD.

 The initial particle creation processes can be described by perturbative
gauge theories, whereas there exists a transition point from perturbative
viewpoints to non-perturbative model constructions. Perturbative QCD
calculations are carried out by expanding with respect to $\alpha_s$, and it is
expected that these expansions can be done within 1\%  level in the energy
range of $500{\rm{GeV}}$.  Higher order calculations can be checked with some
corrections such as exponentiation or resummation included.  However, there
exists uncertainty in choosing the transition point from perturbative formalism
to non-perturbative one,  and we need to study more on this subject.  For the
non-perturbative hadronization processes,  we have to construct models resulting
in jet generating algorithms.  Although there exist many models,  we need
more systematic formalism which can be used to explain many jet system such
as 10 or 20 jets with overlapped configurations. One possibility is the
momentum space flux-tube model which can be applied to explain particle
multiplicities, string effects, angular ordering, and so on.  However, we have
to study more on topics such as color rearrangement, baryon production, and
spin polarization.

%% file: physew/main.tex
\newcommand{\dirwz}{physew}
\chapter{Precision Electro-Weak Physics}
\label{wz-sect}
\section{Introduction}

\begin{figure}[tbhp]
\begin{center}\begin{minipage}[t]{\figurewidth}
\centerline{\epsfysize=20cm \epsfbox{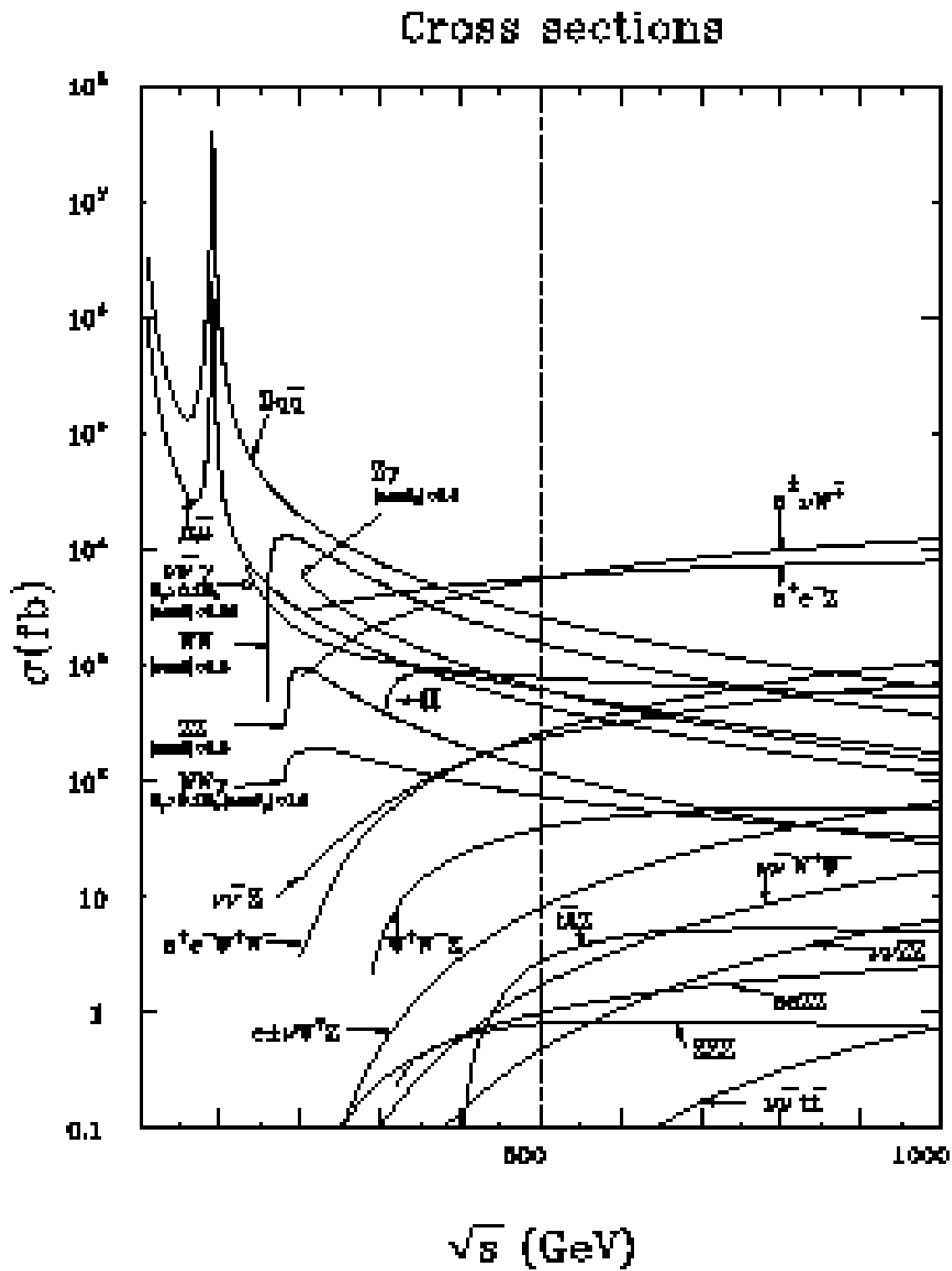}}
\caption{\sl\label{pew-totalcros2.eps}
Cross sections of the standard model processes at energies below 1 TeV.}
\end{minipage}\end{center}
\end{figure}

Total cross sections of standard model processes below
1 TeV are shown in Fig.~\ref{pew-totalcros2.eps}.
As seen in the figure, productions of multiple gauge bosons are 
characteristic feature of experiments at JLC.
Precise measurements of their properties, such as
masses and couplings, allow us to test models at 
higher order and serve as tools to probe physics at high energy
scale.  A clean environment, polarized beams,
variable collision energies  and high luminosities
at JLC are essential ingredients for precise 
measurements.
When operated at the $Z$ pole and just above the $W$ pair threshold,
Giga $Z$ and Mega $W$ samples can be accumulated.
These statistics are two orders of magnitude more than 
LEP-I/II.  At 500 GeV and above, gauge cancellation 
among diagrams of the $e^+e^-\rightarrow W^+W^-$ process
becomes severer, which allows us a direct measurement of
triple-gauge-boson couplings at a precession 
required to probe loop effects.
At higher energies, processes which involve
$t$-channel diagrams increase cross sections with 
energy, from which $WW$ scattering processes
could be measured.  In general, models without
elementary Higgs bosons predict strong interaction among 
longitudinal polarized $W$ bosons.
Such effects would be observed through 
the final-state interaction of $W$'s in the $e^+e^-\rightarrow W^+W^-$
process or resonance production in processes
such as $e^+e^-\rightarrow \nu\bar{\nu}W^+W^-$.

In the following subsections, we discuss topics
on measurements of the $W$ boson mass and triple-gauge-boson couplings.

\section{$W$ Boson Mass Determination}
\label{wz-sect-mass}
\begin{figure}[bhtp]
\begin{center}\begin{minipage}[t]{\figurewidth}
\centerline{\epsfysize=8cm\epsfbox{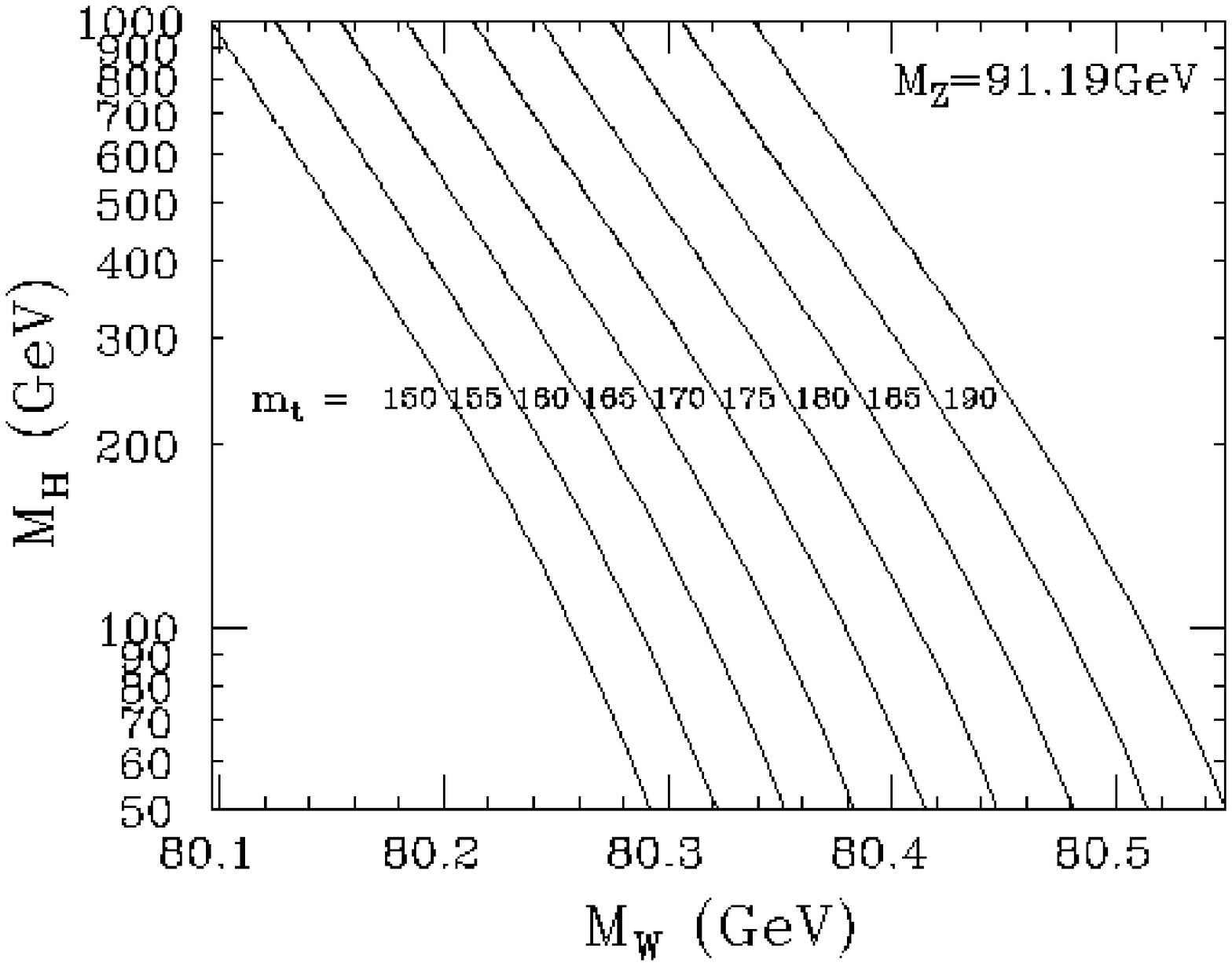}}
\caption{\sl\label{WZFIG-HWTMAS}
The mass of the standard model Higgs boson as a function 
of the $W$ mass for different top quark masses.}
\end{minipage}\end{center}
\end{figure}

\begin{figure}[htbp]
\begin{center}\begin{minipage}[t]{\figurewidth}
\centerline{\epsfxsize=10cm\epsfbox{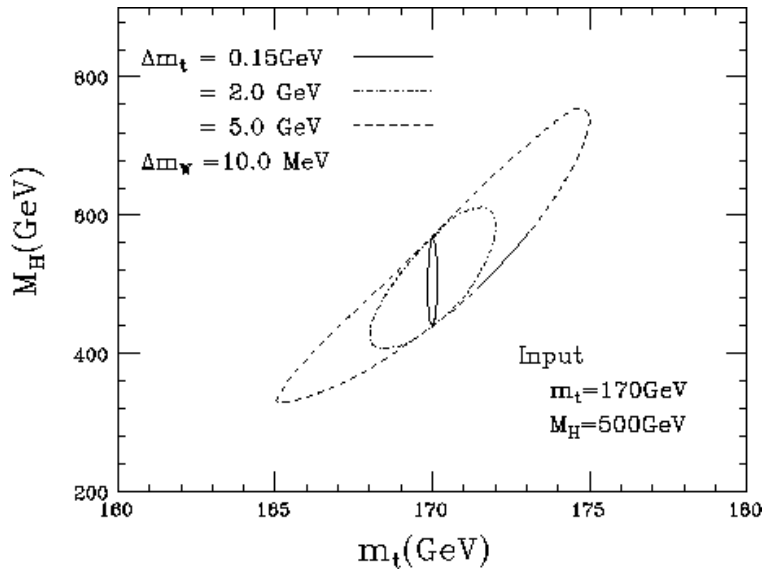}}
\caption{\sl \label{WZFIG-HTMT}
1-$\sigma$ contours in the plane of
$m_t$-$M_H$ for three values of top quark mass error.
We used $m_t=170$GeV and $M_H=500$GeV here.}
\end{minipage}\end{center}
\end{figure}

The masses of $W$, $t$, and Higgs are related through loop corrections.
Their relations are shown in Fig.~\ref{WZFIG-HWTMAS}. As seen 
from the figure, the loop effects of the masses of $W$ and $t$ to that of 
Higgs are quite different in size: if we require the same size,
precisions of the $t$ and $W$ masses should satisfy 
$\sigma_{m_W} \sim 0.7 \times 10^{-2} \sigma_{m_t}$.
At JLC, $\sigma_{m_t} \sim 100 $ MeV is expected for 100 fb$^{-1}$,
which corresponds to $0.7 $MeV for $\sigma_{m_W}$.
Fig.~\ref{WZFIG-HTMT}, on the other hand, shows 1-$\sigma$ contours 
in the plane of $m_t$-$m_H$ for several values of precision for the  
top quark mass measurement.  
From this figure, the importance of precise mass determination 
of the top quark is evident.  
Since the expected precision of the
top quark mass is less than 0.15 GeV, the error on the estimated Higgs mass
is dominated by the error of the $W$ mass measurement.

\begin{figure}[htbp]
\begin{center}\begin{minipage}[t]{\figurewidth}
\centerline{\epsfxsize=10cm\epsfbox{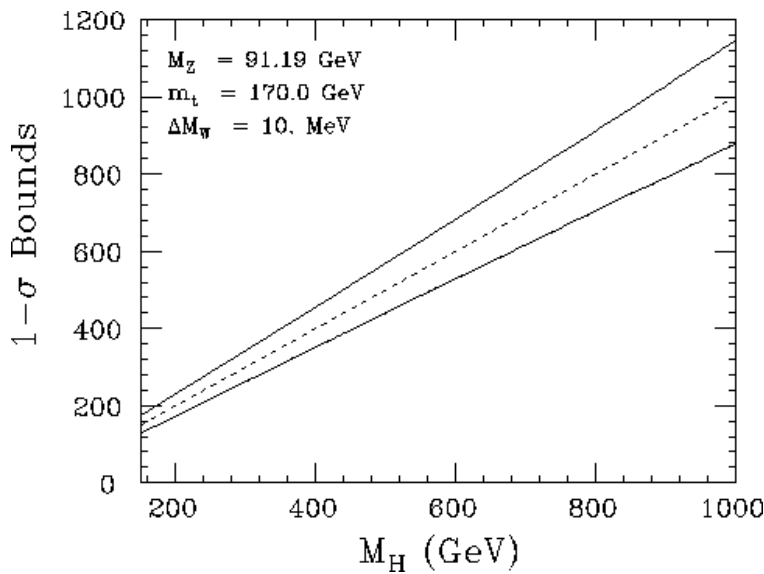}}
\caption{\sl \label{WZFIG-MHDMH}
1$\sigma$ bounds of the mass of Higgs as a function of
the mass of Higgs, 
the precision of $W$ mass is 10 MeV.}
\end{minipage}\end{center}
\end{figure}

The current world average for $m_W$ is $80.448\pm 0.034$ GeV, 
which is the average of
the LEP2 average of 40 MeV precision and the $p\bar{p}$ average of 62 MeV
precision.  The $p\bar{p}$ error is expected to be reduced to 
30 MeV by Tevatron Run-II and to 15 MeV at LHC.  However, the error
is still larger than that required by the expected top quark mass
error as far as the contribution
to the loop effect is concerned.
Further improvement of 
the $m_W$ measurement is desirable.
If no Higgs boson is discovered at any colliders including the JLC,
the precise $m_W$ measurement is of vital importance,
since the loop effect would be the only way to
estimate the Higgs mass.
In Fig.~\ref{WZFIG-MHDMH}, we show the error in the Higgs mass
estimation from the quantum effect as a function of the Higgs mass.
If the Higgs boson is there in the JLC energy region , 
its mass will be measured with a precision similar to 
the mass of the top quark, as described in the Higgs chapter.  
The precise determinations of the $W$, $t$, and Higgs masses 
will enable us to test the standard model
at the loop level, thereby allowing us to probe 
new physics in the loop effect.

At JLC the $W$ mass could be measured at $\sqrt{s}=500$ GeV,
using the process $e^+e^-\rightarrow e \nu W$, since it 
has a large cross section and that there is no
ambiguity in selecting particles from the $W$ decay.
If we require $|\cos\theta|<0.8$ 
to select events well contained in the detector acceptance,
about 80k events are expected to be observed 
for 100fb$^{-1}$\cite{WZREF-LCWS93MIYAMOTO}.
On the other hand, 40k events of 
$e^+e^-\rightarrow \gamma Z$ process can be observed
under the same condition.  This process can be used
for detector calibration.  The invariant mass resolution 
for the Higgs mass of 100 GeV is expected to be less than 4 GeV.
If similar resolution can be achieved for $W$'s, we can expect
the statistical error of the $W$ mass to be about 20MeV.
If the integrated luminosity is 400 fb$^{-1}$,
which is less than two years of running with JLC-Y 
parameters, the statistical error of the $W$ mass becomes about 10 MeV.
In this measurement, there is no systematic error due to 
ambiguity in jet clustering or color recombination.
If the statistics of the $e^+e^-\rightarrow \gamma Z$ process is not 
sufficient for calibration, 
we might run at the $Z$ pole for higher statistics.
We need, however, 
a serious study to find out possible source of detector systematics. 

\section{Triple and Quartic Gauge Boson Couplings}

In the standard model, electroweak
gauge bosons are introduced to preserve the local $SU(2)_L \times U(1)_Y$ 
symmetry.  As a result, there is a universality among 
the couplings of  fermions to the gauge boson, the three gauge bosons, 
and the four gauge bosons.
This universality forms the basis of the success of the standard model
(~see Ref.\cite{HAGIWARA92} for example~). 
So far the fermion-gauge-boson couplings were tested precisely at
various colliders, however the direct measurement of 
the self couplings of the gauge bosons is not precise enough 
to test the standard model at loop level. 

In order to formulate the test of
the self couplings of the gauge bosons, we usually
introduce a Lagrangian containing non-standard-model interactions,
assuming some symmetry for them.  
The most general Lagrangian assuming
the electro-magnetic gauge invariance, the Lorentz symmetry, 
and $C$ and $P$ conservation
is given by\cite{KPZH87}
\begin{eqnarray}
{{\cal L}_{WWV}}/g_{WWV} 
& =&     i g^V_1 ( W_{\mu\nu}^\dagger W^\mu V^\nu 
            - W_\mu^\dagger V_\nu W^{\mu\nu} ) 
 + i\kappa_V W^\dagger_\mu W_\nu V^{\mu\nu} \nonumber \\
& +& { i\lambda_V \over m^2_W } 
     W^\dagger_{\lambda\mu} W^\mu_\nu V^{\nu\lambda}, 
 \label{EQ_CPINV}
\end{eqnarray}
\noindent
where $V\equiv Z$ or $\gamma$ and
$W_{\mu\nu}\equiv \partial_\mu W_\nu - \partial_\nu
W_\mu$,
$V_{\mu\nu}\equiv\partial_\mu V_\nu - \partial_\nu V_\mu$,
$g_{WW\gamma}=-e$, and $g_{WWZ}=-e \cot \theta_W$.
In the standard model, $g_1^V=\kappa_V=1$ and
$\lambda_V=0$. 
The static properties of the $W$, the magnetic dipole moment ($\mu_W$)
and the electric quadrupole moment ($Q_W$), are related to these
couplings as 
\begin{eqnarray}
\mu_W&=&\frac{e}{2M_W} (1+\kappa_\gamma + \lambda_\gamma), \\
Q_W&=&\frac{e}{M^2_W} (\lambda_\gamma - \kappa_\gamma).
\end{eqnarray}

Since $g_1^\gamma=1$ by the electro-magnetic gauge 
invariance, 
five anomalous couplings are involved in
Eq.~(\ref{EQ_CPINV}): 
$\Delta g_1^Z \equiv g_1^Z - 1$,
$\Delta\kappa_\gamma\equiv\kappa_\gamma - 1$,
$\Delta\kappa_Z \equiv\kappa_Z - 1$,
$\lambda_\gamma$, and
$\lambda_Z$.
If we require the global $SU(2)_L$ symmetry for 
Lagrangian~(\ref{EQ_CPINV}), we are left with only one
non-zero coupling: $\lambda\equiv\lambda_\gamma=\lambda_Z$. Since
the $\lambda$ term has an
energy dimension of 6, the requirement of the renormalizability
leads us to the standard model Lagrangian.  
If we require an {\it intrinsic} $SU(2)_L$ symmetry, 
we are left with four independent couplings
with the relation,
\begin{equation}
1 + \Delta g_1^Z = -\tan^2\theta_W
\frac{\Delta\kappa_\gamma}{\Delta\kappa_Z}. \label{EQ_CONST1}
\end{equation}
The number of independent couplings 
becomes two if, in addition, we allow only dimension-4
couplings. If we further require the most divergent quartic one-loop
contribution to the $\rho$ parameter be absent,  we will get an
additional constraint,
\begin{equation}
\Delta g_1^Z \cos^2\theta_W = \Delta\kappa_\gamma,
\end{equation}
\noindent which will lead us to only one coupling of dimension 4.

Lagrangian of Eq.~(\ref{EQ_CPINV}) is adequate for the study of
anomalous triple-gauge-boson couplings at tree level, 
but it is not adequate for the study of loop effects and the 
process involving both triple- and quartic-gauge-boson 
couplings because it does 
preserve local $SU(2)_L \times U(1)_Y$ gauge invariance\cite{RGHM92}.
One approach to overcome this problem is to 
use the most general $SU(2)_L\times U(1)_Y$ gauge invariant 
effective Lagrangian of dimension 6, assuming a Higgs boson,
in the studies of  the possible low energy effects 
of new physics at high energy.
Another approach is to use the gauged chiral Lagrangian
without introducing the Higgs field.

In the locally-$SU(2)_L\times U(1)_Y$-gauge-invariant dimension-6 
effective Lagrangian, the terms involving self-couplings of gauge bosons
at tree level 
can be written in the form\cite{HISZ93}: 
\begin{equation}
{\cal L}= \frac{f_{WWW}}{\Lambda^2} O_{WWW} + 
   \frac{f_W}{\Lambda^2} O_{W} + 
   \frac{f_B}{\Lambda^2} O_{B}, \label{EQ_DIM6}
\end{equation}
\noindent where
\begin{eqnarray}
O_{WWW}&=& Tr\left[ \hat{W}_{\mu\nu} \hat{W}^{\nu\rho} 
     \hat{W}_\rho^\mu \right], \nonumber \\
O_W&=&(D_\mu\Phi)^\dagger \hat{W}^{\mu\nu} 
      (D_\nu\Phi),  \\
O_B&=&(D_\mu\Phi)^\dagger \hat{B}^{\mu\nu}
      (D_\nu\Phi). \nonumber 
\end{eqnarray}
\noindent 
$\Phi$ is a Higgs doublet field,
$D_\mu$ is the covariant derivative
as given by
\begin{equation}
D_\mu\equiv\partial_\mu + \frac{i}{2}g'B_\mu +
  ig\frac{\sigma^a}{2} W_\mu^a,\nonumber
\end{equation}
and the hatted field operators are defined as
$ \left[ D_\mu, D_\nu \right] = \hat{W}_{\mu\nu} + \hat{B}_{\mu\nu}$.
Since the operators in Lagrangian (\ref{EQ_DIM6}) 
have dimension 6, it is 
unrenormalizable and 
the scale $\Lambda$ can be identified as the cut-off energy. 
$f_{WWW}$, $f_{W}$, and $f_{B}$ are the anomalous couplings of
corresponding terms.

In terms of these couplings, the couplings in Lagrangian (\ref{EQ_CPINV})
are expressed as
\begin{eqnarray}
\Delta g_1^Z &=& f_W\frac{m^2_Z}{2\Lambda^2}, 
\label{EQ_DG1ZVSFW} \\
\Delta\kappa_Z&=& \left[ f_W - \sin^2 \theta_W (f_B+f_W) \right]
   \frac{m^2_Z}{2\Lambda^2} ,\\
\Delta\kappa_\gamma&=&\cos^2 \theta_W (f_B+f_W)
   \frac{m^2_Z}{2\Lambda^2}, \label{EQ_DKG}\\
\lambda_\gamma &=& \lambda_Z=\frac{3m^2_W g^2}{2\Lambda^2} f_{WWW}.
\label{EQ_LAMBDA}
\end{eqnarray}
\noindent 
Because of 
Eqs.~(\ref{EQ_DG1ZVSFW}) - (\ref{EQ_LAMBDA}), 
only three of the 
five couplings in Eq.~(\ref{EQ_CPINV}) are 
independent.

Here the above constraints are consequences of the 
electroweak gauge invariance and the restriction to 
the dimension-6 operators\cite{HISZ93}.
Since Lagrangian (\ref{EQ_CPINV}) involves
only the terms relevant to the triple-gauge-boson couplings, 
we shall use Lagrangian~(\ref{EQ_DIM6}) for the study of 
self-couplings in the JLC energy region,
where the quartic-gauge-boson
couplings are equally important at tree level.

In the following discussion, we use $\Delta\kappa_\gamma$,
$\Delta\kappa_Z$, and $\lambda$ 
to parametrize the three
free parameters of Lagrangian~(\ref{EQ_DIM6}) unless otherwise
stated, where $\lambda \equiv \lambda_\gamma = \lambda_Z$ which is  
proportional to $f_{WWW}$ as Eq.~(\ref{EQ_LAMBDA}), and 
\begin{equation}
\Delta g_1^Z = \Delta\kappa_Z + \tan^2\theta_W \Delta\kappa_\gamma,
\label{EQ_DG1ZKZKG}
\end{equation}
\noindent due to constraints (\ref{EQ_DG1ZVSFW}) - (\ref{EQ_DKG}).
In terms of $\Delta\kappa_\gamma$ and $\Delta\kappa_Z$, $f_W$ and $f_B$ 
are expressed as
\begin{eqnarray}
f_W&=& \frac{2\Lambda^2}{m^2_Z}(\Delta\kappa_Z
 + \tan^2\theta_W \Delta\kappa_\gamma), \\
f_B&=& \frac{2\Lambda^2}{m^2_Z}(\Delta\kappa_\gamma
 - \Delta\kappa_Z). 
\end{eqnarray}
\noindent It should be noted that in this convention\cite{HISZ93},
a given set of the anomalous couplings
($\Delta\kappa_\gamma$, $\Delta\kappa_Z$, $\lambda$)
should necessarily contain the $\Delta g_1^Z$ term~(\ref{EQ_DG1ZKZKG})
as well as the quartic couplings in (\ref{EQ_DIM6}).

The standard model processes at JLC  which involve
the triple- and quartic-gauge-boson couplings at tree level are summarized in 
Table~\ref{PROCESS}.
First and second columns show vertices and couplings.
The vertices $WW\gamma Z$, $WWZZ$, and $WWWW$
involve $\Delta\kappa_\gamma$ and $\Delta\kappa_Z$ couplings.
But, since they are proportional to $\Delta\kappa_\gamma
- \Delta\kappa_Z$, they are indicated by $f_W$
( $\equiv\frac{2\Lambda^2}{m^2_Z}(\Delta\kappa_\gamma -  \Delta\kappa_Z$)).
$\bigcirc$'s in the rest of the columns
indicate processes which involve the relevant vertices.
The processes marked by $\bigtriangleup$ involve 
corresponding vertices, but the sensitivities are limited.
The last row of the table  shows the total cross 
sections at $\sqrt{s}=500$ GeV, where $E_\gamma > 25$ GeV 
and $\vert\cos\theta_\gamma\vert<0.85$ are required
for the processes $\nu\bar{\nu}\gamma$ and $WW\gamma$,
and $\vert\cos\theta\vert < 0.9$ is assumed for $\gamma\gamma\rightarrow
W^+W^-$\cite{YWKE92}.

\begin{footnotesize}
\begin{table}[t]
\caption{\label{PROCESS}
{\sl Processes and couplings at tree level.}}
\vglue 4pt
\begin{small}
\centerline{ \begin{tabular}{| c | c | c c c c c c c | c |}
\hline
 & & \multicolumn{7}{c|}{$e^+e^-\rightarrow$} &
$\gamma\gamma\rightarrow$ \\
Vertex & Couplings & $WW$
 & $e\nu W$ & $\nu\bar{\nu}\gamma$ & $\nu\bar{\nu}Z$ 
 & $WWZ$ & $WW\gamma$ & $eeWW$ & $WW$ \\
\hline
$WW\gamma$ & $\lambda,\Delta\kappa_\gamma$ 
   & $\bigcirc$ & $\bigcirc$ & $\bigcirc$ & 
   & $\bigcirc$ & $\bigcirc$ & $\bigcirc$ & $\bigcirc$ \\
$WWZ$ & $\lambda, \Delta\kappa_Z$ 
   & $\bigcirc$ & $\bigtriangleup$ & 
   & $\bigcirc$ & $\bigcirc$ & $\bigcirc$ & $\bigcirc$ & \\
$WW\gamma\gamma$ & $\lambda$ & & & & & & $\bigcirc$
   & $\bigcirc$ & $\bigcirc$ \\
$WW\gamma Z$ & $\lambda, f_W$ & & & & & $\bigcirc$
   & $\bigcirc$ & $\bigcirc$ & \\
$WWZZ$ & $\lambda, f_W$ & & & & & $\bigcirc$
   & $\bigcirc$ & $\bigtriangleup$ & \\
$WWWW$ & $\lambda, f_W$ & & & & & $\bigcirc$
   &  &  & \\
\hline
\multicolumn{2}{|c|}{$\sigma$ at $\sqrt{s}=500$ GeV (fb)}
   & $ 8\times 10^3$ & $ 6\times 10^3$ 
   & $ 10^3$ & $ 350$ & $ 60$ 
   & $ 80$ & $ 400$ & $50\times 10^3$ \\
\hline
\end{tabular} }
\end{small}
\end{table}
\end{footnotesize}

In the approach using the chiral 
Lagrangian\cite{WZREF-APPELQUIST-WU93,WZREF-DAWSON-VALENCIA94,WZREF-TANABASHI95}, 
the physical Higgs field is dropped while the Goldstone fields
are included to give masses to the gauge bosons. To this end, 
the unitary matrix operator, $U(x)$, consisting of
the Goldstone fields, $\omega^a$, is defined:
\begin{equation}
U(x)\equiv \exp\left( \frac{i \tau^a \omega^a}{v} \right)
\end{equation}
Using $U(x)$, the Lagrangian for the kinetic terms of
the gauge fields and the Higgs doublet is given by
\begin{equation}
{\cal L}_{sm} = {v^2 \over 4} {\rm tr}( D_\mu U^\dagger D^\mu U )
- { 1 \over 2 } {\rm tr} ( {\cal W }_{\mu\nu} {\cal W}^{\mu\nu} )
-  { 1 \over 2 } {\rm tr} ( {\cal B }_{\mu\nu} {\cal B}^{\mu\nu} ).
\label{WZEQ:smlagrangina}
\end{equation}
where
\begin{eqnarray}
{\cal W}_\mu &\equiv& { \tau^a \over 2 } W^a_\mu , \nonumber \\
{\cal W}_{\mu\nu} &\equiv& \partial_\mu {\cal W}_\nu
- \partial_\nu {\cal W}_\mu + i g_2 \left[
{\cal W}_\mu, {\cal W}_\nu \right] , \nonumber \\
{\cal B}_\mu &\equiv& { \tau^3 \over 2 } B_\mu , \nonumber \\
{\cal B}_{\mu\nu} &\equiv& \partial_\mu {\cal B}_\nu
- \partial_\nu {\cal B}_\mu , \nonumber \\
D_\mu U &\equiv & \partial_\mu U + i g_2 {\cal W}_\mu U - i g_Y U {\cal B}_\mu.
\label{WZEQ:wzdef}
\end{eqnarray}
$W^a_\mu$ and $B_\mu$ are $SU(2)_L$ and $U(1)$ gauge fields,
respectively.  

If new physics at some high energy scale affects properties
of the gauge boson at low energy, it will produce non-standard
terms in the Lagrangian.  The general terms of dimension
up to 4 is
\begin{eqnarray}
{\cal L} &=&  
 + \beta_1 {v^2 \over 4 } {\rm tr} ( { V_\mu T} ){\rm tr} ({ V^\mu T} )
 + \alpha_1 g_2 {\rm tr} 
\left( {\cal W}^{\mu\nu} U {\cal B}_{\mu\nu} U^\dagger \right) 
\nonumber \\
& +& i \alpha_2 g_Y {\rm tr} 
\left( U^\dagger \left[ V_\mu, V_\nu \right] U { \cal B}^{\mu\nu} \right) 
+ i \alpha_3 g_2 {\rm tr} 
\left( \left[ V_\mu, V_\nu \right] { \cal W}^{\mu\nu} \right)  
\nonumber \\
&+& \alpha_4 {\rm tr} ( V_\mu V_\nu ) {\rm tr} ( V^\mu V^\nu ) 
 + \alpha_5 {\rm tr} ( V_\mu V^\mu ) {\rm tr} ( V_\nu V^\nu ) 
\nonumber \\
&+& \alpha_6 {\rm tr}( V_\mu V_\nu ) {\rm tr} ( T V^\mu ) {\rm tr} ( T V^\nu ) 
 + \alpha_7 {\rm tr} (V_\mu V^\mu) {\rm tr} ( T V_\nu ) { \rm tr} ( T V^\nu )
\nonumber   \\
&+& { 1 \over 4} \alpha_8 g_2^2 {\rm tr} ( T {\cal W}_{\mu\nu} ) 
{\rm tr} ( T {\cal W}^{\mu\nu} ) 
+ { i \over 2 } \alpha_9 g_2 {\rm tr} ( T {\cal W}_{\mu\nu} ) 
{\rm tr} ( T \left[ V^\mu, V^\nu \right] )  
\nonumber \\
&+& {1 \over 2 } \alpha_{10} {\rm tr} ( T V_\mu ) 
{\rm tr} ( T V^\mu ) {\rm tr} ( T V_\nu ) {\rm tr} ( T V^\nu ) 
+ \alpha_{11} g_2 \epsilon^{\mu\nu\rho\lambda}
{\rm tr} ( T V_\mu ) {\rm tr} ( V_\nu {\cal W}_{\rho\lambda} ),
\label{WZEQ:chiral-lagrangian}
\end{eqnarray}
where
$V_\mu \equiv D_\mu U \cdot U^\dagger$, $T \equiv U \tau^3 U^\dagger$,
and $\epsilon_{0123} = - \epsilon^{0123} = 1$.
$\beta_1$ is the chiral coupling of energy dimension 2 and
$\alpha_1 \sim \alpha_{11}$ are dimension-4 chiral couplings.
Among the 12 couplings in Eq.~\ref{WZEQ:chiral-lagrangian},
$\beta_1$ and $\alpha_6\sim \alpha_{10}$ violate
custodial symmetry. $\alpha_{11}$ conserves $CP$ but violates
$P$ invariance.

The triple- and quartic-self-coupling vertices of 
the gauge bosons, the chiral couplings involved, and 
the processes involving these vertices are shown in 
Table~\ref{WZTAB-vertexandprocess}.

\begin{table}[htb]
\begin{small}\begin{center}\begin{minipage}[t]{\figurewidth}
\caption{\sl\label{WZTAB-vertexandprocess}
The chiral couplings involved in the 
triple- and quartic-gauge-boson vertices.
A $\bigcirc$ is shown if the corresponding coupling 
is involved in the vertices.  The processes
which are sensitive to the vertices are shown 
in the right-most column.}
\vspace{6pt}
\begin{center}
\begin{tabular}{| c | c c c c c c c c c c c c | c |}
\hline
vertex & $\alpha_1$ & $\alpha_2$ & $\alpha_3$ & $\alpha_4$ 
 & $\alpha_5$ & $\alpha_6$ & $\alpha_7$ & $\alpha_8$ 
 & $\alpha_9$ & $\alpha_{10}$ & $\alpha_{11}$ 
 & $\beta_1$ & processes\\
\hline
$WW\gamma$ & $\bigcirc$ & $\bigcirc$ & $\bigcirc$
  & & & & & $\bigcirc$ & $\bigcirc$ & & & 
  &  $\rightarrow WW$, $e\nu W$ \\
$WWZ$ & $\bigcirc$ & $\bigcirc$ & $\bigcirc$ 
  & & & & & $\bigcirc$ & $\bigcirc$ & & $\bigcirc$ & $\bigcirc$ 
  & $\rightarrow WW$, $e\nu W$ \\
$ZZWW$ & $\bigcirc$ &  & $\bigcirc$ 
  & &  $\bigcirc$ & &  $\bigcirc$ 
  & & & &  & $\bigcirc$ 
  & $\rightarrow WWZ$ \\
$ZWZW$ & $\bigcirc$ &  & $\bigcirc$ 
  & $\bigcirc$ & &  $\bigcirc$ &
  & & & &  & $\bigcirc$ 
  & $\rightarrow WWZ$ \\
$Z\gamma WW$ & $\bigcirc$ &  & $\bigcirc$ 
  & & & &
  & & & &  & $\bigcirc$ 
  & $\rightarrow WW\gamma$ \\
$ZZZZ$ & & & 
  & $\bigcirc$ & $\bigcirc$ & $\bigcirc$ & $\bigcirc$ 
  & & & $\bigcirc$ &  &
  & $\rightarrow ZZZ$ \\
\hline
\end{tabular}
\end{center}
\end{minipage}\end{center}\end{small}
\end{table}

$\alpha_1$, $\alpha_8$, and $\beta_1$ are included 
in two-point function and expressed in terms of 
$S$, $T$, and $U$ parameters as 
follows\cite{WZREF-APPELQUIST-WU93}:
\begin{equation}
\alpha_1 = -\frac{S}{16\pi}, \;\;
\alpha_8 = -\frac{U}{16\pi}, \;\;
\beta_1=\frac{1}{2} \alpha_{em} T
\end{equation}

Using the PDG values of $S$, $T$, and $U$\cite{ew-PDG2000}, we 
get 
\begin{eqnarray}
\alpha_1 &=& 0.00139 \pm 0.0022 \\
\alpha_9 &=& -0.0022 \pm 0.0030 \\
\beta_1 &=& -0.003\pm 0.0005
\end{eqnarray}
If we neglect the terms including $\alpha_1$, 
$\beta_1$, and $\alpha_8$, the anomalous triple-gauge-boson 
couplings, $\Delta\kappa_\gamma$, $\Delta\kappa_z$, and 
$\Delta g_1^Z$, 
can be expressed in terms of $\alpha_2$, $\alpha_3$, 
and $\alpha_9$ as follows:
\begin{eqnarray}
\Delta\kappa_\gamma&=&{e^2 \over \sin^2\theta_W } (  \alpha_2 + 
\alpha_3 + \alpha_9 ) \nonumber \\
\label{WZEQ:selfcouple}
\Delta\kappa_Z&=& {e^2\over \cos^2\theta_W}\alpha_2
+ {e^2\over \sin^2\theta_W}(\alpha_3 + \alpha_9) \\
\Delta g_1^Z &=& {e^2 \over \sin^2\theta_W \cos^2\theta_W} \alpha_3. \nonumber 
\end{eqnarray}
Since the $\alpha_9$ coupling violates the custodial $SU(2)$ symmetry,
genuine triple-gauge-boson couplings are
$\alpha_2$ and $\alpha_3$, if the custodial symmetry is imposed.

\section{Sensitivity to the Anomalous Gauge Boson Couplings}
\label{wz-sect-ww}
\begin{figure}
\centerline{\epsfxsize=10cm \epsfbox{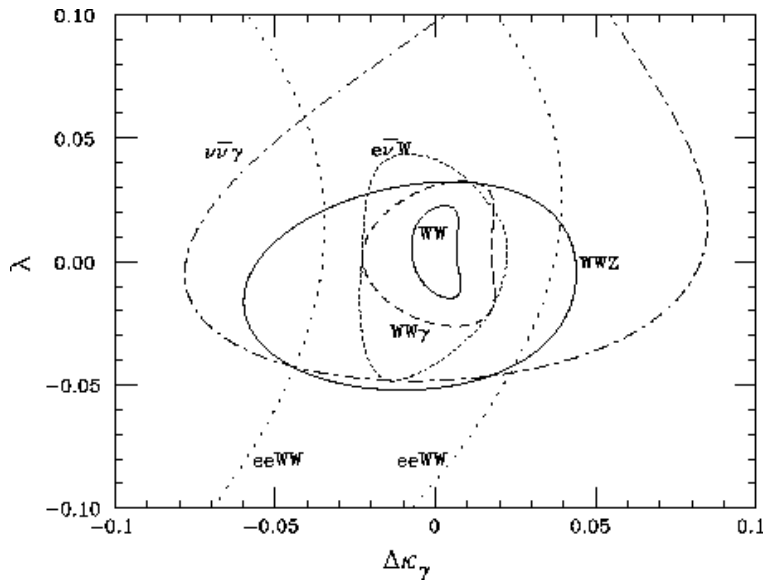}}
\caption{\sl \label{pew-ALLCONT}
Comparison of contours for various processes at the 90\% CL in the
plane of $\Delta\kappa_\gamma$ vs $\lambda$.}
\end{figure}

The sensitivities to the triple and quartic
couplings at JLC were studied in 
Ref.~\cite{WZREF-LCWS93MIYAMOTO}.
Results are summarized 
in Table.~\ref{pew-ALLBOUND}.
In the table, $\sqrt{s}=500$ GeV and the integrated luminosity of 50fb$^{-1}$ 
are assumed.  The limits in the 
plane of $\Delta \kappa_\gamma$ and $\lambda$ are shown in 
Fig.~\ref{pew-ALLCONT}.  In the table and the figure, 
the couplings other than under study are set to zero to derive 
the limits.

Among the processes studied, the process
$e^+e^-\rightarrow W^+W^-$ gives the most stringent bounds.
Nevertheless, since the $WW$ process involves all the three couplings,
the bounds from the processes $e\nu W$, $\nu\bar{\nu}\gamma$, 
and $\nu\bar{\nu}Z$ will provide complementary information. 
Limits obtained from the processes 
$e^+e^-\rightarrow e\nu W$, and 
$e^+e^-W^+W^-$ will improve if polarization information 
of $W$ is used.
In the single parameter case, the bounds on
$\Delta\kappa_\gamma$ and $\Delta\kappa_Z$ are less than 1\%,
and  we may have some sensitivity to new physics 
that gives rise to these effective interactions.

Sensitivities to the chiral couplings, $\alpha_4$ and  $\alpha_5$,
through the process, $e^+e^-\rightarrow W^+W^-Z$,
were studied in Ref.\cite{WZREF-LCWS95MIYAMOTO}.
Assuming the integrated luminosity of 50 fb$^{-1}$
at $\sqrt{s}=500$ GeV, sensitivities of about 0.5
at the 95\% C.L. are expected.  These couplings are sensitive 
to scalar resonances and higher energy is preferred
to get higher sensitivities.

\begin{table}
\caption{\label{pew-ALLBOUND}
{\sl
Summary of sensitivities
of various processes to anomalous couplings at the 95 \% CL,
assuming $\protect\sqrt{s}=500$ GeV and 50 fb$^{-1}$ integrated luminosity.
In the table, $X_W\equiv \cos\Theta_W$, $X_q\equiv \cos\theta_q$, and
$X_l\equiv\cos\theta_l$.}
}
\begin{small}
\begin{center}
\begin{tabular}{| c l | c | c | c | c |} \hline
\multicolumn{2}{|c|}{Process}
 & $\Delta\kappa_\gamma$ & $\Delta\kappa_Z$ & $\lambda$ 
 & method   \\ \hline
\multicolumn{2}{|l|}{$e^+e^-\rightarrow$} & & & & \\
\ & $W^+W^-$ & $-0.0052 \sim 0.0057$  
	            & $-0.0064 \sim 0.0062$ 
             & $-0.012 \sim 0.021$ 
             & $d\sigma/dX_W d\vert X_q\vert dX_l$ \\
\ & $e\nu W$ & $-0.021 \sim 0.020$ & $-$ 
             & $-0.039 \sim 0.038$ 
             & $d\sigma/d\vert X_q\vert$ \\
\ & $\nu\bar{\nu}\gamma$ & $-0.071\sim0.075$ & $-$ 
	            & $-0.044\sim 0.079$ 
	            & $d\sigma/dE_\gamma$ \\
\ & $\nu\bar{\nu}Z$ & $-$ & $-0.29\sim 0.25$ & 
		      $-0.46 \sim 0.17$ & $\sigma_{tot}$ \\
\ & $W^+W^-\gamma$ & $-0.020 \sim 0.016$
             & $-0.018\sim 0.025$ 
             & $-0.025\sim 0.028$
             & $d\sigma/dE_\gamma$ \\
\ & $W^+W^- Z$ & $-0.053 \sim 0.041$ 
               & $-0.071 \sim 0.15$
               & $-0.050 \sim 0.030$
               & $\sigma_{total}$ \\ 
\ & $e^+e^-W^+W^-$ & $-0.032 \sim 0.039$ 
               & $-$
               & $-0.084 \sim 0.12$
               & $\sigma_{total}$ \\ 
\hline
\end{tabular} 
\end{center}
\end{small}
\end{table}

\section{Heavy Higgs and $WW$ Scattering}
\label{wz-strongw}

\begin{figure}[thbp]
\centerline{
\epsfxsize=7cm\epsfbox{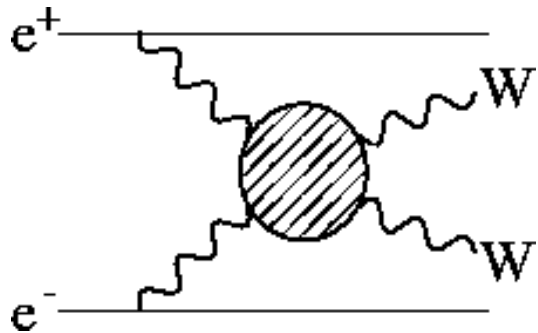}}
\caption{\sl \label{wz-strongh-wfusion.eps}
A Feynman diagram for the $W$ fusion process.}
\end{figure}

Even if no light Higgs boson is observed at JLC 
experiment at the center of mass energy below 500 GeV, 
its mass could be estimated from the precise measurements
of the electroweak interaction.  If the expected mass is less than 
1 TeV,  JLC could increase the energy and the luminosity
to cross the threshold.  If the Higgs boson is heavy, 
the major diagram for Higgs boson production is 
the $W$ fusion process shown in Fig.\ref{wz-strongh-wfusion.eps}.
The background processes to the $W$ fusion process are
\begin{eqnarray}
e^+e^- &\rightarrow& W^+W^-\\
&\rightarrow& \nu_e \bar{\nu_e} Z^0 Z^0 \\
&\rightarrow& e^{\pm}\mathop{\nu_e}\limits^{\mbox{\tiny$(-)$}} W^\mp Z^0\\
&\rightarrow& e^+e^- Z^0 Z^0 \\
&\rightarrow& W^+W^-  .
\end{eqnarray}

The cross sections of these processes were calculated by GRACE\cite{WZREF-GRACE}.
In these calculations, a care has to be taken so as not to violate
the unitarity of the scattering amplitudes,  
as the couplings of heavy Higgs boson are strong.  
We thus enlarge the width of the Higgs boson artificially to keep the amplitude 
below the unitarity limit.  This method leads to a conservative estimate
of the Heavy Higgs signal.

\begin{figure}[htbp]
\begin{center}
\centerline{
\epsfxsize=\figurewidth\epsfbox{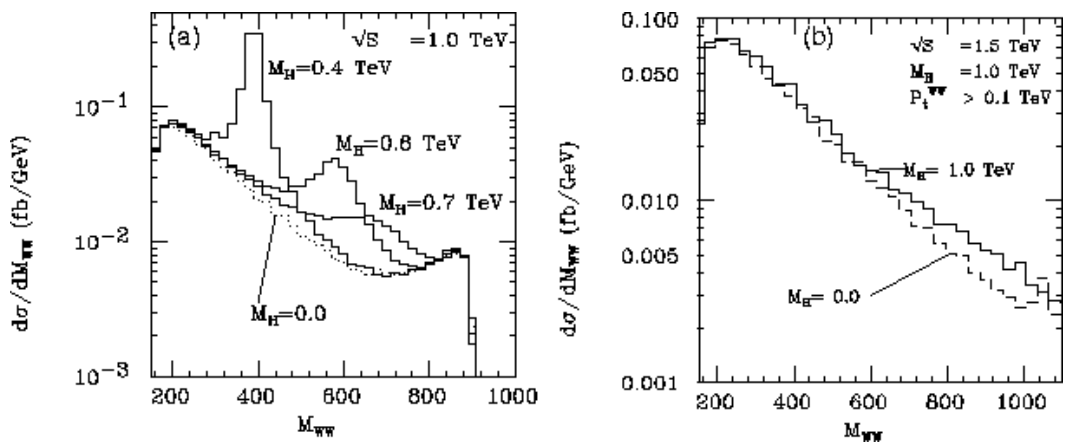}}
\end{center}
\begin{center}\begin{minipage}[t]{\figurewidth}
\caption{\sl \label{wz-strongh-wwmass.eps}
(a) invariant mass distributions of 
$WW$ at $\sqrt{s}$=1 TeV for Higgs masses of 0, 400, 600, and 700 GeV.
(b) similar plots at $\sqrt{s}$=1.5 TeV 
for Higgs masses of 1.0 TeV and 0 TeV.}
\end{minipage}\end{center}
\end{figure}

The Heavy Higgs boson predominantly decays to a $W$ pair,
thus a Higgs resonance is expected to be observed as a peak in 
the $WW$ mass distribution.  The invariant mass distributions of 
$WW$ for several Higgs mass values are shown in Fig.\ref{wz-strongh-wwmass.eps}.
As seen from the figure, the width of Higgs boson increases with mass 
and for the Higgs mass of 1 TeV, no clear peak structure 
in the $WW$ invariant mass
distribution can be seen.

We studied the heavy Higgs signal using the Monte Carlo detector 
simulation\cite{ew-heavyhsim}.  
We studied two cases, one at $\sqrt{s}=$ 1 TeV and a Higgs
mass of 0.7 TeV, the other at $\sqrt{s}=1.5$ TeV 
and a Higgs mass of 1.0 TeV.
We selected hadronic decays of $W$ as follows.
First, we required four jets be observed in the detector 
and both masses of two jet-pairs be consistent with the $W$ mass.
Backgrounds at this stage of selection were 4-jet
events from $e^+e^- \rightarrow e^+e^-W^+W^-$ and $e^+e^-Z^-Z^0$ 
processes.  These processes were rejected by vetoing
energetic $e^\pm$.  When $e^\pm$'s are
undetected as they escaped into the uncovered region around the beam pipe,
the remaining $W$ pair can not have a large missing $p_t$.  
We could thus eliminate most of the background events by
requiring that the $W$ pair system had to have a large missing transverse
momentum.
The $W$'s from Higgs decays have high longitudinal polarization,
while those from other sources do not.  The Higgs signal 
could hence be enhanced if we required jet production angle in 
the $W$ rest frame be large.

\begin{figure}[htbp]
\begin{center}
\centerline{
\epsfxsize=\figurewidth\epsfbox{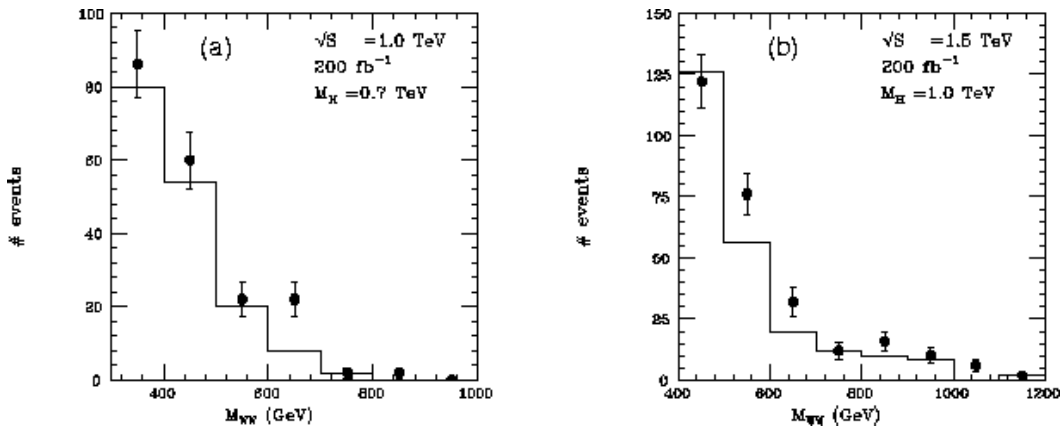}}
\end{center}
\begin{center}\begin{minipage}[t]{\figurewidth}
\caption{\sl \label{wz-strongh-events.eps}
The invariant mass distribution of $W$ pairs at (a) $\sqrt{s}=$1.0 TeV and
the integrated luminosity of 200fb$^{-1}$  and 
(b) $1.5$ TeV.  Histograms are for the Higgs mass of 0 TeV
and data points are for 0.7 TeV(a) and 1.0 TeV(b), respectively.
}
\end{minipage}\end{center}
\end{figure}

The invariant mass distributions of the $W$ pairs in the selected events for 
the integrated luminosity of 200fb$^{-1}$ are shown in 
Fig.~\ref{wz-strongh-events.eps}.
For both of the cases, we could see Higgs effect at statistical significance 
of 3-sigma\cite{wz-strongww}.

The studies of the $WW$ fusion processes are sensitive to 
J=0 and J=1 components of the $WW$ scattering amplitude.
If new strong interaction is responsible for the $W$ mass generation, 
it may produce resonances in J=1 channel, which can be probed 
by a study of the $e^+e^-\rightarrow W^+W^-$ process\cite{WZREF-TANABASHI95}.
Since the center of mass energy of the $WW$ system is larger
than that of the $WW$ fusion process for a given $\sqrt{s}$,
higher sensitivity to the J=1 channel is expected.

A study was performed assuming a vector resonance 
strongly coupled to the longitudinally polarized $W_LW_L$\cite{ew-heavyhsim}.
If its mass is low enough, its effect could be 
observed as anomalous behavior when $W_LW_L$ is produced in 
the $e^+e^-\rightarrow W^+W^-$ process.
For simulation study, we introduce the 
effect by a form factor, $F(q^2)$, in the $W_LW_L$ amplitude.
With this form factor, the cross section is 
expressed as 
\begin{equation}
\sigma \propto | \sum_{(ij)\neq(LL)} M_{(ij)} + M^{J\neq 1}_{(LL)}
+ M^{J=1}_{(LL)}\times F(q^2) |,
\end{equation}
where $M^{J=k}_{(i,j)}$ is the scattering amplitude of 
the $e^+e^-\rightarrow W^+W^-$ process and the subscript 
$(i,j)$ is the polarization of the $W$'s and the superscript $(J=k)$
is the angular momentum of the $W$ pair.  

Two types of form factors were investigated:
the Breit-Wigner type and 
those calculated based on the N/D method\cite{IGIHIKASA}.
The Breit-Wigner-type form factor is expressed as
\begin{equation}
F(q^2) = - {m_V^2 \over q^2 - m_V^2 + i\,m_V\,\Gamma_V\,({q^2\over
m_V^2})},
\end{equation}
where $m_V$ and $\Gamma_V$ are the mass and the width of the vector
resonance, respectively.
It turns out that at $\sqrt{s}=500$ GeV, 
the $\Gamma_V$ dependence of the Breit-Wigner form factor 
is negligible if the mass is greater than 1~TeV
and the width is, say, below 400 GeV.
We therefore fixed the width at 100~GeV in the analysis.
In the case of the N/D method the form factor is expressed as a function of
the masses of the scalar resonance ($m_S$), the vector resonance($m_V$), and
the ratio ($r$) of the contribution of the scalar and vector 
resonances\cite{IGIHIKASA}.  Since the scalar resonance does not
contribute to the $J=1$ channel of the $e^+e^-\rightarrow W^+W^-$ 
process, we fixed $m_S=$ 1~TeV.

With the detector simulation, we studied the $e^+e^-\rightarrow W^+W^-$ 
process where one $W$ decays to $e$ or $\mu$ and the other
decays hadronically.  This mode has a large branching ratio($\sim 27\%$).
$W$'s produced in the backward direction can be detected efficiently
from lepton charge measurement, and angular analysis of quarks 
and leptons maximize the sensitivity to $W$ polarization.

\begin{figure}[htbp]
\centerline{
\epsfxsize=10cm \epsfbox{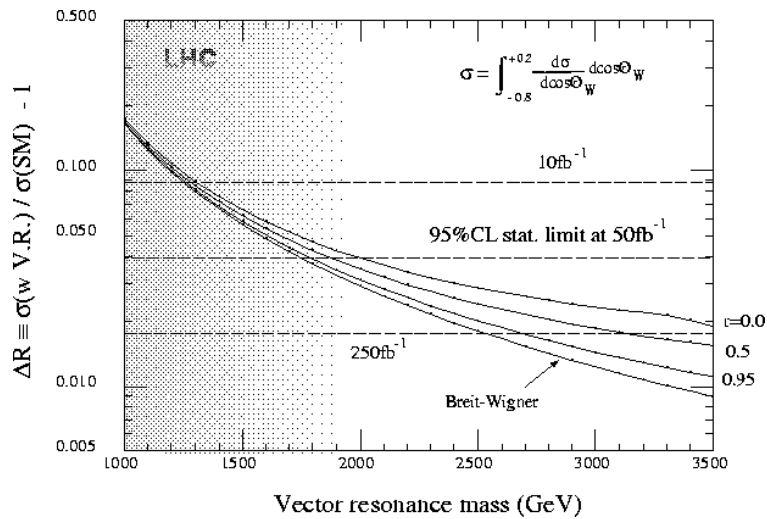}}
\caption{ \sl\label{WWTRRATIO}
Excess of the signal as a function of the vector
resonance mass. The
vertical axis is the excess of the cross section in the given 
angular range normalized to the standard model cross section.
}
\end{figure}

\begin{figure}[htbp]
\centerline{\epsfxsize=10cm \epsfbox{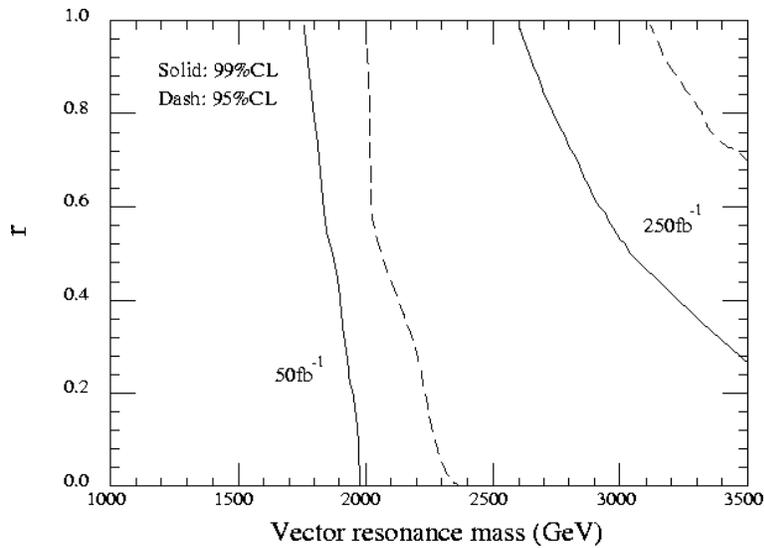}}
\caption{\sl \label{ew-wwtrconta}
Contour plot of the sensitivity to the vector resonance
based on the N/D method, in the plane of
$r$ and the vector resonance mass.}
\end{figure}

The effect of the vector resonance in the $W_LW_L$ amplitude 
is to increase the cross section especially in the 
backward region, as the forward region is dominated
by the transversely polarized $W$ due to the
$t$-channel neutrino exchange diagram.
The increase of the cross section in the angular range
$-0.8 < \cos\theta_W < 0.2$ relative to the 
standard model cross section is shown in Fig.\ref{WWTRRATIO}
as a function of the mass of the vector resonance.
In this figure, the solid lines correspond to the
Breit-Wigner resonance formula and those calculated based on the N/D
method with $r=0.0, 0.5$, and 0.95, respectively.
The statistical limits at the 95\% CL are shown by dashed lines for 
integrated luminosities of 10, 50, and 250~fb$^{-1}$.
The shaded area is the region to be explored by the direct
search at LHC\cite{ew-LHCVR}.
We can see that JLC at $\sqrt{s}=500$ GeV with an integrated
luminosity of 50~fb$^{-1}$ is
sensitive to the vector resonance of mass up to
about 2 TeV, which is similar to the sensitivity expected
with the direct search at LHC.
With five times more luminosity 
we may be sensitive to masses larger than 2.5 TeV.
If we measure polarization of $W$ from angular analysis of decay 
daughters, higher sensitivity is obtained.
Sensitivity based on the N/D method is shown in Fig.\ref{ew-wwtrconta}.

In summary, the vector resonance in the $W_LW_L$ amplitude 
has significant effect at $\sqrt{s}=500$ GeV
even if the mass of the resonance is as high as 
2 TeV, and its effect can be observed at JLC.

{}

%% file: detir/main.tex
\chapter{Interaction Region}
\label{chapter-detir}
\section{Introduction}
\input detir/introduction.tex

\section{Backgrounds}
\input detir/background.tex

\section{Layout}           
\input detir/layout.tex
\section{Estimation of background hits}           
\label{detir-background-section}
\input detir/hits.tex

\section{Background hits in different detector models}           
\input detir/newmask.tex

\section{Pair monitor}
\input detir/pairmonitor.tex

\section{Luminosity Monitor and Active mask}
\input detir/vetosystem.tex

\section{Support system}
\input detir/support.tex

\section{Dump line}
\input detir/dump.tex

\input detir/reference.tex

%% file: detir/introduction.tex
In this chapter we examine various aspects of experimentation that directly interacts with the accelerator design.  Of particular concern in this area is the background to the physics experiments that are caused by the beams passing through the detector. 

The characteristics of  background events at JLC will be very different from those at  typical  $e^+e^-$ colliders, except the SLC.   The features of the background strongly depend on numerous operational parameters of the accelerator,  such as the beam aspect ratio (typically $\sigma_{x}^* / \sigma_{y}^* =0(100)$ ), a high beam intensity ($10^{10}$particles/bunch) and possible tails in the particle distribution  that deviates from a Gaussian distribution. The population of low energy $e^\pm$ pairs that are created during collisions are directly related to the beam aspect ratio, while the tail is mostly responsible for synchrotron radiation and muon background. The optimization of the machine operational parameters must be  considered by taking  this ``interaction" between the experimentation and the machine into account.   

The highest-priority goal here is, of course, to maximize the  luminosity while minimizing the background.   With such a motivation in mind, the effects of $e^\pm$ pairs  have been estimated by detailed Monte Carlo simulations with various configurations of masking system, in addition to simulations of synchrotron radiations and the attenuation of the muon flux that are produced by interactions of the beams with upstream collimator materials. 
Performances of luminosity monitor and active mask are demonstrated under the huge pair background.
Preliminary design of dump line is also presented, where a possibility of beam-energy spectrum is discussed with an accuracy of $\le 0.1$\% and a beam loss is estimated for the major source of neutron-backgrounds at the
interaction point(IP).

Another important issue is a stabilization of final quadrupole magnets to keep "head-on" collisions between nano-meter beams.
A detailed analysis of the support system is presented especially on the oscillation properties.
A real-time measurement of the beam size is vital part of such a stabilization process. As the most promising detector, a pair monitor has been proposed and its expected performances are described.

%% file: detir/background.tex
\begin{figure}[htbp]
\centerline{\epsfxsize=14cm \epsfbox{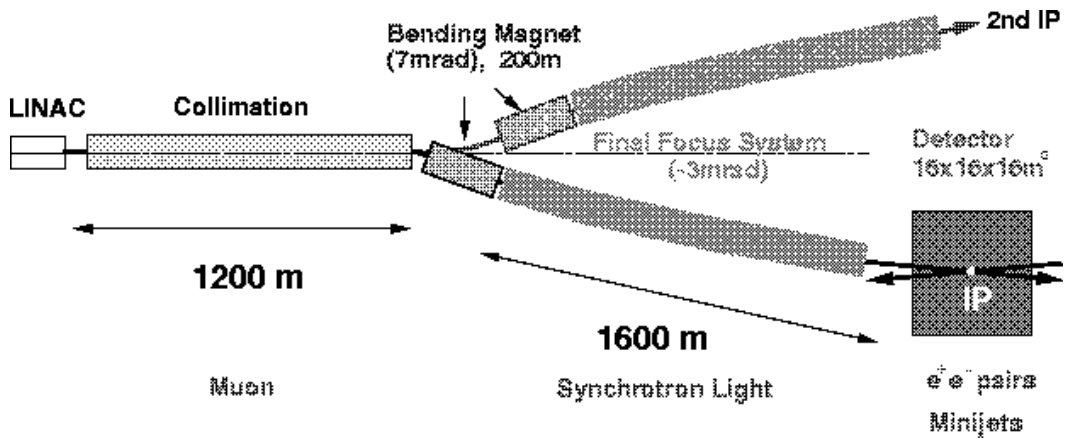}}
\begin{center}\begin{minipage}{\figurewidth}{
\caption{ \sl
Top view of the beam line from the exit of the main linac to the
interaction point (IP) at $E_{cm}= 0.5-1.5$ TeV.
 \label{finalfocus}}
}\end{minipage}\end{center}
\end{figure}

To illustrate where the background particles can originate, the beam line from the exit of the main linear accelerator (linac) to the interaction point (IP) is schematically shown in Figure~\ref{finalfocus}. A $+7$~mrad bending magnet section (200~m
long) downstream of the collimation section is needed in order to create a sufficient amount of separation for two experimental halls and to prevent the background from the upstream linac from directly hitting the detector. In the final focus system, beams are gradually deflected so as to have a horizontal beam crossing angle of $\pm 4$~mrad at IP.
  
There are two major sections for beam collimation (1200~m long) and a final focus system (1800m long) in the beam line to handle beam energies up to 0.75~TeV. While their main purposes are to clip the beam tails, secondary particles are inevitably produced, namely:
(1) muons and (2) synchrotron radiation photons, respectively. In addition at the IP, (3) $e^+e^-$ pairs and (4) mini-jet are created through beam-beam interactions.  They all cause background hits in the detector facility.

In  subsequent sections the first three kinds of background are discussed together with a possible design of the interaction region.

\subsection{Collimation and Muons}

In this section we discuss the production of muons through the interaction of particles in the beam tails when they are collimated at the upstream collimation sections.

Generally, the transverse profile of the beam does not exactly follow a Gaussian distribution at linear colliders. The beams can be accompanied by long tails according to experience from experiments at SLC~\cite{Burke}.  While the origin of these tails is not thoroughly understood at present, we shall conservatively assume that the beam has a flat tail beyond $\pm 3 \sigma_{x(y)}$ in both the horizontal(x) and vertical(y) directions with a relative intensity of $0.1 \sim 1$\%. 

The beam must be collimated within $\pm 6 \sigma_x$ and $\pm 40 \sigma_y$ in order to keep the background due to synchrotron radiation at a manageable level. Since the typical size of the beam core is on the order of a few $\mu$m, collimating such beams is a seriously non-trivial task. A work-around is to expand only the tail part sufficiently by using a non-linear collimation technique, as discussed in Chapter 13.   This is part of the reason why a 1200~m-long collimator section is required for collimating a 0.75~TeV beam.  

One RF pulse will accelerate a bunch train which contains up to 95 bunches separated by 2.8 nsec at a repetition rate of 150Hz. Since each bunch consists of $7.5 \times 10^{9}$ electrons (or positrons) at the IP,  about $10^8$(1\%tail)$\times 10^2$(bunches) electrons may hit collimators at 150~Hz. In the interactions of the beam tails with the collimators a large number of muons are produced through the Bethe-Heitler process, $e^{\pm} N \rightarrow e^{\pm}\mu^+\mu^-N$. Without suitable measures these muons would traverse through the tunnel and create a large amount energy deposit within the detector facility. They would cause serious background problems for conducting high-energy physics experiments.

\begin{figure}[htbp]
\centerline{\epsfxsize=12cm \epsfbox{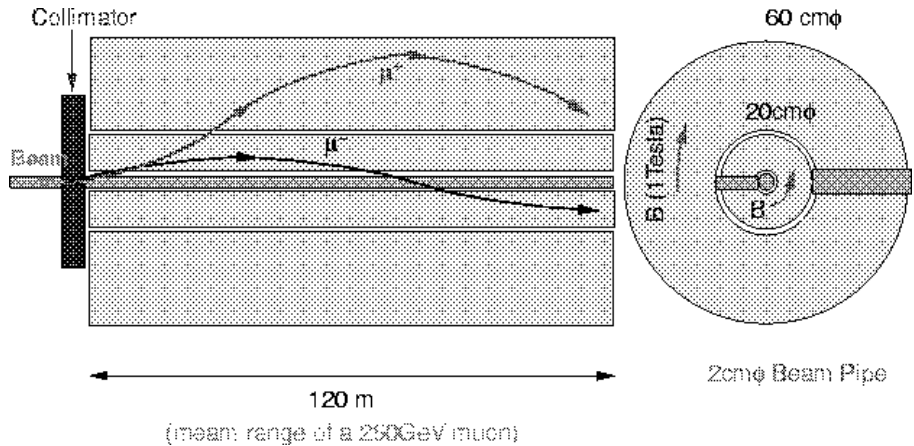}}
\begin{center}\begin{minipage}{\figurewidth}{
\caption{ \sl Original idea of a muon attenuator. Two iron pipes are
magnetized axially in opposite directions for both charged muons, which can be trapped,
where the 120m length of the iron pipe corresponds to a mean range of 250GeV muons. 
\label{attenuator}}
}\end{minipage}\end{center}
\end{figure} 

\begin{figure}[htbp]
\begin{minipage}[t]{5.5cm}
\centerline{\epsfxsize=5.5cm \epsfbox{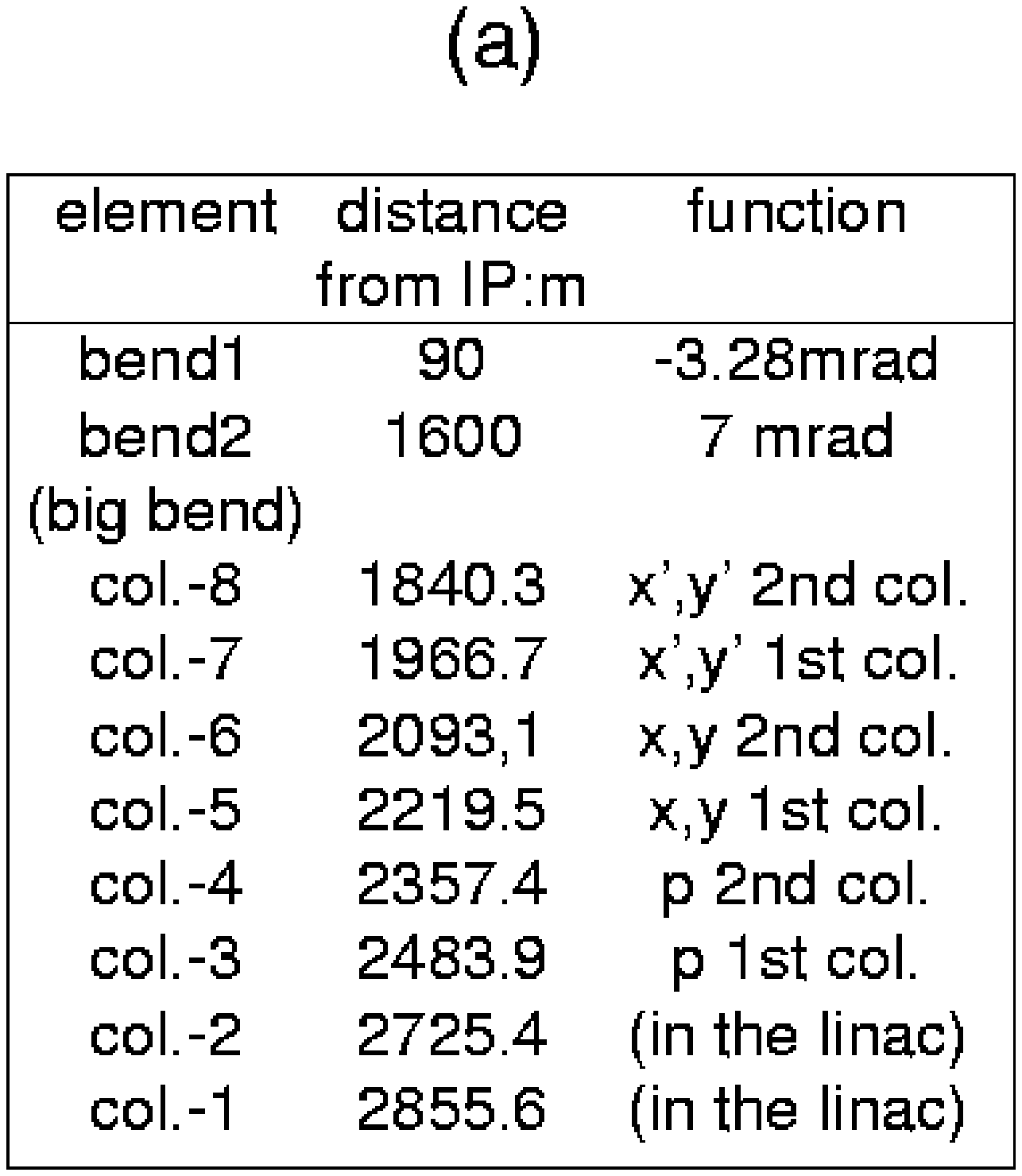}}
\end{minipage}
\hfill
\begin{minipage}[t]{9.8cm}
\centerline{\epsfxsize=9.8cm \epsfbox{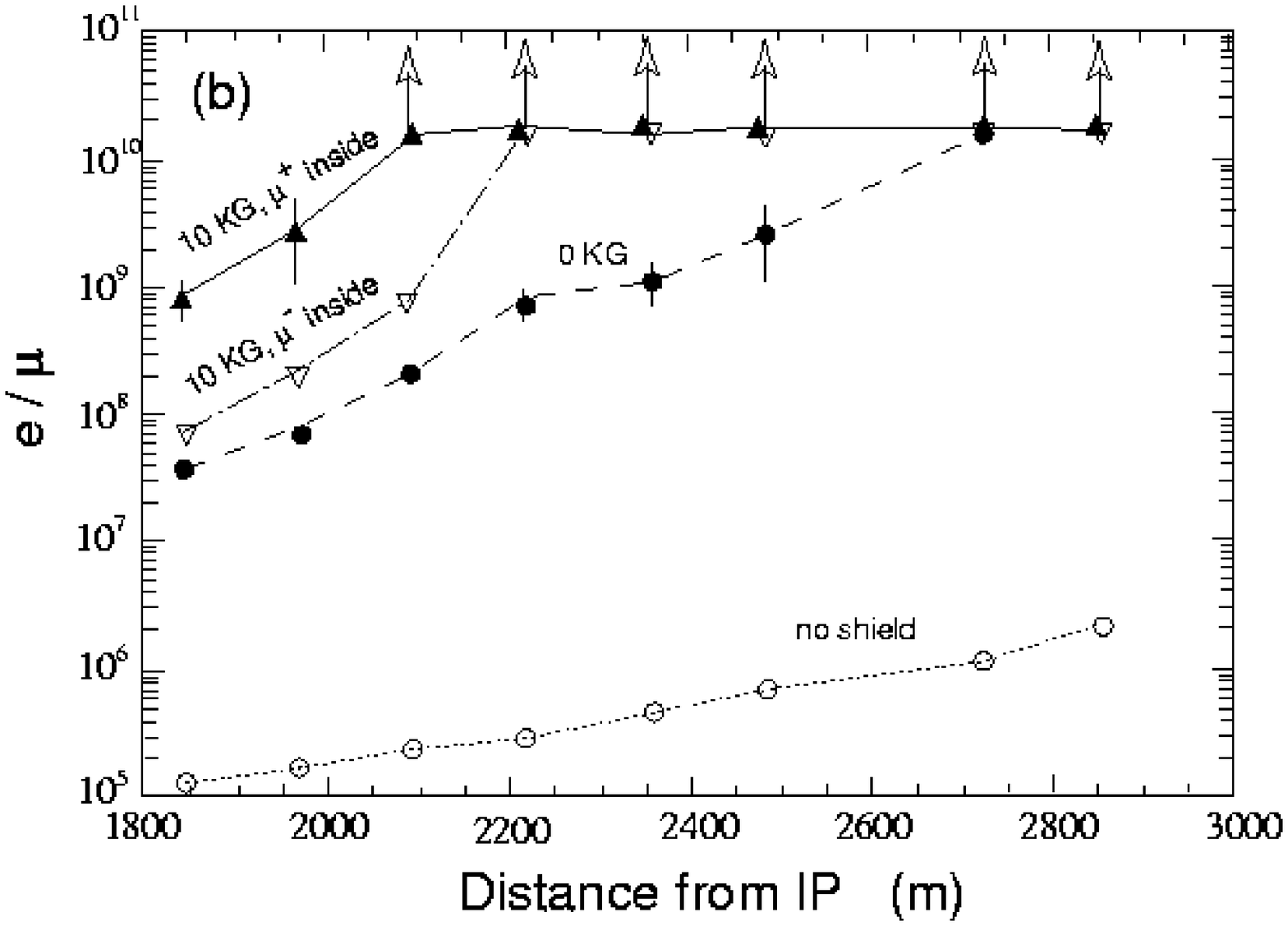}}
\end{minipage}
\begin{center}\begin{minipage}{\figurewidth}{
\caption{\sl (a) Locations of bending magnets and collimators in the beam delivery system. (b) The number of collimated electrons (positrons), which give one muon in a detector of $16 \times 16 \times 16$m$^3$, at eight collimators. Open circles are simulation results with no muon shield, and the others are those with muon attenuators at 1510$\sim$2856m from IP (Y. Namito, Jan.1999).}
\label{mu-background}
}\end{minipage}\end{center}
\end{figure}

There are eight collimators upstream of the big bend, which are located at 
1840.3 $\sim$ 2855.6~m from IP as shown in Fig.\ref{mu-background}~(a).  Two of them (col.-1 and -2) are set in the linac.  They doubly collimate the beams in momentum  space($\Delta p/p < \pm 2$\%) and phase space of transverse profile and divergence ($6 \sigma_x^{(\prime)} \times 40 \sigma_y^{(\prime)}$).
During the collimation, a lot of secondary particles would be produced. Among  them high energy muons would create the most severe backgrounds. A rate of muons depends on a population in a tail of beam.  For the conservative estimation, we assume that the tail is $0.1\sim 1$\% of the bunch population for this simulation study.  

 In the simulation muons were tracked down to the detector of $16 \times 16 \times 16$m$^3$ with various configurations of muon attenuation in a tunnel.  The present simulation does not take account of optical elements except for bending magnets (Fig.\ref{mu-background}(a)) at the beam line.
The muon attenuator is a magnetized iron cylinder of $1< r < 30$cm, which actually comprises two cylinders of $1< r < 10$cm (inner) and $10 < r < 30$cm (outer) as shown in Fig.~\ref{attenuator}. The axis of cylinders corresponds to the beam line.  The two cylinders are magnetized circularly in opposite directions to confine positive and negative muons in each.
The muon attenuator is assumed to cover from the big bend through the whole collimation system, which is 1510 to 2856m from IP.
Figure~\ref{mu-background}(b) shows the number of electrons (e/$\mu$) at 8 collimators to produce one muon in the detector~\cite{namito}.  Open arrows indicate the lower limits since no muon was observed within the simulation statistics.  The best configuration (solid triangles) was the muon attenuator magnetized at 10KG to confine positive muons ( 10 KG, $\mu^+$ inside ) in the inner cylinder, where the beam particles were 250GeV electrons.   It attenuates muons by 4 order of magnitudes  more than  a case  with no muon-shield (open circles).
The oppositely magnetized attenuator( open triangles, 10 KG, $\mu^-$ inside ) has less shielding power than the best case  for col.6-8 since the negative muons can be transported to the detector in the same way as the beam. The magnetization effect is clearly seen  by comparing them with the no magnetized attenuator  (solid circles, 0 KG).
With the best configuration and the $0.1\sim 1$\% beam tail, a few muons would hit the detectors per pulse for the pulse population of $( 0.71 \sim 1.33 ) \times 10^{12}$  corresponding to A$\sim$Y in Table~\ref{JLCparameters}. 
In this case the muons should not be a serious background.

\subsection{Synchrotron Radiation}

Since the intensity of synchrotron radiation is on the same order as the intensity of the beam, itself, it would be very harmful to the experimentation if they are scattered at the pole tips of the nearest quadrupole magnet QC1. It is also very difficult to shield them near to the IP. The amount of synchrotron radiation is determined by the maximum size and angular divergence of the beam profile. Both the maximum beam size $(x ,y)$ and the divergence $(x',y')$ shall be well defined and controlled by the collimators.

The divergences can be expressed by $\sigma_{\theta_{x(y)}} = \sqrt{ \epsilon_{x(y)} / \beta_{x(y)} }$, where $\beta_{x(y)}$ is a (optical) beta function in the final focus system and $\epsilon_{x(y)}$ is an emittance of beams.  The beam size and the divergence are related by $\epsilon_{x(y)}=\sigma_{\theta_{x(y)}}\cdot\sigma_{x(y)}$. 
If we can not control them by some means, we must change the optics to enlarge $\beta_{x(y)}$ so that $\sigma_{\theta_{x(y)}}$ decreases. We may thus even have to sacrifice the luminosity because of $\sigma_{x(y)}= \sqrt{ \epsilon_{x(y)}\cdot\beta_{x(y)} }$.   A similar situation would likely occur at the beginning of operation with a larger emittance than the expected one, as happened in the SLC experiments.

\begin{figure}[tbp]
\centerline{\epsfxsize=14cm \epsfbox{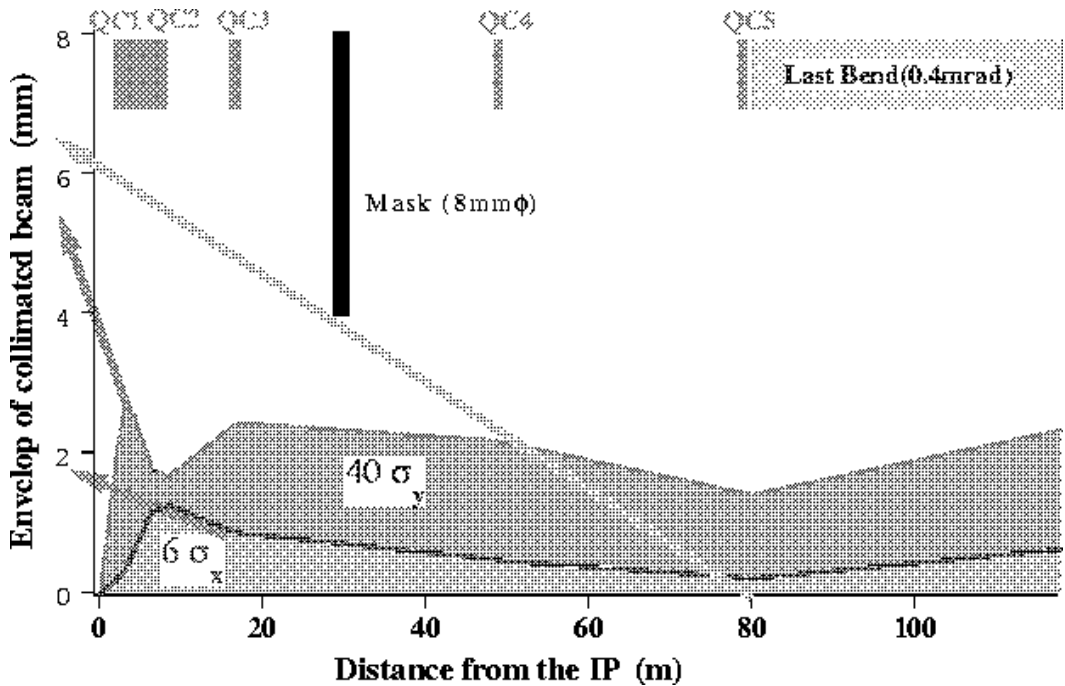}}
\begin{center}\begin{minipage}{\figurewidth}
\caption{\sl Horizontal(6$\sigma_x$) and vertical(40$\sigma_y$) beam envelopes through the last bending magnet and five final focus quadrupole magnets(QC1,QC2,QC3,QC4 and QC5). The maximum divergences of the synchrotron radiation are also drawn by arrows. \label{synrad}  }
\end{minipage}\end{center}  
\end{figure} 
\begin{figure}[htbp]
\begin{minipage}[t]{7.5cm}
\centerline{\epsfxsize=7.5cm \epsfbox{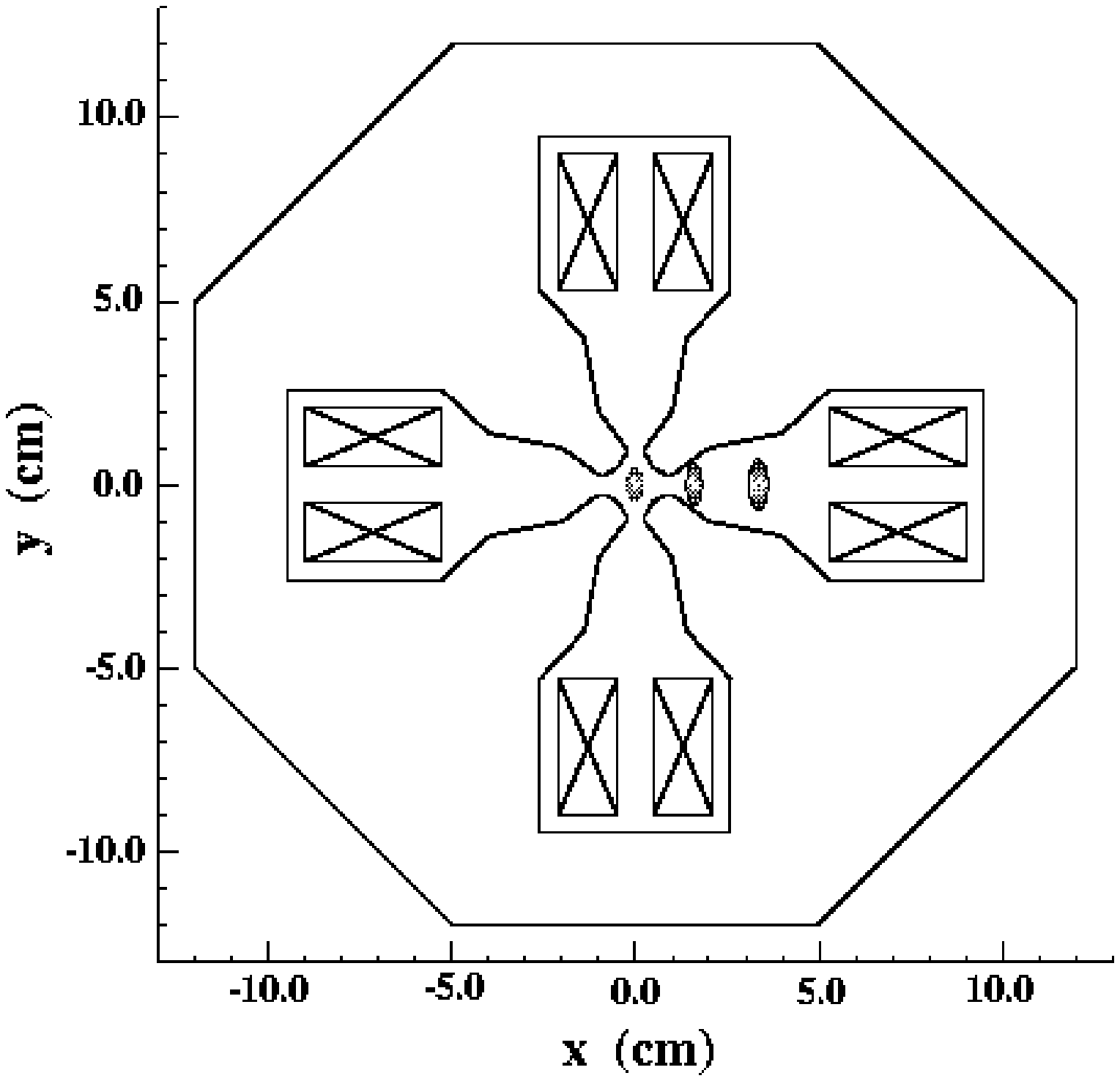}}
\end{minipage}
\hfill
\begin{minipage}[t]{7.5cm}
\centerline{\epsfxsize=7.5cm \epsfbox{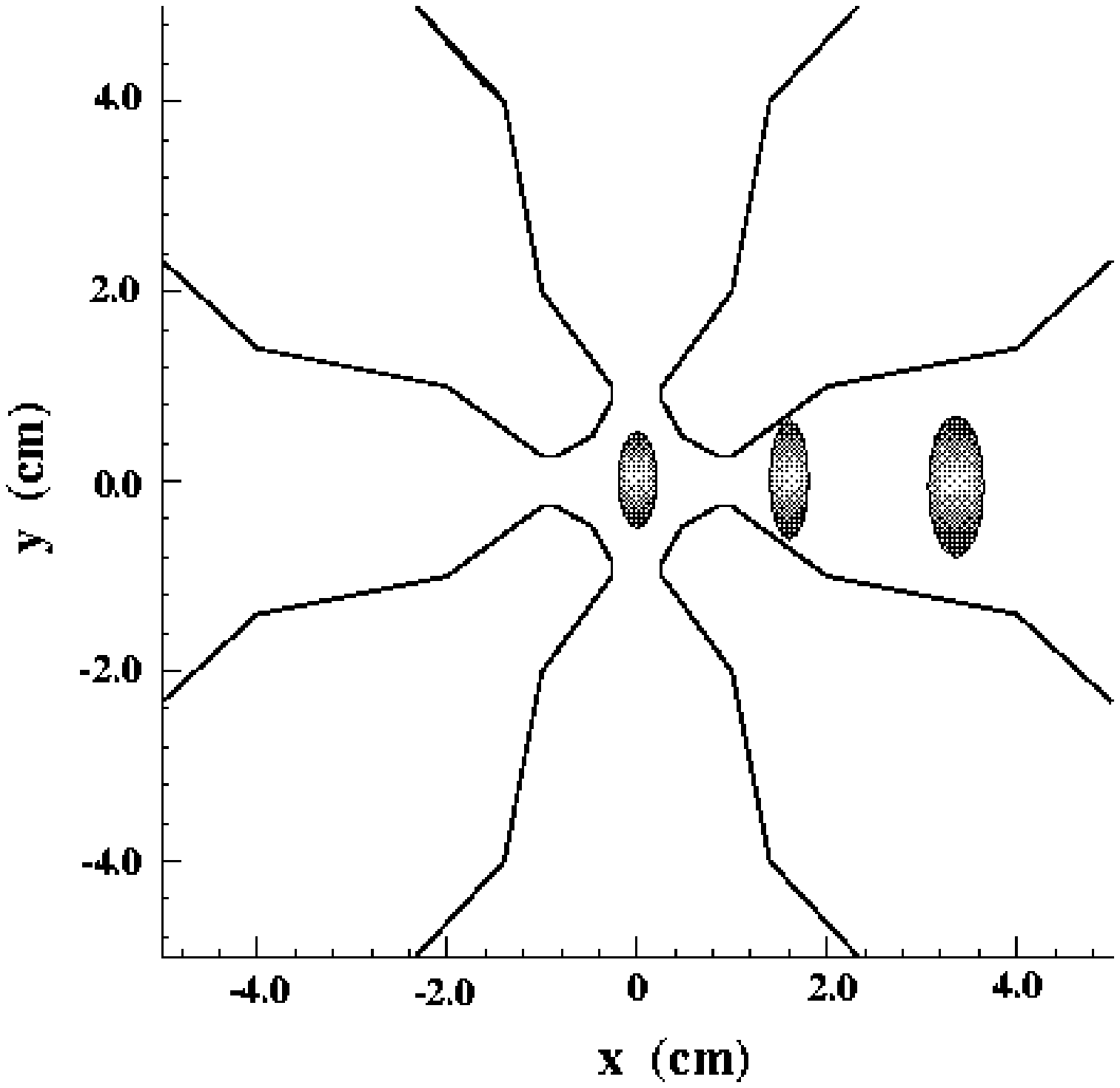}}
\end{minipage}
\begin{center}\begin{minipage}{\figurewidth}{
\caption{\sl Profiles of the synchrotron radiation at QC1;right figure shows the magnified view around the center of QC1. The profile at the center accompanies the in-coming beam. The two right-hand side ones are passing through QC1 of 2.2m long after a collision with a
8mrad horizontal crossing. \label{QC1rad}   }}
\end{minipage}\end{center}
\end{figure}  
Figure~\ref{synrad} shows the development of transverse beam envelops that correspond to $6\sigma_x\times 40\sigma_y$. Figure~\ref{synrad} covers the region from the IP up to the nearest dipole bend magnet.  Here the smearing effect due to collimation is not taken into account, since it was estimated to be very small($10^{-8}$) as mentioned earlier.  

With a mask of 8~mm$\phi$ radial aperture that is located at 30~m from the IP, synchrotron radiation from  upstream magnets beyond the last bending magnet can be completely masked. Any synchrotron radiation that passes through the aperture of  8~mm$\phi$ mask would pass through the final quadrupole magnet (QC1) without scattering. The half aperture of  QC1 is chosen  to be 6.85~mm. As can be clearly seen in Figure~\ref{synrad}, the radiation from QC3
and QC2 provides the maximum divergence at the IP in the horizontal and vertical directions, respectively.  The profiles of the radiation at the QC1 are shown in Figure~\ref{QC1rad}.  The length and inner aperture of QC1 are 2.2~m and 13.7~mm$\phi$, respectively. The front face of QC1 is located 2.0~m from the IP.  The in-coming radiation passes through the central axis of QC1.  After colliding with the opposing beam  at  a horizontal crossing angle of 8~mrad, the out-going radiation passes off-axis through the QC1 magnet on the other side.  The location of the out-going radiation is depicted as two elliptic profiles on the right-hand side of Figure~\ref{QC1rad}.  

As described above, it is expected that there should be no background problems due to the synchrotron radiation if we carefully optimize the collimation and the optics simultaneously. 

\subsection{$e^+e^-$ pairs}\label{eepairs}

An enormous amount of $e^+e^-$ pairs will be created during collisions, for instance a few times $10^5$ pairs per bunch crossing.  While their vast majority are scattered into extremely forward angles, some can be greatly deflected by the strong magnetic field that is produced by the in-coming beam. In such cases they can enter the detector region and can create background noise to the detector facility. 
\begin{figure}[tbhp]
\centerline{\epsfxsize=14cm \epsfbox{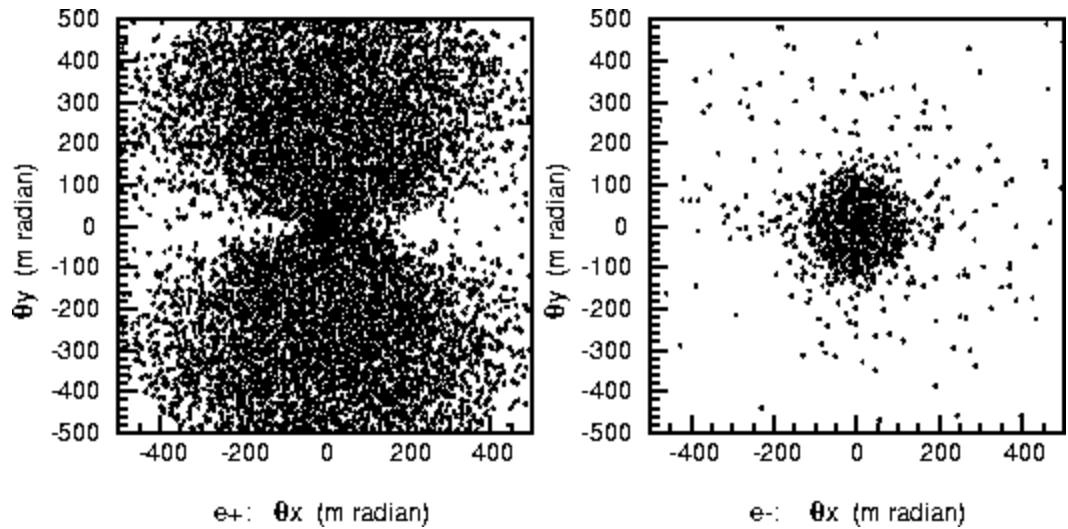}}
\begin{center}\begin{minipage}{\figurewidth}
\caption{\sl \label{angpairs}       
Electrons and positrons scattered by beam-beam interaction at IP.  The vertical and horizontal axes denote the scattering angles in the vertical and horizontal directions, respectively. The left- and right-hand figures show the distributions of the
positrons and electrons, respectively, downstream of the electron beam. Since the positrons have the same sign charge as the in-coming (positron) beam, they are  scattered by larger angles than the electrons.} 
\end{minipage}\end{center}
\end{figure}  
Figure~\ref{angpairs} shows the angular distributions of pair-produced electrons and positrons after the beam-beam interaction. The calculation was made using ABEL\cite{detir-ABEL}. Many particles (electrons or positrons) are scattered at large angles
exceeding 200~mrad.   We can also clearly see an asymmetry in their azimuthal distribution. More particles are deflected in the vertical direction than in the horizontal  because of a very flat transverse beam profile ($\sigma_x^* / \sigma_y^*=260{\rm nm}/3.0{\rm nm}$).  Since the energies of the particles are relatively small, and are at a few hundreds MeV, most of these particles are confined near the beam axis due to the detector solenoid magnetic field.  However, when they hit  QC1 in the out-going side, many secondary photons are back-scattered uniformly into the detector region. They can become serious background.

%% file: detir/layout.tex
\begin{figure}[bp]
\centerline{\epsfxsize=13cm \epsfbox{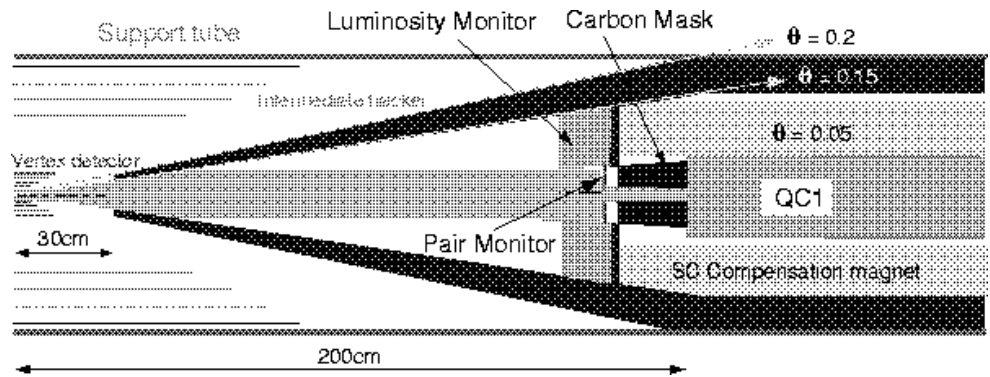}}   
\caption{\sl \label{ip-layout}
Interaction region. }
\end{figure}

IP layout is shown in Fig.\ref{ip-layout}.  Nearest final quadrupole magnet (QC1) is located at 2m from IP, where the QC1 is shielded against a solenoid field of 2 Tesla by a superconducting compensation magnet.   
A huge number of X rays, which are created in electromagnetic showers of $e^\pm$ pairs at QC1,  are absorbed in conical and cylindrical tungsten masks. Angular coverage of the conical mask ranges from 0.15 to 0.2 radian, leaving a small dead cone angle, but the front part of the conical mask could be instrumented as a calorimeter. There is the luminosity monitor covering an angular region between 0.05 and 0.15 radian inside the mask.   The carbon mask in front of the QC1 is very effective to absorb low energy back-scattered electrons.  The 4-layer vertex detector (VTX) is located at radial distance from 2.4 to 6.0cm, followed by an intermediate tracker to link  charged tracks between 
the central drift chamber (CDC) and the VTX. All these instruments are installed in a support tube of 80cm diameter\cite{yamaoka}.

The interaction region has been updated since the 1st ACFA-LC  (Beijing) workshop\cite{first-acfa} in order to further reduce $e^\pm$ pair backgrounds 
in VTX and CDC, especially X-rays in the CDC. 
The major changes are in the beam pipes and the locations of the carbon mask and the pair monitor. The beam pipe inside the VTX is made of 0.5mm thick beryllium, 4cm in diameter, while the major part is a 15cm diameter beam pipe of 2mm thick aluminum in the mask.  These two beam pipes are connected with a conical beryllium pipe of 2mm thickness.  The 15 cm diameter is large enough for $e^\pm$ pair particles to curl inside the beam pipe without interactions.
The pair monitors consisting of active pixel device are now located at 176cm from IP, i.e. inside of luminosity monitor.  
In the previous design, they were closer to IP and were one of the sources of X rays.

%% file: detir/hits.tex
\subsection{$e^+e^-$ pairs}

There is a desire to increase the detector magnetic field ($B$) from 2  to 3 Tesla.  
The motivations are  to reduce backgrounds and  to place a vertex detector closer
 to beam line as well as to reduce the overall detector size.
We study a case of $B=$3Tesla in terms of backgrounds in two major detectors, 
which are CDC filled with 1atm CO$_2$/isobutane 
(90\%/10\%) and VTX of $25 \times 25 \mu$m$^2$ pixels.   
Inner and outer radii of the CDC are 0.45 and 2.3m, respectively, and the 
length is $\pm 2.3$m.  The VTX is assumed to be consisted of three layers 
( r=2.5, 5.0, 7.5 cm and z=$\pm 7.5, 15.0, 22.5$cm, respectively) 
in this study.  Background hits were also simulated with detailed 
geometries by JIM, where photons and electrons(positrons) are tracked 
down to their energies of 10keV and 200keV, respectively.

\begin{figure}[bhtp]
\begin{minipage}[t]{7.5cm}
\centerline{\epsfxsize=7.5cm \epsfbox{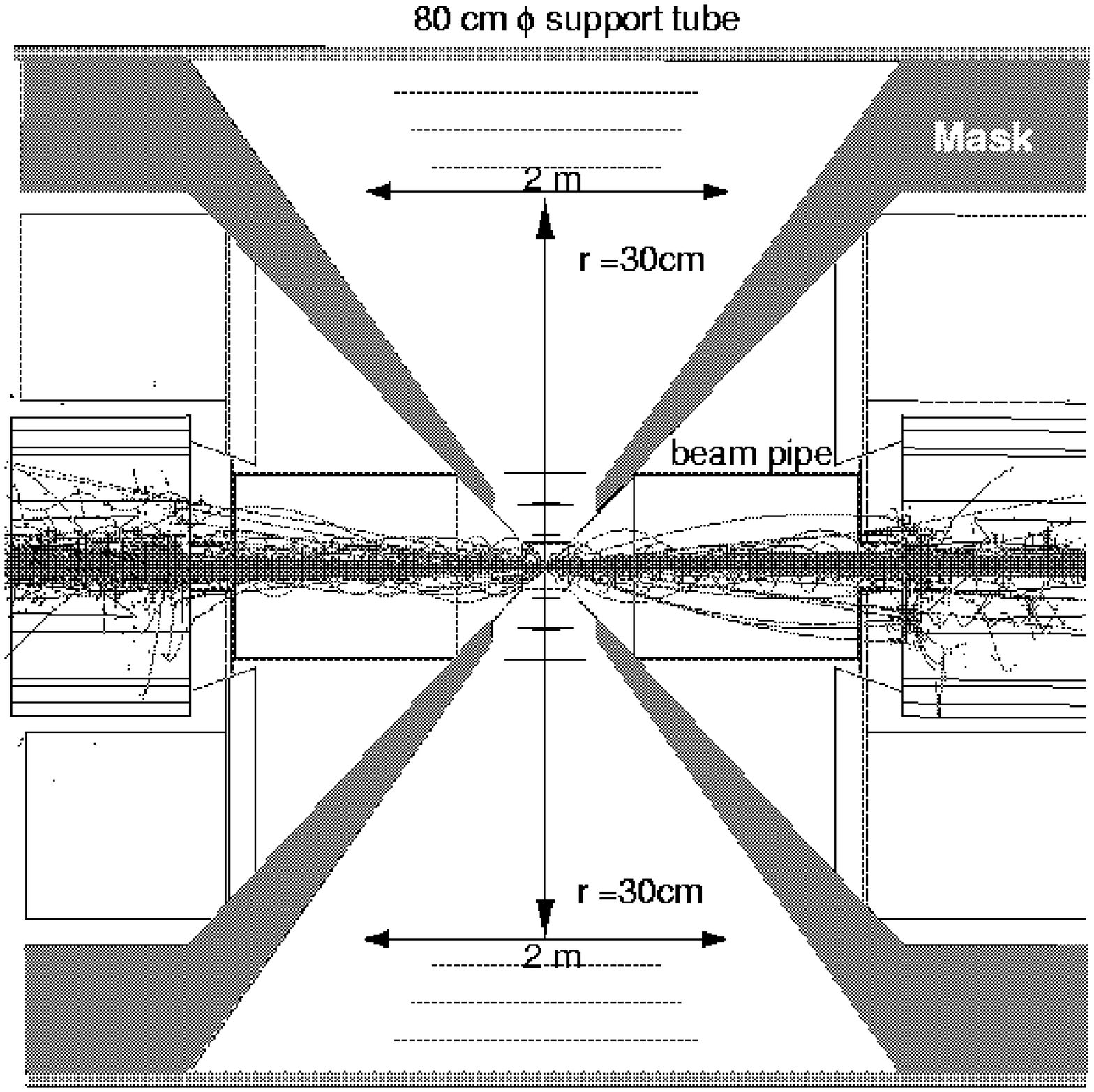}}   
\caption{ \sl \label{oed}
Pair samples in the support tube at $B=$2 Tesla by JIM simulation, 
where photons of $>100$MeV and electrons(positrons) of $>10$MeV are only 
shown for display purpose.}
\end{minipage}
\hfill
\begin{minipage}[t]{7.5cm}
\centerline{\epsfxsize=7.5cm \epsfbox{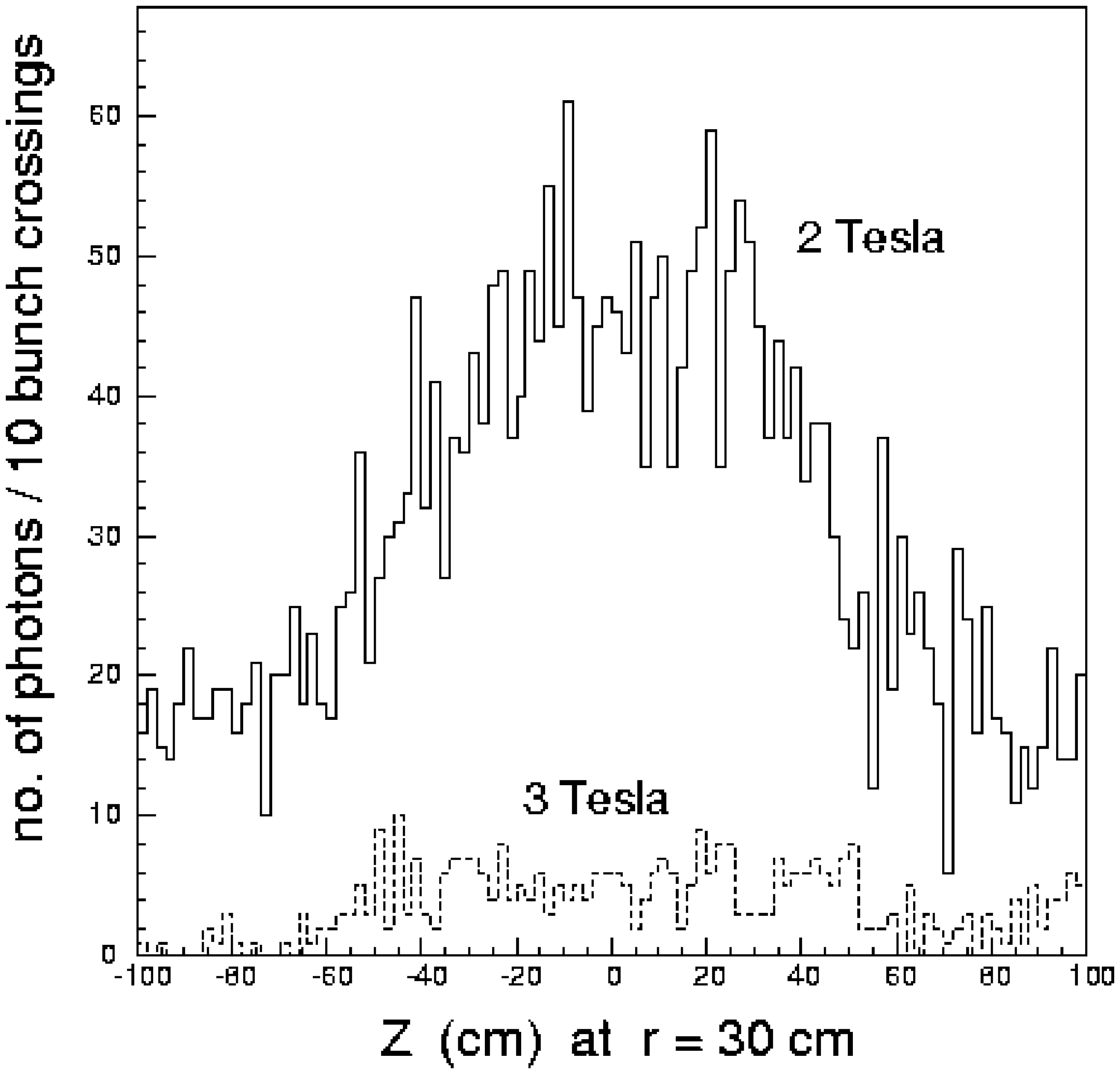}} 
\caption{\sl \label{photons}
 Number of photons traversing a cylinder of 60cm$\phi  \times \pm$
100cm long as a function of z at $B=$2 and 3 Tesla for the CDC background estimation.}
\end{minipage}
\end{figure}

The major backgrounds are secondary photons in the CDC although the CDC 
is shielded against them by the masks.  
For an estimation of the photons entering the CDC,  
we counted the number of photons traversing a surface of 
60cm$\phi \times \pm 1$m long cylinder. Figure~\ref{photons} shows  
the results  at $B=$2  and 3 Tesla for 10 bunch crossings.
Most of the photons are created by low energy electrons (or positrons) 
hitting the 4cm$\phi$ beam pipe very near IP .
We can see an apparent advantage of higher magnetic field in 
Fig.\ref{photons} because of stronger confinement of the electrons(positrons) 
around the beam line at $B=$3 Tesla.
On the other hand, the main backgrounds to VTX are primary electrons(positrons).

\begin{table}[bt]
\begin{center}\begin{minipage}{\figurewidth}
\caption{\sl \label{bkg4cm}
Pair backgrounds for 4cm$\phi$ beam 
pipe at $B=$2 and 3 Tesla for  
beam parameters of the basic JLC-A(95 bunches/train) and the high 
luminosity option of JLC-Y(190 bunches/train), where numbers are 
for 10 bunch crossings if there is no specification.}
\end{minipage}\end{center}
\begin{center} 
\begin{tabular}{ l c c c c c  c }

\hline 
 & r & z &\hfill B =&2 Tesla~~~~~ \hfill&\hfill B =&3  Tesla~~~~~ \hfill\\
 & cm & cm & JLC-A & JLC-Y& JLC-A& JLC-Y\\
\hline 
photons & 30 & $\pm$100 & 3076 & 6082 & 371 & 946\\
vtx-1:hits& 2.5 & $\pm$7.5 & 2186 & 3279 & 900 & 1205 \\
\hfill /mm$^2$/train& & & 0.9 & 2.8 & 0.4 & 1.0 \\
vtx-2:hits & 5.0 & $\pm$15.0 & 720 & 920 & 104 & 306 \\
vtx-3:hits & 7.5 & $\pm$22.5 & 406 & 545 & 34 & 138 \\
CDC:hits(tracks) & 45$\sim$230 & $\pm$230 & 121(101) & 235(194) & 12(9) & 37(28)\\
\hline 
\end{tabular}
\end{center}
\end{table}

Rates of background hits are summarized  both at $B=$2 and 3 Tesla in 
Table~\ref{bkg4cm}.  As mentioned above, the backgrounds hits in CDC 
are proportional to the numbers of photons,  which are one order of 
magnitude less at $B=$3 Tesla than those at $B=$2 Tesla.  
The CDC hit rates were estimated to be 1210 and 120 hits per a train 
crossing at $B=$2 and 3 Tesla, respectively. 
Assuming that the total number of CDC readout channels is 10000, 
corresponding occupancy rates are 12 and 1.2 \% at $B=2$ and 3 Tesla, respectively.
In the VTX the most severe backgrounds were observed at the innermost layer, 
where the hit densities are estimated to be 0.9 and 0.4 /mm$^2$/train 
at $B=$2 and 3 Tesla, respectively.
We also simulated  backgrounds for the case of the high luminosity 
option specified as JLC-Y.   The results are also listed in Table~\ref{bkg4cm}.  
For the JLC-Y, the background rates per bunch crossing  increase by 
1.3$\sim$3 times because of smaller beam sizes as seen in Table~\ref{JLCparameters}.  
Doubling the number of bunches per train crossing at the JLC-Y, the CDC 
occupancy and the VTX hit density can be estimated to be 47 (7.4)\% and 
2.8(1.0)/mm$^2$/train, respectively, at $B=$2 (3) Tesla.

Setting  background tolerances for 10\% occupancy and 1/mm$^2$/train hit 
density in the CDC and VTX, respectively, the backgrounds with $B=2$ Tesla 
are marginal at the JLC-A and they may overwhelm the detectors at the 
luminosity-upgraded JLC-Y.
For the case of $B=3$ Tesla, there seems to be a possibility of placing 
the VTX closer to the beam line at least for the JLC-A.
Therefore two configurations of the VTX and beam pipe have been studied 
by the JIM simulation.  The first comprises the VTX of 1.5cm minimum 
radius ($r_{vtx1}$) and 2cm$\phi$ beam pipe, while the second comprises 
the VTX of  $r_{vtx1}$=1.8cm and 3cm$\phi$ beam pipe.  Radii of the 
second and third layer of the VTX are 2.5cm and 5.0cm, respectively, 
for  both configurations.  
The simulation results are summarized in Table~\ref{bkg2cm}.
From the point view of the background tolerances, the second configuration 
is only allowed for the  CDC occupancy of 7.2\% and the VTX hit density of 1.6/mm$^2$/train.  
At the JLC-Y,  a more elaborate study shall be needed on a detailed 
geometry of beam pipe as well as a tracking algorithm in the CDC with 
the "exceeded" occupancy  even if the innermost layer of the VTX may be overwhelmed by backgrounds.

\begin{table}[bt]
\begin{center}\begin{minipage}{\figurewidth}
\caption{\sl \label{bkg2cm}
Pair backgrounds for 2 and 3 cm$\phi$ beam pipes at $B=$ 3 Tesla 
for the JLC-A, where numbers correspond to 10 bunch crossings if there is no specification. }
\end{minipage}\end{center}
\begin{center} 
\begin{tabular}{ l c c c c }

\hline 
& r & z &\hfill B =&3  Tesla~~~~~ \hfill\\
& cm & cm  & 2cm$\phi$& 3cm$\phi$\\
\hline 
photons & 30 & $\pm$100 & 5857 & 1626 \\
vtx-1:hits & 1.5 & $\pm$4.5 & 3065 & - \\
\hfill /mm$^2$/train& &  & 3.6 & - \\
& 1.8 & $\pm$5.4 & - &  1906 \\
\hfill /mm$^2$/train& & & - &  1.6 \\
vtx-2:hits & 2.5 & $\pm$7.5 & 680 & 804  \\
vtx-3:hits & 5.0 & $\pm$15.0 & 402 & 208  \\
CDC:hits(tracks)& 45$\sim$230 & $\pm$230 & 259(217) & 72(63) \\
\hline 
\end{tabular}
\end{center}
\end{table}

\subsection{Neutrons}

We estimated the neutron background produced through photo-nuclear reaction. 
The pair background particles hit the QC1 and make the EM shower, and the photons 
in the shower cause the photo-nuclear reaction. The neutrons produced here can 
spread over the entire detector inside the iron structure.

The pair background particles which can go through the QC1 cannot be transported 
along the same line as the main extracted beam and will hit the components of 
the extraction beam line resulting the EM shower. The beamstrahlung photons 
have the very small angular divergence but eventually hit the beam-dump and 
also produce neutrons. These neutrons can come back to the IP region 
through the hole for the beam pipe.

In order to study the neutron background produced at the QC1, we used 
a computer code written by Dr.T.~Maruyama at SLAC for the neutron generation 
built into GEANT3. For the neutron transportation in GEANT3, 
FLUKA(for $E_n >20$ MeV) or MICAP(for $E_n < 20$ MeV) was used. 
The cut-off energy for neutral hadrons was 1 keV. We counted the number 
of neutrons passing the z = 0 plane in the simulation. We found 66 neutrons 
passing the plane in r $<$ 10 cm area for 1000 bunch crossing. This number 
corresponds to $3 \times 10^7/$cm$^2$y near interaction point. As the main tracker the central 
drift chamber (CDC) with gas mixture of CO$_2$:Isobutane(C$_4$H$_{10}$)=90:10 
is being considered. Since this gas mixture contains hydrogen atoms, neutrons 
can make background hits by elastic scattering. The number of the background 
hits in the CDC we obtained by the simulation is 800 hits/train.

We have also estimated neutron backgrounds using Fluka98 program.
In this calculation, we used the magnetic field consisted of
the detector solenoid of 2 Tesla, the QCS magnets, and
the compensation magnets to cancel out the solenoid field
in QCS region.  In this configuration, the magnetic field vector
which points along beam direction at IP region changes
the direction to perpendicular to the beam axis.
Since the Z components of the magnetic field which could trap
soft $e^\pm$ was reduced, more $e^\pm$ hit the inner surface 
of QCS.  However, neutrons which returned to IP 
were produced mostly at the front surface
of QC1, the large increase of neutron yield by this effect was
not observed.

\begin{figure}[htb]
\centerline{
\epsfxsize=9cm 
\epsfbox{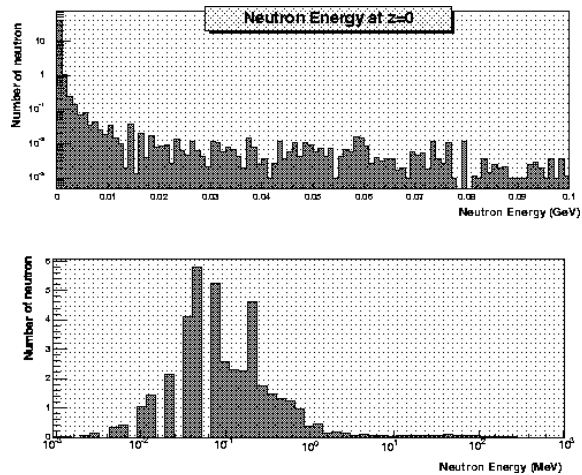}
}
\begin{center}\begin{minipage}{\figurewidth}{
\caption{\sl\label{detir/nenergy.eps}
The energy spectrum of neutron which return to at IP.
Note that the scaling of upper and lower figure is different:
Upper figure is log Y scaling, and lower figure is log X scaling.
}}
\end{minipage}\end{center}
\end{figure}

In FLUKA simulation,  the cut off energy for neutron transport
was 1 keV while those for $e^+/e^-$ and photons were
10.511 MeV an 4 MeV, respectively.  As seen in Fig.~\ref{detir/nenergy.eps},
the only few neutron had energy less than 10 keV.

\begin{figure}[htb]
\centerline{
\epsfxsize=9cm 
\epsfbox{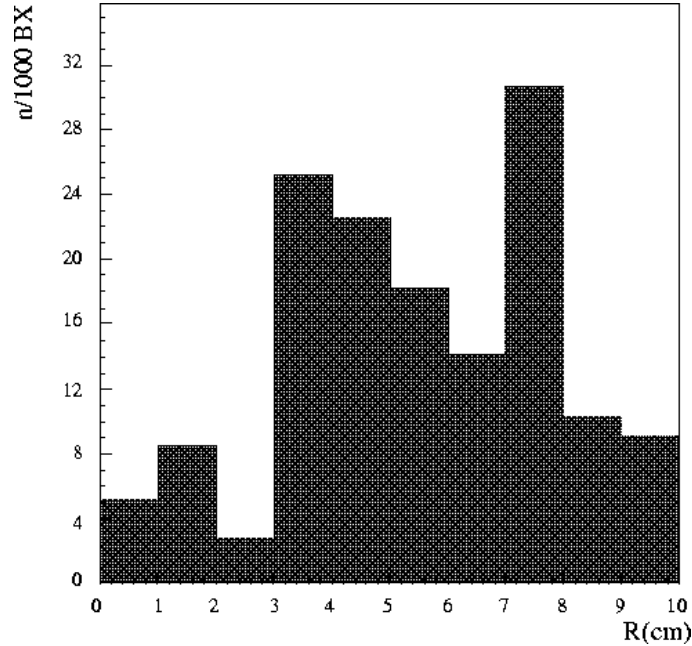}}
\begin{center}\begin{minipage}{\figurewidth}{
\caption{\sl\label{detir/nflux.eps}
The number of neutrons as a function of radius at Z=0 (IP).
The vertical axis is scaled so that it  gives the number of
neutron for 1000 bunch crossing.}
}\end{minipage}\end{center}
\end{figure}

From the Fluka simulation, we obtained the number of neutron 
distribution at Z=0 as shown in Fig.~\ref{detir/nflux.eps}.
From this figure we got the total number of neutron within the 
radius of 10 cm is 292 for 1000 bunch crossing (BX) 
( entries in the figures are doubled to count backgrounds 
by $e^-$ and $e^+$).  If uniform neutron distribution is assumed
the average neutron flux is $0.93\times 10^{-3}$ $n/{\rm cm}^2$ per bunch 
crossing.
For 1 year ( $10^7$sec ), we get $13\times 10^{7}$ $n/cm^2/year$ at IP.
The result is consistent with the GEANT3 simulation within a factor of 4,
if we neglect a difference of magnetic field distribution used for the simulations.

At present it is difficult to make a reliable estimation on the neutron
background from the downstream because we don't have the design of 
the beam extraction line of the JLC yet. So we made a very rough 
estimation on it. By the detector simulation it was found that 
among the total energy of 100 TeV of the pair background particles 
in one bunch crossing, 13 TeV is absorbed in the QC1 and 87 TeV 
goes downstream. By scaling the neutron background from the QC1 
with the deposited energy, we get the upper limit of the neutron 
background produced by pair background particles at the QC1 and 
other beam line components downstream of the QC1 as 
$1.5 \times 10^7 \times 100$~TeV/13~TeV $\sim 1 \times 10^8/cm^2/year$.

For the estimation of the neutron background from the beamstrahlung
photons, we assumed all photons lose their energy in the water beam 
dump located at 300 m from the IP. The electron/positron beam was 
assumed to be bent before the water beam dump and dumped elsewhere 
so that the neutrons from the $e^\pm$ dump can be shielded sufficiently.  

Using the GEANT3 simulation with the neutron generation code we found 
that $6 \times 10^{20}$ neutrons are produced in one year which is 
reduced to $1 \times 10^{17}$ at the surface of the beam dump on 
the IP side owing to the self shielding by the water. If the 
solid angle acceptance at the IP is simply multiplied, 
the resulting neutron yield at the IP is $2 \times 10^7 n/cm^2/year$.

According to the Fluka98 simulation, the neutron yield at the
surface of water dump is $2.8\times 10^{17}$ neutrons per 
year which corresponds to 
$5\times 10^7 n/cm^2/year$ at IP if we take into 
account the acceptance of solid angle and
contribution from two water dumps are added.

%% file: detir/newmask.tex
So far the final focus quadrupole magnets(QC1) are assumed to be located 2~m away from the interaction point($l^*$=2~m).  Recently a new idea of final focus optics is proposed~\cite{panta} which makes longer $l^*$ of 4.3~m possible. With this optics,  the background particles ($\gamma$, $e^{\pm}$, neutron) due to back-scattering from QC1 would be reduced. In this section, we present the results of simulation studies on the pair-background in several detector models which have  different magnetic field, mask design, and $l^*$.

\begin{figure}
 \centering
 \epsfxsize=16cm
\centerline{
 \epsfbox{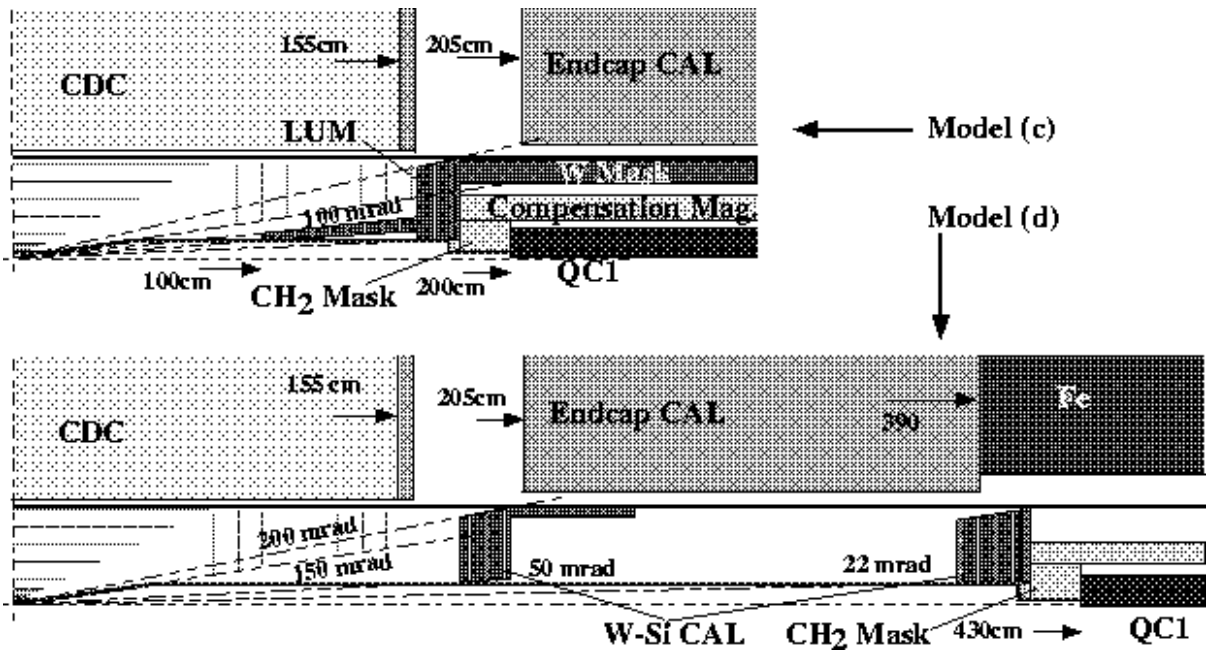}
}
 \caption{\sl  \label{modelcd}
Schematic views of model c) and d) detectors in the 3~T model.}
\end{figure}

Background hit rate has been studied for four detector models (table~\ref{ir-tab1}).
The layout and background estimations in the previous sections are 
based on 2~T model~\cite{JLC-1}. Compared with the 2~T model, the outer radius of the central drift chamber (CDC) and inner radius of the barrel calorimeter of the 3~T models are  reduced in order to keep the minimum $p_t$ of charged particles which reach the barrel calorimeter constant. The length in $z$ direction of the CDC and
the $z$ position of the endcap calorimeter are also reduced in 3~T models.  In models a) and b), large conical tungsten masks are placed to shield the back-scattered particles produced by the pair-background particles hitting the QC1 as shown in Fig.\ref{ip-layout}.  Instead, smaller conical masks are used for the model c). Model d) is for the case of $l^*$=4.3~m. Two sets of forward calorimeters work as active masks in this model. Schematic views of model c) and d) are shown in Fig.~\ref{modelcd}. The pair-background particles were  generated using a program code CAIN~\cite{cain} for the JLC-A.   These particles were put into a full detector simulation program JIM which is based on GEANT3. Magnetic fields  of the QC1 and the compensation magnets were taken into account. The cut-off energy was set to 10~keV for $\gamma$'s and 1~keV for neutrons.

\begin{table}[b]
\begin{center}\begin{minipage}{\figurewidth}
 \caption{\sl Detector models and expected background hits per beam crossing: track density at the innermost layer ($r$=$24$~mm) of vertex detector(VTX), a number of hits in the central tracker(CDC) by $\gamma$ and neutron, and an energy deposit in calorimeter(CAL),
where models a) and b) are called as 2~T model and models c) and d) are called as 3~T model. \label{ir-tab1}}
\end{minipage}\end{center}
 \begin{tabular}{llllllll}
 Model & $B$(T) & $l^*$(m) & Mask & VTX($/{\rm cm}^2$) & CDC($\gamma$)
                             & CDC(n) & CAL(MeV) \\
 \hline
 a  & 2    & 2        & 150 -- 200 mrad & 0.7 & 2 & 30 & 600 \\
 b  & 3    & 2        & 150 -- 200 mrad & 0.4 & 1 & 2  & 900 \\
 c  & 3    & 2        & 70 -- 100 mrad &  0.4 & 2 & 2  & 900 \\
 d  & 3    & 4.3      &  --            &  0.4 & 1 & 0.1 & 30 \\
 \hline
 \end{tabular}
\end{table}

The results of the simulation are given in table~\ref{ir-tab1}.
Background hit rate  of CDC by neutrons in the 2~T model is quite high  because 10~cm thick tungsten mask is not enough to stop
neutrons.  In the 3~T models, the CDC hit rate by neutrons is reduced because the neutron source (QC1) is surrounded by the endcap calorimeter which works as  a neutron shield. With $l^*$=4.3~m optics, the background source is located farther from the detector components and the background hit rate  of the detectors get even less. The simulation was also performed with the JLC-Y.  The result shows about two times more background, while the luminosity is 3 times more than that of the JLC-A.

%% file: detir/pairmonitor.tex
The disk-shaped pair monitor comprises double layers of active pixel sensors whose pixel size and thickness are $100 \times 100 \mu$m$^2$ and 300$\mu$m of silicon, respectively.  The inner and outer radii  are 2 and 8.5cm, respectively.  The pair monitor should measure positions and energy deposits of electrons or positrons and secondary photons. 
\begin{figure}[bp]
\begin{minipage}[t]{7.5cm}
\centerline{\epsfxsize=7.5cm \epsfbox{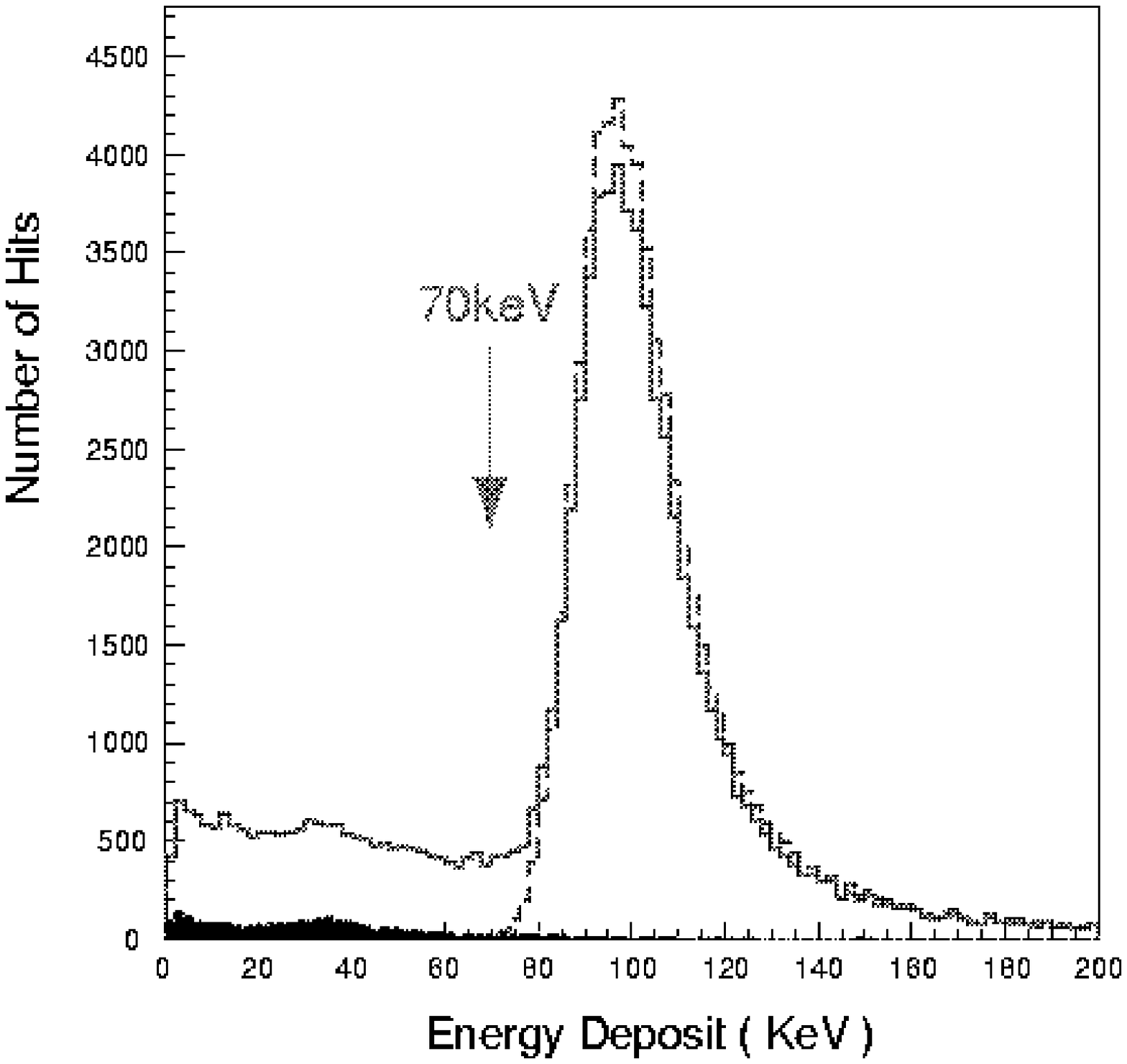}}   
\caption{\sl \label{bm-edep}
 Energy deposits in the pair monitor. The solid and dashed histograms show energy deposits in each cell and those by a track, respectively. The black one show those of secondary backgrounds in each cell. }
\end{minipage}
\hfill
\begin{minipage}[t]{7.5cm}
\centerline{\epsfxsize=7.5cm \epsfbox{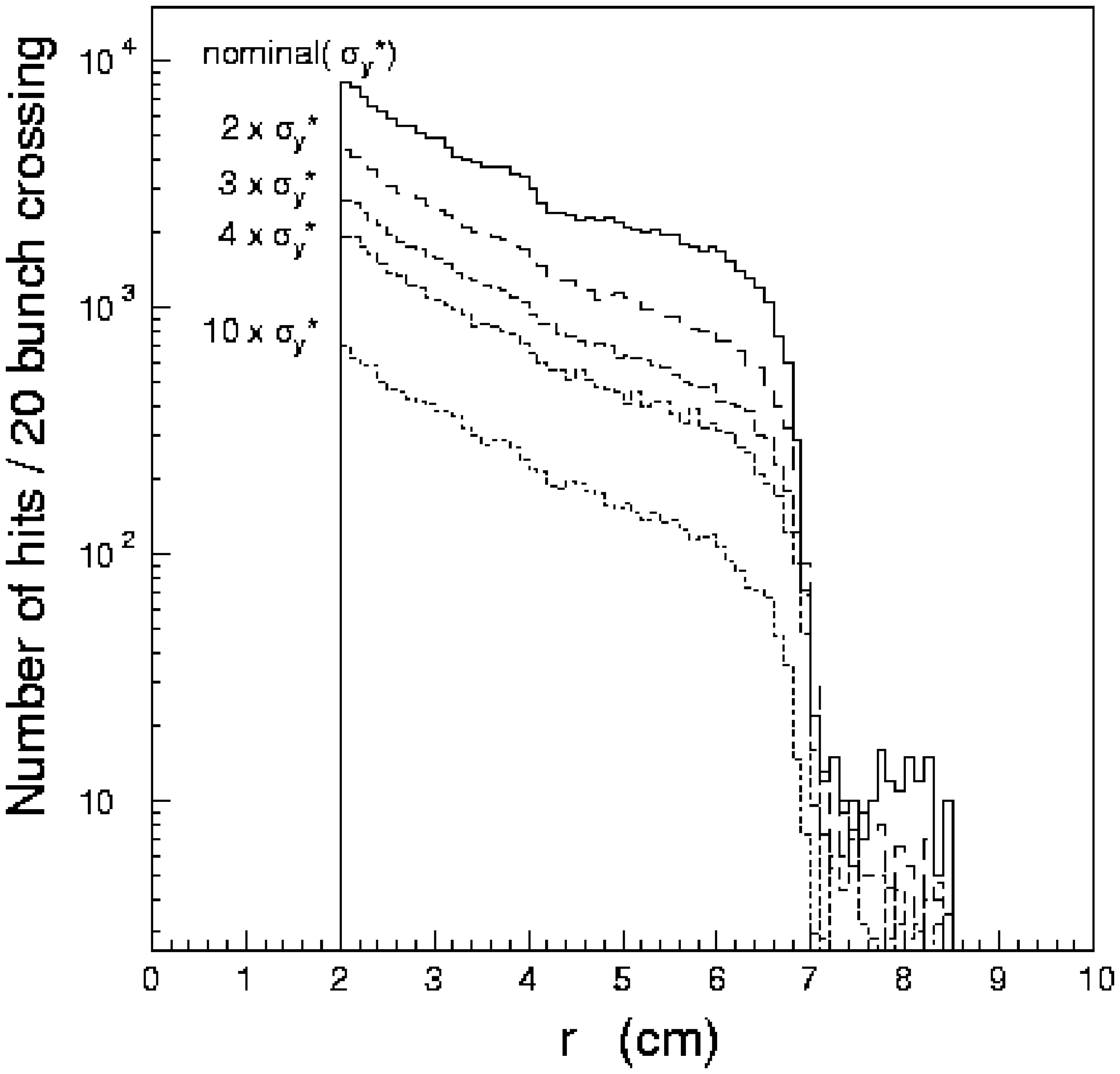}} 
\caption{\sl \label{bm-r}
 Hit distributions normalized at 20 bunch crossings as a function of $r$ for energy deposits of $>$70keV, where histograms correspond to 5 vertical beam sizes at IP; i.e.  basic ($\sigma^o_y$), $2\times,  3\times, 4\times$ and $10\times\sigma^o_y$}.
\end{minipage}
\end{figure}
Figure~\ref{bm-edep} shows the energy deposits simulated with detailed geometries by JIM, where the pairs have been generated with beam parameters of JLC-A by CAIN21d\cite{cain}.   Many of them hit only one cell and,  as clearly seen in this figure,  a very sharp peak appears around 100keV.  Therefore, they can be effectively discriminated from the secondary backgrounds simply by a 70 keV threshold in energy deposits.

In order to see how we can measure the beam sizes, pairs have been generated with five different vertical beam sizes at IP by CAIN\cite{cain}, {\it i.e.} $\sigma_y = \sigma_y^o$(basic), $2 \times \sigma_y^o$, $ 3 \times \sigma_y^o$, $ 4 \times \sigma_y^o$ and $ 10 \times \sigma_y^o$ for the  statistics of 20, 40, 60 50 and 110 bunch crossings, respectively, while the other beam parameters are the same as the basic ones.  The minimum energy of the pair particles was 3 MeV.

Radial distributions of the pairs are shown in Fig.\ref{bm-r}, where the number of hits are normalized at 20 bunch crossings for realistic statistical comparisons. 
All five histograms have sharp shoulders, while negligibly small tails are due to those scattered inherently at large angles. As clearly seen in this figure, positions of the shoulders do not depend on the vertical beam size($\sigma_y$) indeed as described in  ~\cite{nanometer}.  Since they depend on the horizontal beam size($\sigma_x$) as well as well-known beam intensities,  the horizontal beam size can be estimated by the radial distribution.

\begin{figure}[htbp]
\begin{minipage}[t]{7.5cm}
\centerline{\epsfxsize=7.5cm \epsfbox{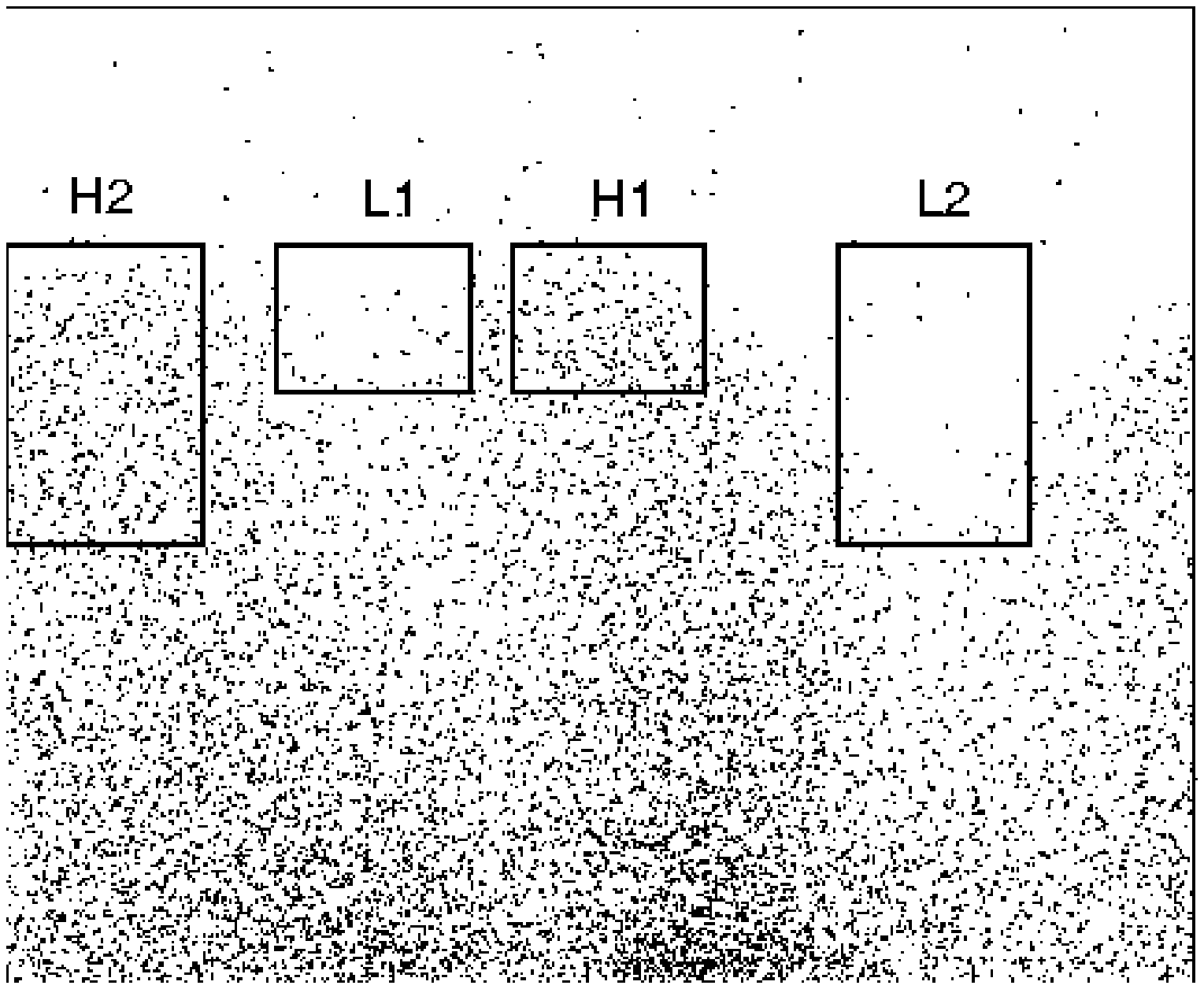}} 
\caption{\sl \label{bm-scatter}
 Scatter plot of hits with $>$70 keV energy deposits in the pair monitor on a plane of azimuthal angle(x) and  radial position (y), where four sensitive areas are also depicted in order to calculate a ratio  of $(L_1+L_2)/(H_1+H_2)$}.
\end{minipage}
\hfill
\begin{minipage}[t]{7.5cm}
\centerline{\epsfxsize=7.5cm \epsfbox{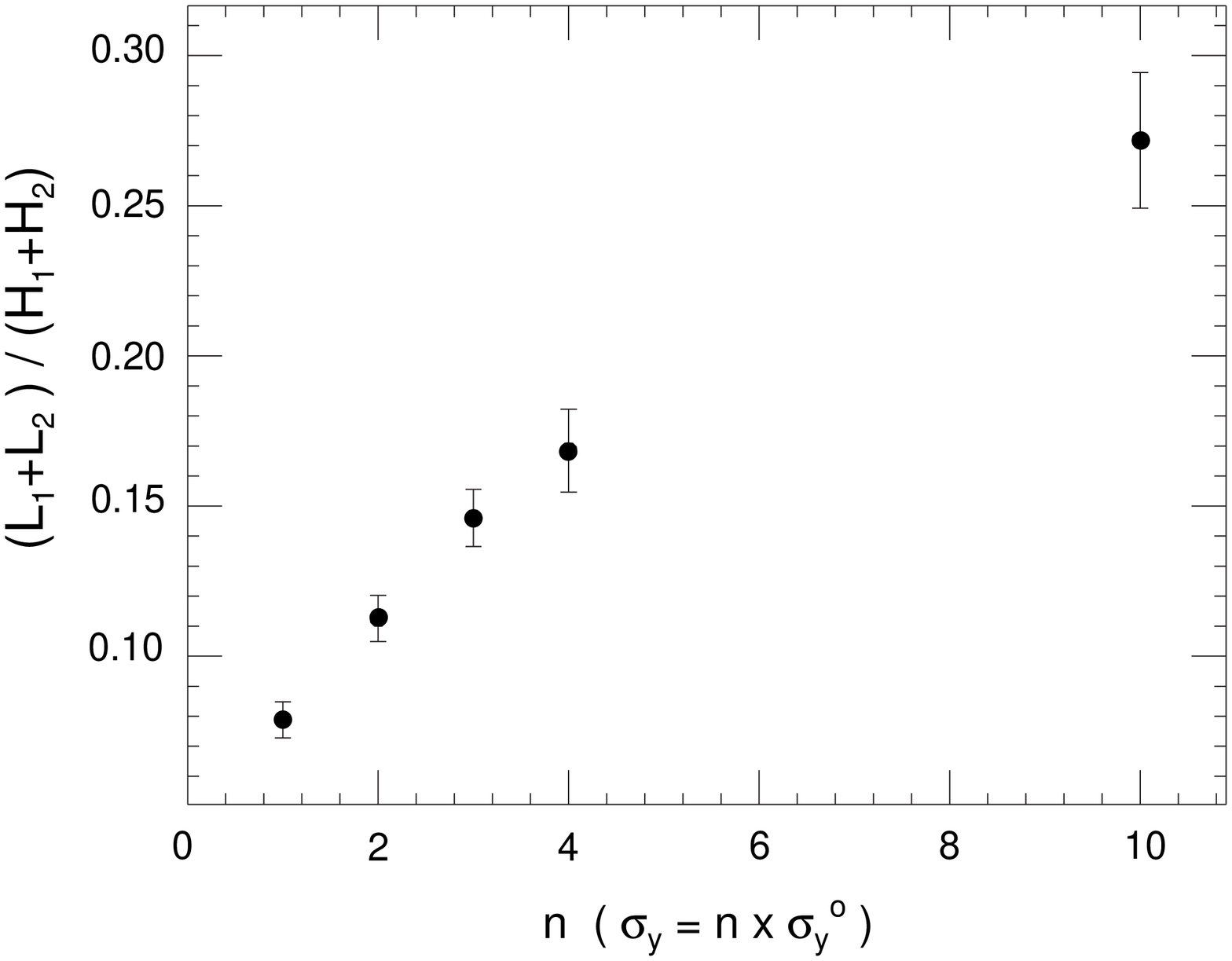}} 
\caption{ \sl \label{bm-ratio}
Ratio of $(L_1+L_2)/(H_1+H_2)$ as a function of vertical beam size ($\sigma_y = n \times \sigma^o_y$) at IP, where error bars are statistical corresponding to 20, 40, 60, 50 and 110 bunch crossings for n=1,2,3,4 and 10, respectively. }
\end{minipage}
\end{figure}

The most interesting observable is the distribution of azimuthal angles, since the angles carry an information of the vertical beam size as a function of aspect ratio,$A \equiv \sigma_x / \sigma_y$. 
Among the pair particles,  particles of the same charge as the incoming beam must be deflected with large angles because of repulsive Coulomb forces.  When the incoming beam has a large aspect ratio, the particles are  scattered asymmetrically  due to the asymmetric Coulomb potential,  that is, more vertically than horizontally.  
Figure~\ref{bm-scatter} shows a scatter plot of hits in the pair monitor on  a plane of azimuthal angle and radial position. 
Two dense bands can be seen, which are made of particles deflected upward or downward  and swum helically in the 2 Tesla solenoid field.  In order to quantify the asymmetry, we set four  regions ($H_2, L_1,H_1, L_2$) as shown in Fig.\ref{bm-scatter}.  Area of $H_2$ and $L_2$ is larger than that of $H_1$ and $L_1$ because of left-right asymmetry with 8 mrad crossing angle.   Then, we defined a ratio, $R \equiv (L_1 + L_2)/(H_1 + H_2)$ as proposed\cite{nanometer}, where $L_1, L_2, H_1,H_2$ represent the number of hits in each region.   It is expected that $R$ becomes smaller as $A$ increases while $R=1$ for round beams.  
The simulation results are shown for beams with five different $\sigma_y$'s in Fig.\ref{bm-ratio}.
 Up to $4 \times \sigma_y^o$, we observed a linear relation between R and the vertical beam size. 
Under the basic beam condition, the vertical beam size could be measured  for 100 bunch collisions with $< 10$\% statistical accuracy.
Since this method, {\it i.e.} counting hit numbers in specific regions with energy deposits of greater than 70 keV,  is very simple, the pair monitor can be a feedback device for beam tuning in realtime. 
With the fast gate, the pair monitor should detect the pairs in specific bunches in order to investigate a variance of beam size in a bunch-train.

%% file: detir/vetosystem.tex
Luminosity monitor is located at 163cm from the interaction point (IP), which is made of tungsten (W) of 15cm thickness (42.86$X_o$) and the (polar) angular coverage ranges from 0.05 to 0.15 radian. It is segmented into 32, 16 and 128 divisions in radius (r), azimuthal angle ($\phi$) and longitudinal coordinate (z), respectively. The physical dimensions are 5mm, 3.2-9.7cm and 1.17mm in r, $\phi$ and z, respectively. At present, all the tungsten segments are treated as sensitive detectors; that is, energy deposits in all segments are calculated in simulations. A real detector must have some sensitive layers made of active material such as scintillator, silicon pad, or crystal (e.g. BGO) instead of
tungsten. 

A front part of the conical mask is instrumented to measure energy deposits in electromagnetic showers. The mask is made of tungsten. The longitudinal position is 30cm from IP and the angular coverage is from 0.15 to 0.2 radian. The active mask consists of 8 layers of silicon pads ( Si(200$\mu$m) + G10(300$\mu$m) ), where the first W layer is 5mm thick and the other W layers are 1cm thick. In simulations, energy deposits in silicon pads are calculated. The active mask is segmented into 8-10 and 32 divisions in r and $\phi$, respectively. The physical dimensions are 2mm and 0.9-1.2cm
in r and $\phi$, respectively.

The $e^+e^-$ pairs were generated with a statistics corresponding to 100 bunch crossings ($\sim$1 train crossing) by cain21d, where 
the nominal JLC-A machine parameters at $\sqrt{s}$=500GeV were used.  They should always overlap on any physics event at JLC. For the signals, electrons of 50 or 250GeV were generated into the front center of the luminosity monitor and the active mask. Their
energy deposits were calculated by the JIM simulation, and they are listed in Tab.\ref{tab-lmam-edep}
\begin{table}[t]
\begin{center}\begin{minipage}{\figurewidth}{
\caption{\sl \label{tab-lmam-edep}
Energy deposits in the luminosity monitor (LM) and active mask (AM) per 100 bunch crossings.}
}\end{minipage}\end{center}
\begin{center} 
\begin{tabular}{ccccc}
\hline 
\multicolumn{1}{c}{$B$=2T}&
\multicolumn{1}{c}{$e^+e^-$ pairs}&
50GeV $e$ & 250GeV $e$ & 250GeV $\mu$ \\
\hline 
AM   &  264MeV &  120.4MeV &  541.0MeV &   0.48MeV \\
LM   &  152GeV &   49.9GeV &  249.5GeV &   1.67GeV \\
\hline 
$B$=3T & $e^+e^-$ pairs & & & \\
\hline 
AM   & 29.3MeV & & & \\
LM   &  46.7GeV & & & \\
\hline 
\end{tabular}
\end{center}
\end{table}
,where energy deposits due to 250GeV muons are also listed. As clearly seen, energy deposits of $e^+e^-$ pairs decreases at higher magnetic field ($B$=3T). The luminosity monitor contains all electromagnetic shower while the active mask has shower leakage of $\sim$10\% in radial direction for a 250GeV electron compared with a 50GeV electron. The longitudinal leakage in the active mask with 21.4$X_o$ was estimated to be $\sim$1\% even for a 250GeV electron.

\begin{figure}[bp]
\begin{minipage}[t]{8cm}
\centerline{\epsfxsize=8cm \epsfbox{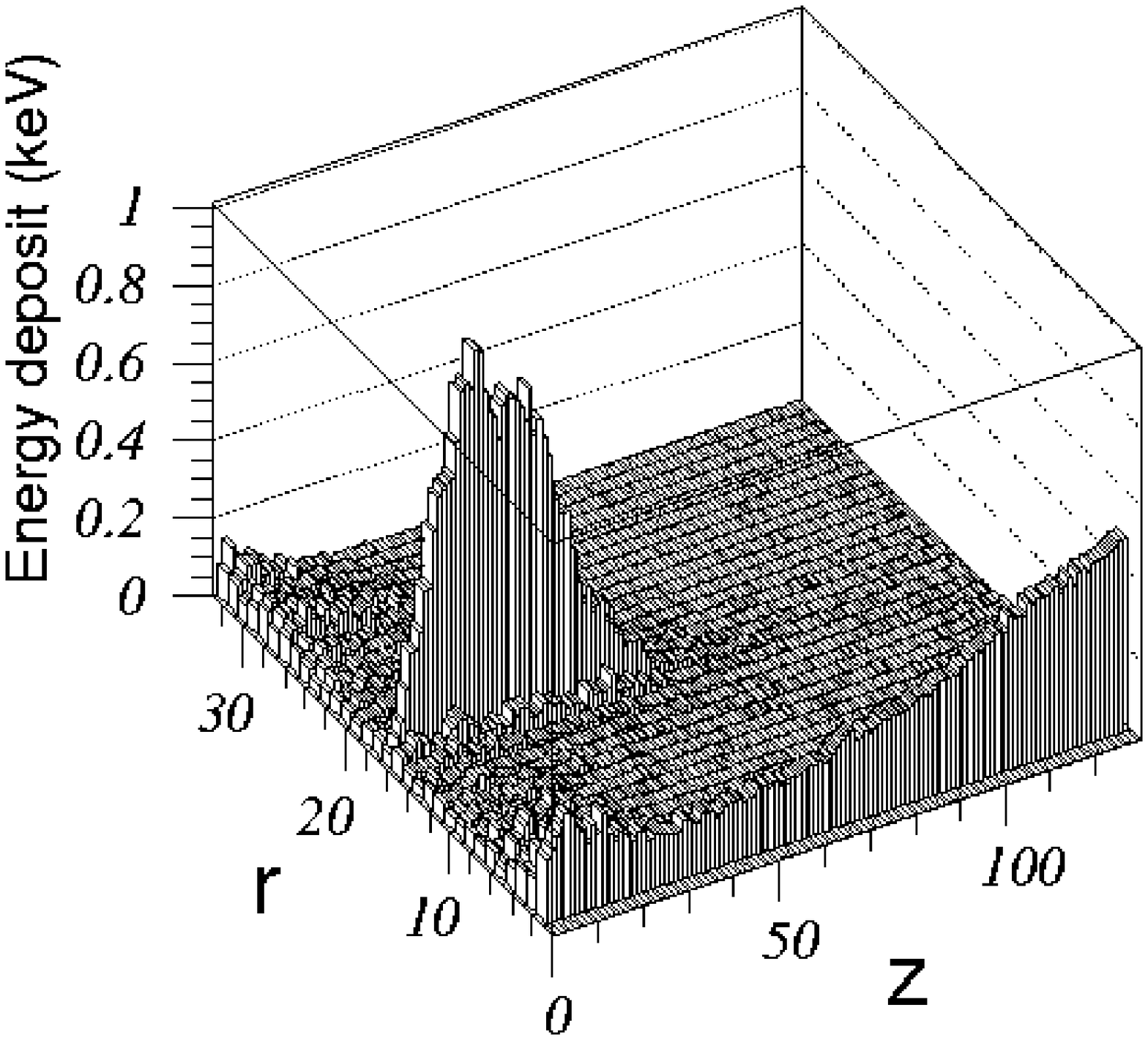}}   
\end{minipage}
\hfill
\begin{minipage}[t]{8cm}
\centerline{\epsfxsize=8cm \epsfbox{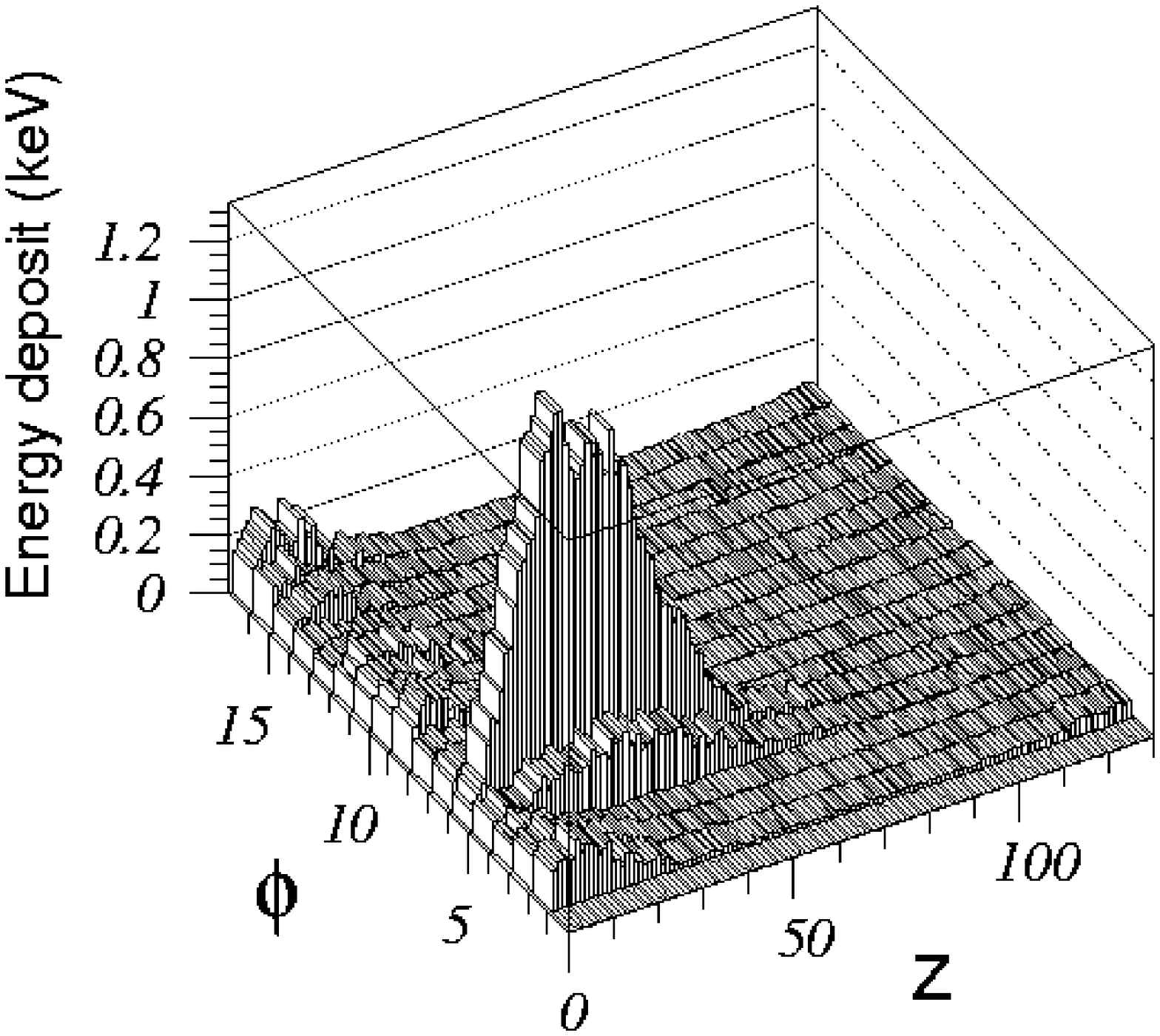}} 
\end{minipage}
\begin{center}\begin{minipage}{\figurewidth}{
\caption{ \sl \label{lm-edep}
Energy deposit of a 50 GeV electron together with $e^+e^-$ pairs whose statistics corresponds to 100 bunch crossings in LM.  Left and right figures are plotted in r-z and $\phi$-z planes, respectively. }
}\end{minipage}\end{center}
\end{figure}

\begin{figure}[bp]
\begin{minipage}[t]{8cm}
\centerline{\epsfxsize=8cm \epsfbox{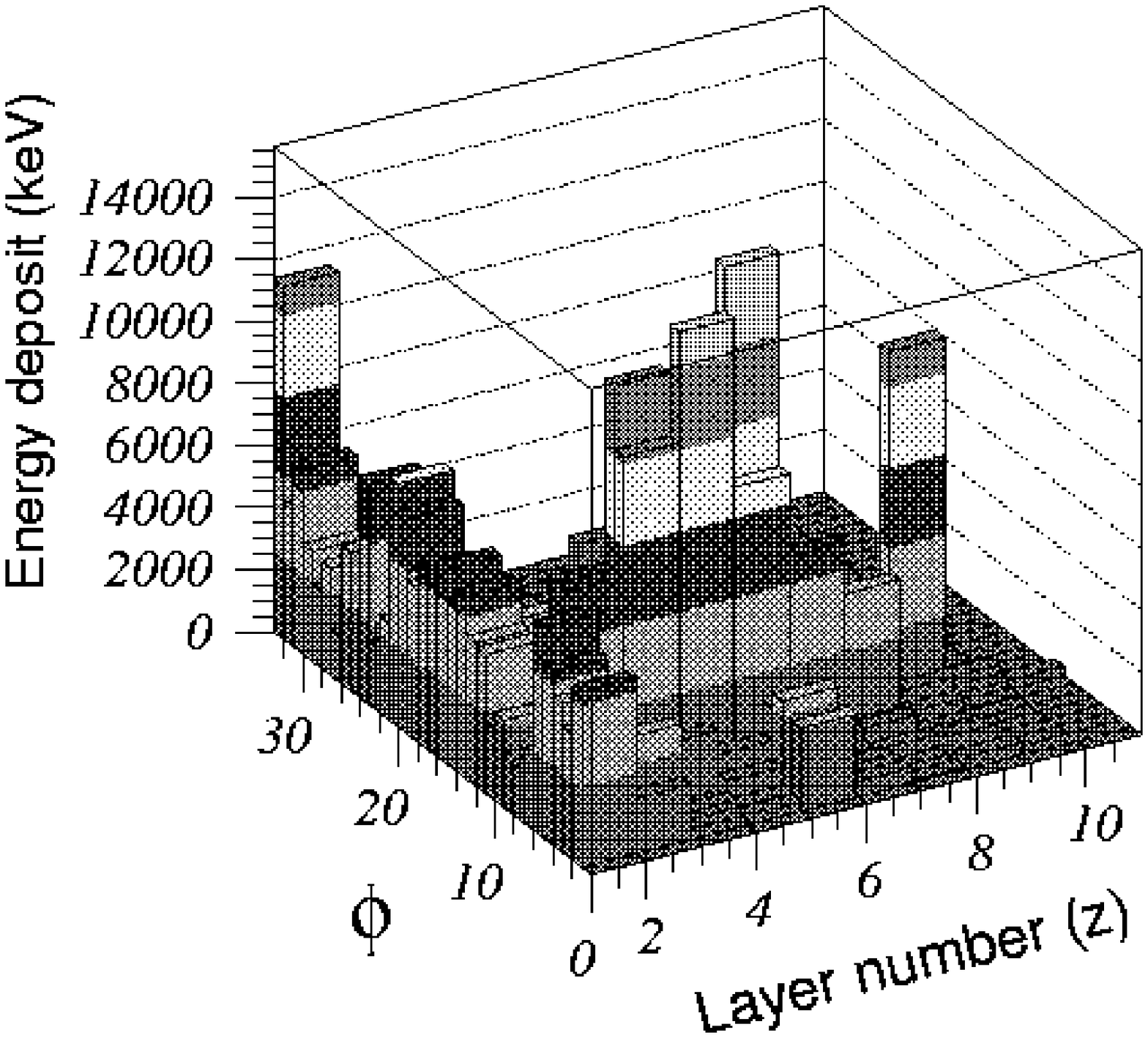}}   
\end{minipage}
\hfill
\begin{minipage}[t]{8cm}
\centerline{\epsfxsize=8cm \epsfbox{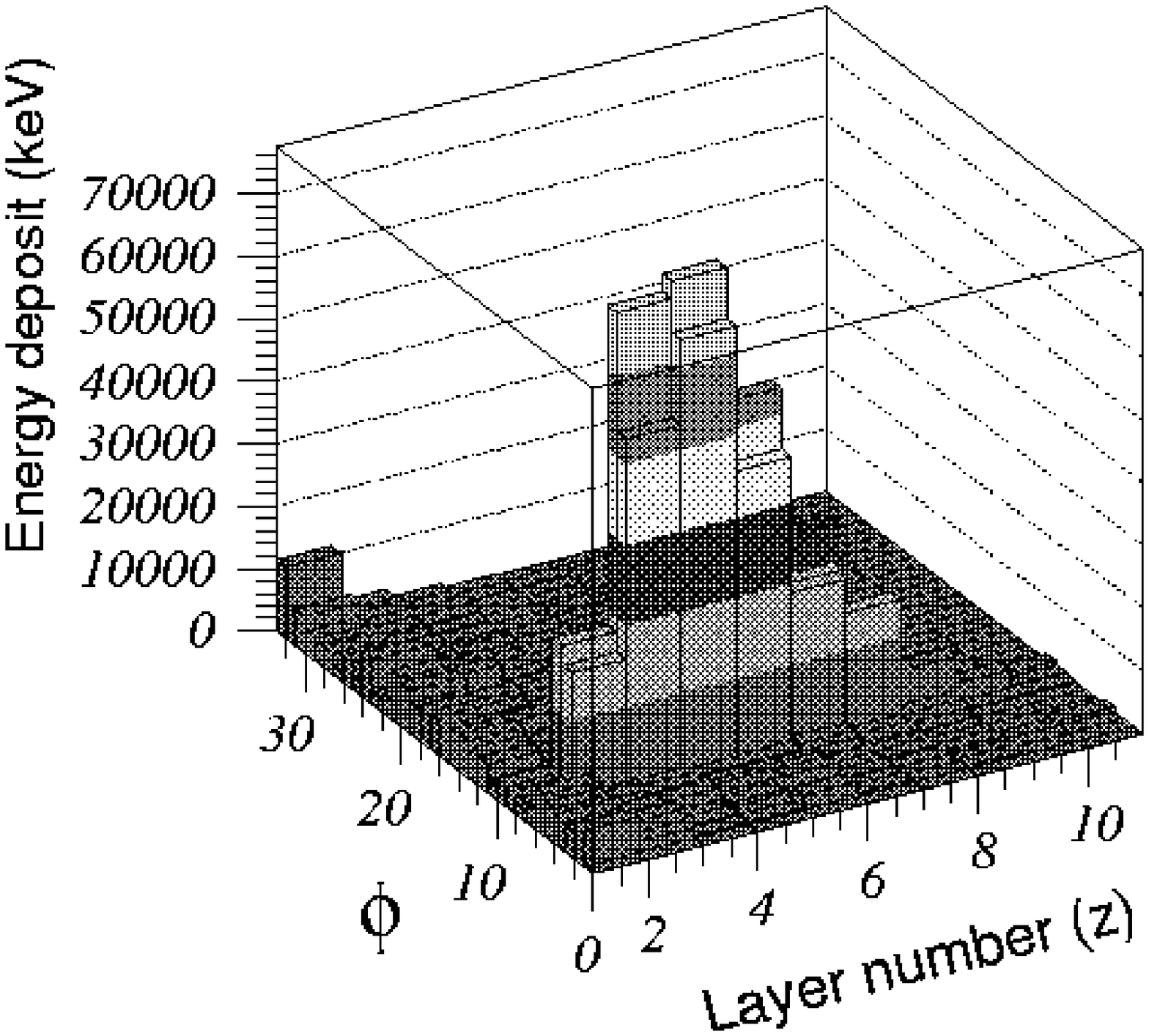}} 
\end{minipage}
\begin{center}\begin{minipage}{\figurewidth}
\caption{ \sl \label{am-edep}
Energy deposit of a high energy electron together with $e^+e^-$ pairs whose statistics corresponds to 100 bunch crossings in AM.  The electron energies are 50 and 250 GeV in left and right figures, respectively. }
\end{minipage}\end{center}
\end{figure}

Although the total energy deposits of $e^+e^-$ pairs are very large, the spatial distributions are very different from those of electrons as seen in Fig.~\ref{lm-edep} and ~\ref{am-edep} for LM and AM, respectively. The simulations show that most of the pairs are absorbed near the surfaces of $\sim 1X_o$ depth because of their low energies while the 50GeV and 250GeV electrons have shower maxima at 5$X_o$ and 8$X_o$, respectively. 

Among the total energy deposit of 152 (46.7) GeV/train in LM due to $e^+e^-$ pairs, only $54^*(14^*)$ GeV comes from the front at B=2(3)T, while most comes from the inner-back. ($^*$ sum of incoming energies). The $\phi$-segmentation (16 div.) is important for angular resolution, and the r-segmentation (32 div.) is desired to determine polar angle with an accuracy of a few milli-radian. So, a fine-segmented W/Silicon calorimeter seems to be ideal. The thickness of tungsten layers must be optimized for energy-resolution.

For AM, first layer (5mm thick W) has $\sim$50\% energy deposit for $e^+e^-$ pairs. Again, $\phi$ segmentation (32 div.) is important for angular resolution. Eight layers of W/Si-pad calorimeter (21.4$X_o$) works very well for vetoing high energy electrons.

%% file: detir/support.tex
To support the components at the interaction region, 80 cm diameter with 10mm thick of support tube made of CFRP was proposed. The support tube is usually acted on self-weight of the components and vibrations from the ground. One of the important requirements is that relative displacement between QC1 separated by 4 m has to be kept less than 1 nm. Therefore, to optimize the support tube configuration, static analysis and dynamic analysis were carried out. And to realize these analyses, the excitation test using prototype model is being done.

At the initial concept of support tube, 12 m long single CFRP tube is provided and it is supported on the both of iron end yokes. Analysis in terms of this configuration was studied and reported at the 2nd ACFA-LC workshop in Seoul. It showed that relative displacement to be excited by ground motions were completely negligible between QC1's. Since then, a new analysis has been executed on a system, where the two final focus doublets are independently supported by two identical cantilevers. If the system is stable in terms of ground motions, it is experimentally more preferable because of less material in detectors as well as easier assembling of the doublets and more straightforward installation.

\begin{figure}[hptb]
\centerline{\epsfxsize=14cm \epsfbox{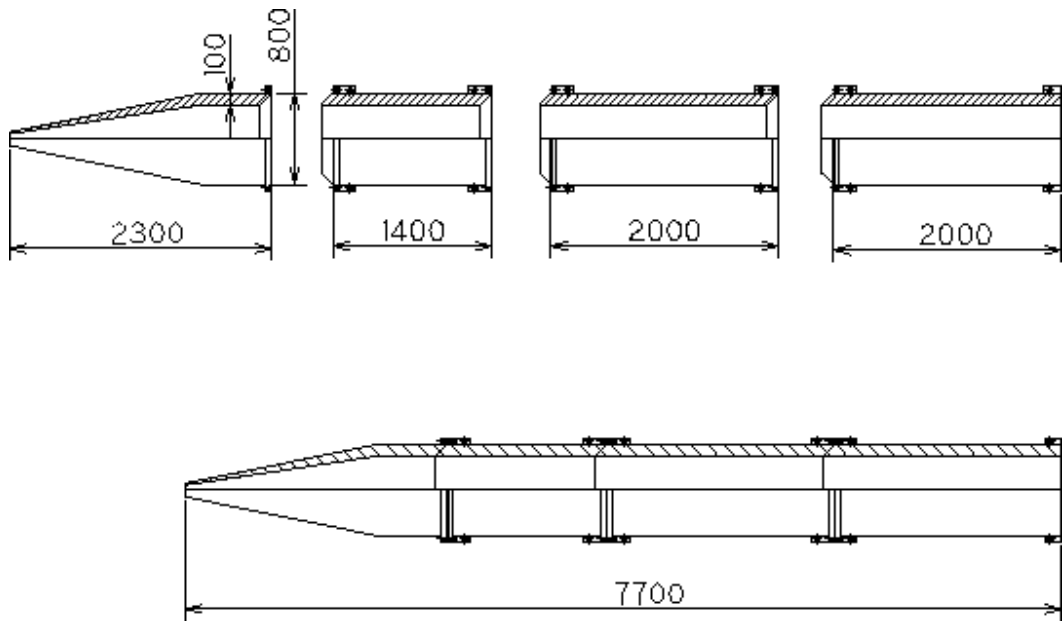}}   
\caption{ \sl\label{wmask}
 Configuration of the tungsten mask. }
\end{figure}

In the calculations for the support tube, stiffness of the support tube was ignored. The bending stiffness can be defined to D=EI, where D is the bending stiffness, E is Young's modulus and I is moment of inertia. Moment of inertia is proportional to $10^4$ of the thickness. Thus, compared to the bending stiffness between 10mm thick of CFRP support tube and tungsten mask, stiffness of the tungsten mask is to be 25 times larger. So, only the stiffness of the 10 cm thick tungsten mask was taken into account. The tungsten mask consists of 4 pieces in the longitudinal direction and 2 pieces in the circumferential direction as shown in Fig.~\ref{wmask}. Overall length after assembling is 8 m and load conditions are self-weight of 32 tons for tungsten mask and other components of 9 tons, respectively. The mechanical properties of tungsten are, Young's modulus is 415 GPa, density is 19.3 and tensile strength is 900 MPa, respectively.

\begin{figure}[hptb]
\centerline{\epsfxsize=14cm \epsfbox{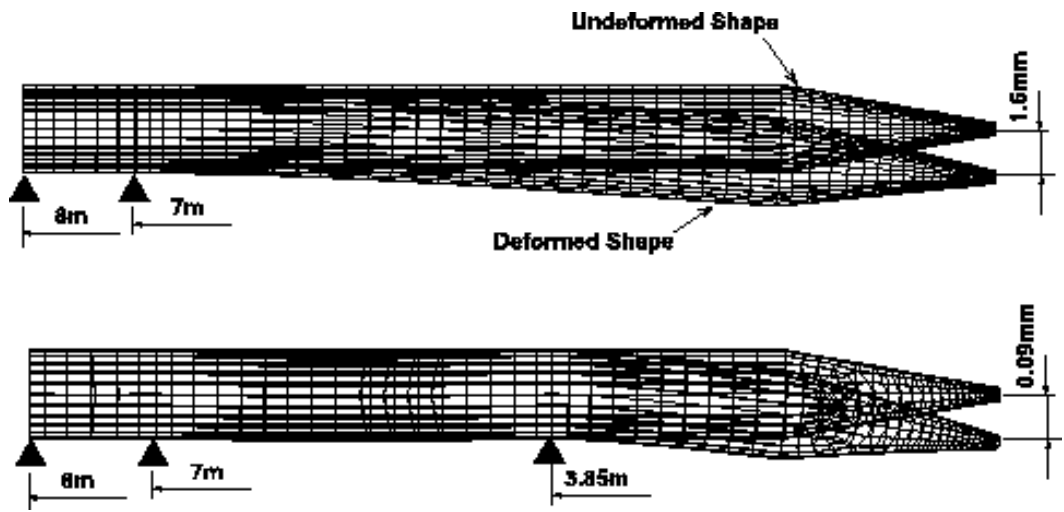}}   
\caption{ \sl \label{deformation}
Deformation of the tungsten mask due to a load of 41 tons. }
\end{figure}

Figure~\ref{deformation} shows the results of static analysis. The gravitational sag due to load of 41 tons was calculated to be 1.6 mm at the top of tungsten mask when it assumed that the tungsten mask is fixed to two points where 7 m far and 8 m far from the interaction point. Stress level was to be 23 MPa that is sufficient low against the tensile strength. This deformation is not accepted. If the mask is additionally fixed at 3.85 m far from the interaction point, the gravitational sag can be improved to 0.09 mm. This indicates that the mask is supported at the back end and the iron yoke.

To know the behavior against vibration such as ground motion, dynamic analysis was carried out. Kinds of dynamic analysis are modal analysis to know the natural frequency and mode shape of structure, harmonic response analysis to know the amplitude at the resonant frequency and spectrum analysis to know the amplitude by input the combined different vibration characteristics. In the dynamic analysis, two analysis models were assumed. One is the support is fixed to the rigid base so called ``rigid case". And the other is the support is fixed to the soft base whose natural frequency is 15 Hz, ``soft case". Usually, natural frequency of large structure is supposed to be relatively low, and from the viewpoint of the seismic design, natural frequency of large structure is to be around 15 Hz. Constraint points are 3 points where 3.85m, 7 m and 8 m from the interaction point, respectively.

\begin{figure}[hptb]
\begin{minipage}[t]{7.25cm}
\centerline{\epsfxsize=7.25cm \epsfbox{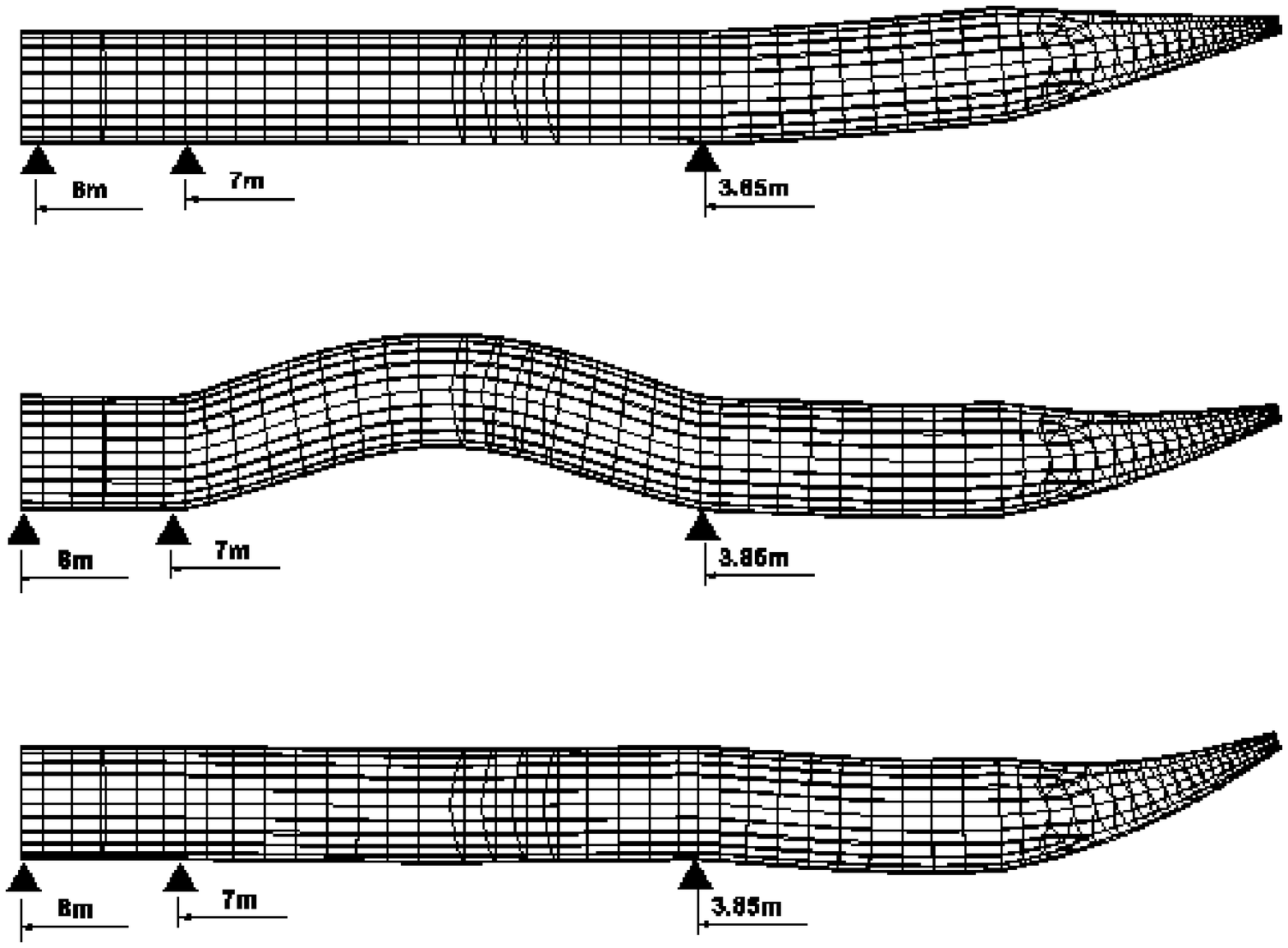}}   
\end{minipage}
\hfill
\begin{minipage}[t]{7.25cm}
\centerline{\epsfxsize=7.25cm \epsfbox{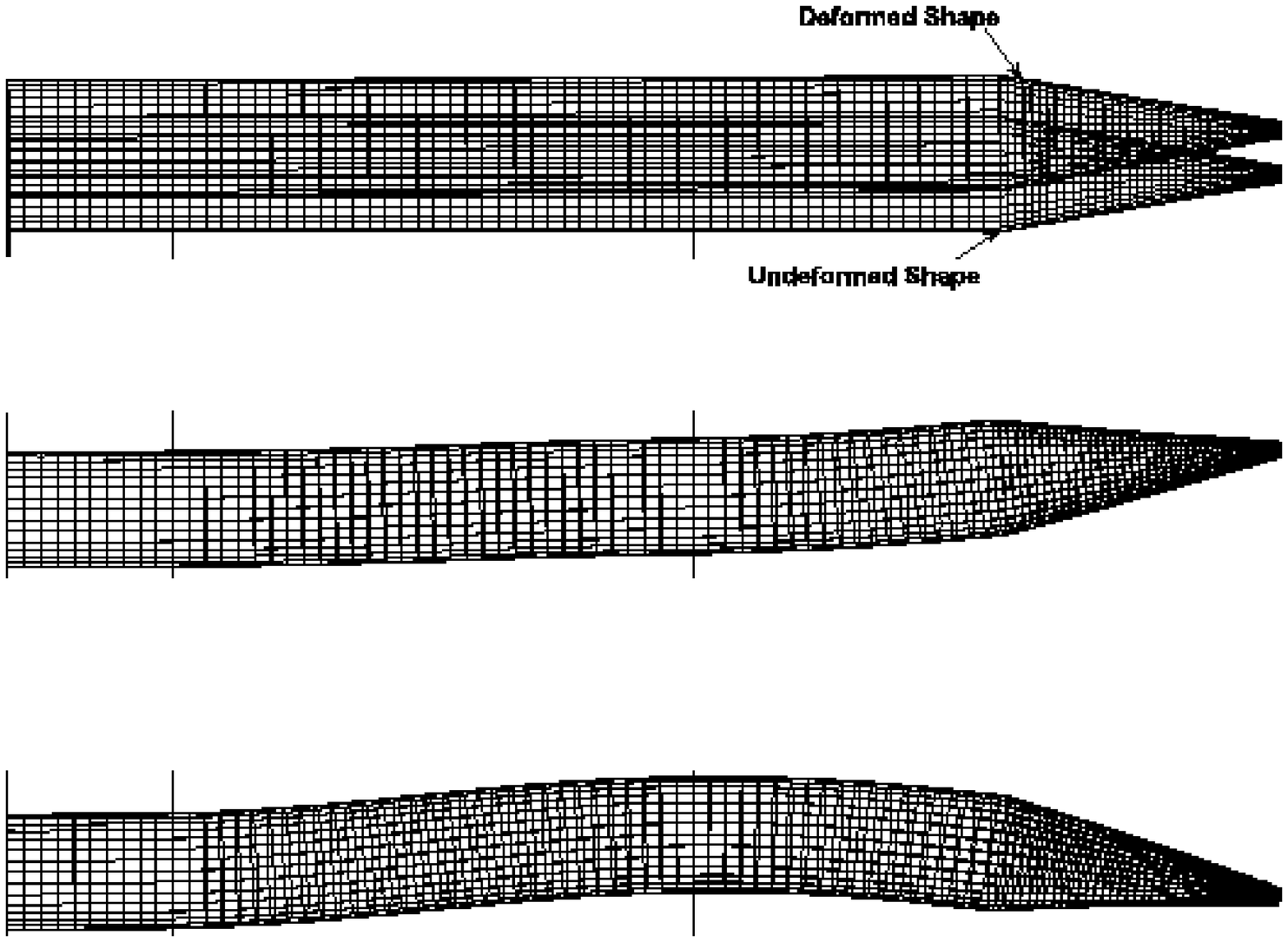}}   
\end{minipage}
\begin{center}\begin{minipage}{\figurewidth}{
\caption{ \sl \label{modal}
Mode shapes by the modal analysis at the first, second and third natural frequencies.  Left and right figures show rigid and soft cases, respectively. }
}\end{minipage}\end{center}
\end{figure}

As shown in Fig.~\ref{modal}, the result of modal analysis on the rigid case is calculated to natural frequency of 71 Hz at first mode, 179 Hz at second mode and 202 Hz at third mode, respectively. In the soft case, natural frequencies are calculated to be 15 Hz at first mode, 54 Hz at second mode and 93 Hz at third mode, respectively.

\begin{figure}[hptb]
\centerline{\epsfxsize=14cm \epsfbox{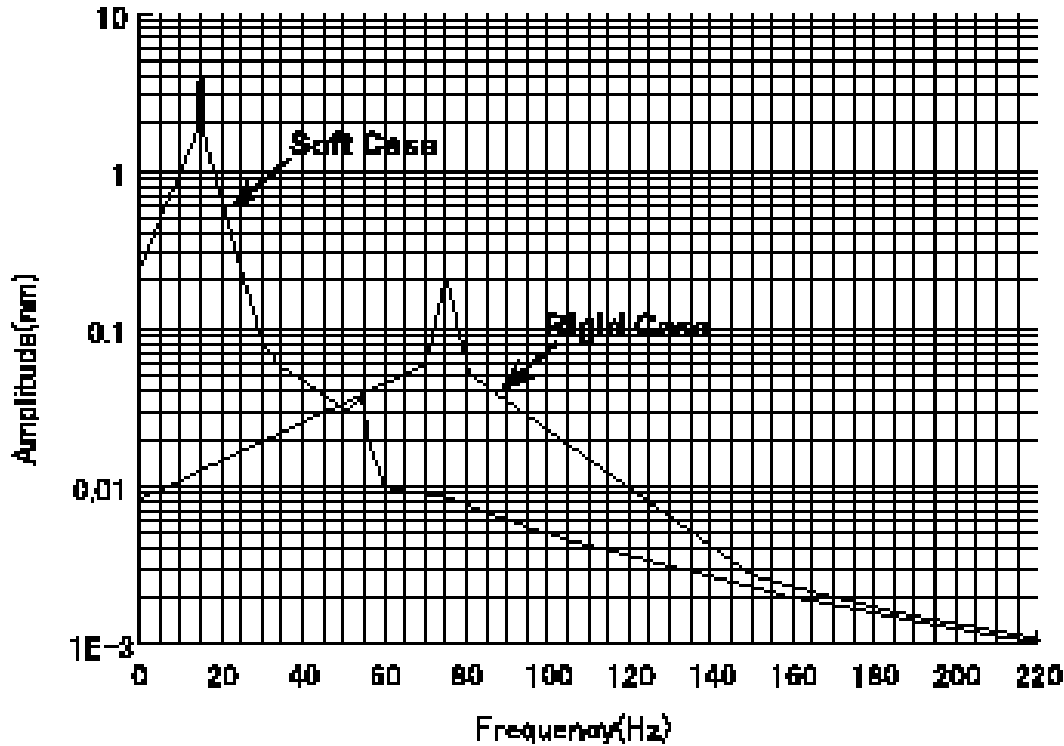}}   
\begin{center}\begin{minipage}{\figurewidth}{
\caption{ \sl \label{harmonic}
Results of the harmonic response analysis , where  peaks are at the first natural frequencies. }
}\end{minipage}\end{center}
\end{figure}

As the next step of dynamic analysis, harmonic response analysis was done. This calculation is a technique used to determine the steady-state response of structure to loads that vary harmonically with time. Applied acceleration was determined from data of grand motion measured in the TRISTAN tunnel. Correlation of amplitude and frequencies are 3 $\mu$m at 3 mHz, 1 $\mu$m at 0.1 Hz, 10 nm at 1 Hz and 5nm at 3 Hz, respectively\cite{sugahara}. From these data, acceleration forces can be converted to $1 \times 10^{-7} , 4 \times 10-^{4}, 4 \times 10^{-5 }$ gal and $2 \times 10^{-4}$ gal, respectively. So acceleration force of $2 \times 10^{-4}$ gal was chosen because of the largest value. Dumping ratio was set to 2 \%. The calculation results of the rigid case is about 0.2 nm as shown in Fig.~\ref{harmonic}. Responded deformation at the soft case is to be 6 nm.

\begin{figure}[hptb]
\centerline{\epsfxsize=14cm \epsfbox{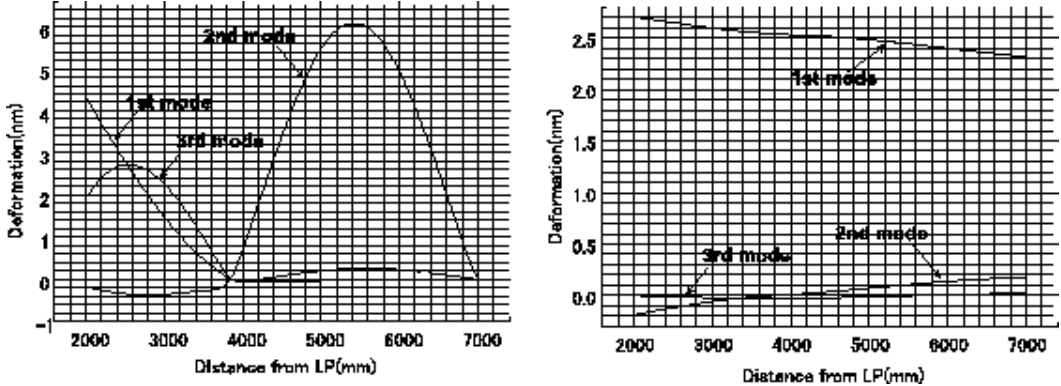}}   
\begin{center}\begin{minipage}{\figurewidth}{
\caption{\sl \label{spectrum}
 Deformations of the tungsten mask between 2 and 7 m from IP by the spectrum analysis.  Left and right figures show rigid and soft cases, respectively.  Mode shapes are similar to those of the modal analysis. }
}\end{minipage}\end{center}
\end{figure}

At the spectrum analysis, three kinds of measured amplitude are input to the constraint points, and then vertical deformations at each resonant frequency are calculated. Figure~\ref{spectrum} shows the calculation results. The deformation of rigid case at each mode is calculated to be from 3 nm to 8 nm. The soft case was turned out to be more stable then the rigid case since the structure was so rigid that it would not be deformed locally. For the soft case, the largest movement was smooth shift by $2.5 \pm 0.2$ nm at the first mode of 15Hz. Good coherency of ground motions has been observed at such short distance between two points in many places. Therefore, the results are encouraging.
Since the coherency generally diminishes at higher frequencies, an oscillation proof must be essential for the support system at $f \ge 10$ Hz.

\begin{figure}[hptb]
\centerline{\epsfxsize=14cm \epsfbox{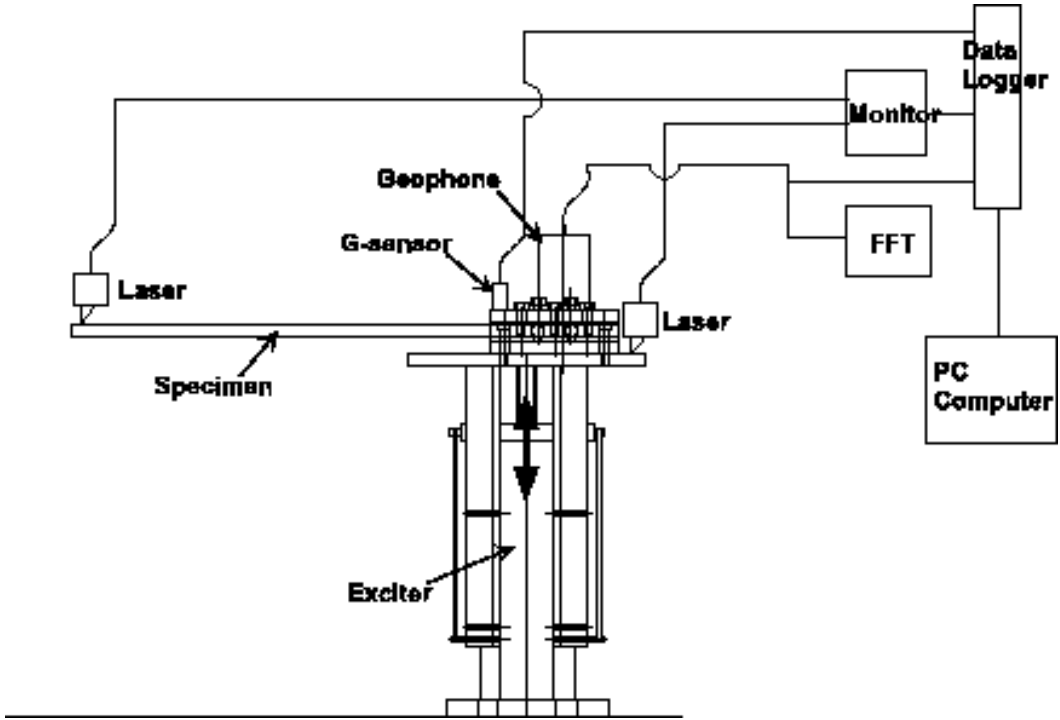}}   
\begin{center}\begin{minipage}{\figurewidth}{
\caption{ \sl \label{prototype}
Test configuration of the prototype in a cantilever case.}
}\end{minipage}\end{center}
\end{figure}

Vibration test using prototype model is just in process. The main purpose is to evaluate a validity of calculations by measuring oscillation properties of the prototype model. In the prototype model design, natural frequency resembles to the real support tube as possible. Two kinds of prototype models are provided those are cantilever type and both ends fixed type. Size of the prototype model is 800mm long and square aluminum bar(100 mm wide and 20 mm thick). They are excited by maximum strength of 10kg long stroke shakers. Input signals are provided by manual sine controller. In the test, some study items are planed; (a) measurement of natural frequencies and amplitudes of the oscillations, (b) measurement of amplitudes with different phases at both ends, (c) comparison between both-ends and cantilever support cases, (d) effects of supporting methods, that is, number of bolts and its sizes (e) measurement of dumping coefficients. Configuration of the data taking as shown in Fig.~\ref{prototype} is two laser position meters are set to the fixed end and free end of the bar, and velocity sensor (Geophone) and G-sensor are placed on the fixed end. Then those data is taken by the data logger and they are sent to the PC computer. Data from Geophone can be analyzed by FFT too.

%% file: detir/dump.tex
\begin{figure}[hptb]
\begin{minipage}[t]{7cm}
\epsfxsize=7cm
\centerline{
\epsfbox{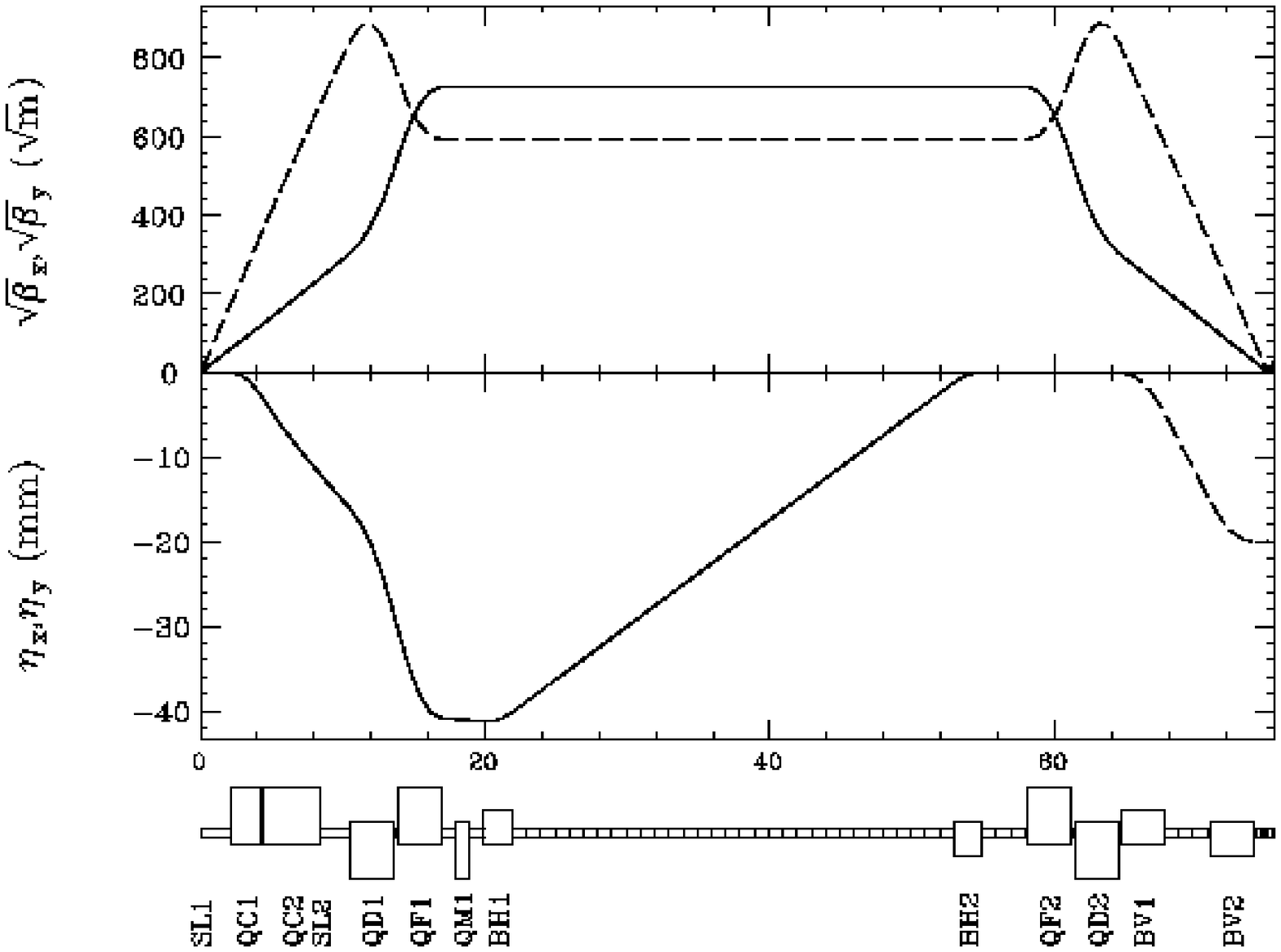}
}
 \caption{\sl  \label{extline}
Optics of the beam extraction  line.}
\end{minipage}
\nolinebreak \hspace{0.5cm}
\begin{minipage}[t]{7cm}
\epsfxsize=7cm
\centerline{
\epsfbox{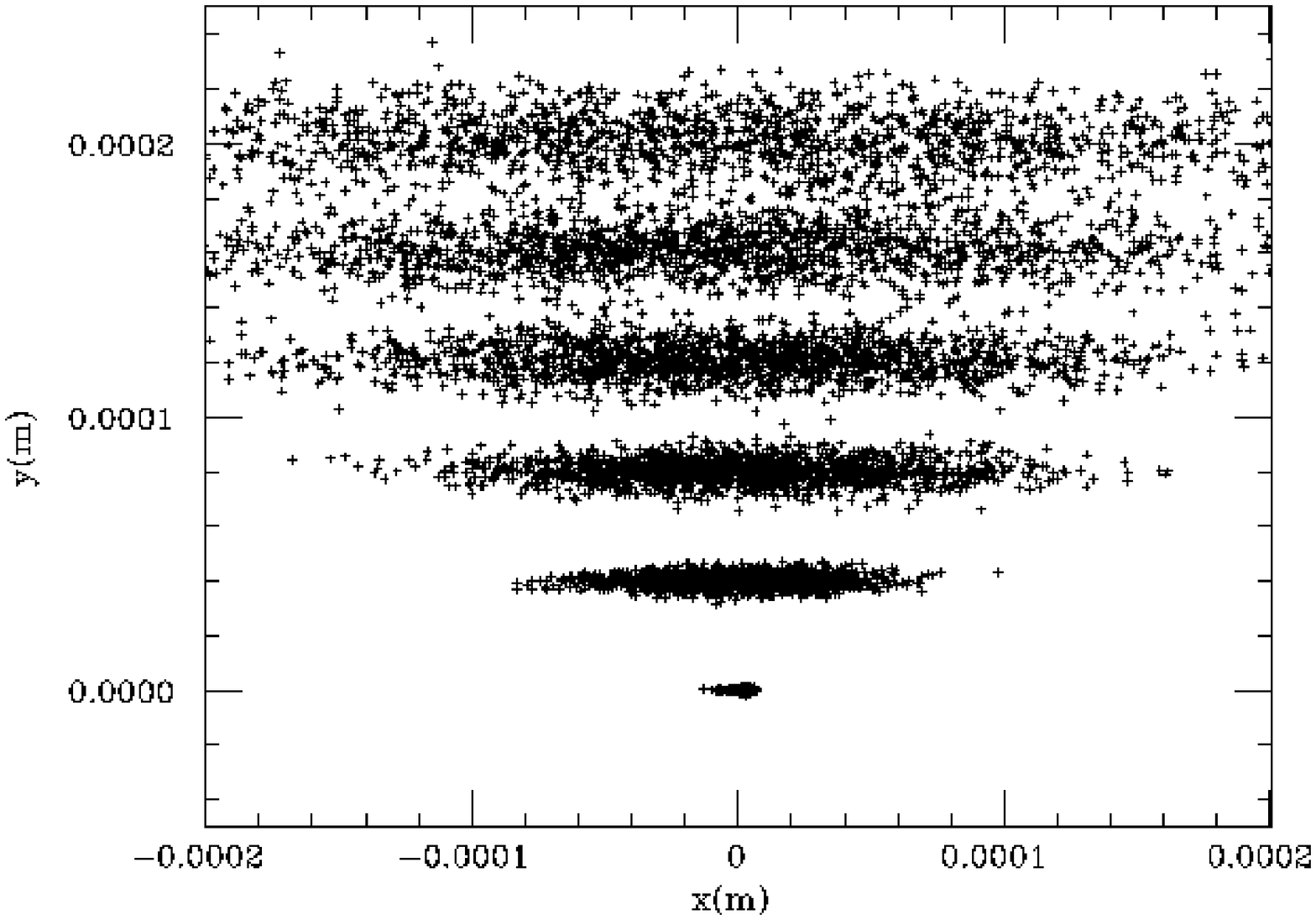}
}
 \caption{\sl   \label{fp2}
Transverse distribution of monochromatic beams at FP2. }
\end{minipage}
\end{figure}

Preliminary design of the dump line, beam line after the
interaction point, was studied. SAD~\cite{sad} was used for this
design. In the beam line, energy distribution of the beam
after the collision should be measured with resolution less
than 0.1\% for $\delta E/E> -1$\% where $\delta E/E$ 
is the relative energy
deviation from the nominal energy. On the other hand, beam
loss in the line should be as little as possible to reduce
background for the detector. In order to measure beam energy
distribution, beam is focused again at 
the second focal point, FP2. And vertical bending is
introduced to make vertical dispersion at FP2. The optics of
the beam line is shown in Fig.~\ref{extline}. 
Four quadrupole magnets, two
sets of doublet, are used for re-focusing. Two vertical
bending magnets are used for introducing dispersion.
Magnetic fields of the two final quadrupole magnets of the
opposite main (incoming)  beam line were also considered. 
Transverse distribution of monochromatic beams at FP2 are
shown in Fig.~\ref{fp2}. Each cluster of the distribution represents
a distribution of electrons in a monochromatic beam of
$-\delta E/E^o_{beam}$=0, 0.2, 0.4, 0.6, 0.8 or 1.0\%, where the beam energy ($E^o_{beam}$) is 250GeV.
Here, we assume the transverse beam parameters after collision as follows; normalized emittance $\epsilon_{x/y}$=1600/20($\times 10^{-8}$~m) and  beta function $\beta_{x/y}$=$4/0.15$(mm), which have been estimated for the JLC-C parameters by CAIN.
Energy distribution will be measured, for
example, scanning a horizontal laser beam with small
vertical size, detecting gamma rays produced by Compton
scattering. And, roughly speaking, the resolution would be
less than 0.1\% for $\delta E/E> -1$\%. 

\begin{figure}[hptb]
\begin{minipage}[t]{7cm}
\epsfxsize=7cm
\centerline{
\epsfbox{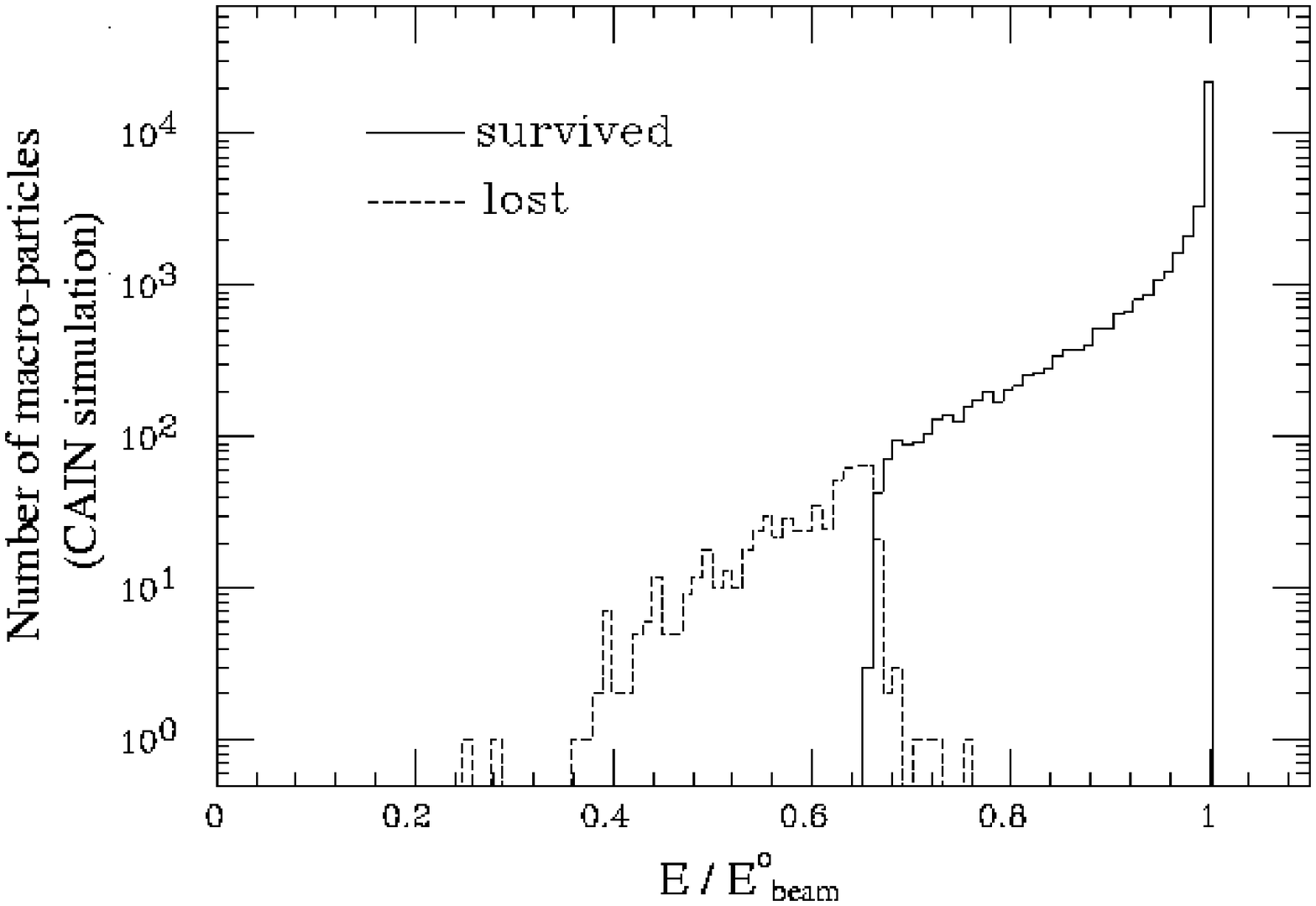}
}
\caption{\sl  \label{edist}
Energy distribution
of survived particles (solid) and  lost particles
(dashed).}
\end{minipage}
\nolinebreak \hspace{0.5cm}
\begin{minipage}[t]{7cm}
 \epsfxsize=7cm
\centerline{
 \epsfbox{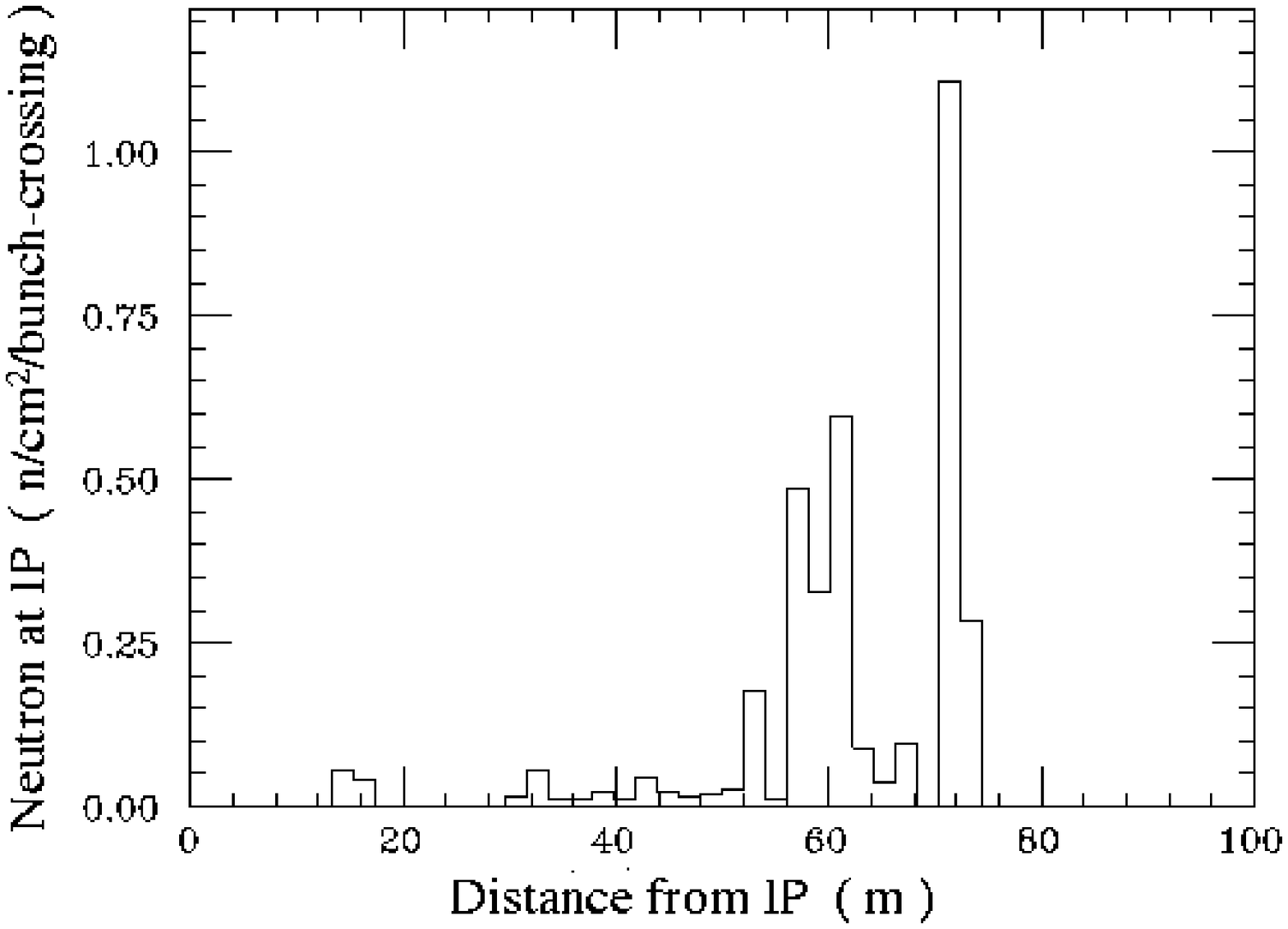}
}
 \caption{\sl   \label{neutron}
Rate of neutron dose at the
interaction point ($/{\rm cm}^2/$bunch crossing).}
\end{minipage}
\end{figure}

To study beam loss, a tracking was performed using SAD, where 
40000 macro particles were used. The initial
conditions of the macro particles were also calculated using
the CAIN. Some simple aperture at each magnet
was assumed and particles went out of it was assigned to be
lost at the position. Figure~\ref{edist} shows the energy distribution
of survived particles (solid) and that of lost particles
(dashed). About 0.9\% of the total particles (which had low
energy) were lost. Figure~\ref{neutron} shows 
the rate of neutron dose at the
interaction point. We assumed that
one neutron would be created per 10~GeV beam loss. The
neutron dose was calculated considering the distance between
the IP and the lost position assuming no shielding between
them.  In total, $4\times10^{11}/{\rm cm}^2$ neutron dose 
is expected in one year.
This result may not be acceptable. 
But it should be noted that 
the self-shielding effect or additional shield 
is not included in this study.
In addition,
more realistic (and complicated) assumption of the
aperture may reduce the beam loss. 
If we loosen the requirement for the resolution of the
energy distribution measurement, the beam line optics may be
modified to reduce the beam loss. 

%% file: dettrk/main.tex
\chapter{Tracking}
\label{chapter-dettrk}

\input{dettrk/vtx/main.tex}

\input{dettrk/itc/main.tex}
\input{dettrk/cdc/main.tex}

\input{dettrk/pfm/main.tex}

%% file: dettrk/vtx/main.tex
\section{Vertex Detector}
\input{dettrk/vtx/ovv/main.tex}
\input{dettrk/vtx/spa/main.tex}

\input{dettrk/vtx/rad/main.tex}

%% file: dettrk/vtx/ovv/main.tex
\newcommand{\dirovv}{dettrk/vtx/ovv/}
\subsection{Overview}
\label{vtxovv}

\subsubsection{Design Criteria}

The primary role of the vertex detector is to reconstruct the
secondary and tertiary decay vertices of B and D meson decays
and tag bottom($b$) and charm($c$) quark jets.  
Distinguishing heavy flavor ($b$ and $c$) jets from
light flavor($u$, $d$, and $s$) jets is relatively easy.
It can be achieved by looking for tracks which have 
impact parameters significantly greater than zero. 
In precision measurements of 
some physics processes  
at the linear collider, discrimination
between $b$ and $c$ quark jets is required.
In this case reconstruction of secondary and
tertiary vertices is necessary for efficient
discrimination of $b$ jets from $c$ jets.
Better spatial resolution and tracking efficiency
are required for the vertex detector to obtain
better tagging efficiency of the jet flavor.
 
The vertex detector  consists of several layers of
sensors surrounding the interaction point concentricaly.
Vertex reconstruction is performed by extrapolating
tracks measured by several layers of the sensors.
In order to get better vertex resolution and better
flavor tagging efficiency, the vertex detector should
have the following features:
\begin{itemize}
\item The length
 of the extrapolation should be as short as possible,
 i.e., the innermost layer should be as close to 
 the interaction point as possible.
\item To minimize the effect of multiple scattering,
 the thickness (in radiation length) of the sensors and
 supporting ladders, particularly that of the inner most
 layer, should be as thin as possible. 
\item The sensors  should be
 two-dimensional (pixel type) sensors with high granularity  
 in order to separate near-by tracks in a collimated jet.
 Of course better spatial resolution is desirable.
\end{itemize}

When we try to satisfy these requirements, we are faced with
several difficulties. As discussed in section 
\ref{detir-background-section}
the track density of the pair background particles 
increases near the interaction point. 
High background hit rate causes the problems of
wrong tracking due to fake hits leading to poor
vertex resolution, and radiation damage of the sensors.
Therefore there is a limit on the minimum radius of
the inner most layer.
The existence of the beam background gives rise to
another requirement for the  vertex detector:
\begin{itemize}
\item The sensors used for the vertex detector should have
 enough radiation immunity so that the first layer
 can be located  in the vicinity of  the interaction point.
\end{itemize}

\subsubsection{Possible Options}

The vertex detector has to consist of pixel type
sensors not only for the separation of tracks 
in a collimated jet but also
for the detection of tracks in the environment 
where a large number of the beam background hits exist.
The following technologies can be candidates for the
sensors used for the vertex detector.
\begin{itemize}
\item {\bf CCD:} CCD (Charge Coupled Device) is an established 
technology widely used also in commercial products.
The thickness of the sensitive region is about 20~$\mu$m.
The wafer thickness can be thinned down to  20~$\mu$m
provided that the mechanical strength is assured with an 
adequate method. Large size chips 
($1.6\times 8\ \mbox{\rm{cm}}^2$ for SLD~\cite{vtxovv:sld}) 
can be made.
Relatively sensitive to radiation damage.
\item {\bf CMOS pixel sensor:} In CMOS pixel sensors, each
pixel has a sensor diode and a read-out amplifier.
Signals from the pixels are read out by x-y addressing.
The thickness of the sensitive region is about 20~$\mu$m.
The radiation immunity is better than that of CCDs. 
\item {\bf Hybrid pixel sensor:} Hybrid pixel sensors have
a sensor diode wafer and a readout wafer connected by
bump bonding technique. This structure inevitably
makes them thicker than CCDs or CMOS sensors.
The radiation hardness is excellent.
\end{itemize}

Among the options listed above, we regard the CCD as
a primary candidate for the sensor used for the 
vertex detector at JLC. 
We have been studying on the properties
of CCD sensors extensively in order
to demonstrate the feasibility for the vertex detector 
at the linear collider experiment. The detail will be
described in subsections \ref{vtxspa} and \ref{vtxrad}.

\subsubsection{Detector Configuration}

\begin{figure}
\centerline{\epsfxsize=10.0cm \epsfbox{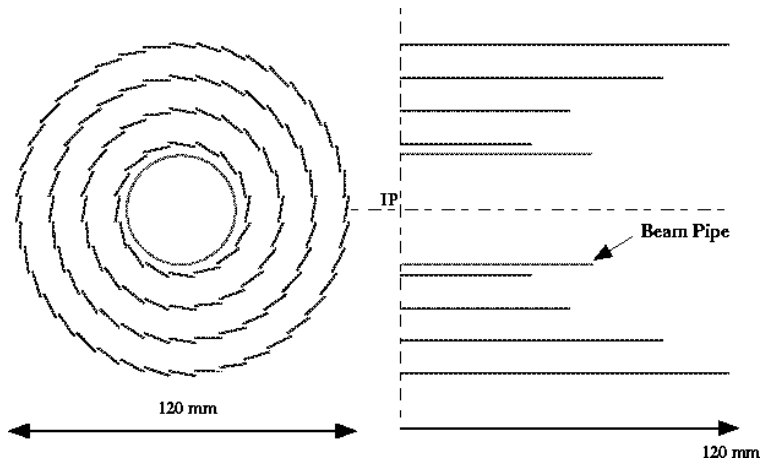}}
\caption{\sl \label{vtxovv:schema}
Schematic design of the CCD vertex detector for JLC.}
\end{figure}

One possible schematic design of the CCD vertex detector,
which is put in the JLC detector full simulator (JIM), 
is shown in Fig.~\ref{vtxovv:schema}. 
Main parameters of the detector are:
\begin{itemize}
\item Four layers of CCDs placed at $r=$24, 36, 48, and 60~mm.
\item Each layer has an angular coverage of 
 $|\cos \theta | < 0.9$.
\item The thickness of each layer is 300~$\mu$m.
\item The pixel size of the CCD is 25~$\mu$m square.
\end{itemize}

Hit density of the vertex detector is quite high
due to collimated jets and the pair-background hits.
In such a situation,
at least four layers are necessary for 
local track finding 
and  self calibration using the vertex detector alone.

In a simple model of the vertex detector which has 
two layers at $r_{in}$ and $r_{out}$ with the spatial resolution
of $\sigma_{in}$ and  $\sigma_{out}$, respectively, the impact parameter
resolution $\sigma_b$ can be expressed as 
\begin{equation}
\sigma_b={{\sigma_{in} r_{out}}\over {r_{out}-r_{in}}} \oplus
 {{\sigma_{out} r_{in}}\over {r_{out}-r_{in}}} \oplus
 { {0.014 r_{in}}\over {p\beta}}\sqrt{X_r \over \sin^3 \theta},
\label{vtx:ovv:eqipres}
\end{equation}
where 
$X_r$ is the thickness of the inner layer in radiation length,  
$p$, $\beta$, and $\theta$ are momentum, velocity, and
polar angle  of the charged particle, respectively. 
Outer radius of the vertex detector has less importance
on the impact parameter resolution as long as
$r_{out} \mathchar12829  r_{in}$ is satisfied. 
In the case of the JLC detector model,  
the impact parameter resolution for high momentum particles
better than  the expression~\ref{vtx:ovv:eqipres} 
is achieved by combining 
the vertex detector and the CDC. 
The reason of this fact is that taking four layers of the 
vertex detector together as the 
``in'' layer and the CDC as the ``out'' layer  
in the expression~\ref{vtx:ovv:eqipres} gives smaller value
for the first two terms in equation~\ref{vtx:ovv:eqipres}.  
Then important parameters for the vertex detector 
are $r_{in}$(radius of the innermost layer), 
$\sigma_{in}$(spatial resolution), and $X_r$(detector thickness). 

The innermost layer  of the vertex detector is 
conservatively located at $r= 24$~mm in the present design. 
With this configuration,  $e^{\pm}$ dose of 
$\sim 1.5\times 10^{11}/\mbox{cm}^2$ is expected 
in one year at the innermost layer with the solenoid
magnetic field of 2~Tesla and with the accelerator 
parameter set of ``A'' described in section~\ref{headers_movv_param}.
As will be shown 
in subsection~\ref{vtxrad},
the CCDs of the innermost layer would stand operation of 
more than 3 years
under this rate of radiation dose. If stronger magnetic field is 
employed, the pair-background particles are confined in the
region of smaller radius and the first layer of the vertex 
detector can be placed closer to the interaction point.
The same background rate is expected at $r=20$~mm with
$B=3$~Tesla. If still smaller radius is required,
or if the luminosity of the machine is increased
(the machine parameter set ``Y'' in section~\ref{headers_movv_param},
for example), 
more frequent replacement of the damaged CCDs, 
still stronger magnetic field, or 
devices more radiation hard are required.

\begin{figure}
\centerline{\epsfxsize=10.0cm \epsfbox{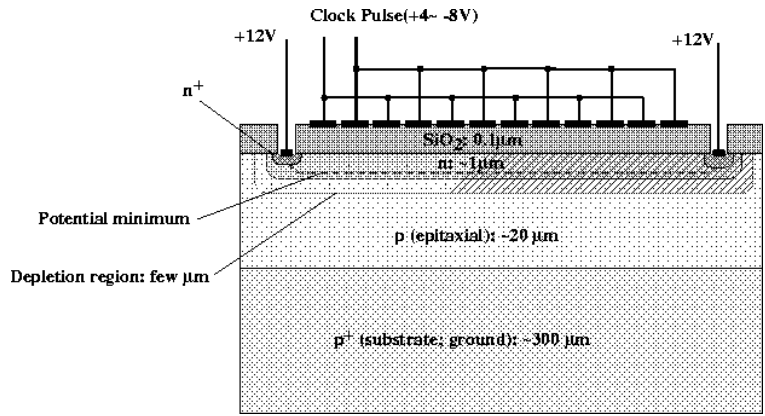}}
\caption{\sl \label{vtxovv:ccd}
Structure of a CCD sensor.}
\end{figure}

One of the advantages of the CCD as a tracking device is
its excellent spacial resolution. 
Structure of a CCD sensor is schematically shown in 
Fig.~\ref{vtxovv:ccd}. The epitaxial layer of
$\sim 20\ \mu$m thick is the sensitive region.
Since most of the epitaxial layer is not depleted,
the signal charge (electrons of e/h pairs) diffuses
thermally in the epitaxial layer and spreads over
several pixels even for the normal incident particles.
Consequently, high precision position measurement
can be achieved by the charge sharing. 

\begin{figure}[hbtp]
\centerline{\epsfxsize=10.0cm \epsfbox{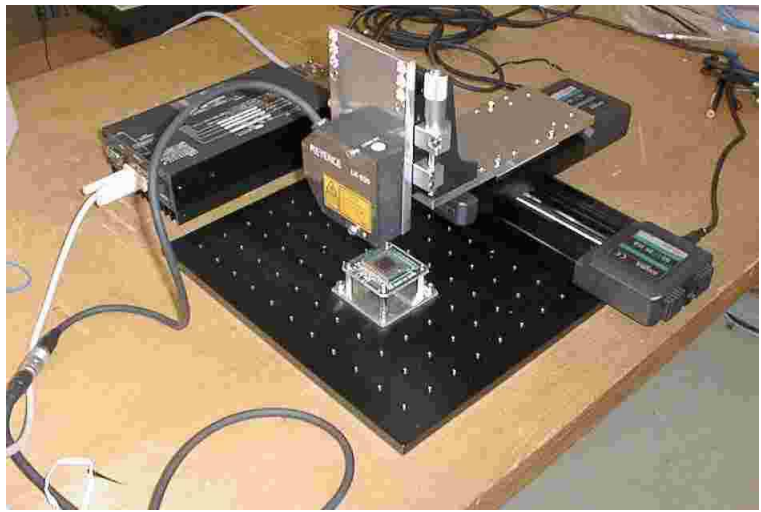}}
\caption{\sl \label{vtxovv:scanner}
Measurement system of CCD flatness.}
\end{figure}

\begin{figure}[hbtp]
\centerline{\epsfxsize=8.0cm \epsfbox{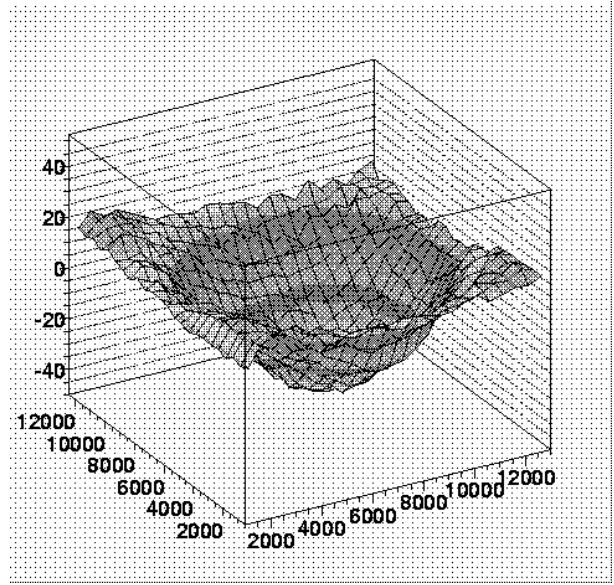}}
\caption{\sl \label{vtxovv:flatness}
Measured surface flatness of a back-thinned CCD sensor.
(units in $\mu$m)}
\end{figure}

As shown in Fig.~\ref{vtxovv:ccd}, normal CCDs 
have thick insensitive substrate. When used for
the vertex detector, this part should be thinned
as much as possible to reduce the multiple 
scattering of charged particles.
On the other hand, it becomes more difficult
to keep thin silicon wafers flat, particularly
operated at low temperature~\cite{vtxovv:sld}. 
We have measured the flatness of CCDs using
a laser displacement meter and an x-y scanner
(Fig.~\ref{vtxovv:scanner}).
We show in Fig.~\ref{vtxovv:flatness} a result
of the flatness measurement of
a commercially available CCD which has thinned
sensitive area for the purpose of the back-illumination.
A sizable bowing presumably due to the temperature
difference between at the fabrication process
and at the room temperature can be seen.
From this point of view, operation of the vertex detector
at near room temperature is desirable to minimize
the distortion of the CCD wafers.
The mechanical structure which can keep the thin CCD
wafers flat has to be studied.

Each CCD wafers will have multiple readout nodes so that
all pixels can be read out during the train crossing
interval of 6.7~msec. If we conservatively 
assume the readout frequency of 10~MHz, 
each readout channel can read out the area of
about $0.4\ \mbox{\rm cm}^2$ between the train crossings.
Since the total area of the CCDs in the vertex detector is
about  $2000\ \mbox{\rm cm}^2$, about 5000 readout channels
are needed. 
Higher readout frequency or longer readout period
by vetoing next collision(s)  
can  reduce the number of the readout channels.  

\subsubsection{R\& D Status}
So far, our R\&D effort has been devoted mainly to the study of
the spatial resolution and the radiation hardness of 
CCD sensors. In the following subsections, we will 
describe  results of the study in detail and 
demonstrate the feasibility of the CCD vertex detector.
In particular, with the conservative design described 
above (machine parameter set ``A'' and $r_{min}=24$~mm),
it will be shown that the operation of CCD vertex detector
near room temperature ($\sim 0^\circ C$) is feasible.

%% file: dettrk/vtx/spa/main.tex
\newcommand{\dirspa}{dettrk/vtx/spa/}

\subsection{Study of Spatial Resolution}
\label{vtxspa}
\subsubsection{Factors Affecting the Spatial Resolution}

The most important parameter for the vertex detectors
is the spatial resolution.
Quite a number of factors can affect the spatial resolution. The most
obvious is the pixel pitch, but other intrinsic effects can be very
important. 
In the case of an analogue readout system, the spatial resolution of CCD
detectors depends on the following parameters:
\begin{enumerate}
\item Signal to Noise ratio (S/N).
\item Charge sharing due to carrier diffusion.
\item Digitization resolution in readout system.
\end{enumerate}

We have carried out test beam to measure the spatial
resolution in CCD detector. 
The influence of those parameters to the spatial resolution
are presented here. 

\subsubsection{Test Beam}

We have tested three types of CCD detectors in Multi Pinned Phase(MPP)
operation~\cite{mpp-mode}.
In the MPP mode, the array clocks are biased sufficiently negative
to invert all phases and the surface potential is ``pinned'' to substrate
potential. As a result, the thermal excitation of electrons is
extremely suppressed. It reduces the dark current by a factor of 20
compared with normal non-inverted operation mode.
Table~\ref{vtxspa:samples} is a list of the test samples with important
 parameters.  Two of the three CCDs
were S5466~\cite{HPK-ref} fabricated 
  by Hamamatsu Photonics K.K. (HPK) , 
but with different depth of epitaxial layers,
 10$\mu$m (HPK10) and 50$\mu$m (HPK50), respectively.
HPK10 is a commercially available, but HPK50 is  specially fabricated
for this experiment.
Those consists of 512 $\times$ 512 pixels and the pixel
size is 24 $\mu$m $\times$ 24 $\mu$m. 
 Another one was CCD02-06SIS~\cite{EEV-ref} fabricated by EEV Company
and is also commercially available.
The CCD02-06SIS consists of 385 $\times$ 578 pixels 
and the pixel size is 22 $\mu$m $\times$ 22 $\mu$m.
It has epitaxial depth of 20$\mu$m (EEV).
HPK devices are two-phase CCD, while EEV device is a three-phase CCD. 

\begin{table}[hbth]
\begin{center}
	\caption{\sl 	\label{vtxspa:samples}
Test Samples.}
\vspace*{6pt}
	\begin{tabular}{lllllll} \hline
Test & CCD & Image  & Pixel  & Pixel  & Epitaxial & Amplifier 	 \\
sample &type & area   & pitch  & format & layer     & sensitivity \\
    &&[mm]    &[$\mu$m]&  (H$\times$V) &[$\mu$m]    & [$\mu$V/e]    \\ \hline
HPK10&S5466&12.29$\times$12.29& 24$\times$24& 512$\times$512& 10 & 2.0 \\
HPK50&S5466(Special)
&12.29$\times$12.29& 24$\times$24& 512$\times$512& 50 & 2.0 \\ 
EEV &CCD02-06&8.5$\times$12.7 & 22$\times$22 & 385$\times$578 & 20 &
1.0 \\ \hline
	\end{tabular}
\end{center}
     \end{table}

\begin{figure}
	\centerline{\epsfxsize=10.0cm \epsfbox{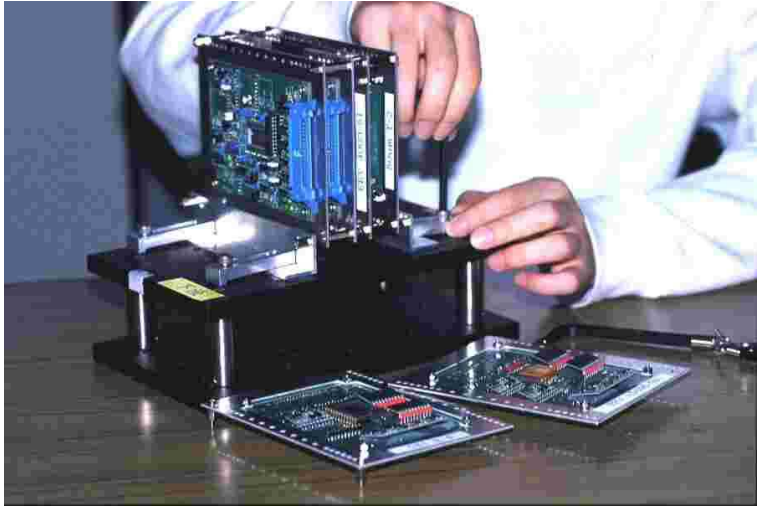}}
\begin{center}\begin{minipage}{\figurewidth}
\caption{\sl \label{vtxspa:setup}
Setup of the CCD tracker.}
\end{minipage}\end{center}
\end{figure}

\begin{figure}
	\centerline{\epsfxsize=10.0cm \epsfbox{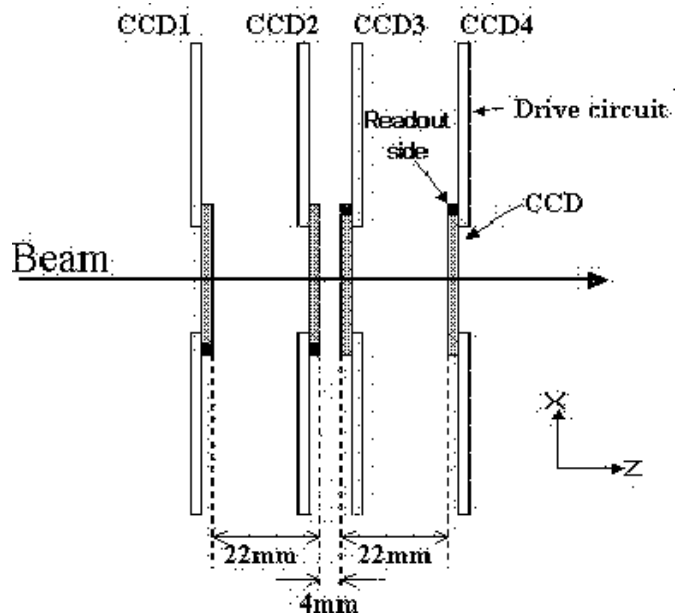}}
\begin{center}\begin{minipage}{\figurewidth}
\caption{\sl \label{vtxspa:set-up}
Schematic view of the setup. 
Both CCD1 and CCD2 were reference CCDs for tracking. CCD3 was the test
sample. CCD4 was used for reduction of the accidental tracks, and the
measurement of the efficiency.
}
\end{minipage}\end{center}
\end{figure}

A CCD tracker (Fig.~\ref{vtxspa:setup}) was set up at
the T1 test beam line of the 12-GeV proton synchrotron (PS) 
at High Energy Accelerator Research Organization (KEK)
as shown in Fig.~\ref{vtxspa:set-up}.
The negative pion beam was used in the experiment.
The momentum of the beam was chosen to be
2.0 GeV/c, 1.0 GeV/c, 0.7 GeV/c, and 0.5 GeV/c.
The normal axis of the CCD tracker was aligned to the beam.
The tracker consisted of four layers of CCDs as shown in
the figure. First and second layers of CCDs are reference
detectors for precise reconstruction of a particle track.
The CCD in third layer is the sensor under test in this experiment.
Another reference CCD was located at the end of the tracker
for the reduction of accidental tracks. The reference CCDs were
HPK/S5466 with epitaxial depth of 10$\mu$m and were same type with
HPK10. All of the CCDs were driven at a rate of 250 kpixels/s, and 
operated in a full-frame readout mode. The CCD's were read out 
every three seconds to synchronize with the PS cycle
using 12-bit analog-to-digital converter(ADC).
The sensitivities of the test samples including the gain of 
amplifier are 61.7$\pm$0.03$\mu$V/electron, 23.99$\pm$0.03$\mu$V/electron,
and 34.63$\pm$0.02$\mu$V/electron, for HPK10, HPK50, and EEV, respectively. 
The details of this experiment are described 
in elsewhere~\cite{beamtest}.

\subsubsection{Signal to Noise Ratio}

  The charge created by a penetrating particle is collected 
generally by several consecutive pixels, which form a cluster.
In order to find the cluster, we first look for a leading pixel
whose charge is higher than the threshold and is highest 
in the adjacent 6$\times$6 pixel region.  
The threshold was defined as 13, 7, and 15 times of the noise level,
for EEV, HPK10, and HPK50, respectively,
in order to reduce the noise contribution with keeping the efficiency good.

\begin{figure}[hbth]
	\centerline{\epsfxsize=7.5cm \epsfbox{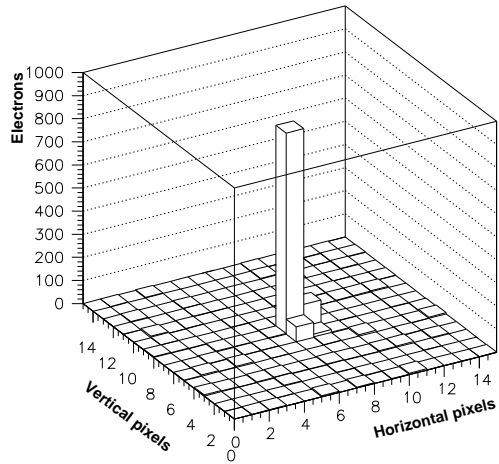}}
\begin{center}\begin{minipage}{\figurewidth}
\caption{\sl \label{vtxspa:typical}
Typical event by a minimum ionizing particle detected in HPK10,
where the height is a collected charge in the unit of 
electrons after the correction
of the sensitivity of the detector including the gain of the amplifier. }
\end{minipage}\end{center}
\end{figure}

Total charge of the cluster is sum of the charges of 
the leading pixel and neighboring pixels.
Fig.\ref{vtxspa:typical}
shows a typical event of a minimum ionizing
particle which was detected in HPK10. 

\begin{figure}[hbth]
	\centerline{\epsfxsize=7.5cm \epsfbox{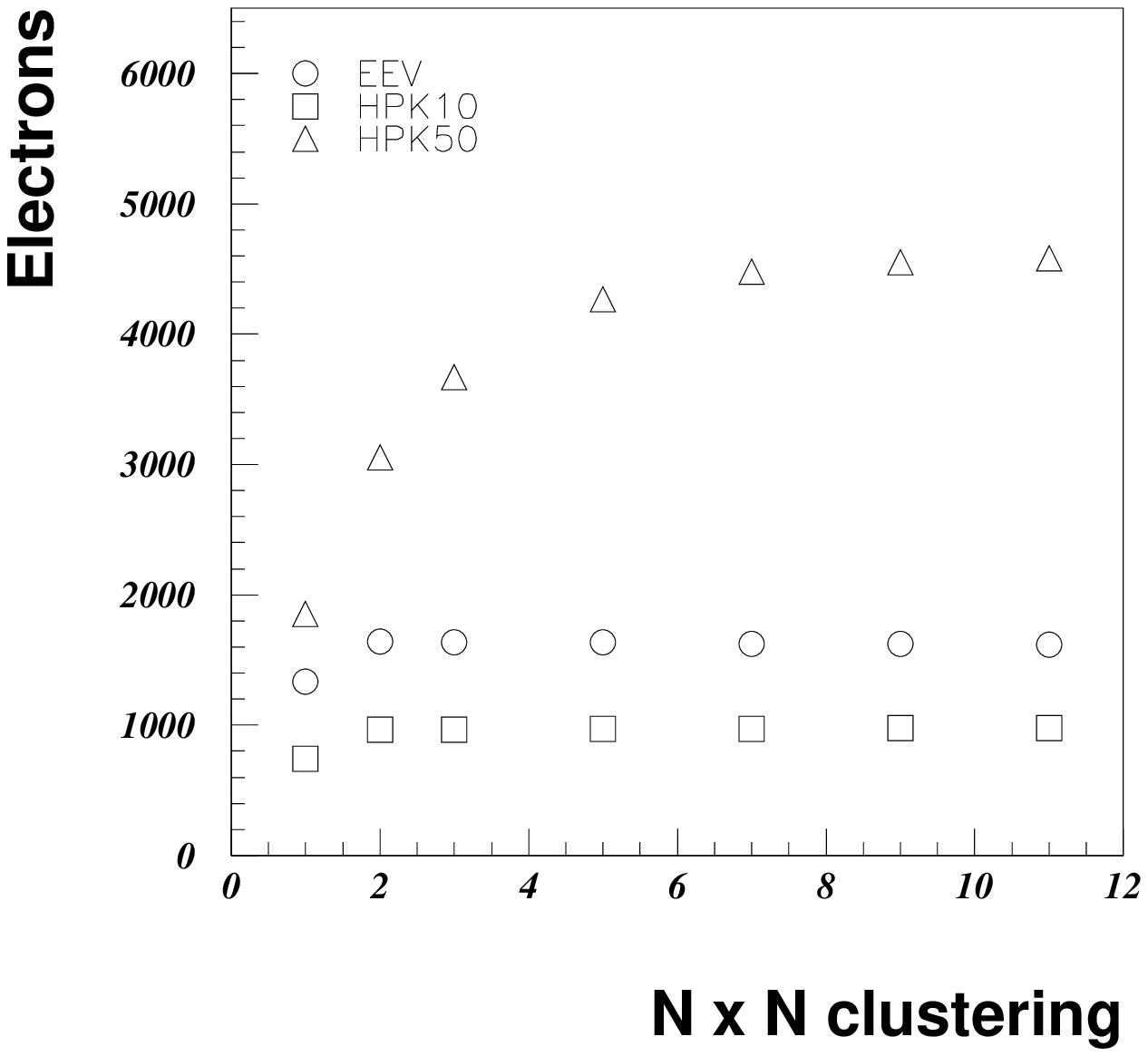}}
\begin{center}\begin{minipage}{\figurewidth}
 \caption{\sl \label{vtxspa:collected}
Collected charge for a minimum ionizing particle as a function
of clustering size, where clustering size was defined as N$\times$N pixels.
The horizontal axis represents a number of N and 
vertical axis is a collected charge in the unit of 
electrons after the correction
of the sensitivity of the detector including the gain of the
amplifier. }
\end{minipage}\end{center}
\end{figure}

Fig.\ref{vtxspa:collected}
shows clustering size dependence of the
total charge at $-$15$^\circ$C, where clustering size is defined as
N$\times$N pixels.
As shown in the figure,
almost all the deposited charges are collected in
the cluster size of  2$\times$2 pixels for 
HPK10 and EEV. For HPK50, however,
more than 5$\times$5 pixels 
are necessary to collect all of the deposited charges.
Approximately 920 electrons/10$\mu$m~\cite{white}
is expected to be produced by a minimum ionizing particle
in silicon, which is consistent with the observed charge
by taking into account the depth of epitaxial layers.

     \begin{figure}[hbth]
	\centerline{\epsfxsize=7.5cm \epsfbox{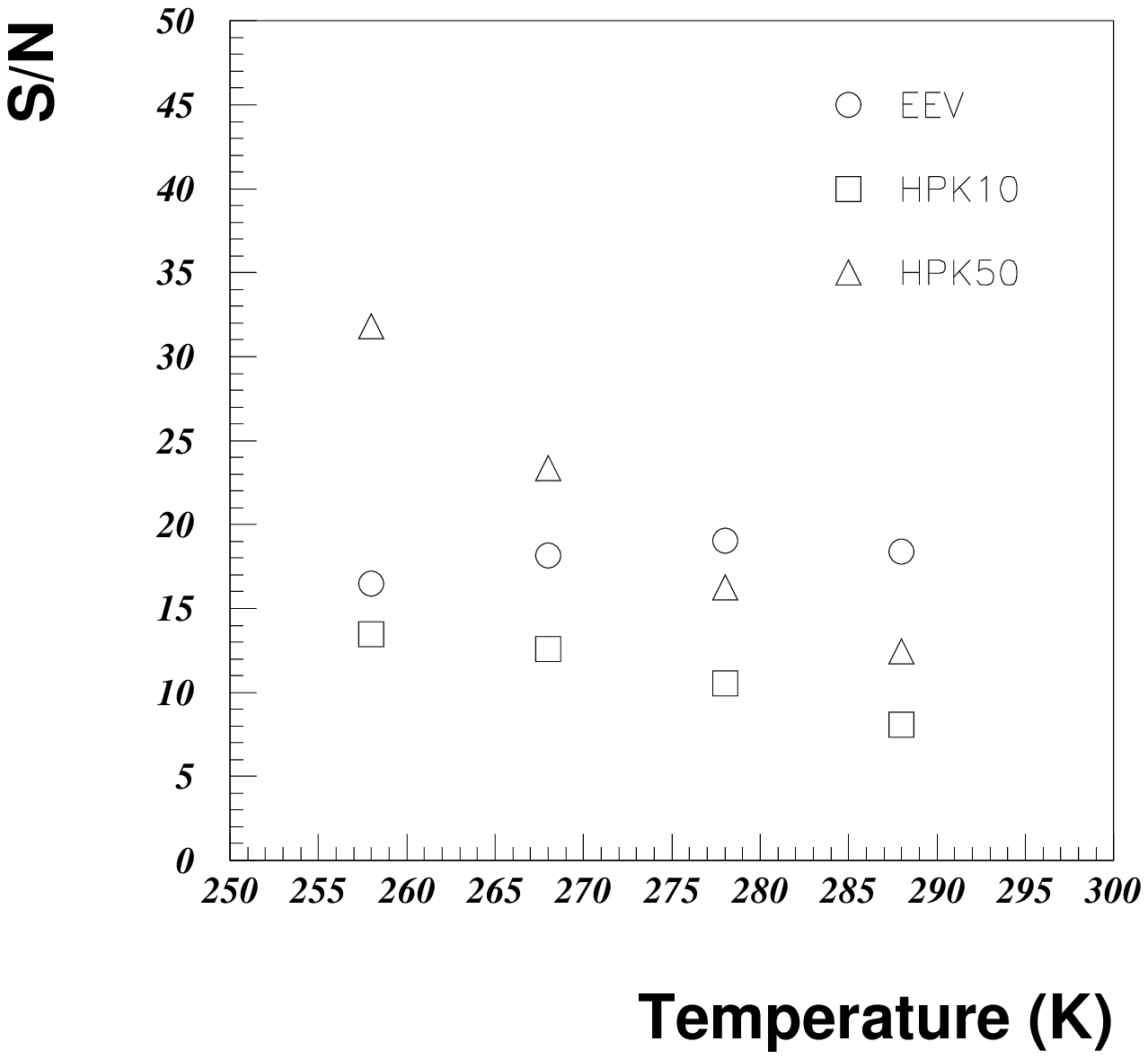}}
\begin{center}\begin{minipage}{\figurewidth}
\caption{\sl 	\label{vtxspa:snratio}
Signal-to-Noise ratio as a function of temperature, where clustering size was chosen to be 2$\times$2 pixels.
In this experiment, the CCD's were read out every three seconds
to synchronize with the PS cycle.
In the JLC project, however, the read out cycle of 6.7ms is planned.
It is consequence of operating faster read out cycle that
the noise due to dark current is suppressed and 
the S/N values are expected to improve significantly. 
}
\end{minipage}\end{center}
     \end{figure}

Fig.\ref{vtxspa:snratio} 
shows the temperature dependence
of S/N for 2$\times$2 pixel clustering.
The S/N is defined as the most
probable value of the collected signal charge distribution
divided by the r.m.s. of the noise charge distribution.
As temperature is raised, the S/N for HPK CCDs significantly
degraded due to the increase of the dark current,
while the S/N of EEV has very small temperature dependence
in the range from $-15^\circ$C to $+15^\circ$C.
This is because the noise caused by the dark current is not a dominant 
source in the total noise for EEV.
The main source is considered to be an external
noise which comes from the cooling system.
The degradation of the S/N for HPK50 in this range was larger
than that for HPK10, since the dark current for HPK50 increases
more rapidly than that for HPK10 as temperature rises.
In the previous paper~\cite{Cool},
 the operational condition at higher
temperature was discussed with the dependence of dark current
on temperature and readout drive frequency.
When CCD is driven at faster rate, the noise due to dark current is
suppressed. Then, the S/N values are expected to improve.

\subsubsection{Charge Sharing and Position Determination}

Since the charge sharing among adjacent pixels depends on 
the hit position of the particle passing through the detector,
it is possible to obtain better resolution by taking into account
a charge sharing characterization which is the effect of the lateral
charge diffusion. The position of the cluster was determined
using the 2$\times$2 pixels.
In this report, we adopted two ways of determination for comparison.

\begin{itemize}
\item {The Analog-Centroid Method}

In the Analog-Centroid-Method (AC-method)\cite{acmethod},
the hit position is reconstructed as the charge weighted average
of the 2$\times$2 pixels as

\begin{equation}
	x = \frac{\Sigma Q_i X_i}{\Sigma Q_i}, \label{vtxspa:XAC}
\end{equation}

where $Q_i$ and $X_i$ represent a charge deposit and
pixel position defined at the center of i-th pixel, respectively. 
In this method we assume a linear relation of charge
deposit to the pixel position.

\item {The Ratio location mapping method}

In the Ratio-Location-Mapping method(RLM-method), 
the relation between the position $x$ of the cluster
and the charge ratio $R_{x}$ of adjacent pixel
was derived from the data, 
where $x$
was calculated by extrapolating the track from the upstream reference
CCDs and normalized by the pixel pitch.
The $R_x$ is defined using the charge deposit of adjacent two pixels as 

\begin{equation}
	R_x = \frac{Q_x}{Q_{max}},\label{vtxspa:RX}
\end{equation}

where $Q_{max}$ and $Q_x$ are a charge deposited in the leading pixel
and in the adjacent pixel to x-direction, respectively.

Since the particle distribution within a pixel is uniform, we can
assume $dN/dx$ is constant, where $N$ is a number of incoming particles.
From this assumption, we can get the relation of $x$ with $R_x$
as $dx/dlog (R_x) \propto dN/dlog(R_x)$.
Taking initial conditions
as $x \rightarrow 0.0$ at $log(R_x)\rightarrow -\infty$ 
and  $x = 0.5 $ at $log(R_x) = 0$, 
we obtain following formula,

\begin{equation}
 x(log(R_x)) = \frac{0.5}{N}\int_{-\infty}^{log(R_x)}
	\frac{dN}{dlog(r)} dlog(r),\label{vtxspa:logRx}
\end{equation}

where 
$dN/dlog(r)$ is a $R_x$ distribution which is obtained from the data.
The initial conditions correspond to the center of leading pixel,
and the middle point of the gap between two pixels, respectively.

Fig.~\ref{vtxspa:rlmfunc}(a) shows data points derived by Eq.\ref{vtxspa:logRx} and 
RLM function obtained by fitting to the data with six order
polynomial function.
Fig.~\ref{vtxspa:rlmfunc}(b) shows a linear relation which is defined as
$x=0.5R_x$ and allowed region of the AC-method with 2$\times$2 clustering
which is obtained from Monte Carlo simulation. 

\begin{figure}[htbh]
	\centerline{\epsfxsize=15.0cm \epsfbox{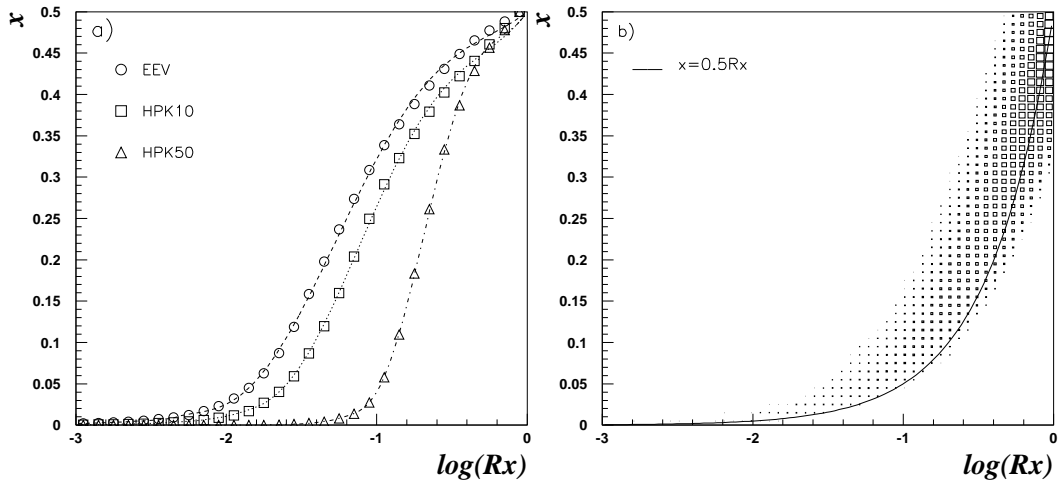}}
\begin{center}\begin{minipage}{\figurewidth}
	\caption{\sl \label{vtxspa:rlmfunc}
The relation between the cluster position $x$ 
and the charge ratio $R_x$, 
where $x$ is the position of the cluster normalized by pixel pitch.
(a)RLM function, where symbols represent measured data and
lines are obtained RLM function.
(b)Linear function, where 
boxes region shows the allowed region of the AC-method
which uses 2$\times$2 pixels.
Solid line represents a linear relation $x=0.5R_x$.
	}
\end{minipage}\end{center}
\end{figure}

The position of the cluster was then calculated from
the measured charge ratio by using the RLM function.
The RLM function represents a spread of charges produced
by incident particle. For example, if a particle
hits the position $x$=0.3, the charge sharing
to the adjacent pixel is about 8\%, 13\%, and 25\% of the leading
pixel, for EEV, HPK10, and HPK50, respectively. HPK50 has more spread
signal than HPK10, since the epitaxial depth of HPK50 is 
larger than that of HPK10. The charge sharing of EEV is smallest
among all samples. This reason is considered  that the EEV has
different structure from HPK type of CCDs.

In the AC-method, 
the linear relation between the position of 
the cluster and the charge ratio of the four pixels is assumed.
However,
the relation between $R_{x}$ and $x$ is not linear as
shown in Fig.~\ref{vtxspa:rlmfunc}.
Therefore, we can expect to obtain  better position resolution
by using the RLM-method.

\end{itemize}

\subsubsection{Measured Spatial Resolution}

\begin{figure}[hbth]
	\centerline{\epsfxsize=15.0cm \epsfbox{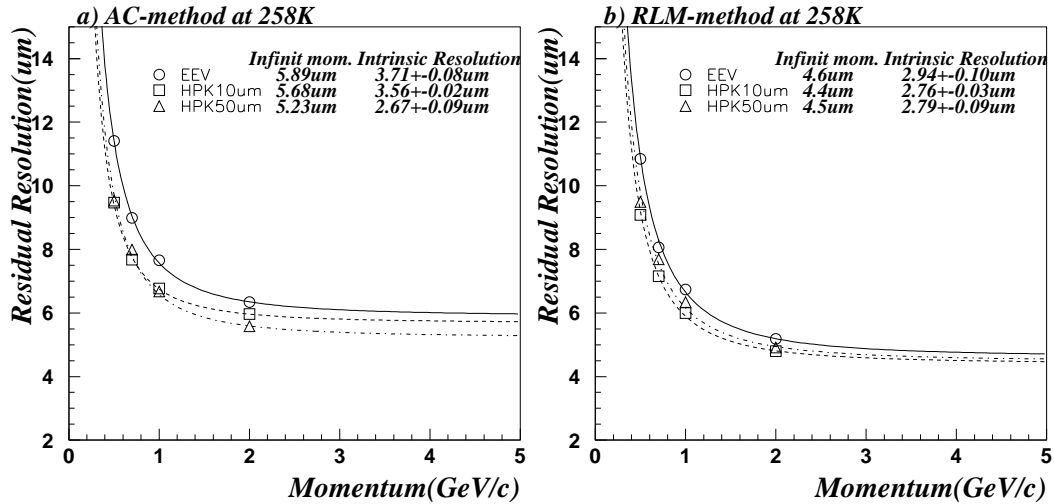}}
\begin{center}\begin{minipage}{\figurewidth}
\caption{\sl 	\label{vtxspa:momdep}
The momentum dependence of the residual resolutions for 
a)AC-method and b)RLM-method.
The lines  are results of fit to data.
}
\end{minipage}\end{center}
     \end{figure}

The spatial resolution was derived from the residual distributions,
which is the deviation of the cluster
in the test sample from the extrapolated hit 
position from the upstream reference CCDs.
The residual resolution ($\sigma_{\mbox{\rm residual}}$) was obtained by
fitting this distribution with the Gaussian distribution.
The $\sigma_{\mbox{\rm residual}}$ contains the contributions 
from multiple Coulomb scattering (MCS) and the finite
resolution of the reference CCDs. 
The MCS effect is suppressed
by the factor $(\beta c p)^{-1}$, where $\beta c$ and $p$ is a velocity 
and momentum of incoming particle. 
We derived the residual resolution at infinite momentum by fitting 
the momentum dependence of residual resolutions with the function
$\sigma_{\mbox{\rm residual}}=\sqrt{A^2+(B/\beta c p)^2}$,
where $A$ and $B$ are fitting parameters, and $A$ represents
residual resolution at infinite momentum.
Fig.~\ref{vtxspa:momdep} shows the momentum dependence of 
residual resolution  with fit result.

Then the contribution from the ambiguity of 
the extrapolated hit position due to the finite resolution of the
reference CCDs are estimated by taking the propagation of errors, 
and subtracted from the data. Here, we assumed spatial resolutions
of reference CCDs were same as HPK10, because the 
reference CCDs were same type with HPK10.

The obtained spatial resolutions for normal incidence
are listed in the Table~\ref{vtxspa:intrinsic},
for the AC-method and the RLM-method, respectively.
At $-$15$^\circ$C, the spatial resolutions for HPK10 and EEV
were improved about 20\% by taking  the nonlinear
charge sharing characteristics by the RLM-method into account. 
When the temperature
was raised from $-$15$^\circ$C to $+$5$^\circ$C,
the spatial resolutions obtained by using the RLM-method were changed
its values according to the variation of the S/N value,
i.e. about 20\% worse for HPK10, and almost unchanged
for EEV. On the other hand, the spatial resolutions
derived from the AC-method were almost same, even if
the temperature was changed from $-$15$^\circ$C to $+$5$^\circ$C.
Since the charge sharing is small in adjacent pixels for HPK10
and EEV, the resolution of $R_x$ is deteriorated by the noise.
This is the reason why the RLM-method is more sensitive to the
S/N value than the AC-method.
This conjecture explains the temperature dependence
for HPK10 and EEV as well.
For HPK50, the spatial resolutions obtained by the RLM-method
and the AC-method were almost same value at both  $-$15$^\circ$C
and $+$5$^\circ$C, because 
the RLM function is similar to
the function in the AC-method as shown in Fig.~\ref{vtxspa:rlmfunc}(b).
The spatial resolution becomes worsened about 30\%
if the temperature was raised from $-$15$^\circ$C to $+$5$^\circ$C.
This is caused by the 50\% degradation of S/N ratio.

\begin{table}[hbth]
\begin{center}
	\caption{\sl 	\label{vtxspa:intrinsic}
Spatial resolution in $\mu$m.}
\vspace*{6pt}
	\begin{tabular}{lllll} \hline
Temperature & Method	    & HPK10 & HPK50 & EEV \\ \hline 
$-15^\circ$C & AC   & 3.56$\pm$0.02 & 2.67$\pm$0.09 & 3.71$\pm$0.08  \\
            & RLM  & 2.76$\pm$0.03 & 2.79$\pm$0.09 & 2.94$\pm$0.10  \\ 
$+5^\circ$C  & AC   & 3.68$\pm$0.03 & 3.34$\pm$0.10 & 3.84$\pm$0.08 \\ 
	  & RLM   & 3.46$\pm$0.04 & 3.67$\pm$0.12 & 2.59$\pm$0.16  \\ 
\hline
	\end{tabular}
\end{center}
     \end{table}

\subsubsection{Efficiency}

Efficiency was measured by looking for the existence of associated
cluster in the interpolated region on
the test sample, for the tracks which have a cluster
in three reference CCDs.
We require that the reference CCD at the end of tracker must have a associated
cluster within 5 times of alignment errors from extrapolated hit position,
for the reduction of the accidental tracks.
The threshold for a leading pixel were applied as 13, 7, and 15 times of
the noise level, for EEV, HPK10, and HPK50, respectively.   
Efficiencies at $-15^\circ$C for 2 GeV pion were
98.8$_{-1.5}^{+1.2}$\%, 99.4$_{-1.6}^{+0.6}$\%,
and 97.8$_{-1.6}^{+2.2}$\%, for HPK10, HPK50, and EEV, respectively.

\subsubsection{Discussion}

\begin{figure}[hbp]
	\centerline{\epsfxsize=9.0cm \epsfbox{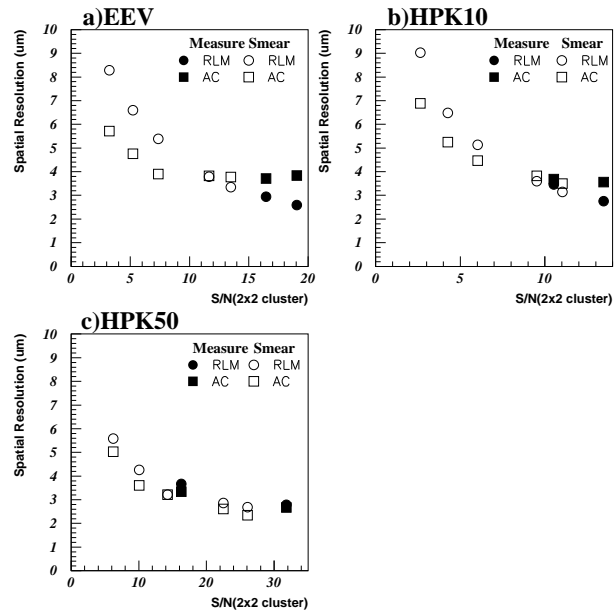}}
\begin{center}\begin{minipage}{\figurewidth}
	\caption{\sl	\label{vtxspa:sndep}
Signal to noise ratio dependence of spatial resolution for
(a)EEV, (b)HPK10, and (c)HPK50, respectively. Black symbols represent
the data measured at -15C$^\circ$ and +5C$^\circ$. Open symbols 
represent the simulated data by smearing the data measured at -15C$^\circ$.
	}
\end{minipage}\end{center}
     \end{figure}

\begin{figure}[htbp]
	\centerline{\epsfxsize=10.5cm \epsfbox{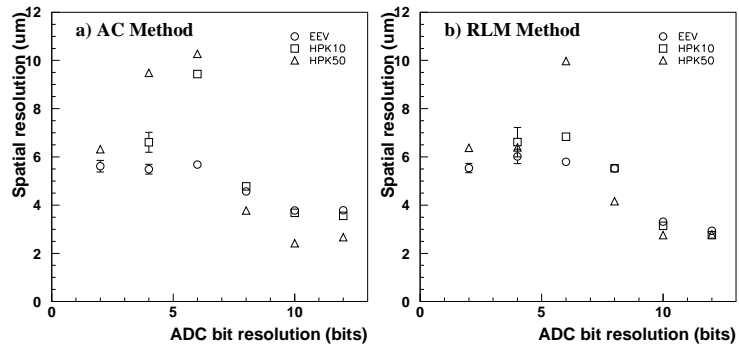}}
\begin{center}\begin{minipage}{\figurewidth}
	\caption{\sl	\label{vtxspa:bitdep}
Digitization accuracy dependence of spatial resolution for
(a)The RLM-method and (b)The AC-method.
	}
\end{minipage}\end{center}
\end{figure}

\begin{figure}[hp]
	\centerline{\epsfxsize=10.0cm \epsfbox{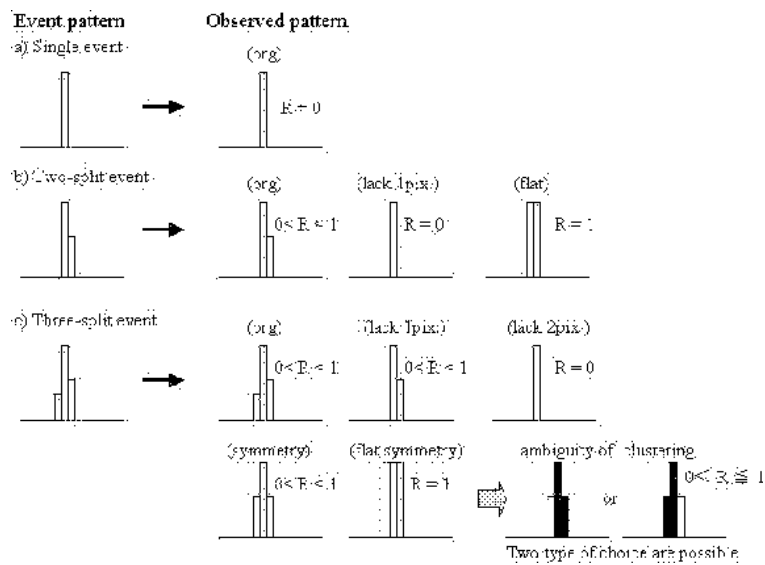}}
\begin{center}\begin{minipage}{\figurewidth}
	\caption{\sl 	\label{vtxspa:adcpat}
Considerable event pattern and its observed pattern.
	}
\end{minipage}\end{center}
\end{figure}

\begin{figure}[hp]
	\centerline{\epsfxsize=10.0cm \epsfbox{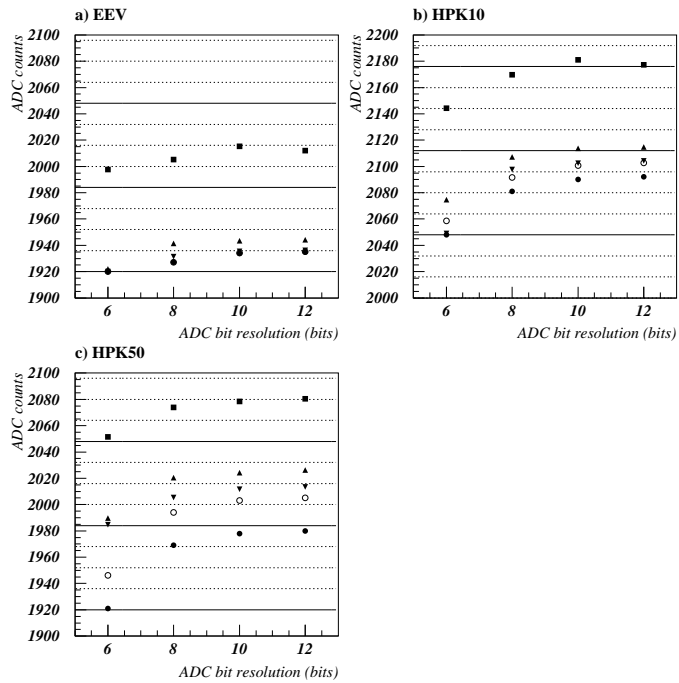}}
\begin{center}\begin{minipage}{\figurewidth}
\caption{\sl 	\label{vtxspa:adcmean}
The mean value of output signals as a function of ADC bit
resolution, for the horizontal over clock(HOC)
(black circle)
which was used for the offset of the output,
dark charge obtained with 12 bit accuracy (open circle),
maximum signal in the cluster (black square), 
the signal used for the calculation of R (black triangle),
and the signal of the pixel in the opposite direction
of R (reversed black triangle).
The solid and dashed lines represent 6 bit and 8 bit
ADC accuracy, respectively.
For the EEV, dark charge is very small, i.e 0.14 adc counts,
the output of dark charge is almost same with 
HOC.
	}
\end{minipage}\end{center}
\end{figure}

\begin{figure}[hp]
	\centerline{\epsfxsize=10.0cm \epsfbox{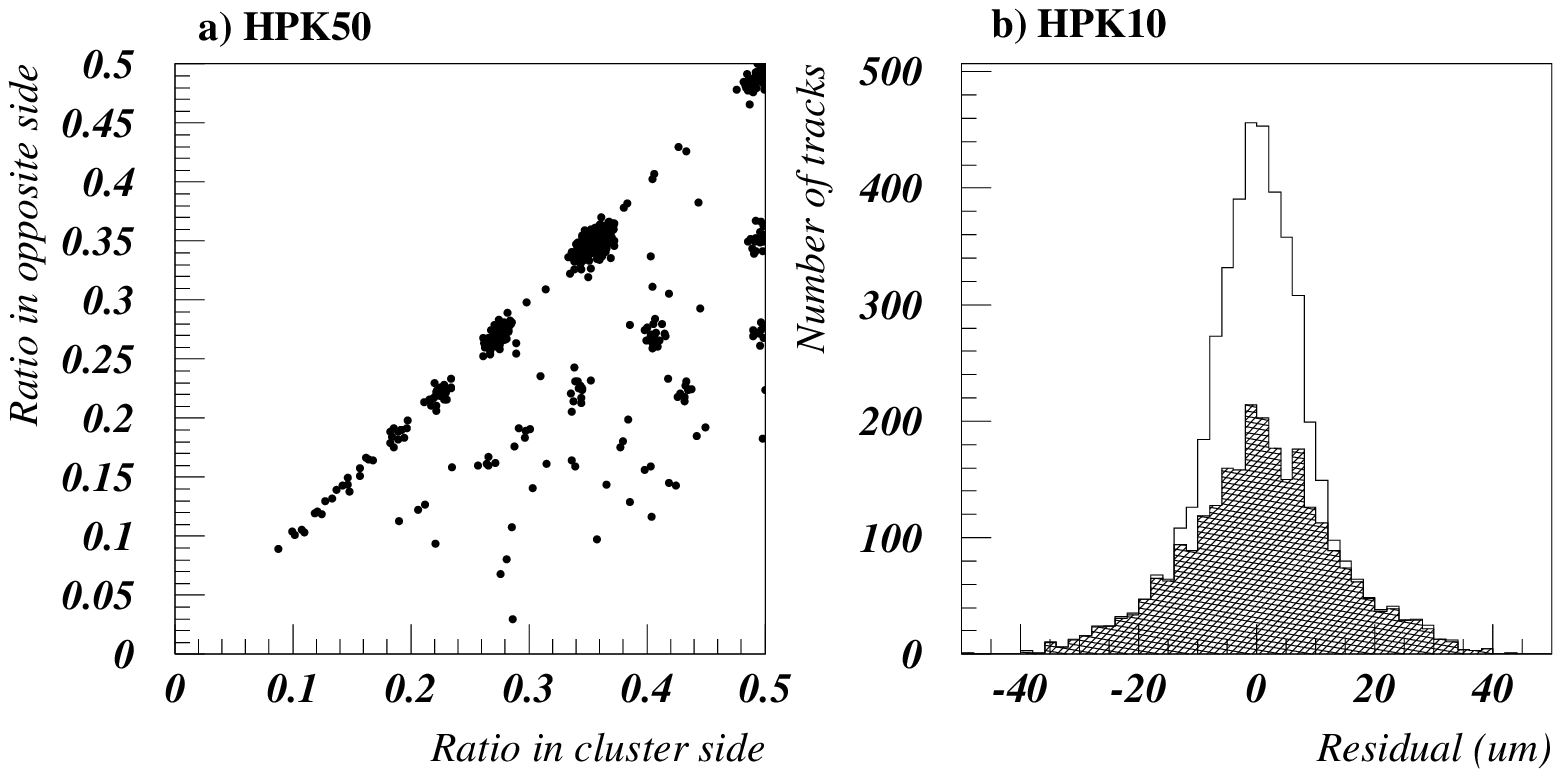}}
\begin{center}\begin{minipage}{\figurewidth}
	\caption{\sl 	\label{vtxspa:acreso}
a) Charge ratio of the  cluster side and opposite
	of the cluster side  for HPK50 at 6 bit ADC resolution.
	 b) The residual distribution for HPK10 at 6 bit ADC resolution,
	where the open area represents  all tracks, 
	and hatched area represents
	the tracks with minus value of the charge ratio.
	}
\end{minipage}\end{center}
\end{figure}

\begin{itemize}
\item{Signal-to-Noise dependence}

The result from the test beam shows the degradation of the spatial
resolution due to the increase of noise.
We have estimated the influence of noise on the
spatial resolution.
The signals from CCDs are smeared with pseudo-noise that
was generated according to the deviates drawn from a Gaussian distribution. 
Fig.~\ref{vtxspa:sndep} shows 
the dependence of the resolution on the 2$\times$2 clustered
signal-to-noise ratio. 
The RLM-method is sensitive to the S/N value, but it can 
improve resolution if the S/N is good enough.
On the other hand, the AC-method is less sensitive to the S/N value than
the RLM-method. In the case of EEV, the AC-method shows that the spatial
resolution
is almost unchanged even if the S/N is decreased from 19 to 7.
The RLM-method requires the S/N about more than 10 for EEV and HPK10,
to obtain better spatial resolution than the AC-method.

\item{Digitization resolution dependence}

An analog readout is a preferred solution for several 
reasons, including
more effective monitoring of the stability of the detector properties
and improvement in the spatial resolution. We need to know how many
bits are required for quick extraction of the digitized information
and reduction of the data size.
We have investigated the spatial resolution by
varying the digitization accuracy. 
Since we have used a 12-bit linear ADC for this experiment,
we varied the digitization accuracy by using only
higher bit information, while the dynamic range was unchanged.

Fig.~\ref{vtxspa:bitdep} shows the digitization accuracy dependence of 
spatial resolutions, for the RLM-method and the AC-method, respectively.
The spatial resolutions are almost unchanged even if the digitization
accuracy is reduced to 10 bits. If the accuracy was reduced
less than 10 bits, the spatial resolution become worse.
Fig.~\ref{vtxspa:adcpat} shows the considerable event pattern and its
observed pattern 
when the ADC accuracy is reduced.
The worst spatial resolution at 6 bits is explained by the
``symmetry'' observed pattern that is described in Fig.~\ref{vtxspa:adcpat}(c).
In this case, there is the ambiguity of choice of clustering.
The cluster can be one of two types of clustering, and loose 
direction information. 
Fig.~\ref{vtxspa:adcmean} shows the mean value of output signals in units of
ADC counts. The points shown in the figure are output level of  
horizontal over clock(HOC),
dark charge which was derived with 12 bit resolution,
the leading pixel in the cluster($P_{max}$), 
the pixel which is the adjacent pixel to x-direction
inside the cluster($P_{x}$), and the pixel which is in the opposite 
side of the $P_{x}$($P_{opp}$).
We used average of HOC output as a baseline for calculation of the
signals so that the HOC also affect to the other signal levels.

As shown in the figure, $P_{x}$ and $P_{opp}$ is almost same as
dark charges for EEV and HPK10, and is disappeared as reduction of
ADC resolution. Then the event type become single event.

Both of $P_{x}$ and $P_{opp}$ are separated even at 
the 8 bit resolution for HPK50.
However, at 6 bit resolution, both signals are recognize as 
identical , while those are separated from dark charge.
This is the case of symmetry event type and loose the direction
information.

 In the case of AC-method, the signal used for position
determination is defined by sum of two pixels. 
 The reason for distortion of the spatial 
resolution at 6 and 4 bit ADC accuracy for HPK50 is same as 
that of RLM-method. But the reason is quite different for
the HPK10. The distortion of the spatial resolution at 6 bit,
is due to the minus signal at the adjacent pixel.
The calculated position become opposite to the correct position
if the adjacent signals were minus values. While the calculated position
constrains to the center of the maximum pixel in RLM-method.
Fig.~\ref{vtxspa:acreso}a) shows the correlation of the ratio
for cluster side and for the opposite side of the cluster for
HPK50, and Fig.~\ref{vtxspa:acreso}b) shows the residual resolution
for HPK10. 

\end{itemize}

\subsubsection{Test with Laser Scanner System}

 The stability and excellent overall response of 
the detector are well known.
Less widely described is the fact that the internal structure of the
pixel gives rise to intra-pixel variations in response.
These response modulations are particularly relevant 
to obtain the best spatial resolution.

We have made detail measurements of the response variations within a CCD, 
using a laser scanner system that gave a 2 micron resolution.
Optical response of the CCD has been determined 
at all sample points within a pixel
at wavelengths of 523 nm and 1064 nm whose absorption length is
about 3 $\mu$m and 300 $\mu$m, respectively.
Since the thickness of the epitaxial layer of CCDs 
is typically 10$\sim$50 $\mu$m,
the pulsed laser of 1064 nm wavelength induces electron-hole  
pairs along the passage of the light nearly uniformly in the epitaxial
layer of CCDs. The Laser light of 1064 nm is suitable for simulating
the signal generated by charged particles in a CCD.

\begin{figure}[htbp]
	\centerline{\epsfxsize=5.0cm \epsfbox{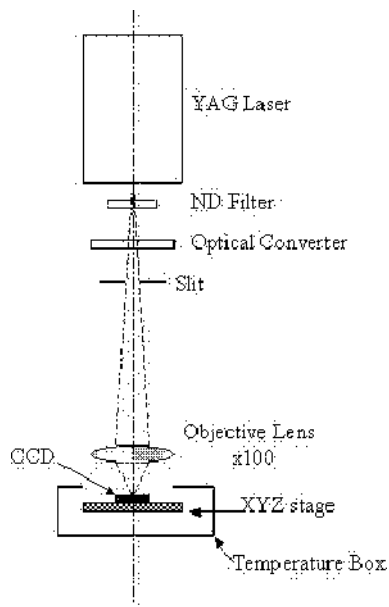}}
	\caption{\sl 	\label{vtxspa:probersys}
The schematic view of the laser scanner system.
	}
\end{figure}

\begin{figure}
	\centerline{\epsfxsize=9.0cm \epsfbox{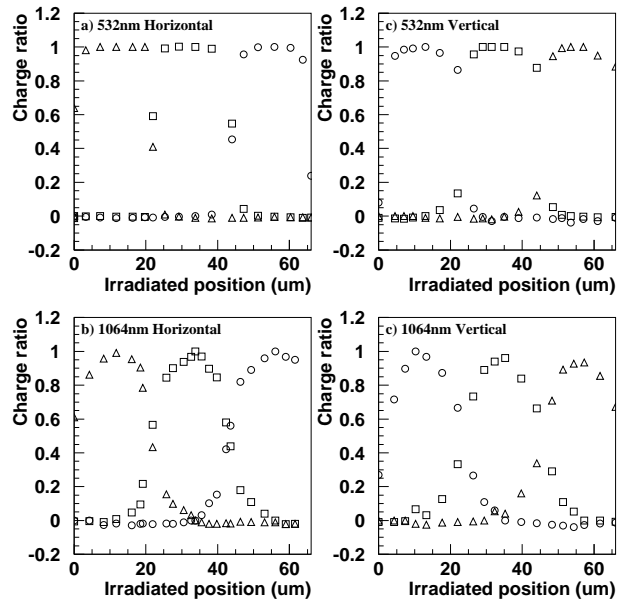}}
\begin{center}\begin{minipage}{\figurewidth}
	\caption{\sl \label{vtxspa:laserdep}
The charge sharing ratio for EEV. (a) and (b) show the
charge ratio in the horizontal scanning with the wavelength of
532 nm and 1064 nm, respectively. (c) and (d) show same figures
in the vertical scanning. 
The three sets of symbols are observed charge ratio on three
adjacent pixels. 
	}
\end{minipage}\end{center}
\end{figure}

Fig.~\ref{vtxspa:probersys} shows the schematic view of the 
experimental arrangement.
The system consists of YAG laser, optics, xyz moving stage,
and temperature box.
A CCD is set on the xyz moving stage that can move with an 
accuracy of about 1$\mu$m. 
The CCD and xyz moving stage are in the temperature
box, and kept at 0$^\circ$C.
The laser power is adjusted to simulate a pulse height of
minimum ionizing particles by using ND filter. 
 The spot size of laser light is about
2$\mu$m$\times$2$\mu$m on the surface of a CCD.
Fig.~\ref{vtxspa:laserdep} shows the result of the measurement fro EEV.
The charge ratio was calculated as signal charge in a pixel
normalized by total charge of the cluster, where cluster 
was defined by adjacent three pixels.
The result of 532 nm shows rather narrow charge sharing region, 
because of the short absorption length which makes 
the effect of the lateral
charge diffusion small.
We observed a broad charge sharing region with a triangle shape
for 1064 nm. Also there is some different property in horizontal and vertical
scanning. 
These results are used to investigate a charge sharing property
and obtain a precise RLM-function. 
Also, this laser scanner system will be used for the calibration of the CCDs.

\subsubsection{Summary}

We have studied the spatial resolution of a CCD. 
 Three types of
CCD detectors which have different 
structures were tested.

The present series of test showed 
that: 

\begin{description}
\item[1)] S/N ratio of more than 10 can be achieved even at
$+5^\circ$C and at accumulation time of 3 s including readout time,
by successful operation in the MPP mode. 
In the JLC project, 
the readout cycle is much faster than this experiment.
We expect the S/N become better, because
noise due to the dark current is largely suppressed.
\item[2)] The spatial resolutions were improved for HPK10 and
EEV by taking the nonlinear charge sharing characteristics
into account by the RLM-method.
\item[3)] The temperature dependence of the spatial resolution 
in the AC-method 
was small for HPK10 and EEV samples, while spatial resolution
become worse about 30\% if the temperature was changed 
from $-15^\circ$C to $+5^\circ$C for HPK50.
\item[4)] The RLM-method needs S/N more than 10 for EEV and HPK10 to keep
the spatial resolution better than the AC-method. 
\item[5)] More than 10 bits digitization accuracy is needed to keep
the spatial resolution better.
\item[6)] The precise measurement of the charge sharing property
was studied using laser scanner system. 
This system will make it easier to measure CCD property. 
\end{description}

%% file: dettrk/vtx/rad/main.tex
\newcommand{\dirrad}{dettrk/vtx/rad/}
\subsection{Radiation Hardness of CCD Sensors}
\label{vtxrad}
The backgrounds at the future LC pose significant challenges to the radiation
hardness of CCD sensors. The main component of the background comes from
$e^{+}e^{-}$ pairs with estimated 10-year fluence of $\approx1.5\times10^{12}$
cm$^{-2}$. Neutron background from ($\gamma$, $n$) reactions is also present,
reaching $\approx5\times10^{9}$ cm$^{-2}$ for the same period. The electron
background is expected to affect mainly the CCDs in the innermost layer,
because it is closest to the beam pipe. The neutron irradiation is almost
uniformly distributed in the volume of the vertex detector and affects all CCD
chips.

Study of commercially available CCDs is required to build a model for the
radiation damage effects, which can be used for estimation of the
device lifetime in the expected environment.  Radiation hardness studies of
CCDs can provide valuable information on the preferable device
architecture, chip geometry and operating conditions, 
which ultimately determine
the vertex detector design, performance and cost. Much of this study was
concentrated on the device application at {\bf near-room} 
temperature because of
the advantages of this operation. Operation at low temperatures is usually
required to suppress the radiation damage effects 
in CCDs \cite{radam:Abe97}. 
By avoiding cooling to cryogenic temperatures it is possible to decrease the
thermal distortions of the supporting CCD ladders and to simplify their design,
which results in better detector geometry and increased measurement precision.
Although attractive from designer's point of view, operation at elevated
temperatures faces problems from radiation damage effects, which are addressed
in this study. \\

Both surface and bulk damage effects are expected to take place in the CCD
sensors because of the type and the energy 
spectrum of the radiation background.
Ionizing radiation creates electron-hole pairs in silicon dioxide, which is
used as gate and field dielectric in CCDs. 
Charge carriers drift in the electric
field (externally applied or built-in) to the 
corresponding electrode. Electrons
quickly reach the positive electrode, 
but some of the holes remain trapped in
the oxide and give rise to radiation-induced trapped positive oxide
charge, which can be stable for long time. 
At any Si-SiO$_{2}$ interface there
are a number of interface traps, which result from the strained or dangling
silicon bonds at the boundary between the two materials. Ionizing radiation
causes the density of these traps to increase, generating radiation-induced
interface traps. 
The formation of radiation-induced trapped oxide charge
(or flat band voltage shift) and interface traps 
is referred under the term {\bf
surface damage}.

Radiation with sufficiently high energy can displace  Si atoms from their
lattice positions, creating displacement damage. 
This process affects
the properties of the bulk semiconductor and is known as {\bf bulk
damage}. 
Low energy electrons and X-rays can deliver only small 
energy to the recoil Si atom
and mainly isolated displacements, or point defects, can be created. 
On the
other hand, heavier particles, such as protons and neutrons can knock out
silicon atoms which have sufficient energy to displace other atoms in the
crystal. 
The secondary displacements form defect clusters, which have high
local defect density and can be tens of nanometers wide. 
Defect clusters often
have complicated behavior and more damaging effect on the properties of
semiconductor devices than point defects. 
The energy threshold for displacement
of a Si atom has been estimated to be about 260~keV for electrons and 190~eV
for neutrons, therefore in the expected radiation environment bulk defects can
be generated by both particles. 
Cluster damage is also expected, because the
threshold for cluster production is known to be $\approx$5~MeV and
$\approx$15~keV for electrons and neutrons \cite{radam:Wunstorf97},
respectively. \\

Trapped holes in the oxides change the parameters of MOS
structures in a way identical to applying an external voltage to the gates.
Small flat band voltage shifts, in the order of few volts, can be accommodated
by adjustment of the amplitude of the gate bias and drive voltages. 
A limitation
may arise from the maximum allowed power dissipation in the
gate drivers and in the CCD chip, because the dissipated power is proportional
to the voltage amplitude squared. 
If the flat band voltage shifts are higher,
the device can stop to function properly because of parasitic charge injection
from the input structures, incomplete reset of the output node or distortion of
the shape of the potential wells \cite{radam:Roy89}. 
However, as long as no
parasitic effects from the increased pulse amplitudes appear, 
the shifts are not
a limitation for the device operation.

Interface traps are the dominant source of dark current in modern CCDs, 
because the generation rate at the Si-SiO$_{2}$ interface is 
usually higher than that in
the epitaxial bulk silicon. Increase of the surface dark current is the main
effect expected from radiation-induced interface defects.

Radiation-induced bulk defects cause the dark current 
in CCDs to increase, which
is well known phenomenon \cite{radam:Janesick89}~\cite{radam:Hopkinson92}.
Dark current is an issue only for near-room temperature operation or long
integration times, because it can be reduced to negligible values by cooling.
Irradiation with heavy particles (e.g. protons, neutrons) often creates large
non-uniformities in the dark current spatial distribution in the CCDs
\cite{radam:Hopkinson96}. 
These non-uniformities, also known as ``dark current
spikes'' or ``hot pixels'' manifest themselves as pixels with much higher dark
current than the average value for the CCD. 
Their presence has been connected
with the high electric fields caused by the device architecture and
field-enhanced emission, the cluster nature of the radiation damage and crystal
strains in the silicon material. 
Dark current spikes have big consequences for
high temperature applications. Additionally, some of them show random
fluctuations of the generated current, or Random Telegraph Signals (RTS)
\cite{radam:Hopkins93}. \\

Another important bulk damage effect is the loss of signal charge during
transfer, or Charge-Transfer Inefficiency (CTI). 
Charge losses occur when bulk
defects capture electrons and emit them at a later moment, 
so that the released
charge cannot join the original signal packet from which it has been trapped.
The basic mechanism can be explained
by the different time constants and temperature dependencies of the electron
capture and emission processes.
In the presence of bulk defects electrons are trapped with a capture time
constant of $\tau_{c}$ and consequently released with an emission time constant
of $\tau_{e}$.  
For a defect at energy position E$_{t}$ below the conduction
band the Shockley-Read-Hall theory gives
\begin{equation}
\tau_{e} = \frac{1}{\sigma_{n}X_{n}v_{th}N_{c}}\mbox{exp}\
\left(\frac{E_{t}}{kT}\right) \label{radam:taue}
\end{equation} and
\begin{equation}
\tau_{c} = \frac{1}{\sigma_{n}v_{th}n_{s}}, \label{radam:tauc}
\end{equation}
where
\begin{center}
\begin{tabular}{l c l}
$\sigma_{n}$ & = & electron capture cross section, \\
$X_{n}$ & = & entropy change factor by electron emission, \\
$v_{th}$ & = & thermal velocity for electrons, \\
$N_{c}$ & = & density of states in the conduction band, \\
$k$ & = & Boltzmann's constant, \\
$T$ & = &  absolute temperature, \\
$n_{s}$ & = & density of signal charge.
\end{tabular}
\end{center}

The capture time constant is typically of the order of several hundred
nanoseconds and has weak temperature dependence, whereas the emission time
constant changes many orders of magnitude because of the exponential
temperature behavior. 
At low temperature the defects can be considered almost
permanently occupied with electrons, because the emission time constant of the
defects can be very large, of the order of seconds. 
The defects cannot capture
signal electrons and the CTI at low temperature is small. 
At high temperature
the emission time constant becomes small and comparable with the charge shift
time. 
Trapped electrons are able to join their signal packet, because most of
them are emitted already during the charge shift time, and charge losses are
small. 
At temperatures between the two extremes these is a peak of the CTI
value.

\begin{figure}[htb]
\centerline{\epsfxsize=5.0cm \epsfbox{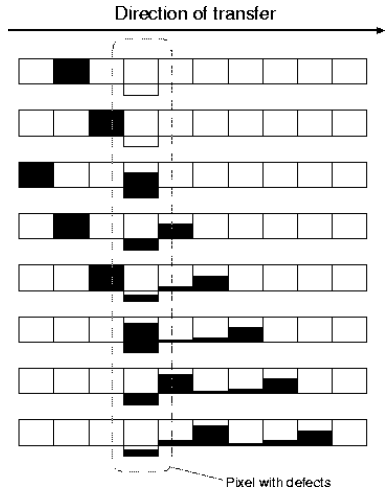}}
\begin{center}\begin{minipage}{\figurewidth}
\caption{\sl \label{fig:radam:ctidemo}
Schematic representation of the mechanism of charge transfer
losses. For simplicity it has been assumed, that only one pixel contains
defects.}
\end{minipage}\end{center}
\end{figure}

The mechanism of charge transfer  losses is illustrated on Fig.
\ref{fig:radam:ctidemo}. 
When the signal packet encounters traps, some of its
electrons are captured and later released. 
Those electrons, which are released
in the trailing pixels do not join their original 
signal packet and account for
the CTI. 
If a charge packet enters a pixel, in which part of the traps are
occupied, less signal electrons can be 
trapped and therefore less charge can be
lost.

We are often interested not only in the CTI value, but in the total
losses the charge suffers after all the transfers 
it takes to reach the output.
For a generated charge $C_{gen}$, transferred $n$ times, the output charge
$C_{out}$ is given by
\begin{equation}
C_{out} = C_{gen}(1 - CTI)^{n}. \label{radam:CTIdef}
\end{equation}
This dependence shows, that CTI and the number of transfers are the two
important parameters that need to be considered for minimizing the charge
transfer losses. \\

\subsubsection{Dark Current and Flat-Band Voltage Shift}
For near-room temperature applications one is particularly concerned with the
dark current of the CCD sensors. Multi Pinned Phase (MPP) mode CCDs
\cite{radam:Saks80a} were used in our experiments because they feature much
lower dark current than standard devices. In MPP CCDs the Si-SiO$_{2}$
interface is biased into inversion and populated with holes, leading
to suppression of the carrier generation from interface defects. With the
surface inverted, only bulk defects contribute to the dark current. Since in
contemporary buried channel CCDs current generation from defects at the
Si-SiO$_{2}$ interface is dominant, MPP mode operation can reduce the device
dark current by more than one order of magnitude.  This method has the ability
to suppress the dark current from the interface states formed during irradiation
and can offer increased radiation hardness \cite{radam:Saks80b}.

The radiation hardness of two types of buried channel, MPP mode CCDs was
studied. The S5466 CCD, manufactured by Hamamatsu Photonics is 2-phase device
with 10$\mu$m epitaxial layer and 512$\times$512 active pixels with size
24$\mu$m$\times$24$\mu$m. Three other types of S5466-based CCDs were
studied, including high speed device (10~Mpix/s), notch CCD and device with
SiO$_{2}$-Si$_{3}$N$_{4}$-SiO$_{2}$ gate (ONO) dielectric. A three-phase CCD,
manufactured by EEV Ltd. was investigated as well. The device CCD02-06 has
20$\mu$m epitaxial layer and 385$\times$578 active pixels with size
22$\mu$m$\times$22$\mu$m. \\

Several devices were exposed to electron irradiation from a $^{90}$Sr source
and 2 CCDs were irradiated by neutrons from a $^{252}$Cf source. The change of
the voltage needed to achieve MPP mode operation was used as a measure of
the flat band voltage shifts. The dark current of the devices was  measured in
the temperature range from $-100^{\circ}$C to $+20^{\circ}$C by averaging the
dark charge over several frames. The CTI of the  horizontal (serial) and the
vertical (parallel) registers of the CCD was measured in the same temperature
range by observing the charge and the position of isolated pixel events (IPE),
created by the 5.9~keV Mn-K$_{\alpha}$ line of a $^{55}$Fe source. After
sufficient averaging, the dark signal of each pixel was stored into memory and
this dark frame was then subtracted pixel by pixel during the X-ray exposure.
The readout speed was usually 250~kpixels/s. For the high-speed Hamamatsu
CCD the speed was raised to 2~Mpixels/s. \\

\begin{figure}[tb]
\begin{minipage}[l]{0.48\linewidth}
\centerline{\epsfxsize=8.0cm \epsfbox{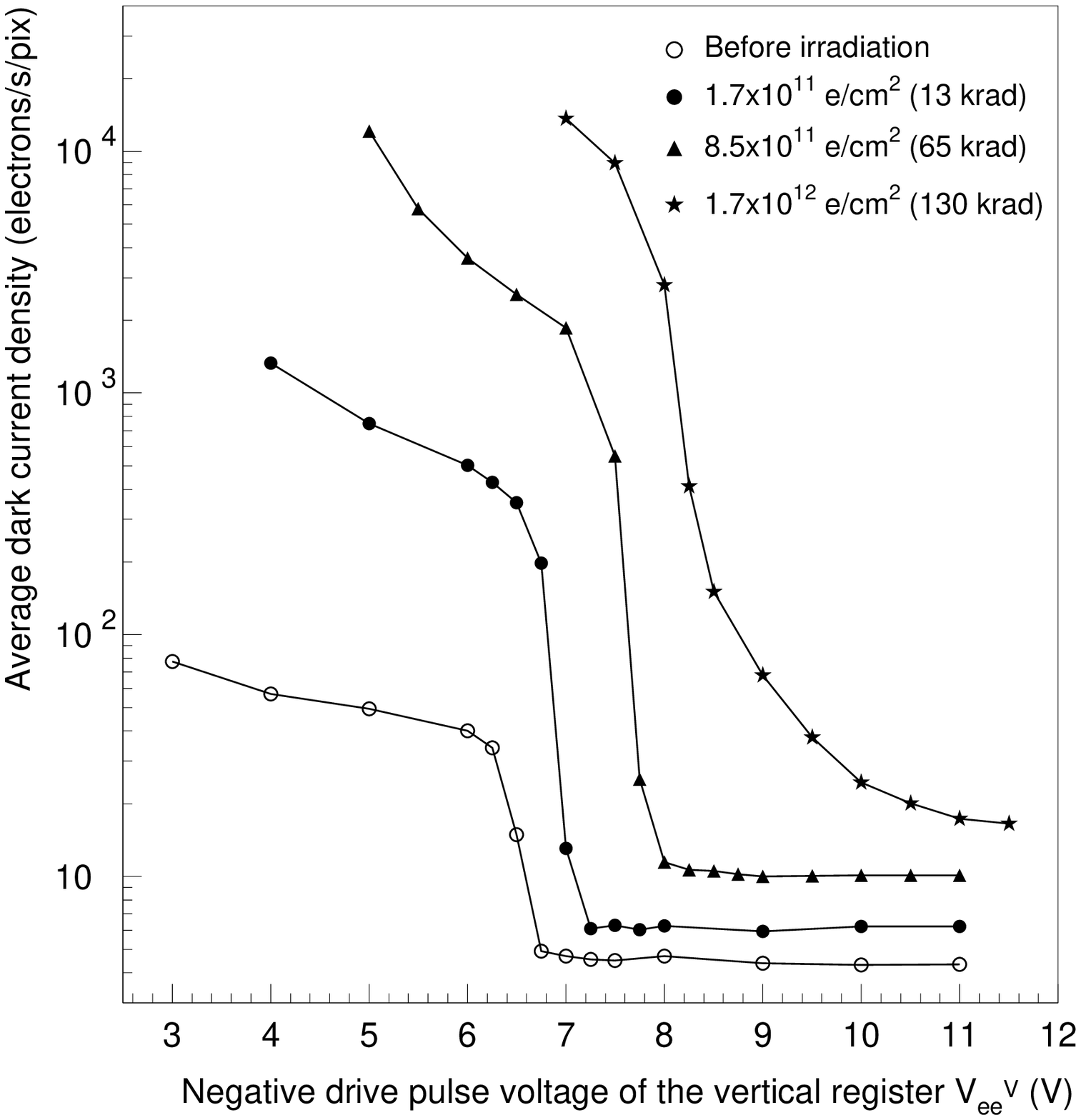}}
\caption{\sl\label{fig:radam:vmppnobias}
Average dark current density as a function of the electron
irradiation fluence and of the negative drive pulse voltage
V$_{ee}^{V}$ at T = $-26^{\circ}$C.  The device \#JS14/026 is irradiated
unbiased.}
\end{minipage}
\hfill
\begin{minipage}[r]{0.48\linewidth}
\centerline{\epsfxsize=8.0cm \epsfbox{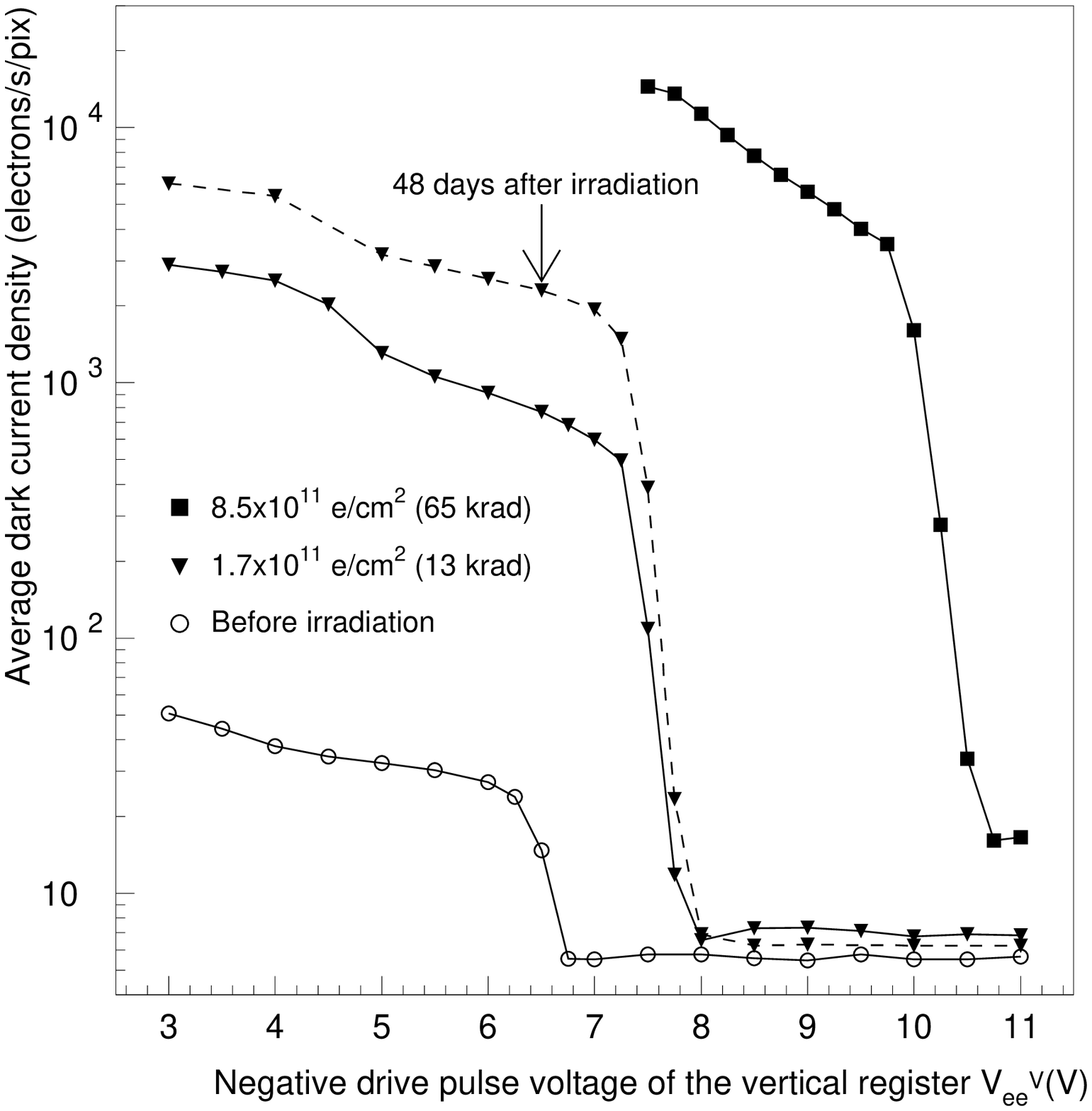}}
\caption{\sl\label{fig:radam:vmppbias}
Average dark current density as a function of the electron
irradiation fluence and of the negative drive pulse voltage
V$_{ee}^{V}$ at T = $-26^{\circ}$C.  The device \#JS8/053 was biased
and clocked during irradiation.}
  \end{minipage}
\end{figure}

The change of the MPP threshold voltage in electron-irradiated S5466 CCD,
irradiated unbiased, is shown in Fig. \ref{fig:radam:vmppnobias}. Due to
the SiO$_{2}$-Si$_{3}$N$_{4}$ gate dielectric in the EEV devices, the change of
the MPP threshold is slightly smaller than than in Hamamatsu CCDs.
In biased and clocked S5466 CCD higher shift was observed
(Fig.\ref{fig:radam:vmppbias}), which is explained by the lower ratio of
recombining holes in biased oxides. The flat band voltage shifts were
estimated to be 0.02 V/krad for unbiased devices and 0.07 V/krad for biased
Hamamatsu CCDs. The flat band voltage shift of electron-irradiated Hamamatsu ONO
CCD was also measured, however it was higher that in standard S5466 devices, on
the contrary to the expectations. In neutron-irradiated devices such shifts were
not expected and were not observed.

The dark current in MPP mode in both electron and neutron irradiated devices
increased after irradiation, which is attributed to creation of bulk
traps. The CCD demonstrates effective suppression of the surface component of
the dark current by almost 3 orders of magnitude (Figs.
\ref{fig:radam:vmppnobias} and \ref{fig:radam:vmppbias}) when
$\mid$V$_{ee}^{V}\mid > \mid$V$_{ee}^{MPP}\mid$ and the interface is
populated with holes. In that condition the dark current is limited by carrier
generation in the bulk. \\

\begin{figure}[htb]
  \begin{minipage}[l]{0.48\linewidth}
     \centerline{\epsfxsize=8.0cm \epsfbox{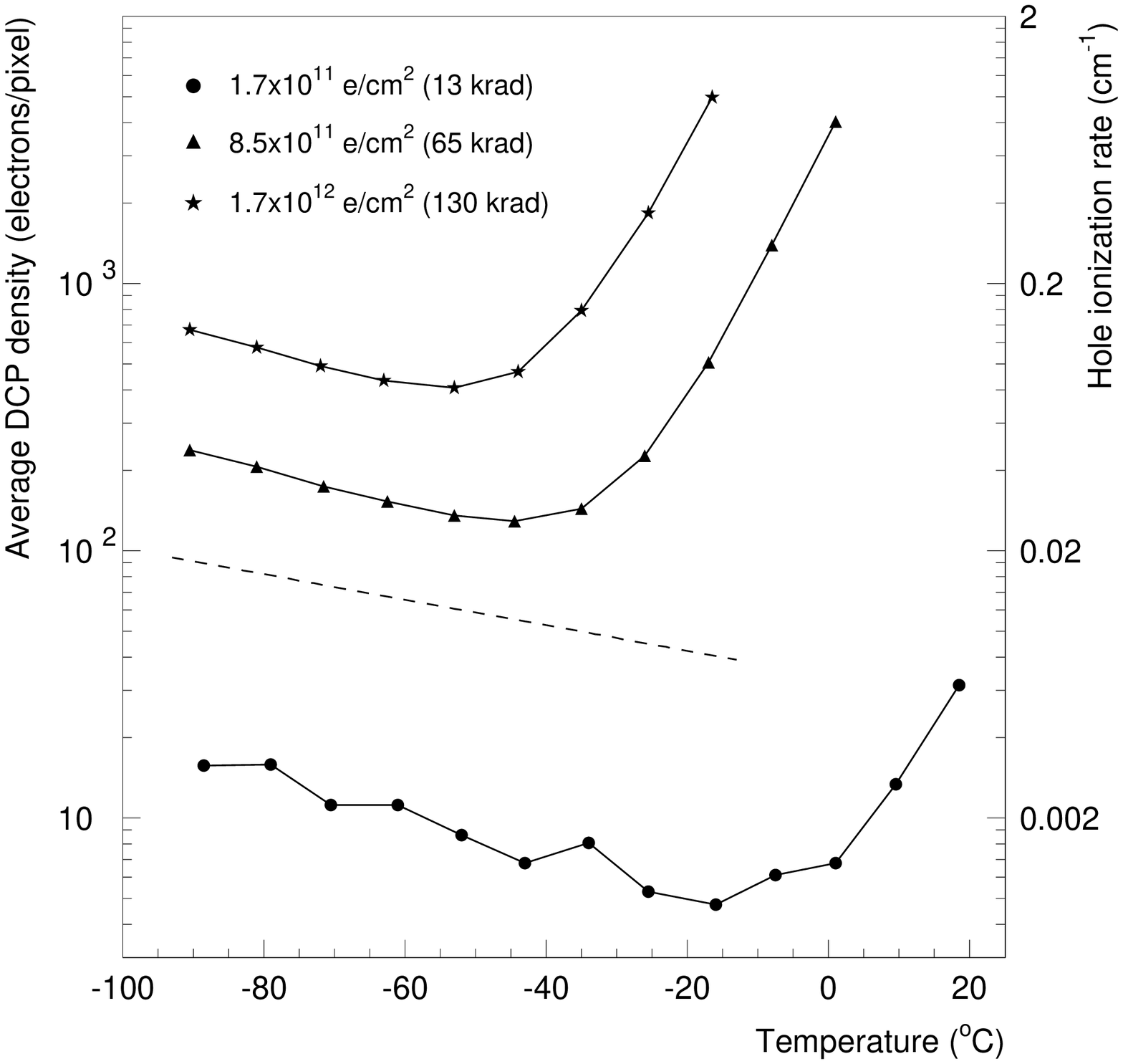}}
     \caption{\sl \label{fig:radam:dcp1}  
Temperature dependence of the average DCP density in device
\#JS14/026 as a function of the electron irradiation. Dashed line shows the
calculated temperature dependence of the hole ionization rate at $10^{5}$ V/cm.}
\end{minipage}
\hfill
\begin{minipage}[r]{0.48\linewidth}
\centerline{\epsfxsize=8.0cm \epsfbox{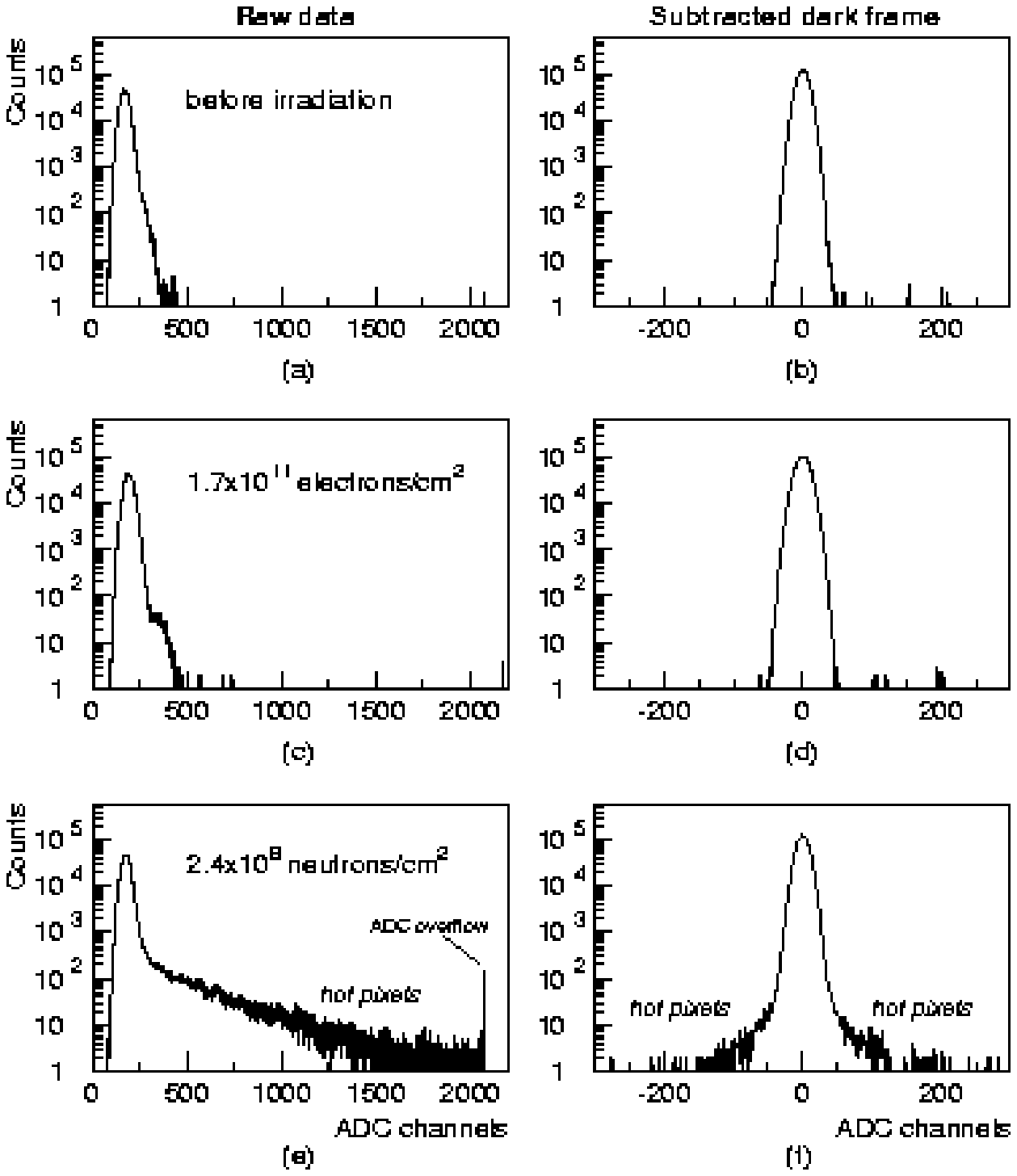}}
\caption{\sl \label{fig:radam:hp1}
Dark current distribution at +10$^{\circ}$C in S5466 CCDs before
irradiation (a), (b); in an electron irradiated to $1.7\times10^{11}$ cm$^{-2}$
CCD (c), (d); and in a neutron irradiated to $2.4\times10^{8}$ cm$^{-2}$ CCD
(e), (f). The gain is $\approx$ 6.0 electrons/ADC channel.}
\end{minipage}
\end{figure}

Spurious dark charge, which was superimposed on the bulk current, was observed
in electron irradiated Hamamatsu CCDs. This spurious current,
which was named ``dark current pedestal'' (DCP) does not depend on the
accumulation time of the readout cycle \cite{radam:Stefanov99}.  Because it
is produced by the clocking, dark current measurements are not affected.
However, the DCP provides background charge in the CCD and decreases the
measured CTI through the ``fat zero'' effect.  It was found, that the spurious
current was caused by impact ionization by holes, returning to the channel stops
and to the adjacent gates after the surface potential has been switched from
inversion to depletion \cite{radam:Stefanov2000a}.  The temperature dependence
of the DCP is shown on Fig. \ref{fig:radam:dcp1}. In the low temperature region
holes are provided by tunneling from near-interface traps, while at higher
temperatures carriers are released by thermal excitation from interface defects.
The DCP severely limits the application of Hamamatsu devices because it is much
larger than the bulk dark current at the same temperature. \\

Investigation of dark current spikes was performed on both Hamamatsu and EEV
devices, which showed very similar results. No increase of the dark current
non-uniformities was observed in electron irradiated devices (Fig.
\ref{fig:radam:hp1}(a)--\ref{fig:radam:hp1}(d)). Neutron irradiated CCDs
however, showed significant creation of dark current spikes (Fig.
\ref{fig:radam:hp1}(e)) after irradiation with fluence as low as
$2.4\times10^{8}$ neutrons/cm$^{2}$. After subtraction of the dark frame pixel
by pixel, the resulting distribution (Fig. \ref{fig:radam:hp1}(f)) is close to
that of a electron-irradiated CCD. This subtraction shows an efficient way to
correct for the large dark current non-uniformities in CCDs, irradiated by heavy
particles. During operation, the dark current map should be updated regularly to
include the contribution of the newly created defects. \\

\begin{figure}[htb]
  \begin{minipage}[l]{0.48\linewidth}
     \centerline{\epsfxsize=8.0cm \epsfbox{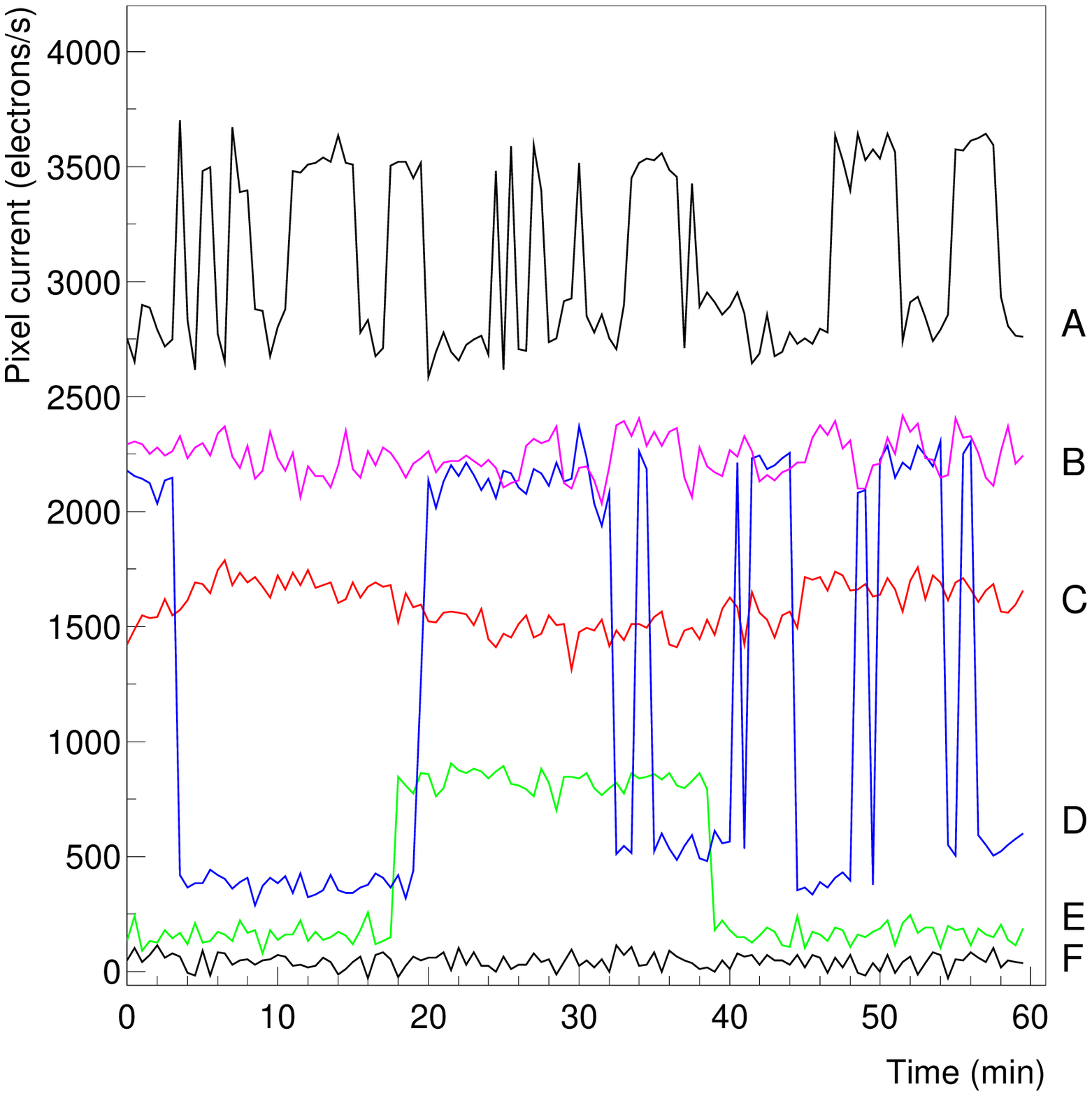}}
     \caption{\sl \label{fig:radam:rts3}
Time variation of the dark current at $-1.6^{\circ}$C in hot
pixels, showing RTS behavior. The measurement was carried out on Notch CCD \#P4
1-5B1-4, irradiated to $5.7\times10^{9}$ neutrons/cm$^{2}$. Pixel F is normal
(not ``hot''), is plotted for reference.} 
\end{minipage}
\hfill
\begin{minipage}[r]{0.48\linewidth}
\centerline{\epsfxsize=8.0cm \epsfbox{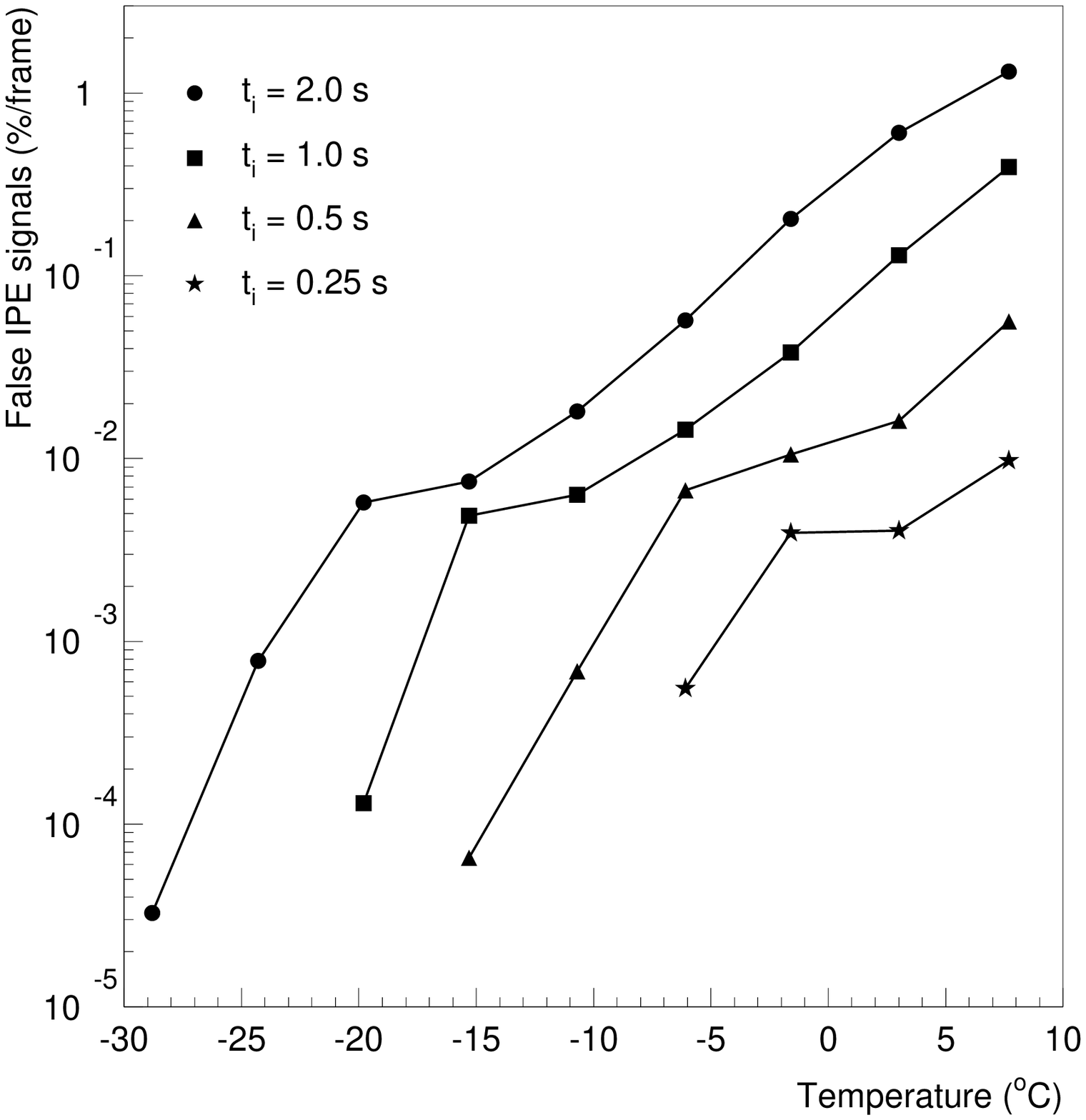}}
\caption{\sl \label{fig:radam:rts4bw}
Ratio of false IPE per CCD frame, caused by RTS, at 4 different
integration times $t_{i}$. The measurement was performed on Notch CCD \#P4
1-5B1-4, irradiated to $5.7\times10^{9}$ neutrons/cm$^{2}$. The threshold is set
at 600 electrons.} 
\end{minipage}
\end{figure}

Random telegraph signals (RTS) were observed in many hot pixels when their
current was recorded for sufficiently long time (Fig. \ref{fig:radam:rts3}).
During the typical time, required for one CTI measurement ($\approx$ 5 min),
these fluctuations are difficult to notice and they have small effect on the
subtraction of the dark frame. In a long time  scale, it is not possible to
correct for the dark current of such pixels as was shown on Fig.
\ref{fig:radam:hp1} because of the random fluctuations of the pixel current. The
RTS can generate false signals, similar to those created by X-rays or MIPs, and
can be a source of erroneous track reconstruction. A measurement to find
isolated pixel events, caused by RTS was performed on Hamamatsu and EEV CCDs,
with similar results. The number of false IPE greatly depends on the threshold,
which is imposed for the search of signal. The same threshold of 600 electrons,
used for the CTI measurements was employed in the measurement. The ratio of
false IPE per CCD readout frame (i.e. the pixel occupancy), caused by RTS is
shown in Fig.~\ref{fig:radam:rts4bw}. The RTS amplitude depends on the
temperature and the integration time $t_{i}$ and these two 
parameters were varied
during the measurement. Although about 40\% of the hot pixels exhibit RTS, most
of them are rejected by the applied threshold for IPE search. When the
integration time is short, a ratio of false IPE signals of less than $10^{-4}$
can be achieved even in the temperature range above 0${^\circ}$C. \\

\subsubsection{CTI}
Much attention was paid to the measurement of radiation-induced CTI,
because it is the main factor that requires low temperature operation
\cite{radam:Abe97}. A model for the charge transfer in 2- and 3-phase CCDs
was developed \cite{radam:Stefanov2000a}, which allows one to calculate the CTI
as a function of the temperature, timing of the transfer, concentration of
background charges, signal size and pixel occupancy.

\begin{figure}[htb]
\begin{minipage}[l]{0.48\linewidth}
\centerline{\epsfxsize=8.0cm \epsfbox{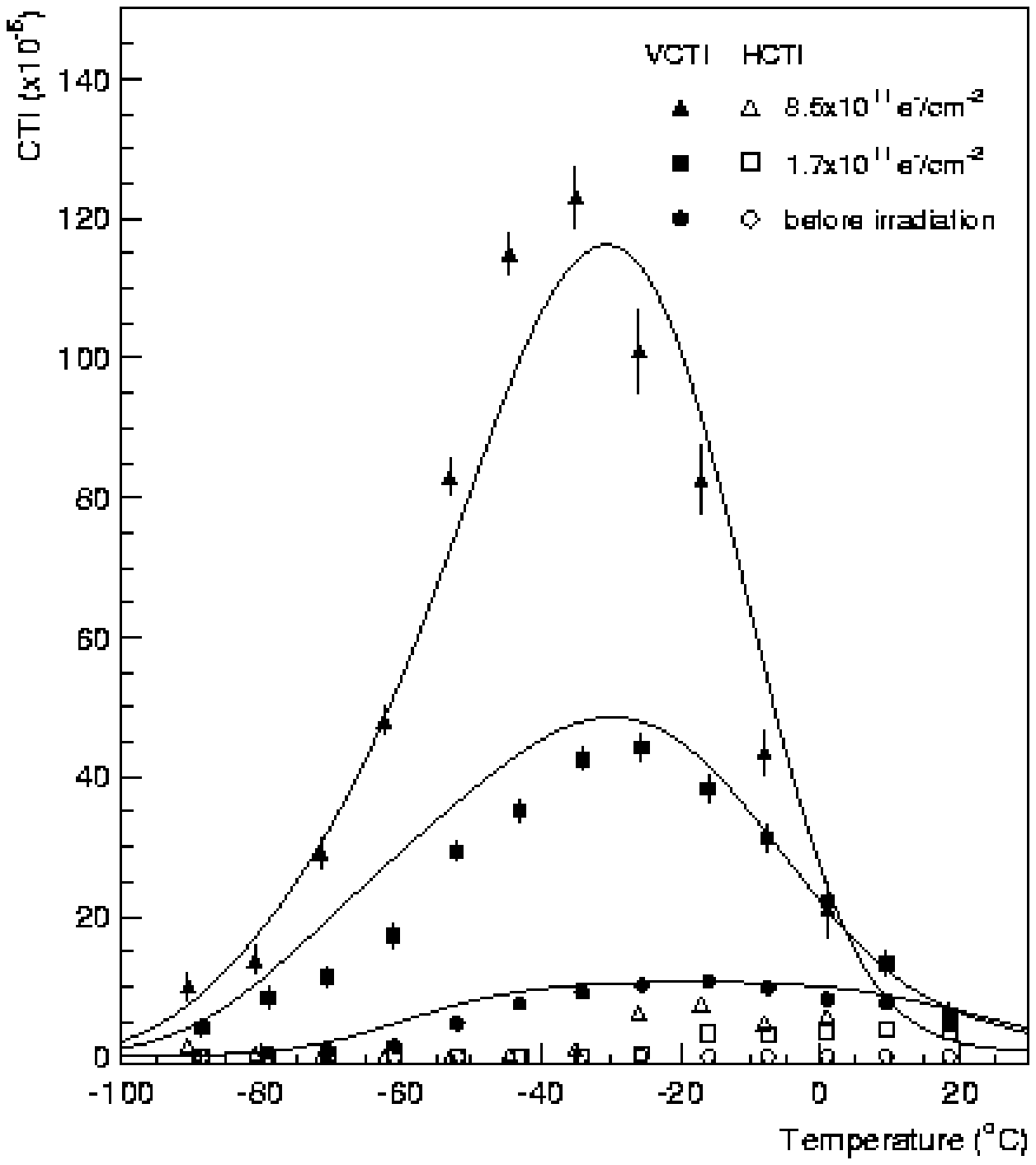}}
\caption{\sl \label{fig:radam:cti_e1}
CTI as a function of the temperature and of the electron irradiation
fluence in S5466 CCD.  Symbols are the experimental data;
solid lines represent the modeled VCTI for the combination of two traps at
E$_{c}-$0.37~eV and E$_{c}-$0.44~eV.} 
\end{minipage}
\hfill
\begin{minipage}[r]{0.48\linewidth}
\centerline{\epsfxsize=8.0cm \epsfbox{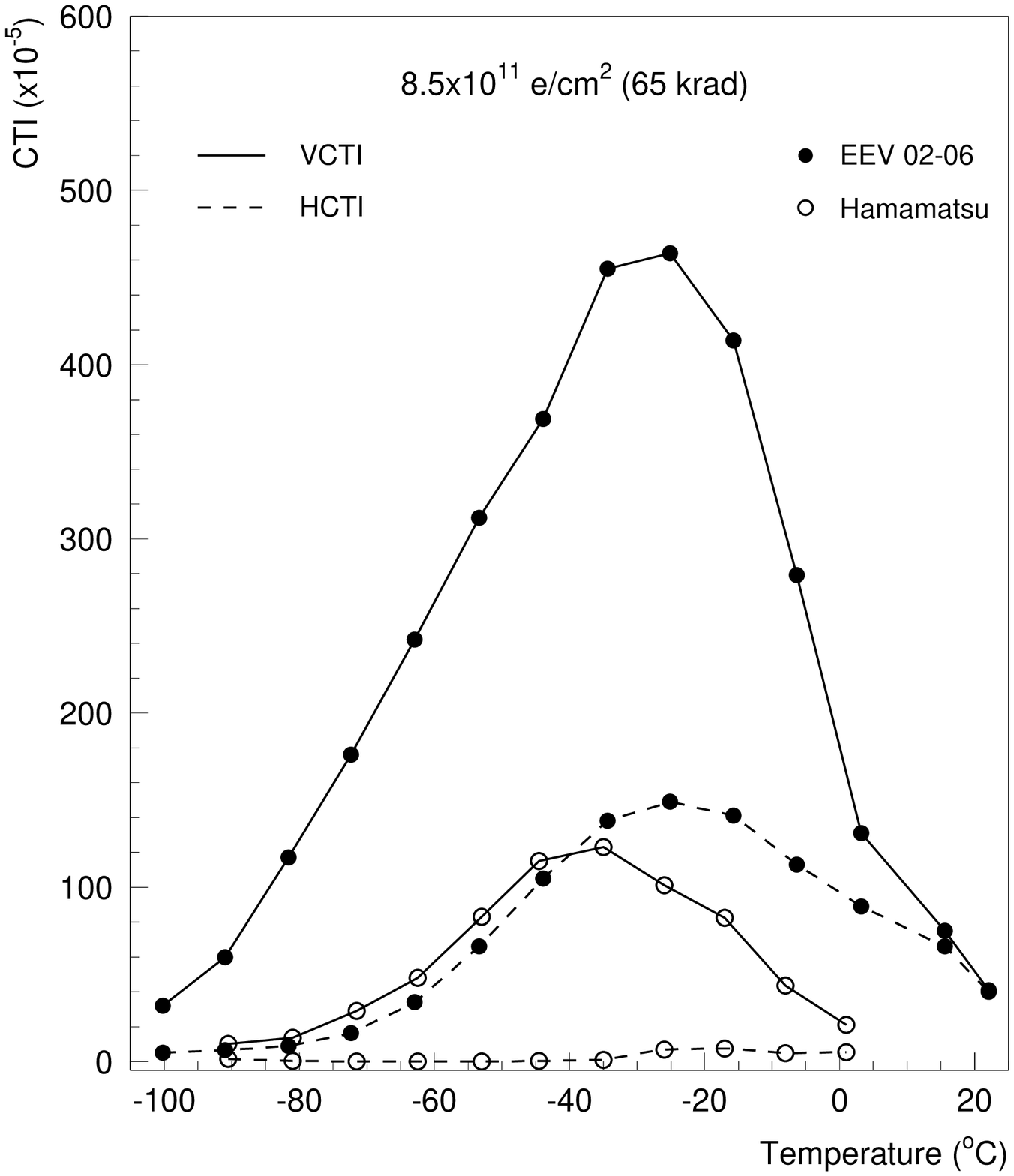}}
\caption{\sl \label{fig:radam:eev_hpk}
CTI of 2-phase (Hamamatsu S5466) and 3-phase (EEV CCD02-06)
electron-irradiated CCDs.} 
\end{minipage}
\end{figure}

The results from the CTI measurements on electron irradiated S5466 CCD,
together with modeled results are shown in Fig.~\ref{fig:radam:cti_e1}.
Modeling was performed using the experimental values of the dark current and of
the DCP, presented in the previous paragraphs.  The modeling on electron
irradiated CCDs showed that a trap at E$_{c}-$0.37~eV with $\sigma_{n}$ =
4$\times10^{-15}$ cm$^{2}$ is responsible for the observed increase of the
vertical CTI. Additional pre-irradiation defect was identified at
E$_{c}-$0.44~eV. The energy position of the main defect, 
derived from the model,
varies between E$_{c}-$0.37~eV and E$_{c}-$0.39~eV in three samples. The
simulated results in Fig.~\ref{fig:radam:cti_e1} are based on the presence of
two traps at E$_{c}-$0.37~eV and E$_{c}-$0.44~eV.  
The concentration of the trap
at E$_{c}-$0.44~eV, obtained from the model increases insignificantly under
irradiation. The CTI of the horizontal register is less than 10$^{-4}$ in the
whole temperature range. As explained by the model, the peak in the HCTI, which
would have occurred at high temperatures is suppressed by the dark current
pedestal and the exponentially increasing dark current.

Similar result was obtained in electron-irradiated EEV CCD, however the VCTI in
that device was about 4 times higher than in S5466 devices. Numerical
simulations showed, that under equal conditions the 3-phase CCDs has about 2.5
times larger CTI than 2-phase devices, because of the higher signal
density in 2-phase CCDs. The greater than expected difference can be explained
mostly by the DCP in Hamamatsu devices, which decreases the measured CTI values
by the ``fat zero effect'' \cite{radam:Stefanov99}. Two-phase devices are
superior to 3-phase CCDs in term of CTI and should be preferred for
applications in radiation environment.

The VCTI of neutron-irradiated to 8.9$\times$10$^{9}$ cm$^{-2}$ S5466 CCD is
about 4$\times10^{-4}$, and VCTI peak of 3$\times10^{-4}$ was measured in an
EEV CCD, irradiated with approximately the same fluence. The same defect at
E$_{c}-$0.37~eV, derived from the model, was found responsible for the CTI
increase in neutron-irradiated devices. In all CCDs the HCTI is significantly
smaller than the VCTI, which is explained by the different timing of the
transfer in the registers.

The defect at E$_{c}-$0.37~eV was identified from annealing studies and
additional measurements as contribution from both $E$ and $V-V^{-}$ centers
\cite{radam:Stefanov2000a}. The $E$ center, which anneals out at $150^{\circ}$C
was found to be dominant, contributing to $\approx$ 60\% of the CTI peak. The
measured energy position of the defects is reduced by the Frenkel-Pool effect.
This effect is also responsible for the thermal activation of the dark current
in most of the hot pixels.  \\

Modeling showed, that in the accumulated 10-year  background from
e$^{+}$e$^{-}$ pairs of 1.5$\times$10$^{12}$ cm$^{-2}$ the VCTI will reach
$1.5\times10^{-1}$ in 3-phase EEV-like CCD. Safe margin is included in this
calculation by considering the bulk damage from e$^{+}$e$^{-}$ pairs to
be 10 times larger than that from $^{90}$Sr electrons. Neutron-induced CTI is
much smaller than those generated by electrons, and only the latter is being
considered. Such a high CTI value is unacceptable, because the charge will be
lost only after a few transfers. Measures have to be taken to decrease the VCTI,
including better CCD design and appropriate device operating conditions, in
order to extend the lifetime of the CCDs. \\

\begin{figure}[htb]
\begin{minipage}[l]{0.48\linewidth}
\centerline{\epsfxsize=8.0cm \epsfbox{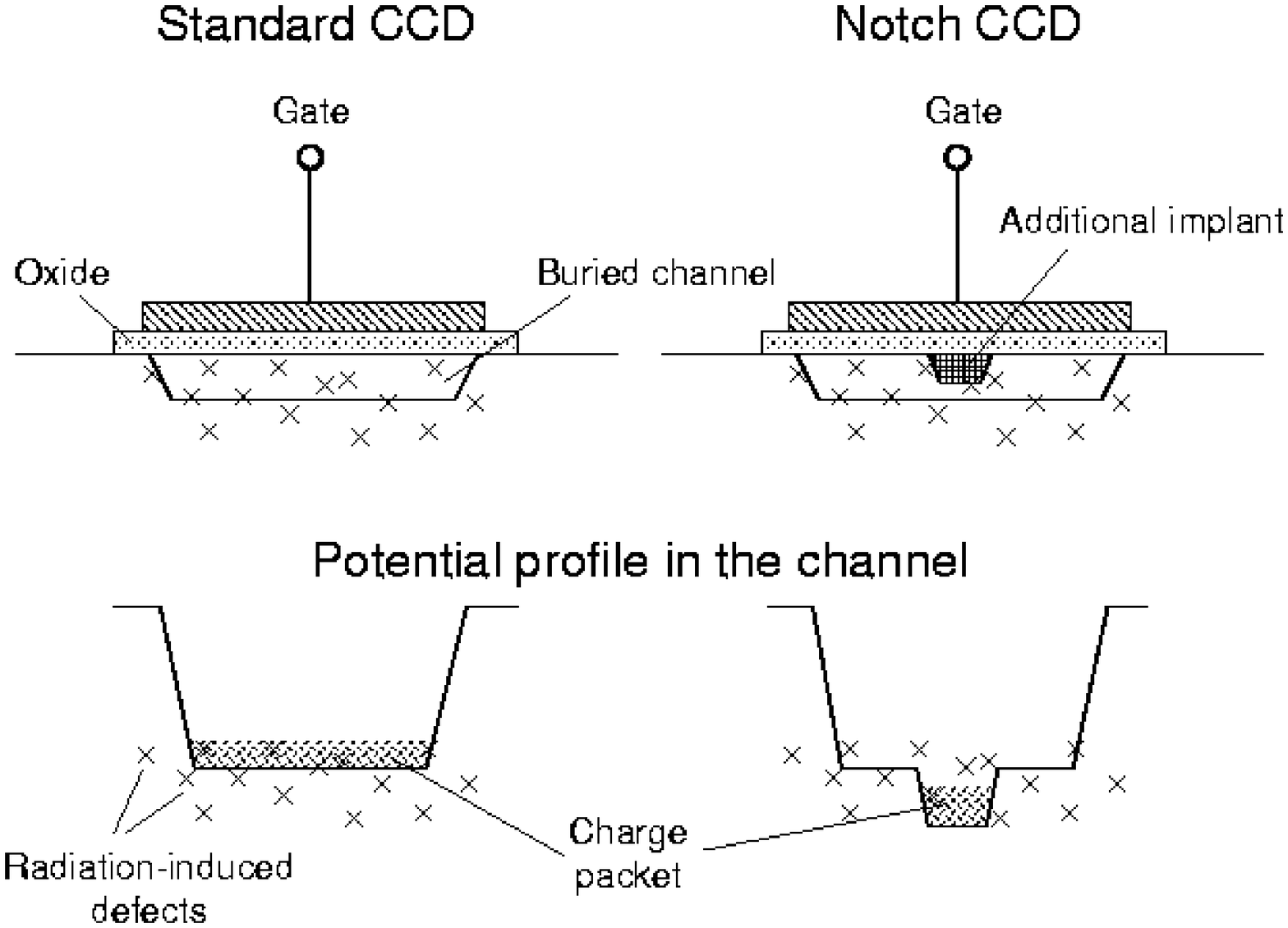}}
\caption{\sl\label{fig:radam:notchfig}
Schematic and potential profile in the channel of standard and
notch CCD.}
\end{minipage}
\hfill
\begin{minipage}[r]{0.48\linewidth}
\centerline{\epsfxsize=8.0cm \epsfbox{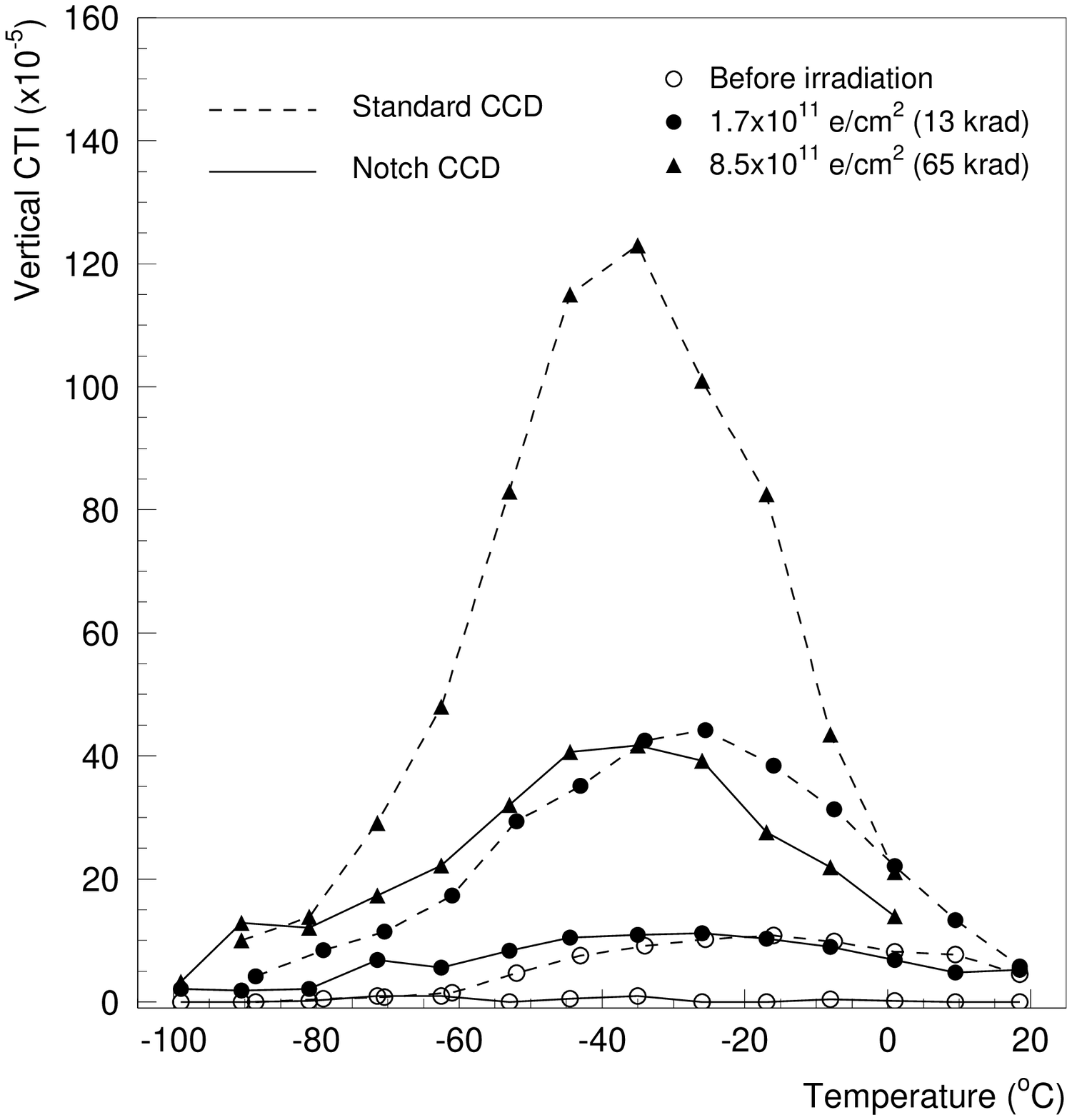}}
\caption{\sl \label{fig:radam:notchcti_e1}
CTI of notch and standard S5466 CCDs.}
\end{minipage}
\end{figure}

One important method to improve the CTI is to use an additional narrow
implant in the CCD channel (Fig.~\ref{fig:radam:notchfig}). This implant forms a
sub-channel, or ``notch'' in the potential profile of the transport channel,
confining the signal charge into a fraction of the pixel volume. CTI depends on
the concentration of signal electrons $n_{s}$ as
\begin{equation}
CTI \propto n_{t}/n_{s}, \label{radam:notchdep}
\end{equation}
where $n_{t}$ is the defect concentration, therefore increasing the signal
density by forcing the same charge into a smaller volume can improve the CTI.

The radiation-induced CTI was measured in electron- and neutron-irradiated notch
CCD, based on the S5466 design. Due to the additional 3$\mu$m-wide notch
channel, its CTI is about 3 times lower than that in conventional S5466 device
(Fig.~\ref{fig:radam:notchcti_e1}). In neutron-irradiated to 5.7$\times$10$^{9}$
cm$^{-2}$ notch CCD the peak CTI was below $10^{-4}$. \\

A calculation of the frequency dependence of the CTI  was carried out to
study the possible improvement by varying the readout speed of the CCDs.
At speeds higher than several Mpix/s, horizontal CTI decreases rapidly because
charge dwell time becomes shorter than the capture time constant of the defects.
Vertical CTI is much less influenced, because charge spends longer time in the
imaging section. Increasing the readout speed has relatively small ability to
control the the VCTI and other methods to reduce it should be chosen. At high
readout speed the horizontal CTI becomes more than an order of magnitude smaller
than the VCTI.

It was not possible to back this calculation by experimental data from
the high-speed Hamamatsu CCD, because of the dependence of the DCP on the width
of the shift pulses \cite{radam:Stefanov2000a}. The frequency dependence of
the CTI was masked by the ``fat zero'' effect from the DCP and a definite
conclusion could not be drawn. \\

\begin{figure}[htb]
\begin{minipage}[l]{0.48\linewidth}
\centerline{\epsfxsize=8.0cm \epsfbox{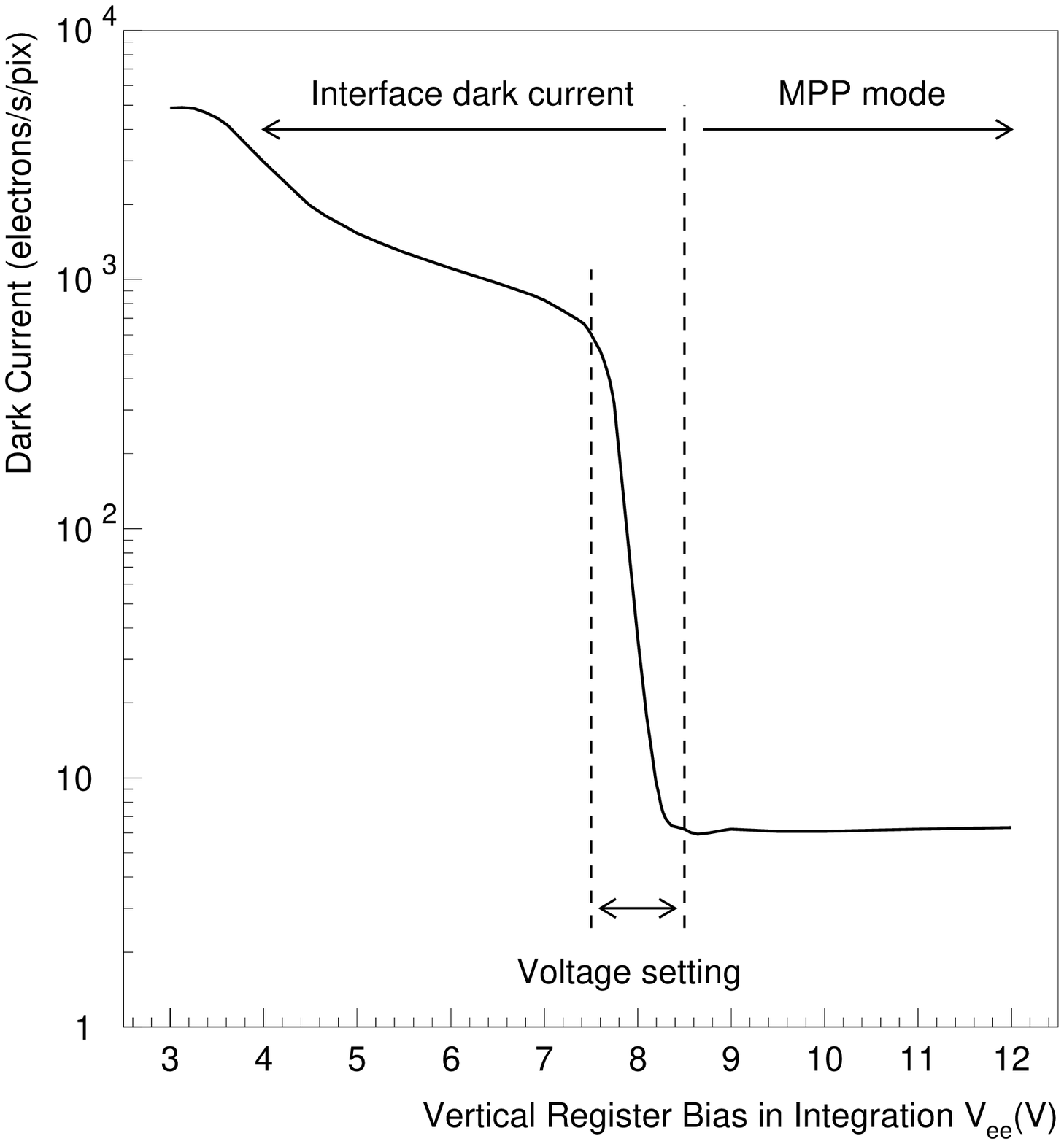}}
\caption{\sl\label{fig:radam:eev1mpp2}
Principle of thermal ``fat zero'' generation in MPP mode CCDs.}
\end{minipage}
\hfill
\begin{minipage}[r]{0.48\linewidth}
\centerline{\epsfxsize=8.0cm \epsfbox{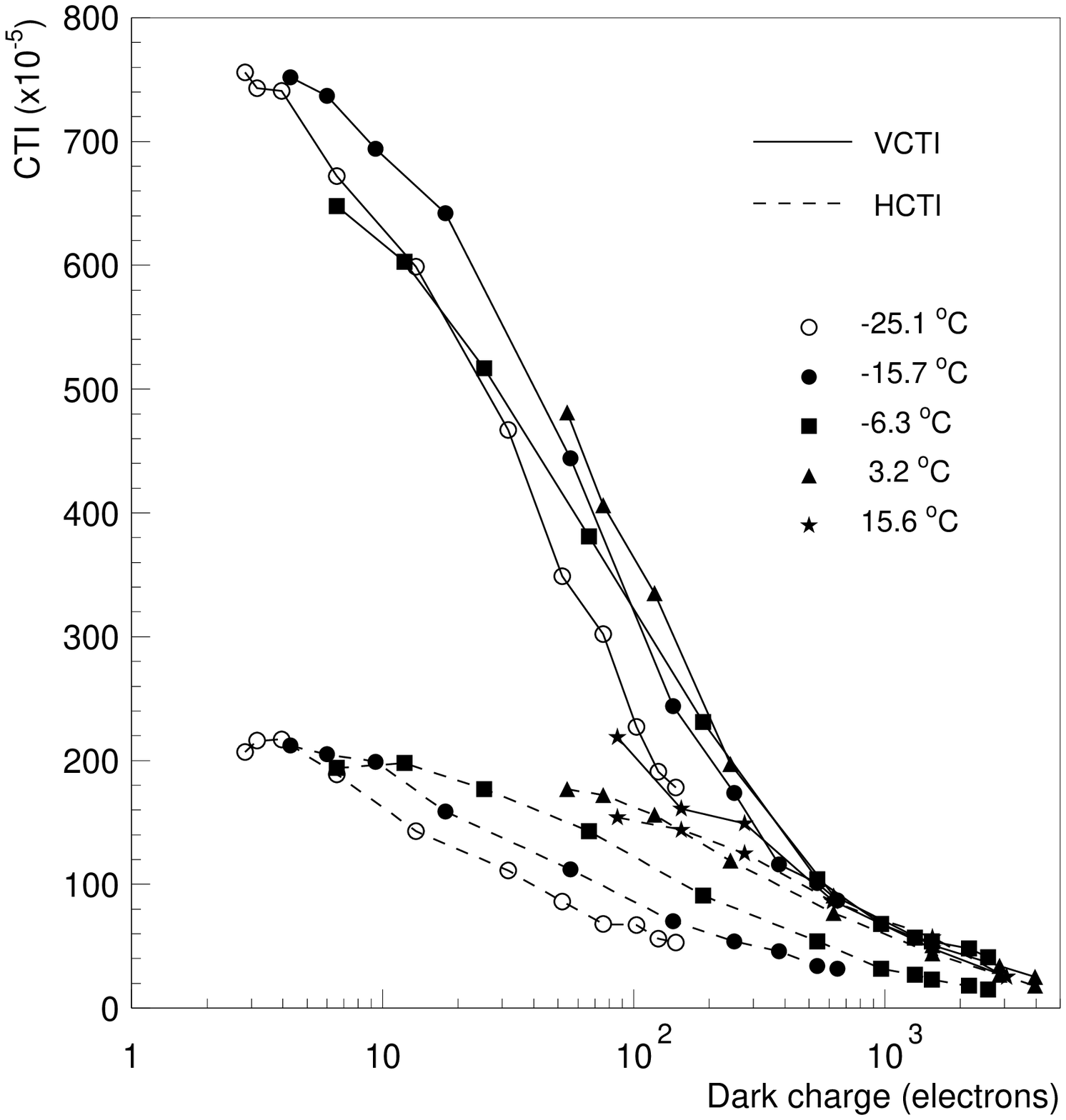}}
\caption{\sl \label{fig:radam:eev1dcti2}
Measurement of CTI reduction by ``fat zero'' effect in EEV
CCD02-06.} 
\end{minipage}
\end{figure}

One of the most powerful methods, that can be used for CTI improvement is
injection of a sacrificial charge, or ``fat zero'' into the CCD registers. This
charge reduces the CTI in two ways: by increasing the signal charge density
(\ref{radam:notchdep}), as it is transported together with the signal; and most
importantly, by filling the traps in front of the signal packet, so that less
electrons can be captured.

Experimental and theoretical studies of several devices have shown, that the
CTI of the vertical register dominates the charge losses in the CCD.
Introducing additional charge in the vertical register through the injection
port is not a trivial task because of the large non-uniformity of the injected
charge \cite{radam:Sugimoto2000}. We have developed another method to inject
``fat zero'', which employs the dark current characteristics of MPP mode CCDs.
By varying the negative gate bias voltage 
of the vertical register V$_{ee}$, the
CCD can be driven into semi-MPP mode and its dark current can be controlled
between the MPP mode value and $\approx$100 times that value
(Fig.~\ref{fig:radam:eev1mpp2}). At high temperatures, this current can be
sufficient to provide charge of the order of 1000 electrons even if the CCD
readout cycle is very short. The proposed method works on any type of MPP CCD,
because the ``fat zero'' is generated thermally. The amount of injected charge
can be controlled either by adjusting the V$_{ee}$, or by setting V$_{ee}$ to
some low value and adjusting the time,  
during which the CCD is out of MPP mode.
Practical implementation of the proposed 
method depends on the irradiation level
and requires precise adjustment.

Another important feature of this method is that it has the ability to
self-adjust for non-uniformities of surface damage. If in some pixels the flat
band voltage shift is higher, corresponding to higher surface and bulk
damage, the interface component of the dark current will be higher than for the
rest of the CCDs at the same setting of V$_{ee}$. 
Therefore, larger dark current
is generated automatically, which compensates for the higher CTI in those
pixels.

The experimental data in Fig.~\ref{fig:radam:eev1dcti2} is obtained by changing
the V$_{ee}$ voltage. It shows, that the CTI can be reduced several times by
injecting ``fat zero'' of approximately 1000 electrons, 
which is almost half the
number of signal electrons. The additional charge increases the noise of the
device, however this does not degrade the performance of the high-speed CCDs,
for which the readout noise is of the order of 100e$^{-}$~RMS. Estimations of
the radiation-induced dark current after receiving 
combined electron and neutron
irradiation at the expected 10-year levels show that dark current generation of
$>1000$ electrons can be achieved by this method at temperatures
$>5^{\circ}$C and integration time of 6.7~ms.\\

As shown in the previous paragraphs, the VCTI is the main point of concern,
because the HCTI can be reduced to less than 10\% of VCTI. 
Using all the options
for CTI improvement described above, 
the VCTI can be reduced more than 60 times,
as shown in Table~\ref{radam:CTIimprove}. However, this may not be sufficient,
and the number of transfers in the vertical direction (i.e., the vertical chip
size) may have to be reduced in accordance with (\ref{radam:CTIdef}) to limit
the charge losses.

\begin{table}[htb]
\caption{\sl \label{radam:CTIimprove}
Vertical CTI improvements.}
\begin{center}
{
\begin{tabular}{|l|c|}	\hline
Option 	  & VCTI improvement \\ \hline
Raise the output speed to $>$5 Mpix/s & $\approx$1.3 times  \\ \hline
Use 2-phase CCD                       & $\approx$2.5 times  \\ \hline
Use notch CCD                         & 3 to 4 times        \\ \hline
Inject ``fat zero'' $>$1000 e$^{-}$   & 6 to 8 times        \\ \hline
Total improvement                     & $\approx$ 60 to 100 times \\ \hline
\end{tabular}
}
\end{center}
\end{table}

Speaking only of the CTI, the CCD lifetime in a radiation 
environment depends on
the chosen value of charge transfer losses. Some example parameters of the CCD
can be calculated if 25\% lost charge is considered acceptable and the HCTI is
taken as 10\% of the VCTI. As was mentioned before, the VCTI of a 3-phase
EEV-like CCD is expected to reach $1.5\times10^{-1}$ 
after 10 years of operation
in the radiation environment of the future linear collider. A CCD, which
contains 2~Mpixels and 12 outputs, each reading 250 vertical and 667 horizontal
pixels, will have 23\% transfer losses after $\approx$3 years of operation.
Although all the measures from Table~\ref{radam:CTIimprove} with improvement
of 60 are applied in this calculation, the main contribution (18\%) comes from
the VCTI. Assuming pixel size of 24$\mu$m, the chip has dimensions
96mm$\times$12mm, and can be read out completely between two pulses (6.7~ms) at
25~Mpix/s. \\

\begin{figure}[htb]
\centerline{\epsfxsize=6.0cm \epsfbox{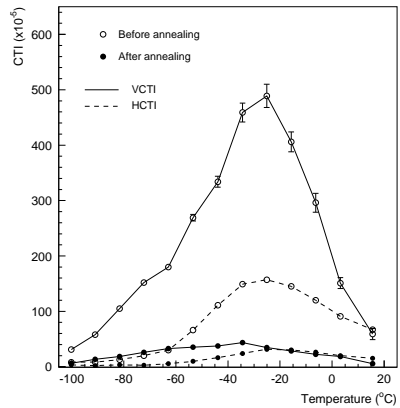}}
\begin{center}\begin{minipage}{\figurewidth}
\caption{\sl \label{fig:radam:CTIanneal}
Temperature dependence of the CTI in electron-irradiated
to $8.5\times 10^{11}$ cm$^{-2}$ EEV CCD02-06 (\#A4003-18) before and
after annealing at $150^{\circ}$C for 2 hours.}
\end{minipage}\end{center}
\end{figure}

Another possible method for reduction of the radiation-induced CTI is device
annealing. It is known, 
that the $P-V$ center, which is the dominant bulk defect
in $n$-type buried channel CCDs \cite{radam:Hopkins93} anneals at about
150$^{\circ}$C. 
By periodically heating of the CCD it is possible to restore the
pre-irradiation CTI, provided that the mechanical support and mounting are
designed to withstand such a temperature cycle.

An annealing experiment was performed on both electron irradiated Hamamatsu
and EEV devices \cite{radam:Stefanov2001}. In the EEV chip the CTI decreased
almost to its pre-irradiation level after 2 hours at 150$^{\circ}$C
(Fig.~\ref{fig:radam:CTIanneal}). Similar experiment on a Hamamatsu device
showed CTI improvement of about 60\%.

\vspace{12pt}
\subsubsection{Conclusions }

The most important results from the radiation damage studies of CCDs
can be summarized as follows: 
\begin{itemize}
\item[$\bullet$] The radiation hardness of 2- and 3-phase MPP mode CCDs from
different manufacturers has been extensively studied; 
\item[$\bullet$]
A model for the charge transfer losses in the CCDs has been
developed and applied in good agreement with the experimental data; 
\item[$\bullet$]
Two-phase, notch CCDs should be used because of the greatly
reduced CTI in comparison to 3-phase and standard devices;
\item[$\bullet$]
Thermal charge injection is the major factor for reducing the CTI and
promises successful near-room temperature operation in 
harsh radiation environment;
\item[$\bullet$]
Device annealing can be used for restoring the CTI to sufficiently
low levels. It can be performed periodically when the detector is not working;
\item[$\bullet$]
Large flat band voltage shifts and spurious dark charge in Hamamatsu
CCDs are the main limitations at the moment. These issuies can be 
solved by more
advanced radiation-hard devices, currently under development.
\end{itemize}

%% file: dettrk/itc/main.tex
\section{Intermediate Tracker\label{section-it}}

The distance between the last layer of VTX and the first layer
of CDC is about 39cm.  The gap is filled by an Intermediate 
Tracker(IT). The intermediate tracker consists of five layers of 
silicon detectors placed at radii of 9.0, 16.0,
23.0, 30.0, and 37.0 cm and covers $|\cos\theta|<0.90$, 
which coincides with the region covered by VTX.
The material thickness of each layer is 0.6\% radiation
lengths, and the spatial resolution is 
40$\mu$m each for $r\phi$ and $z$.

The roles of the intermediate tracker are to improve
\begin{itemize}
\item the linking efficiency of 
a CDC track to the corresponding VTX track,

\item
the reconstruction  efficiency of particles which decay 
between VTX and CDC,
\item
the reconstruction efficiency of low momentum tracks, and
\item
time stamping capability to
discriminate background tracks created by off-timing bunches,
which is impossible with the CCD vertex
detector, since it does not provide signal timing information.
\end{itemize}

\begin{figure}
\centerline{\epsfxsize=10.0cm \epsfbox{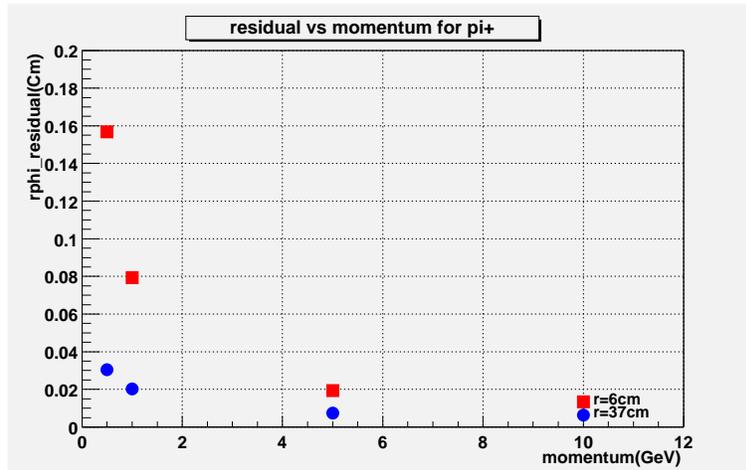}}
\begin{center}\begin{minipage}{\figurewidth}
\caption{\sl \label{imt-rphiresidual}The  r$\phi$ residual
at the last layer of IT(r=37cm) and that of VTX(r=6cm)
for pion momenta: 0.5, 1.0, 5.0, and 10.0 GeV.}
\end{minipage}\end{center}
\end{figure}

To study the efficiency to link a CDC track to 
the corresponding IT and VTX hits, we generated single $\pi^+$ events
at $\cos\theta=0$ and studied the residuals of the CDC tracks 
at the last layer of IT (at $r=37$ cm) 
and the last layer of VTX (at $r=6$ cm),
as shown in Fig.~\ref{imt-rphiresidual}.
Since the residuals at IT are considerably smaller than 
those at VTX, IT will help in extrapolating CDC tracks 
to the VTX region.

We also studied residuals of tracks in Higgs events.
\begin{figure}
\centerline{
\epsfxsize=7.0cm\epsfbox{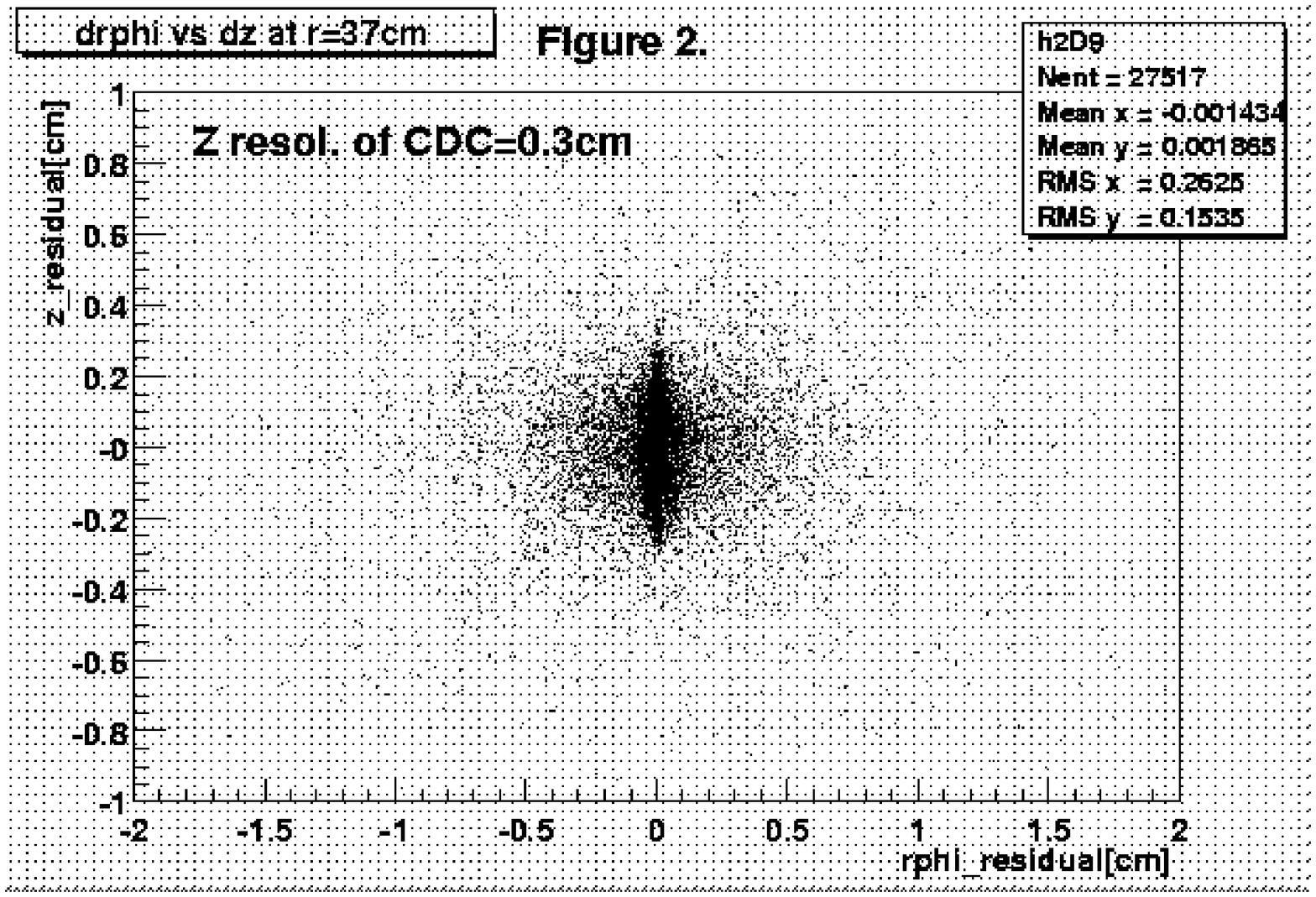}
\epsfxsize=7.0cm\epsfbox{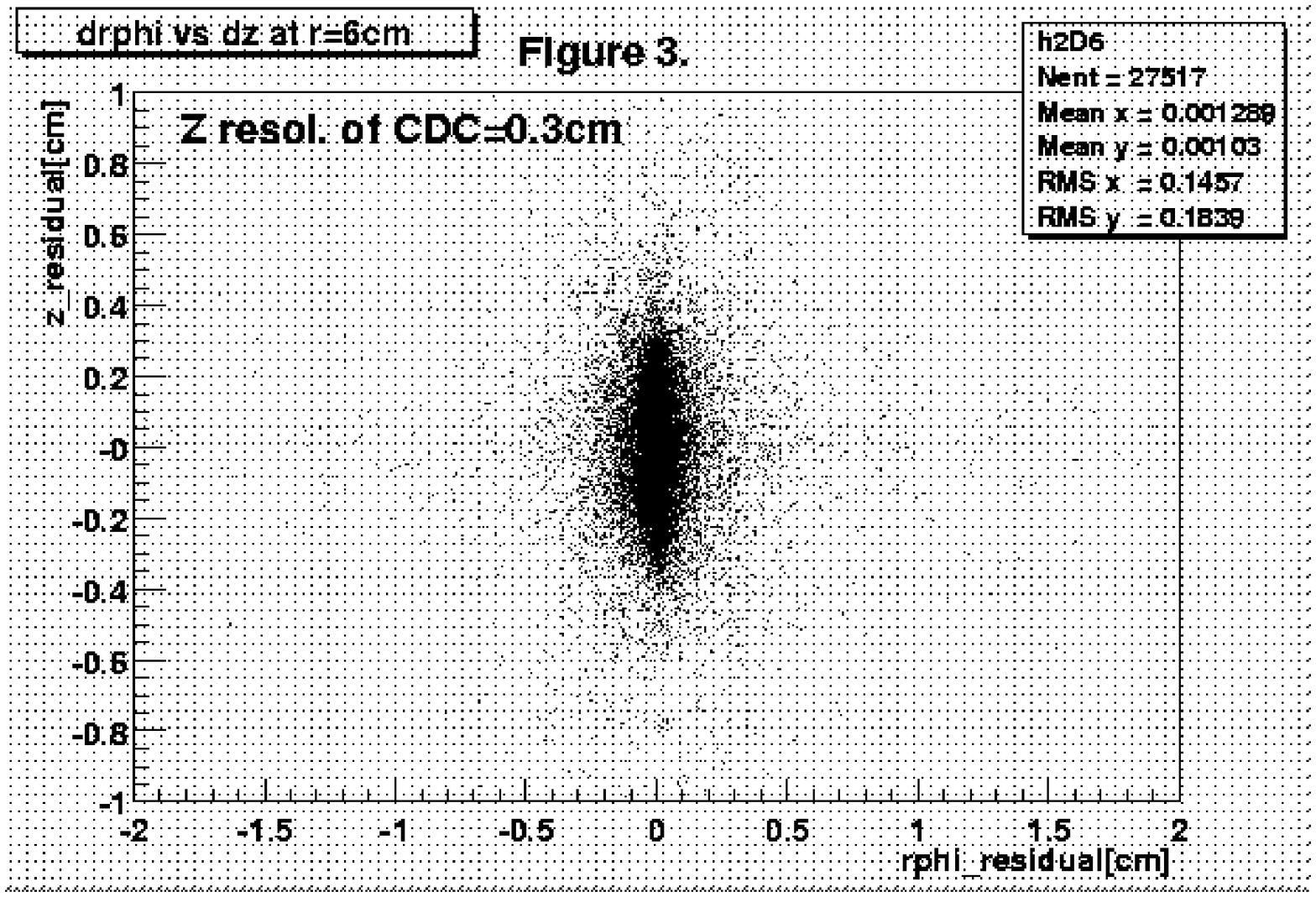}
}
\centerline{
\epsfxsize=7.0cm\epsfbox{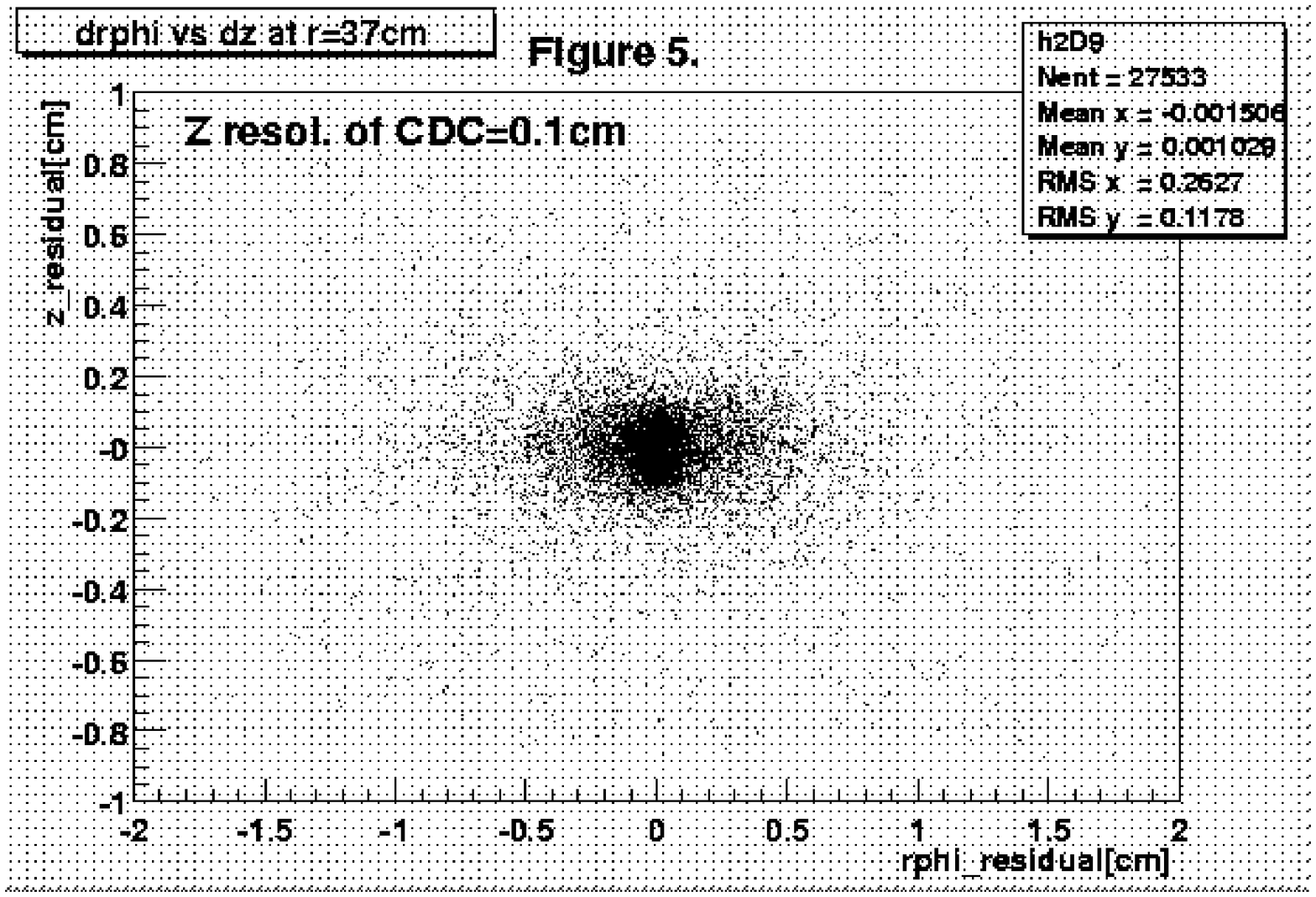}
\epsfxsize=7.0cm\epsfbox{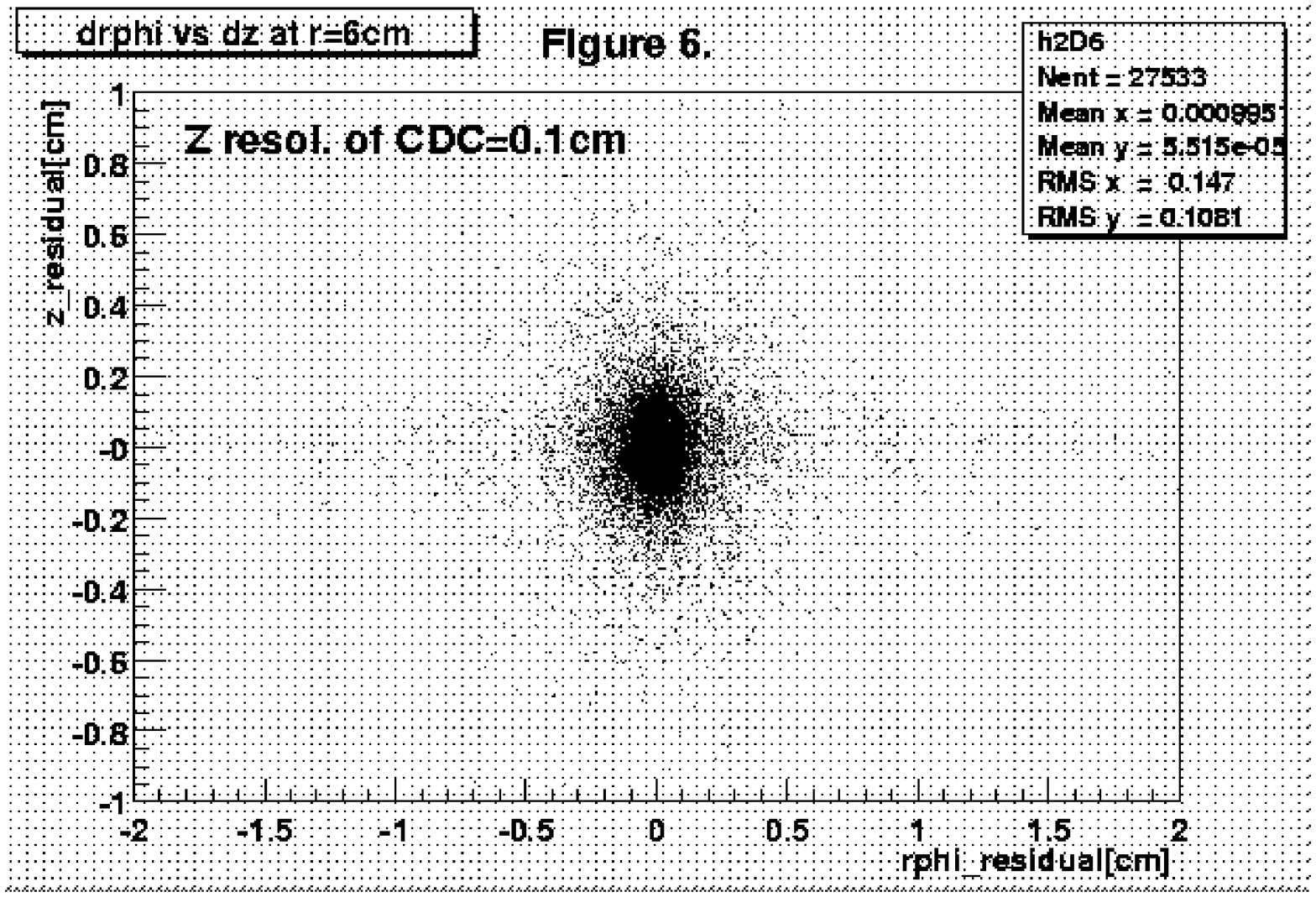}
}
\begin{center}\begin{minipage}{\figurewidth}
\caption{\sl \label{imt-higgsresid.eps}
The residuals of CDC tracks at IT(r=37cm) and VTX(r=6cm) positions,
for Higgs events.  Upper two figures are for the case where the
$Z$ resolution of CDC is 3mm, while the bottom two figures
are for 1mm.}
\end{minipage}\end{center}
\end{figure}
The scatter plots of residuals in the r$\phi$ plane against
those in the $z$ direction are shown in Fig.~\ref{imt-higgsresid.eps}.
The residuals for a CDC 
$z$-resolution of 0.1cm is also shown for comparison.
Again, we see the reduction of residuals at IT compared 
to those at VTX.  The $z$-residual at IT is mainly
due to the $z$-resolution of CDC. It is thus desirable to reduce 
it to 1mm or less.  There are tails
in the distribution at IT.  They are caused
by multiple scattering of low momentum tracks.
By Gaussian-fitting to the central part of the 
distribution, we obtained the standard deviations 
of the residuals as summarized in 
Table~\ref{imt-table1}.
\begin{table}
\begin{center}\begin{minipage}{\figurewidth}
\caption{\sl \label{imt-table1}
Summary of residuals of Higgs events at IT and VTX position.}
\end{minipage}\end{center}
\begin{center}
\begin{tabular}{l l l l }
Position & $\sigma_z$(mm) & $\sigma_{r\phi}$ (mm)& $\sigma_Z$ (mm)\\
IT & 3 & 0.021 & 0.137 \\
     & 1 & 0.023 & 0.055 \\
VTX & 3 & 0.039 & 0.187 \\
     & 1 & 0.038 & 0.073 \\
\end{tabular}
\end{center}
\end{table}

\begin{figure}
\centerline{
\epsfxsize=7.0cm\epsfbox{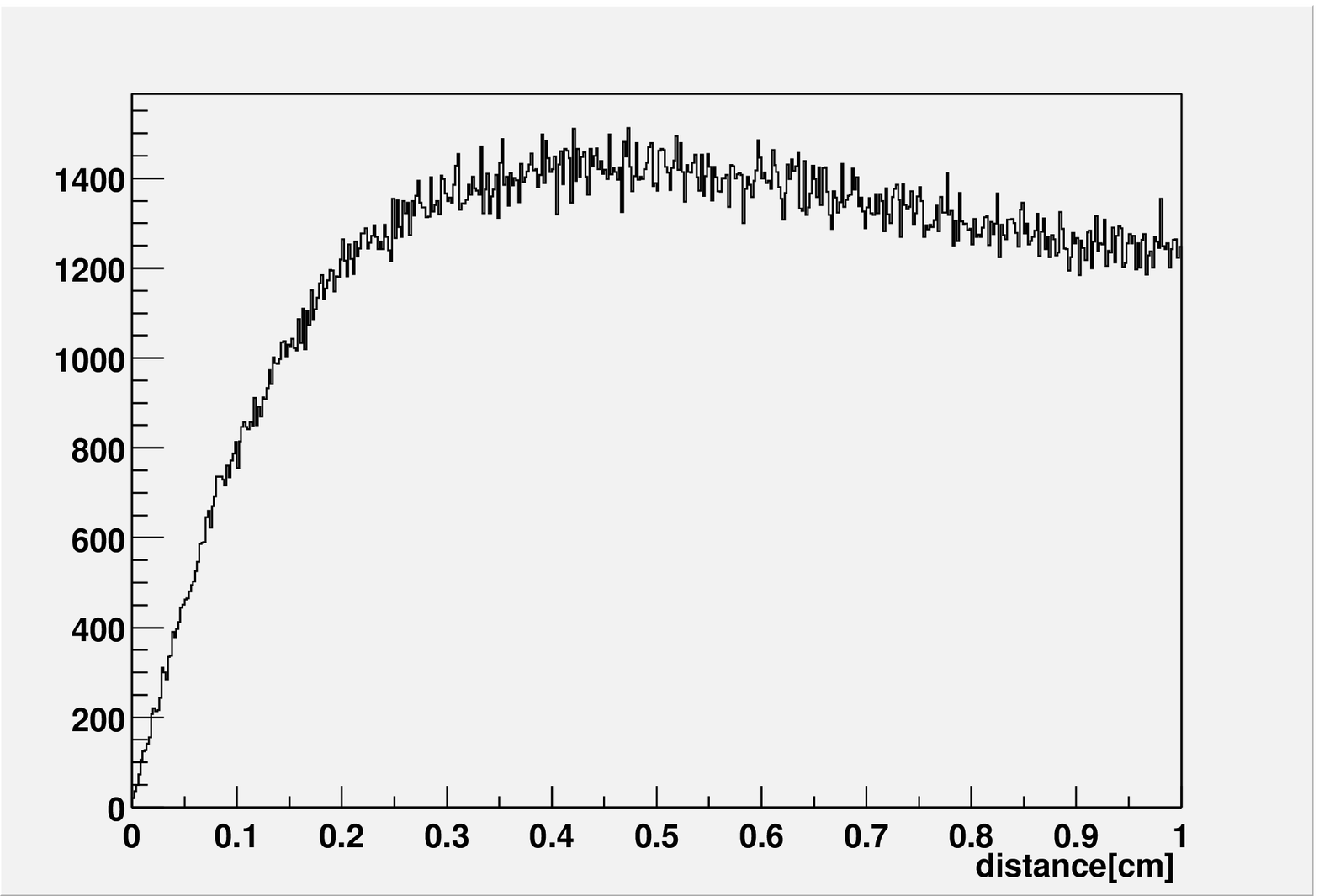}
\epsfxsize=7.0cm\epsfbox{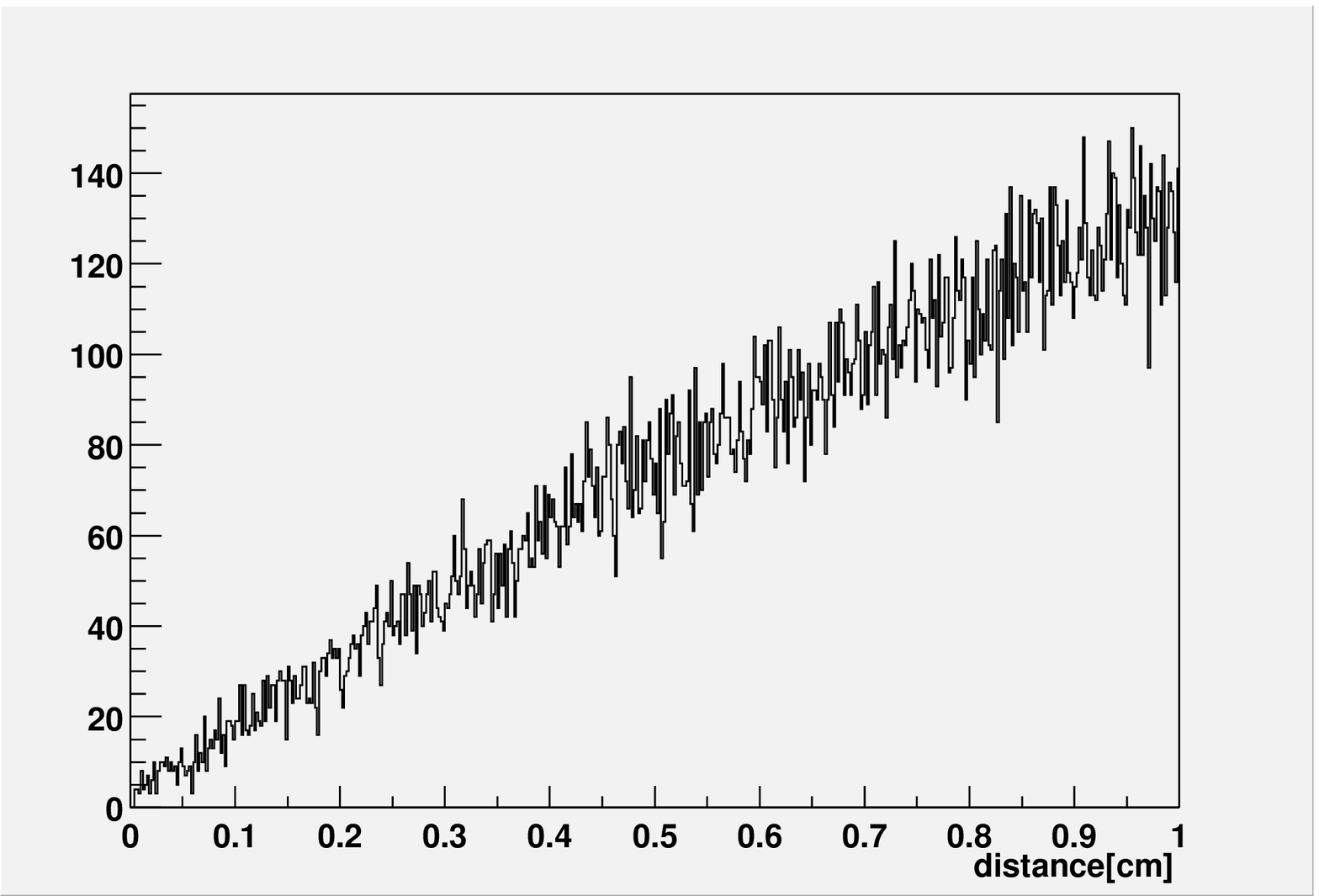}}
\begin{center}\begin{minipage}{\figurewidth}
\caption{\sl \label{imt-hitdensity}
The distance among hits of Higgs events
at (a) VTX position(r=6cm) and (b) IT position(r=37cm).}
\end{minipage}\end{center}
\end{figure}

The distance ($d$) among hits in Higgs events
is plotted in Fig.~\ref{imt-hitdensity},
at the outer-most VTX layer (r=6cm) and 
IT (r=37cm).  The events were generated at $\sqrt{s}=300$ GeV
with a Higgs mass of 120 GeV.
As seen in the figure, 
the numbers of hits near $d$=0 increase linearly with $d$, which 
indicates that the hit density is uniform locally.
From these figures we conclude that the density of hits 
at the IT position ($r=37$cm) is about 0.11 hits/mm$^2$, 
while that at the VTX position ($r=$6cm) is 0.66 hits/mm$^2$.
In the case of $q\bar{q}$ events at $\sqrt{s}$=500 GeV,
those are 0.06 hits/mm$^2$ at IT and 0.34 hits/mm$^2$ at VTX.
If hit densities are high, it is difficult to associate
hits with tracks extrapolated from CDC.  Residuals shown in 
Table~\ref{imt-table1} represent well that extrapolated CDC
tracks can resolve hits in IT or VTX.  
Assuming three times 
the residual value for the size of the window 
for unique hit-to-track association,
the average number of false hits in that window can be obtained
using the hit density.  Those are 0.035(0.016) hits at IT and
0.54(0.21) at VTX when $\sigma_z$ of CDC is 3mm(1mm) in the case 
of Higgs events.  Improvement of hit-track linking is clear
when IT is installed.

%% file: dettrk/cdc/main.tex
\section{Central Tracker\label{section-cdc}}
\newcommand{\figdir}{dettrk/cdc/figs}
\newcommand{\figdiry}{dettrk/cdc/figs}
\newcommand{\figdirgain}{dettrk/cdc/figs-gain}
\newcommand{\figdirlor}{dettrk/cdc/figs-lor}
\newcommand{\figtwotrack}{dettrk/cdc/figs-twotrack}
\input{dettrk/cdc/p1}
\input{dettrk/cdc/p2}

\input{dettrk/cdc/p3}

\input{dettrk/cdc/p4}
\input{dettrk/cdc/p5}

\input{dettrk/cdc/p7}

\input{dettrk/cdc/p8}

\input{dettrk/cdc/p6}

\input{dettrk/cdc/p6baby}

\input{dettrk/cdc/p11}

\input{dettrk/cdc/p12}

\input{dettrk/cdc/p10}

\input{dettrk/cdc/p9}

\input{dettrk/cdc/p13}

%% file: dettrk/cdc/p1.tex
\subsection{Overview of CDC}

We start our discussion with general design principle of a central 
tracker in order to translate the physics requirements to concrete 
chamber parameters.
A possible set of chamber parameters for the central tracker
is given 
according to the general design principle.
We then list up our R\&D items that include
mechanical and operational stability issues,
feasibility studies of required chamber performance,
and effects of space charge and magnetic field.
Finally we summarize our R\&D and discuss future
direction. 

%% file: dettrk/cdc/p2.tex
\subsection{Design principle}
\label{Sec:detector:tracker:cdc:design}

\subsubsection{Outline of Geometrical Parameters Decision Process}

The parameters that determine the geometrical configuration 
of a small-jet-cell cylindrical drift chamber can be classified
into two groups:
(1) global parameters that specify the size and layout of
axial and stereo super-layers, and
(2) local parameters that describe the geometry of a
jet cell in each super-layer.
The first class includes
the length,
the inner and outer radii, and
the stereo angle\footnote{
Strictly speaking, stereo angle,
which is defined as the wire angle measured from
the axial direction, 
slightly varies even within a single jet cell: 
it increases linearly with the radial position of a wire.
As discussed later, the stereo angle of the innermost
wire is the most important in deciding the stereo
geometry.
}
of each super-layer.
If there is a layer of shielding wires between
adjacent super-layers, its radius and stereo angle also
belong to this class.
The parameters in the second class are, essentially, 
the relative locations of wires that comprise a
single jet cell, and determine the local properties
of the chamber.
These parameters are optimized to achieve performance goals
such as required spatial resolution, two-hit separation
capability, etc, together with other non-geometrical
parameters including wire high voltages, gas mixture,
magnetic field, and so on.

\subsubsection{Basic Chamber Parameters in Terms of Physics Requirements}

Now let us outline a typical parameters decision process.
The chamber performance goals required by physics of interest
are often expressed in terms of transverse momentum resolution,
dip angle resolution, two-track separation capability, 
polar angle coverage, etc.
In our case the most stringent requirement comes from
the recoil mass measurement for $e^+e^- \to Zh$
followed by $Z \to l^+l^-$.
In order to confirm the narrowness of the Higgs boson,
the recoil mass against the lepton pair from the
$Z$ decay should be measured to an accuracy only
limited by the natural beam energy spread.
This sets the following performance goal for the
momentum measurement:
$\sigma_{p_{l^\pm}}/p_{l^\pm} \lsim 0.4~\%$ at
$p_{l^\pm} = 50~{\rm GeV/c}$.
Separation of $W$ and $Z$ through jet invariant mass measurements
demands a resolution comparable with
the gauge boson natural widths.
This necessitates good track-cluster matching:
$\sigma_Z \simeq 1~{\rm mm}$
and 2-track separation better than $2~{\rm mm}$.

The transverse momentum resolution $\sigma_{p_T}$ is approximately
given by the following formula\cite{Ref:sigma}:
\begin{equation}
   \frac{\sigma_{p_T}}{p_T} 
	\simeq \sqrt{ \left( \frac{\alpha' \sigma_{r\phi}}{B {l}^2} \right)^2
		 \left( \frac{720}{n+4} \right) {p_T}^2
	       + \left( \frac{\alpha' C}{B {l}} \right)^2
	         \left( \frac{10}{7}\left(\frac{X}{X_0}\right) \right)
	       },
\end{equation}
where 
\begin{eqnarray}
  \left\{
  \begin{array}{lll}
    \sigma_{r\phi} & = & \mbox{spatial resolution in the $r$-$\phi$ plane per wire} \cr
    n & = & \mbox{the number of sampling points} \cr
    \alpha' & =  & 333.56 ~({\rm cm}\cdot{\rm T}\cdot{\rm (GeV/c)}^{-1}) \cr
    C & = & 0.0141 ~({\rm GeV/c}) \cr
    (X/X_0) & = & \mbox{thickness measured in radiation length units } \cr
    {l} & = & \mbox{lever arm length (cm)} \cr
    B & = & \mbox{magnetic field (T)}  
  \end{array}
  \right. .
\end{eqnarray}  
The number of sampling points is practically determined by the lever arm
length and the minimum possible wire-to-wire distance.
The best result is thus expected for the largest possible
$B$ field and lever arm length with the smallest possible
spatial resolution.
Since the required polar angle coverage determines the wire length
when the lever arm length and the radial position of the innermost
wires are given, the lever arm length is limited by the maximum
possible wire length for stable chamber operation\footnote{
There are some external constraints on the radial positions of
the inner- and outermost wires:
the radial position of the innermost wires is constrained,
in the case of the present JLC design, by the size of the support
tube for the final focusing quadrupole magnets
that reside in the detector system.
The radial position of the outermost wires has to be
consistent with the surrounding barrel calorimeter.
}.
The $B$ field essentially determines the Lorentz angle
given a chamber gas mixture\cite{Ref:lorentz} 
and, if a tilted cell
structure is adopted, it imposes a limit on
the two-hit separation capability.
There will also be a technical limit to the maximum
possible $B$ field
attainable with a super-conducting solenoid 
from the mechanical strength
and the stored energy to be released upon quench.

On the other hand, the dip angle resolution $\sigma_{\tan\lambda}$ 
can be written in the form\cite{Ref:sigma}:
\begin{equation}
   \sigma_{\tan\lambda} 
	\simeq \sqrt{ \left( \frac{\sigma_z}{{l}} \right)^2
	              \left( \frac{12}{n} \right)
	       + \left( 1 + \tan^2\lambda \right)
	         \left( \frac{C}{p_T} \right)^2
	         \left( \frac{13}{35} \left( \frac{X}{X_0} \right) \right)
	       },
\end{equation}
where $\sigma_z$ is the spatial resolution per wire in the axial direction
which is given by
\begin{equation}
   \sigma_z = \sigma_{r\phi}/\tan\alpha
\end{equation}
in the case of stereo wires with a stereo angle $\alpha$.
Provided that ${l}$ and $B$ are predetermined by the requirement
on the transverse momentum resolution, the only free parameter
left here is the stereo angle.
Apparently the larger the stereo angle becomes the better the
dip angle resolution.
There are, however, various restrictions to the stereo angle.

The two-hit separation capability is controlled by the width of
a sense wire signal which is determined by the arrival time distribution of 
drift electrons, space charge effects in the avalanche formation,
and the subsequent readout electronics.
Apart from the optimization of the readout electronics,
appropriate choices of the drift line configuration and chamber gas
are thus important.
In any case, the relation of the two-hit separation capability issue
to the stereo geometry is at most indirect and can be treated
separately as a problem concerning 
local properties of the chamber\cite{Ref:two-track}.

\subsubsection{Stereo Wires and Super-layer Structure}

\begin{figure}[htb]
\begin{minipage}[htb]{7cm}
\centerline{
\epsfxsize=6cm 
\epsfbox{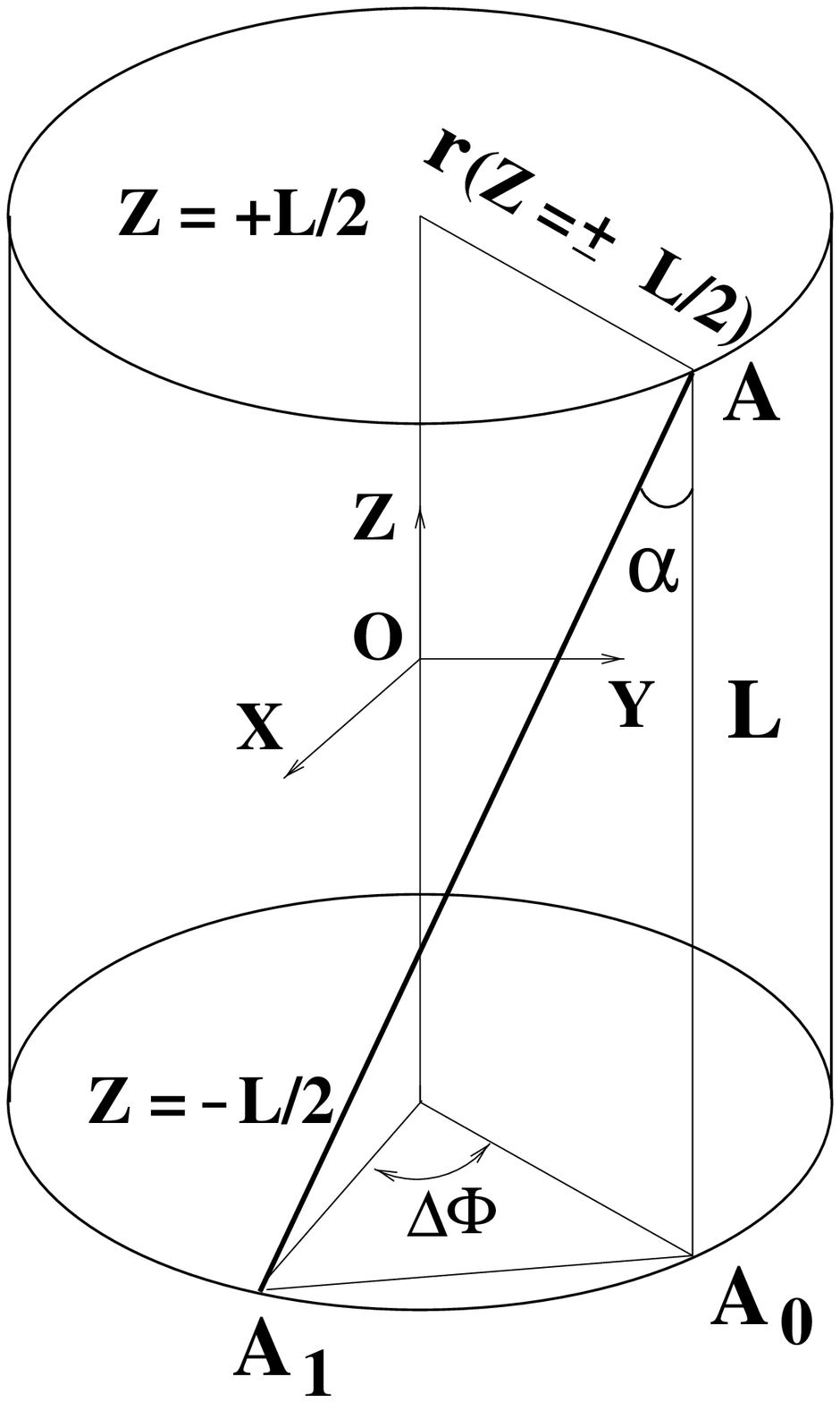}
}
\caption[Fig:3d-stereo-wire]{\label{Fig:3d-stereo-wire}\sl
		3-dimensional view of a single stereo wire.
}
\end{minipage}
\hfill
\begin{minipage}[htb]{7cm}
\centerline{
\epsfxsize=6cm 
\epsfbox{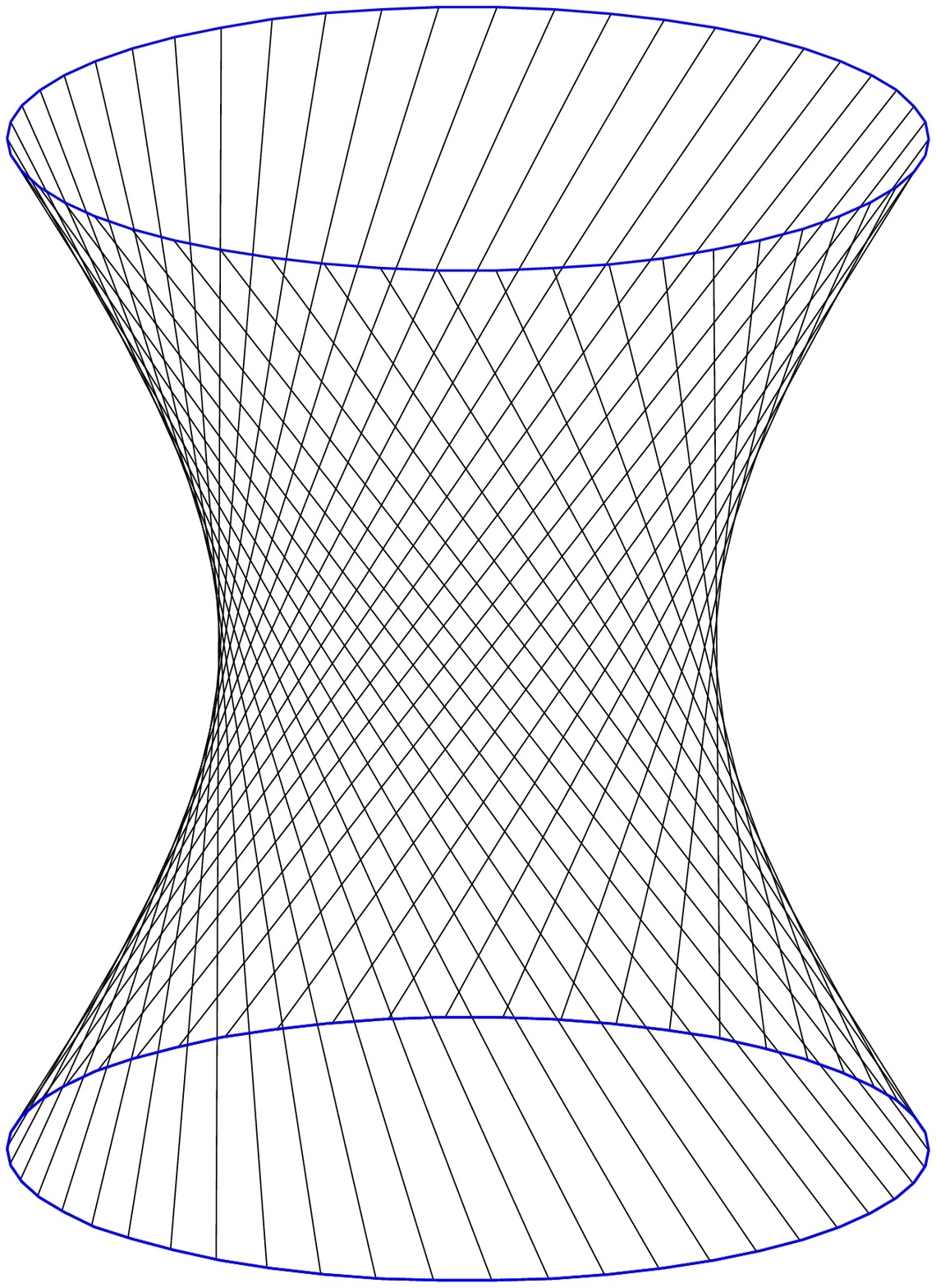}
}
\caption[Fig:hyperboloid]{\label{Fig:hyperboloid}\sl
		Formation of a hyperboloidal surface due to
		stereo wires.
}
\end{minipage}
\end{figure}

At this point we have a rough idea about basic parameters
describing the chamber: the inner and outer radii, the length,
the number of sampling points, and the $r\phi$ spatial resolution 
per wire.
In order to decide the stereo geometry, we further need to know
the super-layer layout of the chamber and problems introduced
by the use of finite stereo angles.
A stereo layer can be formed by rotating the end point
of each wire belonging to this layer by a small angle ($\Delta\phi$)
with the other end point fixed.
As illustrated in Fig.\ref{Fig:3d-stereo-wire},
the stereo angle $\alpha$ is related to $\Delta\phi$ 
and the wire length ($L$) through
\begin{equation}
   \label{Eq:alpha(r)}
   { \everymath{\displaystyle}
   \begin{array}{lll}
   	\alpha & = & \tan^{-1}\left( \frac{2r(z=\pm L/2)}{L} 
   		\sin\left(\frac{\Delta\phi}{2}\right) \right) \cr
   		\rule{0in}{5.0ex}
   	     ~ & \simeq & \frac{r(z=\pm L/2)}{L} \cdot \Delta\phi  ,
   \end{array}
   }
\end{equation}
which suggests $\alpha$ being $r$-dependent even within a single
stereo jet cell as mentioned earlier.

The otherwise cylindrical surface of this wire layer turns
into a hyperboloidal surface as shown in Fig.\ref{Fig:hyperboloid}.
As a result, both the azimuthal angle and the radial position
of the stereo wire becomes $z$-dependent:
\begin{equation}
\label{Eq:phi(z)}
{ \everymath{\displaystyle}
\begin{array}{lll}
   \phi(z) & = & \phi(z=0) 
           + \tan^{-1} \left[ \left(\frac{2z}{L}\right)
                               \tan\left(\frac{\Delta\phi}{2}\right)
                       \right] \cr
   \rule{0in}{5.0ex}
\label{Eq:r(z)}
   r(z) & = & \sqrt{ \left( r(z=0) \right)^2 + \left( z\tan\alpha \right)^2 } \cr
   \rule{0in}{4.0ex}
   ~ & = & \sqrt{ \left( r(z=\pm L/2) \right)^2 
   	+ \left( z^2 - (L/2)^2 \right) \tan^2\alpha },
\end{array}
} 
\end{equation}
where $z$ is measured from the middle of the chamber along the
chamber axis
\footnote{
The azimuthal wire position becomes a linear function,
and the radial wire position a quadratic function of $z$ measured
from the middle of the chamber:
\begin{equation}
  \left\{
  \begin{array}{lll}
   \phi(z) & \simeq & \phi(z=0) + \left( \frac{\alpha z}{r(z=0)} \right) \cr
   \rule{0in}{3.0ex}
   r(z)    & \simeq & r(z=0) + \frac{1}{2}\left(\frac{(\alpha z)^2}{r(z=0)}\right)
  \end{array}
  \right.
\end{equation}
to the lowest order of $\alpha z$.
}.

\begin{figure}[htb]
\centerline{
\begin{minipage}[htb]{12cm}
\centerline{
\epsfxsize=6cm 
\epsfbox{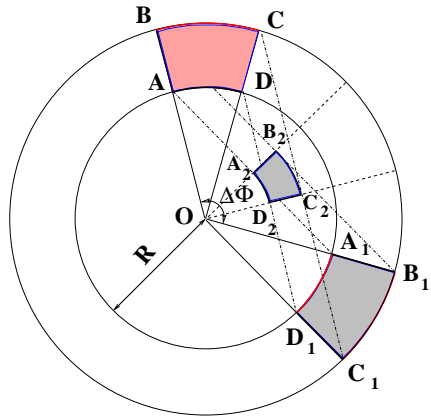}
}
\caption[Fig:stereo-cell]{\label{Fig:stereo-cell}\sl
Exaggerated illustration of the stereo-cell formation through
rotation by an angle $\Delta\phi$ and its deformation along
the wires.
$ABCD$ and $A_1B_1C_1D_1$ are the cross sections of a single stereo cell
at the end plates $(z=\pm L/2)$.
$A_2B_2C_2D_2$ is that of the same stereo cell in-between
$(-L/2~<~Z<~+L/2)$.
}
\end{minipage}
}
\end{figure}

This inhomogeneity distorts the electric field to some extent
and limits the maximum possible stereo angle, thereby
restricting the $z$ resolution $\sigma_z$.
Fig.\ref{Fig:stereo-cell} demonstrates how a single stereo cell
structure deforms as one moves from one endplate to the other.
This $z$-dependent geometrical deformation of the stereo cell
is parametrized by a so called shrink factor and
imposes another important restriction on the
possible range of stereo angles.

The shrink factor $f(z)$ is defined to be
the ratio $A_2B_2(z)/AB(z=\pm L/2)$ in Fig.\ref{Fig:stereo-cell}
and is readily derived from
Eqs.\ref{Eq:alpha(r)} and \ref{Eq:r(z)}:
\begin{equation}
\label{Eq:f(z)}
{ \everymath{\displaystyle}
\begin{array}{lll}
	f(z) & = & \frac{r_B(z)-r_A(z)}{r_B(z=\pm L/2)-r_A(z=\pm L/2)} \cr
   \rule{0in}{7.0ex}
	& = &  \sqrt{1 + \left(\frac{z^2 - (L/2)^2}{L^2}\right)4\sin^2\left(\frac{\Delta \phi}{2}\right)} \cr
   \rule{0in}{7.0ex}
	& = & \frac{\sqrt{R^2+\left(z^2 - (L/2)^2\right)\tan^2\alpha}}{R},
\end{array}
}
\end{equation}
where $R=r(z=\pm L/2)$ and $f(z=\pm L/2) = 1$ by definition.

It is a common practice to use pairs of stereo super-layers ($V$ and $U$)
with alternating stereo angles ($\pm \alpha$) interleaved with 
axial super-layers ($A$)\footnote{
The axial super-layers greatly facilitate track finding in
the $r$-$\phi$ projection, since stereo super-layers alone
cannot provide any absolute coordinate before its $z$ coordinate
is determined through tracking.
There is, however, a logical possibility to do away with
axial layers.
}.
In this case, the inhomogeneity is more serious for the gap between
a stereo super-layer ($V$) and 
the adjacent axial super-layer ($A$) just inside.
The minimum gap attained at the middle point of the chamber
has to be large enough for stable operations of the chamber.

%% file: dettrk/cdc/p3.tex
\subsection{A Possible Central Tracker}
\label{Sec:detector:tracker:cdc:jlccdc}

\subsubsection{Chamber Design}

\begin{figure}[htb]
\begin{minipage}[htb]{7cm}
\centerline{
\epsfxsize=6cm 
\epsfbox{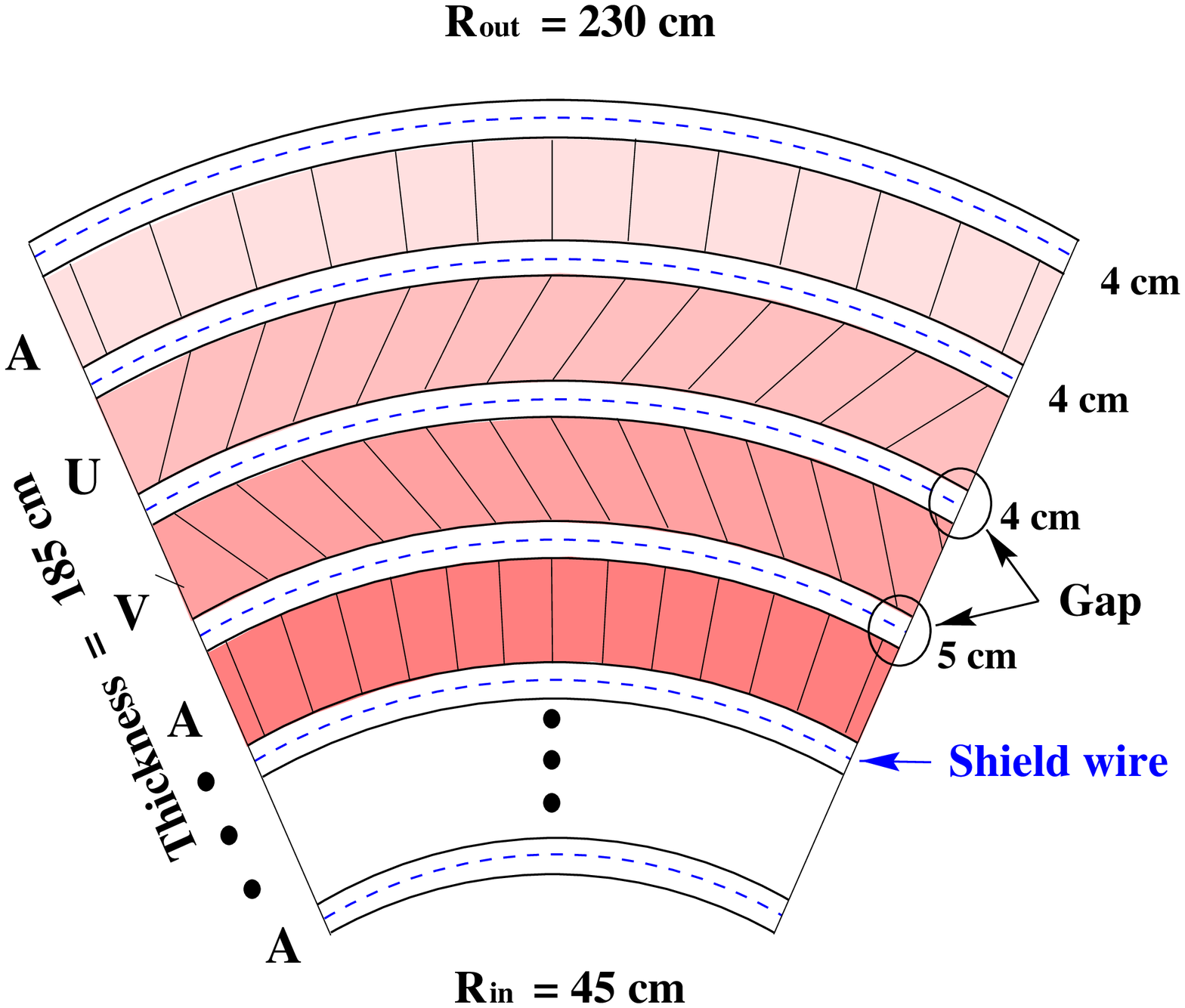}
}
\caption[Fig:jlc-cdc]{\label{Fig:jlc-cdc}\sl
		A possible super-layer layout for the central tracking 
		system of the JLC.
}
\end{minipage}
\hfill
\begin{minipage}[htb]{7cm}
\centerline{
\epsfxsize=6cm 
\epsfbox{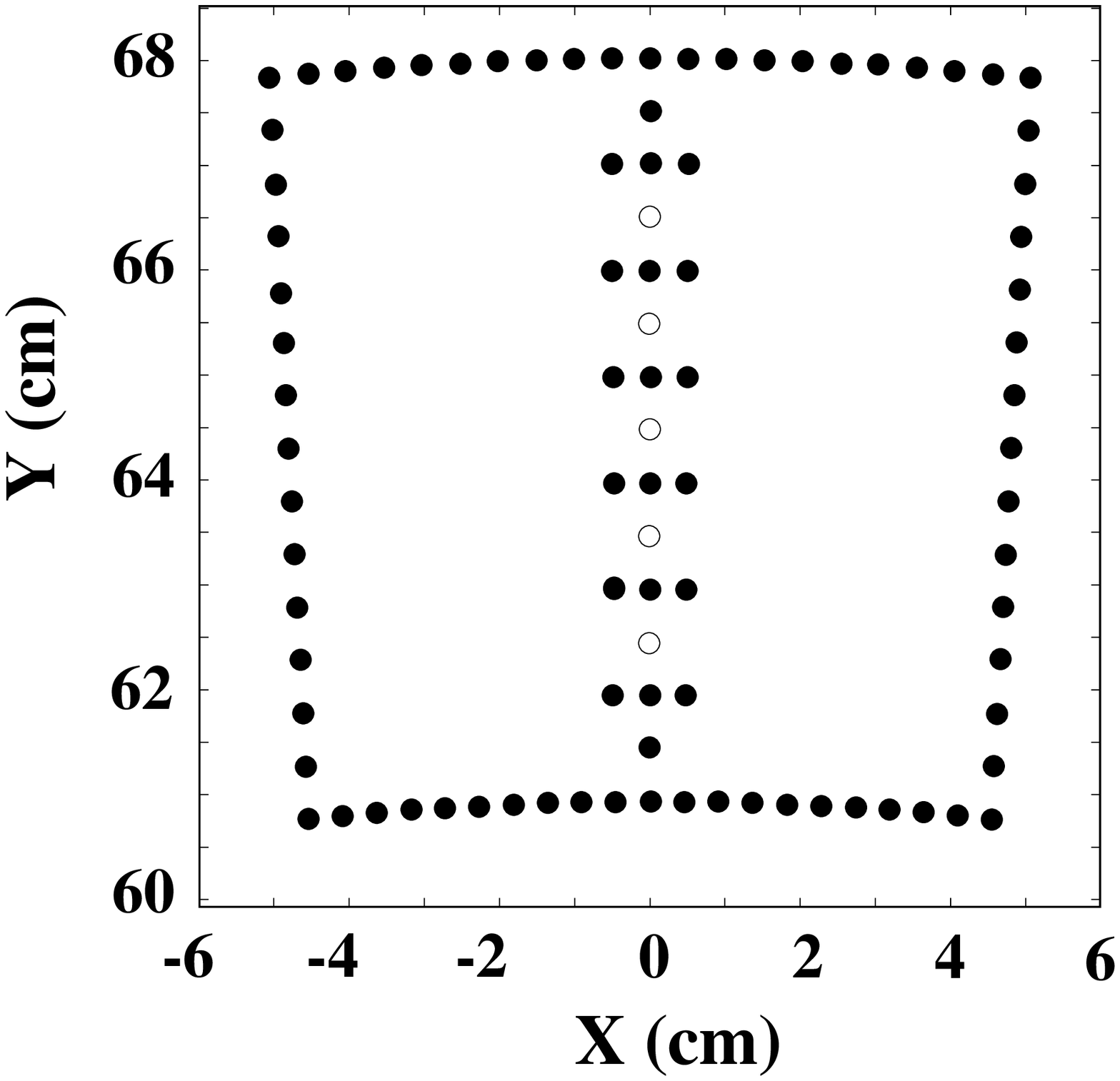}
}
\caption[Fig:jlc-cell]{\label{Fig:jlc-cell}\sl
A typical geometry of a single cell:
the circles ($\circ$) and  ($\bullet$) correspond to sense
and    potential wires, respectively.
}
\end{minipage}
\end{figure}

Applying the general design principle explained above to the JLC case,
we obtained the following 
basic parameters for a possible JLC central tracker:
$B=2~T$, $r_{in}=45~{\rm cm}$, 
$r_{out}=230~{\rm cm}$, $L=460~{\rm cm}$, 
$n~=~80$, 
and $\sigma_{r\phi}=100~\mu{\rm m}$.
The tracker consists of 6 axial and 10 stereo super-layers,
each having 5 anode wires, 
arranged with increasing $r$ in the following order:
$AVUAVUAVUAVUA$,
where $A$, $V$, and $U$ again denote
an axial super-layer, stereo super-layers with
stereo angles of $+\alpha$ and $-\alpha$, respectively.
The gap between adjacent super-layers is
correlated with the stereo angle and is equal to 
either $4.0$ or $5.0~{\rm cm}$ at the endplates, 
depending on its radial location:
the $U$-to-$A$ and the $V$-to-$U$ gaps are equal to $4.0~{\rm cm}$,
while the $A$-to-$V$ is set at $5.0~{\rm cm}$
(see Fig.\ref{Fig:jlc-cdc})\footnote{
We can further optimize the inter-super-layer distances.
}.

In the current design, there is a layer of shielding wires
in each gap between adjacent super-layers,
in order to electrostatically isolate different super-layers.
These shielding wires in a $V$-to-$U$ gap
will have the same stereo angle ($\alpha$) with the $V$ wires.
On the other hand, the shielding wires in
$A$-to-$V$ or $U$-to-$A$ gaps will have half the stereo
angle of the adjacent stereo super-layers ($\pm \alpha/2$).
Note also that the stereo angle varies even within a
single jet cell, according to Eq.\ref{Eq:alpha(r)}.

A possible cell design is shown in Fig.~\ref{Fig:jlc-cell}.
The cell has $5$ sense wires spaced $1~{\rm cm}$, with each
surrounded by ground wires
in the sense wire plane and
grid wires placed $5~{\rm mm}$ away from it.
The two dummy sense wires 
at the both ends of the sense wire plane
are placed to stabilize the upper- and the
lower-most sense wires.
The maximum drift length 
depends on the radial position of each sense wire and has a typical value of 
$5~{\rm cm}$.
 The drift field is produced by $10$ field shaping wires
whose potentials change linearly and cathode wires that cover
the end of the drift region.
The high voltages on the
wires will be chosen so as to get as uniform a drift field as possible
with minimum electrostatic forces,
while keeping a reasonable gas gain for a chamber gas mixture of
$90\%  {\rm CO_2}$ and $10\%$ Isobutane.

%% file: dettrk/cdc/p4.tex
\subsubsection{Questions to be Answered}
The chamber parameters described above lead us to the
following list of questions:
\begin{enumerate}
\item 	Can we control gravitational and electrostatic
sags for 4.6m-long wires?
\item	Is tension drop controllable for $Al$ wires?
\item	Can we stably operate 4.6m-long stereo cells?

\item	Can we achieve a spatial resolution of $100\mu{\rm m}$
everywhere in this big chamber?
\item	Can we achieve 2-track separation better than $2{\rm mm}$?

\item	Is gas gain saturation observed for cool gas mixtures no problem?
\item	Is Lorentz angle small enough to allow straight cells?
\item	Is neutron background 
($\sim$2k hits/train for $B=2$T\cite{Ref:beamBG}) no problem?
How big is the signal charge for a single neutron hit?
Is there any dead time expected?

\end{enumerate}
These questions naturally comprise our R\&D items.
\vskip 1.0cm
Let us begin with the first two questions which have to do with mechanical
stability of the chamber.

%% file: dettrk/cdc/p5.tex
\subsubsection{Mechanical Stability}

\begin{figure}[htb]
\begin{minipage}[htb]{6cm}
\centerline{
\epsfxsize=6cm 
\epsfbox{\figdir/chamb1.epsf}
}
\caption[Fig:chamb]{\label{Fig:chamb}\sl
$4.6~{\rm m}$ CDC test chamber with
wire position measurement system.
}
\end{minipage}
\hfill
\begin{minipage}[htb]{6cm}
\centerline{
\epsfxsize=6cm 
\epsfbox{\figdir/end_cell.epsf}
}
\caption[Fig:cells]{\label{Fig:cells}\sl
Cell layout of our $4.6~{\rm m}$ long test
chamber.
}
\end{minipage}
\end{figure}
\noindent
In order to investigate mechanical stability of a drift chamber with very 
long wires, we have built a $4.6~{\rm m}$ long test chamber 
(see Fig.~\ref{Fig:chamb}).
We  have also been   developing  a  software   program  to  predict  wire
displacement,  knowing  the cell  structure,  wire  tension,  and high
voltage on each wire.
The validity of the prediction has to be proved
experimentally.
We have  developed a wire  tension
and position  measurement  system and
measured  wire  tensions,
gravitational  and electrostatic  sags for the $4.6~{\rm m}$
long test chamber.
The results of the measurements were then compared with
the predictions.

Our $4.6~{\rm m}$ long test chamber has seven
identical cells with the cell parameters which approximate those of the 
aforementioned possible JLC central tracker.
The layout of the seven cells is
shown in Fig.~\ref{Fig:cells}.

\begin{figure}[htb]
\centerline{
\epsfysize=6cm 
\epsfbox{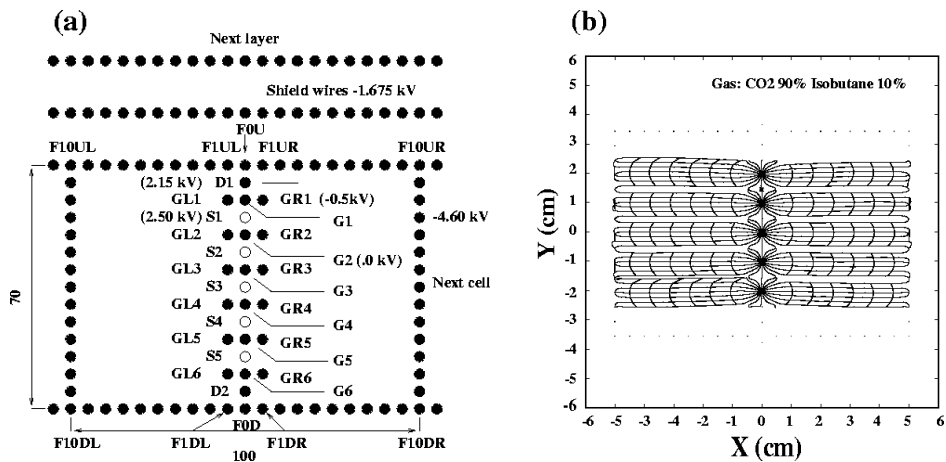}
\hspace{0.5cm}
\epsfysize=6cm 
\epsfbox{\figdir/garf-dr.epsf}
}
\caption[Fig:cell]{\label{Fig:cell}\sl
(a) possible small jet cell structure and
(b) corresponding drift lines ($B$ = 0 T).
}
\end{figure}

The cell design is shown in Fig.~\ref{Fig:cell}-(a)
which also specifies our wire naming convention to be
used hereafter.
The cell has $5$ sense wires ($S1$ to $S5$) held at $2.5~{\rm kV}$ 
and spaced $1~{\rm cm}$, with each
surrounded by ground wires ($G1$ to $G6$)
in the sense wire plane and
grid wires ($GL1$ to $GL6$ and $GR1$ to $GR6$)
at $-0.5~{\rm kV}$ placed $5~{\rm mm}$ away from it.
The two dummy sense wires ($D1$ and $D2$)
at the both ends of the sense wire plane
are held at $2.15~{\rm kV}$ and stabilize the upper- and the
lower-most sense wires ($S1$ and $S5$).
The maximum drift length is $5~{\rm cm}$ where the
drift field is produced by $10$ field shaping wires
($F1**$ to $F10**$) whose potentials change linearly in this order
from $-0.5~{\rm kV}$ to $-4.6~{\rm kV}$, and by 
left and right edge field wires held at $-4.6~{\rm kV}$.
The wire diameter is $30~\mu{\rm m}$ for sense wires and  $125~\mu{\rm m}$
for the others.
The high voltages on the
wires were chosen so as to get as uniform a drift field as possible
with minimum electrostatic forces,
while keeping a reasonable gas gain for a chamber gas mixture of
$90\%  {\rm CO_2}$ and $10\%$ Isobutane.
Fig.~\ref{Fig:cell}-(b) plots
the drift lines and isochrones calculated with
a drift chamber simulation program GARFIELD~\cite{Ref:garfield}.
Unless otherwise stated, wires we are going to deal with belong to 
the central cell ($C4$).

Before stringing wires, we surveyed the endplates
to check their machining precision and
found that the standard deviation of the
actual wire hole positions from the nominal ones
to be less than $8.2~\mu{\rm m}$.
The sense wires are gold-plated tungsten and
all the other wires are gold-plated aluminum.
The reason for this choice of Al 
as the wire material is to minimize the 
total wire tension so as to reduce
the thickness of the chamber endplates as
required by endcap calorimeters. The wire stringing took us three weeks.

\begin{flushleft}
{\bf gravitational and electrostatic sags} 
\end{flushleft}

\begin{figure}[htb]
\centerline{
\epsfxsize=6.cm 
\epsfbox{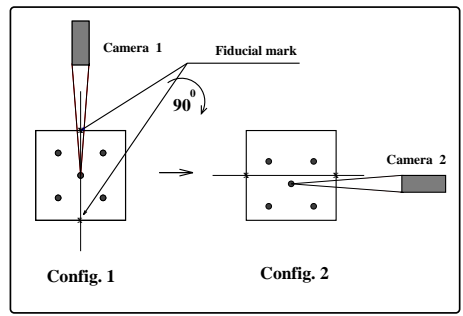}
}
\caption[Fig:grav1]{\label{Fig:grav1}\sl
Principle of  gravitational sag measurements.
}
\end{figure}
\noindent
Fig.~\ref{Fig:grav1} illustrates our measurement scheme of
the gravitational sag of a wire.
The  measurement goes as follows.  
Using the vertical camera, 
we first put fiducial marks on the  chamber  windows
in such a way that the line connecting them passes through
the wire in question (configuration 1 in Fig.~\ref{Fig:grav1}).
Then we rotate the test  chamber by  90$^\circ$
and determine the wire deflection due to  gravitational  sag 
by measuring the wire position relative to
the line  connecting  the fiducial marks,
using the horizontal camera (config.2 in Fig.~\ref{Fig:grav1}).   
We measured the gravitational sags 1.5 years after the wire stringing.  
The  average  gravitational  sags for  field  and
sense wires were 600$\mu$m and 353$\mu$m, respectively.

\begin{figure}[tp]
\centerline{
\epsfxsize=10cm 
\epsfbox{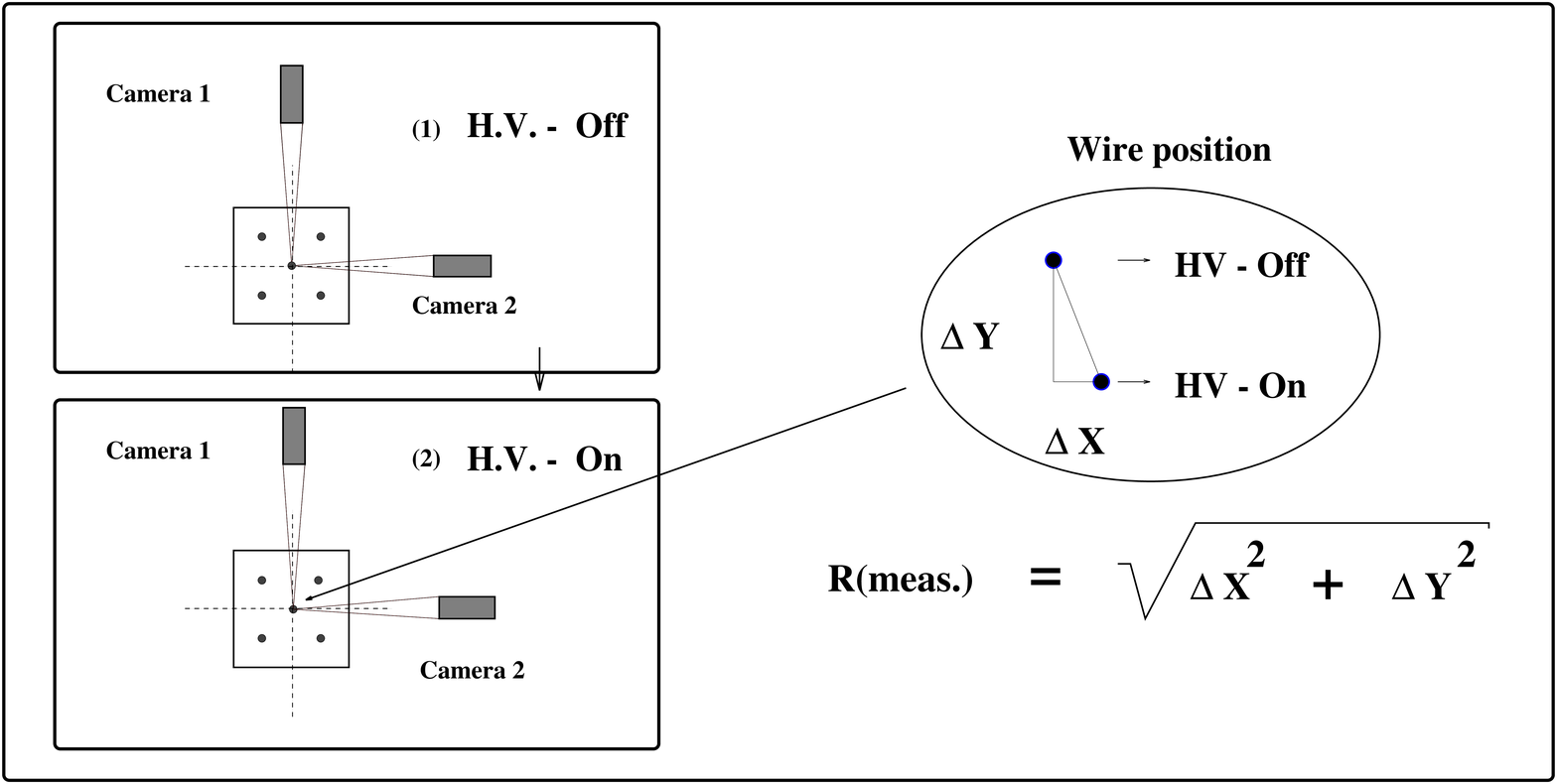}}
\caption[Fig:elsag]{\label{Fig:elsag} \sl
Schematic view of the electrostatic sag measurements.
}
\end{figure}

Our measurement scheme of electrostatic sag
is  sketched  in  Fig.~\ref{Fig:elsag}.  
We first  get a wire in
sight on both the  horizontal  and  vertical  cameras  with high
voltages off.
Then we measure the horizontal ($\Delta$X) and vertical
($\Delta$Y) displacements in the camera screens after turning on the
high voltages.
Notice that the camera stays at rest during the measurement and the 
measurement accuracy is determined completely by the wire image resolutions
on the screens (there is no tilt angle effect here). 
The overall accuracy 
for the electric sag measurement was estimated to be $5~\mu{\rm m}$.
The wire stability in the electrostatic field has been
checked by turning  on and off the high  voltages  several  times.  
We found  no   instability, meaning that the wires always return to the same
positions of equilibrium. 

\begin{figure}[p]
\centerline{
\epsfxsize=10cm 
\epsfbox{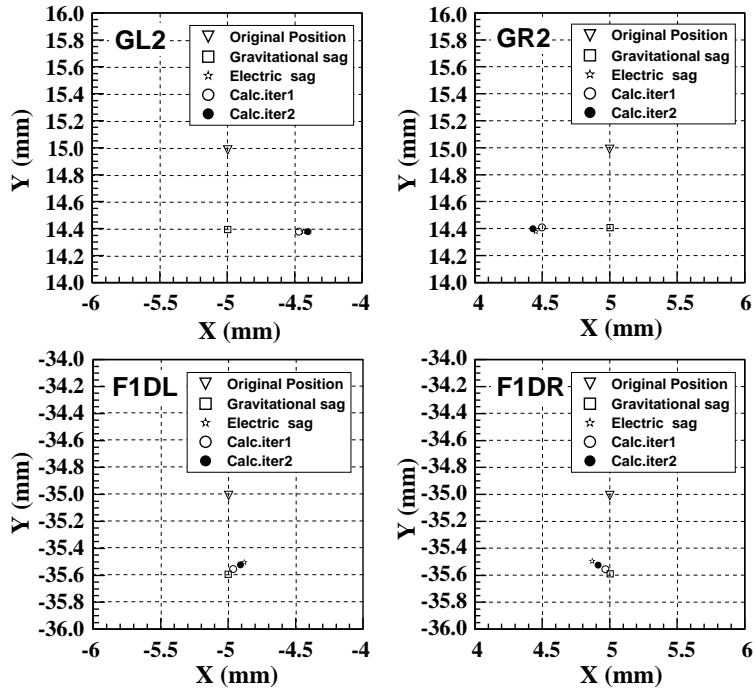}}
\begin{center}\begin{minipage}{\figurewidth}
\caption[Fig:iwate-gf1]{\label{Fig:iwate-gf1} \sl
 Measured positions of some representative field wires 
at the endplates ($\bigtriangledown$), with gravitational sags (\framebox[12pt][
l]{$\;\;$})
before and ($\star$) after electrostatic sags, compared with calculations:
iterations $1$($\circ$) and $2$ ($\bullet$). 
}
\end{minipage}\end{center}
\end{figure}

\begin{figure}[p]
\centerline{
\epsfxsize=10cm 
\epsfbox{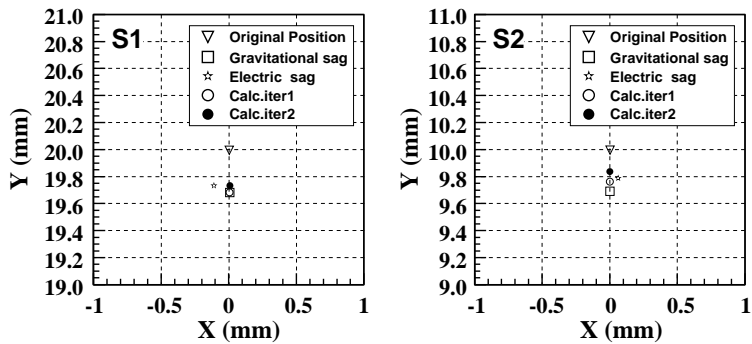}}
\caption[Fig:iwate-s]{\label{Fig:iwate-s} \sl
Same as Fig.~9 for the worst ($S1$) and a typical ($S2$) sense wires.
}
\end{figure}

Figs.\ref{Fig:iwate-gf1}  and \ref{Fig:iwate-s} illustrate
the wire deflections in
our $4.6~{\rm m}$ test chamber
due to gravitational and  electrostatic  forces.
In the figures an {\it original position} ($\bigtriangledown$) is 
the initial position of a wire at the end plates.
On the other hand, denoted as
{\it gravitational sag} (\framebox[12pt][l]{$\;\;$})
and {\it electric sag} ($\star$) are the measured positions
of the wire under the  gravitational  force {\it before}
and {\it after} applying high voltages, respectively.
Called {\t first iteration} ($\circ$) 
is the calculated value of the wire deflection in
the  electrostatic  field when other wires are at the positions  after
the gravitational  sags.
The  input  parameters  for the  second  iteration ($\bullet$) are the 
 new  wire
positions from iteration 1.

In Fig.\ref{Fig:iwate-gf1}  we compare the measurements and predictions
for a few field wires which are on the left and right  sides of the
sense wire plane.  
There is a good agreement not only in the magnitude
of  deflection  but  also  in its  direction for each one of these field wires.  
It is  clear  that  the
prediction from the second iteration is better.

\begin{figure}[htb]
\centerline{
\epsfxsize=5cm 
\epsfbox{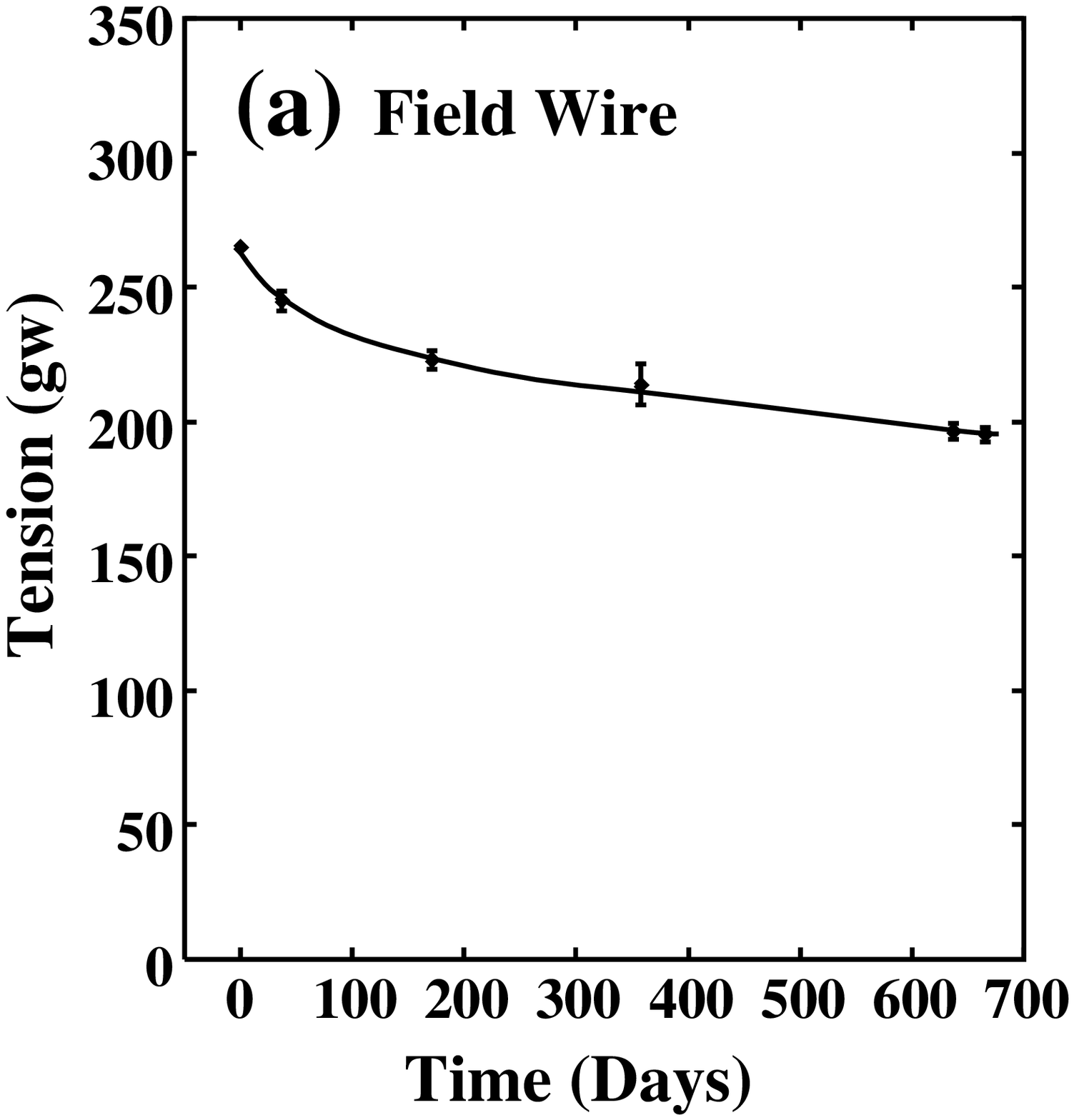}
\hspace{0.5cm}
\epsfxsize=5cm 
\epsfbox{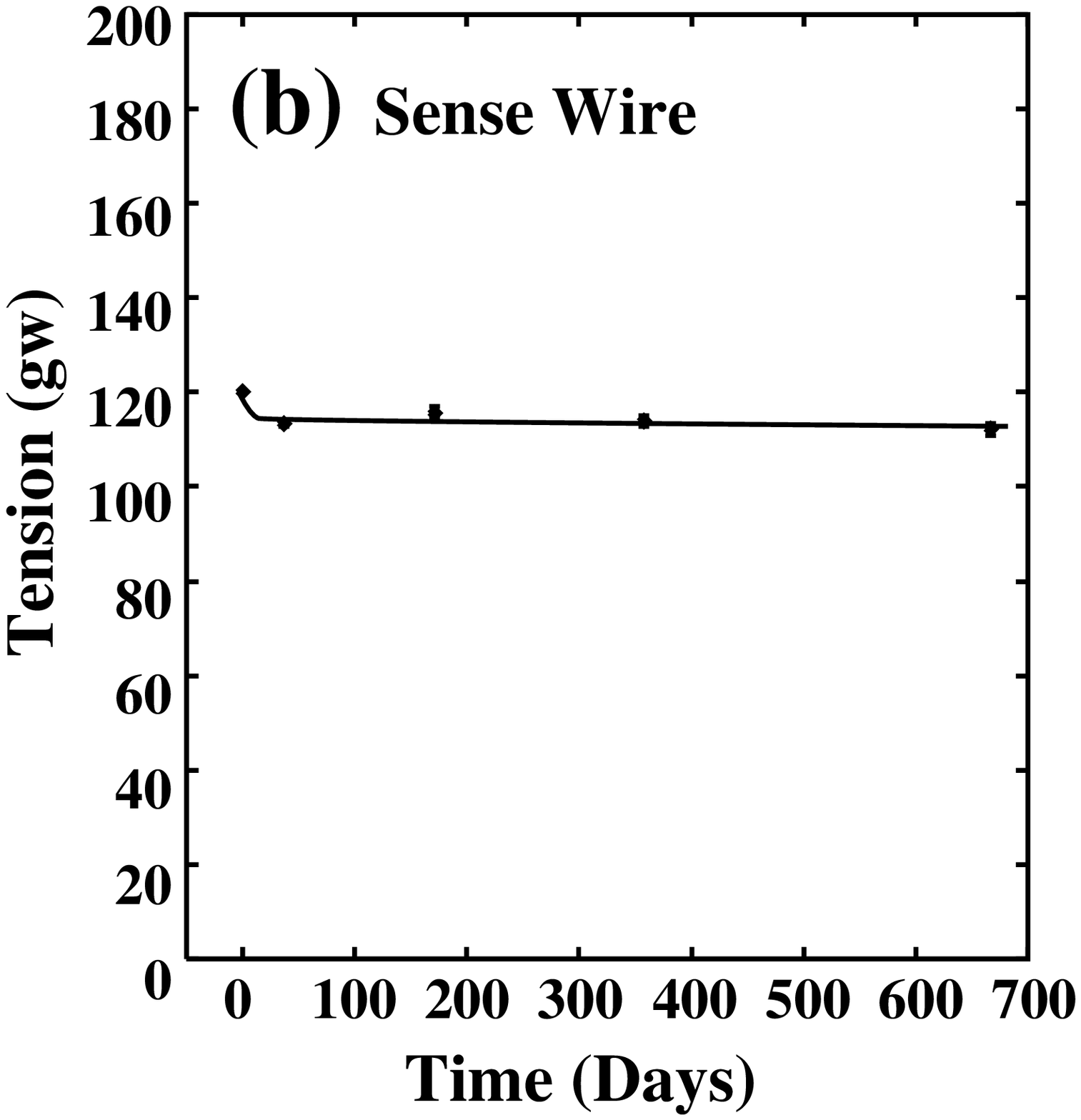}
}
\caption[Fig:tens2]{\label{Fig:tens2} \sl
Tension versus time for (a) field and (b) sense wires.}
\end{figure}

Results of the predictions and  measurements 
for sense wires $S1$ and $S2$
are  shown  in  Fig.\ref{Fig:iwate-s}.  
There  is a noticeable left-right
asymmetry in the measurements
with  respect  to  the   original   sense  wire  plane,
which  could  not  be  reproduced  by the  calculation.
Probably  the source of such an  asymmetry  can be  attributed  to the
influence of the other cells which might have some asymmetry in wire tensions,
since they were strung over a finite period of time.
\begin{flushleft}
{\bf wire tension drop problem}
\end{flushleft}
\label{Sec:tension}
\noindent
Almost all methods of wire tension  measurements are based on the same
basic  principle~\cite{Ref:tension} and ours is no exception.
We excite oscillations  of a wire at its resonant frequency, 
and then determine the tension $T$ by using the following relation:
\begin{equation}
  T =   4 \rho \cdot L^2 \cdot f^2 ,
\label{eq:first}
\end{equation}
where  $f$ is the  fundamental  resonant
frequency.  
The results are shown in Figs.~\ref{Fig:tens2}-(a)
and -(b) for  field  and  sense  wires,  respectively,
as functions of the time measured from the day of
wire stringing.
There is still a
tension drop of about  10\%/year for gold-plated Al field wires unlike
gold-plated W sense wires which now reach a tension plateau.
Since this poses a potential problem and requires a better
wire material, we are continuing wire material studies.
\vskip 1.0cm
We now move on to question 3 which is the operational stability issue 
discussed in detail in~\cite{Ref:stereo} \cite{Ref:gasgain}.

%% file: dettrk/cdc/p7.tex
\subsubsection{Operational Stability}

{\bf surface field  variation due to stereo wires}
\label{Sec:jlc-example}
\\

\begin{figure}[htb]
\begin{minipage}[htb]{7cm}
\centerline{
\epsfxsize=7cm 
\epsfbox{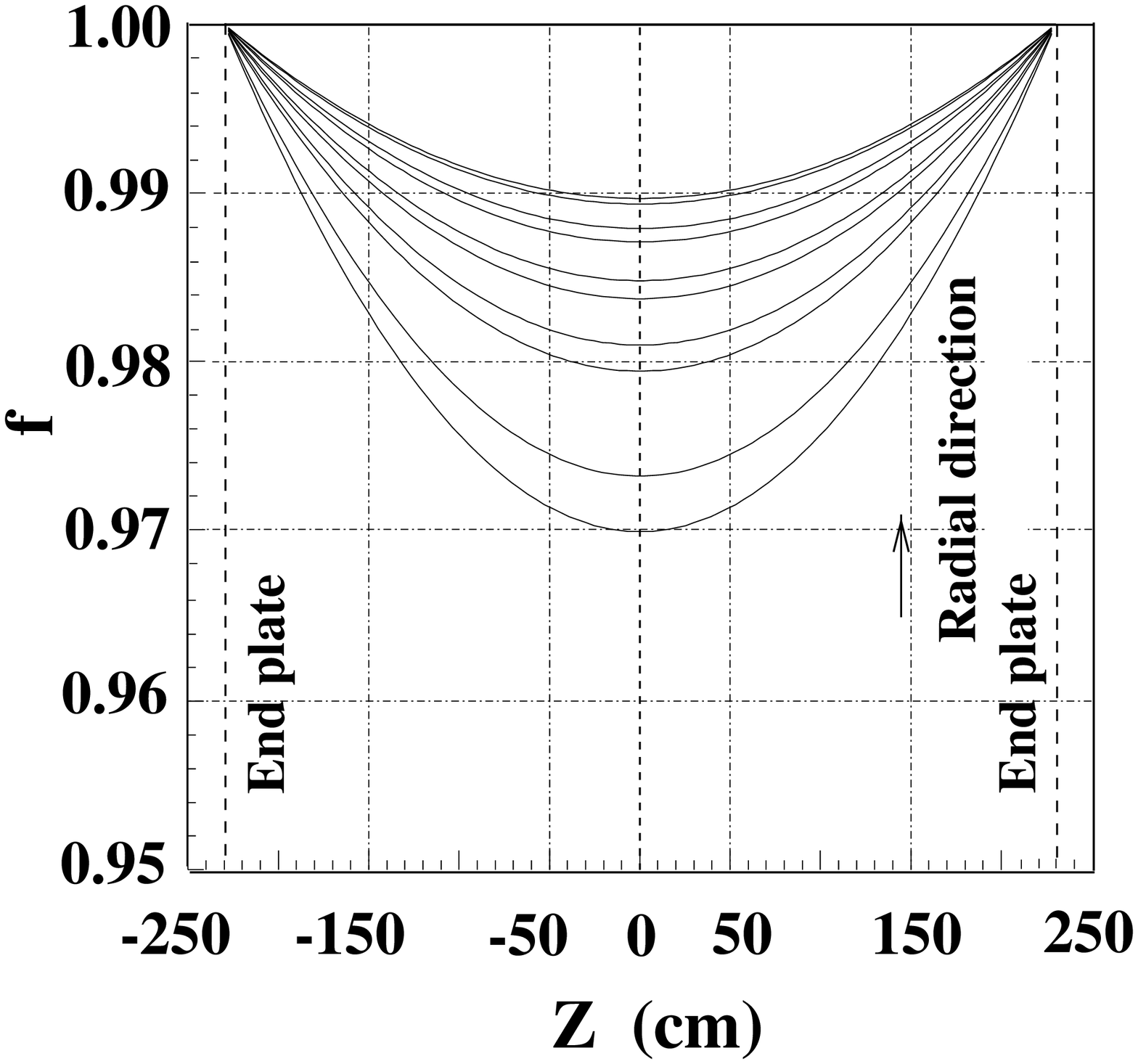}
}
\caption[Fig:f(z)]{ \label{Fig:f(z)}\sl
Shrink factor ($f$) as a function of $z$ for 
all of the 10 super-layers:
from the bottom to the top, the lines correspond to
super-layers ordered in the radial direction.
}
\end{minipage}
\hfill
\begin{minipage}[htb]{7cm}
\centerline{
\epsfxsize=7cm 
\epsfbox{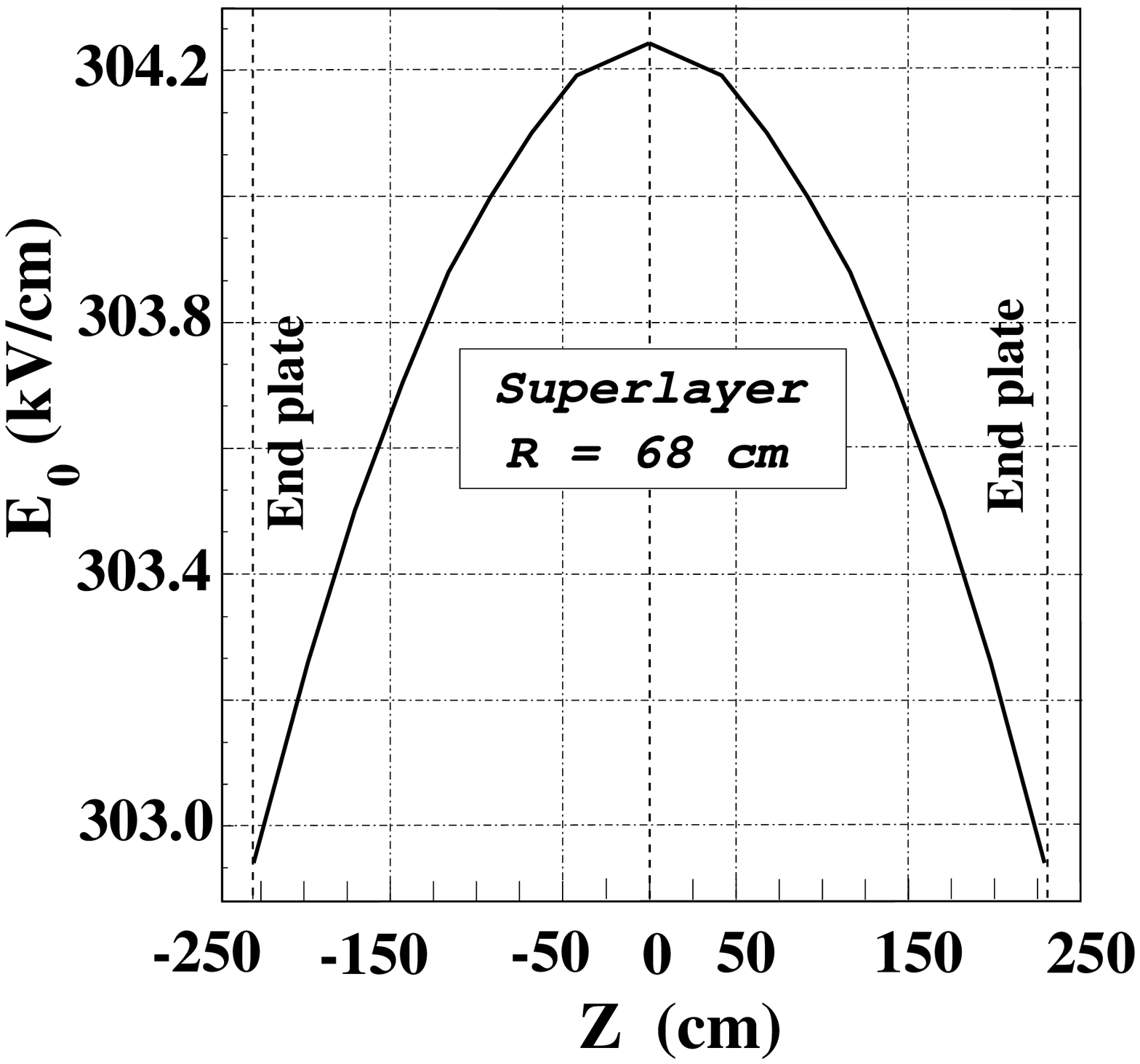}
}
\caption[Fig:E_0(z)]{ \label{Fig:E_0(z)}\sl
		Surface field as a function of $z$ for a sense wire
		in the innermost stereo super-layer at $r = 68~{\rm cm}$.
}
\end{minipage}
\end{figure}

\noindent
As shown earlier, the wires belonging to a stereo layer form 
a hyperboloidal surface.
Consequently, any stereo cell shrinks as one moves
from an endplate to the middle of the wires.
The resultant shrink factor is shown in Fig.\ref{Fig:f(z)}
for all of the 10 ($V$,$U$) stereo super-layers of the possible 
JLC central tracker.

The variation of the
shrink factor $\Delta f/f$ is at most $3~\%$ 
observed for the innermost stereo super-layer which has
$r = 68~{\rm cm}$ and $\alpha = 75~{\rm mrad}$.
The $z$-dependent geometrical deformation of the stereo cell
induces variation of the surface field on the sense wires
in it, thereby introducing gain nonuniformity along the wires.
We calculated the surface field variation on the sense wires,
using the drift chamber simulation program, GARFIELD.
The size of the cell structure depends on the shrink factor $f$ and
varies also with the radial position
or the super-layer that the cell belongs to.
Fig.\ref{Fig:E_0(z)} plots the calculated surface fields against
the $z$ coordinate.
The maximum variation of the
surface field is $\Delta E/E \simeq 0.43~\%$.

\begin{flushleft}
{\bf surface field  variation due to wire sags}
\end{flushleft}

\noindent
In addition to the stereo geometry, there is another
source of geometrical deformation of the cell 
structure that is the wire sags due to gravitational and electrostatic
forces on the wire.
These gravitational and electrostatic sags also contribute
to the surface field variation and are common to
both axial and stereo wires.
It suffices to investigate the effects
of these wire sags for axial wires only.
We, therefore, calculate the
surface field variation with GARFIELD, 
using as input the electrostatic and 
gravitational sags actually measured 
as the difference of the wire coordinates at
$z=\pm L/2$ (wire ends) and $z=0$ (mid point)
for a $4.6~{\rm m}$-long test chamber\cite{Ref:sag}.
The maximum wire-sag-induced variation of the surface field
is $\Delta E_0/E_0 \simeq 1.4~\%$.
It should be emphasized that the fact that
the field variation is comparable with those from
the stereo geometry for our example
is largely due to the use of Aluminum field wires.
We can in principle reduce the field variation
from the gravitational and electrostatic sags,
by optimizing the wire materials and by
stringing these wires with higher tensions.
On the other hand, the limitation coming from
the stereo geometry is inherent in any stereo chamber
and is unavoidable.

%% file: dettrk/cdc/p8.tex
\begin{flushleft}
{\bf gas gain variation along the wires}
\end{flushleft}

\noindent
Now that we know the surface field variation along any
wire, we can study its $z$-dependent gas gain variation
and discuss the operation feasibility of the chamber,
provided that the relation of the surface field
to the gas gain is known.
Conversely, the stereo angle should be so determined to make 
the gain variation acceptable.

We have measured the gas gain for the 
${\rm CO}_2(90\%)$-Isobutane $(10\%)$ mixture,
by using a single wire proportional counter\cite{Ref:gasgain}.
The counter is made of an aluminum cathode tube
with a $10~{\rm mm}$ inner diameter and a gold-plated
tungsten anode wire with a $30~\mu{\rm m}$ diameter.
For the $^{55}$Fe $X$-rays ($5.9~{\rm keV}$)
and $^{90}$Sr $\beta$-rays,
the charges collected on the anode within 
$1~\mu{\rm sec}$ were measured as a function of the
applied voltage.

\begin{figure}
\begin{minipage}[htb]{7cm}
\centerline{
\epsfxsize=7cm 
\epsfbox{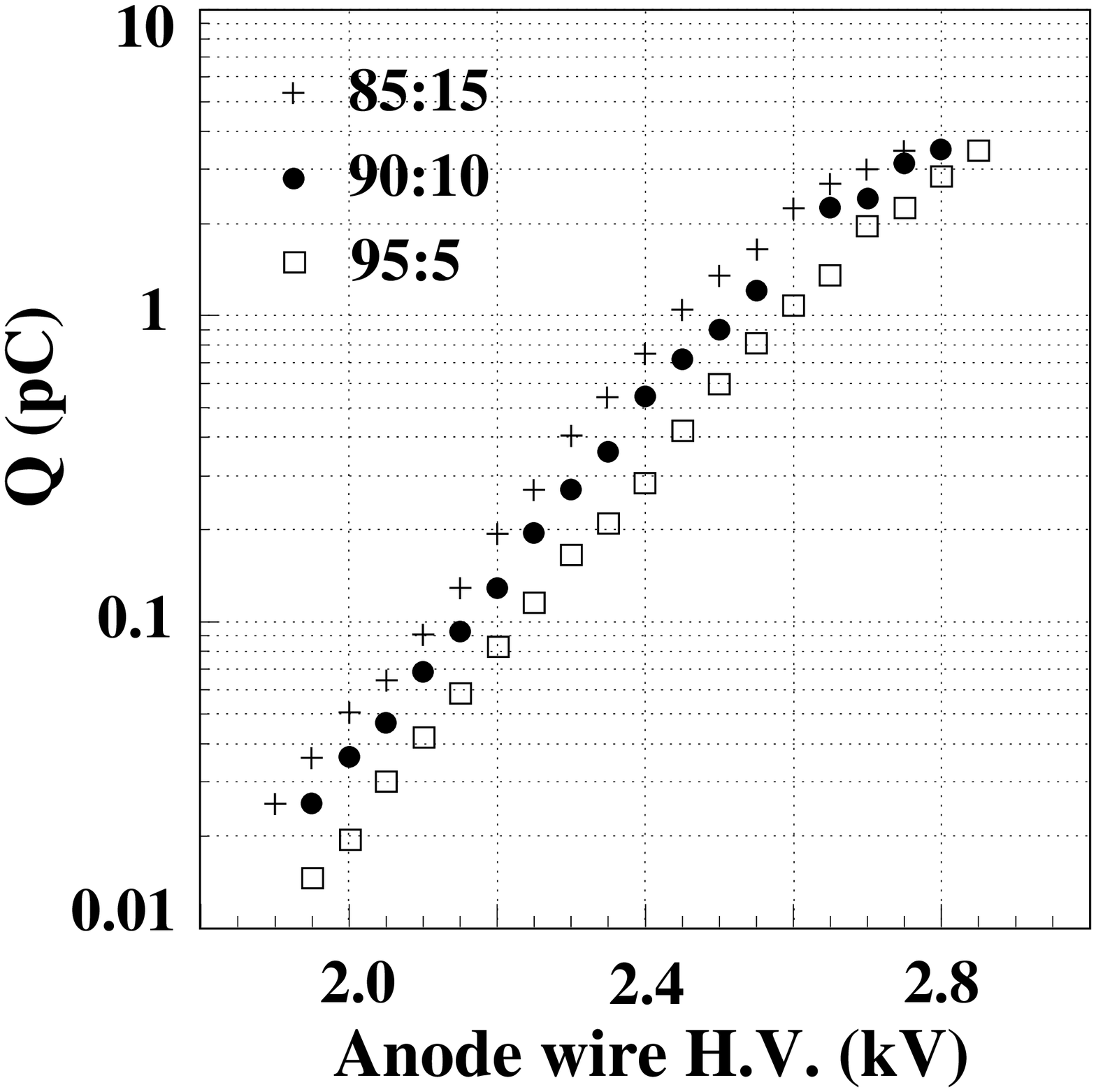}
}
\caption[Fig:fig10]{\label{Fig:fig10}\sl
HV dependence of the charges collected in $1.0~\mu {\rm s}$ 
for  $X$-rays
with three different concentrations of the ${\rm CO}_2$/Isobutane mixture.
}
\end{minipage}
\hfill
\begin{minipage}[htb]{7cm}
\centerline{
\epsfxsize=7cm 
\epsfbox{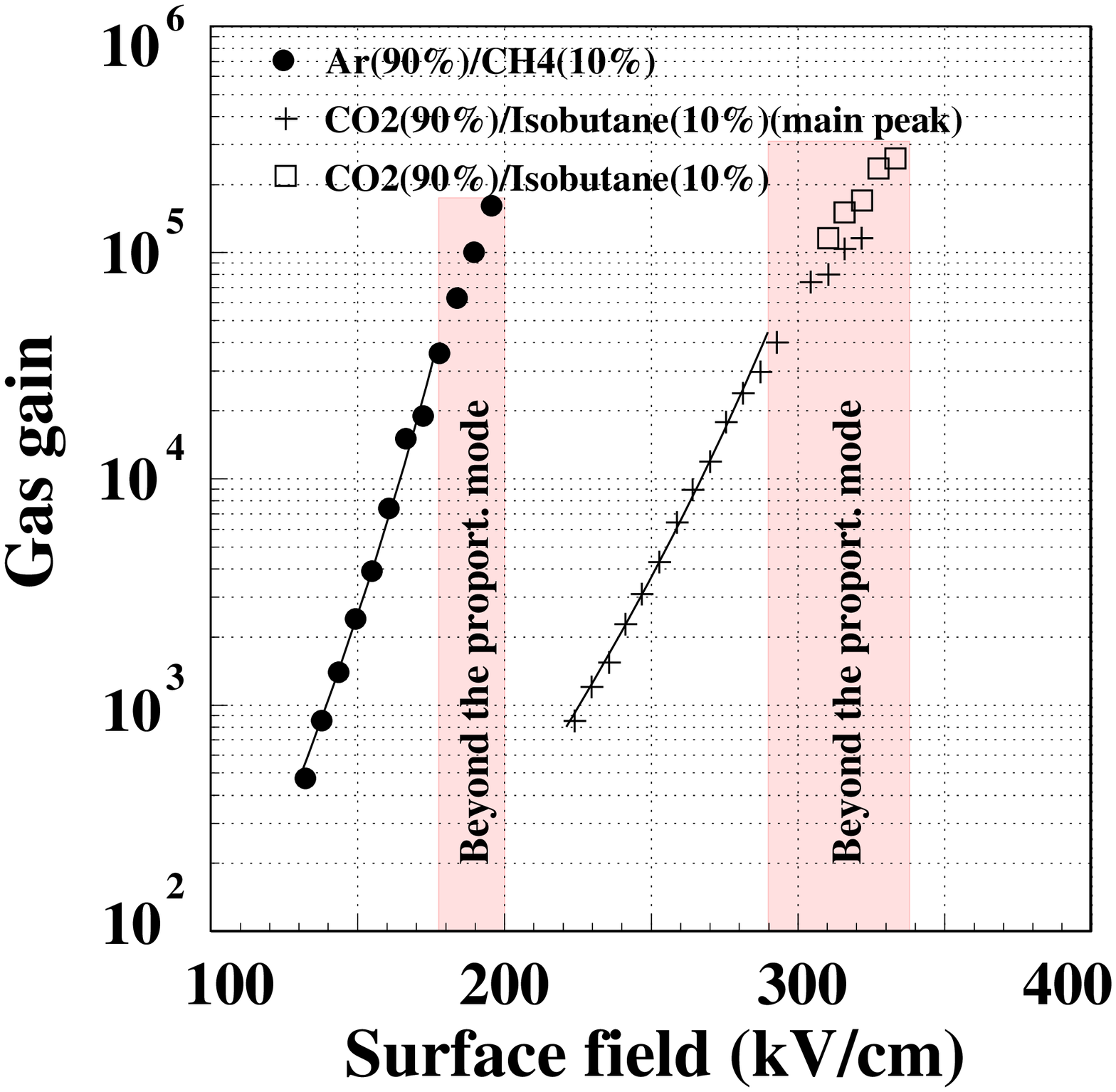}
}
\caption[Fig:fig11]{\label{Fig:fig11} \sl
Gas gain as a function of the surface field for 
the Single Wire Tube with the ${\rm CO}_2(90\%)$-Isobutane $(10\%)$
gas mixture.
}
\end{minipage}
\end{figure}

Gas gain measurements for different Isobutane 
concentrations (5\%, 10\%, and
15\%) have been made. The results for $X$-rays are shown in Fig.\ref{Fig:fig10}.
The gas gain increases as the Isobutane concentration becomes bigger.		

The measured data on the gain variation as a function of the surface field
are fitted in Fig.\ref{Fig:fig11}
to the following empirical formula:
\begin{equation}
	\ln G = a + b E_0,
\end{equation}
where $a$ and $b$ are the parameters to be determined.
From these data, we obtain the following relation:
\begin{equation}
\frac{\Delta G}{G} = (15 \sim 20) \left( \frac{d E_0}{E_0} \right)
\end{equation}
around $E_0 \simeq 300~{\rm kV/cm}$, where
the gas gain is about $10^5$.

By combining these results with the surface field variations
calculated above,
we estimate the gain variation of the stereo wire
due to the stereo geometry
to be at most $8~\%$ which is acceptable.

The surface field variation due to the wire displacements
induced by gravitational and electrostatic forces for
our $4.6~{\rm m}$ test chamber could give rise to some
additional gain variation as large as $25~\%$ at worst.
The wire displacements can, however, be reduced by appropriate
choices of wire materials enduring higher tensions,
a higher gain gas mixture allowing
lower high voltages on sense wires, etc.
On the other hand, the gas gain variation due to the
stereo geometry is irreducible without geometry
redesigning.
\vskip 1.0cm
Finishing up with the stability questions, let us move on to questions 
4 and 5 concerning required chamber performance~\cite{Ref:cosm}.

%% file: dettrk/cdc/p6.tex
\subsubsection{Expected Chamber Performance}

{\bf spatial resolution}
\\

\noindent
Can we actually achieve the
required spatial resolution over the full length of the
$4.6~{\rm m}$-long chamber? 
In order to confirm this, we studied the performance
of the test chamber using cosmic rays~\cite{Ref:cosm}.
Here again we concentrate our attention on the central
cell.
Signals from the 5 sense wires belonging to the
central cell were amplified by a pre-amplifier,
which is mounted on an end-plate of the test chamber,
and by a post-amplifier after passing
through a $5~{\rm m}$-long twisted pair cable.
The gains of the pre-amplifier is about $200~{\rm mV/pC}$ for the short pulse.
The signal from the post-amplifier is fed to 
an 8-bit flash-ADC (REPIC RPC-250), which 
has a $500~{\rm MHz}$ sampling frequency 
over a $16~\mu {\rm sec}$ time window.
Any pulse from $0$ to $-1~{\rm V}$ is 
chopped into $2~{\rm nsec}$ time slots and converted to 
a train of digits in the interval of $0$ to $255$ counts
by this flash-ADC. 
The flash-ADC fully covers
the maximum drift time of the test chamber (about $6~\mu {\rm sec}$)
and provides a high enough sampling frequency 
to ensure a good enough resolution
($2~{\rm nsec}$ corresponding to $16~\mu{\rm m}$ drift equivalent).

The digitized pulse form is read out and recorded on a disk by
a computer (NEC PC98) through a CAMAC system.
Any pulse which continues longer than $120~{\rm nsec}$
above the threshold (pedestal+3 counts)
was regarded as a signal pulse. 
The arrival time of the signal pulse was
determined by its leading edge at which the pulse height
exceeded the threshold.
To convert the drift time to the drift length, 
we use the $x$-$t$ relation calculated
with GARFIELD for each wire.
The validity of the calculated
$x$-$t$ relation was experimentally confirmed by
comparing the drift velocity given by GARFIELD with 
independent drift velocity measurements.

\begin{figure}
\centerline{
\epsfysize=12cm 
\epsfbox{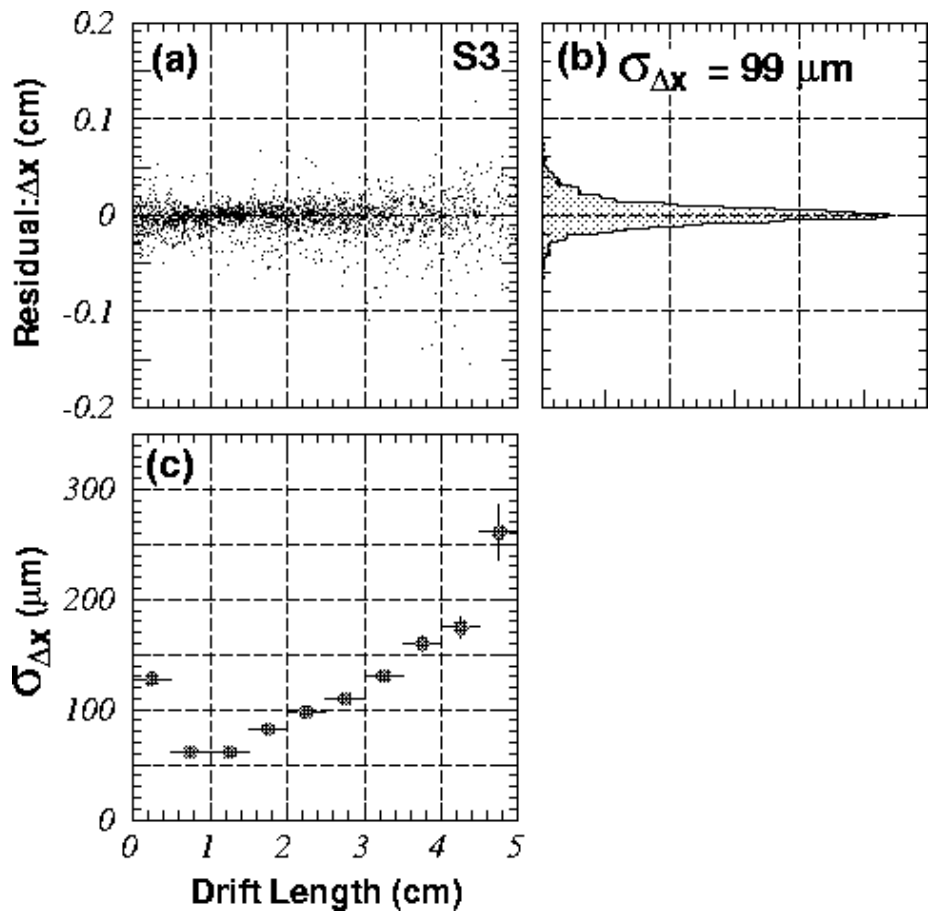}
}
\begin{center}\begin{minipage}{\figurewidth}
\caption[Fig:c-p-a]{\label{Fig:c-p-a} \sl
Residual for $S3$ at the center of the test chamber
(a) plotted against drift length, (b) its projection to
the vertical axis, and (c) the standard deviation of the
residual distribution as a function of the drift length.
}
\end{minipage}\end{center}
\end{figure}

We selected events demanding that the 4 out of the 5 sense wires
excluding the one to study had to have hits and
made a reference track.
We then calculated, on an event-by-event basis,
the residual of each hit on the wire to look at as
the distance from the reference track consisting of
the 4 hits.
The so-obtained residual distribution for the central
wire ($S3$) is plotted in
Fig.~\ref{Fig:c-p-a}-(a) against the drift length
for tracks passing through the central cell ($C4$)
in the middle of the wires, with the two trigger counters
vertically aligned so as to primarily sample tracks perpendicular
to the sense wires (for our naming convention, see Fig.~\ref{Fig:cell}(a)).
Fig.~\ref{Fig:c-p-a}-(c)
shows the one-standard-deviation ($1$-$\sigma$) width
of the residual distribution
as a function of the drift distance.
The width is less than $100~\mu{\rm m}$ for
a drift length less than $2.5~{\rm cm}$,
except for the vicinity of the sense wire\footnote{
The apparent degradation of the resolution near the sense
wire is not only coming from primary ionization
statistics but also from left-right
ambiguity.
}.
It becomes, however, gradually worse
up to about $200~\mu{\rm m}$ towards the edge of the drift volume.
Fig.~\ref{Fig:c-p-a}-(b) is the projection to the
vertical (residual) axis, demonstrating the overall 
residual distribution.

The edge wires ($S1$ and $S5$) have wider distributions.
This is due to the reference track extrapolation error
for $S1$ and $S5$
being much larger than the interpolation error
for $S2$ to $S4$.
In order to estimate the intrinsic spatial resolution
of each wire we calculated the track error as a
function of the drift distance
and subtracted it in quadrature on a bin-by-bin basis.
We then averaged the resultant resolution over the
drift distance to compensate the acceptance non-uniformity.
The so-obtained overall intrinsic spatial resolutions for
individual sense wires are summarized
in Table~\ref{Table:cor_resid}.
Inspection of the table tells us that
the overall resolutions are around $100\mu{\rm m}$, approximately
satisfying our performance goal at least locally.

A similar analysis was carried out for tracks passing
though the central cell near the both ends of the wires,
in order to make sure that the performance goal be
satisfied over the full length of the chamber.
Though the spatial resolutions became slightly worse,
they are still close to $100~\mu{\rm m}$
as listed in Table~\ref{Table:cor_resid}.
\begin{table}[htbp]
\caption{\sl Summary of spatial resolutions.}
\label{Table:cor_resid}
\begin{scriptsize}
\begin{center}
\begin{tabular}{cccc}
\hline
{Wire Number}&{Center}&{Ends}&{Inclined}\\
\hline
S1&$117 \pm 23 (\mu m)$&$163 \pm 75 (\mu m)$&$ 94 \pm 38 (\mu m)$\\
\hline
S2&$103 \pm  7 (\mu m)$&$136 \pm 52 (\mu m)$&$ 95 \pm 15 (\mu m)$\\
\hline
S3&$103 \pm  9 (\mu m)$&$120 \pm 22 (\mu m)$&$102 \pm 19 (\mu m)$\\
\hline
S4&$ 98 \pm  9 (\mu m)$&$119 \pm 22 (\mu m)$&$ 90 \pm 19 (\mu m)$\\
\hline
S5&$109 \pm 16 (\mu m)$&$132 \pm 45 (\mu m)$&$105 \pm 39 (\mu m)$\\
\hline
\end{tabular}
\end{center}
\end{scriptsize}
\end{table}

The intrinsic resolutions may also depend upon the
incident angle of a track to the sense wires.
In order to investigate this effect,
we set the two trigger counters in the middle
of the chamber but slightly displaced so as to primarily
sample inclined tracks having an incident angle
of around $45^\circ$ to the wires.
This incident angle ($45^\circ$) corresponds to
the maximum dip angle with the full number of sense wires
for the possible JLC central tracker.
The width of the residual distributions were
translated into overall intrinsic spatial resolutions
after the track error and acceptance corrections,
and are summarized also in Table~\ref{Table:cor_resid}.
We thus confirmed that
a spatial resolution of about $100\mu{\rm m}$
up to an incident angle of $45^\circ$.

\begin{figure}
\centerline{
\epsfysize=6cm 
\epsfbox{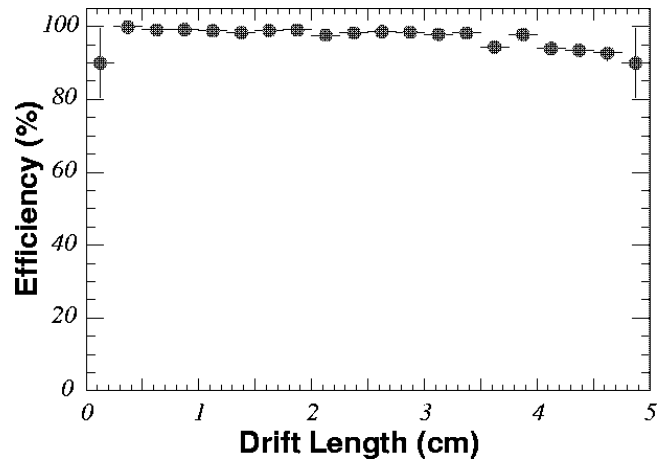}
}
\caption[Fig:c-e]{\label{Fig:c-e} \sl
Efficiency of the central sense wire ($S3$) 
as a function of the drift distance.
}
\end{figure}

The efficiency for each wire was calculated 
as a function of the drift distance to be
the fraction of tracks with
actual hits on the wire in a $5$-$\sigma$ window.
Fig.~\ref{Fig:c-e} is the results of the calculations,
for an oxygen contamination in the chamber gas of
about $60~{\rm ppm}$.
The efficiency stays flat above $95\%$ except in the last
several bins close to the edge of the drift volume.
The slight drop towards the edge is primarily
coming from the drift electron absorption due to
the oxygen contamination.

To investigate the effect of the oxygen contamination in the chamber gas,
we compared the chamber performance for two different oxygen contaminations,
$60~{\rm ppm}$ and $120~{\rm ppm}$.
The oxygen molecules, being electro-negative,
capture some electrons drifting towards a sense wire
and make the signal pulse smaller
as the drift length gets longer.
We chose the chamber gas
with a slow drift velocity to reduce the effect of the diffusion of the
drift electrons.
This choice made the chamber more sensitive to 
a small contamination from electro-negative gases.

\begin{figure}
\centerline{
\epsfysize=8cm 
\epsfbox{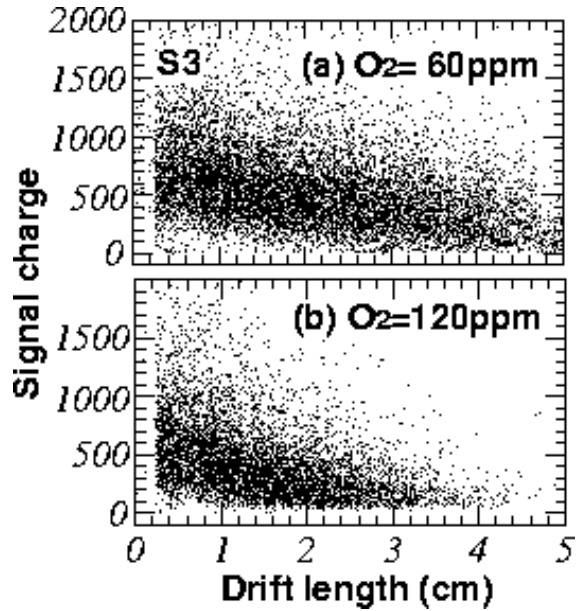}
}
\begin{center}\begin{minipage}{\figurewidth}
\caption[Fig:q(x)]{\label{Fig:q(x)} \sl
(a) Pulse charge on the central sense wire ($S3$)
as a function of the drift distance
for an oxygen contamination of $60~{\rm ppm}$.
(b) Same as (a) but
for an oxygen contamination of $120~{\rm ppm}$.
}
\end{minipage}\end{center}
\end{figure}

The pulse hight becomes smaller with the drift distance
due to the diffusion.
However, the total charge of the pulse must
stay the same, as long as there is no drift electron loss.
Plotted in Fig.~\ref{Fig:q(x)}-(a) is
the integrated signal charge on $S3$ against the drift distance
for an oxygen contamination of $60~{\rm ppm}$.
The signal charge gets smaller as the drift distance
becomes longer, indicating that an electro-negative gas
at work.
In order to confirm this effect, we intentionally
increased the oxygen contamination to $120~{\rm ppm}$
and measured the drift-length dependence of the
signal charge.
As seen in Fig.~\ref{Fig:q(x)}-(b),
the pulse charge rapidly decreases
with the drift distance and
vanishes down below the threshold
at the large drift distance,
suggesting a total loss of drift electrons.

This effect not only diminishes the wire efficiencies but may also
deteriorate the spatial resolutions
in the large drift-length region.
As a matter of fact, Fig.~\ref{Fig:c-p-a}
shows a resolution increment with the drift distance
larger than naively expected from diffusion.
It may be explained by the drift electron loss
due to the oxygen contamination.
A better performance can thus be expected if the oxygen
contamination is reduced.

%% file: dettrk/cdc/p6baby.tex
\begin{figure}[htb]
\centerline{
\epsfysize=6.0cm 
\epsfbox{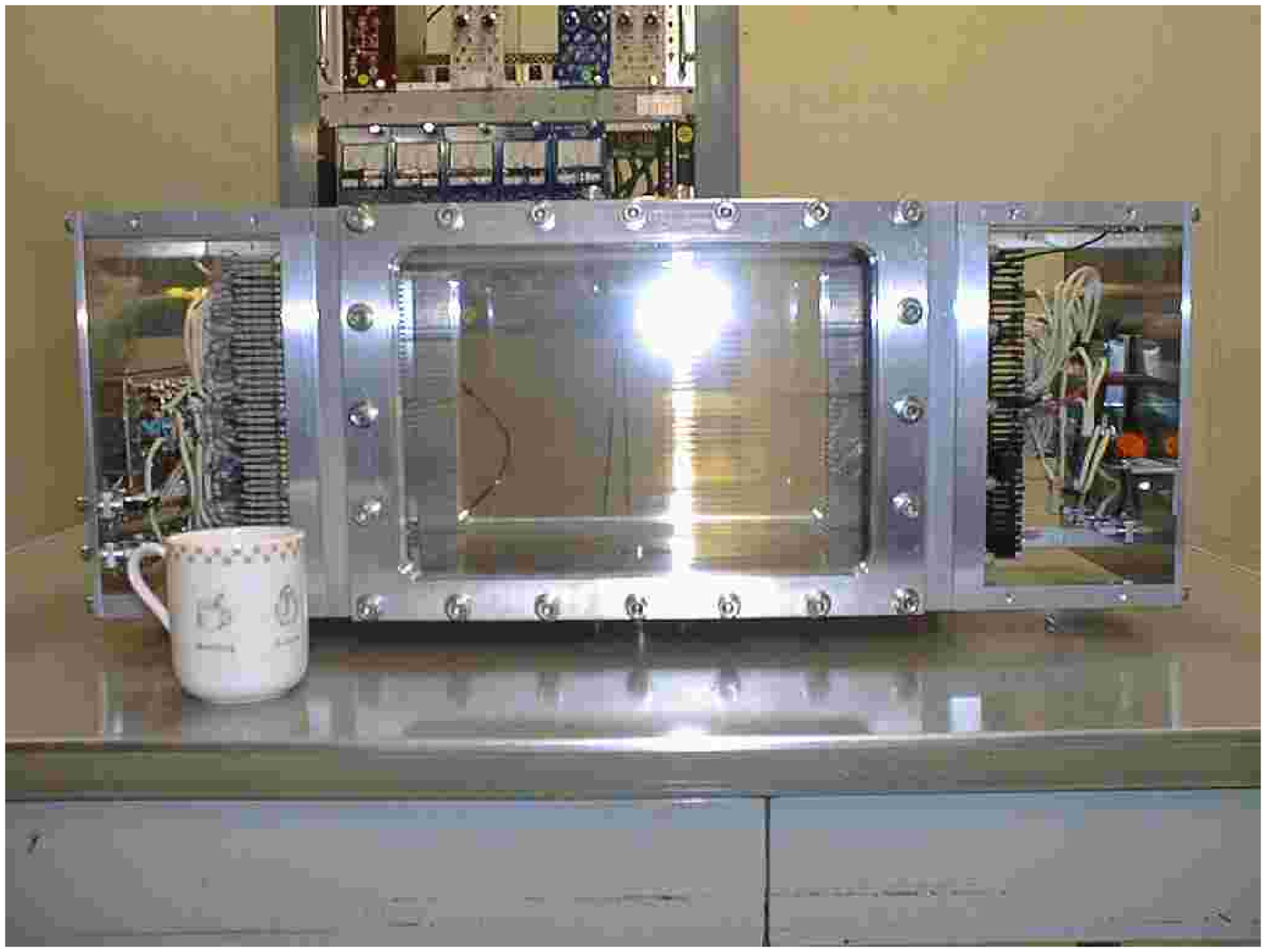}
}
\caption[Fig:babychamber]{\label{Fig:babychamber} \sl
Baby chamber}
\end{figure}

In  order to verify this expectation, 
we studied the effect of the oxygen
contamination with a smaller test chamber
for which further reduction of oxygen contamination is possible.
The smaller test chamber, hereafter
called "baby chamber" has three jet-cells
with the same wire layout as the 4.6~m chamber but with a shorter 
wire length of 40~cm (see Fig.~\ref{Fig:babychamber}). 

\begin{figure}[htb]
\centerline{
\epsfxsize=6.0cm 
\epsfbox{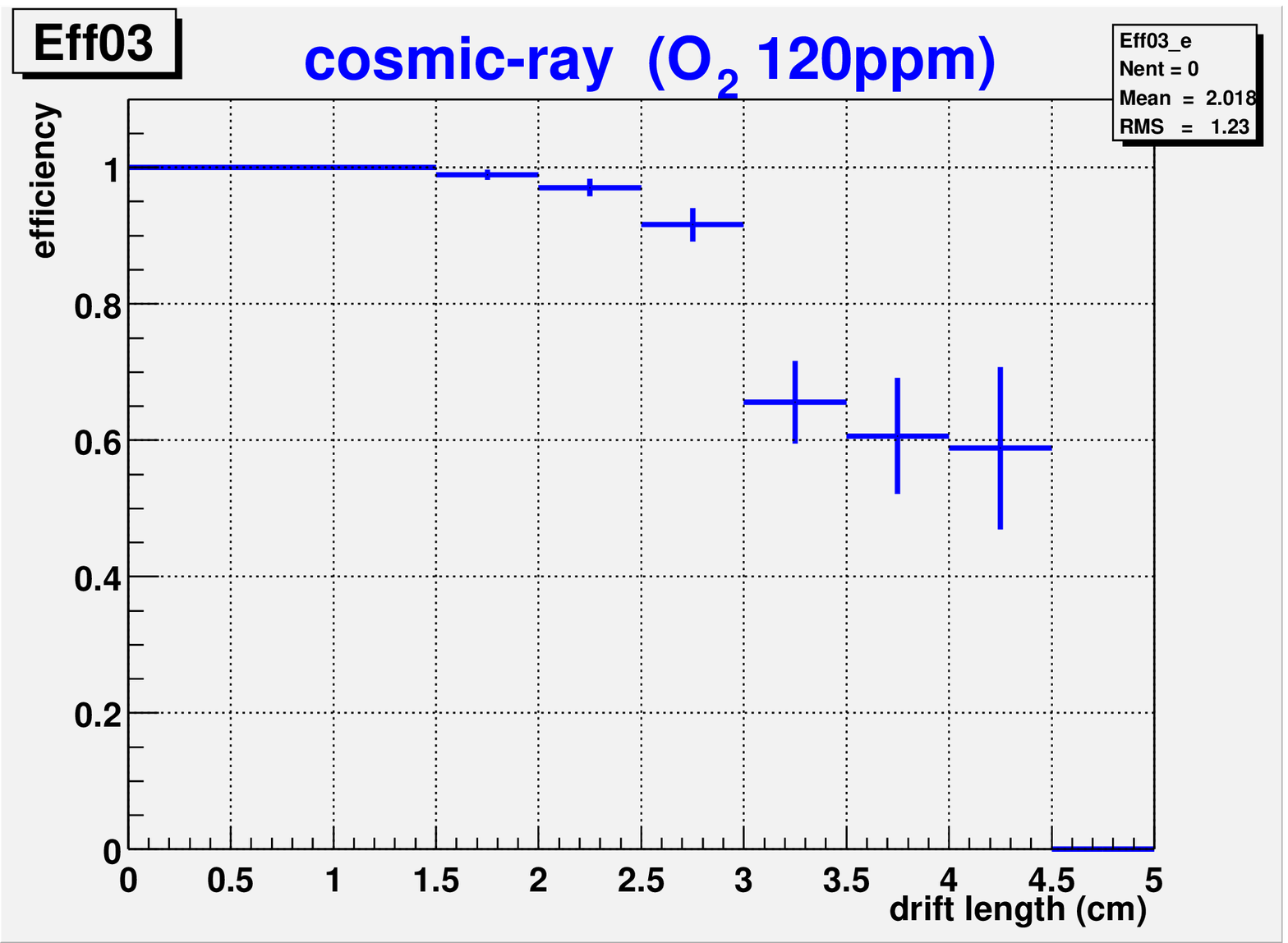}
\hspace{0.5cm}
\epsfxsize=6.0cm 
\epsfbox{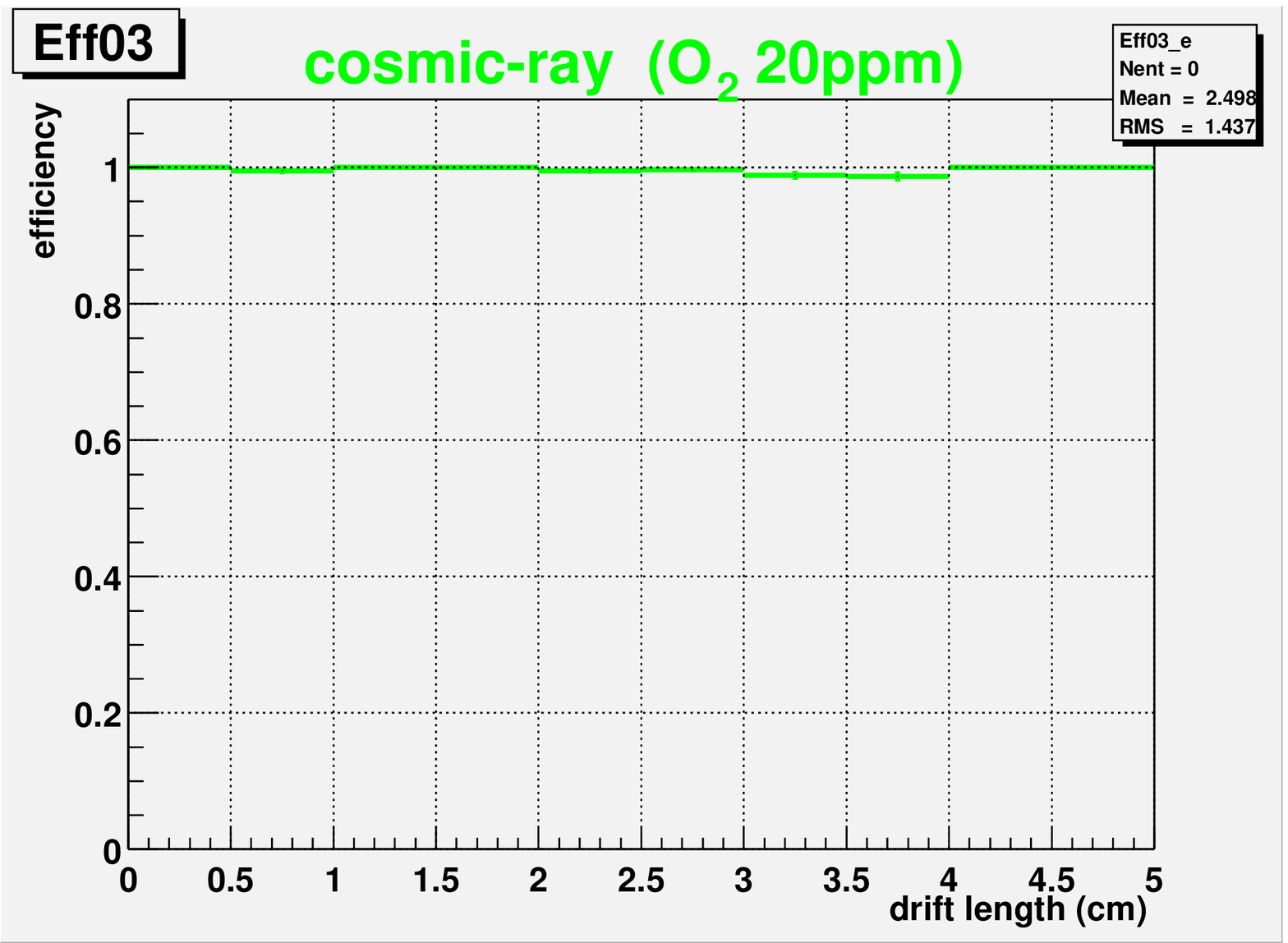}
}
\begin{center}\begin{minipage}{\figurewidth}
\caption[Fig:babyeff]{\label{Fig:babyeff} \sl
Efficiency of the central sense wire ($S3$) 
as a function of the drift distance (baby chamber):
O$_2$ contaminations of 120~ppm (left) and 20~ppm (right).
}
\end{minipage}\end{center}
\end{figure}

\begin{figure}
\centerline{
\epsfxsize=6.0cm 
\epsfbox{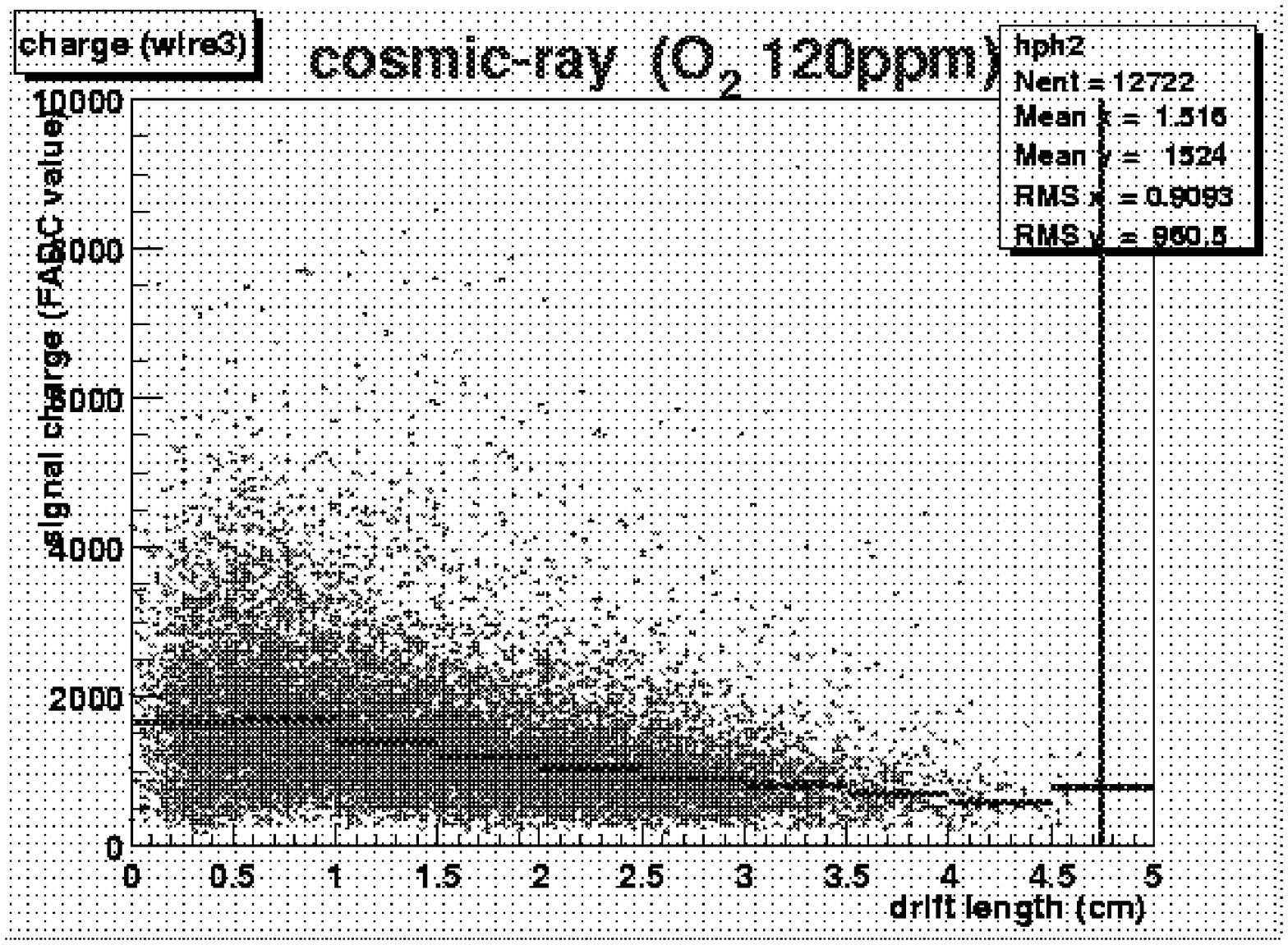}
\hspace{0.5cm}
\epsfxsize=6.0cm 
\epsfbox{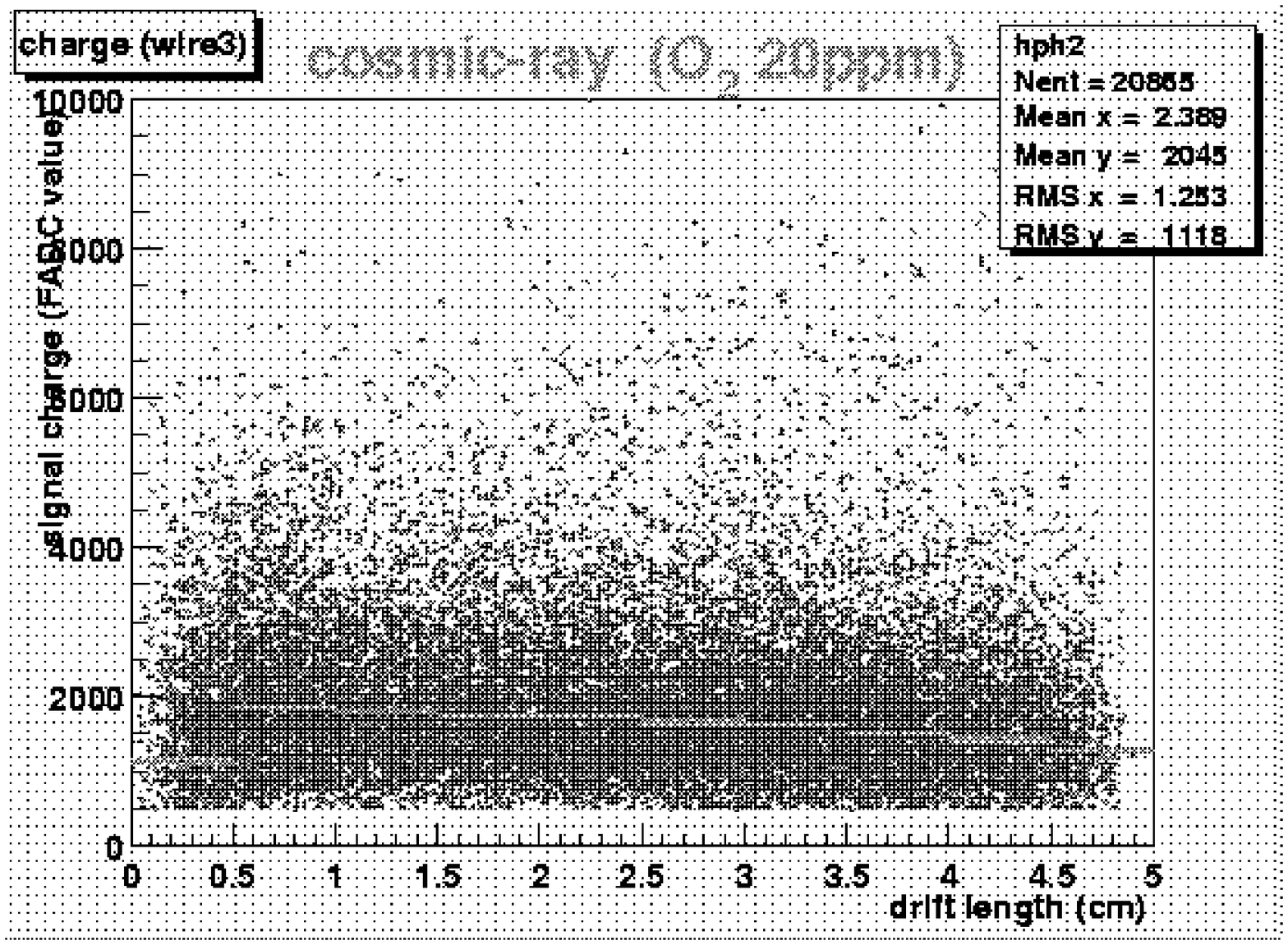}
}
\begin{center}\begin{minipage}{\figurewidth}
\caption[Fig:babycharge]{\label{Fig:babycharge} \sl
Pulse charge on the central sense wire ($S3$)
of the baby chamber
as a function of the drift distance
for oxygen contaminations of $120~{\rm ppm}$ (left)
and $20~{\rm ppm}$ (right).}
\end{minipage}\end{center}
\end{figure}

\begin{figure}
\centerline{
\epsfxsize=10.cm 
\epsfbox{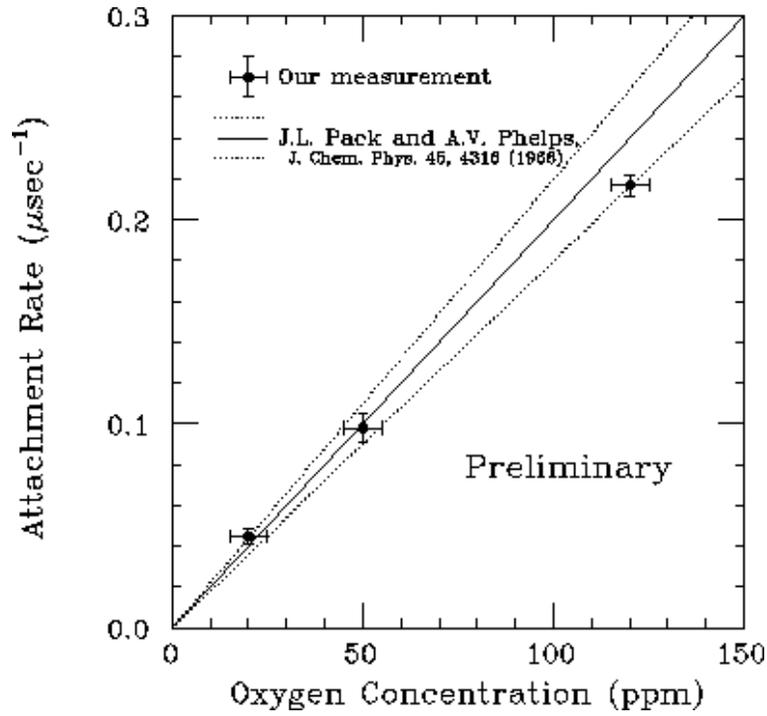}
}
\begin{center}\begin{minipage}{\figurewidth}
\caption[Fig:attach]{\label{Fig:attach} \sl
Attachment rate as a function of oxygen concentration (preliminary).}
\end{minipage}\end{center}
\end{figure}

Using the baby chamber and 
essentially the same data taking system as used
above for the 4.6~m test chamber,
we took cosmic ray data.
We then followed the same data analysis procedure
as above
and calculated the wire efficiency for 
the cosmic ray tracks that passed the chamber
with an incident angle of nearly $90^\circ$.
Fig.~\ref{Fig:babyeff}
shows the efficiency of the central sense wire as a function
of the drift distance for oxygen contaminations of
$120~{\rm ppm}$ and $20~{\rm ppm}$, respectively.
In the case of 
$120~{\rm ppm}$ we see a sharp efficiency
drop towards the edge while for the oxygen contamination
of $20~{\rm ppm}$ the efficiency stays almost $100\%$ everywhere.
The sharp efficiency drop observed for $120~{\rm ppm}$
is coming from the drift electron absorption 
due to the relatively high oxygen contamination, which is obvious 
from Fig.~\ref{Fig:babycharge} showing
the integrated signal charge on the central sense wire as a function 
of the drift
distance: the signal charge gets significantly smaller 
for $120~{\rm ppm}$ than for $20~{\rm ppm}$ as the drift distance becomes
longer.
We obtained similar results for other sense wires.
The attachment rate of drift electrons was evaluated 
from the observed charge attenuation as a function of drift distance
and the known drift velocity.
In Fig.~\ref{Fig:attach} our measurement is compared with that calculated 
using the three-body attachment coefficient measured by J.L. Pack and
A.V. Phelps~\cite{Ref:attach}.   

\begin{figure}
\centerline{
\epsfysize=6.0cm 
\epsfbox{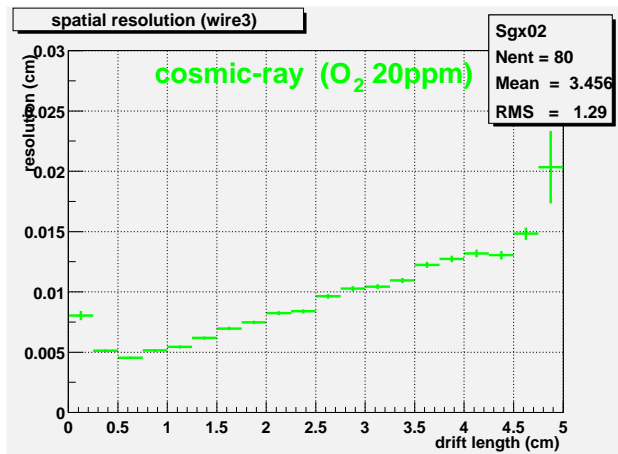}
}
\begin{center}\begin{minipage}{\figurewidth}
\caption[Fig:sigmaX_cosm020]{\label{Fig:sigmaX_cosm020} \sl
Spatial resolution of the central wire of the baby chamber
as a function of the drift length (preliminary).
}
\end{minipage}\end{center}
\end{figure}

Being encouraged by the drift-distance dependence of
the wire efficiency for $20~{\rm ppm}$,
we then studied the spatial resolution per wire.
The analysis procedure was the same as with the
4.6~m test chamber:
we calculated the residual for the wire in question
on the event-by-event basis, using a reference track
drawn with the hits on the other 4 wires.
We then subtracted the track error from the 
standard deviation of the residual distribution
to obtain the spatial resolution.
Fig.~\ref{Fig:sigmaX_cosm020} plots the spatial resolution
of the central sense wire against the 
drift length for $20~{\rm ppm}$,
which shows much more reasonable drift length
dependence compared to Fig.~\ref{Fig:c-p-a}. 

The oxygen contamination of the gas would not be a serious problem
in the real chamber since the oxygen level is planed to be kept lower than
10 ppm with better gas tightness.

%% file: dettrk/cdc/p11.tex
\begin{flushleft}
{\bf two-track separation}
\end{flushleft}

\noindent
Next question is whether we can reproduce
the single track results above
in a multi-track environment.
In order to answer this question,
we have carried out beam tests of the baby chamber,
using $e^+e^-$ pairs produced by gamma conversions at
an $Al$ converter.
The $e^+e^-$ pairs provided closely spaced 2 tracks
at our disposal in the chamber
and their signals were read out
with the same readout electronics as used
for the cosmic ray tests described above.

\begin{figure}
\begin{minipage}[htb]{7cm}
\centerline{
\epsfxsize=6.5cm 
\epsfbox{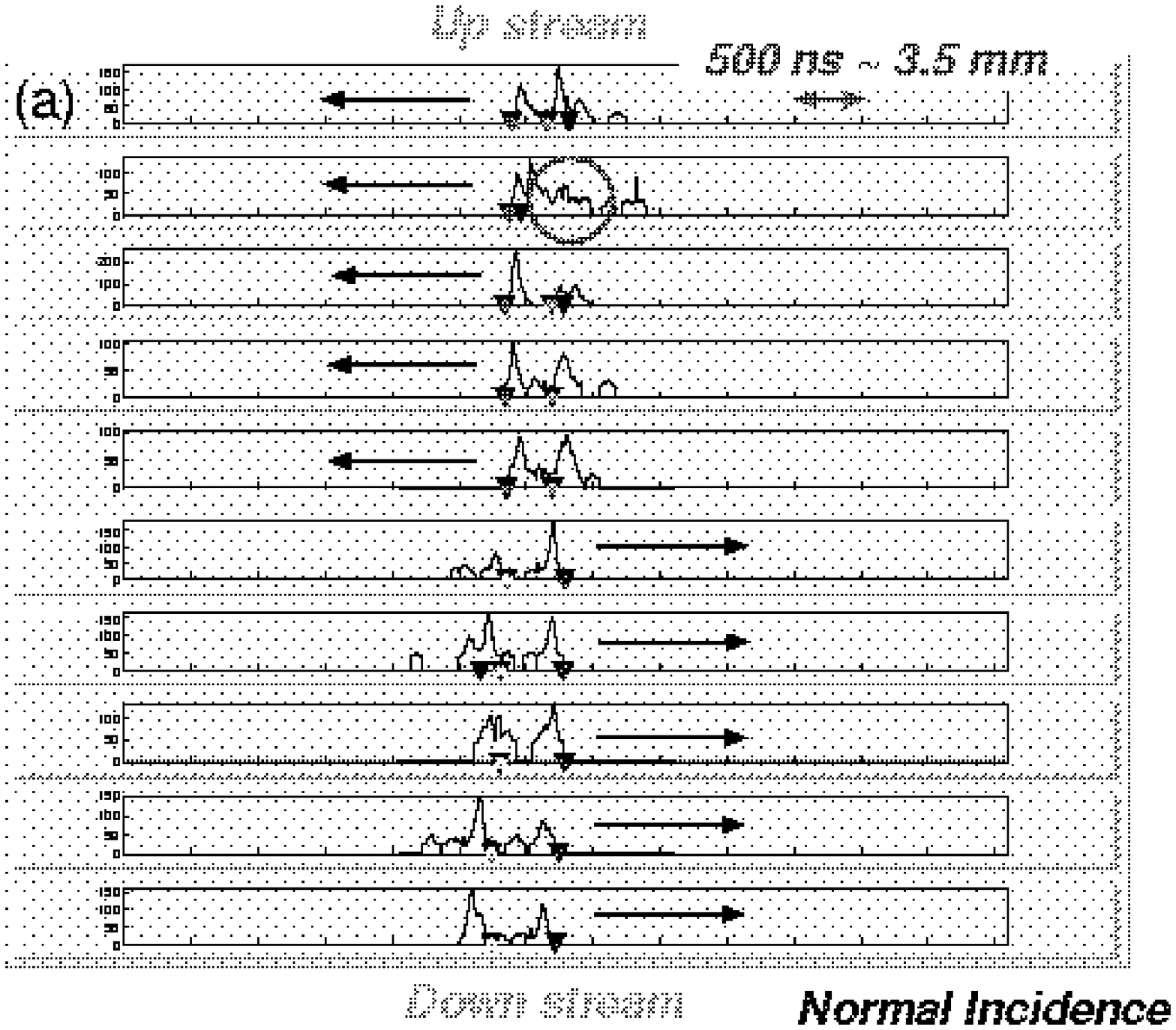}
}
\end{minipage}
\hfill
\begin{minipage}[htb]{7cm}
\centerline{
\epsfxsize=6.5cm 
\epsfbox{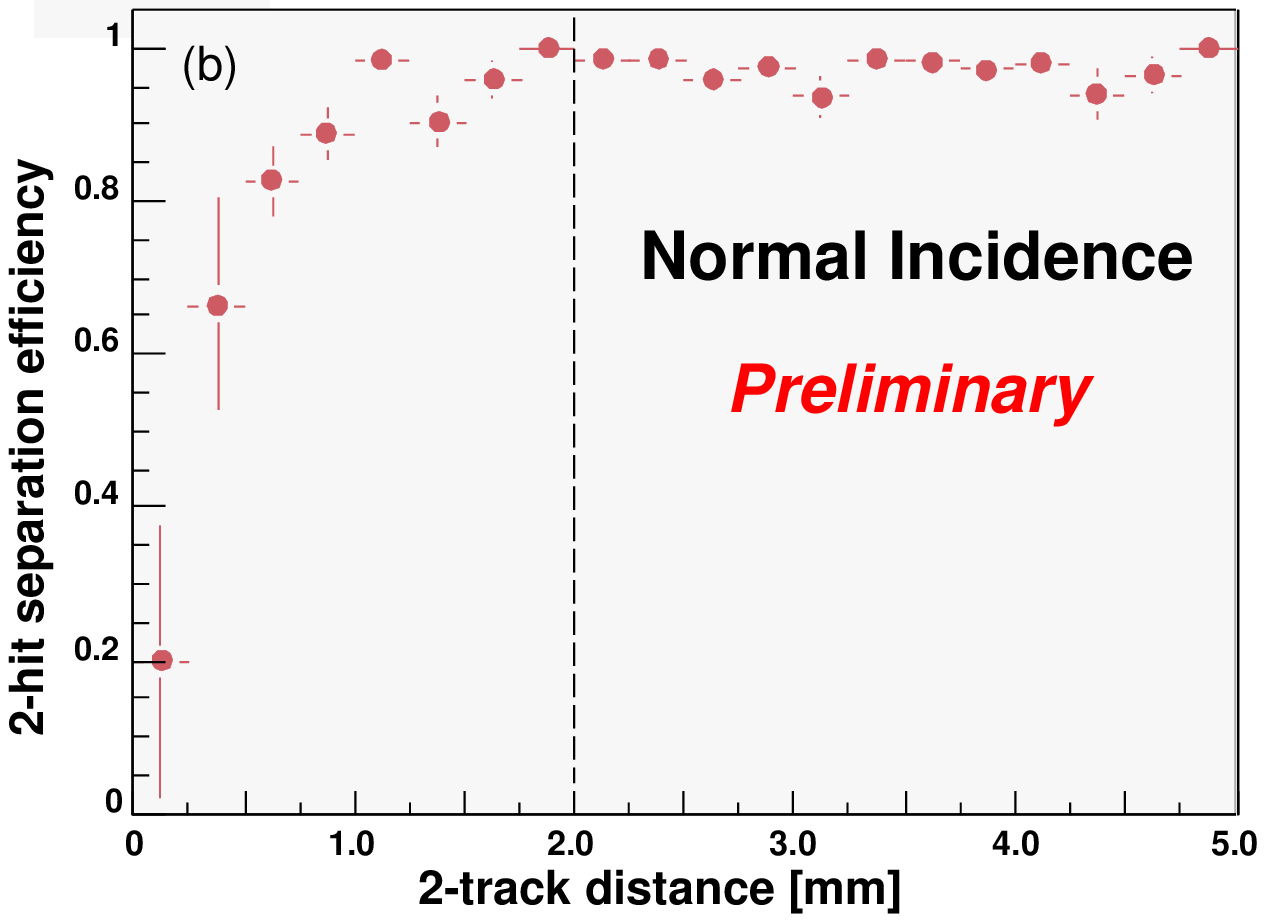}
}
\end{minipage}
\begin{center}\begin{minipage}{\figurewidth}
\caption[Fig:twotrack]{\label{Fig:twotrack}\small \sl
(a) FADC signals as a function of time (500~ns/division)
    for a typical 2-track event.
    The arrows indicate the drift direction.
(b) 2-hit separation efficiency
as a function of 2-track distance (preliminary).
}
\end{minipage}\end{center}
\end{figure}

Fig.~\ref{Fig:twotrack}-(a) shows 
FADC signals from 10 wires, 5 of which belongs to
the upstream and the other 5 to the downstream
cells, for a typical 2-track event.
We can see two closely spaced hits on each wire.
We calculated the 2-hit separation efficiency for 
each wire in a cell,
demanding two reference tracks made of hits on
the 4 reference wires other than the wire
to study.
The two tracks define the 2-track distance
and the expected hit position for the second
track on the wire in question.
Figure \ref{Fig:twotrack}-(b) plots
the efficiency for the second hit
as a function of the 2-track distance
in the case of normal incidence.

Although the result looks promising,
it is still preliminary and
more studies are needed to investigate
possible space charge effects
due to the small diffusion nature of our
gas mixture. 
We will return to this problem later in 
the subsection for space charge effect.

%% file: dettrk/cdc/p12.tex
\begin{flushleft}
{\bf particle identification}
\end{flushleft}

\noindent
Particle identification through energy loss ($dE/dx$) measurements
in a drift chamber is a well established technique and could be 
additional information provided by the JLC-CDC.
However the primary demand to the CDC is high spatial resolution to be 
achieved with a wire gain as high as $10^5$ while $dE/dx$ measurements
require a low gain in order to assure the proportionality of the collected
charge to the total number of ionization acts by a charged particle.
When the wire gain is high this proportionality is lost
because the positive ions created in the avalanche process initiated by
a given drift electron effectively reduce the electric field
near the sense wire surface and then prevent the normal growth of avalanches 
initiated by subsequent electrons (space charge effect).
In addition, the small diffusion of drift electrons (especially along
the wire direction) in a CO$_2$-based gas mixture may cause a severe
space charge effect since the drift electrons get amplified in small 
surface area on the sense wire.  

\begin{figure}
\centerline{
\epsfxsize=6.5cm 
\epsfbox{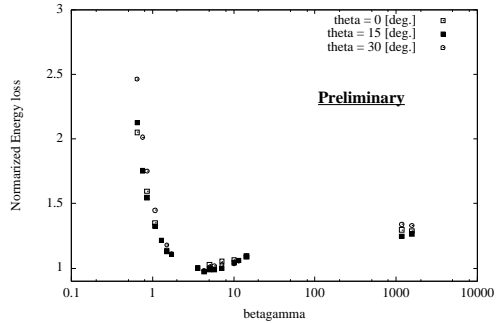}}
\begin{center}\begin{minipage}{\figurewidth}
\caption[Fig:betagamma]{\label{Fig:betagamma}\sl
Relative energy loss as a function of $\beta\gamma$. 
The $dE/dx$ values are normalized to that of 0.5 GeV/c pions and 
corrected for incident angles (preliminary).
}
\end{minipage}\end{center}
\end{figure}

In order to evaluate the particle identification capability
of the JLC-CDC,
we carried out $dE/dx$ measurements by irradiating the baby chamber   
with particle beams ($e$, $\pi$ and $p$) at the 12-GeV KEK proton
synchrotron (KEK-PS). 
Fig.~\ref{Fig:betagamma} shows the normalized energy loss 
as a function of $\beta\gamma$ for different beam incident angles. 
The beam is parallel with the sense-wire plane and the incident
angle ($\theta$) here is defined to be $0^\circ$ when the beam
is  perpendicular to the wire direction.
The energy loss was determined by averaging 80\%-retained truncated means of
10 samples.
The normalized $dE/dx$ decreases for smaller beam angles
especially in the low $\beta\gamma$ region.  
This indicates that the space charge effect does affect the $dE/dx$
measurements for highly ionizing tracks nearly perpendicular to
the sense wires.

\begin{figure}
\centerline{
\epsfxsize=5.5cm 
\epsfbox{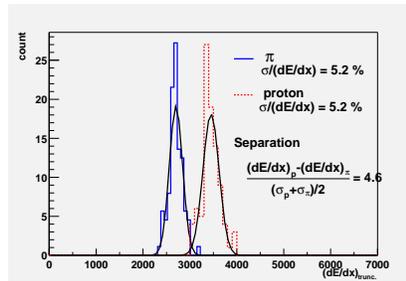}}
\begin{center}\begin{minipage}{\figurewidth}
\caption[Fig:separation]{\label{Fig:separation}\sl
Expected $\pi - p$ separation for 80 samples at 1~GeV/c (preliminary).}
\end{minipage}\end{center}
\end{figure}

Since the present design of the JLC-CDC assumes 16 layers of the 5-sense-wire 
jet cells in the radial direction, 80 samples of 
$dE/dx$ measurement are expected for a single track. 
In order to estimate the particle identification capability with 
this configuration,
we applied the 80\%-retained truncated mean method to the 80 $dE/dx$
measurements in a set of 8 consecutive events (tracks), each with
10 samples, recorded by the baby chamber.    
The result for $\pi - p$ separation at 1 GeV/c and $\theta = 0^\circ$ is
shown in Fig.~\ref{Fig:separation}.
The obtained figure of merit of separation is 
$S \equiv 2\cdot \left|(dE/dx)_A- (dE/dx)_B\right|/(\sigma_A + \sigma_B)
 = 4.6$.

In spite of the possible space charge effect, the $dE/dx$ information 
may still provide a useful means for particle identification.

%% file: dettrk/cdc/p10.tex
\subsubsection{Space Charge Effect}

\begin{flushleft}
{\bf local space charge effect}
\end{flushleft}

\noindent
However, not to mention two-track separation capability, the space charge
effect in multi-track environment may significantly affect the 
chamber performance
including spatial resolution and particle identification by $dE/dx$
measurements.

We thus studied the effect of local space charge on gas amplification 
with parallel laser beams ($\lambda$ = 266 nm) created by a splitter
(quartz plate),
which simulate two closely spaced strings of ionizations
produced by two charged particles in a jet. 
The beam distance was controlled by changing the thickness
of the quartz plate.
The laser beams were injected into a single-wire drift chamber with a
50-mm drift space used originally for the Lorentz angle
measurement~\cite{Ref:lorentz}.
\begin{figure}
\centerline{
\epsfxsize=7.cm 
\epsfbox{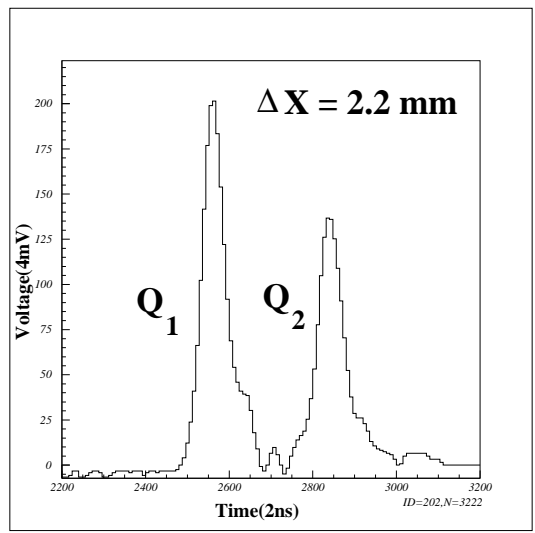}
}
\caption[Fig:spcharge1]{\label{Fig:spcharge1}\sl
Wave forms for double laser beams.}
\end{figure}
In Fig.~\ref{Fig:spcharge1},
typical wave forms for double laser beams at $x_1$ = 10 mm is shown for  
$\Delta x$ = 2.2 mm and $\Delta z$ = 0 mm,
where $x_1$ denotes the drift distance for the first signal,
$\Delta x$ ($\Delta z$) is the beam distance in the drift (wire) direction,
and the beams are parallel with the $y$-axis. 
The two laser tracks are clearly separated. The FWHM of the signal is 80 ns, 
which corresponds to the electron drift distance of 600 $\mu$m.

\begin{figure}
\centerline{
\epsfxsize=7.cm 
\epsfbox{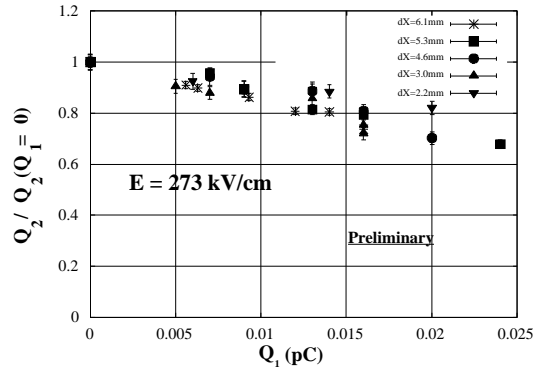}
}
\begin{center}\begin{minipage}{\figurewidth}
\caption[Fig:spcharge2]{\label{Fig:spcharge2}\sl
Gain reduction factor of the second signal as a function of $Q_1$ for
$\Delta z$ = 0 mm and 
$\Delta x$ = 6.1 mm, 5.3 mm, 4.6 mm, 3.0 mm and 2.2 mm,
measured with $x_1$ = 10 mm and $E$ (electric field strength on the
sense wire surface) = 273 kV/cm (preliminary).}
\end{minipage}\end{center}
\end{figure}

\begin{figure}
\centerline{
\epsfxsize=7.cm 
\epsfbox{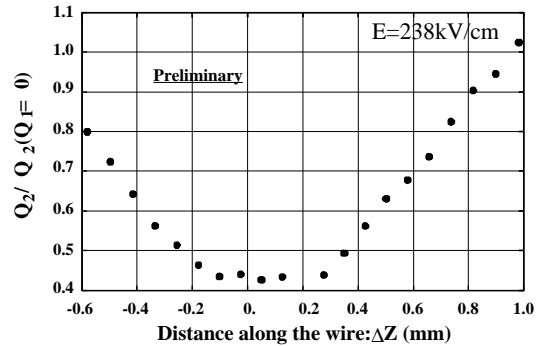}
}
\begin{center}\begin{minipage}{\figurewidth}
\caption[Fig:spcharge3]{\label{Fig:spcharge3}\sl
Gain reduction factor of the second signal as a function of $\Delta z$
measured with  $x_1$ = 10 mm, $\Delta x \sim$ 4 mm and $E$ = 238 kV/cm.
(preliminary)}
\end{minipage}\end{center}
\end{figure}

The signal charges were 
obtained by integrating the wave form. It should be noted that intensities
of the two laser 
beams are different because of the unequal splitting. 
The gain reduction factor of the second signal due to the existence of the 
first one was 
obtained as $Q_2(Q_1)/Q_2(Q_1=0)$, where $Q_1$ and $Q_2$ are the 
integrated charges for the first and the second signals, respectively. 
$Q_2(Q_1=0)$ was obtained by blocking the first laser beam.
The values of $Q_1$ 
and $Q_2$ were measured for each event and the mean values of 300 
events were used to obtain the reduction factor.
In  Fig.~\ref{Fig:spcharge2}, we plot it as a function of the first signal 
charge for different beam distances in the drift direction.    
On the other hand, Fig.~\ref{Fig:spcharge3} shows the reduction factor 
as a function of the distance along the wire.
These figures tell us that
\begin{itemize}
\item The gain reduction due to the space charge depends on the size of
the first signal as naively expected;
\item This gain reduction is a local effect on the sense wire surface
which appears only within a limited distance from the
first avalanches;
\item This local {\it paralysis\/} is sustained for a long time
because of the low mobility of positive ions.
\end{itemize}

The timing quality of the second signal, which is directly reflected
to the spatial resolution, is now under study.       

\begin{flushleft}
{\bf global space charge effect}
\end{flushleft}

\noindent
Another category of the space charge effect may set in 
when the beam background is severe and positive ions are accumulated
in the sensitive volume of the chamber.
Not to mention the wire gain reduction, the positive ions would cause
significant distortion of the otherwise uniform electric field in the
drift space.
In that case the measured coordinates on charged tracks deviate  
from those which would result under circumstances 
free from positive ions
(global space charge).
It should be noted that the drift velocity in a CO$_2$-based gas
is sensitive to the local electric field strength.
These deviations would be dependent on both position and time,
and could be comparable to, or larger than the spatial resolution 
expected of the JLC-CDC.
Therefore it is possible that the global space charge effect 
devastatingly degrade the chamber performance.

In order to study the global space charge effect,
we are now planning a small experiment using the baby chamber and a laser
beam at KEK-PS.
In the experiment the movement of the laser beacon tracks will be observed 
under the controlled intensity of defocused particle beam.

\vskip 1.0cm
So much for the performance issues, let us now move on to question 7,
concerning the magnitude of the magnetic field and how it affects the cell
design.

%% file: dettrk/cdc/p9.tex
\subsubsection{Magnetic Field Effect}

\begin{flushleft}
{\bf 2 tesla design}
\end{flushleft}

\noindent
The Lorentz angle, which is the angle between 
the drift direction of electrons under the influence
of magnetic field and the direction of electric field,
is one of the key parameters to determine
the jet cell design.
The existence of the magnetic field tilts 
drift lines by the Lorentz angle with respect to the
electric field direction.
If this angle is too big, drift lines for the edge
wires hit top or bottom walls of the drift cell,
thereby leading us to loss of detection efficiency.

\begin{figure}
\centerline{
\begin{minipage}[htb]{12cm}
\centerline{
\epsfxsize=10cm  
\epsfbox{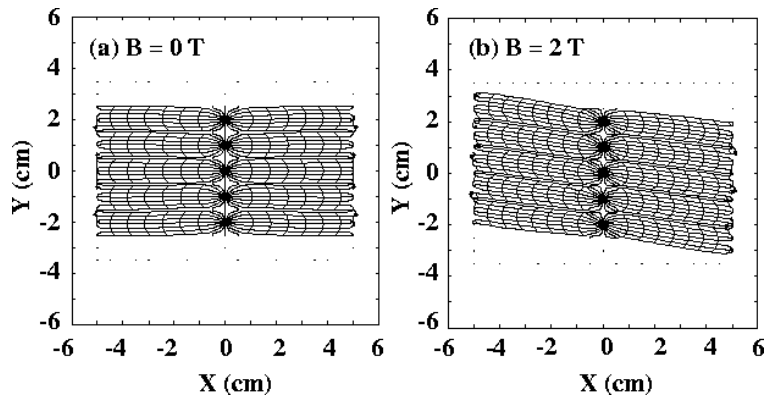}
}
\caption[Fig:drift-magn]
{ \label{Fig:drift-magn}\sl
GARFIELD results of electron drift lines and isochrones 
in the JLC-CDC for the ${\rm CO}_2$/isobutane(90:10) gas mixture.
}
\end{minipage}
}
\end{figure}

Fig.~\ref{Fig:drift-magn} shows electron drift lines in a jet cell
calculated by the chamber simulation program GARFIELD~\cite{Ref:garfield}
(a) without and (b) with the magnetic field. 
The results indicates that the drift lines are
completely contained in the cell
at least up to a magnetic field of $2~{\rm T}$.
There is, however, no systematic experimental data published
for the Lorentz angles of ${\rm CO}_2$/isobutane gas mixtures,
which makes difficult for us
to test the reliability of our calculations 
using the GARFIELD program.
In order to confirm the results of the GARFIELD calculations
and to verify the validity of our cell design, 
we have developed a Lorentz angle measurement system 
for cool gas mixtures, which have small Lorentz angles,
and carried out systematic measurement for 
${\rm CO}_2$/isobutane gas mixtures.
The measurement system is characterized by the use of two
laser beams to produce primary electrons
and flash ADCs to read their signals simultaneously.

\begin{figure}
\centerline{
\epsfxsize=7.0cm 
\epsfbox{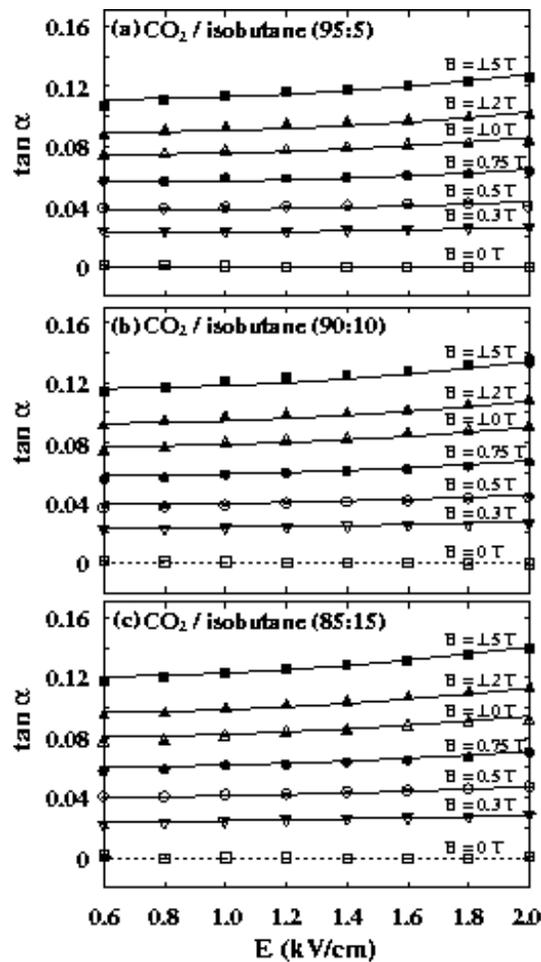}
}
\begin{center}\begin{minipage}{\figurewidth}
\caption[Fig:results]
{ \label{Fig:results}\sl
Tangent of the Lorentz angle
($\tan\alpha$) as a function of the electric field
for ${\rm CO}_2$/isobutane mixtures of
(a) (95:5), (b) (90:10), and (c) (85:15).
Smooth curves are GARFIELD/MAGBOLTZ predictions.
}
\end{minipage}\end{center}
\end{figure}

We measured the Lorentz angles 
for different ${\rm CO}_2$/isobutane gas mixtures.
The results are shown in
Figs.~\ref{Fig:results}-(a), -(b), and -(c)
as a function of the electric field
for three mixing ratios:
$(85:15)$, $(90:10)$, and $(95:5)$.
At each electric field value in each figure,
seven points are plotted, corresponding 
to, from bottom to top, seven magnetic field values:
$0.0$, $0.3$, $0.5$, $0.75$, $1.0$, $1.2,$, and $1.5~{\rm T}$,
respectively.

Curves in Figs.~\ref{Fig:results}-(a), -(b), and -(c)
are predictions obtained with MAGBOLTZ-1 (version 1.16) through
its GARFIELD interface,
although its accuracy for the random velocity distribution of electrons
is known to be limited under certain circumstances\cite{Ref:robson}.
The loss of accuracy is caused by a decomposition of the
velocity distribution function in Legendre polynomials,
in which the lowest two or three terms are retained in the calculation.
Therefore results given by the program may not be precise
enough when the velocity distribution deviates far from
isotropy or it has no axial symmetry as in the case
of crossed electric and magnetic fields.
On the other hand, our experimental condition seems to be
favorable for application of MAGBOLTZ-1:
electrons in ${\rm CO}_2$-based gas mixtures under a low electric
field are nearly thermal, i.e. the velocity distribution is
close to Maxwellian, and the axial symmetry of the velocity
distribution holds to a good extent,
because the momentum transfer cross section is fairly
constant over the main portion of the electron velocity (energy)
distribution.
To confirm this we ran the Monte Carlo version 
(MAGBOLTZ-2, version 2.2\cite{Ref:MAGBOLTZ2}),
which is free from the problems stated above
though time-consuming,
to simulate Lorentz angles for several electric and magnetic
field combinations.
The results were found to be consistent with those
obtained with MAGBOLTZ-1.

\begin{figure}
\centerline{
\epsfxsize=7cm 
\epsfbox{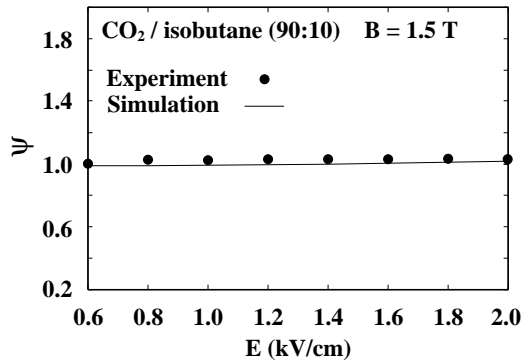}
}
\begin{center}\begin{minipage}{\figurewidth}
\caption[Fig:psi]{\label{Fig:psi}\sl
Magnetic deflection coefficient at $B = 1.5$T
as a function of the electric field
for the ${\rm CO}_2$/isobutane(90:10) mixture.
}
\end{minipage}\end{center}
\end{figure}

\begin{figure}
\centerline{
\epsfxsize=7cm 
\epsfbox{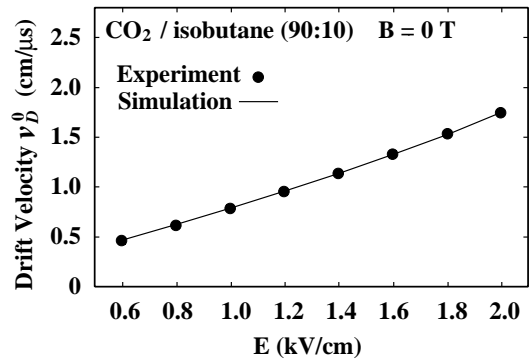}
}
\begin{center}\begin{minipage}{\figurewidth}
\caption[Fig:velocity]{ \label{Fig:velocity}\sl
Drift velocity at $B = 0$
as a function of the electric field
for the ${\rm CO}_2$/isobutane(90:10) mixture.
}
\end{minipage}\end{center}
\end{figure}

\begin{figure}
\centerline{
\epsfxsize=7cm 
\epsfbox{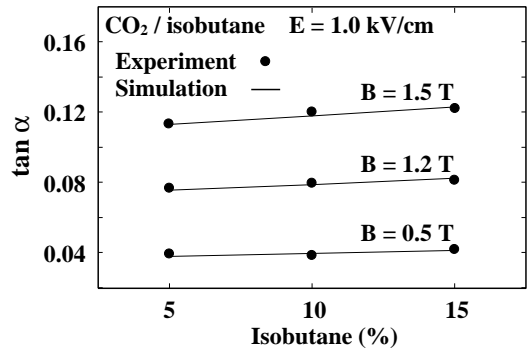}
}
\begin{center}\begin{minipage}{\figurewidth}
\caption[Fig:gasmixtures]
{ \label{Fig:gasmixtures}\sl
Isobutane concentration dependence of $\tan\alpha$
at $E = 1.0~$kV/cm.
The solid lines are GARFIELD/MAGBOLTZ predictions.
}
\end{minipage}\end{center}
\end{figure}

	We calculated the magnetic deflection coefficient ($\psi$)
from the measured Lorentz angles and the drift velocities obtained 
without magnetic field.
Fig.~\ref{Fig:psi} shows the resultant $\psi$ as a function of 
electric field strength
for the CO$_2$/isobutane(90:10) mixture at $B = 1.5$~T,
while Fig.~\ref{Fig:velocity} shows the drift velocity in the 
absence of magnetic field($v_D^0$).
The values of $\psi$ were found to be close to unity within $\pm$ 5\%
for the whole range of the applied electric and magnetic fields
and for all the gas mixtures used.
The gas dependence of the Lorentz angle is shown in Fig.~\ref{Fig:gasmixtures}.
The observed increase of the Lorentz angle with isobutane concentration
is consistent with the increase of drift velocity and with $\psi = 1$.

\begin{figure}
\centerline{
\epsfxsize=7cm 
\epsfbox{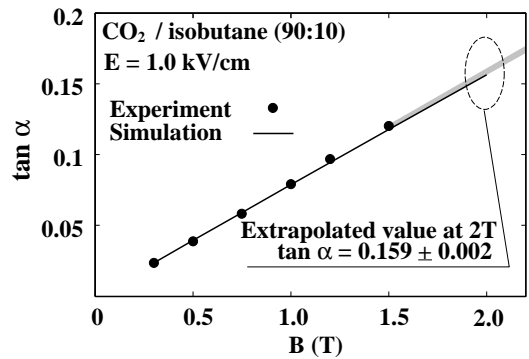}
}
\begin{center}\begin{minipage}{\figurewidth}
\caption[Fig:extra90E10garf]
{ \label{Fig:extra90E10garf}\sl
$\tan\alpha$ at $E = 1.0$~kV/cm plotted 
against the magnetic field for the 
${\rm CO}_2$/isobutane(90:10) mixture.
The shaded band above $B = 1.5$~T is the $1$-$\sigma$
bound for the straight-line extrapolation,
while the solid line is the GARFIELD/MAGBOLTZ simulation.
}
\end{minipage}\end{center}
\end{figure}

The current design of the JLC-CDC 
assumes operation 
under a magnetic field of $2.0~{\rm T}$.
Figure~\ref{Fig:extra90E10garf} plots
the Lorentz angle as a function of the magnetic field.
The Lorentz angle is proportional to the
magnetic field as long as $\psi = 1$.
We thus fit the data points below $1.5~{\rm T}$
to a straight line passing through the origin
and extrapolate the line to $2.0~{\rm T}$,
in order to estimate the Lorentz angle for
the JLC-CDC.
At $E = 1~{\rm kV/cm}$ and $B = 2~{\rm T}$ 
the extrapolated Lorentz angle is
$\tan\alpha = 0.159 \pm 0.002$ for the ${\rm CO}_2$/isobutane(90:10)
mixture.
The shaded band above $1.5~{\rm T}$
indicates $1$-$\sigma$ extrapolation error interval.
The solid line in the figure is the prediction
of GARFIELD/MAGBOLTZ which is consistent with 
the extrapolation for $2~{\rm T}$.

\begin{flushleft}
{\bf higher magnetic field option}
\end{flushleft}

\noindent
Motivated mainly by recent studies
of beam-induced background\cite{Ref:beamBG},
possibility of higher magnetic field
is now under serious considerations.
Our current cell design allows a magnetic
field up to about $3~{\rm T}$.
Since GARFIELD/MAGBOLTZ reproduces
our Lorentz angle data very well,
it is plausible that it continues to work
well at around $3~{\rm T}$, too.
It is, however, desirable to confirm
this experimentally.
We are thus planning to measure Lorentz angles
at higher magnetic fields.

%% file: dettrk/cdc/p13.tex
\subsection{Summary}
\label{Sec:detector:tracker:cdc:conclusion}

We have given affirmative answers
to most of the basic R\&D questions
we raised in the introduction:
in particular, we have demonstrated the possibility
to achieve an average spatial resolution of
$\sigma_{xy} \lsim 100~\mu{\rm m}$.
Still remaining are
those related to the gas gain saturation and space charge effects
inherent in the gas mixtures having small diffusion coefficients,
and the tension drop problem for $Al$ wires.
The space charge effects are being studied using
laser beams as well as charged particles, while
wire material studies are in progress.

As discussed in the beginning of this chapter,
there is a demand for a higher magnetic field
from beam-related background reduction point of view.
We are thus investigating possibilities to increase
the magnetic field from $2$~T to $3$~T.
If we could scale all the chamber parameters that have
the dimension of length by a factor of 2/3,
the chamber performance would stay unchanged.
This is, however, impracticable since
wire spacing, spatial resolution, 
and inner radius are difficult to scale, 
if not impossible.
The expected performance of a $3$~T design,
where these three parameters are kept unchanged,
is discussed at the end of this chapter.
In any case, any design change should be made
consistently to the whole detector system,
and should be justified by physics
simulation studies.
For this purpose, we started developing a full
detector simulator with GEANT4,
while brushing up the one based on GEANT3.

%% file: dettrk/pfm/main.tex
\section{Performances of Tracking System}

\subsection*{Vertexing Performance}
Using a quick simulator, performance of the topological vertexing 
and the mass tagging method was studied.
The topological vertexing algorithm was developed by
the SLD group\cite{ZVTOP}. 
In this method, a tube of probability is defined along a particle
trajectory.  Since the probability is high when trajectories 
overlap, such points are selected as vertices.
When a secondary or tertiary vertex is found,
the $p_t$ corrected mass ($M_{corr}$) is calculated from
the vertex mass ($M_{vtx}$) and the vertex momentum 
transverse to the flight direction  of the vertex ($P_T$):

For the study, we generated quark pair events at the $Z$ pole.
The events were clustered to two jets by using the
JADE clustering algorithm. $Y_{cut}$ was varied so as to 
force clustering to two jets.  
When the production angle of the jet satisfied
$|\cos\theta_{jet}|<0.8$, the tagging method was applied and
efficiencies and purities were studied.

\begin{figure}
\centerline{\epsfxsize=12cm\epsfbox{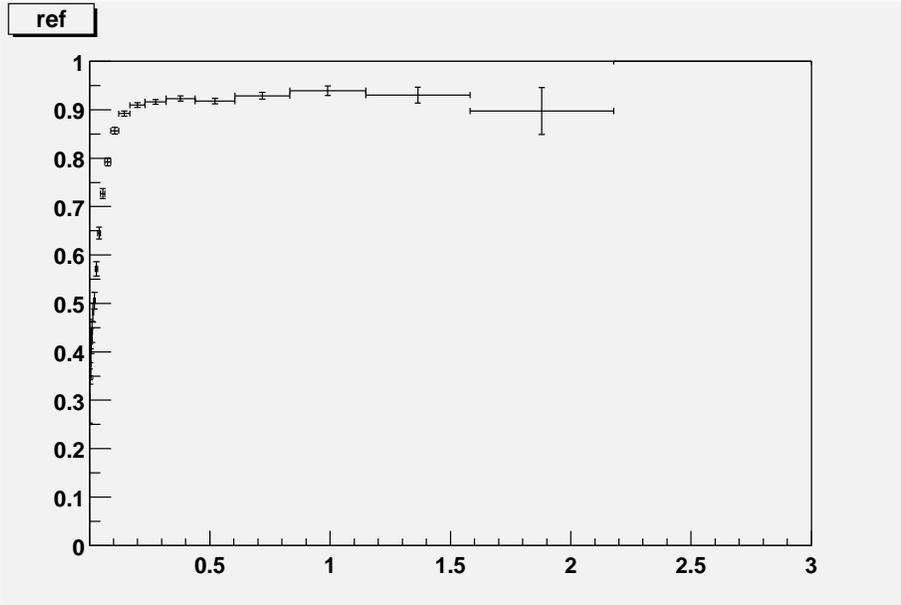}}
\begin{center}\begin{minipage}{\figurewidth}
\caption{\sl\label{dettrk-pfm-beff.eps}
The efficiency to find secondary vertices in $b$-quark jets
as a function of decay length (cm).
}
\end{minipage}\end{center}
\end{figure}

\begin{figure}
\centerline{\epsfxsize=12cm\epsfbox{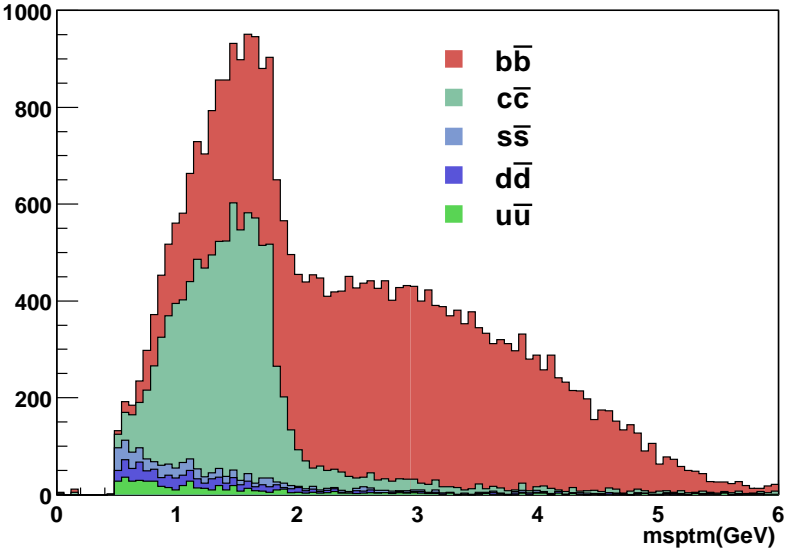}}
\begin{center}\begin{minipage}{\figurewidth}
\caption{\sl\label{dettrk-pfm-msptm.eps}
The $p_t$ corrected secondary mass ($M_{corr}$) 
distribution for each quark.
}
\end{minipage}\end{center}
\end{figure}

Fig.~\ref{dettrk-pfm-beff.eps} shows the probability to find secondary 
vertices for $b\bar{b}$ events.  The efficiency 
was about 90\% and flat for a decay length greater than about 0.1 cm.
The distribution of $M_{corr}$ is shown in Fig.~\ref{dettrk-pfm-msptm.eps}
for jets whose secondary decay lengths are greater than 300 $\mu$m.
We can see clear separation between $b$ quark jets
and $c$ quark jets.

\begin{figure}
\centerline{\epsfxsize=12cm\epsfbox{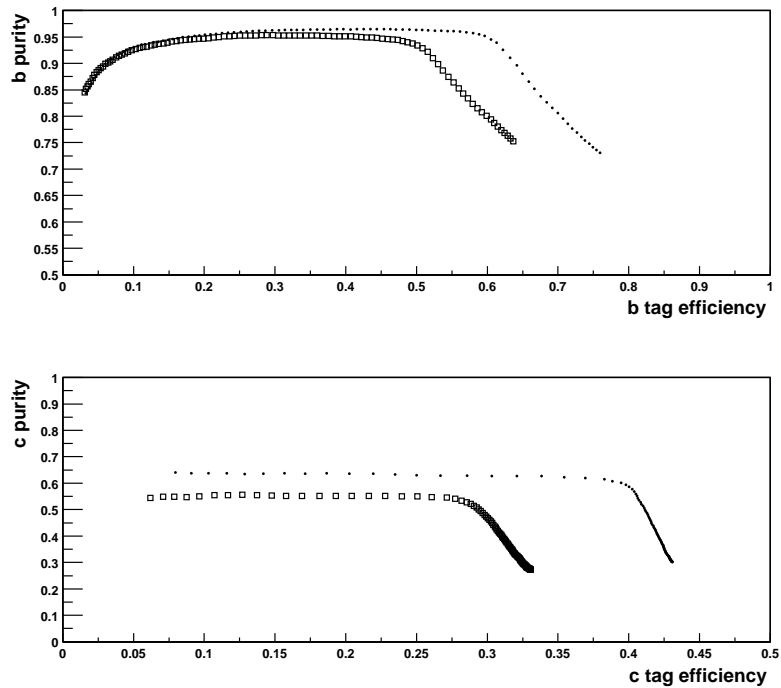}}
\begin{center}\begin{minipage}{\figurewidth}
\caption{\sl\label{dettrk-pfm-effsum.eps}
The upper figure is a
$b$ tag efficiency and purity for the JLC detector (open square)
and for additional CCD layer at 1.2 cm (dots).
The lower one is for $c$ tag.
}
\end{minipage}\end{center}
\end{figure}

We tagged jets as $b$ when the decay length was greater than 
300 $\mu$m and $M_{corr}$ was greater than the cut value.
The purity and the efficiency of the $b$ tagging 
is shown in Fig.~\ref{dettrk-pfm-effsum.eps},
which was obtained by changing the cut value 
from 1.0 GeV to 5.0 GeV. The open-square points are for the 
JLC 3T detector. For comparison, we defined a new geometry
with an additional CCD layer at 1.2 cm and 
the beam pipe at 1.0 cm.  The purity and the efficiency in that
case is shown in the same figure.
In this case, the requirement to the decay length was reduced 
to 150$\mu$m, thanks to the improved vertex resolution. As a result,
the efficiency was increased about 10\% for the selection of purity 
greater than 95\%.

In the case of $c$ tagging, we required the decay length
to be greater than 150$\mu$m and $M_{corr}$ between 0.55 GeV
and the cut value, which was changed from 1.0 GeV to 5.0 GeV.
The result is also shown in Fig.~\ref{dettrk-pfm-effsum.eps} for the
case of the JLC detector and with the additional CCD layer.
The $c$-tagging efficiency is about 30\% with the purity of about 50\%
in the case of the JLC detector.  Both the efficiency and the purity 
will be improved by using additional conditions 
such as vertex momentum, etc. 

\subsection*{The Tracking Performance}

With VTX and CDC combined,
we studied the overall performance of the tracking system,
in terms of the momentum resolution, the impact parameter resolution,
and the missing mass resolution for the $e^+e^-\rightarrow ZH$ process
using the JIM Full simulator.

\begin{figure}
\centerline{\epsfxsize=12.0cm \epsfbox{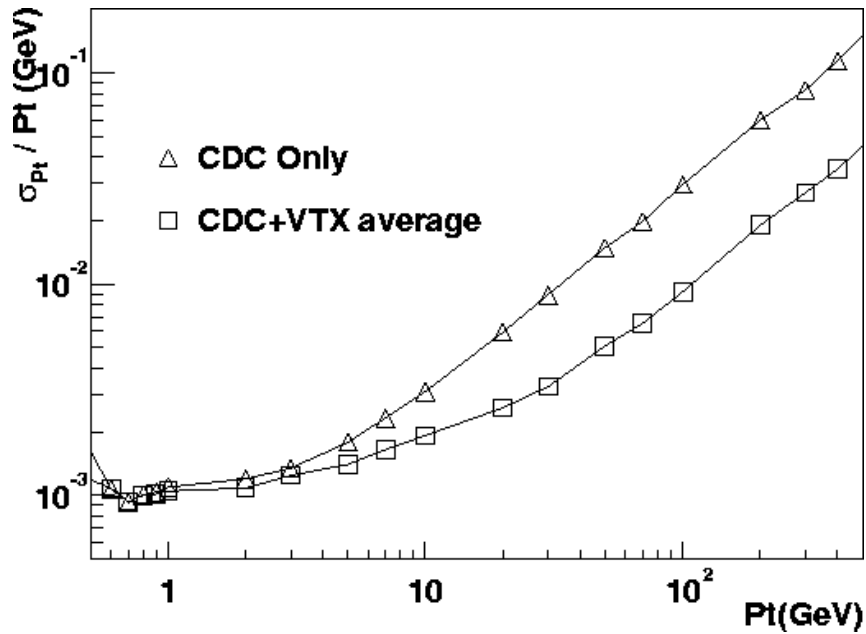}}
\begin{center}\begin{minipage}{\figurewidth}
\caption{\label{ptresol.eps}\sl
The $p_t$ resolution by a fit to CDC hits only and
by a CDC and VTX helix average}
\end{minipage}\end{center}
\end{figure}

To study the momentum resolution, we generated single $\mu$ events
by JIM.  Generated exact hit points were
smeared by assumed resolution.  
In each event, hit points were fitted by a helix
separately for CDC and VTX.  After moving the pivot of 
the CDC helix parameters
to that of the VTX ones, taking into account energy loss and
multiple scattering by materials between them, 
they were averaged with weights of error 
matrices to get a CDC-VTX combined helix parameter vector.
This method does not require any cpu-time-consuming minimization 
procedure such as kink-fitting but yields reasonable results
since both of multiple scattering and energy loss
in the materials can well be approximated by Gaussian.
The obtained $p_t$ resolution is shown in Fig.~\ref{ptresol.eps}.
The $p_t$ resolution was 
$\sigma_{p_t}/p_t = 0.9\times 10^{-4} p_t \mbox{[GeV/c]}$ at high momentum.

\begin{figure}
\centerline{\epsfxsize=12.0cm \epsfbox{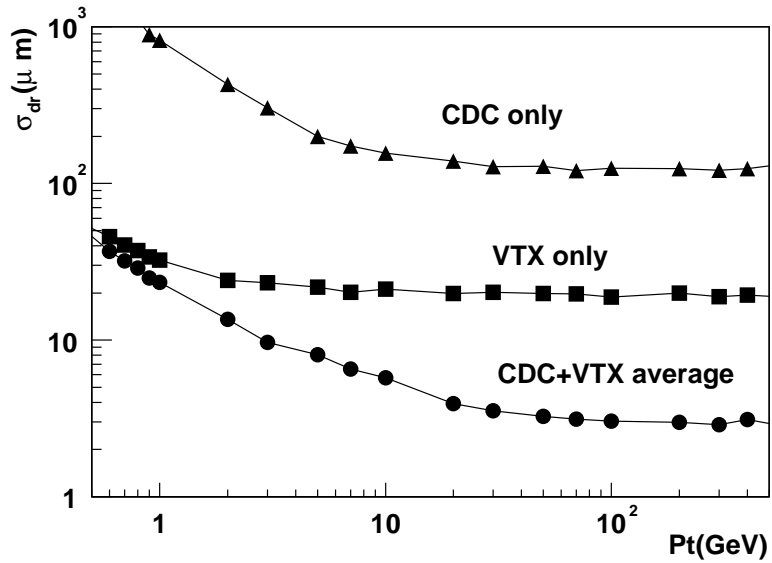}}
\begin{center}\begin{minipage}{\figurewidth}
\caption{\label{ipresol.eps}\sl
The 2D impact parameter resolution as a function of $P_t$
by a fit to CDC hits only, VTX hits only and
CDC-VTX helix average.}
\end{minipage}\end{center}
\end{figure}
The impact parameter resolution as a function of $p_t$ is shown 
Fig.~\ref{ipresol.eps}, which is about 25 $\mu$m at 1 GeV
and about 3$\mu$m at high momentum.

The missing mass resolution was also studied using the
process $e^+e^-\rightarrow \mu^+\mu^- X$.  This is a channel to 
search the Higgs boson independently of its decay mode and   
the width of the Higgs boson can be measured 
if it is large enough.
The beamstrahlung spectrum of 
JLC-I\cite{JLC-I} X-band 300 GeV parameters with 
$\pm$0.5\% initial energy spread was included in the simulation.
The center-of-mass energy was 250 GeV.

\begin{figure}
\centerline{\epsfxsize=12.0cm \epsfbox{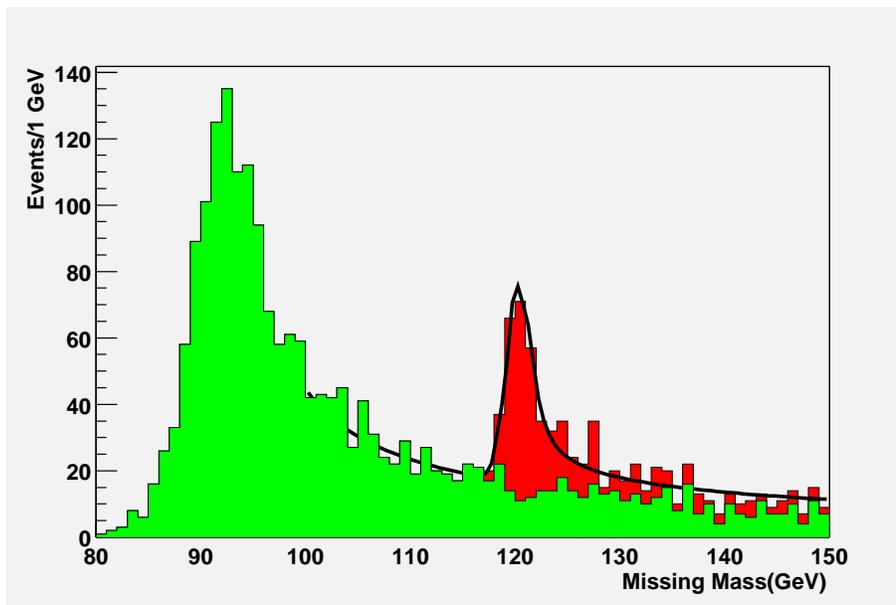}}
\begin{center}\begin{minipage}{\figurewidth}
\caption{\label{mmdis.eps}\sl
The missing mass spectrum of the two tracks for 
$e^+e^-\rightarrow ZH$(red) and $ZZ$(green) processes, where
$Z$ decays to $\mu^+\mu^-$. The center of mass energy is 250 GeV,
and the integrated luminosity is 100 fb$^{-1}$.  The solid line is 
a fit by a function as described in the text.}
\end{minipage}\end{center}
\end{figure}

The events were selected by requiring two good charged tracks
(the number of sampling in CDC is equal to 50)
with an invariant mass between 80 to 100 GeV.
The obtained missing mass spectrum for an integrated luminosity 
of 100 fb$^{-1}$ is shown in Fig.~\ref{mmdis.eps}.
The spectrum was fitted by the curve given by 
\begin{equation}
F(m) = N_H \int F_H(m,t) \exp^{-\frac{t^2}{2\sigma^2}}dt
+ N_Z\cdot F_Z(m) ,
\end{equation}
where $F_Z(m)$ is the scaled missing energy with respect to $Z$(
$F_Z(m)= 1 - \frac{m-M_Z}{\sqrt{s}-M_Z}$),
and $F_H(m,t)$ is that to Higgs ($F_H(m,t)=
Max( 0, 1 - \frac{m-M_H+t}{\sqrt{s}-M_H})$).
From the fit, we obtained the missing mass resolution of 1.0 GeV.
The main contribution to the width is the initial energy spread 
of $\pm 0.5\%$, but the resolution could be slightly improved to 0.9 GeV
if the measurement is performed at 240 GeV.

%% file: detcal/main.tex
\chapter{Calorimetry}
\label{chapter-detcal}
\section{Design Criteria}
\input detcal/Criteria.tex


\section{Baseline Design}
\input detcal/BaselineDesign.tex

\section{Hardware Study}

\subsection{ZEUS-type Test Module}
\input detcal/ZEUS.tex


\subsection{SPACAL-type Test Module}
\input detcal/SPACAL.tex


\subsection{Tile/Fiber Test Module}
\input detcal/TileFiber.tex


\subsection{Preshower and Shower-max Detectors}
\input detcal/PSSM.tex


\subsection{Photon Detectors}
\input detcal/PhotoDetector.tex


\subsection{Engineering R\&Ds}
\input detcal/Engineering.tex


\section{Simulation Study}
\input detcal/Simulation.tex


\section{Future Prospect}
\input detcal/Future.tex

\input detcal/reference.tex

%% file: detcal/Criteria.tex

One of the most important physics to study at linear colliders is
precision study of Higgs boson produced in a reaction
$e^+ e^- \rightarrow Z^0 H$.
Large backgrounds to this process from $W^+ W^-$ and $Z^0 Z^0$ 
pair productions should be
removed by b-tagging and reconstructed 2-jet masses.
Precision study of weak couplings and top properties
also need precise event reconstruction using 2-jet masses
under up to 8-jet environment.
We therefore set a design criteria for the calorimeter system
in combination with the central tracker such that
the 2-jet mass resolution be as good as the natural width
of weak bosons for excellent $W/Z$ identification.
Based on this criteria, the baseline detector model has been
made and simulation studies have been carried out.

In order to achieve the required 2-jet mass resolution,
we set  target energy resolution for calorimeters to be
\vspace{-0.3cm}

\begin{eqnarray*}
    {\sigma_E \over E} &=& {15\% \over \sqrt{E {\rm (GeV)}}} 
        \oplus 1\% ~~~ {\rm for ~ e/\gamma, ~ and} \\
    {\sigma_E \over E} &=& {40\% \over \sqrt{E {\rm (GeV)}}} 
        \oplus 2\% ~~~ {\rm for ~ hadrons,} \\    
\end{eqnarray*}

\noindent
with reasonably fine granularity.
Here $\oplus$ means quadratic sum.
In addition,  the calorimeter should have cluster-position resolution
better than 1 mm to achieve precise track-cluster association
for precise energy reconstruction.

\begin{figure}[tbh]
\centerline{
\epsfysize=5.5cm \epsfbox{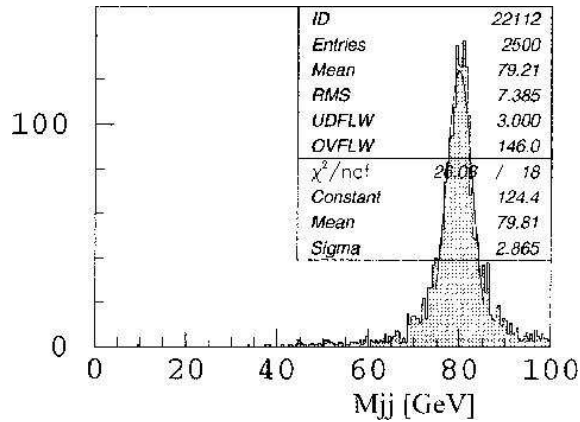}}
\begin{center}\begin{minipage}{\figurewidth}
\caption{\label{WW400}\sl
An example of reconstructed-mass resolution for $W$ with kinematical fit 
for the reaction $e^+ e^- \rightarrow W^+ W^-$ at $\sqrt{S}$ = 400 GeV,
obtained by quick simulation with the baseline detector of 2Tesla version.}
\end{minipage}\end{center}
\end{figure}

Fig.\ref{WW400} shows an example of 2-jet mass resolution
obtained by quick simulation with the baseline detector of 2Tesla design
for the reaction $e^+ e^- \rightarrow W^+ W^-$ at $\sqrt{S}$ = 400 GeV.
Obtained mass resolution of 2.9 GeV with kinematical fit is reasonably good
taking into account that analysis algorithm and parameters are 
not yet optimized very well.
Various origins of the width are decomposed in the table~\ref{WWtable}.
Each contribution was calculated by replacing each information
with generator information.

\vspace{0.5cm}

\begin{table}[h]
\begin{center}\begin{minipage}{\figurewidth}
\caption{\sl
\label{WWtable}
Various contributions to the width of reconstructed $W$ mass for the 
reaction $e^+ e^- \rightarrow W^+ W^-$ at $\sqrt{S}$ = 400 GeV.}
\end{minipage}\end{center}
\begin{center}
\begin{tabular} { l l }
\hline
\hline
Natural width of W        & $\sim$ 1.6 GeV ( in terms of $\sigma$ ) \\
Neutrino escape           & $\sim$ 0.8 GeV \\
CDC momentum resolution   & $\sim$ 1.3 GeV \\
CAL energy resolution     & $\sim$ 1.2 GeV \\
Cluster-track association & $\sim$ 1.9 GeV \\
Jet clustering            & $\sim$ 1.1 GeV \\
\hline
Total Width               & $\sim$ 3.3 GeV ( without kinematical fit ) \\
\hline
\hline
\end{tabular}
\end{center}
\end{table}


It has been widely known since the early days of $e^+ e^-$ collider
experiments that
in order to achieve the best event reconstruction,
tracking information should be used for charged particles,
while calorimeter information should be used only for neutral particles.
For this purpose, calorimeter hits generated by charged particles
should be precisely identified and deleted.
Therefore, excellent clustering followed by precise track-cluster 
association plays essential role in analysis.
This conventional method has been used for analysis of quick-simulation data.
Existence of neutral objects are identified solely by
energy-momentum unbalance of associated cluster and track for
overlapping clusters,
which does not not require very fine granularity.

From the table, it is seen that more improvements in track-cluster 
association algorithm, and consequently in hadron clustering,
should further improve this result.
Needless to say, however, verification with full simulation
is indispensable to validate/optimize the baseline parameters.

%% file: detcal/BaselineDesign.tex

The baseline calorimeter system is characterized by :
\begin{itemize}
\item excellent hadron energy resolution;
\item reasonably fine granularity;
\item whole calorimeter system inside of the superconducting solenoid.
\end{itemize}
Required hadron energy resolution of $40\%/\sqrt{E}$ can only
be achieved with compensation, which means that calorimeter
signals for EM particles and for hadrons of the same energy 
have the same amplitude.
There are two ways to achieve the compensation;
one is hardware compensation and the other is software compensation.
Software compensation, first adopted by H1 $liq.$Ar calorimeter \cite{H1},
requires very fine granularity to apply to linear collider detectors 
aimed at precision physics, and thus results in very high cost.
On the other hand, hardware compensation can be achieved by
simply adjusting the calorimeter composition, and requires no
additional effort/cost.
We therefore chose hardware-compensation with tile/fiber technique
as the baseline design.
By adopting hardware compensation, we can de-couple energy resolution
and granularity ; the best energy resolution can be achieved regardless
the granularity, and granularity can be determined only by the
criteria from topological reconstruction.
 
\vspace{0.5cm}
\noindent
The tile/fiber technique, which will be described
in detail in the following sections, 
has following features:
\begin{itemize}
\item Essentially crack less hermeticity;
\item Capability for very fine longitudinal segmentation;
\item Easy to design projective tower layout;
\item Easy to integrate pre-shower and shower-max detectors;
\item Reasonable transverse segmentation;
\item Less expensive than crystals and liq.Ar.
\end{itemize}
In addition, the hardware compensation results in following features:
\begin{itemize}
\item Excellent hadron energy resolution and linearity;
\item High density;
\item Reasonable EM energy resolution;
\item Relatively low light yield.
\end{itemize}

Our analysis based on quick simulations indicates that
very fine granularity is not necessary for event reconstruction.
Therefore  present baseline design is considered to be almost optimum 
for the hadron calorimeter (HCAL).
For the EM calorimeter (EMC), on the other hand,
very fine granularity might be required for precise event topology
reconstruction, 
even though quick simulation tells us that
present granularity shows high performance.
In that case software compensation would be another option.
Further simulation analysis will answer this question.

We also need excellent pre-shower detector (PSD) 
and shower-max detector (SMD) for
$e/\gamma/\pi^{\pm}/\pi^0$ separation, two-cluster
identification, and track-cluster association.
To have plural layers of shower position detectors (SPD)
in EMC enables precise measurement of off-vertex photon direction.

All the calorimeter system, including the photon detectors,
are designed to be located inside of the 3Tesla (or 2Tesla) magnetic field.
We therefore need high-gain high-sensitivity photon detectors
operational in the strong magnetic field, since
hardware compensation results in relatively poor
photon yield compared to the usual sampling calorimeters.
We discuss such photon detectors in a separate section.

\begin{figure}[h]
\centerline{
\epsfysize=9cm \epsfbox{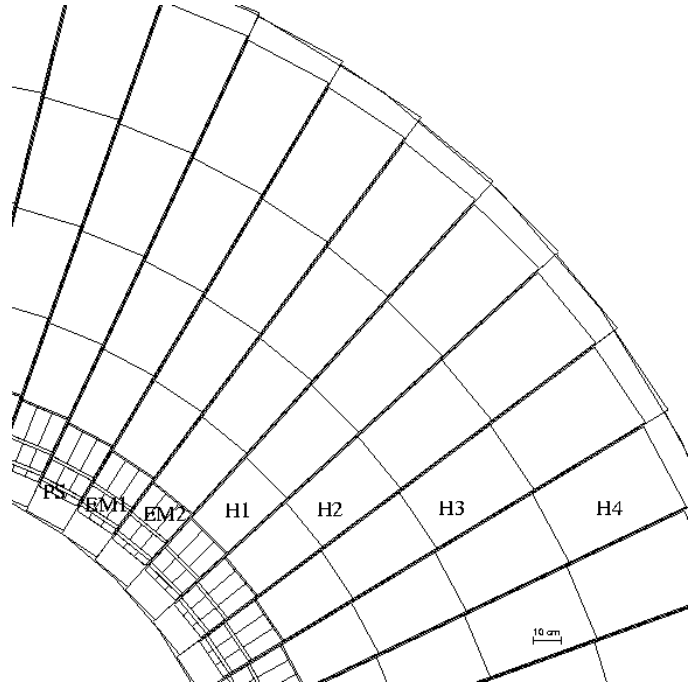}}
\begin{center}\begin{minipage}{\figurewidth}
\caption{\label{JIMcal}\sl
Configuration of the baseline design of 
barrel calorimeter system (x-y cross section)
implemented into a full simulator.}
\end{minipage}\end{center}
\end{figure}

The GEANT3 drawing of the baseline-design
barrel calorimeter is shown in Fig.\ref{JIMcal}.
There are 2Tesla option and 3Tesla option. 
Sizes of the calorimeter are smaller for the latter,
but configuration is almost the same.

In the case of the 3Tesla design,
inner and outer radii of the barrel calorimeter are 160 cm and 340 cm, 
respectively, and the length is $\pm$180 cm.
Inner radius of the endcap calorimeter is 50 cm, and the front face
locates at $z=\pm190$ cm.
Angular coverage of the endcap calorimeter with full-thickness
therefore extends to $\mid$cos$\theta\mid<0.966$, 
while partial-thickness coverage extends to $\mid$cos$\theta\mid<0.991$.

The calorimeter is an array of super-towers with pointing 
geometry with small offset.
One super-tower is composed, from the inner-radius to the outer-radius, 
of PSD (3$\times$3 in a super-tower), SMD, EMC-1(3$\times$3), 
SPD, EMC-2(3$\times$3), HCAL-1, HCAL-2, HCAL-3, and HCAL-4.
There are about 2000 super-towers in the barrel calorimeter,
and about 3000 in total in the endcap calorimeters.
One super-tower has readout channels of 31 and 48
for calorimetric measurement and for position measurement, respectively.

Parameters of these sub-detectors are summarized in Table~\ref{CALtable}.
All the parameters in the table are the first trials, and
optimization with full simulation is needed.

\begin{table}[htb]
\begin{center}\begin{minipage}{\figurewidth}
\caption{\label{CALtable}\sl
Parameters of the calorimeter system implemented into a full simulator.
2Tesla option is a revision of the previous large detector option, 
and 3Tesla option is a new detector design.}
\end{minipage}\end{center}
\centerline{
\epsfxsize=15cm \epsfbox{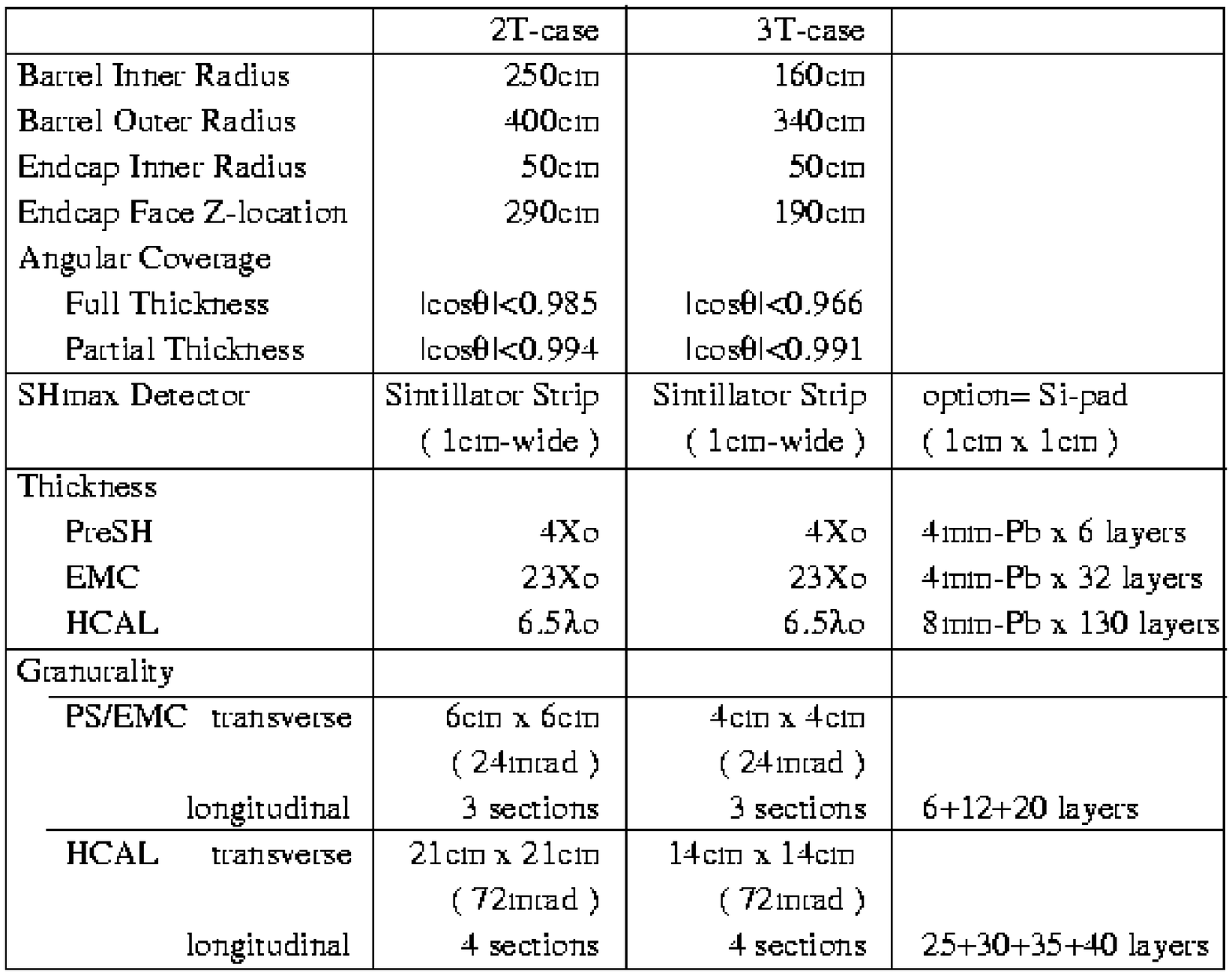}}
\end{table}

\subsection{EM Calorimeter (EMC)}

EMC is a sampling calorimeter composed of 4mm-thick lead plates
and 1mm-thick plastic scintillator plates.
Photons from the scintillator plates (tiles) are read out
via wavelength shifting fibers (WLS fibers) embedded in the tiles.
Each WLS fiber is connected to a clear fiber at the exit
of the tile, and photons are transfered
through the clear fibers to photon detectors (tile/fiber technique).
There are several ways to layout the clear fibers.
Here we adopted CDF-type\cite{CDF} ;
additional 1mm-thick acryl plates are stacked on the
tile assembly to accommodate clear fibers.
This sampling frequency is expected to achieve stochastic
term of the EM energy resolution of 15$\%/\sqrt{E}$,
provided that photon statistics does not contribute.

EMC is longitudinally divided into PSD, EMC1 and EMC2,
to improve hadron/electron identification.
Thicknesses of these sections are 4.3 $X_0$ (radiation length),
8.6 $X_0$, and 14.3 $X_0$, respectively.
We expect pion rejection better than 1/1000 with electron
efficiency of 98\% when used with HCAL information.
Transverse cell size of EMC is 4cm$\times$4cm in the baseline design
of 3Tesla-version, and 3$\times$3 EMC cells cover one HCAL cell.
This transverse size is limited by the cost and technical feasibility
of small tiles, or to be more precise, bending radius
of WLS fibers at corners.

\subsection{Hadron Calorimeter (HCAL)}

HCAL is composed of 8mm-thick lead plates, 2mm-thick plastic 
scintillator plates, and 2mm-thick fiber-routing acryl plates.
HCAL also has tile/fiber configuration.
This sampling frequency was expected to achieve stochastic
term of the hadron energy resolution of 40$\%/\sqrt{E}$,
provided that photon statistics did not contribute.

HCAL is longitudinally divided into four sections with 
thicknesses of 1.25 $\lambda_0$ (interaction length),
1.5 $\lambda_0$, 1.75 $\lambda_0$, and 2.0 $\lambda_0$.
Total thickness including EMC is 7.5 $\lambda_0$.
Transverse segment size of HCAL is
72 mrad at $\theta=90^{\circ}$ seen from the interaction point,
which corresponds to 14cm at the front face of HCAL section
in the case of 3Tesla-design.

Number of longitudinal segmentation is limited only
by number of photon-detector channels.
Improvement in multi-channel photon detectors enables
much finer longitudinal segmentation, and thus enables
more sophisticated clustering and two-cluster separation algorithms.
Transverse segment size, on the other hand,
is limited by cost, and it seems difficult to make it much smaller.

\subsection{Shower-Max Detector(SMD) and Shower-Position Detector(SPD)}

The purpose of SMD is $\pi^0/\gamma$ identification,
precise measurement of cluster position for track-cluster
association, and two-cluster recognition. 
SMD also improves $e/\pi^{\pm}$ separation by transverse signal distribution.
For this purpose, fine segmentation is necessary for SMD.
Also needed is wide dynamic range to detect minimum-ionizing particles (MIP)
and developing EM showers simultaneously.

The baseline design for SMD is an arrays of plastic scintillator strips.
Strip size of 1cm-wide and 5mm-thick is assumed at present.
An array of 12 strips covers one super-tower in the case of 
3Tesla design, and 18 strips in the case of 2Tesla design. 
Two orthogonal layers give $\theta-\varphi$ position information
with accuracy of better than 1 mm for $e/\gamma$ of energy greater
than a few GeV.
Ghosts are removed, though not completely, by pulse-height analysis.

In the baseline design, 
photons from the scintillator strips are read out by WLS-fibers,
similar to the tile/fiber technique.
Another option is under study to read out photons
using photo-diodes directly attached on the strips.

\vspace{0.3cm}

Another shower position detector (SPD) is installed between EMC-1
and EMC-2.
SPD has the same configuration as SMD.
The purpose of SPD is to measure the direction
of the isolated off-vertex photons.
SPD also helps hadron shower position measurement
by measuring early stage of hadron shower development,
and thus improves track-cluster association for charged hadrons.

The baseline design has only one set of SPD.
Whether we need more layers or not should be investigated.

\subsection{Photon Detectors}

Photon detectors for scintillating light
are designed to sit at the end of the calorimeter assembly
in the magnetic field, rather than stretching the clear fibers
to the outside of the iron yoke of the whole detector.
Therefore high-sensitivity photon detectors operational
inside of 3Tesla (or 2Tesla) magnetic field are necessary.
At present, we assume that;
\begin{itemize}
\item Both EMC and HCAL are read out by multi-channel HPDs;
\item PSD are read out either by multi-channel HPDs or multi-channel HAPDs;
\item SMD and SPD are read out either by multi-channel HAPDs or EBCCDs.
\end{itemize}
Here HPD, HAPD and EBCCD stand for Hybrid Photo-Diode, Hybrid Avalanche
Photo-Diode, and Electron-Bombarded CCD, respectively.
Though some fine-mesh photo-multiplier tubes show quite high gain
even in 2.5 Tesla, we decided not to use them.
Detailed R\&Ds are further needed to decide which option is the optimum.

\vspace{0.3cm}

There are 125k , 19k, and 223k channels of EMC, HCAL, and SMD/PSD
to read out, respectively.
Therefore, it is impractical to use single-channel photon detectors.
With multi-channel HPDs and EBCCDs under study at present,
only one multi-channel HPD and one EBCCD can read out of all the 
calorimetric signals and position detector signals
from one super-tower, respectively.

\subsection{Other Detector Options}

There are several calorimeter schemes other than the
baseline-design tile/fiber scheme. 
The features of various detector schemes
are summarized in Table~\ref{CALoptions}.
The listed features are just potentials, and whether constructed detectors 
really have such features or not is a different question.
For the energy resolution, resolutions of typical EM calorimeters are quoted.
Hadron energy resolution strongly depends on the detailed structure
and analysis algorithm, and thus is not quoted.
It should be stressed that longitudinal segmentation is essentially
important for $e/\pi$ separation.
\begin{table}[bh]
\caption{
\label{CALoptions}
\sl Properties of various detector schemes for EM calorimetry.}
\begin{small}
\begin{tabular}{|l||c|c|c|c|c||c|}
\hline
\hline
\rule[-12pt]{0pt}{30pt}
\begin{minipage}[c]{2cm}
Detector \\ Scheme 
\end{minipage}
& Crystal & SPACAL & Shashlik & W/Si & Liq/Ar & Tile/Fiber \\
\hline
\hline
\rule[-12pt]{0pt}{30pt}
\begin{minipage}[c]{2cm}
EM Energy \\
Resolution 
\end{minipage}
& Excellent & Good & Good & Good & Good & Reasonable \\
\hline
\rule[-12pt]{0pt}{30pt}
\begin{minipage}[c]{2.2cm}
Transverse \\ Segmentation
\end{minipage}
& Good & Good & Good & Excellent & Excellent & Reasonable \\
\hline
\rule[-12pt]{0pt}{30pt}
\begin{minipage}[c]{2.2cm}
Longitudinal \\
Segmentation 
\end{minipage}
& Poor & Poor & Poor & Excellent & Excellent & Excellent \\
\hline
\rule[-4pt]{0pt}{18pt}
Hermeticity 
& Good & Good & Good & Excellent & Reasonable & Excellent \\
\hline
Cost 
& Expensive & Expensive & Expensive 
& 
\rule[-12pt]{0pt}{30pt}
\begin{minipage}[c]{1.8cm} 
\begin{center} Very \\ Expensive \end{center}
\end{minipage}
& Expensive & Reasonable \\
\hline 
\hline
Examples
& \begin{minipage}[c]{1.6cm} 
\begin{center} L3, Belle \\ CMS \end{center} \end{minipage}
& \begin{minipage}{1.2cm} 
\begin{center} H1,\\  KLOE \end{center} \end{minipage}
& \begin{minipage}{1.6cm} 
\begin{center} Phenix, \\ Hera-B  \end{center} \end{minipage}
& \begin{minipage}{1.6cm} Luminosity \\ monitors  \end{minipage}
& \rule[-18pt]{0pt}{45pt}
\begin{minipage}{1.4cm} 
\begin{center} SLD, \\ D0, H1, \\ ATLAS \end{center} \end{minipage}
& \begin{minipage}{1.2cm} 
\begin{center} CDF, \\ STAR \end{center}
\end{minipage} \\
\hline
\hline
\end{tabular}
\end{small}
\end{table}

Besides the properties in the table, cryo-liquid calorimeters have 
additional features inadequate for precision physics such as:
\begin{itemize}
\item Cryogenic pipes and walls tend to behave as anomaly 
(or a hole at the worst case);
\item Materials and gap between inside tracker and EMC deteriorate 
track-cluster association and EMC performance.
\end{itemize}
Nevertheless, $liq$.Ar calorimeter has been adopted by ATLAS EMC
due to its high resistance against radiation damage,
which is not the case for linear collider environment.
For large-scale general-purpose detectors at linear colliders, therefore, 
tile/fiber scheme has the best-balanced features.

%% file: detcal/ZEUS.tex

\begin{figure}[tb]
\centerline{
\epsfxsize=10cm \epsfbox{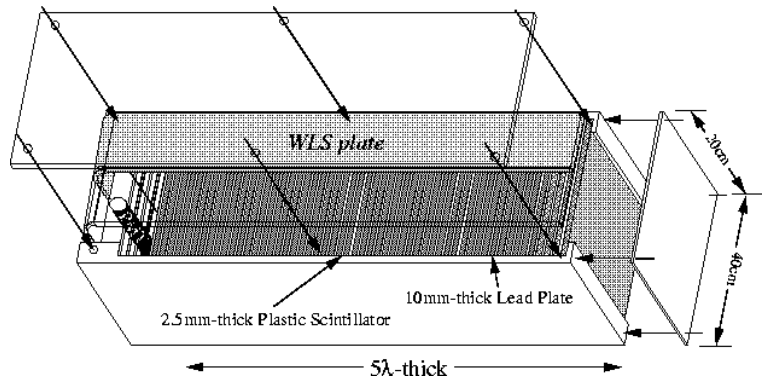}}
\caption{\label{m93sw}\sl
Schematical drawing of ZEUS-type test module.}
\end{figure}

We first investigated the possibility of ZEUS-type sandwich calorimeter
because it was the best-operating hadron 
calorimeter in the collider detectors when we started R\&Ds.
We built four hadron calorimeter test modules with similar configuration
to the ZEUS test modules\cite{ZEUS}.
It was composed of 10mm-thick lead plates and 2.5mm-thick
plastic scintillator plates as shown in Fig.\ref{m93sw}. 
This volume ratio was expected to achieve hardware compensation.
Photons were read out by WLS plates attached on the both sides of the stack.
Photon collection efficiency was higher than the tile/fiber scheme.
Number of layers was 80, which corresponds to 
the total thickness of 5$\lambda_0$.
This simple structure realizes low-cost and easy assembling,
but at the same time results in a massless gap between modules.

\begin{figure}[bt]
\centerline{
\epsfxsize=7.5cm \epsfbox{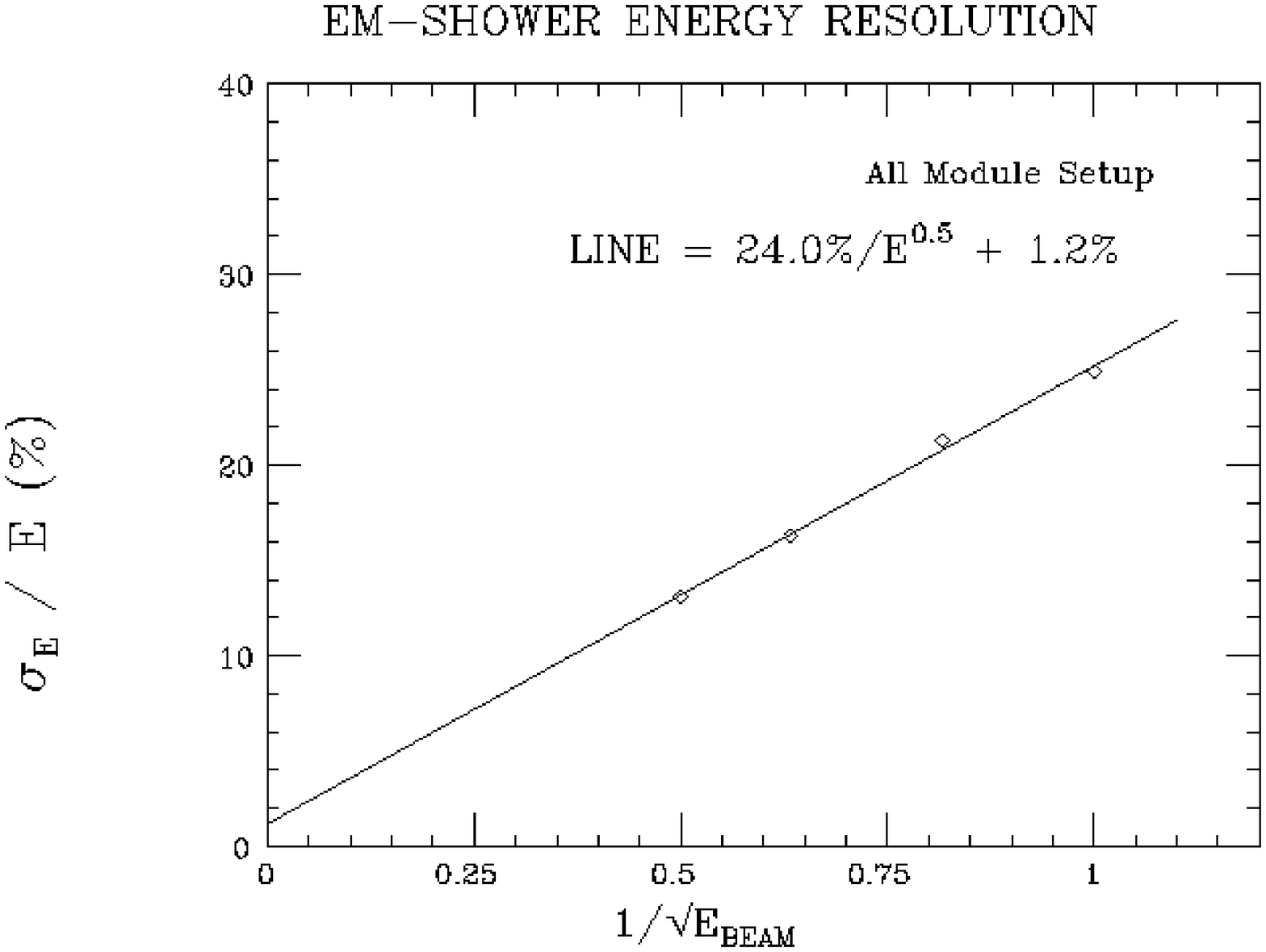}
\hspace{0.5cm}
\epsfxsize=7.5cm \epsfbox{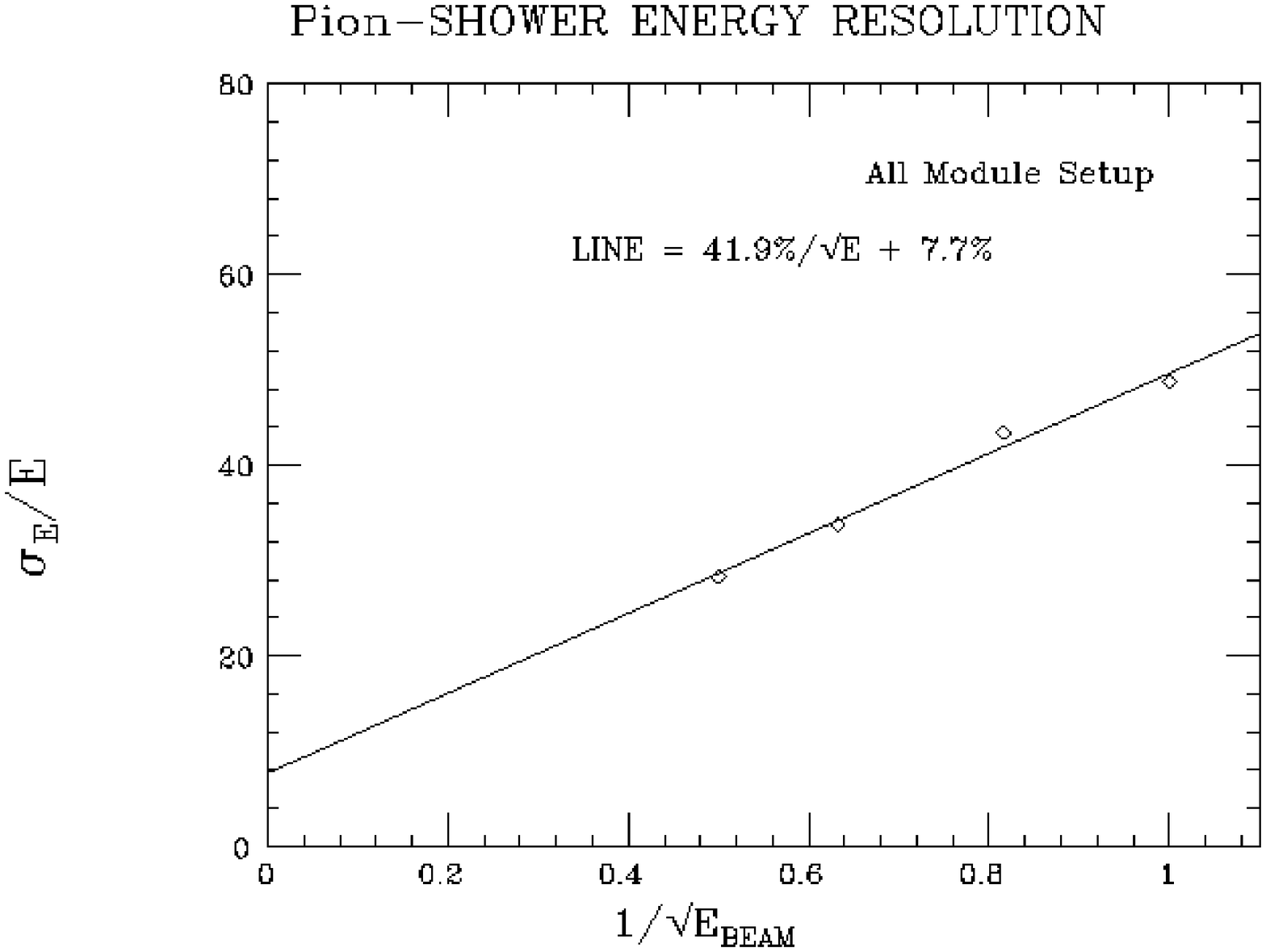}}
\begin{center}\begin{minipage}{\figurewidth}
\caption{\label{SWEres}\sl
Energy resolutions of ZEUS-type hadron calorimeter test modules 
for electrons (left) and for pions (right).}
\end{minipage}\end{center}
\end{figure}

\begin{figure}[bt]
\centerline{
\epsfxsize=7.5cm \epsfbox{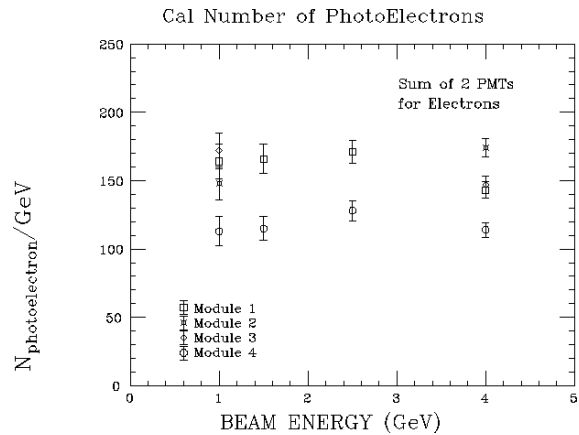}}
\begin{center}\begin{minipage}{\figurewidth}
\caption{\label{SWNpe}\sl
Photo-electron yield of ZEUS-type hadron calorimeter test modules 
for electrons. The same amount is expected for pions since the
module was hardware-compensating. }
\end{minipage}\end{center}
\end{figure}

\vspace{0.3cm}

A test beam measurement was carried out at KEK $\pi$2 beamline\cite{T305} 
using 1-4 GeV beams with a PSD, an SMD, and a SPACAL-type test
module, which is described in the next section.
Measured energy resolution is shown in Fig.\ref{SWEres}.
The results were as expected except the large
constant term for pions, which was due to transverse shower leakage
caused by small detector assembly cross section of 60cm$\times$60cm.
An e/$\pi$ ratio, a measure of compensation, was obtained to be 1.01.
Also obtained was photo-electron yield of the module to be 
160 p.e./GeV on average as shown in Fig.\ref{SWNpe}.
This photo-statistics worsens the pion energy resolution
from 40.0\%/$\sqrt{E}$ to 40.8\%/$\sqrt{E}$.
This is not a problem for hadron calorimetry.

\begin{figure}[tb]
\centerline{
\epsfxsize=7.5cm \epsfbox{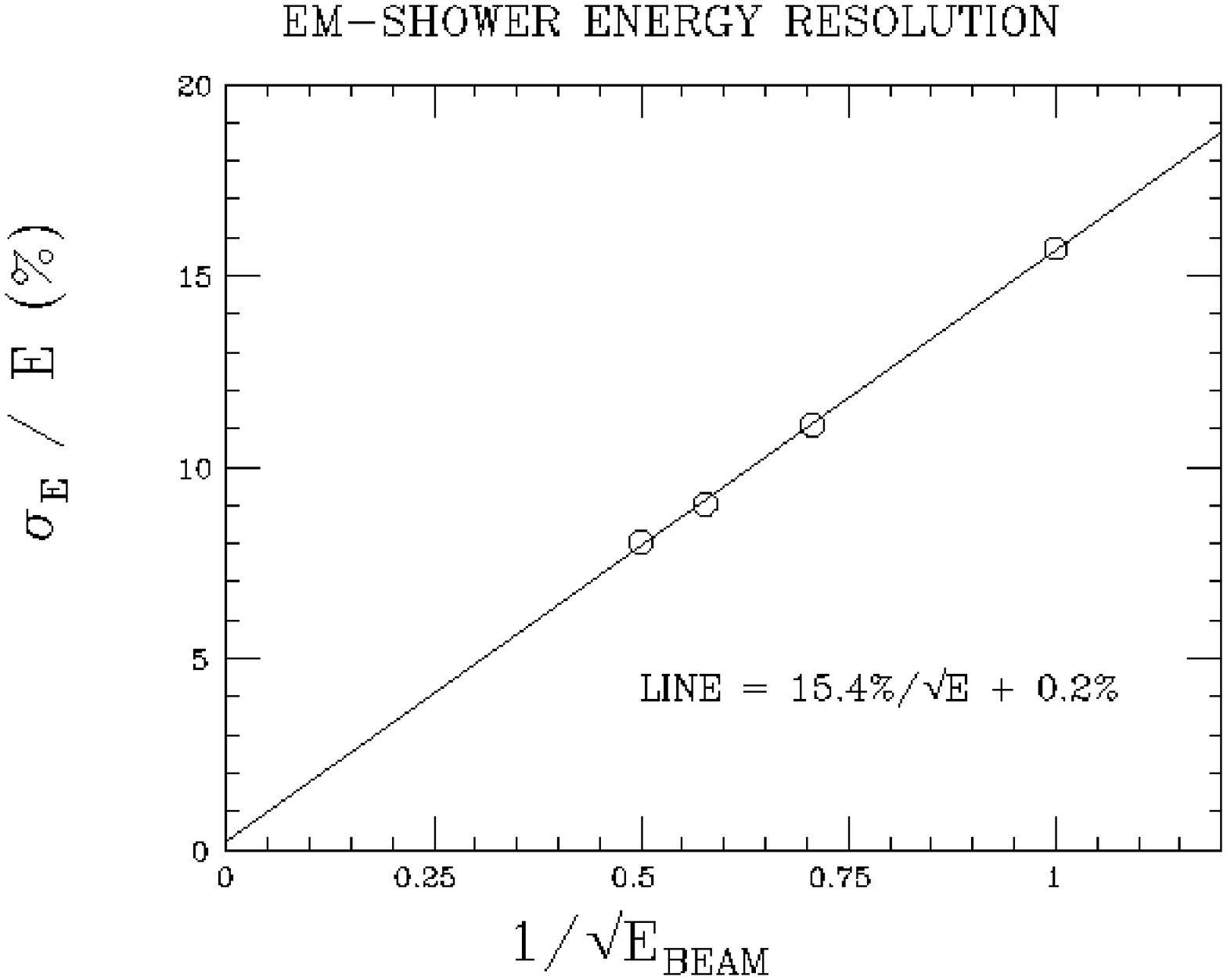}
\hspace{0.5cm}
\epsfxsize=7.5cm \epsfbox{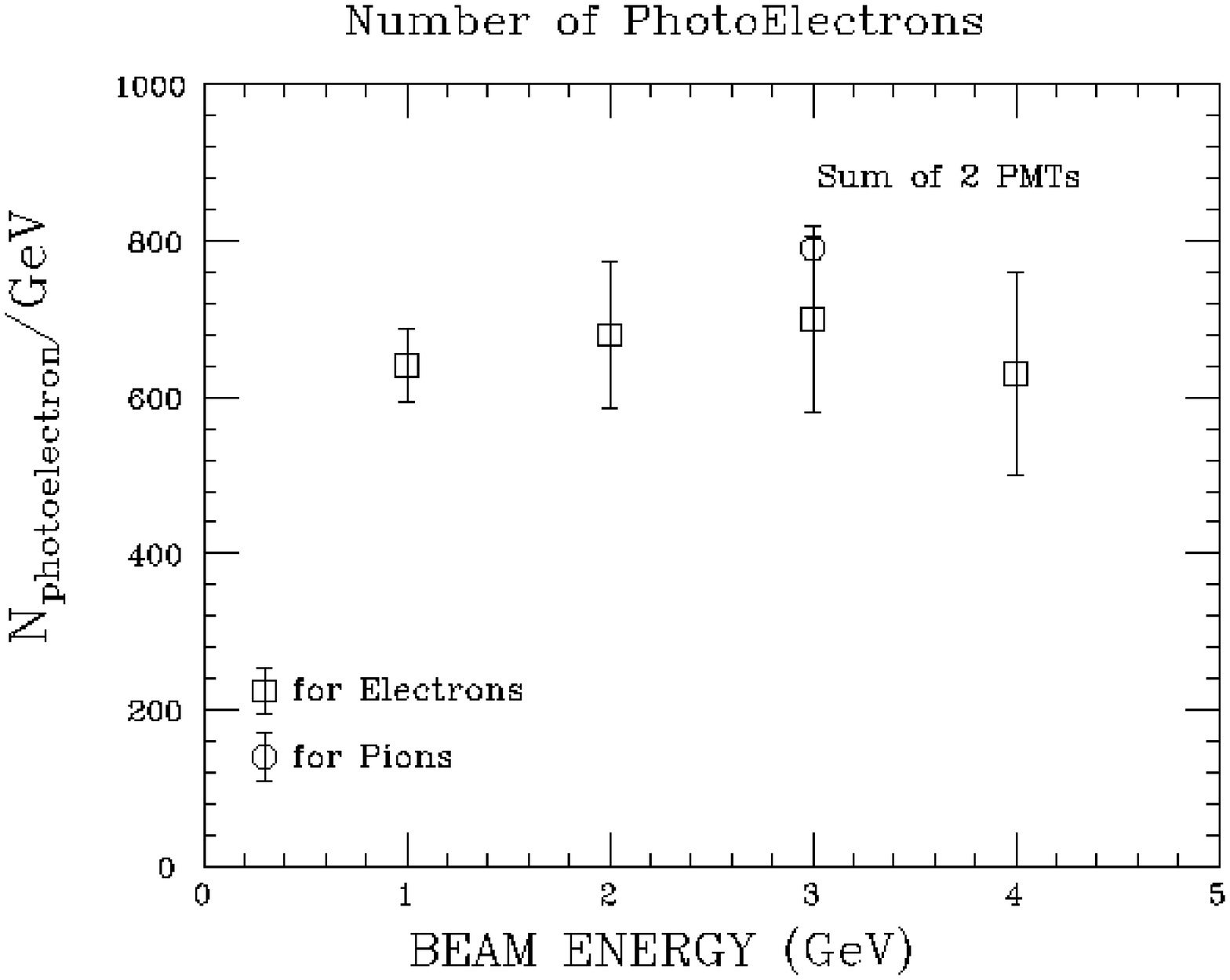}}
\begin{center}\begin{minipage}{\figurewidth}
\caption{\label{EMSW}\sl
Energy resolution (left) and photo-electron yield (right) 
of ZEUS-type EM test module.}
\end{minipage}\end{center}
\end{figure}

In order to test the response to electrons, 
another ZEUS-type EM calorimeter test module was made with 
4mm-thick lead plates and 1mm-thick plastic scintillator plates.
Beam test was carried out at KEK $\pi$2 beamline with 1-4 GeV beams.
Measured energy resolution and photo-electron yield are 
shown in Fig.\ref{EMSW}.
Obtained energy resolution of $15.4\%/\sqrt{E} + 0.2\%$
satisfies expected resolution.
The measured photo-electron yield of 660 p.e./GeV has effect 
on the stochastic term of the energy resolution
to worsen it from 15.0\%/$\sqrt{E}$ to 15.5\%/$\sqrt{E}$.
Though this would be acceptable, photon readout scheme with
better photon collection efficiency might be preferred for EMC.

Though measured energy resolutions were quite satisfactory,
we decided not to continue R\&Ds of ZEUS-type sandwich 
calorimeter due to following reasons:
\begin{itemize}
\item Longitudinal segmentation is difficult;
\item Massless gap made by WLS plates running radially 
	introduces significant response anomaly.
\end{itemize}

%% file: detcal/SPACAL.tex

\begin{figure}
\centerline{
\epsfysize=9cm \epsfbox{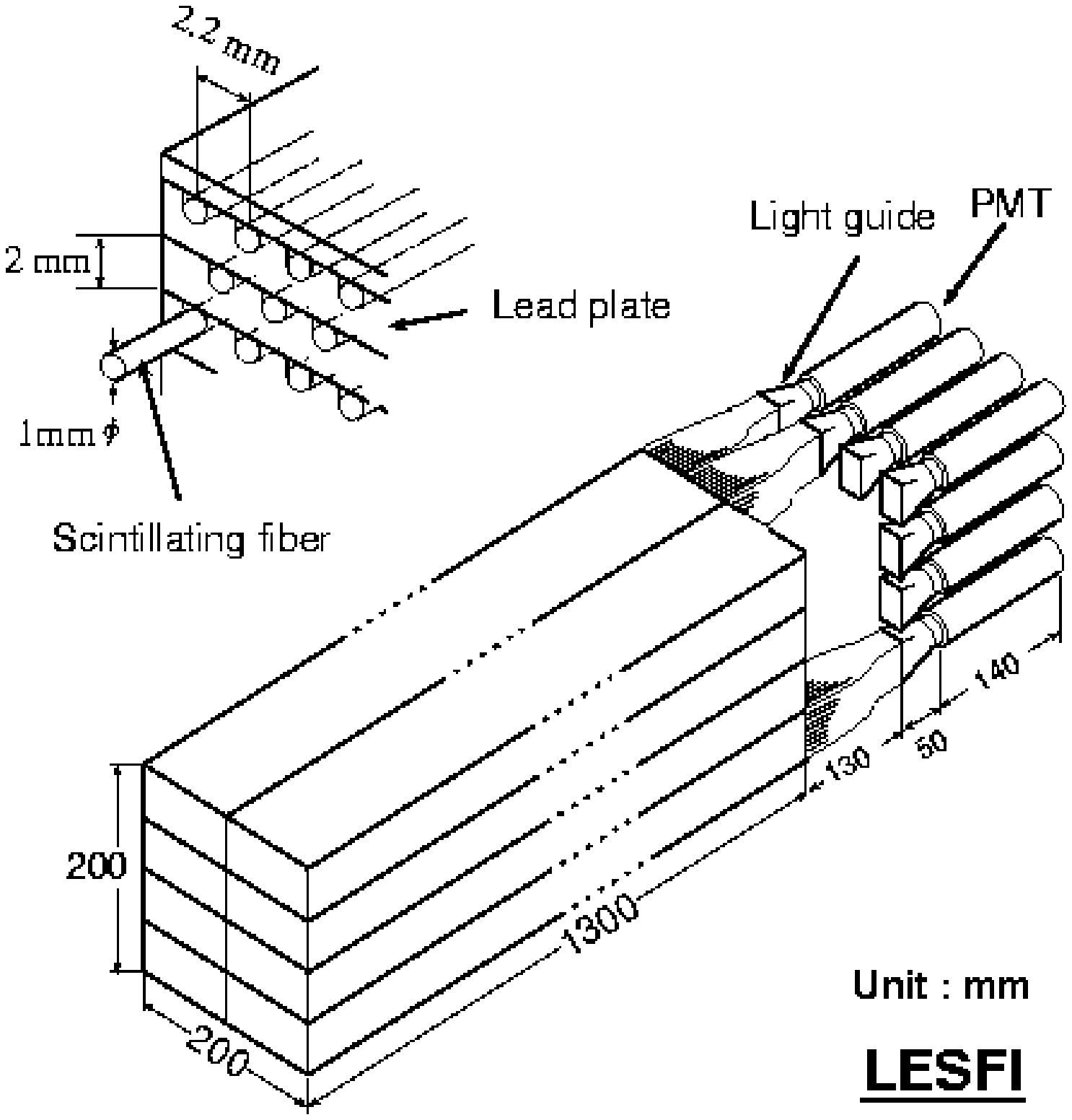}
\hspace{0.2cm}
\epsfysize=9cm \epsfbox{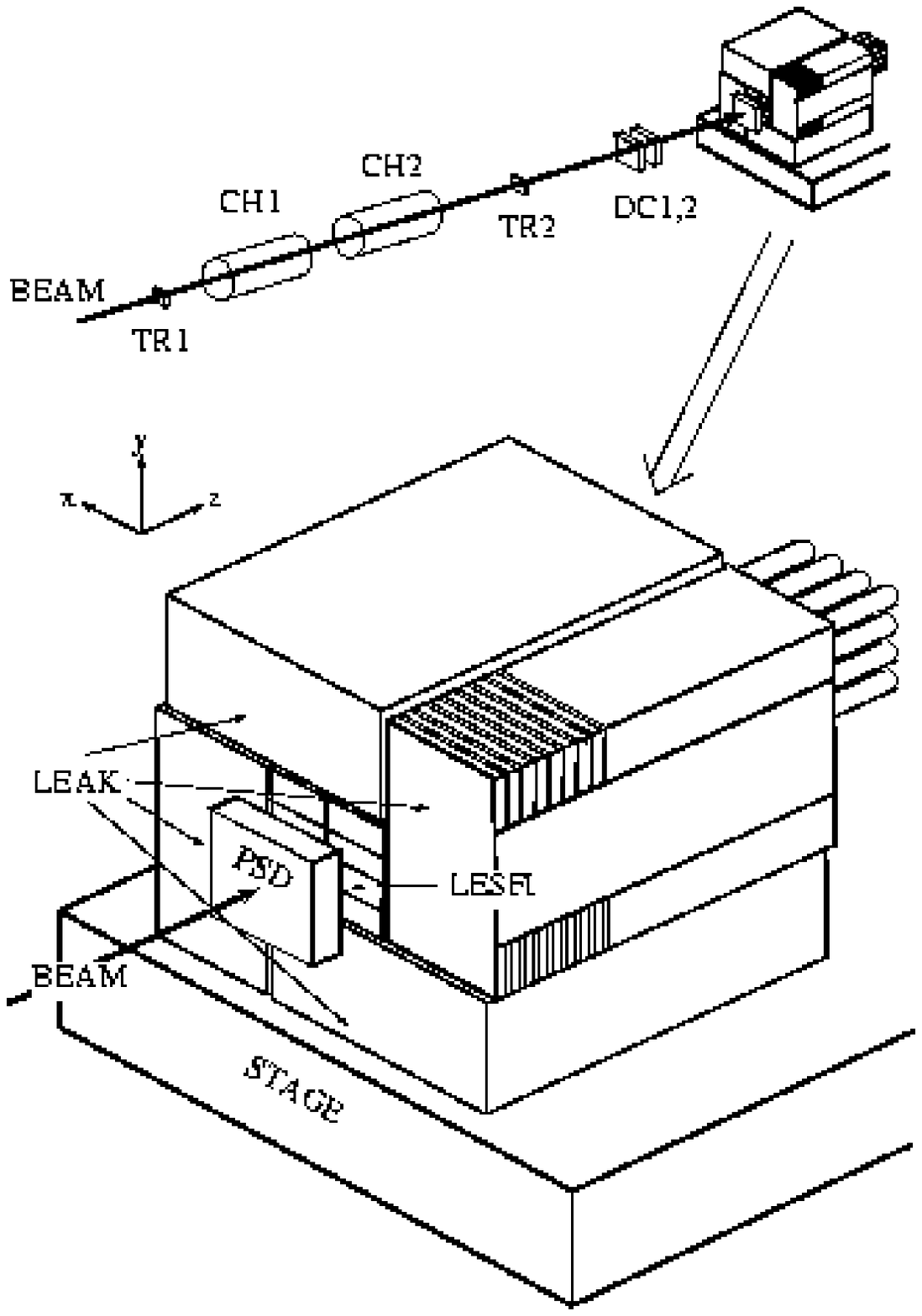}}
\begin{center}\begin{minipage}{\figurewidth}
\caption{\label{SPACAL}\sl
Schematical drawing of the SPACAL-type test module (left) and
setup of the combined beam test of ZEUS-type and SPACAL-type
test modules at KEK-$\pi2$ beamline (right).}
\end{minipage}\end{center}
\end{figure}

\begin{figure}
\centerline{
\epsfxsize=7.5cm \epsfbox{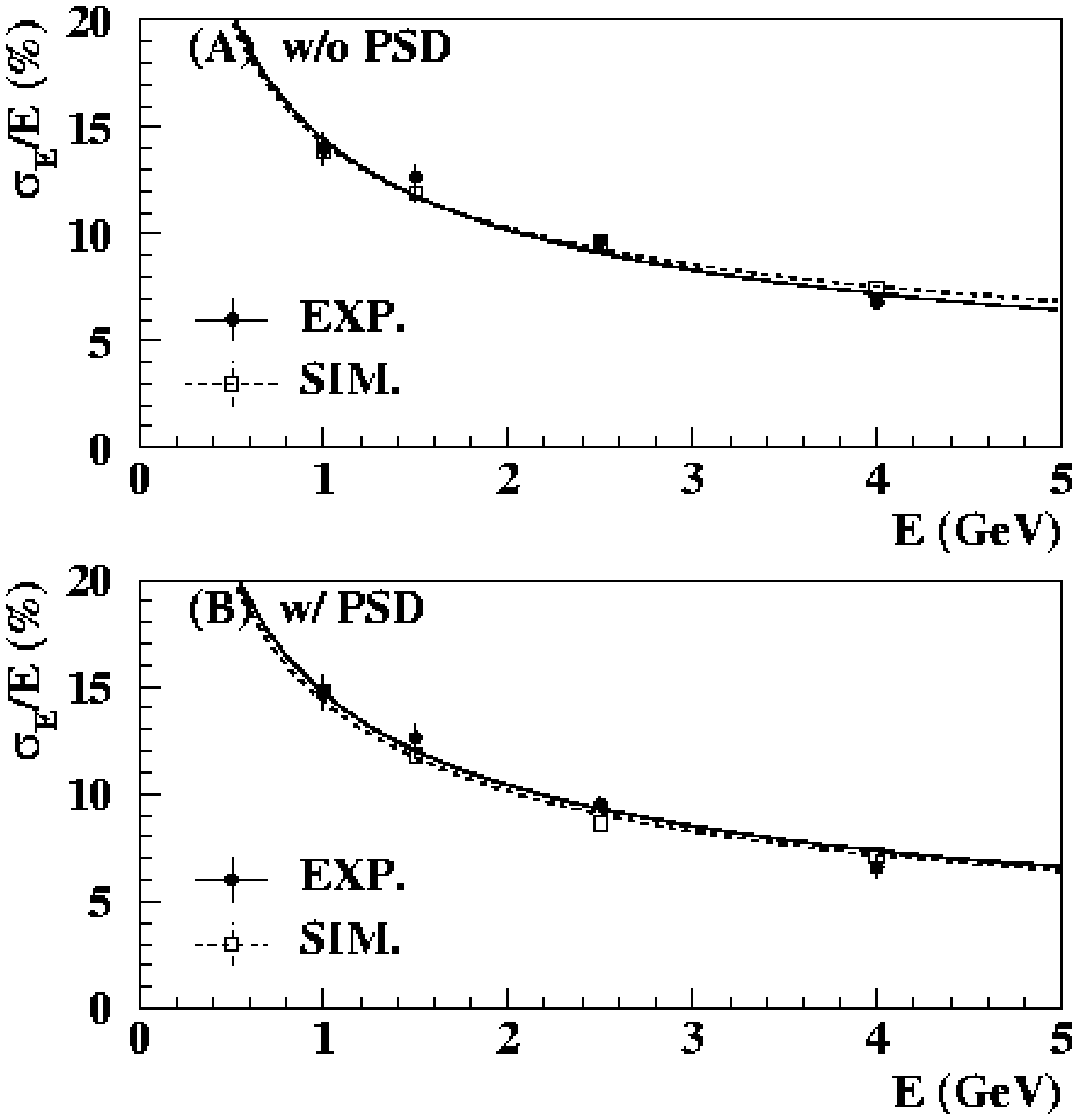}
\hspace{0.5cm}
\epsfxsize=7.5cm \epsfbox{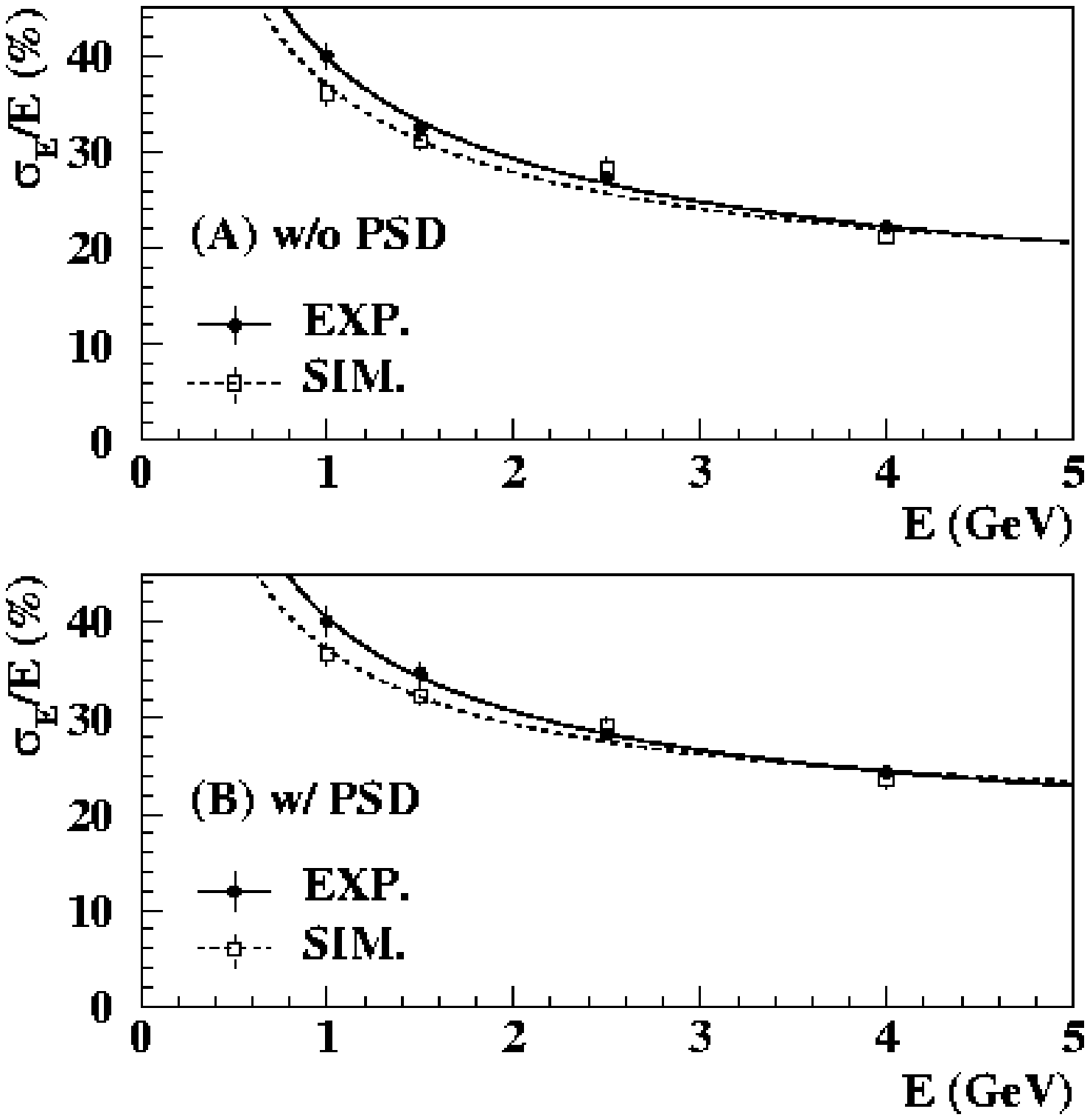}}
\begin{center}\begin{minipage}{\figurewidth}
\caption{\label{SFEres}\sl
Energy resolutions of SPACAL-type test module for electrons (left)
and for pions (right).
Bottom figures are for measurements with PSD in front,
and top ones are without PSD.
Dotted lines are results of GEANT simulation.}
\end{minipage}\end{center}
\end{figure}

We studied SPACAL-type calorimeter in parallel with
the ZEUS-type calorimeter at the first stage of the R\&D 
because the SPACAL-type calorimeter was the
best-performance calorimeter that time, 
even though there was no running experiment using SPACAL.
Since SPACAL-type calorimeter was quite expensive
and extremely elaborating, we built only one module with sizes 
of 20cm$\times$20cm$\times$130cm as shown in Fig.\ref{SPACAL}.
Pure lead plates with grooves were stacked, and
1 mm$\phi$ scintillating fibers were embedded in the grooves.
The volume ratio of lead to scintillator was 4:1 for hardware compensation.
One module was read out by 16 PMTs to have $4\times4$ sub-tower structure
for transverse shower profile measurement.

Beam test was done at KEK $\pi$2 beamline using 1-4 GeV beams\cite{T305}.
The setup is also shown schematically in Fig.\ref{SPACAL}.
One SPACAL-type module at the center was
surrounded by four ZEUS-type calorimeter modules, 
and a PSD was set in front of the SPACAL module for several measurements.

Measured energy resolutions are shown in Fig.\ref{SFEres}.
The results are parametrized as

\begin{eqnarray*}
    {\sigma_E \over E} &=& {( 14.4 \pm 0.3 )~\% \over \sqrt{E {\rm (GeV)}}} 
        \oplus ( 0.0 \pm 1.4 )~\% ~~~ {\rm for ~ electrons,} \\
    {\sigma_E \over E} &=& {(38.1 \pm 1.3)~\% \over \sqrt{E {\rm (GeV)}}} 
        \oplus (11.6 \pm 1.4)~\% ~~ {\rm for ~ pions.} \\    
\end{eqnarray*}

\noindent
These resolutions were very good except the large constant
term due to transverse shower leakage.
We also obtained pion rejection factor of 1/200 with electron
efficiency of 98\% using PSD pulse height and transverse shower profile.
This rejection score is not very high because
it is difficult to separate $e/\pi$ with transverse shower profile
at low energies.

In conclusion of the series of R\&Ds, we decided not to continue R\&Ds 
of SPACAL-type calorimeter due to following reasons:
\begin{itemize}
\item Longitudinal segmentation is very difficult;
\item Pre-shower and shower-max detectors can not be integrated naturally;
\item Really expensive and elaborating.
\end{itemize}

%% file: detcal/TileFiber.tex

   Tile/fiber structure is characterized by a wavelength shifting 
fiber embedded in a plastic scintillator tile.
This enables compact optics for light collection and transfer from 
scintillators to photon detectors without sacrificing photon yield so much.
This scheme was extensively studied by SDC group\cite{SDC}, 
and then by CDF group\cite{CDF} with slightly different fiber layout.
The tile/fiber calorimeters have been widely adopted 
by CDF, STAR\cite{STAR}, CMS\cite{CMS}, and so on.
For the JLC calorimeter, the tile/fiber structure with hardware 
compensation was adopted as a revised baseline design in 1996, 
and systematic R\&Ds were initiated.
Following bench tests of each component, test modules were 
constructed for generic shower study and realistic performance study,
and series of beam tests were carried out at KEK and at FNAL.
In the following sections, detail of the R\&Ds are described.

\begin{figure}[bht]
\centerline{
\epsfysize=7.5cm \epsfbox{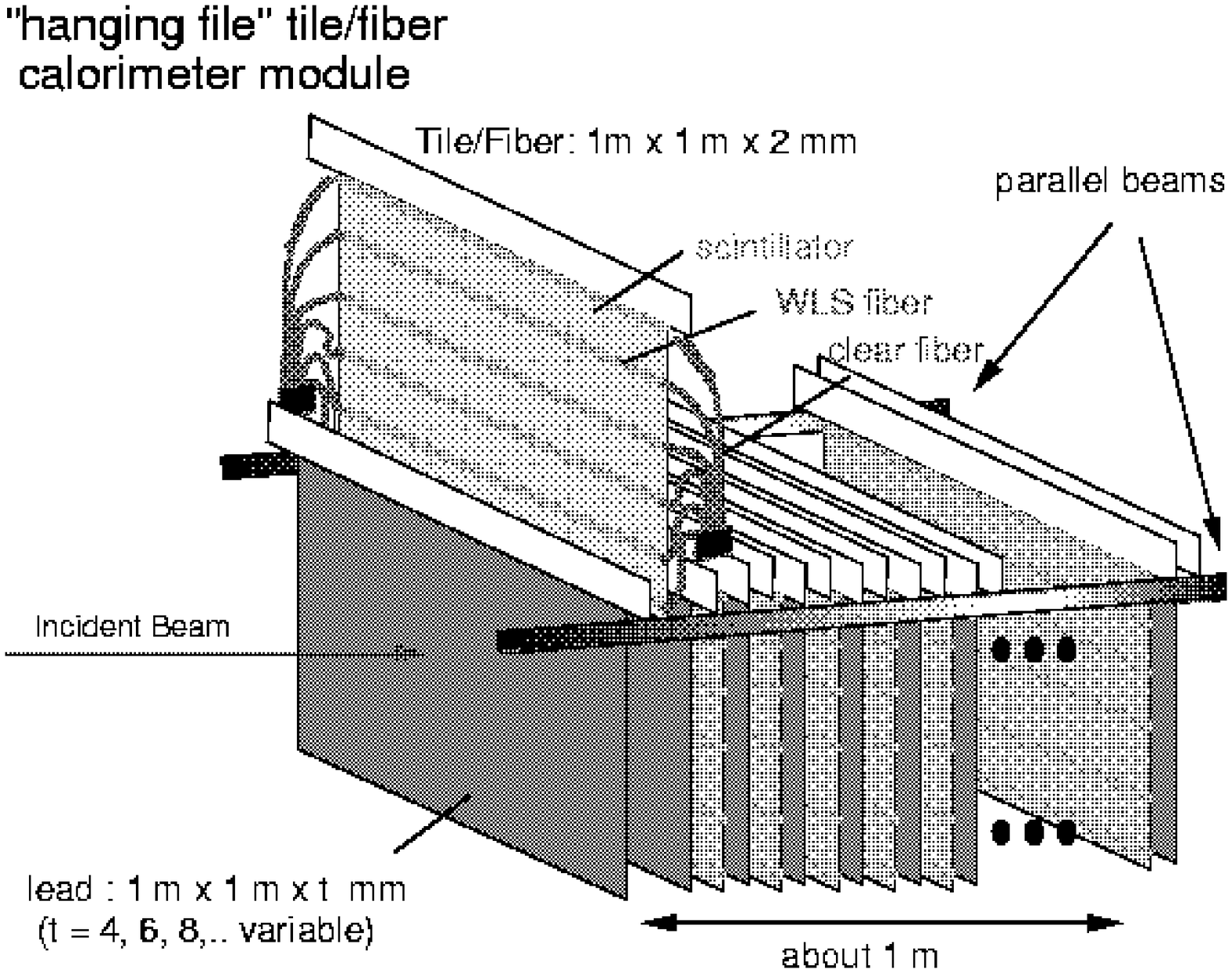}
\hspace{0.5cm}
\epsfysize=8.0cm \epsfbox{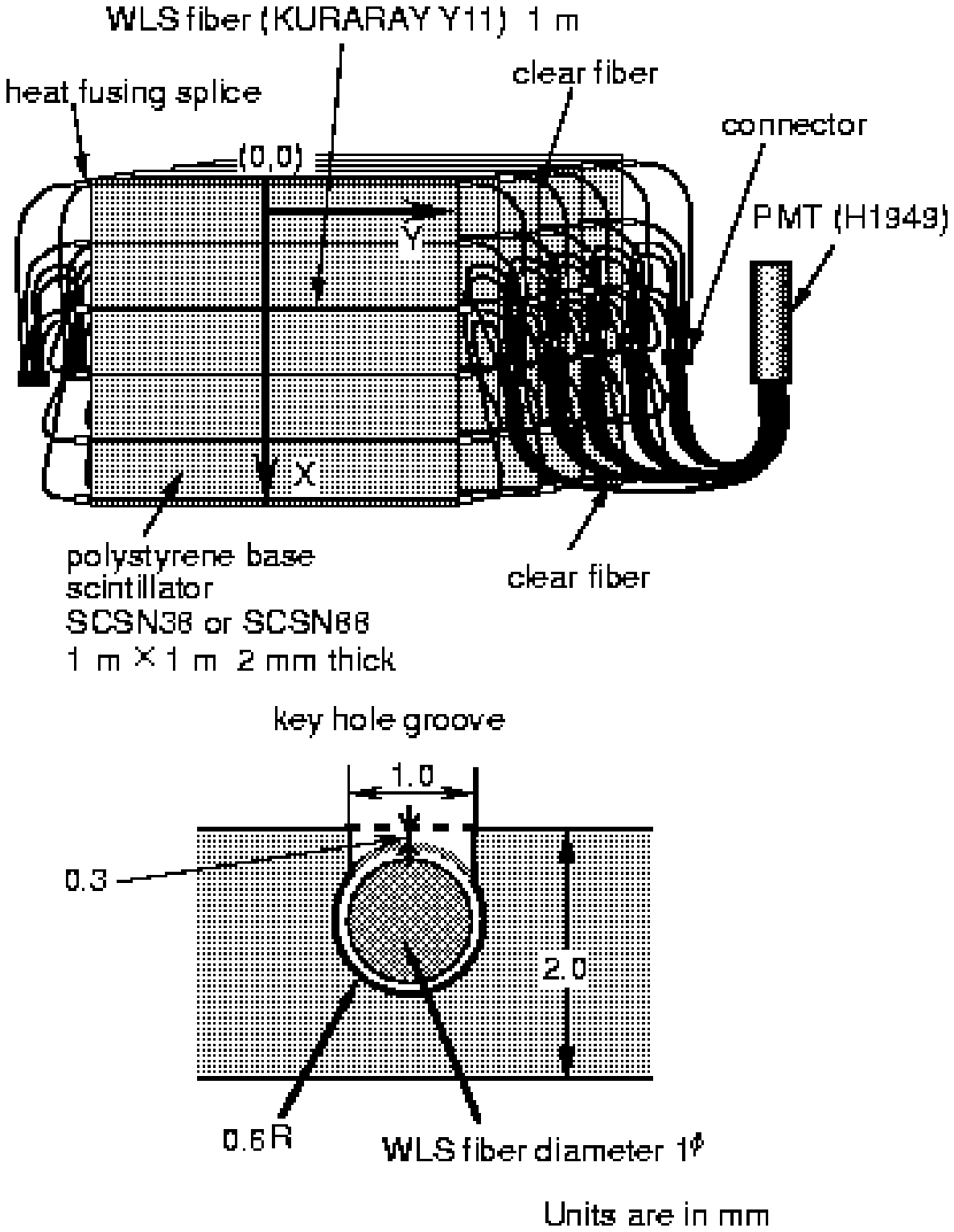}}
\begin{center}\begin{minipage}{\figurewidth}
\caption{\label{RCAL}\sl
Schematical view of straight-groove hanging-file test module (left),
and optical readout scheme (right).}
\end{minipage}\end{center}
\end{figure}

\subsubsection{1) Generic Studies with Straight-groove Module}

Though extensive R\&Ds had been carried out by SDC/CDF group before 1996,
combination with hardware compensation raised two open questions:
\begin{itemize}
\item Possibility of hardware compensation itself 
with tile/fiber configuration must have been established.
There were several calculations and measurements on compensation 
with lead and plastic scintillator sampling 
calorimeters\cite{T305,Wigmans,Acosta}.
However those results had significant discrepancies with each other.
Addition of fiber-routing plates for tile/fiber configuration
was another unknown factor on compensation.
This might affect the compensation condition.
Those problems must have been examined before making 
calorimeter module of tile/fiber design.
\item In order to realize required energy resolution with 
hardware-compensating composition using lead absorber,
very thin scintillator plates of 1mm-thickness 
should be used for EM calorimeter, and 2mm for hadron calorimeter.
Photon yield, uniformity, and mechanical feasibility must have been
examined for such thin scintillator plates.
\end{itemize}

A test module with hanging-file structure was constructed 
to examine possibility of hardware compensation.
The structure is schematically shown in Fig.\ref{RCAL}.
Lead plates and scintillator plates, and acryl plates if necessary,
were hung over a pair of supporting beams
to enable re-configuration for generic studies.
As absorbers, 4mm-thick lead plates and 2mm-thick lead sheets were
used to change the lead thickness from 4 mm to 16 mm by 2mm-step.
The thicknesses of plastic scintillator plates and acryl plates, 
on the other hand, were fixed to be 2 mm.
Transverse size of the plates were 1m$\times$1m for good 
shower containment.

\begin{figure}[bth]
\centerline{
\epsfxsize=12cm \epsfbox{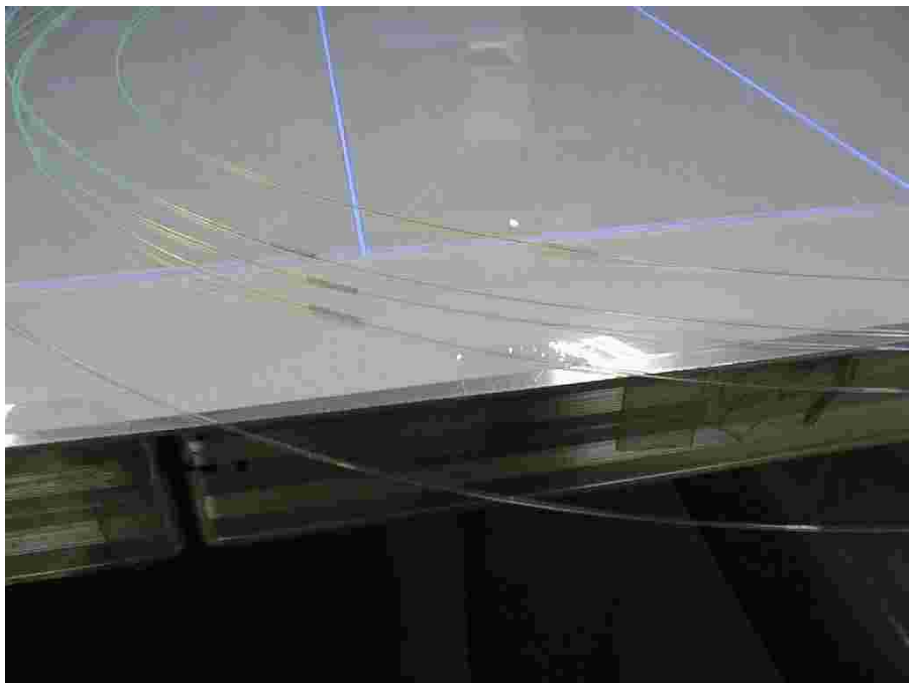}}
\begin{center}\begin{minipage}{\figurewidth}
\caption{ \label{TileWLS} \sl
A scintillator plate with WLS fibers.
Blue lines on the scintillator are the grooves.
Green part of the fibers are WLS fibers to be embedded in the grooves.
A clear fiber is connected to each WLS fiber by heat-fusing 
with a protective sleeve.}
\end{minipage}\end{center}
\end{figure}

Photon-readout scheme is shown in Fig.\ref{RCAL} 
and in Fig.\ref{TileWLS}.
The scintillator plates had six straight grooves of key-hole cross section,
where 1 mm$\phi$ WLS fibers were embedded.
Distance between the grooves, 20cm, was determined
to re-use the scintillator plates for the tile/fiber module.
Fibers from five successive scintillator plates on each side
were ganged to form a super layer, and were read out by one PMT.
There were 42 super layers in the configuration with 4mm-thick lead plates,
and detailed study on longitudinal shower profile was carried out.
In the case of 8mm-thick lead plates, there were 24 super layers,
resulting in the total thickness of 6$\lambda_0$. 

\begin{figure}[bth]
\centerline{
\epsfxsize=7.5cm \epsfbox{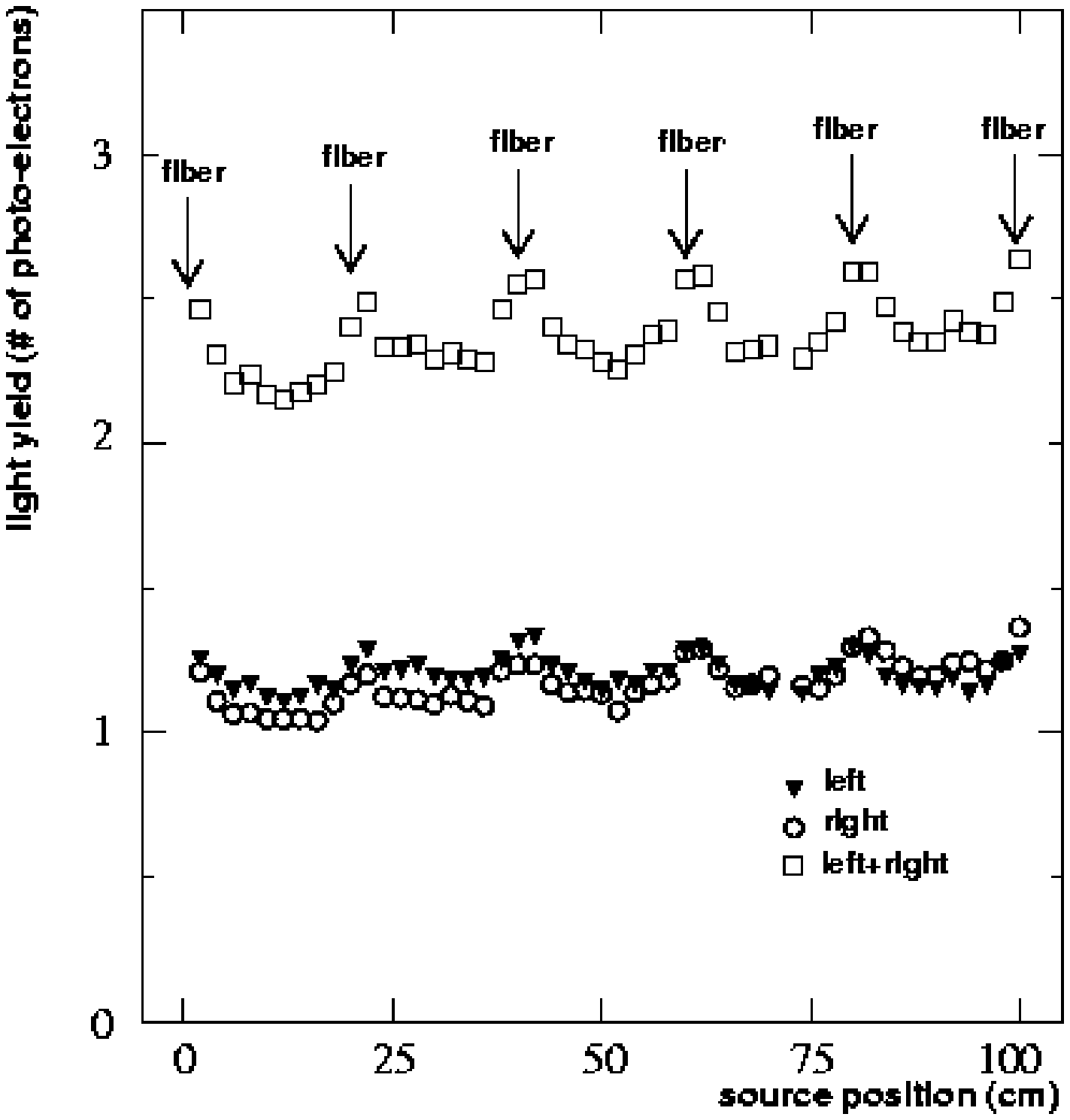}
\hspace{0.5cm}
\epsfxsize=7.8cm \epsfbox{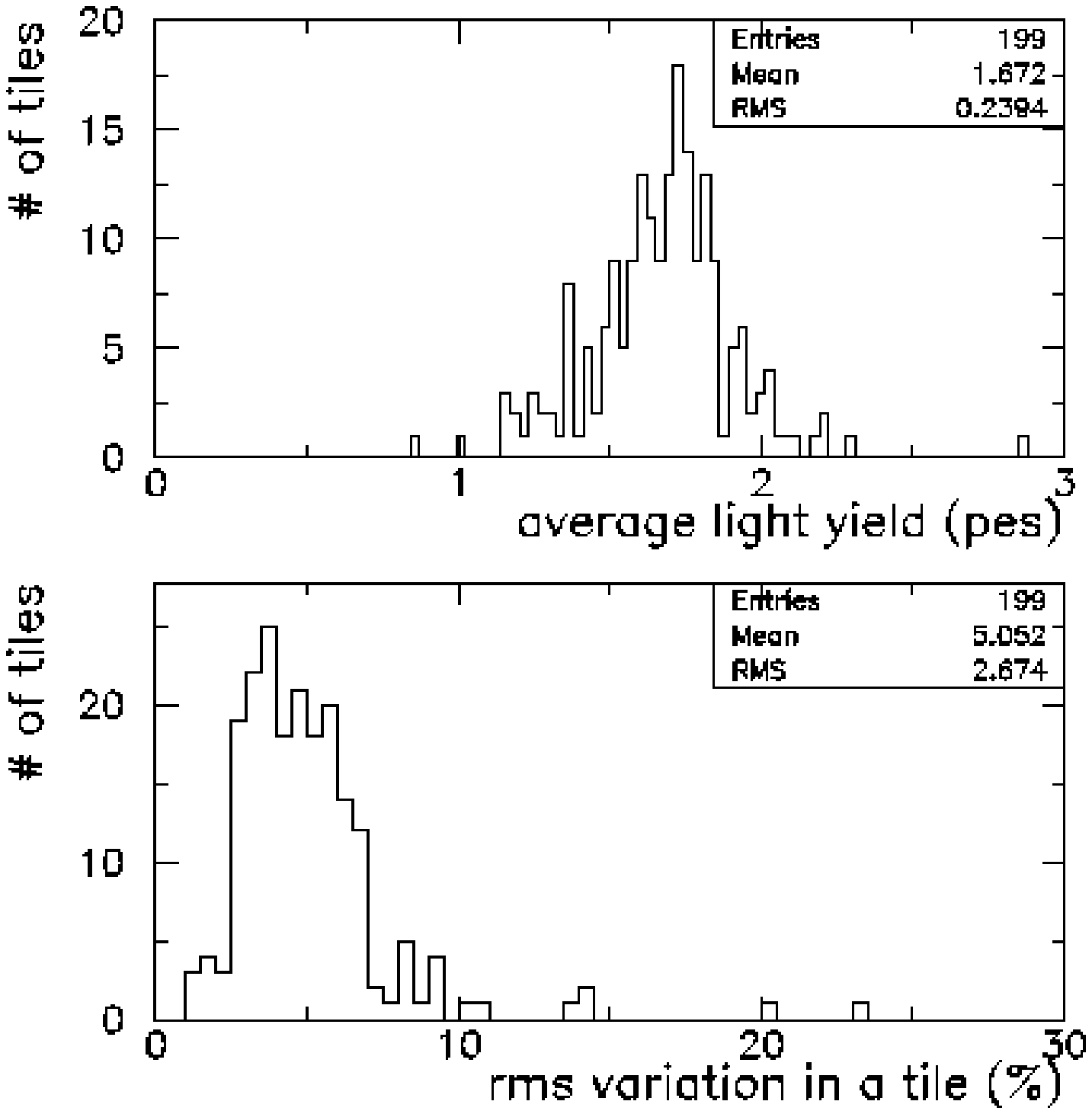}}
\begin{center}\begin{minipage}{\figurewidth}
\caption{\label{T405Npe}\sl
Non-uniformity of photo-electron yield over one scintillator plate (left),
distributions of its average (right-top) 
and RMS (right-bottom) for 199 plates.}
\end{minipage}\end{center}
\end{figure}

Response of scintillator plates were measured with an RI source
before assembling.
Non-uniformity of photo-electron yield over one scintillator plate and
distributions of its average and RMS for 199 plates 
are shown in Fig.\ref{T405Npe}.
Since the WLS-fiber distance is rather large, average photo-electron
yield is only 1.6 p.e./MIP.
Measured response map shows prominent peaks at the embedded WLS locations.
Effects of both measured non-uniformity and 
distribution of average photo-electron yield 
on calorimeter responses were examined by GEANT simulation,
and concluded not to be significant for hadrons\cite{T405}.


\vspace{0.3cm}
\noindent
{\bf Energy Measurement}
\vspace{0.2cm}

\begin{figure}
\centerline{
\epsfxsize=12cm \epsfbox{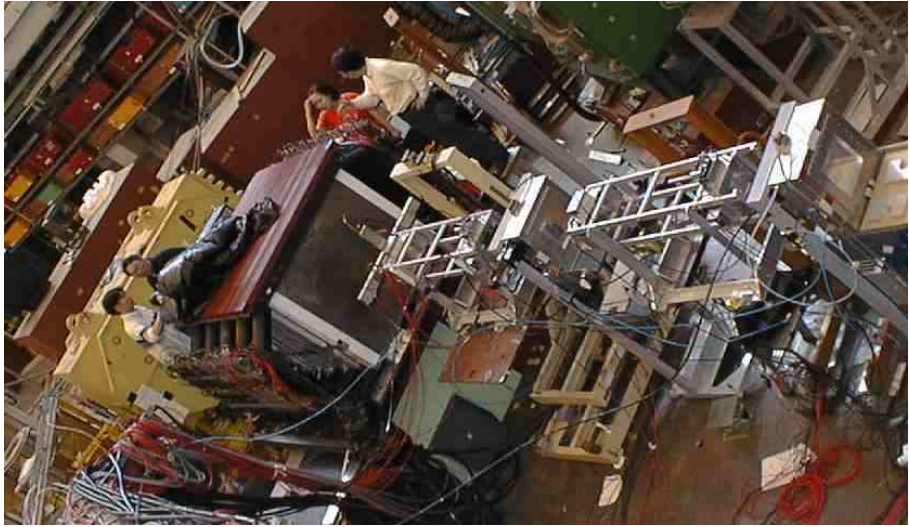}}
\begin{center}\begin{minipage}{\figurewidth}
\caption{
\label{T411photo}\sl
Setup of the beam test of hadron calorimeter test module at KEK.}
\end{minipage}\end{center}
\end{figure}

\begin{figure}
\centerline{
\epsfxsize=8.0cm \epsfbox{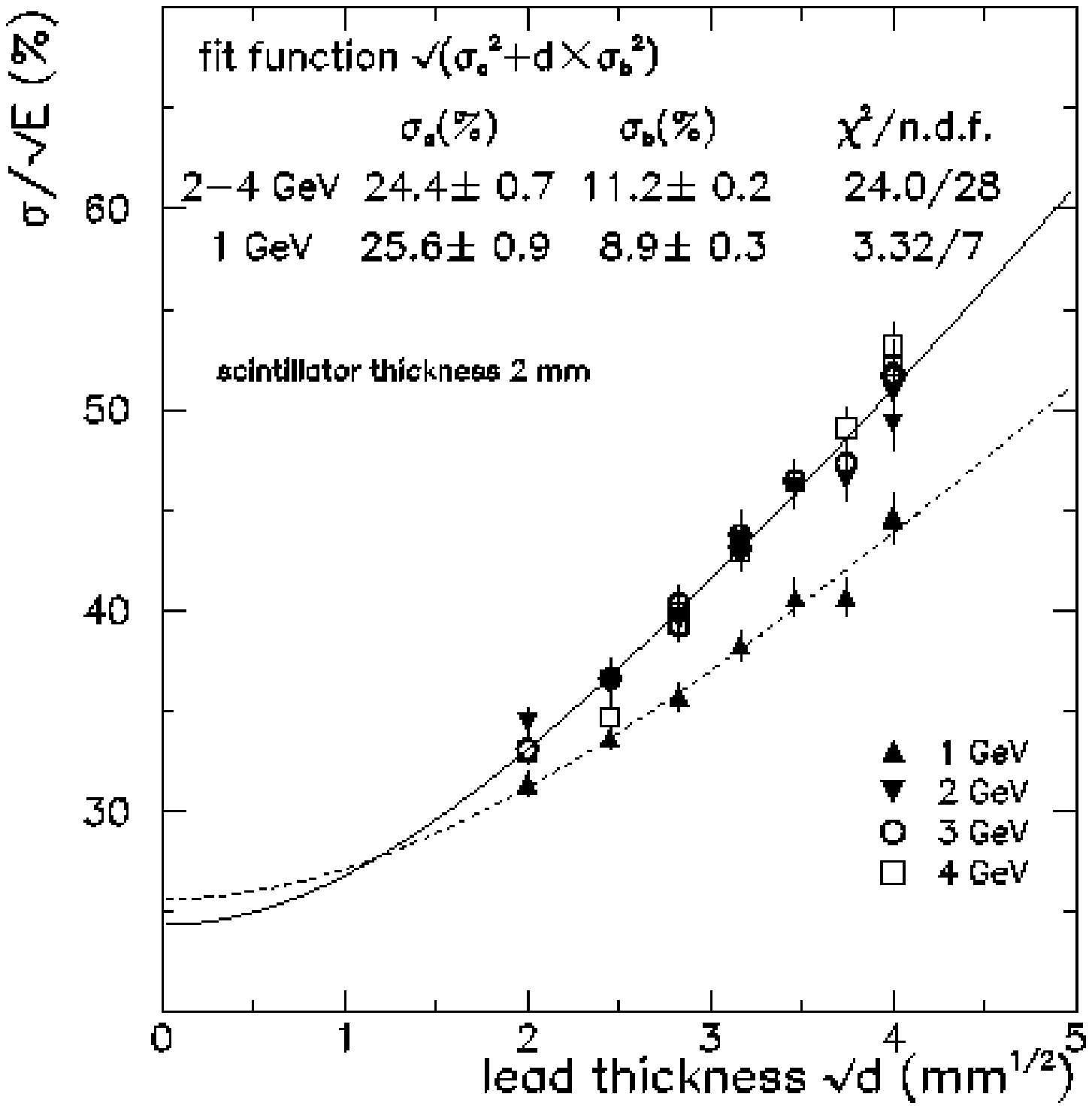}
\epsfxsize=8.0cm \epsfbox{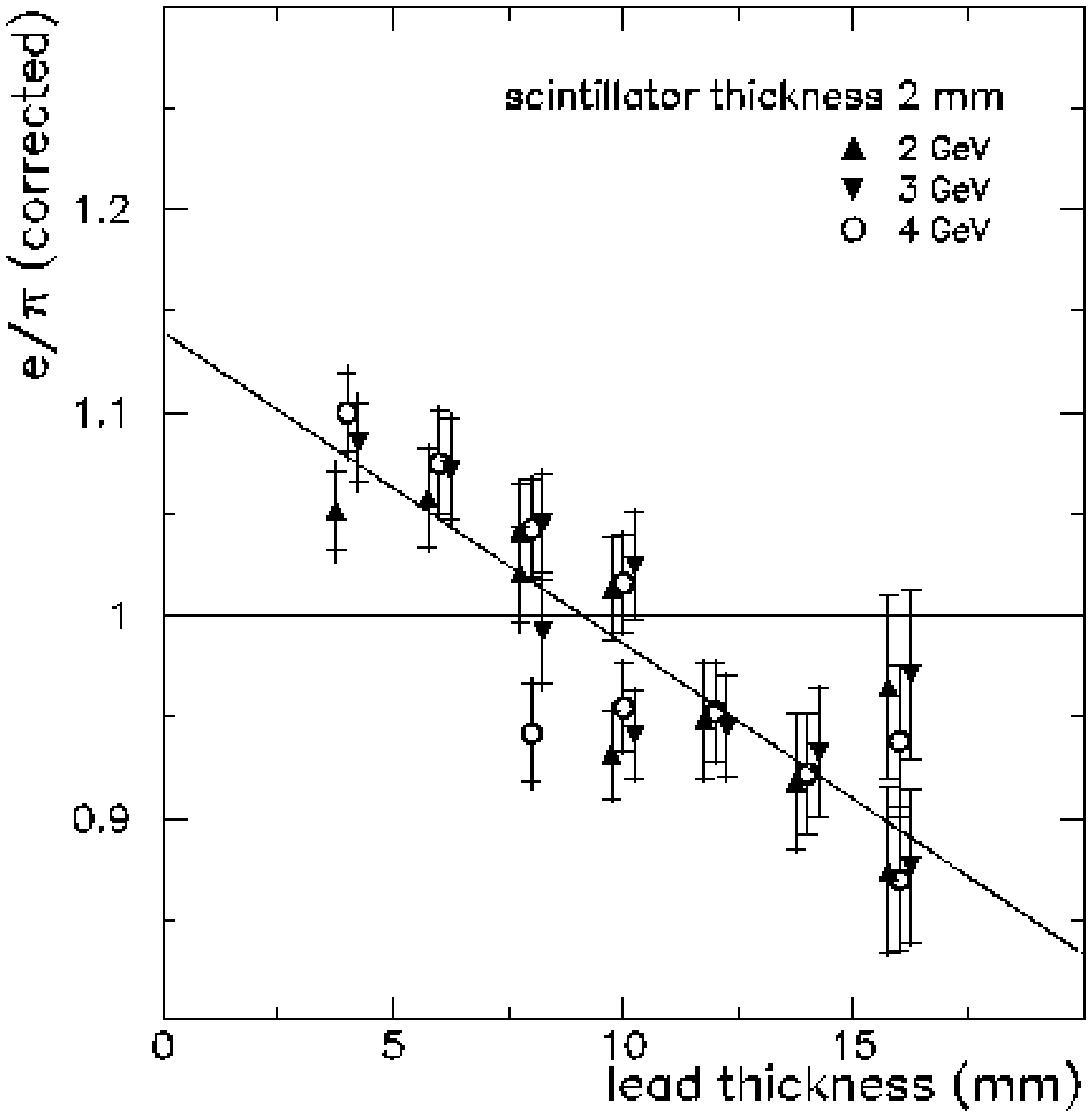}}
\begin{center}\begin{minipage}{\figurewidth}
\caption{
\label{T405Eres}\sl
Energy resolution for pions (left) and e/$\pi$ ratio (right).}
\end{minipage}\end{center}
\end{figure}

Series of beam tests were carried out at KEK $\pi$2 beamline 
with beam energies from 1 GeV to 4 GeV\cite{T405}.
The setup of the beam test is shown in Fig.\ref{T411photo}.
Fig.\ref{T405Eres} shows energy resolution for pions and 
e/$\pi$ response ratio versus lead plate thickness.
Target energy resolution is achieved with lead plates of
8mm-thick or thinner.
The hadron energy resolution of 33.6\%/$\sqrt{E}$ achieved 
with 4mm-thick lead plates are one of 
the best hadron energy resolution achieved so far.
In Fig.\ref{T405Eres}, dependence of energy resolution
on lead thickness $d$ is parametrized as
$\sigma_E = \sqrt{\sigma_{intrinsic}^2 + 
(\sigma_{sampling}^2+\sigma_{photostat}^2)/d} $,
neglecting the constant term. 
Intrinsic hadron shower fluctuation was
obtained to be 24.4$\pm$0.7\% as a result of the fitting.
This is significantly larger than previously reported value of
13.4$\pm$4.7\% by ZEUS \cite{Drew},
which was derived only from two sampling frequencies.
However this could simply be due to the difference of incident energy.

It is concluded from the fitting of $e/\pi$ ratio in Fig.\ref{T405Eres}
that hardware compensation should be 
achieved with 9mm-thick lead plates 
in the case of 2mm-thick plastic scintillator plates.

\vspace{0.3cm}

In order to check effect of fiber-routing acryl plates on compensation,
three configurations are tested in the case of 8mm-thick and 16mm-thick
lead plate configurations:
\begin{itemize}
\item no acryl plates were used (lead and scintillator only);
\item an acryl plate was put upstream-side of every scintillator plate;
\item an acryl plate was put downstream-side of every scintillator plate.
\end{itemize}
The motivation of measurement with 16mm-thick lead plates was such that
compensation might be achieved with 16mm-thick lead plates + 2mm-thick 
scintillator plates + 2mm-thick acryl plates, resulting in
the volume ratio of lead to total plastic to be 4:1. 

Effects of acryl plates on the energy resolution and e/$\pi$
ratio is summarized in Table~\ref{acryleffect}.
It is seen from the results of 8mm-thick lead-plate case 
that acryl plates does not destroy the hardware compensation
if they are placed downstream-side of scintillator plates.
Hadron energy resolution, however, gets slightly worse by their presence
regardless of their location.
These behaviour is not clear in the case of 16mm-thick lead-plate
case due to large error. 
However it is clear that compensation was not achieved, meaning that
acryl plates do not contribute to the volume-ratio counting.

\begin{table}
\begin{center}
\begin{minipage}{\figurewidth}
\caption{
\label{acryleffect}\sl
Energy resolutions and $e/\pi$ ratios with and 
without acryl plates for 4~GeV/c electrons and pions.}
\end{minipage}
\end{center}
\begin{small}
\begin{center}
\begin{tabular}{|l|l|c|c|c|}
\hline
lead thickness & position of acryl & \small $\sigma_{E}/E$ for electrons & \small $\sigma_{E}/E$ for pions & $e/\pi$ ratio \\
\hline
 8mm & No acryl plates        & (12.0$\pm$0.5)\% & (20.5$\pm$0.4)\% & 1.03$\pm$0.02 \\
     & Acryl plates upstream  & (11.6$\pm$0.5)\% & (22.7$\pm$0.4)\% & 1.07$\pm$0.02 \\
     & Acryl plates downstream& (12.0$\pm$0.5)\% & (22.8$\pm$0.4)\% & 1.01$\pm$0.02 \\
\hline
16mm & No acryl plates        & (18.1$\pm$1.1)\% & (27.1$\pm$0.7)\% & 0.90$\pm$0.04 \\
     & Acryl plates upstream  & (18.1$\pm$1.1)\% & (28.1$\pm$0.7)\% & 0.95$\pm$0.04 \\
     & Acryl plates downstream& (18.1$\pm$1.1)\% & (28.8$\pm$0.7)\% & 0.87$\pm$0.04 \\
\hline
\end{tabular}
\end{center}
\end{small}
\end{table}


\vspace{0.3cm}
\noindent
{\bf Shower Fluctuation Analysis}
\vspace{0.2cm}

\begin{figure}
\centerline{
\epsfxsize=7.5cm \epsfbox{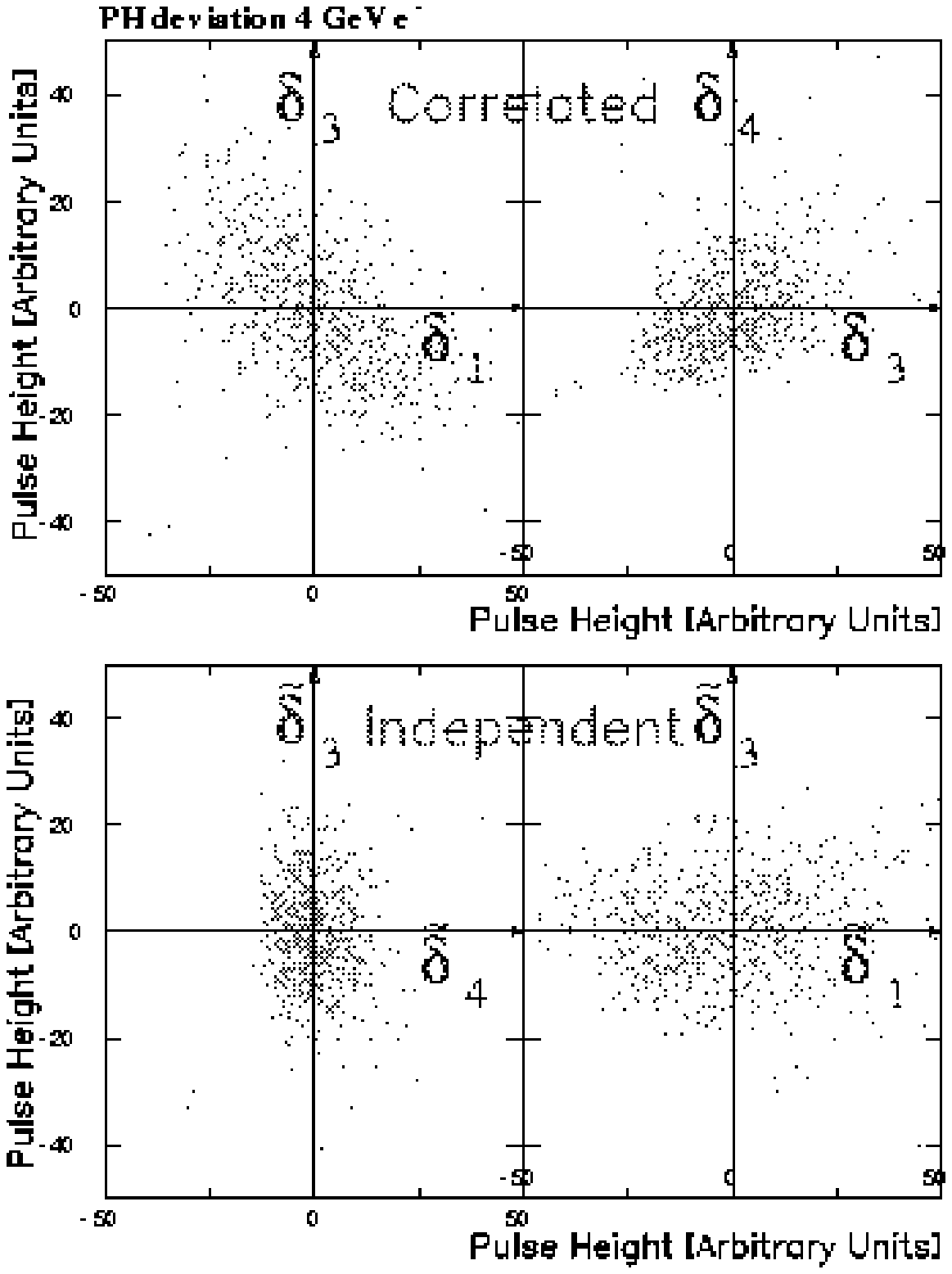}
\hspace{0.5cm}
\epsfxsize=7.5cm \epsfbox{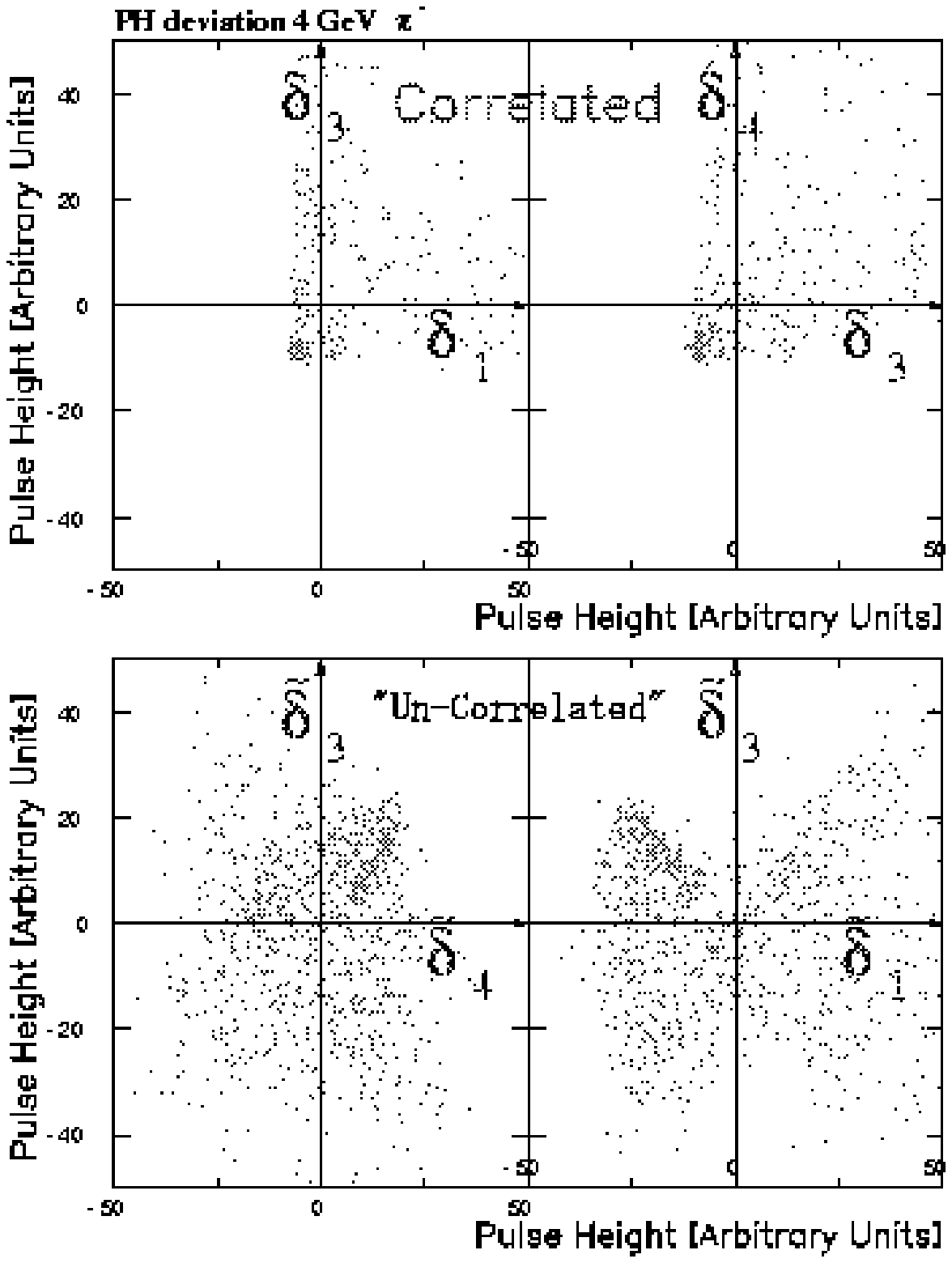}}
\begin{center}\begin{minipage}{\figurewidth}
\caption{\label{fluccorr}\sl
Correlation between pulseheight fluctuations for different super layers
for electrons (left) and for pions (right).
Upper distributions are for raw fluctuations, and
lower ones are after orthogonalization.}
\end{minipage}\end{center}
\end{figure}
        
\begin{figure}
\centerline{
\epsfxsize=8cm \epsfbox{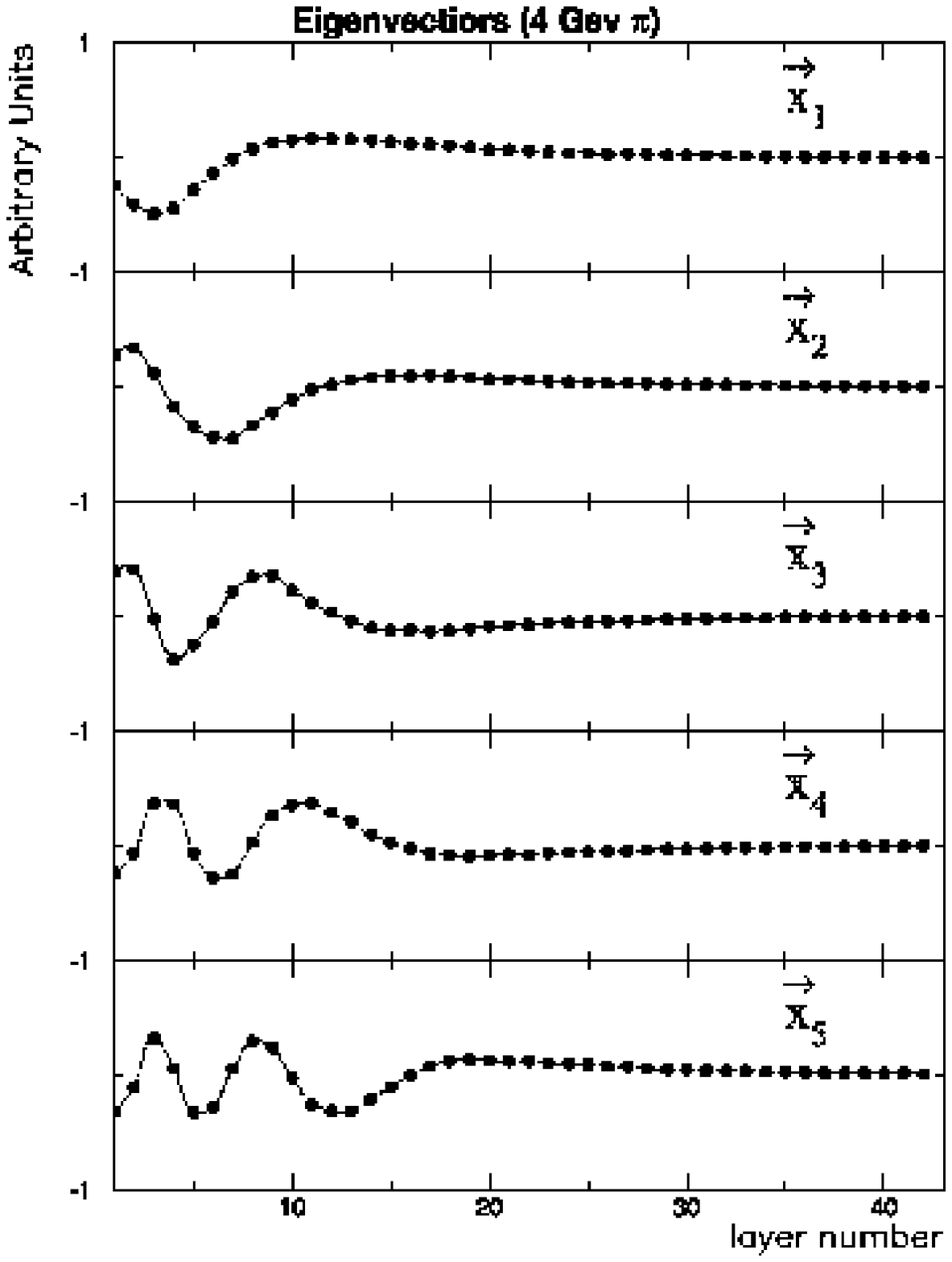}}
\begin{center}\begin{minipage}{\figurewidth}
\caption{\label{eigenvec}\sl
Shape of the orthogonalized fluctuation mode (eigen vectors)
for 4 GeV pion shower. Five major modes are shown. 
Horizontal axis is super layer number,
and vertical axis is amplitude of deviation (arbitrary unit).}
\end{minipage}\end{center}
\end{figure}

Longitudinal shower profile and its event-by-event fluctuation
have been studied with configuration of 4mm-thick lead plates,
which had 42 longitudinal samplings (super layers).
The purposes of this study are;
\begin{itemize}
\item Improve hadron energy resolution (software weighting);
\item Construct parametric hadron shower generator with realistic fluctuation;
\item Obtain $e/\pi$ separation capability as a function of
the longitudinal segmentation of the calorimeter.
\end{itemize}
Analysis has been done for both EM and hadron showers
following the work by CDF for EM shower analysis\cite{miyata}.
Correlation between pulseheight fluctuations for different super layers
are shown in Fig.\ref{fluccorr}.
Upper plots are for raw fluctuations $\delta_i$ from average
shower shape, where $i$ means $i$-th super layer.

Orthogonalization is done by diagonalizing correlation matrix 
$$ 
C_{ij} = \frac{1}{N} \sum_k \delta_i^k \delta_j^k 
$$
\noindent
where $k$ means $k$-th events.
As the eigen vectors and eigen values of the diagonalized matrix,
we obtain orthogonalized fluctuation modes and 
their amplitude $\tilde{\delta}$, respectively.
Fig.\ref{eigenvec} shows five major eigen vectors (orthogonalized
fluctuation modes) for 4GeV pion shower.
Horizontal axis is super layer number.
Lower plots in Fig.\ref{fluccorr} are correlation between
fluctuation of amplitudes $\tilde{\delta_i}$ and $\tilde{\delta_j}$ 
for $i$-th and $j$-th orthogonalized fluctuation mode.
In the case of electrons (left), no correlation is seen between amplitudes
of the orthogonalized eigen vectors.
On the other hand, $\tilde{\delta_i}$ for pions (right) have correlation
even after orthogonalization procedure, meaning that 
orthogonalization of hadronic fluctuation is not yet successful.
This is because hadronic fluctuation is not Gaussian due to
its two-component nature; global hadronic fluctuation
and local $\pi^0$ generation.
An attempt has been under study to decompose two fluctuations.
Therefore parameterization of hadron shower fluctuations with
orthogonal parameter set has not been completed yet.
       
\begin{figure}
\centerline{
\epsfxsize=7cm \epsfbox{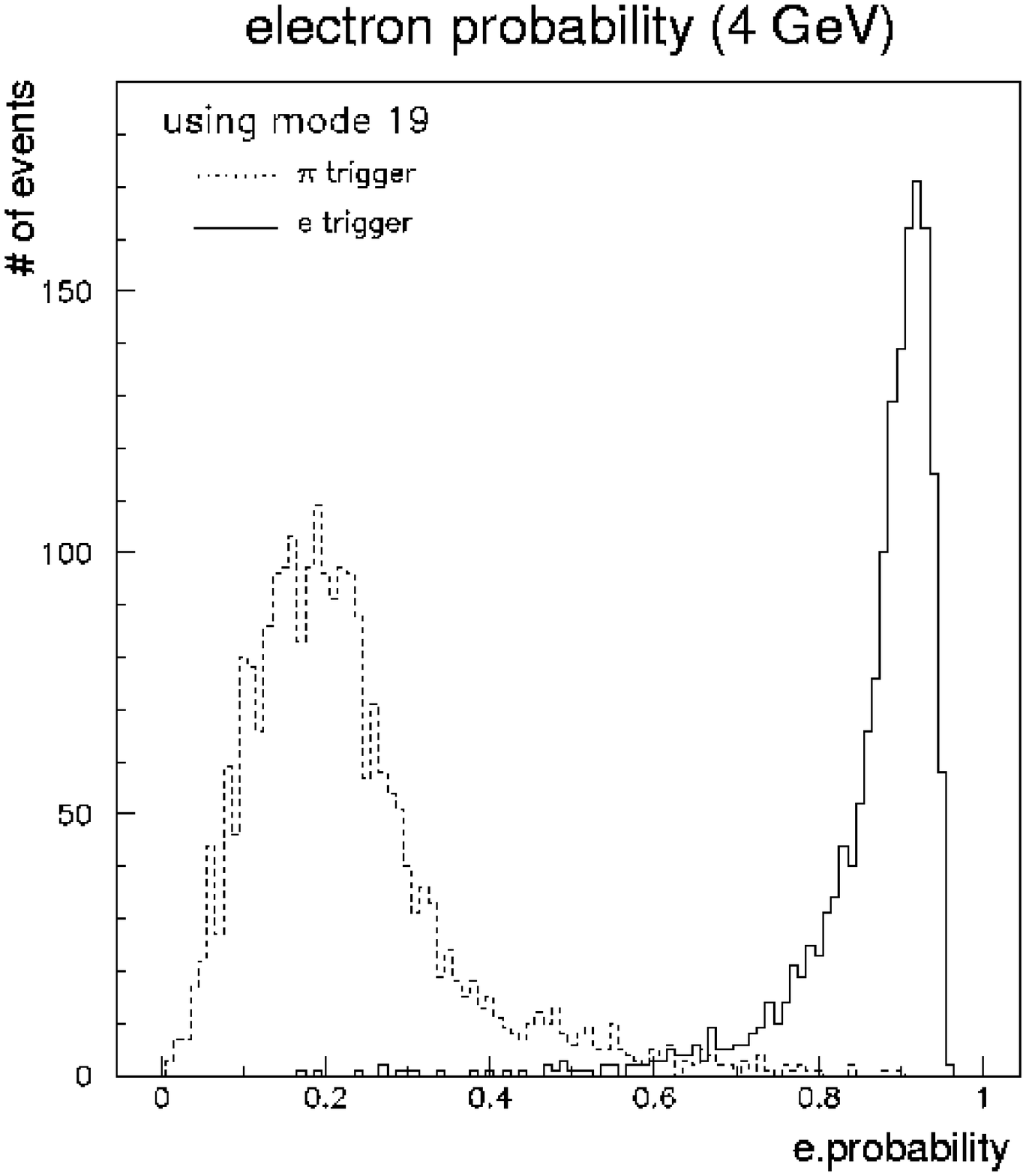}
\hspace{0.5cm}
\epsfxsize=7cm \epsfbox{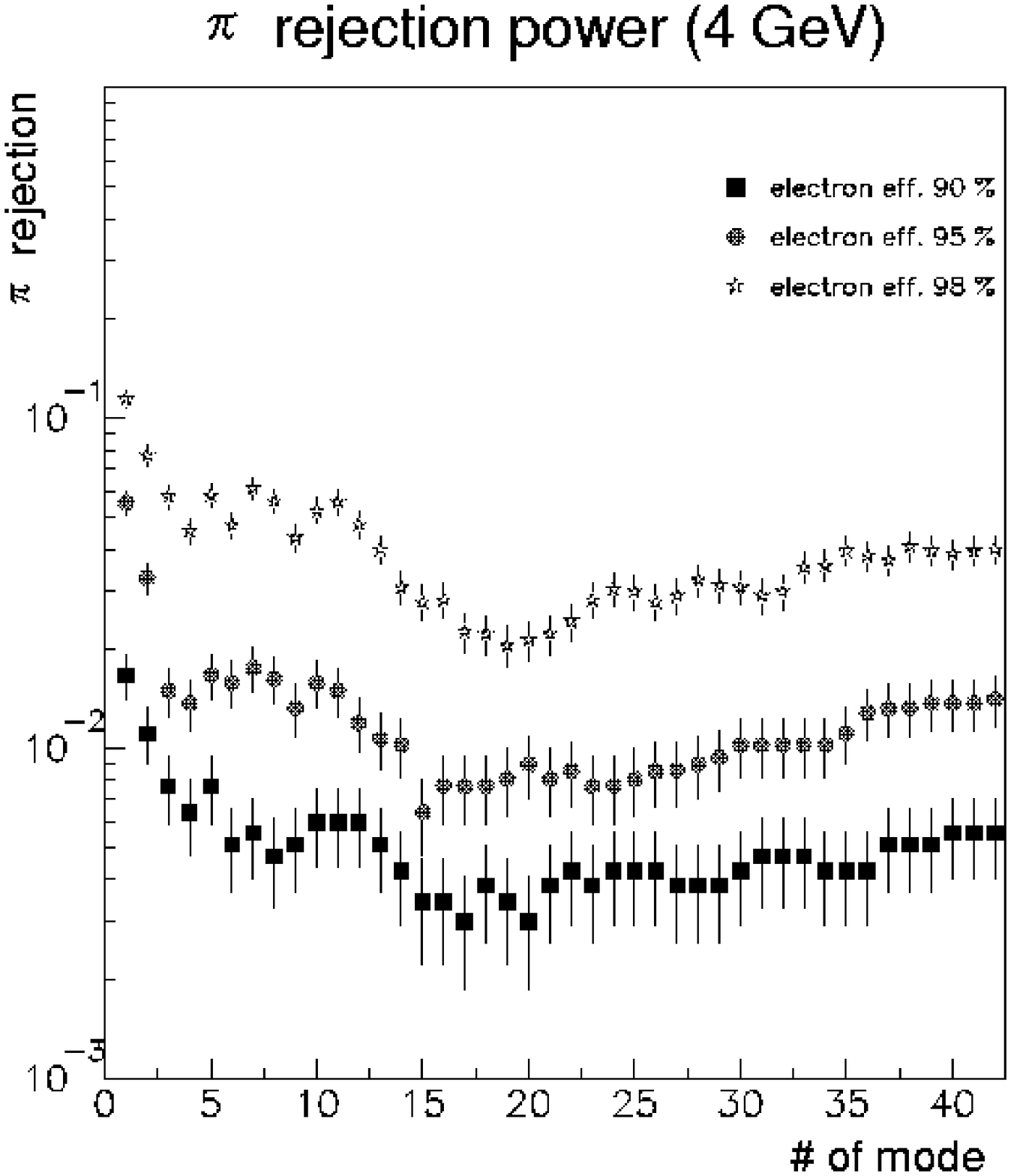}}
\begin{center}\begin{minipage}{\figurewidth}
\caption{\label{flucepi}\sl
$e/\pi$ separation by deviation of pulseheight fluctuation from template 
fluctuation distributions for electrons and pions.
Left histogram is electron probability distribution for
electron and pion samples, and right plot is pion rejection
factor as a function of number of modes (used eigen vectors).}
\end{minipage}\end{center}
\end{figure}

Pulseheight fluctuation can be used to improve 
$e/\pi$ identification, which is usually done by shower profile only.
Deviations of measured pulseheights from electron profile template and from
pion profile template are normalized by typical fluctuation for
each particle, and likelihood to each particle are calculated.
Distribution of electron-ness and pion rejection capability
are shown in Fig.\ref{flucepi}.
Number of mode in the right figure is number of eigen vectors
of pulseheight fluctuation used in the analysis.
Since there are 42 super layers, there are 42 orthogonal eigen
vectors to represent pulseheight fluctuation.
Using six or seven major eigen vectors achieves almost
the best score ; higher minor eigen vectors are noisy,
and using them does not improve the score.
For single-particle $e/\pi$ separation, therefore,
six or seven longitudinal segmentation should be good enough
if they are segmented in optimum way.


\subsubsection{2) Study of Tile/Fiber Module}  

\begin{figure}
\centerline{
\epsfysize=6.5cm \epsfbox{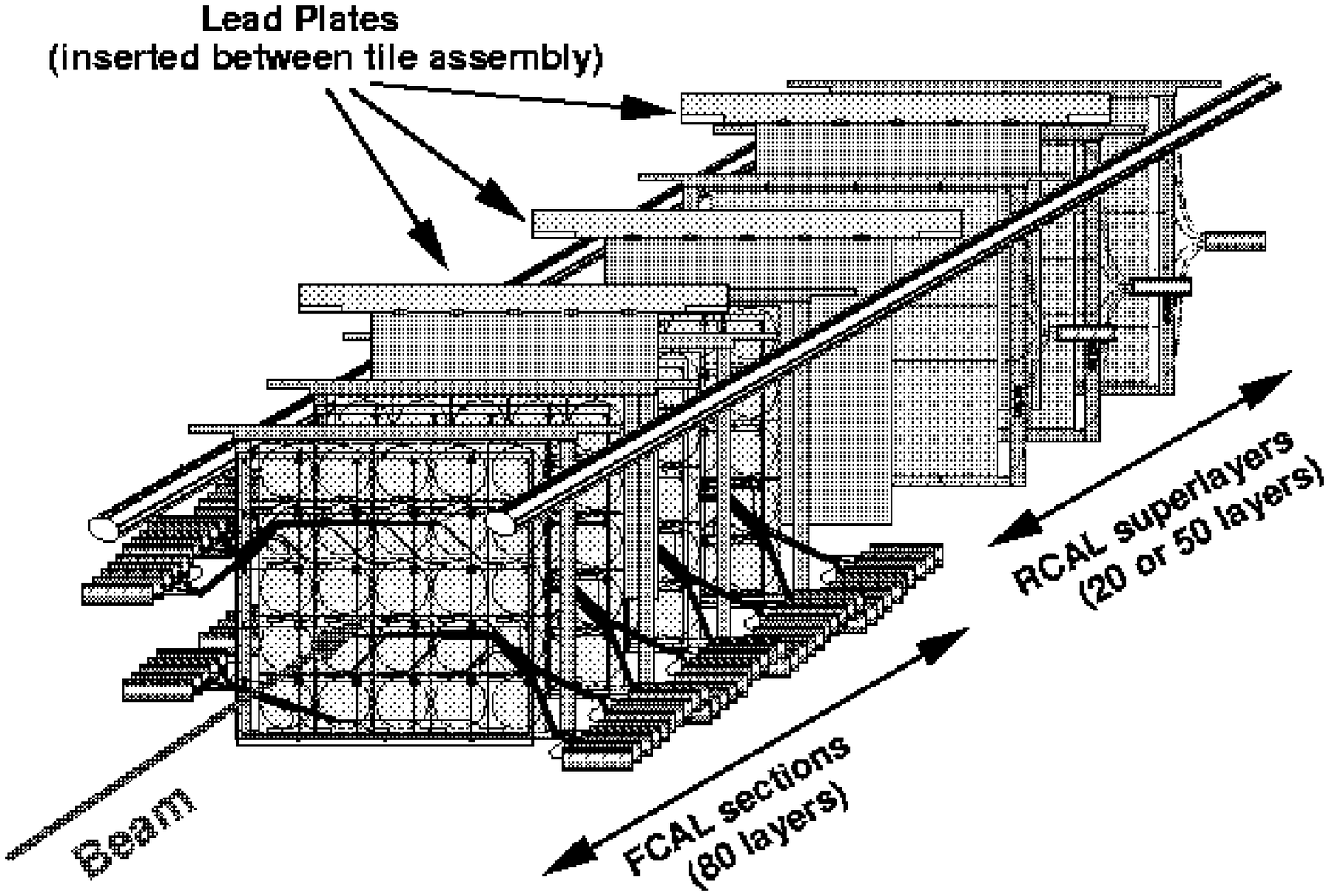}
\epsfysize=7.0cm \epsfbox{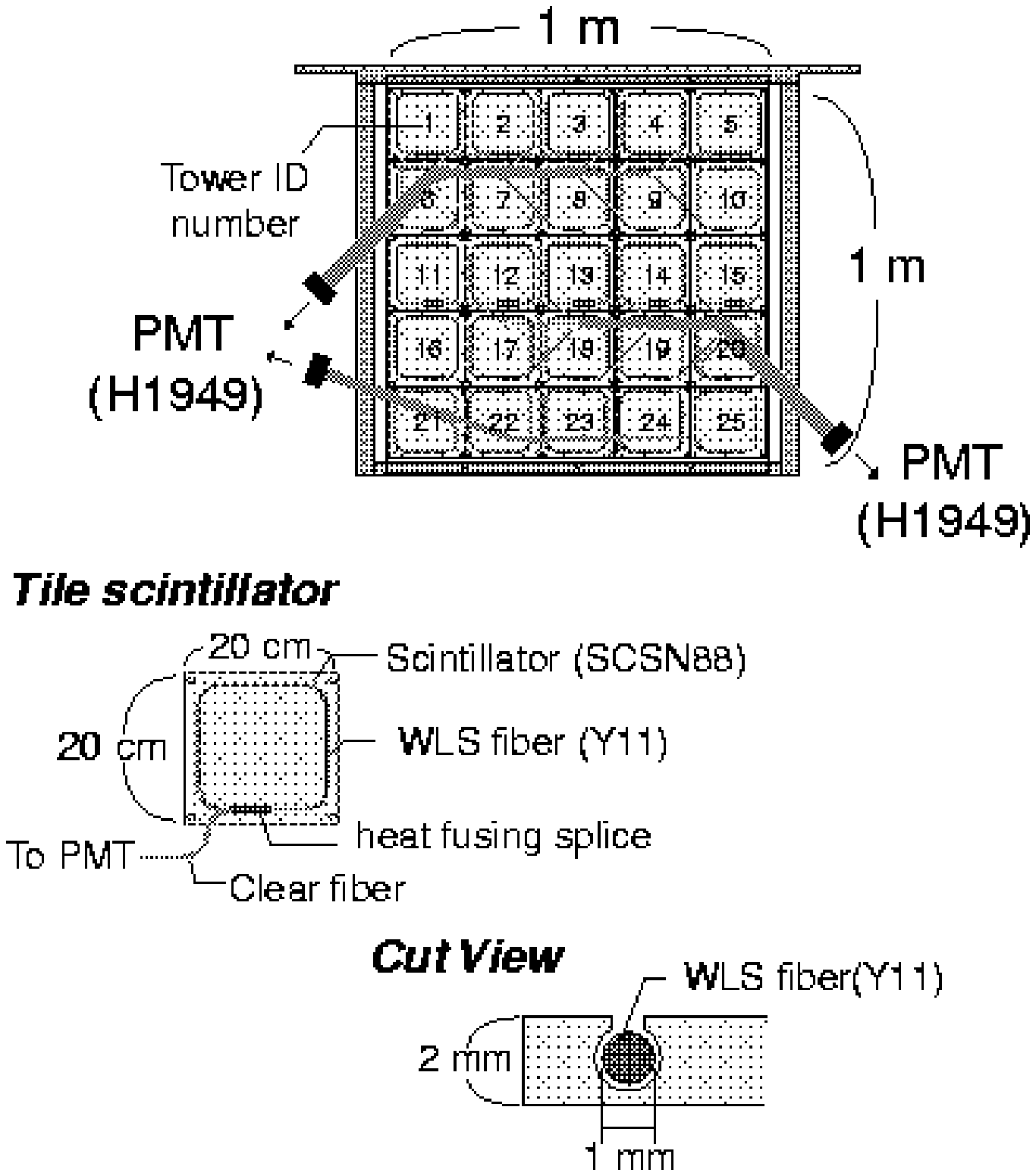}}
\begin{center}\begin{minipage}{\figurewidth}
\caption{\label{TFmodule}\sl
Schematic view of the tile/fiber test module (left) and
layout of fibers and tiles (right).}
\end{minipage}\end{center}
\end{figure}

Based on the results of the generic studies and tile bench tests, 
the straight-groove module was rebuilt to be a tile/fiber calorimeter module.
The design of the module is schematically shown in Fig.\ref{TFmodule}.
The module was composed of a front section (FCAL) and a rear section (RCAL).

\begin{figure}
\centerline{
\epsfxsize=12cm \epsfbox{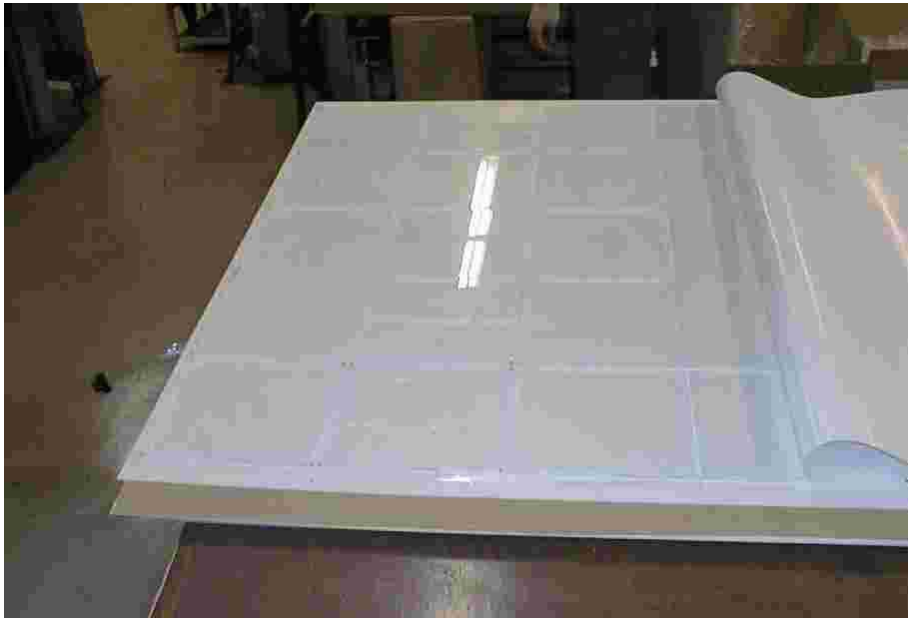}}
\begin{center}\begin{minipage}{\figurewidth}
\caption{ \label{tile5x5} \sl
Scintillator tiles arranged as $5 \times 5$ on the fiber-routing plate,
and covered with white PET films.}
\end{minipage}\end{center}
\end{figure}

In the FCAL part,
the 1m$\times$1m plastic scintillator plate was modified
to be a scintillator tile assembly as shown in Fig.\ref{TFmodule}.
Twenty-five tiles with size of 20cm$\times$20cm$\times$2mm-thick were 
arranged on a fiber-routing acryl plate to form 5$\times$5 tower structure.
Four sides of the tiles were painted white with TiO$_2$-based emulsion,
and the assembly was covered with white PET films 
for better light collection efficiency
as shown in Fig.\ref{tile5x5}.
The tile had a $\sigma$-shaped groove with a key-hole cross section,
where a WLS fiber was embedded.
The WLS fiber was connected to a clear fiber by heat splicing
at the exit from the tile.
The clear fiber then ran in the groove on the fiber-routing acryl plate
to exit from the detector assembly, and was connected to a photon detector.

\begin{figure}
\centerline{
\epsfxsize=12cm \epsfbox{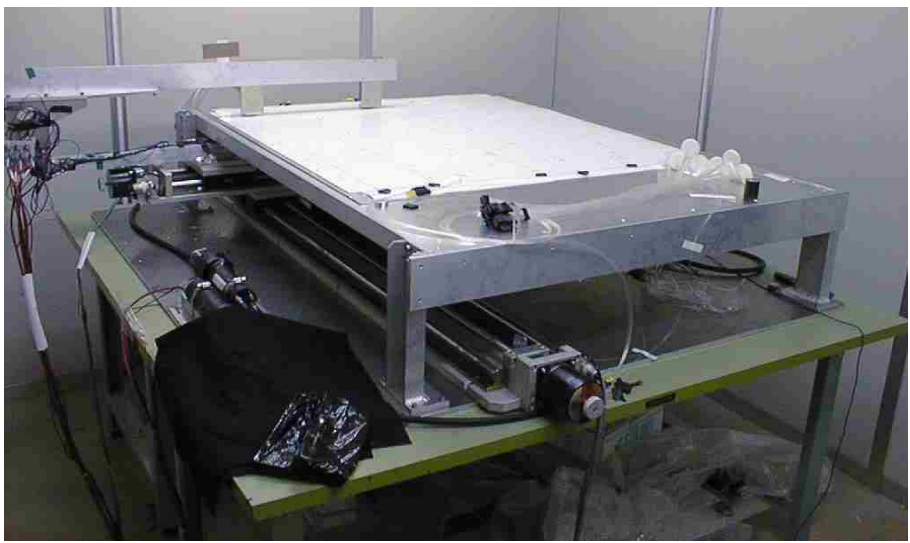}}
\begin{center}\begin{minipage}{\figurewidth}
\caption{ \label{tilestage} \sl
Automatic measurement stage for photo-electron yield uniformity.}
\end{minipage}\end{center}
\end{figure}

\begin{figure}
\centerline{
\epsfysize=10.0cm \epsfbox{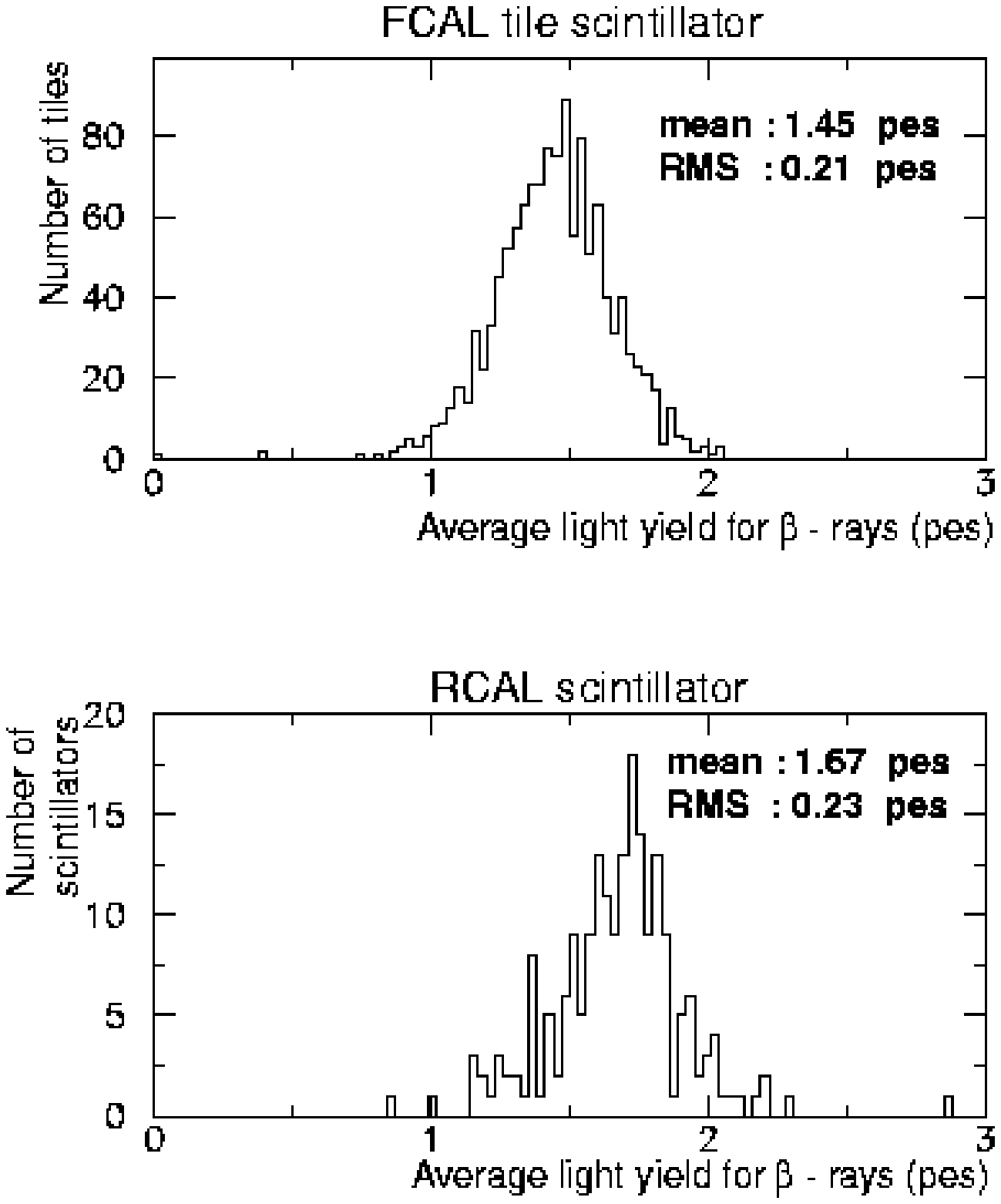}
\hspace{0.5cm}
\epsfysize=10.0cm \epsfbox{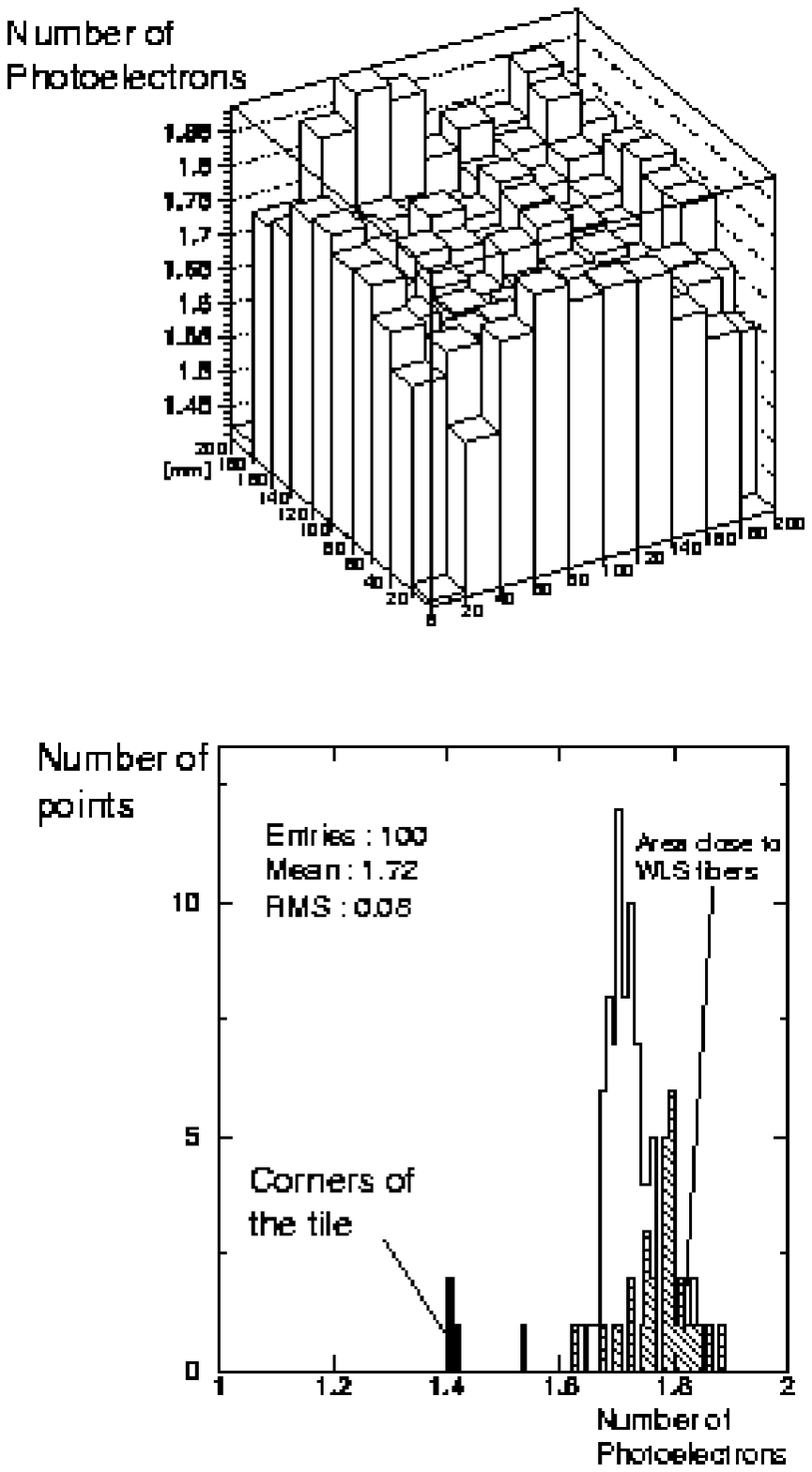}}
\begin{center}\begin{minipage}{\figurewidth}
\caption{\label{tilepes}\sl
Average photo-electron yield for tiles (left-top),
average photo-electron yield for straight-groove plates (left-bottom),
uniformity over one tile (right-top),
and distribution of photo-electron yield for 100 sampling points
over one tile (right-bottom).}
\end{minipage}\end{center}
\end{figure}

Prior to assembling to 5$\times$5 matrix, 
photo-electron yield uniformity over a tile were measured with 
$\beta$ rays from an RI source for several tiles.
Then after assembling to $5 \times 5$ matrix,
photo-electron yield at the center of each tile
was measured for all 2000 tiles.
The setup of the measurement system is shown in Fig.\ref{tilestage}, 
and the results are shown in Fig.\ref{tilepes}.
Obtained photo-electron yield is translated to the calorimeter response of
83 p.e./GeV for 8:2 configuration,
about the half of the ZEUS-type modules.
This slightly worsens the stochastic term of hadron energy resolution from 
40\%/$\sqrt{E}$ to 41\%/$\sqrt{E}$, 
which is not a problem for hadron calorimetry.
The measured non-uniformity was estimated using GEANT3 simulation
to introduce systematic uncertainty of 0.3\% on average on 
energy resolution measurement.
This, again, is not significant.

\begin{figure}
\centerline{
\epsfxsize=8.0cm \epsfbox{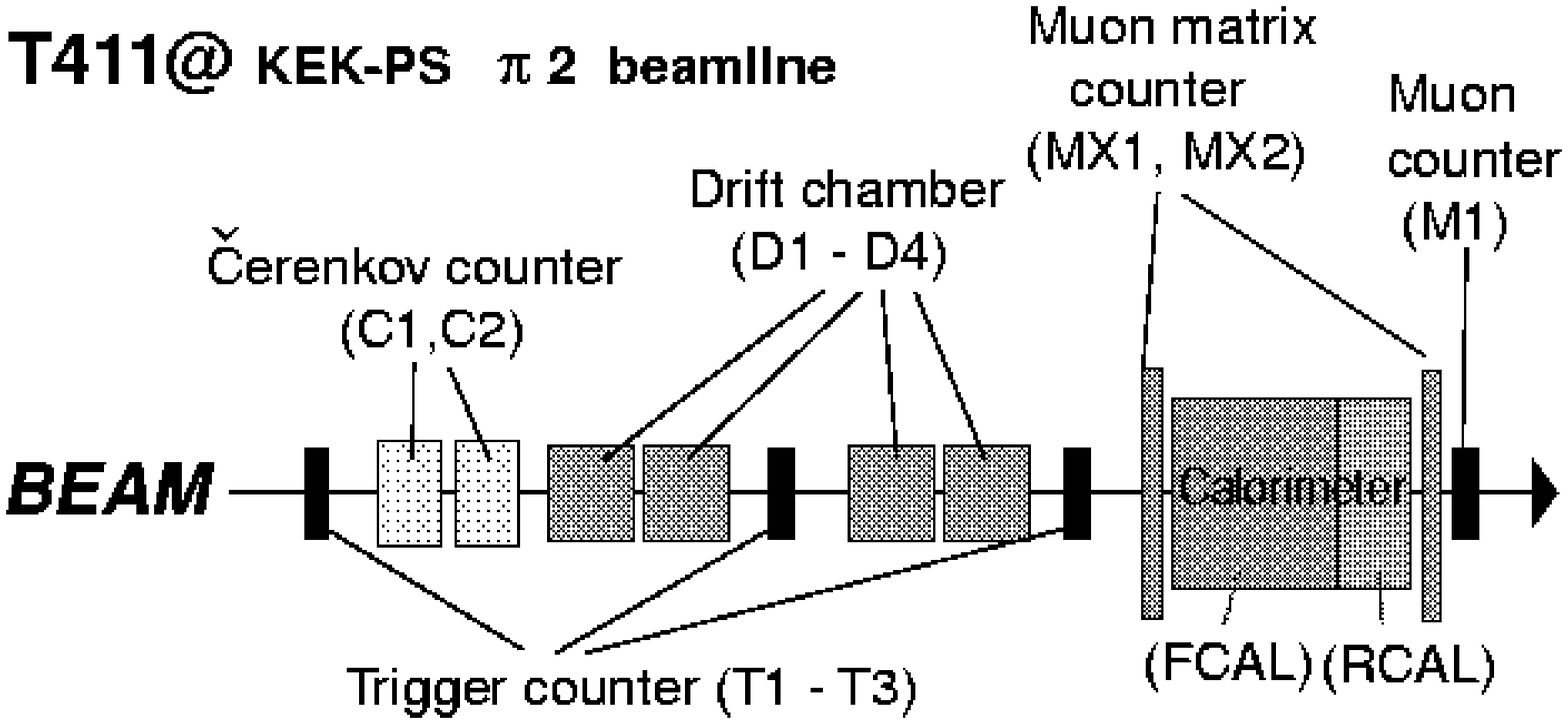}
\hspace{0.5cm}
\epsfxsize=7.0cm \epsfbox{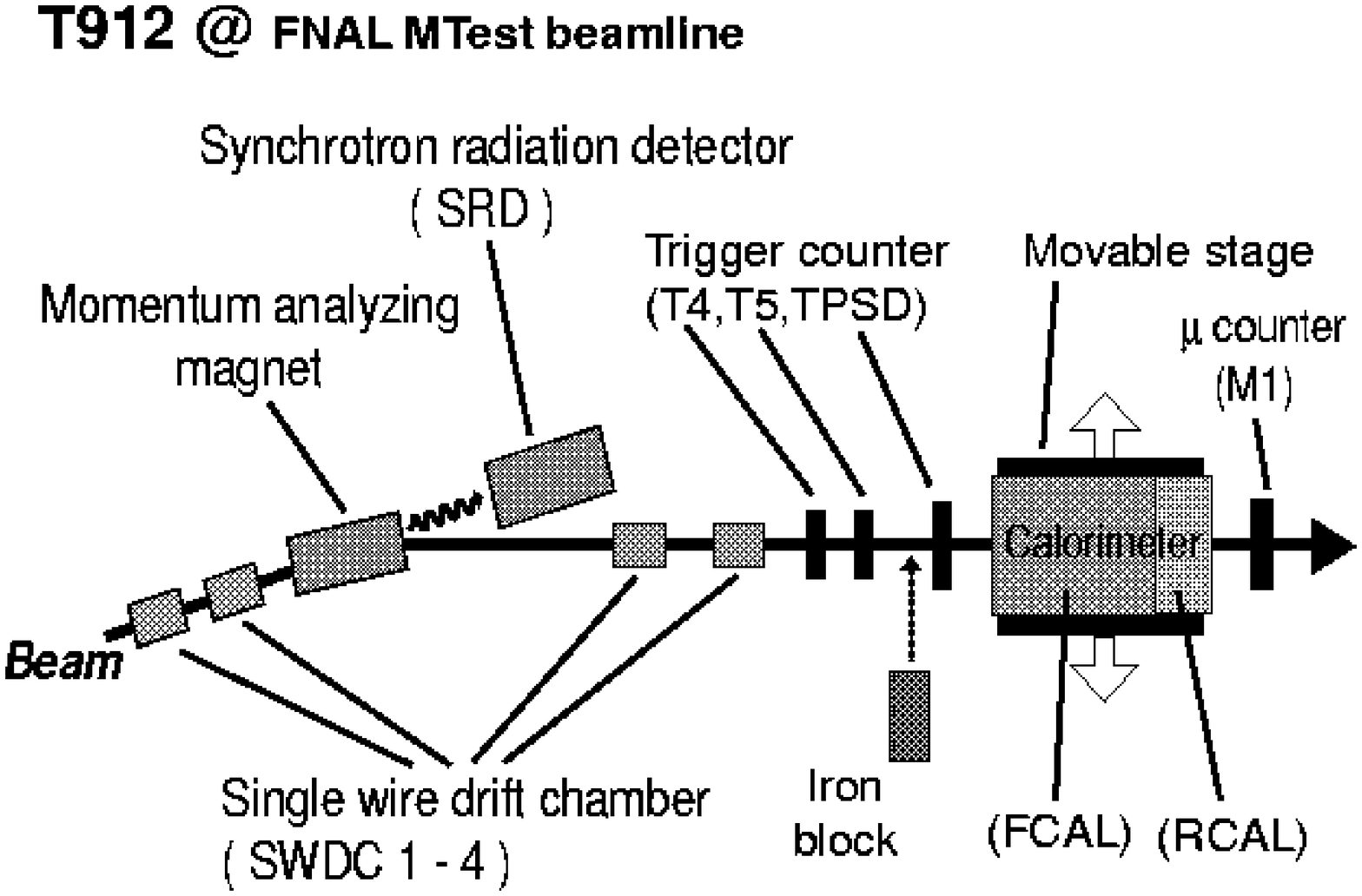}}
\begin{center}\begin{minipage}{\figurewidth}
\caption{\label{T411setup}\sl
Setup of the beam test at KEK (left) and at FNAL (right).}
\end{minipage}\end{center}
\end{figure}

These tile assemblies were interleaved with 8mm-thick lead plates.
There were 80 layers of the tile assemblies in FCAL in total.
They were divided into four sections longitudinally,
each of which contained 20 layers.
One quadrant of a tower was read out by one PMT.

Following the tile/fiber part, unmodified straight-groove section
remained as RCAL. 
There were two super layers in RCAL 
in the case of low energy beam test at KEK.
In the case of high energy beam test at FNAL,
ten super layers were installed, the last of which did not have
lead absorbers to be used for muon tagging.
RCAL also had 8mm-thick lead plates as absorber.

The layouts of the beam tests at KEK and at FNAL are schematically 
shown in Fig.\ref{T411setup}.
Both setups are quite similar except for that there were 
momentum-analyzing devices at FNAL beam test,
and that SRD was used for electron identification at FNAL
instead of Cherenkov counters.
Details of the setup and analysis are given elsewhere\cite{T405,T912}.
Energy of the beam was 1-4 GeV and 10-200 GeV at KEK and at FNAL, respectively.
The beam was unseparated, and particle identification was done off-line.
In the case of FNAL test, however, electron-rich and pion-rich conditions
were realized by selecting appropriate converters/absorbers on the beamline.
Tower-to-tower gain calibration was done using penetrating muons.


\vspace{0.3cm}
\noindent
{\bf Energy Measurement}
\vspace{0.2cm}

\begin{figure}
\centerline{
\epsfysize=9cm \epsfbox{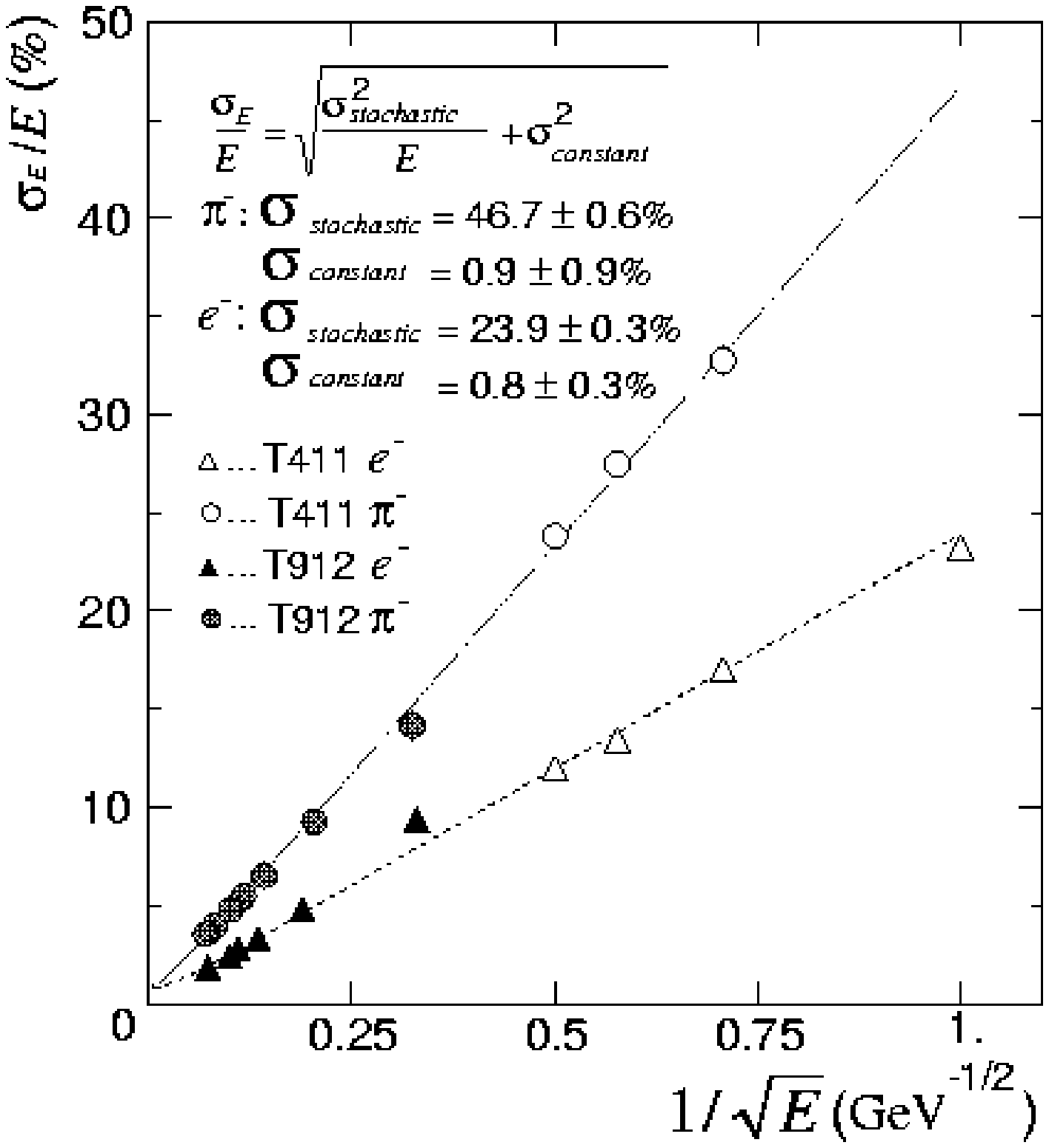}
\hspace{0.5cm}
\epsfysize=9cm \epsfbox{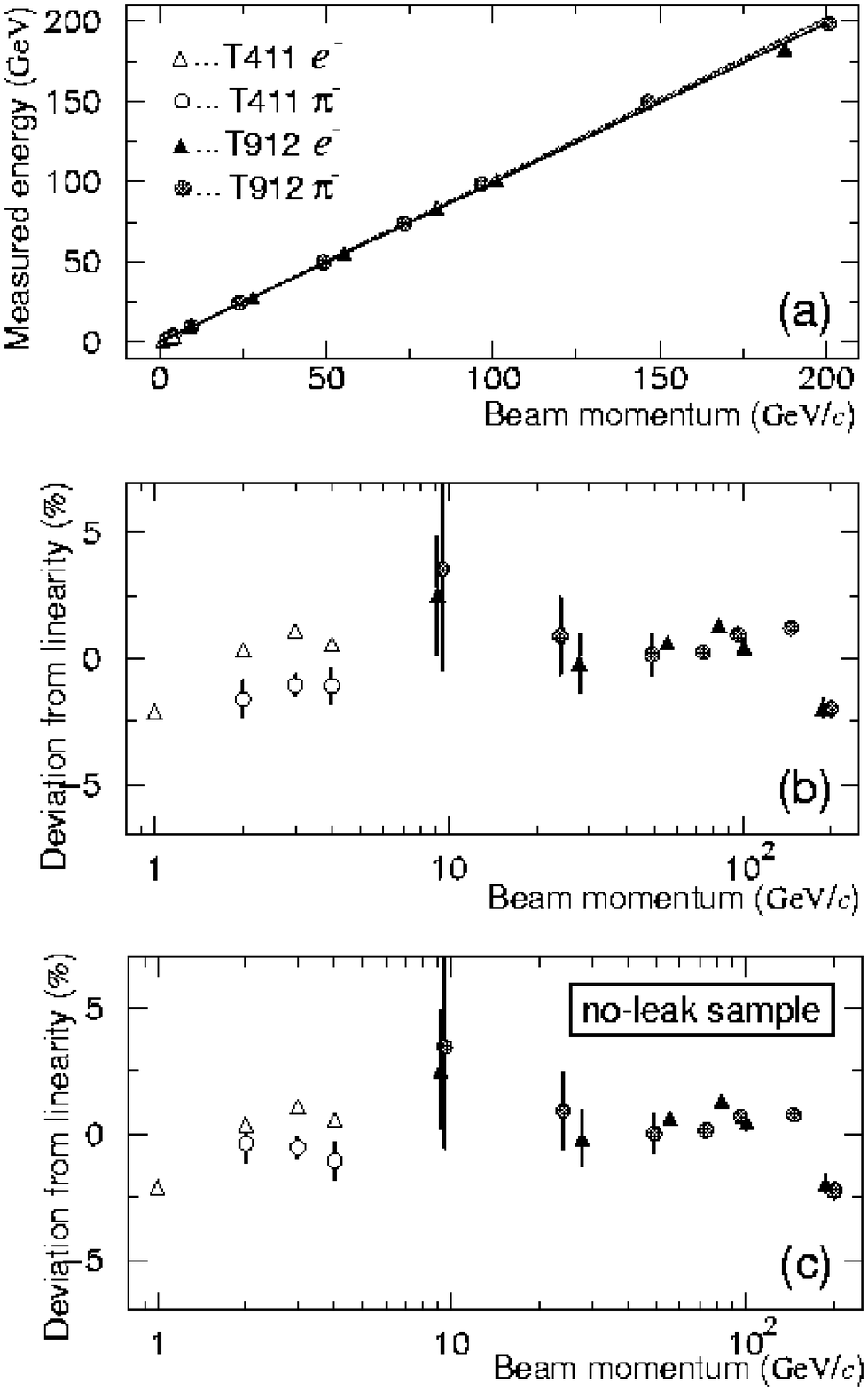}}
\begin{center}\begin{minipage}{\figurewidth}
\caption{\label{T912Eres}\sl
Energy resolution (left) and linearity (right) of the tile/fiber test module.}
\end{minipage}\end{center}
\end{figure}

\begin{figure}
\centerline{
\epsfxsize=7.5cm \epsfbox{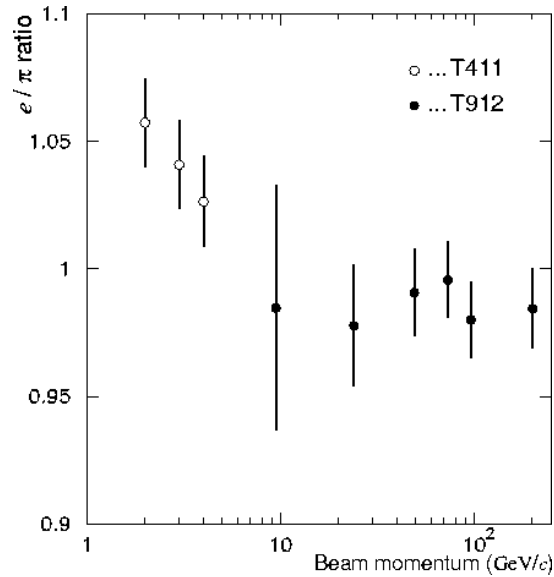}}
\begin{center}\begin{minipage}{\figurewidth}
\caption{\label{T912epi}\sl
e/$\pi$ response ratio of the tile/fiber test module.
T411 and T912 denote beam test at KEK and at FNAL, respectively.}
\end{minipage}\end{center}
\end{figure}

Measured energy resolution, linearity, and $e/\pi$ ratio are shown
in Figs.\ref{T912Eres} and \ref{T912epi}.
The energy resolution is slightly worse than those 
shown in Fig.\ref{T405Eres},
and is consistent with that of the straight-groove module 
with acryl plates interleaved shown in the Table~\ref{acryleffect}.
We can therefore conclude that acryl plates used for fiber-routing do
deteriorate the energy resolution, regardless their location.
On the other hand, the $e/\pi$ ratio is consistent with 1.0, and
hardware compensation is established to be retained in the
case of the tile/fiber structure if each acryl plate is
located downstream-side of each scintillator assembly.

If we stick to the hadron energy resolution of 40\%/$\sqrt{E}$,
sampling frequency of the HCAL should be finer than 
8mm-thick lead plates plus 2mm-thick plastic scintillator.
However measured hadron energy resolution of 46\%/$\sqrt{E}$ 
may be acceptable for physics analysis. 
This should be examined by further simulation study.
  

\vspace{0.3cm}
\noindent
{\bf Tower Property}
\vspace{0.2cm}

\begin{figure}
\centerline{
\epsfxsize=7.0cm \epsfbox{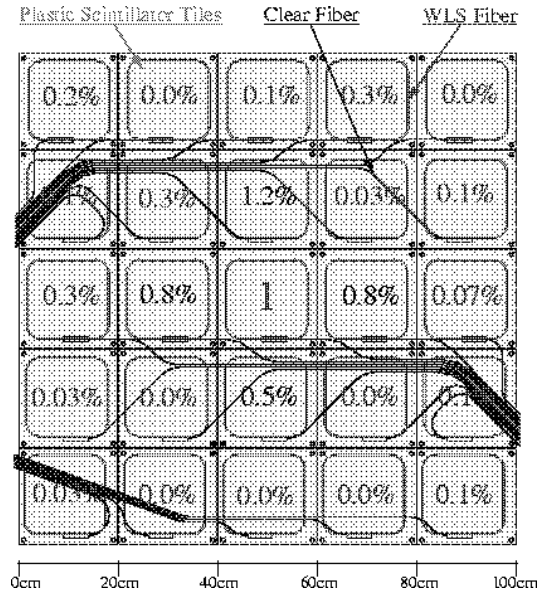}}
\begin{center}\begin{minipage}{\figurewidth}
\caption{\label{crosstalk}\sl
Cross talk between towers for the tile/fiber test module.}
\end{minipage}\end{center}
\end{figure}

Cross-talk between towers measured with electron central injection
are shown in Fig.\ref{crosstalk}.
Amount of cross talk to adjacent towers are 1\%-level.
This is consistent with the optical crosstalk at the bench test
measured with an RI-source.
Cross talk of this level is negligible to hadron shower analysis.
For EMC of finer-granularity option, however, cross talk may
have significant impact, and needs to be examined.

\begin{figure}
\centerline{
\epsfxsize=6.0cm \epsfbox{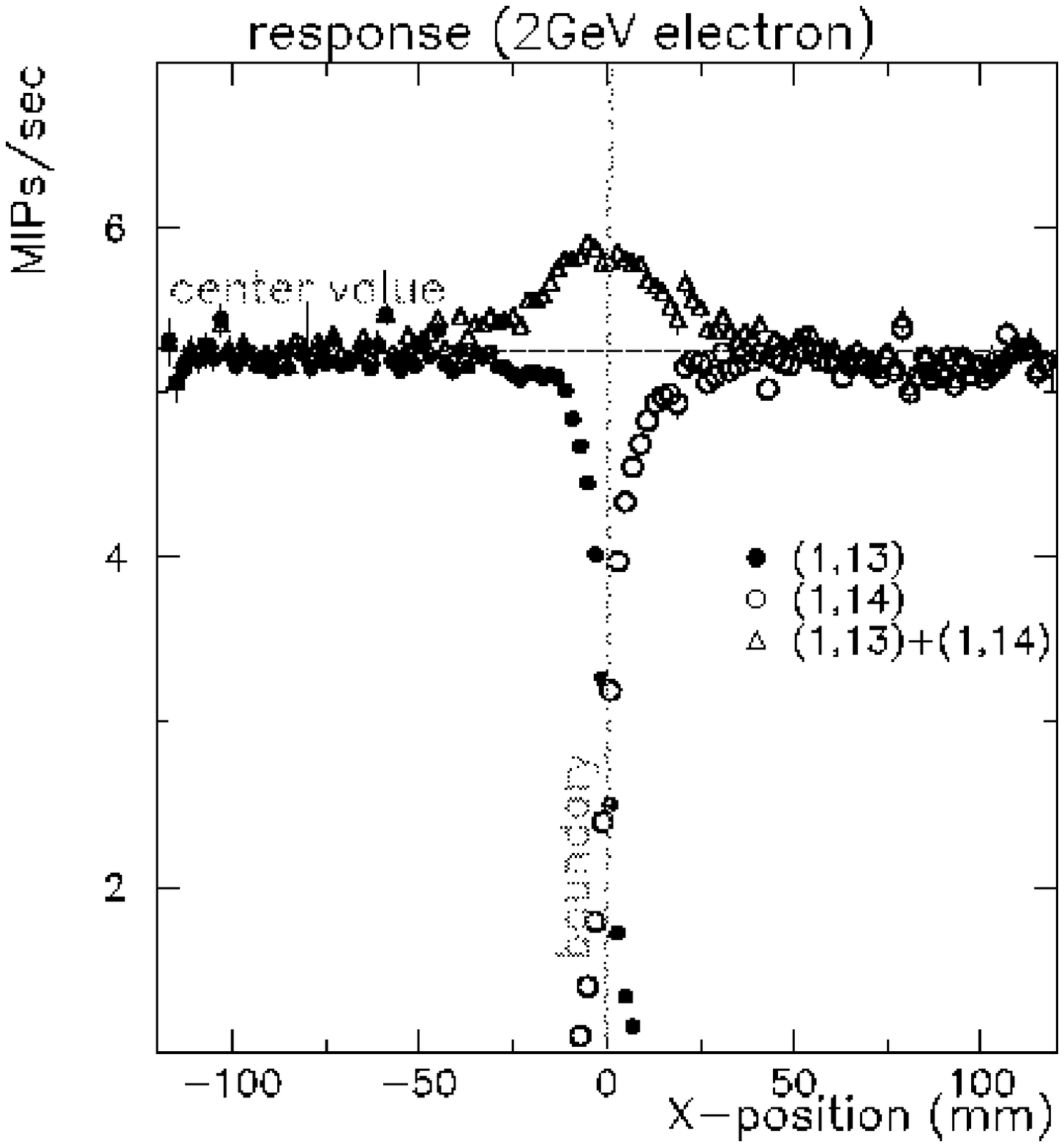}
\hspace{1cm}
\epsfxsize=6.0cm \epsfbox{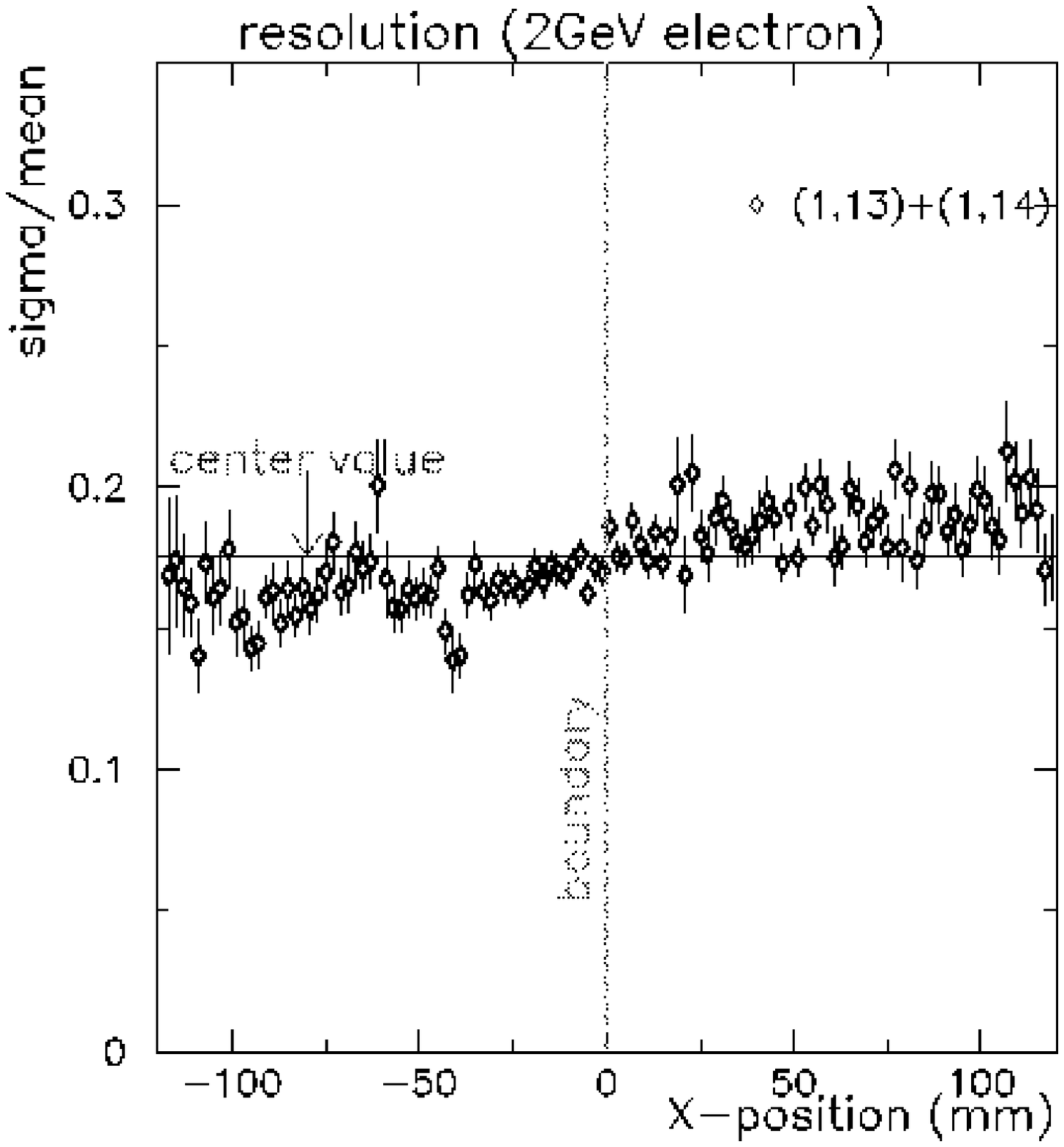}}
\vspace{0.2cm}
\centerline{
\epsfxsize=6.0cm \epsfbox{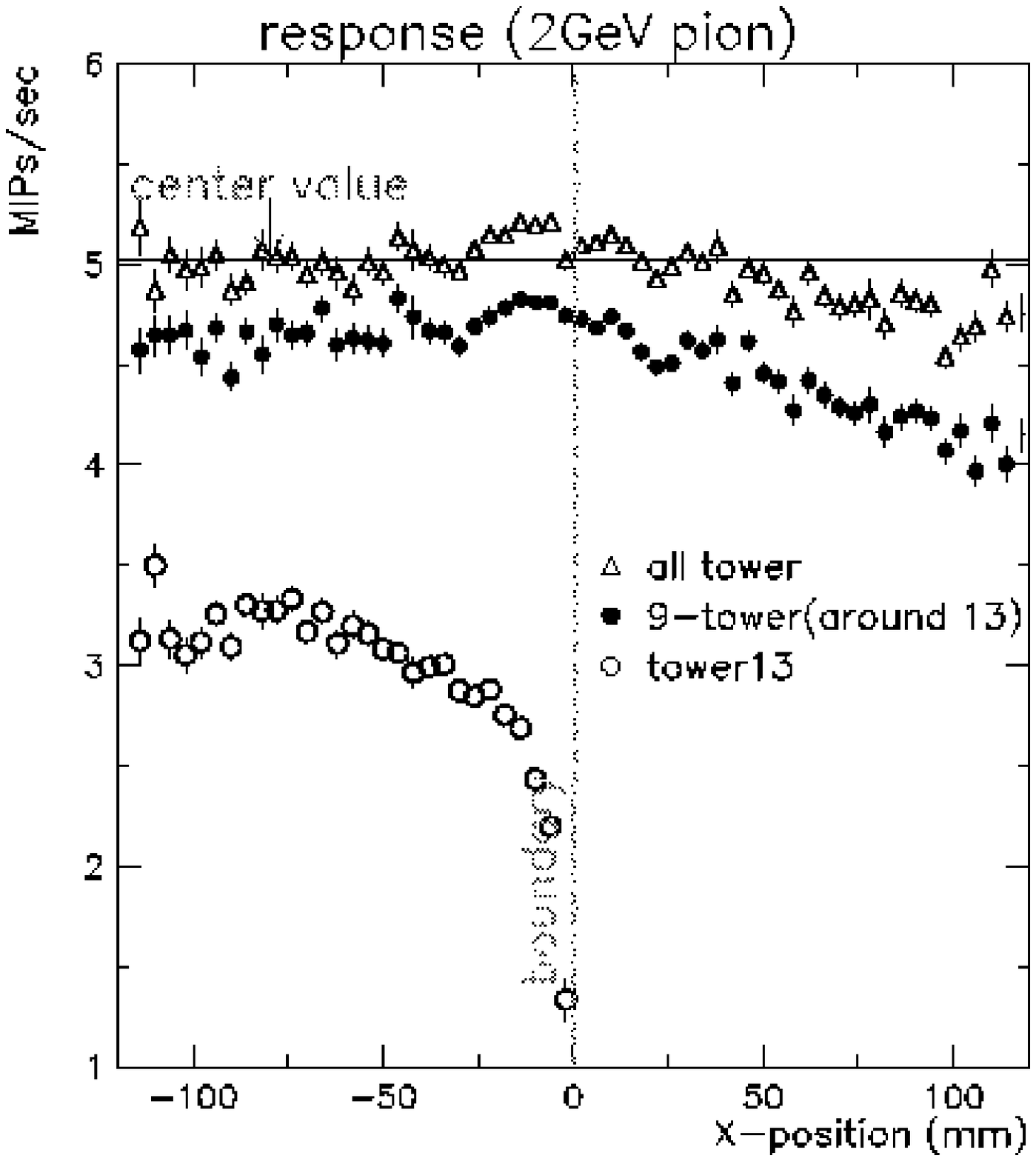}
\hspace{1cm}
\epsfxsize=6.0cm \epsfbox{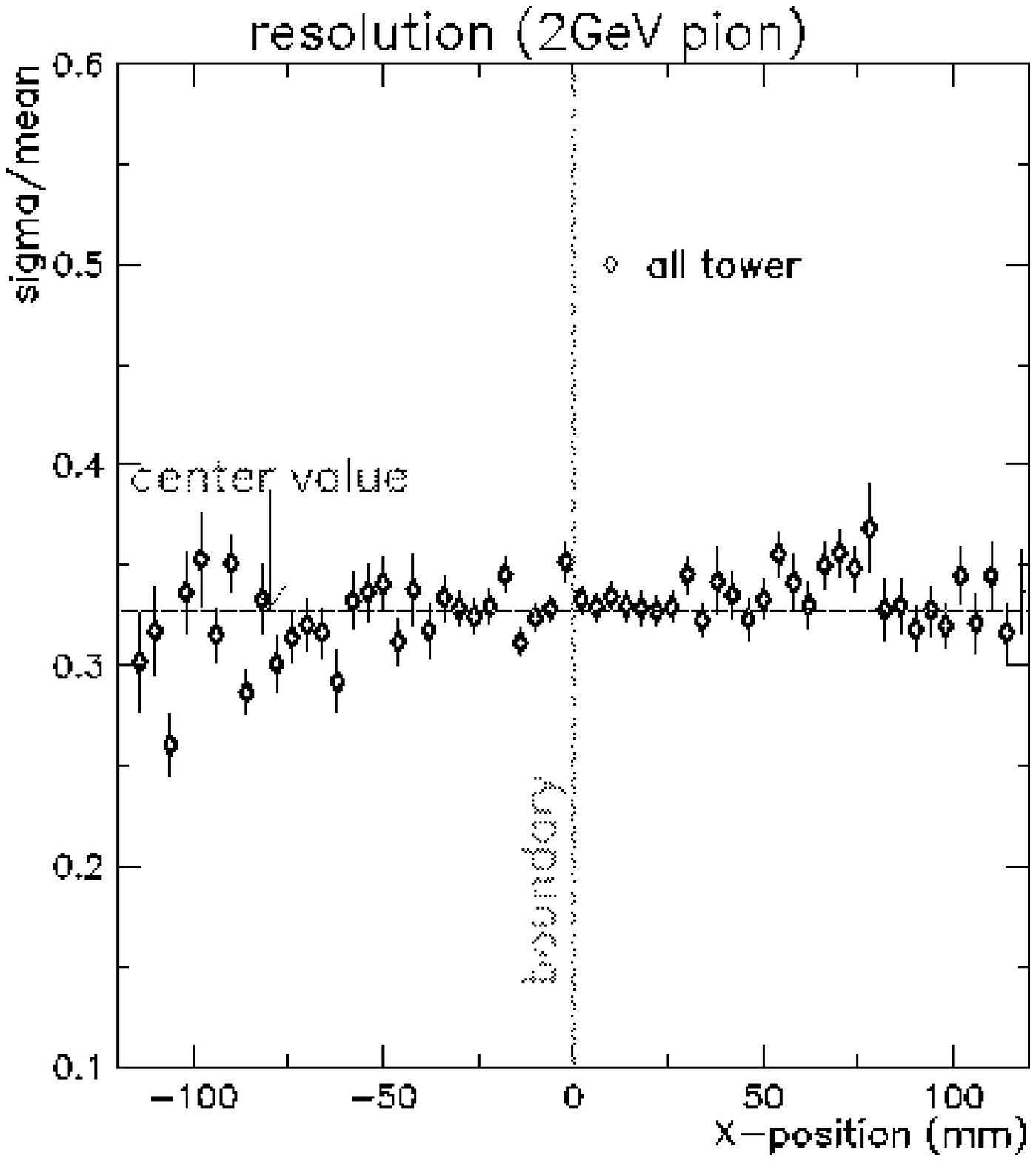}}
\begin{center}\begin{minipage}{\figurewidth}
\caption{\label{boundary}\sl
Tower boundary response of the tile/fiber test module.
Top figures are responses for electrons, and bottom ones are for pions.
Left figures are energy measurement, while right ones are energy resolutions.}
\end{minipage}\end{center}
\end{figure}

Response at the tower boundary are shown in Fig.\ref{boundary}.
Energy measurement for pions may have enhancement of about 3\%
at the region where WLS fibers are embedded.
This is consistent with the non-uniformity at the bench test
shown in Fig.\ref{tilepes}.
This enhancement is not significant when EMC is located
before the HCAL and response smears out.
On the other hand, energy measurement for electrons has
prominent enhancement at the boundary.
This must be equalized by adjusting reflection index of tile
surface around the WLS when EM modules are constructed.

\begin{figure}
\centerline{
\epsfxsize=8.0cm \epsfbox{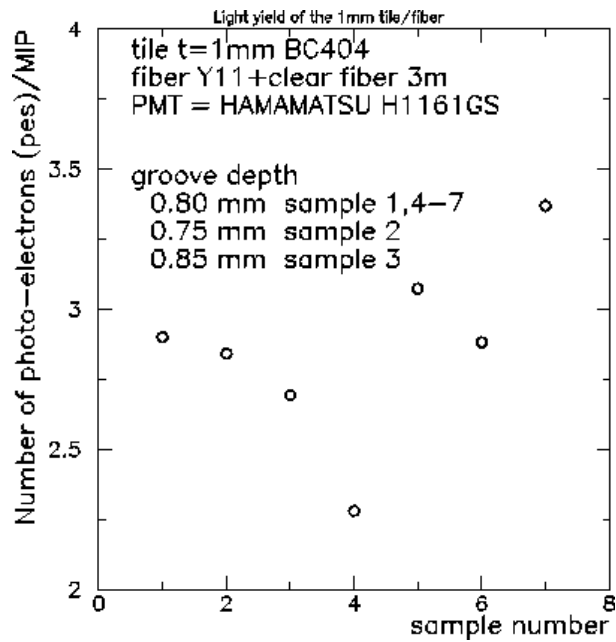}}
\begin{center}\begin{minipage}{\figurewidth}
\caption{\label{tile1mm}\sl
Photo-electron yield of 10cm$\times$10cm$\times$1mm-thick tiles.}
\end{minipage}\end{center}
\end{figure}


\vspace{0.3cm}
\noindent
{\bf 3) EMC R\&Ds at Testbench}
\vspace{0.2cm}

\begin{figure}
\centerline{
\epsfysize=8.0cm \epsfbox{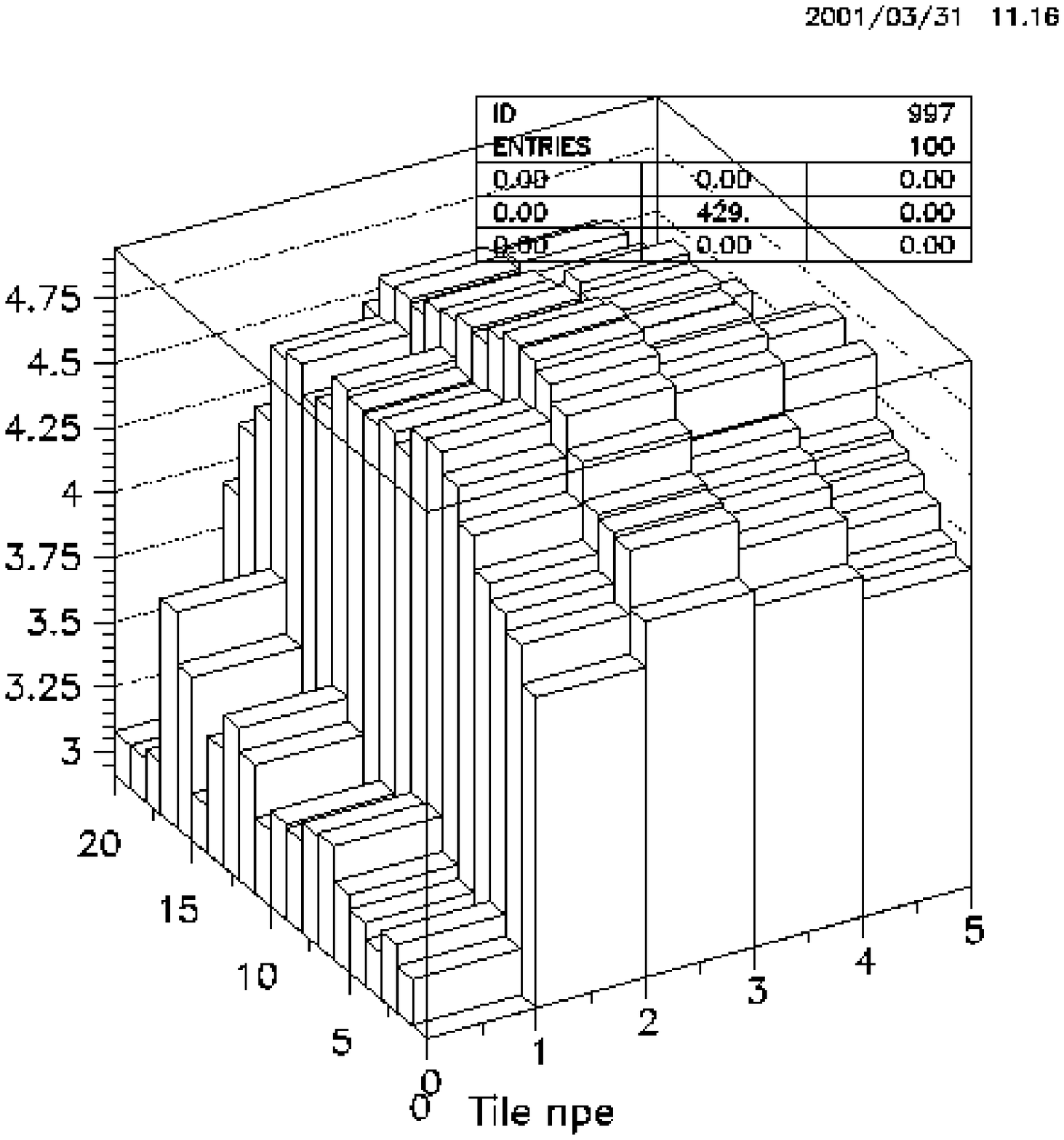}}
\begin{center}\begin{minipage}{\figurewidth}
\caption{\label{EMCstrip}\sl
Photon yield uniformity over a 1cm-wide strip.}
\end{minipage}\end{center}
\end{figure}

Photo-electron yield of 1mm-thick scintillator for EM calorimeter
was measured for 10cm$\times$10cm tiles.
It also had a $\sigma$-groove, and 0.7 mm$\phi$ WLS fiber was used.
The average photo-electron yield was obtained to be three
as shown in Fig.\ref{tile1mm}.
This corresponds to 350 p.e./GeV for compensating calorimeter,
and is not sufficient for EM calorimetry.
Comparison between 20cm$\times$20cm$\times$2mm-thick tiles and
10cm$\times$10cm$\times$1mm-thick tiles suggests that
photo-electron yield is inversely proportional to the area of tiles,
since we know from independent measurement that 
photo-electron yield is proportional to the tile thickness.
Smaller tiles such as 5cm$\times$5cm is considered to have
sufficient photo-electron yield.
This, together with non-uniformity, should be established
by further measurement.

\vspace{0.3cm}
R\&Ds on another option of finer-granularity EMC, made of
stacks of scintillator strip arrays like SMD, are in progress.
This option requires huge number of photo-detector channels,
and has become possible by recent advances in multi-channel HPD and EBCCD.
Whether we can remove ghosts efficiently or not must be
investigated with a full simulator.

Bench tests on strip properties were carried out for various
strip sizes with thickness of 2 mm.
Non-uniformity of photo-electron yield over a strip was
measured to be 4.8\% at most for 1cm-wide strips as shown in
the Fig.\ref{EMCstrip}, and is acceptable for EMC.
Average photo-electron yield, on the other hand, is 4.6 p.e./MIP.
This corresponds to 260 p.e./GeV in the case of compensating EMC,
and needs further improvement.
However, if software compensation is possible even with the orthogonal 
$\theta-\varphi$ strip layout,
thicker strips can be used, and photo-electron yield will not be a problem.
Detailed full simulation studies are again needed to examine
applicability of software compensation for this scheme.

%% file: detcal/PSSM.tex

   The purpose of PSD and SMD 
are to identify $e/\pi^{\pm}/\gamma/\pi^0$, and improve cluster separation
and track-cluster association.
For this purpose, fine segmentation is necessary for SMD.
First we studied SMD of Si-pad array as HES\cite{HES} of ZEUS.
Test modules were made with Si-pad array firstly with pad size of 
1cm$\times$1.5cm and array size of 18$\times$12, 
then secondly with pad size of 1cm$\times$1cm and array size of 
16$\times$16 with improved readout electronics.
Beam tests were carried out in combination with
PSD and hadron calorimeter modules\cite{Sipad}.

Later on, we moved to the scintillator-strip option since:
\begin{itemize}
\item Si-pad SMD was thought to cost too much;
\item Scintillator-strip SMD used the same technique as the tile/fiber
	calorimeter, and was easily integrated with calorimeter.
\end{itemize}
Since the size of the 3Tesla-version baseline JLC calorimeter 
is now much smaller than the 
previous one, the first reason may be no more true.
However re-examination of the possibility of Si-pad option
has not yet started.

\begin{figure}
\centerline{
\epsfysize=7.0cm \epsfbox{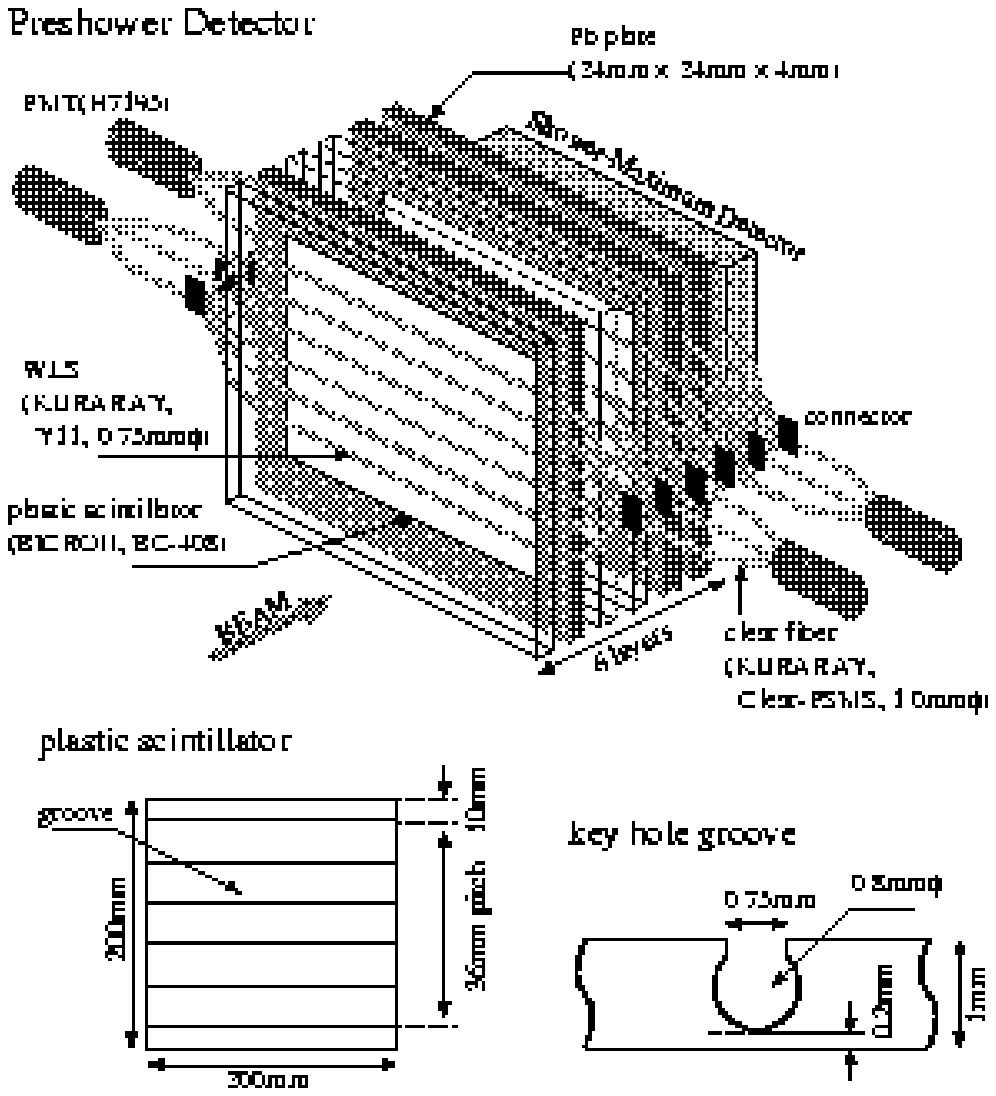}
\hspace{0.3cm}
\epsfysize=7.0cm \epsfbox{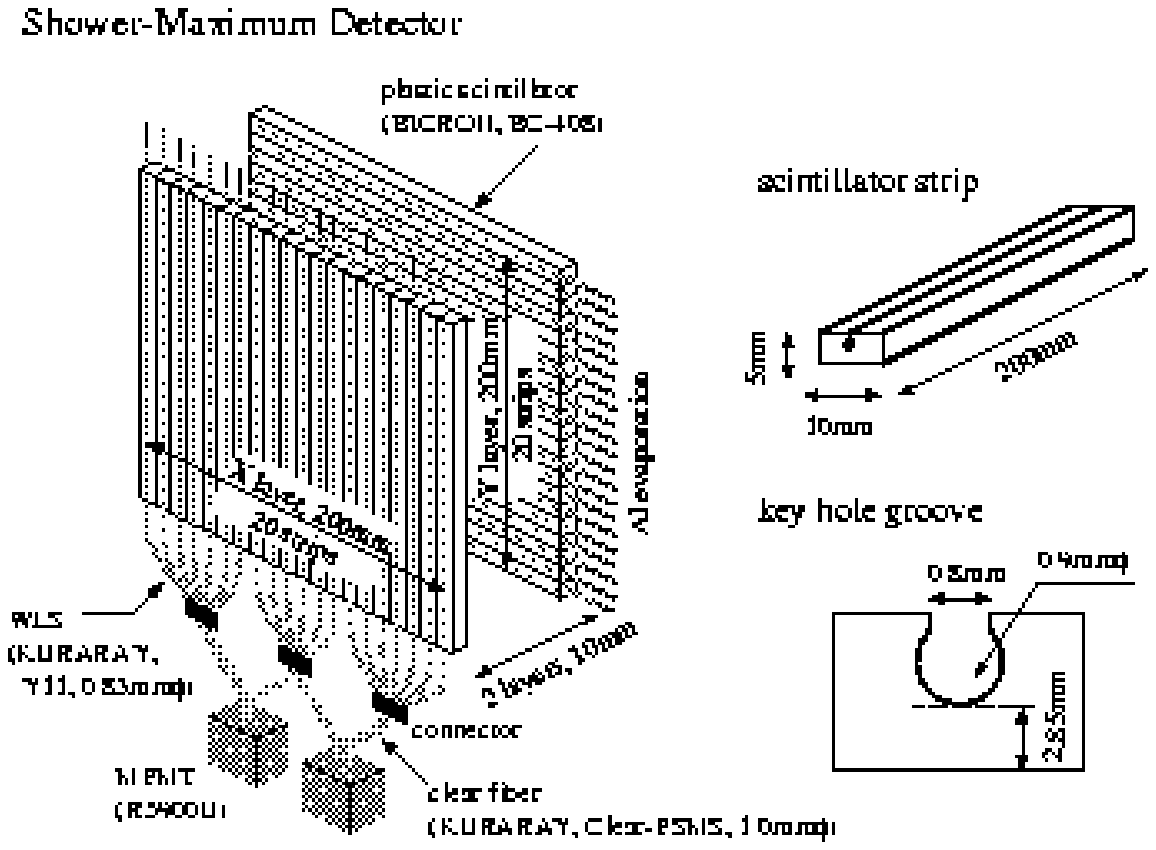}}
\begin{center}\begin{minipage}{\figurewidth}
\caption{\label{PSSMD}\sl
Schematical drawing of PSD (left) and SMD (right).}
\end{minipage}\end{center}
\end{figure}

\begin{figure}
\centerline{
\epsfxsize=15.0cm \epsfbox{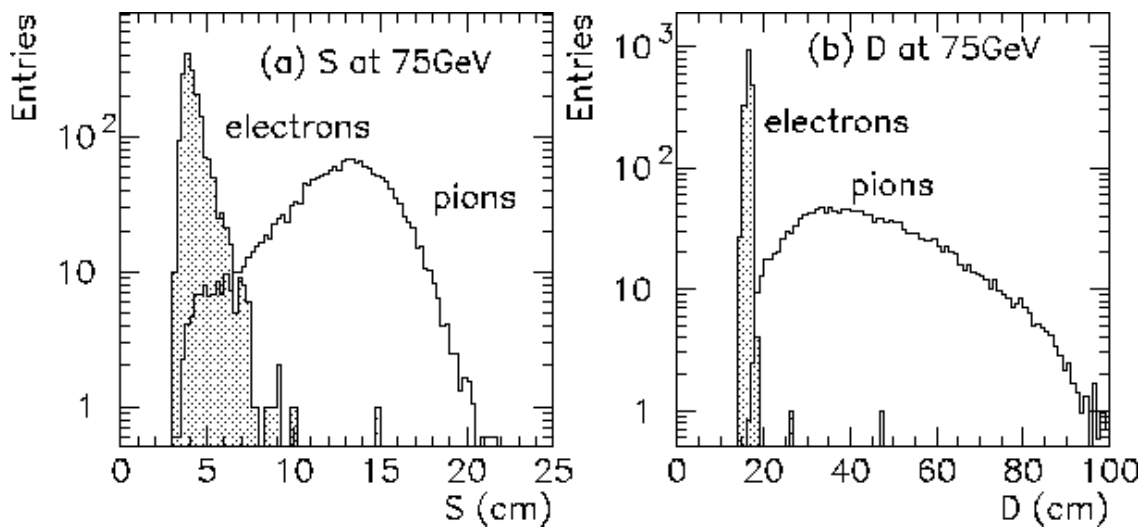}}
\begin{center}\begin{minipage}{\figurewidth}
\caption{\label{SDscore}\sl
e/$\pi$ separation capability of $S$-variable (left) 
and $D$-variable (right).}
\end{minipage}\end{center}
\end{figure}

According the the baseline design, we built test modules
of PSD and SMD, and carried out a beam test at FNAL
in combination with the tile/fiber hadron calorimeter\cite{PSSMFNAL}.
The designs of PSD and SMD modules are schematically shown in 
Fig.\ref{PSSMD}.
Photons from scintillators were read out by WLS fiber assembly
in both PSD and SMD.
In the case of SMD, fibers were connected to multi-channel PMTs,
and each strip was read out separately.

In order to separate electrons from pions, three variables are defined:
\begin{itemize}
\item $S$ is a measure of lateral shower spread defined as
      $ S = \sqrt{ \sum_{i} d_i^2 E_i / E_{tot} } $
\item $D$ is a measure of longitudinal shower distribution defined as
      $ D = \sum_j z_j E_j / E_{tot} $
\item $R$ is a ratio of energy in PSD to total energy.
\end{itemize}

\begin{figure}
\centerline{
\epsfxsize=12.0cm \epsfbox{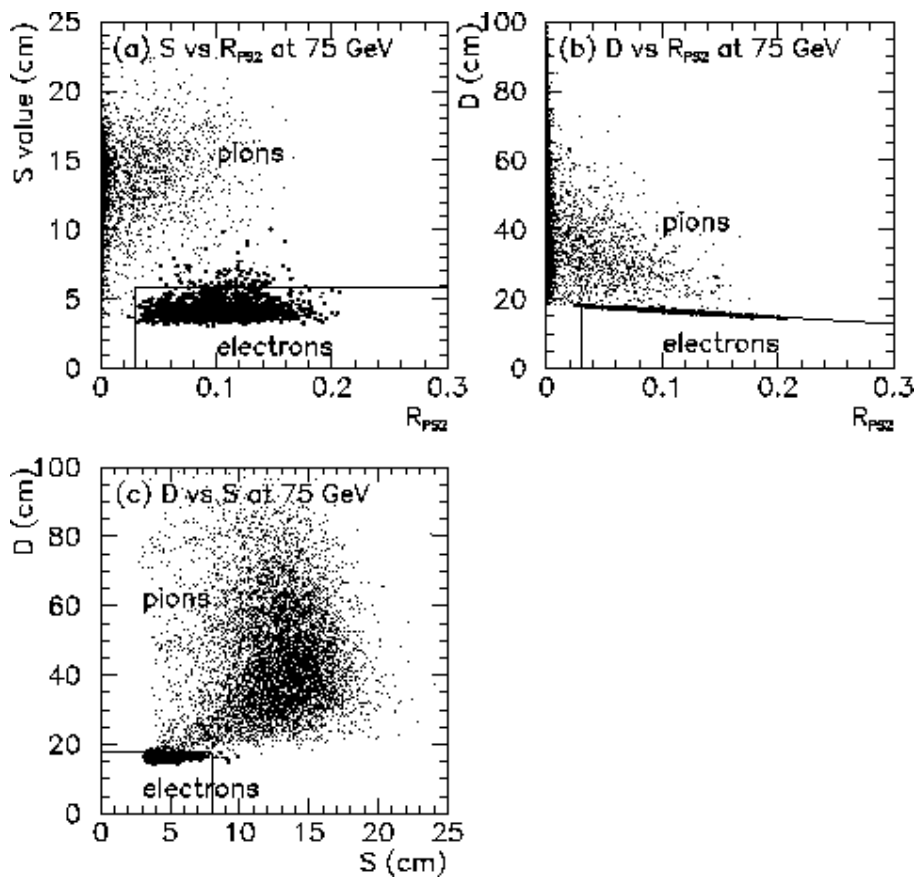}}
\begin{center}\begin{minipage}{\figurewidth}
\caption{\label{PSSMepi}\sl
e/$\pi$ separation capabilities using combinations of
$S$, $D$, and $R$ variables..}
\end{minipage}\end{center}
\end{figure}

Examples of $e/\pi$ separation capability of $S$ and $D$ variables
are shown in in Fig.\ref{SDscore}.
It is seen that $D$-variable has excellent performance in $e/\pi$ separation,
and thus that longitudinal segmentation is quite important.
Measured $e/\pi$ separation capabilities are shown in Fig.\ref{PSSMepi}
and summarized in Table~\ref{PSSMepitable}.
Obtained score of mostly better than 1/1000 are comparable 
to other measurements~\cite{SPACALepi}, and is satisfactory.

Hit multiplicity of SMD can also be used to identify electrons.
By requiring more than 10 hit-strips out of 40 strips,
pion rejection factor of about 10 can be achieved with
electron efficiency of 98\%.
However this has strong correlation with R-value, and
is not included in the table~\ref{PSSMepitable}.

\begin{figure}
\centerline{
\epsfxsize=8.0cm \epsfbox{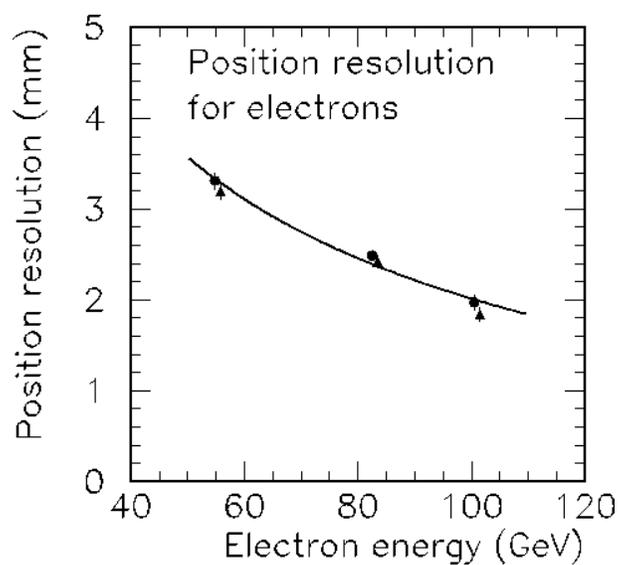}}
\begin{center}\begin{minipage}{\figurewidth}
\caption{\label{PSSMposition}\sl
Position resolution of SMD.}
\end{minipage}\end{center}
\end{figure}

\begin{table}
\begin{center}\begin{minipage}{\figurewidth}
\caption{\sl Pion rejection factors with the combinations of
$R_{PS2}$, $S$ and $D$ values.}
\label{PSSMepitable}
\end{minipage}\end{center}
\begin{center}
\begin{tabular}{|c|c|c|c|c|c|c|c|c|c|}\hline
Method & 
\multicolumn{3}{c|}{$S$ value and $R_{PS2}$} &  
\multicolumn{3}{c|}{$D$ value and $R_{PS2}$} &  
\multicolumn{3}{c|}{$D$ and $S$ values} \\ \hline
$\epsilon_{e}$ (\%) & 90 & 95 & 98 & 90 & 95 & 98 & 90 & 95 & 98 \\ \hline
$r_{\pi}$ at 50~GeV  & 
473 & 312 & 180 & 1274 & 1104 &  720 & 1183 &  828 & 753 \\
$r_{\pi}$ at 75~GeV  & 
657 & 487 & 343 & 1258 & 1162 &  944 & 1510 & 1162 & 888 \\
$r_{\pi}$ at 100~GeV & 
984 & 656 & 281 & 1640 & 1406 & 1406 &  984 &  820 & 656 \\ \hline
\end{tabular}
\end{center}
\end{table}

Measured position resolution is shown in Fig.\ref{PSSMposition}.
Present result is significantly worse than the baseline requirement of
better than 1 mm for energies higher than a few GeV.
This is due to saturation and cross talk of multi-channel PMTs, 
and should be solved before going to the next step.

\vspace{0.3cm}

The WLS fiber assembly costs significant amount of
the total cost of scintillator-array SMD. 
In order to reduce the total cost of SMD, an option which use 
photo-diodes (PDs) instead of WLS fiber assembly are being studied.
Silicon PIN PDs or APDs are attached at the both end of
the strips, and photons are directly read out by the PDs.
Direct punch-through of charged particles will be identified
by double-layer scheme; a blind PD is pasted behind the detector PD.

Studies so far indicates that PIN PDs do not have enough
sensitivity, and studies with APDs are in progress.
Much more studies are needed to get some conclusion on this option.

%% file: detcal/PhotoDetector.tex

The whole calorimeter system is designed to be located
inside the superconducting solenoid.
Thus photon detectors operational in strong magnetic field are needed.
In the case of crystal calorimeter, the light yield is large 
and popular PIN silicon photodiode or APD can be used.
On the other hand, in the case of sampling calorimeters,
especially compensating calorimeters,
the light yield is relatively poor and high-gain high-sensitivity
photo-detector is needed.

 One conventional solution is fine-mesh 
photomultiplier tubes (FMPMTs). 
This has been widely used for magnetic field below 1 Tesla.
Possibility to improve its performance at higher magnetic
field was first investigated.

 Another option is hybrid devices such as Hybrid
Photodiodes (HPDs) or Hybrid Avalanche Photodiodes (HAPDs).
Performances of such devices have been extensively studied so far. 
As the result, these hybrid devices are thought 
to be the best option at present.

\subsubsection{1) FMPMT}

\begin{figure}
\centerline{
\epsfxsize=8cm 
\epsfbox{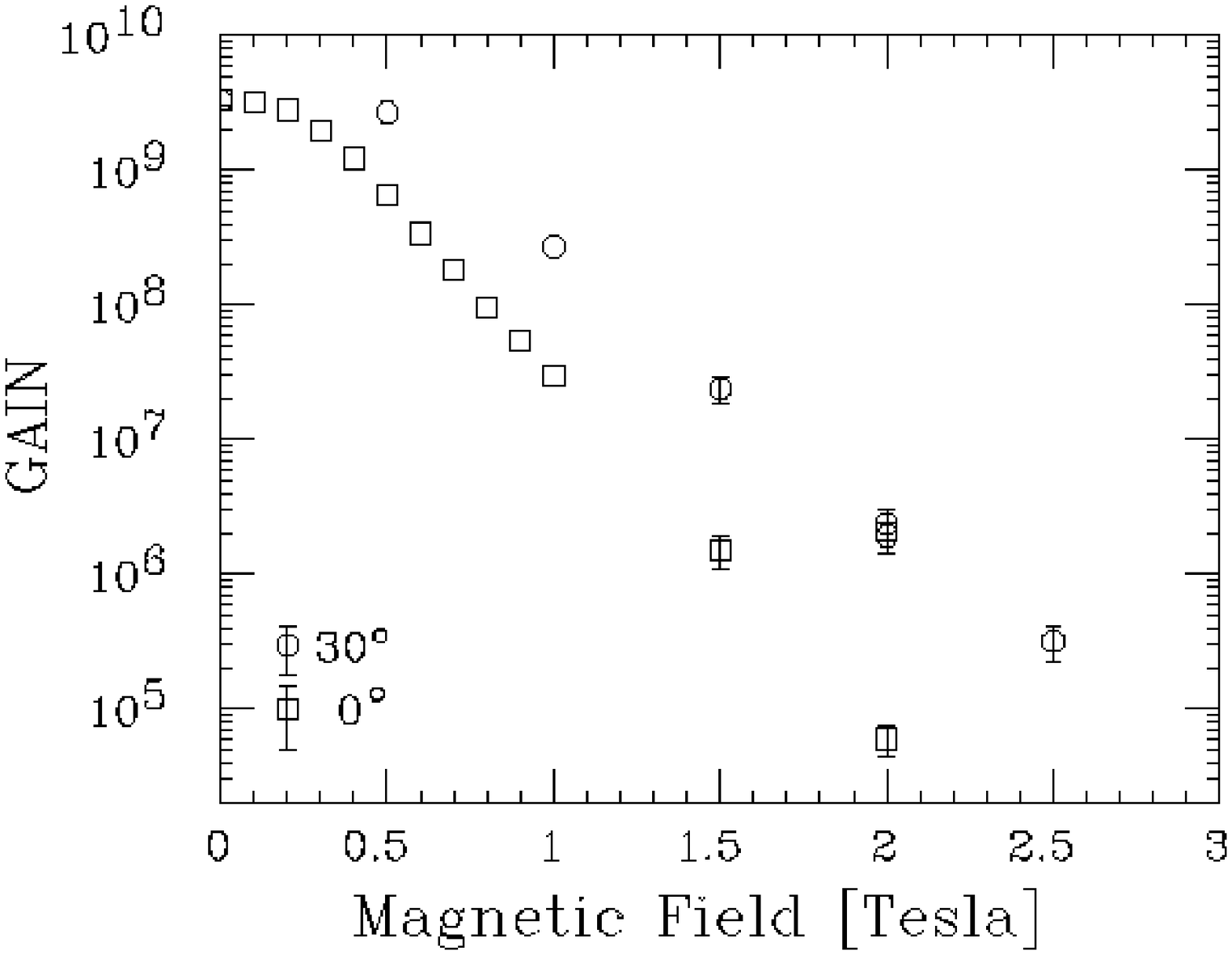}
\epsfxsize=8cm 
\epsfbox{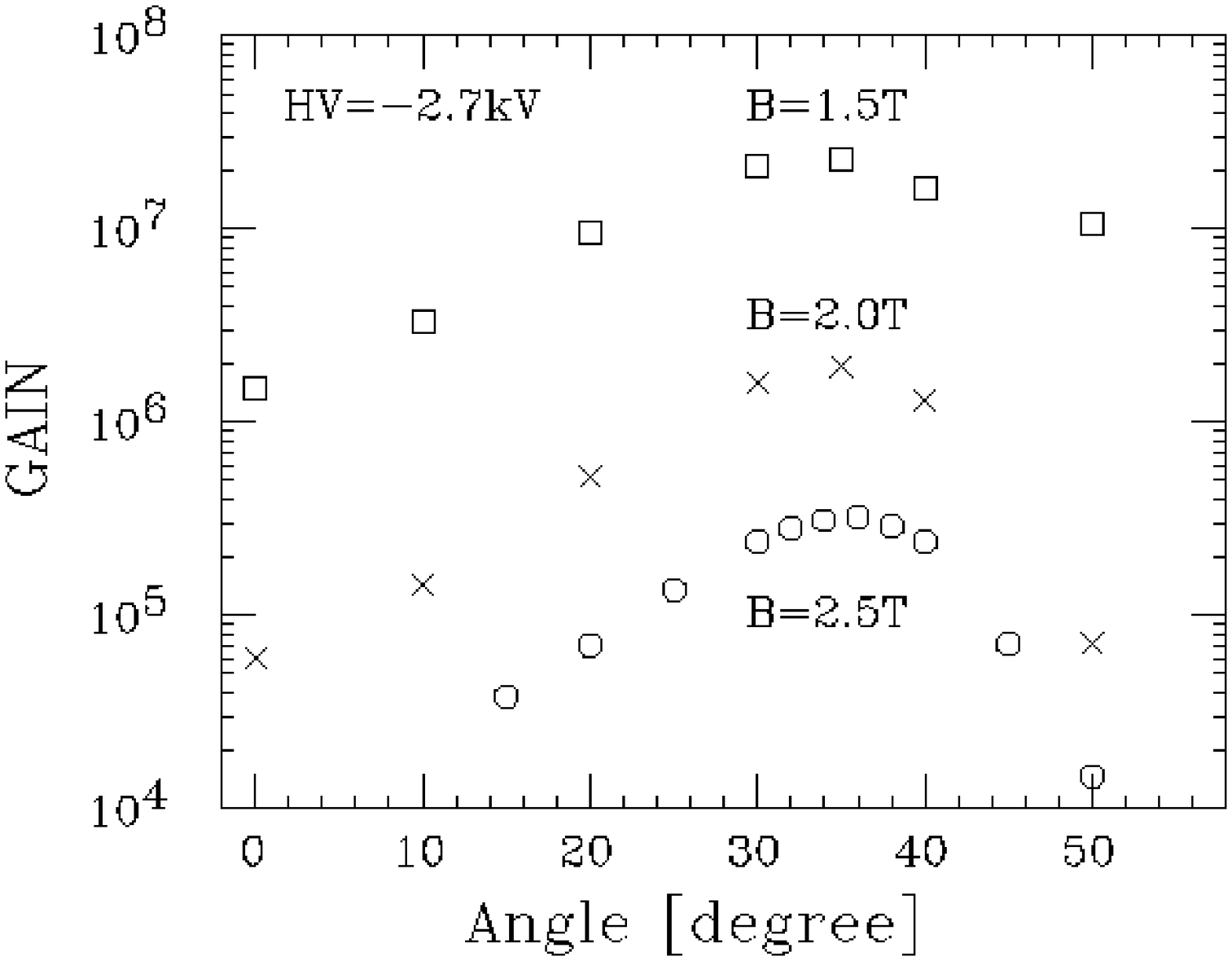}
}
\begin{center}\begin{minipage}{\figurewidth}
\caption{\label{FMPMT}\sl
Magnetic field dependence of FMPMT gain. Horizontal axes are
magnetic field strength (left) and angle between magnetic field
and PMT axis (right).}
\end{minipage}\end{center}
\end{figure}

In order to keep high gain even in the
magnetic field of higher than 1 Tesla,
high-gain FMPMTs with 24-stages of dynodes (Hamamatsu H2611SXA) were 
made and tested with magnetic field up to 2.5 Tesla 
using SKS spectrometer at KEK \cite{HPD95}.
Fig.\ref{FMPMT} shows magnetic field dependence of the gain.
It still had gain of $3 \times 10^5$ at 2.5 Tesla
when the angle between the PMT axis and magnetic field was 30$^{\circ}$
even after rapid drop of gain with magnetic field.
By extrapolation, it was expected to have gain of  $3 \times 10^4$ at 3 Tesla.
However the gain variation with respect to field strength and 
PMT angle to the field direction is quite steep.
This feature is not desirable, though is not fatal.
We therefore concluded that FMPMT can not be the primary solution.

\subsubsection{2) Hybrid Devices}

\begin{figure}
\centerline{\epsfysize=6cm 
\epsfbox{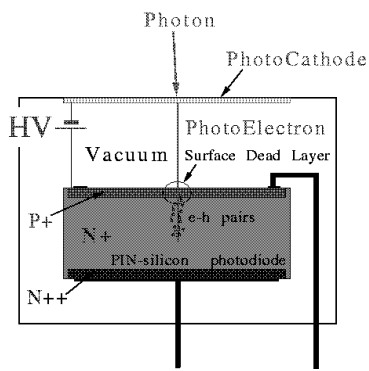}}
\caption{
\label{HPDstructure}\sl
Schematic view of the HPD structure.}
\end{figure}

 HPD consists of a photo-cathode and a PIN photodiode
facing to each other with a narrow vacuum gap in between.
This is essentially insensitive to the axial magnetic field
as is easily seen from the structure shown in Fig.\ref{HPDstructure}.
High-voltage is applied between the photocathode and the PIN silicon diode.
Emitted photoelectrons are accelerated by this field,
and injected to the PIN diode.
Those photoelectrons deposits their energy when they pass
the depletion layer and create electron-positron pairs.
Roughly speaking, since excitation energy of one 
electron-position pair is 3.6 eV,
gain of 3000 is expected with photo-cathode voltage of -11 kV.
Actual gain is slightly lower due to the surface layer of the diode.

HAPD uses APD instead of PIN photodiode, and has
higher gain due to the gain of APD itself.
With an APD of gain 100, HAPD would achieve total gain comparable to PMTs.


\begin{figure}
\centerline{
\hspace{1.5cm}
\epsfysize=6.5cm \epsfbox{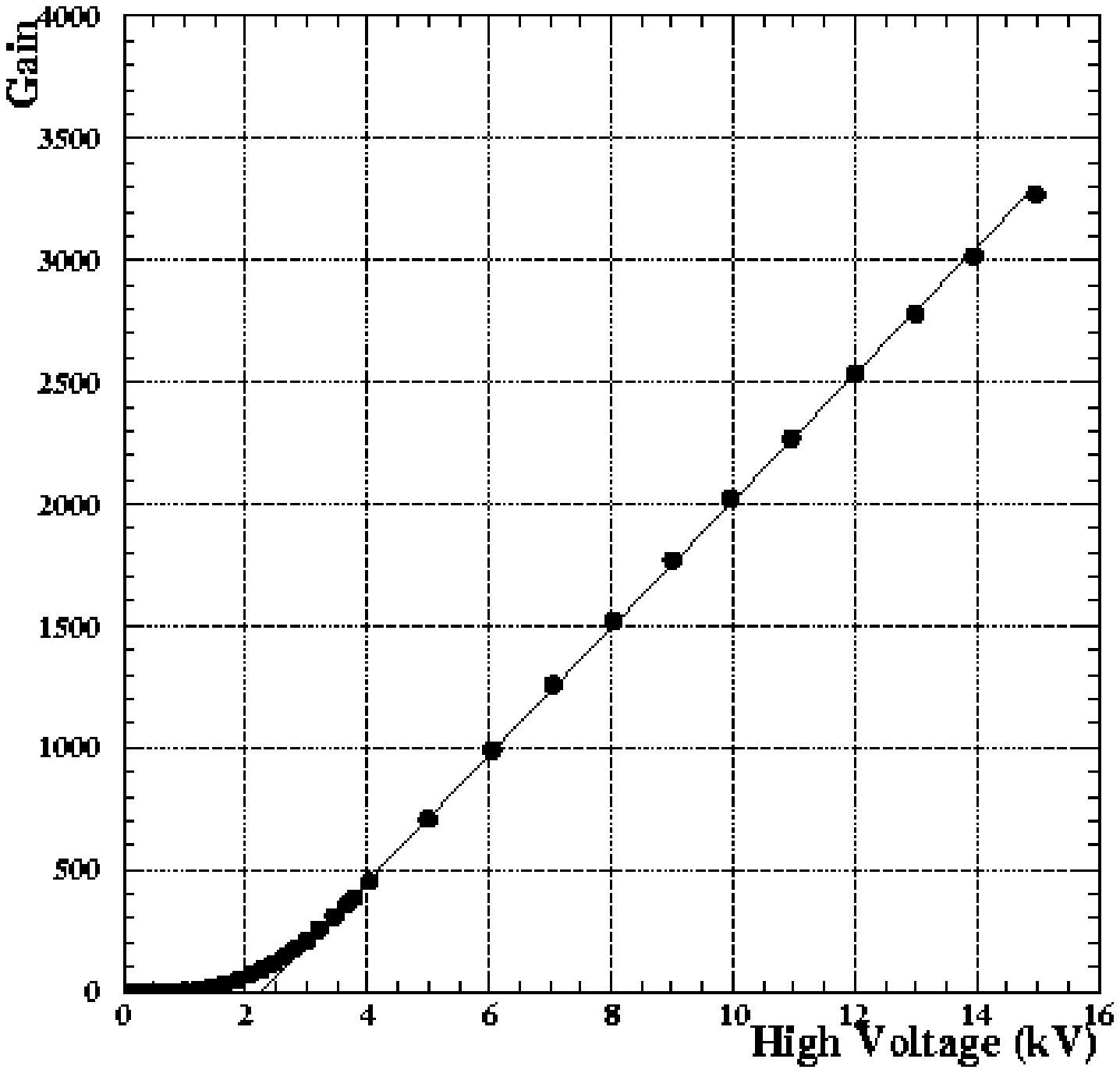}
\epsfysize=6.0cm \epsfbox{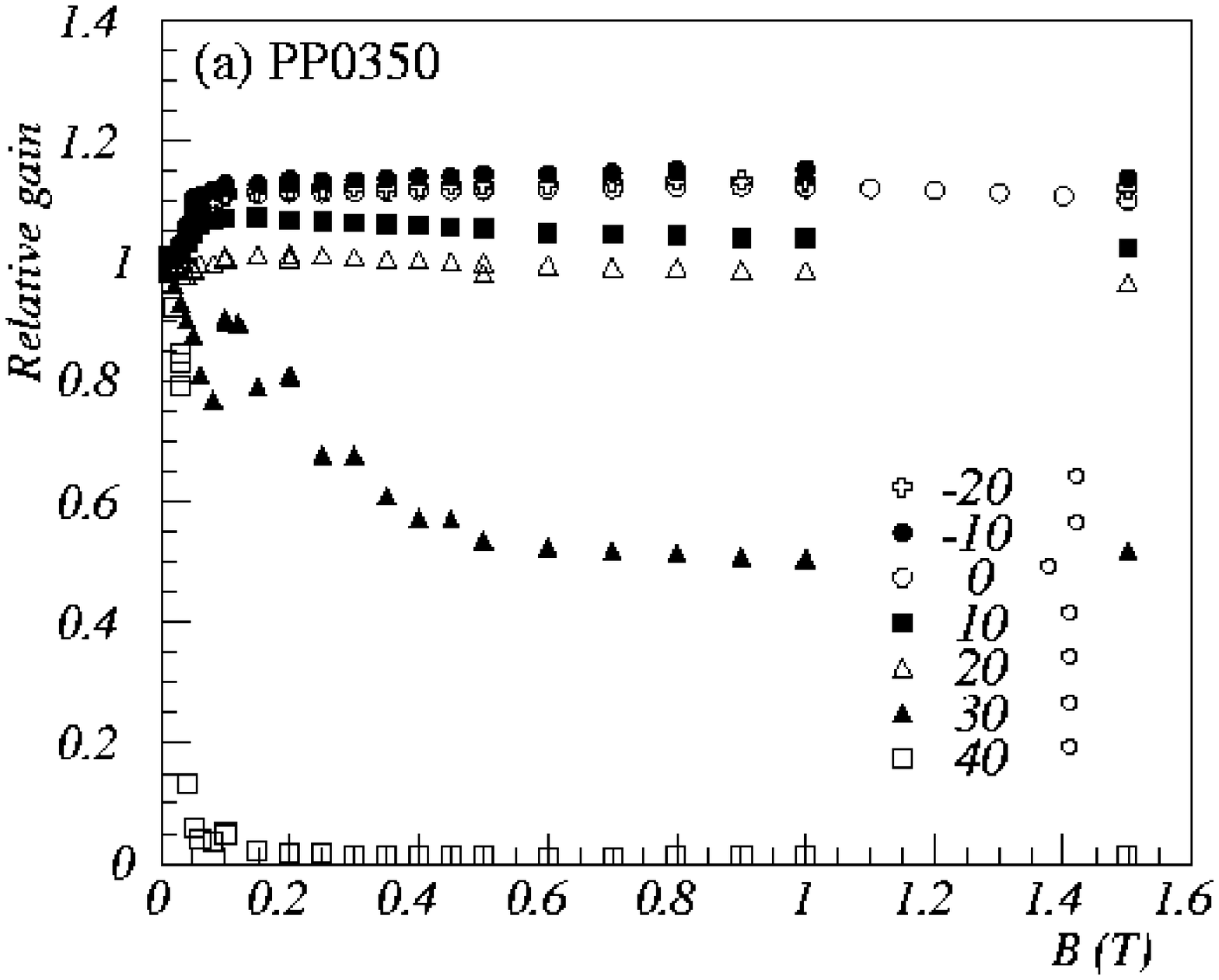}}
\begin{center}\begin{minipage}{\figurewidth}
\caption{\label{HPDgain}\sl
Gains of high-gain HPD out of (left) and in (right) a magnetic field.}
\end{minipage}\end{center}
\end{figure}

\begin{figure}
\centerline{
\hspace{-0.5cm}
\epsfysize=5.5cm \epsfbox{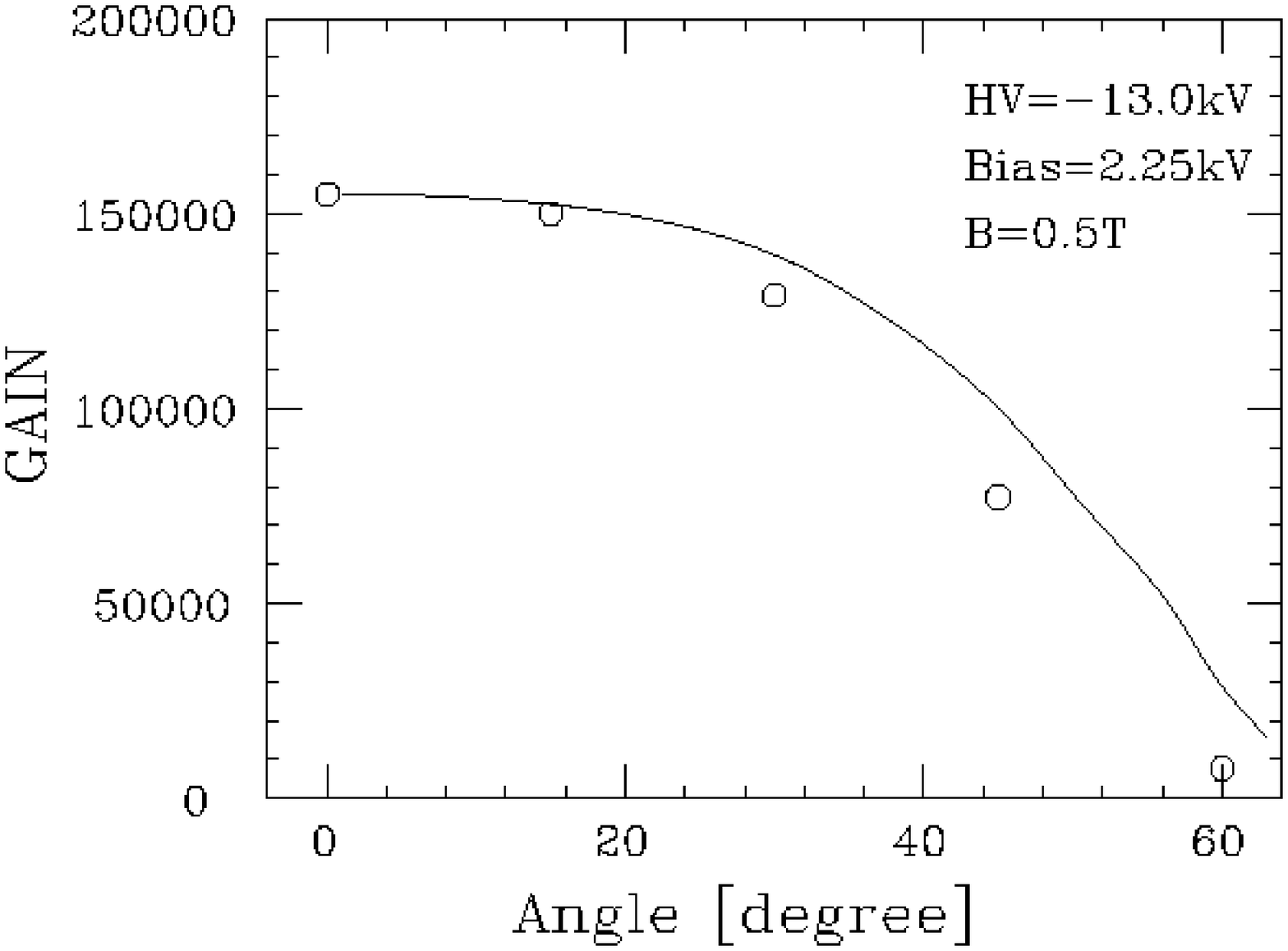}
\hspace{0.5cm}
\epsfysize=5.5cm \epsfbox{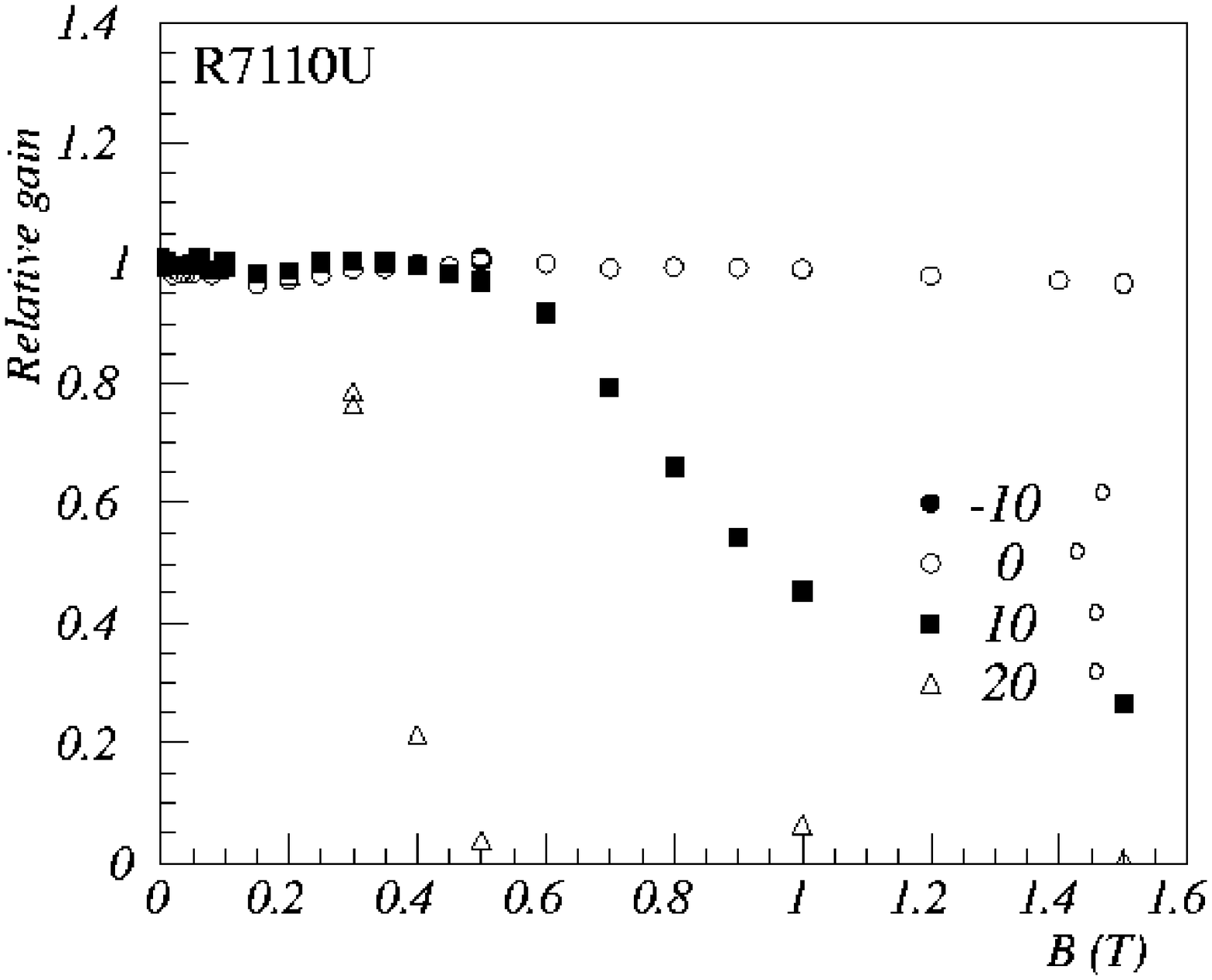}}
\begin{center}\begin{minipage}{\figurewidth}
\caption{\label{HAPDgain}\sl
HAPD gain in a magnetic field for API 748-73-75-631 (left)
and for Hamamatsu R7110U.  
Solid line in the left figure is result of calculation.}
\end{minipage}\end{center}
\end{figure}

\begin{figure}
\centerline{
\epsfysize=6.5cm \epsfbox{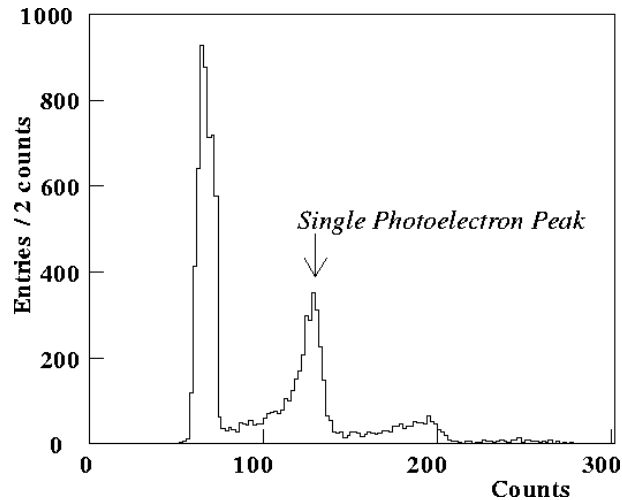}}
\begin{center}\begin{minipage}{\figurewidth}
\caption{\label{HAPDspp}\sl
Single photo-electron peak measured outside of magnetic field by 
Hamamatsu R7110U.}
\end{minipage}\end{center}
\end{figure}

Properties of HPDs and HAPDs have been extensively studied so 
far \cite{HPD95,HPD99}. 
Gains of an HPD (Delft PP350F) in and out of magnetic field 
are shown in Fig.\ref{HPDgain}.  
The gain in the magnetic field with $V_{PC}$=-15 kV 
was measured to be 4000.
If we assume the photo-electron yield of EMC to be 260 p.e./GeV,
preamp noise to be 1000 $e^-$, and HPD gain of 4000,
we can achieve readout noise of about 1MeV.

Though the gain of HPDs is high enough for calorimetry,
it is not high enough for PSD and SMD.
We therefore studied performances of HAPDs.
The gains of HAPDs were measured for 
748-73-75-631 made by Advanced Photonix 
and for Hamamatsu R7110U in magnetic field.
The results are shown in Fig.\ref{HAPDgain}. 
It is clearly seen that HAPDs are
operational in a magnetic field with high gain.
It was also established that HAPDs had sensitivity to measure 
single photon as shown in Fig.\ref{HAPDspp}.
Therefore HAPDs can be a good candidate for PSD readout.


\begin{figure}
\centerline{
\epsfxsize=9cm
\epsfbox{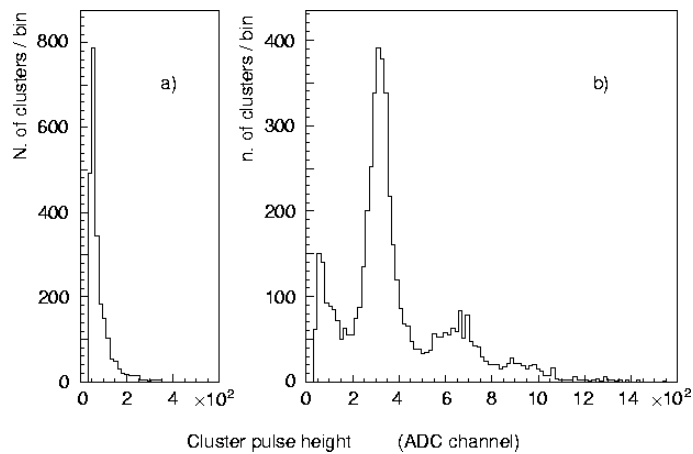}}
\begin{center}\begin{minipage}{\figurewidth}
\caption{\label{EBCCDSPP}\sl
Single-photo-electron peak measured with electro-focus EBCCD by RD46.}
\end{minipage}\end{center}
\end{figure}

\begin{figure}
\centerline{
\epsfxsize=8cm
\epsfbox{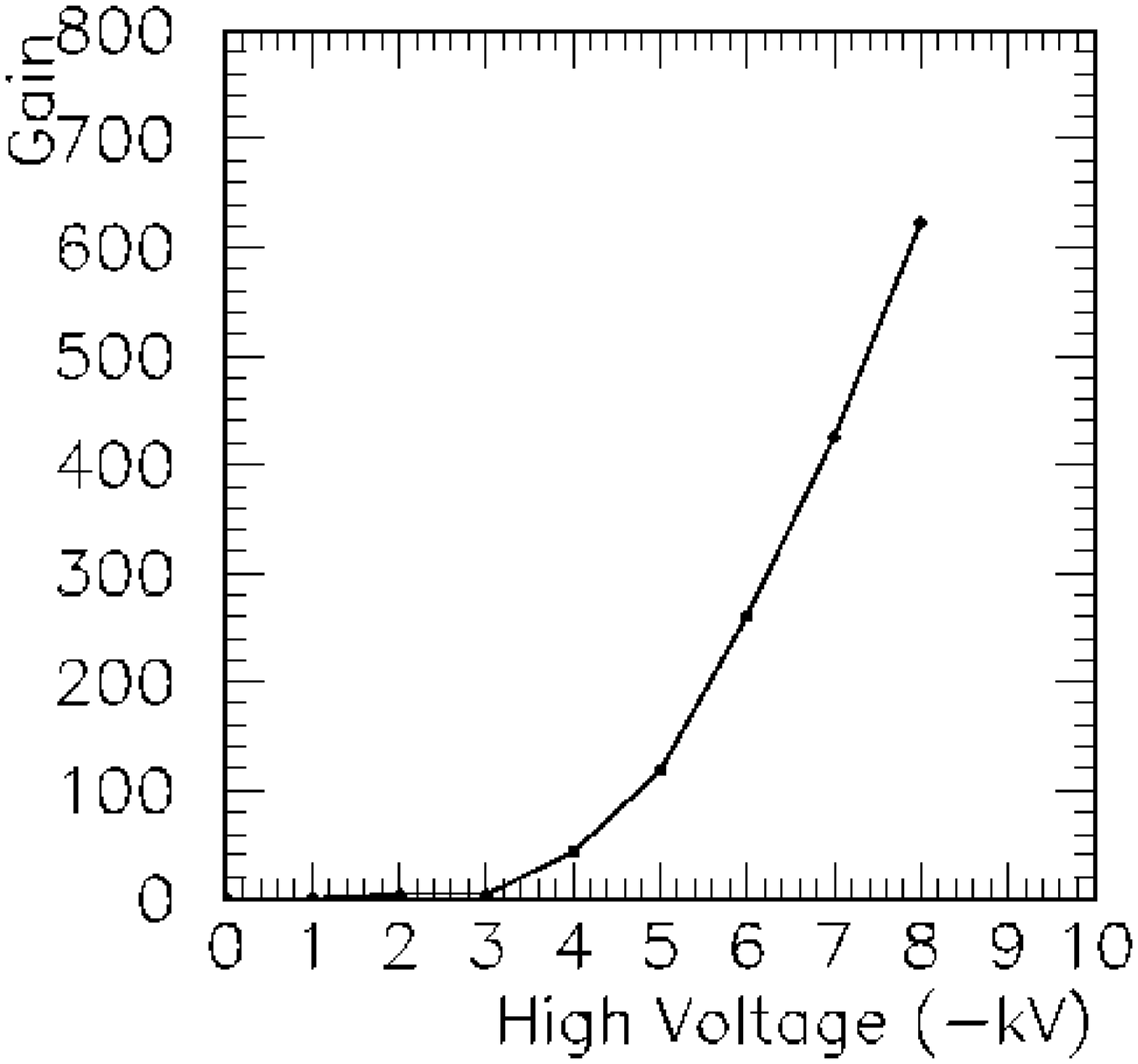}
\epsfxsize=8cm
\epsfbox{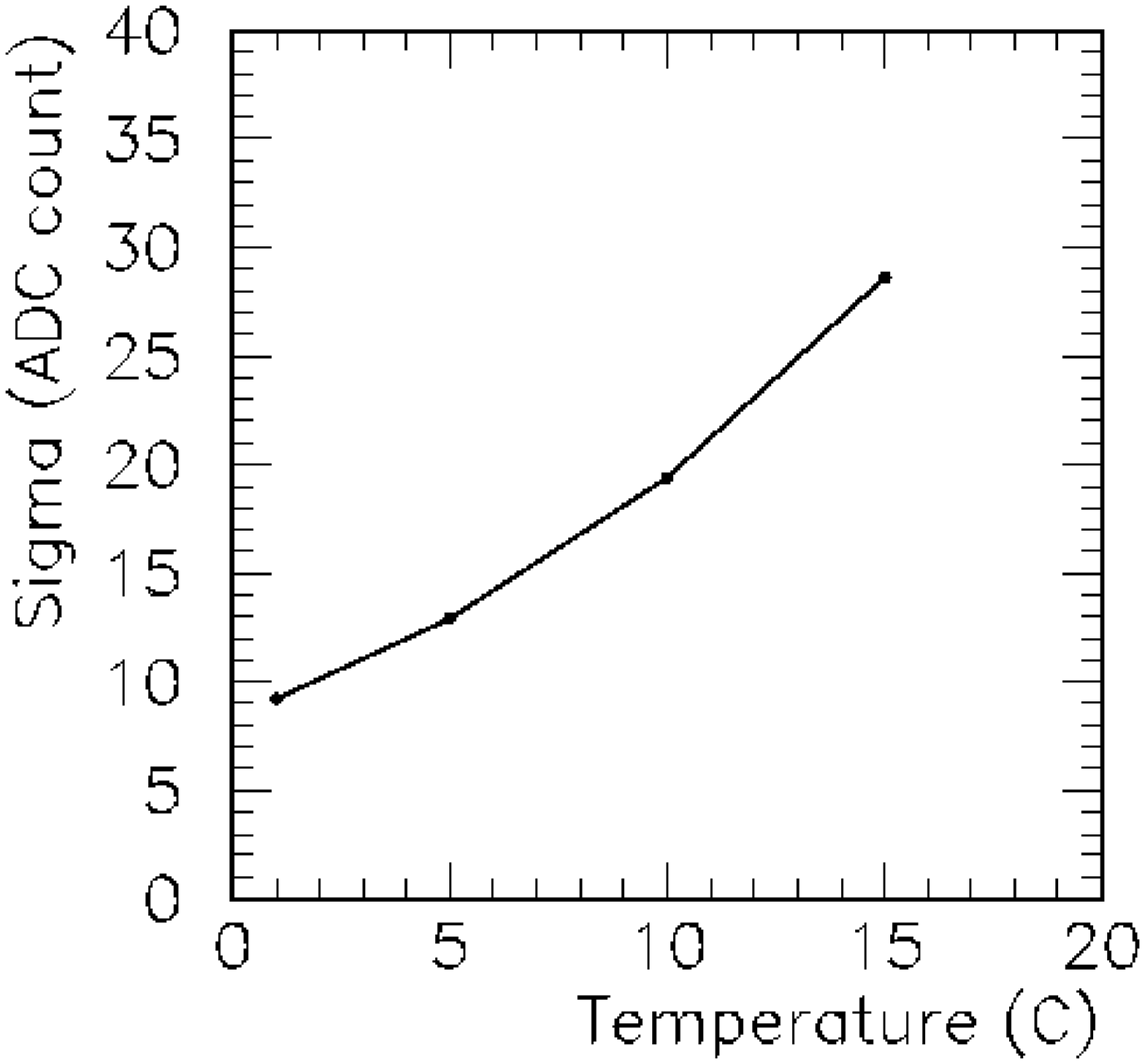}
}
\begin{center}\begin{minipage}{\figurewidth}
\caption{\label{EBCCD-GV}\sl
Gain vs. photo-cathode voltage (left)
and noise vs. temperature (right) for proximity-focused EBCCD N7220.}
\end{minipage}\end{center}
\end{figure}

Scintillator-strip SMD requires 
ultra-multi-channel photon detectors.
Though once Advanced Photonix made 9-pixel HAPDs,
it is not so easy to make multi-pixel HAPDs.
Multi-channel HPDs are available with order of
100 channel/device at present.
However sensitivity of HPDs are not high enough to use for SMD.
Recently developed EBCCDs are ultra-multi-pixel
devices naturally, and can achieve single-photon 
sensitivity~\cite{EBCCDHPK,EBCCDRD46}.
This is thought to be the best candidate for SMD, 
and basic surveys have been initiated.
RD46 collaboration has succeeded to observe single photo-electron
peak with an EBCCD as shown in Fig.\ref{EBCCDSPP}.
However it is an electro-focus type like image intensifiers,
and can not be used in a magnetic field.
We therefore started performance study on proximity-focused 
EBCCDs\cite{EBCCDHPK}.
Structure of proximity-focused EBCCD is quite similar to
that of HPDs shown in Fig.\ref{HPDstructure}, replacing
PIN silicon photodiode by a CCD.
Photo-electrons bombard back-side of CCD, where substrait
is thinned to enable photo-electrons to reach to the epitaxial layer.
Fiber bundle with appropriate fiber spacing is attached to the
photocathode window, and hits are reconstructed by
clustering algorithm.
Fig.\ref{EBCCD-GV} shows gain curve with respect to the
applied photo-cathode voltage and noise vs. temperature.
Measured sensitivity is yet unable to detect single photon,
and further studies are in progress.

%% file: detcal/Engineering.tex

There are several engineering issues related to the actual
construction of the calorimeter system:
\begin{itemize}
\item Rigid and strong lead alloy;
\item Mass production scheme and cost of tiles;
\item Mass production scheme and cost of WLS assembly;
\item Overall structure.
\end{itemize}
Most of these are yet open questions.

\vspace{0.3cm}

\noindent
{\bf Lead Alloy}

\vspace{0.3cm}

Our baseline design is to use lead alloy as absorber material
because of its high-$Z$, density and cost.
However lead is very soft metal, and development of
rigid and hard lead alloy is indispensable.

\begin{figure}
\centerline{
\epsfxsize=8cm \epsfbox{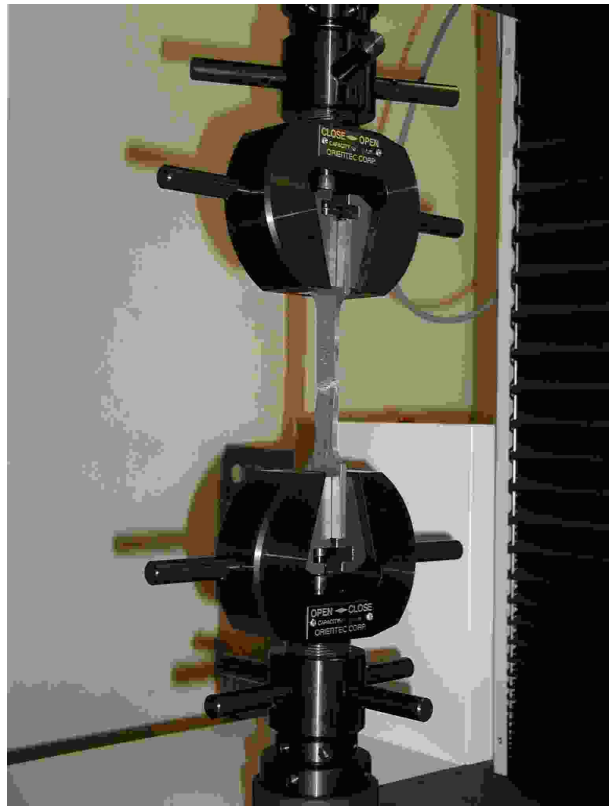}}
\begin{center}\begin{minipage}{\figurewidth}
\caption{\label{Pbsetup} \sl
A photo of the setup to measure mechanical properties of lead alloys.
Broken test piece after measurement is seen.}
\end{minipage}\end{center}
\end{figure}

\begin{figure}
\centerline{
\epsfxsize=8cm \epsfbox{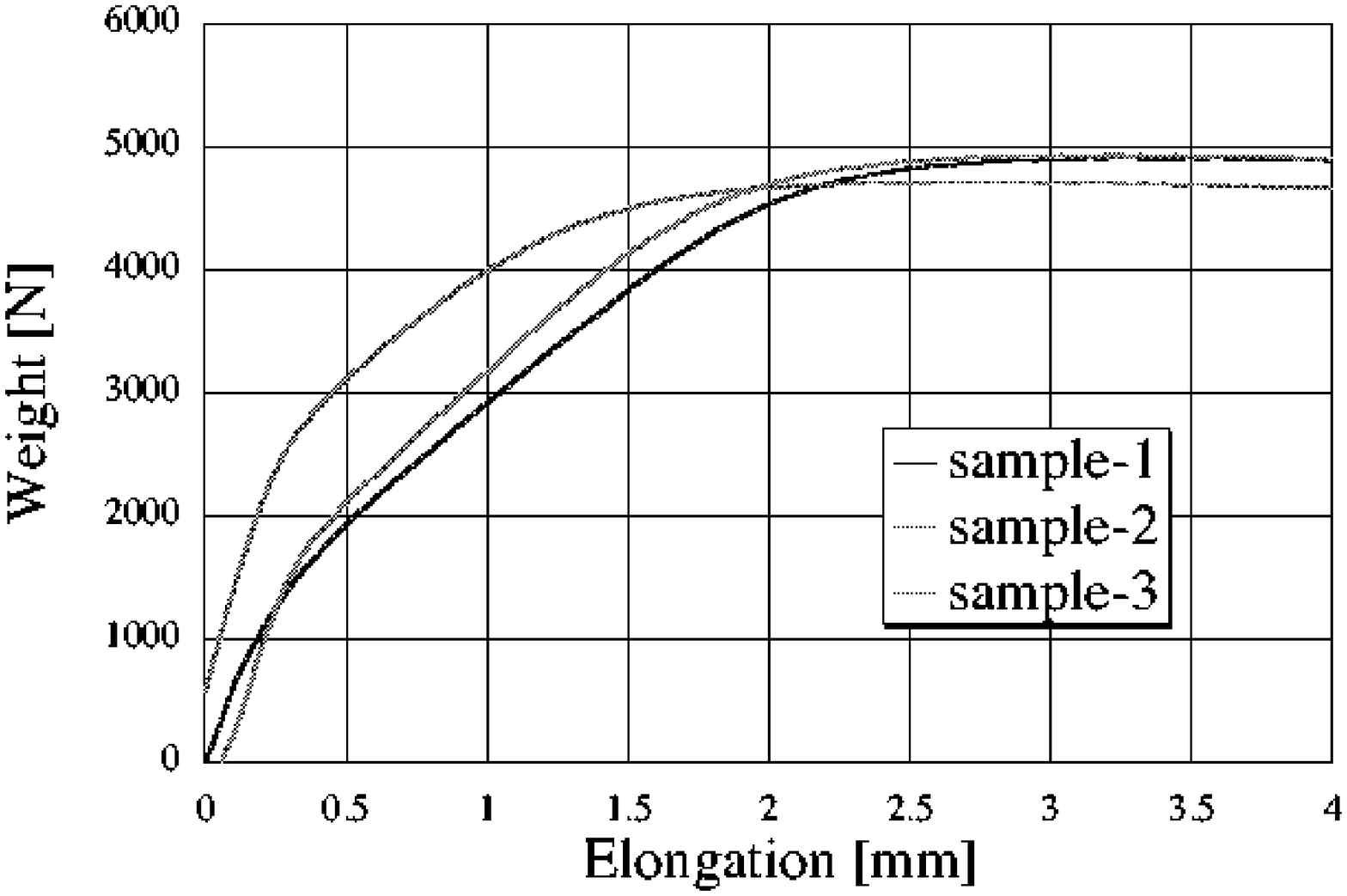}
\epsfxsize=8cm \epsfbox{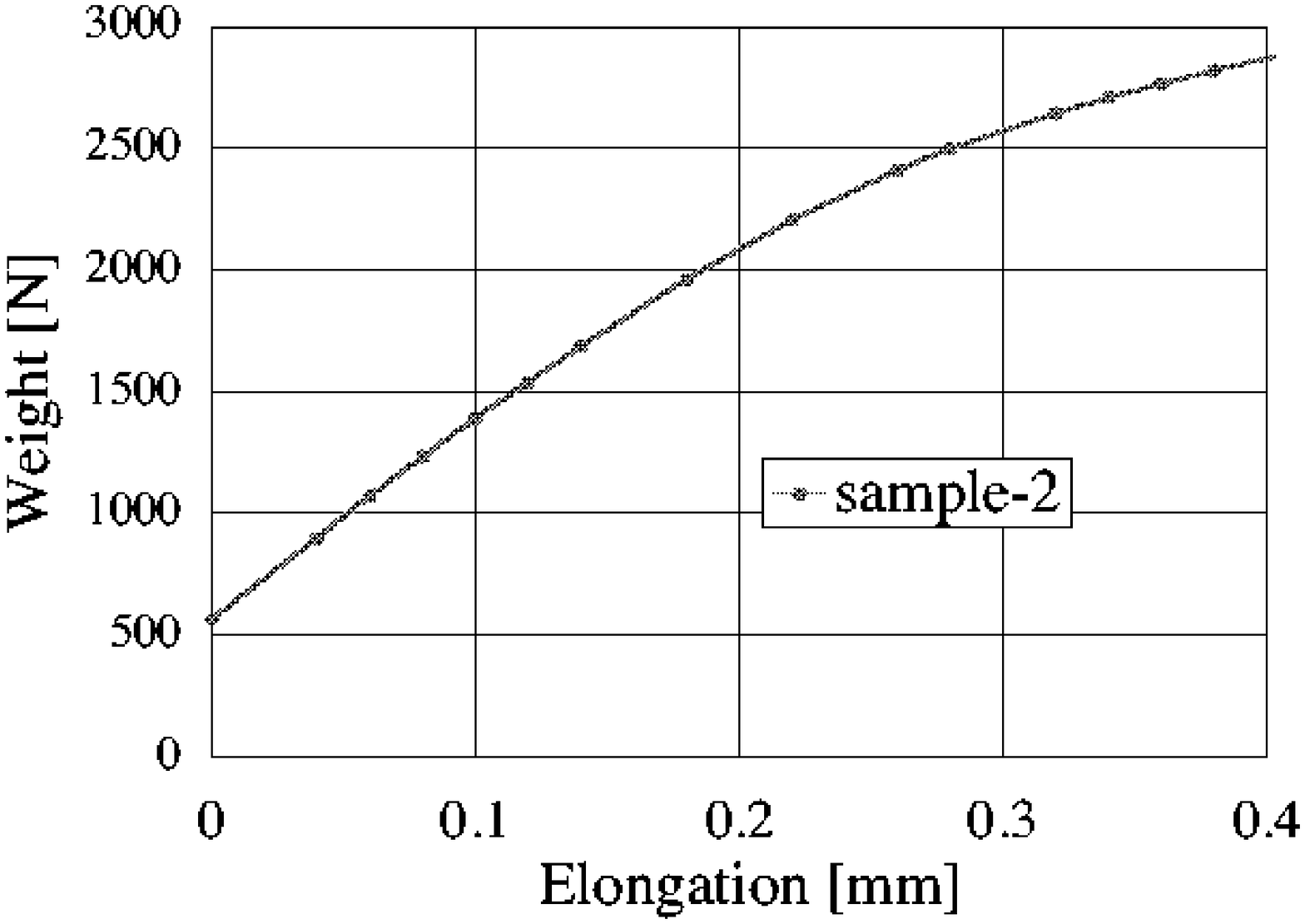}}
\begin{center}\begin{minipage}{\figurewidth}
\caption{\label{PbCa} \sl
Elongation vs. applied weight for a Ca-doped lead alloy.
Left figure shows whole curve which gives yield tensile
strength, and right figure shows rising region which gives
Young modulus.}
\end{minipage}\end{center}
\end{figure}

Lead with Sb or with Ca/Sn are the most popular hard lead alloys.
It is also well known that
heat treatment and/or mechanical treatment improves
mechanical feature for some lead alloys.
In order to study these properties,
test pieces made of Pb(Ca/Sn) and Pb(Sb) alloys 
of several different concentrations have been made,
and measurement of strength, Young's modulus, and creep are
in progress.
Properties of heat treatment and mechanical treatment
are also under study.

Fig.\ref{Pbsetup} and Fig.\ref{PbCa} show the measurement system and 
a result of strength measurement for a Ca-doped lead alloy, respectively.
Tensile strength (yield) of this sample is measured
to be 49MPa, which is 7 times stronger than pure lead.
However measured Young modulus of 13GPa is almost the
same as that of pure lead.

\begin{figure}
\centerline{
\epsfxsize=8cm \epsfbox{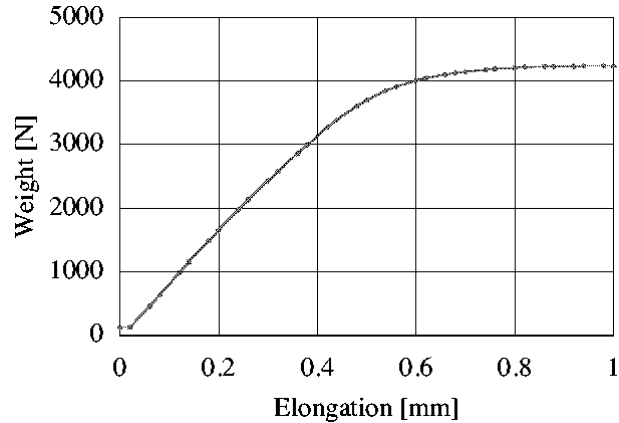}}
\begin{center}\begin{minipage}{\figurewidth}
\caption{\label{PbCaMT} \sl
Elongation vs. applied weight for mechanicaly-treated lead alloy.
Elastic region extends to higher tension region, 
while slope itself (Young modulus) stays the same.}
\end{minipage}\end{center}
\end{figure}

Improvement of tensile strength itself is not an important result
because one can not use material beyond elastic region.
Fig.\ref{PbCaMT} shows how mechanical treatment
improves elastic region of lead alloy.

Studies on heat treatment is in progress.

\vspace{0.3cm}

Feasibilities of other options such as tungsten, copper, stainless
steel or hybrid material such as CFRP-sandwiched lead
are yet open questions. 

\vspace{0.3cm}

\noindent
{\bf Mass-Production of Tiles and WLS Assemblies}

\vspace{0.3cm}

Machining of $\sigma$-grooves on tiles costs
about \$10/tile for fabrication scale of a few thousand pieces.
Since there are 2.2 million tiles in total,
machining cost is significantly high and should be decreased.
This can be solved by making tiles with casting.
However, casting of tiles with $\sigma$-grooves seems very
complicated, though is not thought to be impossible.
On the other hand, the strip-EMC option enables quite easy casting.
Drastic cost reduction is expected by mega-strip casting ;
casting of the strip array as a whole.
This possibility should seriously be pursued.

\vspace{0.3cm}

Cost reduction of WLS assembly is yet an open question.
Direct readout by attached photo-diode,
which is under study as a part of the shower-max detector
R\&D, may enable us to avoid this problem.

\vspace{0.3cm}

\noindent
{\bf Structural Study}

\vspace{0.3cm}

\begin{figure}
\centerline{
\epsfxsize=10cm \epsfbox{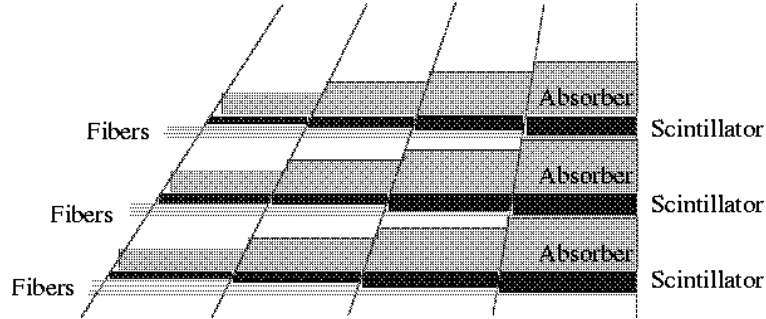}}
\begin{center}\begin{minipage}{\figurewidth}
\caption{\label{CALlayout}\sl
Conceptual layout of constant-sampling barrel calorimeter.}
\end{minipage}\end{center}
\end{figure}

There are two major ways to build the barrel calorimeter assembly.
One is a stack of doughnuts, and the other is an assembly of orange sectors.
Orange sector assembly is a natural solution for z-insertion of mega-tiles. 
In this case, however, absorber plate orientation can not be
perpendicular to the polar angle direction.
Therefore sampling frequency,
and accordingly energy resolution, becomes $\theta$-dependent.
One way to avoid this problem is to make a structure as shown
in Fig.\ref{CALlayout}.
This layout enables constant sampling frequency and at the same time
reserving room for clear fibers, which increases as $\theta$ increases.
However mechanical strength should be a very difficult problem.
Simulation study is needed to estimate the effect of $\theta$-dependent
resolution on physics capabilities.

%% file: detcal/Simulation.tex

A full simulator (named JIM) based on GEANT3 was constructed for
detector parameter optimization and performance study.
Calorimeter geometry and parameters listed in the Table~\ref{CALtable}
are implemented, together with all other detectors and structural
components.
Calorimeter signal normalization, response mapping, linearity and
energy resolution examinations have been done by comparing
with the beam test results. 
Development of hadron-shower clustering algorithm is in progress

\vspace{0.3cm}

\noindent
{\bf Simulator Tuning}

\vspace{0.3cm}

\begin{figure}
\centerline{
\epsfysize=10cm \epsfbox{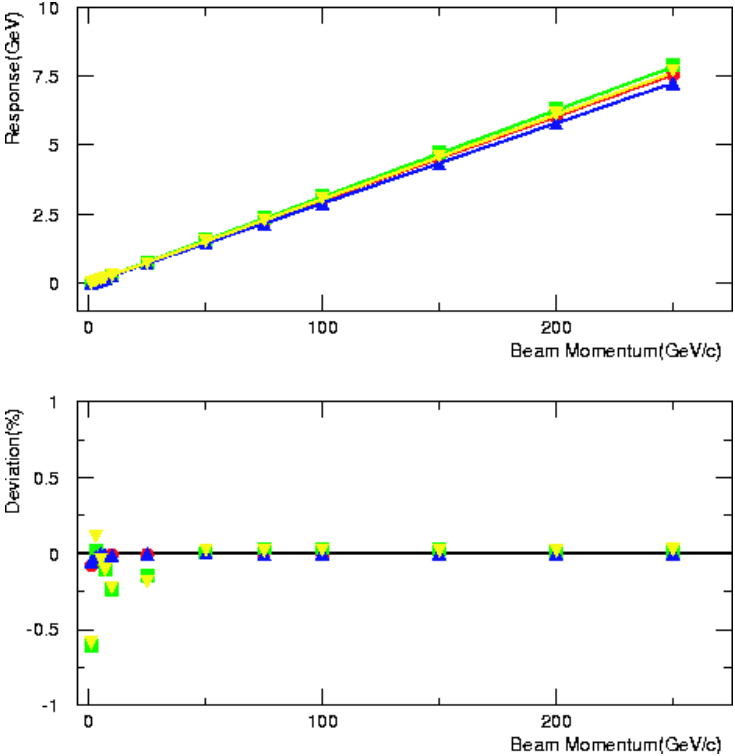}}
\begin{center}\begin{minipage}{\figurewidth}
\caption{\label{BClinear} \sl
Linearity of the calorimeter implemented in a full simulator.
Red, green, blue, and yellow lines indicates barrel CAL response
to electrons, barrel CAL response to pions, endcap CAL response
to electrons, and endcap CAL response to pions, respectively.}
\end{minipage}\end{center}
\end{figure}

Linearity of calorimeter response is shown in Fig.\ref{BClinear}.
It has an excellent linearity of less than 0.3\% from 2GeV up to 250GeV,
while 1GeV data has slightly large deviation probably due to 
cross section of hadronic reaction and low multiplicity of shower
at low energy.

\begin{figure}
\centerline{
\epsfysize=10cm \epsfbox{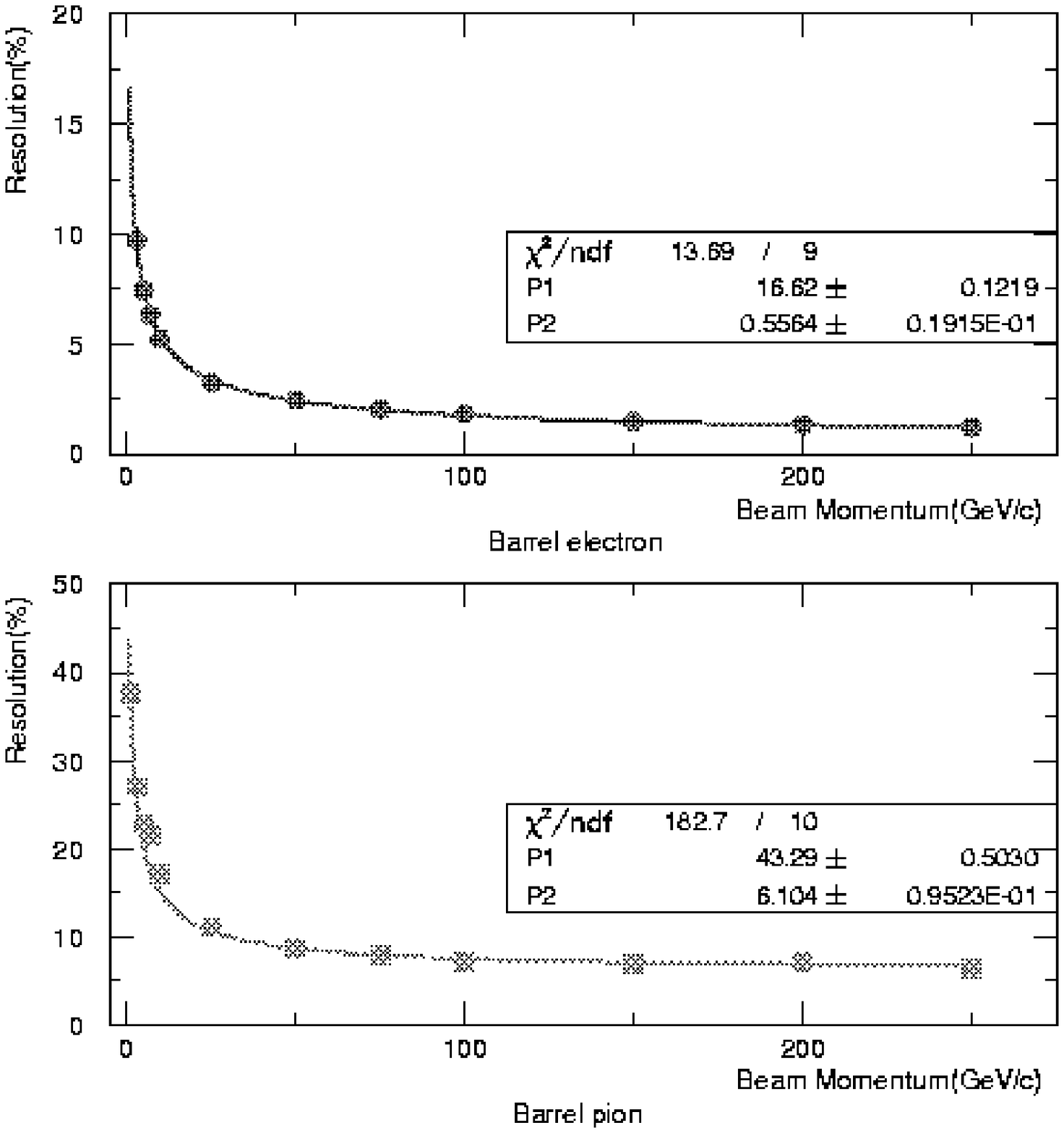}}
\begin{center}\begin{minipage}{\figurewidth}
\caption{\label{BCresolution} \sl
Energy resolution of barrel calorimeter implemented in a full simulator.
Top is for electrons, and bottom is for pions}
\end{minipage}\end{center}
\end{figure}

Fig.\ref{BCresolution} shows full simulation result of
the energy resolution of barrel calorimeter
for electrons and for pions.
Obtained hadron energy resolution of 
$ 43.3\% / \sqrt{E} \oplus 6.1\% $ for pions 
has much worse constant term
than our beam test results shown in Fig.\ref{T912Eres}.
It has already been shown that this is not due to shower 
leakage in the simulation.
More tuning of detailed geometry and/or 
cross section of physics processes are necessary
to reproduce measured results.

\begin{figure}
\centerline{
\epsfxsize=8cm \epsfbox{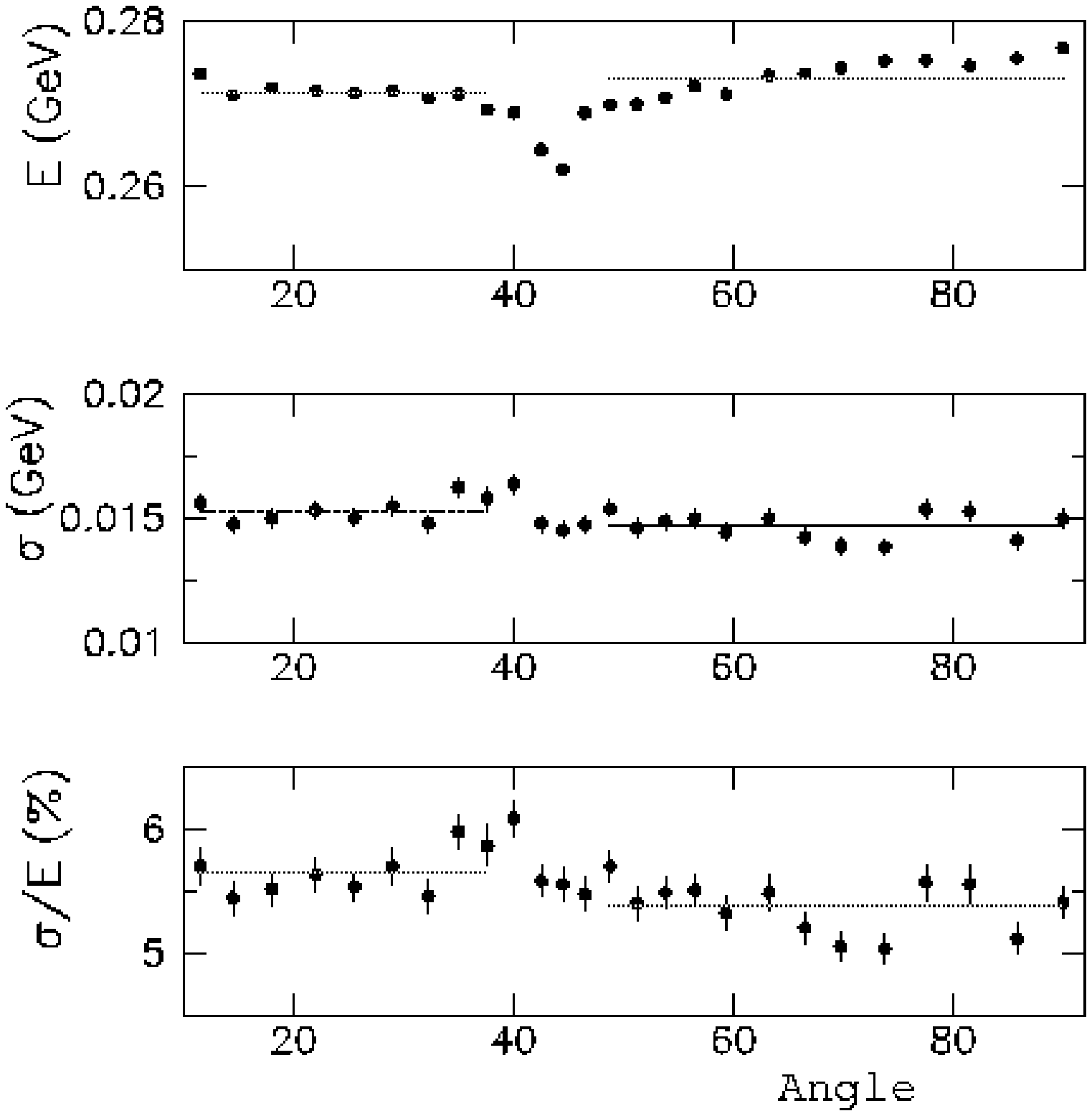}
\hspace{0.1cm}
\epsfxsize=8cm \epsfbox{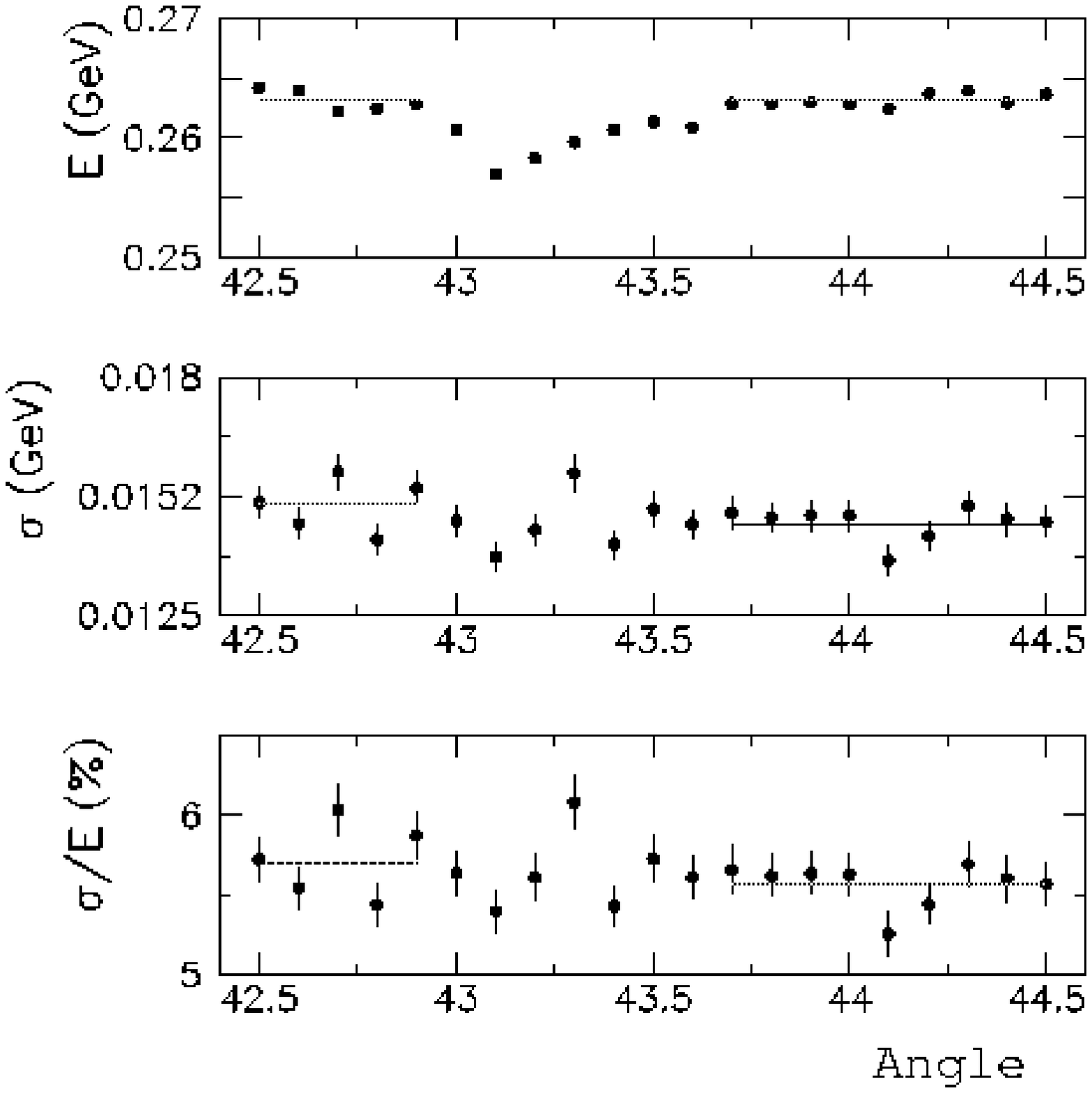}}
\begin{center}\begin{minipage}{\figurewidth}
\caption{\label{CALresponse} \sl
Calorimeter response vs polar angle obtained by
10 GeV electron injection.
Dips in the left and right figures show boundary between the endcap 
and barrel calorimeters and boundary between barrel super-towers,
respectively.}
\end{minipage}\end{center}
\end{figure}

Response mapping was also done using electrons
with respect to the polar angle 
as shown in Fig.\ref{CALresponse},
and to the azimuthal angle.
We observe a dip in energy measurement at the super-tower boundary.
The beam test data, on the other hand, shows enhancement rather than dip
as shown in Fig.\ref{boundary}.
This difference is considered to be caused by two reasons:
1)non-uniformity of photo-electron yield over a tile 
is not implemented into the full simulator, and
2)there exists unexpected gap between supertowers in the geometry definition.
The latter was found by difference of response map between
electron injection and positron injection, and should be fixed.
Impact of non-uniformity on physics analysis should be investigated
to know necessity to implement non-uniformity into the full simulator.

\vspace{0.3cm}

\noindent
{\bf Hadron Shower Clustering}

\vspace{0.3cm}

\begin{figure}
\centerline{
\epsfxsize=8cm \epsfbox{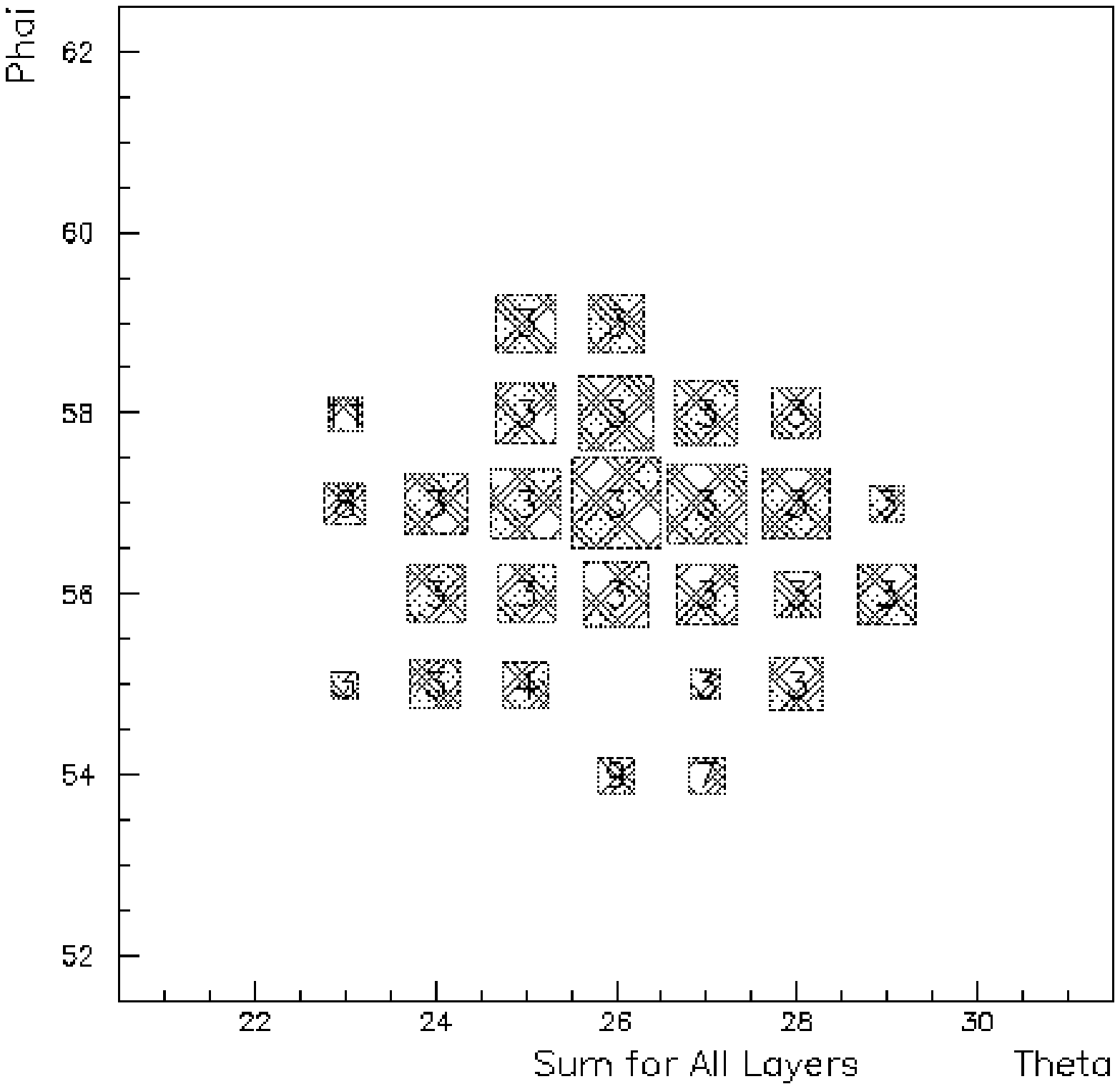}
\hspace{0.1cm}
\epsfxsize=8cm \epsfbox{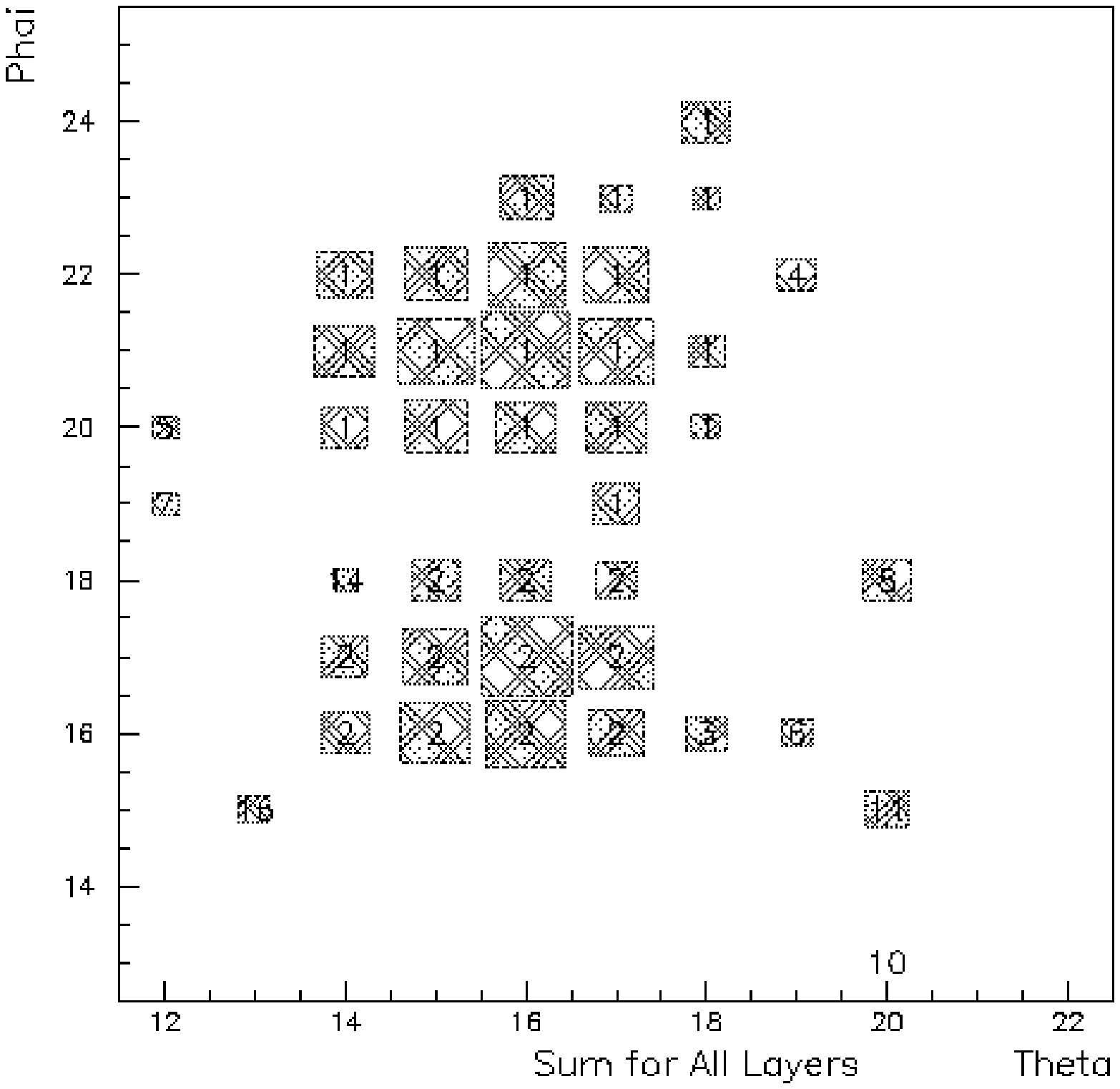}}
\begin{center}\begin{minipage}{\figurewidth}
\caption{\label{CALcmap} \sl
Hit-cell distribution for one-pion injection (left) and for
two-pion injection (right), and results of 3-D contiguous clustering.
One box corresponds to one calorimeter cell, but longitudinal layers
are projected to 2-D.
The size of the boxes correspond to the energy in log-scale.
Overlapped numbers indicate serial cluster number.}
\end{minipage}\end{center}
\end{figure}

Clustering of hadron shower is essentially important for
precise track-cluster association.
Conventional clustering (JADE algorithm) for EM shower
does not work very well for hadron showers because
it tends to split one shower into plural clusters,
and thus leaves excessive neutral energy undeleted.

There are two possibilities to avoid this problem;
one is to make large contiguous cluster with plural
peaks in it, and the other is to make 'super-cluster',
a cluster of single-peak clusters.
The former method with two-dimensional clustering
is very successful for quick-simulation results,
but does not work well for full-simulation data.
Extension of this method to 3-dimensional clustering
is first tried.
The latter, which has been studied by TESLA group,
has not yet been tried.

Fig.\ref{CALcmap} shows how this clustering works
on a hadron shower induced by one 100GeV-pion (left),
and two 100GeV-pions (right).
Boxes indicate a tower, and the sizes represent the energy. 
Numbers overlapped on boxes are cluster ID numbers.
Some contiguous towers have different cluster ID,
which means that individual cells in the towers 
are not contiguous in 3-D space.

\begin{figure}
\centerline{
\epsfxsize=10cm \epsfbox{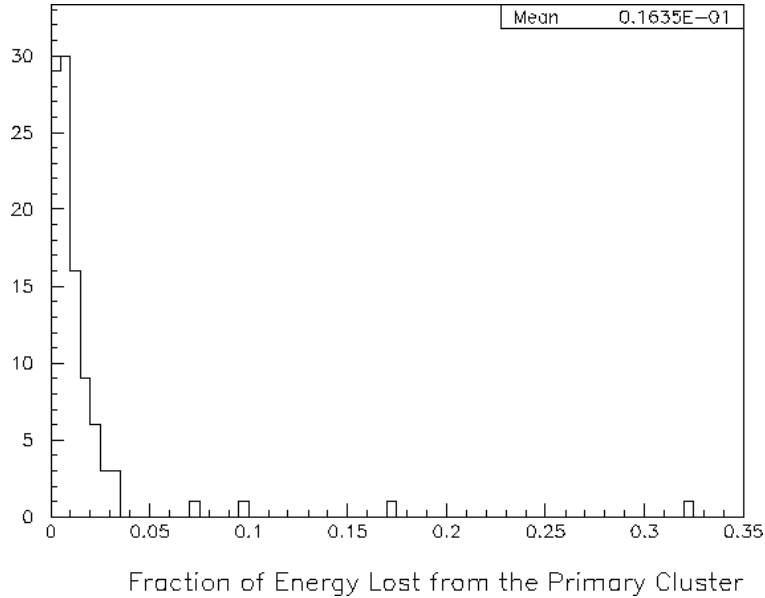}}
\begin{center}\begin{minipage}{\figurewidth}
\caption{\label{CALc1frac} \sl
The fraction of the energy summed over satellite clusters
to the total measured energy.}
\end{minipage}\end{center}
\end{figure}

Several satellite clusters are made around/apart from
the main cluster. 
These satellite cluster are not deleted at the track-cluster
association stage, and thus result in excessive neutral energy.
The fraction of the energy summed over satellite clusters
to the total energy are shown in Fig.\ref{CALc1frac}
for one-pion events.
There are four events with large satellite energy.
However total satellite energy is only 1.6\%
of the total energy even with the four events,
and most of the energy are collected by the main cluster.

\begin{figure}
\centerline{
\epsfxsize=10cm \epsfbox{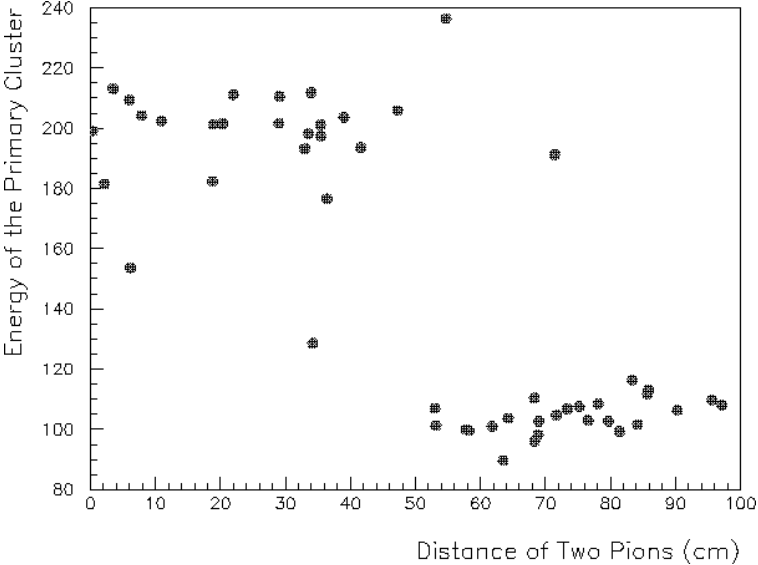}}
\begin{center}\begin{minipage}{\figurewidth}
\caption{\label{2pidist} \sl
Energy of the primary cluster as a function of distance
between two pions of 100GeV.
200GeV means two pions formed one connected cluster.}
\end{minipage}\end{center}
\end{figure}

Right figure of Fig.\ref{CALcmap} shows merit of contiguity
check in 3-D space.
Two clusters looks contiguous in 2-D projection,
but are not in 3-D space,
and thus correct clustering is performed.
Figure \ref{2pidist} shows separation capability
of two pion clusters made by 100GeV pions.
Two clusters can be reconstructed separately down to about 
50cm-distance, which corresponds to four hadron cells.
It means at least one vacant cell is needed between
two $3 \times 3$ clusters.
Thus present score is almost at the geometrical limit.

Two pions with less distance mostly result in one cluster
in this 3-D contiguous clustering.
In this case, existence of two pions must be identified
by shower-max detectors,
and energy decomposition may be done
by momentum-energy subtraction.
This, however, gives worse energy resolution
than geometrical cluster decomposition.
To avoid this deterioration,
conventional JADE algorithm can be tried only
for such cases.
This method is yet under development.

Study of effect on physics sensitivity be forthcoming.

%% file: detcal/Future.tex

Results of quick simulation for various physics processes
have demonstrated that the baseline calorimeter system
can realize excellent physics sensitivity.
However, validation with a full simulation is essentially important, 
especially for optimization of granularity.
This must be completed urgently.

Extensive hardware studies have proven the technical feasibility
and performance of tile/fiber scheme with hardware compensation.
There are still open questions in high-granularity EMC option 
and in photon detector options.
These issues are also related to the total cost,
and thus should be solved before moving to the proto-type R\&D stage.
Engineering studies including the heavy metal choice should also be
finished before the proto-type stage.

%% file: detmuon/main.tex
\chapter{Muon detector}
\label{chapter-detmuon}

\section{Introduction}

The muon momentum can be precisely measured 
by the inner tracking detectors.
Therefore the muon detector should only have
many position measurements
to obtain good matching with the tracks,
where modest position resolution is required.
Modest timing resolution is also required 
to reject cosmic ray backgrounds.
As the requirements are much looser than
those for the LHC experiments,
we think that we can construct the muon detector 
with existing detector technologies.
We may apply new technologies developed for LHC
if required.

\begin{figure}[htbh]
\centerline{
\epsfxsize=10cm
\epsfclipon
\epsfbox{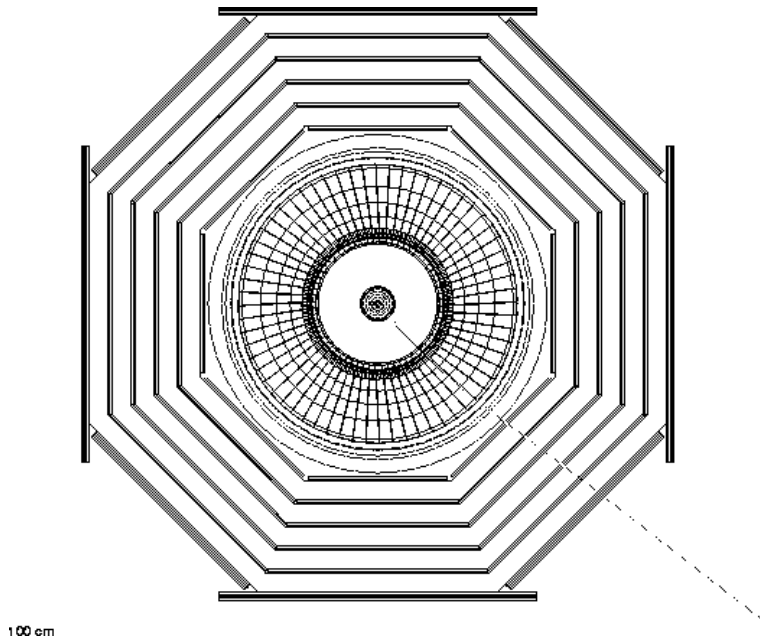}
}
\caption{\sl Layout of the muon detector.
\label{detmuon:layout}}
\end{figure}

Fig.~\ref{detmuon:layout} is
a cross sectional view of the standard detector,
where the muon detector layout is shown.
In the barrel region 
the solenoid magnet is surrounded by one layer of plastic scintillator
counters, followed by six muon detector layers;
one layer just inside the return yoke, four layers in the return yoke,
and one layer just outside the return yoke.
The endcap muon detectors have a similar detector configuration.
The total area of the muon layers
amounts to about 4,000~m$^2$.

\section{Material Effect}

We studied the effect of the detector material to the muon detector.
Table~\ref{tab:thickness} summarizes 
the amount of material of the standard JLC detector
in front of the return yoke 
and in the return yoke itself at $\theta = 90^{\circ}$.
The muon momentum is required to be more than 2.2~GeV
to reach the return yoke,
and to be more than 5.2~GeV to fully penetrate the return yoke.

\begin{table}[htbh]
\caption{\sl Material thickness of the detector at $\theta = 90^{\circ}$. 
Those in front of the return yoke, in the return yoke itself, 
and the total thickness are listed. 
\label{tab:thickness}}
\vspace{0.4cm}
\begin{center}
\begin{tabular}{|c|c|c|c|} \hline
&In front & Return Yoke & Total \\ \hline
Weight & 
1.09 kg/cm$^2$ & 1.51 kg/cm$^2$ & 2.60 kg/cm$^2$ \\ \hline
Radiation length & 
158 $X_0$ & 109 $X_0$ & 267 $X_0$ \\ \hline
Interaction length &
5.6 $\lambda_0$ & 11.2 $\lambda_0$ & 16.8 $\lambda_0$ \\ \hline
\end{tabular}
\end{center}
\end{table}

Multiple scattering due to the material in front of the muon detector
is simulated by JIM~\cite{detmuon:JIM}, 
the full simulation program of the JLC detector.
In the simulation, 
muons are generated at the interaction point with a momentum of
$(p_x , 0, 0)$ where the $+z$ direction is along the electron beam.
The hit positions 
at the inner surface
of the return yoke ($x=500$~cm) are recorded.

\begin{figure}[tbh]
\centerline{
\epsfxsize=11cm
\epsfbox{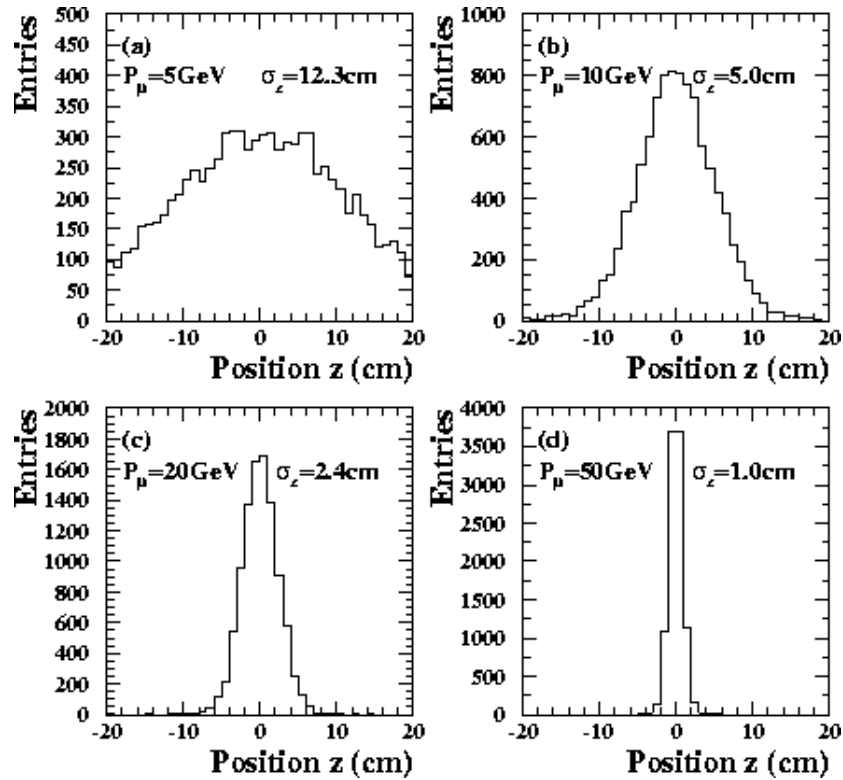}
}
\begin{center}\begin{minipage}{\figurewidth}
\caption{\sl Distributions of the $z$-coordinate
at the return yoke surface 
(a)~$p_{\mu}=5$~GeV,
(b)~$p_{\mu}=10$~GeV,
(c)~$p_{\mu}=20$~GeV,
and (d)~$p_{\mu}=50$~GeV.
\label{detmuon:ms_plot}}
\end{minipage}\end{center}
\end{figure}

\begin{figure}[htbh]
\centerline{
\epsfxsize=12cm
\epsfbox{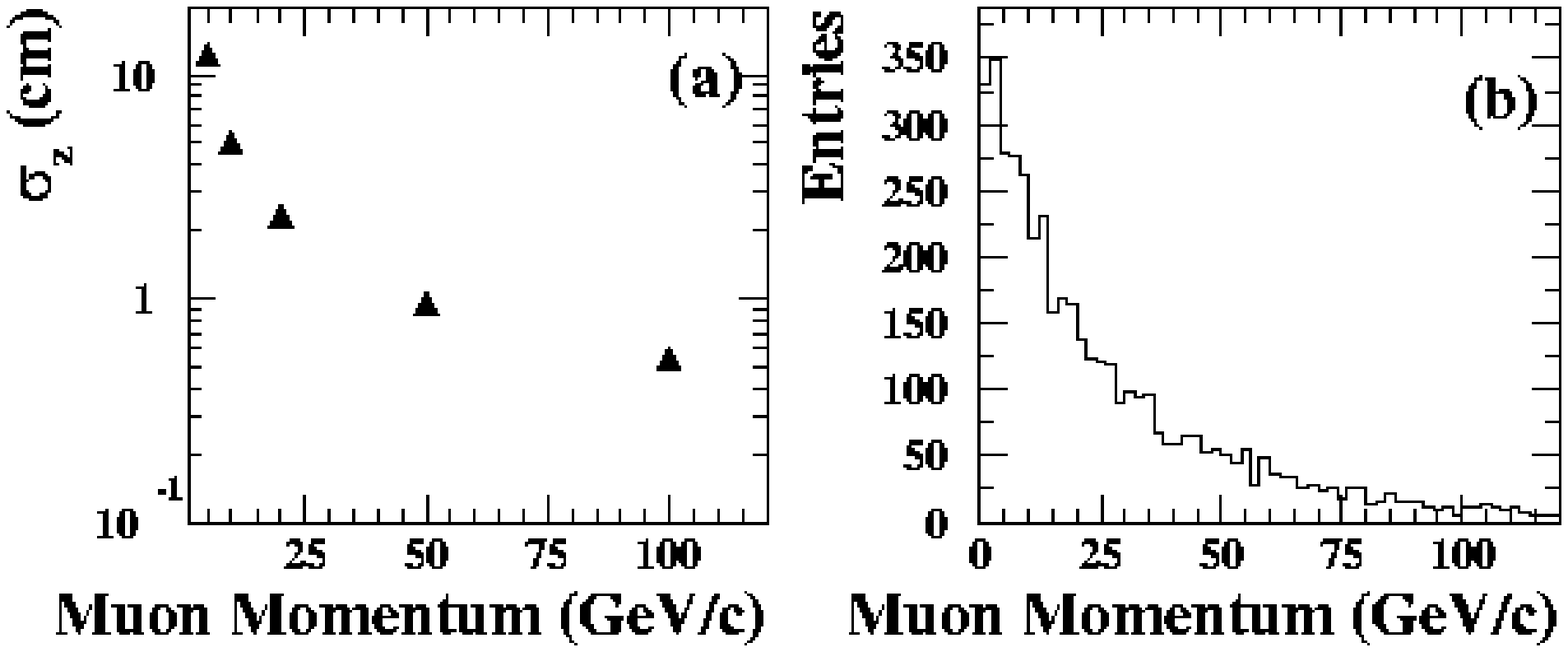}
}
\begin{center}\begin{minipage}{\figurewidth}
\caption{\sl
(a) The standard deviation of the $z$-coordinate at the return yoke surface
as a function of the muon momentum,
and (b) the distribution of the muon momentum in the $b$-quark pair
production at ${s}=500$~GeV.
\label{detmuon:ms_func}}
\end{minipage}\end{center}
\end{figure}

Fig.~\ref{detmuon:ms_plot} shows the distributions of the $z$-coordinate 
for muons with initial momentum of 5, 10, 20, and 50~GeV.
The distributions of the $y$-coordinate
are very similar to those of the $z$-coordinate.
Fig.~\ref{detmuon:ms_func}(a) shows 
the standard deviation of the $z$-coordinate ($\sigma_z$) 
due to multiple scattering
as a function of the muon momentum.
This result is similar to that obtained for the TESLA detector
\cite{detmuon:Tesla}.
Fig.~\ref{detmuon:ms_func}(b) shows the momentum distribution
of muons
in the $b$-quark pair production at $\sqrt{s}=500$~GeV.
As seen in this histogram the majority of muons have momentum below
50~GeV, for which $\sigma_z$ is larger than 1~cm.
We thus conclude 
that the position resolution of about 1~cm is sufficient
for the muon tracking device.

\section{Detector Options}

The JLC muon detector 
should have position resolution of 1~cm or better,
and should be constructed with a well-established technology.
In addition, as the muon detector covers a very large area,
the technology must be inexpensive.
Single-cell drift chambers, resistive plate chambers,
and thin gap chambers are currently considered as the candidates.

\subsection{Single Cell Drift Chambers}

The first candidate, single-cell drift chambers (SCDCs),
is described in the JLC-I report~\cite{detmuon:JLC-I}.
A SCDC super-layer has four layers with xx'yy' wire configuration
to measure both $\phi$ and $z$ coordinates.
The cell size of the SCDC is $10 \times 5$~cm$^2$,
and the wire length is $10 \sim 15$~m.
A wire support is located at the middle of the wire
to reduce the wire sag.
The expected position resolution is about 500~$\mu$m
dominated by the gravitational wire sag.
Once the hit coordinate along the wire is known,
the position resolution can be much improved.
The number of readout channels is relatively small,
about 10,000 in total.

\subsection{Resistive Plate Chambers}

The second candidate is resistive plate chambers (RPCs).
Large area RPC systems are used 
by experiments at B-Factories 
\cite{detmuon:Belle,detmuon:BaBar},
and will also be used by experiments at LHC 
\cite{detmuon:ATLAS,detmuon:CMS}.
Such a large area RPC system can be constructed 
with a very low cost,
because of their simple structure with no wires.
The RPC can be very thin, for example,
the total thickness of an RPC doublet
for the BELLE KLM detector is only 3.2~cm.
The position resolution is determined 
by the dimensions of readout cathode strips, 
and a resolution of 1~cm can easily be achieved.
RPC is a very promising candidate for the JLC muon detector.

\subsection{Thin Gap Chambers}

The third candidate is thin gap chambers (TGCs),
a kind of MWPC with a very thin gap ($\sim 3$~mm).
TGC is highly efficient except for the dead space due to wire supports.
The signal is fast enough for our purpose.
The position resolution of 1~cm is easily achieved,
similar to the case of RPC.
TGC can be operational at a very high particle rate of 1~kHz/cm$^2$.
The only demerit is the
higher construction cost than that of RPC,
as TGC uses a huge number of wires.

The endcap muon trigger system for the ATLAS experiment
uses the TGC technology~\cite{detmuon:ATLAS}.
To this purpose intensive R\&D works 
were performed in Israel and Japan.
Production of a total of 1,000 large TGC doublets/triplets 
are in progress at KEK.
According to the ATLAS TGC schedule,
the production will be completed in 2003, and
then installed at the LHC experimental hall by the end of 2004.
Through the TGC production we will at least gain an excellent experience
to construct a large area muon detector system.

\section{Summary}

A simulation study is made to investigate the effect of the
detector material to the muon detector.
The position resolution required for the muon tracking device
is found to be about 1~cm. 

The detector technology is still to be determined,
We have currently three possible candidates for the muon tracking device;
single-cell drift chambers, resistive plate chambers,
and thin gap chambers.
The technology will be determined by further simulation studies,
together with experiences obtained at B-factories and LHC.

%% file: detmag/main.tex
\chapter{Detector Magnet}
\label{chapter-detmag}
\section{Superconducting coil}
\input detmag/coil.tex
\section{Iron structure}
\input detmag/iron.tex

%% file: detmag/coil.tex
\begin{figure}[hptb]
\centerline{\epsfxsize=12cm \epsfbox{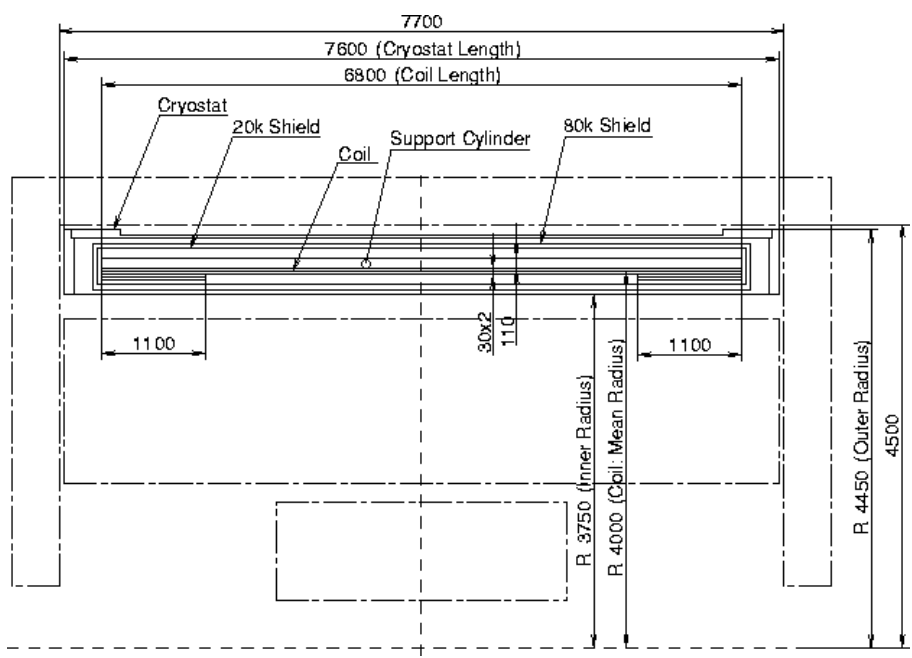}}   
\caption{\sl Configuration of the JLC superconducting solenoid magnet.
\label{solenoid} }
\end{figure}

A superconducting solenoid magnet has been studied for the JLC detector. The JLC superconducting solenoid magnet is shown in Fig.~\ref{solenoid}. The magnet provides a central magnetic field of 3T at nominal current of 4725A in a cylindrical volume of 8m in diameter and 6.8m in length. It placed outside the calorimeter to achieve good hermetic. The coil consists of double layer aluminum-stabilized superconductor wound around the inner surface of an aluminum support cylinder made of JIS-A5083. At the both of the end part, four layers of coil will be wounded to improve field uniformity in the CDC volume. Length of the end part is 1.1m. Indirect cooling will be provided by liquid helium circulating through a single tube welded on the outer surface of the support cylinder. Since the magnet is located outside of the calorimeter, it is not necessary to design as a thin superconducting solenoid magnet. Therefore, a thickness of outer/inner wall of cryostat can be thick as much as sub-detectors are necessary.

\begin{figure}[hptb]
\centerline{\epsfxsize=12cm \epsfbox{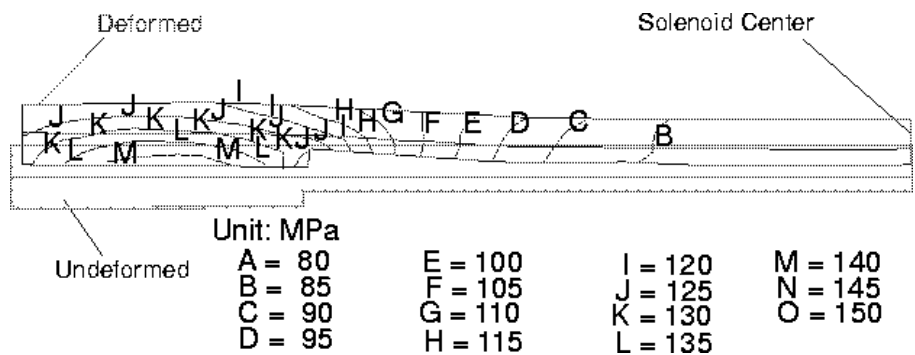}}   
\caption{\sl Stress and deformation shape due to magnetic force in the coil. 
\label{solenoid-stress}}
\end{figure}

It was considered about a superconducting wire for the JLC solenoid based on the ATLAS superconducting wire. The size is 4.3mm width and 30mm height. On design of a superconducting solenoid, it is required to be reliable two main mechanical stresses, hoop stress and axial stress in a coil. The evaluation is done by calculated combined stress. Even though the yield strength of superconducting wire is achieved to around 250MPa at recent R\&D of high strength superconducting wire, it assumed to yield strength of 150MPa in our design. The combined stress, Von Mises, in the coil when the central magnetic field of 3T is shown in Fig.~\ref{solenoid-stress}. Since the calculation result, required thicknesses to keep within yield strength in the coil are to be 110mm for support cylinder and 60mm for coil, respectively. From these required thickness, the cold mass is approximately to be 95tons. The stored energy is calculated to be 985MJ.

\begin{figure}[hptb]
\centerline{\epsfxsize=12cm \epsfbox{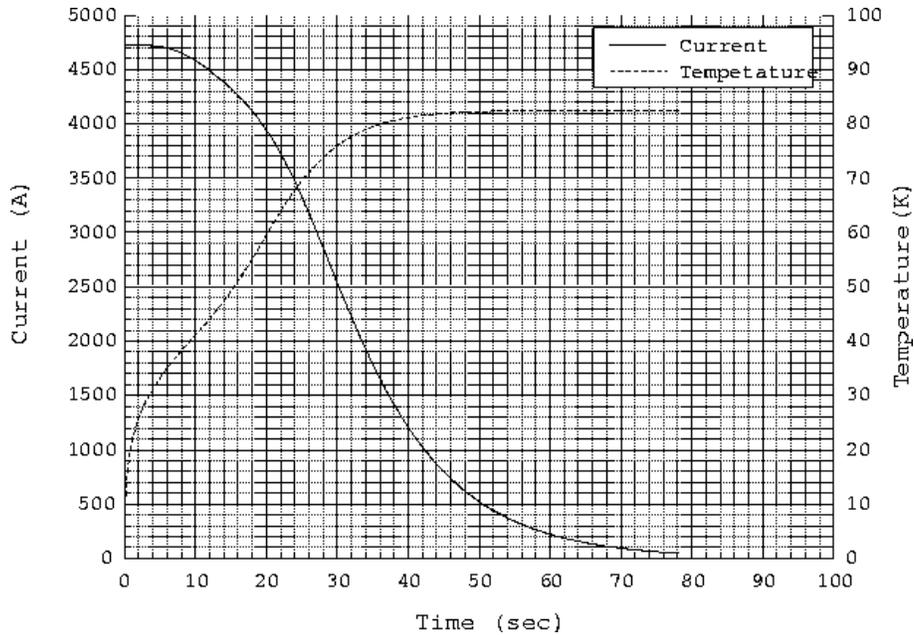}}   
\caption{\sl Quench behavior of the coil with pure aluminum strips. 
\label{solenoid-quench}}
\end{figure}

To protect the coil due to the magnet quench is one of the most important issues for solenoid design. Figure~\ref{solenoid-quench} shows the temperature rise and current decay after magnet quench in case of mounting pure aluminum strips. In this calculation, total magnetic energy is dumped entirely into the superconducting wire, and it assumed that the pure aluminum strips are mounted on the inner/outer surface of the solenoid. By mounting pure aluminum strips, the quench propagation velocity can be increased drastically. As the calculation results, the maximum temperature rise after quench is to be around 80K, and the magnet current is decayed within 80 second from the nominal current of 4725A.

Radiation shields consist of 20K-shield and 80K-shield are placed between the coil and the vacuum vessel. These shields have to be decoupled electrically from both the coil and the vessel walls in order to avoid the effects of the eddy current induced by the fast current discharge of the coil. The vacuum vessel for this solenoid magnet consists of inner and outer coaxial cylinders that are connected by flat annular bulkheads at each end. The solenoid magnet will be supported at the innermost layers of barrel iron yoke by fixing the support structure on the outer vacuum vessel. The coil support system has to be transmitting both the weight of cold mass and the magnetic de-centering forces. If the geometrical center of the solenoid and the iron yoke are deviated each other, the magnetic force is generated. This force has to be taken into account on the coil support design.

%% file: detmag/iron.tex
\subsection{Introduction}

\begin{figure}[hptb]
\centerline{\epsfxsize=12cm \epsfbox{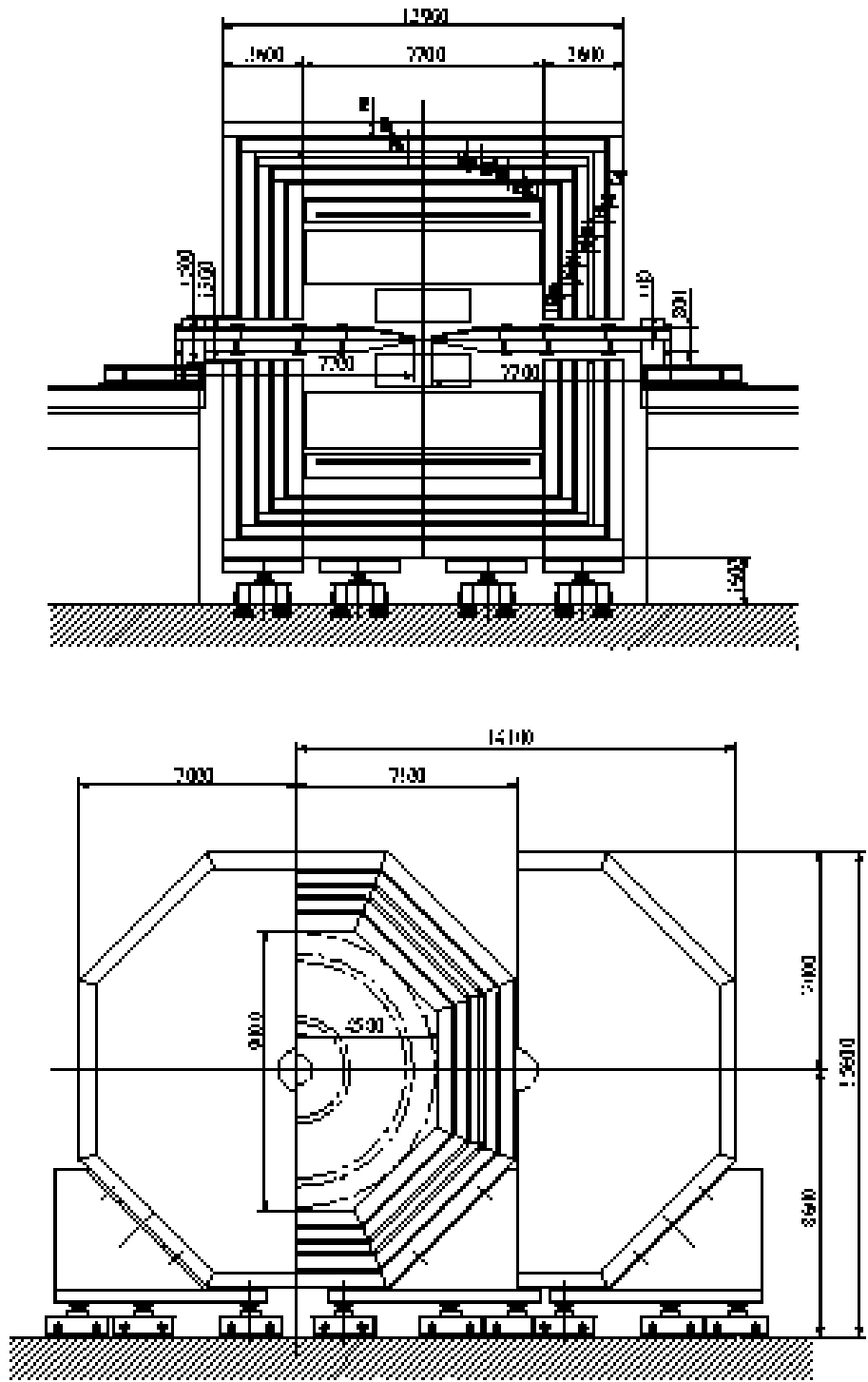}}   
\caption{\sl Basic configuration of the JLC iron yoke.\label{iron-yoke} }
\end{figure}

An iron structure for the JLC detector has been studied and designed. Figure~\ref{iron-yoke} shows the basic configuration of the JLC iron yoke structure. The structure consists of the barrel and two end-yoke sections. The overall height is about 16m from the floor level; the depth is about 13m. The iron structure will be made from low-carbon steel (JIS-S10C: 0.008 Ð 0.012 wt\% carbon). The total weight except for sub-detectors is approximately 11,000tons, and it will be placed on a transportation system for roll-in/out. Assembling of the iron yoke and installation of the sub-detector will be carried out at the roll-out position and then move to the roll-in position for the experiment. The end-yoke separates at its center and then opens to access to the sub-detectors.

Some kinds of design studies are required to optimize the iron yoke configuration. Those are the influence of the magnetic field, stability against the acted forces such as self-weight of iron yoke (mechanical design) and assembling/maintenance consideration. In the magnetic field study, the iron yoke is acted on the magnetic field of 3T from the JLC solenoid magnet, amount of iron have to be determined for improving the field uniformity in the CDC volume, and leakage field outside of the iron yoke is kept minimize by absorbing the flux return. In the mechanical design, it is required to study on the deformation and stress level against heavy self-weight of iron yoke and strong magnetic force. As the other important issue for the mechanical design, an earthquake resistant design is required. In these studies, way of iron yoke assembling, access to the sub-detectors and cables pass have to be taken into account.

\subsection{Magnetic Field Design}

\begin{figure}[hptb]
\begin{minipage}[t]{7.75cm}
\centerline{\epsfxsize=7.75cm \epsfbox{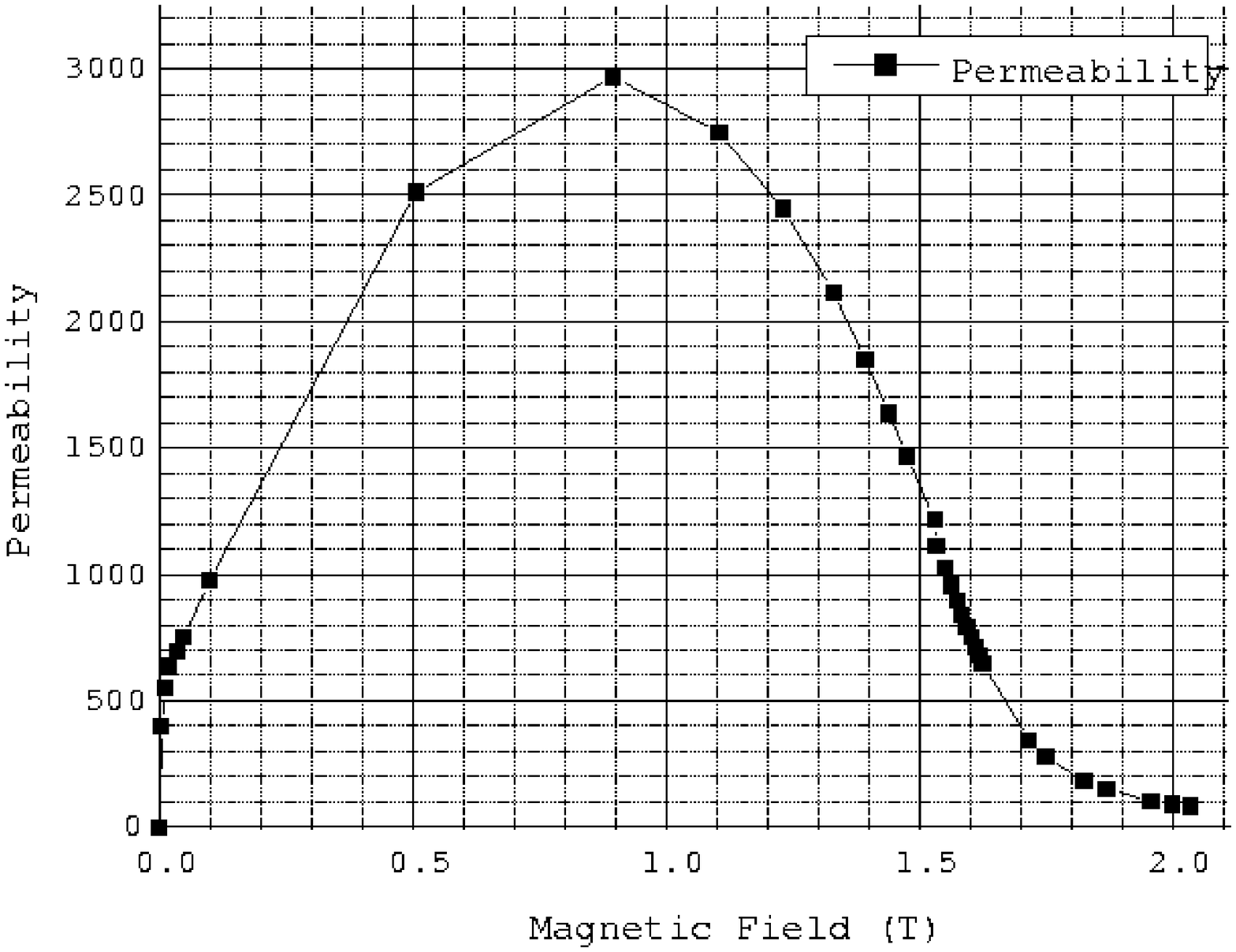}}
\caption{\sl Permeability of iron material (JIS-S10C), which was used for the magnetic field calculation.\label{permeability}}
\end{minipage}
\hfill
\begin{minipage}[t]{7.75cm}
\centerline{\epsfxsize=7.75cm \epsfbox{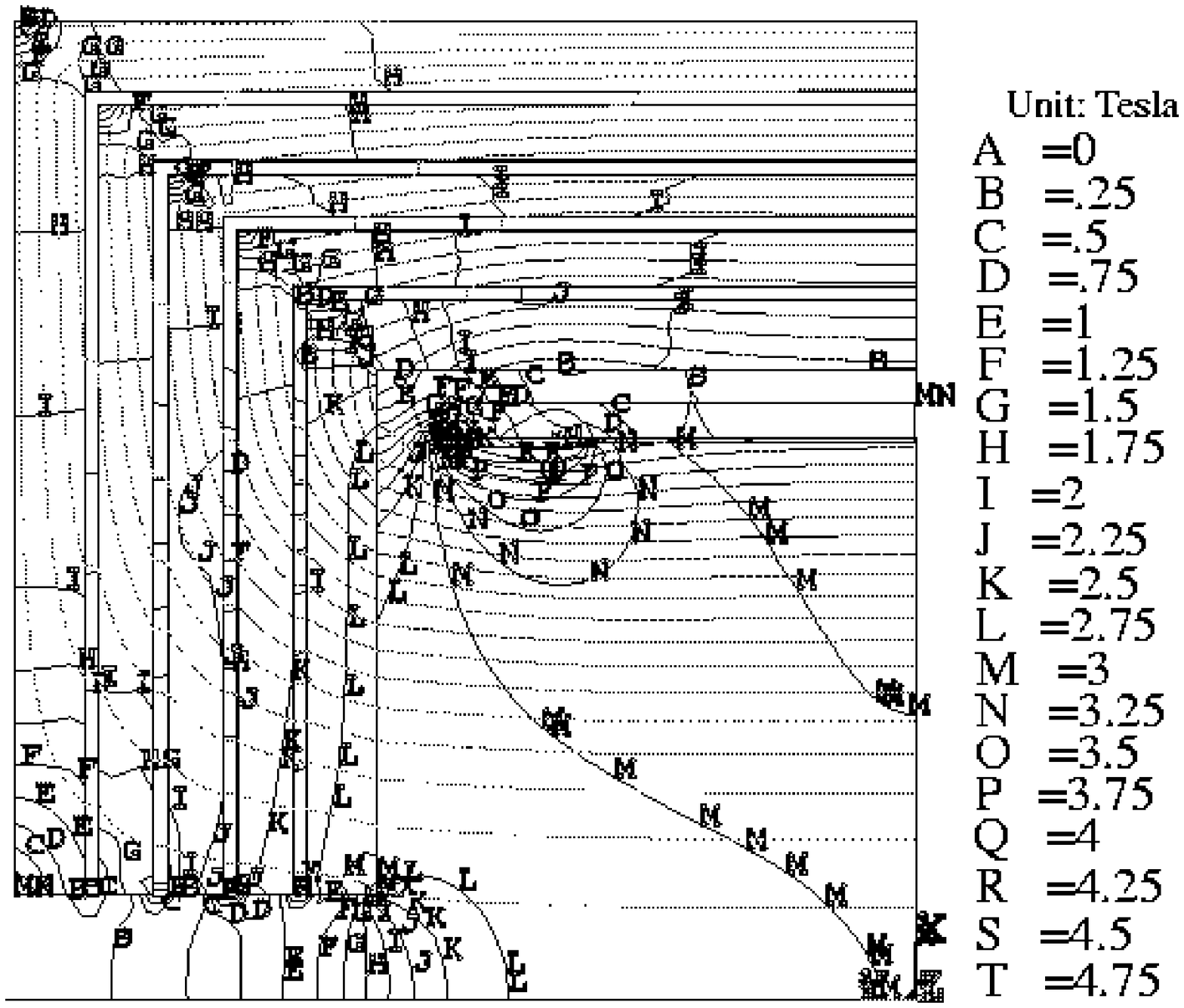}}   
\caption{\sl Calculated magnetic field at the central magnetic field of B=3~T.\label{magnet3t}}
\end{minipage}
\end{figure}

\begin{figure}[hptb]
\begin{minipage}[t]{7.75cm}
\centerline{\epsfxsize=7.75cm \epsfbox{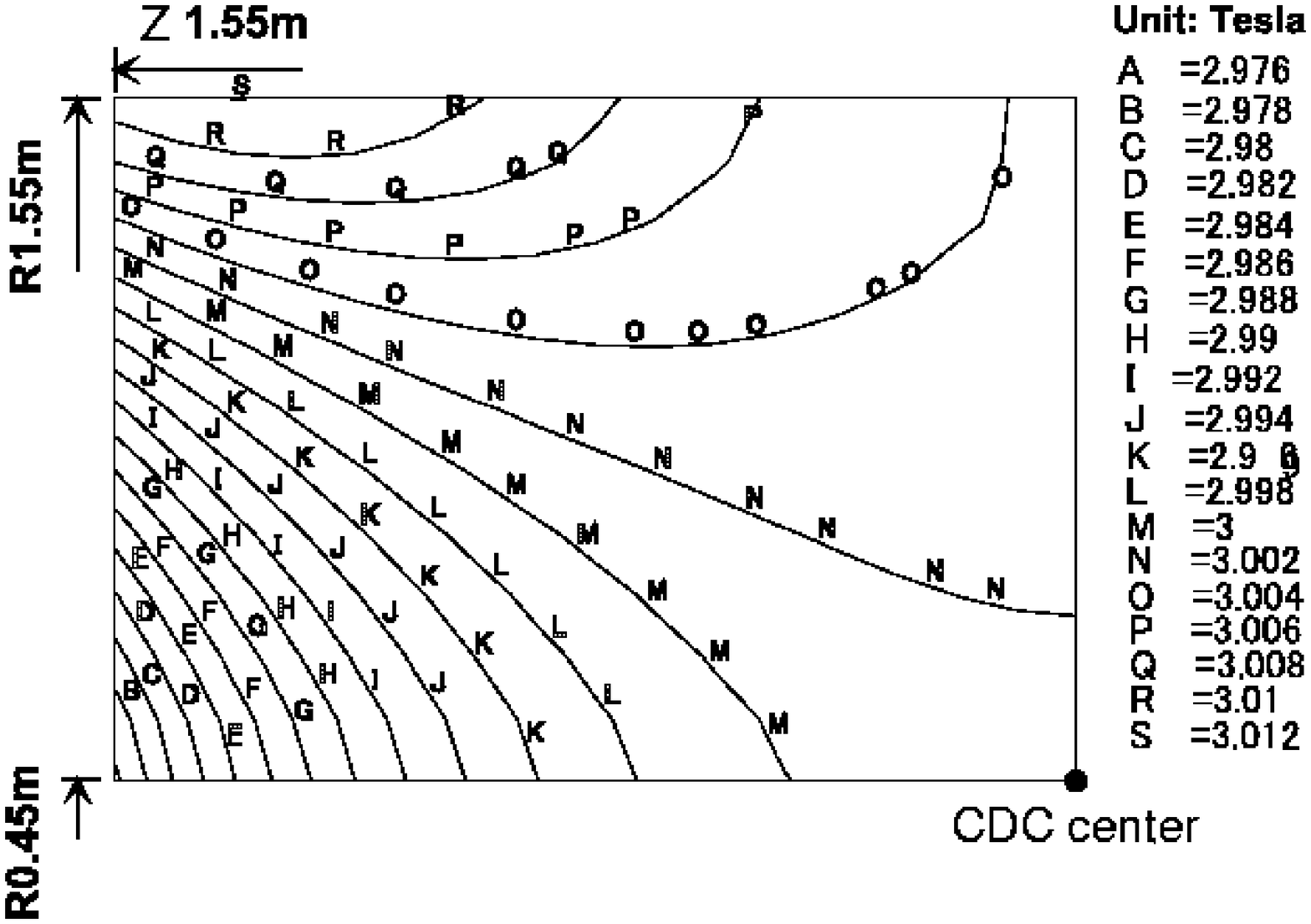}} 
\caption{\sl Field uniformity in the CDC volume at B=3T. The uniformity is simply calculated from the ratio of the minimum and maximum fields. \label{field-uniformity}}
\end{minipage}
\hfill
\begin{minipage}[t]{7.75cm}
\centerline{\epsfxsize=7.75cm \epsfbox{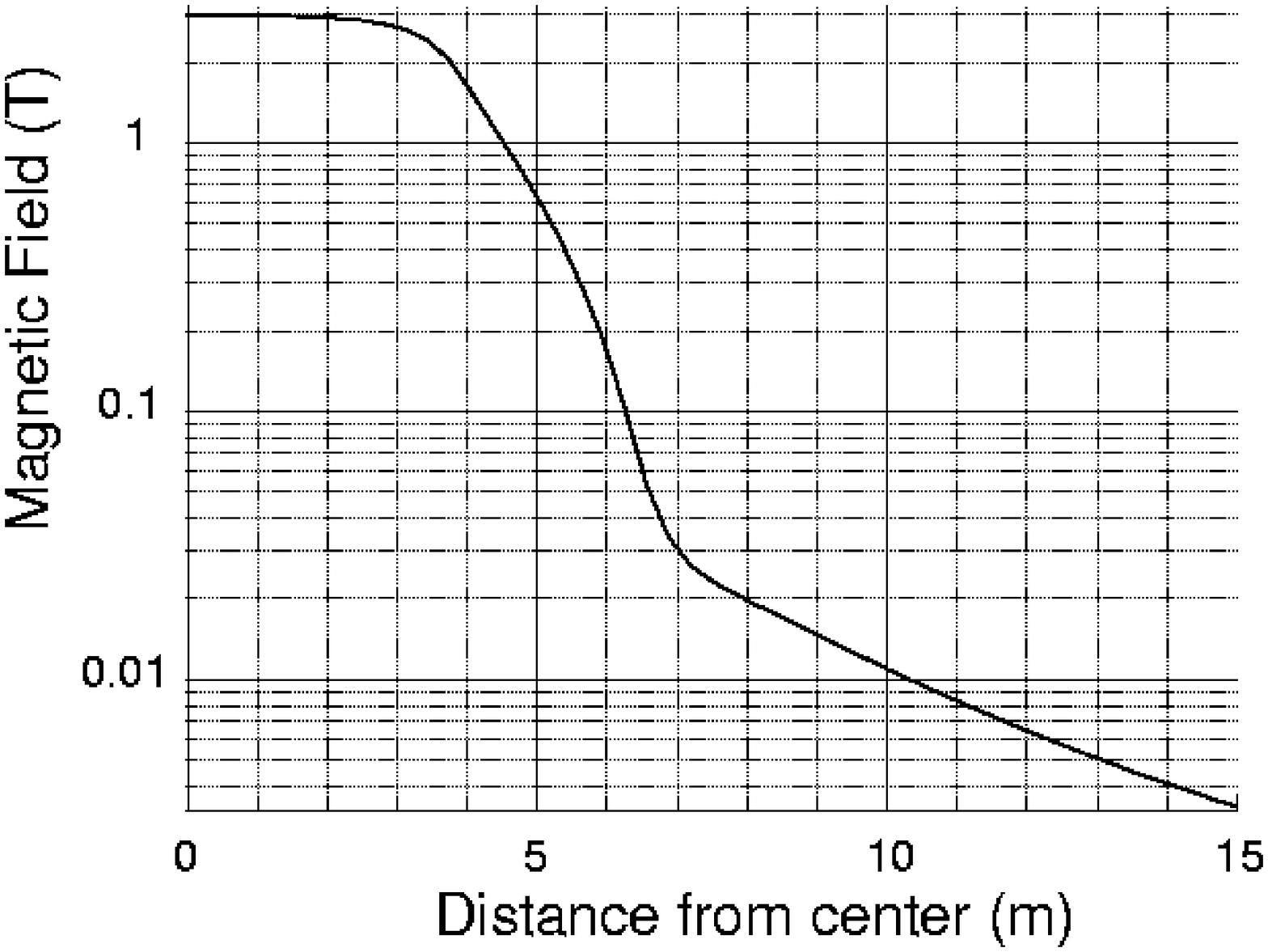}}
\caption{\sl The magnetic field along the beam axis at B=3~T. 
\label{field-z} }
\end{minipage}
\end{figure}

The criteria for the magnetic design to determine the iron yoke structure are around field uniformity of 1\% in the CDC volume, and an allowable leakage field at 10m far from the solenoid center is less than 100 gauss. To set the tolerance against leakage field, an important criterion is to work an electromagnetic valve. With keeping these requirements, thickness of the iron plates were tried to minimize as possible. Permeability of iron plate for calculation is shown in Fig.~\ref{permeability}. This is the measurement data of the BELLE iron yoke, which is the same material as the planning JLC iron yoke. The ways to improve the field uniformity without increasing the amount of iron are coil thickness at the end of the solenoid is thicker, and the inner diameter of the end-yoke is smaller. The results of magnetic field calculation when the central magnetic field of 3T is shown in Fig.~\ref{magnet3t}. In this calculation, the end part of the coil is two times thicker than the other part. The maximum magnetic field in the iron yoke is to be 3.3T. The field uniformity is to be 1.2\% as show in Fig.~\ref{field-uniformity}. The magnetic field along the beam line is shown in Fig.~\ref{field-z}. The leakage field was calculated to be 110 gauss.

From this calculation, iron plate thicknesses of barrel and end-yoke were optimized. Innermost and outermost layers of yokes are understood to require a thicker iron plate of 50cm-thick than other layers plates of 30cm-thick and 40cm-thick.

\subsection{Mechanical Design}

\begin{figure}[hptb]
\begin{minipage}[t]{8.25cm}
\centerline{\epsfxsize=8.25cm \epsfbox{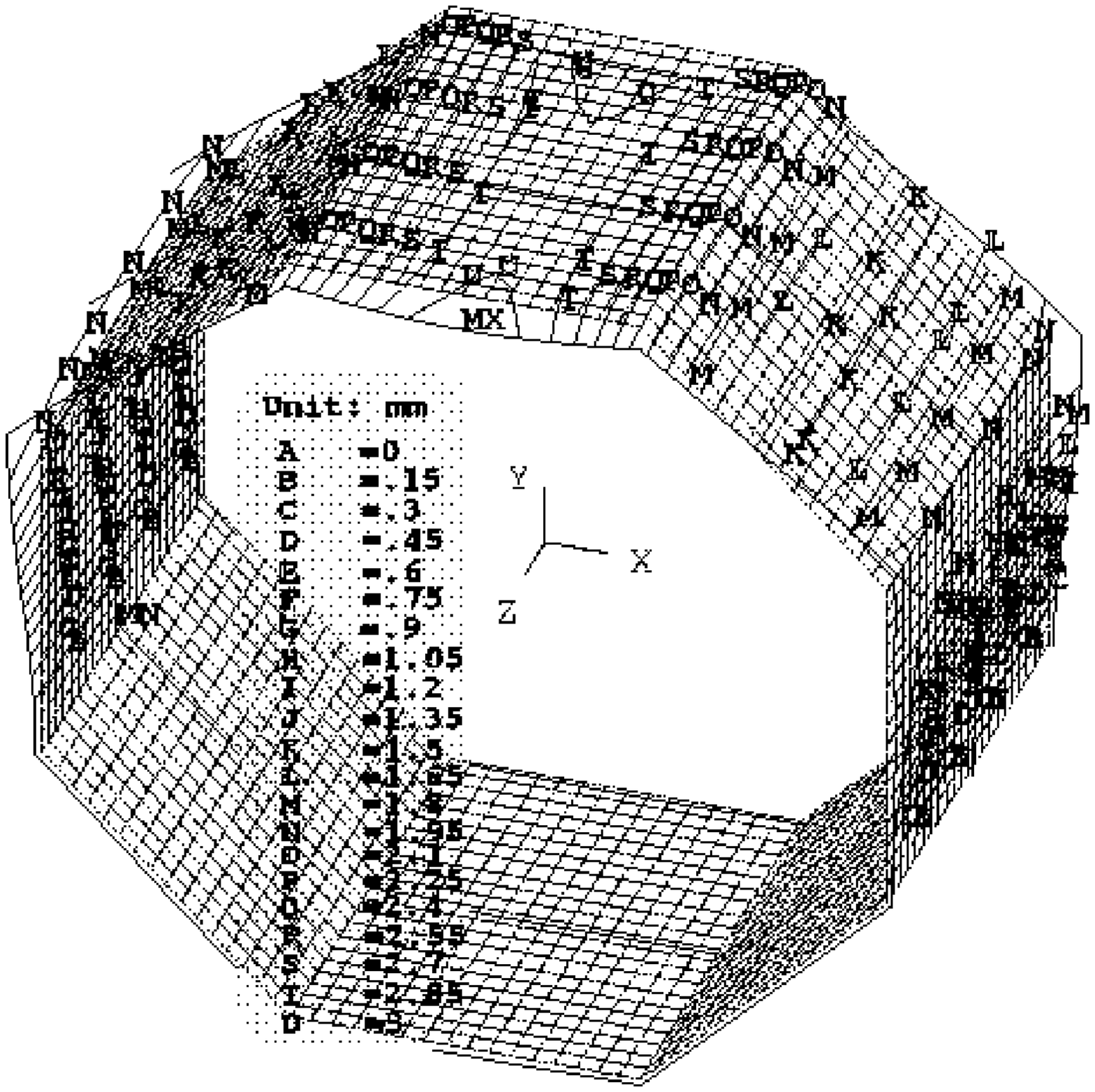}}
\caption{\sl Deformation of the outermost layer of the barrel-yoke due to self-weight of 1,352 tons. 
\label{self-deformation}}
\end{minipage}
\hfill
\begin{minipage}[t]{7.25cm}
\centerline{\epsfxsize=7.25cm \epsfbox{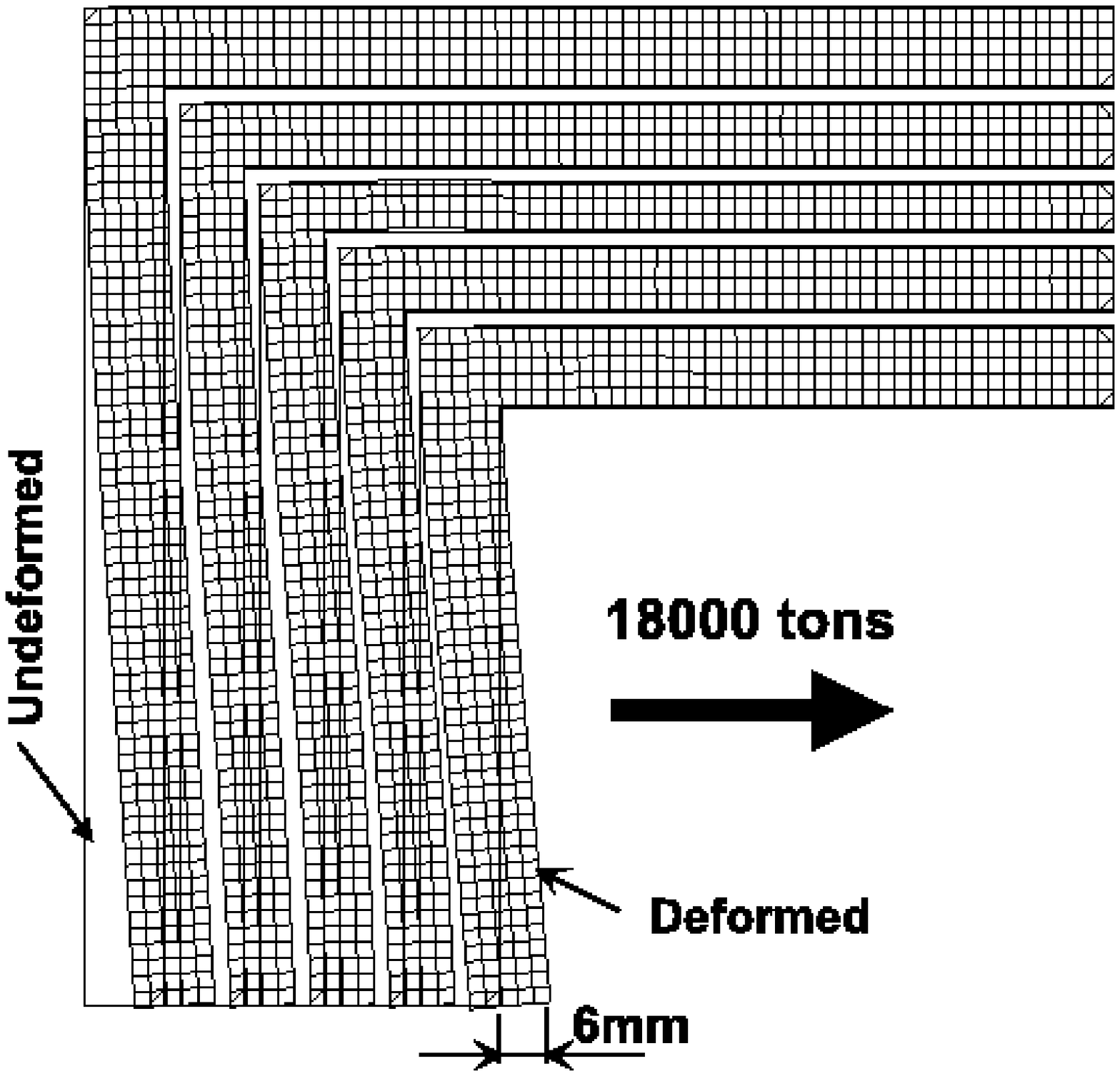}}   
\caption{\sl Deformation of the end-yoke due to magnetic force of 18,000 tons.\label{magnet-deformation}
 }
\end{minipage}
\end{figure}

Load conditions of iron yoke are self-weight of 11,000tons. The barrel-yoke is about 5,000~tons and the end-yoke is 6,000~tons, respectively. The magnetic force was calculated to be 18,000~tons. The mechanical design will be followed to the design guideline for construction of nuclear power plant in Japan. In this reference, the safety margins for various stresses are indicated and load conditions are also indicated at the various situations. At the earthquake resistant design, an input acceleration will be 0.2G or 0.3G. Earthquakes of this magnitude are expected to occur in the Tsukuba area with a frequency of once per year of 30 years.
Figure~\ref{self-deformation} shows the deformation of the outermost layer of the barrel-yoke due to self-weight. The self-weight of the outermost layer is approximately 1,352 tons. The maximum deformation was calculated to be 3.1mm. However, it should be noticed that this assumption is ideal conditions. The practical situation is each octagonal section will be assembled by bolted joints, so its stiffness must be lower than the complete octagonal shape. It is important to assemble with keeping the rigid structure. The gravitational sag of end-yoke due to the self-weight will be neglected. Because the self-weight is acted on the vertical direction, the end yoke is the rigid body in this direction.

The deformation of the end-yoke due to the magnetic force of 18,000 tons is shown in Fig.~\ref{magnet-deformation}. The maximum deformation was calculated to be 6mm. The end yoke will be assembled from 4 sections from the viewpoint of crane capacity and easy assembling, it will not be able to expect to the stiffness as the completed octagon.
At the earthquake resistant design, the natural frequency of iron yoke is calculated as the first step. If this value is lower than the 10Hz, the reinforcement of the iron yoke structure is required due to avoid to resonant with earthquake. As the next step, various earthquake waves those maximum magnitudes are corresponding to 0.2G or 0.3G are input to the iron yoke, then get the responded magnitude. If the magnitude is 0.4G, this value is applied to the horizontal direction as the static load. At the final step, the deformation and stress level will be calculated using 0.4G.

\subsection{Configurations}

The barrel-yoke structure is constructed from eight flux-return and eight Muon detector modules. Each barrel Muon module consists of 5 layers with 100mm thick instrumental gaps. Thicknesses of steel plates are 500mm, 400mm, 300mm, 400mm and 500mm from innermost layer, respectively. Few millimeters of flatness of steel plate has to be taken into account on the Muon detector design. To improve the stiffness of the barrel-yoke, some support structures will be necessary.

Each end-yoke is planned to separate into four quadrants containing five iron plates. Each thickness is the same as the corresponding plate of the barrel-yoke. The inner diameter of 1,300mm is required for th			e support tube. The end-yoke must slide out to provide access to the inner detector. An end-yoke transport system has to be studied.

%% file: detsim/main.tex
\chapter{Monte Carlo Simulation Tools}
\label{chapter-detsim}

\section{Overview}

Monte Carlo simulation comprises an essential part of
any modern experimental high energy physics.
In the designing stage of a large scale project such as 
the JLC, its primary goal is to identify
important physics targets and then set
machine parameters such as beam energy, luminosity,
beam energy spread, beamstrahlung, beam related background, etc.  
and detector parameters such as
momentum resolution for charged particle tracking,
calorimetric energy resolution, 
impact parameter resolution,
minimum veto angle,
particle identification, and so on.
These performance requirements constrain
the machine and the detector designs
eventually to be integrated into a TDR.
The simulation studies thus link three major components of
the project: physics, machine, and detector.
It is therefore very important to simulate,
with sufficient accuracy and speed,
all of key features of linear collider experiments.

\begin{figure}[htb]
\centerline{
\epsfxsize=12cm 
\epsfbox{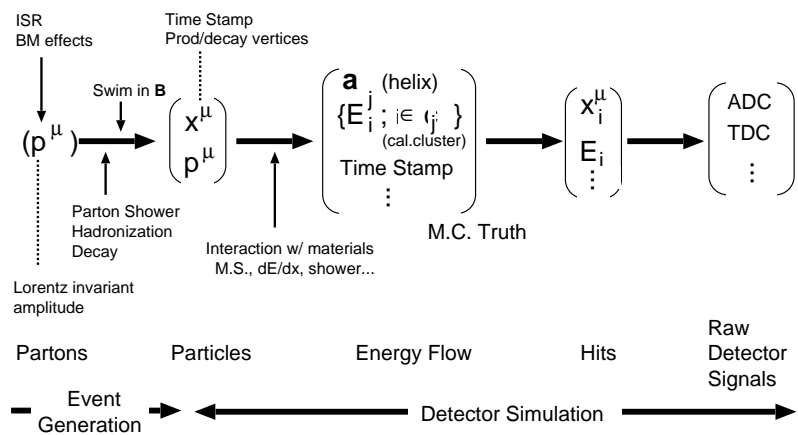}
}
\caption[Fig:sim]{\label{Fig:sim}\sl
Flow chart for Monte Carlo simulations
}
\end{figure}

Monte Carlo simulation that connects physics, 
machine, and detector parameters has the structure shown in
Fig.\ref{Fig:sim}.
We start from 4-momenta of initial partons generated
with initial state radiation (ISR) and beam effects.
After parton showering, hadronization, and decays
we are left with pairs of two 4-vectors $(x^\mu,p^\mu)$
for final state particles, where $x^\mu$'s carry
information on production and decay vertices as well as
time stamps.
These particles are swum in the detector volume,
interacting with materials, leave tracks in
tracking devices and, depending on the nature of the
particles, shower in calorimeters.
These tracks are usually parametrized as helices
and individual showers as calorimeter clusters,
resulting in hits in the tracking devices and
energy deposits in calorimeter cells.
If necessary we can then convert these into
raw detector signals such as ADC or TDC values, etc.

\begin{figure}[htb]
\centerline{
\epsfxsize=12cm 
\epsfbox{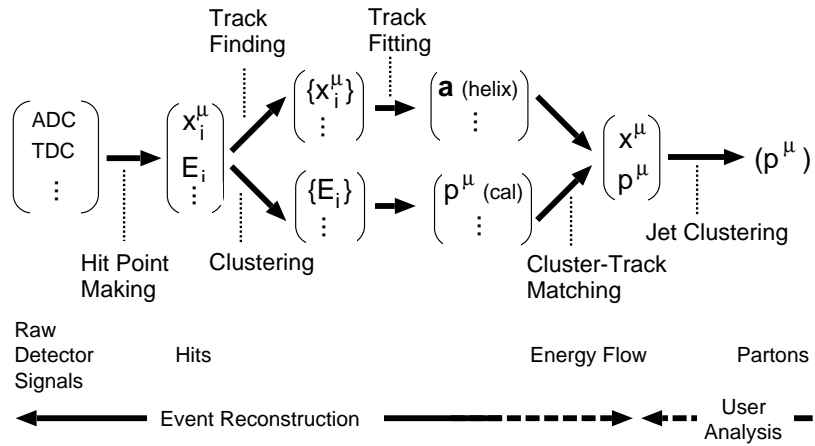}
}
\caption[Fig:anl]{\label{Fig:anl}\sl
Flow chart for event reconstruction and analysis.
}
\end{figure}

There is a mirror image of this flow diagram, which
is called event reconstruction and analysis,
and is shown in Fig.\ref{Fig:anl}.
From the raw data signals we make hit points (hit point making).
In the case of hits in trackers, the map from the set of individual hits
to subsets of hits (tracks) is called track finding, while
in the case of calorimeter hits, the map is called clustering.
The map from the tracks to helix parameters is called track fitting.
These tracks and calorimeter clusters are combined through
track-cluster matching and yield energy flow information.
Together with vertex information, these comprise a set of
pairs of 4-vectors $(x^\mu,p^\mu)$ of the final state particles.
Jet clustering follows this step and hopefully
leads us to reconstructed initial parton 4-momenta.

Depending on how we short cut these processes from
the 4-momenta given by the event generator
to the reconstructed 4-momenta,
there can be many levels of simulation.
It is very useful to do it in various levels, since
by doing so we will know the effects of what we
skipped or included, thereby leading us to
deeper understanding of the detector
as well as to fast but sufficiently accurate
way of simulation.

In order to facilitate such multi-level simulations,
it is desirable to have a general simulation study framework
which allows easy switching from one level to another
without affecting end users in the analysis stage.
Considering the recent advance of object-oriented programming (OOP)
paradigm, we are developing a framework
called the JLC Study Framework (JSF) which is sketched in
the next section.
Section \ref{Sec:detsim:evgen} then lists up
various Monte Carlo event generators that can be input
to the JSF.
Brief description of
detector simulators we have used throughout this report
is given in Section \ref{Sec:detsim:sim}.
Section \ref{Sec:detsim:future} then discusses future direction.

\section{JSF: JLC Study Framework}
\label{Sec:detsim:jsf}

\input detsim/jsf.tex

\section{Event Generators}
\label{Sec:detsim:evgen}

\input detsim/generator.tex

\section{Detector Simulators}
\label{Sec:detsim:sim}

\input detsim/sim.tex

\section{Future Direction}
\label{Sec:detsim:future}

As was mentioned in \ref{Sec:detsim:jsf}, the current JSF system
uses, as its part of core libraries, legacy software packages
written mainly in FORTRAN.
One such example is the helicity amplitude subroutine package (HELAS)
and differential cross section functions composed with HELAS.
We are planning to develop a C++ version of HELAS (HELAS++) to
facilitate object-oriented implementation of event generators.
Another example is the detector simulators:
a substantial part of our fast detector simulator QuickSim
and essentially all of our GEANT3-based full simulator JIM
are written in FORTRAN.
We are now developing a new detector simulation system (JUPITER)
based on GEANT4.

\input detsim/ref.tex

%% file: detsim/jsf.tex
\subsection{Philosophy}

The JLC Study Framework (JSF) is written in C++ and is
based on ROOT\cite{Ref:root}.
As its name suggests,
the JSF sets a general framework for event-based data processing,
scope of which covers all the JLC-related software works
that extend from various Monte Carlo simulations to
online data taking and monitoring for beam tests and
their off-line analyses.
Basic concept of the JSF is to organize an event-based 
data processing program as consisting of modules.
Each module comprises a program unit for a certain task
and is implemented as derived from a base class called {\tt JSFModule}, 
which specifies a common interface for processing
begin run, event, end run data, etc. 
Any module that inherits {\tt JSFModule}
is automatically included in the sequence
of the program execution, which is controlled
by a kernel routine, an instance of a class called {\tt JSFSteer}.
The kernel routine also takes
care of inputs from and/or outputs to files,
and provides a graphical user interface
as well as an event display facility.

For Monte Carlo simulation studies,
the JSF provides a common interface 
to various event generators 
both fast and full detector simulators,
event reconstruction, and analysis.

\begin{figure}
\centerline{\epsfxsize=10.0cm \epsfbox{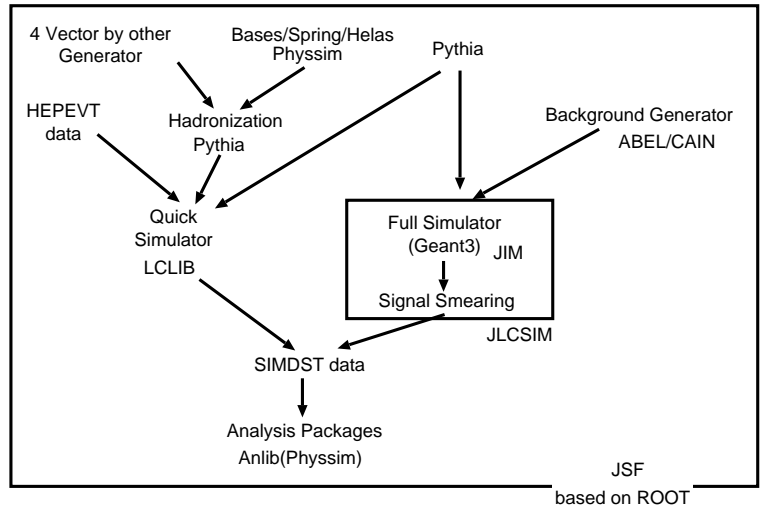}}
\caption{\label{detsim/figs/overview.eps}
Software packages and their relations for JLC studies.}
\end{figure}

Several modules are thus included in the standard JSF distribution:
a base class for event generators using BASES/SPRING\cite{Ref:bases}, 
an interface to PYTHIA\cite{Ref:pythia},
a module to handle parton showering and hadronization using JETSET, 
and modules to control a quick simulator (QuickSim)
and a full simulator (JIM).
There are also interface modules to read 
external event data files such as ASCII text files of parton four momenta,
data created by JIM, etc.
Their relations are
shown in Fig.~\ref{detsim/figs/overview.eps}.

\subsection{Interfaces to Event Generators}

There are interfaces to several types of event generators
implemented in the JSF.

\subsubsection{Pythia}
The Pythia generator module uses the ROOT's interface to
Pythia\cite{Ref:pythia} and is built in the JSF
as the standard event generator.
Event generation with this module can thus be controlled
by a macro file in a standard format or a built-in GUI
or both, which specify event types, decay modes,
the number of events to generate, etc.

\subsubsection{Bases/Spring-based Generators}
The Bases/Spring module provides an interface mainly for
those generators that utilize HELAS\cite{Ref:helas}
for helicity amplitude calculations.
The current version of the JSF comes with a C++ version
of BASES/SPRING, where BASES performs integration
of amplitudes squared for the process in question and
SPRING carries out generation of weight-one events
according to the integration result.
Resultant partons are passed to a hadronizer
described below.
Many event generators have been developed using this interface,
of which a package called Physsim is described 
in the next section.

\subsubsection{Other Generators}
For other event generators which are not integrated into the JSF,
two types of interface modules are prepared.  
One is to read generator data in HEPEVT format.
In this case, the 4-momenta of stable particles 
are directly fed into detector simulators.
The other is to read 4-momenta of quarks and leptons,
which are prepared in an ASCII text file.
This category includes various generators
based on GRACE\cite{Ref:grace}.
As with the Bases/Spring-based generators,
the generated partons
are hadronized using the standard hadronization module,
being ready for consumption by a detector simulator.
This provides a very handy 
interface to a toy Monte Carlo.

\subsubsection{Beamstrahlung}

Proper account of beamstrahlung
is indispensable for realistic simulations
of experiments at the JLC.
The JSF thus provides a standard package that generates 
average electron and
positron energy during collision
under the influence of beamstrahlung.

The beamstrahlung spectrum, $F(Z)$, used there is given by 
\begin{equation}
F(z)=\int^1_0e^{-n_\gamma\tau}\frac{e^{-\eta}H(n_1\eta^{1/3}\tau)}{1-z}d\tau
\end{equation}
where $z=E/E_0$ is the fractional energy of the electron
and 
\begin{equation}
\xi=\frac{r_e^2E_0N}{\frac{1}{2}m_e\alpha\sigma_z(\sigma_x+\sigma_y)},\;\;
n_{cl}=\frac{\alpha r_e N}{\frac{1}{2}(\sigma_x+\sigma_y)},\;\;
n_\gamma=n_{cl}\left(\frac{1-0.598\xi+1.061\xi^{5/3}}{1+0.922\xi^2}\right),
\end{equation}
\begin{equation}
\eta=\frac{1-z}{\xi z}, \; \;
n_1=\frac{n_{cl}+\xi\eta n_\gamma}{1+\xi\eta}.
\end{equation}
$r_e$, $m_e$, and $\alpha$, are classical electron radius, electron 
mass, and fine structure constant, respectively. 
$E_0$ is the nominal beam energy and 
$N$ is the number of electrons in a bunch.
A function $H(x)$ is given by
\begin{equation}
H(x)=\sqrt{3}{8\pi}\left( \frac{\sqrt{x/3}}{1+0.53x^{-5/6}}\right)^{3/4}
e^{4(x/3)^{3/4}} .
\end{equation}
In this formula, the energy spectrum is specified by a set
of parameters: $(E_0, N, \sigma_x, \sigma_y, \sigma_z)$, 
in addition to the initial energy spread of 
electron and positron beams.
For efficient spectrum generation, 
the beamstrahlung spectra for different energies were tabulated
for JLC-I accelerator parameters and included
in a standard routine for beamstrahlung generation.

\subsection{Tau Polarization, Parton Showering, and Hadronization}

As already stated, there are many event generators
based on BASES/SPRING that output 4-momenta of 
final-state partons.
If they are colored, they should be hadronized before
being passed to a detector simulator.
The parton-showering and fragmentation of
colored partons are taken care of by 
a module called {\tt JSFHadronizer} that
provides a C++ interface to JETSET7.4.
It should be noted that, in this hadronizer module,
gluon emissions from the daughter partons
after the decays of heavy intermediate states
are treated properly, 
so that the daughter $b$- and $\bar{b}$-quarks from 
a $t\bar{t}$ pair, for instance, radiate independently.
Another important feature of {\tt JSFHadronizer}
is that it takes proper account of tau polarization.
This is crucial since the polarizations of
tau leptons play an essential role in
various physics analyses.
The standard JETSET, however, ignores the tau polarizations.
In order to overcome this problem, 
{\tt JSFHadronizer} exploits TAUOLA\cite{Ref:TAUOLA},
when it find a tau in the input list of partons.

\subsection{Detector Simulation}
The JSF supports both quick (QuickSim) and
full (JIM) simulators, of which
QuickSim is included in the standard distribution
as a default detector simulator.
Though the JIM itself is not distributed with the JSF,
the JSF comes with not only 
a module to read JIM detector simulator data 
but also an interface to run the JIM full detector simulator 
within itself.
Both of QuickSim and JIM use basically the same
format to save tracking and calorimeter cluster information, 
so as to make end-users' easily switching from one
to the other.
Further description of detector simulator will be given in 
Section~\ref{Sec:detsim:sim}.

%% file: detsim/generator.tex
\subsection{Physsim Generator}
Physsim\cite{detsim:physsim}
is a collection of various event generators
and works within the JSF.
Currently the following processes are included
in the package:
\begin{center}
\begin{tabular}{l l l}
Higgs &:& $e^+e^-\rightarrow ZH$ \\
SUSY &:& $e^+e^-\rightarrow \tilde{f}\bar{\tilde{f}}, 
\tilde{\chi}\bar{\tilde{\chi}}$ \\
TOP &:& $e^+e^-\rightarrow e^+e^-t\bar{t}, \nu\bar{\nu}t\bar{t},
t\bar{t}h, t\bar{t}, t\bar{t}Z$\\
Two photon &:&  $e^+e^-\rightarrow e^+e^- f\bar{f} ( f\equiv e, \mu, \tau, q)$ \\
$W/Z$ &:&  $e^+e^-\rightarrow e^+e^-W^+W^-, e^+e^-Z, e\nu W$\\
  & & $\nu\bar{\nu}W^+W^-, \nu\bar{\nu}Z, W^+W^-, W^+W^-Z, ZZ$
\end{tabular}
\end{center}
all of which are based on BASES/SPRING.
The amplitudes squared for these processes are
calculated with HELAS and include full helicity amplitudes,
thereby properly reproducing various angular
correlations among decay daughters of unstable partons,
such as $W$, $Z$, $top$, and $H$, that
appear in intermediate states.
It should also be noted that beam polarization
can be specified for all of the above processes.
The polarizations of taus are retained and
fed to the hadronizer described in Sec.~\ref{Sec:detsim:jsf},
which decays them with TAUOLA.
Note also that, when appropriate, 
as with 
$e^+e^- \to e^+e^-t\bar{t}$,
$\nu\bar{\nu}t\bar{t}$,
$t\bar{t}Z$,
$e^+e^- W^+W^-$,
$\nu\bar{\nu}W^+W^-$,
and 
$W^+W^-Z$,
diagrams with a Higgs boson in intermediate states are included,
so that they can serve as Higgs generators.
One can readily switch on and off beam effects including
beamstrahlung, whose spectrum is 
given in Sec.~\ref{Sec:detsim:jsf}.
The effect of initial state radiation is taken into account using 
an exponentiated formula.

Physsim also provides a class library called Anlib, which
include jet finders, event shape routines, a rather primitive
vertex tagger, etc., together with some examples showing
their usage.

\subsection{Other Generators}
Being an $e^+e^-$ collider, the JLC should be able to
provide an ideal laboratory for not only discovery physics
but also various precision measurements.
Precision measurements, however, require
event generators with matching precision,
which in turn necessitates exact calculations
wherever possible.
Such generators inevitably become very huge
and often rather slow, therefore being
left outside of the standard JSF distribution.
This is why they are called 'external generators'.
There are a lot of activities in the field of external generators, 
which can be categorized into the following groups: 
1) generators of generators,
2) generators for precision measurements, and
3) generators for the MSSM.

\subsubsection{Generators of Generators}

In the JLC energy region, we expect final states with
many partons and the JLC experiment there
is going to provide us with data of high accuracy.
This means that 
the cross sections for complex processes must be
calculated including higher-order corrections. 
Such a huge computation sometimes
exceeds limit of desk work by theorists. 
Since perturbative calculations
of quantum field theory consist of a well established algorithm, 
it is natural to try to invent some automated system on computer 
to solve the problem.
Such an automated system generates 'event generators',
and is sometimes called a 'generator of generators'.

An example of such large scale computation is the four-fermion
production in $e^+e^-$ collision at LEP-II. 
It has been demonstrated that automated systems are quite efficient 
for the purpose among many programs
contributed to its calculation. 
In this case
one needed to generate all the 76 channels of four-fermion final 
states and the maximum number of Feynman diagrams reached 
144 for two electron pairs. 
Automated systems, for example, {\tt grc4f}\cite{grc4f}
spawned from {\tt GRACE}\cite{grace}, made such computation possible 
even keeping the fermion masses non-zero.

The automated systems obviously take important role to 
calculate the radiative corrections in one-loop and beyond 
because of the number of diagrams that contribute 
and also the complexity of the calculations. 
Such an example is
the one-loop corrections to $e^+e^- \rightarrow W^+W^-$ in the MSSM.
This requires the computation of $\sim$1,000 Feynman diagrams\cite{eeww1,eeww2},
certainly beyond the scope of manual calculations.

In the ACFA working group, there are two activities:
\begin{itemize}
\item
{\tt GRACE} has a generator for Feynman diagrams to any order
({\tt grc})\cite{grc}. In the tree level calculation, 
it produces {\tt FORTRAN} source code for each of them using the
{\tt CHANEL}\cite{chanel} 
package which performs
the calculation of the helicity amplitudes on a purely numerical basis. 
In the case of multi-particle final states, the integration
over their phase space is performed with the help of the multi-dimensional
integration package {\tt BASES}\cite{bases} and the event generator 
{\tt SPRING}\cite{bases} which generates unweighted events. 
In the one-loop calculation, there is an 
additional step of the algebraic manipulation 
to treat Feynman integrals, which
is also available for the processes of the MSSM.
 
\item
{\tt FDC}\cite{fdc} is written in RLISP+REDUCE. 
It also aims at automatic one-loop computation and
provides a function to deduce Feynman
rules from the input Lagrangian.
One of its example is equipped with
the MSSM.
\end{itemize}

Outside the ACFA working group, there are several other activities:
\begin{itemize}
\item
{\tt CompHEP}\cite{comphep} provides an interactive user interface. 
It can generate events and has been applied to many physical processes. 
Though restricted to the tree-level at present, 
the extension to the one-loop is in progress. 
There are several built-in models including the MSSM.
Using {\tt LanHEP}\cite{lanhep}, the Feynman
rules are deducted from the Lagrangian in {\tt CompHEP} format.
\item
{\tt MadGraph}\cite{madgraph} is also for the tree calculation 
where the amplitude is evaluated by the {\tt HELAS}\cite{helas} library.
\item
{\tt FeynArts/FeynCalc/FormCalc/LoopTool}\cite{feyn} is constructed on 
Mathematica and calculates one-loop amplitudes. 
Several processes have been calculated extensively not only 
in the SM but also in the MSSM.
\item
{\tt ALPHA}\cite{alpha} algorithm is for tree calculations, 
which is unique in the sense that it does not rely on
the perturbation by Feynman diagrams. 
The method is based on the fact that only discrete number of
momenta defined by external particles appear in the intermediate
states at the tree level. 
The field operators are expanded in terms of these
discrete modes and then the scattering amplitude becomes 
directly calculable with a generating functional. 
This method has shown to be efficient to
save time for the amplitude computation.
\item
{\tt O'mega}\cite{omega} is an optimizing compiler for tree amplitudes
inspired by the {\tt ALPHA} algorithm. 
It reduces the growth in calculational effort 
from a factorial of the number of particles to an exponentiation.

\item
{\tt HELAC}\cite{helac} is a {\tt FORTRAN} based package 
to efficiently compute helicity amplitudes in the SM. 
The algorithm exploits 
the virtue of the Dyson-Schwinger equations as compared to Feynman 
diagram approach.
\end{itemize}

\subsubsection{Generators for Precision Measurements}
\par
Through the LEP-II MC-workshops\cite{yr96,yr2k}, it is recognized that 
the following 3 points are indispensable to make precision measurements:
1) keeping masses in the exact matrix elements, 
2) implementation of the loop corrections wherever possible, and 
3) treatment of the QED higher order effects. 

The first point was already noticed in the first LEP-II workshop in 1995. 
A remarkable example is the analysis of 
the process of $e^+e^- \rightarrow e^- \bar{\nu}_e u \bar{d}$\cite{enud}.
When the mass of the electron is neglected, 
the results easily become wrong by a factor. 
Since it is complicated to keep mass terms, a lot of generators
are written by means of the automatic systems 
described in the previous section
\cite{yr96}.

{\tt GENTLE}\cite{gentle} produced a CC03 cross section, typically in the 
$G_F$-scheme, with universal ISR QED 
and non-universal ISR/FSR QED corrections, 
implemented with the so-called current-splitting technique.
At the second LEP-II workshop in 1999, it was pointed out that a new
electroweak $O(\alpha)$ CC03 cross-section in the framework of
the double pole approximation (DPA), showing a result that is 2.5$\sim$
3\% smaller than CC03 cross section from {\tt GENTLE}. 
This is a big effect and a good example to 
show the effects of the loop corrections.
The DPA of the lowest-order cross-section emerges from
CC03 diagrams on projecting the W-boson momenta in the matrix elements to
their on-shell values. 
The important fact is that this approximation keeping
a gauge invariance because DPA is based on the residue of the double 
resonance. 
{\tt BBC}\cite{bbc}, {\tt RacoonWW}\cite{racoonww}, and
{\tt YFSWW}\cite{yfsww}
are written based on this DPA.
One may be able to calculate the full 1-loop corrections to the 4-fermion
production, but it is not trivial to keep the gauge invariance. 
Another idea to keep the gauge invariance is so-called fermion-loop scheme,  
where one includes the minimal set of Feynman diagrams 
that is necessary to compensate
the gauge violation caused by the self-energy graphs. 
{\tt WTO} and
{\tt WPHACT} are written according to this scheme.

In order to further the precision measurements,
the inclusion of initial state radiative corrections (ISR)
that go beyond $O(\alpha)$ is inevitable in the generators. 
As tools for such higher order ISR,
the structure function(SF)\cite{sf} and 
the QED parton shower model\cite{qedps}
and YFS exclusive exponentiation\cite{yfs}
are widely used for $e^+e^-$ annihilation processes.

Single-$W$-production processes present an opportunity to study
the anomalous triple-gauge-couplings (TGC) at the JLC.
The main contribution, however, comes from the non-annihilation
type diagrams.
Therefore, the universal factorization method used for 
the annihilation processes is obviously inappropriate. 
The main problem lies in the determination
of the energy scale of the factorization.
A general method to determine the energy
scale to be used in SF\cite{sf} and ${\tt QEDPS}$\cite{qedps}
is established\cite{qedpst}.                                       
The numerical
results of testing  SF and ${\tt QEDPS}$ for the processes of
$e^- e^+ \rightarrow e^- {\bar \nu}_e u {\bar d}$ and
$e^- e^+ \rightarrow e^- {\bar \nu}_e \mu^+ \nu_\mu$
are given. 
 
Total and differential cross sections of semi-leptonic process of
$e^- e^+ \rightarrow e^- {\bar \nu}_e u {\bar d}$ and 
leptonic one,
$e^- e^+ \rightarrow e^- {\bar \nu}_e \mu^+ \nu_\mu$,
are calculated with the radiative correction by SF or ${\tt QEDPS}$.
Fortran code to calculate amplitudes of the above processes has been 
produced using ${\tt GRACE}$ system\cite{grace}. 
All fermion-masses are kept finite in the calculations.
Numerical integrations of the matrix element squared in the four-body
phase space are done using ${\tt BASES}$\cite{bases}.
For the study of the radiative correction for the single-$W$ productions,
only $t$-channel diagrams (non-annihilation diagrams) are taken into 
account.

\begin{figure}
\centerline{\epsfxsize=14cm \epsfbox{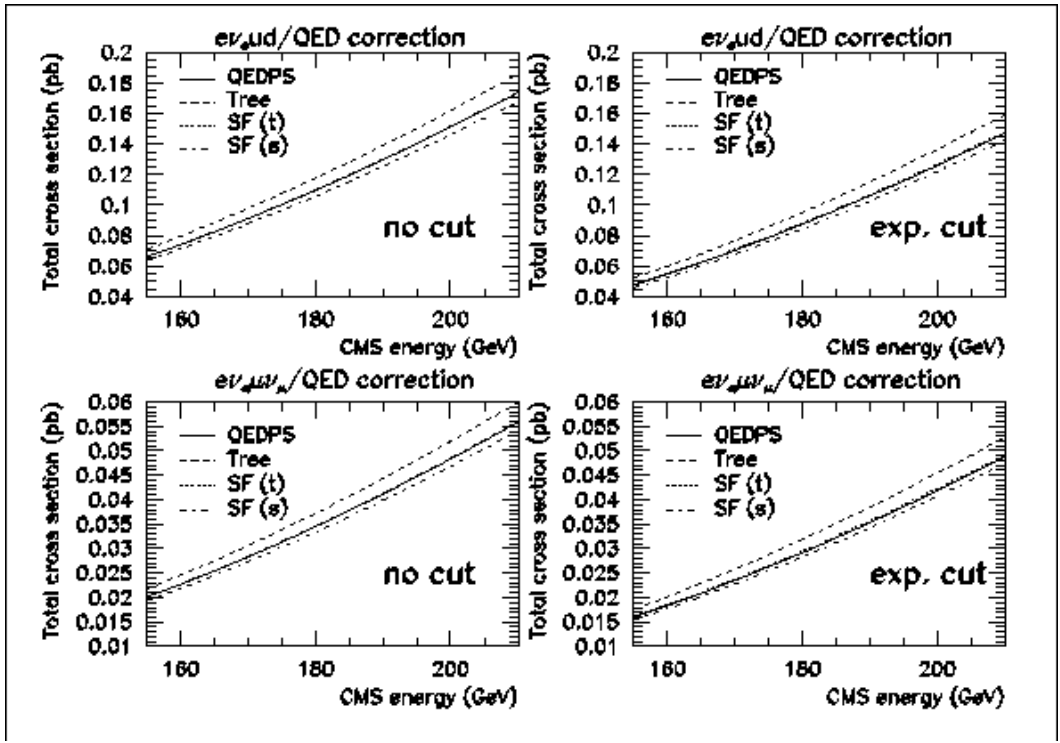}}
\begin{center}\begin{minipage}{\figurewidth}
\caption{\sl
Total cross sections for $e\nu_e\bar{u}d$ and $e\nu_e\mu\nu_{\nu}$
processes 
without and with the experimental cuts described in the text.
SF(t) denotes SF with correct energy scale and SF(s) with wrong
energy scale ($s$).
\label{qedpst}}
\end{minipage}\end{center}
\end{figure}

Total cross sections as a function of the CM energy at LEP-II with
and without experimental cuts are shown in 
Fig.\ref{qedpst}. 
The experimental cuts applied here are
\begin{enumerate}
\item $M_{q {\bar q}}>45 \hbox{GeV}$,
\item $E_l>20 \hbox{GeV}$.
\end{enumerate}
The effects of the QED
radiative corrections on the total cross sections are obtained to be
7 to 10\% at LEP-II energies. 

If one uses a wrong energy scale, $s$, in SF, 
the ISR effect is overestimated by about 4\% 
both with and without the cuts.
For the no-cut case SF with the correct energy scale 
is consistent with ${\tt QEDPS}$ to around 0.2\%. 
On the other hand with the experimental cuts, 
SF with the correct energy-scale deviates by about 1\%
from ${\tt QEDPS}$.

In the JLC region, processes with more particles in the final state,
such as $e^+e^- \rightarrow 6$ fermions become important 
because it is expected for the top-quark pair production. 
The experience at LEP-II tells us that 
one needs a generator equipped with
the exact matrix elements that include all the diagrams that lead
to the same 6-fermion final state, 
not only the ones via the resonant $t\bar{t}$ pair
but also other non-resonant background diagrams.
Already several activities appeared for this process:
Montagna {\it et al.}\cite{mont}
reported a full calculation of the 6-fermion process
at $e^+e^-$ linear colliders,
where the Higgs boson  with an intermediate mass is produced,
by the program package {\tt ALPHA}\cite{alpha} for 
the calculations of the matrix elements and
{\tt HIGGSPV/WWGENPV}\cite{higgspv} for Monte Carlo event generation.
Accomando {\it et al.}\cite{ball} have been calculating a semi-leptonic process
such as
$e^+e^- \to b\bar{b}qq'l\nu$ by program package {\tt PHACT}\cite{phact}. 

In Ref.\cite{yuasa} the process
$e^+e^- \rightarrow b \bar{b} u \bar{d} \mu \bar{\nu} _{\mu} $ is discussed, 
which has been calculated by means of the parallel
{\tt GRACE} combining the parallel {\tt BASES} and PVM {\tt GRACE}
with MPI and the load balancing\cite{pvm}. 
It shows scalability up to 16 processors.
Among the total of 232 diagrams (in unitary gauge) of the process,
the $t \bar{t}$ production diagrams are dominant
over the others.
We have divided the {\it major background diagrams} into three categories:
the diagrams with $W^+ W^- \gamma$ (hereafter $WW\gamma$),
those with $W^+ W^- Z$ (hereafter $WWZ$),
and those with single-$t$ through $W^+W^-$ pair production (hereafter $tWW$).
\par
\begin{figure}
\centerline{\epsfxsize=14cm \epsfbox{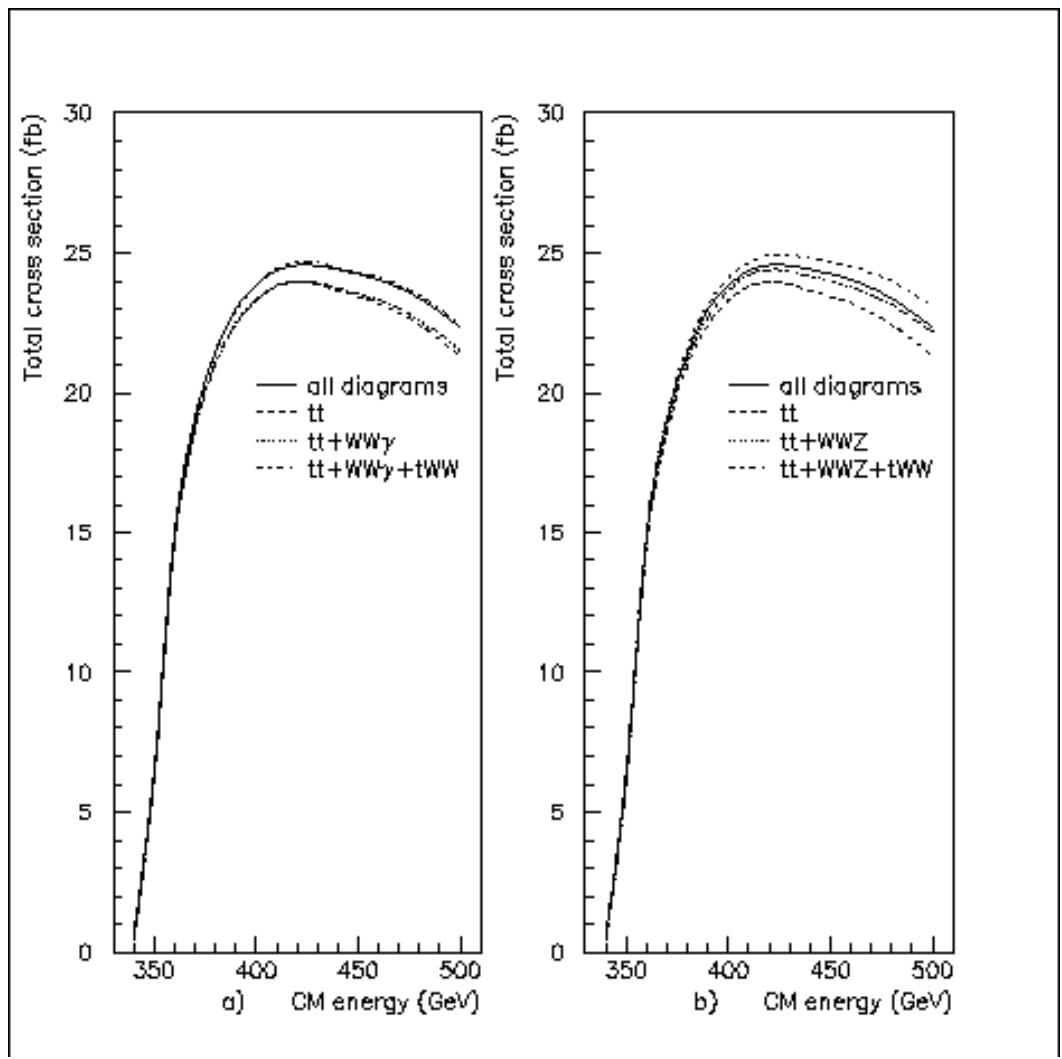}}
\begin{center}\begin{minipage}{\figurewidth}
\caption{\label{ee6f}
{\tt a)} The total cross section with all the diagrams (solid line),
with the $t \bar{t}$ diagrams alone (dashed line), 
with the $t \bar{t}$ and $WW\gamma$ (dotted line), 
and with the $t \bar{t}$, $WW\gamma$, and $tWW$ (dot-dashed line).
{\tt b)} The total cross section with all diagrams (solid line),
with the $t \bar{t}$ diagrams alone (dashed line), 
with the $t \bar{t}$ and $WWZ$ (dotted line), 
and with the $t \bar{t}$, $WWZ$, and $tWW$ (dot-dashed line).
}
\end{minipage}\end{center}
\end{figure}

The total cross sections with various sets of the diagrams
in the CM energy range 340-500 GeV are plotted in 
Figs.\ref{ee6f} {\tt a)} and {\tt b)}, 
where the solid line is the
numerical result of the total cross section with all the diagrams,
while the dashed line is the result with only the dominant
$t \bar{t}$ diagrams.
Besides the result with all the diagrams and that with the
$t \bar{t}$ diagrams alone, 
those with the $t \bar{t}$ and the major
background diagrams are also shown for comparison.
The dotted line in {\tt a)} shows the result with the 
$t\bar{t}$  and $WW \gamma$ diagrams,
while that in {\tt b)} is with the $t \bar{t}$ and $ WWZ$.
The dot-dashed lines in {\tt a)} and {\tt b)} show the result with
the $t \bar{t}$, $ WW \gamma$, and $tWW$ diagrams
and that with the $t \bar{t}$, $ WWZ $, and $tWW$ diagrams, respectively.
These results shown by dotted and dot-dashed lines include the
interferences among the selected diagrams.
\par
As shown in Fig.\ref{ee6f}, cross sections with both the $t \bar{t}$ 
and the major background diagrams 
show different behaviors from that with all the diagrams.
Their difference is about 3\% at $\sqrt s$ = 500 GeV.
This means that the effect from the interference between $t \bar{t}$
and the rest of the diagrams except $WW\gamma$, $WWZ$, and $tWW$ 
is also non-negligible and important.
The contribution from the background diagrams to the total
cross section, ${(\sigma_{all} - \sigma_{t \bar{t}})}/{\sigma_{t\bar{t}}}$,
is less than 5\% in total above
the $t\bar{t}$ threshold.

\subsubsection{Generators for the MSSM}

A lot of 2-to-2 reactions in the models beyond the SM are implemented
in {\tt SPYTHIA} and {\tt ISAJET}.
For precision measurements, however,
we need generators with full matrix elements that lead
to the same multi-body final states as expected for these
2-to-2 reactions after decays.
In the MSSM, there appear 55 particles and 3,553 vertices.
The number of Feynman diagrams for such a multi-body
final state naturally becomes huge. 
Therefore, again, an automatic system plays important role.
The generators for the processes 
$e^+e^- \rightarrow \tilde{\chi}_1^+ \tilde{\chi}_1^- 
\rightarrow 
\tilde{\chi}_1^0 \tilde{\chi}_1^0 q \bar{q'} q'' \bar{q}''' $\cite{lafage},
and $e^+e^- \rightarrow \tilde{t}_1 \tilde{t}_1^* H$\cite{belanger} are
available.

About 1-loop corrections, there exist only some calculations, 
no generator at this moment because calculations of the cross sections
cost too much CPU time.
In the full MSSM, the following 
one-loop corrections are available:
$H^+ \rightarrow t \bar{b}$\cite{mssm-htb},
$e+e^- \rightarrow t \bar{t}$\cite{mssm-eett}, 
$e+e^- \rightarrow W^+ W^-$\cite{eeww1,eeww2}, 
$e+e^- \rightarrow Z^0 h^0$\cite{mssm-eezh}, and
$e+e^- \rightarrow H^+ H^-$\cite{mssm-eehh}.
For $e^+e^- \rightarrow \tilde{t} \tilde{t}^*$, 1-loop corrections in
QCD-MSSM has been calculated\cite{mssm-eestst}.

Among the processes above, 
$e+e^- \rightarrow W^+ W^-$ has a huge cross section
in the JLC energy region and 
the deviation from the SM is important.
For the analysis here, we use the following values 
for the input parameters:

\begin{center}
\vspace{3mm}
\noindent{
\(M_Z=91.1867\)~GeV, \(M_W=80.35\)~GeV, \(m_t=174\)~GeV, \(m_b=4.7\)~GeV,}

\vspace{1mm}
\noindent{
\(m_e=0.51099906\)~MeV, \(m_{\mu}=105.658389\)~MeV, 
\(m_{\tau}=1.7771\)~GeV, }

\vspace{1mm}
\noindent{
\(m_u=58\)~MeV, \(m_d=58\)~MeV, 
\(m_s=92\)~MeV, \(m_c=1.5\)~GeV,}

\vspace{1mm}
\noindent{
\(\tan\beta=15\), \(\mu=-300\)~GeV, \(M_2=200\)~GeV,}

\vspace{1mm}
\noindent{
\(m_{\tilde{e}_1}=
  m_{\tilde{e}_2}= 300\)~GeV,~~\(\theta_e=0.1\),~~
 \(m_{\tilde{\nu_e}}= 289.14\)~GeV,}

\vspace{1mm}
\noindent{
\(m_{\tilde{\mu}_1}=
  m_{\tilde{\mu}_2}= 300\)~GeV,~~\(\theta_{\mu}=0.1\),~~
\(m_{\tilde{\nu_\mu}}= 289.14\)~GeV,}

\vspace{1mm}
\noindent{
\(m_{\tilde{\tau}_1}=
  m_{\tilde{\tau}_2}= 300\)~GeV,~~\(\theta_{\tau}=0.1\),~~
\(m_{\tilde{\nu_\tau}}= 289.14\)~GeV,}

\vspace{1mm}
\noindent{
\(m_{\tilde{u}_1}=
  m_{\tilde{u}_2}=
  m_{\tilde{d}_1}= 300\)~GeV,~~\(\theta_u=\theta_d=0.1\),~~
  \(m_{\tilde{d}_2}= 396.25\)~GeV,}

\vspace{1mm}
\noindent{
\(m_{\tilde{c}_1}=
  m_{\tilde{c}_2}=
  m_{\tilde{s}_1}= 300\)~GeV,~~\(\theta_c=\theta_s=0.1\),~~
  \(m_{\tilde{s}_2}= 396.25\)~GeV,}

\vspace{1mm}
\noindent{
\(m_{\tilde{t}_1}=200\)~GeV,~~
\(m_{\tilde{t}_2}=400\)~GeV,~~
\(m_{\tilde{b}_1}=200\)~GeV,~~\(\theta_t=0.85,\theta_b=0.95\),~~
\(m_{\tilde{b}_2}= 221.8\)~GeV.}

\vspace{1mm}
\noindent{
\(m_{h_0}=90.258\)~GeV,~~
\(m_{H_0}=250.337\)~GeV,~~
\(m_{H_A}=250\)~GeV,~~
\(m_{H_{\pm}}=262.595\)~GeV,~~
}

\vspace{1mm}
\noindent{
\(m_{\tilde{\chi}^0_1}=94.66\)~GeV,~~
\(m_{\tilde{\chi}^0_2}=184.23\)~GeV,~~
\(m_{\tilde{\chi}^0_3}=309.89\)~GeV,~~
\(m_{\tilde{\chi}^0_4}=326.97\)~GeV,~~
}

\vspace{1mm}
\noindent{
\(m_{\tilde{\chi}^+_1}=184.41\)~GeV,~~
\(m_{\tilde{\chi}^+_2}=330.01\)~GeV,~~
}
\end{center}
\vspace{5mm}

This set was chosen based on the following criteria:
1) the value of $\tan\beta$ is not too small to have already been 
experimentally excluded,
2) the lighter chargino can be pair-produced at $\sqrt{s}=$500 GeV,
3) all sfermion masses are less than $\sim$ 1 TeV but heavier than 250 GeV, 
so that they are not pair-produced at $\sqrt{s}=500$ GeV, and
4) all the Higgs bosons except for the lighter CP-odd Higgs boson are not
pair-produced at $\sqrt{s}=$500 GeV.

The result of the calculation has been confirmed through the following checks:
\begin{itemize}
\item The UV divergence is canceled among the 1-loop diagrams. 
\item The infrared divergence is canceled between the 1-loop diagrams and
the soft photon corrections. 
\item The results are independent of the renormalization schemes
\cite{hollik} or \cite{kyoto} adopted in the computation. 
\end{itemize}

\begin{table}
\begin{center}\begin{minipage}{\figurewidth}
\caption{\sl \label{detsim-generator-one}
The total cross section at tree level $\sigma_{tree}$, 
$O(\alpha)$ corrected total cross sections 
in the SM($\sigma^{O(\alpha)}_{SM}$) and in the 
MSSM($\sigma^{O(\alpha)}_{MSSM}$), respectively, 
at $\sqrt{s}=$200 and 500GeV. $\delta$ shows the deviation 
of the MSSM results from the SM one in \%. $k_c$ is set $0.05\sqrt{s}$.
}
\end{minipage}\end{center}
\begin{center}
\begin{tabular}{|c|c|c|c|c|}\hline
$\sqrt{s}$(GeV)& $\sigma_{\mbox{tree}}$(pb)&
$\sigma^{O(\alpha)}_{\mbox{SM}}$(pb)& 
$\sigma^{O(\alpha)}_{\mbox{MSSM}}$(pb)&
$\delta(\%)$ \\ \hline
200&18.001&15.928&15.785&$-$0.897\\ \hline
500&6.765&5.789&5.700& $-$1.69\\ \hline 
\end{tabular}
\end{center}
\end{table}

\begin{figure}
\centerline{\epsfysize=7cm \epsfbox{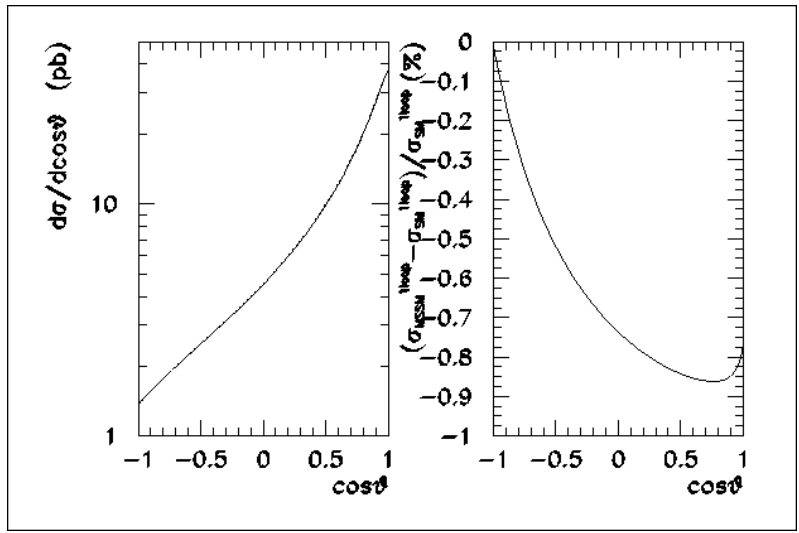}}
\begin{center}\begin{minipage}{\figurewidth}
\caption{\label{fw200}
$O(\alpha)$ corrected $d\sigma/d\cos\theta$ in the SM(the dashed line) and
in the MSSM(the solid line) and the deviation $\delta$ in \% at $\sqrt{s}
=$200 GeV with $k_c=0.05\sqrt{s}$.
}
\end{minipage}\end{center}
\end{figure}

\begin{figure}
\centerline{\epsfysize=7cm \epsfbox{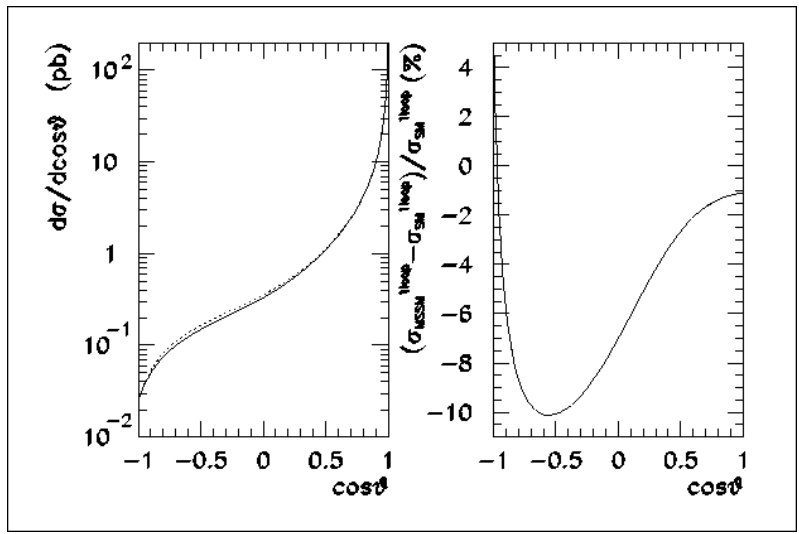}}
\begin{center}\begin{minipage}{\figurewidth}
\caption{\label{fw500}
$O(\alpha)$ corrected $d\sigma/d\cos\theta$ in the SM(the dashed line) and
in the MSSM(the solid line) and the deviation $\delta$ in \% at $\sqrt{s}
=$500 GeV with $k_c=0.05\sqrt{s}$.
}
\end{minipage}\end{center}
\end{figure}

The results are shown in Table~\ref{detsim-generator-one}.  
The soft photon cut parameter $k_c$ is set to $0.05 \sqrt{s}$. 
For the set of the above input parameters,
the deviation from the SM value is less than 1\% 
in the LEP-II energy region,
and still small at $\sqrt{s}=$500 GeV in the JLC
energy region. 
Figs. \ref{fw200} and \ref{fw500} show the angular dependence of the
differential cross sections with $k_c=0.05\sqrt{s}$ 
and of the deviation, $\delta$. 
The solid and dashed lines of the differential cross sections correspond to 
the MSSM case and to the SM case, respectively. 
At $\sqrt{s}=$200 GeV, the two curves are visually indistinguishable.

On the other hand, at $\sqrt{s}=$500 GeV GeV,
the deviation becomes large in the backward region. 
It might not be, however,
so easy to see it in real experiments because of its small cross sections
in that angular region.

\begin{figure}
\centerline{\epsfysize=7cm \epsfbox{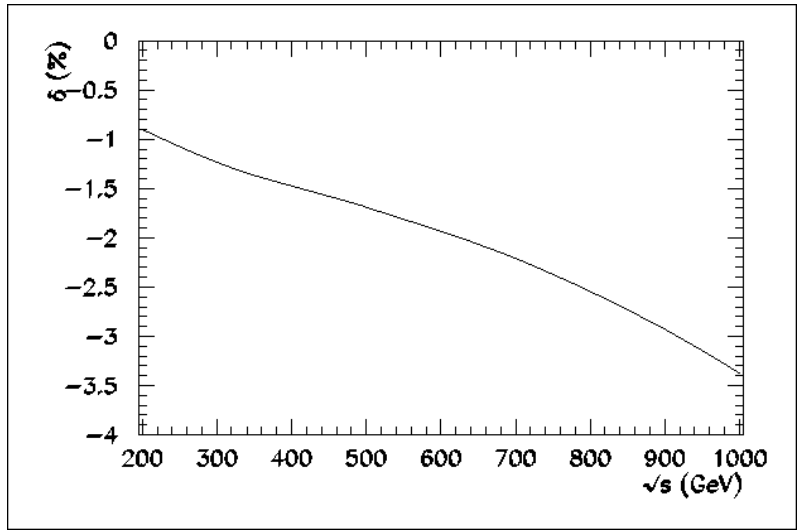}}
\begin{center}\begin{minipage}{\figurewidth}
\caption{\label{delta}
The center of mass energy dependence of the deviation $\delta$ 
in \% with $k_c=0.05\sqrt{s}$.
}
\end{minipage}\end{center}
\end{figure}

Fig.\ref{delta} shows the energy dependence of the deviation $\delta$ with the
above set of the input parameters. 
Up to $\sqrt{s}=$1000 GeV, the deviation
stays small and is less than around 3\%. 

%% file: detsim/sim.tex
\subsection{QuickSim: a Fast Simulator}

There are two approaches to quick simulations.
The first approach that is adopted for tracking part of the JSF
relies on calculations of various error matrices.
Consider a track parameter vector (${\bf a}$)
and its error matrix ($E_{\bf a}$).
Linearization reduces the $\chi^2$ of its measurement
into a parabolic form:
\begin{equation}
\chi^2 = \frac{1}{2} \Delta{\bf a}^T 
\cdot E_{\bf a}^{-1} \cdot \Delta{\bf a}.
\end{equation}
Being symmetric, $E_{\bf a}$ can be diagonalized using an orthogonal
matrix $O$ as
\begin{equation}
E_{\bf b}^{-1} = O^T \cdot E_{\bf a} \cdot O
=\left( \begin{array}{ccc}
1/\sigma_1^2 & ~  & 0   \cr
~  & \ddots & ~ \cr
0 & ~ & 1/\sigma_n^2
\end{array}\right).
\end{equation}
Using this, we can rewrite the $\chi^2$ as
\begin{equation}
\chi^2 = \frac{1}{2} \Delta{\bf b}^T 
\cdot E_{\bf b}^{-1} \cdot \Delta{\bf b},
\end{equation}
where $\Delta{\bf b}$ is defined by
\begin{equation}
\Delta{\bf b} = O^T \cdot \Delta{\bf a}.
\end{equation}
Since the components of $\Delta{\bf b}$ are independent,
we can now smear them as
\begin{equation}
\Delta {\rm b}_i 
= \sigma_i \cdot (\mbox{Gaussian random number with unit width})
\end{equation}
from which we can obtain a smeared ${\bf a}$ as
\begin{equation}
{\bf a} = {\bf a}_{\rm true} + O \cdot \Delta{\bf b}.
\end{equation}
Notice that in this way it is possible to properly take into account
any correlation among components of the original parameter
vector ${\bf a}$.
Since one can approximately calculate error matrices,
knowing basic detector parameters such as geometry and
resolutions, this approach is useful to see
how the overall detector performance varies with these
basic parameters.

The second approach 
is based on a parametric smearing and is
more empirical and practical, which is
used for calorimetry.
Smearing parameters are tuned to reproduce results
from corresponding full simulators, and,
once tuned, the simulation can be quick and
fairly accurate.
This approach is therefore useful for physics simulations
that require high statistics.

Components we have in the Quick Simulator are as follows:
a beam pipe, Vertex Detector (VTX), Intermediate Tracker (IT), 
Support Tube, Central Drift Chamber (CDC), 
Electromagnetic (EM) and Hadron (HD) Calorimeters.
Parameters of them, such as the number of layers, thickness,
resolution, etc. are all controlled by input cards.

In QuickSim, charged particle smearing goes
as follows.
\begin{enumerate}
\item A particle information is obtained from generator data and 
a track starts.
\item The track is swum in the magnetic field 
towards the beam pipe and, if it reaches there, its direction 
gets smeared due to multiple scattering.
\item
The track is further swum to the inner-most VTX layer, 
and its direction gets smeared again by multiple scattering.
\item
From the position and momentum of the track at the inner-most layer of VTX,
a VTX track parameter vector is formed and smeared 
by the error matrix method as described above.
\item
The track is then swum to
the next VTX layer and its direction is smeared again
due to multiple scattering.
This procedure is repeated all the way through the rest of the VTX layers, 
all of the IT layers, and the support tube to the inner-most 
sense-wire layer of CDC.
\item
From the position and momentum of the track
at the inner-most CDC layer, 
a CDC track parameter vector is calculated and smeared
with the error matrix method.
\item
The track is further swum to the calorimeter and creates cell signals.
Electrons leave signals in the EM section
and hadrons in the HD section of the calorimeter.  
Currently, muons do not generate signals in the calorimeter.
\item
The pivot of the CDC track parameter vector is moved back to that of 
the VTX track parameter vector taking into account the effects of
multiple scattering on the error matrix.
They are then averaged
to get a combined VTX-CDC track parameter vector.
\end{enumerate}
The energy loss of particles is not considered in QuickSim.

The smearing of neutral particles is more straight-forward 
as we just extrapolate them to the calorimeter and generate
smeared signals.  
To simulate the effects of the lateral spread of the
calorimeter signals, 
a projected shower distribution given by,
\begin{equation}
f(x)=a_1 e^{-|x|/\lambda_1} + a_2 e^{-|x|/\lambda_2}.
\end{equation}
is used, where
$a_i$ and $\lambda_1(i=1,2)$ are tunable 
parameters for shower shape.
The energy deposit in each of the hit cells is
calculated by integrating this distribution over the cell size.

To use the calorimeter signals for physics analysis, 
the signals distributed to cells have to be clustered,
which is done in QuickSim.
The algorithm we adopted to cluster EM calorimeter signals is as follows:
\begin{enumerate}
\item
Find the highest energy cell BL1 among those with an energy larger than ETH. 
\item
Attach neighbor cells BL2 to the seed cell BL1.
\item
Check BL3 that is a neighbor cell of BL2 and 
include it in the cluster if 
one of the following two conditions is satisfied:
\begin{itemize}
\item[(a)] $E(BL2) > c1\times E(BL1)$
\item[(b)] $E(BL3) > c2\times E(BL1)\; \mbox{~and~} \;E(BL3)< c3\times E(BL2)$,
\end{itemize}
where $c1$, $c2$, and $c3$ are parameters to be optimized,
depending on the calorimeter geometry.
\end{enumerate}

\begin{figure}[tb]
\centerline{\epsfxsize=7.0cm \epsfbox{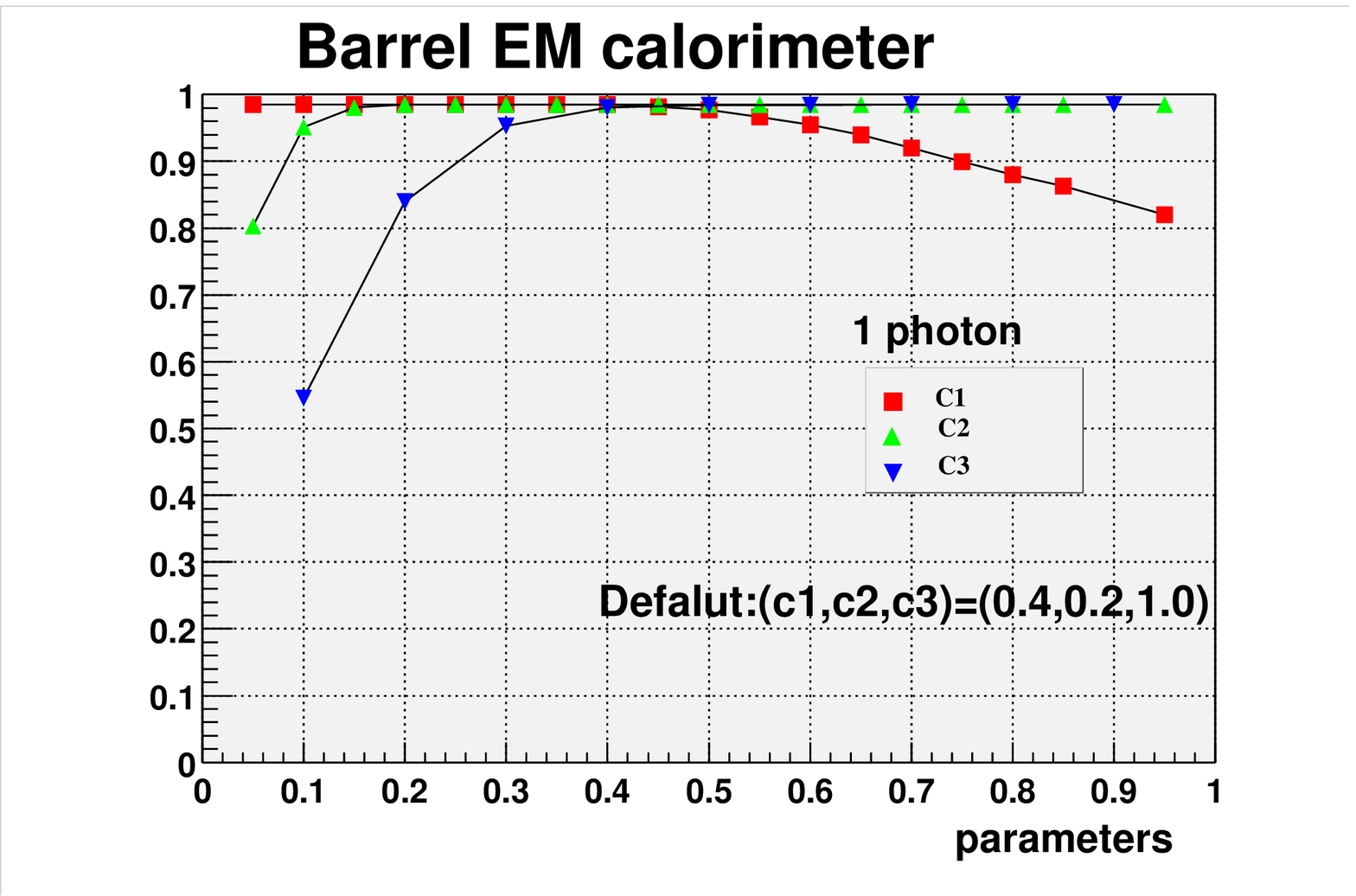}
\hspace*{12pt}
\epsfxsize=7.0cm \epsfbox{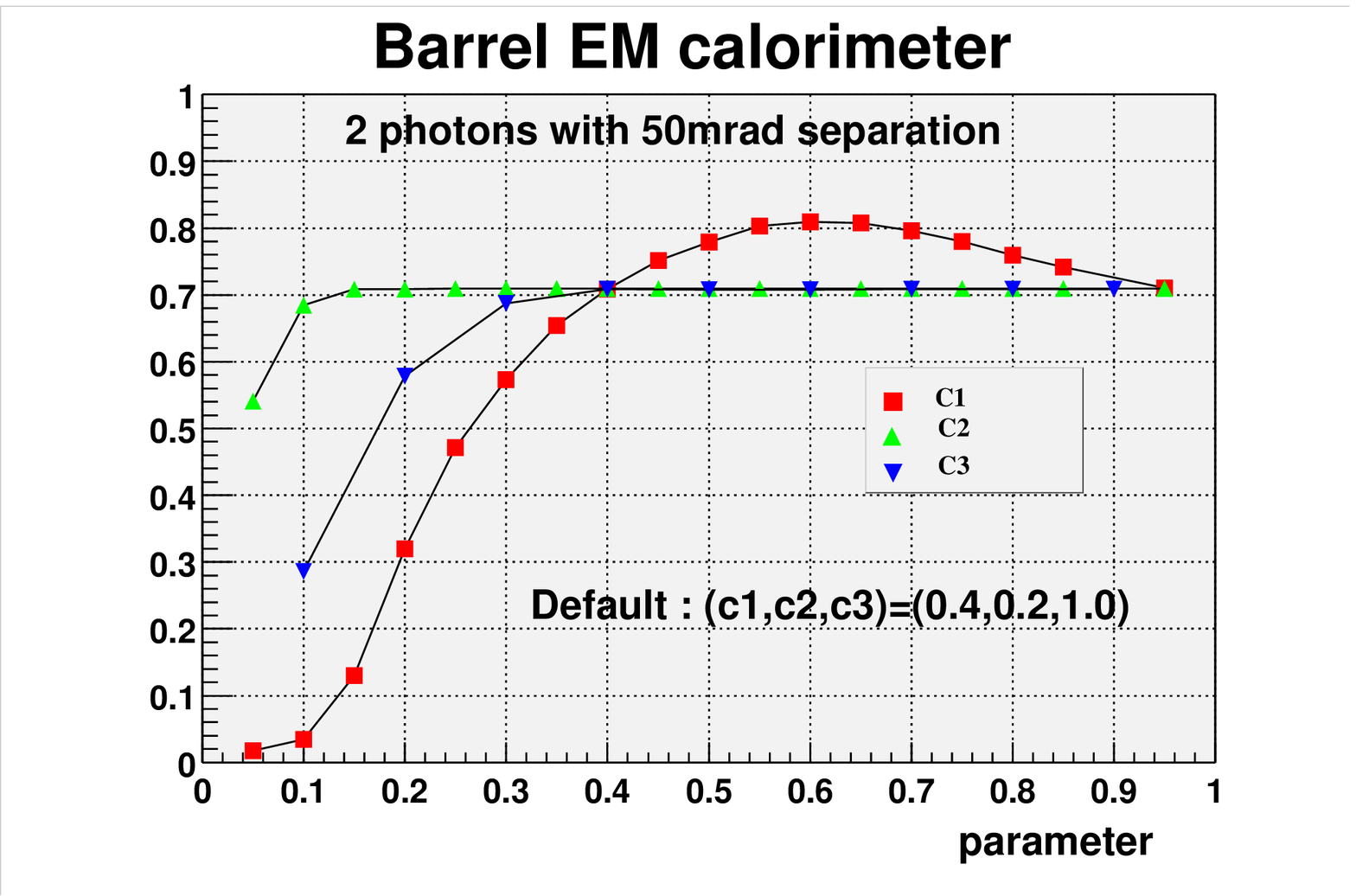}}
\vspace*{-4.5cm}\hspace*{1.0cm}a)\hspace*{7cm}b)\vspace*{4.5cm}
\begin{center}\begin{minipage}{\figurewidth}
\caption{\sl \label{det:calshower-sep}
The efficiency to cluster (a) a single photon as a single photon,
and (b) two photons as two photons.  
The photon energy is varied from
1 GeV to 50 GeV, randomly, while the opening angle in the case of 
two-photon generation is fixed at 50 mrad, while their directions
are randomly varied among Barrel calorimeter region.
Red squares, green up, and blue down triangles
are efficiencies when only $c1$ or $c2$ or $c3$ is modified, respectively.
}
\end{minipage}\end{center}
\end{figure}

In order to optimize the clustering parameters, 
we generated a single photon or two photons with a fixed opening angle
but with a varying energy between 1 GeV to 50 GeV,
and tuned the parameters so that the number of reconstructed photons
matches the generated one.  
The opening angle of the photons in the case
of the two-photon generation is fixed at 30mrad, 50mrad, and 100mrad.
The clustering efficiencies are 
shown in Fig.~\ref{det:calshower-sep}
as a function of the parameters.

A charged track is flagged as muon or electron
with the help of generator information.
If there is no electron matched with an EM cluster, 
it is considered as a photon cluster.  
The cluster-track
matching is performed based on their position and energy differences.
If a track and an EM cluster match in position but differ in energy
by more than 2-$\sigma$ of their resolutions, 
the energy  excess in the EM cluster is 
attributed to a photon cluster overlapping with the electron cluster.

In the case of the HD clusters, we used 
a method to link multiple tracks to a connected set of HD cells,
instead of relying on clustering using
the HD section alone\cite{LINK-TRACK-CLUSTER}.
In this method, we first form a connected set, called an island
hereafter, of hit HD cells without trying to separate multiple peaks 
within the island.
Charged tracks are then extrapolated to the island.
If the sum of the track energies matches 
with the total energy deposit in the island
within $2\sigma$ of the calorimeter resolution,
the island is simply deleted and each of the matched tracks are
flagged as a combined track from a purely charged particle.
If they differ by more than $2\sigma$,
we subtract the track energies from the energy deposit
in the island and attribute the excess to
a possible neutral hadron.
Each of the tracks that matched in position with the island
is then flagged as a charged hadron in a mixed HD cluster. 

Results of the cluster-track matching are stored 
in a class called {\tt JSFLTKCLTrack}, which serves 
as a starting point of the most of physics analyses.

\subsection{JIM and JLCSIM: a Full Simulator}

\begin{figure}[hb]
\begin{minipage}[c]{7cm}
\centerline{\epsfxsize=6.0cm \epsfbox{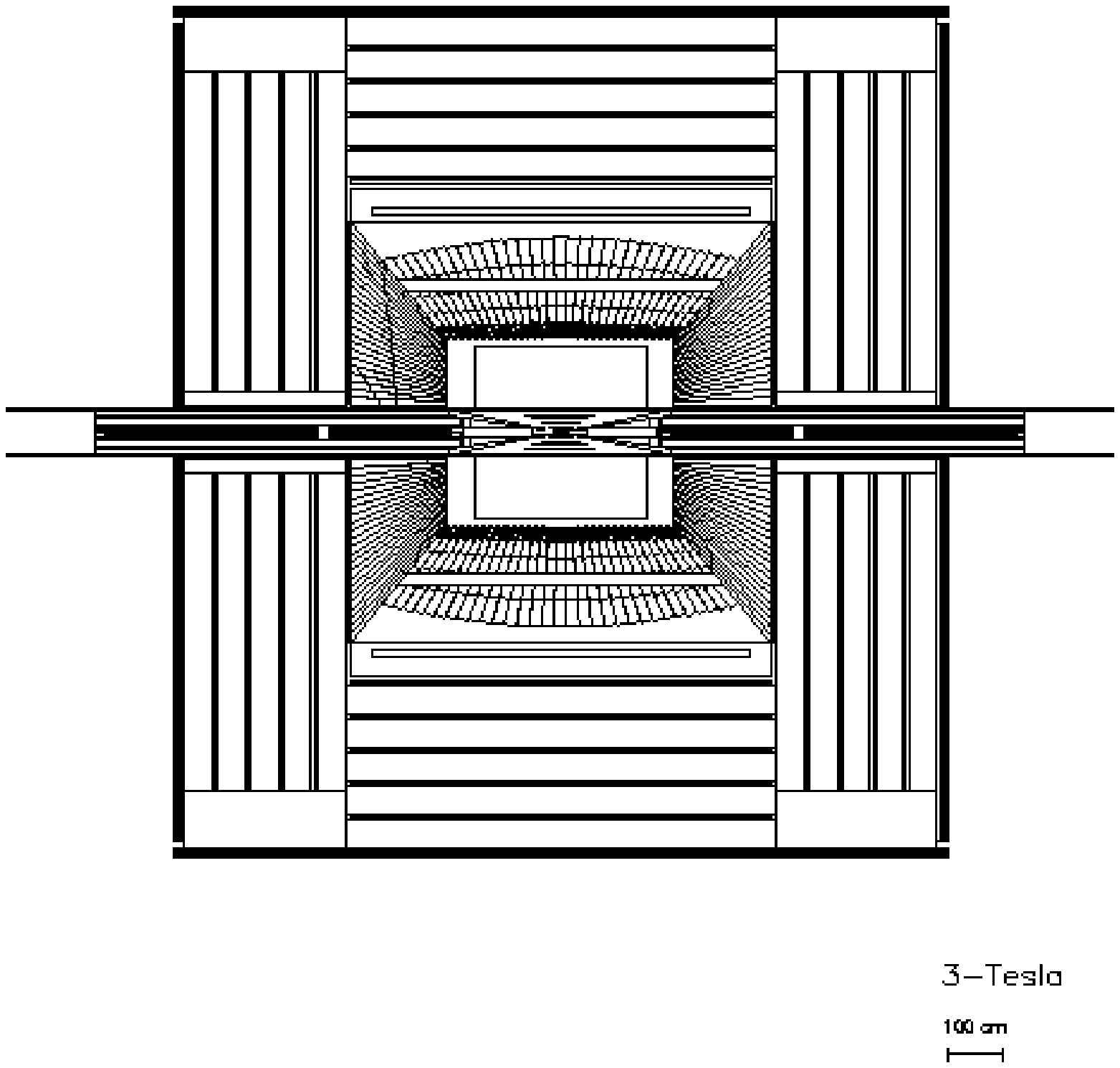}}
\end{minipage}
\begin{minipage}[c]{7cm}
\centerline{\epsfxsize=6.0cm \epsfbox{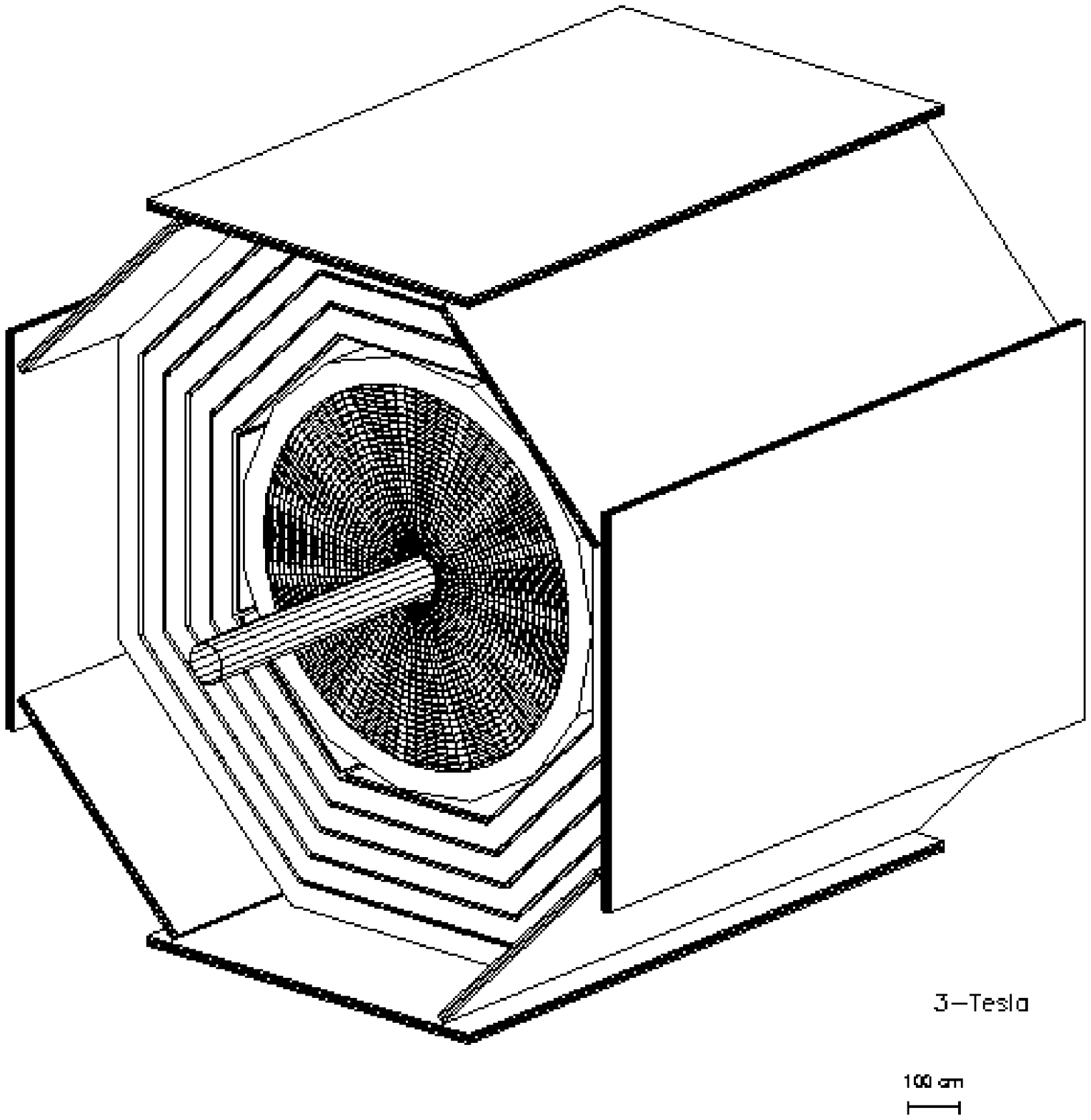}}
\end{minipage}
\caption{\sl \label{jim-3tesla.eps}
The JLC detector geometry of 3 Tesla model.
The End Cap Muon system and Calorimeters are removed in the 3-D view
to show the structure inside.}
\end{figure}

As for full simulators, we have to implement
materials and physics processes as realistic as possible,
since their primary usage is to check the validity of
the corresponding quick simulators.
JIM is a GEANT3-based full detector simulator
being developed for this purpose and
is distributed as the JLCSIM package together with
utilities to analyze JIM data.
It includes two versions of geometry models: one for
2 Tesla magnetic field\cite{JLC-I} and the other for
3 Tesla. 
The 3 Tesla model was introduced recently
to accommodate a higher luminosity.
The geometry of the 3 Tesla model is shown in 
Fig.\ref{jim-3tesla.eps}.  
In both models, realistic shapes of the
detector and structural components are implemented
with the exception of the 3-Tesla version of CDC
and the Intermediate Tracker in both models.
Geometry components and its status are summarized 
in Table.~\ref{tab:jimgeometry}.

\begin{table}[hb]
\begin{center}\begin{minipage}{\figurewidth}
\caption{\label{tab:jimgeometry}\sl
Detector components and its status defined in JIM.}
\end{minipage}\end{center}
\begin{center}
\begin{tabular}{l l l }
Name & 2-Tesla & 3-Tesla \\
Central Tracker & Full & Cylinder \\
Intermediate Tracker & Cylinder & Cylinder \\
Calorimeters & Full & Full \\
Vertex & Full & Full and Cylinder \\
Muon & Full & Full \\
Mask and beam pipe & Full & Full \\
Magnet and return yokes & Full & Full 
\end{tabular}
\end{center}
\end{table}

Hit positions of trackers created by JIM are exact hit information of particles
and their smearing is performed by the analysis routines
provided by JLCSIM.  
There are also programs for tracking and clustering,
results of them are stored as a ntuple file with a format 
that is the same as with QuickSim.  
When JIM data are analyzed by the JSF,
the JSF calls these utilities to store their data as 
instances of the objects
in the JSF.

\subsection{Comparison of QuickSim and JIM}
Since QuickSim is an independent program of JIM, 
their performances are compared in terms of momentum resolution, 
impact parameter resolution, and calorimeter resolution,
for the case of the 3 Tesla detector parameters.

To compare the momentum resolution, we generate a 
single muon track at 90$^\circ$ with various momenta
and the momentum resolution with and without VTX constraint 
are compared.  
Since the CDC geometry in JIM is an idealized cylindrical one, 
momentum resolutions of QuickSim and JIM should be
essentially the same except for 
the effect of the ionization energy loss and
large angle multiple scattering both of which 
are not considered in QuickSim.  

\begin{figure}
\centerline{\epsfxsize=12.0cm \epsfbox{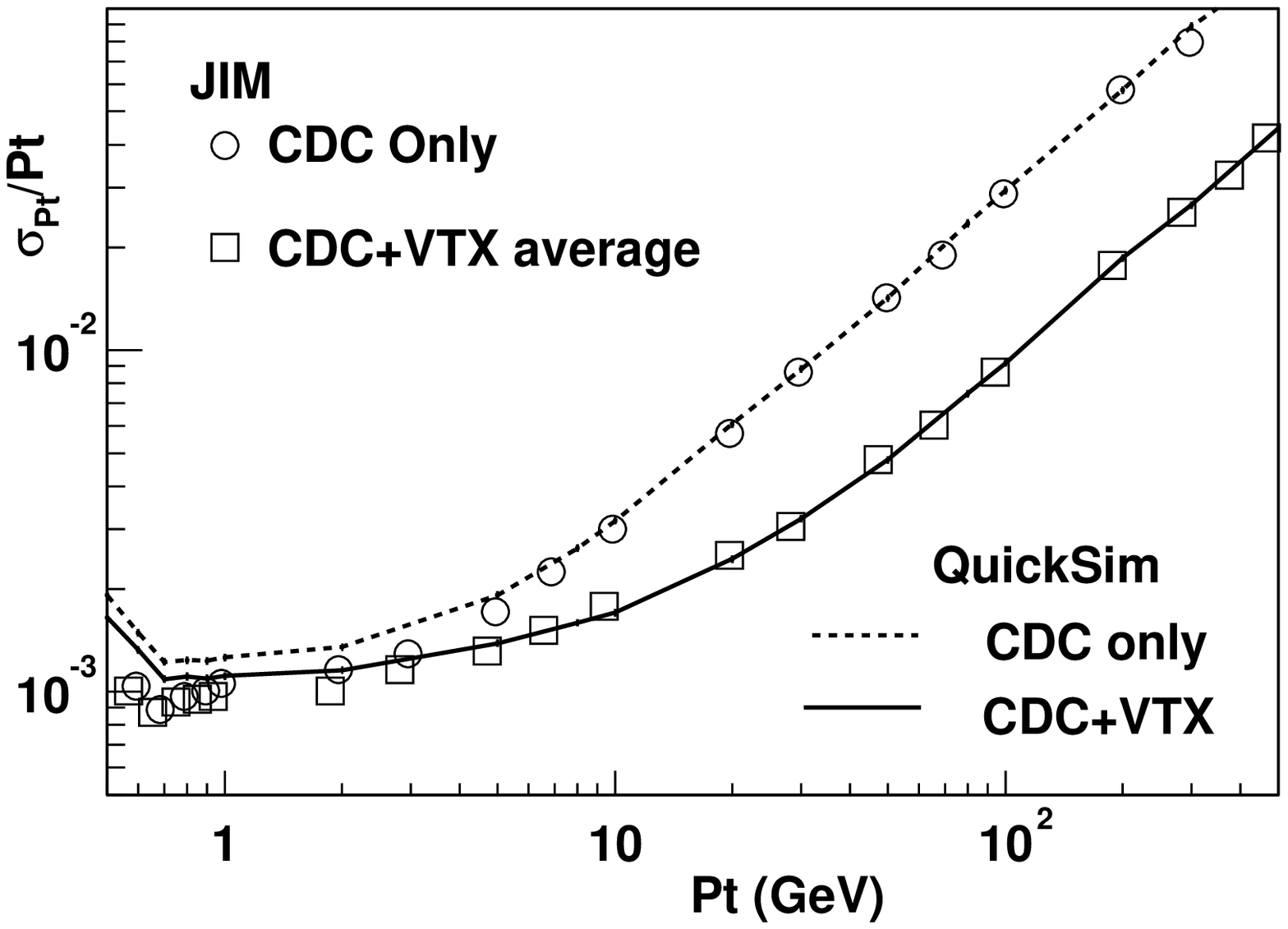}}
\begin{center}\begin{minipage}{\figurewidth}
\caption{\sl\label{ptres-compare.eps}
Momentum resolutions obtained by QuickSim and JIM.}
\end{minipage}\end{center}
\end{figure}

To get the reconstructed momentum of each CDC track
simulated by JIM, we made a five-parameter helix fitting
to the CDC hit points. 
There are materials of 5\% radiation lengths in front of CDC, 
which causes energy loss of about 2 MeV for a 10 GeV muon.
When the momentum resolution of CDC is compared,
this shift is ignored.  
For the combined track fitting, however, this shift cannot be 
neglected especially in the low momentum region.
For the CDC-VTX combined track fitting of JIM data, 
we shift the momentum by amount of the average energy loss 
and the error matrix of the helix parameters is enlarged
to include the effect of multiple scattering.
The so obtained CDC helix parameter vector is then
averaged with the corresponding VTX helix parameter vector.
Resolutions are compared as a function of the track momentum
in Fig.~\ref{ptres-compare.eps}, which shows good agreement
both in the cases of CDC only and CDC+VTX average.
In the momentum region below 1 GeV, the resolution of JIM 
is better than QuickSim.  
At low momenta, tracks curl up in CDC volume and 
produces a large number of hit points. 
In QuickSim, the number of CDC hits
are, however, restricted to the number of wires
hit by the first radial segment
for conservative simulation,
while in JIM this restriction is absent.  
The better performance of JIM in the low momentum region 
can be attributed to this difference,
but further verification is necessary.

\begin{figure}
\centerline{\epsfxsize=12.0cm \epsfbox{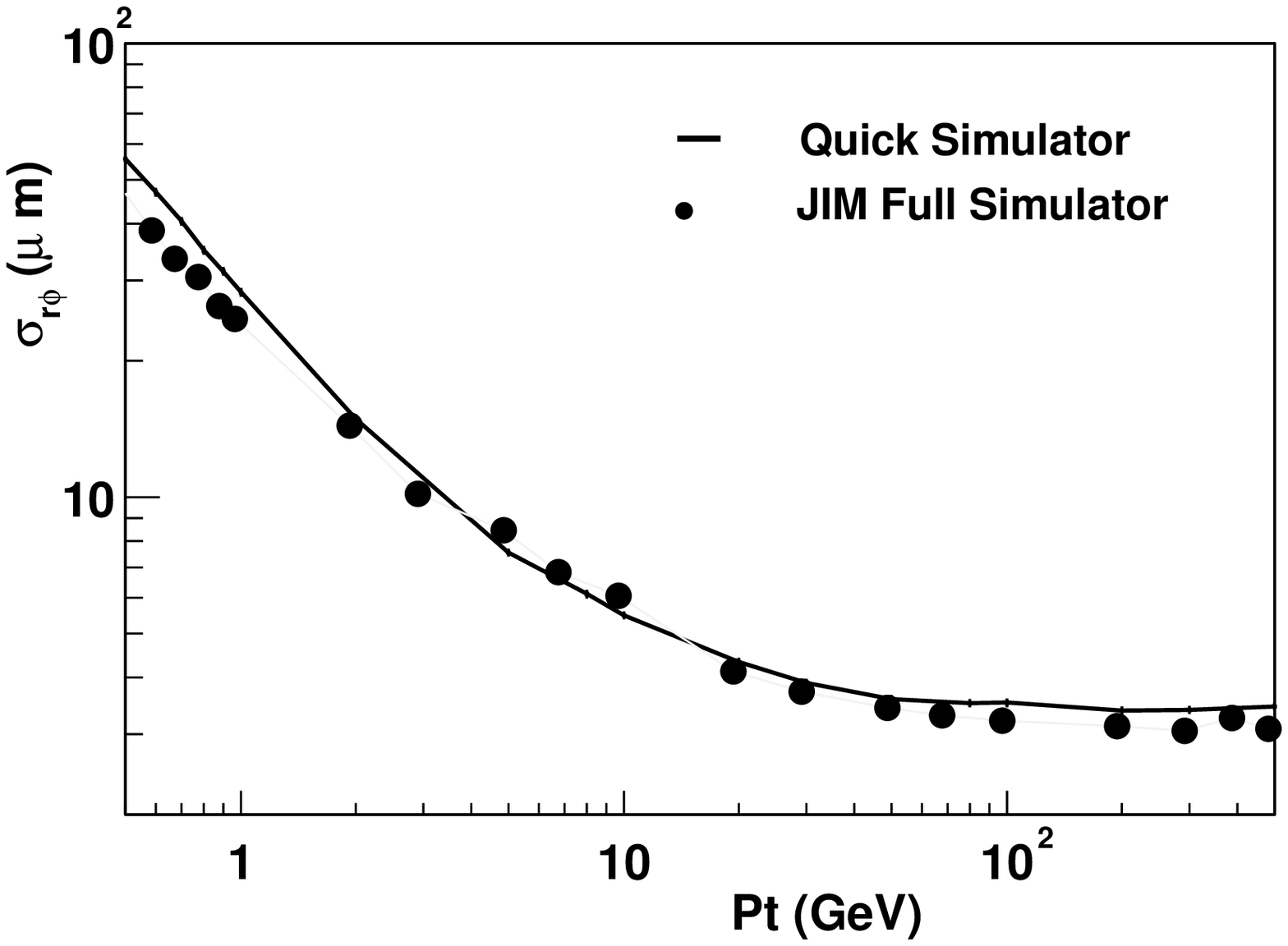}}
\begin{center}\begin{minipage}{\figurewidth}
\caption{\sl\label{ipres-compare.eps}
2-D impact parameter resolutions obtained by QuickSim and JIM.}
\end{minipage}\end{center}
\end{figure}

The 2-D impact parameter resolutions are also compared between 
QuickSim and JIM in Fig.~\ref{ipres-compare.eps}, showing good 
agreement.

\begin{figure}
\centerline{\epsfxsize=14.0cm \epsfbox{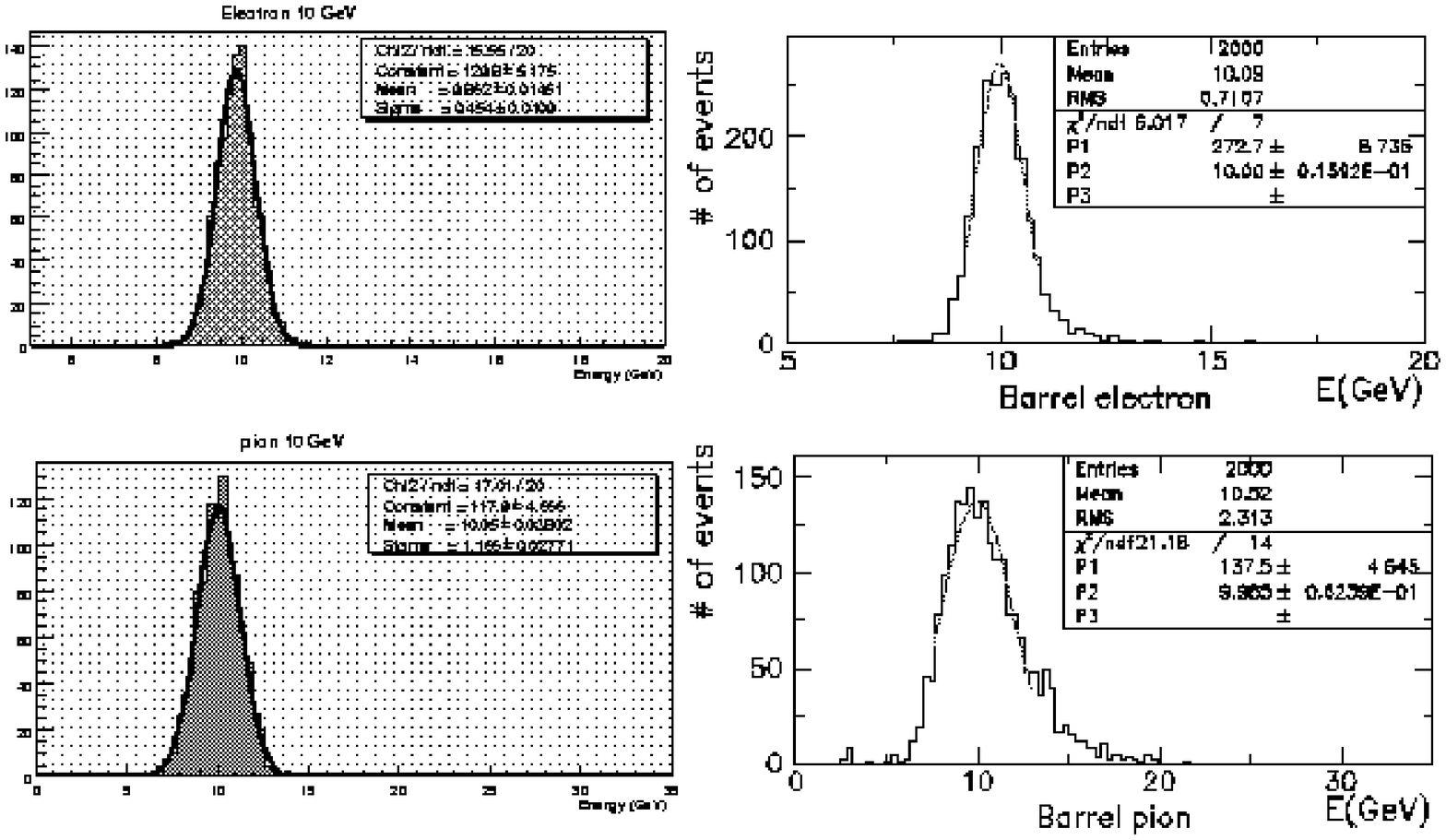}}
\begin{center}\begin{minipage}{\figurewidth}
\caption{\sl\label{calresol.eps}
The energy distribution of calorimeter
by QuickSim compared with that of JIM.}
\end{minipage}\end{center}
\end{figure}

The energy resolutions of the calorimeter 
as simulated by QuickSim 
and JIM are compared in Fig.~\ref{calresol.eps}
for 10 GeV electrons and pions.  
The energy resolutions for the the electrons are mutually consistent, 
but that for the pions in JIM is larger than the QuickSim result.  
The energy resolution of the hadron calorimeter in
JIM is even worse than beam test results,
which suggests necessity of 
further tuning of the calorimeter
simulation in JIM.  

\subsection{JUPITER : a GEANT4-based Simulator}
As a new approach to full detector simulation, 
we are developing a GEANT4-based detector simulator called
JUPITER (JLC Unified Particle Interaction and Tracking EmulatoR).
One of the design goals of JUPITER is to allow easy modification 
of geometrical parameters that describe
various detector components, 
since its currently conceived most important usage is 
to optimize the detector parameters for the JLC.
To this end, we decided to organize base classes of JUPITER
to mimic detector building process: assembling of each device
and its subsequent installation,
thereby facilitating easy replacement,
uninstallation and reinstallation, of any detector component.
In our design, every geometrical component, 
even the World Volume, must be a derived class
from a base class called {\tt J4VDetectorComponent}.
This pure virtual base class
specifies basic interfaces that
any detector component must be equipped with:
an {\tt Assemble()} method
to define a shape and material of the object as
a {\tt G4Solid} and a {\tt G4LogicalVolume},
an {\tt InstallIn()} method to install it into its
mother volume, thereby appending positional
information to the object as a {\tt G4VPhysicalVolume}.
The base class also provides an automatic naming system
for the hierarchy of the detector components
as well as some handy interfaces to frequently used shapes
including tubes.
For parallel and distributed development of various 
detector components, JUPITER also provides a base
class called {\tt J4VMaterialStore} for material definitions.
By instantiating a class object derived from this base class,
a developer of each detector component can define materials
without conflicting with developers of the other parts.

\begin{figure}
\centerline{\epsfbox{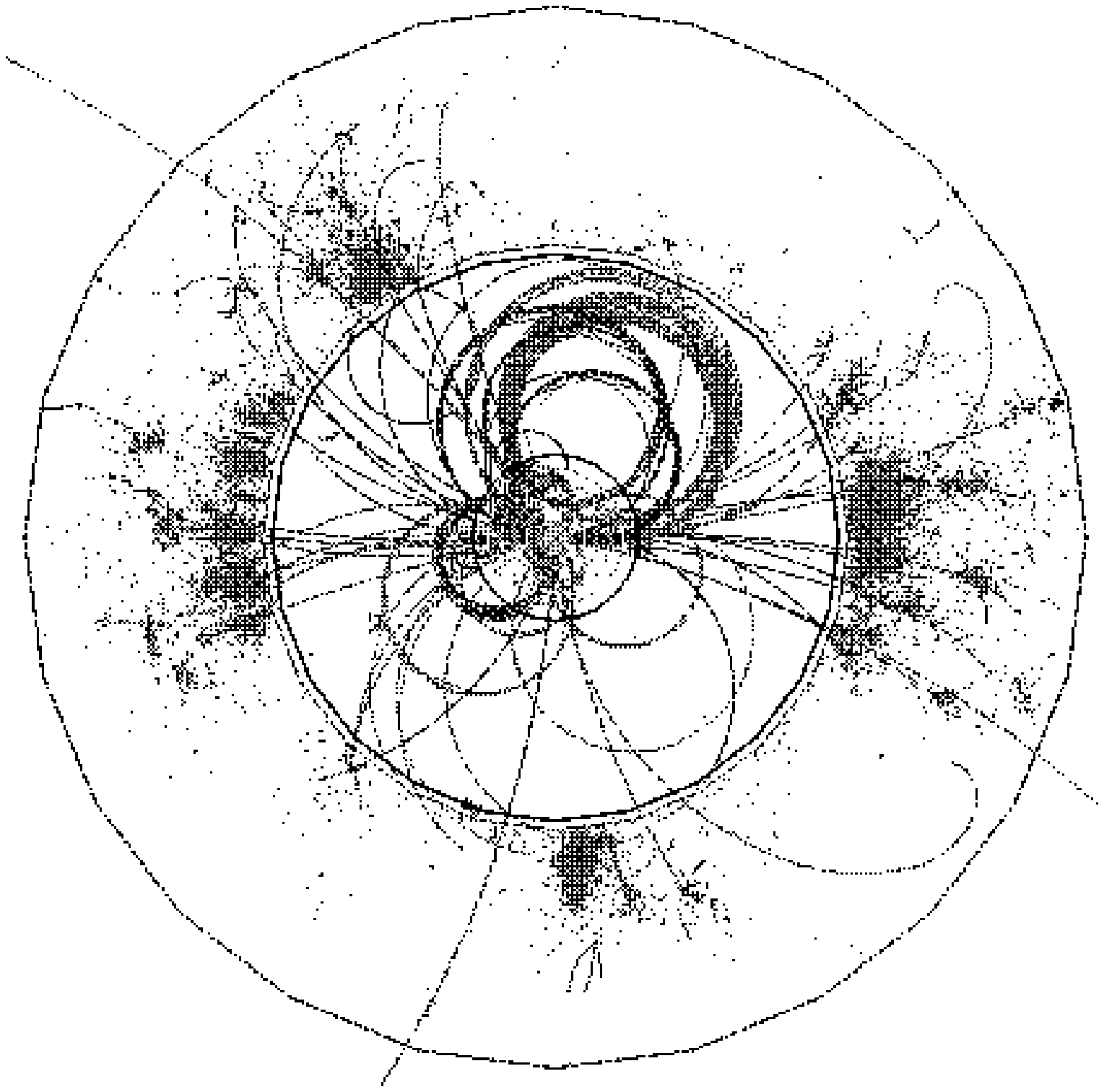}}
\begin{center}\begin{minipage}{\figurewidth}
\caption{\sl\label{det-sim-j4event}
Event display of $e^+e^-\rightarrow ZH$ 
at $\sqrt{s}=300$ GeV simulated by JUPITER.
Only charged tracks are shown.}
\end{minipage}\end{center}
\end{figure}

Using these classes, we started with the development 
of the CDC part of JUPITER,
which can now handle multi-particle final states as from
the process $e^+e^-\rightarrow ZH$.
A sample event display is shown in Fig.\ref{det-sim-j4event}
for this process generated at $\sqrt{s}=300$ GeV.
The implementation of the VTX detector has begun recently,
but those of the other detector components, not to mention
the analysis packages for simulated data, 
are yet to be developed.
It should also be noted that the validity of GEANT4
itself should be checked out, which needs dedicated studies.

%% file: options/main.tex
\chapter{$\gamma\gamma$, $\lowercase{e}^-\gamma$, $\lowercase{e}^-\lowercase{e}^-$}
\label{chapter-options}
\newcommand{\gamgam}{$\gamma \gamma$}
\newcommand{\eg}{$e \gamma$}
\newcommand{\epm}{$e^+e^-$}
\newcommand{\emm}{$e^-e^-$}
\newcommand{\x}[1]{$\times 10^{#1}$}
\def\lsim{\:\raisebox{-0.5ex}{$\stackrel{\textstyle<}{\sim}$}\:}
\def\gsim{\:\raisebox{-0.5ex}{$\stackrel{\textstyle>}{\sim}$}\:}

\input options/intro.tex

\section{Physics}\label{options/chapter:physics}
\subsection{Higgs Bosons}\label{options/sect:higgs}
\input options/higgs.tex

\subsubsection{$H$-$A$ interference in MSSM}
\input options/haint.tex

\subsection{W Boson}\label{sect:W}
\input options/w.tex

\subsection{Top Quark}
\input options/top.tex
\input options/top-techni
\subsection{Supersymmetric Particle Productions}
\input options/susy.tex
\subsection{Excited Leptons in $e\gamma$ Collision}
\input options/estar.tex
\subsection{CP Violation Studies by Linearly Polarized Beams} \label{sec:cp}
\input options/linear-pol.tex

\subsection{Hadronic Cross-sections in \gamgam\ collisions}
\input options/hadron.tex
\subsection{Luminosity Measurement}
\input options/lum-meas.tex
\section{Technical Issues}
\input options/tech.tex

\subsection{Interaction Region}
\input options/ir.tex
\subsection{Lasers}
\input options/laser-koba

\input options/laser-llnl
\section{Summary and Future Prospect}
\input options/summary.tex

\input options/ack.tex

%% file: options/intro.tex
\section{Introduction}

At the time of its operation, the \epm~ linear collider will be 
 source of the highest energy 
and the most intense electron beams.
Using these electron beams, it is also possible to construct 
facilities to study high energy interactions other than 
the \epm~ collisions.

The simplest case is an \emm~ interaction which is
realized by just replacing the positron beam with the
electron beam. 
The other possibility is to convert the electron beam to the 
photon beam by the backward Compton scattering, 
providing facilities of 
the \gamgam~ interaction and/or the \eg~ interaction. 

In the JLC working group, 
feasibility of these 'optional' interaction has been studied
and the activity was mainly focused on the \gamgam~ and
the \eg~ interactions.
A summary of the recent study was published as a KEK Report\cite{options/JLCgg}.
The activities are also found in the
North America and the European LC working group and 
their recent works  were described  in 
refereces\cite{options/JLCgg,options/NLC,options/TESLAgg,options/e-e-99}.

As results of these studies, the overall picture of the physics 
opportunities with the options is fairly understood 
and we see many interesting and important subjects to be 
explored with them. 
They have potential to give an additional view  or to give a complementary 
information to the \epm~ interactions. 
Therefore, full understanding the feasibility of 
the \gamgam~, \eg~, and \emm interactions both in physics and in technical aspects 
are important as a part of the linear accelerator project.

On the other hand, since the linear collider communities have 
paid their attention mainly to the \epm collision so far,  
the understanding of the feasibility of these optional interaction is not the same
level as those in the \epm~ interaction.

This chapter aims to describe overall view of the optional interaction mainly 
of \gamgam~ and \eg~ interaction with the current knowledge, i.e, 
\begin{itemize}
\item Overview of the \gamgam~ and \eg~ interaction.
\item Machine parameters and luminosities
\item Physics opportunities
\item Technical issues
\item Future prospects
\end{itemize}

\section{\gamgam~ and \eg~ Collider}

\begin{figure}[tbp]
\centerline{\epsfxsize=14cm \epsfbox{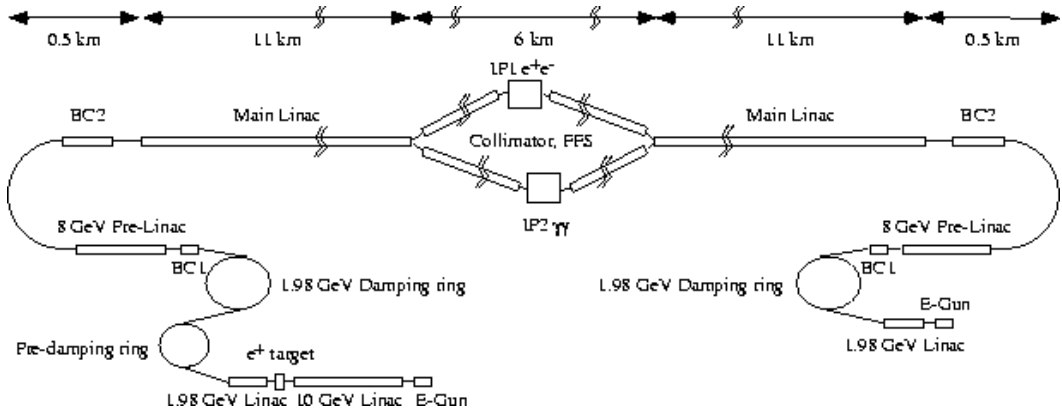}}
\caption{\sl \label{options/fig:jlc}
Schemtic of the JLC with the second interaction region.}
\end{figure}

The JLC accelerator scheme with the second interaction region is illustrated
in Fig. \ref{options/fig:jlc}.

\begin{figure}[tbp]
\centerline{\epsfxsize=9cm \epsfbox{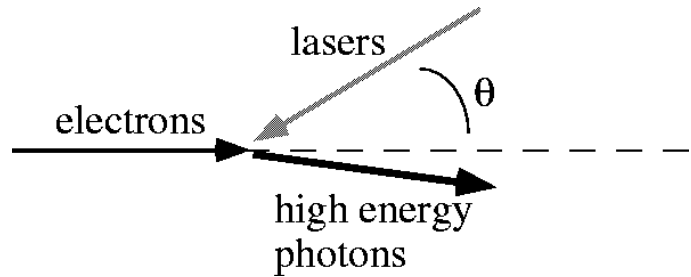}}
\caption{\sl \label{options/fig:compton}
Kinematics of inverse-Compton scattering.}
\end{figure}

\begin{figure}[tbp]
\centerline{\epsfxsize=12cm \epsfbox{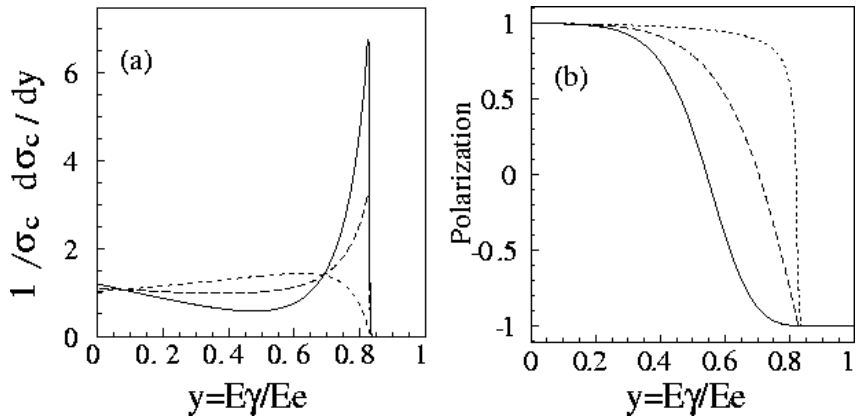}}
\begin{center}\begin{minipage}{\figurewidth}
\caption{\sl \label{options/fig:spectrum}
Calculated energy and polarization
of back-scattered photons by 
laser-Compton scattering for different combination of the laser and 
the electron beam polarization, i.e., $P_L P_e = -1$ (solid), 
$P_L P_e = 0$ (dashed) and $P_L P_e = 1$ (dotted), respectively.  }
\end{minipage}\end{center}
\end{figure}

In  the \gamgam~ and the \eg~ colliders,  photon beams are
generated by the backward Compton scattering of electron- and 
laser-beams just before the interaction point. 
The maximum energy of the generated photon is expressed as;
\begin{equation}
E_\gamma\big |^{max} ={x \over {x+1}}E_e.
\label{options/eqn:e-gam}
\end{equation}
Here, $x$ is a kinematics parameter of the Compton scattering, 
defined as
\begin{equation}
x={{4E_e\omega _L} \over {m_e^2}}\cos ^2(\theta / 2),
\label{options/eqn:x}
\end{equation}
where $E_e$, $\omega_L$ and $\theta$ are the
electron energy, laser photon energy and angle between 
the electron beam and the laser beam (see Fig.~\ref{options/fig:compton}).
A calculated photon energy spectrum generated by the backward
 Compton scattering is
shown in Fig.~\ref{options/fig:spectrum}.
As  seen from the figure, energy distribution of the
photon depends on the polarization of the electron and the laser beam,
{\it i.e.},
one can obtain a broad energy distribution 
or peaked distribution
by controlling the polarization.

According to equation (\ref{options/eqn:e-gam}), the maximum photon energy 
increases as $x$ becomes higher. 
However, when  $x$ exceeds $2(1+\sqrt{2})\approx 4.83$, the
energy of a Compton-scattered photon and a laser photon   
system exceeds the threshold of \epm pair creation.
This pair-creation process could waste generated high-energy photons
and could be an additional source of detector background
so that, at least as a first assumption, $x$ parameter is 
restricted to be smaller than 4.83, corresponding to 
the maximum photon energy of about 80\% of the electron energy. 

Since a precise estimation of the luminosity of \gamgam ~colliders requires
a detail consideration of the laser-Compton scattering and  geometry 
of the interaction region, which need a simulation of the laser-Compton
scattering as well as beam-beam interaction,
we estimate here the luminosity of \gamgam ~colliders by a simple
analysis.
With the typical parameters of the photon collider,
the Compton-conversion probability for an electron, 
{\it i.e.}, 
mean number of Compton interactions of an electron in a laser pulse, 
is assumed to be 1. 
With this assumption, the conversion efficiency $k$, 
is calculated as;
\begin{equation}
k = \sum\limits_{n = 1}^\infty  {P_n^1  = 1 - e^{ - 1} }  \approx 0.63
\label{options/eqn:k}
\end{equation}
where $P_n^1$ is a probability that an electron encounters $n$ laser photons 
when its average is 1.
Thus the number of scattered photons ($N_\gamma$) is $0.63 N$, 
with $N$ being the number of electrons in a bunch.
Usually, we are interested in the high energy part of the spectrum 
and if we take high energy part ($E_{\gamma} > 65\%$) of the spectrum,
the effective conversion efficiency $k'$ can be regarded as $k' \approx 0.3$.

The photon beam spot size at the interaction point is more or less 
same as that of the electron beam since the scattered photons are boosted
toward the electron beam direction. 
Particularly, spot size of the high energy part can be assumed to be the
same as the electron beam. 
Therefore, the luminosity of the \gamgam ~collider is expressed 
approximately as;
\begin{equation}
L_{\gamma \gamma} = k'^2 L_{ee} \approx 0.3^2L_{ee} \approx 0.1L_{ee},
\label{options/eqn:lum_est}
\end{equation}
where  $L_{ee}$ is the geometric luminosity of 
$e^- e^-$ collisions defined as;
\begin{equation}
L_{ee}  = {{N^2 f} \over {4\pi \sigma _x^e \sigma _y^e }}
\label{options/eqn:geom}
\end{equation}
with $N$, $f$, $\sigma _x^e $, $\sigma _x^e $ and $\sigma _x^e $ 
being the number of electrons in a bunch, the bunch repetition, 
the horizontal and the vertical bunch size at the interaction point,
respectively.
It should be noted that the geometric luminosity is not necessary
to be the same as the \epm collider. 
Since the effect of the beam beam interaction is less serious then
\epm ~interaction, 
the geometric luminosity  is possibly higher that the \epm collision.

More precise estimation of the luminosity spectrum can be performed using
numerical simulation with a specific parameters of the electron and 
the laser beam.
As an example, the  parameters for the 
JLC \gamgam ~collider is shown in table \ref{options/tbl:parameters}.
\begin{table}[tbp]
\begin{center}\begin{minipage}{\figurewidth}
\caption{\sl \label{options/tbl:parameters}
Major parameters of $\gamma \gamma$~ collisions at JLC-I.
}
\input options/parameter_tbl.tex

\end{minipage}\end{center}
\end{table}
Parameter sets are one 
for the high energy operation with 250 GeV electron beam 
and two for low energy operation. 
Difference of two parameter sets for low energy operation is the wave length of the laser, i.e., one with $1 \mu m$ and one with $0.37\mu m$. 
For both case, the maximum photon energy is set at 60 GeV with 120 GeV Higgs 
boson in mind.

\begin{figure}[tbp]
\centerline{\epsfxsize=9cm \epsfbox{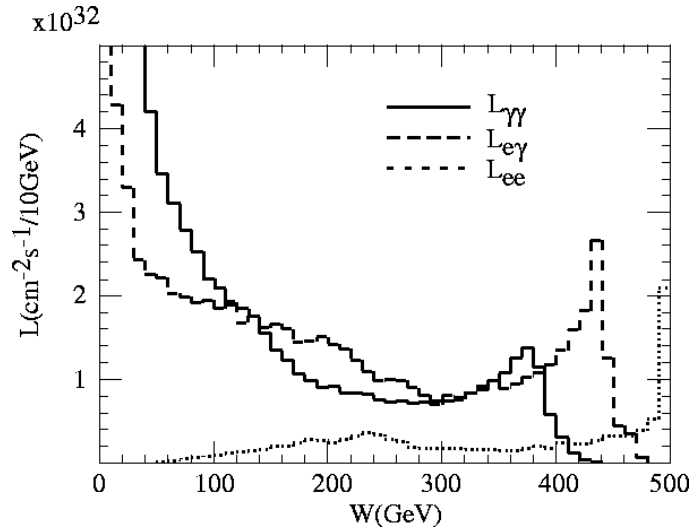}}
\begin{center}\begin{minipage}{\figurewidth}
\caption{\sl \label{options/fig:lumgg}
Luminosity distributions simulated by CAIN for collisions,
the solid, dashed and dotted lines are luminosities of
\gamgam~, \eg ~ and \emm~collisions, respectively, and $W$ is their
center-of-mass energies.}
\end{minipage}\end{center}
\end{figure}

The luminosity distributions simulated using CAIN\cite{options/cain}, 
a simulation program for the laser-electron and the beam-beam 
interaction,
 for the  high energy operation is
shown in Fig. \ref{options/fig:lumgg}. 
We have \gamgam, 
\eg~  and  $e^-e^-$~ interaction 
simultaneously as well as low energy interaction caused by 
beamstrahlung photons.

The other feature of the \gamgam ~interaction is that it is possible prepare
polarized photon beam both circularly and linearly.
For example, the Higgs bosons are produced with the $J_z=0$ combination 
of the initial photons while the background process as 
$\gamma \gamma  \to f\bar f$ 
are suppressed with this helicity configuration.

\begin{figure}[tbp]
\centerline{\epsfxsize=13cm \epsfbox{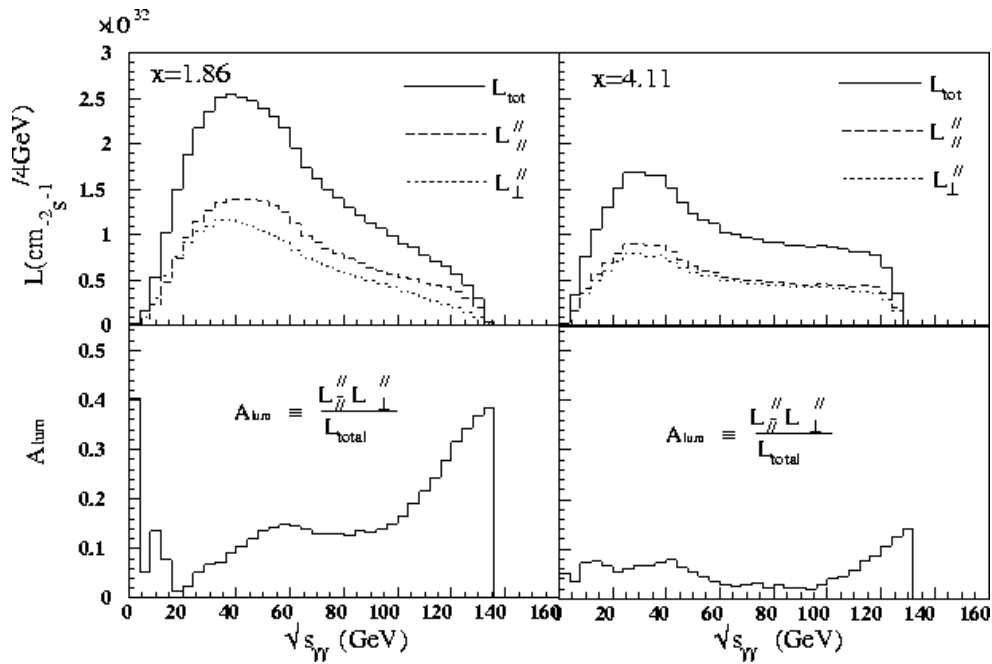}}
\begin{center}\begin{minipage}{\figurewidth}
\caption{\sl \label{options/fig:lum-lin}
Linearly polarized luminosities(top) and the asymmetries(bottom) at
120-GeV \gamgam~ collisions, which are simulated by CAIN, where
$L^\parallel_{\parallel}$(dashed),
$L^\parallel_{\perp}$ (dotted) and  $L_{\rm{total}} \equiv
L^\parallel_{\parallel}+ L^\parallel_{\perp}$(solid). The left and right
figures correspond to the parameter sets of (b) and (c) in
table~\ref{options/tbl:parameters}, respectively. }
\end{minipage}\end{center}
\end{figure}

 It is also pointed out  that the linearly polarized photon beam is 
useful to investigate
CP nature of the Higgs boson
\cite{options/Yang50,options/GF92,options/KKSZ94,options/GK94,options/CK95}.
Fig. \ref{options/fig:lum-lin} shows luminosity distribution
(CAIN simulation)
 when parameters are set for linearly polarized photons.(see \ref{options/sect:higgs}) 

\subsection{\eg ~Collider}

\begin{figure}[tbp]
\centerline{\epsfxsize=9cm \epsfbox{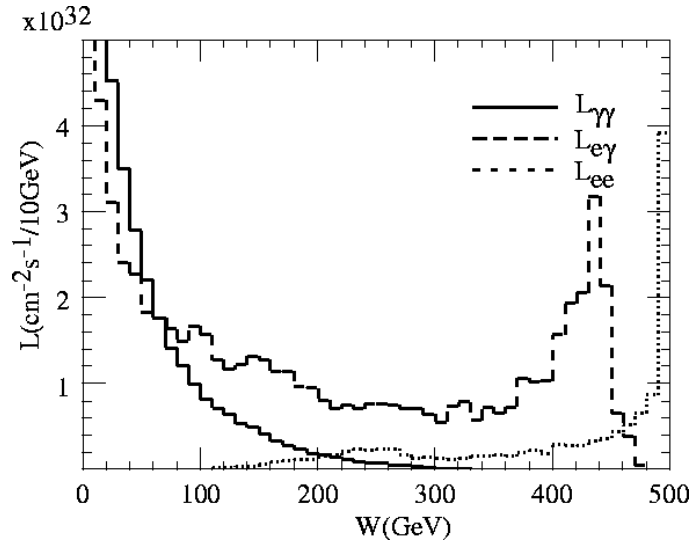}}
\begin{center}\begin{minipage}{\figurewidth}
\caption{\sl \label{options/fig:lum-eg}
Luminosity distributions for \eg~ collider mode 
simulated by CAIN, where
the solid, dashed and dotted lines are luminosities of
\gamgam,~ \eg~ and \emm~ collisions, respectively, and $\sqrt{s}$ is their
center-of-mass energies.}
\end{minipage}\end{center}
\end{figure}

The \eg ~collider mode can be obtained by turning off the laser for 
one side at the \gamgam ~collider operation. 
If we use the same discussion on the luminosity of the \gamgam ~collider, 
the luminosity of the high energy part of the \eg ~luminosity distribution 
can be estimated as;
\begin{equation}
L_{e \gamma} = k' L_{ee} \approx 0.3L_{ee}.
\label{options/eqn:lum_eg}
\end{equation}
The result of the of numerical simulation is shown in Fig.\ref{options/fig:lum-eg}.

\subsection{\emm Collider}

\begin{figure}[tbp]
\centerline{\epsfxsize=9cm \epsfbox{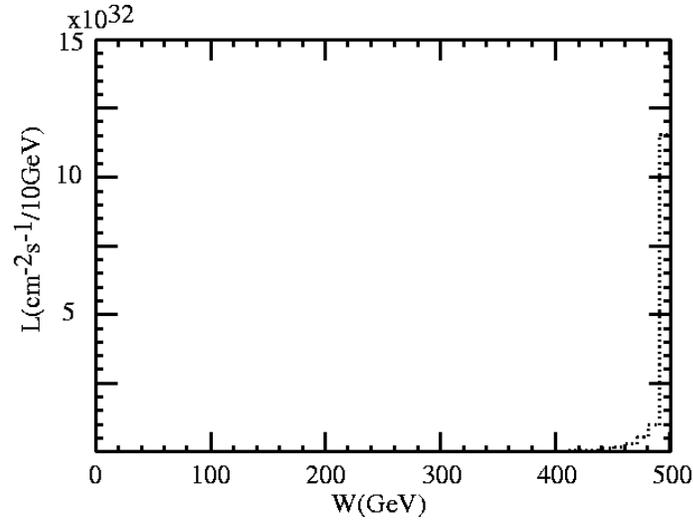}}
\begin{center}\begin{minipage}{\figurewidth}
\caption{\sl \label{options/fig:lumee}
Luminosity distributions for \emm~ collider mode 
simulated by CAIN}
\end{minipage}\end{center}
\end{figure}

The collision of the same sign beam 
suffers from the 
anti-pinch effect due to the repulsive force between them.
It requires a careful tuning of the beam parameters because, 
small beam size increases the beamstrahlung but may not gain the luminosity. 
The discussion on the optimization of the luminosity of the \emm collider
is found in ref. \cite{options/e-e-kath}.
According to the discussion, the typical luminosity of the 
\emm~ collider is around 30\% of the \epm collider. 
A simulated luminosity distribution is show in Fig.\ref{options/fig:lumee}.
As an example of the parameter, a parameter set for 
JLC/NLC (ISG-B)\cite{options/ISG}
\epm ~collider were used with replacing a positron beam to the electrons.
In this particular case, the luminosity was about 20\% of the those of 
the \epm ~luminosity with the same parameters.

\subsection{Summary of the luminosities}

The summary of typical luminosity of \gamgam,~ \eg,~ and \emm collider is 
shown in table \ref{options/tbl:lum}.
\begin{table}[tbp]
\begin{center}\begin{minipage}{\figurewidth}
\caption{\sl \label{options/tbl:lum}
Summary of luminosities of various LC options based on the JLC parameters.
}
\end{minipage}\end{center}
\begin{center}
\input options/lum-tbl.tex

\end{center}
\end{table}
Roughly speaking, typical luminosity of the options are:
\begin{itemize}
\item $L_{\gamma \gamma}^{eff} \approx 0.1 L_{geom}^{e^-e^-}$
\item $L_{e \gamma}^{eff} \approx 0.3 L_{geom}^{e^-e^-}$
\item $L_{e \gamma}^{eff} \approx 0.3 L^{e^+e^-}$
\end{itemize}

It has to be noted that the optimization of these luminosities are still 
possible and we could expect higher luminosity without major change in 
accelerator parameters. 
More detail discussion on the optimization can be found in 
\cite{options/e-e-kath, options/gg2000}.

%% file: options/parameter_tbl.tex
{\tabcolsep=3pt \footnotesize
\begin{tabular}{|llllll|}
\hline
Parameter set & & & (a) & (b) & (c) \\
\hline
\bf{$e^-$} & & & & & \\
Beam energy & $E_e$ & {\rm GeV} & 250 & 103 & 80 \\    
Particles/bunch & $N$ & & 0.95\x{10}& $\leftarrow$ &$ \leftarrow$ \\   
Repetition rate & $f_{rep}$ & {\rm Hz} & 150& $\leftarrow$ &$ \leftarrow$ \\
bunches/pulse & $n_b$ & & 95 & $\leftarrow$ &$ \leftarrow$ \\
Bunch length & $\sigma_z$ & $\mu$m & 120 & $\leftarrow$ &$ \leftarrow$ \\
Bunch sizes at IP & $\sigma_x^*/\sigma_y^*$ & nm  & 680/12.8 & 77/12.2 & 88/10.7 \\
Beta func. at IP & $\beta_x^*/\beta_y^*$ & mm  & 0.5/0.8 & 0.3/0.5 & 0.3/0.3 \\
Norm. emittance & $\varepsilon_{xn}/\varepsilon_{yn}$ & 
nm$\cdot$r& 4500/100 & 4000/60 &$ \leftarrow$ \\
Vertical offset & & $\sigma_y^*$ & 1 &$ \leftarrow$ &$ \leftarrow$ \\
CP-IP distance & $d$& mm& 7.0&3.3 & 2.8 \\
Geom. luminosity & L$^{ee}_{geom.}$  &
cm$^{-2}$s$^{-1}$ & 11.8\x{33} & 6.78\x{33} & 6.79\x{33}\\
\bf{Laser } & & & & & \\
Wave length & $\lambda_L$    & $\mu$m & 1.053 &$\leftarrow$& 0.37 \\
Pulse energy & $E_L$     & J   & 1.0 & $\leftarrow$ & $\leftarrow$ \\
Pulse length & $\sigma_z^L$   & $\mu$m & 230 & $\leftarrow$ & 120 \\
r.m.s spot size & $\sigma_0^L$   & $\mu$m & 3.17 & $\leftarrow$ & 1.53 \\
$x$ parameter &$4\omega_LE_e/m_e^2$  &    & 4.51 & 1.86 & 4.11 \\
\hline
\end{tabular}}

%% file: options/lum-tbl.tex
{\tabcolsep=3pt \footnotesize
\begin{tabular}{|llllll|}
\hline
Luminosity in \x{33} cm$^{-2}$s$^{-1}$ & \gamgam (a)& \gamgam (b) &\gamgam (c) & \eg & \emm \\
\hline
$e^- - e^-$ geometric & 11.8 & 6.8 & 6.8 & 11.8 & 6.3 \\
$\gamma - \gamma$ total & 10.5 & 9.3  & 5.3 & 5.5 & N/A \\
$\gamma - \gamma$ [eff] & 1.4 & 1.6 & 1.0  & 0 & N/A  \\
$e^- - e^-$ total & 1.3  & 0.5  &0.4  &  1.2   &  1.4\\
$e^- - e^-$ [eff] & 0.9  & 0.4  &0.2  & 1.0  & 1.4  \\
$e - \gamma$ total & 8.1  & 5.1   & 3.4  & 6.8  & N/A \\
$e - \gamma$ [eff] & 2.5  & 2.1   & 1.1  & 2.3 & N/A  \\
\hline
\end{tabular}}

%% file: options/higgs.tex
  The search for and study of the Higgs boson, the last missing
member of the standard model family, are among the most important
tasks for future colliders, such as the Large Hadron
Collider (LHC) and \epm linear colliders.

  The interaction of high-energy photons at a \gamgam~
collider~\cite{options/gin1,options/gin2,options/gin3,options/tel1,options/tel2,
options/watanabe,options/JLCgg,options/gold} provides an
opportunity to study the Higgs boson. 
As described later, the number of events expected in the \gamgam~ 
interaction is expected to be 5000 per 10fb$^{-1}$ 
for a Higgs boson of 120GeV, while it is
approximately 1000 per 10fb$^{-1}$ in an \epm collision.
The Higgs-boson physics at the
\gamgam~ collider has been studied by several authors, and it has been
shown that studies of  the intermediate-mass Higgs in the mass
region $M_{W} < M_{H} <2M_{W}$ through the $\gamma\gamma \to H \to
b\bar{b}$ process is complementary to an $e^{+}e^{-}$ linear collider
or a hadron collider, such as study of the CP property, or
exploring physics beyond the standard model through 
measurements of the two-photon decay width
\cite{options/bor92,options/bor93,options/bor94,options/boudjema,
options/belanger,options/wata94,options/wata95,options/jik95,options/jik96,options/NLC}.

\begin{figure}[tbp]
\centerline{\epsfxsize=5cm \epsfbox{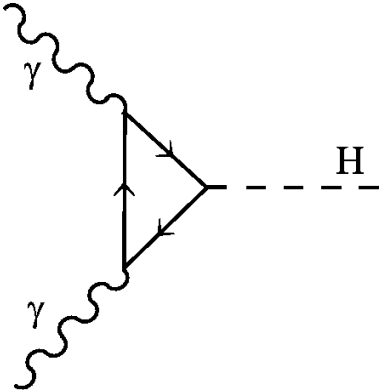}}
\caption{\sl \label{options/fig:cou}
Coupling of the Higgs boson to two photons.}
\end{figure}

Figure~\ref{options/fig:cou}
shows a schematic diagram of the coupling of the Higgs boson to two
photons.
Since two photons do not directly couple to the Higgs boson, but
do only through a loop diagram of massive charged particles,
any kind of massive charged particles contributes to the two-photon
decay width of the Higgs
boson \cite{options/gun90,options/oku82}.
In the standard model, the dominant contributions come from 
the top and W-boson loop. 
It should be noted that since the contribution of bosons and fermions
in the loop interfere destructively,
it is expected to observe this destructive interference
between the top and W-boson loop for the Higgs of 500-700 GeV.
If the gauge coupling of the Higgs and W boson is measured
in the \epm interaction with good precision, a measurement of
the two-photon decay width of the Higgs boson could provide
an information concerning the Yukawa coupling of the Higgs boson and
the top quark.

Regarding physics beyond the standard model,
a deviation of the two-photon
width from the standard model prediction indicates an additional
contribution from unknown particles, and is a signature of physics
beyond the standard model.
For example, the supersymmetric
extension of the standard model (MSSM) has additional charged
particles, such as scalar fermions, charged Higgs bosons and
charginos. Since the masses of these particles partly originate
from the Higgs mechanism of the electroweak symmetry breaking,
the presence of theses particles results in a shift of the
two-photon decay amplitude of the Higgs boson
from its value of the SM.
In fact, a minimal extension of the standard model(MSSM)
predicts  the ratio of two-photon decay widths ($\Gamma(h^{0} \to
\gamma\gamma,\mbox{MSSM})/\Gamma(H \to \gamma\gamma, \mbox{SM})$)
to be as much as 1.2 for the lightest Higgs boson having a mass of 
120 GeV Higgs \cite{options/bor93}.

\begin{figure}[tbp]
\centerline{\epsfxsize=9cm \epsfbox{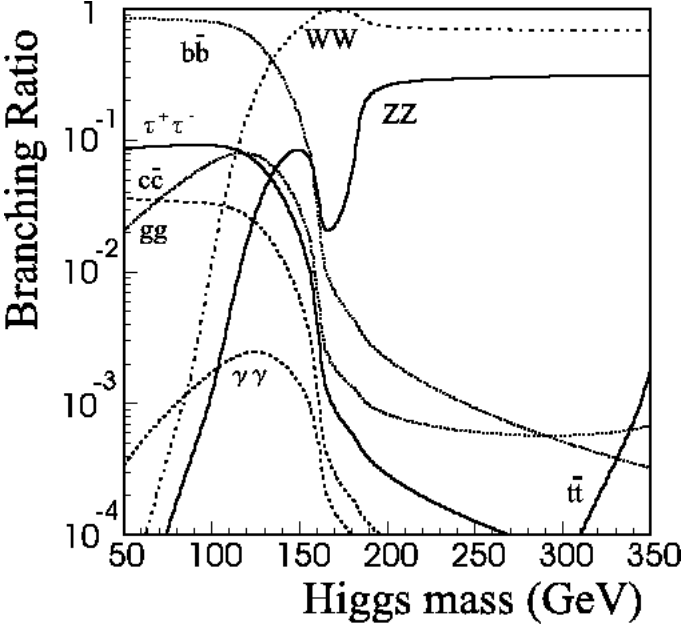}}
\begin{center}\begin{minipage}{\figurewidth}
\caption{\sl \label{options/fig:bra}
Branching ratios for the standard model Higgs boson for a
176-GeV top-quark mass.}
\end{minipage}\end{center}
\end{figure}

\subsubsection{SM Higgs Boson}

 The intermediate-mass Higgs boson in the SM mainly decays into
a $b \bar{b}$ pair, as is shown in Fig.~\ref{options/fig:bra},
and the daughter b-flavored hadrons will be easily identified
due to their long lifetime; therefore, $b \bar{b}$ events are
the best signal of the intermediate-mass Higgs bosons.
Although the main background may be the continuum
$\gamma \gamma \to q \bar{q}$ processes, 
the background events are dominantly produced by the initial photon
collision in the $J_z = \rm 2$ angular momentum state, and it can be
suppressed by controlling the polarization of the colliding photon
beams. Simultaneously, beam polarization enhances the Higgs boson
signals, which are only accessible by the  $J_z = 0$ collisions
\cite{options/gun93, options/bor93}.
The feasibility of a measurement of the two-photon decay
width of the Higgs boson in this mass region has been studied by
Borden {\it et al.}\cite{options/bor92,options/bor94}.

  Recently, several authors reported that the effect of QCD
corrections to $\gamma\gamma \to q\bar{q}$ is large,
since the helicity suppression in the $J_z=0$ channel
does not work due to gluon emission.
It could be a
serious source of background in the
intermediate-mass Higgs boson detection,
since some of the three-jet events from the $J_z=0$ state may mimic
two-jet events \cite{options/bor94,options/jik95,options/jik96}.

In this work, we simulate a measurement of the two-photon
decay width of the Higgs boson with a mass of 120 GeV at a
\gamgam~ collider, including the effect of QCD correction for
the  $\gamma\gamma \to q\bar{q}$ process in the manner of
Jikia and Tkabladze \cite{options/jik96}.
To perform a realistic evaluation, the Monte-Carlo programs
CAIN\cite{options/CAIN0, options/ohgaki96, options/cain}, JETSET7.3\cite{options/sjo94} and
JLC-I detector simulator\cite{options/JLC-I} were applied to find the
luminosity distribution of a \gamgam~ collider,
hadronization and selection performance in the detector,
respectively.

Assuming the Higgs boson mass of 120 GeV, a set of parameters
suitable for a study was prepared, as listed
in tabel~\ref{options/tbl:higgs_lum}.
\input options/param.tex

\begin{figure}[tbp]
\centerline{\epsfxsize=9cm \epsfbox{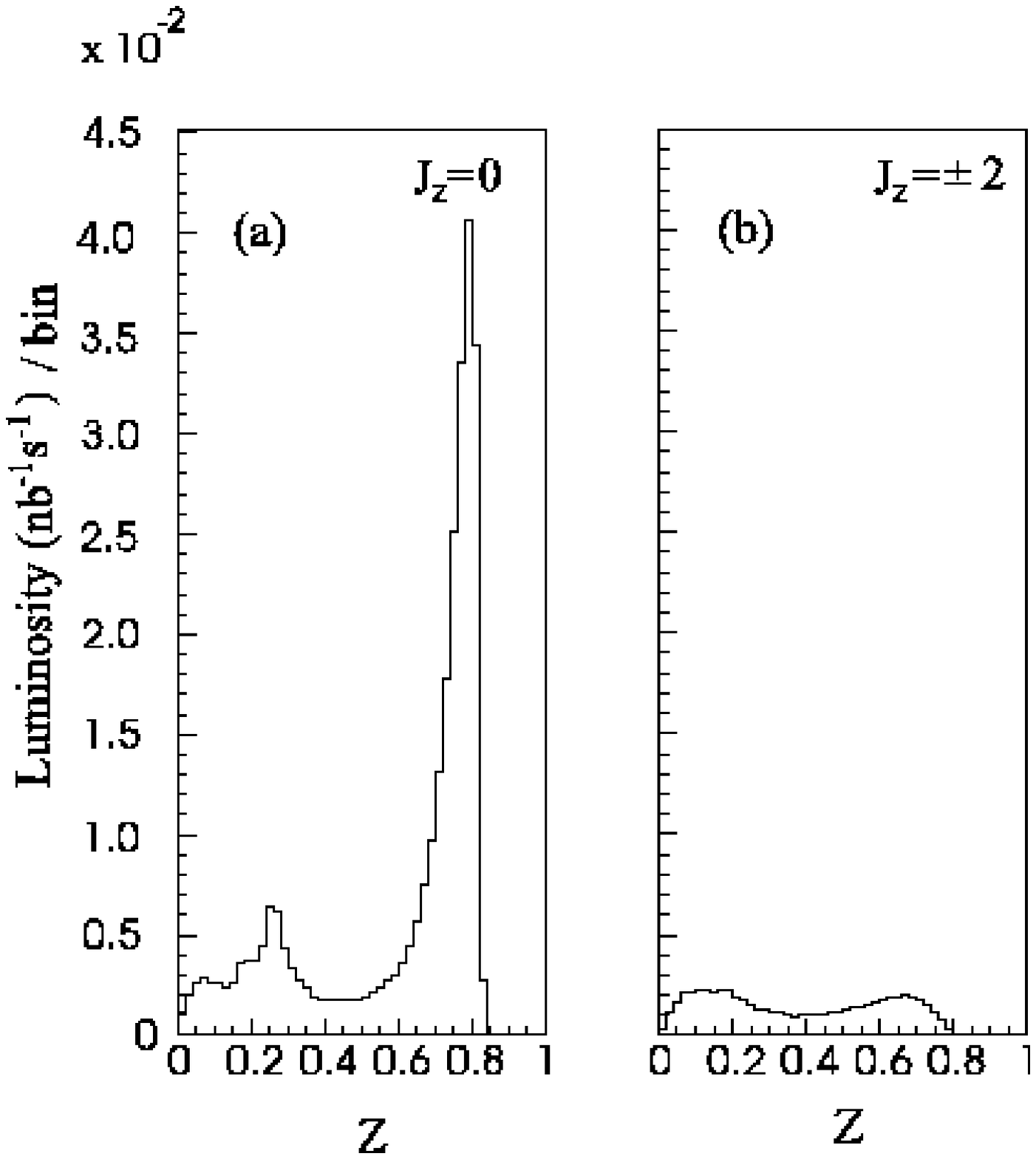}}
\begin{center}\begin{minipage}{\figurewidth}
\caption{\sl \label{options/fig:hlum1}
Luminosity distribution used in the Higgs analysis, where
$z=\sqrt{s_{\gamma \gamma}}/2E_{beam}$. a) and b) are for the spin of
two colliding photon system, $J_z=0$ and $J_z=\pm 2$, respectively.
}
\end{minipage}\end{center}
\end{figure}

\begin{figure}[tbp]
\centerline{\epsfxsize=9cm \epsfbox{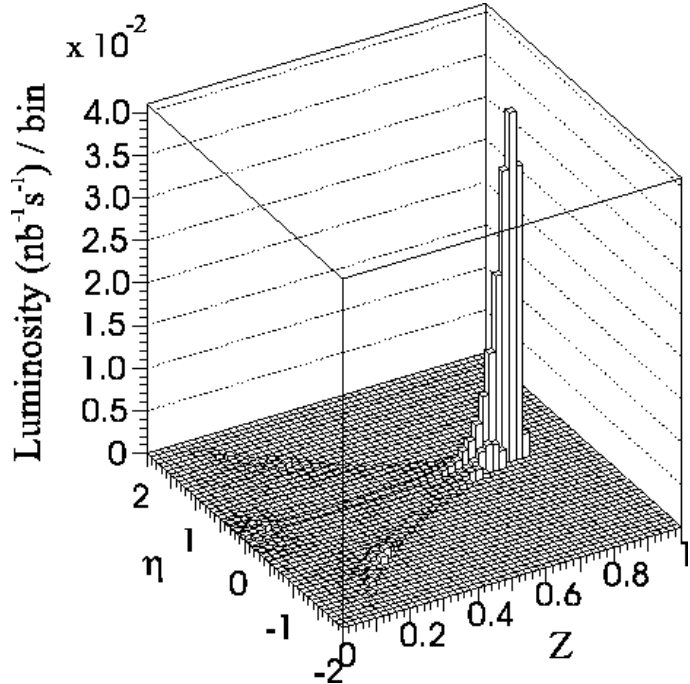}}
\begin{center}\begin{minipage}{\figurewidth}
\caption{\sl \label{options/fig:hlum2}
Rapidity versus the center-of-mass energy of two colliding
photons.}
\end{minipage}\end{center}
\end{figure}

The electron and laser beam energy are 75~GeV and 4.18~eV,
respectively, resulting in the maximum photon energy of
60~GeV.
The combination of the
polarization of the laser ($P_L$) and the electron ($P_e$) are chosen
to be $P_{L}P_{e}=-1.0$ to make the generated photon spectrum
peaked at its maximum energy.
In order to enhance the Higgs-boson production
while suppressing any background from the production of quark pairs,
in the tree level at least, the helicity
combination of two high-energy photons is arranged
so that  $J_{z}=0$ is dominant.
The parameters of the electron and the laser beam are
essentially identical to the
one which is described in a later chapter, except for a treatment of
the spent electron.
In this analysis we assume a CP-IP distance of 1~cm and
that spent electrons are swept  away by a 3~T
external magnetic field.
Thanks to the sweeping of electrons from
the IP, the effect of the electron beam
on the luminosity distribution, such as
electron-electron collisions and the collisions of beamstrahlung
photons and the electron/photon, is negligible,  and a
relatively clean \gamgam~ luminosity distribution is obtained.
Estimating a realistic luminosity distribution,
we performed a
numerical simulation using CAIN\cite{options/CAIN0,options/cain}.
Figure~\ref{options/fig:hlum1} shows the simulated
differential luminosity distribution.
Since the cross section of the Higgs boson
and the background $q \bar{q}$ production depends
strongly on the helicity combination as well as on
the center-of-mass energy of the two-photon system,
information about the differential luminosity,
$$
\left. {{{dL} \over {d\sqrt {s_{\gamma \gamma }}}}} \right|_{J_z},
$$
as shown in Fig.~\ref{options/fig:hlum1}, is very important for
a realistic simulation.
In addition to the center-of-mass energy and spin combination of
the colliding two photons, the rapidity of the system must be
taken into account as well, since the
energy of the colliding two photons
is not always equal, while they are nearly equal for \epm colliders.
Figure~\ref{options/fig:hlum2} shows the center-of-mass energy versus
rapidity of the two-photon system.
In this figure, the rapidity ($\eta$) is defined as
$$
\eta = {1 \over 2}ln{w_1 \over w_2},
$$
where $w_1$ and $w_2$ are the energy of the two photons.
As expected from characteristics of the backward Compton scattering,
the rapidity is smaller at a higher collision energy, and becomes
larger in a lower energy region,
meaning that the Higgs bosons are produced at rest, while background
in lower energy  region tends to escape into a small angle.

The total $\gamma\gamma$
luminosity was calculated to be 3.4$\times
10^{32}\mbox{cm}^{-2}\mbox{s}^{-1}$
over the entire energy region.

  For the intermediate-mass Higgs, the cross section of $\gamma\gamma
\to H \to b\bar{b}$
near to the mass pole can be described by a Breit-Wigner
approximation,
\begin{eqnarray}
\label{options/eq:3}
  \sigma_{\gamma\gamma \to H \to b\bar{b}}=8\pi\frac{ \Gamma(H \to
\gamma\gamma) \Gamma(H \to b\bar{b}) }{ (s_{\gamma\gamma}-M^{2}_{H})
^{2} + M^{2}_{H}\Gamma^{2}_{H} }(1+\lambda_1\lambda_2),
\end{eqnarray}
where $M_H$ is the Higgs-boson mass,
$\Gamma(H \to \gamma\gamma)$, $\Gamma(H \to b\bar{b})$ and
$\Gamma_{H}$ are two photons, $b$ quark pair and total
decay width of the Higgs boson. $\lambda_1$ and $\lambda_2$ are
the initial photon helicities.

  For  a realistic estimation of the Higgs production,
the luminosity distribution calculated by CAIN was convoluted so
as to estimate the number of generated events as
$$
N =\sum\limits_{J_z} {\int {\sigma
_{\gamma \gamma \to H}^{J_z}(s_{\gamma \gamma })}L_{J_z}(s_{\gamma \gamma})
ds_{\gamma \gamma }},
$$
where N and $s_{\gamma \gamma}$ are the number of events and
center-of-mass energy of the \gamgam~ system.
$J_z$ is the longitudinal component of the
spin of the \gamgam~ system and the summation runs
over the $J_z =0,2$ state.
Throughout the analysis, we adopt quark masses of
$m_{b}$=4.3 GeV, $m_{c}$=1.3 GeV, and $m_{t}$=176 GeV.
The branching
ratios ($Br(H \to b\bar{b})$, $Br(H \to c\bar{c})$, and $Br(H \to
gg)$) are taken to be 64\%, 2.7\%, and 8.3\%, respectively, which
were calculated by the HDECAY program\cite{options/spi96}.
In HDECAY program, a $O(\alpha_s^3)$ radiative correction
by the $\overline{MS}$ scheme to the
$\Gamma(H \to b\bar{b})$, was taken into account.
The QCD correction affects the Higgs-boson production
significantly and reduces the expected number of events by almost
50\% of the tree level calculation
around a Higgs-boson mass of 120 GeV.

\begin{table}[tbl]
\begin{center}\begin{minipage}{\figurewidth}
\caption{\sl \label{options/tbl:crs}
Effective cross sections and generated events with the
 luminosity distribution of the photon-photon collider.}
\end{minipage}\end{center}
\leavevmode
\centering
\vspace*{3mm}
{\tabcolsep=3pt \footnotesize
\begin{tabular}{lccc}
\hline
       & $\sigma_{\vert\cos\theta\vert<0.95}$ (pb) &
Events & Simulated events\\
       & (with Luminosity distribution) & (10 $\mbox{fb}^{-1}$) & \\
\hline
Signal events                               & & & \\
  \hspace*{3mm} $\gamma\gamma \to H \to b\bar{b}$  &  0.508 &   5080 &
10000 \\
Backgrounds ($\sqrt{s}>$ 75 GeV)          & & & \\
  \hspace*{3mm} $\gamma\gamma \to H \to c\bar{c}$  & 0.0210 &    210 &
10000 \\
  \hspace*{3mm} $\gamma\gamma \to H \to gg      $  & 0.0633 &    633 &
10000 \\
  \hspace*{3mm} $\gamma\gamma \to b\bar{b}(g)$     &  0.727 &   7270 &
10000 \\
  \hspace*{3mm} $\gamma\gamma \to c\bar{c}(g)$     &   15.1 & 151000 &
50000 \\
\hline
\end{tabular}}
\end{table}

  A convolution with the luminosity distribution
is performed by a Monte-Carlo integration package,
BASES \cite{options/bases}.
Kinematical cuts of $|\cos\theta|<0.95$ and
$\sqrt{s}_{\gamma\gamma}>75$ GeV are imposed for effective event
generation.
The effective cross sections calculated with BASES are
listed in table~\ref{options/tbl:crs}.
The expected number of events of $b\bar{b}$ pairs
from Higgs-boson decay is 5080 with an integrated luminosity of 10
$\mbox{fb}^{-1}$.
For a further analysis,
four-momenta of
$b\bar{b}$ pairs from the Higgs boson decay are generated by the
SPRING\cite{options/bases} program.
Subsequent parton evolution and hadronization is simulated
using the JETSET 7.3 \cite{options/sjo94} program
with a parton shower and a string fragmentation prescription.

  The $\gamma\gamma \to q\bar{q}$ background events are generated in a
similar way as in Higgs production, except that the production
amplitude is calculated by the helicity-amplitude calculation program
HELAS\cite{options/helas}, and that only events with
$\sqrt{s_{\gamma \gamma}}>75$ GeV are generated.
As mentioned in the previous section, QCD corrections have to be
taken into account for a reliable  background estimation.
We use a calculation in \cite{options/jik96} for this purpose,
which includes the soft/hard gluon emission
and virtual corrections.
To simulate a $q \bar{q} g$ event, it is necessary to calculate the
differential cross section of the process; however, it is
reported in \cite{options/jik96} that there is a difficulty to
calculate the differential cross section with a small $y_{cut}$
value, where $y_{cut}$ is the minimum 'distance' between two partons
in terms of the invariant-mass squared to be
regarded as two separated jets.
The definition of the $y_{cut}$ is
$$
y_{cut} \equiv { p_ip_j(1-cos\theta) \over s_{\gamma \gamma} },
$$
where $p_i$, $p_j$ are the momentum of the $i$ and $j$th jet
and $\theta$ is the angle between them.
Thus we used the parton shower evolution in the JETSET program for
event topology and scaled with the  total cross section calculated
in \cite{options/jik96}. We discuss this point again in a later section.

\begin{figure}[tbp]
\centerline{\epsfxsize=9cm \epsfbox{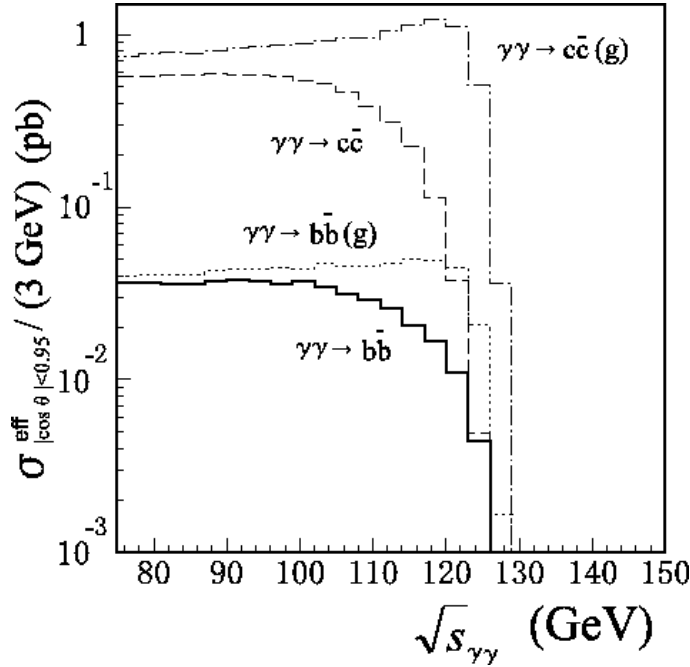}}
\begin{center}\begin{minipage}{\figurewidth}
\caption{\sl \label{options/fig:parton}
Effective cross sections with the luminosity distribution of
the photon-photon collider at
$P_{L}P_{e}=-1.0$. The solid line corresponds to $\gamma\gamma \to
b\bar{b}$, the dashed line to $\gamma\gamma \to c\bar{c}$, the
dotted line to
$\gamma\gamma \to b\bar{b}$ with QCD corrections, and the
dash-dotted line to $\gamma\gamma \to c\bar{c}$ with QCD
corrections. The bin size is 3 GeV.}
\end{minipage}\end{center}
\end{figure}

  The effective cross sections and the number of generated events
with the QCD corrections are listed in table~\ref{options/tbl:crs}. In this
table,
$\gamma\gamma \to q\bar{q}(g)$ indicates the process $\gamma\gamma \to
q\bar{q}$, taking into account of the QCD corrections.
 Figure~\ref{options/fig:parton} shows the effective cross sections
for the $J_z=0$ channel as a function of $\sqrt{s_{\gamma \gamma}}$.
In this figure,
in order to see the effect of the radiative correction,
the effective cross sections with a tree-level calculation is shown
as well. The suppression for $c\bar{c}$ and $b\bar {b}$ production
works as the energy of \gamgam~ system increases, and even
more for a lighter particle at the tree level. 
Once the suppression is removed by gluon emission,
the effect is indeed very large.
Thus, the effect of the radiative correction
is large
around $\sqrt{s}_{\gamma\gamma}$=120 GeV,
and is much more significant for the
$\gamma\gamma \to c\bar{c}$ process than for $\gamma\gamma \to
b\bar{b}$.

  In order to evaluate the performance  of the Higgs-boson
identification at a \gamgam~ collider,
we performed a Monte-Carlo
simulation using the JLC detector simulation program \cite{options/JLC-I}.
The main components used in the analysis are a
vertex detector, a central drift chamber(CDC) and calorimeters. The
$b$-quark tagging by the vertex detector(VTX) is crucial in this
simulation. A CCD (charge coupled device) detector for the VTX is
assumed in the current JLC-I design and its impact parameter
resolution is
\begin{eqnarray}
\label{options/eq:4}
  \sigma_{d}^2=11.4^{2}+(28.8/p)^{2}/\sin^{3}{\theta} \
(\mu\mbox{m}^2),
\end{eqnarray}
where $p$ is the momentum of the charged particle in GeV/c and
$\theta$ is the angle with respect to the beam axis.
\footnote{After this study, the impact parameter resolution 
in the JLC detector was studied and updated which is used
in the top quark analysis described later.
An improvement for the Higgs study is expected but yet to be
estimated.}
The new parameter was used in the top  found to be able to be improved 
The energy resolution of calorimeters, which plays the main role in
the two-jet mass reconstruction, are
$15\%/\sqrt{E}+1\%$ and $40\%/\sqrt{E}+2\%$ for
electromagnetic and hadronic calorimeter, respectively.

  The analysis requires a reconstruction of the two-jet final state
from the Higgs-boson decay.
 In the analysis, the calorimeters are used for jet-mass
reconstruction
and the tracking detectors are used exclusively for b-quark tagging.

First of all, well reconstructed tracks and clusters in tracking devices
and in calorimeters, respectively,
are selected, and only these tracks and clusters are used
in the analysis.
A `good track' requires $|\cos\theta|<0.95$,
$P_{t}>0.1$ GeV/c; also the CDC track and VTX space points
must be successfully linked.
A `good cluster' is
defined as that of  $E>0.1$ GeV and $|\cos\theta|<0.99$ among the
clusters reconstructed in electromagnetic and hadron calorimeters.

The number of good tracks is
required to be greater than 10 in order to choose multi-hadronic
events. JADE clustering \cite{options/jade} is applied 
with $y_{cut}$=0.02 using good clusters
to choose two-jet events.
A cut of $|\cos\theta_{jet}|<0.7$ is applied, where
$\theta_{jet}$ is the angle of the jet with respect to
the beam axis, to ensure that the event is well contained
in the fiducial volume.

To identify the $b\bar{b}$ final state,
a $b$($\bar{b}$) jet is selected by requiring five or
more tracks which have the normalized impact parameter
$d/\sigma_{d}>2.5$ and $d<1.0 \ \mbox{mm}$ in each jet, where $d$ is the
impact parameter.
Events of which both jets are identified as
b-quark jets are regarded as being Higgs-boson candidates.
 With this requirement, the tagging efficiency of $b\bar{b}$
events is
\begin{eqnarray}
\label{options/eq:5}
  \varepsilon_{tag}=\frac{\# \ \mbox{of tagged events}}{\# \ \mbox{of
two-jet events}}.
\end{eqnarray}
Table~\ref{options/tbl:tag} summarizes the tagging efficiencies and the number of
events.
\begin{table}[tbl]
\begin{center}\begin{minipage}{\figurewidth}
\caption{\sl \label{options/tbl:tag}
Tagging efficiencies and the number of events with
10fb$^{-1}$.}
\end{minipage}\end{center}
\leavevmode
\centering
\vspace*{3mm}
{\tabcolsep=3pt \footnotesize
\begin{tabular}{lcc}
\hline
                    & Tagging efficiencies (\%) &  Events  \\
                    & $b\bar{b}$ tag & $b\bar{b}$ tag  \\
\hline
Signal events                 &  & \\
  \hspace*{3mm} $H \to b\bar{b}$            & 59.5 & 582  \\
Backgrounds                   &  & \\
  \hspace*{3mm} $H \to c\bar{c}$            & 19.7 & 7.85 \\
  \hspace*{3mm} $H \to gg      $            & 6.17  & 1.58 \\
  \hspace*{3mm} $\gamma\gamma \to b\bar{b}(g)$ & 60.0  & 278 \\
  \hspace*{3mm} $\gamma\gamma \to c\bar{c}(g)$ & 14.9 & 1320  \\
\hline
\end{tabular}}
\end{table}

In order to enhance the signal, the mass region of the Higgs
signal is chosen in such a way that the statistical significance of
the signal,
$(N_{obs}-\langle N_{bg}\rangle )/\sqrt{N}_{obs}$,
is maximized, where $N_{obs}$
is assumed to be the number of observed events, while
$\langle N_{bg} \rangle $ being the number of expected background events.
As a result,
events in two-jet mass regions of 106
GeV $< M_{jj} <$ 126 GeV are selected.
The selection efficiency is defined as
\begin{eqnarray}
\label{options/eq:6}
  \varepsilon_{sel}=\frac{\# \ \mbox{of selected events}}{\# \
\mbox{of generated events}}.
\end{eqnarray}
They are listed in table~\ref{options/tbl:sel2}.

\begin{figure}[tbp]
\centerline{\epsfxsize=10cm \epsfbox{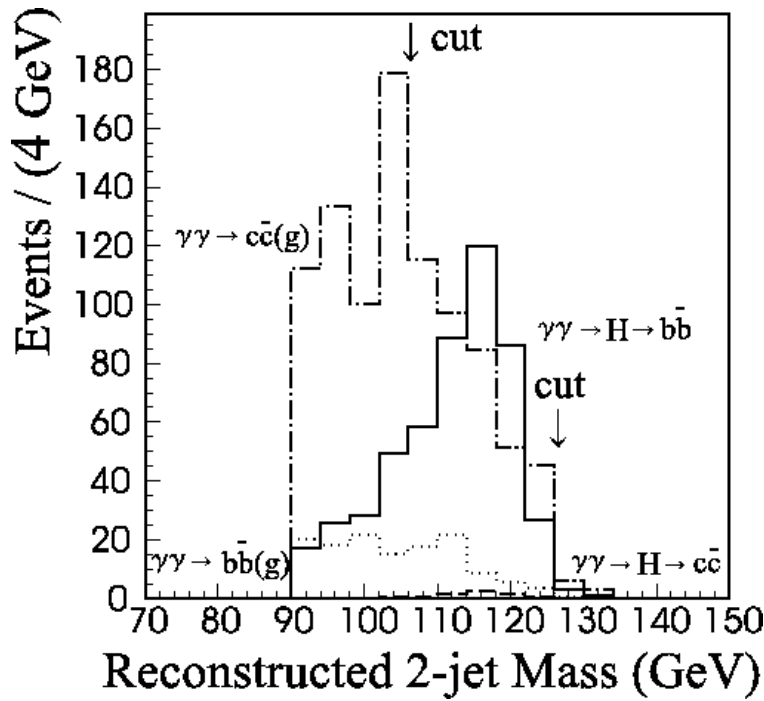}}
\begin{center}\begin{minipage}{\figurewidth}
\caption{\sl \label{options/fig:higgs2}
Mass distribution of two-jets with applying b-tagging
requirements. An integrated luminosity of 10 $\mbox{fb}^{-1}$ and
standard model branching fractions for the Higgs boson are
assumed. The effects of the QCD corrections to $\gamma\gamma \to
q\bar{q}$ as background process are included.}
\end{minipage}\end{center}
\end{figure}

\begin{table}[tbl]
\begin{center}\begin{minipage}{\figurewidth}
\caption{\sl \label{options/tbl:sel2}
Selection efficiencies and the number of events. The effects
of the QCD corrections to $\gamma\gamma \to q\bar{q}$ as background
process are included.}
\end{minipage}\end{center}
\leavevmode
\centering
{\tabcolsep=3pt \footnotesize
\begin{tabular}{lcc}
\hline
   & $\varepsilon_{sel}$ \ (\%) & Events \\
\hline
Signal events                                & & \\
  \hspace*{3mm} $\gamma\gamma \to H \to b\bar{b}$   & 7.48 & 380 \\
Backgrounds                                  & & \\
  \hspace*{3mm} $\gamma\gamma \to H \to c\bar{c}$   & 3.16 & 6.65 \\
  \hspace*{3mm} $\gamma\gamma \to H \to gg$         & 0.230 & 1.46 \\
  \hspace*{3mm} $\gamma\gamma \to b\bar{b}(g)$      & 0.790 & 57.4 \\
  \hspace*{3mm} $\gamma\gamma \to c\bar{c}(g)$      & 0.260 & 394 \\
\hline
Signal to Background with QCD corrections & & 380 / 459 \\
\hline
\end{tabular}}
\end{table}

The
statistical errors,
$\sqrt{N}_{obs}/(N_{obs}-\langle N_{bg}\rangle )$,
of the two-photon
decay width of the SM Higgs boson at $M_{H}$=120 GeV is
found to be 7.6\%, assuming an integrated luminosity of
10${\rm fb}^{-1}$ with S/N=0.83.
Any systematic error related to an estimation of background
is not included in the result.
As a reference, the statistical error
with the tree-level calculation for $q \bar{q}$ was evaluated
to be 6\%.

Toward a further improvement of Higgs boson identification,
it is crucial to reduce any contamination of c-quark jet within the
b-quark jet candidates.
Recently, a sophisticated method for b-quark identification was
developed by the SLD group using the topological vertexing method
\cite{options/vtx}.
It is reported that by using the 
topological vertexing method, the efficiency of b-quark jet
identification and the purity are 70\% and 0.7\%, respectively, for
equal mixing of c- and b-quark jets of 50 GeV within
$|\cos \theta | < 0.7$.
Therefore, tagging efficiency of $b\bar{b}(g)$ and $c\bar{c}(g)$
events is 49\% and 0.005\% with double tagging, and 
91\% and 1.4\% with single tagging method, respectively.
Assuming that these numbers can be kept for
Higgs-boson decay and $b\bar{b}/c\bar{c}$ production
in the $\gamma \gamma$ collision, 
we can estimate the number of signal and the background.

For an estimation, we adapt the number of events surviving 
multi-hadron selection with 
$|\cos \theta_{jet} | < 0.7$.
The restriction for $|\cos \theta_{jet} | < 0.7$ is applied to
simulated events to fit the condition used in a topological 
vertexing analysis in \cite{options/vtx}.
The tagging efficiencies assumed in the estimation and 
the number of estimated events for signal and background 
are tabulated in table~\ref{options/tbl:newvtx}.
\begin{table}[tbl]
\begin{center}\begin{minipage}{\figurewidth}
\caption{\sl \label{options/tbl:newvtx}
Tagging efficiencies and the number of events
 by the topological vertexing method.}
\end{minipage}\end{center}
\leavevmode
\centering
\vspace*{3mm}
{\tabcolsep=3pt \footnotesize
\begin{tabular}{lcc}
\hline
  Tagging efficiencies                   &  Double tag &  Single tag  \\
\hline
b jet  & 0.70 & 0.70 \\
c jet  & 0.007 & 0.007 \\
 \hspace*{3mm} $H \to b\bar{b}$  & 0.49               &  0.91 \\
\hspace*{3mm} $H \to c\bar{c}$   & $5 \times 10^{-5}$ & 0.014 \\
  \hspace*{3mm} $H \to gg $      & 0                 & 0 \\
  \hspace*{3mm} $\gamma\gamma \to b\bar{b}(g)$ & 0.49  & 0.91 \\
  \hspace*{3mm} $\gamma\gamma \to c\bar{c}(g)$ & $5 \times 10^{-5}$  & 0.014  \\
& & \\    
Number of expected events &  Double tag &  Single tag  \\ 
within  $106 GeV < m_{jj} < 126 GeV$ & & \\
\hline
  \hspace*{3mm} $H \to b\bar{b}$ & 832 & 1545  \\
  \hspace*{3mm} $H \to c\bar{c}$ & 0 & 1.3 \\
  \hspace*{3mm} $H \to gg      $ & 0  & 0 \\
  \hspace*{3mm} $\gamma\gamma \to b\bar{b}(g)$ & 121 & 223 \\
  \hspace*{3mm} $\gamma\gamma \to c\bar{c}(g)$ & 0.3 & 101  \\
\hline signal/background & 832/121=6.9 & 1545/325=4.7 \\
\hline
Statistical accuracy &   3.7\% & 2.8 \% \\  
\hline
\end{tabular}}
\end{table}
$\gamma \gamma \to c\bar{c}(g)$ contamination is now
negligibly small, and only the $\gamma \gamma \to b\bar{b}(g)$
process is the major background.
As a result, the statistical accuracy of the two-photon decay width
is  3.7\% and 2.8\% by the double-tagging and single-tagging
condition, respectively, which shows a good improvement from the
conventional b-quark identification method.
It is noted that
the signal-to-background ratio(S/N) in this case is
drastically improved as
$S/N=812/121=6.9$ and $1545/325=4.7$ with the double and single
tagging method, respectively, while it is 0.83 by the previous
analysis.  Since the background from $\gamma \gamma \to c\bar{c}$
is most serious and  affected largely
by an ambiguous radiative correction,
an improvement of the signal-to-background ratio indicates 
a significant reduction of systematic errors coming from the
background estimation.

This result shows that the \gamgam~ collider will be sufficient to
distinguish the Higgs boson of MSSM from one of the standard model
for an intermediate-mass Higgs boson when the ratio of the
two-photon decay widths ($\Gamma(h^{0} \to
\gamma\gamma,\mbox{MSSM})/\Gamma(H \to
\gamma\gamma, \mbox{SM})$) is about 1.2 for a 120 GeV Higgs boson
\cite{options/bor93}.

Recently, G.~Jikia, M.~Melles and their collaborators 
\cite{options/Jikia00,options/Melles} have performed aggressive studies on 
evaluating the effects of the radiative corrections with 
the resummation technique of double-logarithmic corrections, 
as well as the realistic Monte Carlo simulations with optimized event 
selection criteria.  
Applying some appropriate cuts, the statistical error of the 
measurement of the Higgs production cross section is estimated to be 
better than 2{\%} for the Higgs mass between 100 and 140 GeV 
\cite{options/Melles}.  
Such a high precision of this measurement can only be done at the 
photon-photon mode of the linear collider.

The heavy SM Higgs whose mass is heavier than the $W$-threshold 
mainly decays into a $W^+W^-$ pair.  
Since there is no tree diagram of the continuum background process, 
$\gamma\gamma \rightarrow ZZ$ mode is the promising channel to 
detect the heavy Higgs boson if its mass is heavier than 
$2m_Z$ \cite{options/Jikia93,options/Berger93}.  
For the mass range between 140 GeV and 160 GeV, $\gamma\gamma 
\rightarrow WW^*$ mode is found to be available with some appropriate 
experimental cuts \cite{options/BIKOPSS98}.  

The detection methods for the SM Higgs boson is illustrated in the 
middle column of Table \ref{options/T:H-detect}.  

\begin{table}[t]
\begin{center}\begin{minipage}{\figurewidth}
\caption[Detection methods of Higgs bosons]{
\sl \label{options/T:H-detect}
The detection methods of Higgs bosons, which are strongly 
correlated with the dominant decay mode \protect\cite{options/Watanabe99}.  
}
\end{minipage}\end{center}
\vspace{3mm}
\begin{center}
\begin{tabular}{|c||l|l|}
\hline
 & \multicolumn{1}{|c|}{SM} & \multicolumn{1}{c|}{MSSM} \\
\hline \hline
 light & \multicolumn{2}{|c|}{} \\
 {\boldmath $h^0$} 
       & \multicolumn{2}{|c|}%
         {\bf \boldmath $b\overline{b}$ mainly} \\
       & \multicolumn{2}{|c|}{} \\
\hline
 heavy & $WW$ mainly & $WW$,$ZZ$ suppressed \\
 {\boldmath $H^0$} 
       & {\bf \boldmath $ZZ$ useful} 
       & {\bf \boldmath $t\overline{t}$ mainly} \\
       & $t\overline{t}$ ? 
       & $b\overline{b}$ for large $\tan\beta$ \\
\hline
 CP-odd & & no $WW$,$ZZ$ \\
 {\boldmath $A^0$} 
       & \multicolumn{1}{|c|}{---} 
       & {\bf \boldmath $t\overline{t}$ mainly} \\
       & & $b\overline{b}$ for large $\tan\beta$ \\
\hline 
\end{tabular}
\end{center}
\end{table}

\subsubsection{More Complicated Higgs Models}

If the Higgs sector has more complicated structure than that in the SM, 
one has to devise some other schemes to detect the various Higgs 
bosons.  
The minimal supersymmetric extension of the SM (MSSM) is one of the 
typical example of such new model which has a complicated Higgs 
sector.  

In MSSM, there are three neutral Higgs bosons:  
the light CP-even Higgs $h$, the heavy CP-even Higgs $H$ and 
the CP-odd Higgs $A$.  
If we assume a large value of the $A$ mass, 
the light CP-even Higgs boson $h$ is similar to the light SM Higgs 
boson, and its detection can be done by the $b\overline{b}$ decay mode, 
just as the SM Higgs.  
However, for the heavy Higgs bosons, the situation is different.  
The $WW$ and $ZZ$ decay modes of $H$ are suppressed for heavy $A$ 
case, and these decays are forbidden for $A$ boson. 
Instead of them, $t\overline{t}$ decay mode may be useful, if the Higgs 
boson masses are heavier than $2m_t$, and if $\tan\beta \ll 10$.  
For the moderate and large values of $\tan\beta$, the decay mode to 
$b\overline{b}$ (and to $\tau^+\tau^-$ in some cases) is available 
\cite{options/Muhlleitner}.  
The situation is also illustrated in Table \ref{options/T:H-detect}.  
Furthermore, in the wide range of the MSSM Higgs parameter space, 
$H$ and $A$ often have similar values of mass.  
Then a question arises:  
Can we separate two signals of $A$ and $H$ at the photon-photon 
collider?  
Or, how do these signals of heavy Higgs bosons look like at a 
$\gamma\gamma$ collider?  

The separation of $H$ and $A$ signals can clearly be done 
by using the linear polarization of the colliding photons.  
In the frame of the linear polarization of the initial photons, 
$H$ and $A$ states can only be produced from collisions of the parallel 
and the perpendicularly polarized photons, respectively 
\cite{options/Yang50,options/GF92,options/KKSZ94,options/GK94,options/CK95}.  
That is the answer to separate the $H$ and $A$ signals.

%% file: options/param.tex
{\tabcolsep=2pt \footnotesize
\begin{table}[htbp]
\begin{center}
\begin{tabular}{|cccc|}
\hline
{\sl Electron beam parameters} & & & \\
\hline
Number of electrons per bunch & $N_e$  &
$0.63\times10^{10}$ & \\
Number of bunches per pulse & $n_b$  & 85 & \\
Repetition rate    & $f_{rep}$ & 150 & Hz \\
Normalized emittance 
& $\gamma\epsilon_{x,e}$ & $3.3\times10^{-6}$ & m
\\
& $\gamma\epsilon_{y,e}$ & $4.8\times10^{-8}$ & m
\\
R.m.s.~bunch length & $\sigma_{z,e}$ & 90 & $\mu$m \\
Beta functions at I.P. & $\beta^{*}_{x,e}$ & 0.30 & mm \\
& $\beta^{*}_{y,e}$ & 10.0 & mm \\
Beam size at I.P. without conversion 
& $\sigma^{*}_{x,e}$ & 82 & nm \\
& $\sigma^{*}_{y,e}$ & 57 & nm \\
Beta functions at C.P. 
& $\beta^{CP}_{x,e}$ & 0.33 & m \\
& $\beta^{CP}_{y,e}$ & 20 & mm \\
Beam size at C.P. 
& $\sigma^{CP}_{x,e}$ & 2.7 & $\mu$m \\
& $\sigma^{CP}_{y,e}$ & 81 & nm \\
\hline
{Laser beam parameters} & & & \\
\hline
Wavelength  & $\lambda_L$ & 0.297 & $\mu$m \\ 
Photon energy & $\hbar\omega_L$ & 4.18 & eV \\
R.m.s.~pulse length & $\sigma_{z,L}$ & 300 & $\mu$m (1ps) \\
Laser beam size at C.P.
&$\sigma^{CP}_{x,L}$ & 5 & $\mu$m \\
&$\sigma^{CP}_{y,L}$ & 5 & $\mu$m \\
Number of laser photons in a pulse & $N_L$ &
$1.1\times10^{19}$ & \\
Energy per pulse & $\hbar\omega_L N_L$ & 7 & Joule \\
Laser peak power (effective rectangular pulse) & $P$ & 2.0 
& TW \\
Maximum electric field (Gaussian peak) & ${\cal E}_{L,max}$
& $2.2\times10^{12}$ & V/m \\
Nonlinear QED parameter at Gaussian peak & $\xi_{peak}$ & 0.20 &\\
\hline
{$\gamma$ beam} & & & \\
\hline
Number of photons per electron bunch & $N_{\gamma}$ &
$0.41\times10^{10}$ &\\
Beam size at I.P. & $\sigma^{*}_{x,\gamma}$ & 107 & nm \\
&$\sigma^{*}_{y,\gamma}$ & 89 & nm \\
$\gamma$-$\gamma$ luminosity & ${\cal L}_{\gamma\gamma}$ & 
$3.4\times10^{32}$ & cm$^{-2}$s$^{-1}$ \\
Distance between C.P. to I.P. & $L$ & 1.0 & cm\\
\hline
\end{tabular}

\end{center}
\caption{\sl Parameters of the photon-photon collider based on JLC 
for $M_H$=120 GeV.}

\label{options/tbl:higgs_lum}

\end{table}
}

%% file: options/haint.tex
With the circular polarization of the photon beams, 
there is another interesting story \cite{options/AKSW00}.  
The circular polarization of the photon can be expressed in the  
45-degree mixture of the two linear polarizations.  
It means that the collision of the circularly polarized photon 
beams is accessible to both the $H$ and $A$ amplitudes simultaneously.  
A similar situation can be seen in the decay of a Higgs state into 
a fermion pair in the definite helicity states, 
and the helicity of the top quark can be measured statistically 
by looking its decay angle distribution \cite{options/JK89,options/HMW91,options/MP96}.  
Then, the process of $\gamma\gamma \rightarrow t\overline{t}$ 
with circular photon polarization and top helicity detection 
has an interference effect of $H$ and $A$ amplitudes.  
The effect can be never observed if measuring of top quark helicity is 
not performed, because the interference term for $\gamma_+ \gamma_+
\rightarrow t_R \overline{t}_R$ cancels the one for $\gamma_+ \gamma_+
\rightarrow t_L \overline{t}_L$. 
\begin{table}[h]
\begin{center}\begin{minipage}{\figurewidth}
\caption{\sl \label{options/tab:exp}
Helicity dependence of the amplitudes for
$\gamma \gamma \rightarrow H, A \rightarrow t \bar{t}$}
\end{minipage}\end{center}
\vspace{0.4cm}
\begin{center}
\begin{tabular}{|c|c|c|}
\hline
&
${\large t_R \bar{t}_R}$ &
${\large t_L \bar{t}_L}$
\\ 
\hline
{\large $\gamma_+ \gamma_+$}&
\begin{minipage}{1.0in}
\begin{center}
${\cal M_H} $\\
${\cal M_A}$
\end{center}
\end{minipage}
&
\begin{minipage}{1.0in}
\begin{center}
$-{\cal M_H} $\\
${\cal M_A}$
\end{center}
\end{minipage}
\\ 
\hline
\end{tabular}
\end{center}
\end{table}

\begin{figure}[tbp]
\centerline{\epsfxsize=9cm \epsfbox{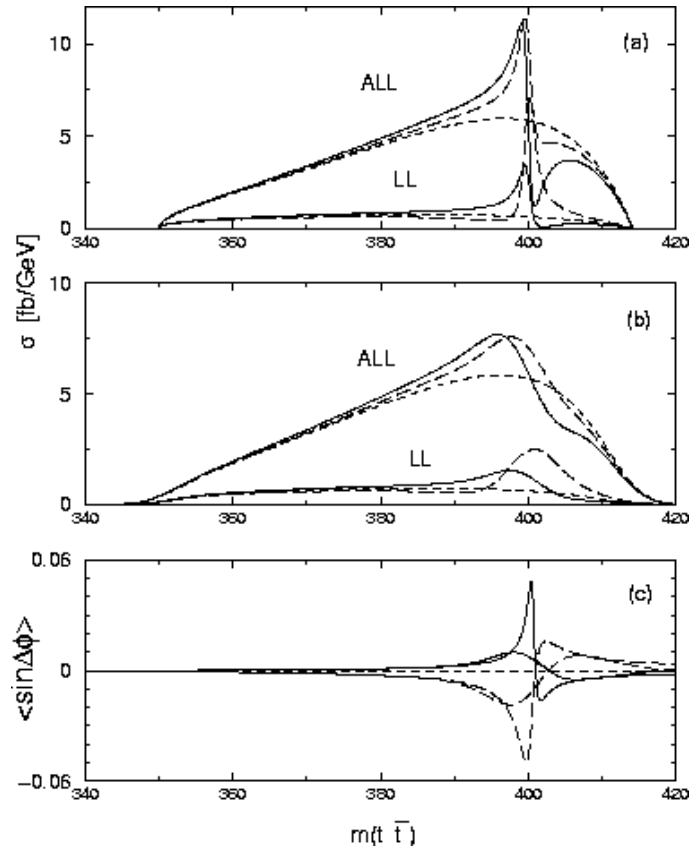}}
\begin{center}\begin{minipage}{\figurewidth}
\caption{\sl \label{options/fig:haint}
The $m(t \overline{t})$ dependence of the cross sections
for $\gamma \gamma \rightarrow t \overline{t}$.
The energy and polarization dependence of the $\gamma \gamma$
luminosity has been considered (we set the highest laser frequency,
$x = 4.83$ at $\sqrt{s}_{ee} = 500$ GeV, and $P_e P_L = -1$.).
Short-dashed curves show QED predictions, while solid (long-dashed)
curves show predictions when $A$ ($H$) of $400$ GeV is produced.
In (a) and (b), the thick lines are for total events and the thin lines
are for events where final top-pairs are left-handed.
Gaussian smearing with $\Delta m (t \overline{t}) = 3$ GeV
is applied in (b). Azimuthal decay angular correlation is shown in (c)
with (without) the smearing by thick (thin) lines.} 
\end{minipage}\end{center}
\end{figure}

In the process $\gamma \gamma \rightarrow t \overline{t}$,
the above amplitudes interfere with the continuum
amplitudes. By observing the interference effect, it is possible to 
judge whether a resonance is from $H$ or $A$.
Denoting the helicity amplitudes for Higgs resonance 
$M_{\phi}^{LL {\rm or} RR}$ 
($\phi = H$ or $A$) and of continuum process 
$M_{cont}^{LL {\rm or} RR}$,
the helicity dependent interference terms are written by
\begin{eqnarray}
sgn\left[ 
{\cal M}_{\phi}^{RR}
{\cal M}_{cont}^{RR}
\right]
&=& - sgn \left[ 
{\cal M}_{\phi}^{LL}
{\cal M}_{cont}^{LL} 
\right] \hspace{5mm} {\rm for} ~~H,
\nonumber \\
sgn \left[ 
{\cal M}_{\phi}^{RR}
{\cal M}_{cont}^{RR}
\right]
&=& ~~~sgn \left[ 
{\cal M}_{\phi}^{LL}
{\cal M}_{cont}^{LL} 
\right] \hspace{5mm} {\rm for} ~~A.
\end{eqnarray}
The difference appear in the cross section with circular polarized photons.
Numerical results are shown 
in Fig.~\ref{options/fig:haint} (a,b).
For definiteness,
we use the MSSM prediction for 
the total and partial widths for $A$ adopted in ref.~\cite{options/AKSW00}.
It is found that little interference is observed
for $H$ (long-dashed curves) because the interference effects for
$t_L \overline{t}_L$ and $t_R \overline{t}_R$ events
cancel each other. On the other hand, the effects for $A$
(solid curves) can be large due to additive interference
for both events.

G.~Jikia and his collaborator calculated the QCD Next-to-Leading order 
correction to this process in an {\it arbitrary} photon beam 
polarization \cite{options/Jikia}.

%% file: options/w.tex
\subsubsection{Anomalous Coupling of the W bosons}

The cross section of W pair
production in the \gamgam 
collision is about 90pb for $\sqrt{s_{\gamma \gamma}} > 200$ 
GeV and is almost independent of the \gamgam center-of-mass energy.
This cross section is O($10-10^{2}$) higher than the fermion pair
production,
and is also 
O($10-10^{2}$) higher than the W pair production in \epm collisions.
It has been pointed out that the W pair production cross section in
the \gamgam 
interaction is sensitive to the anomalous coupling of W bosons to photons,
$\delta  \kappa_\gamma$\cite{options/boudjema,options/choi}.
In this study, the feasibility of a measurement of the W pair
production cross  section in the \gamgam collision and its sensitivity
to  the anomalous coupling of W bosons to 
photons have been investigated\cite{options/takahashi95}.
We studied a measurement of the W pair production cross 
section in a realistic condition, {\it i.e.}, 
using a luminosity distribution simulated by CAIN, while
taking the detector effect into account by simulation.

The laser and machine parameters used are almost identical to
the 400GeV  \gamgam  collider parameters listed in
table~\ref{options/tbl:parameters}. The difference is that the parameter
used in this section assumes a relatively large CP-IP distance of
2cm and the spent electrons coming out from the CP are  swept away
from the IP by an external magnetic field.
The \gamgam luminosity was 
$6.9\times 10^{32}{\rm cm^{-2}s^{-1}}$
and 
$3.7\times 10^{32}{\rm cm^{-2}s^{-1}}$
in total and 
above the threshold of the W pair production, respectively.
In a following study, the total integrated luminosity was assumed to
be $10{\rm fb^{-1}}$, which corresponds to about one year of
experimental operation.

The helicity amplitude for the 
$\gamma \gamma \to W^+W^- \to f\bar{f}f'\bar{f'}$
was calculated using the helicity amplitude calculation program 
HELAS\cite{options/helas}.
The phase-space integration and event generation were performed by 
a Monte-Carlo integration (BASES) and event generation (SPRING) 
program\cite{options/bases}, respectively.
For a detector simulation, we adopted the JLC-I
detector\cite{options/JLC-I}; its performance, mainly related with this
analysis, is the resolution of calorimeters, which are
$15\%/\sqrt{E}+1\%$ and $40\%/\sqrt{E}+2\%$ for 
electromagnetic and hadronic calorimeter, respectively.

The W pair events were selected by 4-jet events, {\it i.e.}, 
for the case that both the
W bosons decayed into quark pairs.
For each event, more than, or equal to, 10 charged tracks were 
required for the central 
tracking detector to be chosen as a multi-hadron event.
After multi-hadronic event selection, JADE clustering\cite{options/jade}
with $y_{cut}=0.007$  was applied and only 4-jet events were selected.
28\% of the W pairs (60\% of hadronic decay) survived the cuts.
For remaining 4-jet events, $\chi_W^2$ was defined as
$$
\chi _W^2\equiv {{\left( {m_{ij}-m_W} \right)^2} \over {\sigma _m^2}}+{{\left( {m_{kl}-m_W} \right)^2} \over {\sigma _m^2}},
$$
where $m_{ij}$ is an invariant mass of the i-th and j-th jet
combination and 
$m_W$ was the W mass.
$\sigma_m$ is the resolution of the W-mass  reconstruction, and was
estimated to  be 5 GeV by the simulation.
One out of the possible three combinations which minimize $\chi_W^2$
was assigned to be the correct combination of jets from W decay.
For the assigned jet combination, both W masses
 were required to be greater than  65~GeV and less than 95~GeV and 
$\left| {\cos \theta _W} \right|<0.9$,
where $\theta_W$ is the angle of the W boson 
with respect to the beam axis.
The overall detection efficiency by this selection was 15\%.

\begin{figure}[tbp]
\centerline{\epsfxsize=9cm \epsfbox{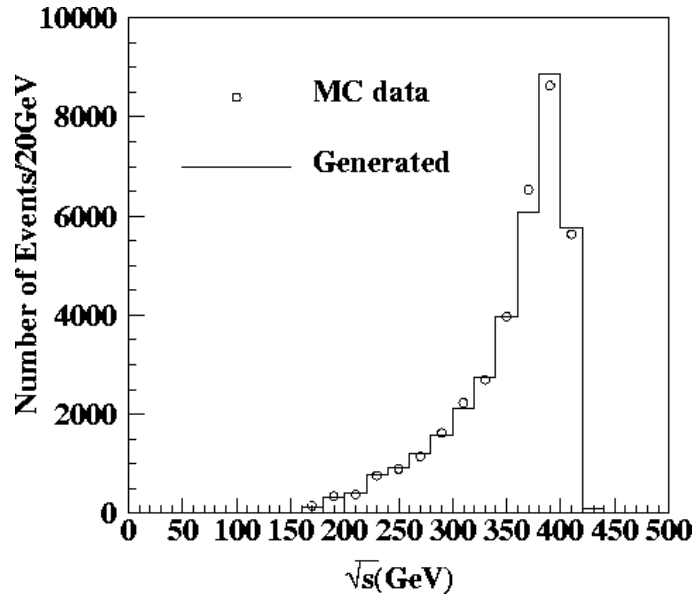}}
\begin{center}\begin{minipage}{\figurewidth}
\caption{\sl \label{options/fig:wevent}
Number of W pair events as a function of the center-of-mass
energy.  The solid line is the number of generated and circles are
the efficiency-corrected number of detected events. 
The normalization corresponds to a total 
integrated luminosity of 10${\rm fb^{-1}}$. }
\end{minipage}\end{center}
\end{figure}

The production cross section for W pairs weighted by luminosity, 
$$
\mathord{\buildrel{\lower3pt\hbox{$\scriptscriptstyle\frown$}}\over \sigma } _{WW}\equiv \sum\limits_{J=1,2} {\int {{{dL_{\gamma \gamma }^J} \over {d\sqrt s}}\sigma _{WW}^J}}d\sqrt s
$$
was approximately 50pb.
Thus, the expected number of event with an integrated luminosity of 
10${\rm fb}^{-1}$ is about 75000.
In order to obtain the production cross section of W pairs, 
the center-of-mass energy of each event was calculated and put into
20 GeV bins.  For each bin, the detection efficiency was estimated 
by a simulation to correct the cross section.
Figure~\ref{options/fig:wevent} shows corrected number of 
W pairs as a function of their center-of-mass energy.

\begin{figure}[tbp]
\centerline{\epsfxsize=9cm \epsfbox{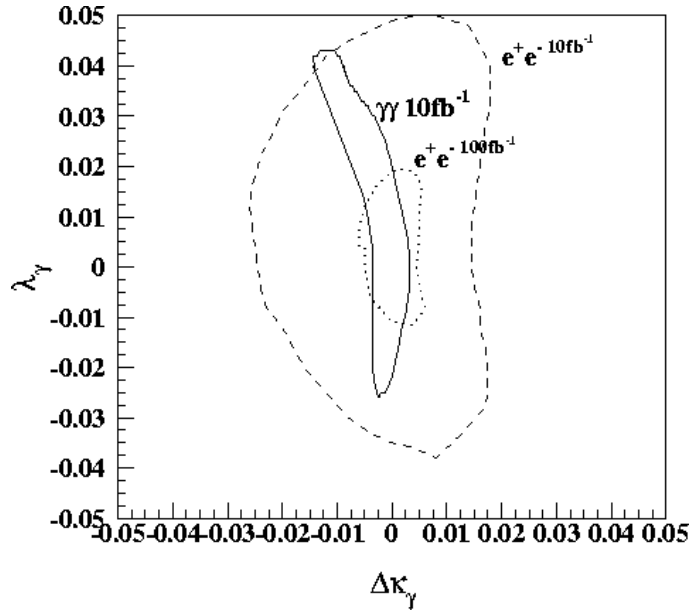}}
\begin{center}\begin{minipage}{\figurewidth}
\caption{\sl \label{options/fig:anom}
Expected 90\% limit in W's anomalous coupling parameter
$\delta \kappa _\gamma$, $\lambda_\gamma$.
An estimation with an \epm collision is also shown by the dotted 
and  dashed lines.
The total integrated luminosity is assumed to be $10{\rm fb^{-1}}$.}
\end{minipage}\end{center}
\end{figure}

For the obtained number of events, the sensitivity to the anomalous
coupling,
$\delta \kappa _\gamma$ and $\lambda_\gamma$, were examined.
Using the cross-section formula including the anomalous coupling
parameter\cite{options/boudjema},  the $\chi^2$ was defined as
\begin{equation}
\chi ^2\equiv \sum\limits_i {\left( {{{N^i-L_0^i\sigma
_0^i-L_2^i\sigma _2^i} \over {\delta N^i}}} \right)^2},
\label{options/eqn:chi2}
\end{equation}
where $N^i$, $L^i_0$, $\sigma^i_0$, $L^i_2$\, $\sigma^i_2$ are
the number of  W pairs, the luminosity and the cross section for J=0
and 2 in the i-th bin, respectively.
$\delta N^i$ is the error of $N^i$, and was estimated as
\begin{equation}
\delta N^i=\sqrt {N^i+\left( {{dL_0^i}\sigma _0^i} \right)^2+
\left( {{dL_2^i}\sigma _2^i} \right)^2}.
\label{options/eqn:delta_N}
\end{equation}
The $dL_k^i$ in (\ref{options/eqn:delta_N}) is the error of the luminosity
determination of the J=k component in the i-th bin, which will be 
described later. From the  defined $\chi^2$ using (\ref{options/eqn:chi2}),
the expected 90\% confidence limit in the
$\delta \kappa _\gamma$-$\lambda_\gamma$
plain was calculated(Fig.~\ref{options/fig:anom}).

As can be seen from the figure, the feasibility for a 
measurement of the anomalous coupling of W bosons by \gamgam colliders
is comparable to that from \epm colliders. 
It should be pointed out that this feasibility is derived 
only from measurement of the total cross section, and improvements
are expected by a comprehensive analysis using the decay property of
W bosons.

\subsubsection{Extra Dimensions}

Arkani-Hamed, Dimopoulos and Dvali \cite{options/extradim1} have proposed that the 
existence of extra dimensions in which the gravity propagates can explain why 
the gravitational interaction is much weaker than the other elementary forces. 
The extra $n$ dimensions are compactified with some space scale and the 
corresponding mass scale is assumed to be a few TeV. In the scenario, 
the exchange of Kaluza-Klein graviton towers can contribute to various 
scattering processes at collider experiments. The Feynman rules have been 
developed by using the effective Lagrangian method. 
The graviton exchange could give rise to effective contact interactions 
between the SM particles in high energy scattering experiments \cite{options/extradim2}.  
In particular, the W-boson production at the $\gamma\gamma$ colliders 
could be affected sensitively by the graviton exchange contributions 
\cite{options/extradim3}.  
The precise measurement of the $W$-pair production cross section with 
appropriate initial beam polarization will give us the largest reach for 
the mass scale of the extra dimensions. The sensitivity on the scale of 
the cross sections is much better than those of other colliders as well 
as the other processes at $\gamma\gamma$ colliders such as the 
$\gamma\gamma$ and $ZZ$ productions \cite{options/extradim3, options/extradim4}.

%% file: options/top.tex
Physics with top quarks in the \epm ~linear collider has been studied and possibilities of rich physics with them has been demonstrated.(see chapter \ref{chapter-phystop}.)
It has also been suggested that there are many topics of top quark physics to 
be explored in \gamgam colliders.
In particular, the \gamgam collider has a unique feature that the polarization of 
initial state photons are controllable. 
The controllability of the polarization allows us to study aspects of physics 
which are virtually impossible with the \epm colliders.
However, there has
been no realistic simulation analysis on the top quark pair detection in \gamgam
colliders so far. 
The estimation of detection efficiency and purity of the events is important 
since W boson 
pair production cross section is very large (about 90 pb independent of energy)
and can be a serious background source. 

In this study, we estimated the number of top quark events detected  in a \gamgam 
collider with a realistic luminosity distribution and with 
simulation for the JLC-I detector. 

\subsubsection{Production Cross Section}

The effective top quark production cross section was 
calculated by convoluting luminosity
distribution with the cross section as:
\begin{equation}
\hat \sigma _{t\bar t}  \equiv {1 \over {L_{tot} }}\sum\limits_{J = 0,2} {\int {{{dL_{\gamma \gamma }^J } \over {d\sqrt {s_{\gamma \gamma } } }}} } \sigma _{t\bar t} \left( {\sqrt {s_{\gamma \gamma } } } \right)d{\sqrt {s_{\gamma \gamma } } } 
\label{options/eq:sigma}
\end{equation}
where the integration ran over the entire energy spectrum.
The effective cross section defined in Eq.~(\ref{options/eq:sigma}) was 0.17 pb which 
corresponds to the number of events as:
$$
\displaylines{
N_{\exp } /year = \hat \sigma ( \approx 0.17pb) 
\times L_{\gamma \gamma } /year( \approx 22.4fb^{ - 1} ) \cr 
\approx 4000 \cr} 
$$
where a year of operation was assumed to be $10^7sec$.

The most serious background source for top quark events is W boson pair
production. With the same definition as Eq.(\ref{options/eq:sigma}), production cross
section of the W boson pairs is about 45 pb corresponding number of events
in a year of 1.2 milion.\cite{options/takahashi95} 
It is 300 times larger than that of top quark pair
production. Suppression of this background is the most important
issue for the analysis.

\subsubsection{Event selection}

The events were selected for two decay modes from top quark pairs.
\begin{itemize}
\item lepton + 4 jets; one top(anti-top) quark decays into a 
lepton, a neutrino and a b(anti b) quark jet, and the other goes to three jets.
\item 6 jets; both top and anti-top quarks decay into three jets.
\end{itemize}
29\% of top quark pairs decay into lepton + 4 jets while 46\% decay into 
6 jet mode. In both cases, final state jets include at least two b-quark jets.
Thus, an efficient b-quark tagging is crucial to discriminate top quark events from 
W boson pairs.

\begin{figure}[tbp]
\centerline{\epsfxsize=9cm \epsfbox{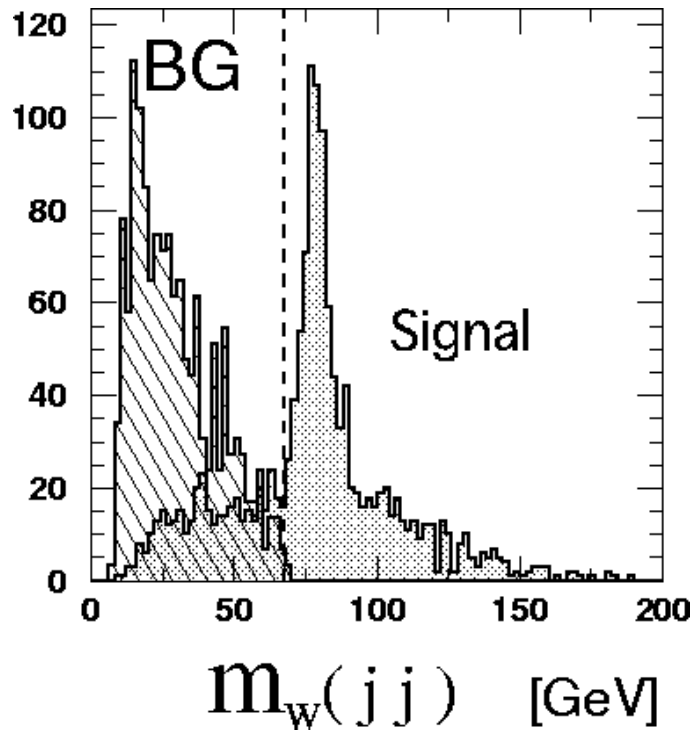}}
\begin{center}\begin{minipage}{\figurewidth}
\caption{\sl \label{options/fig:cut}
Jet-jet mass distribution for W boson candidate.}
\end{minipage}\end{center}
\end{figure}

The lepton + 4 jets events selection first required an isolated lepton in a hadronic event.
The isolation required no particle within $20^ \circ$ around the lepton. 
The JADE clustering algorithm~\cite{options/jade} was applied for remaining 
events with varying $y_{cut}$
from 0.004 to 0.04 and the event had to have four hadronic jets at some 
$y_{cut}$.
After selecting 4 jet events, b-quark jets were searched using 
the impact parameter method. 
The impact parameter resolution was assumed as:
$$
\sigma _b (\mu m) = 4.4 + 5.5/(p\sin \theta )^{2/3},
$$
where p is the momentum of a track in GeV/c and $\theta$ is angle of the track 
with respect to the beam axis.
In each jet, $N_{sig}$ was defined as the number of tracks having 
the impact parameter three times larger than $\sigma_b$.
Then, two or more jets had to be $N_{sig}$ greater than one, and at least
one jet had to be  $N_{sig}$ greater than three.
Finally, the jet-jet invariant masses were calculated for three possible 
combination and the one closest to the W boson mass was assigned to the W boson candidate.
We required the jet-jet mass of the W boson candidate to be greater than 
70 GeV.
Fig. \ref{options/fig:cut} shows jet-jet mass for the W boson candidate.
We obtained a detection efficiency of 9\% with signal-to-background 
ratio of 10, corresponding to about 360 events per year.

For the 6 jet decay mode, an event must be classified as a 6 jet event by 
the same jet clustering algorithm
 as lepton + 4 jets analysis.
For selected events, the b-quark tagging was applied in the same way 
as the lepton + 4 jet mode.
15.6\% of top pair events survived the cuts with signal-to-background ratio 
of 10, which corresponds to 620 top quark events per year. 

\subsubsection{Toward Full Event Reconstruction}

In order to apply selected top quark events to physics analysis, 
one has to 
reconstruct events and obtain, for example, angular 
distribution of top quarks. 
Here, we tried to reconstruct top quark angular distribution 
in the lepton + 4 jet mode.

To reconstruct top quarks from selected jets, we have to choose the right
combination of W boson and b-quark jets. The W boson was already 
reconstructed in the event selection procedure.
The remaining task is to choose the right b-quark jet out of 
two possibilities.
In this analysis, both possibilities were tested and the combination 
where  
the reconstructed mass closer to the top quark mass was assumed to be
the right one.

\begin{figure}[tbp]
\centerline{\epsfxsize=14cm \epsfbox{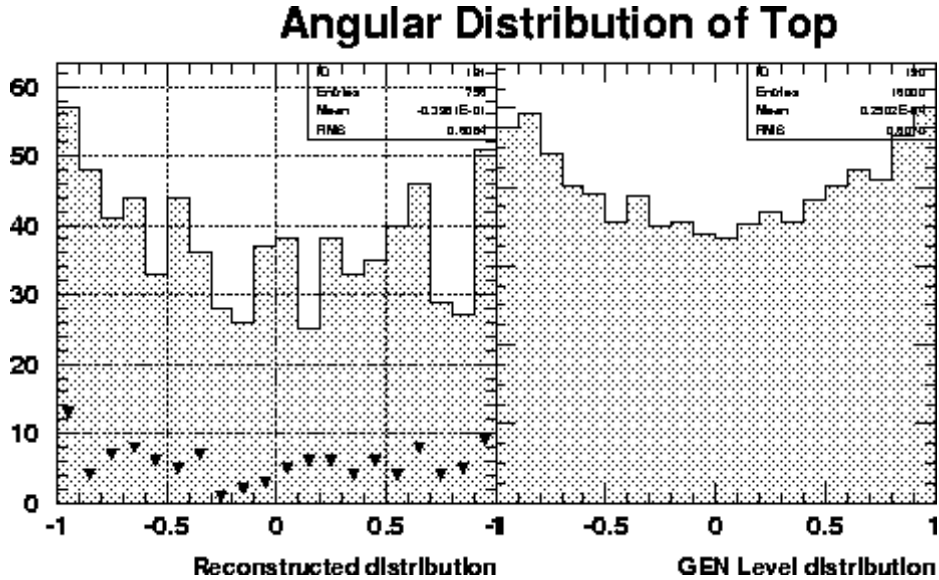}}
\begin{center}\begin{minipage}{\figurewidth}
\caption{\sl \label{options/fig:top-ang}
Angular distribution of the top quark at reconstructed (left) and 
at generated (right) level. Triangle points are contamination 
of miss-reconstructed events.}
\end{minipage}\end{center}
\end{figure}

Fig. \ref{options/fig:top-ang}-a shows the angular distribution of reconstructed
top quarks. It appears to reflect real angular distribution which is shown 
in Fig. \ref{options/fig:top-ang}, However, a more detailed study is necessary. 
Particularly, we still had order of 10\% contamination from 
a wrong assignment of W 
boson and b-quark jet. 

\subsubsection{Summary on the SM top quark}
We studied the feasibility of detection of top quark events in a \gamgam collider
as an option of the JLC. It was found we could detect about 1000 events with 
the signal-to-background ratio of 10 in a year of operation, i.e.,
\begin{itemize}
\item detection efficiency of 9\%~(360events/year) for lepton + 4 jets.
\item detection efficiency of 15.6\%~(620events/year) for 6 jets.
\end {itemize}

A test of reconstruction of the angular distribution showed that it 
appeared to be possible to obtain information of the physical property of 
the top quark from the detected events. 
However more detailed simulation study is 
necessary to obtain quantitative estimation.

%% file: options/top-techni.tex
\subsubsection{Top Quark Production in Technicolor Model}

One of our most important tasks in the present particle physics 
is to understand the mechanism of electroweak symmetry breaking.  
Roughly speaking, the proposed models which go beyond the standard model are 
classified into two categories: supersymmetric models and 
composite models.  

The composite models assume some new kinds of fermions which feel 
some additional strong interactions.  
A pair of new fermion and its anti-particle compose a vacuum 
condensate, which acts just like a Higgs field in the standard 
model.  
For example, in the technicolor model, new fermions which belong 
to SU(2)$_L \times$U(1)$_Y$ are assumed:  
\begin{equation}
Q_L = \left( {U \atop D} \right)_L, \quad
U_R, \quad D_R.  
\end{equation}
To cancel the anomaly by themselves, the hypercharges of $Q_L$, 
$U_R$ and $D_R$ may be chosen as $0$, $1/2$ and $-1/2$, respectively.  
The above fermions also belong to the fundamental representation 
of the SU($N_{TC}$) gauge group of the new strong interaction, the
so-called the technicolor, which is an analogue of QCD color.  
Due to technicolor, pairs of technifermion and anti-technifermion 
make some tight bound-states, just like mesons of quarks in QCD; 
however, the energy scale of the condensation is assumed to be 
at the weak scale.  
Among of the technimesons, technipions are consumed by $W$ and $Z$
bosons to acquire their masses.  

If the energy of future colliders is high enough to create 
the technifermions or technimesons, of course we can examine 
new physics directly and in detail.  

Since the top quark is known to have a heavy mass as the weak scale, 
it is natural in the technicolor model to expect that 
techniforce has some role to create the top-quark mass.  
It is often introduced as an effective four-fermion operator, 
\begin{equation}
\frac{G}{M^2} \, ({\bar Q}_L U_R) ({\bar t}_R q_L) + {\rm h.c.}, 
\label{options/4f}
\end{equation}
where $G$ is the dimensionless coupling constant, $M$ the 
intrinsic mass scale of the operator, and $q_L$ the ordinary quark 
doublet of the top and bottom.  
The top-quark mass from the techniquark condensate can be 
estimated as 
\begin{equation}
m_t = \frac{G}{M^2} \, \langle \bar{U_L} U_R \rangle .   
\end{equation}

The four-fermion interaction (\ref{options/4f}) can also affect 
the $\gamma\gamma \rightarrow t\bar{t}$ amplitude.  
The additional $\gamma\gamma\bar{t}t$ vertex function due to 
the techniforce can be parametrized by two form factors 
($A$ and $B$) as 
\begin{eqnarray}
\Gamma^{\mu\nu}_{NEW}(k,\bar{k}) & = & Q_U^2 \frac{G}{M^2} 
\{ A(k,\bar{k}) 
[ (k \cdot \bar{k}) g^{\mu\nu} - \bar{k}^\mu k^\nu ] 
\nonumber \\
& & \qquad \qquad 
+ B(k,\bar{k}) 
[ k^2 \bar{k}^2 g^{\mu\nu} 
- k^2 \bar{k}^\mu \bar{k}^\nu 
- \bar{k}^2 k^\mu k^\nu 
+ (k \cdot \bar{k}) k^\mu \bar{k}^\nu ] \} , 
\end{eqnarray}
where $Q_U$ is the electric charge of the techni-$U$ fermion, 
and the two photons have four-momenta $k$ and $\bar{k}$, 
and Lorentz indices $\mu$ and $\nu$, respectively 
\cite{options/asaka}.  
Here the $B$ term vanishes after contacting the 
on-shell photon wave functions.  
The remaining $A$ form factor is assumed to have a pole-type 
function, 
\begin{equation}
A(k,\bar{k}) = \frac{r_S}{(k + \bar{k})^2 - M_S^2} , 
\label{options/poleansatz}
\end{equation}
to represent the contributing scalar channel \cite{options/asaka}.  
Here, $M_S$ is the pole mass, which is taken to be the lowest-lying
scalar technimeson mass.   The value of $r_S$ can be estimated by
computing 
$\Gamma^{\mu\nu}_{NEW}$ in Euclidean momentum space, 
and extrapolating it to the physical region with the ansatz 
(\ref{options/poleansatz}).  

The additional amplitude from $\Gamma^{\mu\nu}_{NEW}$ interferes 
destructively with the standard-model amplitude.  
Assuming the one-generation SU(3) technicolor model at the 
scale $M$ = 3 TeV, the cross section of the $\gamma\gamma 
\rightarrow t\bar{t}$ process for the same sign photon 
helicities is found to be suppressed for several percent 
at $\sqrt{s}_{\gamma\gamma}$ = 1 TeV \cite{options/asaka}.  
The deviation from the standard model grows rapidly when 
the collider energy increases.  

%% file: options/susy.tex
The supersymmetric (SUSY) standard model is the most promising
extension of the standard  model as it could naturally give a 
solution to the gauge hierarchy problem. 
The most phenomenologically important consequence of the model 
is in the fact that existence of the SUSY particles (sparticles) 
with masses $O$(100GeV$\sim$1TeV) are predicted. 
The search for these new particles and the study on 
their interactions must be an important 
purpose of the photon colliders as well as the LC and hadron 
colliders. 

In estimation of the production rates for the sparticles at 
collider experiments, the automatic calculation codes could be 
very efficient since we must treat a large number of particles and 
unknown model parameters. 
{\tt GRACE}\cite{options/grace} system is a code which generates 
automatically the matrix element in terms of helicity amplitudes 
for any process in the MSSM 
once the initial and the final states have been specified. 
We have developed an interface between {\tt GRACE} and {\tt CAIN} 
\cite{options/cain}, 
which will be useful in extensive studies of the sparticle 
production at the photon colliders. 

\subsubsection{Chargino and Sfermion Production in $\gamma\gamma$ Collisions}

In $\gamma\gamma$ collisions, the charged sparticles,  
the charginos ${\widetilde{\chi}_i^{\pm}}$ 
\cite{options/JLCgg,options/susy_gg1,options/susy_gg2,options/chargino_gg1,
options/chargino_gg2} and the sfermions 
${\widetilde{f}^{\pm}}$ \cite{options/JLCgg,options/susy_gg1,options/susy_gg2,
options/sfermion_gg}, 
can be produced in pairs as far as it's mass is below the 
kinematical bound, 
\begin{eqnarray}
\gamma\gamma &\to& {\widetilde{\chi}_i^{+}}{\widetilde{\chi}_i^{-}} \\
\gamma\gamma &\to& {\widetilde{f}} {\overline{\widetilde{f}}}
\end{eqnarray}
As they are pure SUSY QED processes, 
the cross sections to leading order depend only on the mass, 
the electric charge and the color degree of freedom of the sparticles. 
In other words, they do not have complicate dependence on the mixing 
angles of the sparticles as well as the weak gauge-boson contribution, 
which are inevitably involved in $e^+e^-$ collisions. 
This property could provide us complementary information about the 
models, e.g., 
universality of the masses for sleptons and squarks in the 
1st and 2nd generations. 

\begin{figure}[tbp]
\centerline{\epsfxsize=9cm \epsfbox{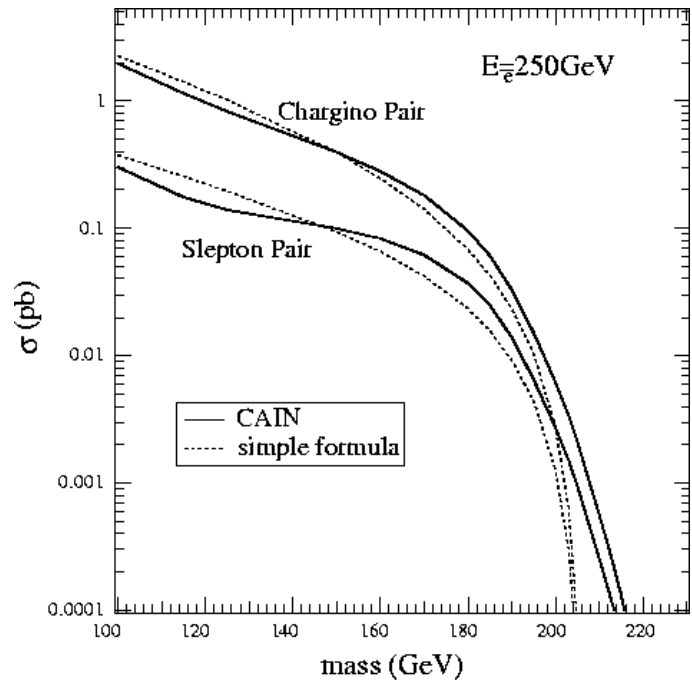}}
\begin{center}\begin{minipage}{\figurewidth}
\caption{\sl \label{options/fig:ggcs}
mass dependence of the total cross sections 
for chargino and slepton production. 
We take $E_e$ $=$ 250GeV. 
Solid and dashed line respectively corresponds to calculation with 
{\tt CAIN} outputs and the simple formula for photon spectrum. 
}
\end{minipage}\end{center}
\end{figure}

The mass dependence of the total cross sections with 
{\tt CAIN} outputs for the photon spectrum  
is shown in Fig.~\ref{options/fig:ggcs}, where we set $E_e$ $=$ 250GeV. 
For a comparison, we also plot those with 
the simple photon spectrum formula \cite{options/gin2,options/gin3}. 
We find the cross sections become non-zero even above the naive 
kinematical bound, 
\begin{displaymath}
m < 0.8 E_e. 
\end{displaymath}
In the calculation with {\tt GRACE} and {\tt CAIN}, 
the initial photon polarization has been appropriately included. 

It should be emphasized that, as the $\gamma\gamma$ cross sections 
involve an $s$-wave contribution, they will be much 
larger than that of $e^{+}e^{-}$ 
if $\sqrt{s}$ is large compared to the mass threshold. 
This property not only provides large production rates  
compared to $e^{+}e^{-}$ collisions but also 
enables us to see the 
{\it squarkonium} production. 
We could detect a stoponium \cite{options/kon93b,options/stoponium} 
(sbottomonium) 
resonance peak in the invariant mass distribution 
for 
$\gamma \gamma \to \widetilde{t}_1 \overline{\widetilde{t}}_1$ 
($\gamma\gamma \to \widetilde{b}_1\overline{\widetilde{b}}_1$). 
From the peak structure, we could extract valuable information 
about the SUSY parameters as well as on the SUSY QCD. 

\subsubsection{Sfermion Production in $e\gamma$ Collisions}

In $e\gamma$ collisions, a charged particle can be singly produced 
accompanied by a neutral one. 
Consequently, if there is large mass difference between the 
charged and the neutral particle, the kinematical limit for the 
production 
in $e\gamma$ collisions could be larger than that of 
pair production processes of each particle in $e^+e^-$ collisions. 

In the framework of the MSSM we can assume ${\widetilde{Z}}_{1}$ as the LSP. 
In this case we find that the process 
\begin{equation}
  e\gamma \to \widetilde{e}_{R}{\widetilde{Z}}_{1}. 
\end{equation}
has a lower mass threshold of 
$m_{\widetilde{e}_{R}}+m_{{\widetilde{Z}}_{1}}$ than 
that of $2 m_{\widetilde{e}_{R}}$ for the selectron pair production at 
$e^{+}e^{-}$ colliders. 
It is expected that $e\gamma$ colliders 
will be efficient in searching for the heavy selectron with a mass 
larger than half of $\sqrt{s}$ in $e^{+}e^{-}$ colliders 
\cite{options/JLCgg,options/susy_eg1,options/susy_eg2,options/susy_eg3}. 

\begin{figure}[tbp]
\centerline{\epsfxsize=9cm \epsfbox{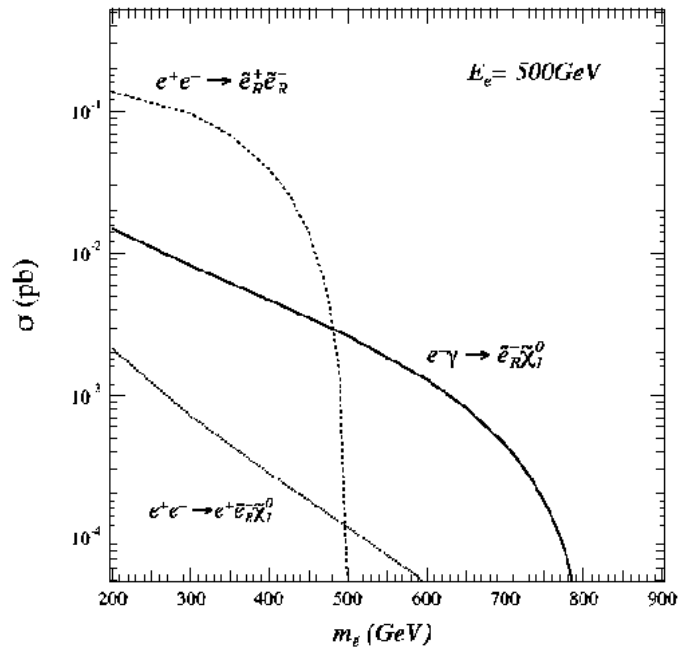}}
\begin{center}\begin{minipage}{\figurewidth}
\caption{\sl \label{options/fig:egsesz1} 
Selectron mass dependence of the total cross sections. 
We take $E_e$ $=$ 500GeV, $\tan\beta=12$, 
$\mu=-300{\rm GeV}$ and $M_2=200{\rm GeV}$, which correponds to 
$m_{{\widetilde{Z}}_{1}}=96{\rm GeV}$. 
}
\end{minipage}\end{center}  
\end{figure}

Figure~\ref{options/fig:egsesz1} shows the selectron mass dependence of the
total cross section. 
The selectron could be discovered up to the kinematical limit of 
$e\gamma$ collisions, 
\begin{displaymath}
m_{\widetilde e} < 1.8 E_e - m_{{\widetilde{Z}}_{1}}. 
\end{displaymath}
We can obtain a large cross section for 
$e\gamma$ $\to$ $\widetilde{e}_{R}{\widetilde{Z}}_{1}$, 
even for the heavy selectron, 
$m_{\widetilde{e}_{R}}\mathrel{\vcenter
     {\hbox{$>$}\nointerlineskip\hbox{$\sim$}}} E_e$. 
Note that the initial beam polarization will be efficient 
to enhance the signal cross section \cite{options/JLCgg}. 

Another important property of the process is the simple dependence
of the cross section on the SUSY parameters.
The cross section is proportional to the bino component ($N_{11}$) 
of the lightest neutralino, 
${\widetilde{Z}}_{1} = 
N_{11}\widetilde{B}+N_{12}{\widetilde{\rm W}}+
N_{13}\widetilde{H}^{0}_{1}+N_{14}\widetilde{H}^{0}_{2}$. 
Arbitrary SUSY parameters appearing in the cross section 
are only $m_{\widetilde{e}_{R}}$, $m_{{\widetilde{Z}}_{1}}$ 
and $|N_{11}|$. 
This contrasts with the situation 
for the selectron pair production at $e^{+}e^{-}$ colliders, 
to which all masses and mixing angles of the neutralinos 
contribute. 
If we know $m_{\widetilde{e}_{R}}$ and $m_{{\widetilde{Z}}_{1}}$ 
from an analysis of the process 
$e^{+}e^{-}$ $to$ $\widetilde{e}_{R}^+\widetilde{e}_{R}^-$, 
we could determine $|N_{11}|$ by a measurement of the total cross section 
for $e\gamma$ $\to$ $\widetilde{e}_{R}{\widetilde{Z}}_{1}$. 

In the framework of the MSSM with $R$-parity (lepton number) violating 
superpotential ($i \sim m$ denote generation indices) \cite{options/rbphys}, 
\begin{equation}
W_{\slash\hspace{-6pt}R}=\lambda_{ijk}L_iL_j{\overline{E}_k}
   + \lambda'_{1lm}L_1Q_l{\overline{D}_m}, 
\qquad \qquad (i\neq j \quad {\rm or} \quad k) =1
\end{equation}
sleptons or squarks can be singly produced with ordinary matter 
fermions in $e\gamma$ collisions, e.g., through 
\begin{eqnarray}
e\gamma &\to& \ell_i \widetilde{\nu}_j, \quad 
   {\widetilde{\ell}}_i \nu_j
 \qquad \qquad (\lambda_{ij1} \neq 0), \\
e\gamma &\to& d_m \widetilde{u}_l, \quad 
   {\widetilde{d}}_m u_l
 \qquad \qquad (\lambda'_{1lm} \neq 0).
\end{eqnarray}
In this case the kinematical limit becomes 
\footnote{
One of authors, T.K., is thankful Prof.I.Ginzburg for his 
valuable suggestion. 
}
(except $e\gamma \to {\widetilde{d}}_m t$)
\begin{displaymath}
m \mathrel{\vcenter
     {\hbox{$<$}\nointerlineskip\hbox{$\sim$}}} 1.8 E_e. 
\end{displaymath}

\begin{figure}[tbp]
\centerline{\epsfxsize=9cm \epsfbox{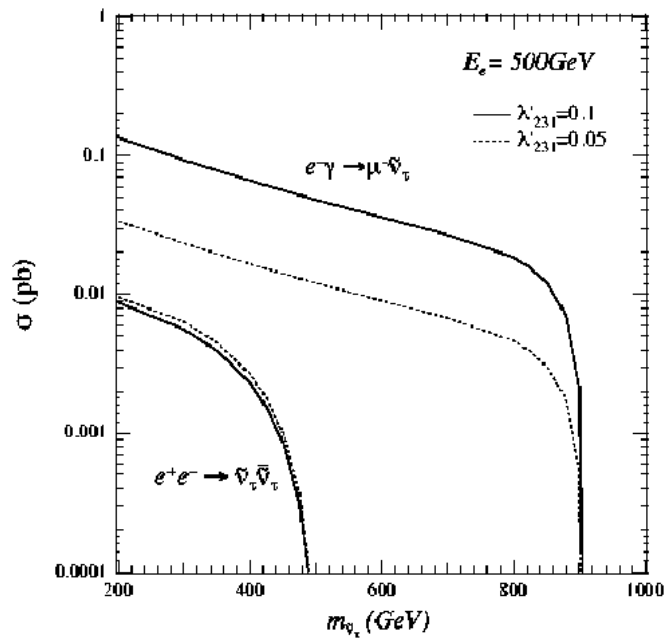}}
\begin{center}\begin{minipage}{\figurewidth}
\caption{\sl \label{options/fig:egmsn}
Sneutrino mass dependence of the total cross sections. 
We take $E_e$ $=$ 500GeV. 
}
\end{minipage}\end{center}   
\end{figure}

As an example we show the sneutrino mass $m_{\widetilde{\nu}_\tau}$ 
dependence of the
total cross section for 
$e\gamma \to \mu \widetilde{\nu}_\tau$ 
in Figure~\ref{options/fig:egmsn}. 
We find that the mass reach improves considerably in 
$e\gamma$ collisions in comparison with the pair 
${\widetilde{\nu}_\tau}$ production at $e^+e^-$ colliders. 
Moreover, the SUSY parameter dependence of the cross section 
is simple in $e\gamma \to \mu \widetilde{\nu}_\tau$ 
becuase it is proportional to $|\lambda'_{231}|^2$. 
It means that we can easily extract information on the 
${\slash\hspace{-6pt}R}$ 
strength from the cross section measurements. 
On the other hand, it is not so easy in 
$e^+e^- \to \widetilde{\nu}_\tau {\overline{\widetilde{\nu}_\tau}}$, 
since the sensitivity on the coupling strength will be smeared 
owing to the $Z$-boson contribution, which does not depend on 
$\lambda'_{231}$. 

%% file: options/estar.tex
The $e\gamma$ collider is a suitable option 
for the search for excited electrons 
$e^{\ast}$\cite{options/gin4,options/konEG,options/djouadi}. 
If $m_{e^{\ast}}<\sqrt{s_{e\gamma}}$, 
a clear signal of $e^{\ast}$ as an $s$-channel resonance 
is expected. 
The production cross section is much larger than 
that of $e^{\ast}$ single production in $ee$ collision. 
The present lower limits of excited lepton masses are 
obtained as $\sim 90 {\rm GeV}$ by LEP2 experiment
\cite{options/aleph,options/opal,options/L3}. 
$e\gamma$ colliders can extend the searchable mass range 
up to $\sqrt{s_{e\gamma}}$ 
($\sim$ twice of $\sqrt{s_{ee}}$). 
Even if $m_{e^{\ast}}$ is beyond the energy range of 
the $e\gamma$ collider, 
its existence may be detectable by observing the interference 
with the standard model process.

By operating $e\gamma$ collider 
in such a mode to enhance the polarization of 
$\gamma$ and $e$ beams, 
we can get information on the coupling of $e^{\ast}$ to $e$ 
or more generally $e^{\ast}$ to leptons. 
Let us write the polarized cross section 
for the process $e\gamma\rightarrow e\gamma$ 
$$
\begin{array}{rl}
d\hat{\sigma}(\lambda,\xi_2) 
=&\frac{1}{4}(1+\xi_2)\{
 (1+2\lambda)d\hat{\sigma}[e_R\gamma_{+}]
+(1-2\lambda)d\hat{\sigma}[e_L\gamma_{+}]\}\\
+&\frac{1}{4}(1-\xi_2)\{
 (1+2\lambda)d\hat{\sigma}[e_R\gamma_{-}]
+(1-2\lambda)d\hat{\sigma}[e_L\gamma_{-}]\},
\end{array}
$$
where $\lambda$ and $\xi_2$ are 
the mean helicity of electron 
and the Stokes parameter of $\gamma$, respectively. 
If $e^{\ast}_L$ couples radiatively to $e_L$ 
as suggested in some model \cite{options/cabibo,options/kuehn,options/hagiwara}, 
then the $s$-channel resonance is seen 
solely in the case $\lambda=-1/2, \xi_2=-1$. 
Conversely, 
if $e^{\ast}$ is observed as an $s$-channel resonance 
in $e_R\gamma_+$ mode, 
$e^{\ast}_R$ couples radiatively to $e$. 
Furthermore, 
if $e^{\ast}$ has spin 3/2, 
resonance in $e_R\gamma_-,e_L\gamma_+$ mode are expected. 

The value of $\xi_2$ cannot be tuned exactly 
but is determined by the polarization $P_c$ of 
initial laser beam and $\lambda'$ of electron scattered by laser. 
For instance, by adjusting the polarization as 
$(P_c,2\lambda')=(1,-1)$, 
the energy of $\gamma$ beam concentrates on 
its kinematically allowed limit 
and $\xi_2$ is almost $-1$. 
Thus the $\gamma$ beam has a suitable polarization 
for the search for $e^{\ast}$ suggested by some model. 

A sharp peak in the $z$ spectrum of the final states 
locates at $z_{\rm peak}=m_{e^{\ast}}^2/s_{ee}$ 
and $m_{e^{\ast}}$ is definitely determined. 
Background $\gamma$ due to QED process are produced in 
backward direction, 
an appropriate angular cut can make the signal clearer. 

%% file: options/linear-pol.tex
One of the most important research subjects in the current 
particle physics is to explore the origin of the CP violation,
which realizes the present universe filled with matter.  

There are two techniques to observe the CP-odd quantities 
in high-energy collider experiments: one is to analyze the 
correlation of the spins and momenta of the final particles; 
the other one is to observe the angular distributions 
of the final particles, which are produced from the polarized 
initial beams.  
By tuning the polarizations of both the laser and the electron 
beam, one can obtain back scattered photon beam in almost 
arbitrary polarization demanded.  
Thus, a \gamgam collider can be a unique probe 
to look for a different aspect of  the CP 
violation through Higgs bosons, $t\bar{t}$ and $W^+W^-$ productions.

\subsubsection{CP properties of Higgs bosons}

As is already discussed in the section~\ref{options/sect:higgs}, 
the polarization of the photon beams is a very effective tool to 
analyze the CP properties of the Higgs bosons. 
One of such interest is on the $A$-$H$ intereference~\cite{options/AKSW00} 
which is already discussed in the above.  
Moreover, the method to determine the CP property of the heavy neutral 
Higgs bosons is recently discussed in the model without definite 
CP-parity~\cite{options/ACHL00}.  
It is a great advantage of the $\gamma\gamma$ collider that both of 
the circular and the linear polarizations of the photon beams can be 
available in high degree.  

\subsubsection{Top-Quark Electric-Dipole Moment(EDM) measurement }

The top quark EDM bears an additional interaction between the 
top quark and the photon by a T-odd effective Lagrangian, 
\begin{equation}
{\cal L}_{EDM} = -i e Q_t \left( \frac{d_t}{2 m_t} \right) 
\bar{t} \sigma_{\mu\nu} \gamma_5 t F^{\mu\nu},  
\end{equation}
where $Q_t$ = 2/3, $m_t$ is the top-quark mass, $F^{\mu\nu}$ 
the photon field strength and $\tilde{\mu_t}$ $\equiv 
e Q_t d_t / 2 m_t$ the top-quark EDM.  
By observing the process $\gamma\gamma\rightarrow t\bar{t}$ with 
linearly polarized photon beams, one can extract the 
electric dipole moment (EDM) of the top quark.  
Detailed analyses can be found in \cite{options/choiT,options/Baek97}.  
The CP-odd contribution to the differential cross section 
is proportional to 
\begin{equation}
 \eta_1 \eta_2 \, ( 1 - \beta_t^2 \cos^2\theta ) 
 \sin [2(\phi_1 - \phi_2)] \, {\rm Re}(d_t), 
\end{equation}
where $\eta_1$ and $\eta_2$ are the degrees of the linear 
polarization of the two photon beams, $\phi_1$ and $\phi_2$ 
are the azimuthal angles of the photon linear polarizations, and
$\beta_t$ and $\theta$ are the velocity and the scattering 
angle of the top quark in the center-of-mass frame.  
It was found that only the number-counting method without 
the information of decaying top quarks is required here 
for the measurement of $d_t$.   
By adjusting the Compton-scattering conversion parameter ($x$), 
one can explore the top-quark EDM up to 
${\rm Re}(\tilde{\mu}_t)$ $= 1.8$ and $0.2 \times 10^{-17}$ 
$e\cdot {\rm cm}$ in the one-sigma level at $\sqrt{s}$ = 
0.5 and 1.0 TeV machines with a 20 fb$^{-1}$ integrated 
luminosity, respectively \cite{options/Baek97}.  
This limit at 1.0 TeV is much smaller than that of $e^+e^-$ 
machines \cite{options/Cuypers95}.
Poulose and Rindani also studied the dipole coupling using circularly 
polarized photon\cite{options/rindani1,options/rindani2}.
A detail description is found in chapter \ref{chapter-phystop}.   

\subsubsection{W-boson Productions}

The general form of the $\gamma WW$ vertex has seven independent 
terms, which characterize their Lorents structures
\cite{options/Hagiwara87}.   Out of them, three terms violate CP
symmetry.   Due to the QED gauge invariance, one form factor of the
three  should vanish at $\gamma\gamma$ or $e\gamma$ collisions with 
the initial on-shell photons, and the remaining two, often called 
$f_6$ and $f_7$ in the literature, can be surveyed at the 
$\gamma\gamma$ and $e\gamma$ colliders.  
By adopting only the standard-model terms and the CP-odd terms, 
the effective $\gamma WW$ coupling would be expressed as 
\begin{eqnarray}
\Gamma^{\mu\nu\lambda}(q,\bar{q},k) & = & 
-ie \left[  (q-\bar{q})^\lambda g^{\mu\nu} 
+ (\bar{q}-k)^\mu g^{\nu\lambda}
+ (k-q)^\nu g^{\lambda\mu} \right. 
\nonumber \\
& & \qquad \left. 
+ f_6 \epsilon^{\mu\nu\lambda\rho} k_\rho 
+ \frac{f_7}{m_W^2} (q-\bar{q})^\lambda 
\epsilon^{\mu\nu\rho\sigma} k_\rho (q-\bar{q})_\sigma 
\right] , 
\end{eqnarray}
where $q$, $\bar{q}$ and $k$ are the $W^-$, $W^+$ and $\gamma$
momenta,  and $\mu$, $\nu$ and $\lambda$ are the Lorentz indices of
their  external bosons, respectively.  

The $f_6$ depencence of the differential cross section in the 
process $\gamma\gamma \rightarrow W^+W^-$ vanishes in the leading 
order of $f_6$, when the directions of the linear polarizations of 
the colliding two photon beams are parallel.  
However, it maximizes and is proportional to 
${\rm Re}(f_6) \, \sin 2\phi$ when the two beam polarizations 
are perpendicular, where $\phi$ stands for the azimuthal angle of 
the scattered $W$ boson with respect to the direction of the linear 
polarization of one beam \cite{options/boudjema}.  

With an integrated luminosity of 20 fb$^{-1}$ at $\sqrt{s}$ = 
500 GeV, a $\gamma\gamma$ collider can measure the ${\rm Re}(f_6)$ 
up to an accuracy of $2.3 \times 10^{-2}$ and up to 
$0.6 \times 10^{-2}$ with 100 fb$^{-1}$ at 1 TeV 
\cite{options/boudjema}.  
These bounds are given at the $3 \sigma$ level.  

The imaginary part of $f_6$ is accessible up to $1.2 \times 10^{-3}$ 
at 0.5 TeV with 20 fb$^{-1}$ by using the circular photon polarization 
\cite{options/boudjema}.  

The process $e^-\gamma \rightarrow \nu W^-$ is also effective to 
measure the CP-odd form factor $f_6$, only if the helicity of the 
final $W^-$ is understood \cite{options/boudjema}.

%% file: options/hadron.tex
An accurate knowledge of hadronic cross-sections in \gamgam
collisions ($\sigma_{\gamma \gamma}^{\rm had}$) at high energies is of  two fold
importance. It is very interesting from a theoretical view point
of acquiring an almost `first principle' understanding
of a nonperturbative quantity such as total/inelastic cross-sections
in a QCD based picture. On a more pragmatic note it is essential
to be able to estimate the hadronic backgrounds at the future 
Linear colliders, both in the \epm~ and the \gamgam~
mode. It should be pointed out right at the beginning  that
even though the quantity $\sigma_{\gamma \gamma}^{\rm had}$ can be measured  
at an \epm~ collider as well, the Compton colliders will play  a 
very important role in realizing  the first of the above stated  aims,  
due to the 'unfolding' procedures that have to be used to extract the 
\gamgam~ cross-section from the measured hadronic cross-sections
for the process 
$e^+e^- \to e^+e^- \gamma \gamma \to e^+e^-  +$  
hadrons.

\subsubsection{Different theoretical models for the hadronic \gamgam~
cross-sections and discrimination among them at the \gamgam~ colliders}

It is now well established that all the hadronic cross-sections 
$pp / p\bar p$, $\gamma p~$ and \gamgam~  rise with the energy
of the colliding beams. The latter two, involving the hadronic
structure of the photon, have become available only recently from
HERA~\cite{options/1,options/2} and LEP~\cite{options/3,options/4}. 
A variety of models~\cite{options/5},
some using  factorization along with a  generalization of the 
phenomenological ideas introduced to  describe the rise with energy
observed for the $pp/p\bar p$ cross-sections~\cite{options/6}, 
treating the photon like a proton  and some based on QCD, using
the information  on the parton structure of the photon, have been 
put forward.  In all these descriptions, the extra parameters required 
for the case of the photon are mostly determined from the $\gamma p$
case and then cross-sections are predicted for the \gamgam~ case.
Present $\gamma p$ and \gamgam~ data seem to indicate that the energy rise
of the cross-sections involving photons might be faster than that for 
the $pp/p\bar p$ case, which is expected generally in the `eikonalised
minijet' models~\cite{options/8,options/7}.  However, the values of 
$\sigma_{\gamma p}^{\rm had}$, 
$\sigma_{\gamma \gamma}^{\rm had}$ at HERA  and LEP respectively 
are extracted from a  study of the reactions $ep \to e \gamma p \to eX$ 
and $e^+ e^- \to e^+ e^- \gamma \gamma \to e^+ e^- X$, respectively. 
 As a result the errors in 
the extracted experimental values are substantial. 
Within those errors, 
predictions of some of these other models also tend to be consistent with 
the data. 
Thus a \gamgam~ collider which does not suffer from the 
`unfolding' problems can serve very well indeed for discriminating
between these various theoretical models and achieving a better 
understanding of the theoretical situation.

\begin{figure}[htb]
\begin{center}
\centerline{
\epsfxsize=7cm \epsfbox{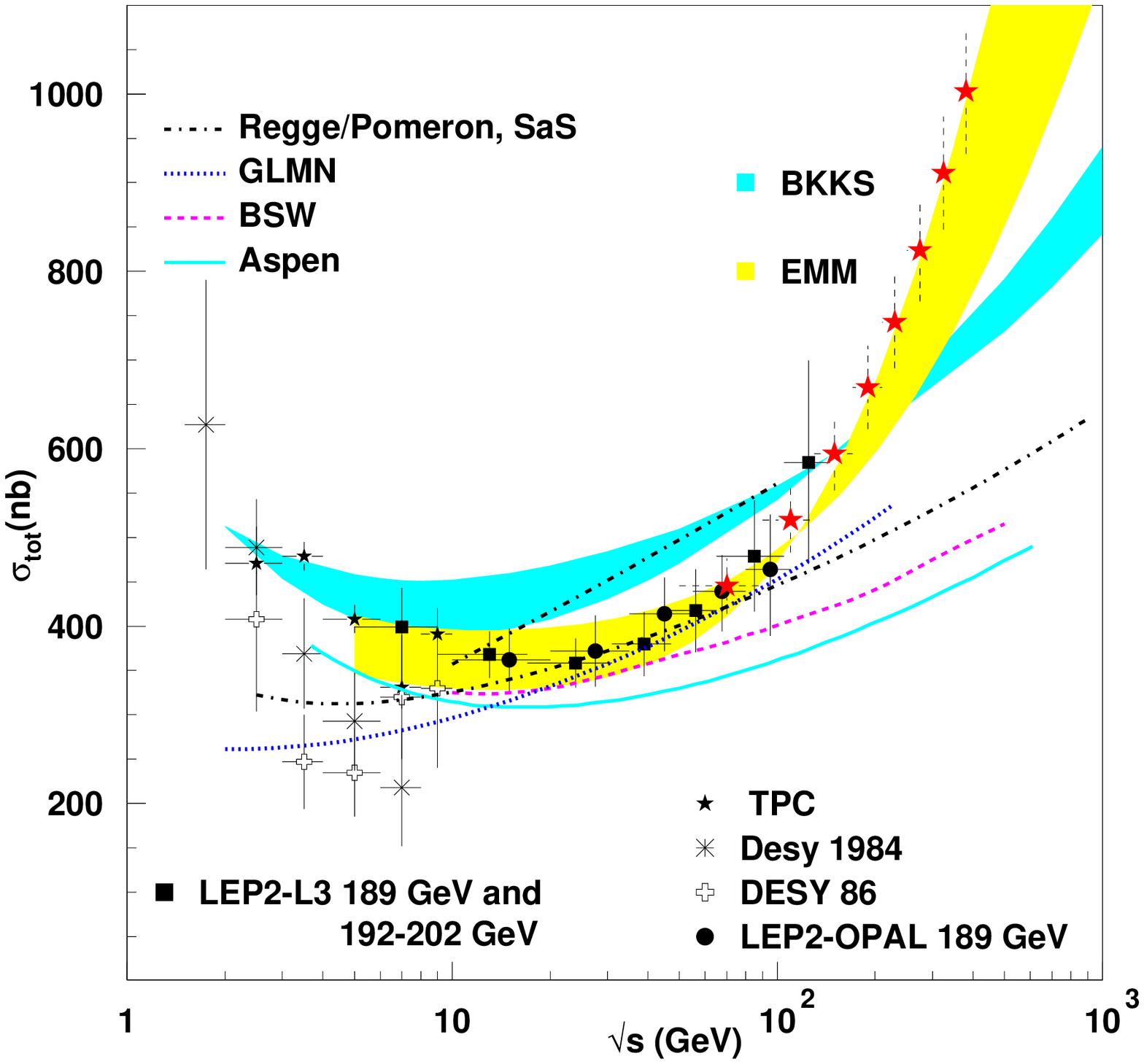}
\epsfxsize=7cm \epsfbox{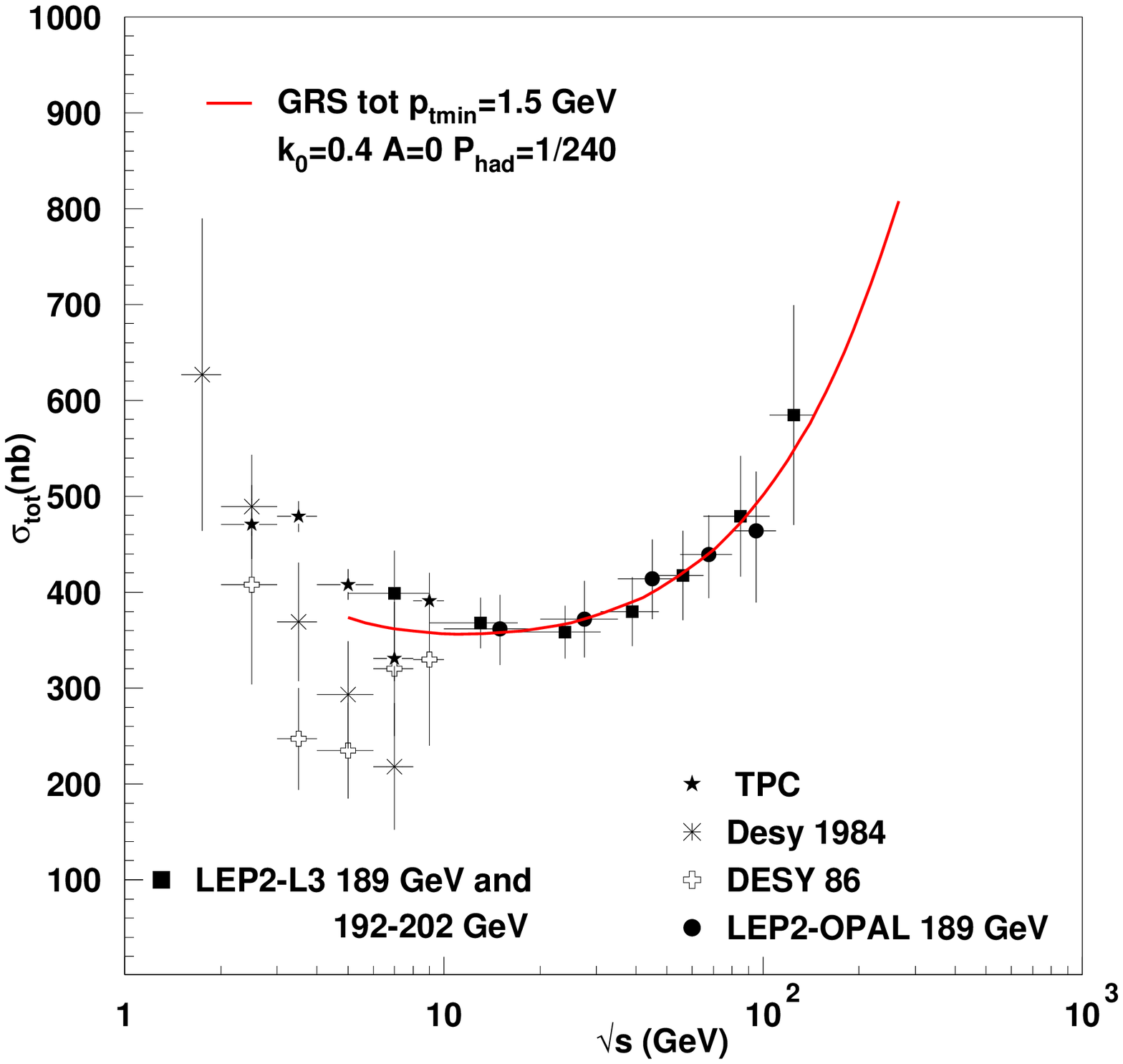}
}
\begin{center}\begin{minipage}{\figurewidth}
\caption{\sl \label{options/one}
The predictions from factorization models, 
Regge-Pomeron exchange and a QCD structure function models together with 
those from the EMM and a comparison with present data in the left panel
and the EMM description of the latest LEP data for parameters extracted from 
$\gamma p$ with $10\%$ changes\protect~\cite{options/7} in the right panel as an 
illustration.}
\end{minipage}\end{center}
\end{center}
\end{figure}

The left panel of Fig.~\ref{options/one}
shows a compilation of the latest data on the \gamgam\ cross-sections and the 
predictions of the different models. A parameterization of the various  model 
predictions in the form
\begin{equation}
\sigma_{\gamma \gamma}^{~\rm had} (s_{\gamma \gamma}) 
= a~s_{\gamma \gamma}^{\epsilon}  + 
b~ s_{\gamma \gamma}^{-\eta}. 
\label{options/eq1}
\end{equation}
gives $\epsilon$ values between 0.09 (Aspen model)\cite{options/9}, 
0.13 (BKKS)~\cite{options/10} and 0.30(Eikonalised minijet models 
EMM)\cite{options/7,options/8}. The QCD based models~\cite{options/7,options/10} 
seem to describe the current
data somewhat better.  The figure shows also the expected \gamgam~ 
cross-sections at a Compton collider for a 500 GeV Linear Collider with 
TESLA design,  in the `minijet' models, for a choice of parameters 
which produces the `best fit' to the LEP data and is consistent with the 
$\gamma p$ HERA data~\cite{options/7,options/11}, 
as `pseudo data points' expected to be
measured at such a collider.  
These are indicated by the  red 
stars, with estimated errors~\cite{options/12} on them indicated by the vertical bars. 
In the right panel of the same figure the EMM prediction in its total
formulation is shown along with the latest L3 data. The values of the
parameters of the EMM used here are consistent with the $\gamma p$ data within
$10 \%$ uncertainty. 

\begin{figure}[htb]
\centerline{\epsfxsize=8cm \epsfbox{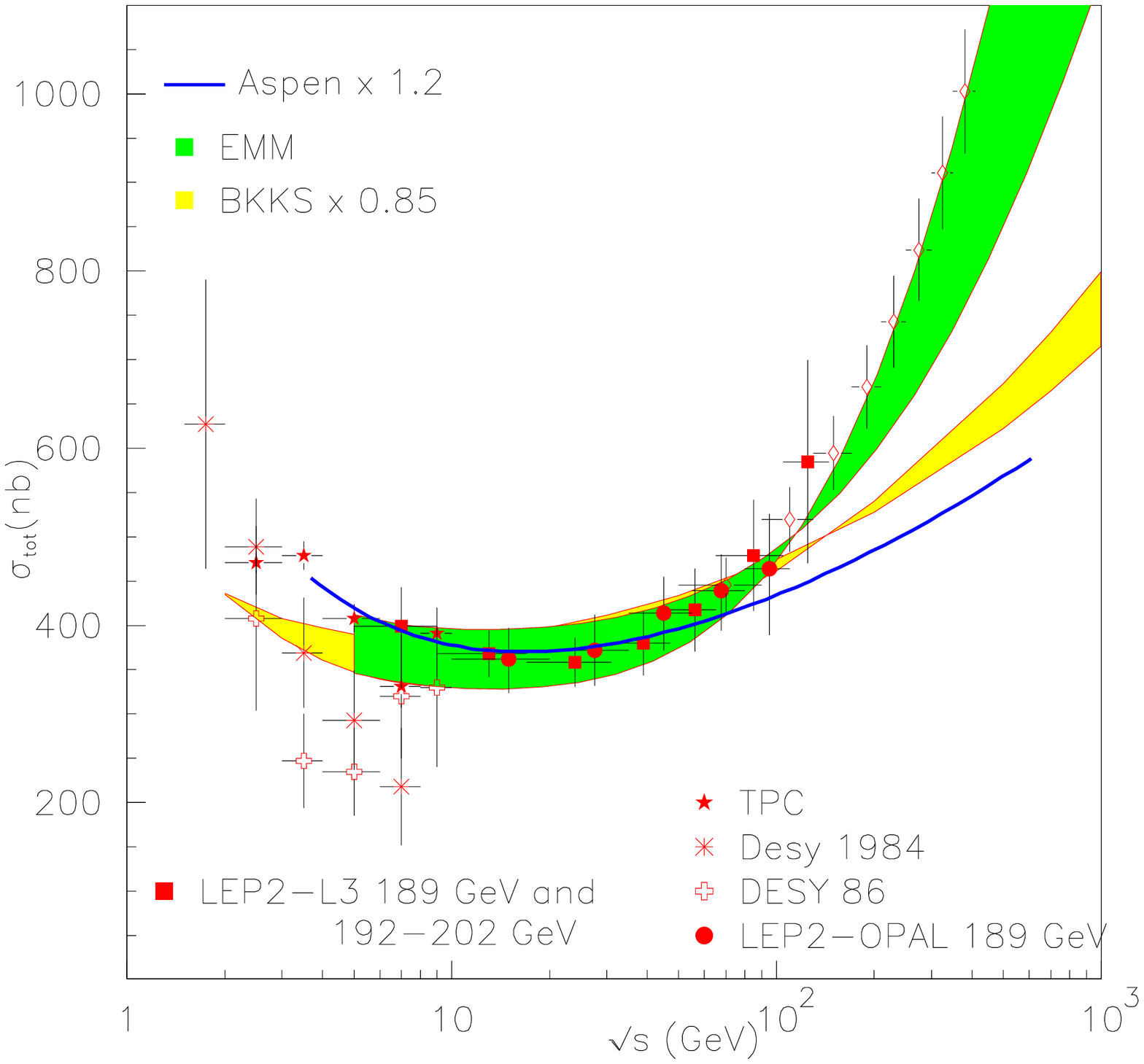}}
\begin{center}\begin{minipage}{\figurewidth}
\caption{\sl \label{options/two}
The total \gamgam~ cross-section as a function of the collision 
energy compared with some model calculations: BKKS band\protect~\cite{options/10}
band corresponds to different partonic densities for the photon, EMM band 
corresponds to the different choices of parameters in the EMM
model~\protect\cite{options/8,options/7}, the solid line corresponds to a proton like 
model~\cite{options/9}. The Aspen and BKKS results have been normalized to the data.}

\end{minipage}\end{center}
\end{figure}

The difference in the rate of the  rise with energy of the predicted \gamgam\ 
cross-sections, is seen more clearly in Fig.~\ref{options/two}
where the predictions of the Aspen and BKKS model have been normalized 
to the data.  This figure again has the same `pseudo data points' as shown 
in the Fig.~\ref{options/one}. This figure shows very clearly that \gamgam~ collider 
will be able to indeed discriminate between these models and clarify our 
theoretical ideas about the high energy behavior of the total hadronic 
cross-sections.

In Table~\ref{options/tabone} we show the expected total $\gamma \gamma$ cross-sections 
for three different models  which treat a photon more or less like a proton
~\cite{options/9,options/13,options/14}. The last column shows the 1$\sigma$ level 
precision needed to discriminate between Aspen\cite{options/9}
and BSW\cite{options/13} models. The difference between DL\cite{options/14} 
and either Aspen or BSW is bigger than between Aspen and BSW at  
each energy value.
\begin{table}[htb]
\begin{center}\begin{minipage}{\figurewidth}
\caption{\sl \label{options/tabone}
Precision required for the measurement of \gamgam~
cross-sections to distinguish between the different `proton' like models}
\end{minipage}\end{center}
\vspace{0.5cm}
\begin{center}
\begin{tabular}{|c||c|c|c|c||}
\hline 
$\sqrt{s_{\gamma \gamma}} (GeV)$ & Aspen &  BSW & DL & $1 \sigma$ \\ \hline
\hline
 20    & 309 nb & 330 nb & 379 nb &  7\%  \\ \hline
 50    & 330 nb & 368 nb & 430 nb &  11\%   \\ \hline
 100   & 362 nb & 401 nb & 477 nb &  10\%   \\  \hline
 200   & 404 nb & 441 nb & 531 nb &  9\%   \\  \hline
 500   & 474 nb & 515 nb & 612 nb &  8\%   \\  \hline
 700   & 503 nb & 543 nb & 645 nb &  8\%   \\ \hline
\end{tabular}
\end{center}
\end{table}

A similar table can be drawn for distinguishing between the total and 
inelastic formulation of the minijet models~\cite{options/7} and the BKKS 
model\cite{options/10}, for instance.
\begin{table}[hbt]
\begin{center}\begin{minipage}{\figurewidth}
\caption{\sl \label{options/table2}
Precision required for the measurement of \gamgam~ cross-sections
to distinguish between different formulations of the 
EMM and BKKS~\protect\cite{options/10}}
\end{minipage}\end{center}
\vspace{0.3cm}
\begin{center}
\begin{tabular}{|c||c|c|c|c||}
\hline 
$\sqrt{s_{\gamma \gamma}} (GeV)$ &EMM, Inel,GRS &EMM, Tot,GRV & BKKS& $1 \sigma$ \\ 
& ($p_{\rm tmin}$=1.5 GeV)& ($p_{\rm tmin}$=2 GeV)              & GRV & \\ \hline
\hline
 20    &399  nb & 331 nb      & 408 nb &   2 \%  \\ \hline
 50    &429  nb & 374 nb      & 471 nb &   9\%   \\ \hline
 100   &486  nb & 472 nb      & 543 nb &   11\%   \\  \hline
 200   & 596 nb & 676 nb      & 635 nb&   6\%   \\  \hline
 500   & 850 nb & 1165 nb      & 792 nb &  7  \%   \\  \hline
 700   & 978 nb & 1407 nb     & 860 nb &   13 \%   \\ \hline
\end{tabular}

\end{center}
\end{table}
The last column in Table~\ref{options/table2} now gives the percentage 
difference between the two models which bear closest results, 
i.e. EMM with GRS~\cite{options/15} densities and inelastic formulation on 
the one hand and BKKS, as well as EMM with GRV~\cite{options/16}densities and total 
formulation on the other. From the two tables one sees that a measurement
to  $7\%$ accuracy, will be able to distinguish within a class of models 
of a particular type. Figs.~\ref{options/one},\ref{options/two} shows that for a distinction 
between the different classes even a precision of $\lsim 20 \%$ is sufficient,
for the \gamgam~ energies of up to $\sqrt{s} < 500 $ GeV.

\subsubsection{Hadronic backgrounds at the future colliders}

One of the ways of estimating the hadronic backgrounds will be indeed  to
use the measured values of the hadronic \gamgam~ cross-sections from LEP in a 
parameterized form. Due to the somewhat larger errors in the highest energy data 
points as well as the uncertainties caused by having to extrapolate
the measured two photon hadronic cross-section outside the kinematic region
covered by measurements, the current data  are consistent with pretty different 
behavior of the values of positive exponent of $\sqrt{s_{\gamma \gamma}}$ in the 
parameterization of the data. In Table~\ref{options/table3} we give the fitted 
values for the parameters $\epsilon,\eta$ of Eq.~\ref{options/eq1}, 
for different model  predictions.
Note here that this is just a parameterization of the curves drawn in
the left panel of Fig.~\ref{options/one}. Only representative values of the
parameters in each case have been given. A more complete job by giving 
the dependence of these powers on the parameters of the model,
{\em e.g.} the minimum $p_t$ used $p_{t}^{min}$ or $k_0$ the parameter 
determining the overlap function in the case of the `minijet' models,
would indeed be welcome. In the  table numerical values obtained from the fit
to the lower most and upper most curves of the EMM and BKKS bands as well as 
those for Aspen model are given.
\begin{table}[htb]
\begin{center}\begin{minipage}{\figurewidth}
\caption{\sl \label{options/table3}
Values of the parameters a,b,$\eta,\epsilon$ of Eq.~\ref{options/eq1}
obtained by numerical fits to the various model predictions.}
\end{minipage}\end{center}
\vspace{0.5cm}
\begin{center}
\begin{tabular}{|c||c|c|c|c||}
\hline 
Model & a(nb) & $\epsilon$ & b(nb) & $\eta$ \\ \hline
\hline
BKKS (upper edge)  & 166.5  & 0.13 & 538.2 &  0.38 \\ \hline
BKKS (lower edge)  & 180.6 & 0.11 & 356.5 &  0.18  \\ \hline
Aspen  & 145.7 & 0.094 & 517.5 &  0.39   \\  \hline
EMM (lower edge) &14.01  &0.34  &475.4 & 0.14    \\  \hline
EMM (upper edge) & 19.9  &0.29  &475.3 & 0.084   \\  \hline
\end{tabular}
\end{center}

\end{table}
Note here that this parametrization should be used  only for 
$\sqrt{s_{\gamma \gamma}} >  10 $ GeV  or so. It should be mentioned here that
the `minijet' models in their current formulation do have certain 
lacuna when confronted with the $pp, p\bar p $ along with the
$\gamma p$ , \gamgam~ data~\cite{options/17p,options/17pp}. This indicates the need to modify the
modeling of the overlap function which is not strictly calculable from 
first principles. A QCD model for this has been proposed~\cite{options/17p}. 
Work is in progress to apply the Bloch-Nordsieck approach  to photon-induced 
processes. This is expected to tamper the high energy rise somewhat. What would
be even more interesting is to study the implications of the 
Bloch-Nordsieck formalism to multiplicity distributions or for multiple 
parton processes to determine the parameters of the EMM model using these.
The values given in Table~\ref{options/table3} represent the spread of our predictions
at present.

Yet another 'rule of thumb' measure of the `messiness' at the Compton
colliders can also be obtained by looking at the quantity 
$\sigma^{\rm jet}_{\gamma \gamma}$ defined 
\begin{equation}
 \sigma^{\rm jet}(s,p_{\rm tmin})=
 \int_{p_{\rm tmin}}
 d^2{\vec p_t} {{d\sigma_{jet}}\over{d^2{\vec p_t}}}.
\label{options/minjets}
\end{equation}
where the $d\sigma_{jet}/d^2{\vec p_t}$ is calculated using
perturbative QCD and photonic parton densities measured experimentally.
This rises steeply with increasing energy and decreasing $p_{\rm tmin}$.
While it is true that only part of this rise
with $\sqrt{s}$ is  reflected in  the energy dependence 
of $\sigma^{\rm had}$, the quantity is still a good measure of the   
messiness caused by the hadronic backgrounds at the JLC due to beamstrahlung.
Here we  give a new parameterization of the `minijet' 
cross-sections in  \gamgam~ collisions which can be used in estimating the
hadronic backgrounds at the JLC's by folding it with appropriate beamstrahlung
spectra. This supersedes the corresponding parameterization that was given 
earlier~\cite{options/17}.

The 'minijet' cross-section, for the  GRV~\cite{options/16} and SAS~\cite{options/18} 
densities, is given (in nb) 
\begin{eqnarray}
\sigma_{minijet}& = &
\left[222 \left( {2 \;{\rm GeV} \over p_{\rm tmin}}\right)^2 - 161 
\left({2\; {\rm GeV} \over p_{\rm tmin}}\right) 
+ 36.6 \right] \left({\sqrt s\over 50}\right)^{1.23} \label{options/grv}\\
& =&\left[77.6 \left({2\; {\rm GeV} \over p_{\rm tmin}}\right)^2 - 45.9 
\left({2\; {\rm GeV} \over p_{\rm tmin}}\right) + 9.5 \right] 
\left({\sqrt s \over 50}\right)^{1.17}
\label{options/eq:sas}
\end{eqnarray}
by Eqs.~\ref{options/grv} and \ref{options/eq:sas} respectively. Here $\sqrt{s}$ is the 
\gamgam~ c.m. energy in GeV.

It is essential to fix  $pt_{\rm tmin}$  due to the very strong dependence of the 
\gamgam~ `minijet' cross-section on it.  From our earlier discussions it 
is clear that this value will be $\sim  2$ GeV.

\subsubsection{Is discrimination still possible with only an \epm collider?} 

\begin{figure}[htb]
\centerline{\epsfxsize=8cm \epsfbox{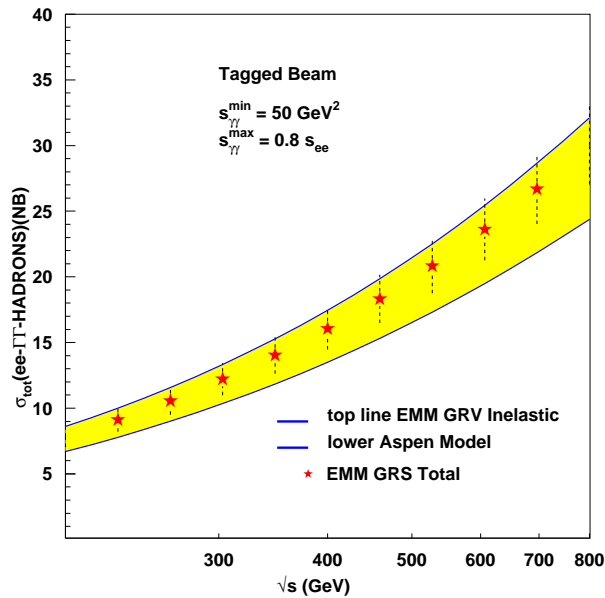}}
\begin{center}\begin{minipage}{\figurewidth}
\caption{\sl \label{options/three}
Cross-sections for hadron production due to \gamgam~ interactions 
in \epm reactions. The red stars indicate the expected cross-sections in
the EMM model, in total formulation with GRS densities with error bars 
representing the estimated errors.}
\end{minipage}\end{center}
\end{figure}

While it will be difficult to measure the $\sigma^{\rm had}_{\gamma \gamma}$ in the
\epm~ mode with sufficient accuracy some information can still be gained
by measuring $\sigma(e^+ e^-  \to e^+ e^- ~{\rm hadrons})$. This can be calculated
by convoluting the \gamgam~ hadronic cross-sections with the spectrum of
the photons. This spectrum is given by the Weiz\"acker Williams
approximation for the bremsstrahlung photons. The approximation has to be
improved to include the effect of the virtually of the photons.
The expected hadronic cross-section due to the two photon 
processes at an \epm~ collider is shown in Fig.~\ref{options/three}.
The integration over $s_{\gamma \gamma}$ is restricted to a minimum value such that
corrections due to losses along the beam direction can be made with precision
and errors due to the unfolding are reduced. An upper limit is imposed to
reduce the contamination by the annihilation events. The figure corresponds
to a choice of $50\ GeV^2 < s_{\gamma \gamma} < 0.64 s_{ee}$. The top curve here 
corresponds to the inelastic formulation of the EMM model~\cite{options/7} with
GRV densities  and the lower curve to Aspen model~\cite{options/9}. 
The 'pseudo data points' expected for the total formulation of the EMM 
with GRS densities with the choice of the parameters used in the 
figure in the right panel of Fig.~\ref{options/one}. The vertical error bars are the
errors expected at TESLA~\cite{options/12}. The figure thus clearly shows that even a
\epm~ collider will be able to shed some light on the issue of hadronic 
\gamgam~ cross-sections.

\underline{Summary}

Thus in conclusion we can say the following
\begin{enumerate}
\item 'Photon is like a proton' models predict a rise of $\sigma_{\gamma
\gamma}$ with $\sqrt{s_{\gamma \gamma}}$, slower than shown by the data,
albeit compatible with them within 2 $\sigma$,   
i.e. value of predicted $\epsilon$ is lower than what the data seem to show.
\item The extrapolated $\gamma p$ data seem to show similar trends.
\item The predictions of the EMM model show good agreement with the data.
\item Even in the EMM formulations use of Bloch Nordsieck ideas to calculate
the overlap function $A(b)$ seems to slow down this rise.
\item An obvious improvement in the EMM models is to try and determine
$A(b)$ by more refined `theoretical' ideas or determine it in terms of the
multiple parton interactions measured  at the HERA/Tevatron collider.
\item However, extraction of $\sigma_{\gamma \gamma}$ and $\sigma_{\gamma p}$ 
from $\sigma_{e^+e^-}$ and $\sigma_{ep}$ respectively, is no mean task and 
has large uncertainties.
Moreover, a difference of about a factor two in the predicted values of 
$\sigma^{\rm tot}_{\gamma \gamma}$ in different models, gets reduced to only
about $30\%$ when folded with the photon spectrum expected in the 
WW approximation in $e^+e^-$ collisions. While the good part is that it reduces
the uncertainty in our predictions of the hadronic background at the $e^+e^-$
linear colliders due to the corresponding uncertainties in 
$\sigma^{\rm tot}_{\gamma \gamma}$, the studies of  two-photon hadronic 
cross-sections at $e^+e^-$ colliders, will not be very efficient in 
shedding  much light on the theoretical models used to calculate them.
\item Therefore measurements of total cross-sections at a 
$\gamma \gamma$ collider with its monochromatic photon beam, in the 
energy range $300 < \sqrt{s_{\gamma \gamma}} < 500$ GeV, 
can play a very useful role 
in furthering our understanding of the 'high' energy photon interactions.
A precision of $\lsim 7-8\%(8-9\%)$ 
is required to distinguish among the different formulations of the EMM models 
(models which treat photon like a proton), whereas a precision of 
$\lsim 20\%$ is required to distinguish between these two types of models.
\end{enumerate}

%% file: options/lum-meas.tex
At the \epm collider,  Bhabha scattering has large cross section and 
is a good channel to 
monitor the luminosity\cite{options/miller,options/kawabata}. 
In the \gamgam ~collider, there is no
 such processes having a large cross section 
as Bhabha scatting in the \epm ~interaction. 
In addition, the luminosity spectrum of the \gamgam ~collision is far from the 
monochromatic and,  information of 
the polarization  is essential for physics study. 
Therefore, to pursue good physics at the \gamgam~ collider, 
measurement of differential luminosity as:
\begin{equation}
{{dL} \over {dE_\gamma ^1 dE_\gamma ^2 d\xi _\gamma ^1 d\xi _\gamma ^2 }}
\label{options/eqn:general-diff}
\end{equation}
where $E_\gamma ^{1(2)}$ and $\xi _\gamma ^{1(2)}$ are the energy 
and polarization of the first(second) photon.
It can be equivalently expressed in away which is more suitable for physics
analysis as 
\begin{equation}
{{dL^{J = 0,2} } \over {d\sqrt {s_{\gamma \gamma } } d\eta _{\gamma \gamma } }}
\label{options/eqn:diff-hel}
\end{equation}
for helicity based analysis or 
\begin{equation}
{{dL^{\parallel , \bot } } \over {d\sqrt {s_{\gamma \gamma } } d\eta _{\gamma \gamma } }}
\label{options/eqn:diff-lin}
\end{equation}
Here, $\eta_{\gamma \gamma}$ is the rapidity of the \gamgam~ system defined as
$$
\eta _{\gamma \gamma }=\log \sqrt {E_{1}/ E_{2}},
$$
, $J$ is  (z component of) total spin of the photon photon system and 
$\parallel$($\bot$) stands for that the electric field of two photons are
in parallel(perpendicular) each other. 

 For a good measurement of the luminosity, 
processes 
must have large cross sections and can be selected with high efficiency with
small background contamination. 
The possible candidate processes are;  
$\gamma \gamma \rightarrow e^+e^-$,
$\gamma \gamma \rightarrow \mu^+\mu^-$, 
$\gamma \gamma \to W^+W^-$.
The lepton pair production are good processes, however, 
it is only sensitive to the J=2 component of the 
luminosity so that it is necessary to measure total 
luminosity by other processes.
The total cross section of 
$\gamma \gamma \rightarrow \mu^+\mu^- \mu^+ \mu^-$,
is as large as about 150pb independent of energy, however, 
most of muons are radiated in to the beam direction.
For, example, if we require that at least two like sign muons are
$\left| {\cos \theta  < 0.98} \right|$ and $E > 1GeV$, the 
effective cross section goes down to about 2pb at 100GeV and 0.1pb at
400 GeV. \footnote{The cross sections has been calculated with 
CompHep\cite{options/comphep} and
GRACE\cite{options/grace} program and the results agree with each other.}
The radiative electron positron pairs, i.e., 
$e^ +  e^ -   \to e^ +  e^ -  \gamma $ could be useful for 
the luminosity measurement\cite{options/asakawa-lum}.
Though the cross section does not appear to be large enough for
high energy but could be till useful for luminosity measurement 
below W boson threshold.
If we assume that W is  the standard model gauge boson, 
the W pair production is a good candidate for a luminosity
measurement at a center-of-mass energy greater than 200 GeV.
 
A possibility of measuring polarization dependent luminosity 
using lepton pairs were
proposed\cite{options/tel-lum}.
When the \gamgam ~collider is running with $P_L P_e=-1$, 
the helicities of two photon beams are usually chosen to have the luminosity
distribution of either the J=0 or J=2 dominant distribution.
These two modes can be easily exchanged if
both $P_L$ and $P_e$ are flipped simultaneously for one of the photon beams.
This operation keeps $P_L P_e=-1$ but changes 
the sign of the helicity of the generated 
photons, making it possible to 
 exchange between
J=0 and J=2 mode without, in ideal condition, changing 
luminosity distributions.
 Thus, with this helicity-flipping method, the J=0 luminosity can
be directly given by the measured J=2 luminosity.

\begin{figure}[tbp]
\centerline{\epsfxsize=13cm \epsfbox{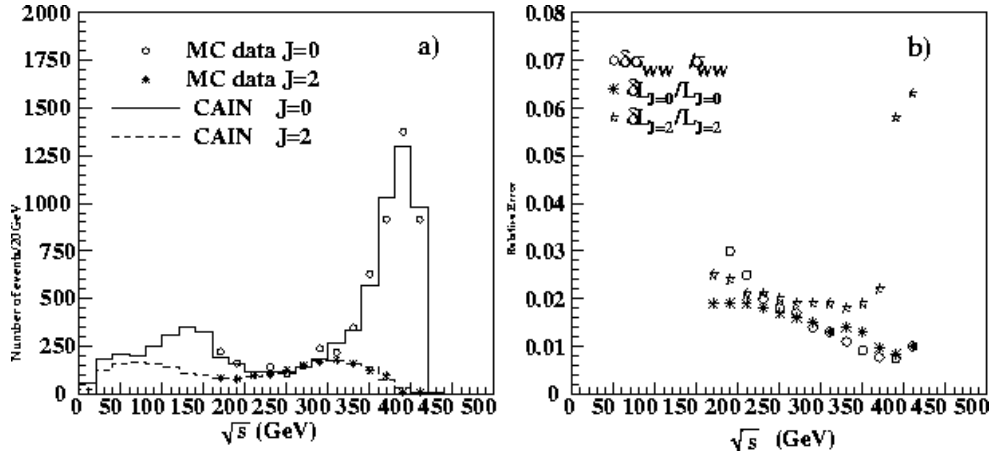}}
\begin{center}\begin{minipage}{\figurewidth}
\caption{\sl \label{options/fig:lum_meas}
a) Reconstructed events at the process $\gamma \gamma
\rightarrow e^+ e^-$,where the
circles and asterisks are the measured luminosities with J=0 and J=2,
while the solid and dashed lines are the generated luminosities
with J=0 and J=2, respectively.
b) Estimated statistical error for the luminosity measurement, assuming 
the integrated luminosity of $10fb^{-1}$.}
\end{minipage}\end{center}
\end{figure}

A feasibility study for a luminosity measurement was performed for the
process $\gamma \gamma \rightarrow e^+ e^-$  and for
$\gamma \gamma \rightarrow W^+ W^-$ \cite{options/takahashi95,options/JLCgg}. 
The reconstructed luminosity distribution and the statistical errors on 
the luminosity measurements are shown in  
Fig.~\ref{options/fig:lum_meas}.

As shown in this figure, for both electron and W pairs, 
the statistical errors are expected to be   $1 \sim 2$\%
for each energy bin.
This measurement assumes, however,  the use of sophisticated helicity 
flipping to obtain polarized luminosities which potentially 
has systematic effects. 
More detail study on the systematic effect of the measurement 
is necessary on this point.

%% file: options/tech.tex
In a current design of the \gamgam~ and \eg~ collider, all particles from the 
conversion point are brought to the interaction region and 
experience the beam beam interaction. 
In addition,  
the \gamgam ~ and the \eg ~collider needs TW laser pulses 
synchronized with the electron beam at the conversion point which is 
about a half $cm$ from the interaction point.
It adds technical issues to be solved in addition to the \epm~ collider, 
i.e., 
\begin{itemize}
\item Development of high-repetition/high-power lasers.
\item Optical system for laser pulses in the interaction region.
\end{itemize}
In following sections, the status of the design of the interaction region and the 
laser system will be described.

%% file: options/ir.tex
The issue on the interaction region are:
\begin{itemize}
\item Protect detectors from the beam-beam backgrounds.
\item Extract spent electron beam.
\item Take TW laser pulses in and out from the interaction region 
without hitting anywhere.
\end{itemize}
 
In the sense of  the beam related background, 
situation is more or less similar to the 
\epm~ collider except energy spectrum of the spent electron.
Since the electron beam already interacted with the laser pulse at the 
conversion point, energy of the electron beam is widely spread and the 
energy could be as low as O(GeV) which are disrupted away 
to the detector volume.

\begin{figure}[htbp]
\centerline{\epsfxsize=13cm \epsfbox{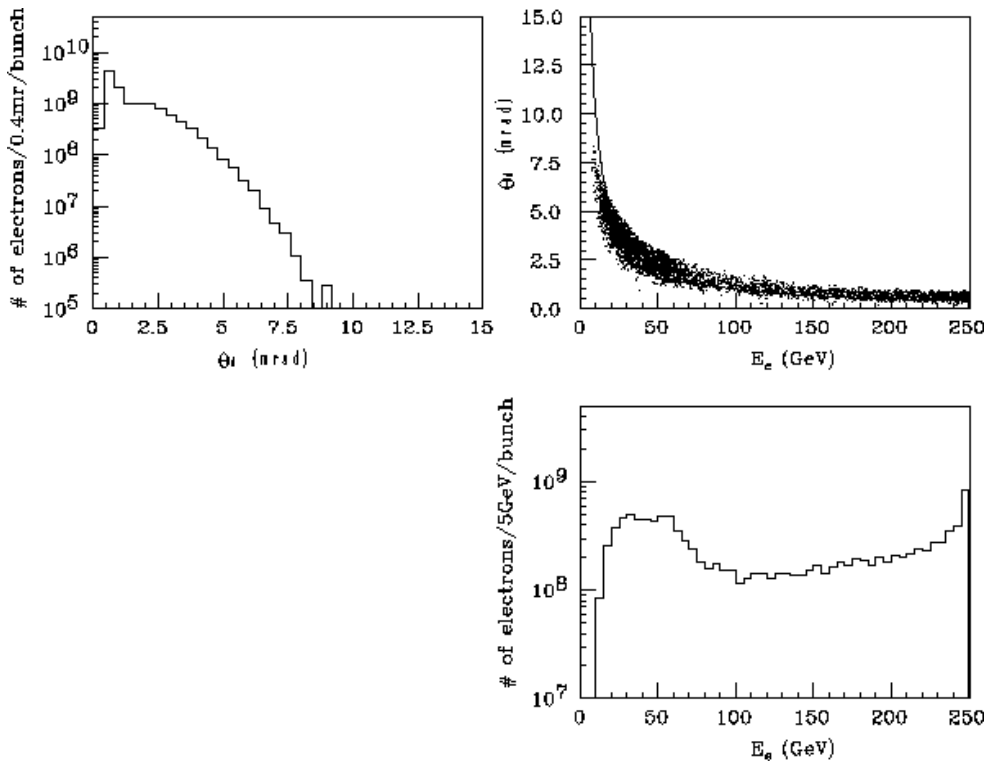}}
\begin{center}\begin{minipage}{\figurewidth}
\caption{\sl \label{options/fig:e_angle}
Scatter plot of disrupted electrons at $\sqrt{ s_{e^-
e^-} }=500$GeV, where the horizontal and vertical axes are the
energy (E$_e$) and disruption angle ($\theta_d$), respectively.  The
solid line is a calculation by equation~\ref{options/eqn:disp1}. Histograms projected
on both axes are also shown.
}
\end{minipage}\end{center}
\end{figure}

To estimate the  beam-beam effect,
CAIN was employed to simulate the interaction of electron 
bunches when they are passing by.
Figure~\ref{options/fig:e_angle} shows a plot of disrupted electrons in
a plane of the disrupted angle and the energy, together with two
histograms projected on both axes.
(parameter (a) in table~\ref{options/tbl:parameters})
The disruption angle can also be approximately expressed by
\begin{equation}
\theta_d \approx { N r_e \over{ \sqrt{2} \gamma ( \sigma^*_x +
\sigma^*_y ) }} \approx  { 94 [{\rm mr}]\over{E_e[{\rm GeV}]}},
\label{options/eqn:disp1}
\end{equation}
where $N$ and $r_e$ are the electron-beam intensity and the electron
classical radius, respectively, and $\sigma^*_{x(y)}$ is the
electron-beam size at the IP (see table~\ref{options/tbl:parameters}).  
An
analytic calculation a little bit overestimates the disruption angle 
compared to the
simulation at low energy, while they agree with each other in
the high-energy region, as shown in Fig.~\ref{options/fig:e_angle}. The maximum
disruption angle is about $10mr$, which corresponds to the electron  energy
of 10GeV. 

Because of the large disruption angle,  the crossing angle should be 
large for the
disrupted electrons to pass through the nearest final focus magnet
without any interaction. 
Therefore a crab-crossing scheme has to be
necessary to avoid reduction of the luminosity.

The other source of backgrounds is low-energy 
\epm~ pairs produced by incoherent productions, such as
Breit-Wheeler ($\gamma \gamma \to e^+e^-$), 
Bethe-Heitler ($\gamma e^\pm \to e^\pm e^+e^-$), 
and
Landau-Lifshitz ($e^+e^-\to e^+e^-e^+e^-$).
Since the mechanism of the pair creations is the same as that of the \epm
collision, the background in detectors should be similar to that at
\epm collision, where a detailed simulation result has been
presented\cite{options/JLC-I}.  

The low-energy particles are moving along the helical
trajectories with a radius of $R = 2 \rho \sin \phi/2$, where $\rho =
p_t /0.3 B$ and $\phi = 0.3 B z /p_z$ in a detector-solenoidal magnetic
field of $B=$2T. Here, $z$ is the distance along the beam axis.  Since
they are created at very forward angles ($m_e/E_e$), they obtain their
transverse momenta ($p_t$) mainly from deflections due to a strong
electromagnetic field of the opposing beam.  Their trajectories make
a bundle along the beam axis. The boundary of the bundle is given by a
particle with the maximum $p_t$, that is the maximum deflection angle. 
Any detectors and masks must be located outside of the bundle in order to
avoid a huge background. 
There are also particles with inherently large scattering angles outside
of the bundle; however since the number is relatively small, they
can be tolerable, as mentioned below.  The maximum deflection angle
($\theta_{max}$) can be expressed by
\begin{equation}
\theta_{max}= \theta_\circ \sqrt{ 2 \ln (4 \sqrt{3} D_x/\epsilon) /
\sqrt{3} \epsilon D_x }~,
\label{options/eqn:thmax}
\end{equation}
where $\theta_\circ = D_x \sigma^*_x/\sigma^*_z$, 
$D_x= 2 r_e N \sigma^*_z / ( \gamma \sigma^*_x
(\sigma^*_x+\sigma^*_y) )$ and $\epsilon = p / E_{beam}$.
For the high energy \gamgam collision,
$D_x=$1.68 and
$\theta_\circ =1.08 mr$. Since $\theta_{max}$ is less than 0.2 for our
cencerned particles, we can approximate $p_z \approx p$. Thus, $\phi$ is a
function of $p$ alone.  Apparently, the maximum radius is $R_{max}= 2
\rho$ at $\phi=\pi$, where 
$p_{max} = 0.3 B z / \pi$.   
We therefore obtain $R_{max}$ as a function of $z$, 
\begin{equation}
R_{max} \approx {2 p_{max} \theta_{max} \over{ 0.3 B}}
= {2 z \theta_{max} \over{\pi}}. 
\label{options/eqn:rmax}
\end{equation}
For the innermost layer of the vertex detector with $z = \pm 5.2$ cm
($| \cos \theta | <$0.90 ),
$R_{max}$ was calculated to be 1.6cm, which is smaller than the present
$r_{vtx}$=2.5cm.
\begin{figure}[tbp]
\centerline{\epsfxsize=9cm \epsfbox{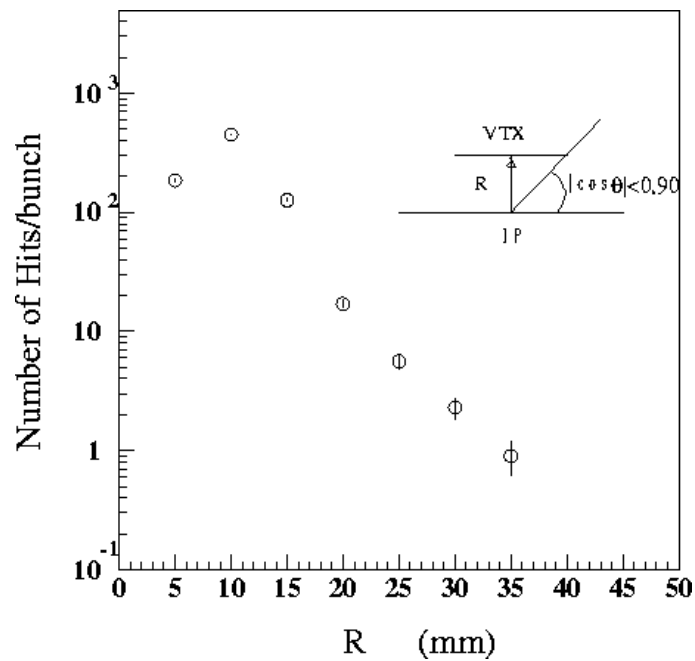}}
\begin{center}\begin{minipage}{\figurewidth}
\caption{\sl \label{options/fig:vtx}
Expected number of background hits in tracking detectors
per bunch crossing. }
\end{minipage}\end{center}
\end{figure}

The number of background hits was actually calculated
as a function of the radial distance (R) from the beam axis in the angular
region $| \cos \theta | <$0.90 with no secondary interaction.  The
result is shown in Fig.~\ref{options/fig:vtx}.
In the estimation, multiple hits were registered by a helical
track. At $r_{vtx}$=2.5cm, the number of hits is about 10 per bunch
crossing, that is  O($10^3$) hits per pulse train.  Since the number is
indeed similar to that of the
\epm collision, a very similar environment of backgrounds is expected 
due to the pairs.

Secondary particles, which are back-scattered at the final focus magnet, are
prevented from going into the detectors by a conical masking system.

\begin{figure}[tbp]
\centerline{\epsfxsize=9cm \epsfbox{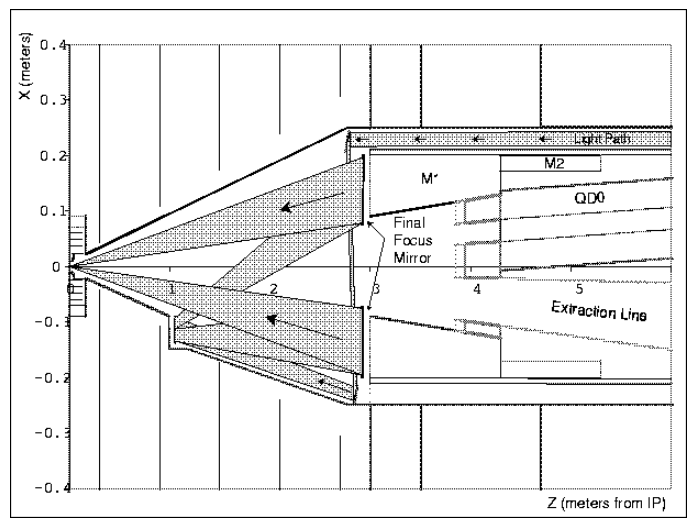}}
\begin{center}\begin{minipage}{\figurewidth}
\caption{\sl \label{options/fig:nlc-ir}
A schematic of the interaction region of the \gamgam~ collider
\protect~\cite{options/nlc-orange}.}
\end{minipage}\end{center}
\end{figure}

The engineering level design of the interaction region including the laser optics
and masking system are still under consideration. 
A design closest to the engineering level at this moment
is shown in 
Fig. \ref{options/fig:nlc-ir} which is  found in \cite{options/nlc-orange}.
In this design, the aperture of the electron beam exaction line is about $10mr$ to 
accommodate the spent electrons and the beam crossing angle is enlarged to 
$30mr$ to avoid interference between the extraction line and the final focus magnet.
The laser path is also shown in Fig. \ref{options/fig:nlc-ir}. 
The final focus mirror is mounted on the tungsten mask(M1) which is 3 m from the 
interaction point. The mirror is 38cm in diameter with a hole of 15 cm diameter for 
the electron lines.
The detail description of this design is found in \cite{options/nlc-orange}.

%% file: options/laser-koba.tex
 Requirements for the laser system are shown in table \ref{options/Laser-Requirements}.  It is difficult to generate such a high-energy pulse and a complicated time structure by a single laser. One of the promising solutions is to build 95 lasers( 190 lasers in total for two linacs), which is a 190 laser system. In this case, each laser generates 1 J/pulse at 150 Hz. Key issues are 
\begin{enumerate}
\item 1J pulse with 1 ps width
\item Synchronization of each pulse
\item Collision with beam-bunches and pulses
\item Focus and pointing stability
\end{enumerate}

In this section, current and future technologies for the above issues are described.

\begin{table}[tbp] 
\begin{center}  
\caption{\sl  \label{options/Laser-Requirements} Requirements for the laser system}  
\begin{tabular}{|l|l|} 
\hline   
Time structure & a train of 95 pulses with 2.8 ns intervals \\ 
Repetition rate & 150 Hz \\
Pulse energy & 1 J \\
Pulse width & 1 ps \\
Wavelength & 1 $\mu$m \\
focusing spot size in diameter & 3 $\mu$m (r.m.s) \\
\hline  
\end{tabular} 
\end{center}
\end{table}

\subsubsection{1 J laser pulse with 1 ps pulse width} 

The rapid and remarkable progress of short pulse lasers has led to the demonstration of  $>$ 1J femto-(pico-)second pulse. 

A Nd:glass laser has been often used for high energy short pulse generation. However, its repetition rate is too low ($\leq$ 1 Hz) because of the poor thermal property. The other Nd-doped materials have been also used for high energy pulse lasers, but the pulses are not short enough for our purpose.  Even in the low energy mode-locked oscillators, the shortest pulse has been obtained to be 2 ps (Nd:YLF) or 5 ps (Nd:YAG)~\cite{options/Laser-Svelto}. 

One of realistic choices is to use Ti:sapphire crystals with a Chirped Pulse Amplification(CPA) systems~\cite{options/Laser-Strickland}. The mode-locked oscillator can generate pulses as short as tens of femtosecond. These short pulses must be stretched in hundreds of picosecond or in a nanosecond, then a pulse is picked up by a Pockels cell to be amplified. Finally, the amplified pulse is compressed.  Generally in the usual CPA system, it is compressed into the initial or slightly longer pulse. However, for a $\gamma \gamma$ collider, it is intended to be compressed insufficiently for the requirement of  1 ps pulse-width. 

Problematic issues of the Ti:sapphire laser system are high cost, low reliability, short wavelength, and energy chirp in a pulse.  
High cost and low reliability come from the same reason.  
The pumping wavelength is around 500 nm, where Ar$^+$ ion lasers or a second harmonics of Neodium doped lasers (e.g. crystals of Nd:YLF or Nd:YVO$_4$) are used as a pump source. 
Since Ar$^+$ ion lasers are pumped by a discharge, the stability is too poor for a $\gamma \gamma$ collider. 
On the other hand, the Nd doped lasers are sufficiently stable since they can be pumped by laser diodes(LDs). 
Wavelength of the Nd lasers must be reduced to half in the second harmonics.  In this case, a pumping scheme is much more complicated than that without the wavelength conversion. Generally, a complicated scheme has high cost and low reliability which may cause a serious problem in the 190 laser system.
Wavelength of Ti:sapphire laser is around 800 nm which is shorter than the required wavelength of $1 \mu$m. The insufficient compression leaves residuals of energy chirp. Therefore the wavelength varies about 5\% (40-50 nm in 800 nm) in a pulse. 

Nevertheless, the Ti:sapphire CPA system is an influential candidate, since the technology has been well established. The required specifications, except for the wavelength, for the $\gamma \gamma$ collider can be constructed using existing technologies. Additionally, due to the wide spectral bandwidth, square temporal profiles can be generated using a spatial light module (SLM). The laser power can  be constant in the pulse~\cite{options/Laser-Takasago}.

Another possibility is to use a crystal of Ytterbium doped with Sr$_5$(PO$_4$)$_3$F (Yb:S-FAP)~\cite{options/Laser-Payne}. The laser wavelength is 1047 nm consistent with the required specification. The spectral bandwidth is sufficiently wide and seems to be good for sub picosecond or 1 ps pulses. It can be directly pumped by LDs. Furthermore, the upper state lifetime is 1.2 ms, which is five times longer than Nd:YAG and 2.5 times longer than Nd:YLF.  Fewer LDs are required to produce the same energy output because the pumping energy can be stored in the crystal over a longer period of time. 
Comparing to the Ti:sapphire crystal with Nd doped laser(second harmonic), the Yb:S-FAP crystals have advantages of reduced cost and improved reliability.  Yb:S-FAP crystals do not require another laser as a pumping source nor frequency conversion. The total system is much simpler and the required pumping energy is one order of magnitude less than the Ti:sapphire laser. One the other hand, the Yb:S-FAP crystals have not been well investigated.  Further research is required to determine if this crystal can generate 1 ps pulses.

\subsubsection{Timing synchronization of each laser pulse}

The synchronization of 190 lasers is one of the most important technologies in the 190 laser system, where each pulse must be synchronized to the electron beam bunch. 
It is not realistic to measure the electron timing for a feedback to the laser system because the electron pulses travel at almost the speed of light and the time to feedback is too short. The laser pulses must be synchronized to the RF signal which drives the linacs. Then the laser pulses can be synchronized to the electron pulses. In order to achieve the synchronization, three stages will be built.

The first stage is the synchronization of a mode-locked master oscillator. Either for Ti:sapphire or Yb:S-FAP, a single oscillator will be used as an injection seeder for all amplifiers. 
The pulse timing of mode-locked oscillator can be adjusted by the optical length of the laser cavity. So, the oscillator can be synchronized to the RF signal with a conventional phase locking technique by adjusting the position of the cavity mirror. The synchronization of less than 100 femtosecond has been already demonstrated~\cite{options/Laser-Kobayashi}.

\begin{figure}[tbp]
\centerline{\epsfxsize=10cm \epsfbox{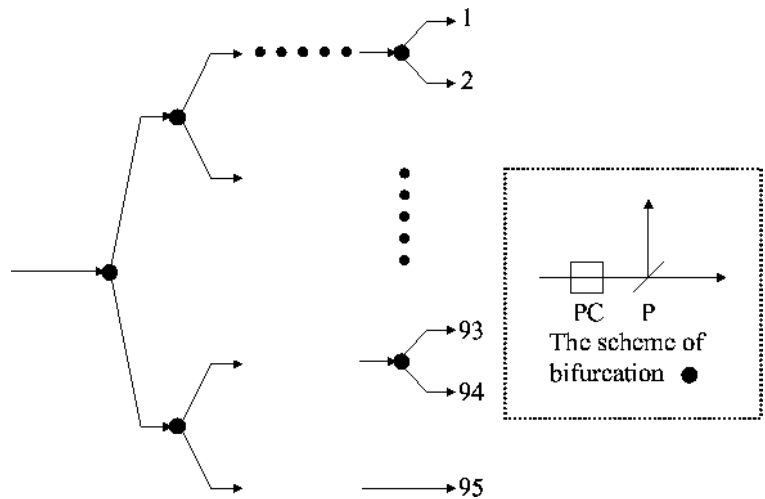}} 
\begin{center}
\begin{minipage}{\figurewidth} 
\caption{\sl  \label{options/Laser-PC} Pulse divider, consist of the series of Pockels cells.  The scheme of bifurcation is shown in the dotted box. PC: Pockels cell,  P: thin film polarizer or Grand Thomson prism.}
\end{minipage}
\end{center}
\end{figure}

\begin{figure}[tbp]
\centerline{\epsfxsize=10cm \epsfbox{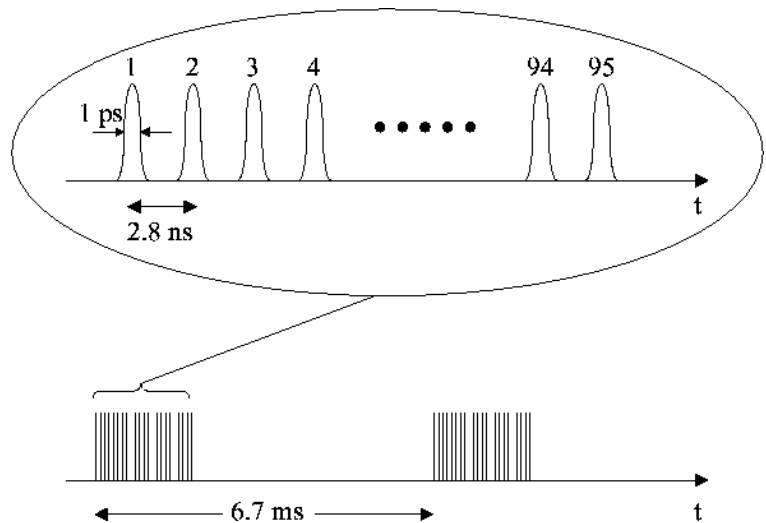}}
\begin{center}
\begin{minipage}{\figurewidth} 
\caption{\sl  \label{options/Laser-timing} The timing structure of the laser pulses.}
\end{minipage}
\end{center}
\end{figure}

The second stage is the pulse divider. The repetition rate of the oscillator should be 357 MHz (2.8 ns interval). Using high-speed Pockels cells, 95 laser pulses are picked up and divided into 95 beams. The timing of the drive signal for each Pockels cell is not critical because it is used only for the "gate signal". The accuracy is determined by the mode-locked oscillator. Finally, each picked pulse is divided into two by a beam splitter and delivered to the amplifiers for two linacs. The series of Pockels cells and the timing structure are shown in Fig.\ref{options/Laser-PC} and Fig.\ref{options/Laser-timing}, respectively.

The third stage is the stabilization of the amplified pulses. The amplified pulses are automatically synchronized in principle because the injected seed pulses are synchronized to the RF signal. However, a stabilization is still required because additional disturbances may induce fluctuations. One example of such disturbances is the thermal expansion. Since the laser pulses travel in a long distance, the arriving time to the colliding point may fluctuate. The other is the mechanical vibration. Vibrations of the beam delivering mirrors cause a change in the optical pass length. The repetition rate of each amplifier is 150 Hz. A statistical technique cannot be applied to measure the fluctuation because of the low repetition rate, while it can be applied for the mode-locked oscillator. An elegant method to measure and stabilize the amplified pulses has already been demonstrated~\cite{options/Laser-Miura}. In this method, stabilized oscillator pulses are used as a reference and the timing lag between the oscillator pulse and the amplified pulse is measured by a modified cross-correlation technique. Using this technique, sub femtosecond stabilization has been achieved. 

\subsubsection{Collision between the beam-bunch and the laser pulse}

Since 95 optical amplifiers will be required for each linac, Collision between the beam-bunch and the laser pulse is an important issue.  The laser pulses shall be focused by 95 different angles to a colliding point, where 95 optical mirrors must be placed on the base of a cone whose tip is the collision point. Instead of the 95 mirrors, a single parabola reflector can be used too. The advantage of this method is a simplicity. We can align the path of each laser pulse independently. On the other hand, it occupies a large space. Also, each amplifier requires a pulse compressor.

Another solution is to combine the pulses using Pockels cells. Reverse procedure of pulse divider for the oscillator can be applied. But this time, the energy of the amplified pulse is much higher than that of the oscillator. To avoid optical damage, pulse diameter should be very large. It is necessary to develop a large aperture, high damage threshold, and high-speed ($\geq$ 357 MHz) Pockels cell. If such a Pockels cell can be developed, every pulse can be co-axially overlapped. After overlapping, only one compressor and one focusing mirror are required. 

\subsubsection{Tight focus and pointing stability}

The tight focus and  pointing stability of the amplified laser pulses is a key issue for the stable collisions. In general, the thermal problem is the biggest issue in high energy laser amplifiers. Inhomogeneous thermal distributions result in wave-front distortions~\cite{options/Laser-Mourou,options/Laser-Spielmann} and then the focus becomes poor. Also, the fluctuations of the thermal distributions lead to pointing instabilities.

There are two choices to compensate for the inhomogeneous thermal distributions. One is to use a phase-conjugation mirror such as SBS cell. But the reflectivity is not very high, and it requires much bigger amplifiers to obtain sufficient energy. 

\begin{figure}[tbp]
\centerline{\epsfxsize=10cm \epsfbox{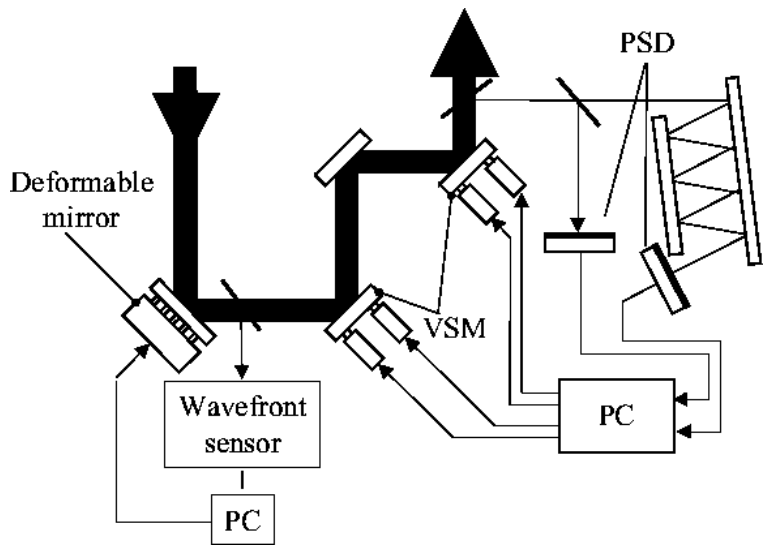}}
\begin{center}
\begin{minipage}{\figurewidth} 
\caption{\sl  \label{options/Laser-optics} Wave-front correction and the pointing stabilization. PC: personal computer,  VSM: vector scanning mirror, PSD: position sensitive detector.}
\end{minipage}
\end{center}
\end{figure}

The another choice is to use adaptive optics (deformable mirrors). First order wave-front distortions can be removed using a telescope~\cite{options/Laser-Ito}. Remaining distortions will be measured using a Shack-Hartmann type wave-front sensor ~\cite{options/Laser-Shack-Hartmann}, and adaptive optics will be used to remove the distortion. To reduce the pointing instability, a set of mirrors will be used as a vector scanning system. Figure \ref{options/Laser-optics} shows the set of adaptive optics and vector scanning mirrors.

\subsubsection{Summery of the laser system}

\begin{figure}[tbp]
\centerline{\epsfxsize=11cm \epsfbox{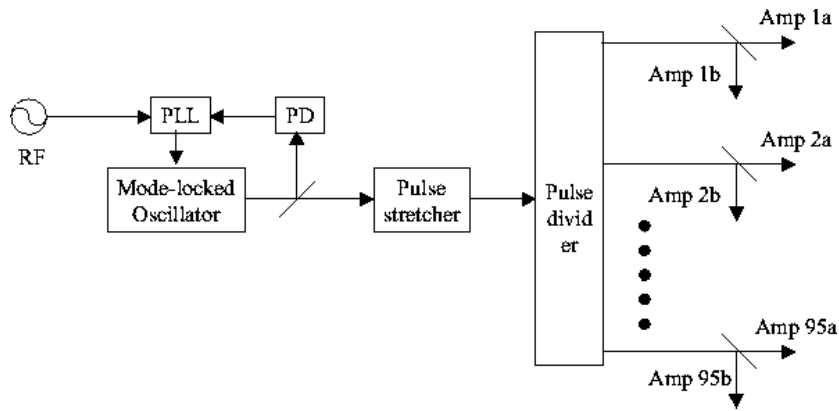}}
\begin{center}
\begin{minipage}{\figurewidth} 
\caption{\sl \label{options/Laser-oscillator} The oscillator stage of the laser system.}
\end{minipage}
\end{center}
\end{figure}

\begin{figure}[tbp]
\centerline{\epsfxsize=11cm \epsfbox{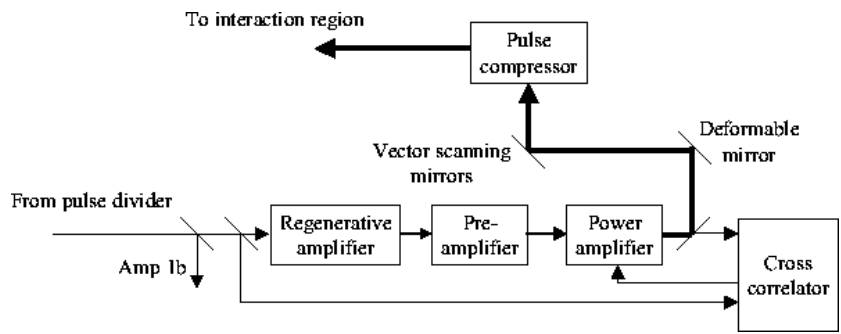}}
\begin{center}
\begin{minipage}{\figurewidth} 
\caption{\sl \label{options/Laser-amplifier} The amplifier stage of the laser system.}
\end{minipage}
\end{center}
\end{figure}

In this section, a laser system for the $\gamma \gamma$ collider is proposed. The basic idea is to build an amplifier for each pulse, and combine the pulses with synchronizing technique. This system can be built, in principle, with the state of the art Ti:sapphire laser technology. However, the cost is expected to be very high, and the reliability may not be sufficient. We have enough time to develop a next generation laser system. The key technologies are picosecond pulse amplification with Yb:S-FAP crystals and the beam combination with large aperture Pockels cells.  Figures \ref{options/Laser-oscillator} and \ref{options/Laser-amplifier} show the proposed laser system.  Figure \ref{options/Laser-oscillator} is the oscillator stage, consisting of a mode-locked oscillator, synchronization circuit to RF reference signal, pulse stretcher, and pulse divider. Figure \ref{options/Laser-amplifier} is the amplifier stage. 95 amplifier stages (total 190 amplifiers for two linacs) will be built. The amplified pulses will be delivered to the interaction region with 95 different paths or a single path after beam combination by a series of Pockels cells.

%% file: options/laser-llnl.tex
One of the most unknown part of above laser system is 
the way of laser pulse combination, i.e., how it is possible to 
deliver 95 laser path from deferent path to each electron bunch. 
A method is to install 95 mirrors at the interaction region but seems to be 
impossible because there is no space available. 
The other way is to combine pulses from different paths to pulse train by Pockels
cell before entering the interaction region. 
However, the development of required high power and fast Pockels cell are issue to be 
solved. 

The NLC group proposed different idea for the laser scheme based on Mercury laser system\cite{options/laser-jim}. 
The Mercury laser is 100J/pulse, 10Hz with 2-10ns pulse length which is now 
under development at LLNL.
The idea to avoid the problem of pulse combination is that 
a pulse train consists of 95 pulses with 2.8 ns spacing is formed before the amplification. 
Each pulse is chirped and overlapped temporarily so that the envelop is a 
$\approx 300ns$ pulse. 
However each pulse can still be identified because they are not overlapped in 
frequency.
After amplification by the Mercury amplifier, the pulses are compressed to 95 of 
1ps pulses. 
Since the Mercury laser is design to operate 10Hz, it only needs 12 systems to provide
the laser pulses to ${\rm 95 \times 120 Hz}$ electron bunches.

%% file: options/summary.tex
According to the studies in the last decade, the characteristics of the beams provided 
by the \gamgam and the \eg collider and overall view of the physics opportunities have 
been clear. 
It has been shown that \gamgam~ and the \eg~ colliders using backward Compton scattering 
could provide comparable luminosity with the \epm collider as well as have unique feature
of providing polarized photon beams.
It also has many good physics opportunities, some of which 
are unique and only possible with 
these facilities and some of which  are complementary to the \epm interaction.

Regarding these aspects, the R\&D efforts are now tuning in to reveal feasibility of 
the \gamgam~ and \eg~ collider as realistic as possible 
in both physics and technical aspect.
In the physics study, we need simulation study by taking into account the realistic 
luminosity distribution and the detector performance.
In the technical aspect, continuous effort for the development of the laser and 
design of the interaction region is the key issue for the realization of the 
\gamgam~ and \eg~ colliders.
However, looking at  the manpower available, we understand  the world wide level collaboration 
is very importance to perform these R\&D effort effectively. 
For this purpose, we are now organizing an international working group including Asia, 
North America and European region and started a real collaborative work with 
the North America group for the design of the interaction region. 

We believe that the feasibility  of the \gamgam~ and \eg~ collider will be  clear as the 
same level as is in the \epm collider for now in a couple of years
and is possible to  put it on a part of 
the linear collider project.

%% file: options/ack.tex
{\bf Acknowledgement}

Many topics described in this chapter are based on discussions and/or works
in the international collaborative group on the \gamgam~ and \eg~ collider.
The author would like to thank all participants of the collaboration in the 
North America and Europe. 
Particularly, we thank colleagues in the Lawrence Livermore National Laboratory for 
providing us materials on the interaction region and the laser sections.

%% file: headers/acknowledgements.tex
\begin{Large}
\begin{center}
{\bf 
Acknowledgements
}
\end{center}
\end{Large}

This work is supported in part by the following organizations/programs:
Japan Society for the Promotion of Science (JSPS),
-JSPS-CAS Scientific Cooperation Program under the Core University System,
-JSPS Japanese-German Cooperative Program,
-Japan-US Cooperative Program in High Energy Physics;
Foundation for High Energy Accelerator Science;
The Australian Research Council (ARC) and
The Department of Industry, Science \& Resources (DISR);
Commission on Higher Education of the Philippines.

%% file: headers/acfastatement.tex
\chapter{ACFA Statement\label{chapter-acfastatement}}

\begin{large}\begin{center}
Statement Of Physics Study Group On The $e^+e^-$ Linear Collider
\end{center}
\end{large}
\vspace{12pt}

On the 2nd plenary meeting, ACFA announced its endorsement of 
the e+e- linear collider as one of the major future
facilities in the Asia-Pacific region. 
In fact, recent world-wide research at existing facilities has 
enabled us to form a more and more concrete picture of 
"TeV-scale physics" and, consequently, 
has made the $e^+e^-$ linear collider's role 
more and more crucial in its exploration.
According to the recent picture, the linear collider is 
expected to produce very important,
decisive physics outputs even in the initial stage 
(in the energy region below 500GeV) of its energy upgrading program ; 
for instance, a top quark study at threshold, which is very 
important in its own right, can be a key to new physics and, more
importantly, the Higgs particle will almost certainly 
manifest itself there or the SUSY/GUTS scenario will be disapproved. 

\vspace{12pt}

In addition to its role as an energy frontier machine for 
High Energy Physics, the linear collider has a facet which can be
shared with a new means for materials science. 
The ultra-low emittance beam essential to the linear collider 
is also an indispensable element of the next-generation, 
coherent x-ray source. In order to efficiently and 
effectively promote accelerator science in the region,  
one should start seriously thinking about the possibility 
of integrating both into a single project. 

\vspace{12pt}

Turning our attention to activities in 
Asian region, we see significant progress in high 
energy and synchrotron radiation 
experiments at various domestic facilities. 
Not only that, many researchers from ACFA member nations are actively
participating in large-scale experiments 
such as at LEP-II, Tevatron collider, HERA and PEP-II/KEKB. The Asian
physics community on which ACFA is based has grown 
significantly and has set a firm enough foundation to prepare for
further advancement. 

\vspace{12pt}

In response to the ACFA statement issued in the last year, 
considering the importance of the linear collider project and the
potential of our community to realize it, 
we agreed to set up a study group under ACFA. 
The charge of the group should be to elucidate
physics scenario and experimental feasibilities and 
to write up a report to ACFA within two years. Taking account of the
scale of and the world-wide interests in such project, 
actual studies shall hopefully be carried out in a more global scope in
spite of the regional nature of ACFA's initiative. 

\vspace{12pt}

\noindent
July 22, 1998\\
Zhipeng Zheng \\
Chairman, Asian Committee for Future Accelerators\\
Director, Institute of High Energy Physics,\\
Chinese Academy of Sciences\\